\documentclass[structabstract]{aa}  
\usepackage{graphicx}
\usepackage{txfonts}
\usepackage{natbib}
\usepackage{verbatim} 
\usepackage{enumitem} 
\setlist{nolistsep}
\bibpunct{(}{)}{;}{a}{}{,} 
\begin{document}
   \title{Listening to galaxies tuning at $z\sim 2.5-3.0$: \\The first strikes of the Hubble fork}

\author{M. Talia
\inst{1}
\and
A. Cimatti\inst{1} \and
M. Mignoli\inst{2} \and
L. Pozzetti\inst{2} \and
A. Renzini\inst{3} \and
J. Kurk\inst{4} \and
C. Halliday\inst{5}
}
          
\offprints{M. Talia\\
\email{margherita.talia2@unibo.it}}

\institute{Dipartimento di Fisica e Astronomia, Universit\`a di Bologna,
Via Ranzani 1, I-40127, Bologna, Italy.
\and  
INAF- Osservatorio Astronomico di Bologna,
Via Ranzani 1, I-40127, Bologna, Italy
\and
INAF- Osservatorio Astronomico di Padova,
vicolo dell'Osservatorio 5, I-35122, Padova, Italy
\and
Max Planck Institut f\"ur extraterrestrische Physik,
Postfach 1312, 85741 Garching bei M\"unchen, Germany
\and
23, rue d\'Yerres, 91230 Montgeron, France
}

   \date{}

 
  \abstract
   {}
   {We investigate the morphological properties of 494 galaxies selected from the Galaxy Mass Assembly ultra-deep Spectroscopic Survey (GMASS) at $z > 1$, primarily in their optical rest frame, using \emph{Hubble} Space Telescope (HST) infrared images, from the Cosmic Assembly Near-IR Deep Extragalactic Legacy Survey (CANDELS).}
   {The morphological analysis of Wield Field Camera (WFC3) H$_{160}$ band images was performed using two different methods: a visual classification identifying traditional Hubble types, and a quantitative analysis using parameters that describe structural properties, such as the concentration of light and the rotational asymmetry. The two classifications are compared. We then analysed how apparent morphologies correlate with the physical properties of galaxies.}
   {The fractions of both elliptical and disk galaxies decrease between redshifts $z \sim 1$ to $z \sim 3$, while at $z > 3$ the galaxy population is dominated by irregular galaxies. The quantitative morphological analysis shows that, at $1<z<3$, morphological parameters are not as effective in distinguishing the different morphological Hubble types as they are at low redshift.

No significant morphological \emph{k}-correction was found to be required for the Hubble type classification, with some exceptions. 

In general, different morphological types occupy the two peaks of the $(U-B)_{rest}$ colour bimodality of galaxies: most irregulars occupy the blue peak, while ellipticals are mainly found in the red peak, though with some level of contamination. Disks are more evenly distributed than either irregulars and ellipticals. We find that the position of a galaxy in a \emph{UVJ} diagram is related to its morphological type: the ``quiescent'' region of the plot is mainly occupied by ellipticals and, to a lesser extent, by disks.

We find that only $\sim$33$\%$ of all morphological ellipticals in our sample are red and passively evolving galaxies, a percentage that is consistent with previous results obtained at $z < 1$. Blue galaxies morphologically classified as ellipticals show a remarkable structural similarity to red ones.

We search for correlations between our morphological and spectroscopic galaxy classifications. Almost all irregulars have a star-forming galaxy spectrum. In addition, the majority of disks show some sign of star-formation activity in their spectra, though in some cases their red continuum is indicative of old stellar populations. Finally, an elliptical morphology may be associated with either passively evolving or strongly star-forming galaxies.}
   {We propose that the Hubble sequence of galaxy morphologies takes shape at redshift $2.5{<}z{<}3$. The fractions of both ellipticals and disks decrease with increasing lookback time at $z{>}1$, such that at redshifts $z{=}2.5-2.7$ and above, the Hubble types cannot be identified, and most galaxies are classified as irregular.}  
   {}

   \keywords{Galaxies: high-redshift, Galaxies: structure, Galaxies: evolution}


   \maketitle
%

\section{Introduction}\label{sec:Introduction}
In the local Universe, galaxies can be classified according to their appearance along the so-called Hubble sequence.
At $z > 1$, in contrast, the majority of the galaxy population consists of objects with irregular morphologies that are very dissimilar from their low-redshift descendants \citep{abraham1996, cassata2005, conselice2008}. 

How and when did these objects turn into the ellipticals and spirals that we see today? 
The shape of a galaxy can be modified by many processes: star-formation activity, the effects of an active nucleus, interactions with neighbouring galaxies, galaxy merging, or infall of cold gas from the inter-galactic medium \citep{conselice2003}. Both morphological classification at all redshifts, to study how galaxies change shape through cosmic time, and the determination of how galaxy morphologies are related to their other physical and spectral properties are fundamental to the understanding of galaxy formation and evolutionary processes. 
The assignment of a galaxy to a particular morphological class is strongly affected by human subjectivity, especially for faint high-redshift galaxies. The Hubble types are still used in morphological classification at high redshift, even though galaxy shapes often do not fit perfectly into these traditional classes. In the past few years, some authors have developed simpler visual classification schemes based on the apparent nucleation of galaxy light profiles and the number and identification of distinct clumps \citep{law2007, law2012a, law2012b}.

To make the classification process more objective and automatic, non-parametric approaches have also been introduced. The most widely used systems are CAS and G-M$_{20}$\footnote[1]{CAS: concentration - asymmetry - clumpiness; G-M$_{20}$: Gini coefficient - second-order moment of the light distribution.} \citep{abraham1994, schade1995, abraham1996, bershady2000, conselice2003, lotz2004}. In high-redshift galaxies, the clumpiness is the least well-defined parameter, owing to the small size and faintness of the galaxies and to the limited resolution of the images \citep{lotz2004, law2012a}. 
A principal component analysis of the aforementioned parameters combined with a parametric description of the galaxy light is used, for example, in the Zurich Estimator of Structural Types (ZEST) classification scheme \citep{scarlata2007}.
A simpler approach to quantitative morphological galaxy classification is to examine the position of a galaxy in the planes defined by the four cited parameters. 
At low redshifts, galaxies belonging to different morphological classes occupy well-defined regions in the C-A and G-M$_{20}$ planes \citep{abraham1996, bershady2000, conselice2003, lotz2008}, within which even merger candidates are effectively distinguished. At high redshifts, the typical parameter values for the various morphological types widely differ. 
Previous studies at high-redshift have shown that, in the C-A plane, ellipticals have the highest concentrations and the lowest asymmetries \citep{conselice2008, cassata2005}. In addition, ellipticals are confined to a particular region in the G-M$_{20}$ plane, with M$_{20} \sim -1.7$, and G values spanning a $\sim 0.2$ range between 0.5 and 0.8, depending on the study. 
Galaxies with both disky and irregular morphologies are more scattered across both planes, but are more tightly confined in the G-M$_{20}$ one. 
In general, spheroids have higher G and lower M$_{20}$ values, while most irregulars have lower G and high M$_{20}$ \citep{lotz2008, law2012a, wang2012}. There are, however, some discrepancies between different studies \citep[see, for example,][]{conselice2008}.
The primary aim of morphological studies is to determine the robustness of non-parametric statistics in differentiating between morphologies at all redshifts, and develop a reliable classification tool for cases when the traditional method ``by eye'' becomes too uncertain.

The redshift evolution of galaxy morphologies, regardless of the way in which they are defined, has to be studied in the same rest-frame band for all redshifts. Since at different wavelengths different gas phases and stellar populations are probed, the appearance of a galaxy may also change, depending on the wavelength at which the galaxy is seen. Hence, it is important to determine whether there is any need for a ``morphological $k$-correction'' as a function of redshift \citep{conselice2011}. 

At high redshift, we need to search for relations between the morphological classification and other properties, such as spectral type, mass, colours, and star-formation history to understand the interplay between physical processes and merger events in the shaping of a galaxy. At low redshift, quiescent galaxies usually have spheroidal structures, while star-forming galaxies are mostly either irregulars or disks. In general, this is also true at high redshift \citep{cassata2011, wuyts2011, bell2012, szomoru2011, szomoru2012}, but not always. \citet{conselice2011} find that a significant fraction of galaxies visually classified as early-types have star-forming spectral types, which suggests that very few purely passive massive galaxies exist at $1 < z < 3$, at least within the \emph{Hubble Ultra Deep Field}. 
For a sample of IRAC-selected Extremely Red Objects (IEROs), however, \citet{wang2012} find that quiescent and dusty star-forming galaxies, classified according to their IR colours, can be clearly separated in the G-M$_{20}$ plane, thus implying that there is a relation between the morphology and star-forming status of IEROs. In particular, this may indicate that the quenching process for star formation occurs with (or is the cause of) an increase in the galaxy concentration. 
	\begin{figure}[h!]
	\centering
	\includegraphics[scale=0.45]{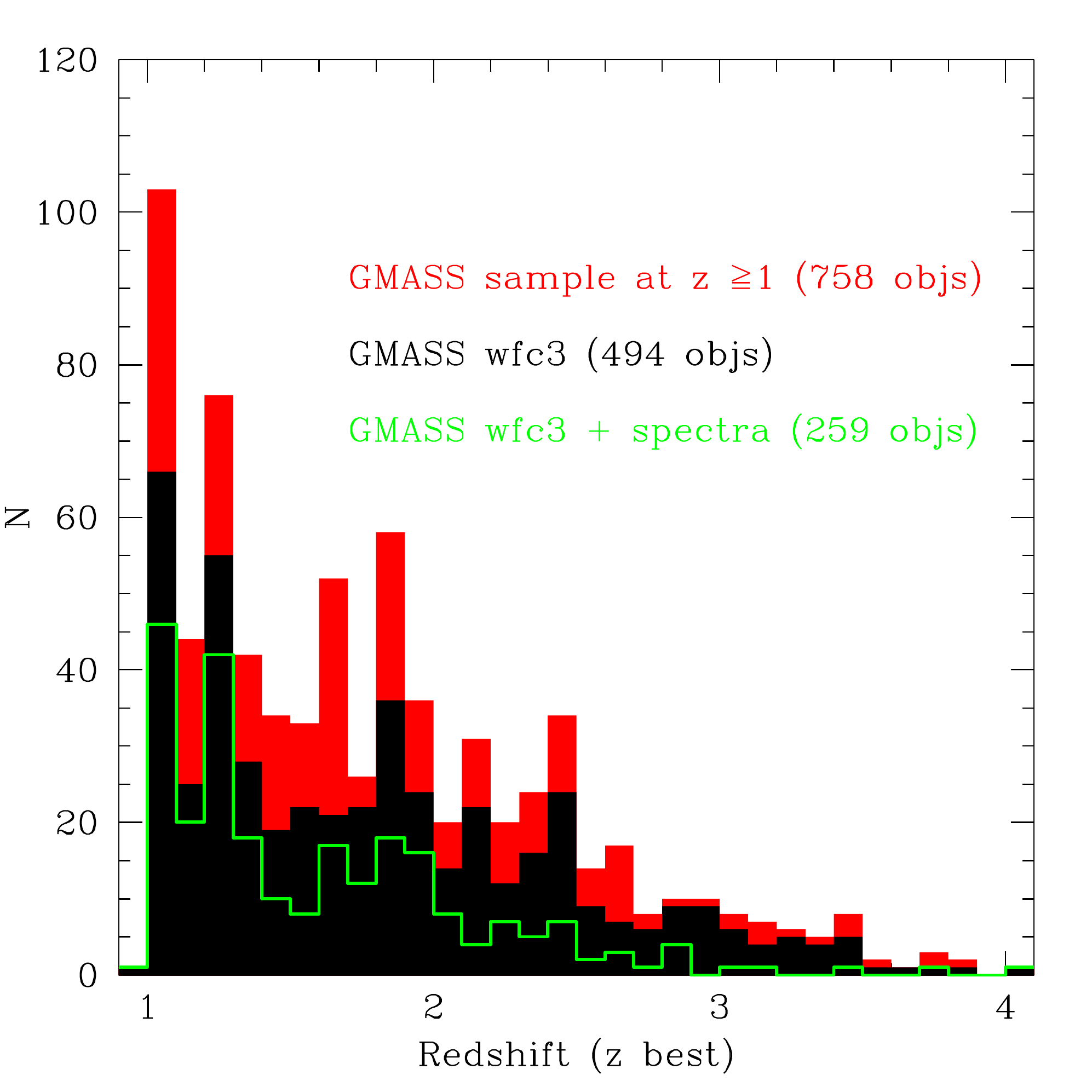}
	\caption{Redshift distribution. The red histogram represents best-redshift (spectroscopic or photometric) distribution of the total GMASS sample at $z \geq 1$; the black histogram represents best-redshift distribution of the 494 galaxies from the \emph{GMASS-WFC3 sample}; the green histogram represents the distribution for the 259 galaxies with a spectroscopic redshift.}
	\label{hist_z}
	\end{figure}

The aim of this paper is to investigate the morphological properties of galaxies at intermediate to high redshift, and how they correlate with physical properties such as the star-formation activity. We decided to use recent images acquired using the H$_{160}$ band filter of the Wide Field Camera 3 (WFC3), mounted on the \emph{Hubble} Space Telescope, because the observed-frame near-infrared wavelengths are needed to resolve the location of most stellar mass in galaxies, at high redshifts \citep{conselice2011}. This paper is structured as follows. First, the results of our morphological classification are presented, which are based on both Hubble types and non-parametric statistics. The need for morphological $k$-correction is also investigated by comparing rest-frame ultraviolet (UV) and optical images. We then study the connections between the morphologies and physical properties of $z > 1$ galaxies, examining the colour-stellar mass diagram and the distributions of galaxies with different morphologies in the well-established bimodal colour distribution of galaxies. Finally, we focus on a spectroscopic sub-sample of galaxies to search for possible correlations between our morphological and spectroscopic classifications.

\section{The sample}\label{sec:The sample}
We analyse rest-frame optical morphologies of a sample of $z \geq 1$ galaxies. We started from the photometric catalogue of the \emph{Galaxy Mass Assembly ultra-deep Spectroscopic Survey} - GMASS \citep{kurk2012}, and searched for counterparts to galaxies at $z \geq 1$ in the H$_{160}$ band mosaics (v0.5) of the \emph{Cosmic Assembly Near-IR Deep Extragalactic Legacy Survey} - CANDELS \citep{grogin2011, koekemoer2011} in the GOODS-South. 
We found 494 matches, which we refer to as the \emph{GMASS-WFC3 sample}. We note that 259 of the selected galaxies have a spectroscopic redshift, which was collected primarily from the GMASS spectroscopic survey, but also from the European Southern Observatory (ESO) public database\footnote[2]{The compilation of GOODS/CDF-S spectroscopy master catalogue v2.0 was used for this work. Spectroscopic redshifts were taken in particular from the following surveys: the ESO-GOODS/FORS2 v3.0 \citep{vanzella2008} and ESO-GOODS/VIMOS v2.0 \citep{popesso2009, balestra2010}, the VVDS v1.0 \citep{lefevre2005}, the K20 \citep{mignoli2005}, and the spectroscopic follow-up programme of X-ray sources in the Chandra Deep Field South \citep{szokoly2004}.}.
The redshift distribution is shown in Fig.\ref{hist_z}.

\subsection{The GMASS survey}\label{sec:The GMASS survey}
The GMASS survey \citep{cimatti2008, kurk2012} is an ESO Very Large Telescope (VLT) large programme project based on data acquired using the FOcal Reducer and low dispersion Spectrograph (FORS2). 
The project's main objective is to use ultra-deep optical spectroscopy to measure the physical properties of galaxies at redshifts $1.5 < \emph{z} < 3$. 
The \emph{GMASS photometric catalogue} is a purely magnitude-limited selection of sources detected in the GOODS-South public image taken at $4.5 \mu m$ with the \emph{Infrared Array Camera} (IRAC) mounted on the \emph{Spitzer Space Telescope}, the limiting magnitude being $\emph{m}_{4.5} < 23.0$ (AB system).
The photometric catalogue contains 1277 objects, for 131 of which it was possible to determine secure spectroscopic redshifts. The GMASS spectra have a spectral resolution of $R = \lambda/\Delta\lambda \sim 600$, which was chosen as the best compromise between extensive wavelength coverage and the ability to resolve and identify spectroscopic features for redshift determination. 
Physical properties, such as stellar mass and star-formation rate, were estimated by fitting broad-band photometry, from the \emph{U} band to the IRAC $5.8\mu m$ band, with the synthetic spectra of the \citet{maraston2005} evolutionary population-synthesis models. Star-formation histories were parametrized by exponentials ($e^{\frac{-t}{\tau}}$) with e-folding time-scales $\tau$ between 100 Myr and 30 Gyr, plus the case of constant SFR. A Kroupa initial mass function \citep{kroupa2001}, fixed solar metallicity, and a Calzetti law for dust extinction \citep{calzetti2000} were assumed.

\subsection{The CANDELS survey}\label{sec:The CANDELS survey}
The CANDELS survey \citep{grogin2011, koekemoer2011} is a 902-orbit Multi-Cycle Treasury program on the HST. The project aims to investigate galaxy evolution up to redshift $z \sim $ 8 using deep imaging data collected with the infrared (IR) and UVIS channels of the WFC3, as well as the Advanced Camera for Surveys (ACS), targeting five existing survey fields on the sky. Science drizzled images, along with weight images, are being publicly released as the observations are carried out. In this work, images acquired within the Deep program with the WFC3/IR F160W filter were used, which have an exposure depth of 1100s and a points spread function (PSF) full width at half maximum (FWHM) of 0".18. 

\section{Visual morphological analysis}\label{sec:Visual morphological analysis}
The morphological analysis of H$_{160}$ images was done using two different methods: a visual classification based on traditional Hubble types, and a quantitative analysis using parameters that quantify morphological properties, such as the concentration of light and the galaxy asymmetry. 
	\begin{figure}[h!]
	\centering
	\includegraphics[trim=100 40 100 400, clip=true, width=27mm]{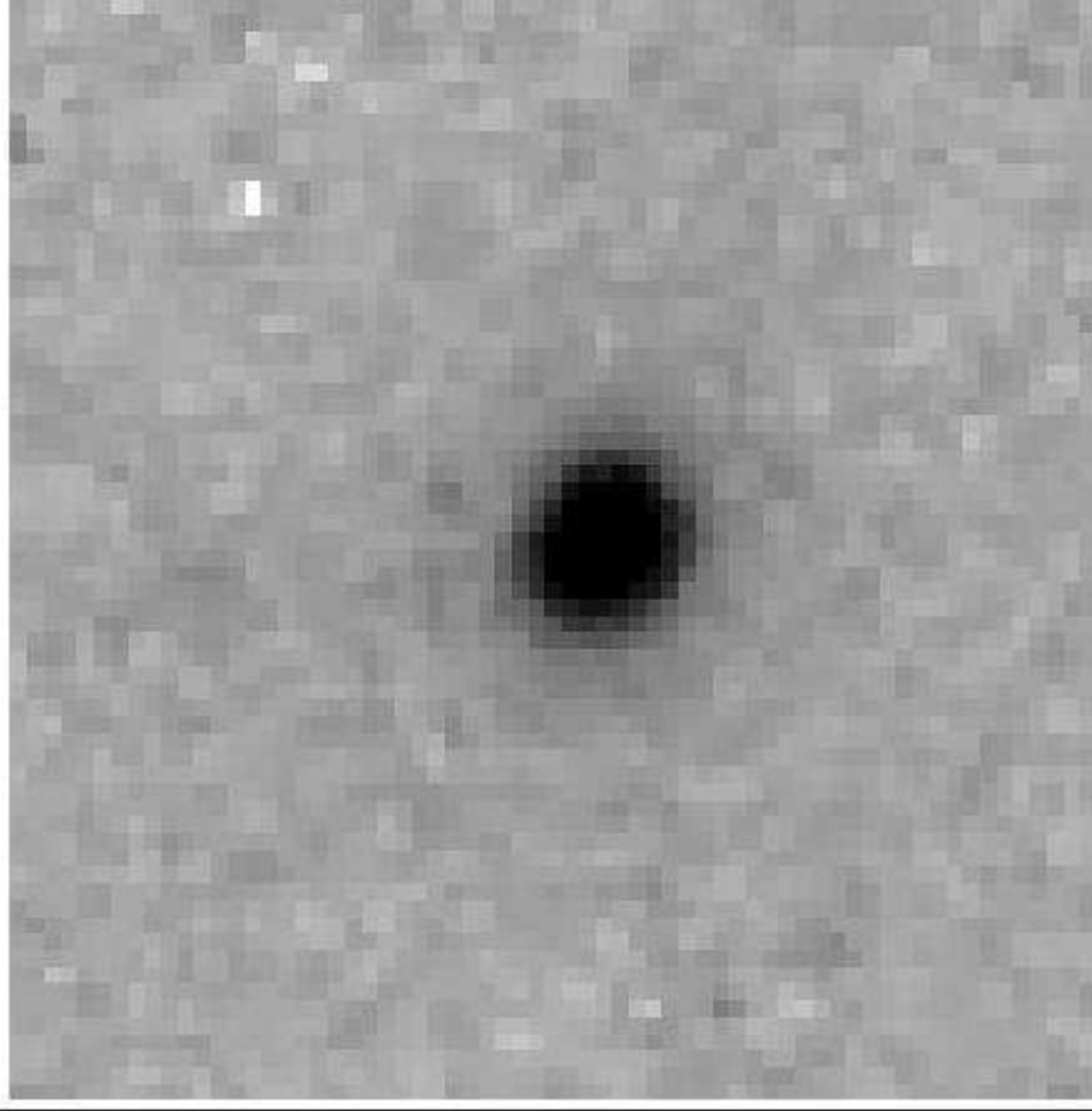}
	\includegraphics[trim=100 40 100 400, clip=true, width=27mm]{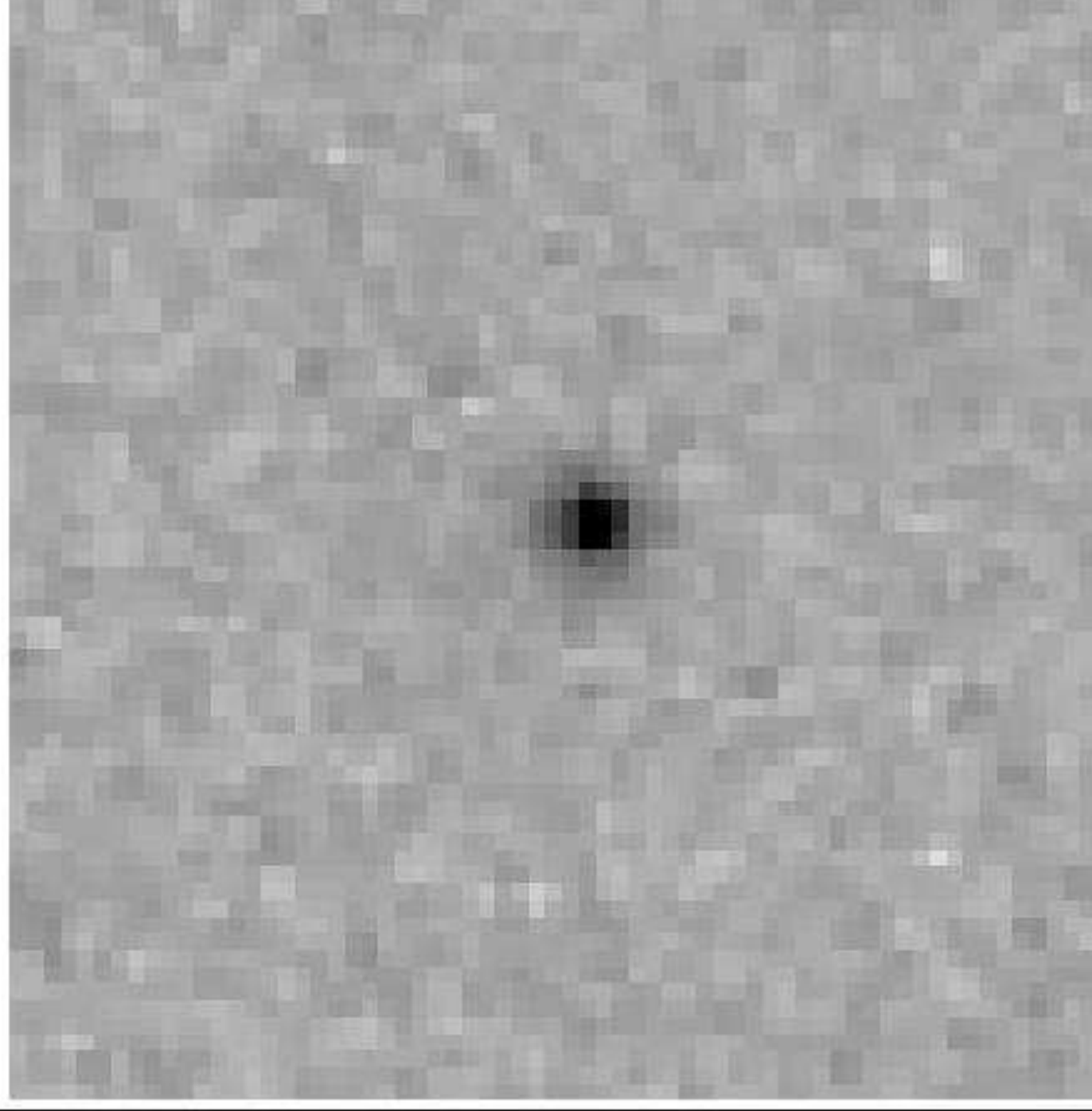}
	\includegraphics[trim=100 40 100 400, clip=true, width=27mm]{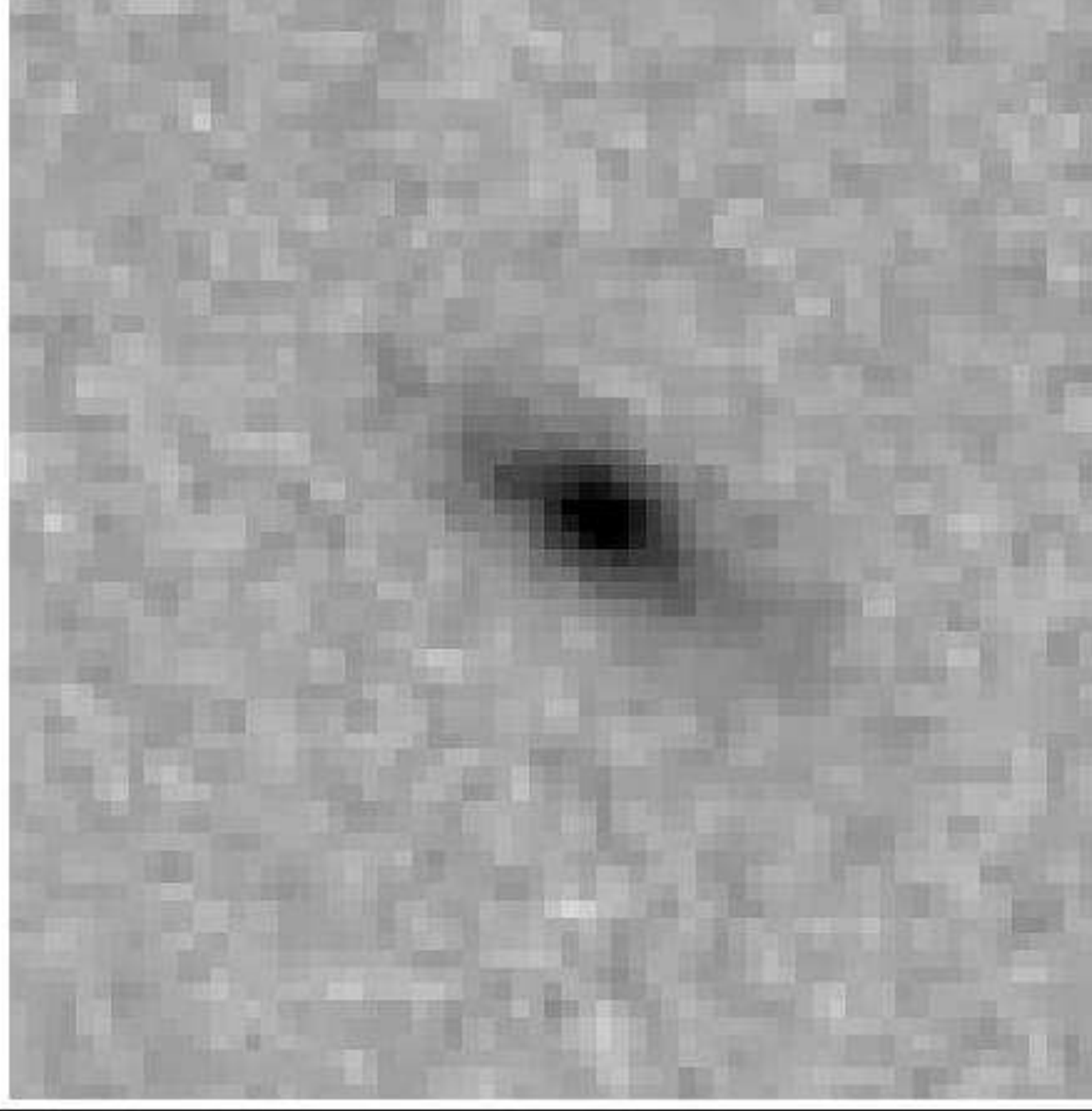}
	\includegraphics[trim=100 40 100 400, clip=true, width=27mm]{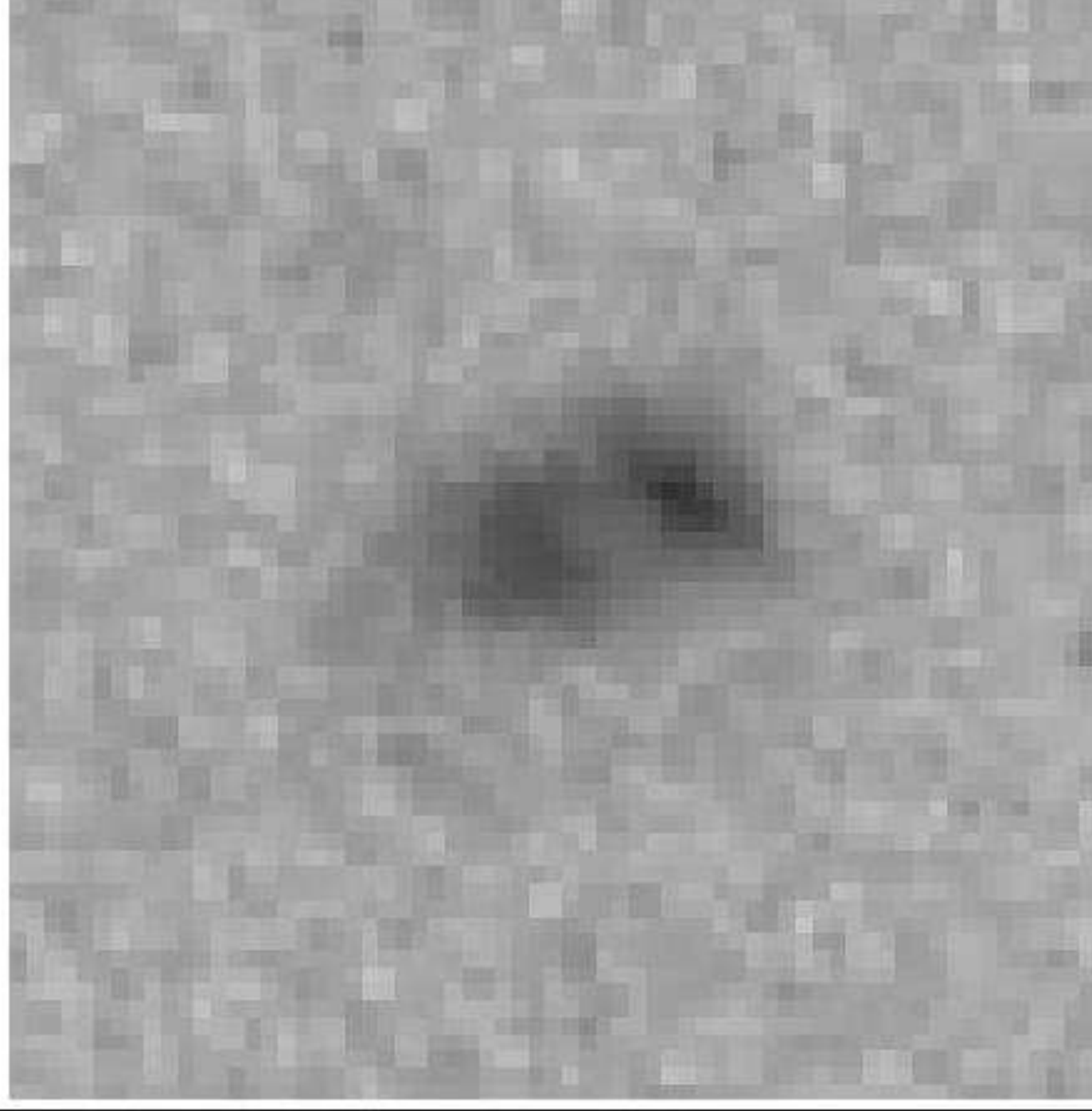}
	\includegraphics[trim=100 40 100 400, clip=true, width=27mm]{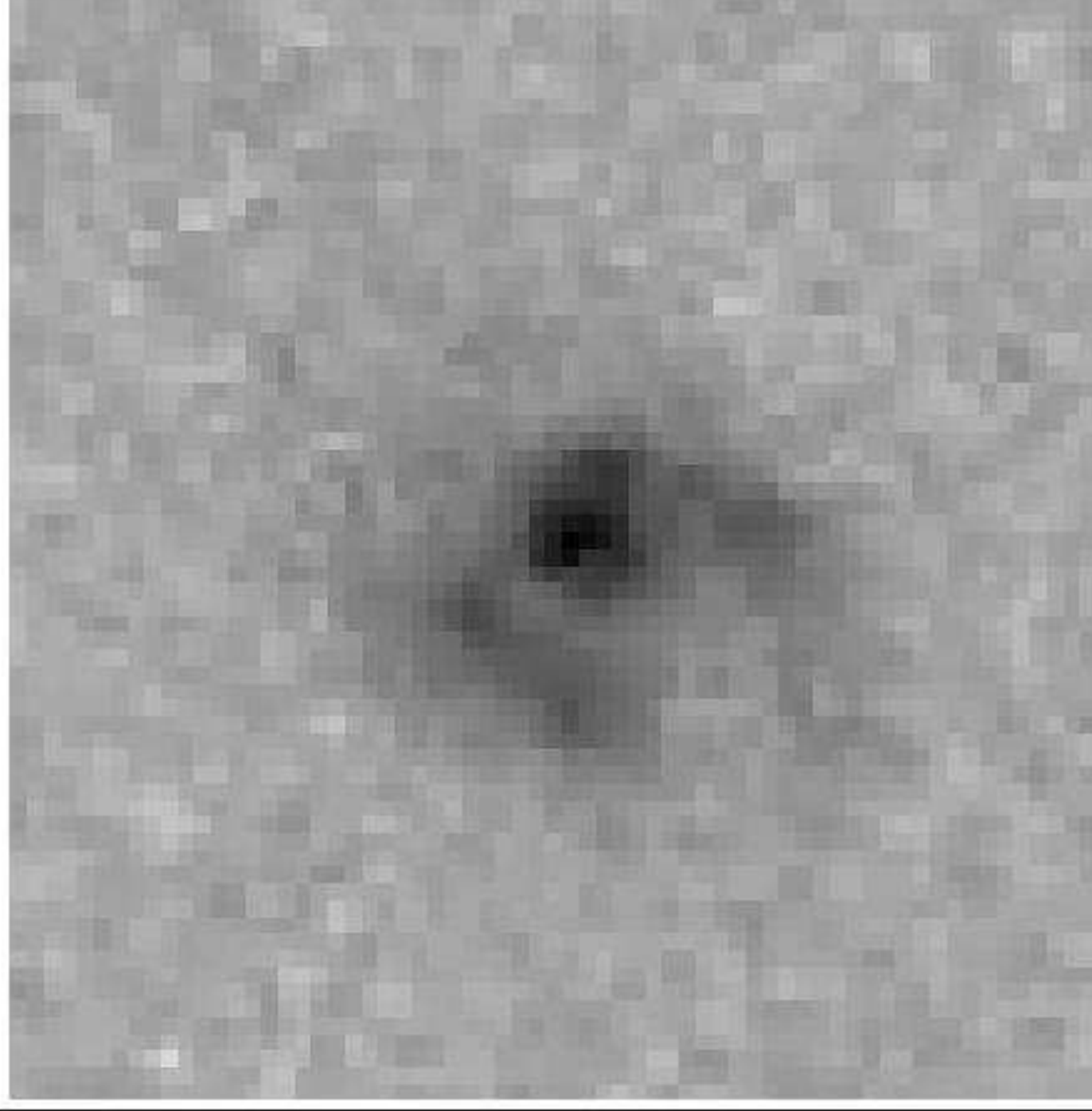}
	\includegraphics[trim=100 40 100 400, clip=true, width=27mm]{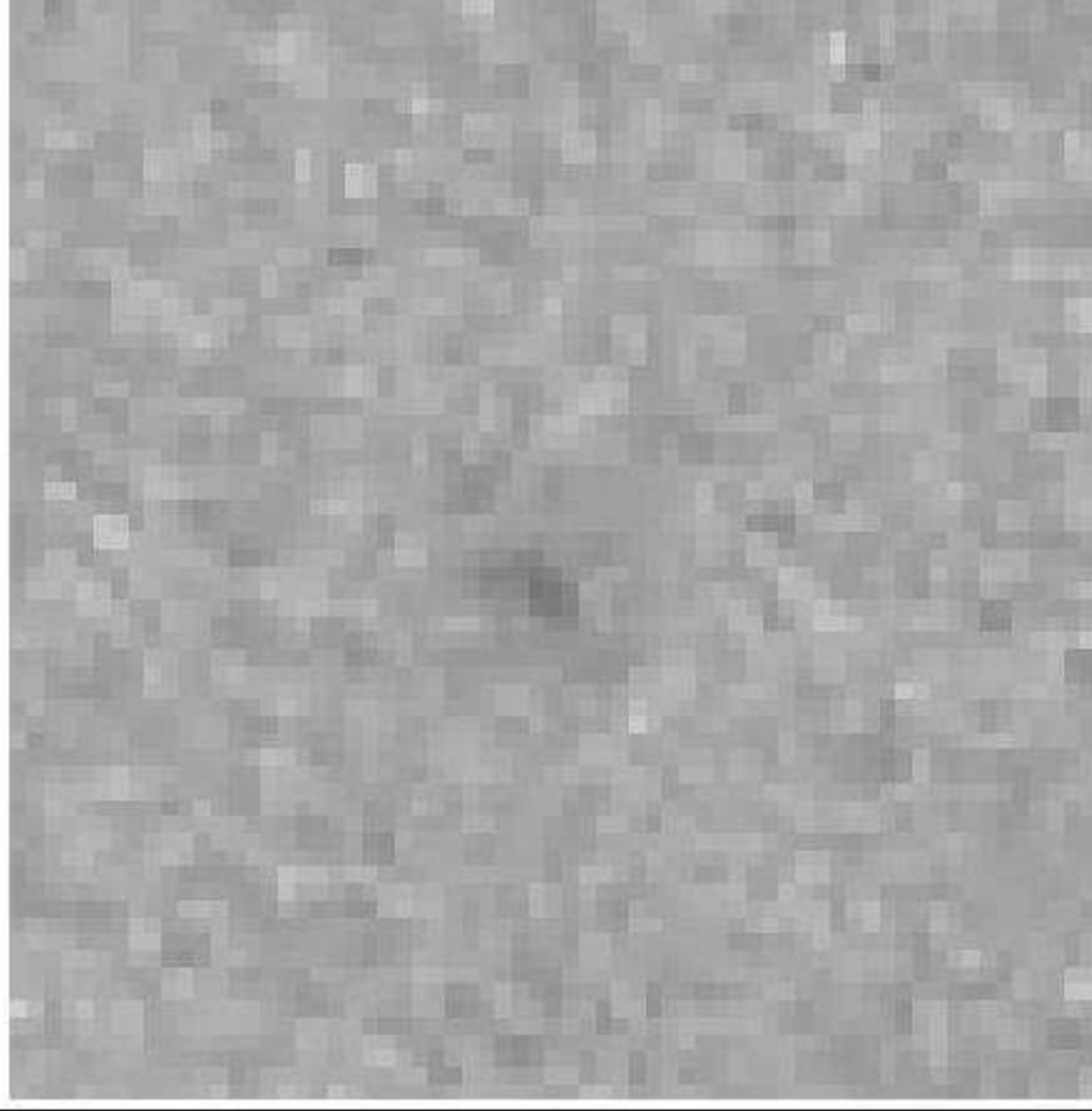}
	\caption{Sample snapshots (4``$\times$4``) of galaxies to illustrate the visual classification scheme. From top left: elliptical ($z{=}1.09$), compact ($z{=}3.19$), disk-like ($z{=}1.22$), irregular ($z{=}1.22$), irregular ($z{=}1.99$), \emph{faint object} ($z{=}2.30$).}
	\label{snapshot}
	\end{figure}

The galaxy shapes were classified into five main types based on the appearance of the galaxies in the H$_{160}$ band images:
\begin{itemize}
	\item\emph{Ellipticals} for single centrally concentrated sources with no evidence of outer structures. 

	\item\emph{Compact} for single centrally concentrated sources that are very smooth, symmetric, and display no evidence of any substructure. It differs from the elliptical classification in that a compact galaxy contains no features such as an extended light distribution or a light envelope.

	\item\emph{Disk-like} for undisturbed sources with disk-like shapes.

	\item\emph{Irregulars} where the galaxy shapes are none of the above. These systems are possibly in some phase of a merger \citep{conselice2003}. Some of them show evidence of two or more distinct nucleated sources of comparable magnitude.

	\item\emph{Faint objects} where the objects are too faint to allow any reliable classification. 
\end{itemize}

	\begin{table}[h!]
	\caption[]{Number of galaxies of each morphological type.}
	\label{vis_class}
	\centering                          
	\begin{tabular}{l l l l l}        
	\hline\hline                 
	Ellipticals & Compact & Disk-like & Irregulars & Faint\\
	\hline                      
	75 & 36	& 157 & 199 & 27 \\
	\hline\hline
	\end{tabular}
	\end{table}

Galaxies were, at first, classified by three authors (MT, AC, and MM) independently. All classifiers were allowed to adjust the stretch and contrast of the images. The three classifications were then compared and any disagreements were discussed before assigning the single galaxy classifications used throughout this paper.
We emphasize that our visual classification is based only on the appearance of the galaxy as seen through the observed H$_{160}$ filter: no additional information such as colour, spectrum, mass, or star-formation rate (SFR) was used to assign a morphological type to a galaxy.
In Fig. \ref{snapshot}, sample snapshots of galaxies are provided to illustrate the visual classification scheme, while Table \ref{vis_class} summarizes the number of galaxies in each class. A morphological atlas of all 494 galaxies of our sample is presented in the Appendix.
	\begin{figure*}[t!]
	\centering
	\includegraphics[scale=0.45]{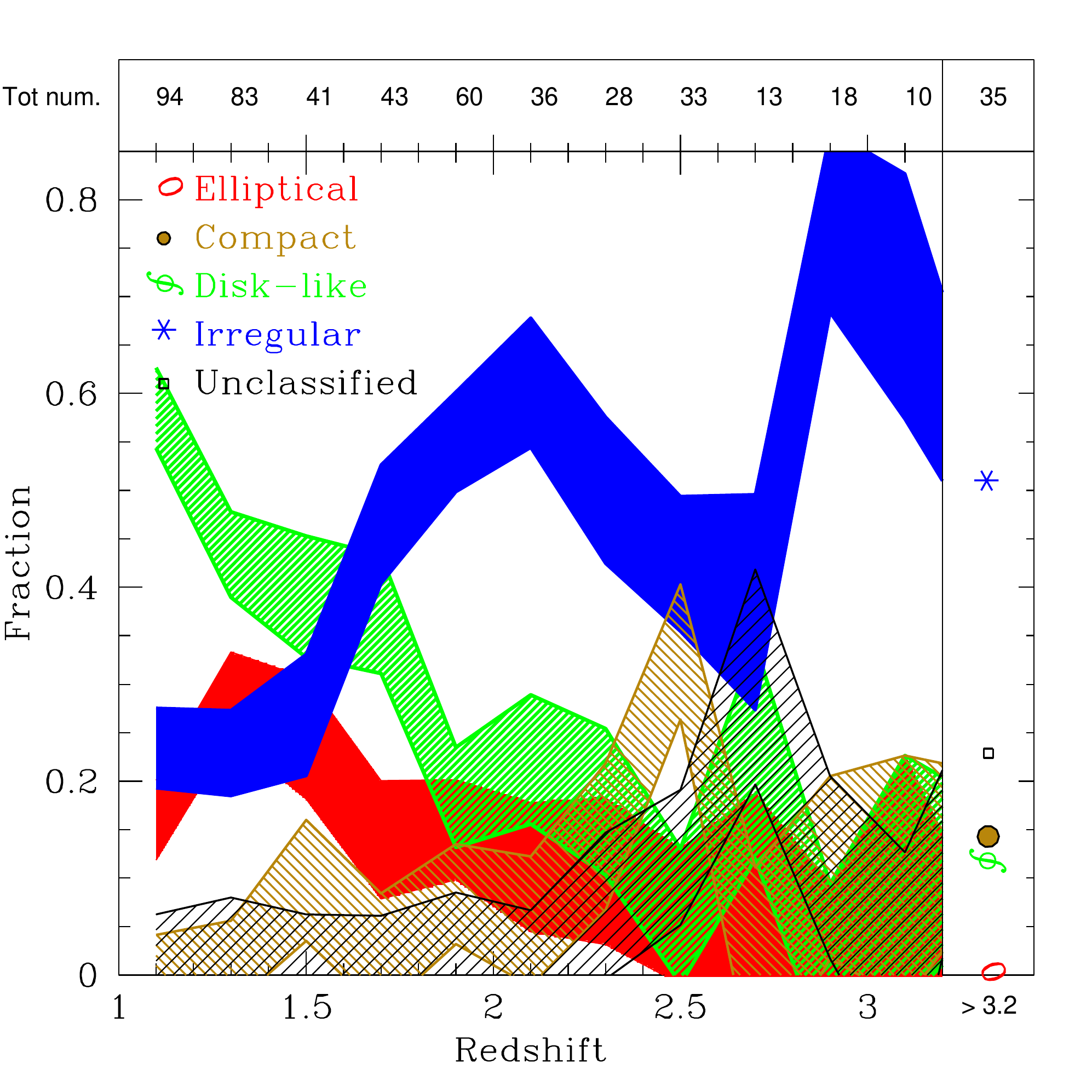}
	\includegraphics[scale=0.45]{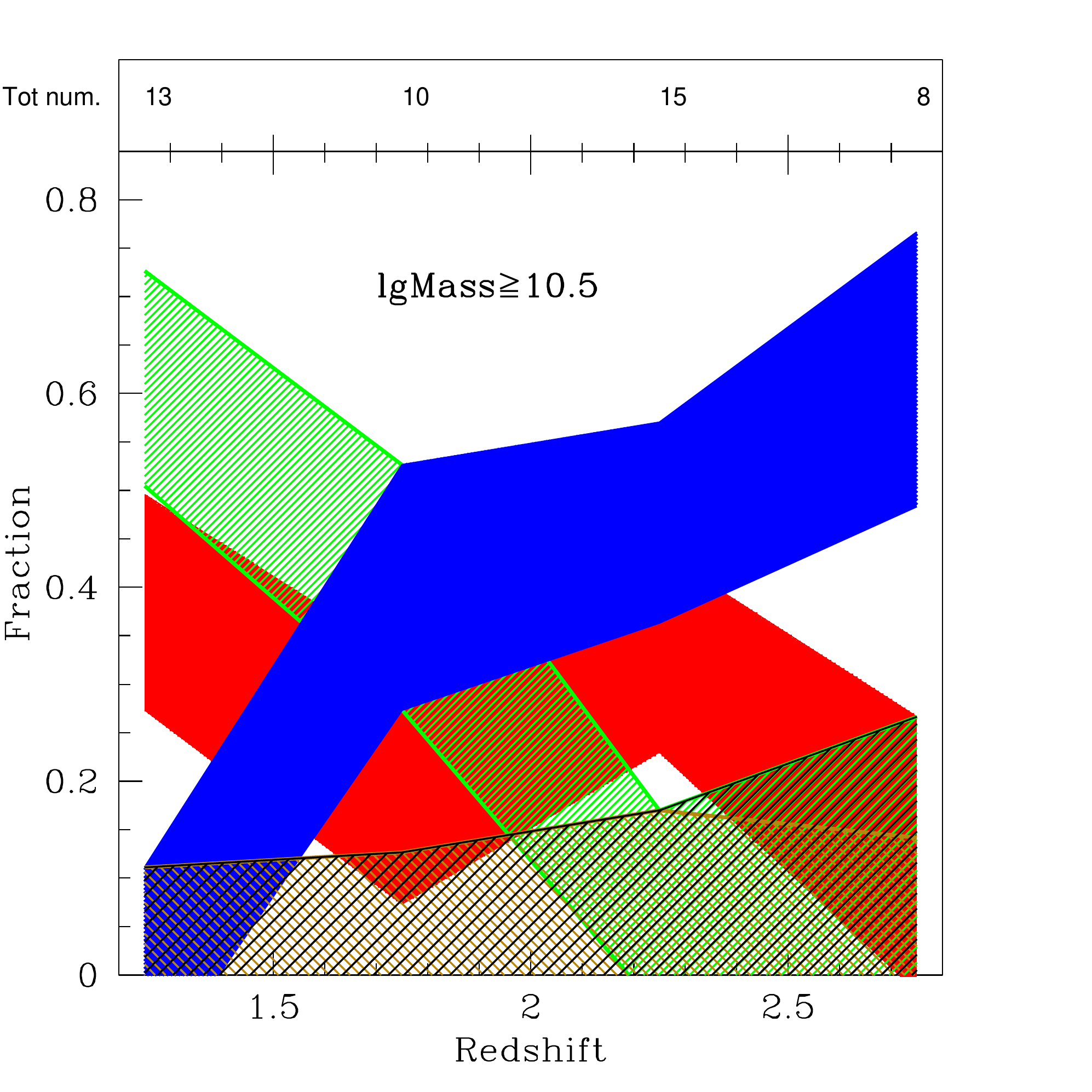}
	\caption{Left plot: the relative fractions of the different galaxy morphological types as a function of redshift. The total number of galaxies in each redshift step ($\Delta z = 0.2$) between $z = 1$ and $z = 3.2$ is indicated. The last step includes all galaxies at $z > 3.2$. The confidence regions reflect both the random uncertainty and the differences between classifications made by three different authors. See the text for more details about the classification procedure. Right plot: same as left, but for a mass complete sample ($log(M_{\odot}lim)=10.5$). In this plot the redshift step is $\Delta z = 0.5$.}
	\label{viscl_fct_z}
	\end{figure*}
The relative fraction of the different visual morphological types, as a function of redshift, is shown in Fig. \ref{viscl_fct_z}. The confidence regions account for both the random uncertainty and the differences between the three original classifications, before they were reconciled into the definitive one.
We find that the fractions of ellipticals and disks decrease with increasing lookback time and that the most galaxies at redshift $z \sim 2$ and above are either irregular or too faint to be classified. The fraction of compact galaxies is almost constant ($\sim 0.1$) between redshift 1 and 3, but has a peak around $z \sim 2.5$, as also observed by \citet{conselice2008}. Morphological type fractions were computed also for a sample selected to be complete in terms of stellar mass. We applied a minimum-mass threshold of $log(M_{\odot}lim)=10.5$, which is the mass completeness limit in the redshift range considered in this paper \citep{kurk2012}. The plot, presented in Fig. \ref{viscl_fct_z}, shows that the general trends found in the complete sample are preserved.

Using the observed H$_{160}$ band, we observe a rest-frame wavelengths extending from $\lambda_{rest}{\sim}8000$ at $z{\sim}1$ to $\lambda_{rest}{\sim}4000$ at $z{\sim}3$. To check whether this variation in galaxy rest-frame wavelength could bias our results, and whether a morphological $k$-correction was required by instead performing morphological classifications at approximately identical rest-frame wavelengths for all galaxies. We used ACS z-band images for galaxies at $z \leq 1.7$, WFC3 J-band (from CANDELS) at $1.7<z \leq 2.5$ and H-band at $z>2.5$\footnote[3]{All images at shorter wavelengths were degraded to match the pixel size and PSF of H-band images. See Sec. \ref{sec:Morphological k-correction}.}. The fractions of morphological types, at all redshifts, are almost identical to those presented in Fig. 3. The only differences are a slightly higher fraction of irregulars and a slightly lower fraction of disks at $z<1.5$, but the general trends remain unchanged. A thorough discussion about morphological $k$-correction is presented later in the paper.

Comparing our trends with the literature, we are able to explore the full range of redshifts, down to $z \sim 0$. Below the minimum redsfhit of our sample, i.e. $z < 1$, disks and spheroids are the dominant population, with only about 20$\%$ of galaxies being irregulars. At redshift $z \sim 1-1.5$ the fractions of irregulars and compact are higher, while the fractions of both disks and ellipticals are lower than at redshift $z<1$ \citep{conselice2005, conselice2008}. 

Our results allow us to identify the origin of the Hubble sequence at redshift $2.5 < z < 3$. At these redshifts, 20$\%$ of galaxies have elliptical or disk morphologies, whereas at redshifts $z > 3$ almost all galaxies are classified as irregulars.

\begin{figure}[h!]
	\centering
	\includegraphics[scale=0.45]{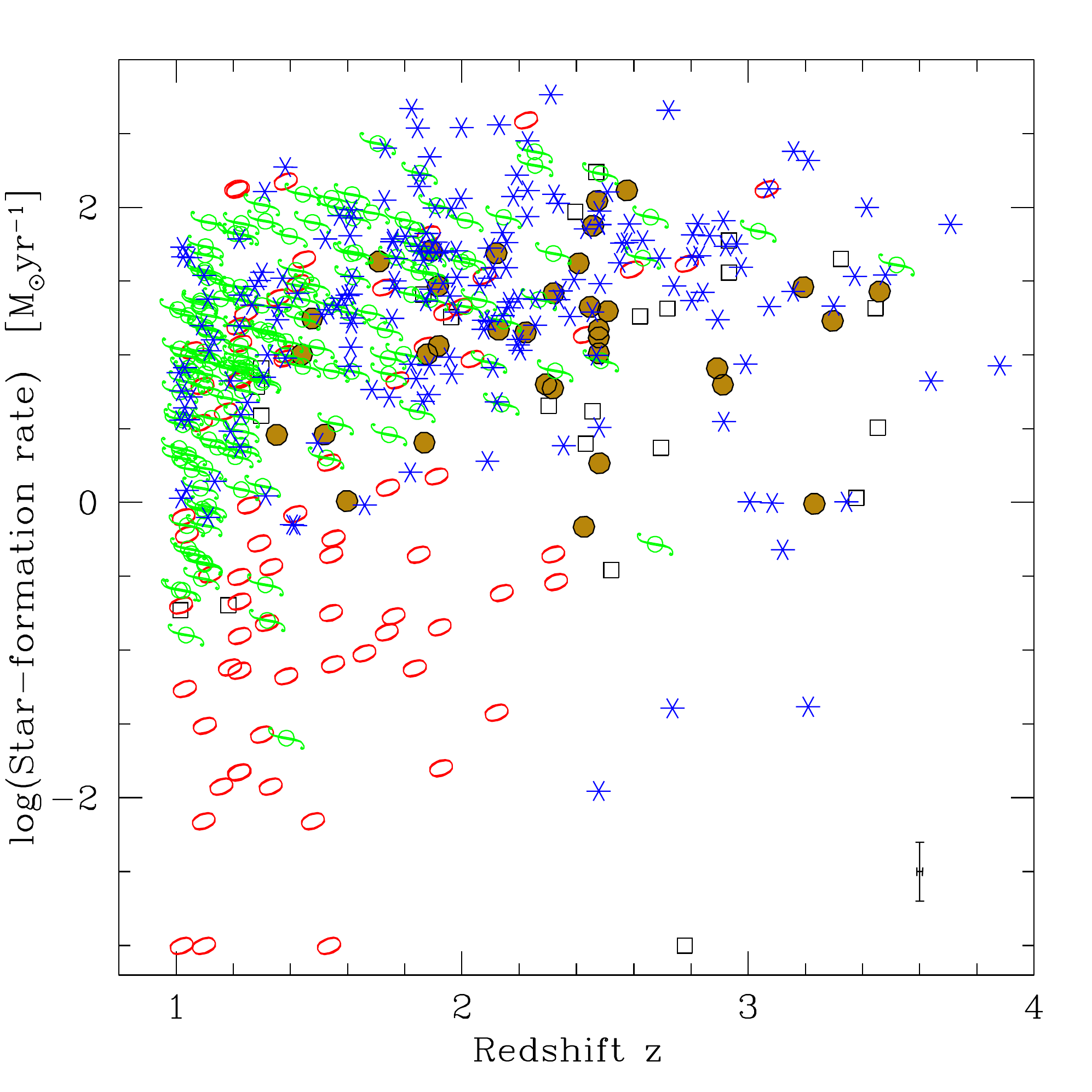}
	\includegraphics[scale=0.45]{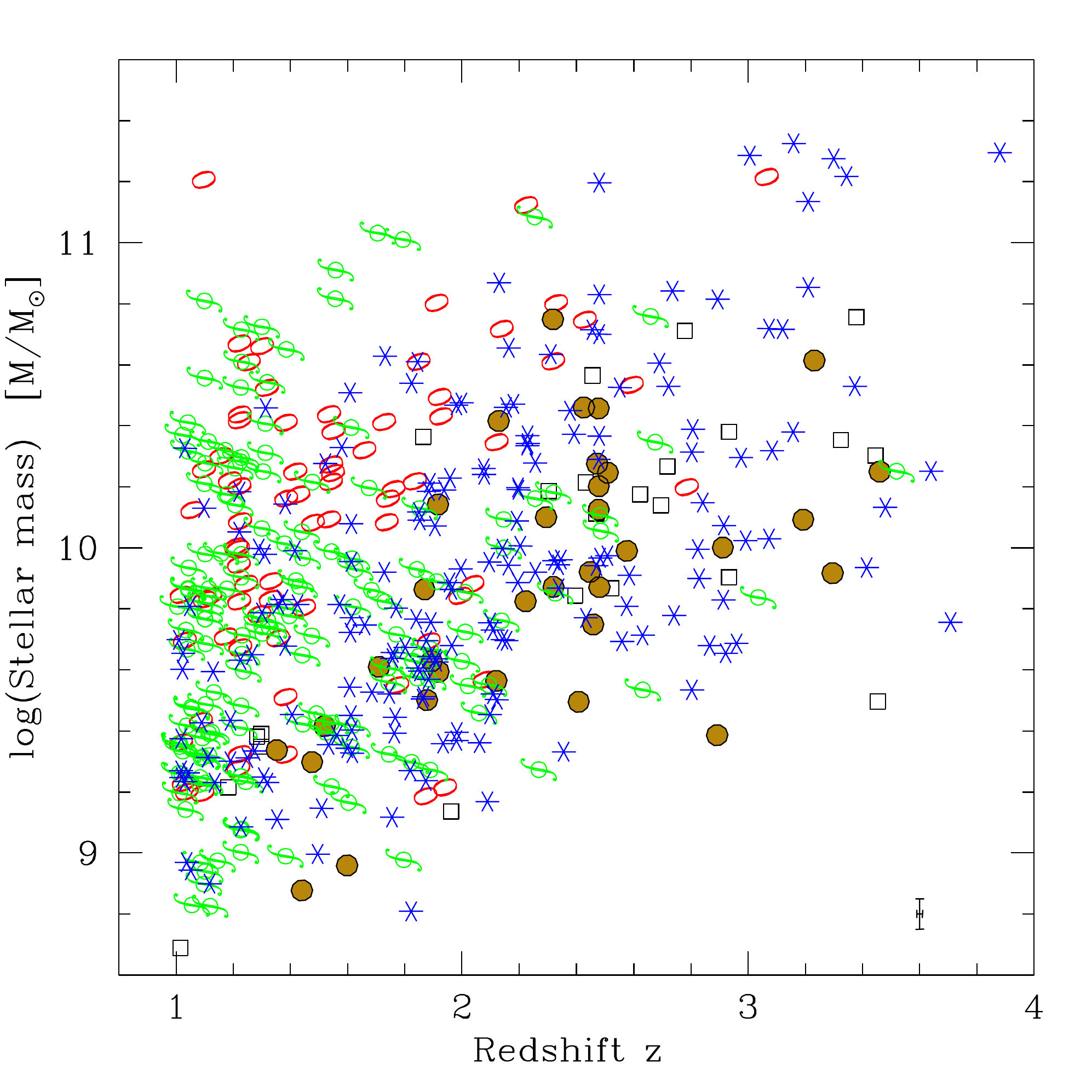}
	\caption{SFR (upper panel) and stellar mass (bottom panel) as a function of redshift for the 494 galaxies in the \emph{GMASS-WFC3 sample}. Points are shape- and colour-coded according to the morphological visual class (as in Fig. \ref{viscl_fct_z}).}
	\label{prop_vs_zbest}
	\end{figure}

In Fig. \ref{prop_vs_zbest} we plot SFRs and stellar masses as a function of redshift of all the galaxies in our sample. 
Galaxies hosting little or no star-formation are mainly ellipticals, though a small fraction are disks. On the other hand, galaxies with higher SFRs display a wider range of morphologies. In particular, almost all irregular galaxies have $SFR \gtrsim 10$ $M_{\odot}yr^{-1}$. We can conclude that non-SFGs at $z > 1$ are generally ellipticals, while an irregular morphology is associated with active star-formation. It is interesting to notice that most faint objects are star-forming galaxies, with ${<}SFR{>} \sim 10$ $M_{\odot}yr^{-1}$, which are probably heavily obscured by dust or of low surface brightness. 
There is also a trend between morphological types and stellar masses: at each redshift, faint objects and both irregular and compact galaxies have, on average, lower stellar masses than ellipticals. 

\section{Quantitative morphological analysis}\label{sec: Quantitative morphological analysis}
Another approach to the morphological classification of galaxies is the calculation of non-parametric statistics. We concentrated on the four most widely used parameters: concentration \emph{C}, rotational asymmetry \emph{A}, Gini coefficient \emph{G}, and second-order moment of the light distribution M$_{20}$, which were computed using the code MORPHEUS \citep{abraham2007}.  
	\begin{table*}[t!]
	\caption[]{Morphological parameters. For the total sample, we provide the minimum, maximum, and mean values (with r.m.s.) of the morphological parameters, for each morphological type. In parentheses, the number of galaxies for which the image $S/N > 50$, for each morphological type, is also indicated.}
	\label{morph_par}
	\centering   
	\begin{tabular}{l c c | c c c c}        
	\hline                 
	& Min & Max & Ellipticals (75)& Compact (24)& Disk-like (151)& Irregulars (138)\\
	\hline 
	concentration 	& 0.12 & 0.40& 0.29 $\pm$ 0.04 & 0.28 $\pm$ 0.04 & 0.25 $\pm$ 0.04 & 0.23 $\pm$ 0.04 \\
	asymmetry 	& 0.00 & 0.58& 0.19 $\pm$ 0.08 & 0.17 $\pm$ 0.07 & 0.12 $\pm$ 0.06 & 0.17 $\pm$ 0.10 \\
	Gini	 	& 0.23 & 0.49& 0.41 $\pm$ 0.02 & 0.39 $\pm$ 0.03 & 0.37 $\pm$ 0.04 & 0.34 $\pm$ 0.04 \\
	M$_{20}$	&-2.29 &-0.78&-1.64 $\pm$ 0.17 &-1.63 $\pm$ 0.19 &-1.67 $\pm$ 0.19 &-1.51 $\pm$ 0.26\\
	\hline\hline
	\end{tabular}
	\end{table*}

\subsection{Parameter definitions}\label{sec: Parameter definitions}
The concentration index \emph{C} is the ratio of the intensity of light contained within a central region to that within one larger radius. 
A higher value of \emph{C} indicates that a larger amount of light in a galaxy is contained within its central region.

The asymmetry index \emph{A} was first defined by \citet{schade1995} as a measure of the asymmetry of a galaxy to a 180$^{\circ}$ rotation. 
Lower values of \emph{A} imply that a galaxy is largely symmetric, while higher values of \emph{A} indicate an asymmetric light distribution.

The Gini coefficient \emph{G} \citep{abraham2003, lotz2004} is a statistical tool that measures the cumulative flux distribution of a population of pixels and is insensitive to the actual spatial distribution of the individual pixels. 
Higher \emph{G} values indicate that the majority of the total flux is concentrated in a small number of pixels, while lower values represent a more uniform distribution of flux.

Finally, M$_{20}$ \citep{lotz2004} is defined as the second-order moment of the brightest pixels that constitute 20\% of the total flux in the segmentation map, normalized by the second-order moment of all the pixels in the segmentation map. 
Increasingly negative values of M$_{20}$ correspond to more regular objects, whereas less negative values are found for more irregular objects often with multiple clumps. 

The precise definitions of the parameters can vary from author to author. Here we use the concentration defined by \citet{abraham1994}, and the definitions of Gini, M$_{20}$, and asymmetry given in \citet{law2012a}, although we use instead the correction term defined by \citet{lauger2005} to remove the background asymmetry. We refer the readers to the cited papers for more details.
To define the segmentation map (i.e. decide which pixels belong to the galaxy), the quasi-Petrosian method of \citet{abraham2007} was adopted\footnote[4]{The pixels in a preliminary segmentation map are sorted in decreasing order of flux into an array $f_i$, which is then used to construct a cumulative flux array. The so-called quasi-Petrosian isophote is then set by determining the pixel index at which the pixel flux is equal to some fraction (called the quasi-Petrosian threshold $\eta$) of the cumulative mean surface brightness.}. We chose a threshold of $\eta = 0.3$ following the prescriptions of \citet{law2012a}, who found this to be the optimal way of maximizing the advantages of the method. 

In this part of the analysis only images with $S/N > 50$ were used\footnote[5]{$S/N=F/sqrt(A)*rms$, where $F$ is the total flux of the galaxy, $A$ is the total number of pixels, and $rms$ is the average r.m.s. of the background over all sky pixels.}, since images of a too low a $S/N$ did not allow a correct galaxy mapping to be made over all pixels \citep{law2012a}. Some tests showed that parameters computed on images with $S/N < 50$ assume unrealistic values, for example negative Asymmetries, due to a background over-correction.
This $S/N$ cut, which roughly corresponds to a cut in H-band magnitude of H$_{AB} < 24$, removes 106 objects, including the galaxies visually classified as faint objects. Our quantitative morphological analysis was conducted on the remaining sample of 388 galaxies.

\subsection{Exploration of the parameter space}\label{Exploration of the parameter space}
For each morphological type, mean values of all four parameters C, A, G, and $M_{20}$ are given in Table \ref{morph_par}. 

At low redshift, different morphological types can be clearly distinguished using the four aforementioned parameters \citep{abraham1996, conselice2003, lotz2004}. In contrast, at high redshift the parameters cannot be effectively used to separate the different visual morphologies. 
\\

\begin{figure}[h!]
	\centering
	\includegraphics[scale=0.45]{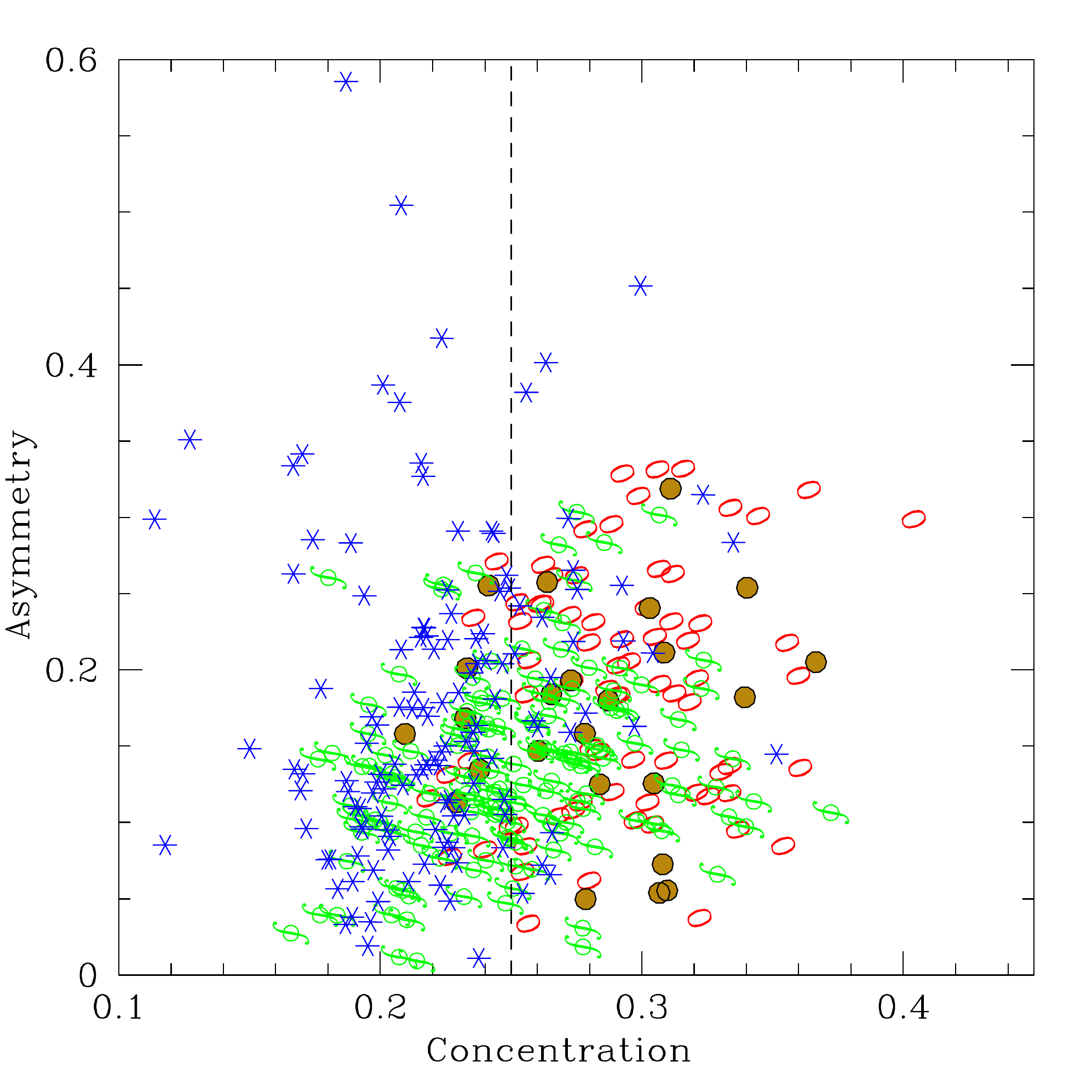}
	\caption{Concentration vs. asymmetry. The dashed line separates the regions typically populated by ellipticals and irregulars. Points are shape and colour coded with respect to the morphological visual class (see Fig. \ref{viscl_fct_z}).}
	\label{cvsa_plot}
	\end{figure}

{\bf The \emph{C} vs. \emph{A} plane} (Fig. \ref{cvsa_plot}) shows that ellipticals and compacts tend to have the highest concentrations, while the majority of irregular galaxies have $C \lesssim 0.25$. There is, however, a large overlap between the regions occupied by different types, hence it is impossible to distinguish an elliptical or irregular from a disk based solely on its value of \emph{C}.
Asymmetry cannot be used alone to distinguish the morphological types, although we note that a tail in the \emph{A} distribution towards higher values is populated only by irregular galaxies.  

Comparing our results with other high-redshift studies, in the rest-frame optical regime, we found general qualitative agreement \citep{conselice2011, law2012a}, although in the \citet{conselice2011} sample ellipticals show, on average, lower asymmetries than irregular galaxies.  
Our results were also compared with studies at lower redshift \citep{bershady2000, conselice2003, conselice2011, scarlata2007}. The greatest difference was found in terms of asymmetry. At low-z, the \emph{A} parameter is as effective as the concentration in distinguishing morphologies, at odds with what is seen at high-z. In addition, all the morphological types at high-z have a higher average level of asymmetry than their local siblings. This is particularly true for elliptical galaxies, whose mean asymmetry at $z \sim 2$ measured in this work is ${<}A{>} \sim 0.2$, about one dex higher than measured at lower redshifts.
\\

\begin{figure}[t!]
	\centering
	\includegraphics[scale=0.45]{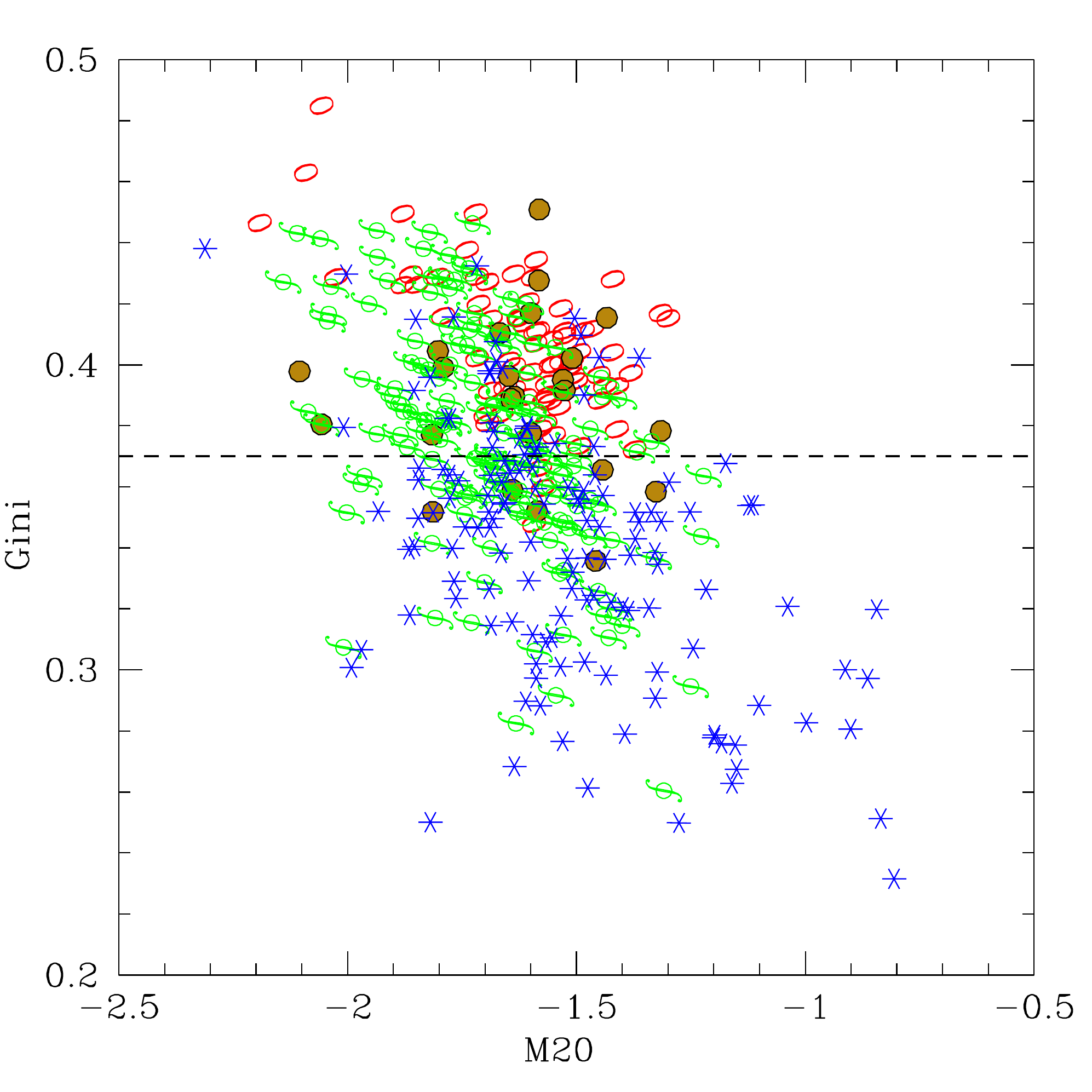}
	\caption{M$_{20}$ vs. Gini coefficient. The dashed line separates the regions typically populated by ellipticals and irregulars. Points are shape and colour coded according to the morphological visual class (see Fig. \ref{viscl_fct_z}).}
	\label{gvsm20_plot}
	\end{figure}

{\bf The \emph{M$_{20}$} vs. \emph{G} plane} (Fig. \ref{gvsm20_plot}) shows that, at $z \geq 1$, ellipticals and compacts have $G \gtrsim 3.7$, while most of the irregulars are characterized by smaller values $G \lesssim 3.7$. However, as found in the \emph{C vs. A} plane, the region of the plane occupied by disks overlaps with both of the other two types.
There is no clear separation of different morphologies in terms of $M_{20}$, the majority of galaxies having $M_{20}$ values in a short range between $\sim -1.8$ and $\sim -1.4$. Beyond these boundaries, however, we note that the tail of the $M_{20}$ distribution towards higher values is populated only by irregular galaxies, while ellipticals and disks are found in the opposite tail. 
The $M_{20}$ distribution in our sample is in close agreement with other high redshift studies \citep{law2012a, wang2012}. 
We found more discrepancies, however, when looking at the Gini parameter. The sample of SFGs studied by \citet{law2012a} have a similar range of Gini values to the galaxies analysed here, although galaxies cannot be distinguished based on \emph{G}. On the other hand, \citet{wang2012} report much higher values for their IERO sample ($0.3 < G < 0.8$), but they observe a clear separation between spheroids and irregular galaxies. 
The discrepancies between the Gini values reported in various works likely depend on the different selection criteria adopted. This suggests that \emph{G} is strongly related to the physical properties of galaxies, such as their stellar mass or star-formation status.
The overall appearance of the \emph{G} vs. $M_{20}$ plane does not change much with decreasing redshift. From a quantitative point of view, however, some differences emerge. In particular, low-z ellipticals have far more negative $M_{20}$ values ($M_{20} < -2$; \citet{lotz2004}), though some discrepancies exist across the literature (see, for example, \citet{conselice2008}). 
\\

	\begin{table*}[t!]
	\caption[]{Visual morphological classification in WFC3-IR $H_{160}$ vs. ACS$_{deg}$ $z_{850}$ images. In parentheses the total number of galaxies belonging to each morphological type, in each band, is reported. Groups of galaxies with the same morphological classification at both wavelengths are marked in bold-face.}
	\label{vis_class_k_corr}
	\centering                          
	\begin{tabular}{l l|c c c c c}        
	\hline\hline  
		  &           		&          	   &           	   & $z_{850}$		&           		&\\	
		  &           		& Ellipticals (59) & Compact (22)  & Disk-like (117)  	& Irregulars (208)  	& Faint objects (88)\\		
	\hline     	
		  & Ellipticals (75)  	& {\bf 53} 	   &   6       	   &   5        	&   8       		&   3\\
		  & Compact (36)  	&  3       	   & {\bf 16}  	   &   1        	&   7       		&   9\\
	$H_{160}$ & Disk-like (157) 	&  0       	   &   0       	   & {\bf 104}  	&  49       		&   4\\
		  & Irregulars (199)  	&  3      	   &   0       	   &   7        	& {\bf 138} 		&  51\\
		  & Faint objects (27) 	&  0       	   &   0       	   &   0        	&   6       		& {\bf 21}\\     
  	\hline\hline
	\end{tabular}
	\end{table*}

From these aforementioned results, we can conclude that, at $1 < z < 3$, quantitative morphological parameters cannot be used alone to effectively characterize the galaxy morphologies. Even if \emph{G} and \emph{C} have proven to be good in separating ellipticals from irregulars, it is impossible to distinguish these two morphological types from disks based solely on the position of a galaxy in the two above-mentioned planes. Additional galaxy-structure measurements are required. Nevertheless, it is important to study whether and how morphological parameters are related to the physical properties of galaxies, since at $z \gtrsim 3$ it becomes difficult to apply the Hubble classification, and structural parameters are the only way to characterize galaxy morphologies. 

We therefore searched for possible correlations between morphological parameters and the specific SFR (sSFR) of galaxies, which are independent of the morphological Hubble class. Only a mild trend was found, where galaxies with $log(sSFR){\gtrsim}-1$ [$Gyr^{-1}$] span the entire range of all parameters, while galaxies with $log(sSFR){\lesssim}-1$ [$Gyr^{-1}$] only have the highest values of Gini and concentration parameters, and the lowest values of $M_{20}$ ($G \gtrsim 0.35$, $C \gtrsim 0.25$, and $M_{20} \lesssim -1.5$).

\section{Morphological $k$-correction}\label{sec:Morphological k-correction}
Before analysing in greater detail the correlations between the morphological and physical properties of galaxies, we consider how the morphological features depend on the rest-frame band in which the galaxies are observed. 

To check whether a morphological $k$-correction should be applied to our sample, the images of galaxies acquired in WFC3-H$_{160}$ band were compared with those in ACS imaging in the $z_{850}$ band \citep{giavalisco2004}, which probes the rest frame from B-band ($\lambda \sim 4300\AA$ at $z\sim1$) up to the UV regime ($\lambda\sim 2000\AA$ at $z\sim3$) for galaxies in our chosen redshift range.
Owing to the differences in resolution and PSF, ACS images were first degraded to match the pixel size and PSF of those obtained with the WFC3. In the resulting images (ACS$_{deg}$), a complete morphological analysis was performed, similar to that outlined in the previous sections.

\subsection{Visual appearance}\label{sec:Visual appearance}
\begin{figure}[h!]
	\centering
	\includegraphics[trim=100 40 75 390, clip=true, width=37mm]{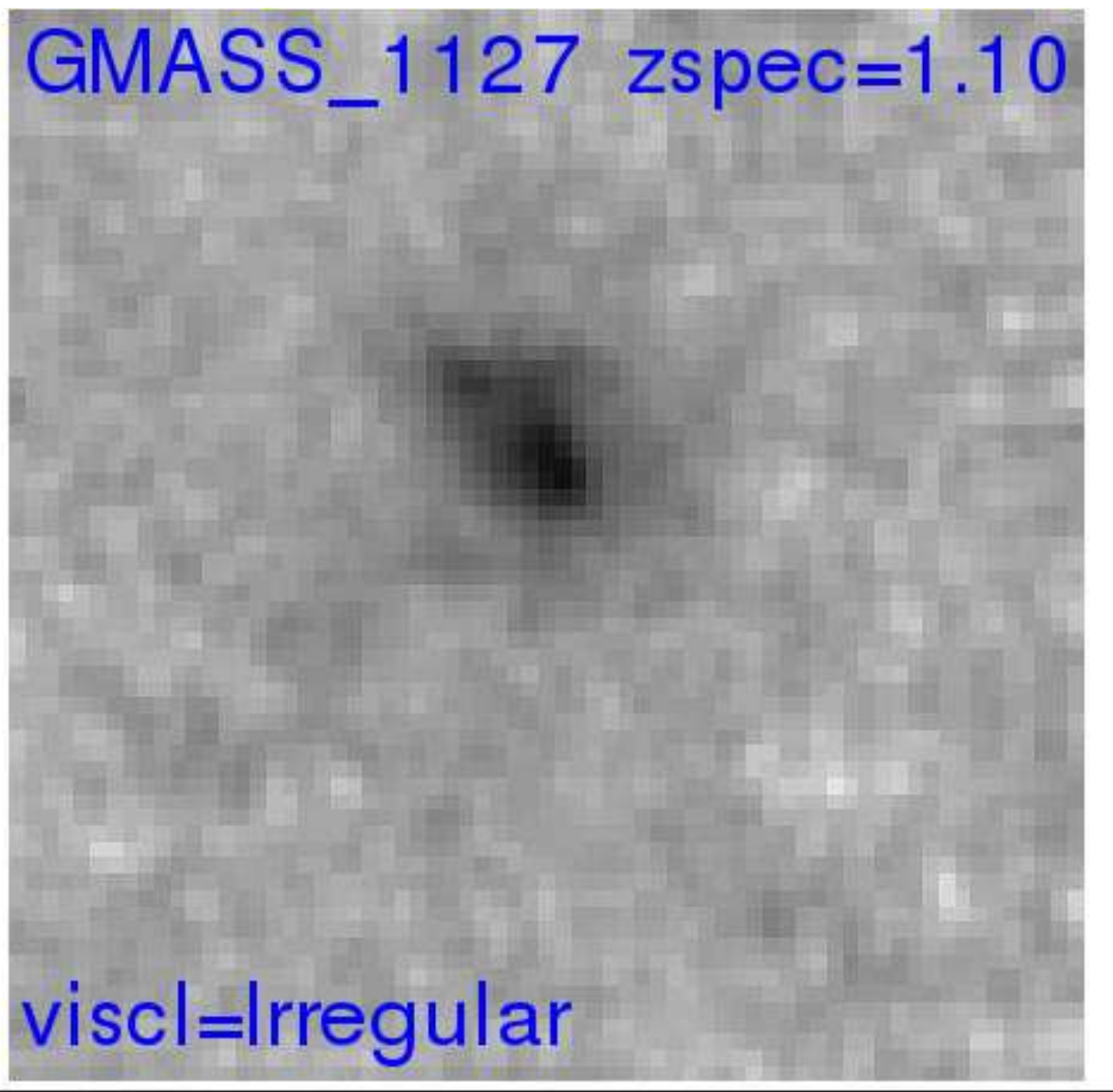}
	\includegraphics[trim=100 40 75 390, clip=true, width=37mm]{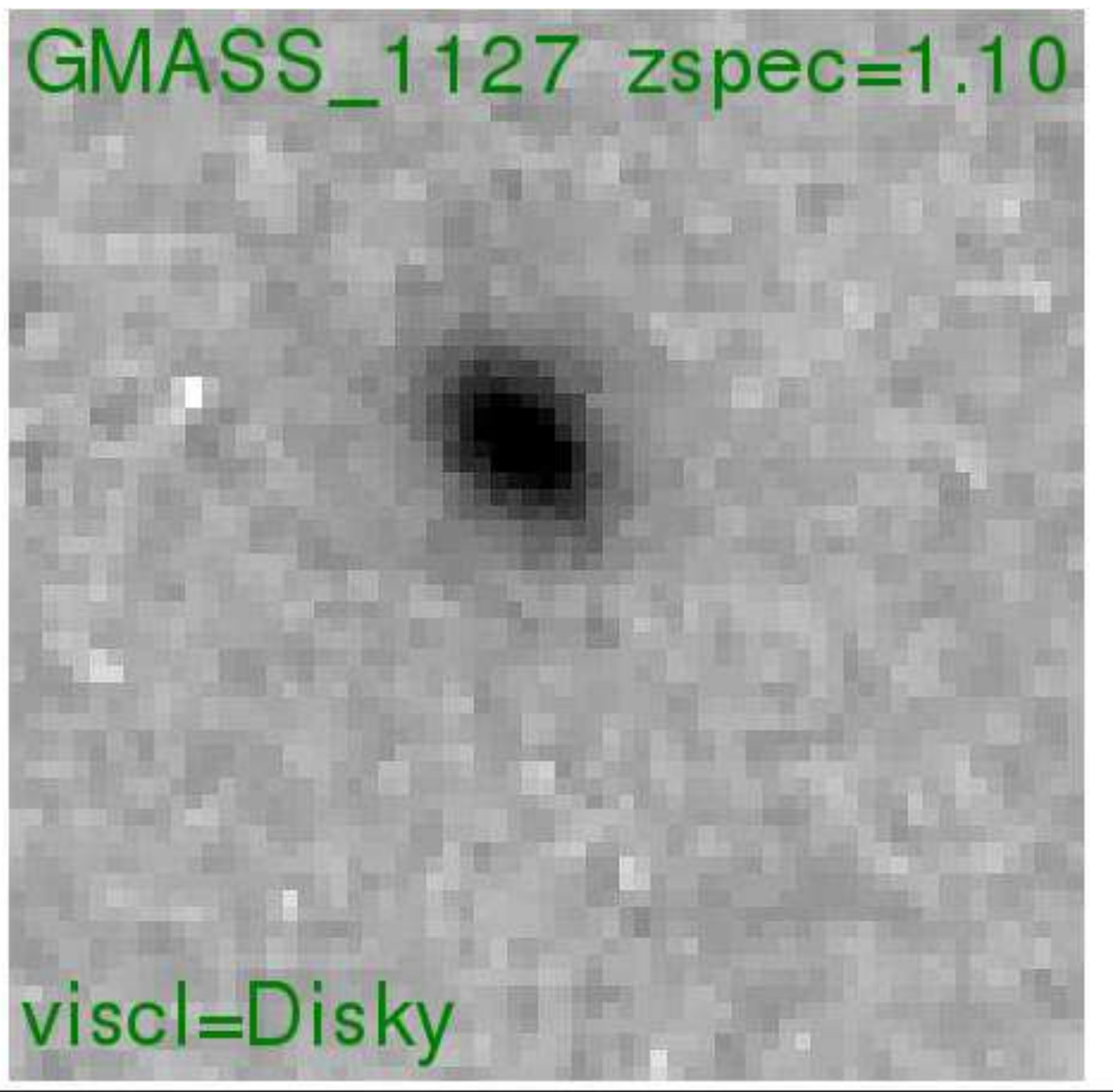}
	\includegraphics[trim=100 40 75 390, clip=true, width=37mm]{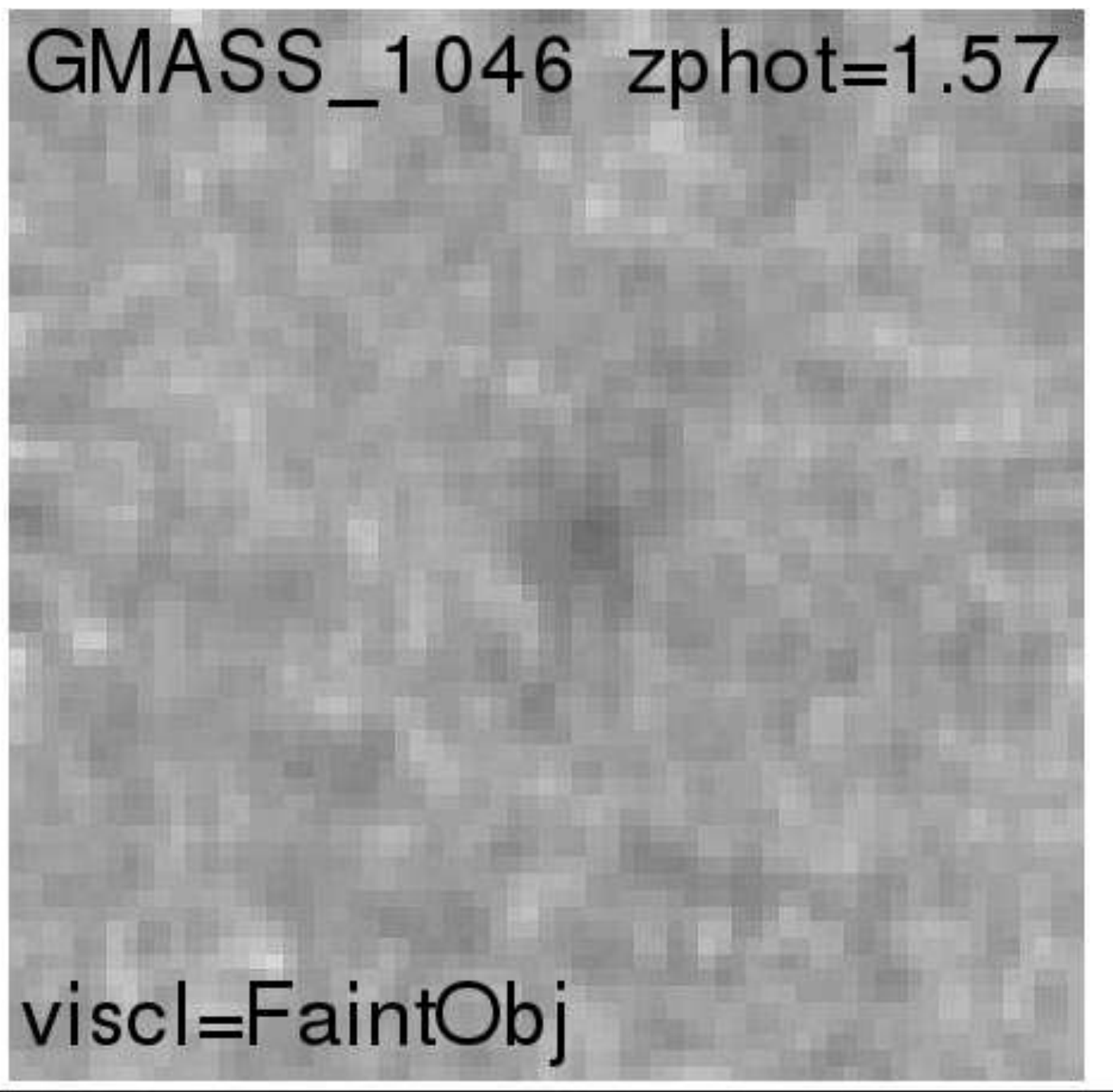}
	\includegraphics[trim=100 40 75 390, clip=true, width=37mm]{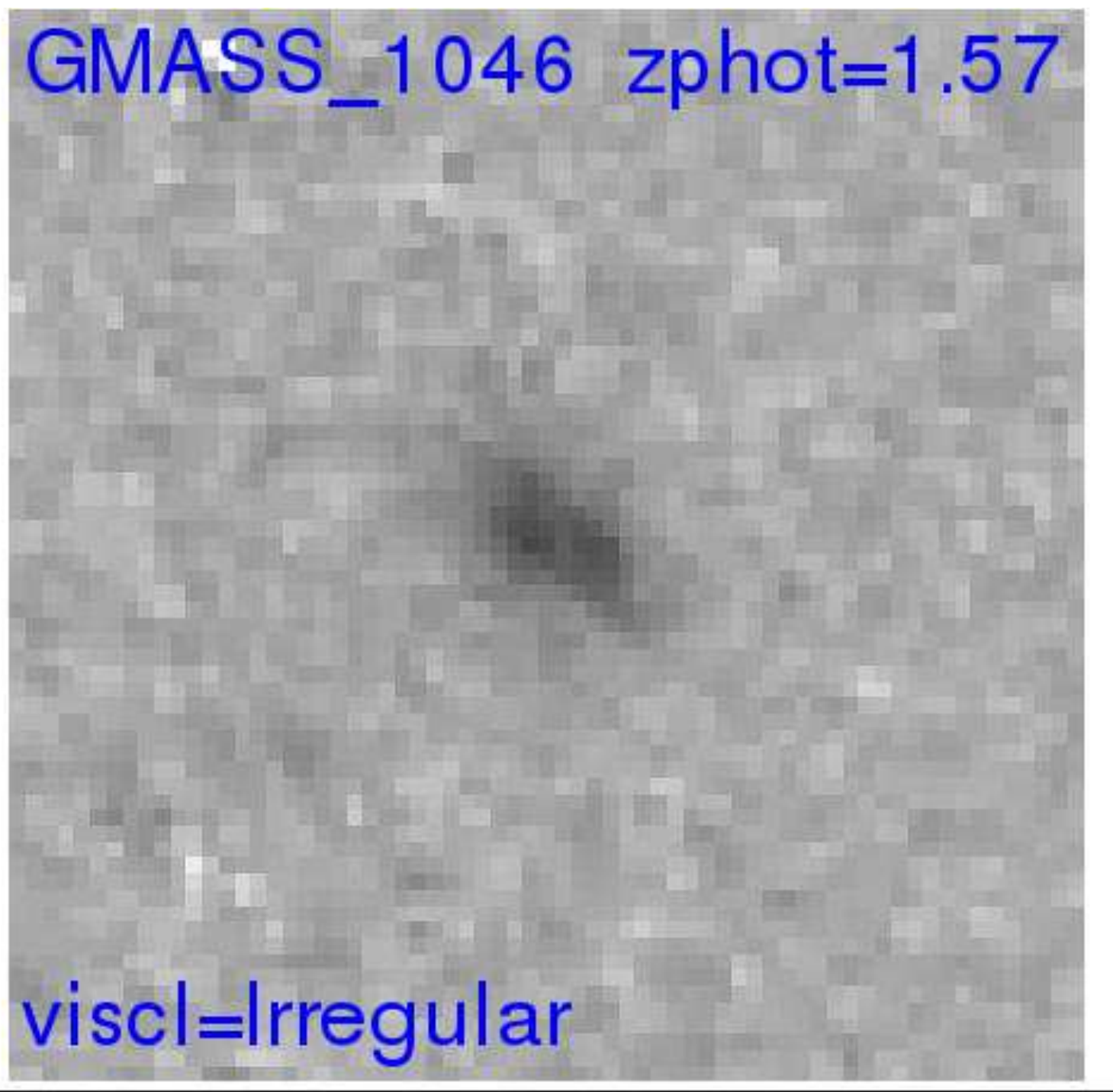}
	\caption{Two examples of galaxies for which a morphological $k$-correction is required between the optical and UV band. For each galaxy, two images are shown: WFC3-H$_{160}$ (right) and ACS$_{deg}$-$z_{850}$ (left). In each snapshot, galaxy ID, redshift and classification are reported.}
	\label{strong_kcorr}
	\end{figure}

The comparison between morphological classifications in the ACS-z$_{850}$ and WFC3-H$_{160}$ is reported in Table \ref{vis_class_k_corr}. We find that the majority of galaxies have the same visual morphology at both wavelengths. Among the galaxies with different morphological classifications, 49 galaxies classified as disks in H$_{160}$ images were re-classified as irregulars in the $z_{850}$ band. We compare the two images for two example galaxies in Fig. \ref{strong_kcorr}. The difference in morphology between the two wavelength bands may reflect some differences in the distributions of the various stellar populations that inhabit the galaxy.  

We found that 51 objects with irregular morphologies in the H$_{160}$  band, are almost undetectable in the $z_{850}$ imaging (see Fig. \ref{strong_kcorr}). In this case, we cannot strictly speak of ``morphological \emph{k}-correction'', since the faintness of these galaxies at shorter wavelengths makes it impossible to classify them. We note that these objects have ongoing star-formation and intermediate-to-high stellar masses relative to the total sample, i.e. $log({<}SFR{>}) = 1.59 \pm 0.67$ [$M_{\odot}yr^{-1}$] and $log{<}M_{\star}{>} = 10.48 \pm 0.51$ [$M/M_{\odot}$]. We assume that they could be dusty star-forming galaxies in which the UV light from young stars is almost entirely absorbed. 

\subsection{Quantitative changes}\label{sec: Quantitative changes}
\begin{figure}[h!]
	\centering
	\includegraphics[scale=0.45]{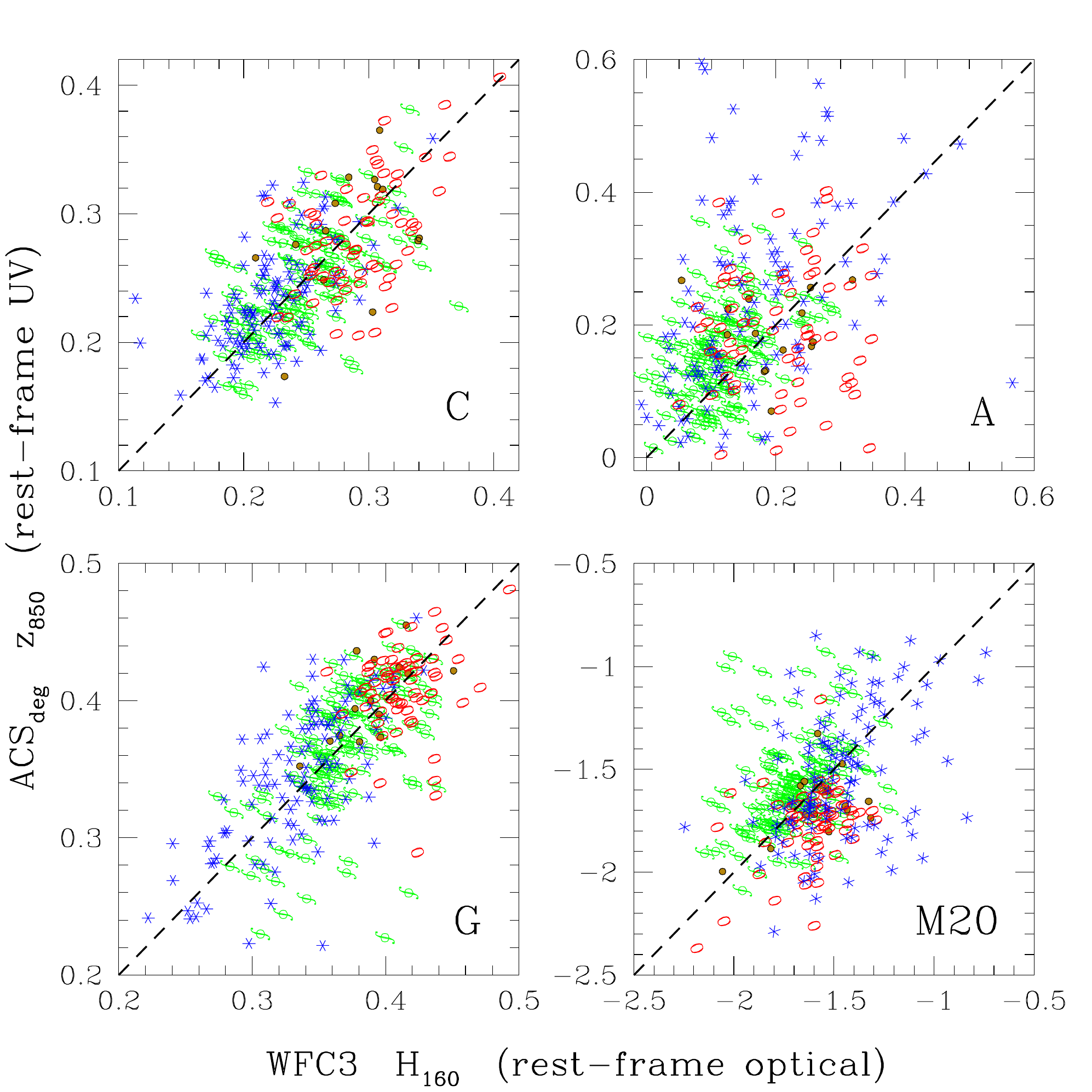}
	\caption{Comparison between morphological parameters computed in the WFC3-H$_{160}$ and ACS$_{deg}$-$z_{850}$ images. From top left, clockwise, concentration, asymmetry, $M_{20}$, and the Gini coefficient. Points are shape and colour coded with respect to the morphological classification in the H$_{160}$ band (see Fig. \ref{viscl_fct_z}).}
	\label{k_corr}
	\end{figure}

In our visual Hubble classification, we find that, for galaxies at $z > 1$, no significant morphological $k$-correction is required. This finding is however challenged by our quantitative analysis. 
In Fig. \ref{k_corr}, we compare the morphological parameters computed in the WFC3-H$_{160}$ and ACS$_{deg}$-$z_{850}$ images. For all morphological types, we detect no difference between the concentrations and the Gini coefficients measured for the two different filters. 
The largest differences are found for the asymmetry and $M_{20}$ parameters, where disks in particular have higher values of \emph{A} and $M_{20}$ for the shorter-wavelength filter. This confirms our previous results for high-redshift disk galaxies, where young stars have a more irregular spatial distribution than the older stellar populations generating the light at optical wavelengths. Irregulars are more asymmetric at shorter wavelengths, as expected because they are actively star-forming galaxies \citep{conselice2011}, but for $M_{20}$ there are no significant differences, on average, between the two wavelength ranges. Compact and elliptical galaxies have, on average, less negative $M_{20}$ values (hence more irregular morphologies) at longer rest-frame wavelengths. 
	\begin{table*}[t!]
	\caption[]{Median and semi-interquartile range (in parenthesis) of \emph{C}, \emph{A}, \emph{G}, $M_{20}$, stellar mass, and SFR for blue ($(U{-}B)_{rest}<1$) and red ($(U{-}B)_{rest}>1$) galaxies, in each morphological class. The percentage of galaxies belonging to each colour group, for each morphological type, is also provided.}
	\label{blue_vs_red}
	\centering                          
	\begin{tabular}{l|c c|c c|c c|c c}        
	\hline\hline  
			\multicolumn{9}{c}{Total \emph{GMASS-WFC3 sample}}\\
	\hline
	                & \multicolumn{2}{c|}{Elliptical} & \multicolumn{2}{c|}{Compact} & \multicolumn{2}{c|}{Disk} & \multicolumn{2}{c}{Irregular}\\
			& Red ($60\%$) & Blue ($40\%$) & Red ($25\%$) & Blue ($75\%$) & Red ($22\%$) & Blue ($78\%$) & Red ($15\%$) & Blue ($85\%$)\\
	\hline				
	Mass $log(M/M_{\odot})$	     & 10.25 (0.20) & 9.71 (0.34) & 10.46 (0.11) & 9.82 (0.25) & 10.40 (0.37) & 9.66 (0.27) & 10.81 (0.43) & 9.78 (0.31)\\
	SFR $log(M_{\odot}yr^{-1})$  & -0.84 (0.58) & 1.29 (0.37) &  1.07 (0.30) & 1.17 (0.33) &  0.50 (1.05) & 1.25 (0.36) &  0.90 (0.98) & 1.46 (0.34)\\
	\hline\hline  
			\multicolumn{9}{c}{Galaxies with image $S/N > 50$}\\
	\hline
	                & \multicolumn{2}{c|}{Elliptical} & \multicolumn{2}{c|}{Compact} & \multicolumn{2}{c|}{Disk} & \multicolumn{2}{c}{Irregular}\\
			& Red ($60\%$) & Blue ($40\%$) & Red ($25\%$) & Blue ($75\%$) & Red ($22\%$) & Blue ($78\%$) & Red ($10\%$) & Blue ($90\%$)\\
	\hline    	
	concentration   	     & 0.29 (0.03) & 0.28 (0.02) & 0.29 (0.04) & 0.28 (0.02) & 0.27 (0.02) & 0.25 (0.03) & 0.23 (0.02) & 0.22 (0.02)\\
	asymmetry  		     & 0.22 (0.06) & 0.12 (0.05) & 0.17 (0.07) & 0.18 (0.05) & 0.13 (0.04) & 0.11 (0.03) & 0.10 (0.05) & 0.15 (0.06)\\
	Gini  			     & 0.41 (0.01) & 0.40 (0.01) & 0.40 (0.02) & 0.39 (0.01) & 0.38 (0.02) & 0.37 (0.02) & 0.34 (0.03) & 0.34 (0.02)\\
	M20   			 & -1.60 (0.06) & -1.63 (0.06) & -1.64 (0.11) & -1.59 (0.09) & -1.75 (0.12) & -1.66 (0.12) & -1.56 (0.17) & -1.55 (0.15)\\
	\hline\hline
	\end{tabular}
	\end{table*}

\section{Colour bimodality}\label{sec: Colour bimodality}
It has been well-established by several studies that the galaxy rest-frame \emph{(U-B)} colour has a bimodal distribution \citep{strateva2001, hogg2002, cassata2007, cassata2008, bell2004, weiner2005, whitaker2011, bell2012}. 

Here, we determine the distribution of the different morphological types in the colour bimodality, to study the correlation between the morphologies and physical properties of high-redshift galaxies. 
	\begin{figure}[b!]
	\centering
	\includegraphics[scale=0.45]{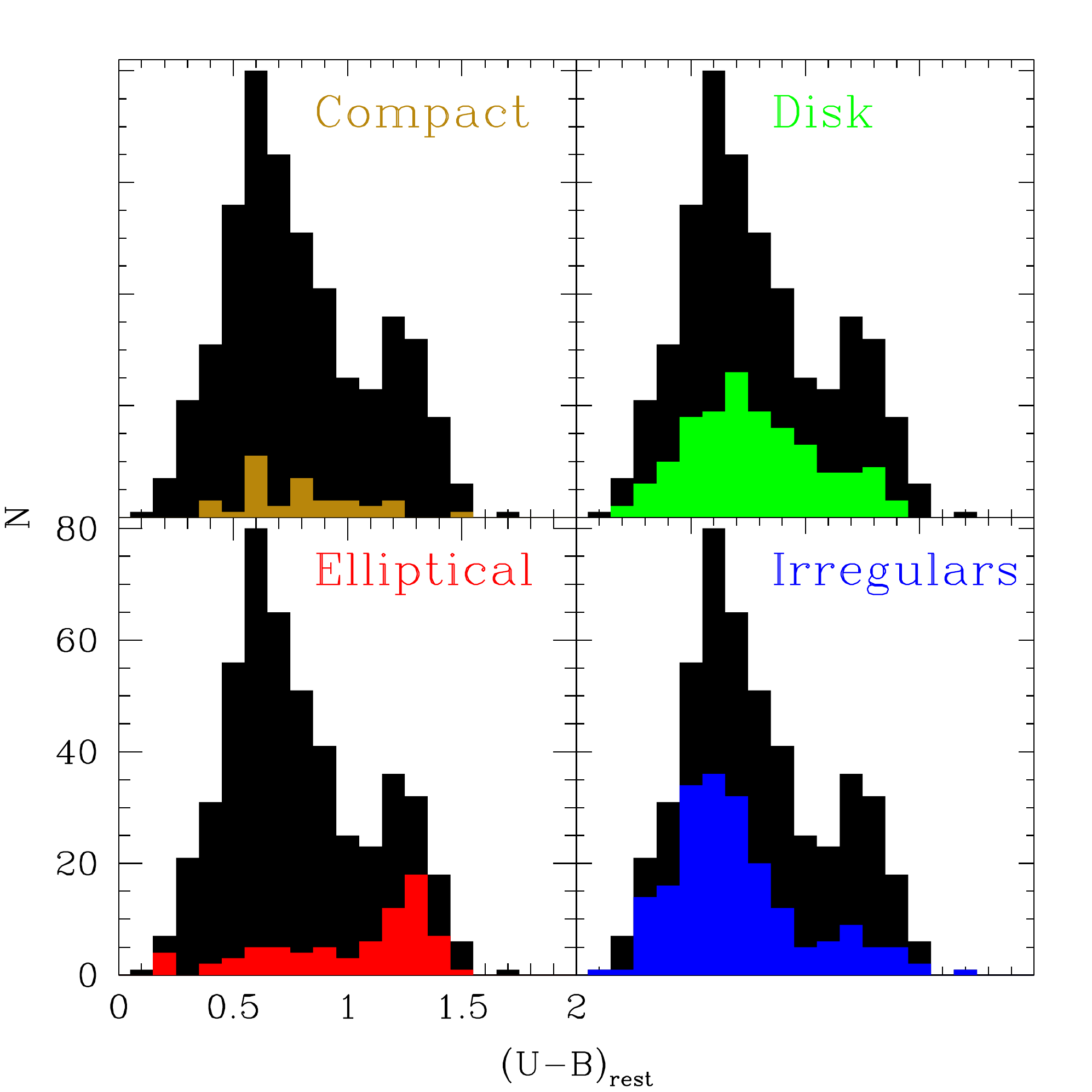}	
	\caption{Rest-frame (U-B) colour distributions of each morphological visual class (coloured histograms) vs. that of the full \emph{GMASS-WFC3 sample} (black histogram).}
	\label{col_hist}
	\end{figure}

Figure \ref{col_hist} compares the $(U{-}B)_{rest}$ colour\footnote[6]{The absolute magnitude in a given filter X was computed using the observed apparent magnitude in the filter Y, which had been chosen to be the closest to $\lambda(Y) \sim \lambda(X) * (1+z)$.} distribution for the \emph{GMASS-WFC3 sample} with the distribution for each morphological type. The WFC3 sub-sample has a similar colour bimodality to that found for the GMASS total sample by \citet{cassata2008}. 
We divided the the colour distribution into four redshift bins. We found that the bimodality can be clearly seen up to redshifts as high as $2.0 < z < 2.5$, whereas at higher redshift the colour distribution is far smoother because a larger fraction of galaxies populates the region between the red and the blue peaks. The bin in-between the red and blue peaks containing the least galaxies, $(U{-}B)_{rest}{\sim}1$, was chosen as the separation value between the red sequence and the blue cloud: this is consistent with threshold values reported in the literature \citep{cassata2008, kriek2009}.

\clearpage
	\begin{figure*}[t!]
	\centering
	\includegraphics[trim=20 0 0 150, clip=true, scale=0.85]{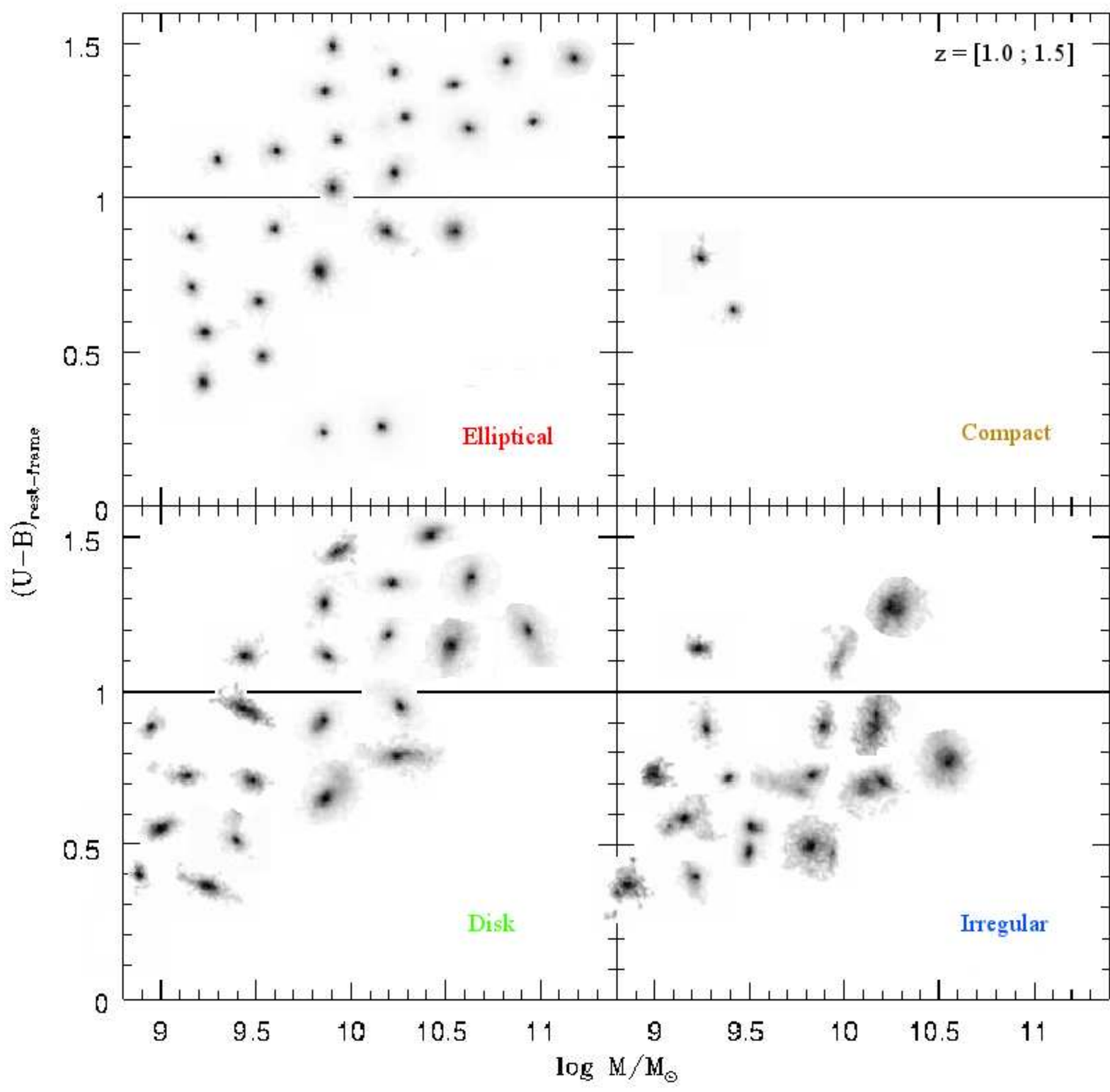}
	\caption{Rest-frame colour vs. stellar mass for each morphological type in the low-redshift bin $1.0 < z < 1.5$. Instead of symbols, we plot randomly selected representative H$_{160}$ cutouts from our sample.}
	\label{snap_1}
	\end{figure*}
\clearpage
	\begin{figure*}[t!]
	\centering
	\includegraphics[trim=20 0 0 150, clip=true, scale=0.85]{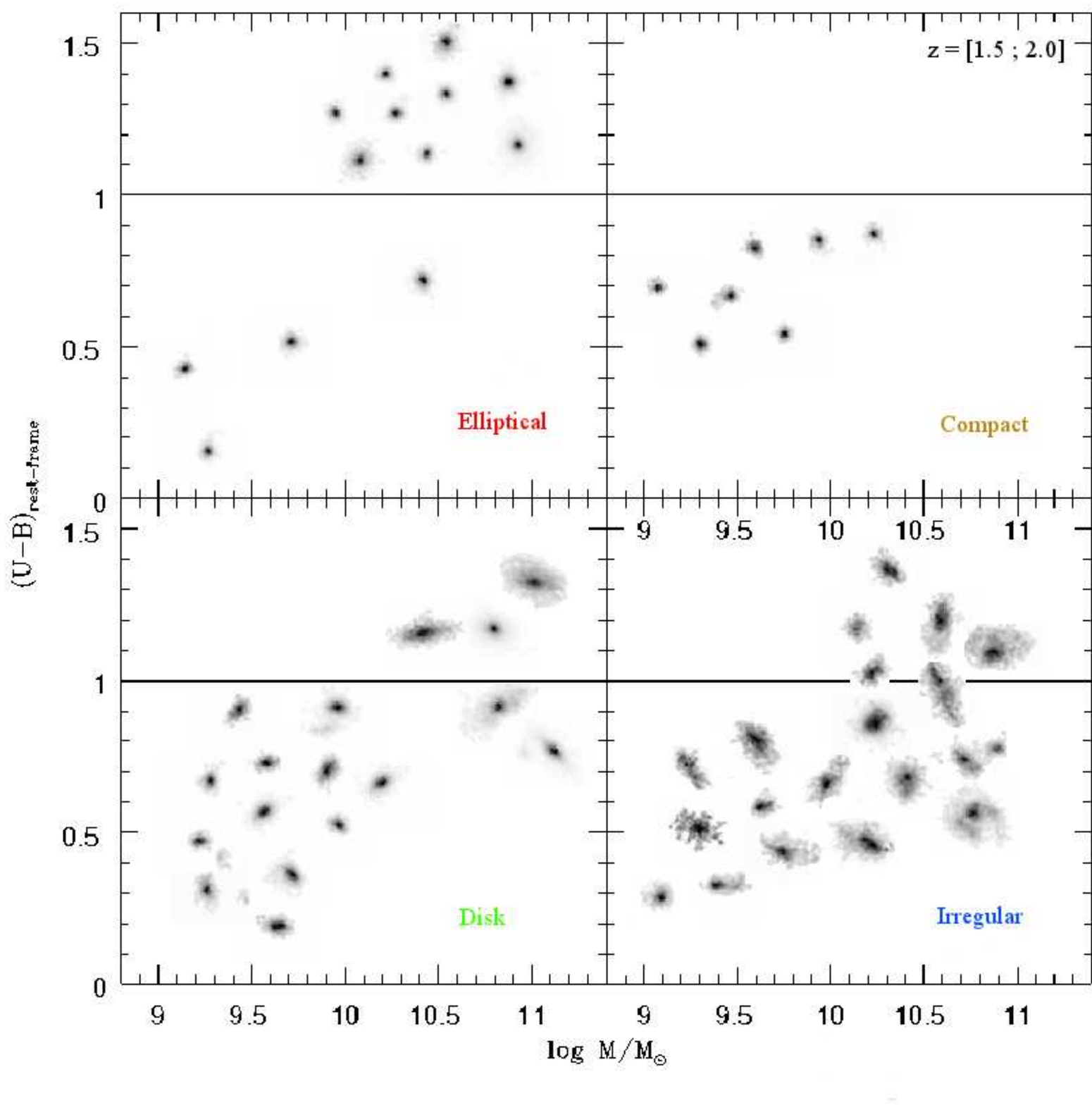}
	\caption{Rest-frame colour vs. stellar mass for each morphological type in the intermediate-redshift bin $1.5 < z < 2.0$. Instead of symbols, we plot randomly selected representative H$_{160}$ cutouts from our sample.}
	\label{snap_2}
	\end{figure*}
\clearpage
	\begin{figure*}[t!]
	\centering
	\includegraphics[trim=20 0 0 150, clip=true, scale=0.85]{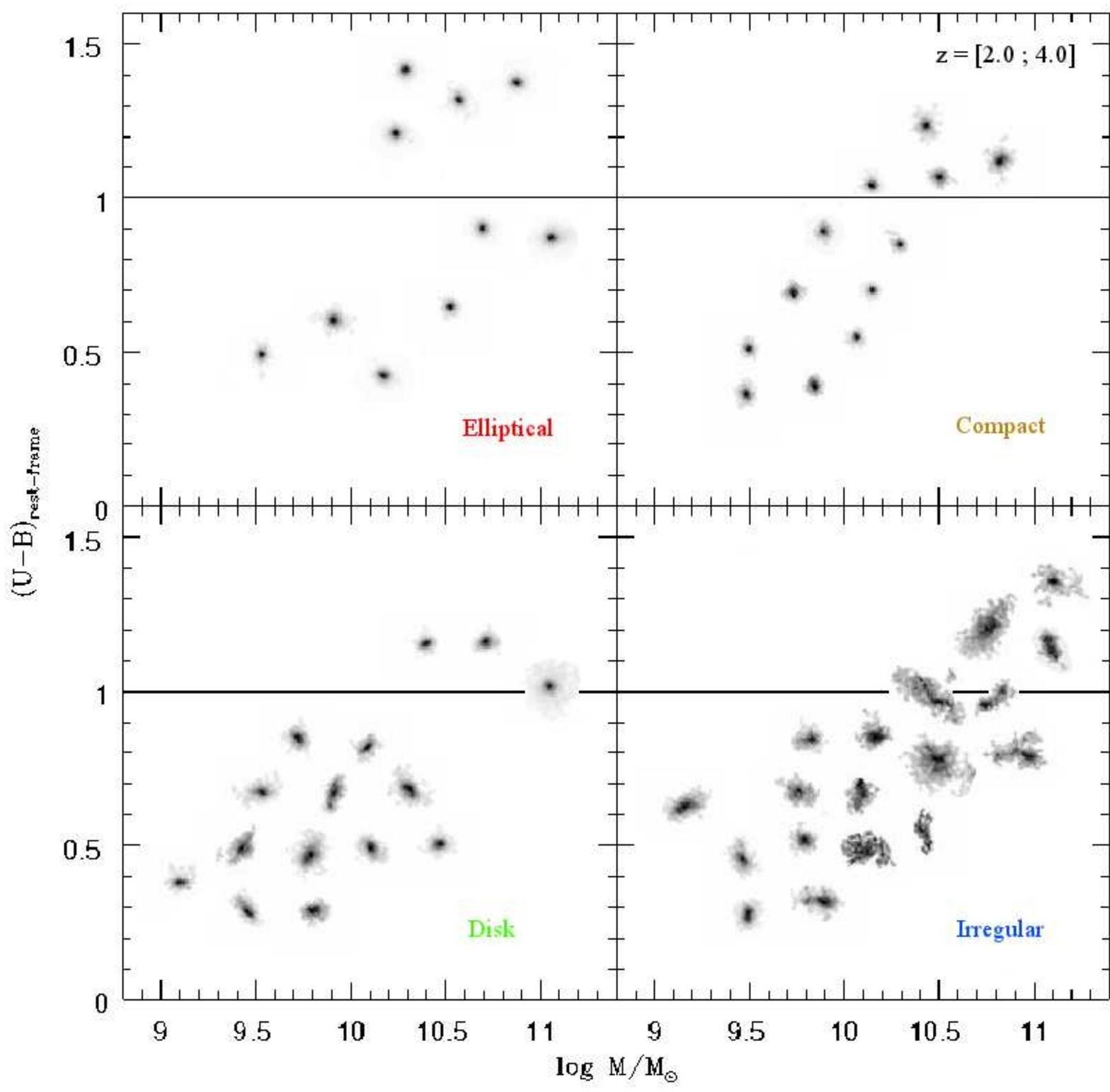}
	\caption{Rest-frame colour vs. stellar mass for each morphological type in the high-redshift bin $2.0 < z < 4.0$. Instead of symbols, we plot randomly selected representative H$_{160}$ cutouts from our sample.}
	\label{snap_3}
	\end{figure*}
\clearpage

The colour of a galaxy clearly correlates with its morphological type: we find that (i) most irregulars (85$\%$) occupy the blue peak (${<}(U{-}B)_{bluepeak}{>}{=}0.59$); (ii) the red peak is mainly populated with ellipticals (60$\%$, with ${<}(U{-}B)_{redpeak}{>}{=}1.26$); (iii) disks have a smoother colour distribution than the other types and are the main type occupying the region between the blue and red peaks.
In Table \ref{blue_vs_red}, we indicate the median of the distribution of some physical and morphological properties, for both red and blue galaxies, and each morphological type. The median values of stellar mass and SFR correspond to the entire sample, whereas the median values of the morphological parameters are only for images with $S/N > 50$. We point out that the percentages of blue and red objects in the total sample are insensitive to the S/N cut. 

There is a strong correlation between the $(U{-}B)_{rest}$ colour of a galaxy and its stellar mass. This result is confirmed by the diagrams shown in Fig. \ref{snap_1}, \ref{snap_2}, and \ref{snap_3} where red galaxies are more massive than blue ones, regardless of their morphological type. 

In terms of star-formation activity, there is little difference between red and blue objects for all morphological classes apart from ellipticals. Almost all galaxies classified as compact, disks, and irregulars have ongoing star-formation, regardless of their colour. Blue galaxies classified as ellipticals are also star-forming, but red ellipticals, in addition to a small fraction of red disks, display no signs of star-formation. This confirms a previously presented result in this work that quiescent galaxies generally have an elliptical morphology (Sect. \ref{sec:Visual morphological analysis}; see also Sect. \ref{sec:The UVJ diagram}). While the red colours of elliptical depend mainly on the old age of their stellar populations, red galaxies belonging to the other morphological classes are in most cases star-forming objects obscured by dust that reddens their colours \citep{scarlata2007, franzetti2007}. The physical properties of \emph{red-sequence} galaxies in the GMASS total sample have already been extensively studied in \citet{cassata2008}. 

No correlation is found between colour and the quantitative morphological parameters, the parameter distributions of red and blue galaxies being quite consistent for morphological Hubble class. In contrast, the value of $M_{20}$ is lower for red disk galaxies than blue ones. The most remarkable variation is that of the asymmetry of elliptical galaxies, with blue ellipticals appearing less asymmetric than red ones.

\subsection{The UVJ diagram}\label{sec:The UVJ diagram}
\begin{figure}[h!]
	\centering
	\includegraphics[scale=0.45]{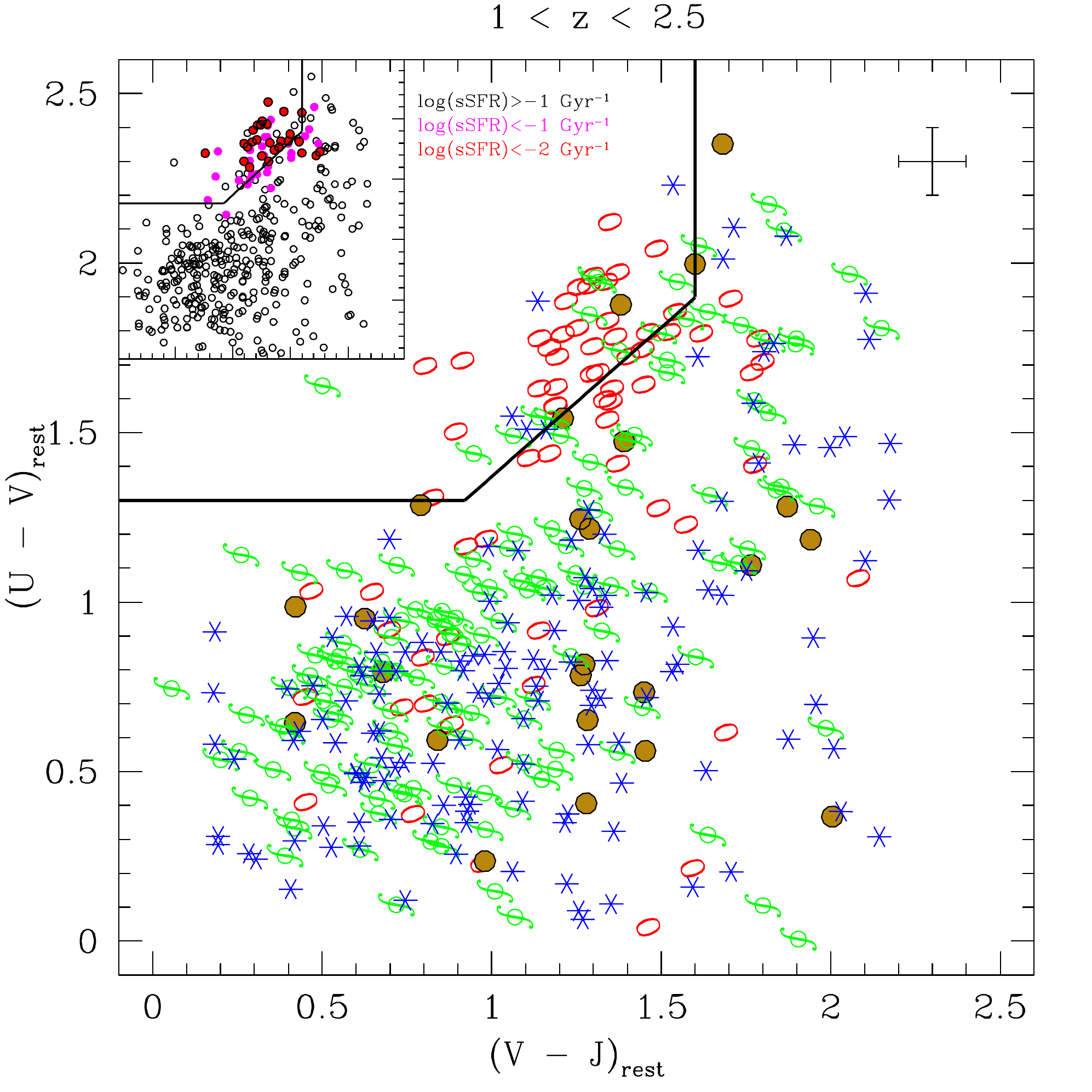}	
	\caption{Rest-frame \emph{U-V} vs. \emph{V-J} colours for galaxies at $1<z<2.5$ in our sample. In the big plot, points are shape and colour coded with respect to the morphological visual class (see Fig. \ref{viscl_fct_z}). In the small insert, points are colour coded with respect to their sSFR. In both plots, the ``quiescent'' region, defined by \citet{williams2009} for the chosen redshift range, is also marked.}
	\label{uvj_plot}
	\end{figure}

The galaxies in our sample were also plotted in a rest-frame \emph{U-V} vs. \emph{V-J} colour diagram, which is shown in Fig. \ref{uvj_plot}. 
\citet{williams2009}, and previously \citet{wuyts2007} found that star-forming and quiescent galaxies occupy two distinct regions in the plane. 
The \emph{UVJ} diagram allows us to distinguish red star-forming dust-obscured galaxies from red, quiescent ones, which would be impossible using the \emph{U-B} colour alone. In particular, \citet{williams2009} defines the region where quiescent galaxies fall, in different redshift bins. 
The quiescent region identified at $1<z<2$ is indicated in Fig. \ref{uvj_plot}. We plot only galaxies up to $z \sim 2.5$, from our sample, since this is the limit up to which the bimodality in the \emph{UVJ} diagram is still visible \citep{williams2009}.

Fig. \ref{uvj_plot} shows that the quiescent region of the diagram is mainly populated by ellipticals and, to a lesser extent, by disks, reinforcing the results reported in the previous section. In addition, these galaxies have the highest values of concentration and Gini (see end of Sec. \ref{Exploration of the parameter space}).
The quiescent state of these galaxies is also consistent with their sSFR, $log(sSFR)< -1$ [$Gyr^{-1}$]. In Fig. \ref{uvj_plot}, we also label passive galaxies, i.e. with $log(sSFR)< -2$ [$Gyr^{-1}$], following the definitions given by \citet{ilbert2010} and \citet{pozzetti2010}. We note that, while ``quiescent'' galaxies have either elliptical or disk morphologies, ``passive'' galaxies are mainly elliptical.

\subsection{An in-depth look at ellipticals}\label{sec:An in-depth look at ellipticals}
\begin{figure}[b!]
	\centering
	\includegraphics[scale=0.45]{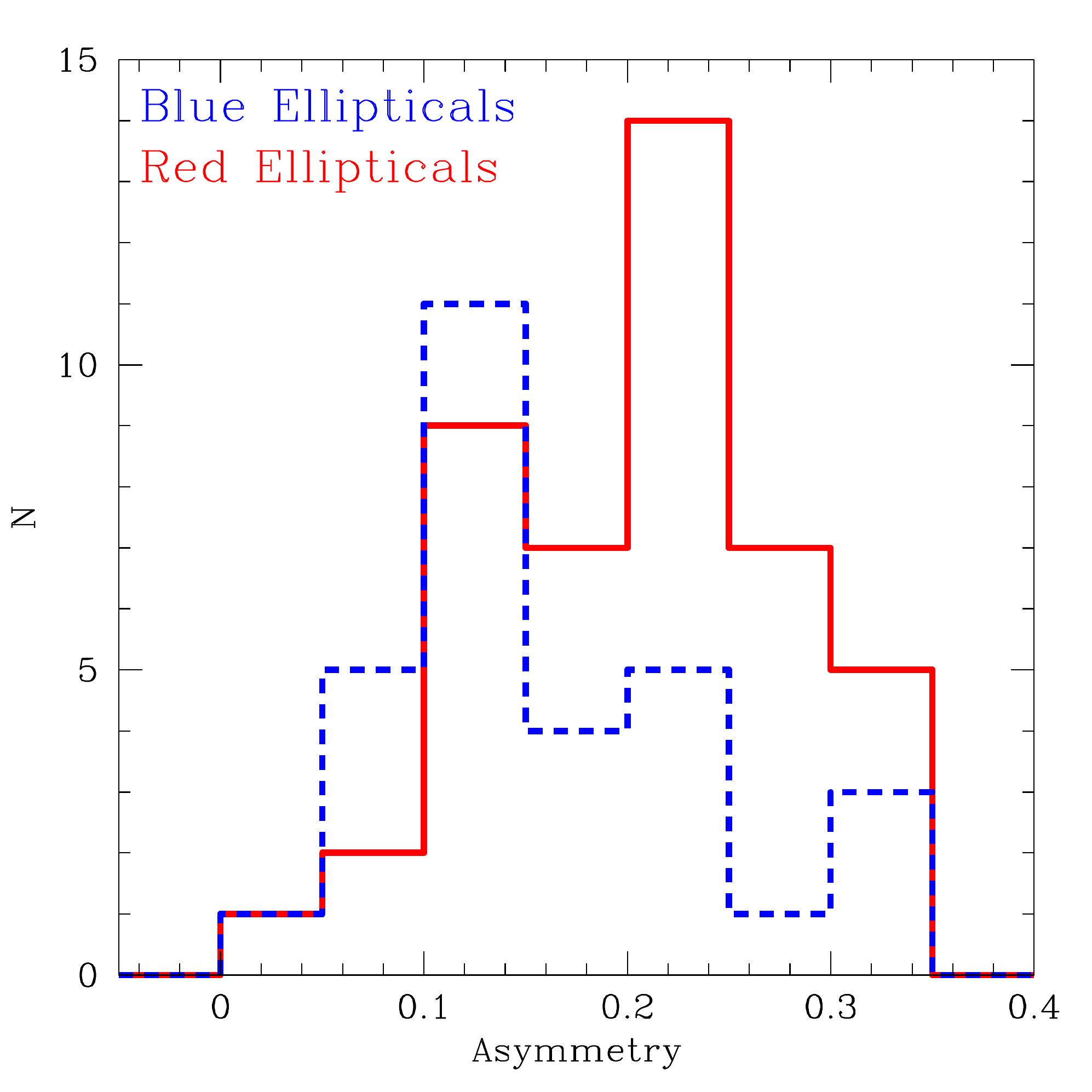}	
	\caption{asymmetry distribution for blue ($(U{-}B)_{rest}{<}1$) and red ($(U{-}B)_{rest}{>}1$) ellipticals.}
	\label{asy_hist}
	\end{figure}

Although most ellipticals have a red rest-frame \emph{(U-B)} colour and occupy the ``quiescent'' region of the \emph{UVJ} diagram, a significant fraction (40$\%$) are found in the blue peak of the \emph{(U-B)} distribution. 
In general, the physical properties of these two groups of galaxies are very different: red ellipticals have high stellar masses and low to null star-formation activity, while blue ones occupy the low-mass end of the stellar mass distribution and have SFRs extending from tens to hundreds $M_{\odot}yr^{-1}$ (see Table \ref{blue_vs_red}). 
Apart from sharing the same Hubble classification, red and blue ellipticals have, however, remarkably similar structure, having similar distributions of all morphological parameters except asymmetry. Figure \ref{asy_hist} shows the distributions asymmetry for both red and blue ellipticals. Although the blue peak is clearly evident at low asymmetries, there is no sharp separation between blue and red ellipticals, because red galaxies span a wide range of asymmetry values. 

Red and blue ellipticals also differ in terms of their S\'ersic index\footnote[7]{The S\'ersic profile: $\Sigma(r) = \Sigma_{e}exp(-\kappa[(r/r_{e})^{1/n}-1])$, where $r_{e}$ is the effective half light radius and \emph{n} is called the S\'ersic index. $n{>}2$ is commonly used as a criterion to select spheroids, while disk-dominated galaxies are characterized by $n{<}2$ \citep{buitrago2008}.} (n). We took S\'ersic index values from the catalogue of structural parameters released by the CANDELS team \citep{vanderwel2012}. We refer the reader to the cited paper for the details about the parameter's computation. We found blue ellipticals to have ${<}n{>}{=}3.1{\pm}1.3$ while red ellipticals show ${<}n{>}{=}4.5{\pm}1.9$. On one hand, this result strengthens the classification of blue ellipticals; on the other hand, it highlights another possible structural difference between blue and red ellipticals. However, the difference between the two groups is not significant, therefore not even the S\'ersic index can be used to separate ``a priori'' red from blue ellipticals.
We conclude that it is impossible to distinguish between red and blue ellipticals on the basis of morphological properties alone.
 
The true importance of searching for structural differences between red and blue ellipticals is the need to establish some suitable morphological criteria for selecting galaxies that are no longer forming stars.
	\begin{figure}[t!]
	\centering
	\includegraphics[scale=0.45]{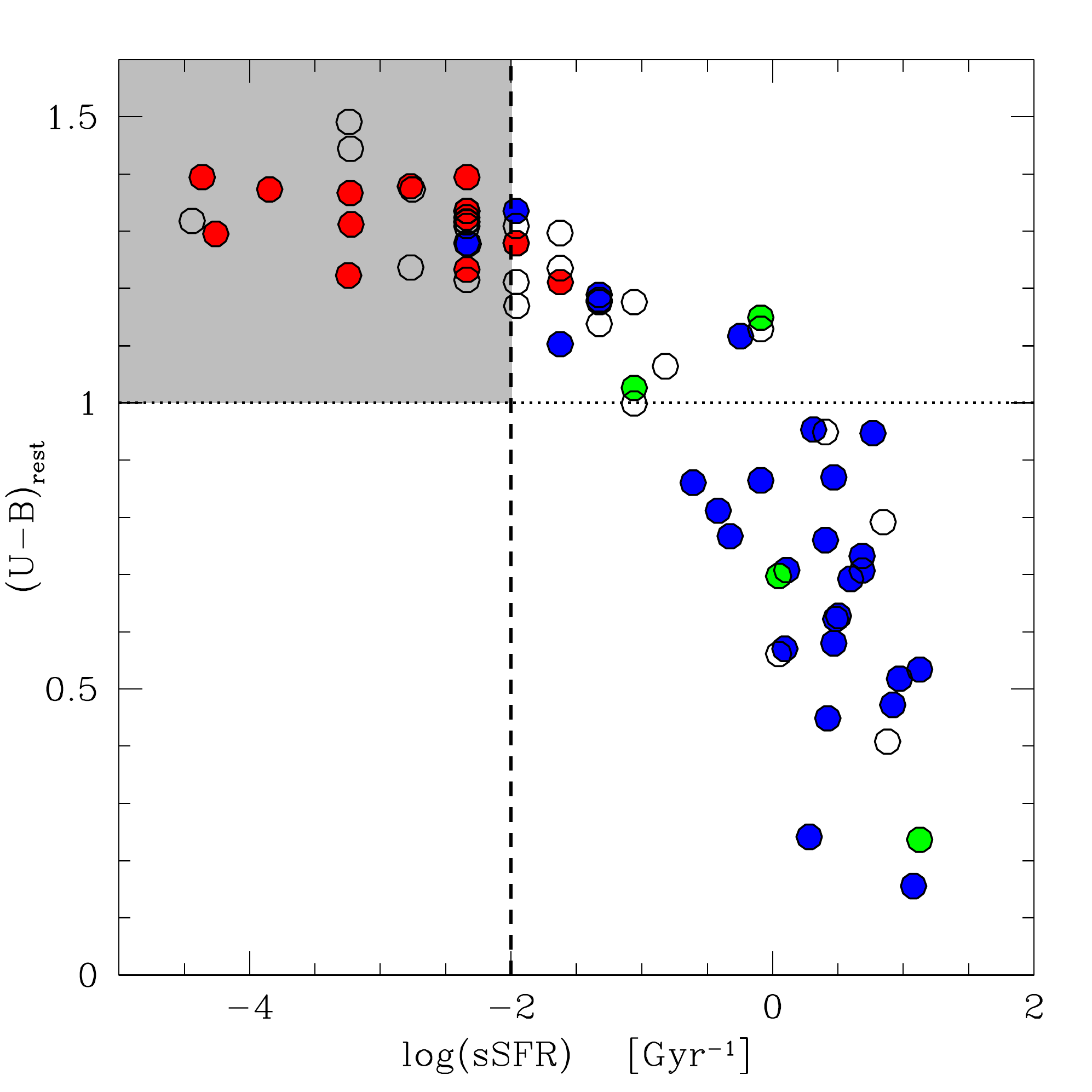}	
	\caption{sSFR vs. $(U{-}B)_{rest}$ colour for morphological ellipticals in the sample. Empty black points represent galaxies with no spectrum. Filled points indicate galaxies with a spectrum, and are colour-coded with respect to the spectral classification: red for \emph{early-types}, blue for \emph{star-forming}, and green for \emph{AGNs}. The vertical line marks the sSFR cut to select passive galaxies. The horizontal line marks the separation between red and blue galaxies. The grey area identifies the locus in the plane where ``red and passive'' galaxies fall.}
	\label{ssfr_vs_col}
	\end{figure}
Although the properties of passively evolving, early-type galaxies are quite established, \citep{renzini2006}, different criteria have been used across the literature to select ``pure'' samples, i.e. those containing the smallest numbers of star-forming contaminants. 
The best way of performing such a selection is, of course, to use a combination of morphological, photometric and spectroscopic data \citep{moresco2010}. 
This type of selection, however, is highly demanding in terms of information used. \citet{moresco2013} demonstrates that a selection based on sSFR is the closer to a combined criterion, in terms of percentage of contaminants. 

In the previous section we have seen that almost all passive galaxies at $z>1$ are ellipticals. We now determine the fraction of passive galaxies among all galaxies with morphological elliptical classifications in our sample. Fig. \ref{ssfr_vs_col} shows the distribution of sSFR vs. $(U{-}B)_{rest}$ colour for all ellipticals in our sample. Only $\sim$33$\%$ of all ellipticals are passive galaxies. It is interesting to note that this percentage is consistent with the result of \citet{moresco2013} for their sample of $z < 1$ galaxies.
The sSFR selection of passive galaxies in our sample is also consistent with the spectroscopic classification. In Fig. \ref{ssfr_vs_col}, galaxies from the spectroscopic sub-sample are indicated by different colours depending on their spectral class: galaxies with an \emph{early-type} spectrum occupy the ``red and passive'' locus of the plane. In the remaining sub-sample, the majority of galaxies have a \emph{star-forming} spectrum, while four have spectra typical of active galactic nuclei (AGNs). 
We refer the reader to \citet{cimatti2008} for a thorough analysis of the spectral properties of early-type galaxies performed on spectra from the GMASS spectroscopic survey. Unfortunately, almost all galaxies analysed in that paper fall outside the CANDELS field and could not be included in this work.

\section{Morphological vs. spectroscopic classification}\label{sec:Morphological vs. spectroscopic classification}
\begin{figure}[b!]
	\centering
	\includegraphics[scale=0.45]{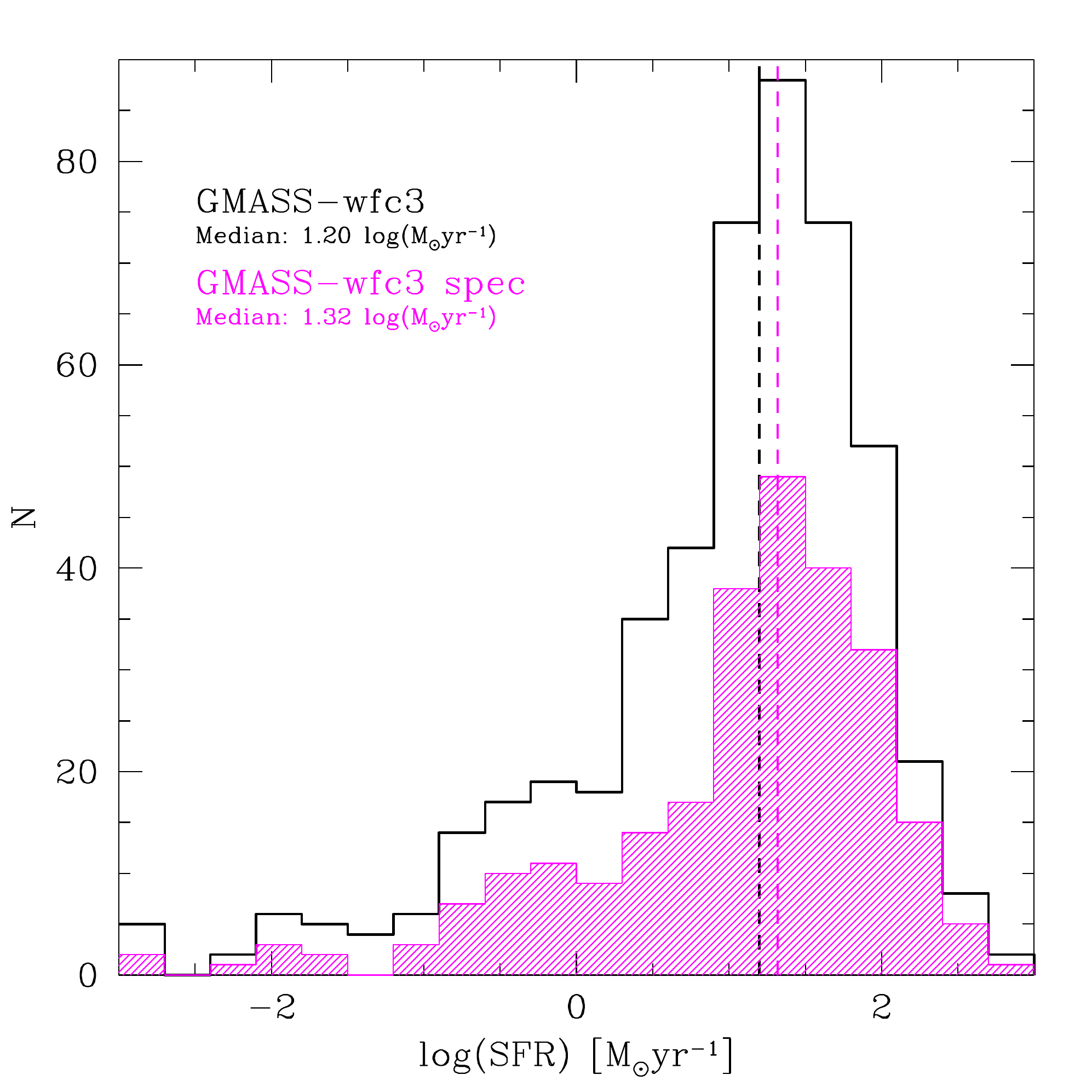}
	\includegraphics[scale=0.45]{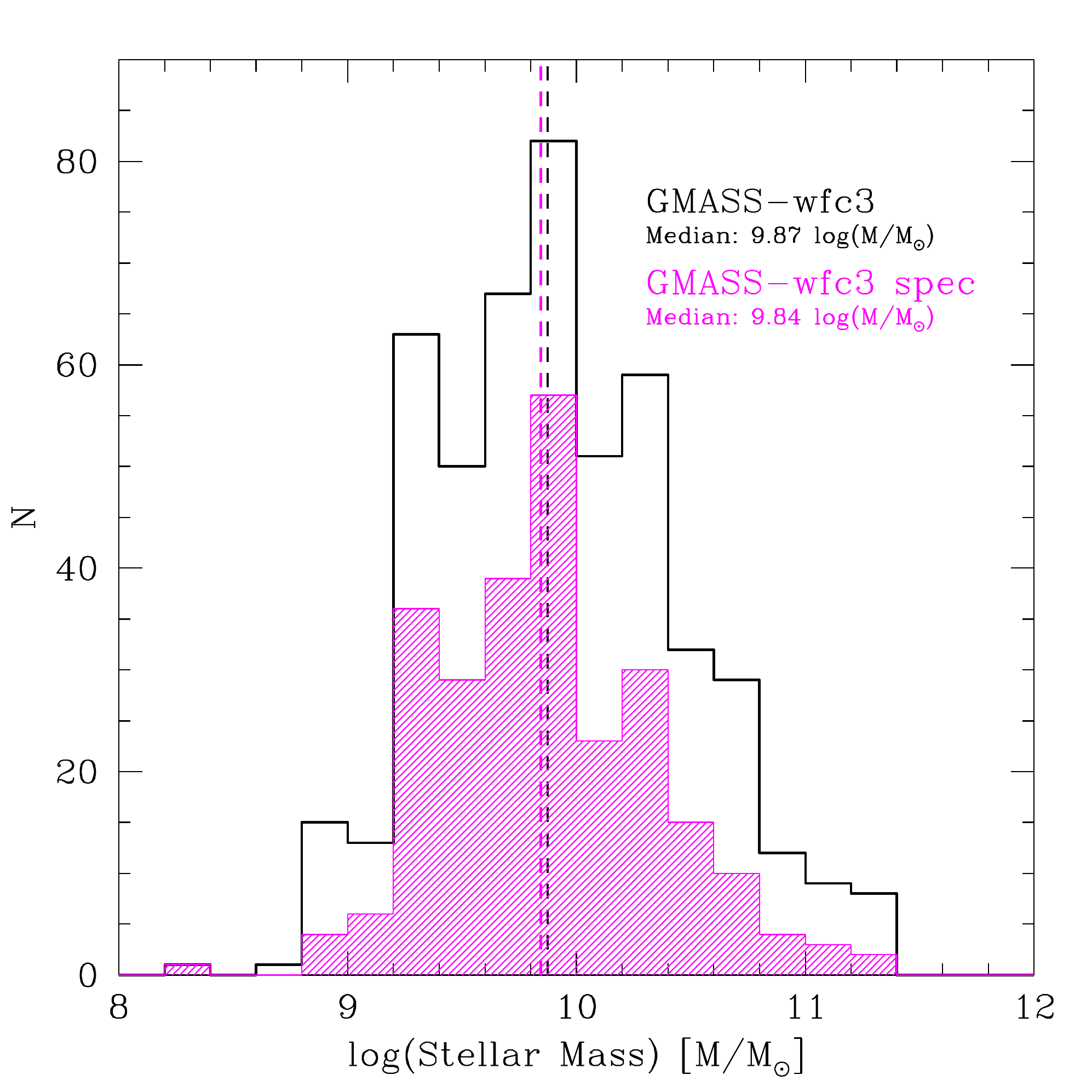}	
	\caption{Distribution of SFR (top panel) and stellar mass (bottom panel) for the \emph{GMASS-WFC3 sample} (black histogram) and the \emph{GMASS-wfc3 spec sample} (magenta histogram). Median values for each sample are also marked by vertical dashed lines.}
	\label{phot_vs_spec}
	\end{figure}

Spectra were collected for a sub-sample of 259 galaxies (the \emph{GMASS-wfc3 spec sample}) to investigate the correspondence between morphological and spectroscopic classifications of high-redshift galaxies. 
We know that the requirement of spectroscopy feasibility may introduce a bias in our analysis, since the spectra were collected from different surveys with different selection criteria and depths. To establish how representative the spectroscopic sub-sample is with respect to the parent sample, we compared the distributions of SFR and stellar mass. Fig. \ref{phot_vs_spec} shows that the \emph{GMASS-wfc3 spec sample} is slightly biased towards higher SFRs.
We also considered whether there was any bias toward any particular morphological Hubble type. We verified whether the relative percentages of different morphological types in the parent sample were conserved in the spectroscopic sub-sample. None of the galaxies classified as faint objects are present in the \emph{GMASS-wfc3 spec sample}, as we would expect. We found that $\sim$70$\%$ of the ellipticals and disks are in the spectroscopic sample, in contrast to only $\sim$50$\%$ of the irregulars and $\sim$20$\%$ of compact. Most irregulars with no spectroscopic information are galaxies with image $S/N<50$. Therefore, we conclude that, in terms of morphological classification, the spectroscopic sub-sample is biased slightly more towards elliptical and disk morphologies than irregulars, compared to the parent sample of galaxies. 

Galaxies in the \emph{GMASS-wfc3 spec sample} were divided into four classes depending on their spectroscopic features \citep{mignoli2005, talia2012}: (i) \emph{early-type} (red continuum, metal absorption lines, no nebular emission lines); (ii) \emph{star-forming} (strong [OII]$\lambda$3727 emission, strong inter-stellar absorption lines); (iii) \emph{intermediate} (red continuum plus [OII]$\lambda$3727 emission); and (iv) \emph{AGN} (type 1 having broad emission lines, type 2 narrow CIV$\lambda$1549$\AA$ emission line). 
	\begin{table}[t!]
	\caption[]{Morphological type vs. Spectral class for the 259 galaxies in the \emph{GMASS-wfc3 spec sample}.}
	\label{vis_type_spec_class}
	\centering                          
	\begin{tabular}{c l| c c c c c}        
	\hline\hline    
				&				& \multicolumn{5}{c}{Spectral class\tablefootmark{b}} \\
				&       			& Early & Interm & SF & AGN & Uncl \\		
	\hline  	   		
				&	\emph{Ell} 	   	& 16 & 0  & 29 & 5 & 0 \\
				&	\emph{Comp}	  	& 0  & 1  & 6  & 1 & 0 \\
	Morph\tablefootmark{a}  & 	Disk	 	& 4  & 12 & 91 & 1 & 0 \\
	type    		&	\emph{Irr} 	    	& 1  & 1  & 86 & 2 & 2 \\
				&	\emph{Faint}	  	& 0  & 0  & 1  & 0 & 0 \\   
  	\hline\hline
	\end{tabular}
	\tablefoot{
	\tablefoottext{a}Legend: Ell = Elliptical; Comp = Compact; Disk = Disk-like; Irr = Irregular; Faint = Faint Objects.\\
	\tablefoottext{b}Legend: Early = Early-type; Interm = Intermediate; SF = Star-forming; AGN = AGN; Uncl = Unclassified.\\
	}
	\end{table}
The spectroscopic classification was compared to the morphological one and the results are shown in Table \ref{vis_type_spec_class}. There is a close correspondence between the morphological and spectroscopic galaxy classifications. Almost all galaxies with an early-type spectrum have an elliptical appearance. These elliptical galaxies are also passively evolving following a definition based on their sSFR, as reported in the previous section of this work. Our so-called intermediate spectra appear instead to be linked to a disk-like morphology. Galaxies with star-forming spectra do not tend to have any particular morphology: this class contains irregulars as well as disks, compact, and ellipticals. Conversely, we find that irregular galaxies always have a star-forming spectrum. Disks usually display some sign of star-formation activity in their spectra, though in some cases their red continuum represents older stellar populations. In addition, at $z > 1$ an elliptical morphology is associated with either a passively evolving or strongly star-forming spectrum. Finally, we note that, in our sample, spectroscopically confirmed AGNs are preferentially hosted by elliptical galaxies.

\section{Summary and conclusions}\label{Summary and conclusions}
We have analysed the rest-frame optical morphology of galaxies at $z \geq 1$ observed as part of GMASS, using IR images from HST/WFC3 to help us relate galaxy morphologies to their physical properties.

The main result of this paper is that the fraction of ellipticals declines going back with cosmic time from $z = 1$, as does the fraction of disks. Up to $z \sim 2.5-2.7$, we found that Hubble morphological types are still recognizable, while at higher redshifts the galaxy population is dominated by irregular galaxies. 
This result implies that the build-up of the Hubble sequence occurred at redshifts $2.5<z<3$.

Other important results of our analysis can be summarized as follows. 
Galaxies in our selected sample were first classified among the traditional Hubble types, finding that:
\begin{itemize}
\item Galaxies with little or no sign of star formation generally have elliptical morphologies, while an irregular morphology is linked to intense, ongoing star-formation. Galaxies too faint to allow a morphological classification ``by eye'' are mainly SFGs that are likely to be heavily obscured by dust or characterized by low surface brightness. There is a trend between morphological types and stellar masses, such that at each redshift, faint objects, irregulars and compact galaxies have, on average, lower stellar masses than ellipticals. \\ 
\end{itemize}
A quantitative morphological analysis was also performed using the four most commonly used morphological parameters: concentration, asymmetry, Gini, and $M_{20}$. 
\begin{itemize}
	\item Comparing the results of our quantitative analysis to our Hubble classification, we found that at $1<z<3$ morphological parameters cannot be used to clearly separate different morphological types. Even if \emph{C} and \emph{G} were able to differentiate ellipticals from irregulars, with ellipticals having the highest values of both parameters and irregulars having the lowest ones, it would be impossible to distinguish between different morphologies, because there are severe overlaps in the distributions of all parameters, for different Hubble types. No significant difference between the various morphological types was found in terms of asymmetry or $M_{20}$.
	\item The comparison of our results with studies at lower redshift showed that high-z galaxies have a higher general level of asymmetry than their local siblings. This is true in particular for ellipticals, whose mean asymmetry at $z \sim 2$ is about one dex higher than measured at lower redshifts. High-z ellipticals also have less negative $M_{20}$ values than at low-z.
	\item Finally, we found that galaxies with $log(sSFR){\gtrsim}-1$ [$Gyr^{-1}$] span the entire range of all quantitative morphological parameters, while galaxies with $log(sSFR){\lesssim}-1$ [$Gyr^{-1}$] have the highest values of G and C, and the lowest $M_{20}$ ($G \gtrsim 0.35$, $C \gtrsim 0.25$, and $M_{20} \lesssim -1.5$). \\
\end{itemize}
We assessed whether any morphological k-correction was required by comparing the morphological classifications made for each galaxy in WFC3-H$_{160}$ and ACS-$z_{850}$ images. These images sample, respectively, rest-frame optical wavelength and the rest-frame B-band ($z\sim1$) to UV ($z\sim3$). We found that:
\begin{itemize}
	\item No significant \emph{k}-correction was required when the Hubble type classification was considered. A notable exception were a group of galaxies classified as disks in optical images, which were found to be irregular at UV rest-frame wavelengths. 
	\item A fraction of irregular galaxies, in WFC3-H$_{160}$ band, are very faint in the ACS-$z_{850}$ images, though in this case we cannot strictly speak of ``morphological \emph{k}-correction'', since the faintness of these galaxies in their UV rest frame makes their classification impossible.
	\item From a quantitative point of view, few differences were found. In particular, disks have higher \emph{A} and $M_{20}$ in ACS-$z_{850}$ images, and irregulars are much more asymmetric at shorter wavelengths, as expected because they are actively star-forming galaxies. Finally, compact and elliptical galaxies have, on average, less negative values of $M_{20}$ in their WFC3-H$_{160}$ than ACS-$z_{850}$ images. \\
\end{itemize}
To investigate the correlations between the morphologies and physical properties of high-z galaxies, we analysed how different Hubble types are distributed in the well-known colour bimodality of galaxies.
The colour distribution was checked in different redshift bins: the bimodality may still be seen clearly up to $2.0 < z < 2.5$, while at higher redshifts the distribution is smoother and a larger fraction of galaxies populates the region in-between the blue and red peaks. 
The following results refer to the complete redshift range covered by our sample as a whole. 
\begin{itemize}
	\item The morphological type of a galaxy correlates with its colour: the majority of irregulars occupy the blue peak of the colour distribution, while ellipticals mainly populate the red peak, with some exceptions. Disks have a smoother colour distribution and are the morphological type that occupy most of the region in-between the blue and red peaks.  
	\item The comparison between the physical properties of blue and red galaxies showed that there is a positive correlation of colour with stellar mass, regardless of the morphological type. In terms of star-formation activity, almost all galaxies classified as compact, disks, and irregulars have ongoing star-formation, independently of their colour. Blue galaxies classified as ellipticals are also star-forming, while only red ellipticals, and a small fraction of red disks, show no sign of star formation. 
	\item There is strong similarity between the parameter distributions of red and blue galaxies within each morphological Hubble class. There is, however, a difference in the asymmetry within the elliptical class, with blue ellipticals appearing less asymmetric than red ones. \\
\end{itemize}
Examining the position of different morphological Hubble types in a rest-frame \emph{U-V} vs. \emph{V-J} colours diagram, we found that:
\begin{itemize}
	\item Galaxies inhabiting the ``quiescent'' region of the plot are almost all ellipticals and, to a leser extent, disks.
	\item There is a correspondence between the position of a galaxy in the plot and its sSFR: almost all galaxies in the ``quiescent'' region of the diagram have $log(sSFR)< -1$ $Gyr^{-1}$.
	\item Galaxies classified as ``passive'' according to their sSFR: $log(sSFR)< -2$ [$Gyr^{-1}$], have an elliptical morphology. \\
\end{itemize}
Galaxies with an elliptical morphology were examined in greater detail. We found that:
\begin{itemize}
	\item In our sample, elliptical galaxies can be separated into two groups: 40$\%$ have blue colours, relatively low stellar masses, and SFRs extending from tens to hundreds of $M_{\odot}yr^{-1}$, while 60$\%$ are red, massive galaxies with low to null star-formation activity. However, from a structural point of view, there is a remarkable similarity between red and blue ellipticals: apart from sharing the same Hubble classification, they also have similar distributions of all morphological parameters. Differences were found only in terms of asymmetry and S\'ersic index, with blue ellipticals being characterized by lower values of both parameters. However, given the relative distributions of both parameters, it is impossible to distinguish between the two groups on the basis of morphological properties alone.
	\item To establish the percentage of ``pure'' passively evolving galaxies in our sample of morphological ellipticals, we adopted a cut in sSFR: $log(sSFR)< -2$ [$Gyr^{-1}$]. We found that only $\sim$33$\%$ of all morphological ellipticals are ``passive'' galaxies: this percentage is consistent with the result of \citet{moresco2013} for their sample of $z < 1$ galaxies. \\
\end{itemize}
Finally, spectra collected for a sub-sample of galaxies were used to investigate the correspondence between the morphological and spectroscopic classifications of high-redshift galaxies. We found that almost all irregular galaxies have a star-forming spectrum, and that disks usually have some sign of star-formation activity in their spectra, even though some have red continuum indicative of old stellar populations. Finally, galaxies with an early-type spectrum have an elliptical shape, but an elliptical morphology may be associated with either passively evolving or highly star-forming galaxies.

\begin{acknowledgements}
This work is based on observations taken by the CANDELS Multi-Cycle Treasury Program with the NASA/ESA HST, which is operated by the Association of Universities for Research in AStronomy, Inc., under NASA contract NAS5-26555. AC and MM acknowledge the grants ASI n.I/023/12/0 ``Attivit\`a relative alla fase B2/C per la missione Euclid'' and MIUR PRIN 2010-2011 ``The dark Universe and the cosmic evolution of baryons: from current surveys to Euclid''. Part of the work has been supported also by an INAF grant ``PRIN-2010''. The authors would like to thank the anonymous referee for his/her valuable comments, and P.Nair and R.Abraham for sharing the code MORPHEUS. MT wishes to thank Fabio Bellagamba for useful discussions and suggestions. 
\end{acknowledgements}

\bibliographystyle{aa} 
\bibliography{references} 

\appendix
\section{The morphological atlas}
\label{sec:Appendix A}

\begin{figure*}
\centering
\includegraphics[trim=100 40 75 390, clip=true, width=30mm]{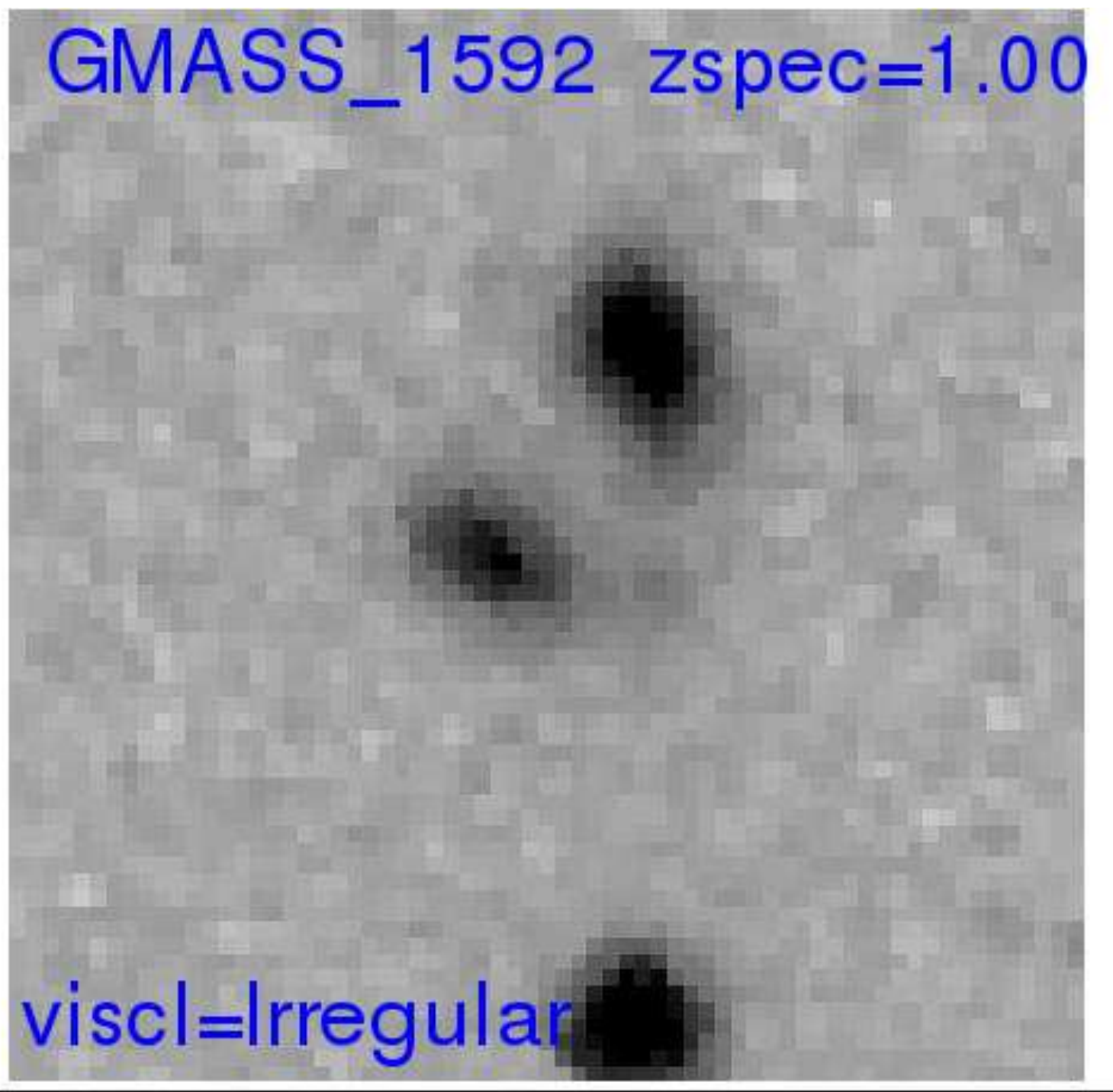}	
\includegraphics[trim=100 40 75 390, clip=true, width=30mm]{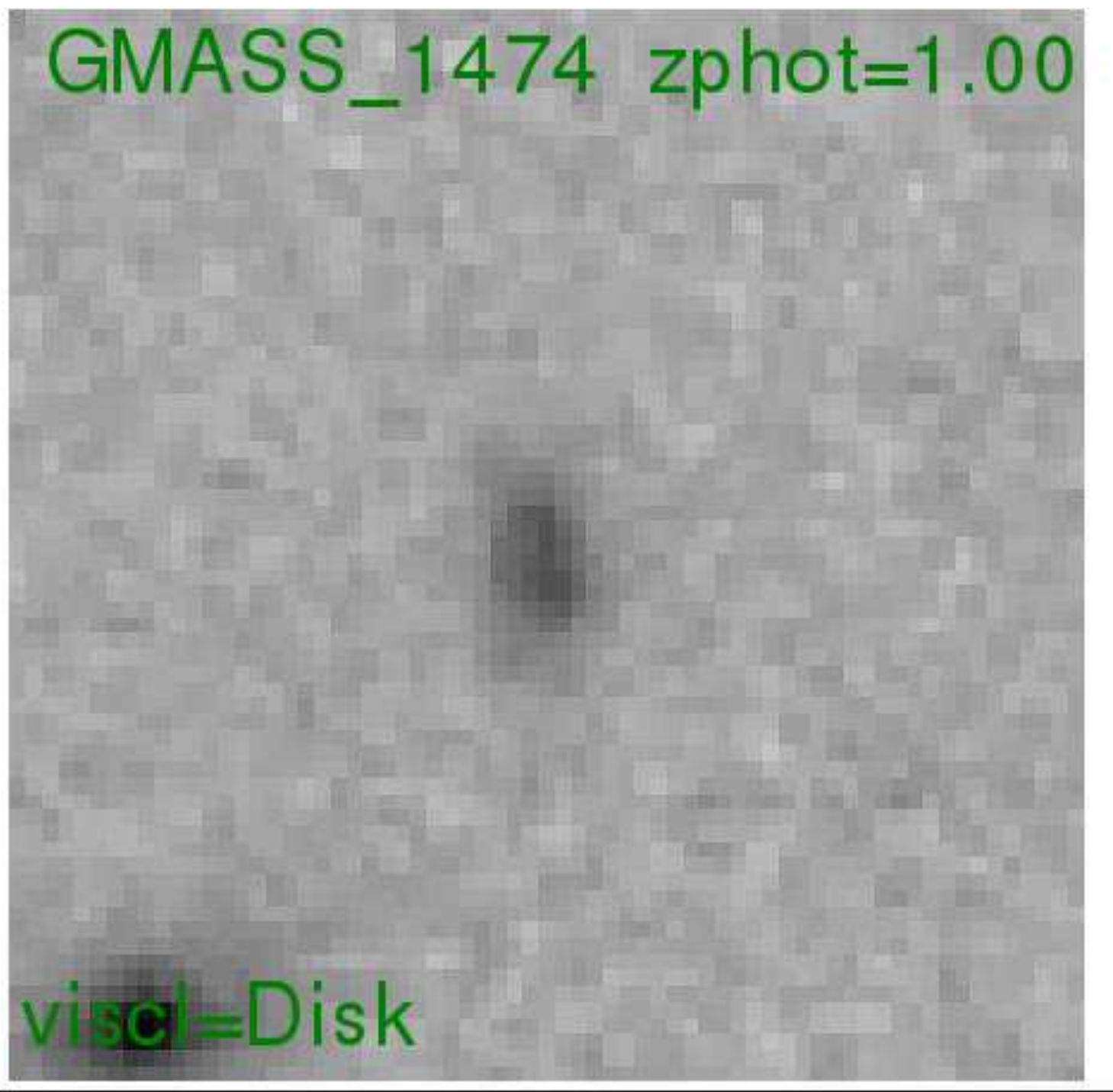}			     
\includegraphics[trim=100 40 75 390, clip=true, width=30mm]{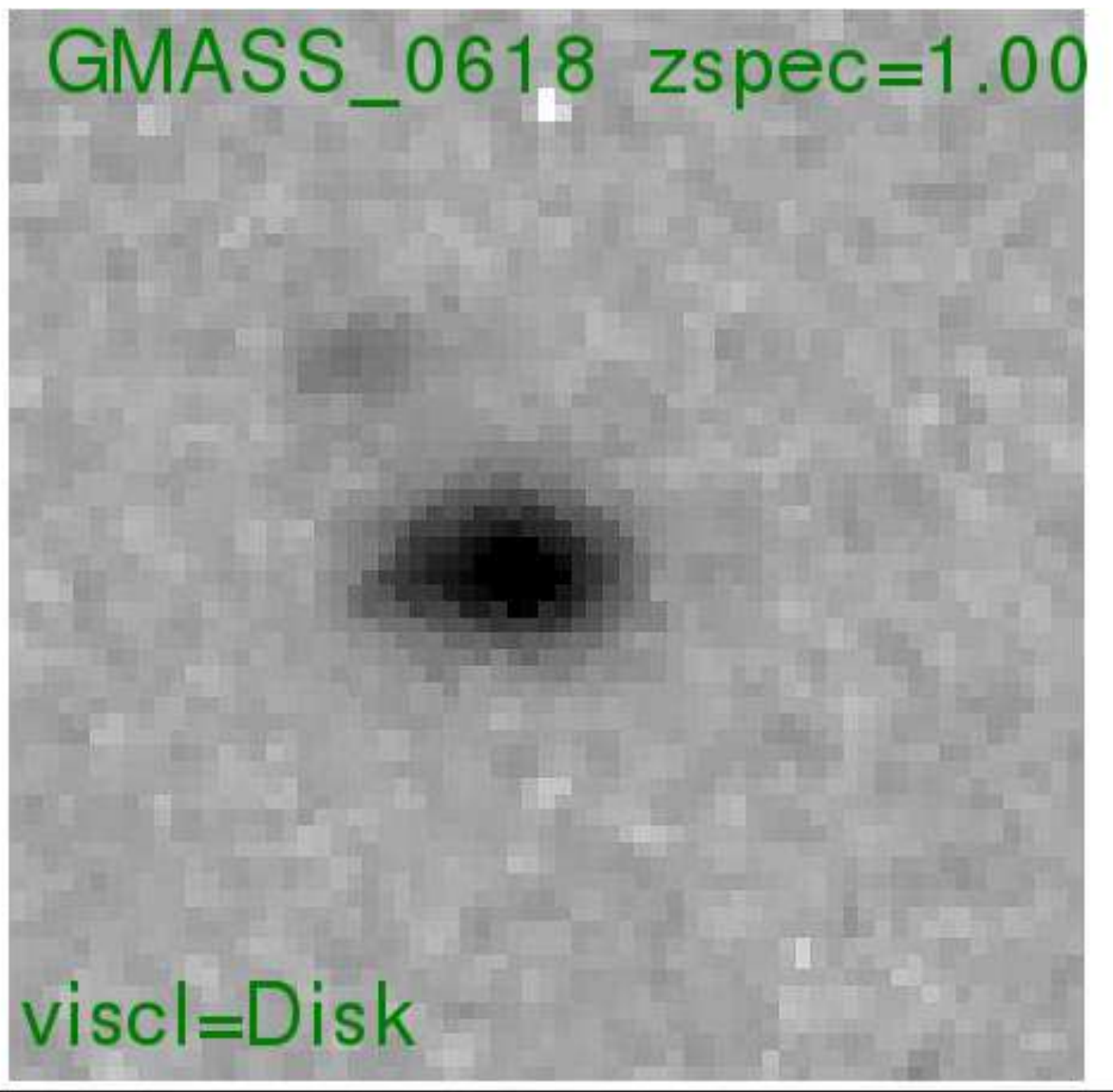}		     
\includegraphics[trim=100 40 75 390, clip=true, width=30mm]{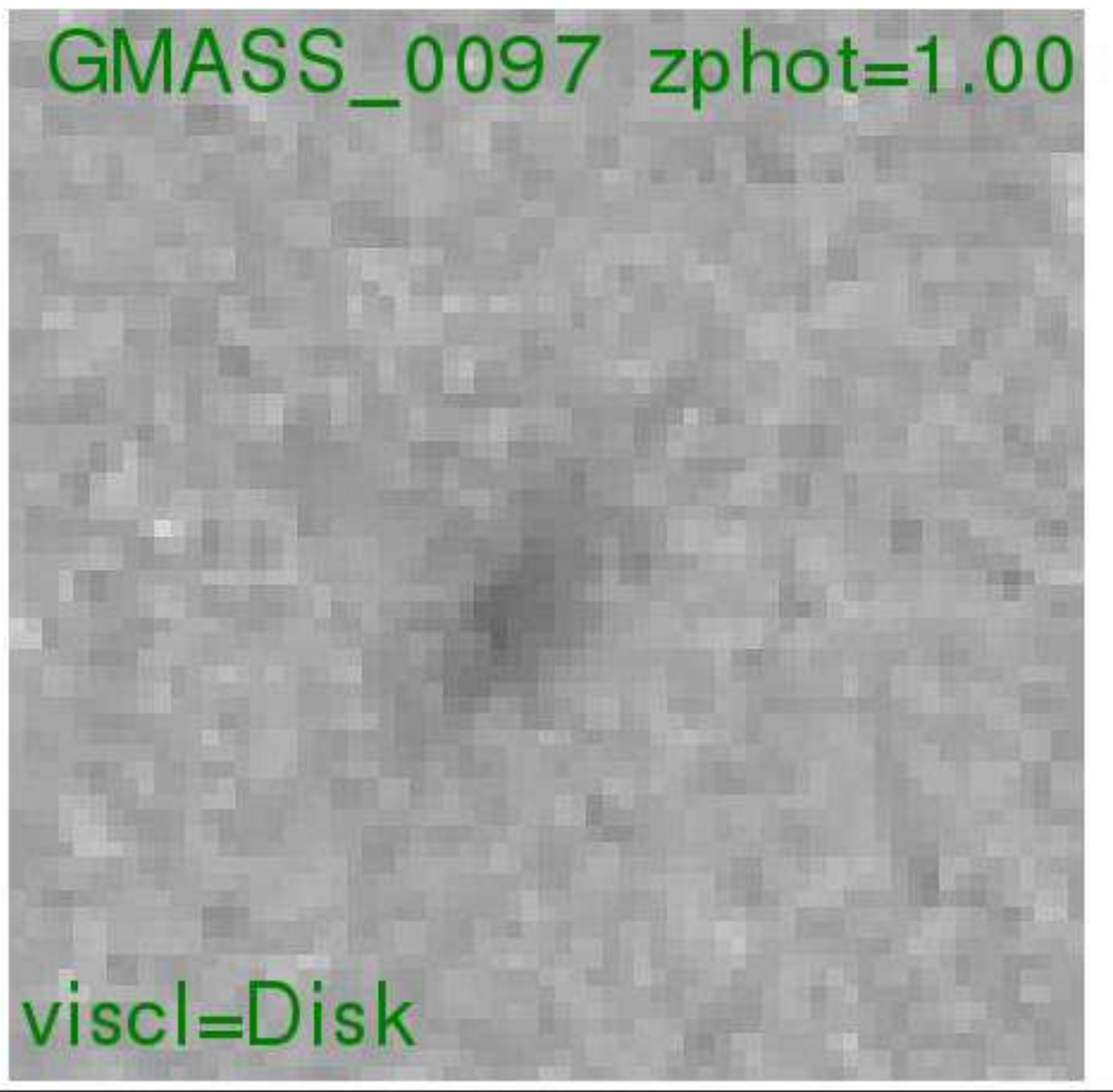}			     
\includegraphics[trim=100 40 75 390, clip=true, width=30mm]{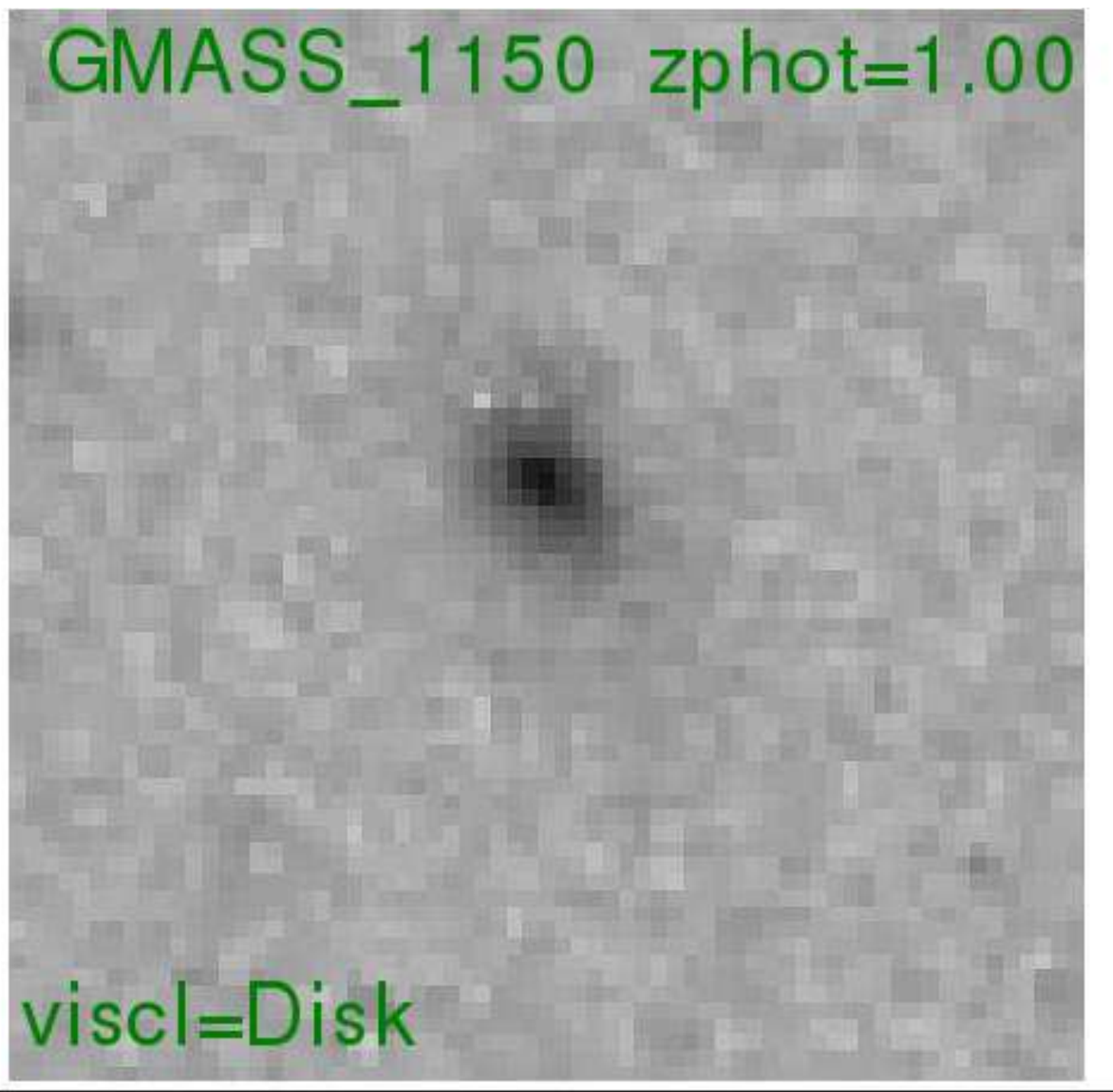}			     
\includegraphics[trim=100 40 75 390, clip=true, width=30mm]{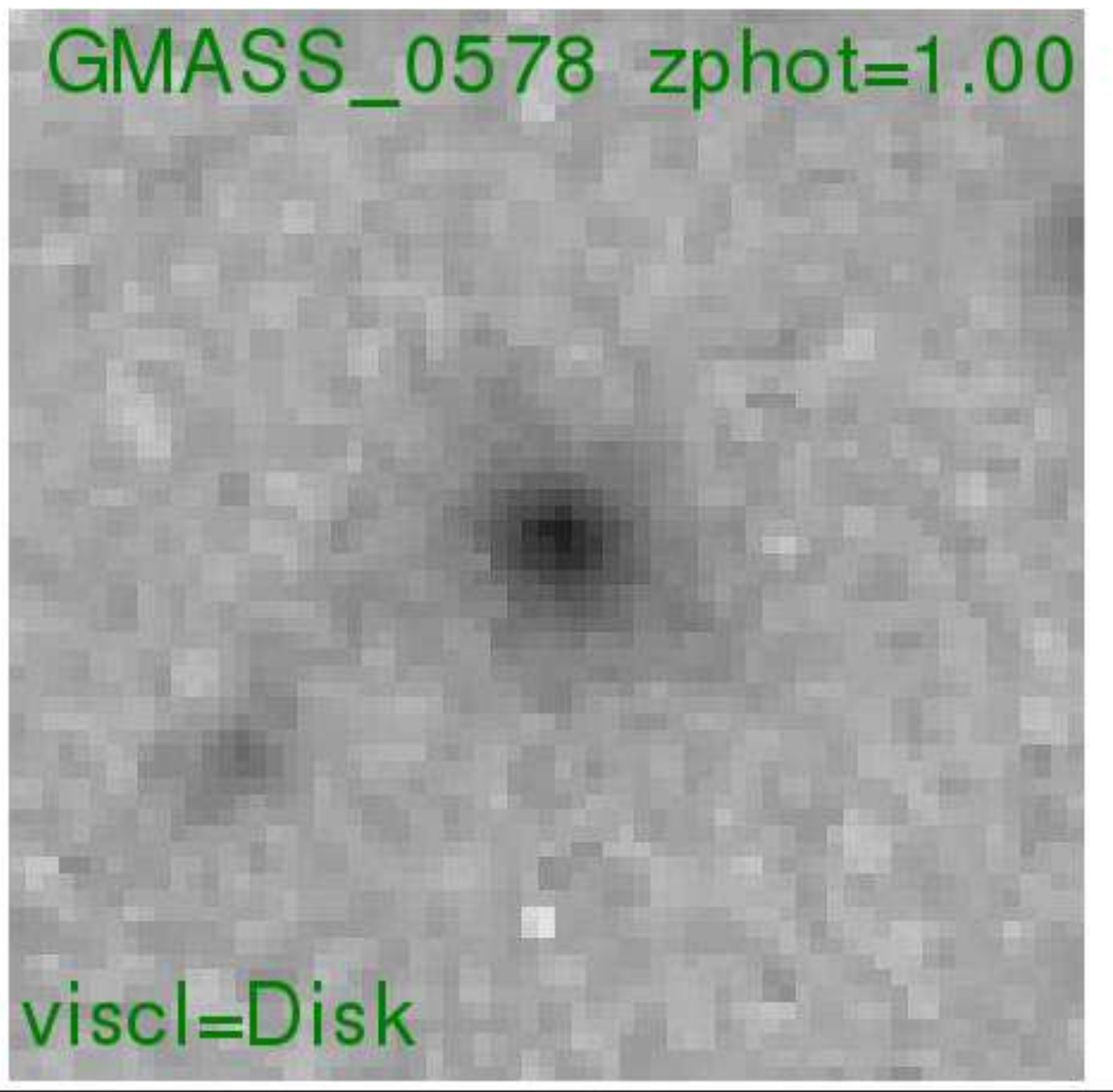}
			     
\includegraphics[trim=100 40 75 390, clip=true, width=30mm]{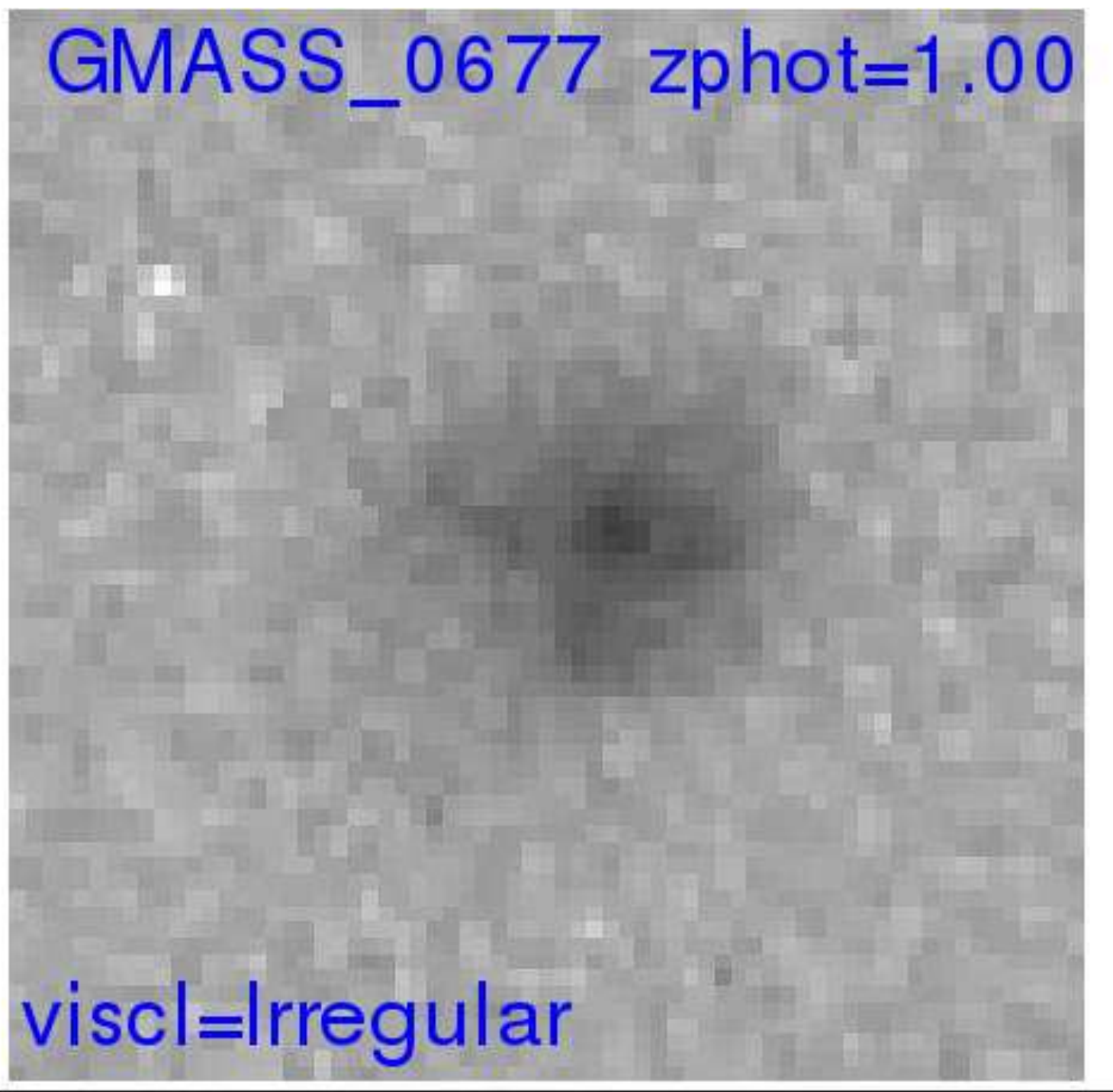}			     
\includegraphics[trim=100 40 75 390, clip=true, width=30mm]{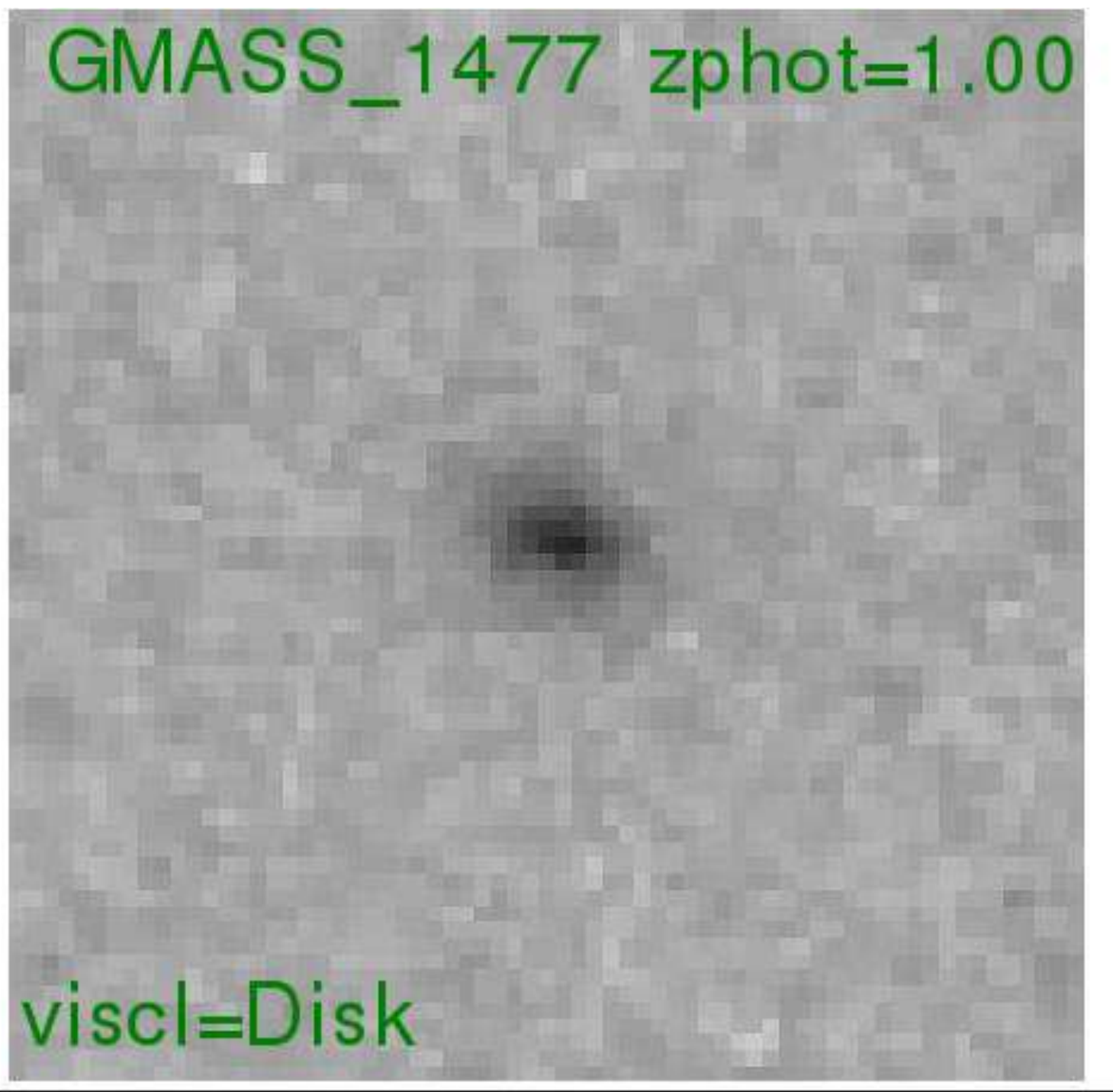}			     
\includegraphics[trim=100 40 75 390, clip=true, width=30mm]{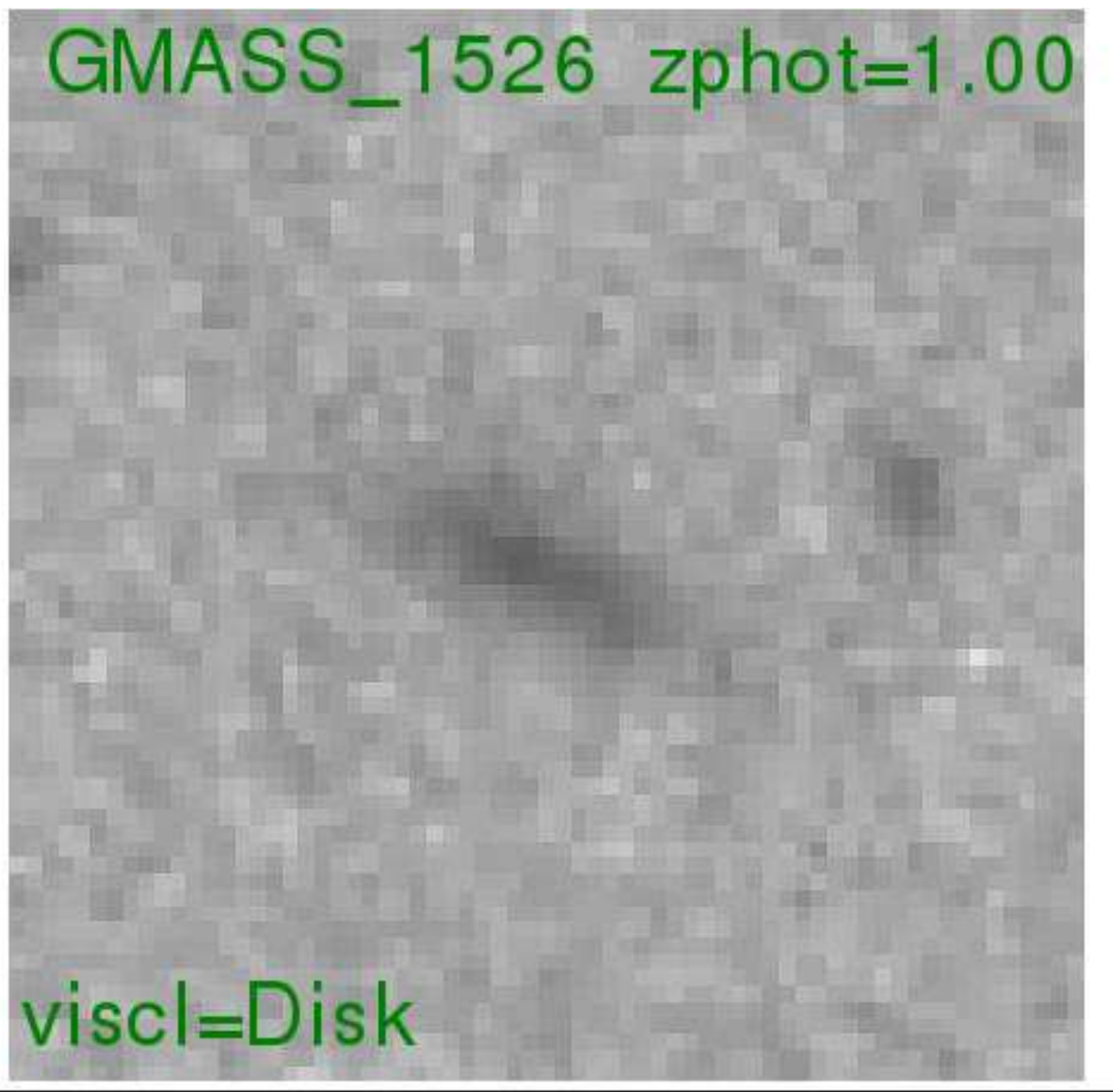}			     
\includegraphics[trim=100 40 75 390, clip=true, width=30mm]{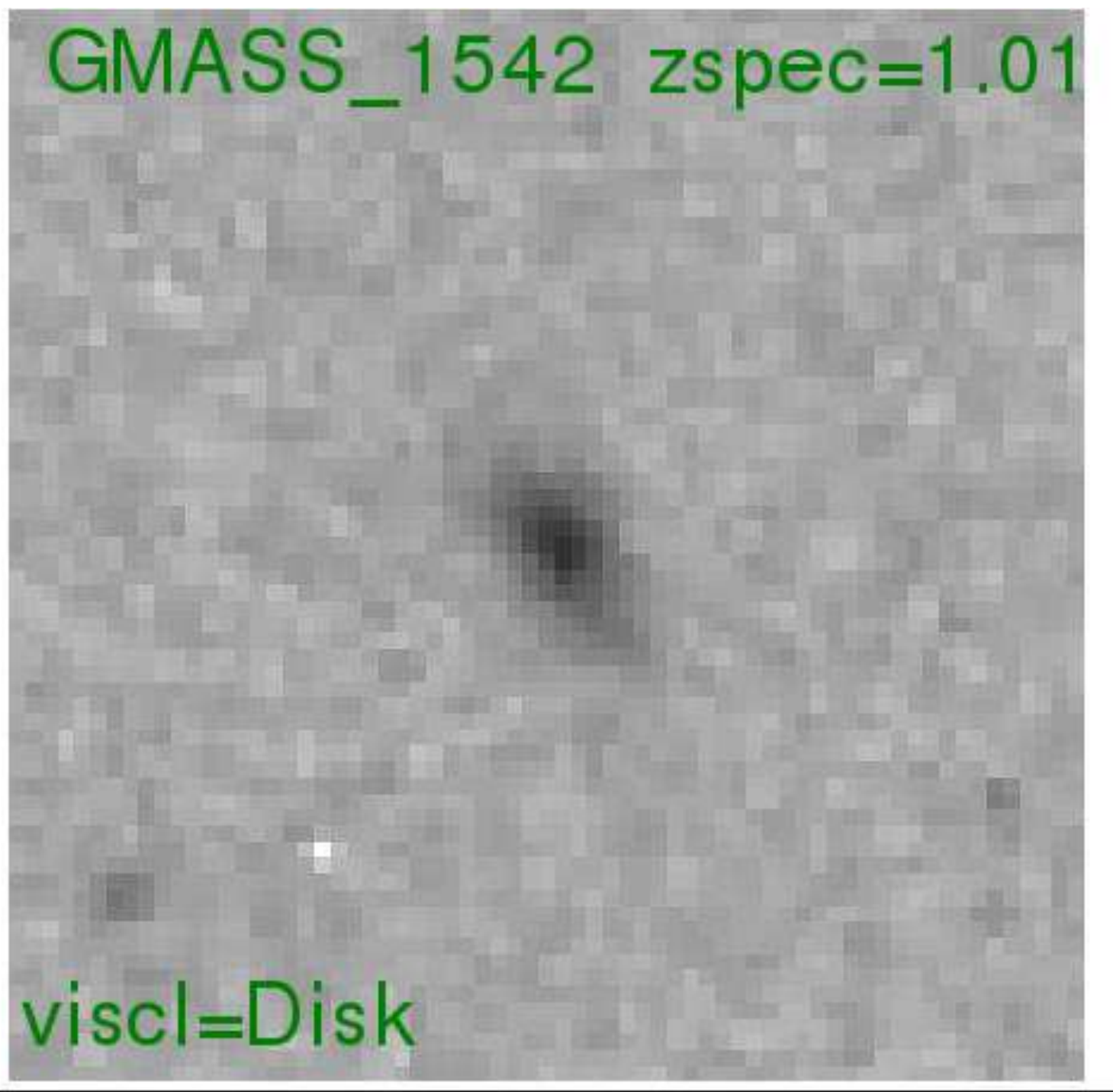}			     
\includegraphics[trim=100 40 75 390, clip=true, width=30mm]{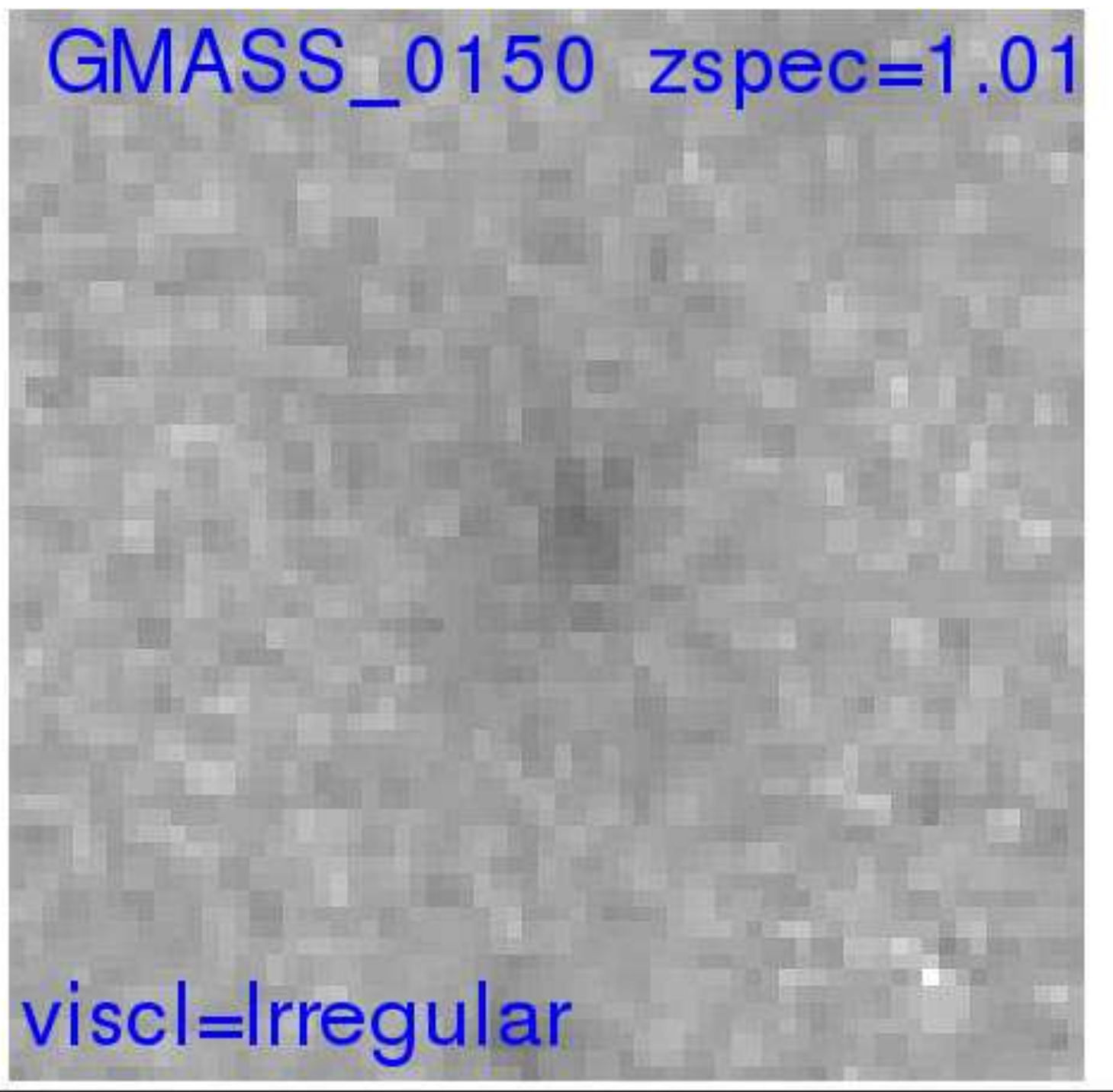}			     
\includegraphics[trim=100 40 75 390, clip=true, width=30mm]{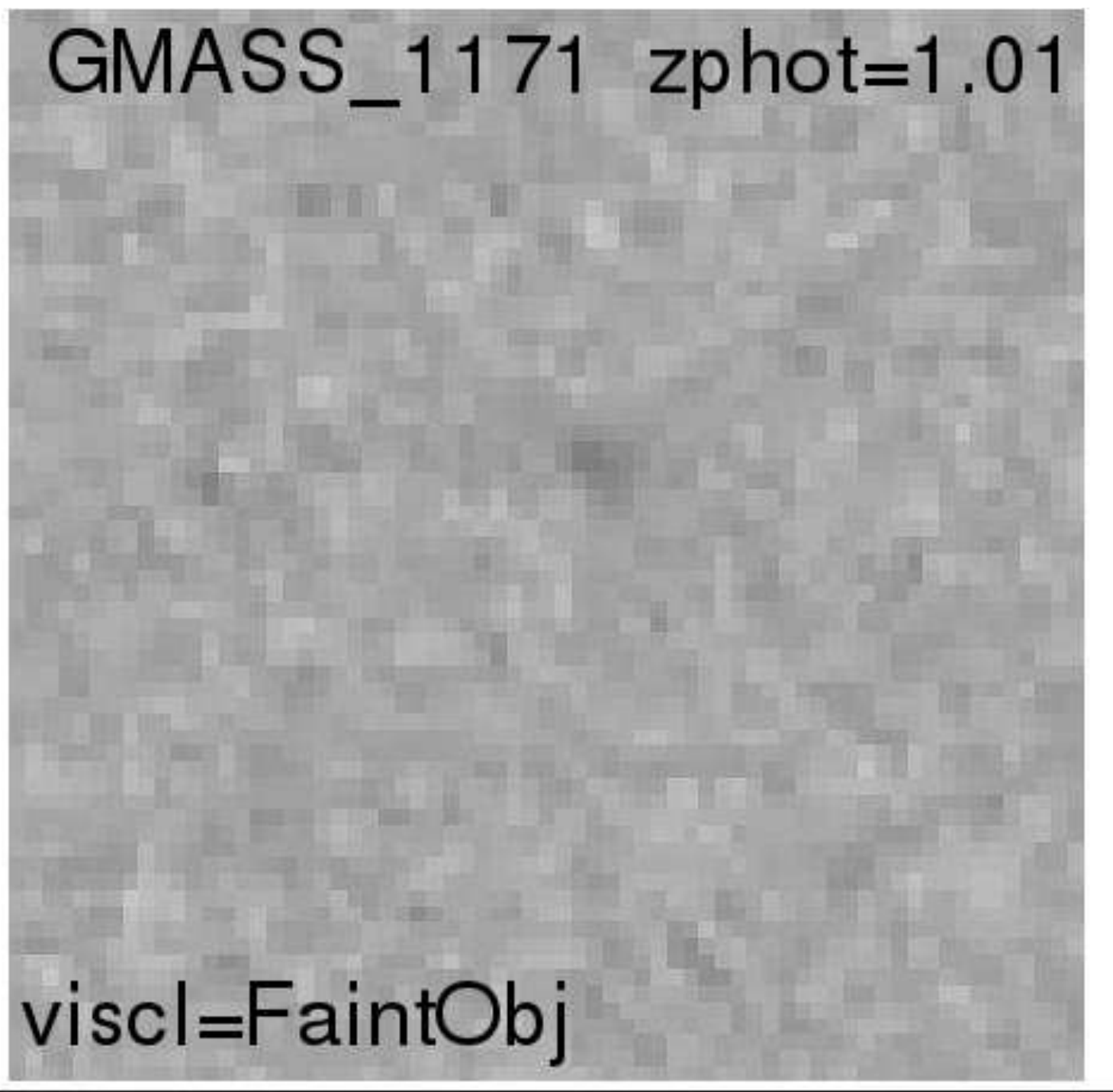}		     

\includegraphics[trim=100 40 75 390, clip=true, width=30mm]{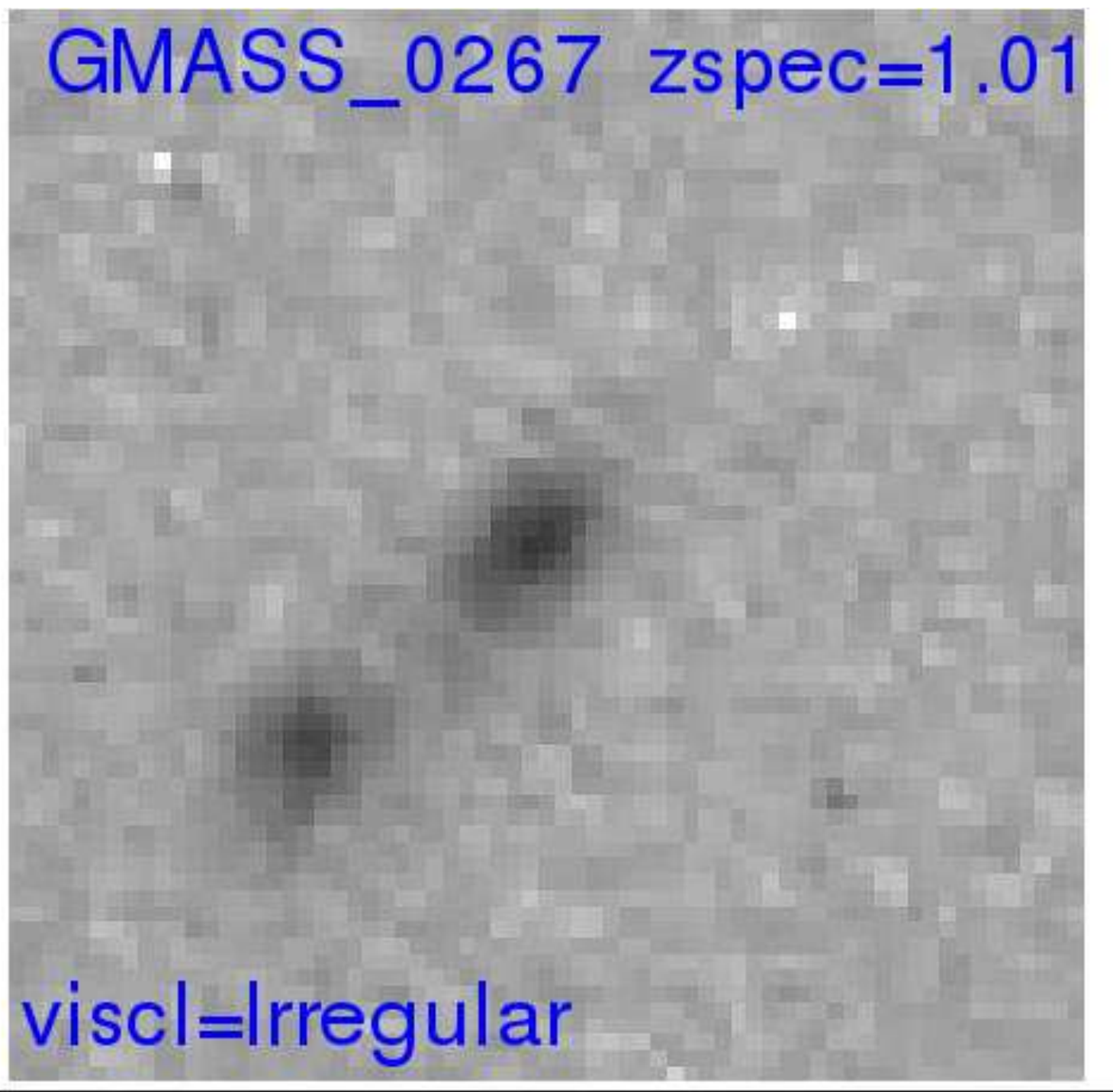}			     
\includegraphics[trim=100 40 75 390, clip=true, width=30mm]{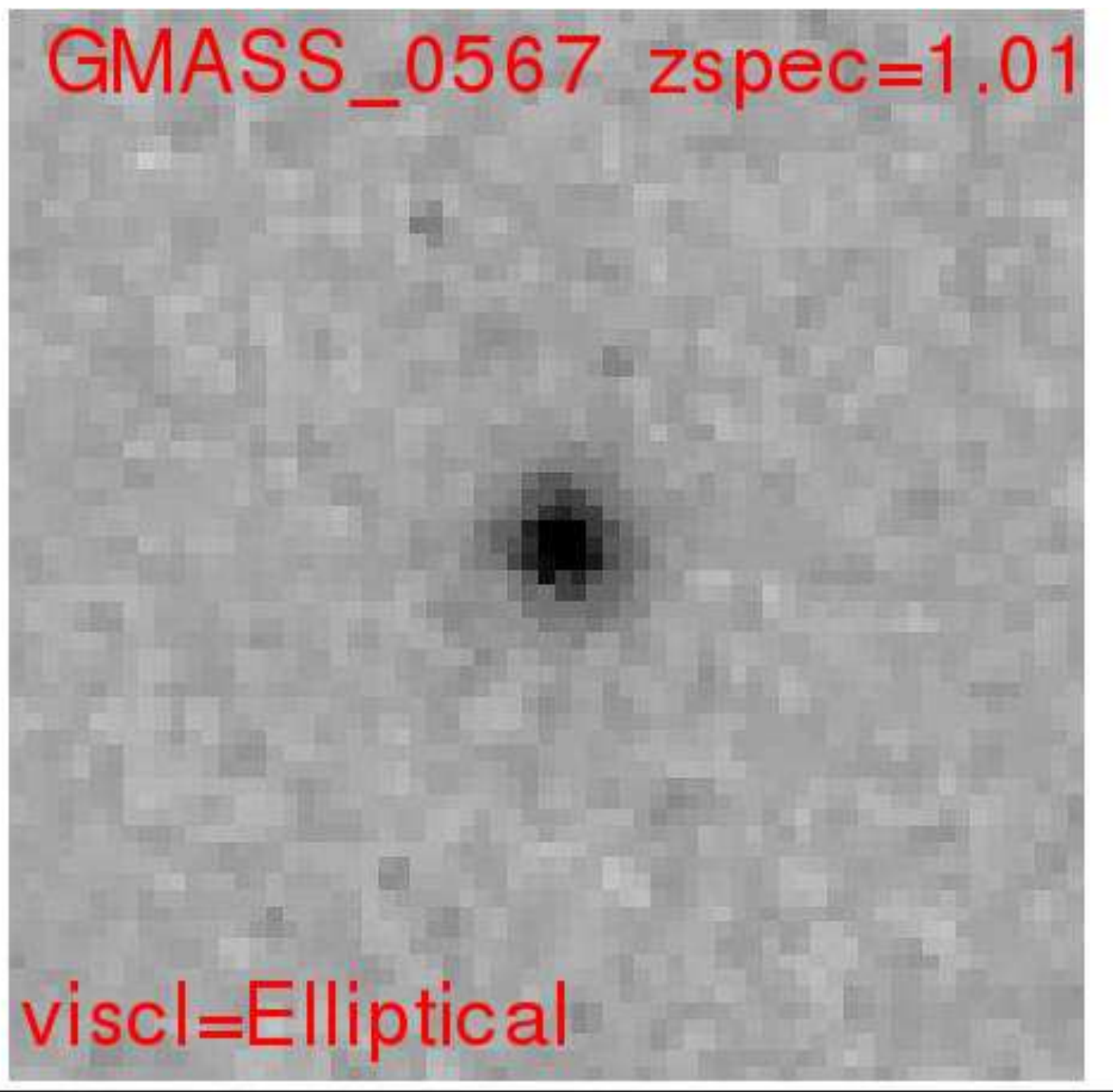}			     
\includegraphics[trim=100 40 75 390, clip=true, width=30mm]{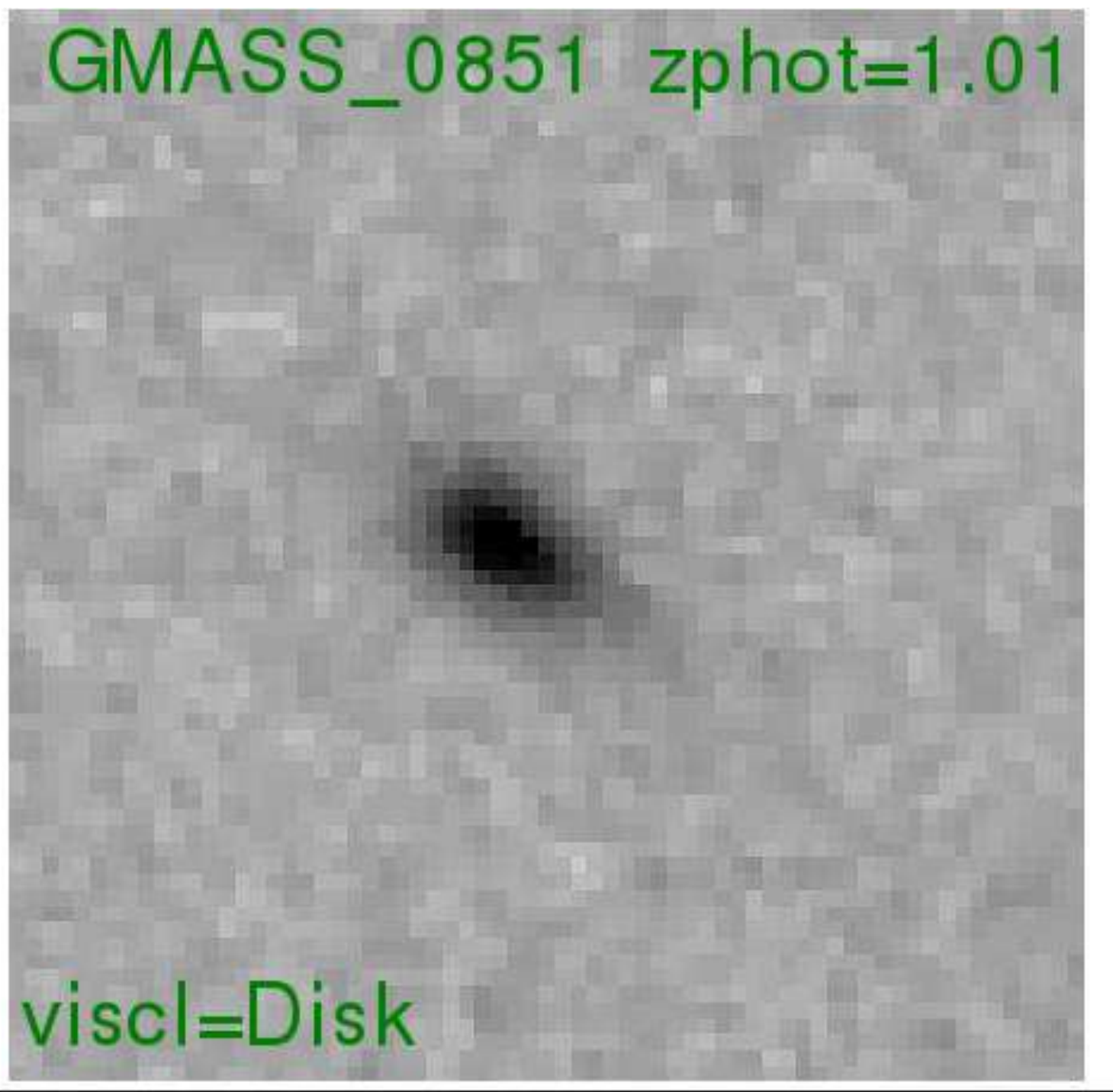}			     
\includegraphics[trim=100 40 75 390, clip=true, width=30mm]{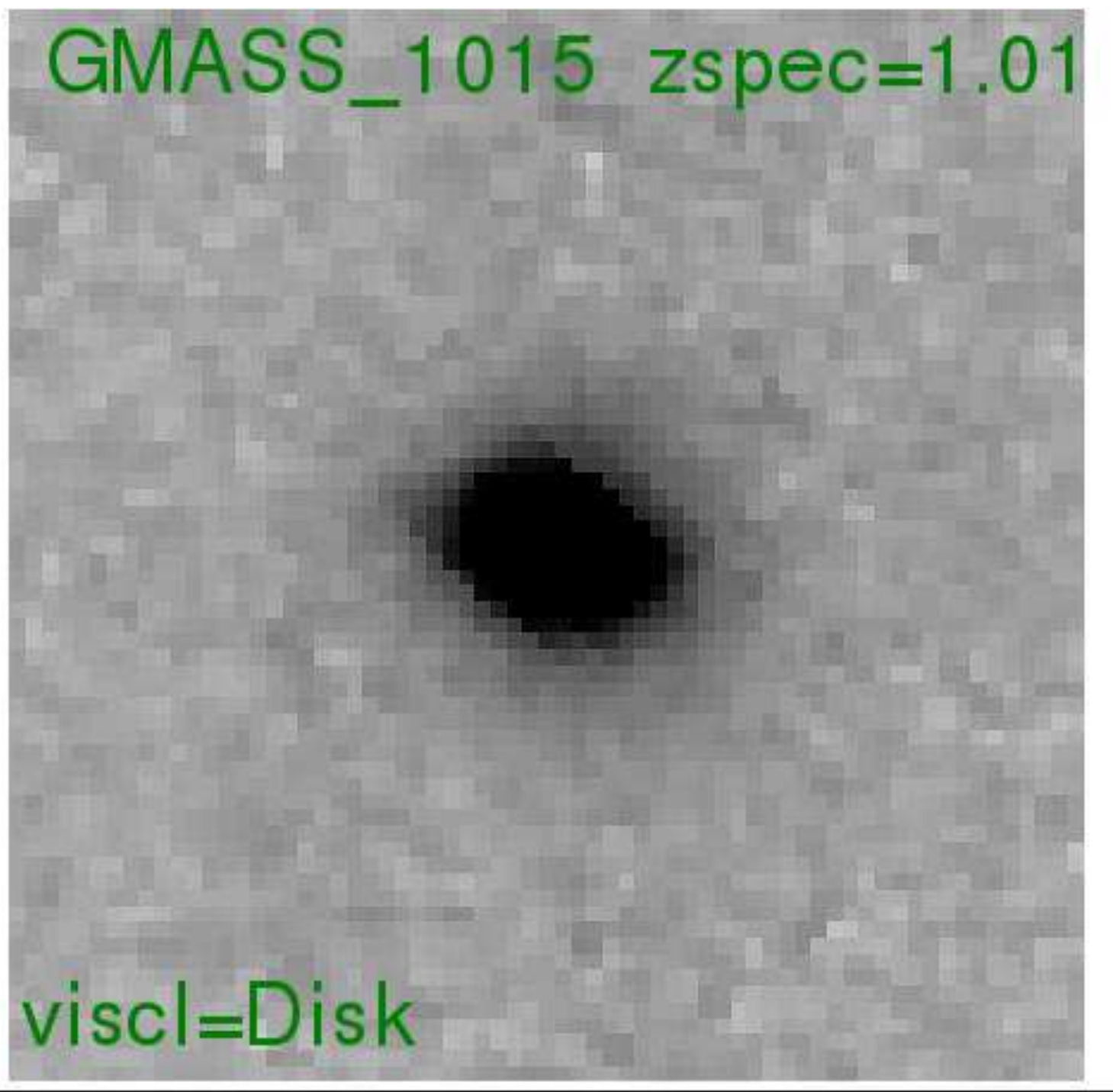}			     
\includegraphics[trim=100 40 75 390, clip=true, width=30mm]{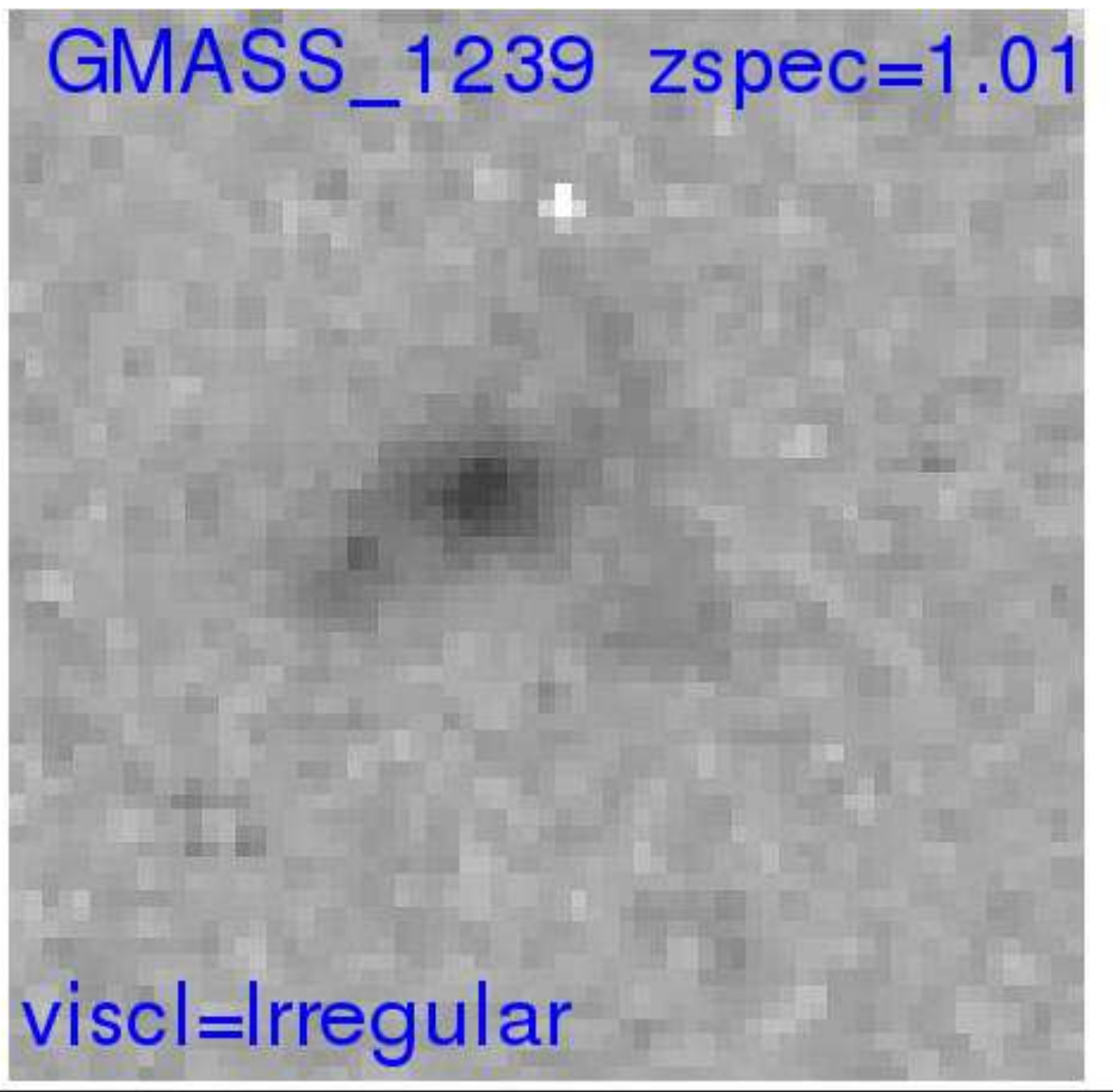}			     
\includegraphics[trim=100 40 75 390, clip=true, width=30mm]{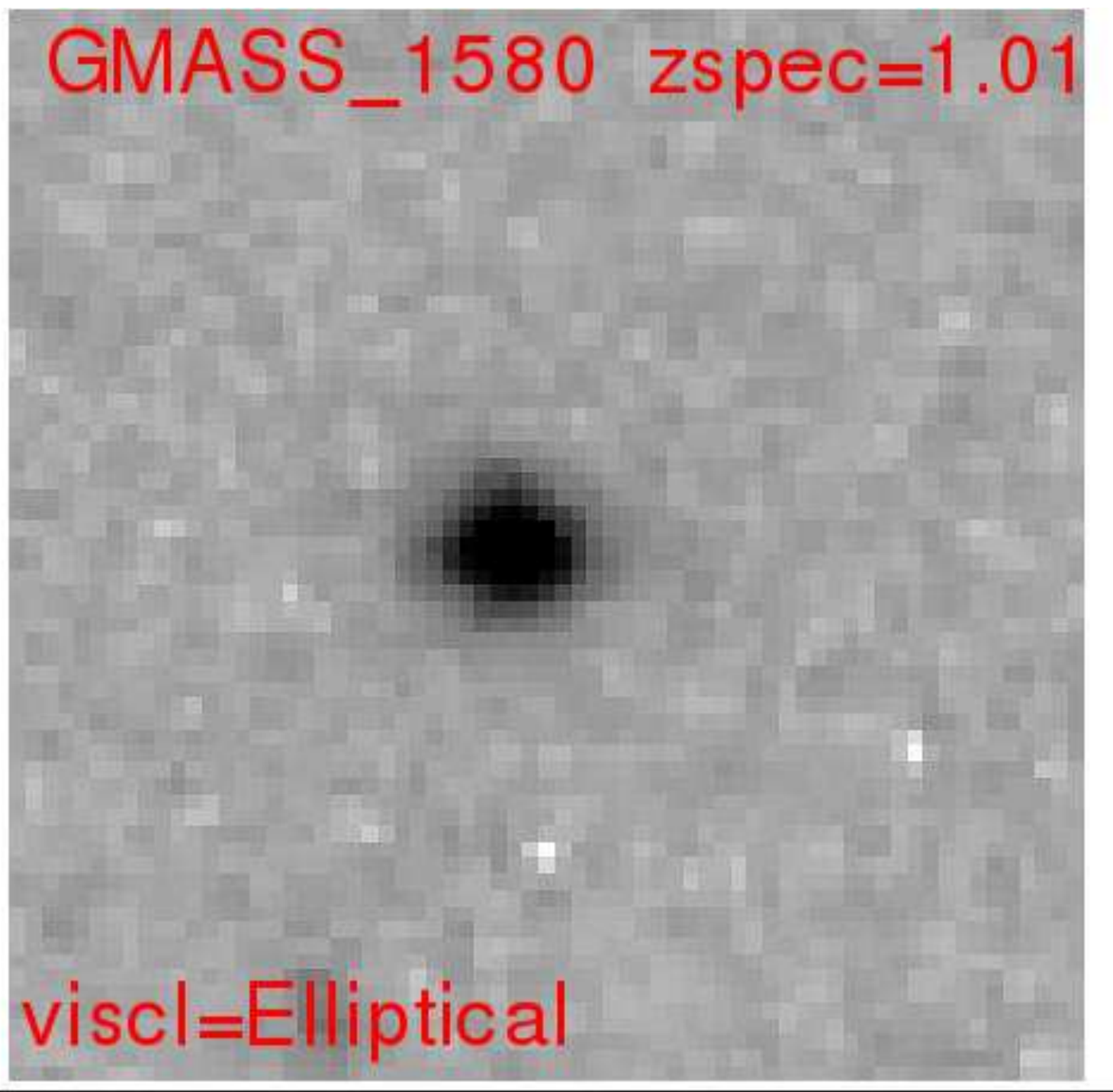}			     

\includegraphics[trim=100 40 75 390, clip=true, width=30mm]{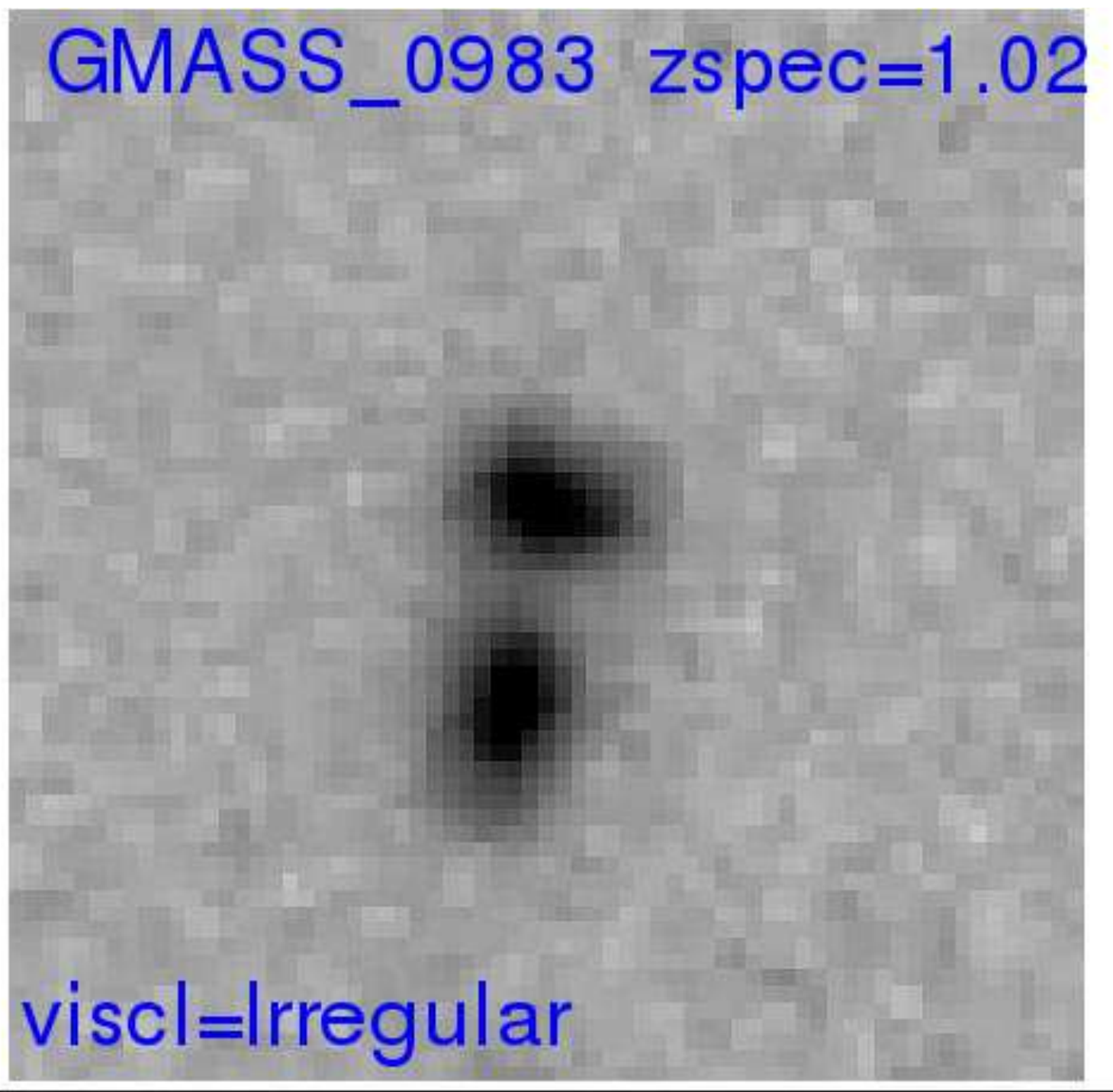}			     
\includegraphics[trim=100 40 75 390, clip=true, width=30mm]{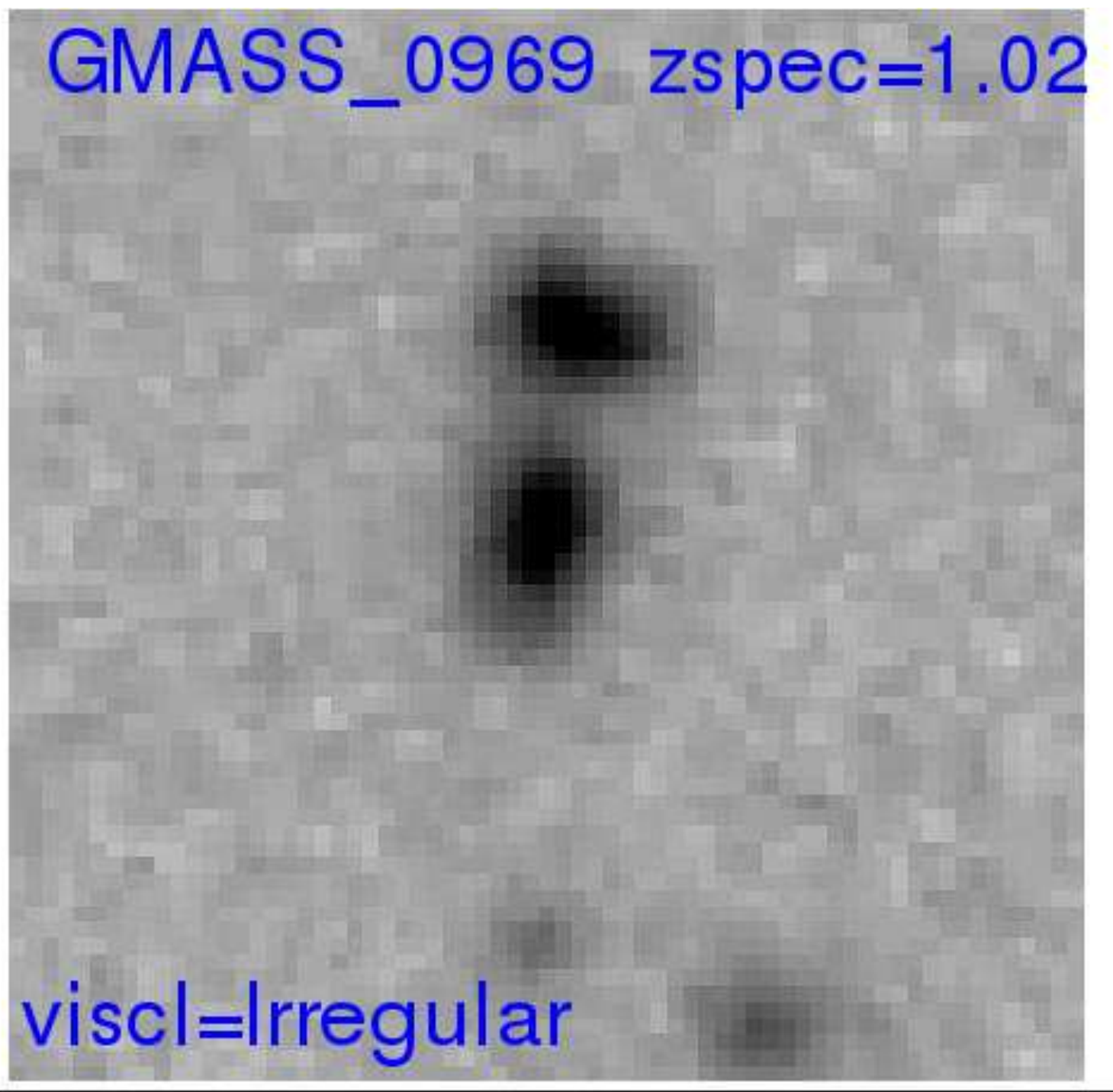}			     
\includegraphics[trim=100 40 75 390, clip=true, width=30mm]{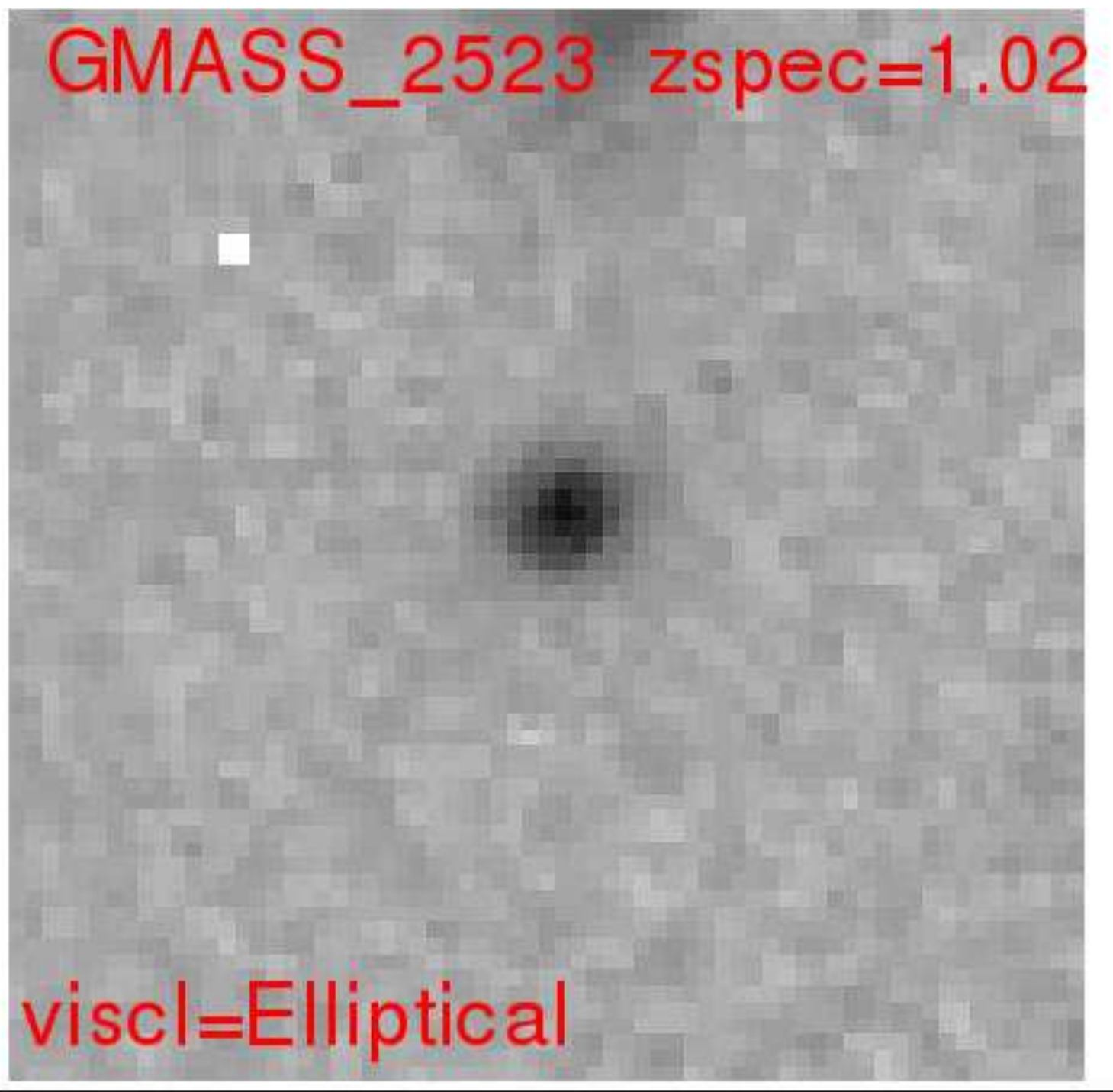}		     
\includegraphics[trim=100 40 75 390, clip=true, width=30mm]{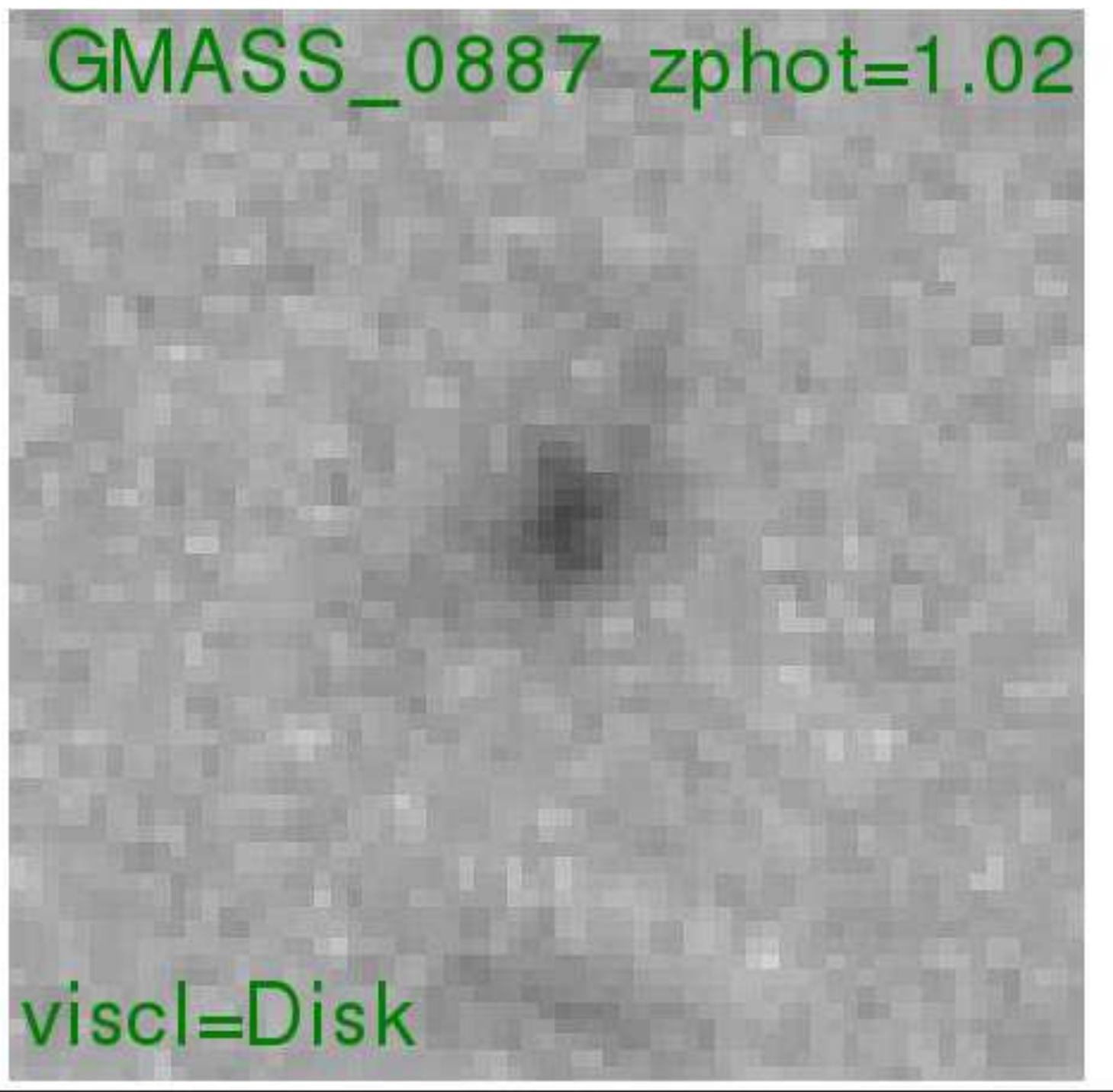}			     
\includegraphics[trim=100 40 75 390, clip=true, width=30mm]{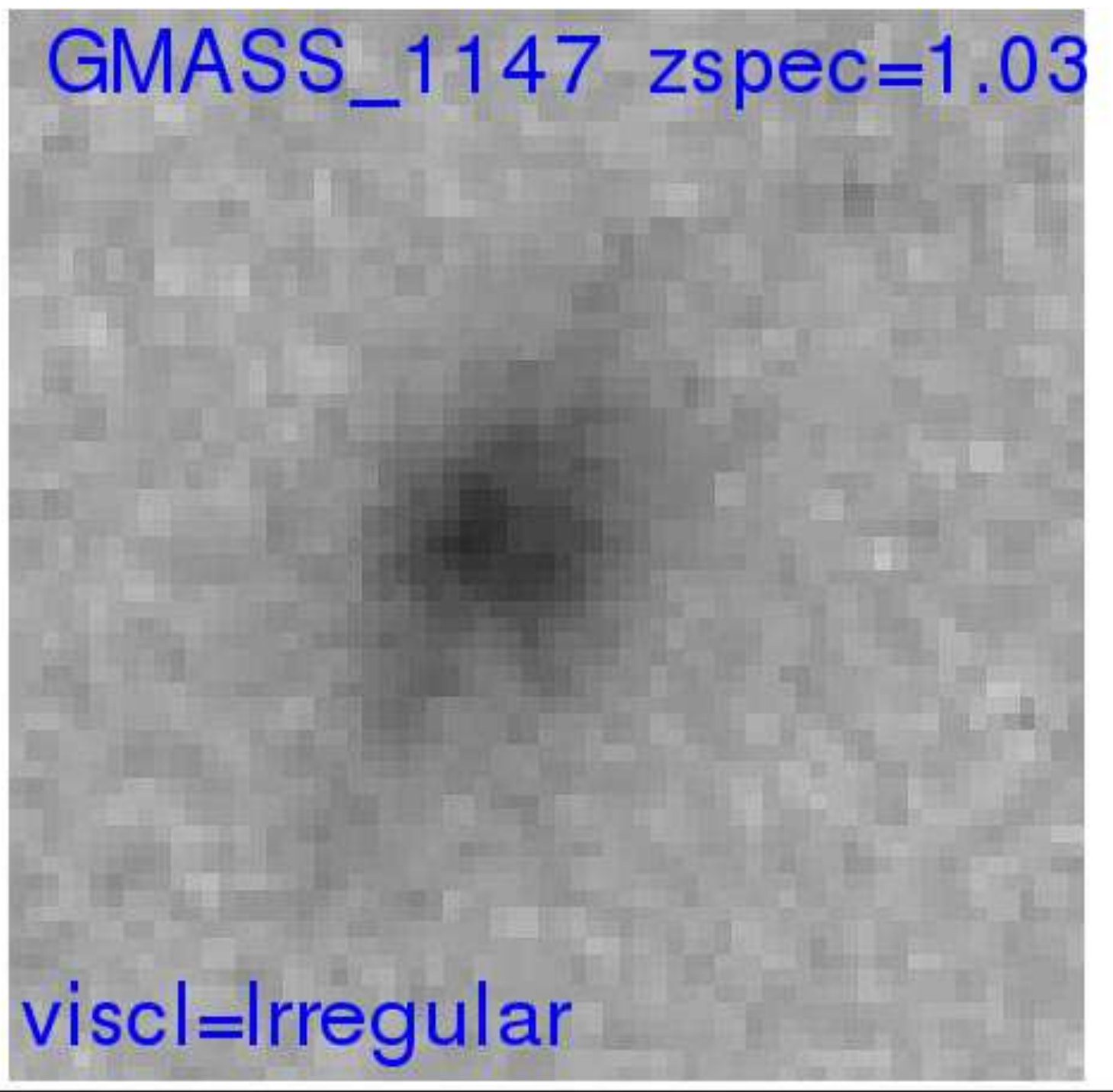}			     
\includegraphics[trim=100 40 75 390, clip=true, width=30mm]{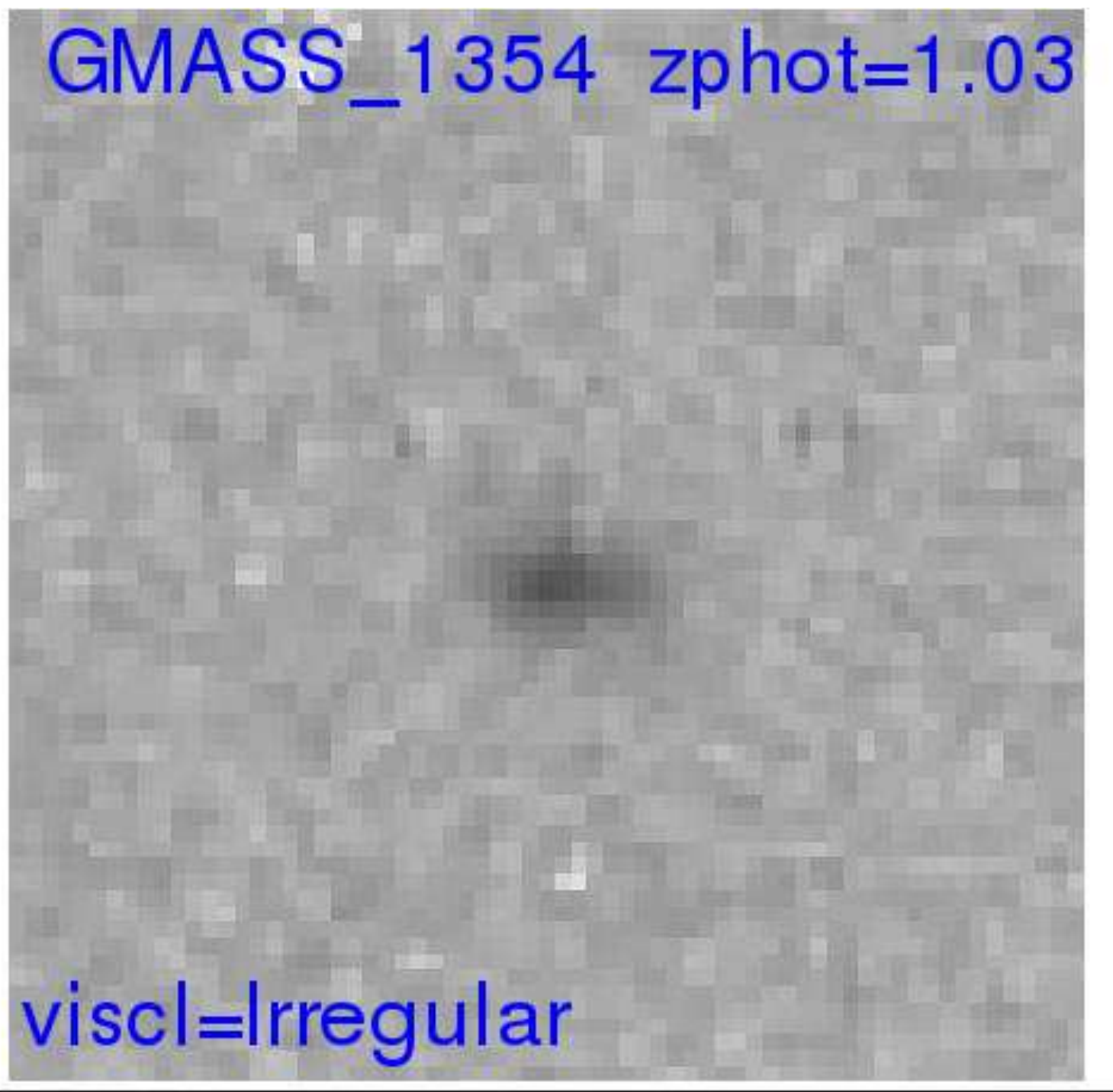}			     

\includegraphics[trim=100 40 75 390, clip=true, width=30mm]{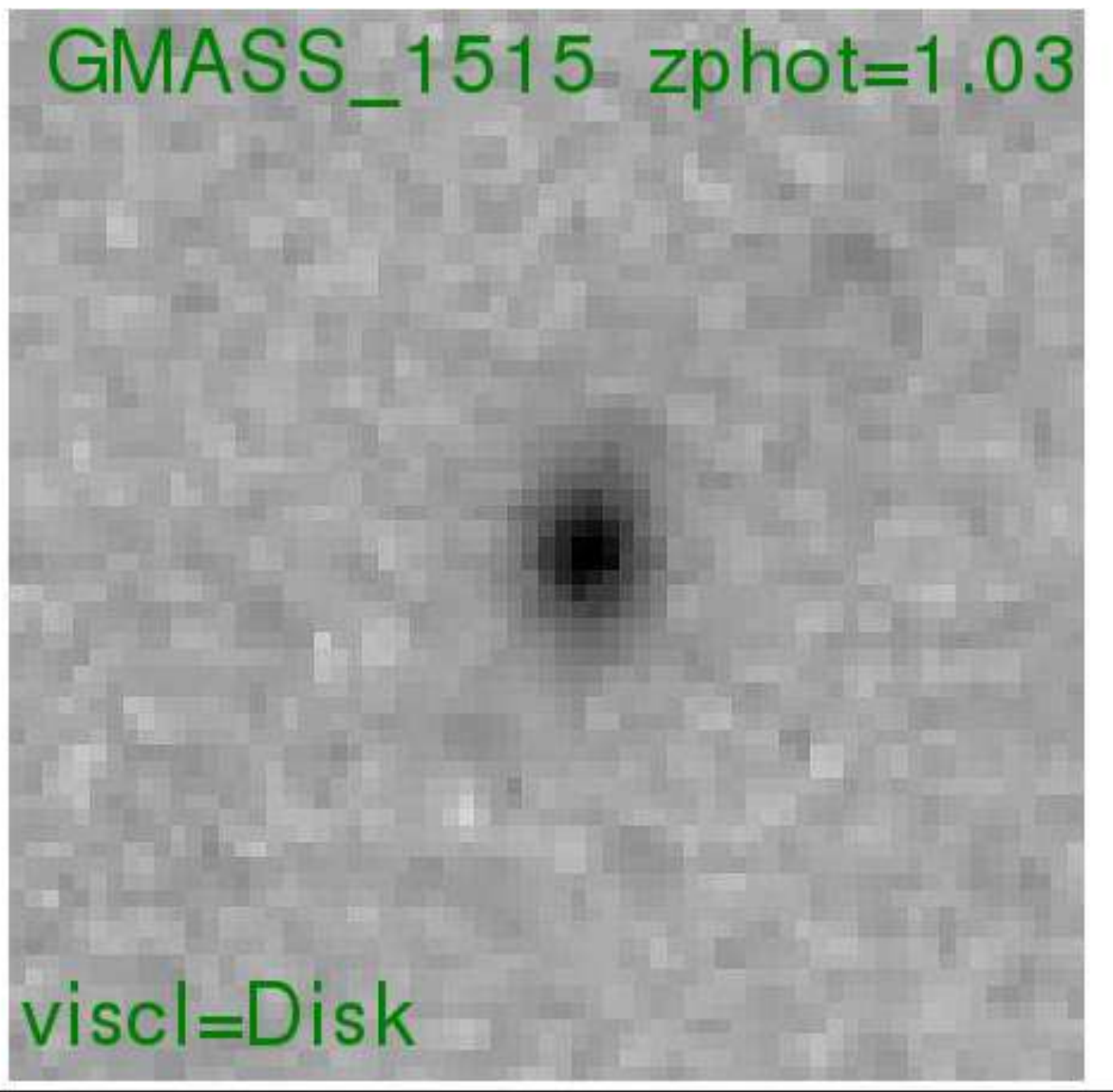}			     
\includegraphics[trim=100 40 75 390, clip=true, width=30mm]{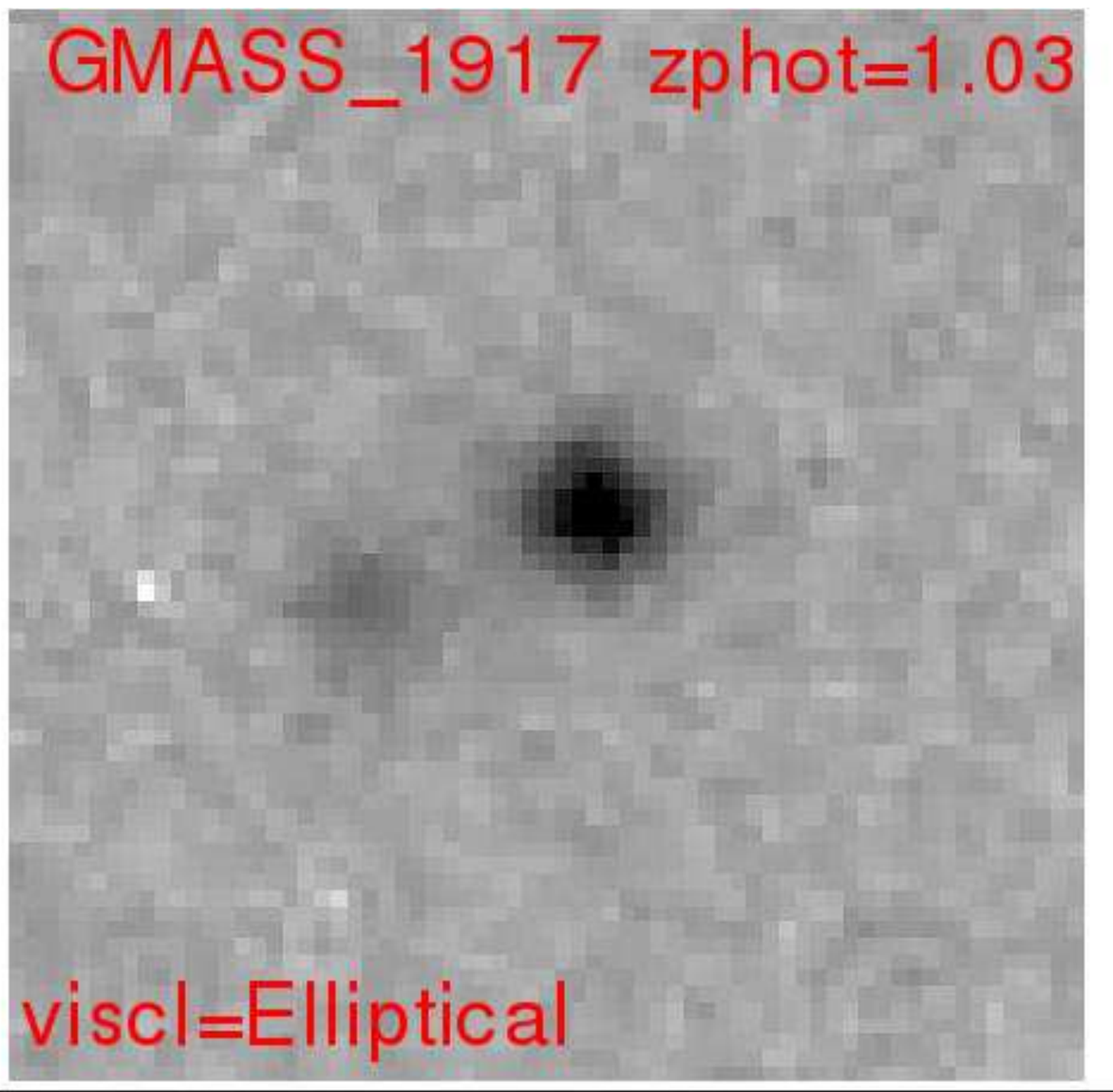}			     
\includegraphics[trim=100 40 75 390, clip=true, width=30mm]{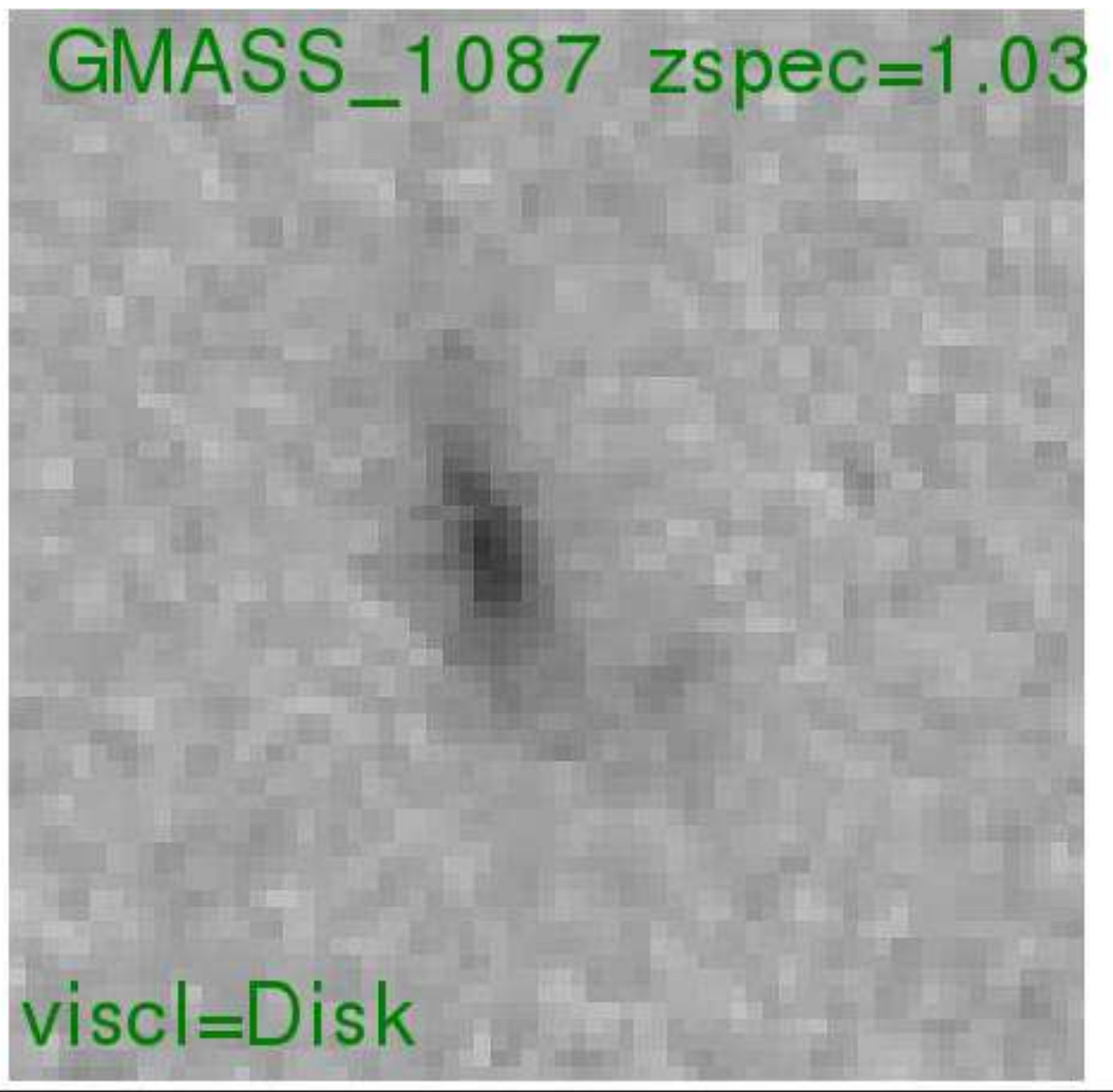}			     
\includegraphics[trim=100 40 75 390, clip=true, width=30mm]{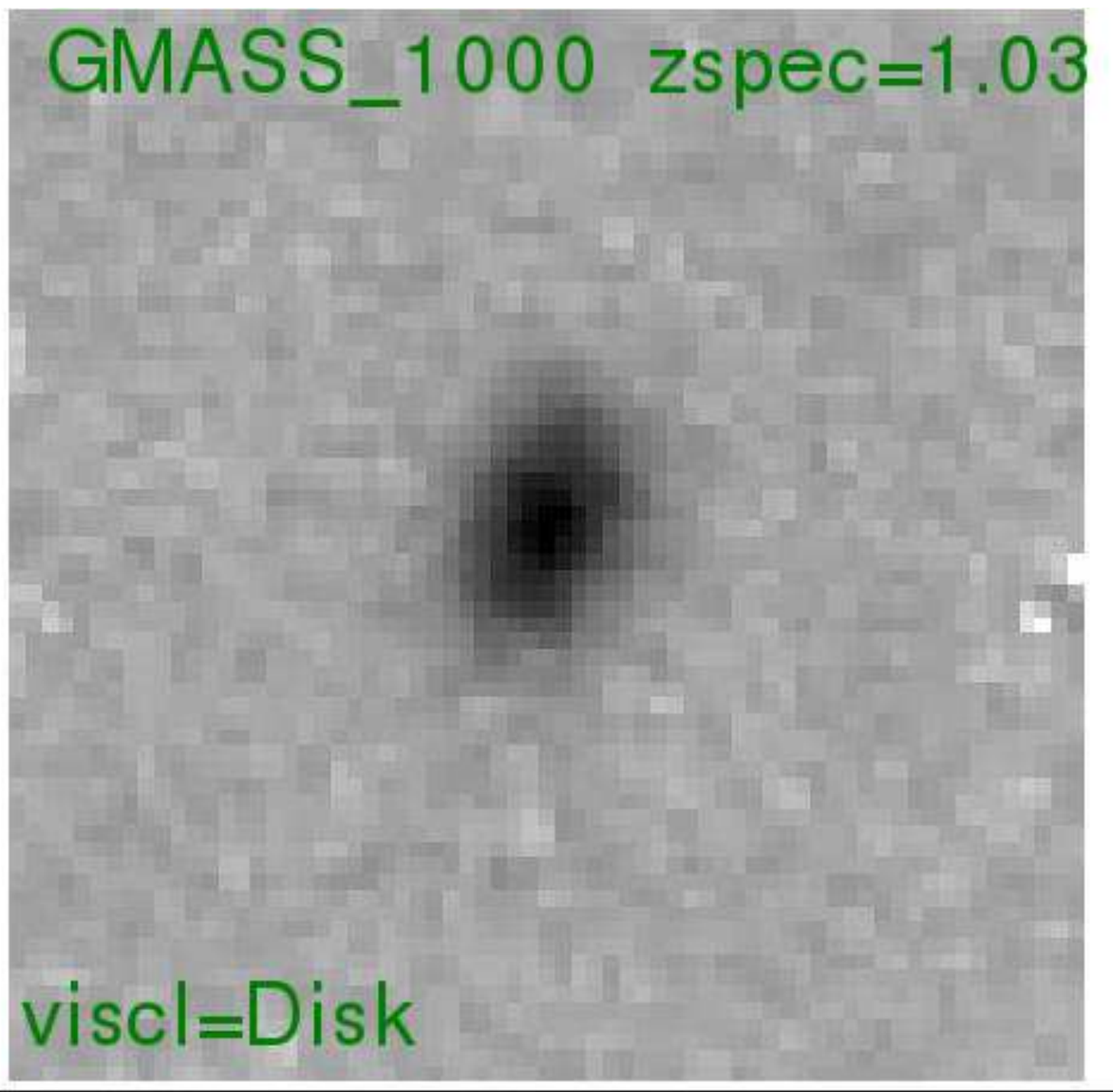}			     
\includegraphics[trim=100 40 75 390, clip=true, width=30mm]{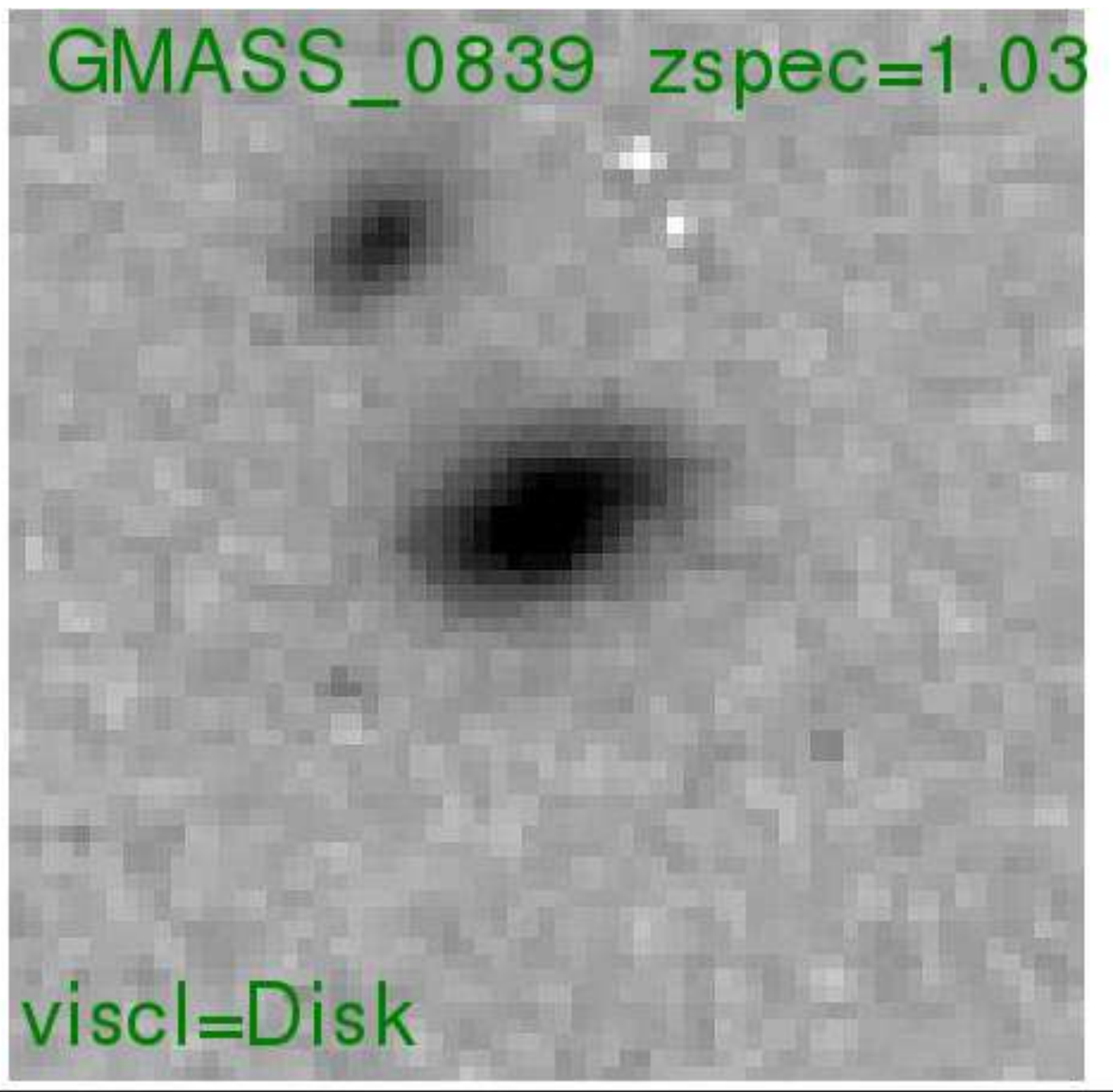}			     
\includegraphics[trim=100 40 75 390, clip=true, width=30mm]{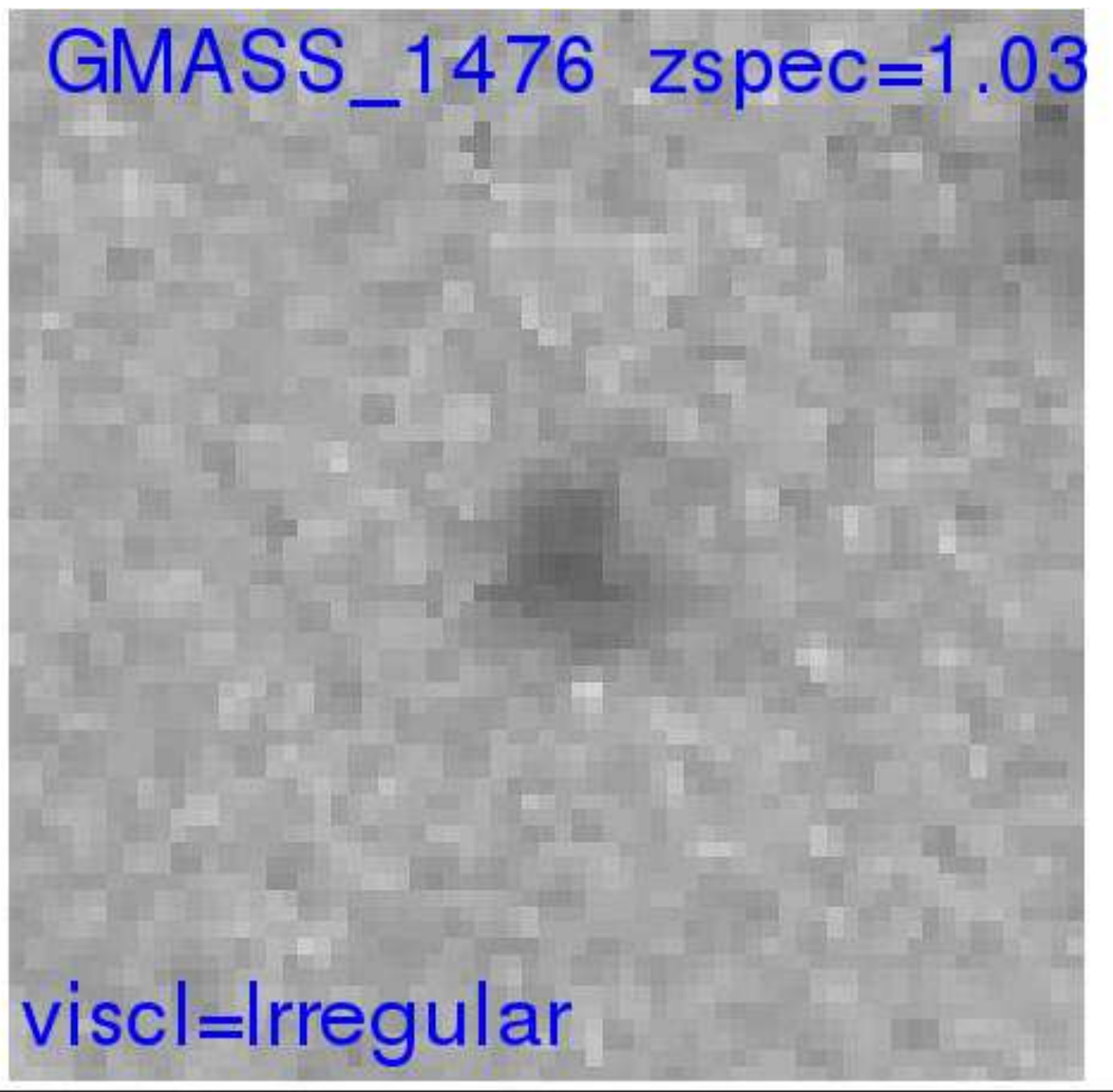}		     

\includegraphics[trim=100 40 75 390, clip=true, width=30mm]{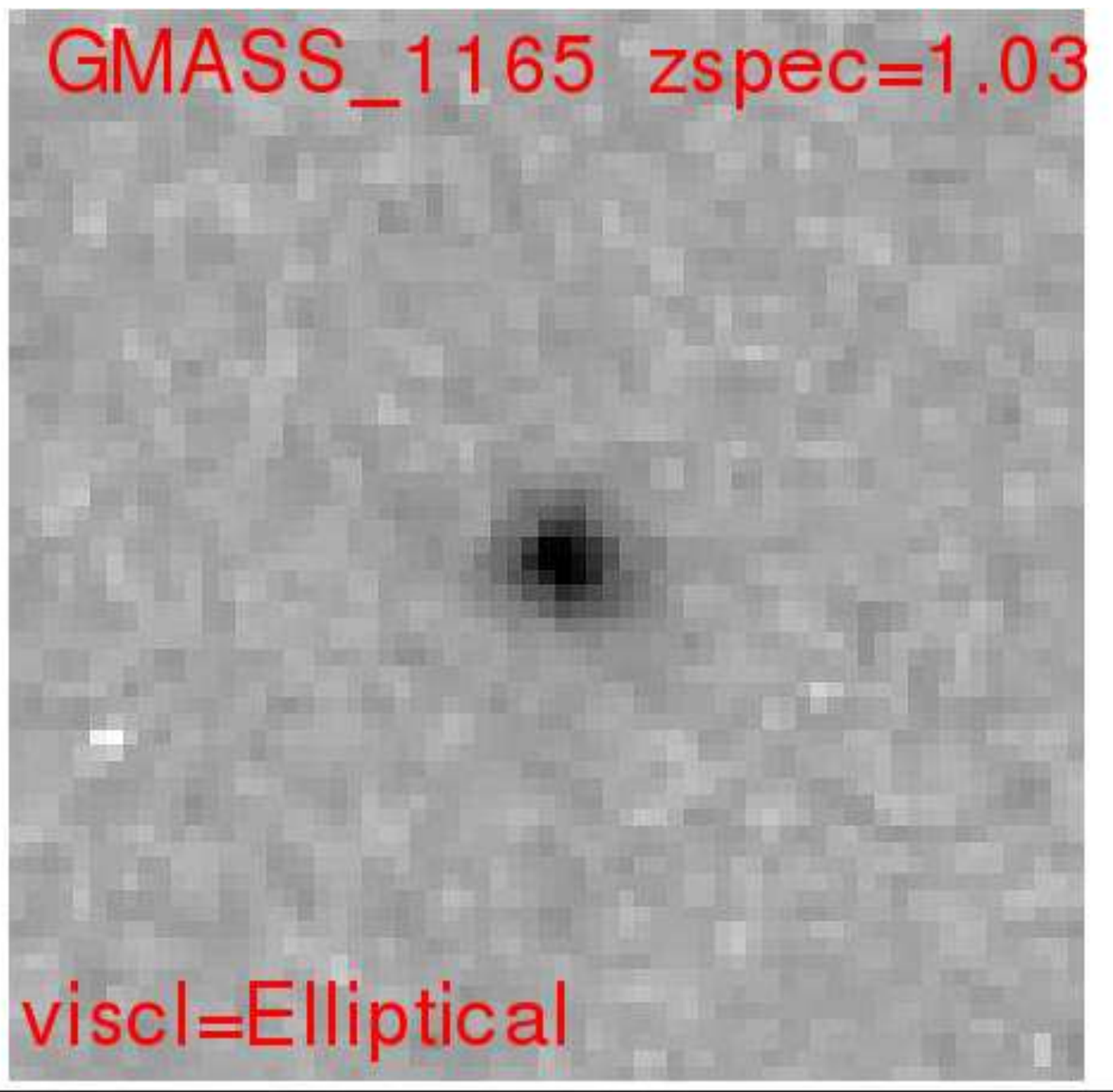}			     
\includegraphics[trim=100 40 75 390, clip=true, width=30mm]{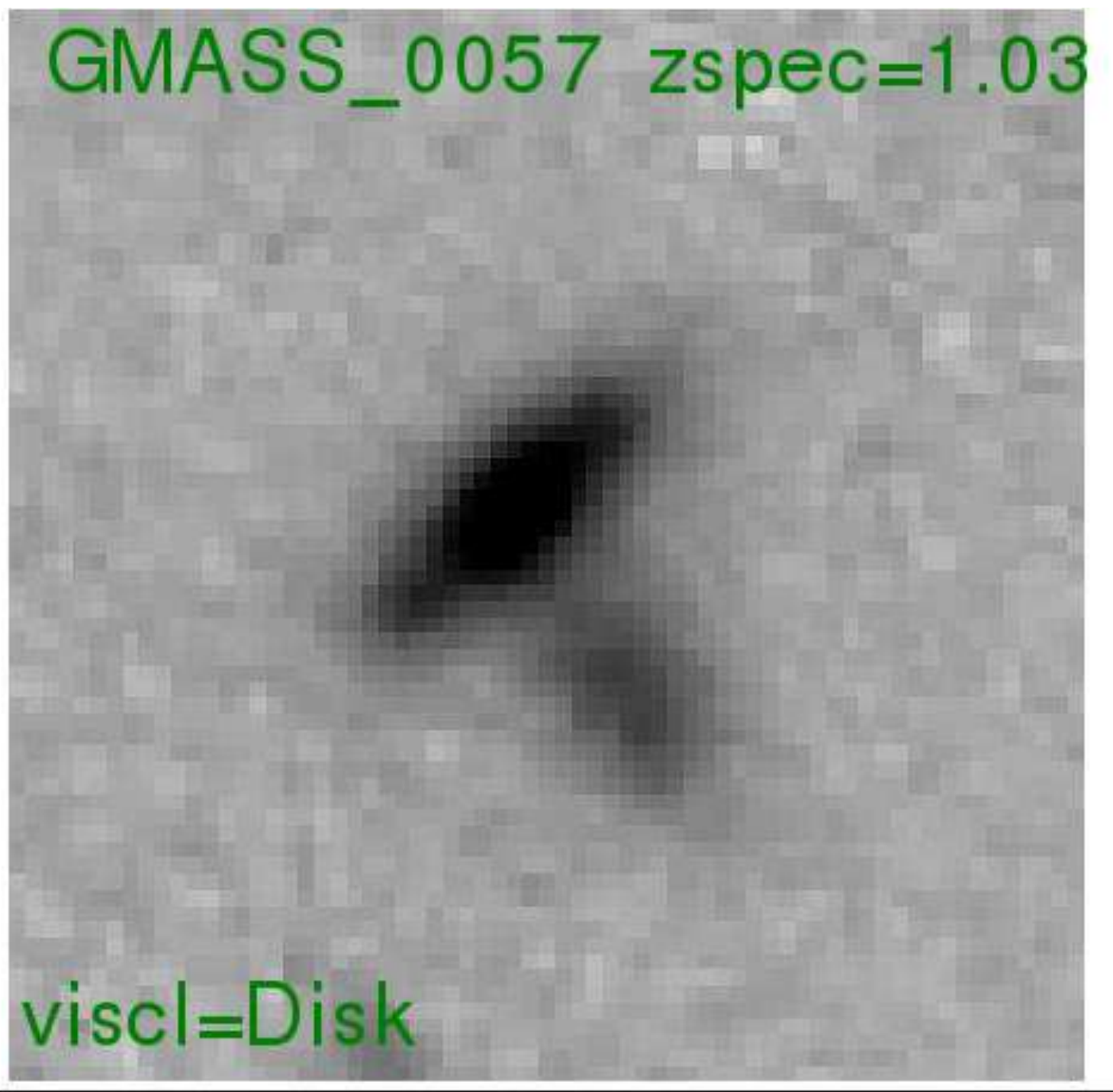}			     
\includegraphics[trim=100 40 75 390, clip=true, width=30mm]{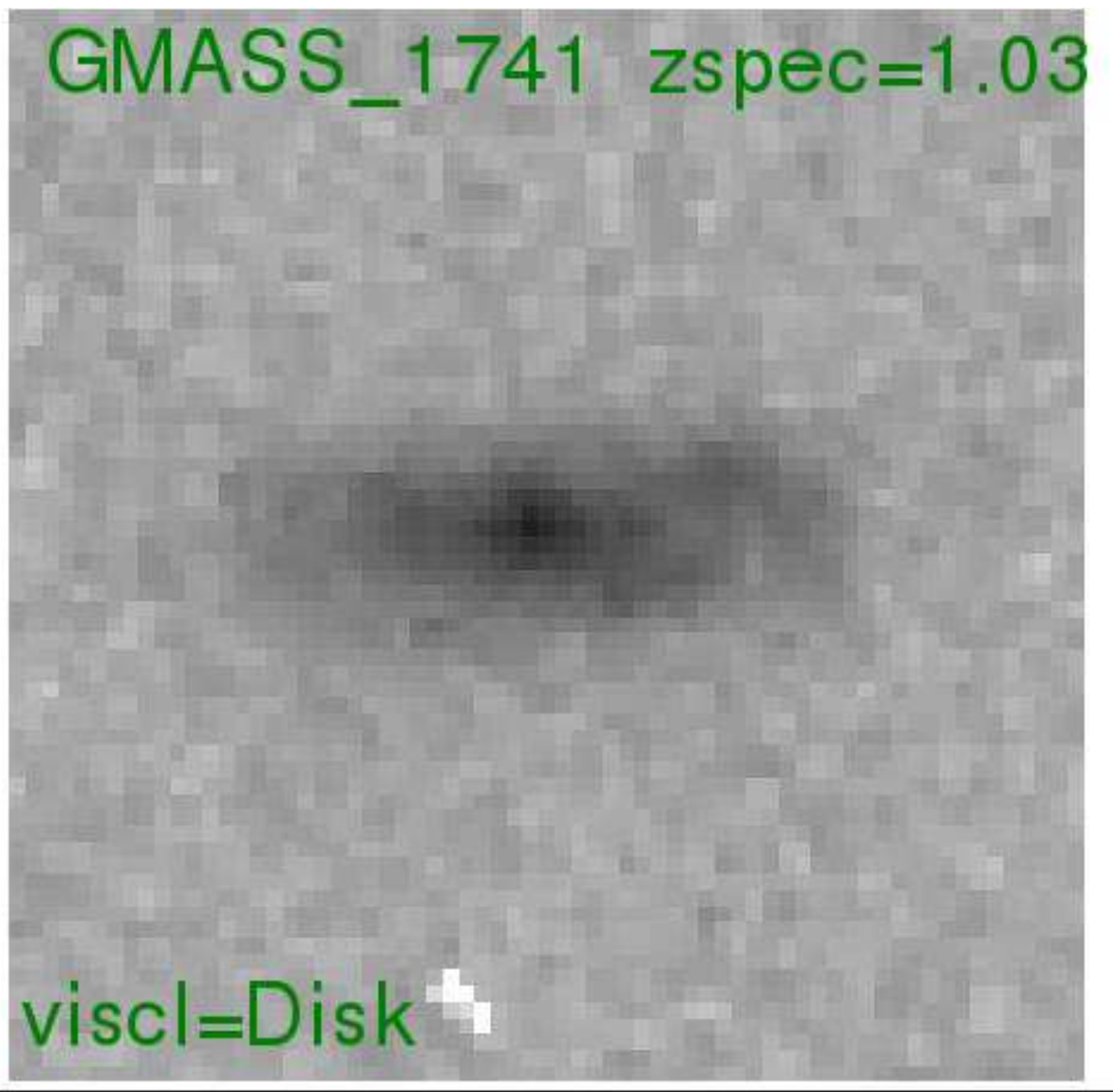}			     
\includegraphics[trim=100 40 75 390, clip=true, width=30mm]{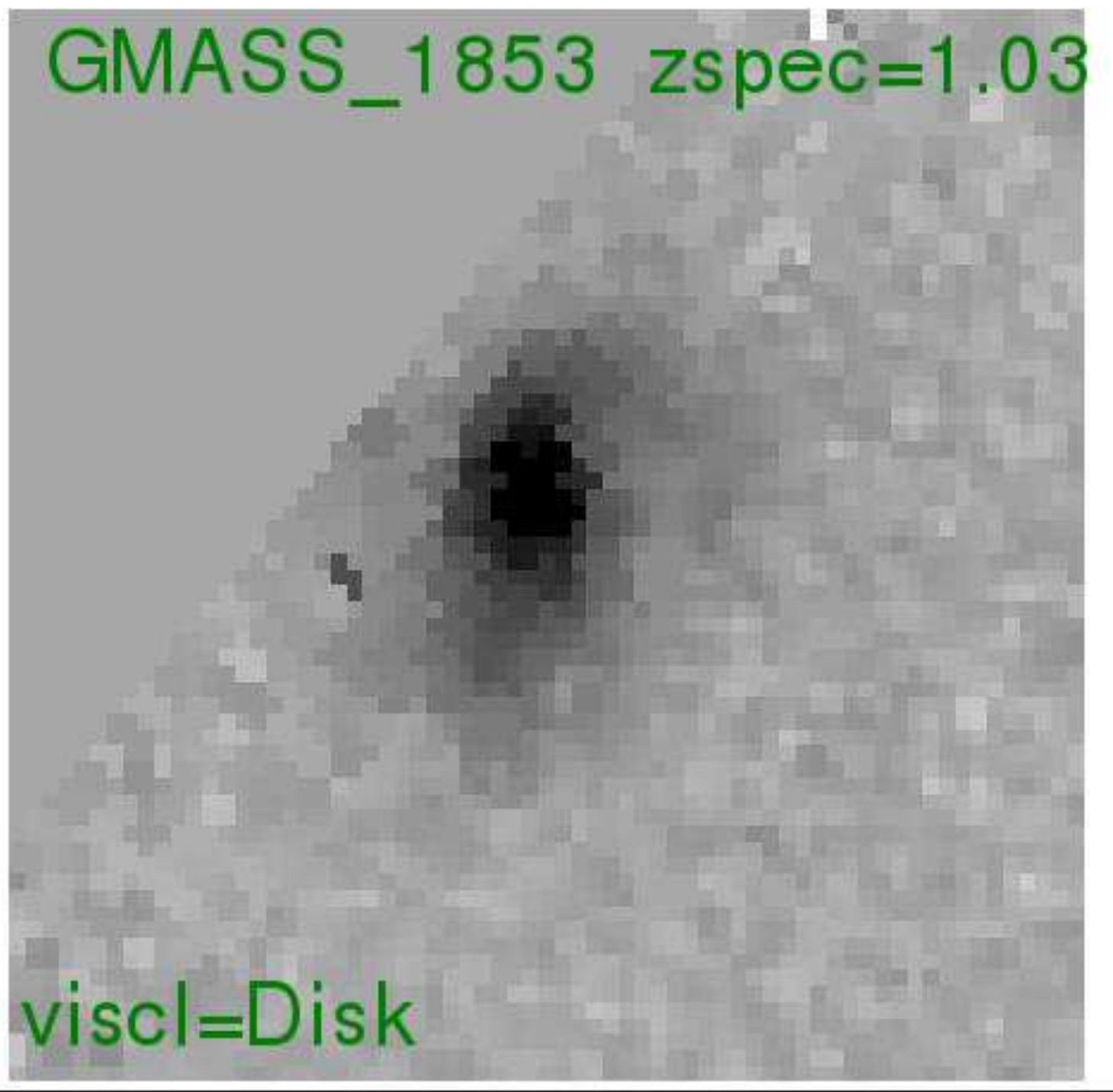}		     
\includegraphics[trim=100 40 75 390, clip=true, width=30mm]{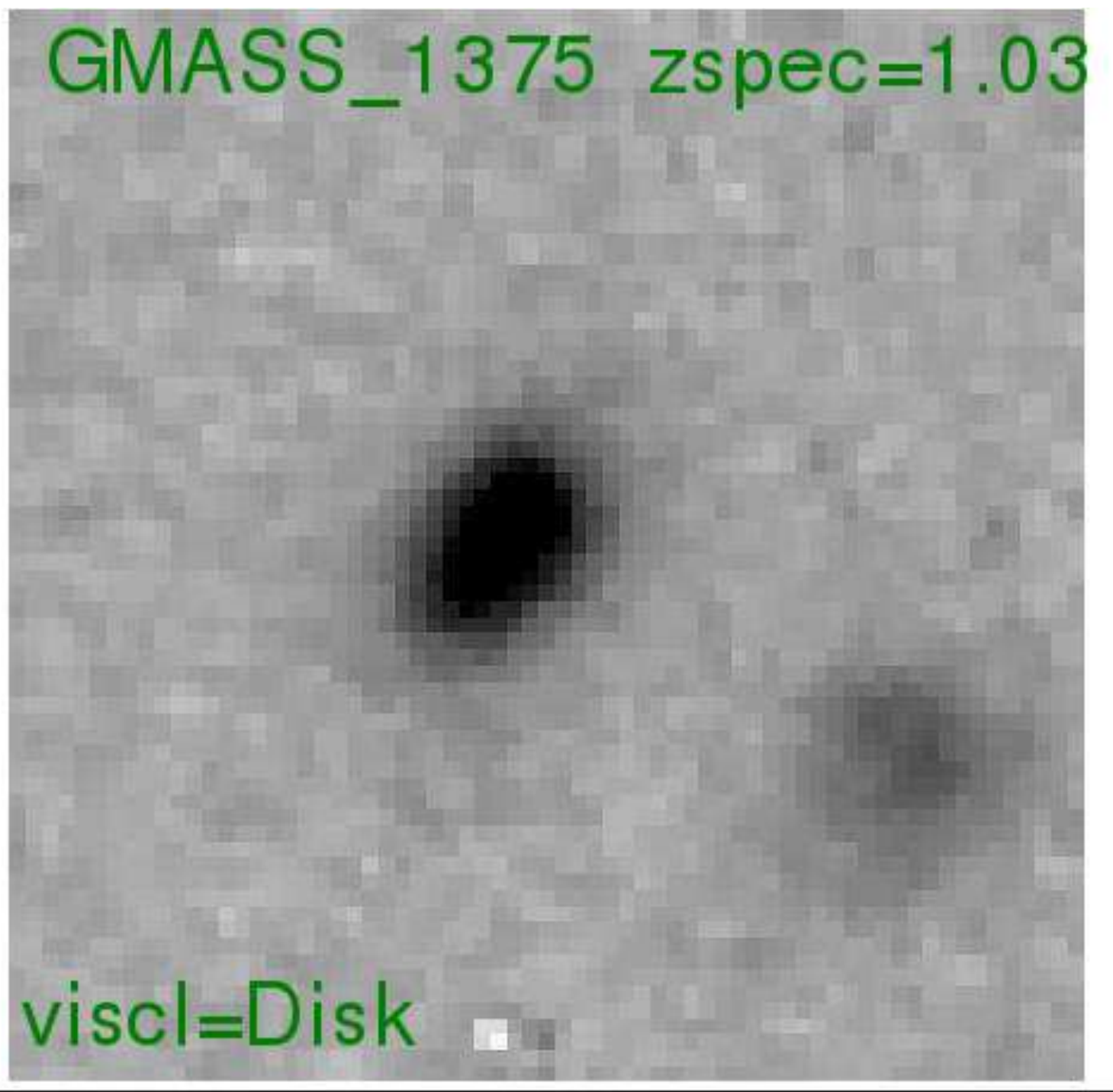}			     
\includegraphics[trim=100 40 75 390, clip=true, width=30mm]{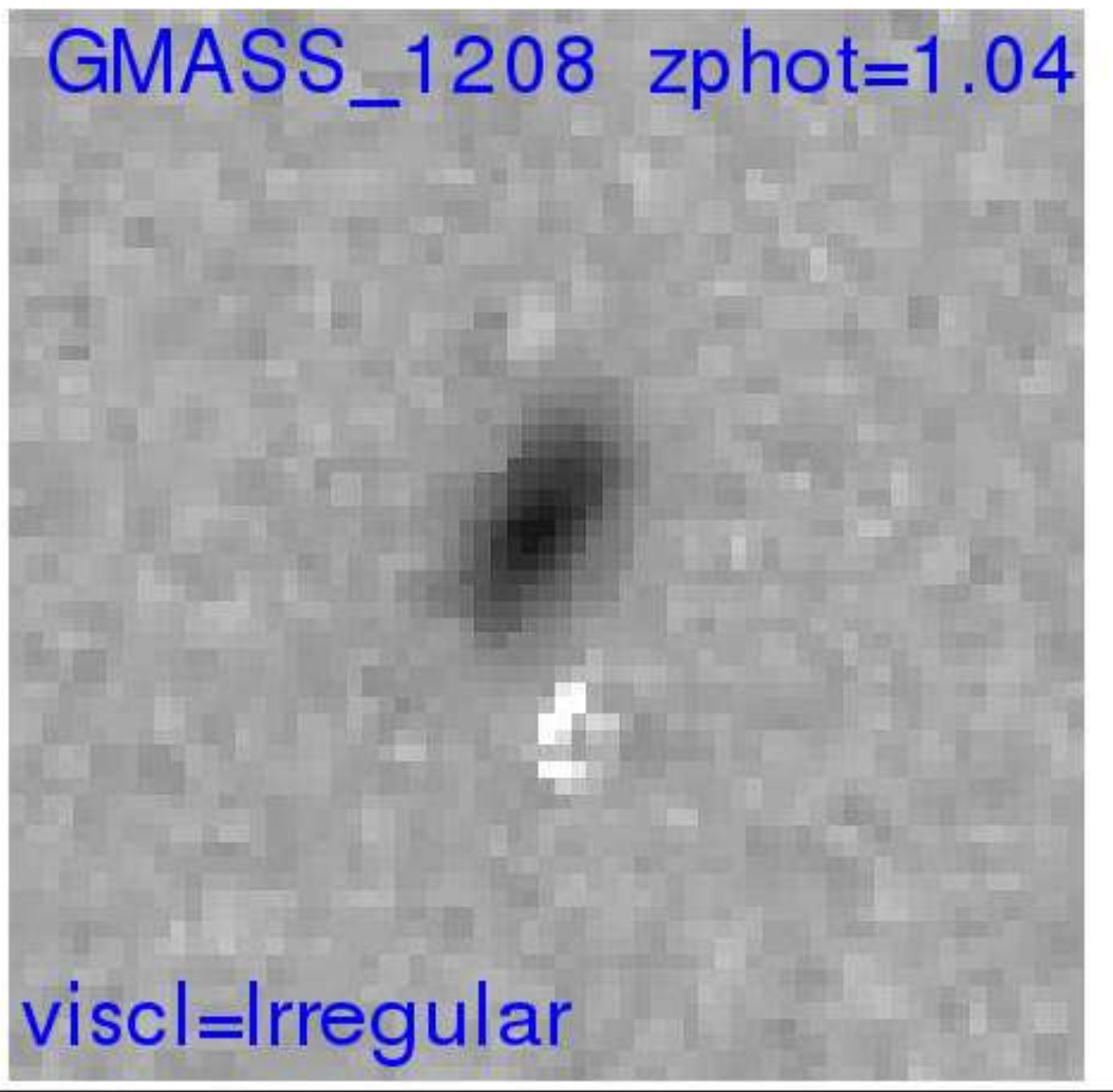}			     

\includegraphics[trim=100 40 75 390, clip=true, width=30mm]{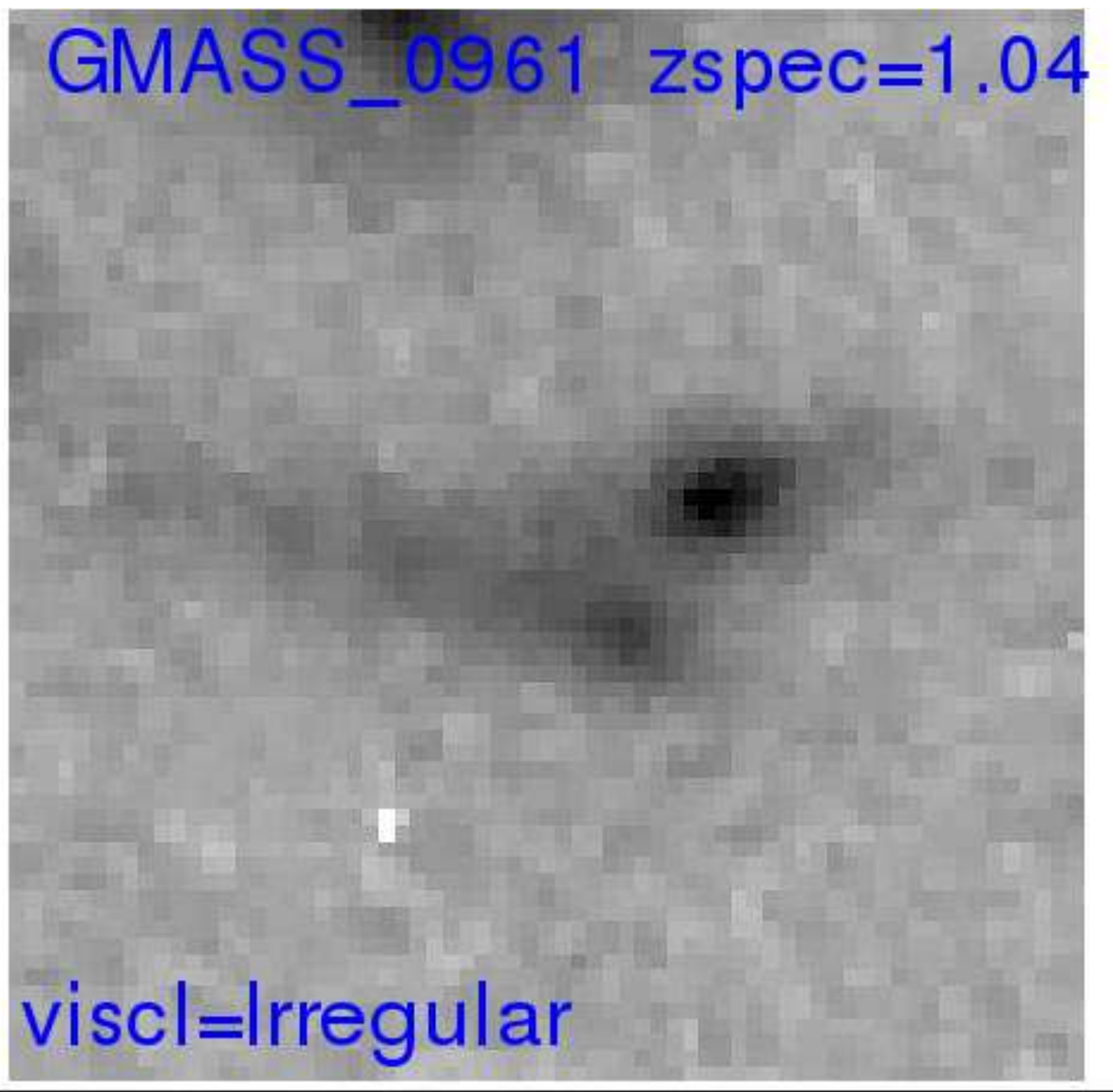}			     
\includegraphics[trim=100 40 75 390, clip=true, width=30mm]{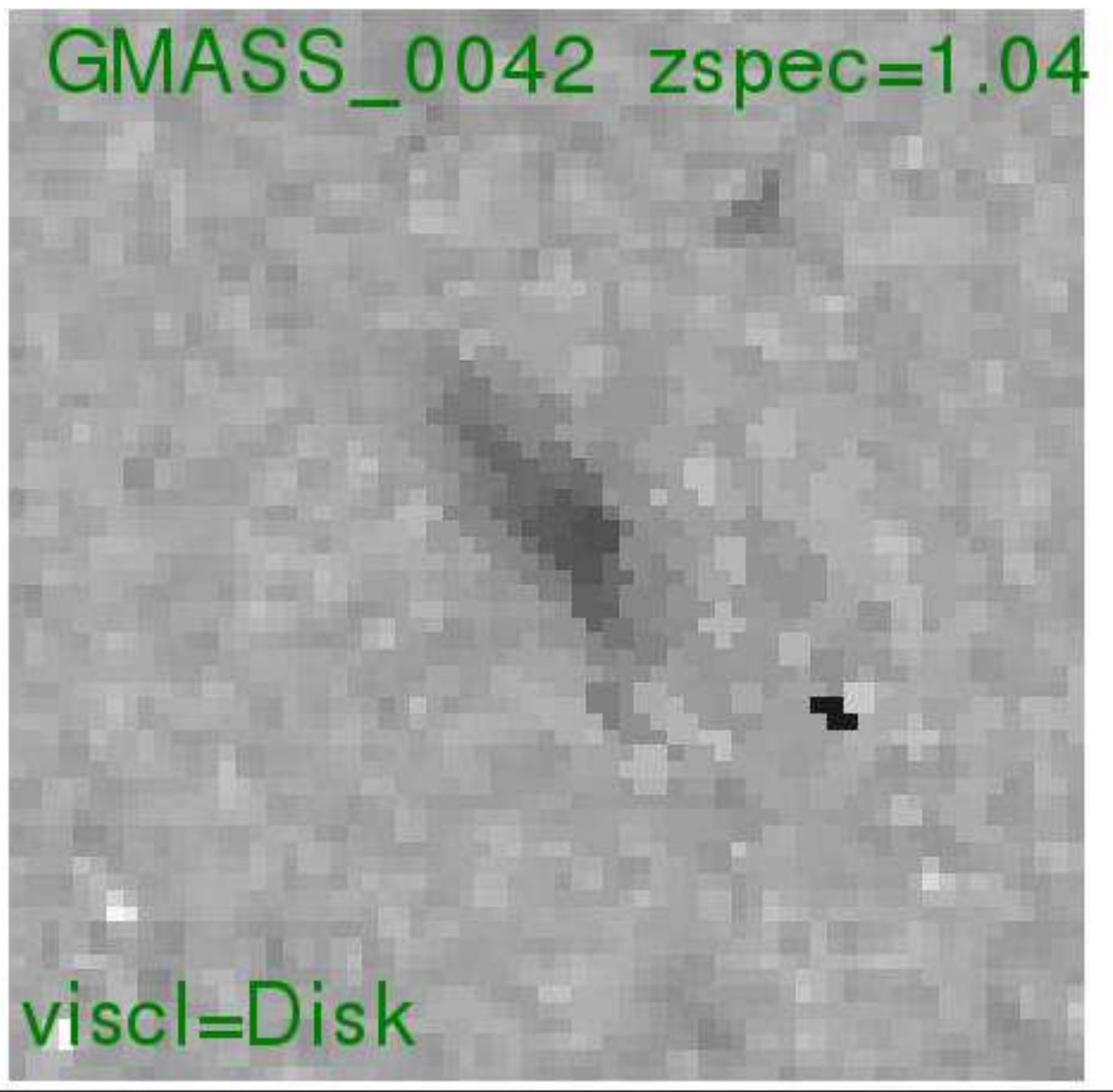}			     
\includegraphics[trim=100 40 75 390, clip=true, width=30mm]{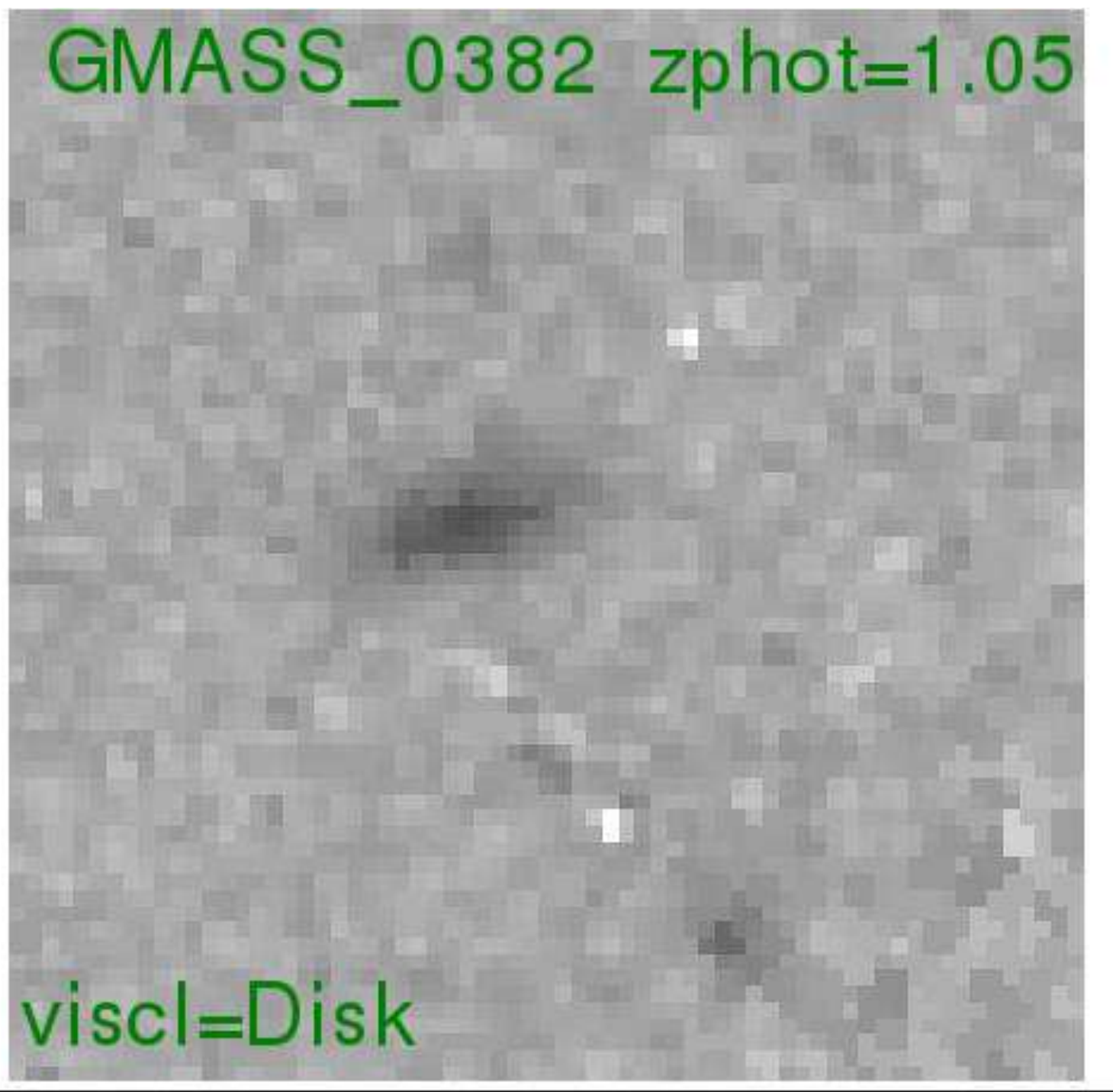}			     
\includegraphics[trim=100 40 75 390, clip=true, width=30mm]{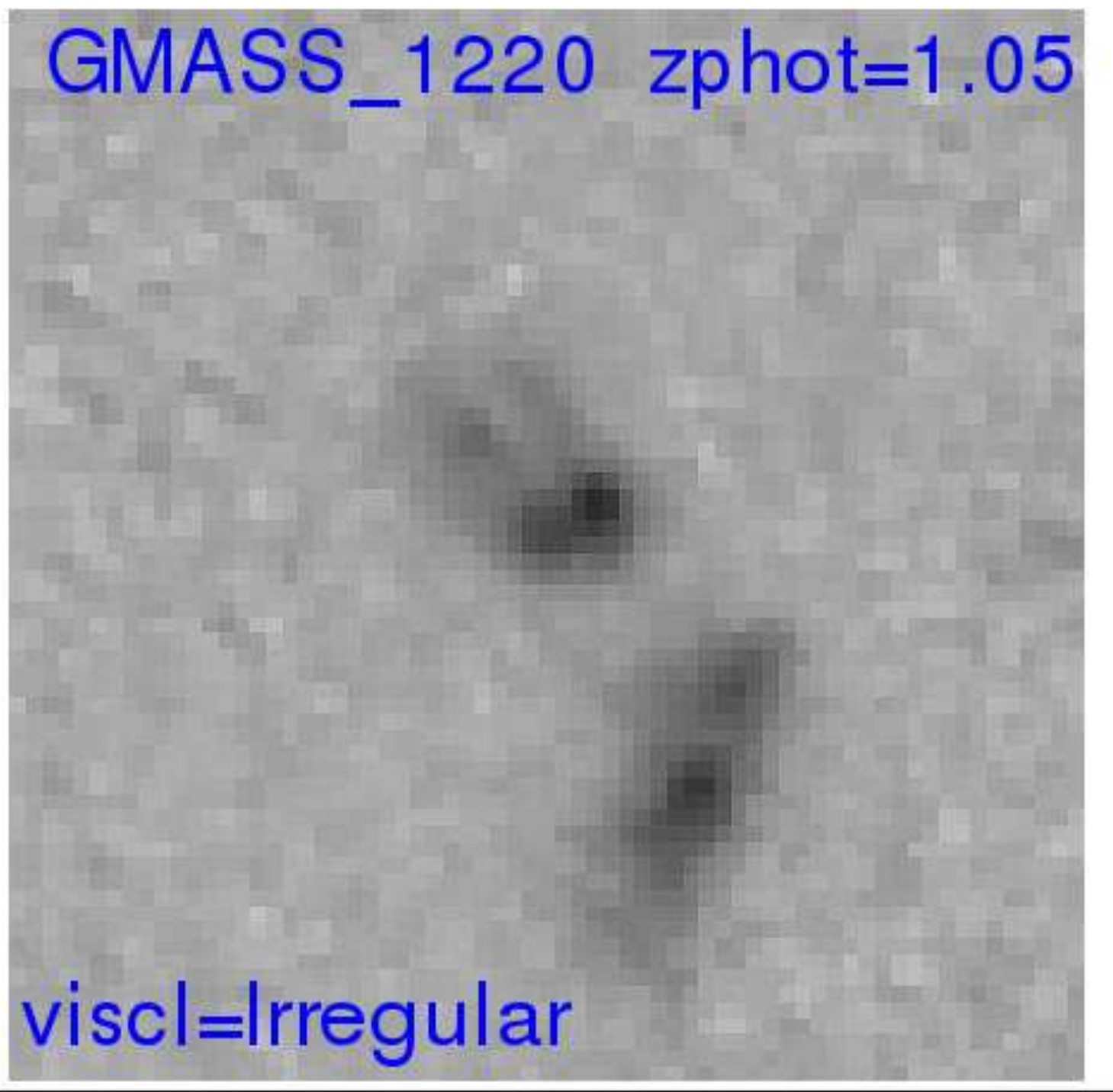}			     
\includegraphics[trim=100 40 75 390, clip=true, width=30mm]{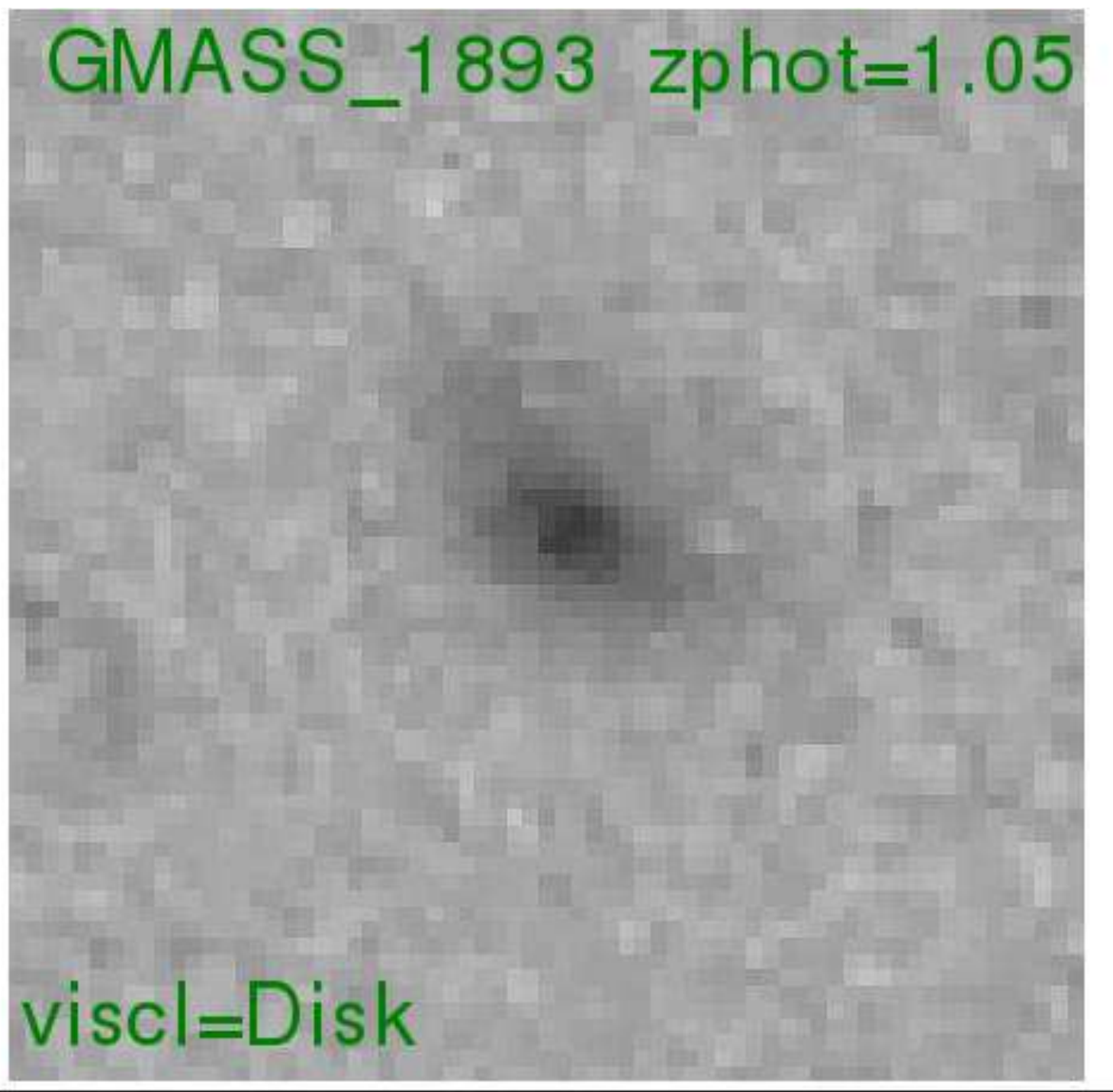}			     
\includegraphics[trim=100 40 75 390, clip=true, width=30mm]{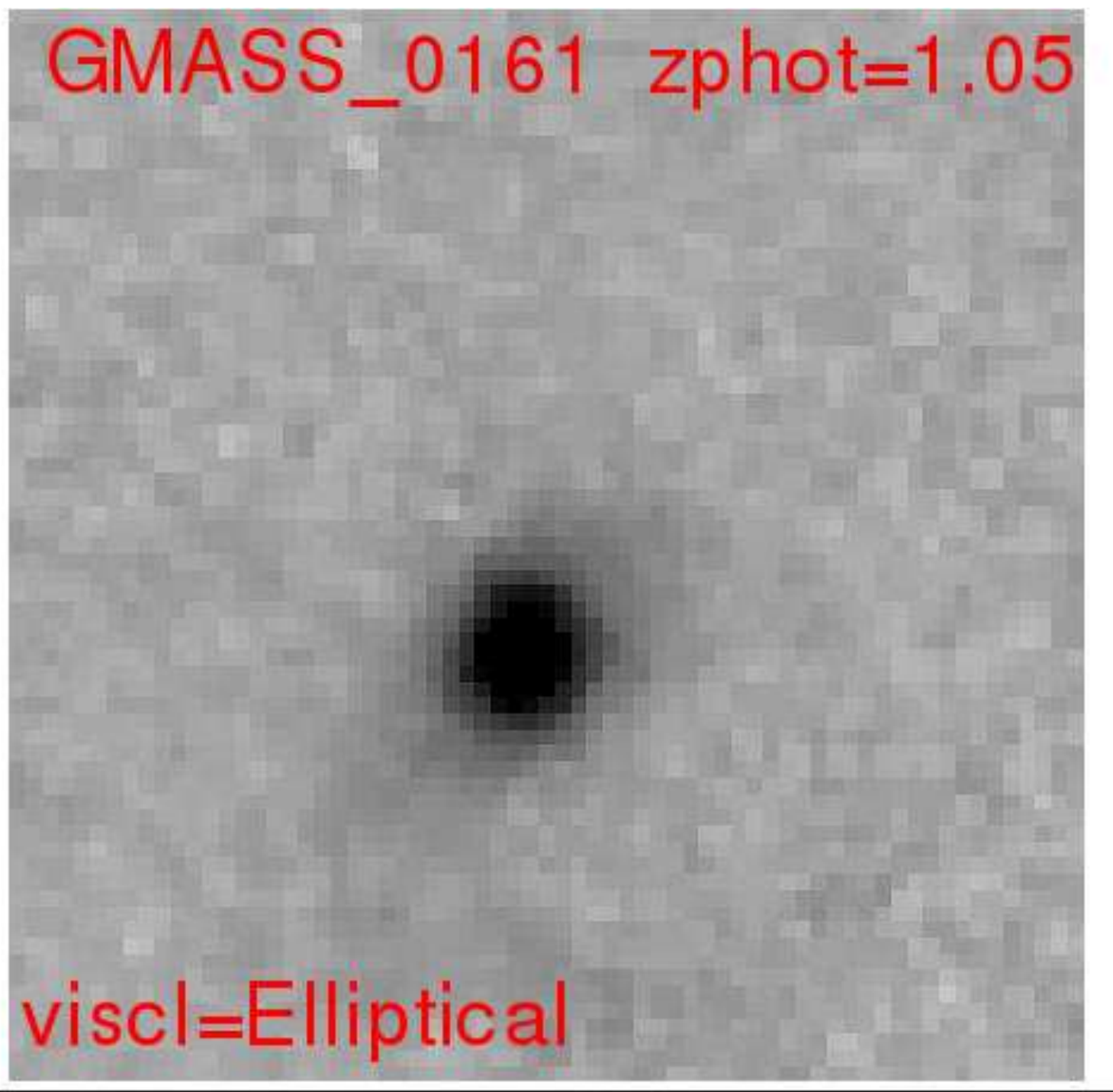}		     

\includegraphics[trim=100 40 75 390, clip=true, width=30mm]{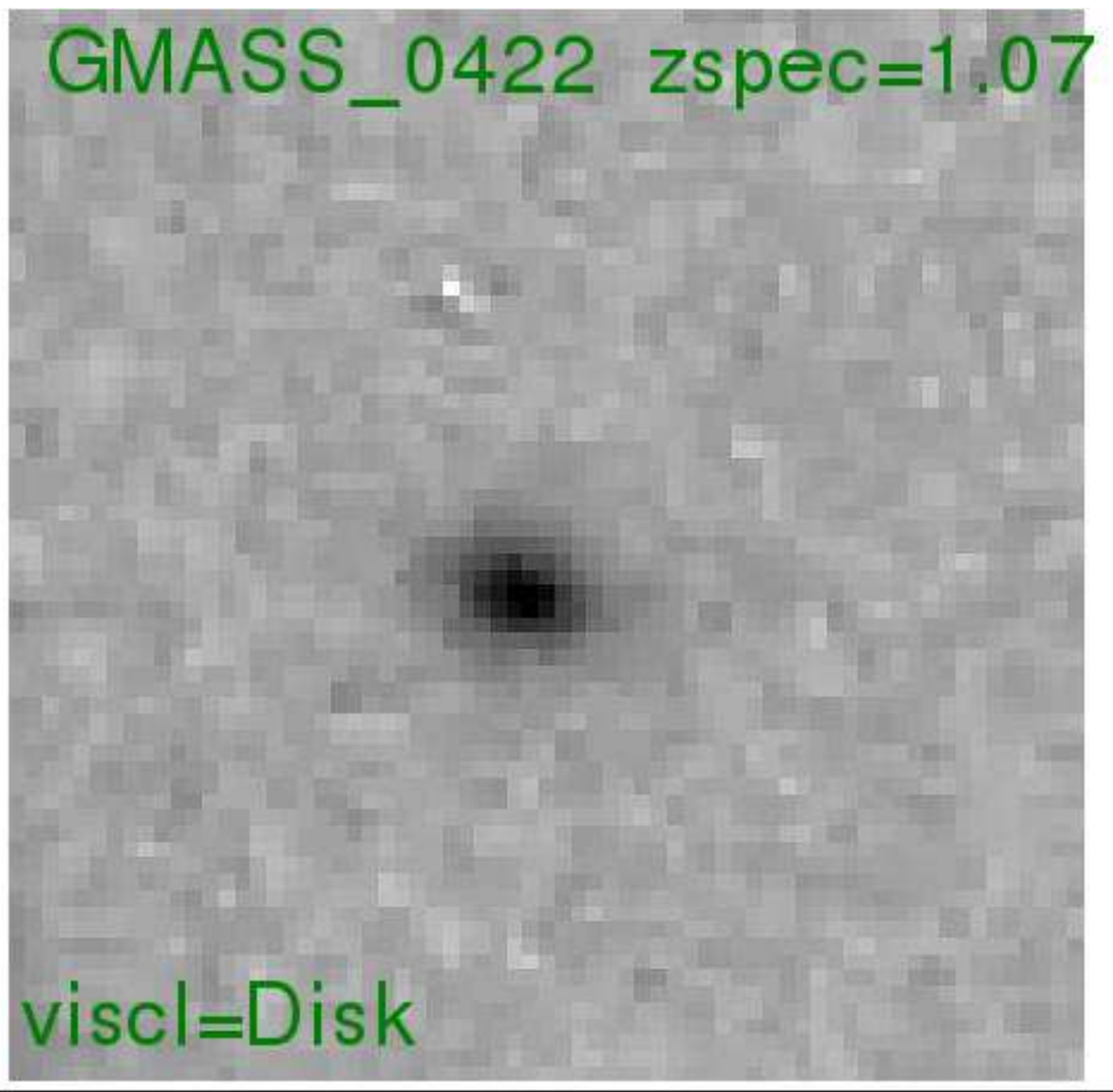}			     
\includegraphics[trim=100 40 75 390, clip=true, width=30mm]{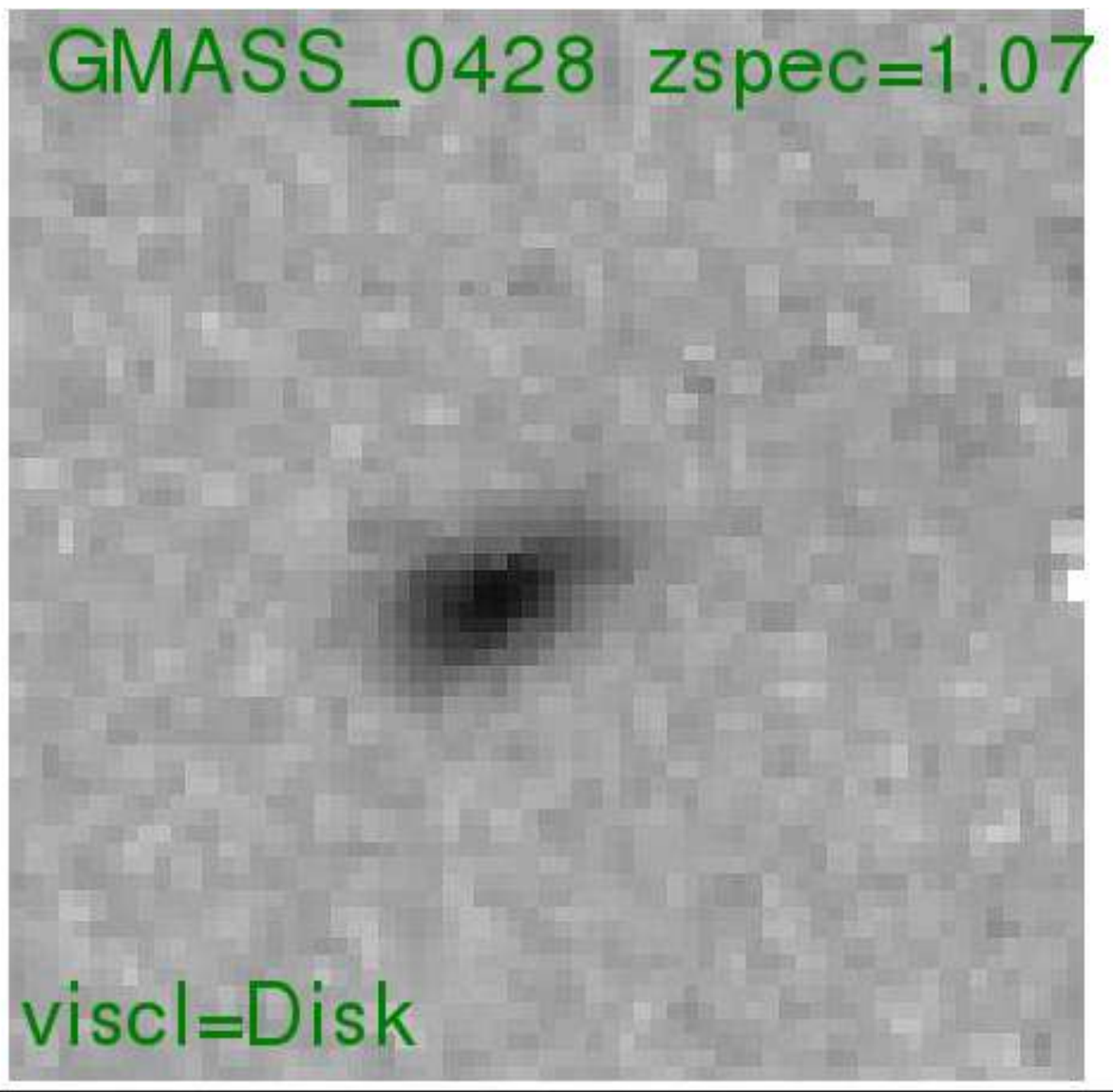}			     
\includegraphics[trim=100 40 75 390, clip=true, width=30mm]{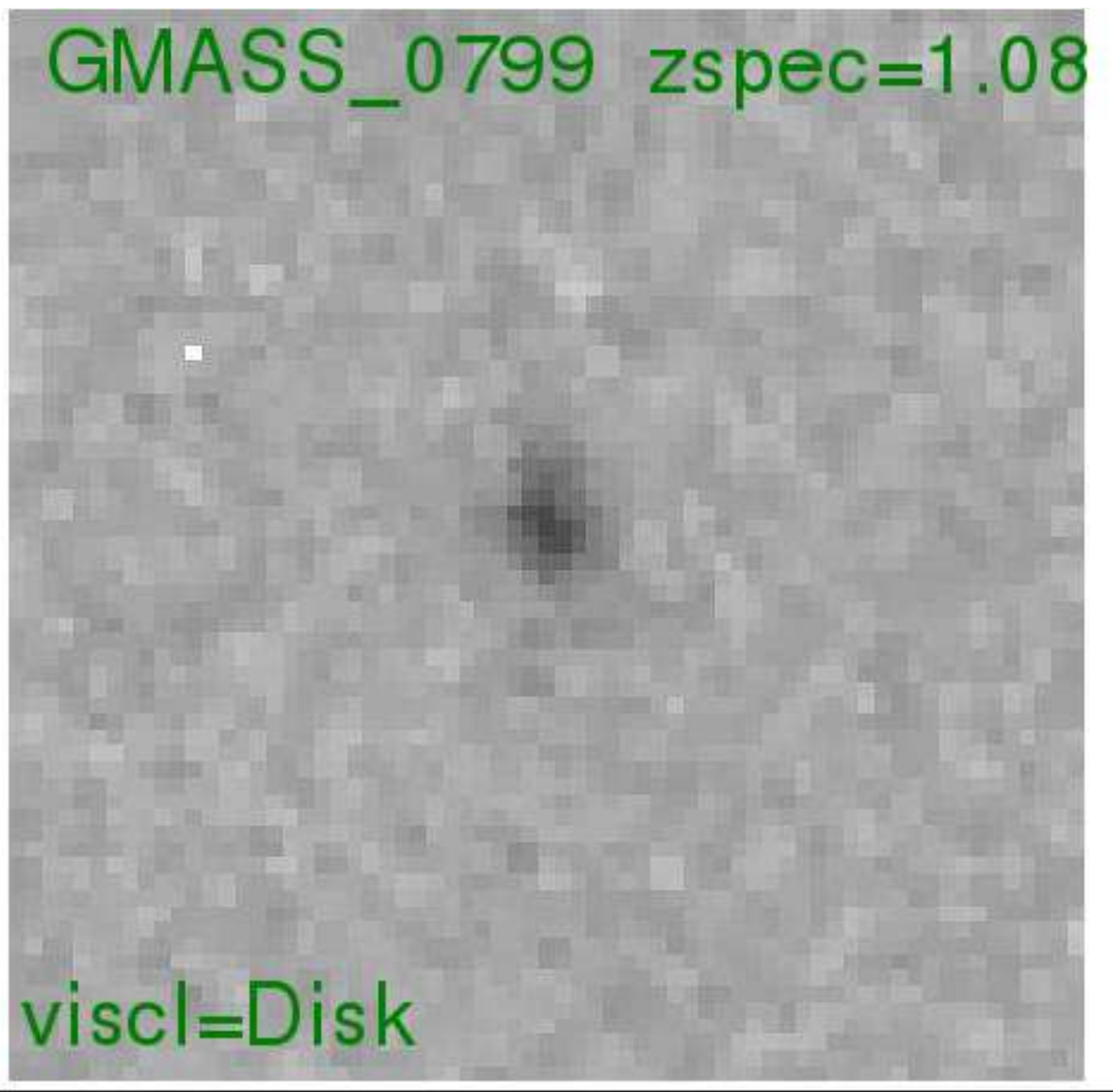}			     
\includegraphics[trim=100 40 75 390, clip=true, width=30mm]{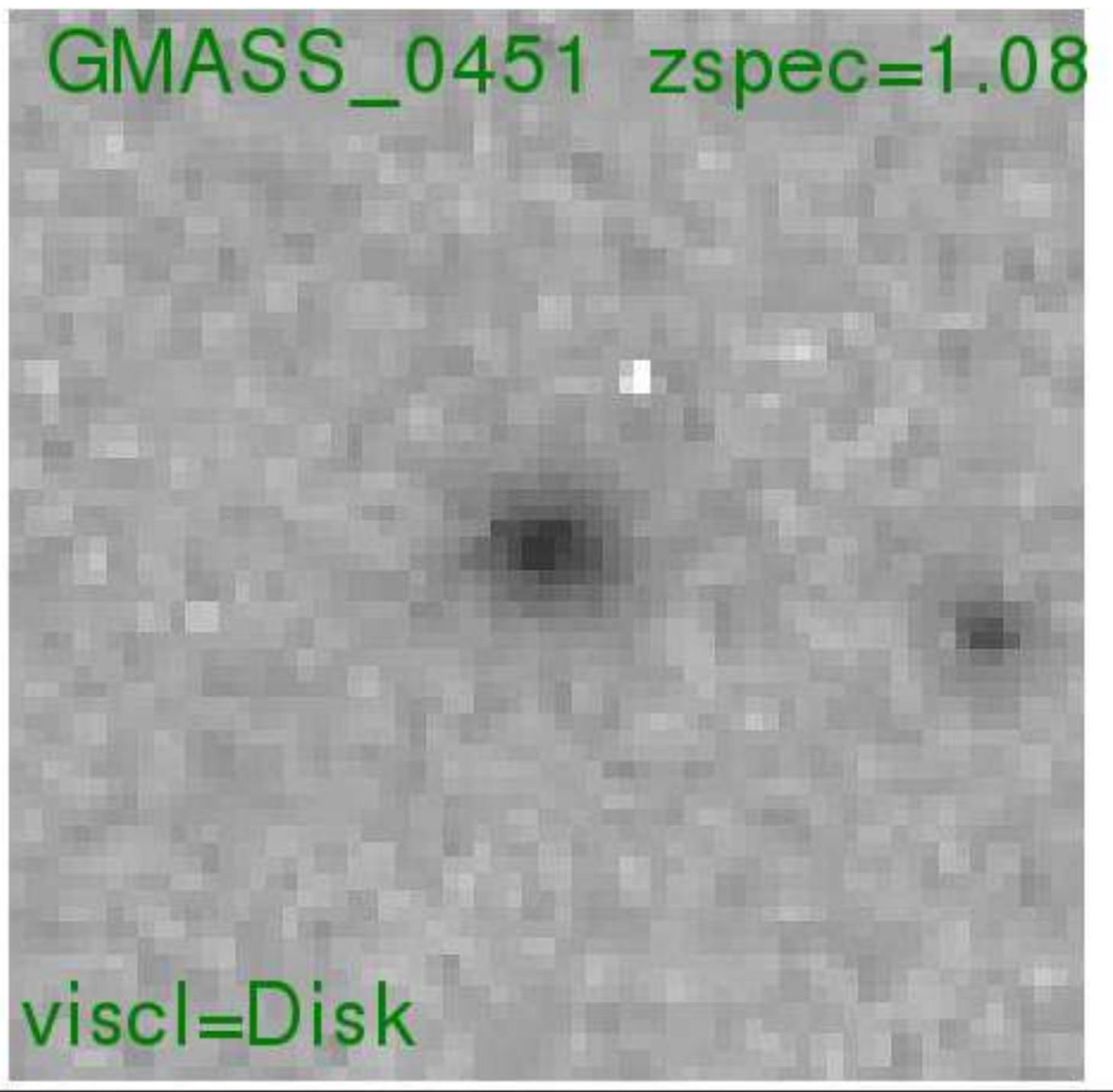}			     
\includegraphics[trim=100 40 75 390, clip=true, width=30mm]{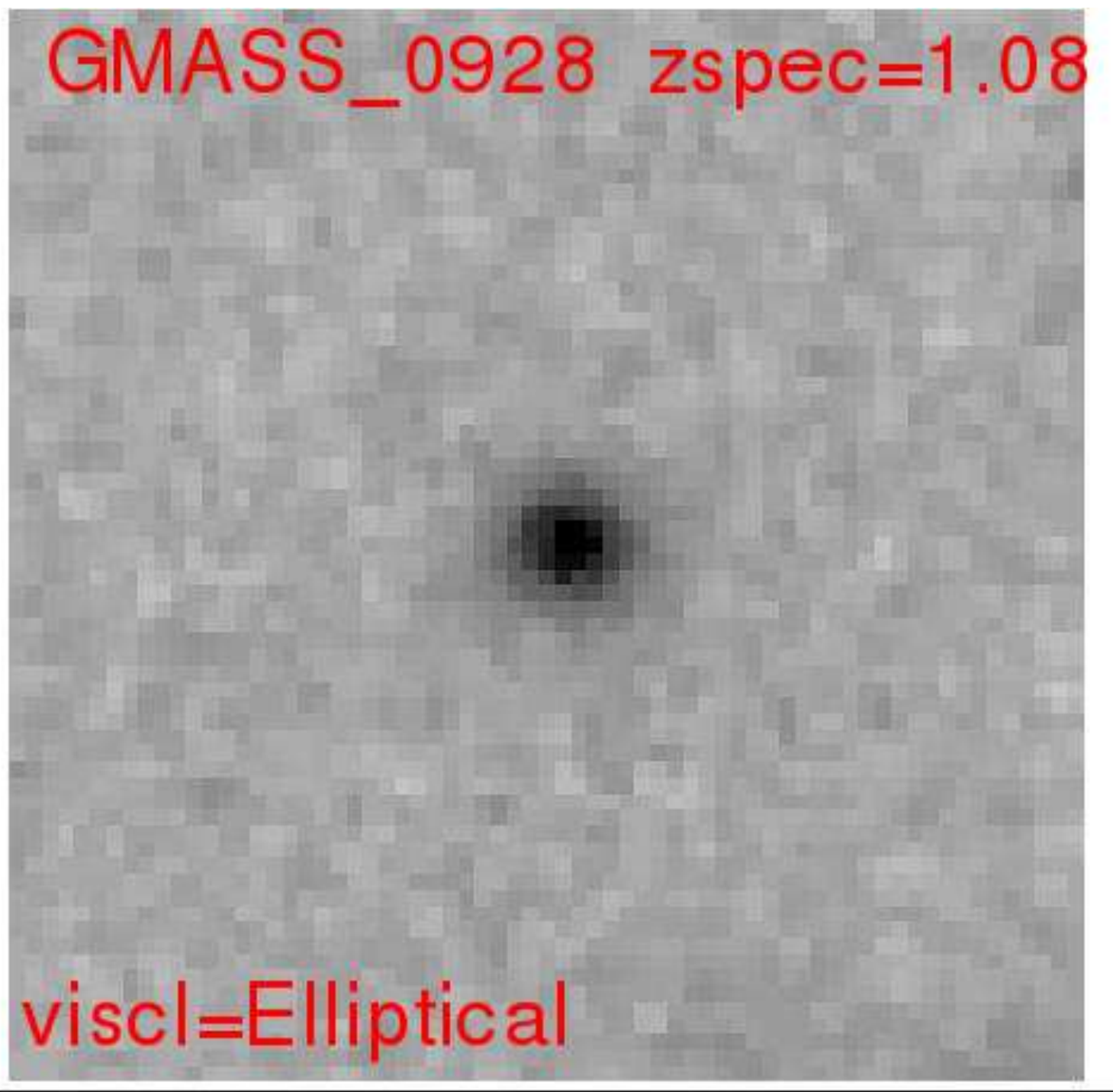}			     
\includegraphics[trim=100 40 75 390, clip=true, width=30mm]{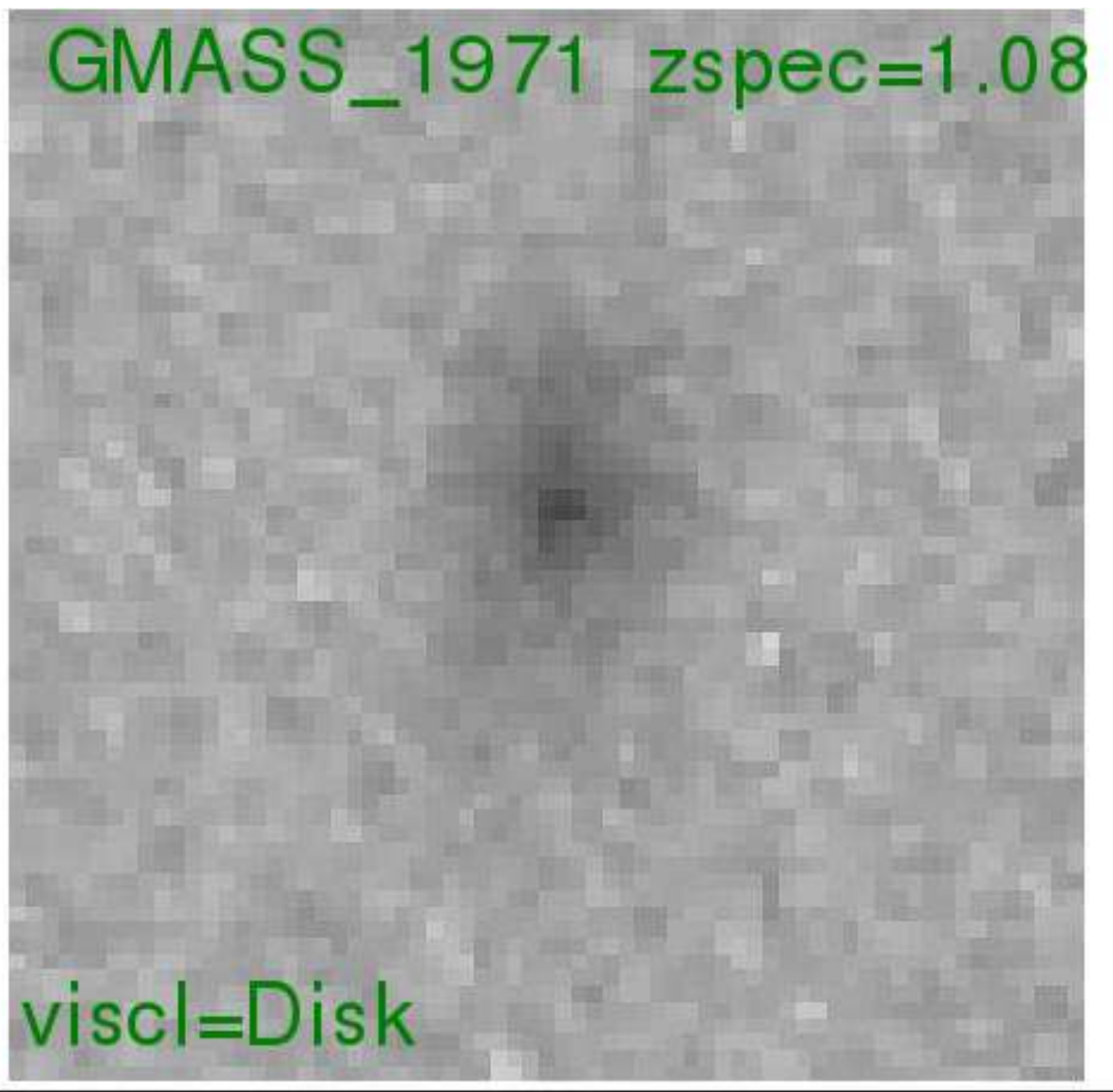}	
\caption{WFC3-IR H$_{160}$ cutouts of all the galaxies of our sample.}
\label{atlas}
\end{figure*}

Fig. \ref{atlas} shows WFC3-IR H$_{160}$ cutouts of all the galaxies of our sample, arranged in order of increasing redshift. The chosen display for the 4``$\times$4`` cutouts is a square root scale with fixed minimum and maximum pixel count limits, to ensure a uniform background. This means that some galaxy images may appear to be saturated. We point out, however, that the visual classification was done by looking at the images display in a linear scale too. On each snapshot, the GMASS identification number, galaxy redshift and visual classification are also indicated, using different colours according to the morphological type.

\newpage
\begin{figure*}
\centering
\includegraphics[trim=100 40 75 390, clip=true, width=30mm]{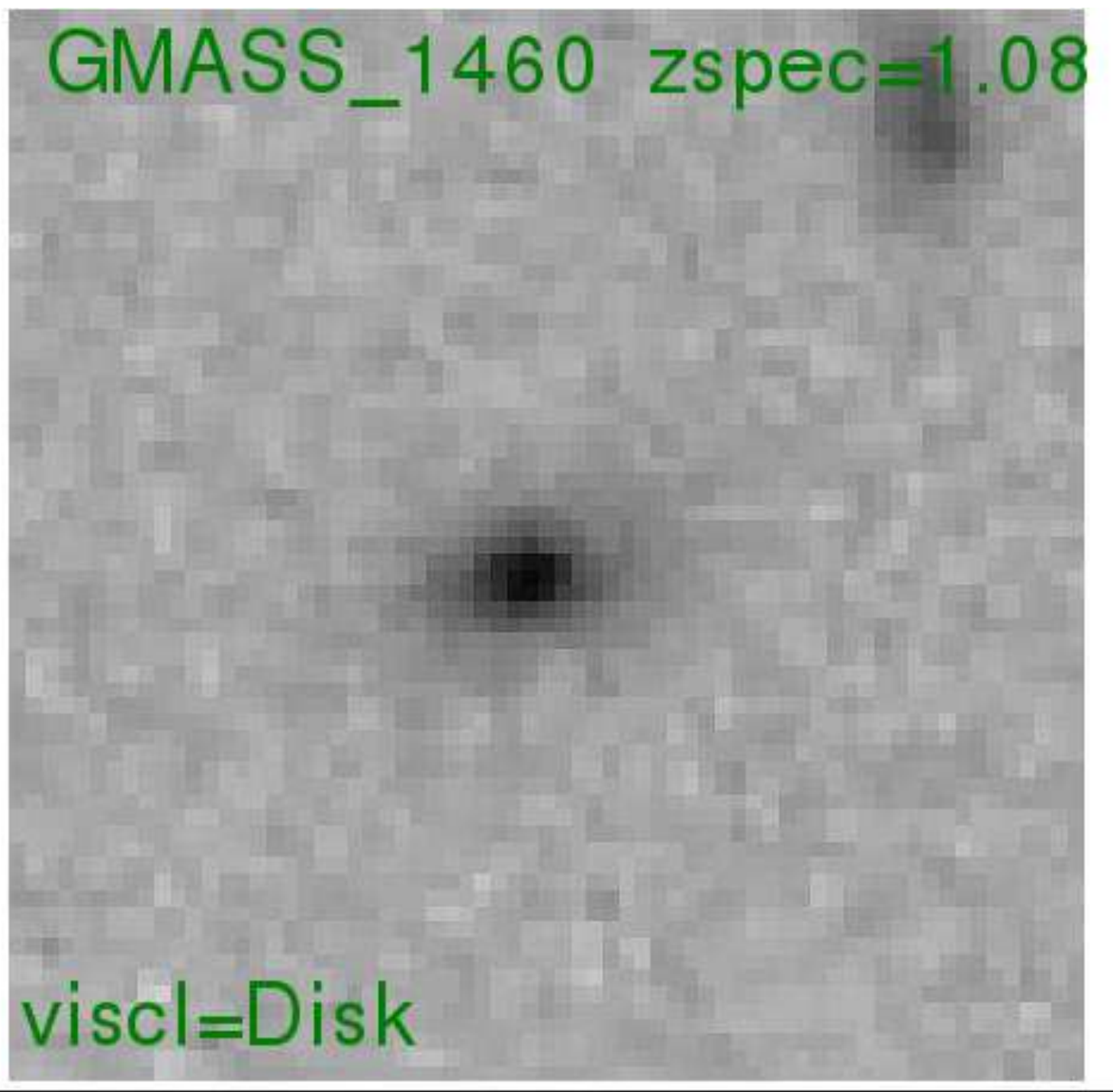}		     
\includegraphics[trim=100 40 75 390, clip=true, width=30mm]{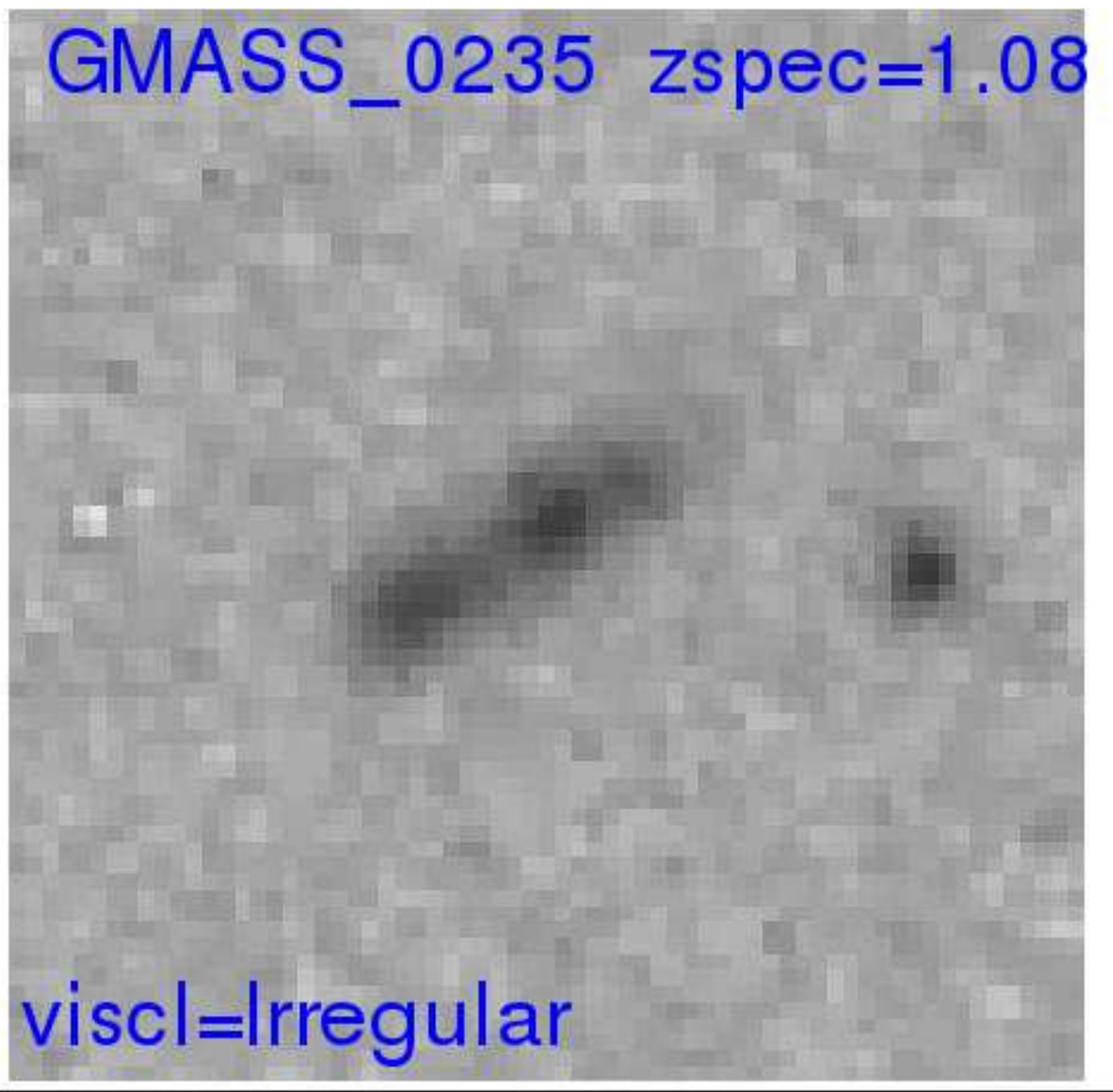}			     
\includegraphics[trim=100 40 75 390, clip=true, width=30mm]{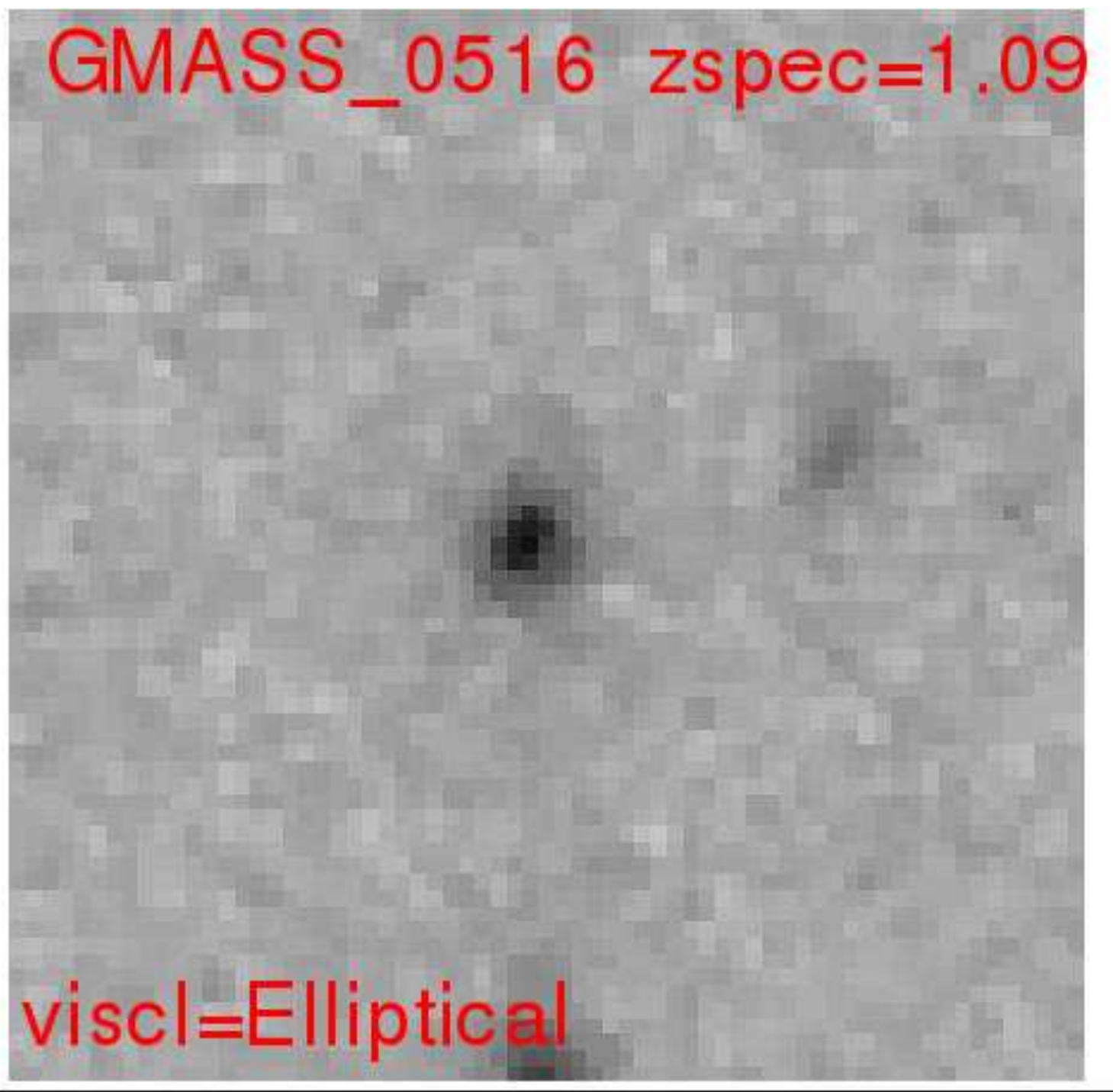}		     
\includegraphics[trim=100 40 75 390, clip=true, width=30mm]{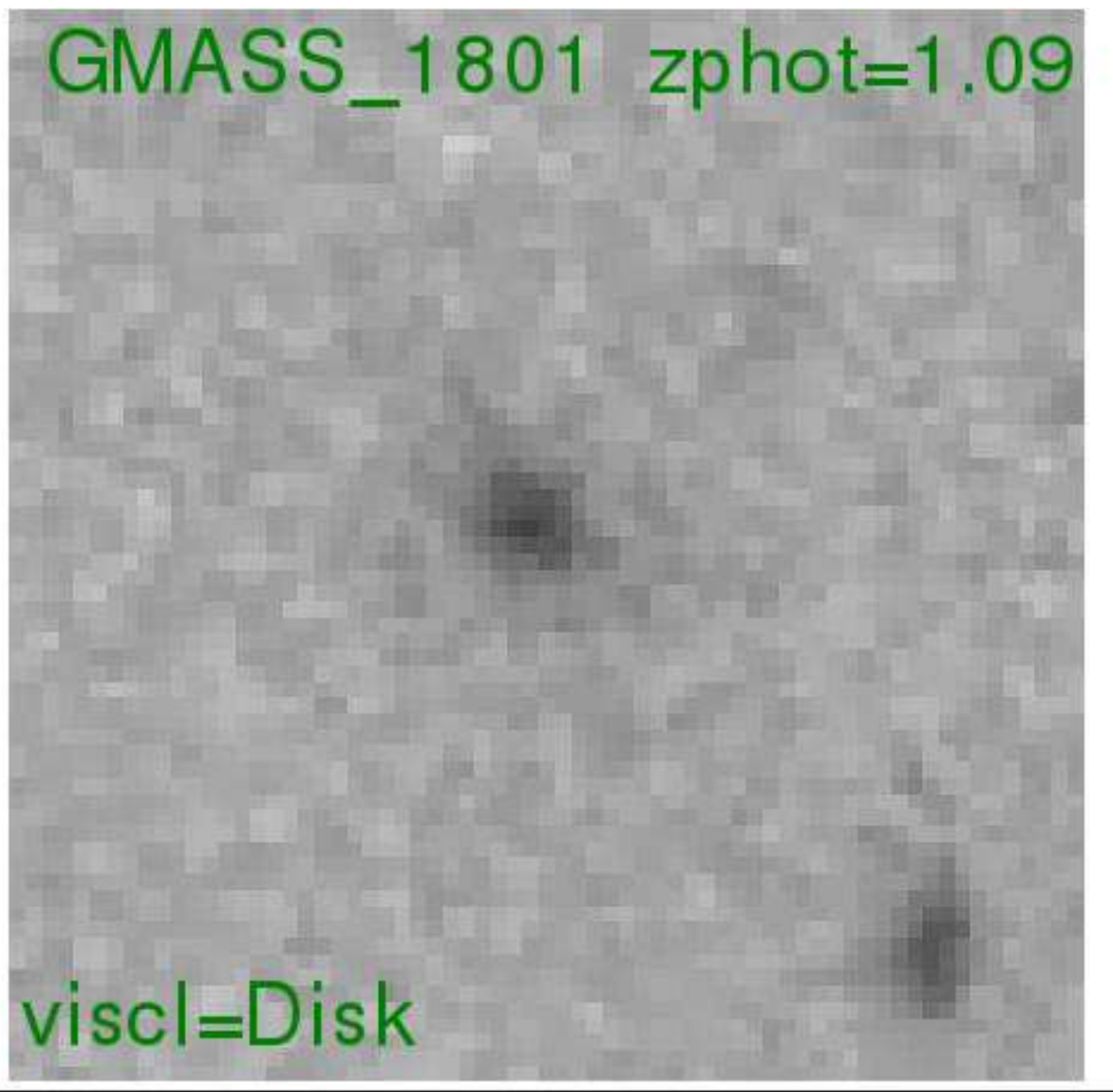}			     
\includegraphics[trim=100 40 75 390, clip=true, width=30mm]{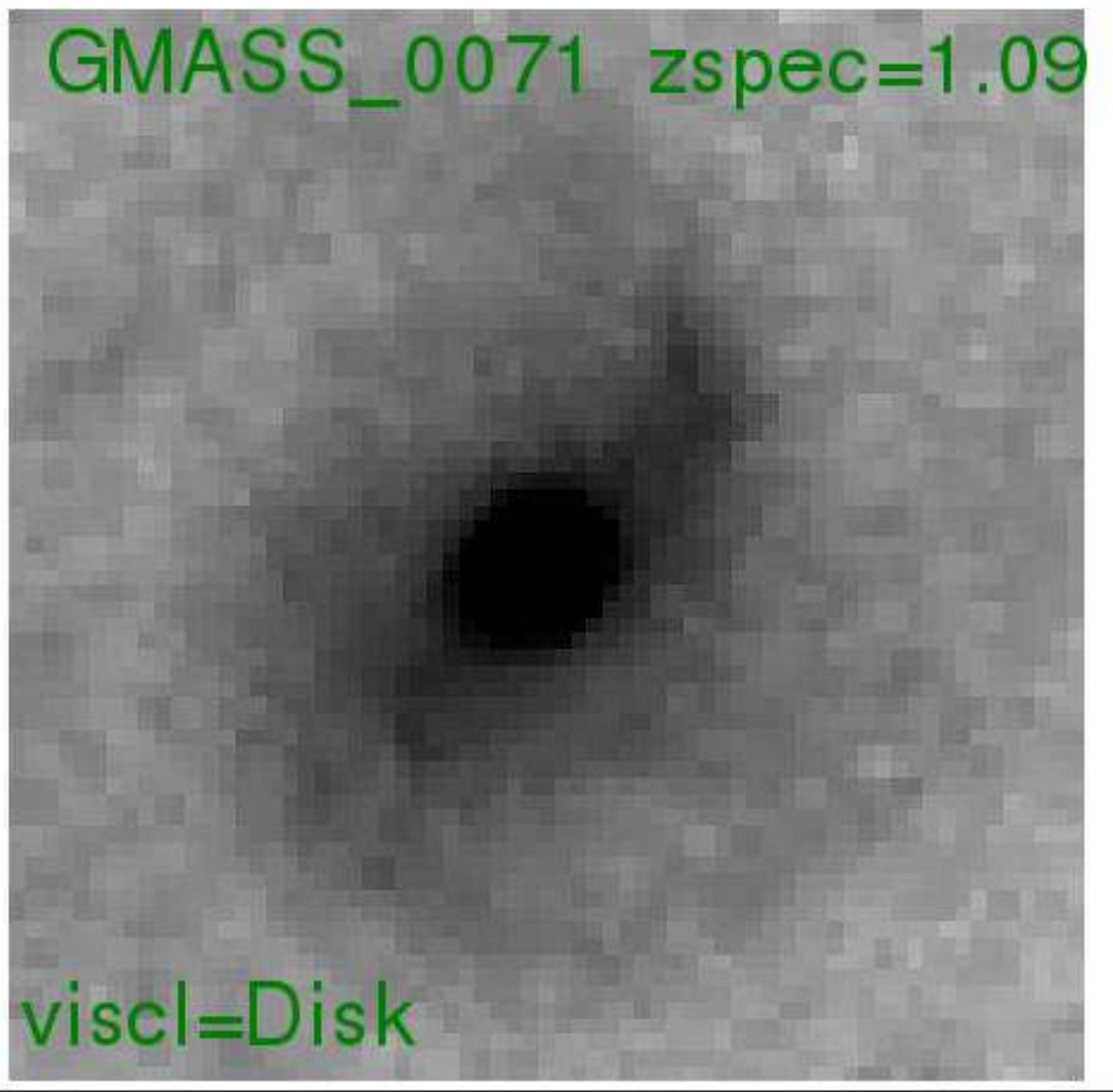}			     
\includegraphics[trim=100 40 75 390, clip=true, width=30mm]{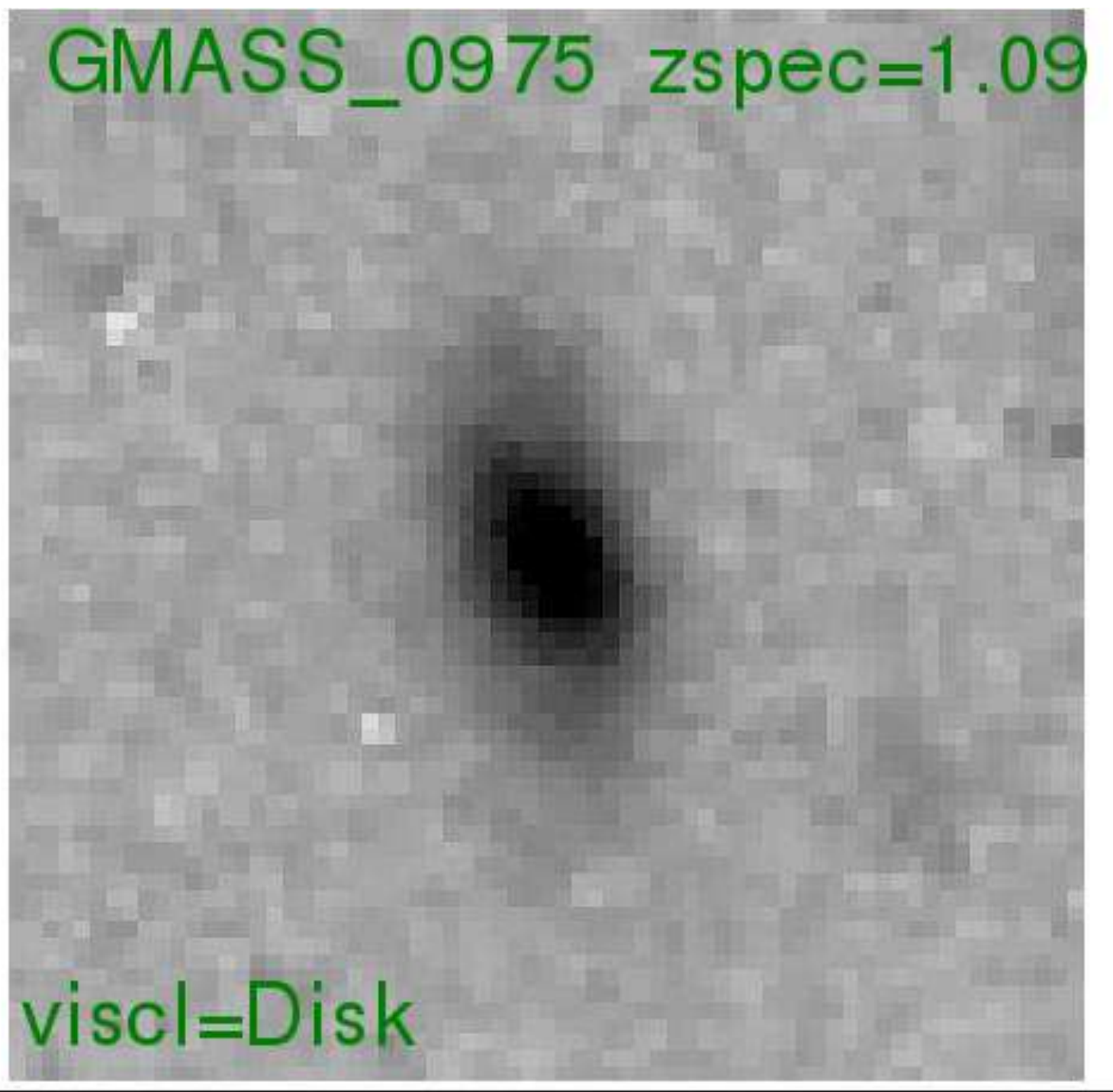}			     

\includegraphics[trim=100 40 75 390, clip=true, width=30mm]{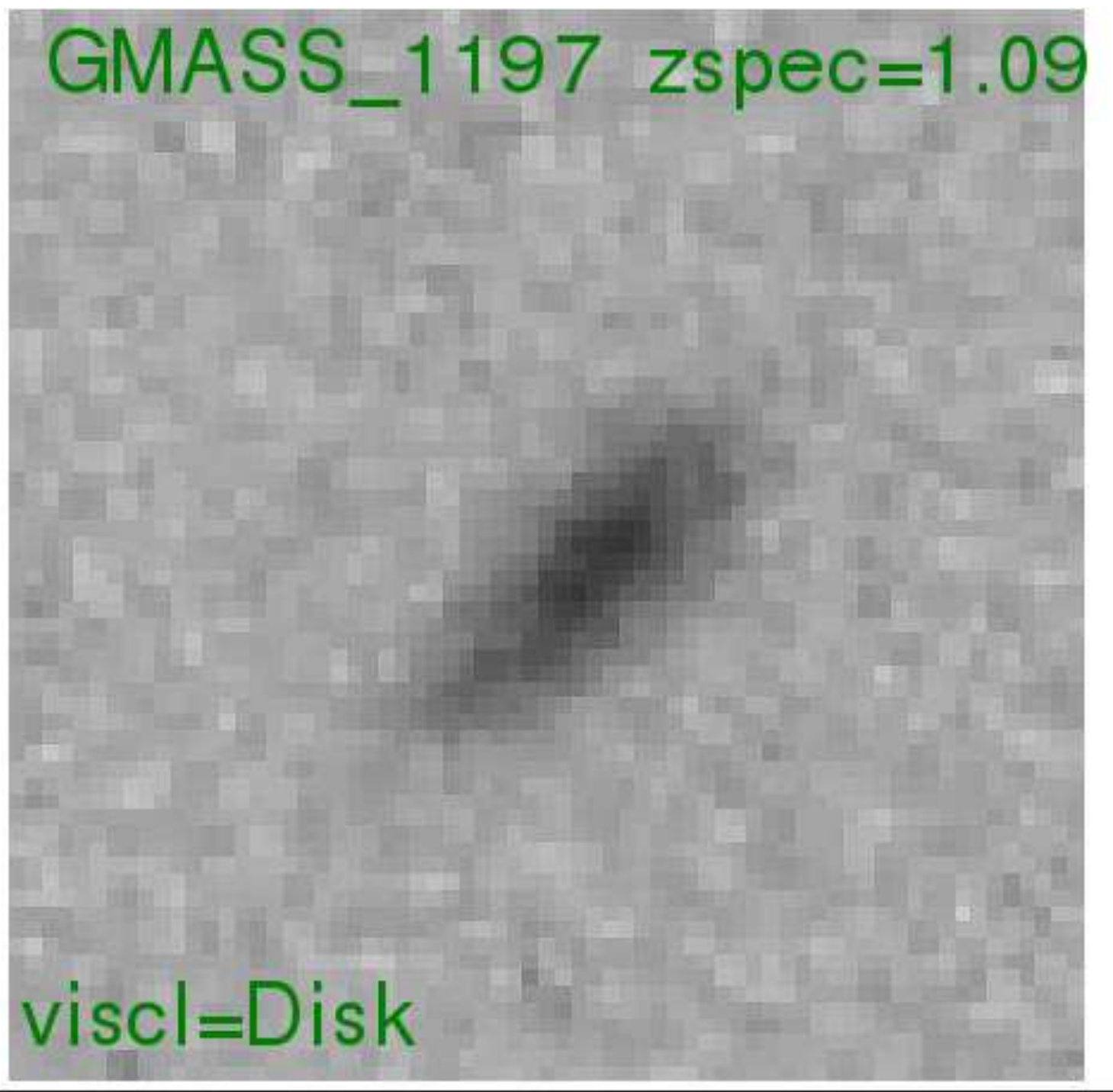}			     
\includegraphics[trim=100 40 75 390, clip=true, width=30mm]{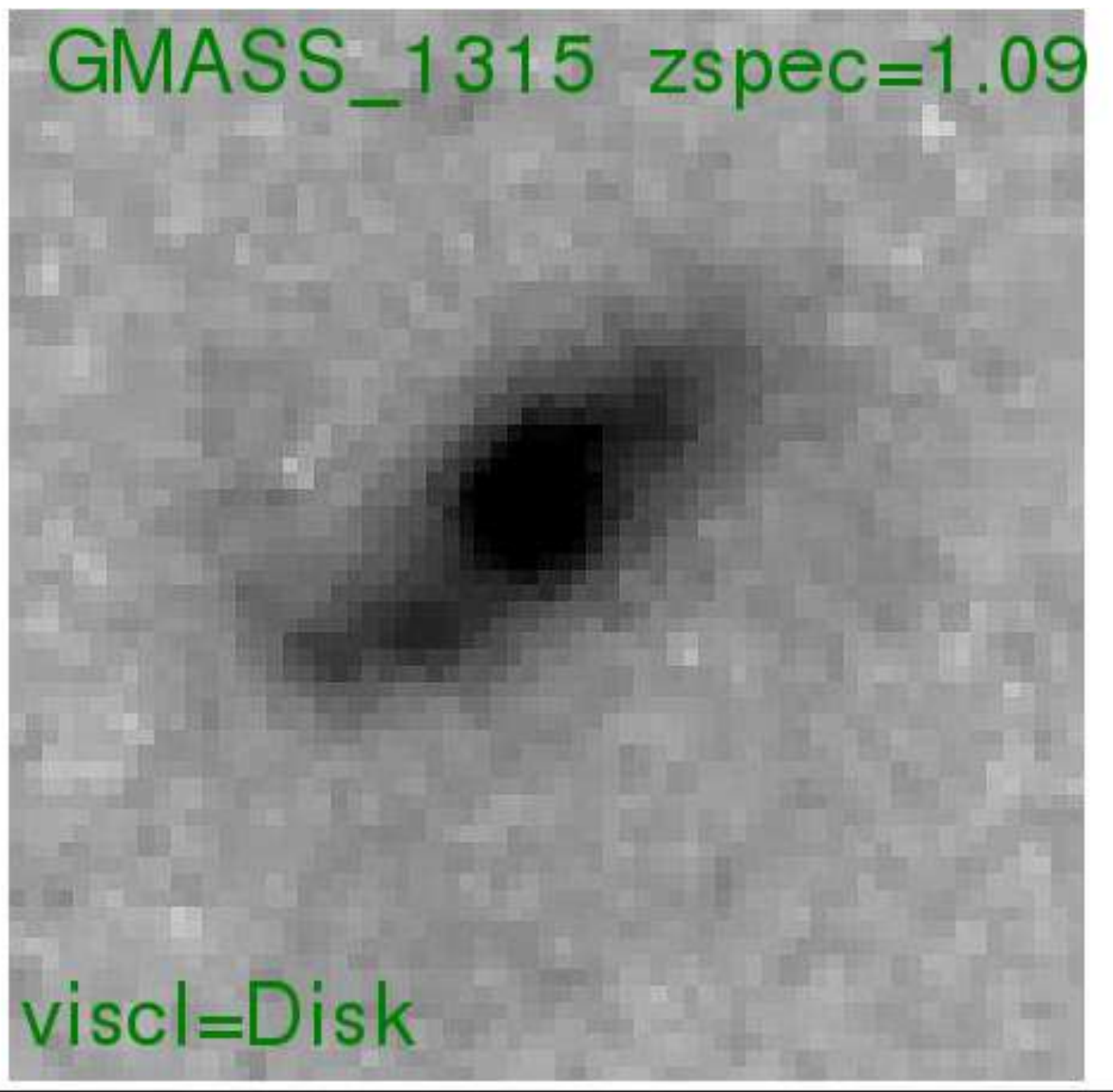}			     
\includegraphics[trim=100 40 75 390, clip=true, width=30mm]{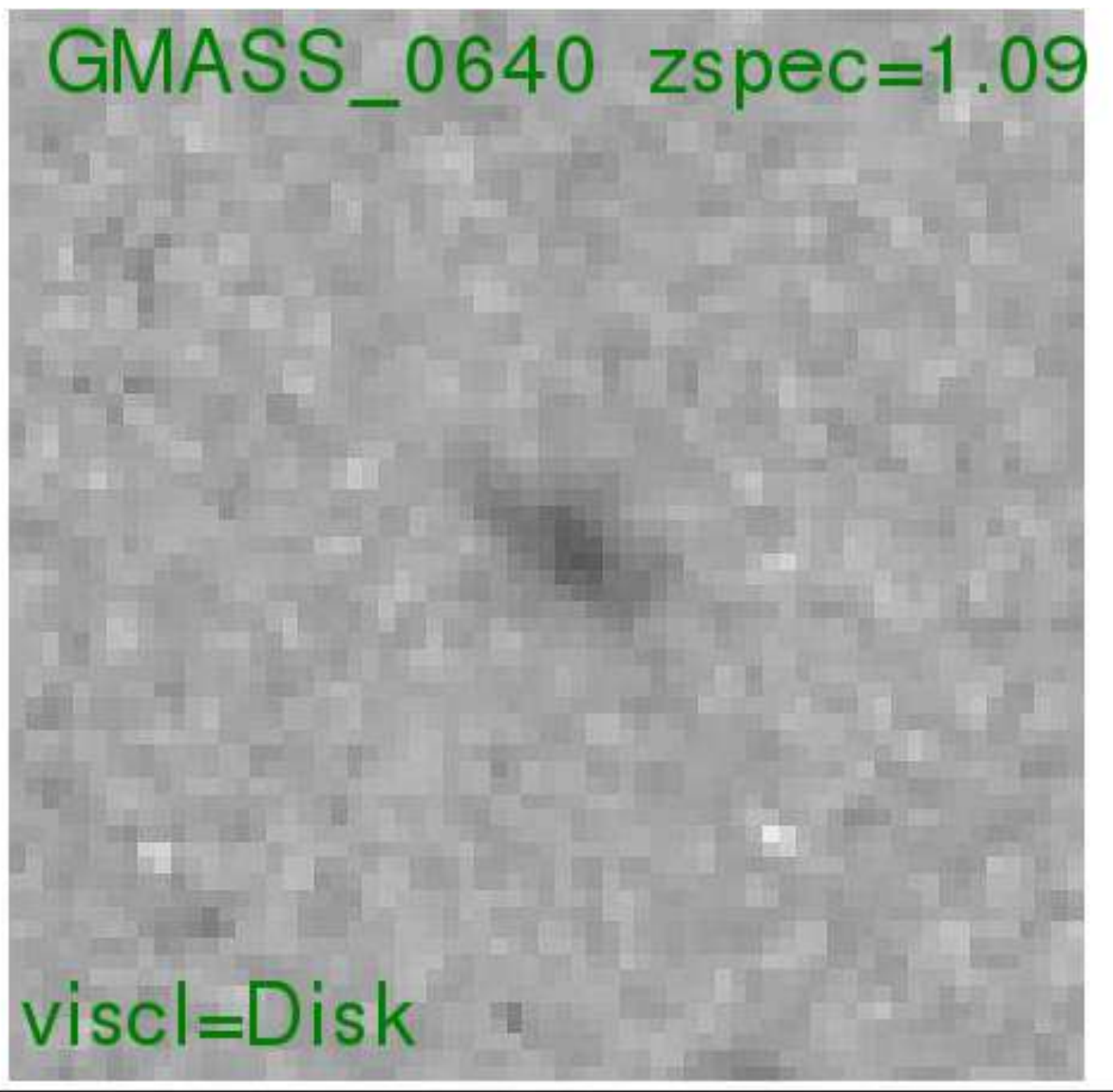}			     
\includegraphics[trim=100 40 75 390, clip=true, width=30mm]{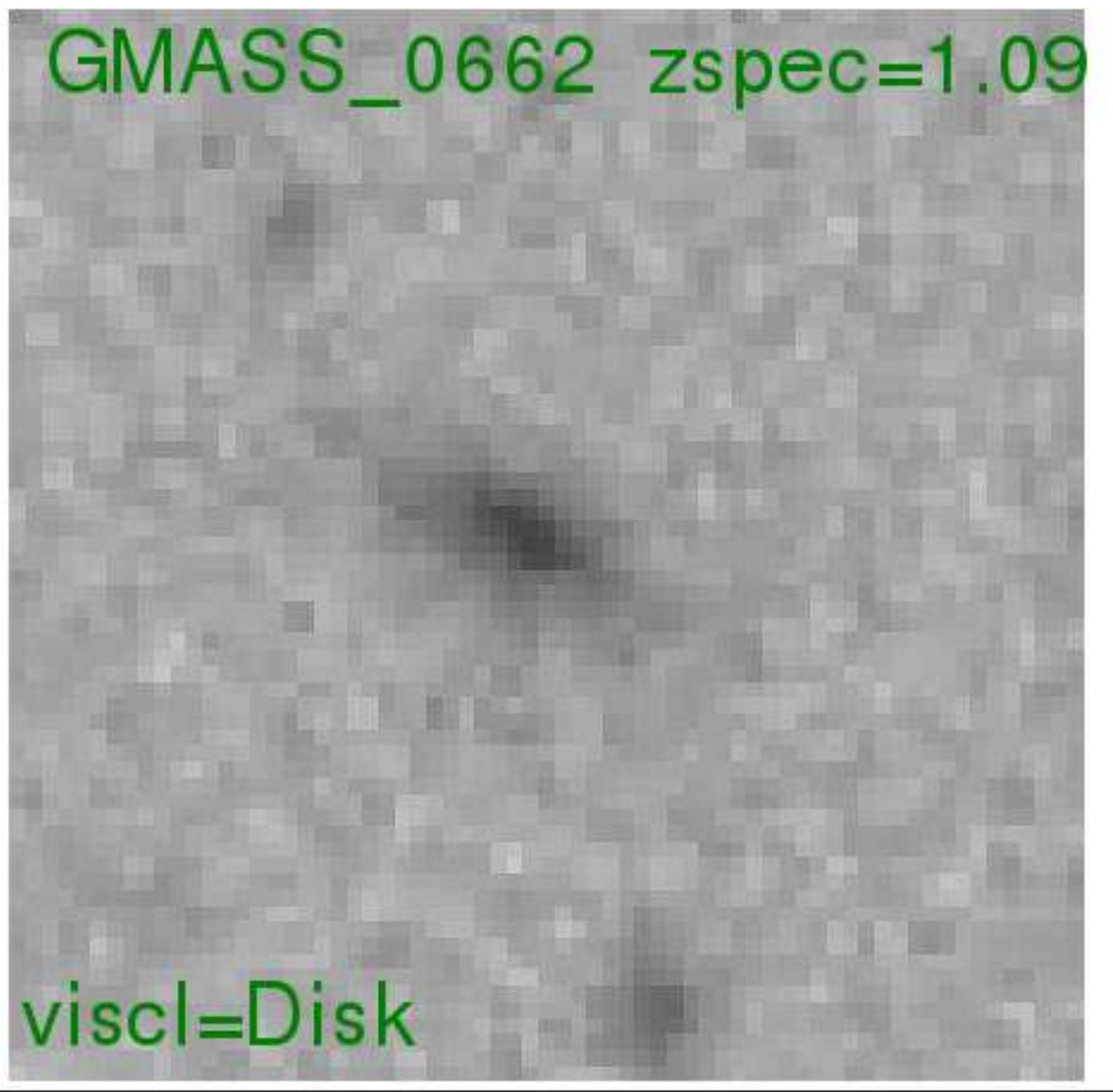}			     
\includegraphics[trim=100 40 75 390, clip=true, width=30mm]{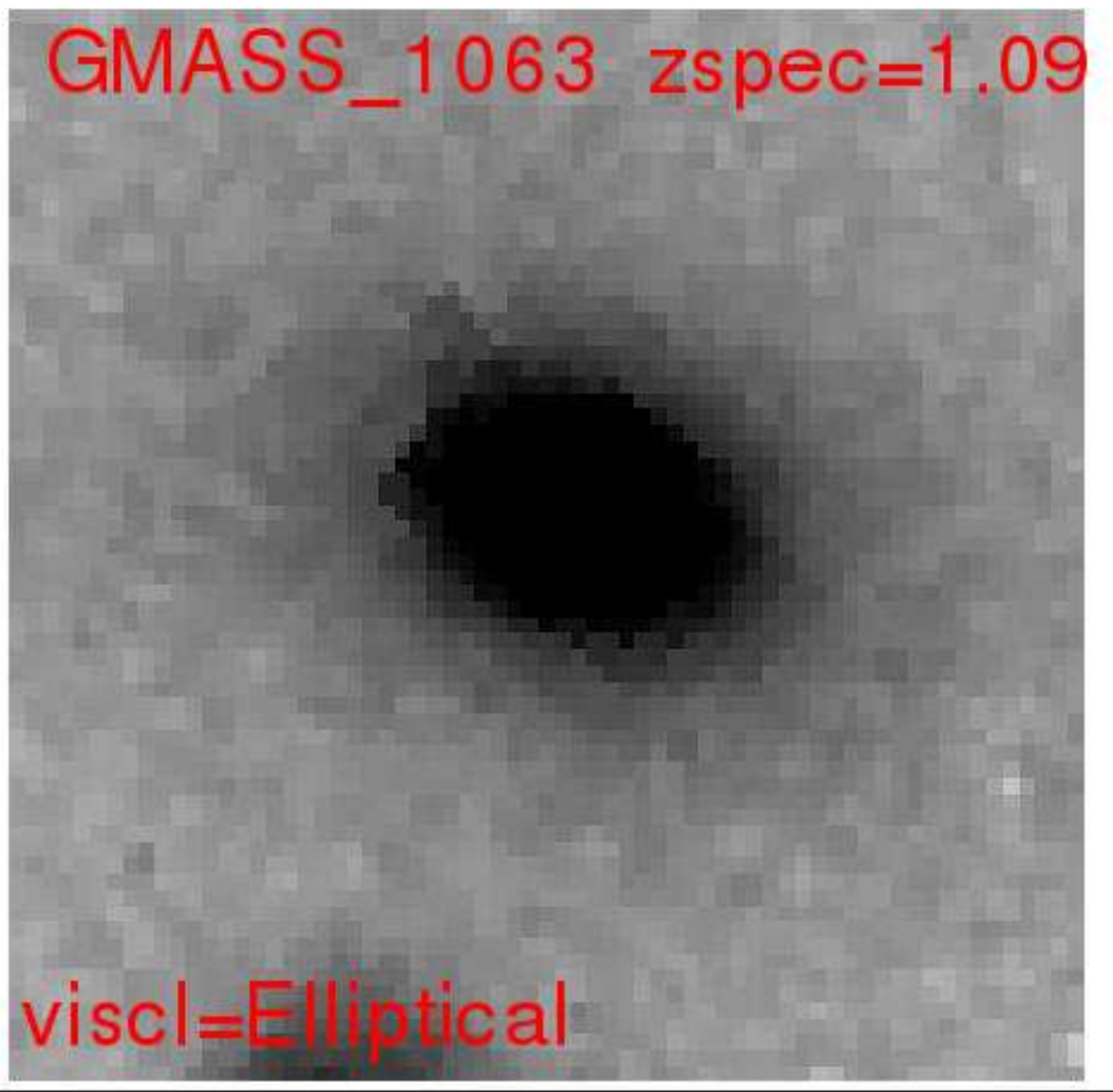}			     
\includegraphics[trim=100 40 75 390, clip=true, width=30mm]{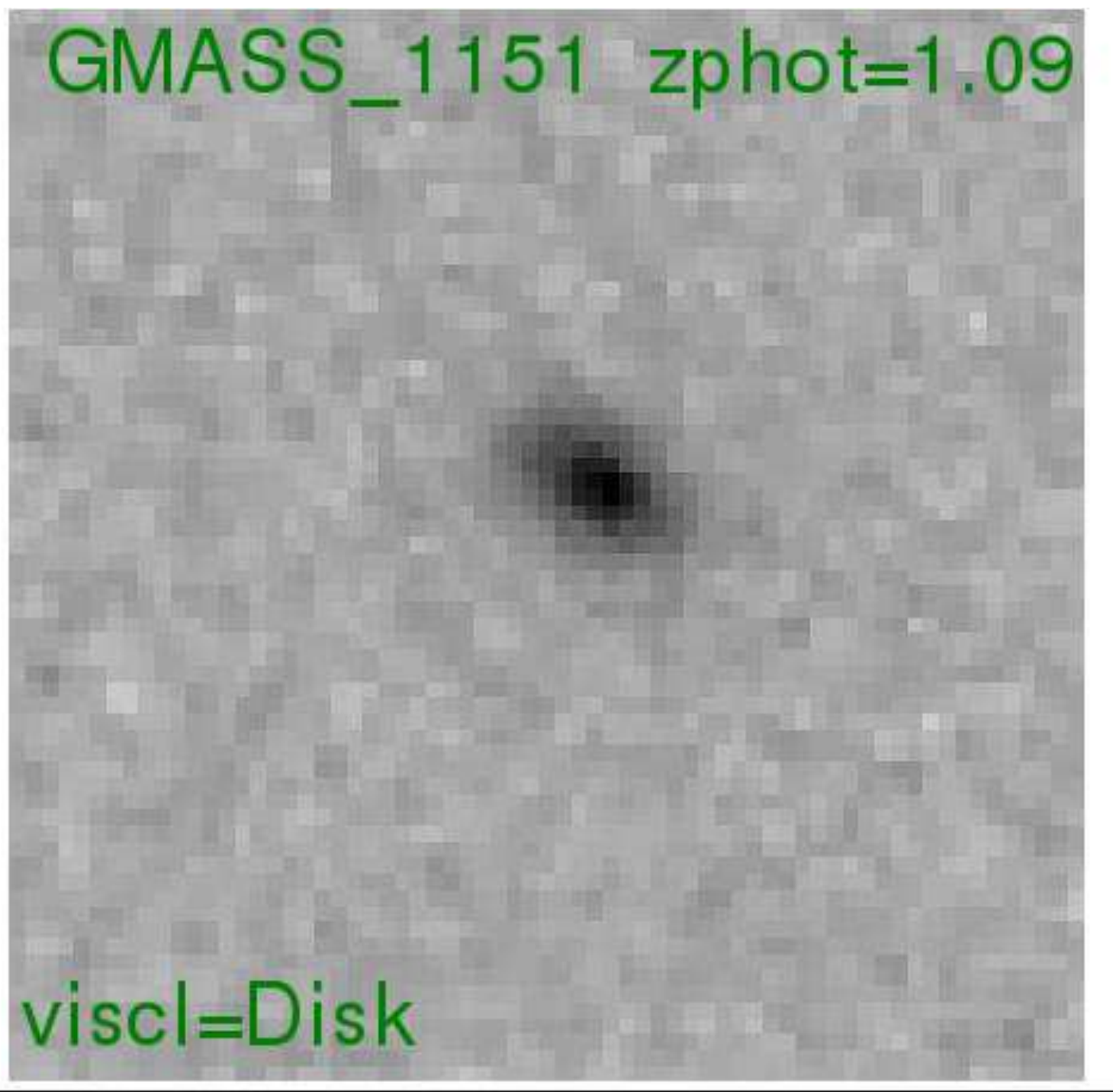}			     

\includegraphics[trim=100 40 75 390, clip=true, width=30mm]{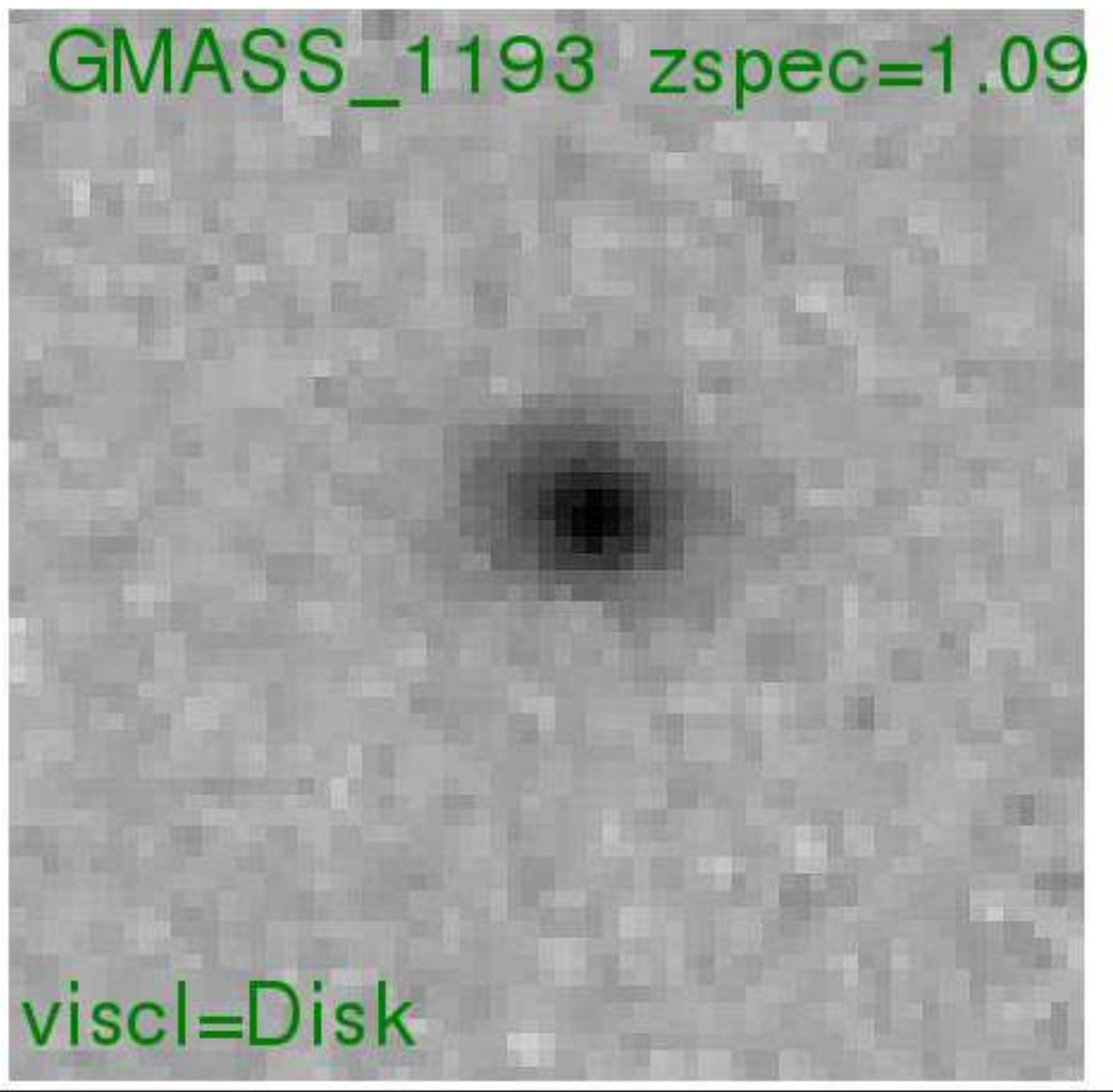}			     
\includegraphics[trim=100 40 75 390, clip=true, width=30mm]{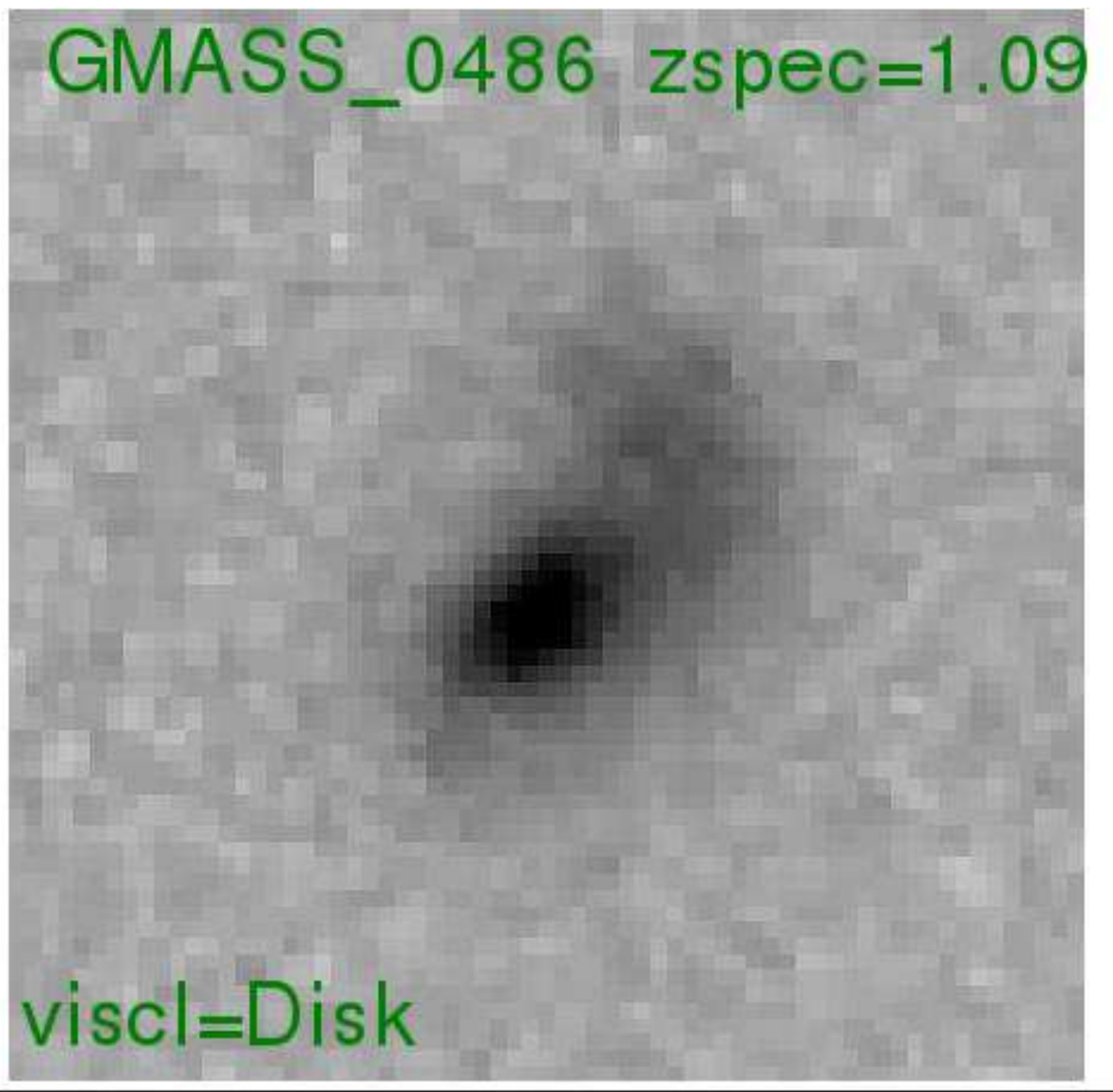}			     
\includegraphics[trim=100 40 75 390, clip=true, width=30mm]{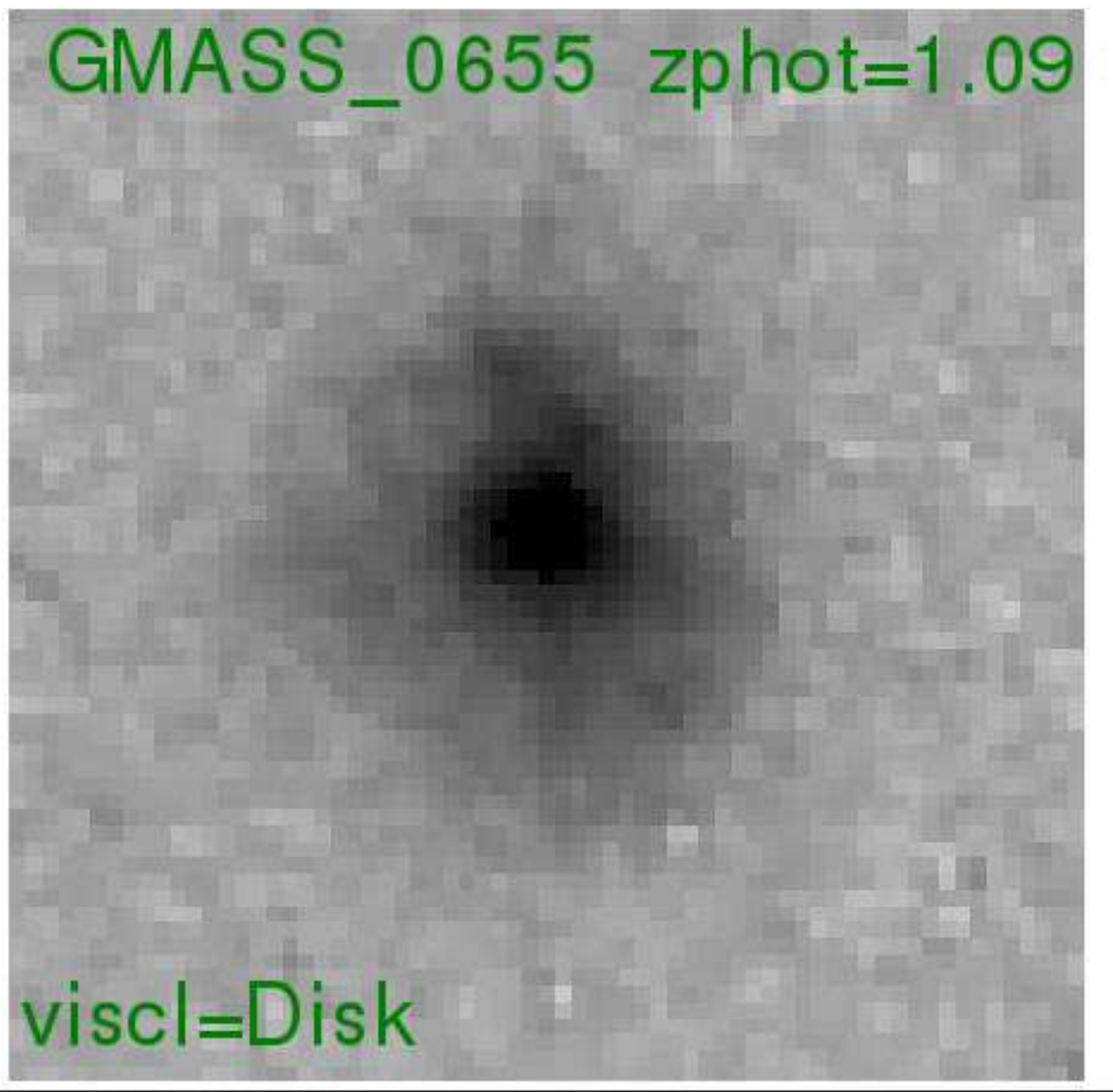}			     
\includegraphics[trim=100 40 75 390, clip=true, width=30mm]{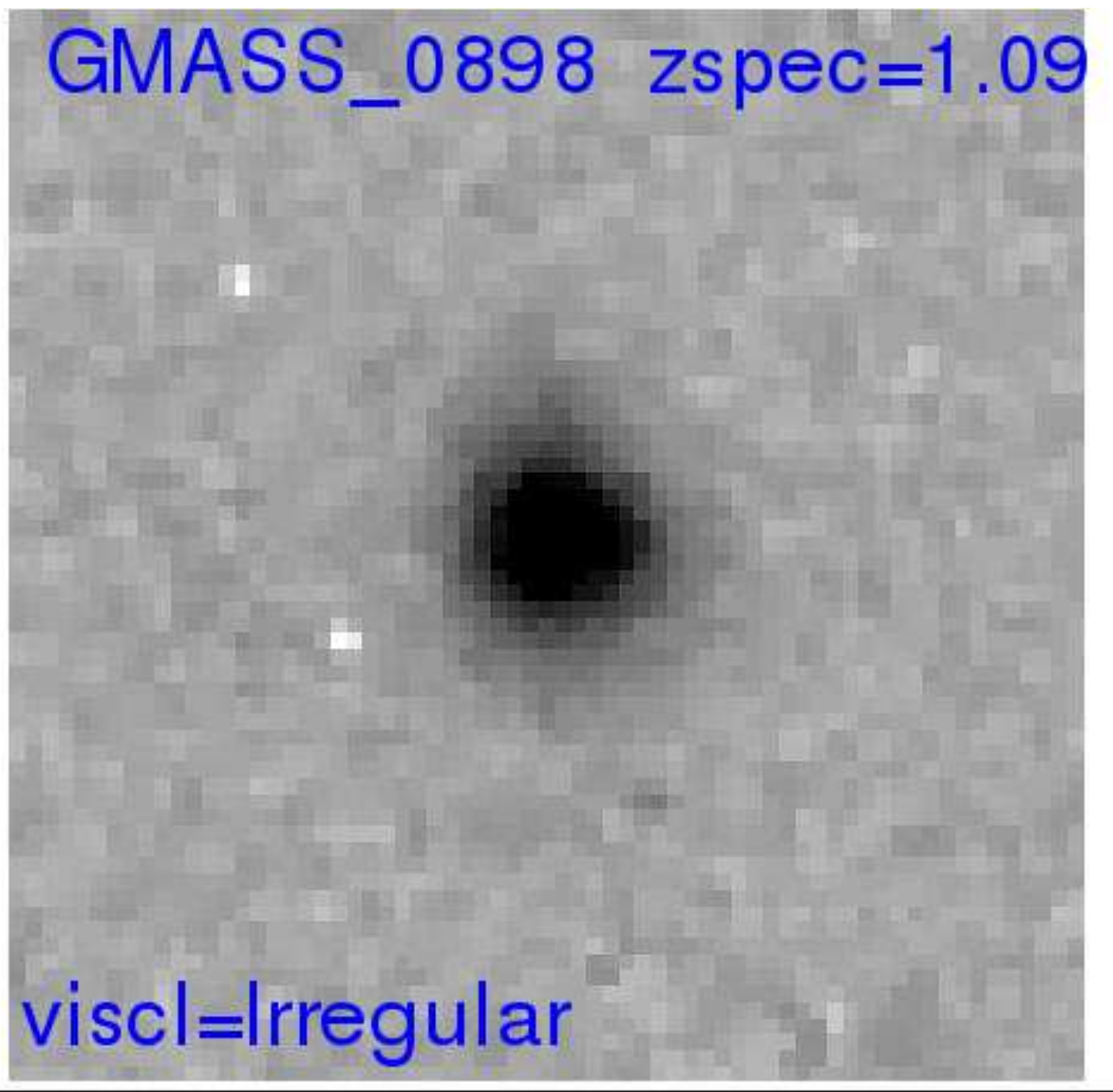}			     
\includegraphics[trim=100 40 75 390, clip=true, width=30mm]{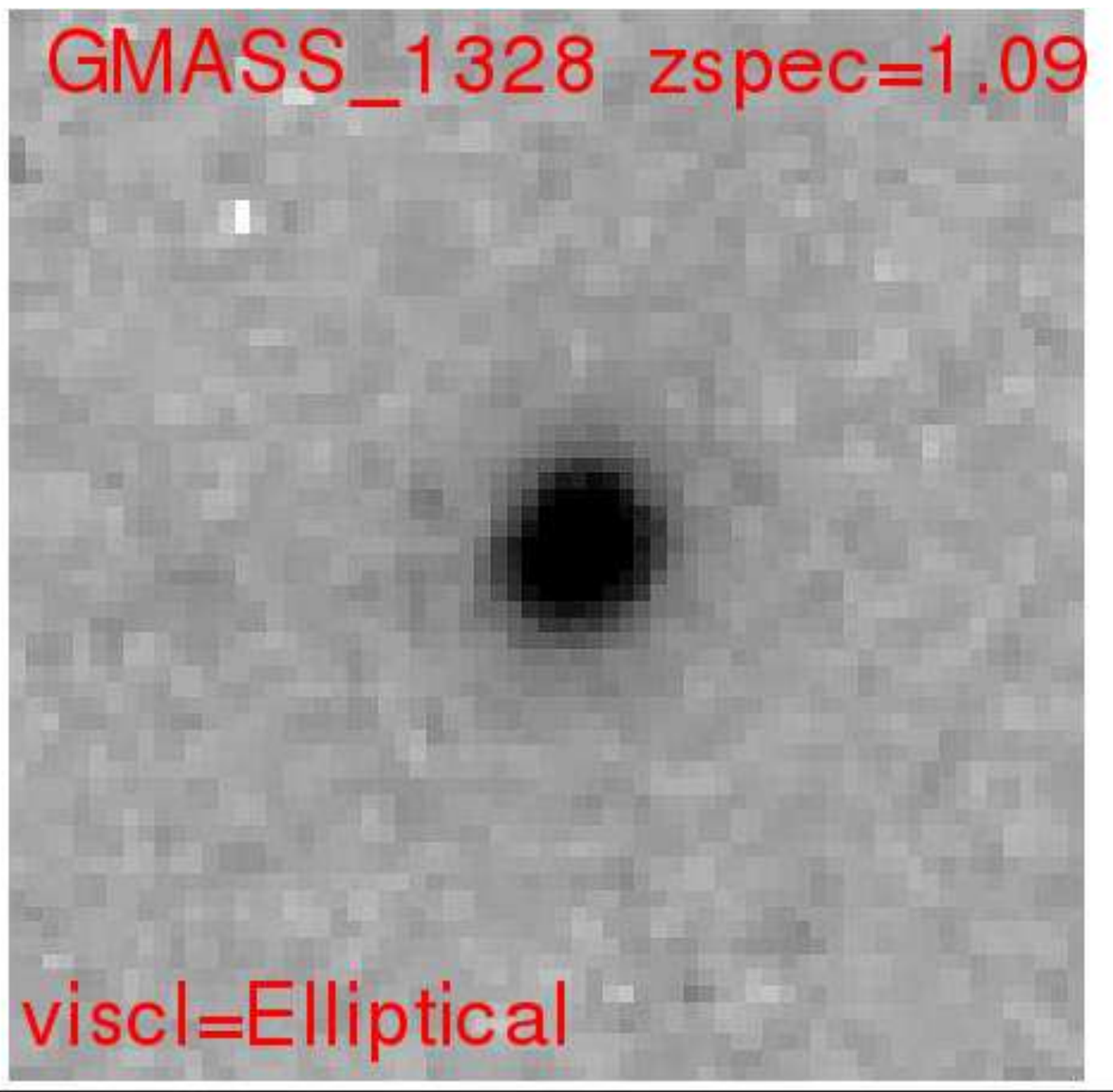}			     
\includegraphics[trim=100 40 75 390, clip=true, width=30mm]{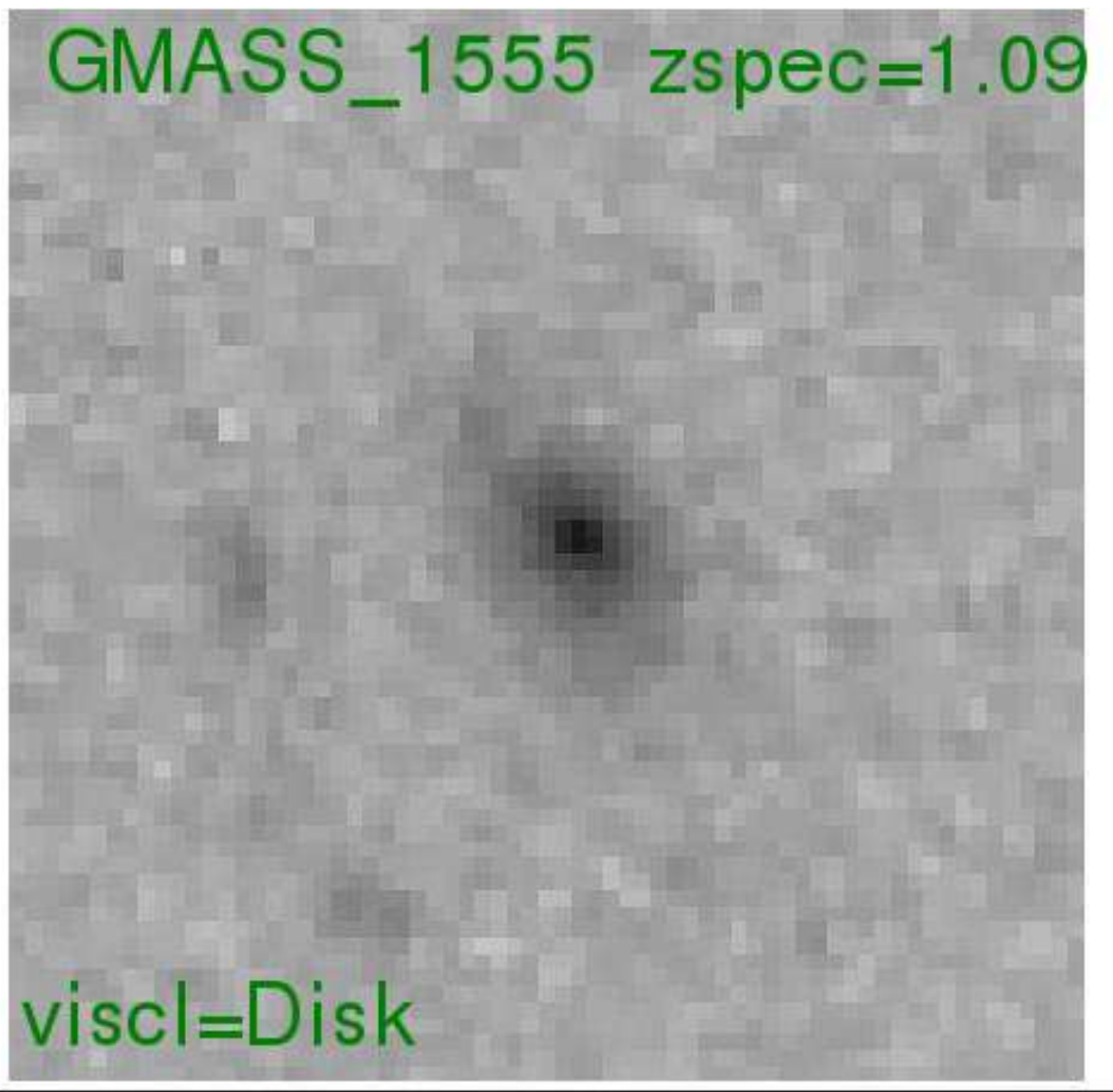}			     

\includegraphics[trim=100 40 75 390, clip=true, width=30mm]{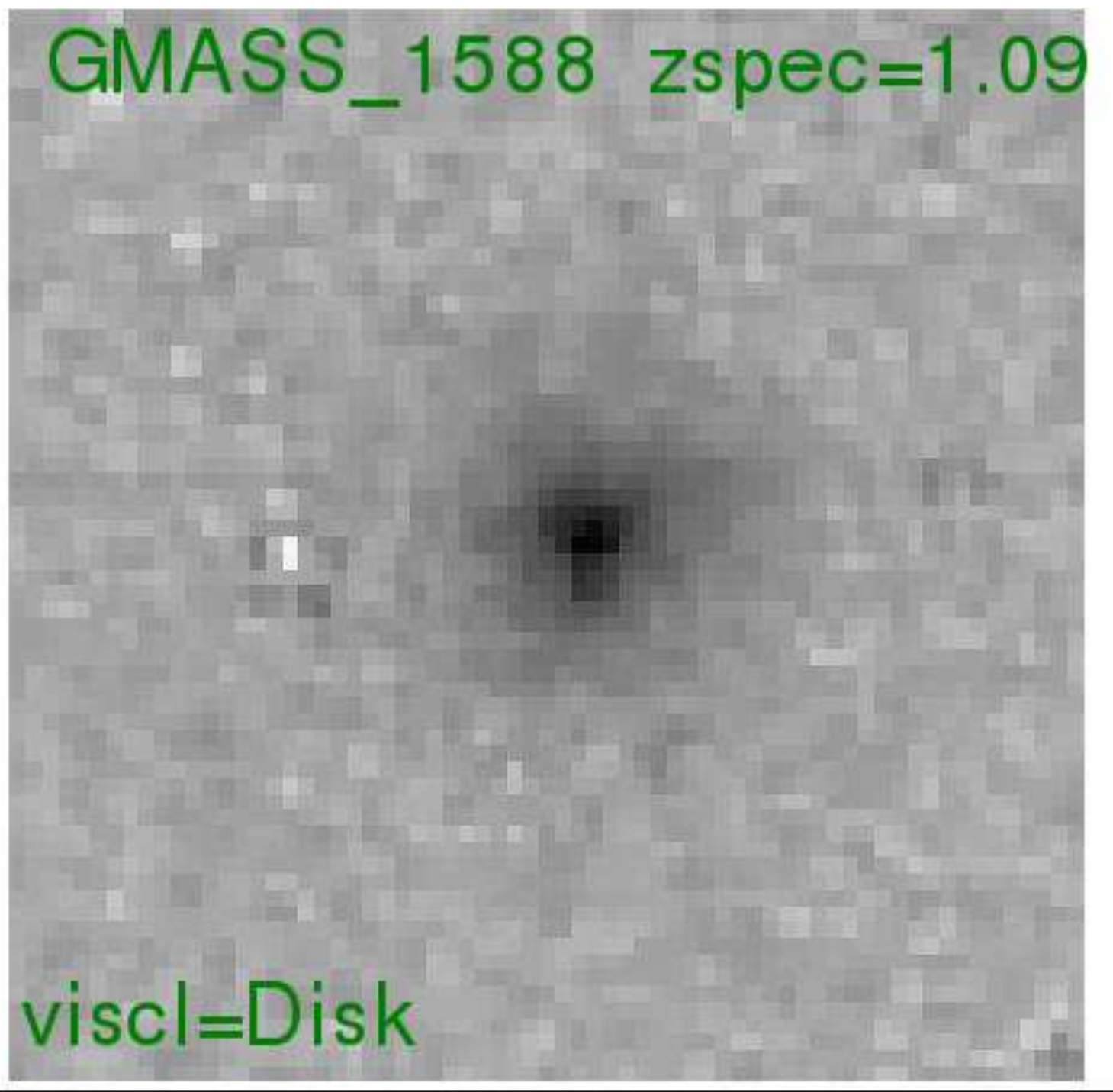}			     
\includegraphics[trim=100 40 75 390, clip=true, width=30mm]{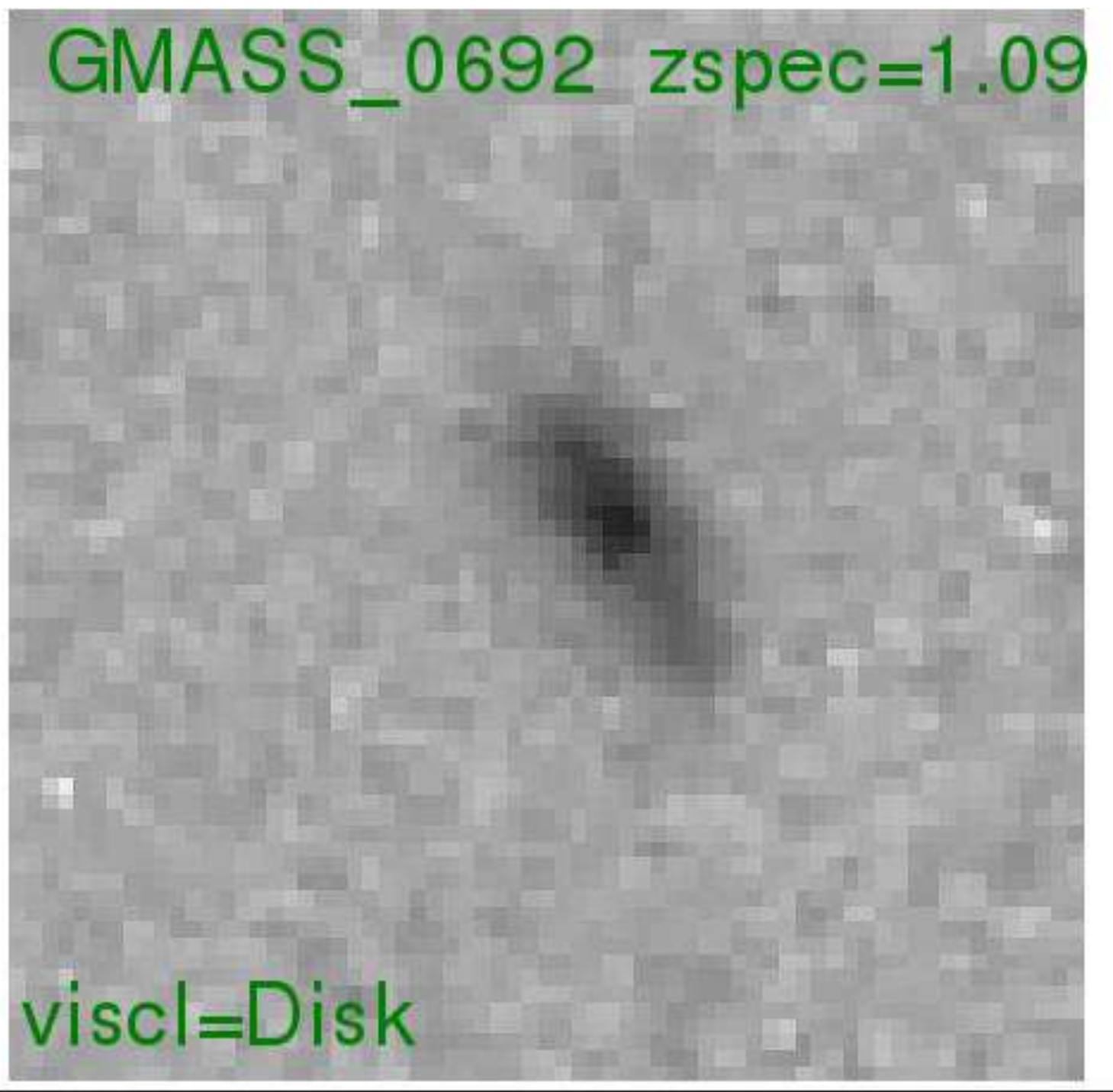}			     
\includegraphics[trim=100 40 75 390, clip=true, width=30mm]{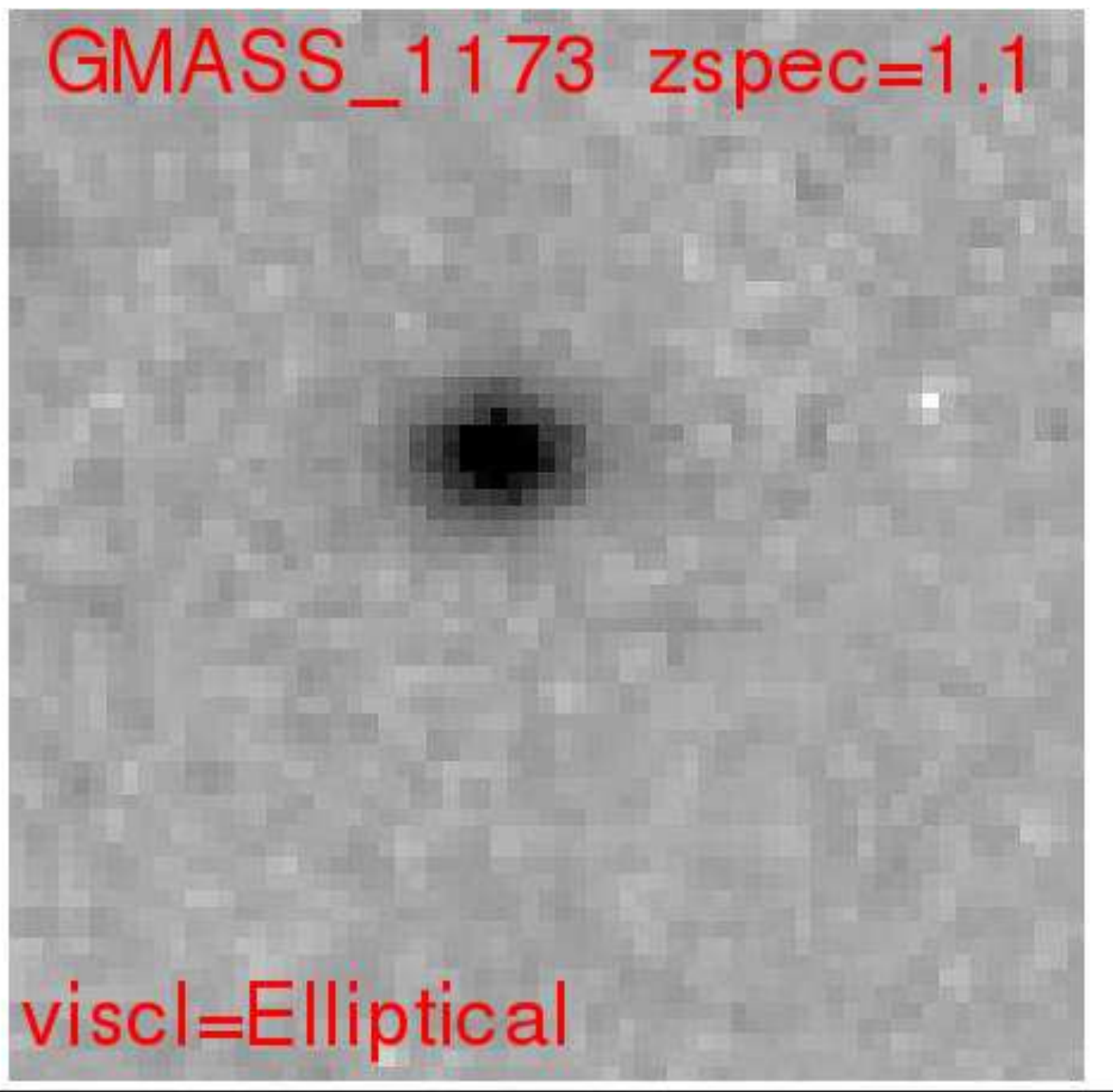}			     
\includegraphics[trim=100 40 75 390, clip=true, width=30mm]{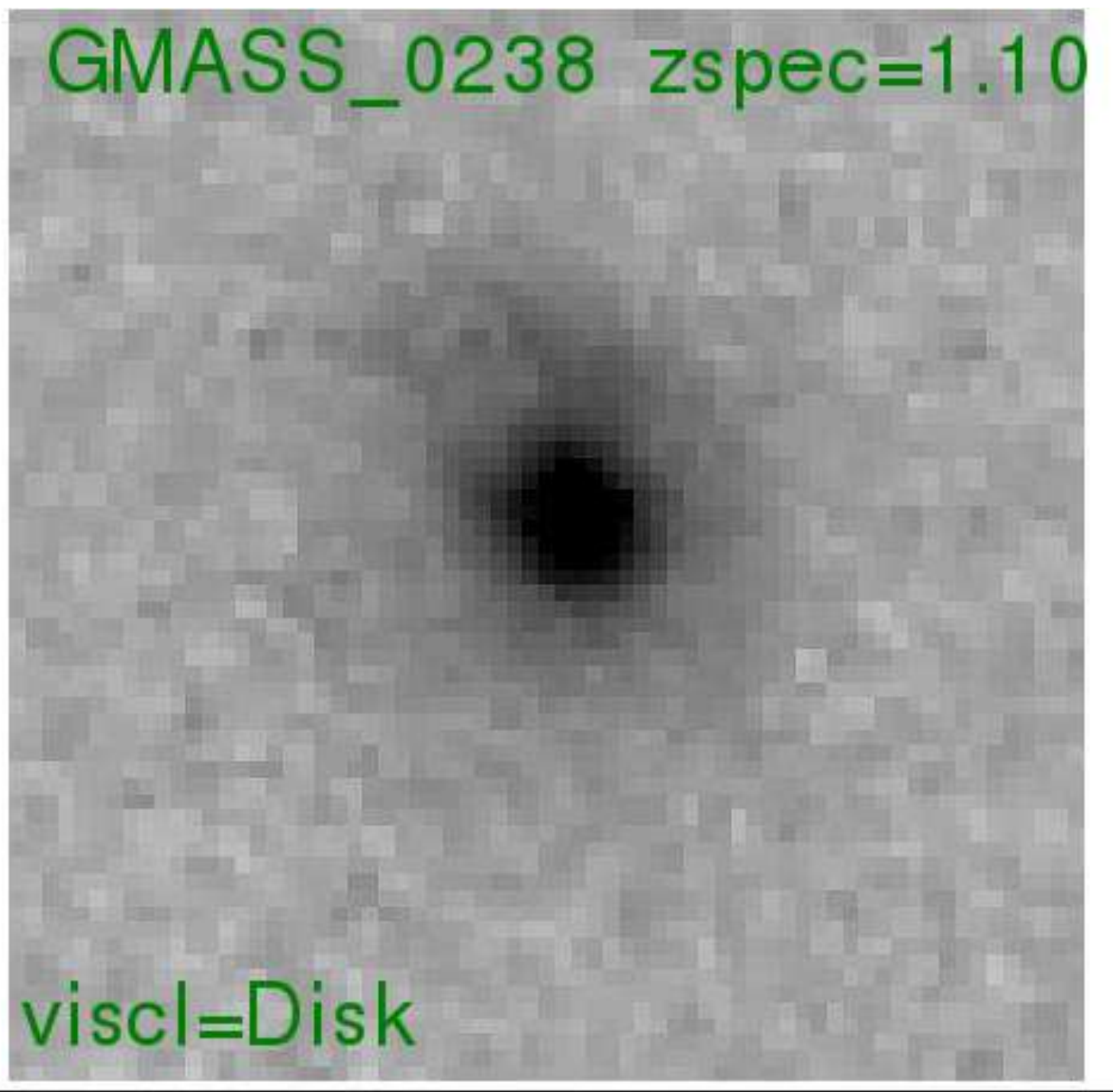}			     
\includegraphics[trim=100 40 75 390, clip=true, width=30mm]{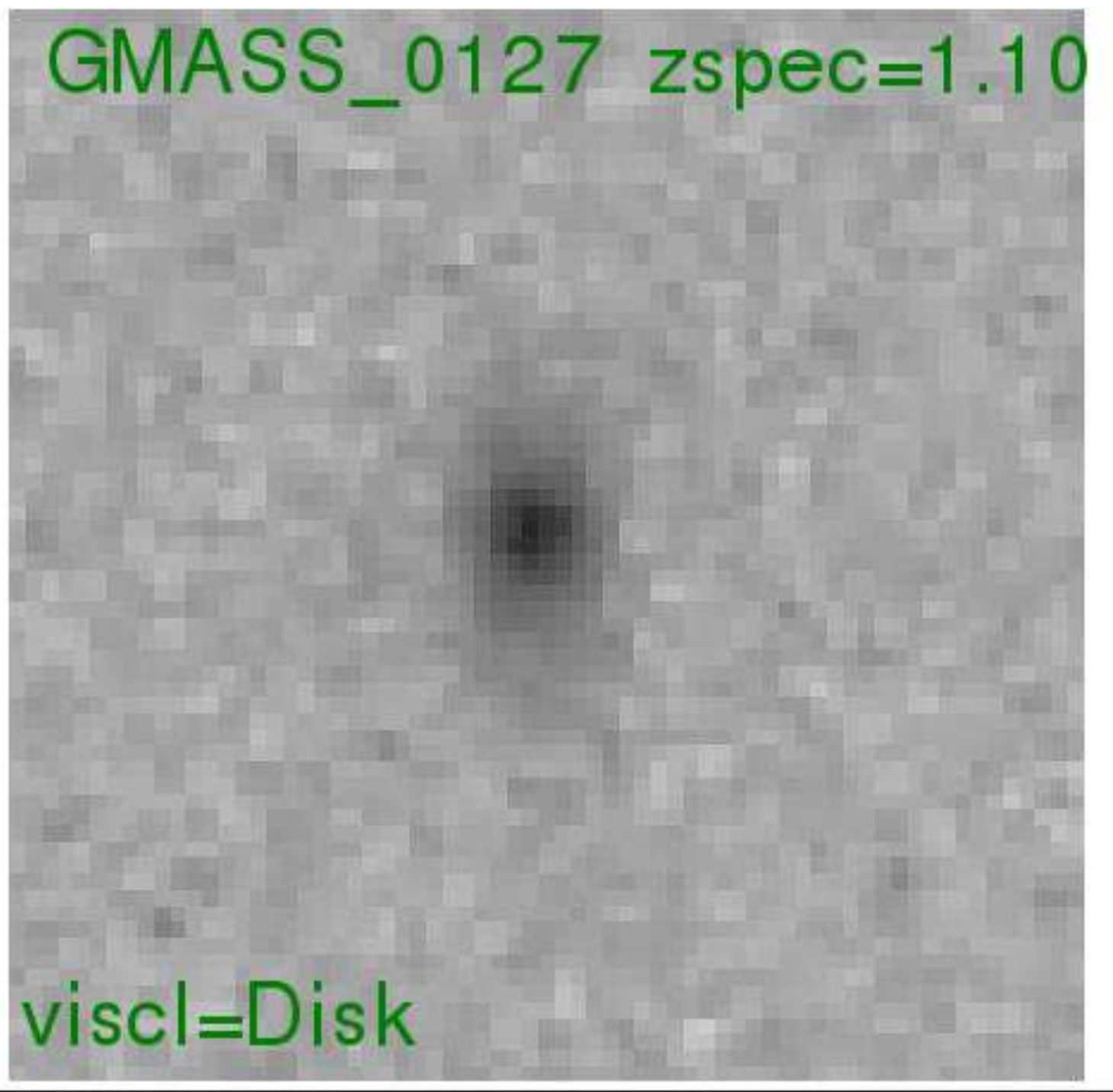}			     
\includegraphics[trim=100 40 75 390, clip=true, width=30mm]{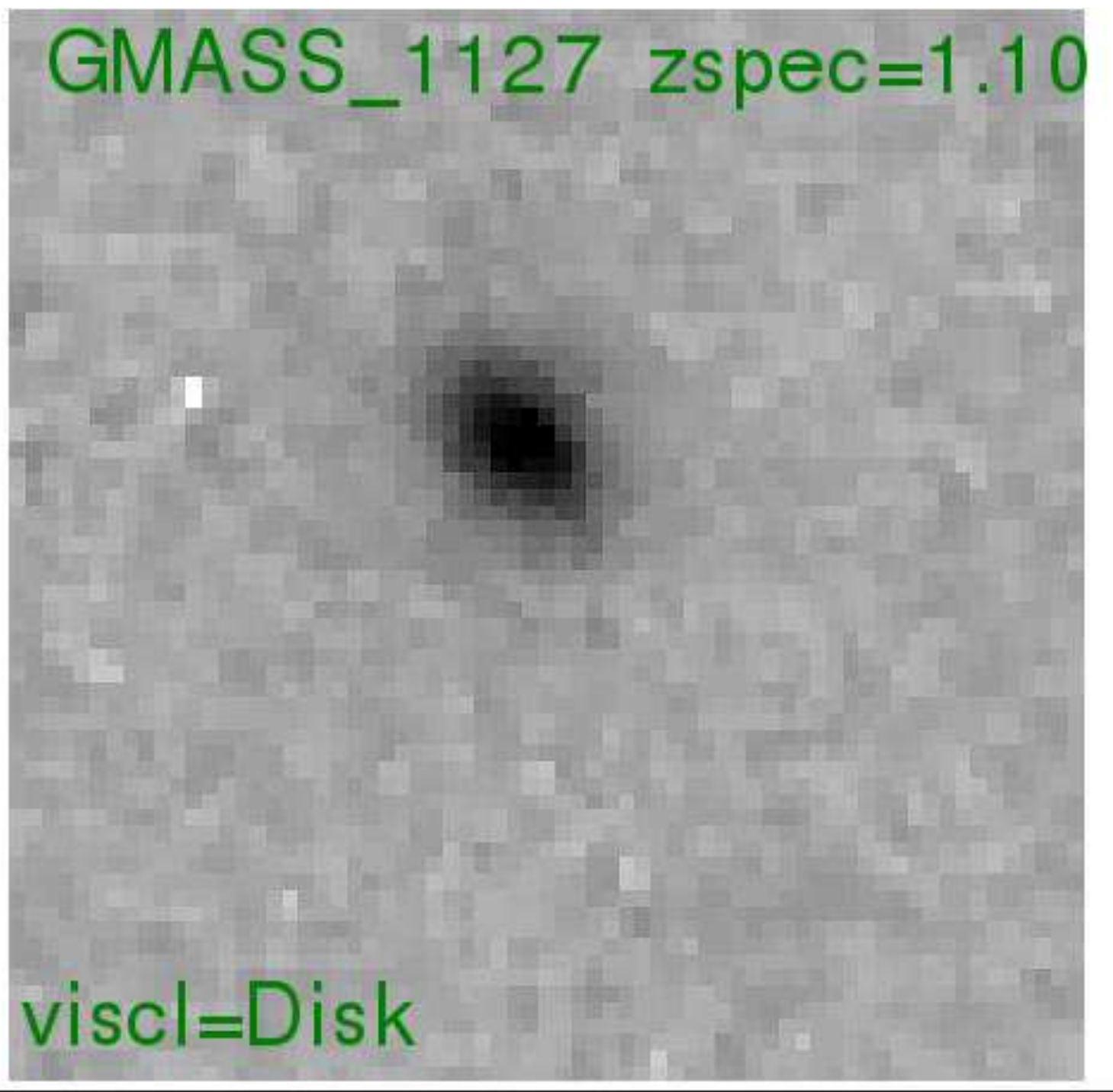}			     

\includegraphics[trim=100 40 75 390, clip=true, width=30mm]{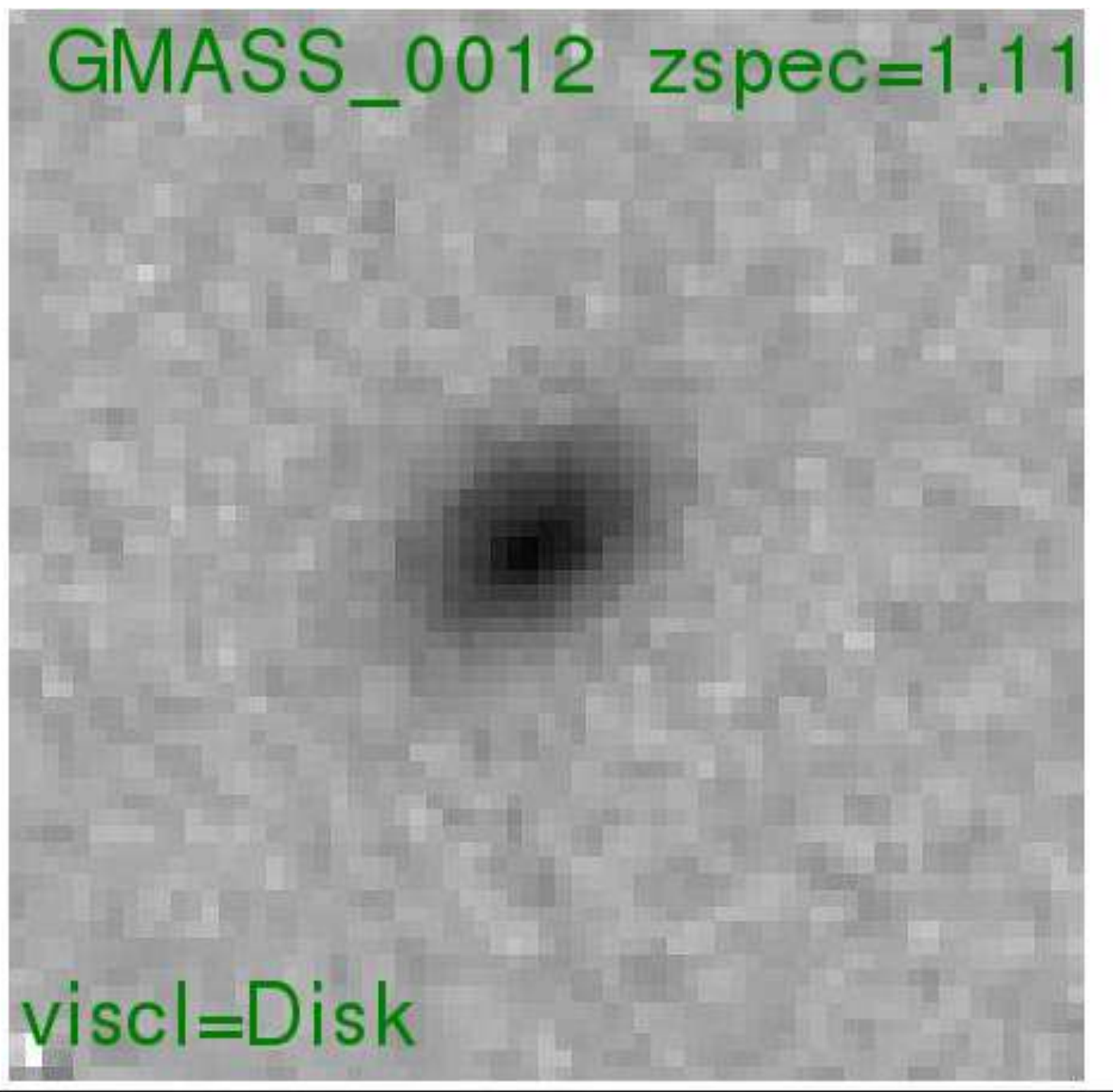}			     
\includegraphics[trim=100 40 75 390, clip=true, width=30mm]{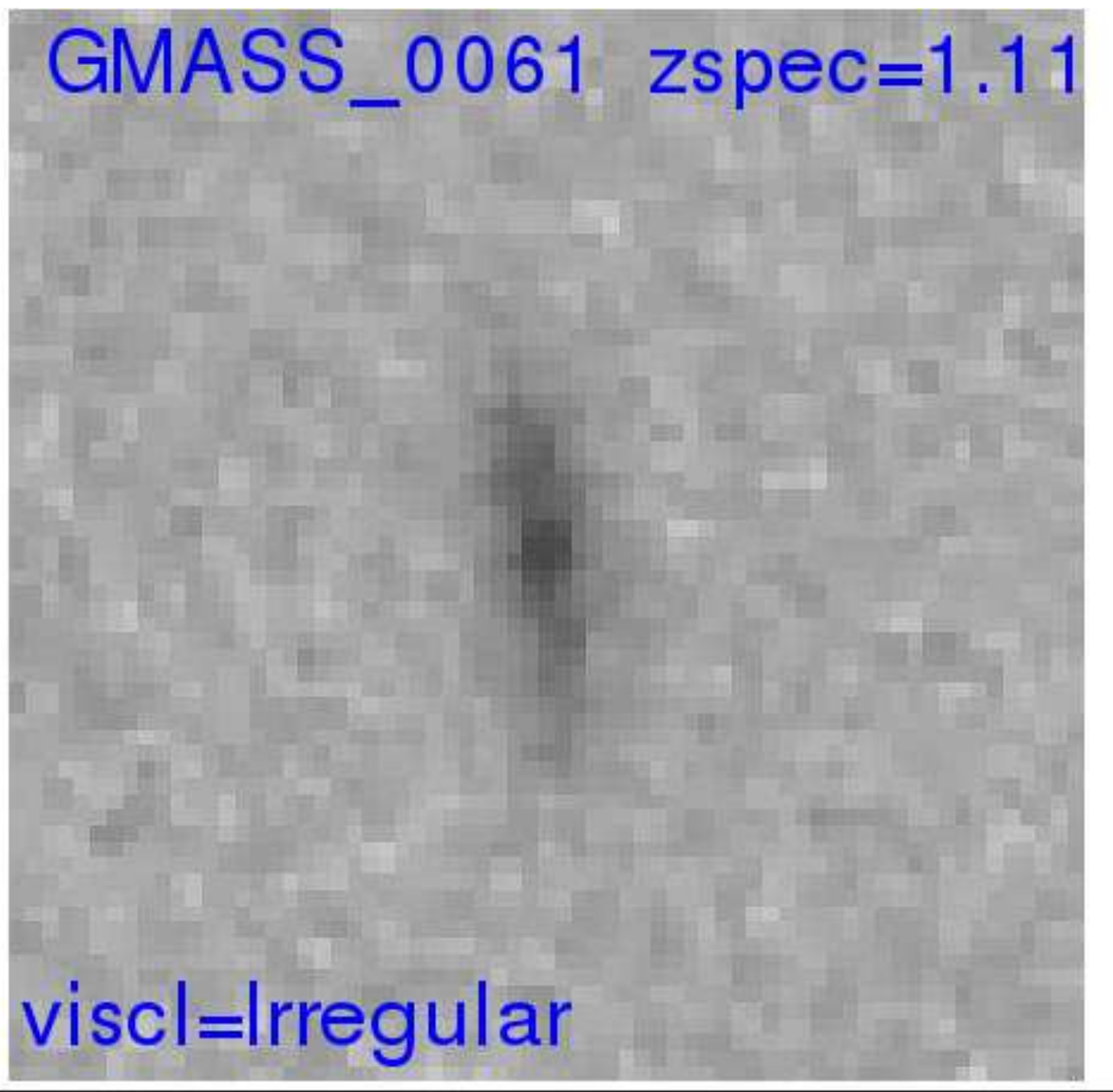}		     
\includegraphics[trim=100 40 75 390, clip=true, width=30mm]{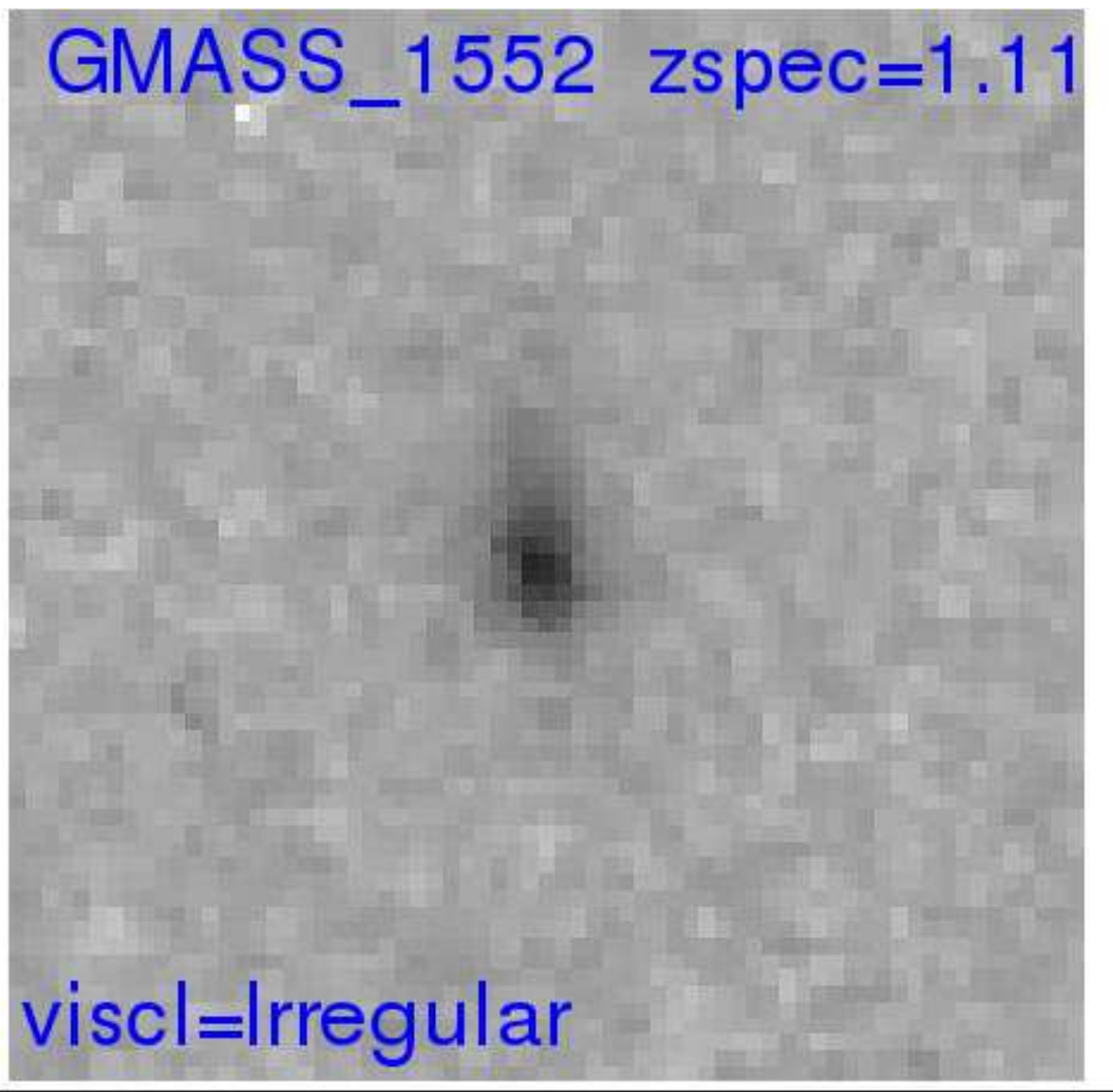}			     
\includegraphics[trim=100 40 75 390, clip=true, width=30mm]{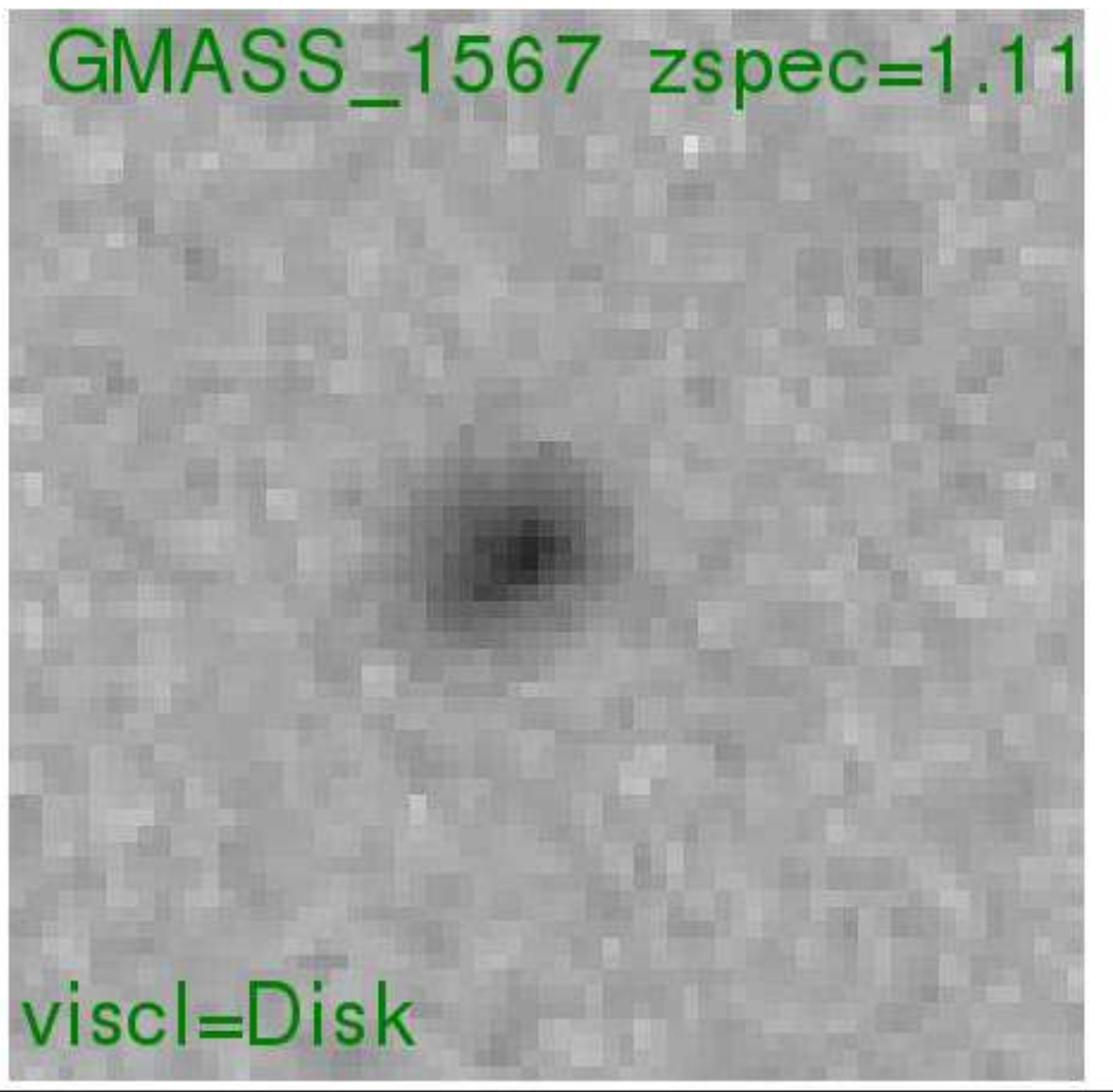}			     
\includegraphics[trim=100 40 75 390, clip=true, width=30mm]{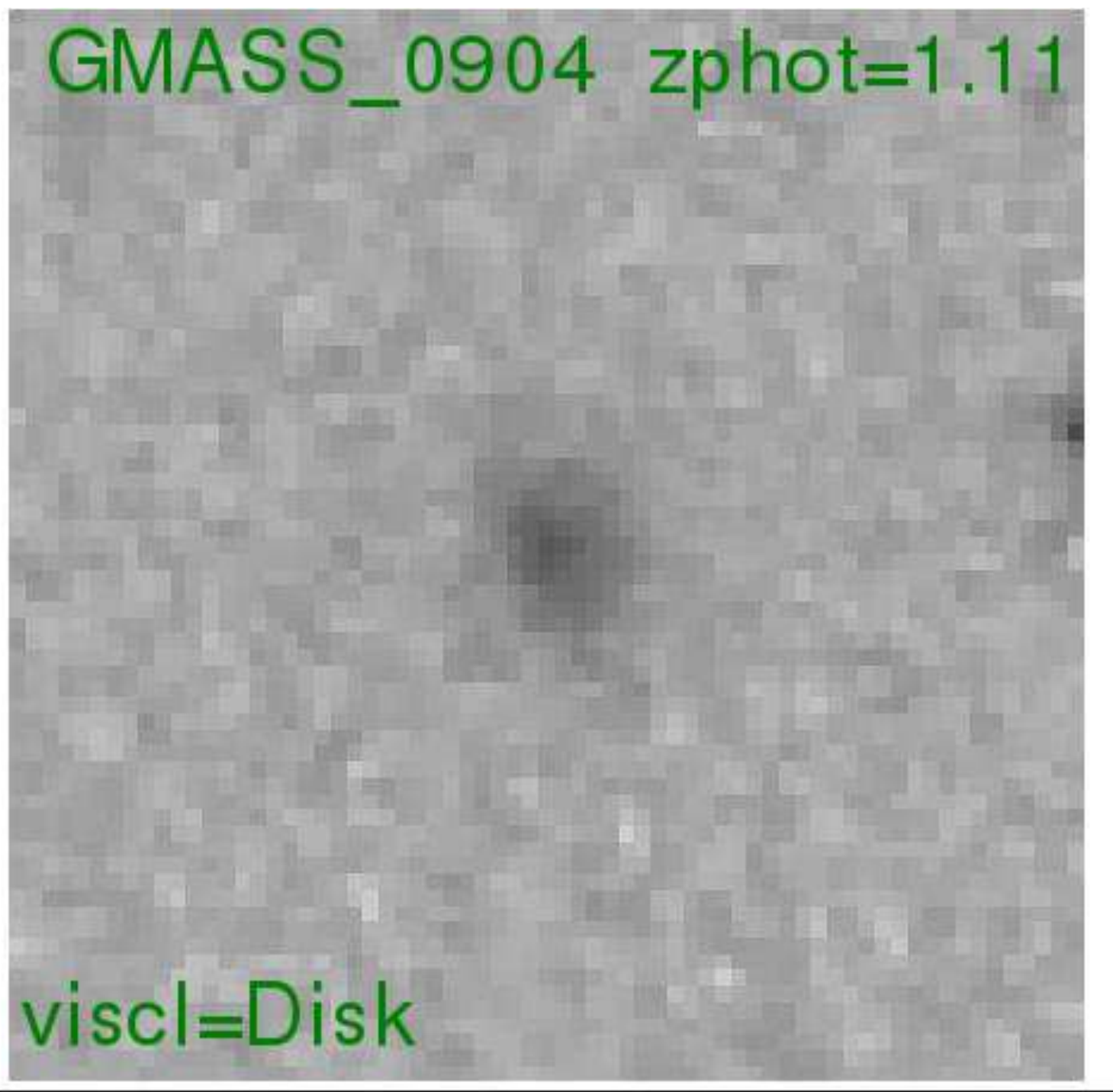}			     
\includegraphics[trim=100 40 75 390, clip=true, width=30mm]{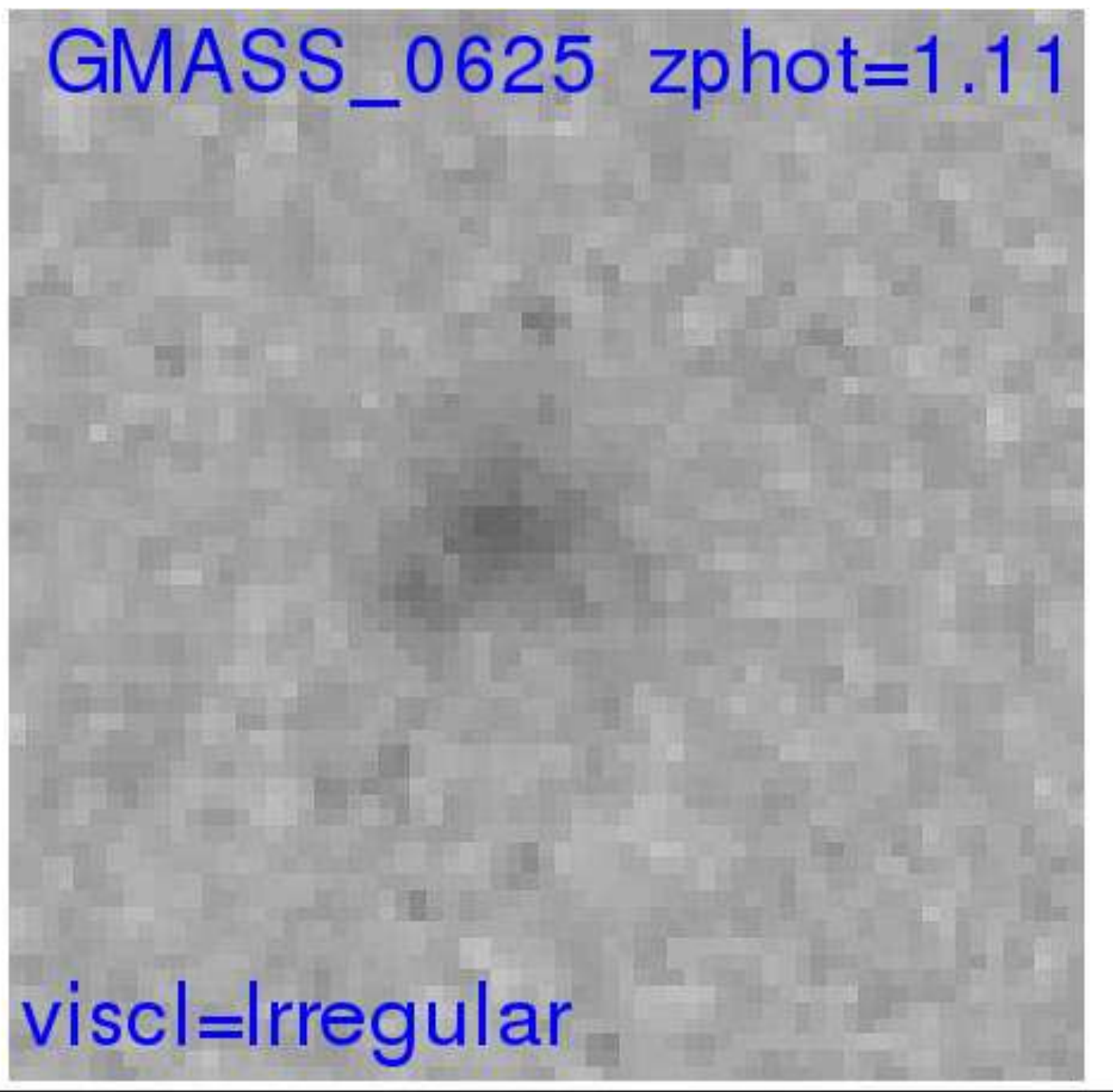}			     

\includegraphics[trim=100 40 75 390, clip=true, width=30mm]{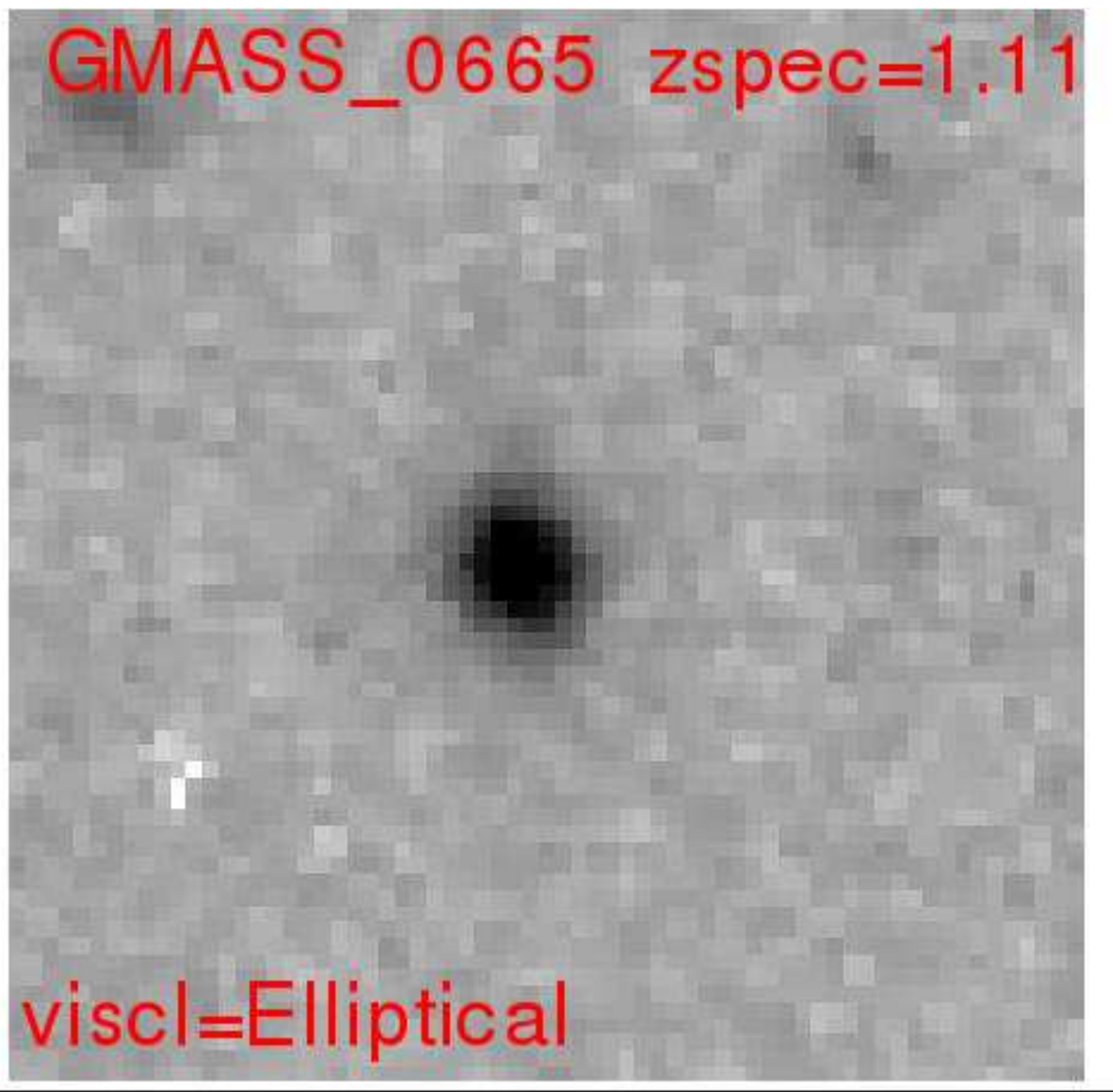}			     
\includegraphics[trim=100 40 75 390, clip=true, width=30mm]{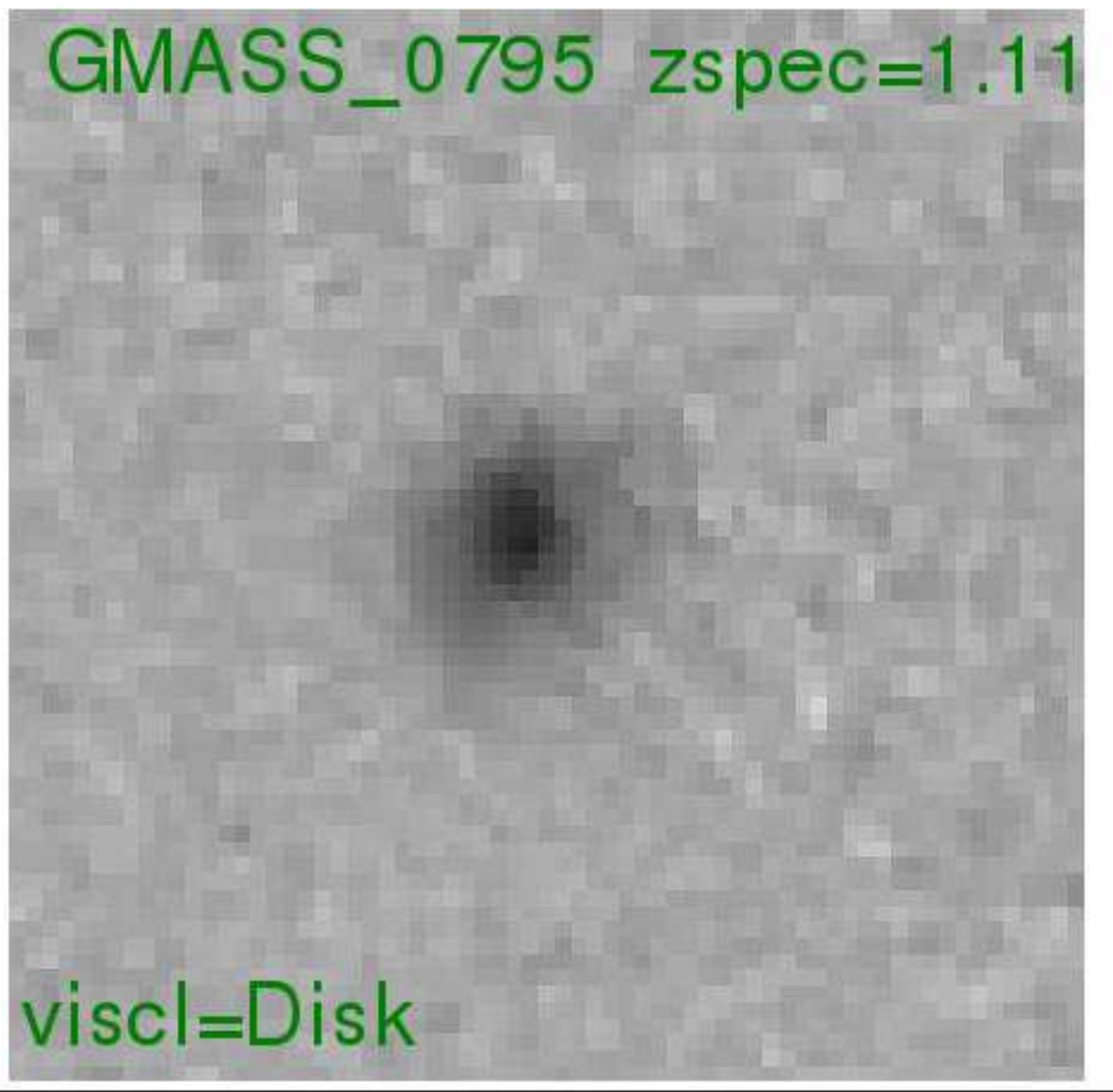}			     
\includegraphics[trim=100 40 75 390, clip=true, width=30mm]{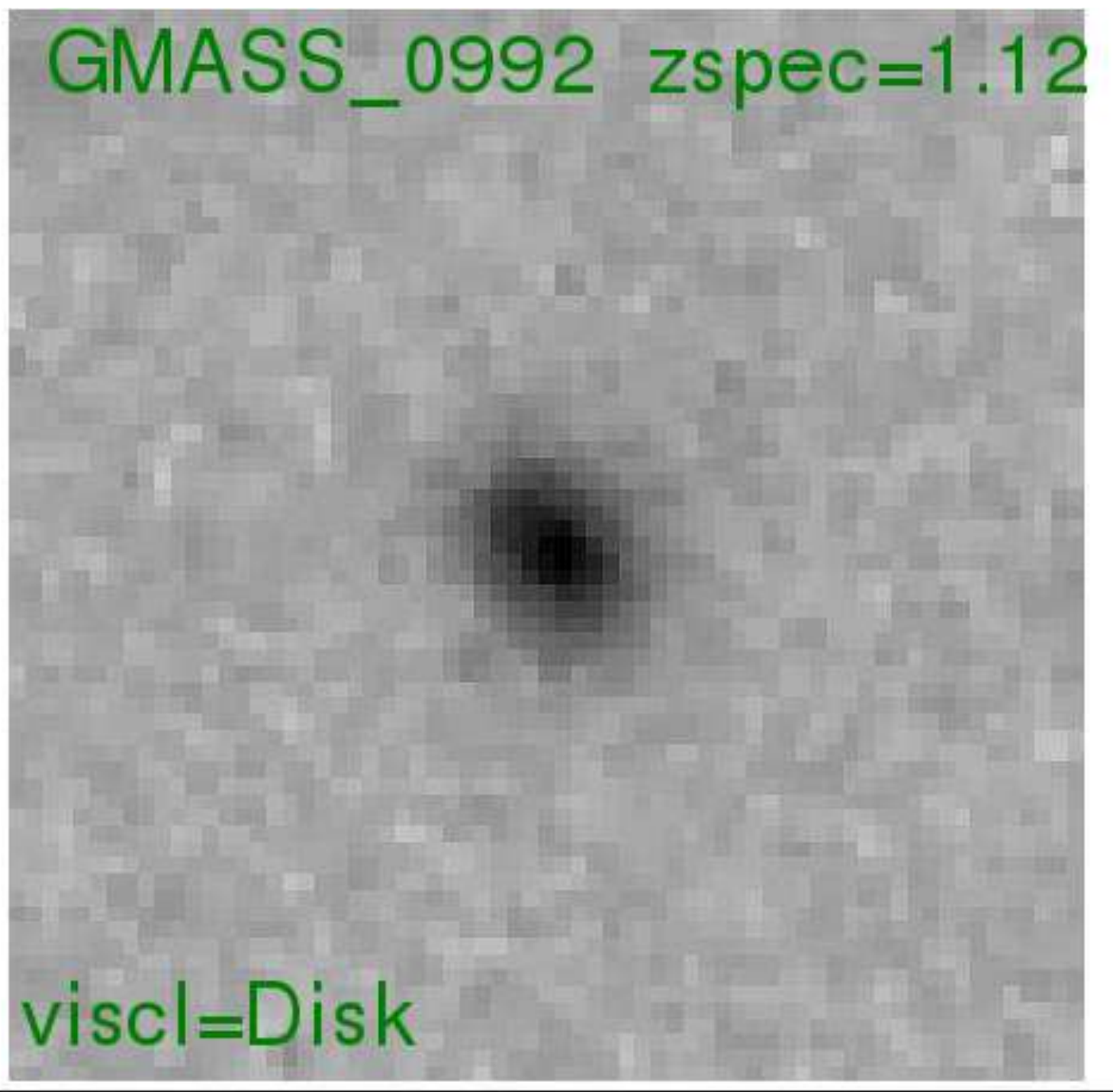}			     
\includegraphics[trim=100 40 75 390, clip=true, width=30mm]{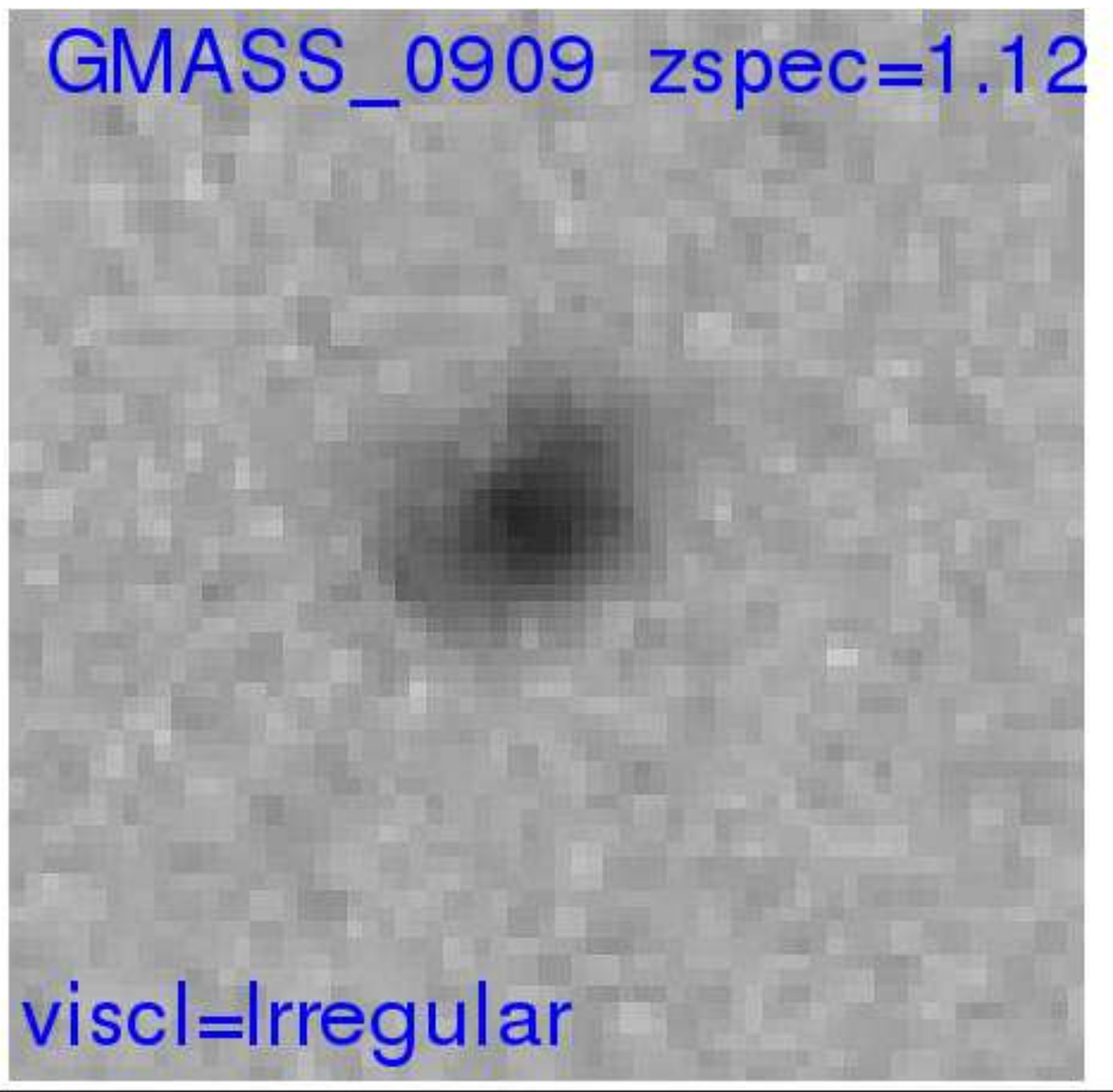}			     
\includegraphics[trim=100 40 75 390, clip=true, width=30mm]{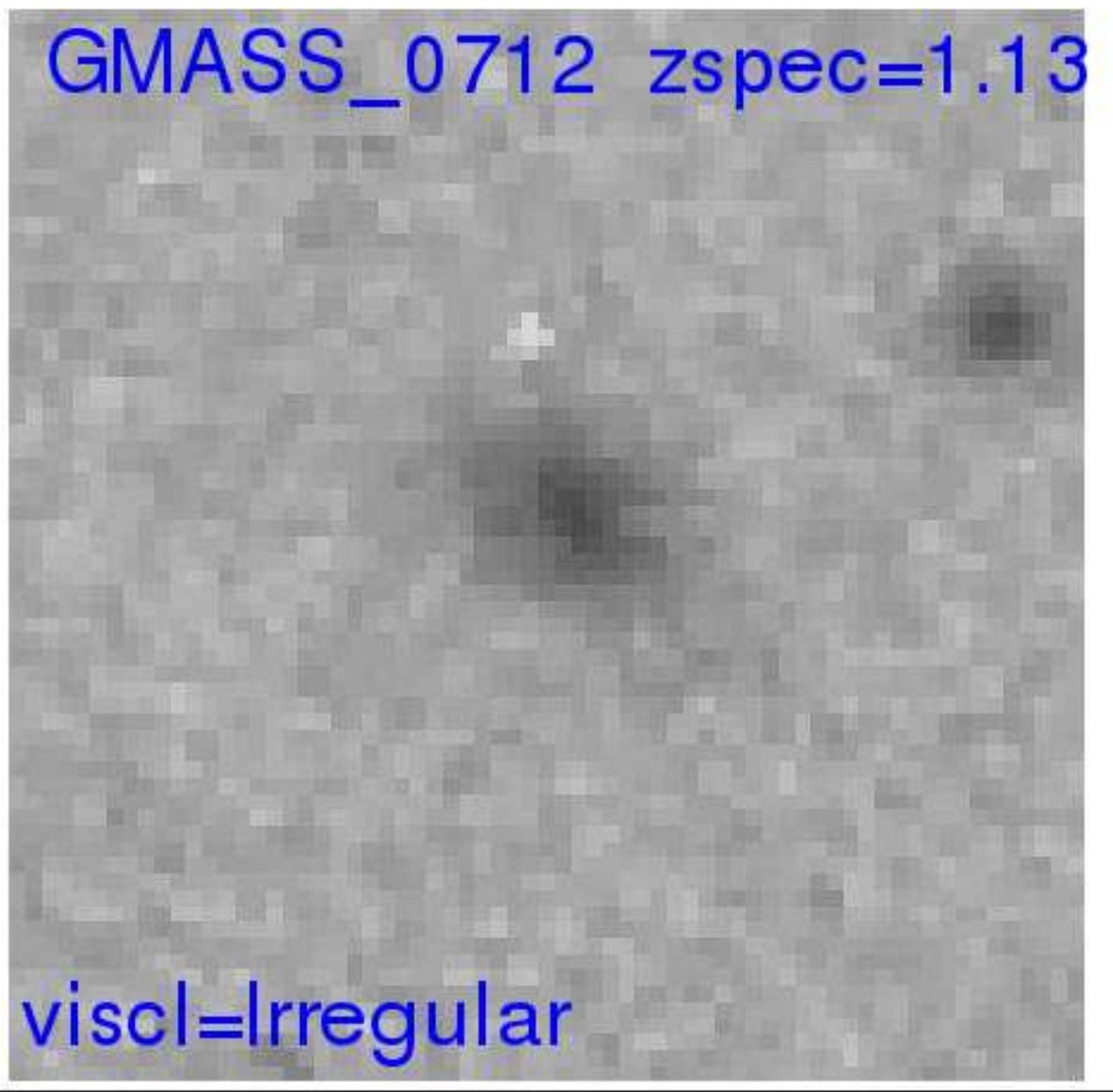}			     
\includegraphics[trim=100 40 75 390, clip=true, width=30mm]{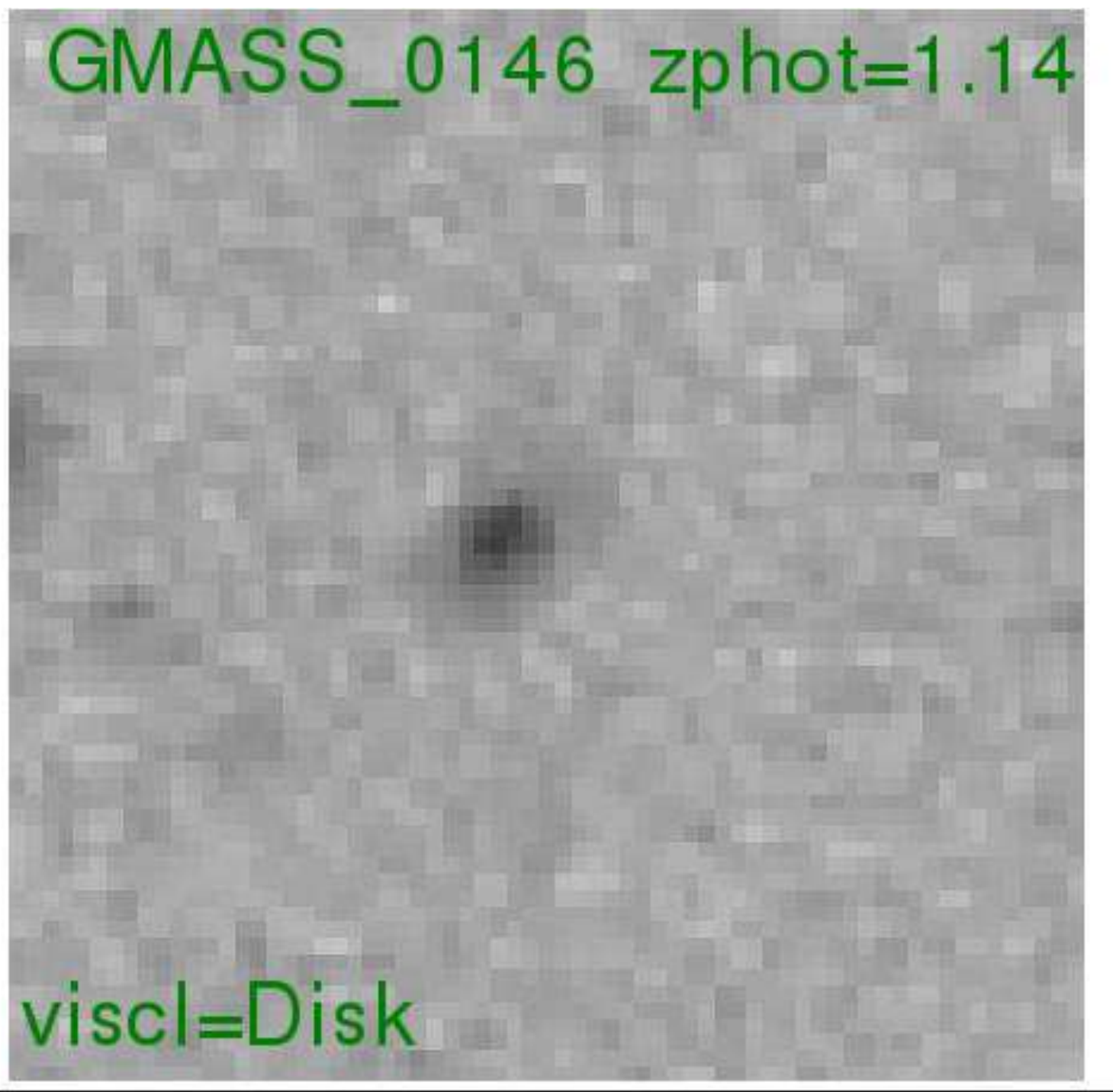}		     

\includegraphics[trim=100 40 75 390, clip=true, width=30mm]{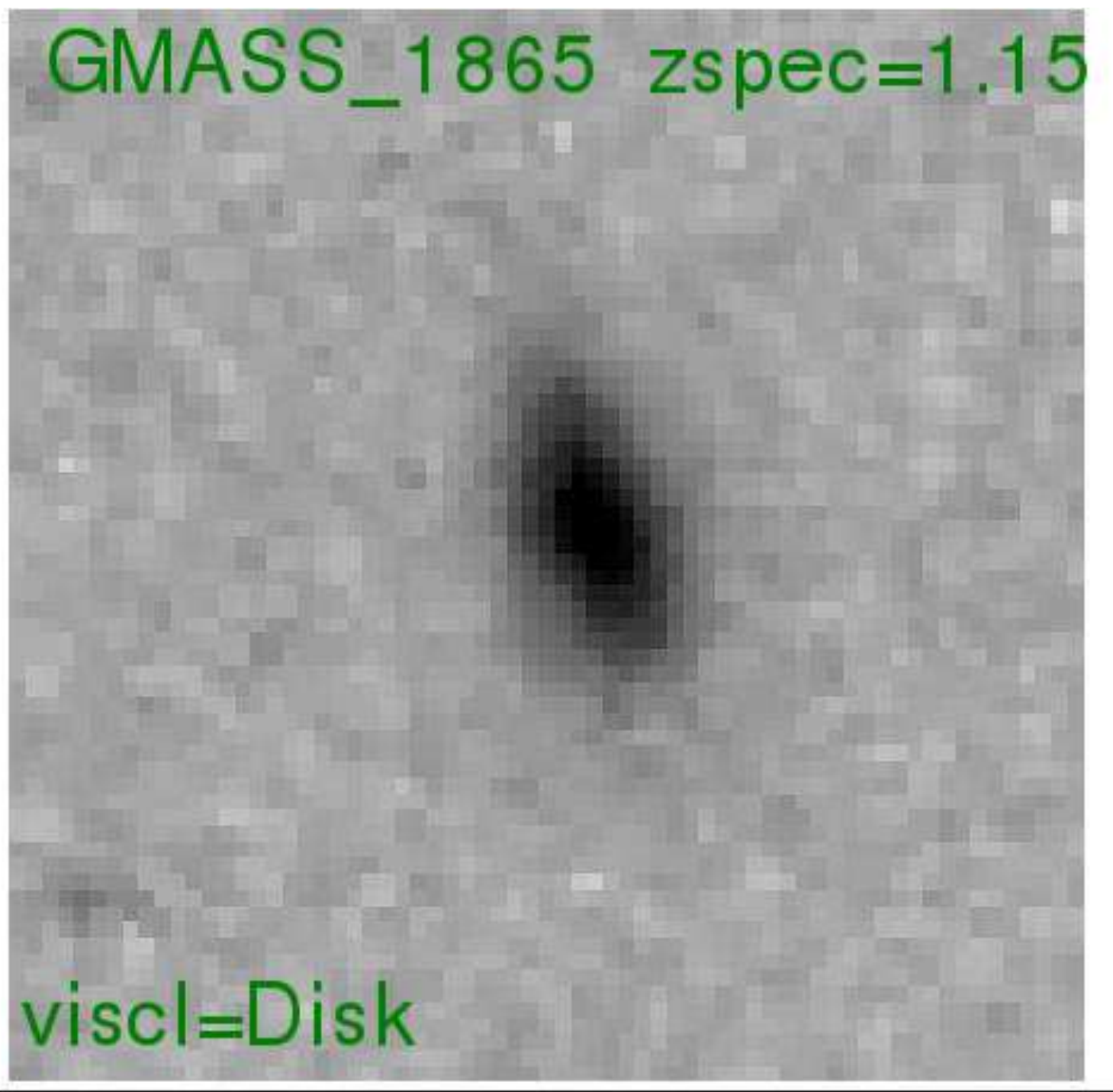}			     
\includegraphics[trim=100 40 75 390, clip=true, width=30mm]{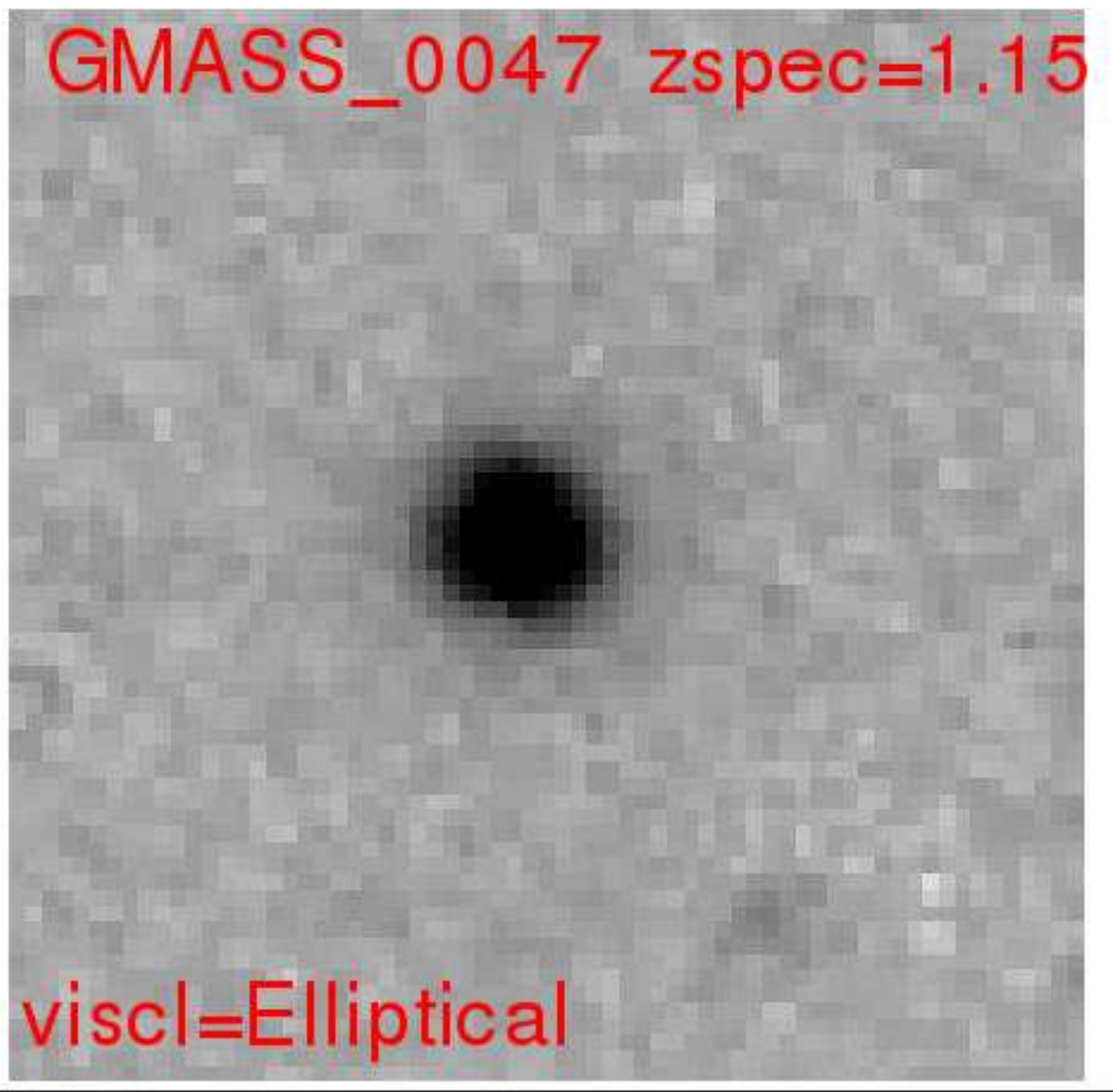}			     
\includegraphics[trim=100 40 75 390, clip=true, width=30mm]{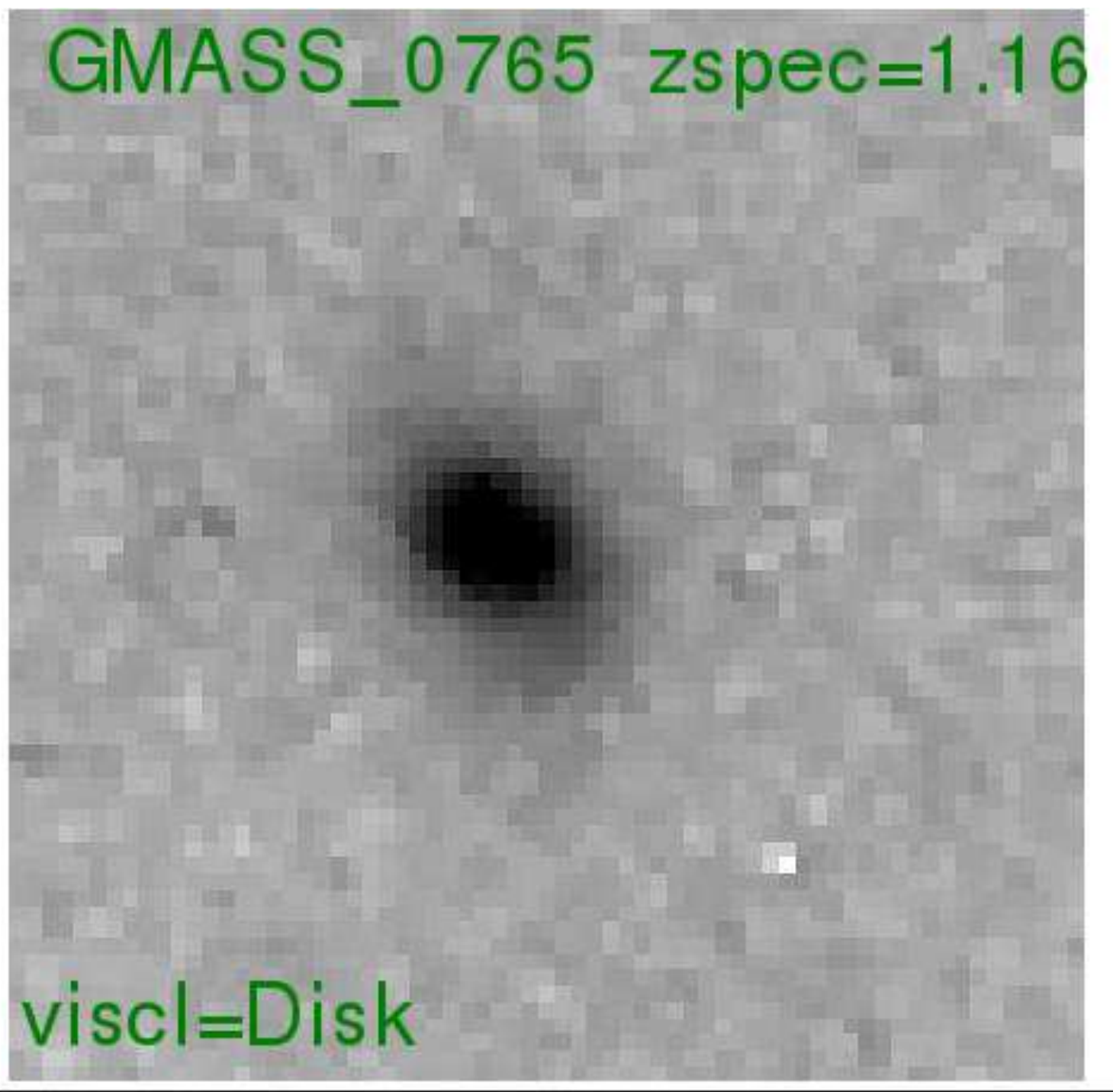}			     
\includegraphics[trim=100 40 75 390, clip=true, width=30mm]{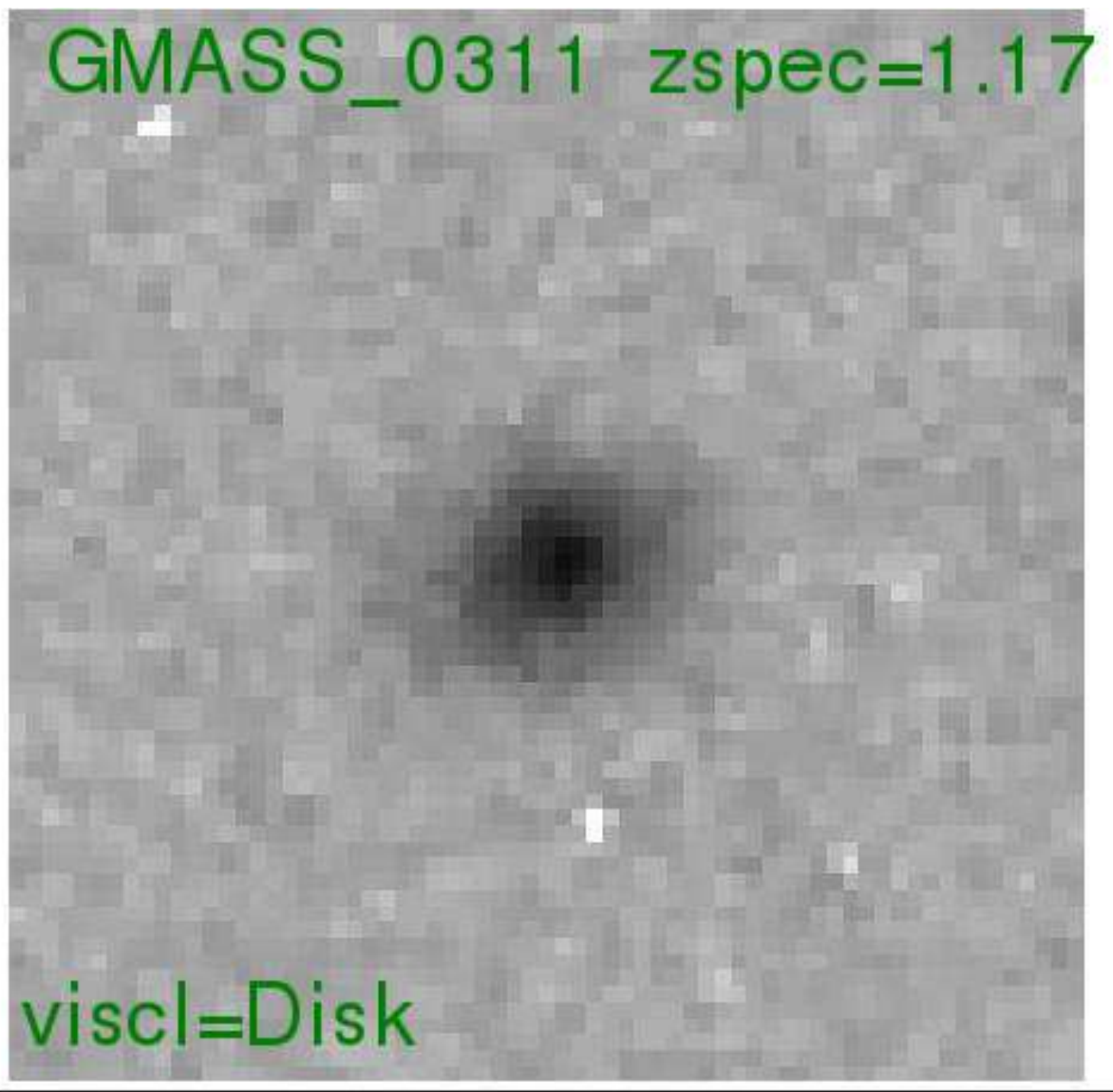}			     
\includegraphics[trim=100 40 75 390, clip=true, width=30mm]{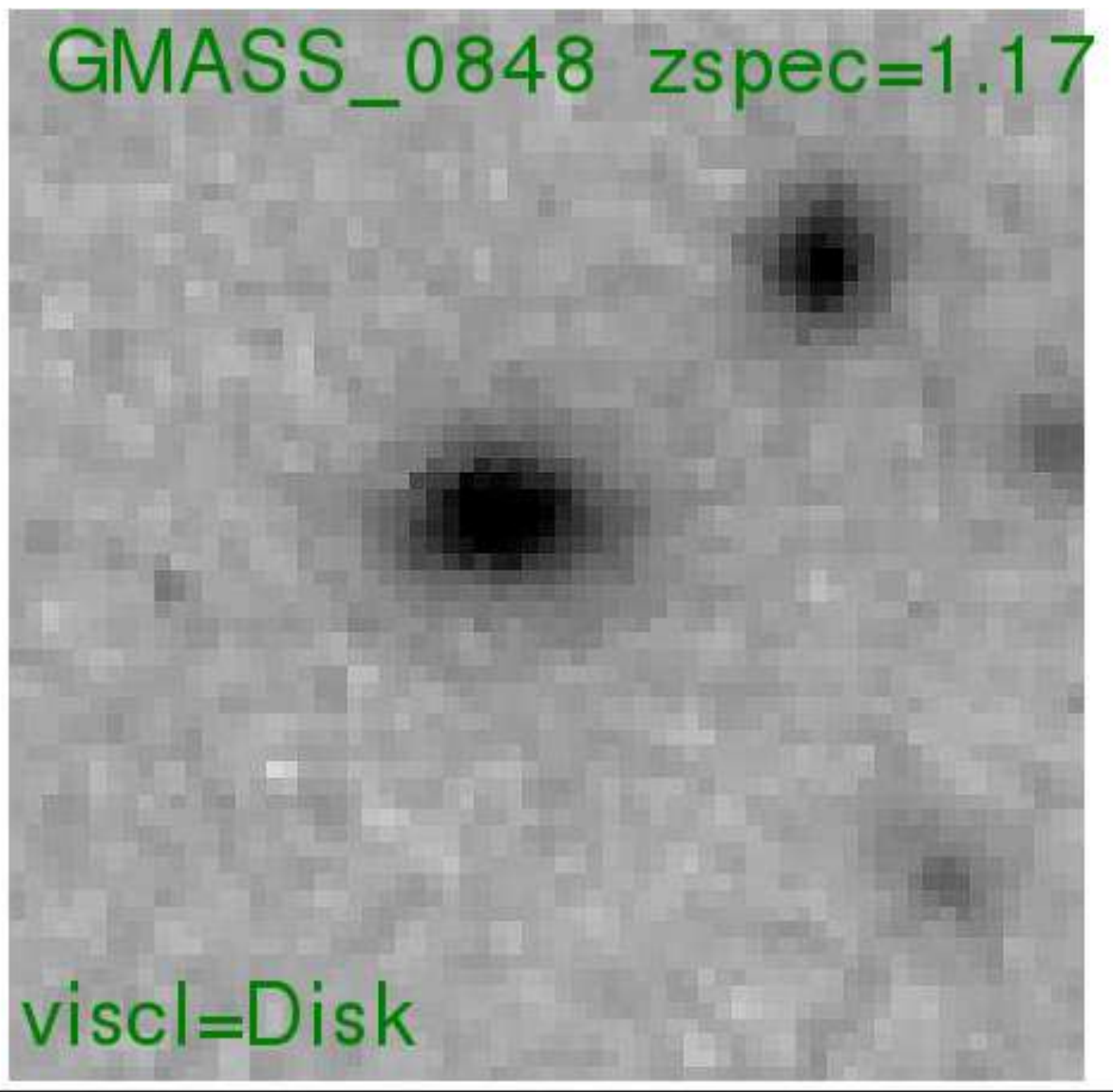}			     
\includegraphics[trim=100 40 75 390, clip=true, width=30mm]{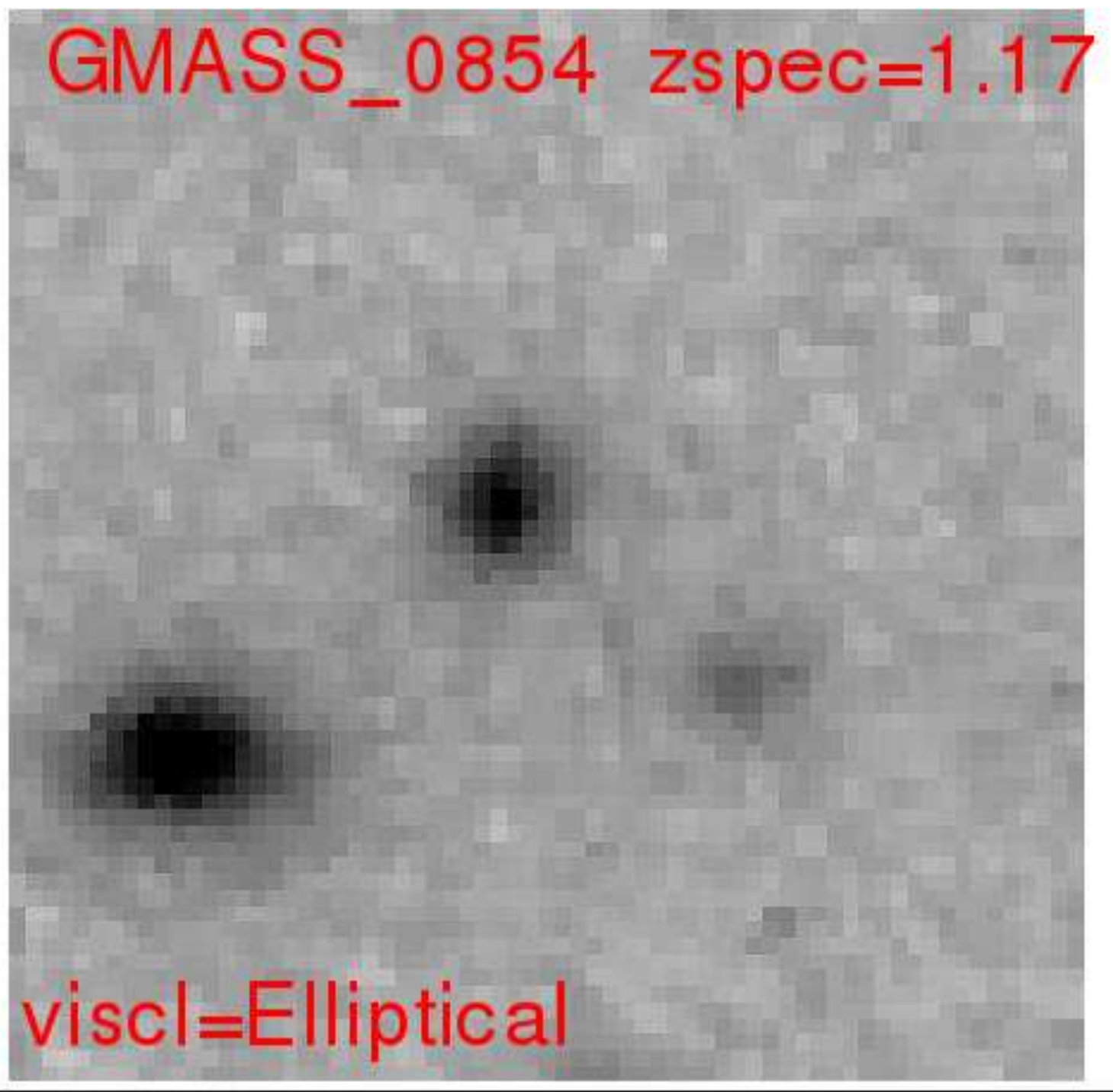}		     

\includegraphics[trim=100 40 75 390, clip=true, width=30mm]{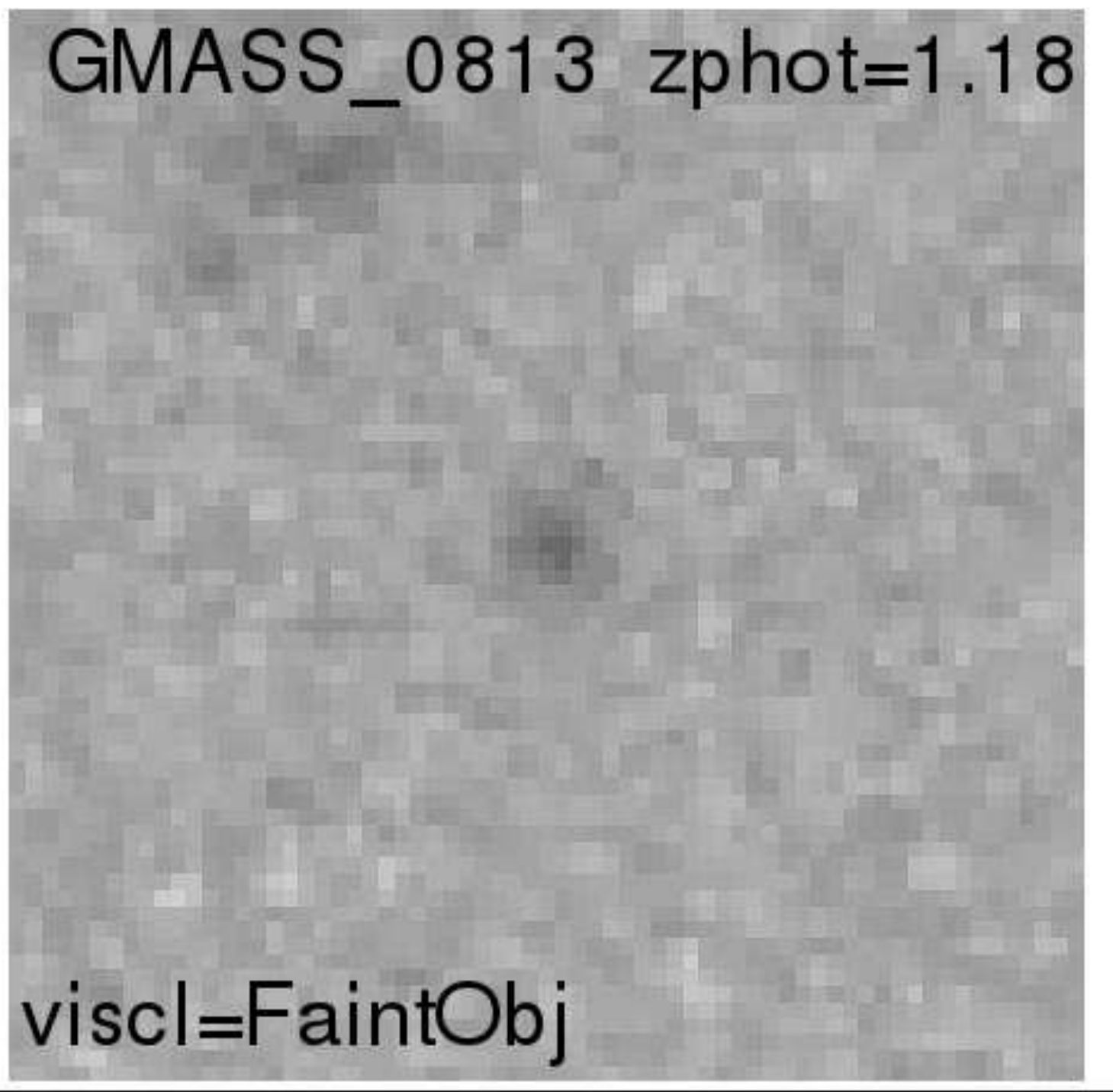}			     
\includegraphics[trim=100 40 75 390, clip=true, width=30mm]{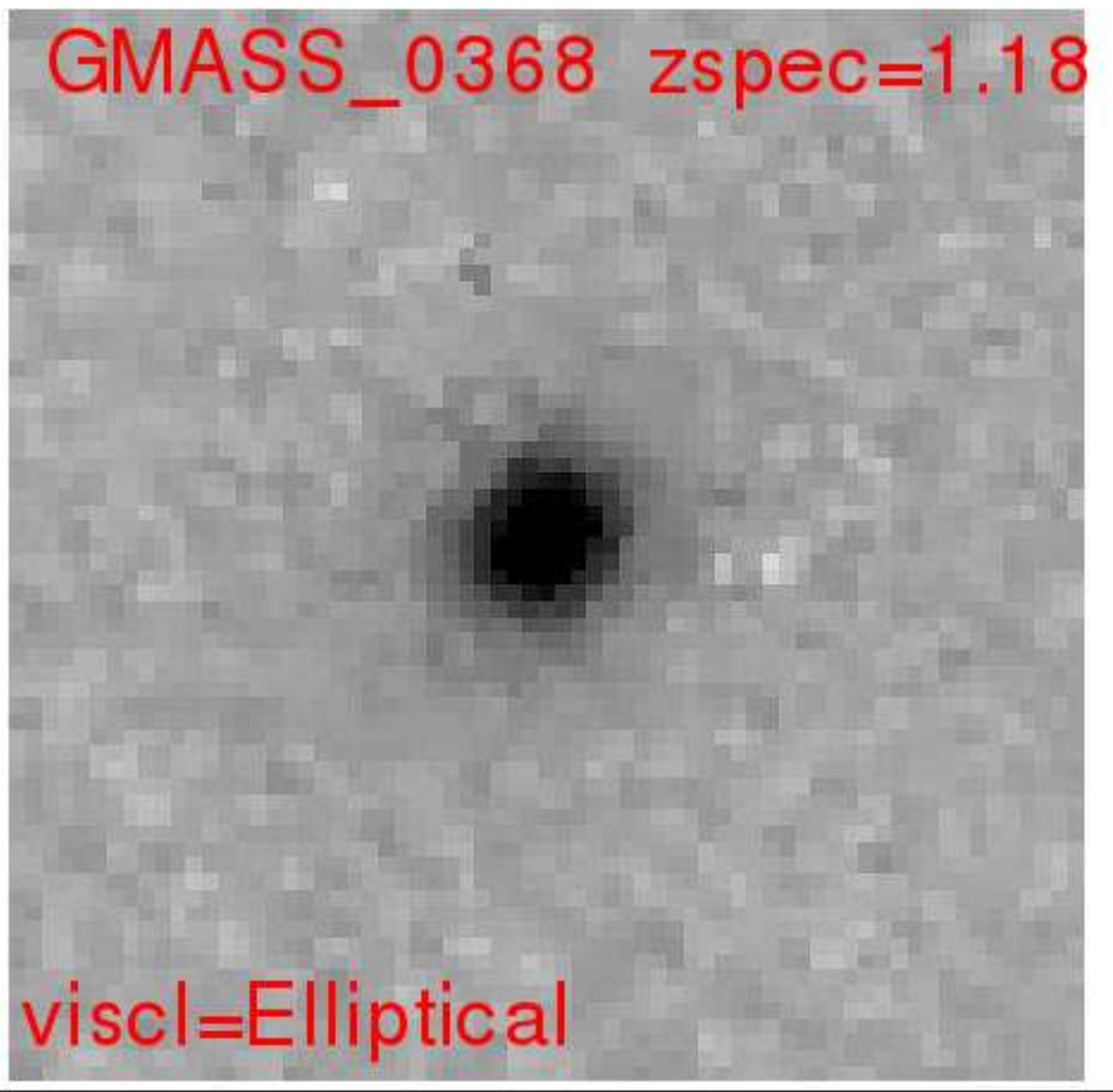}			     
\includegraphics[trim=100 40 75 390, clip=true, width=30mm]{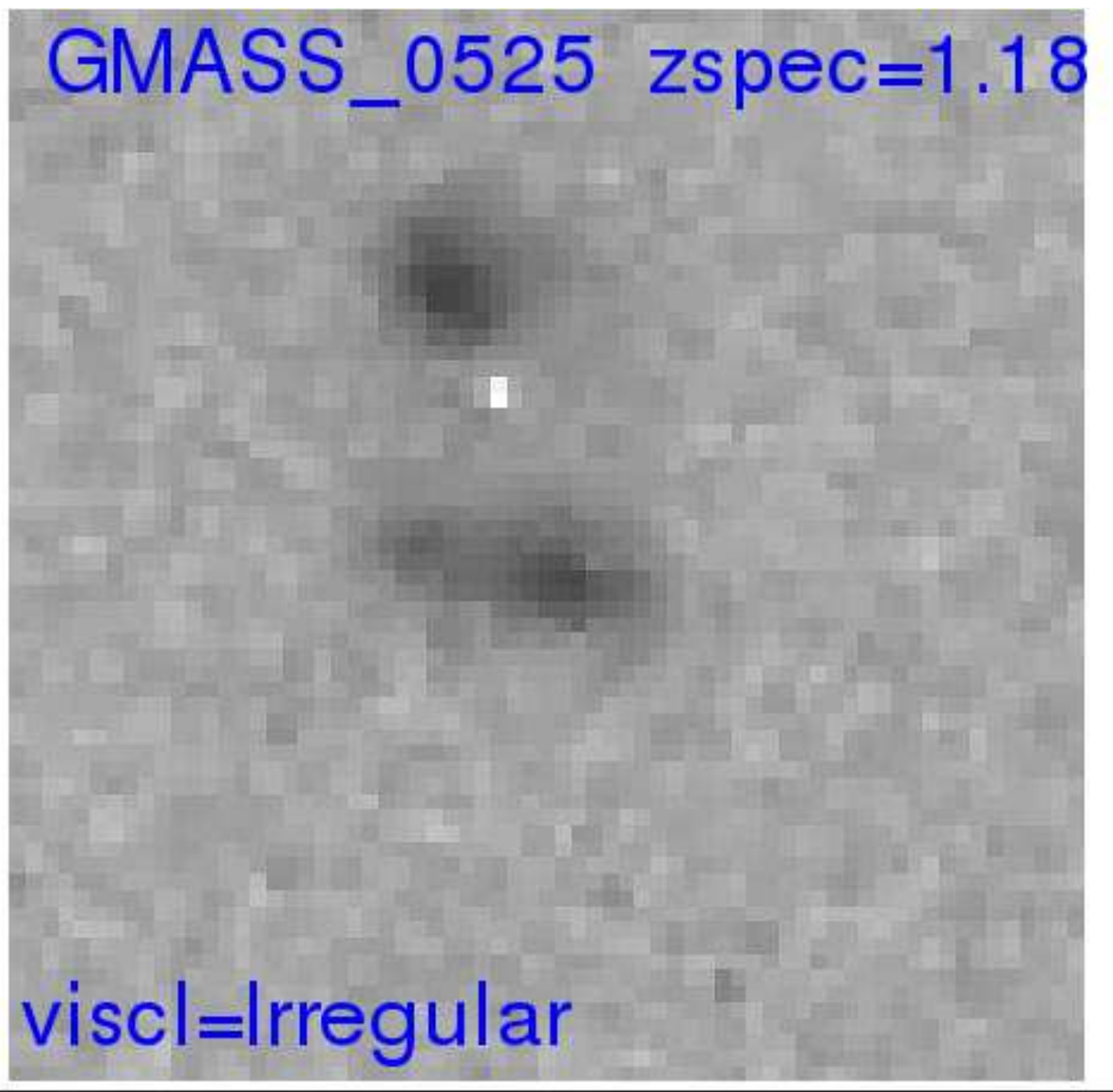}			     
\includegraphics[trim=100 40 75 390, clip=true, width=30mm]{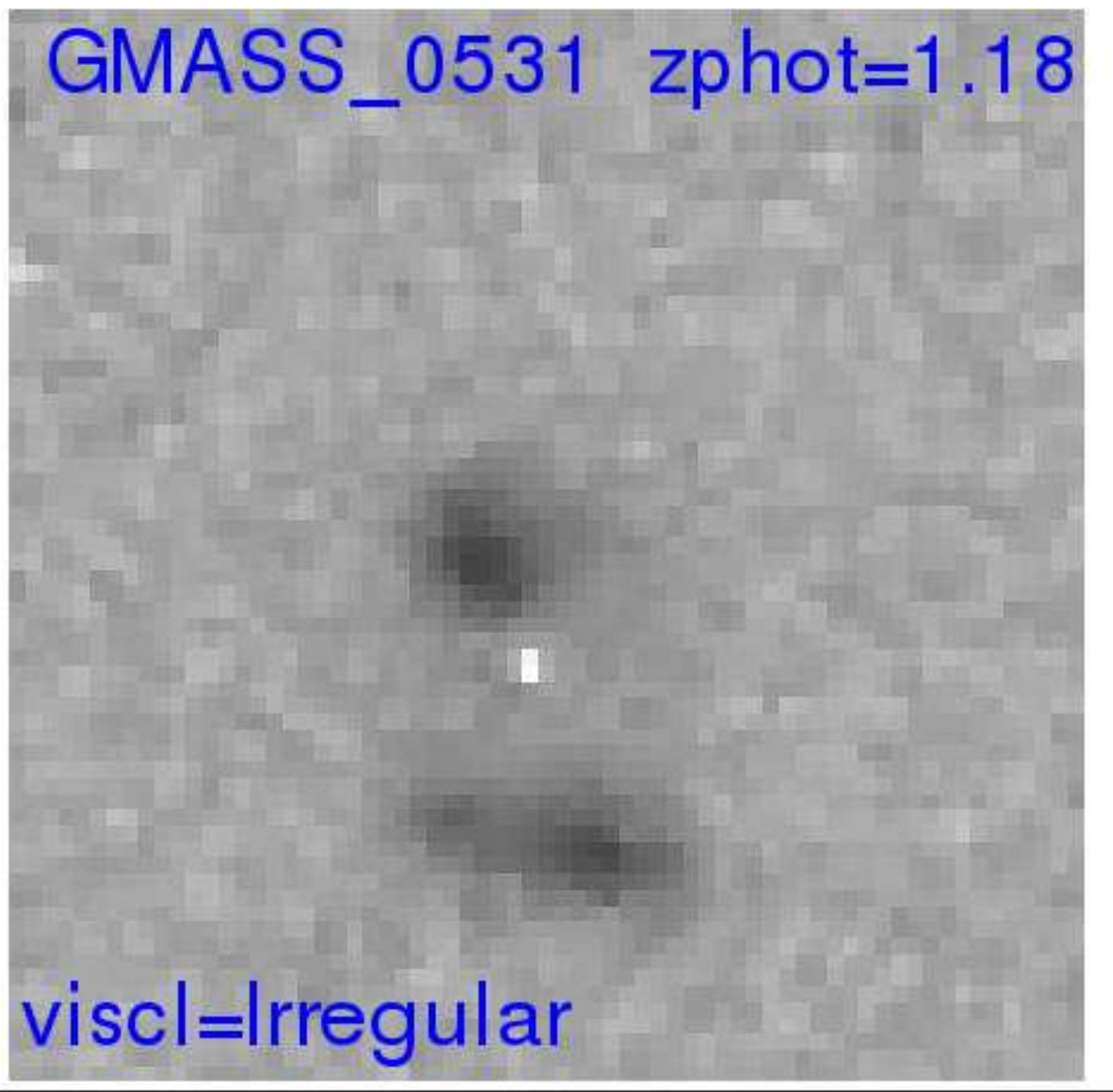}			     
\includegraphics[trim=100 40 75 390, clip=true, width=30mm]{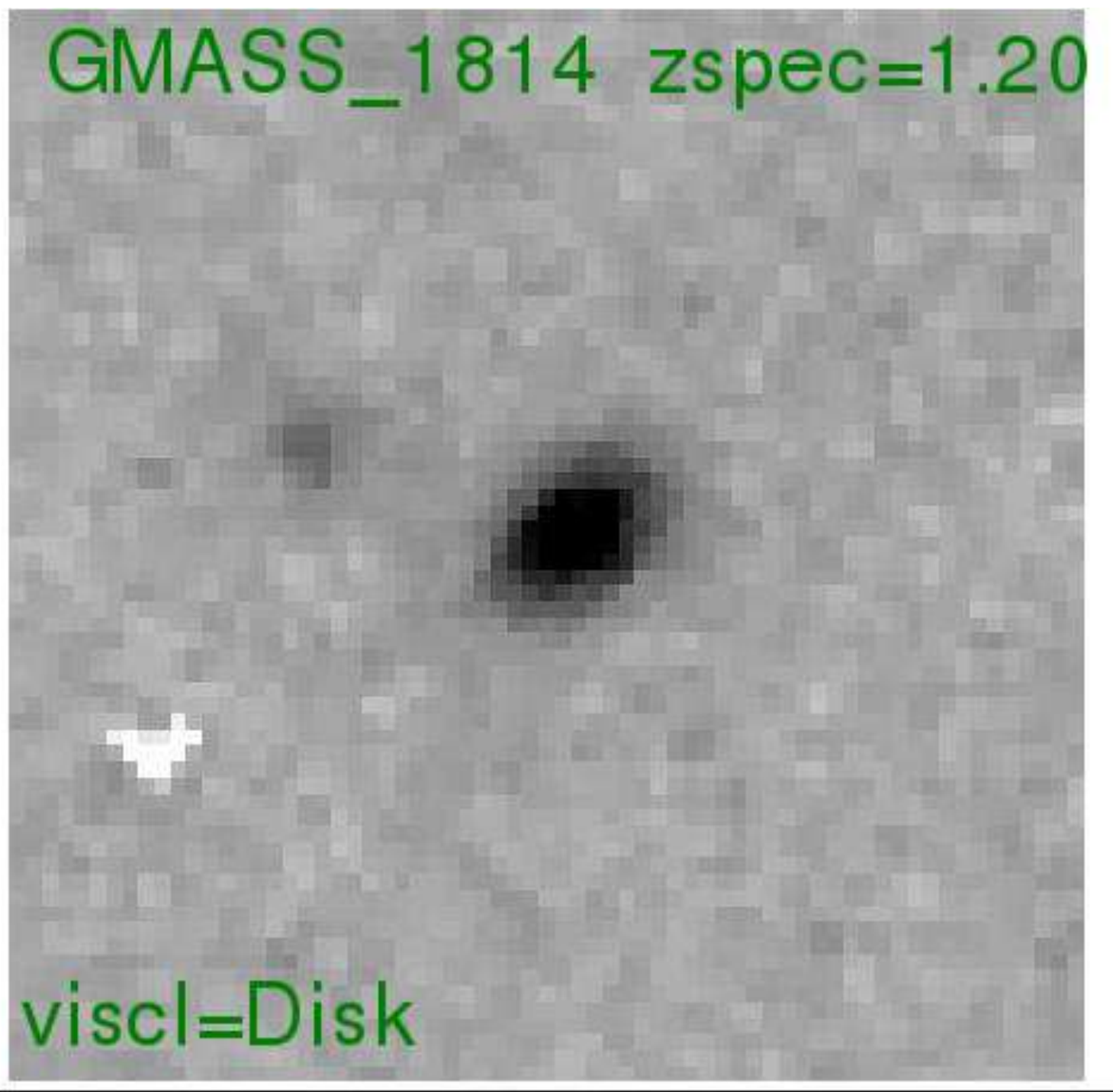}			     
\includegraphics[trim=100 40 75 390, clip=true, width=30mm]{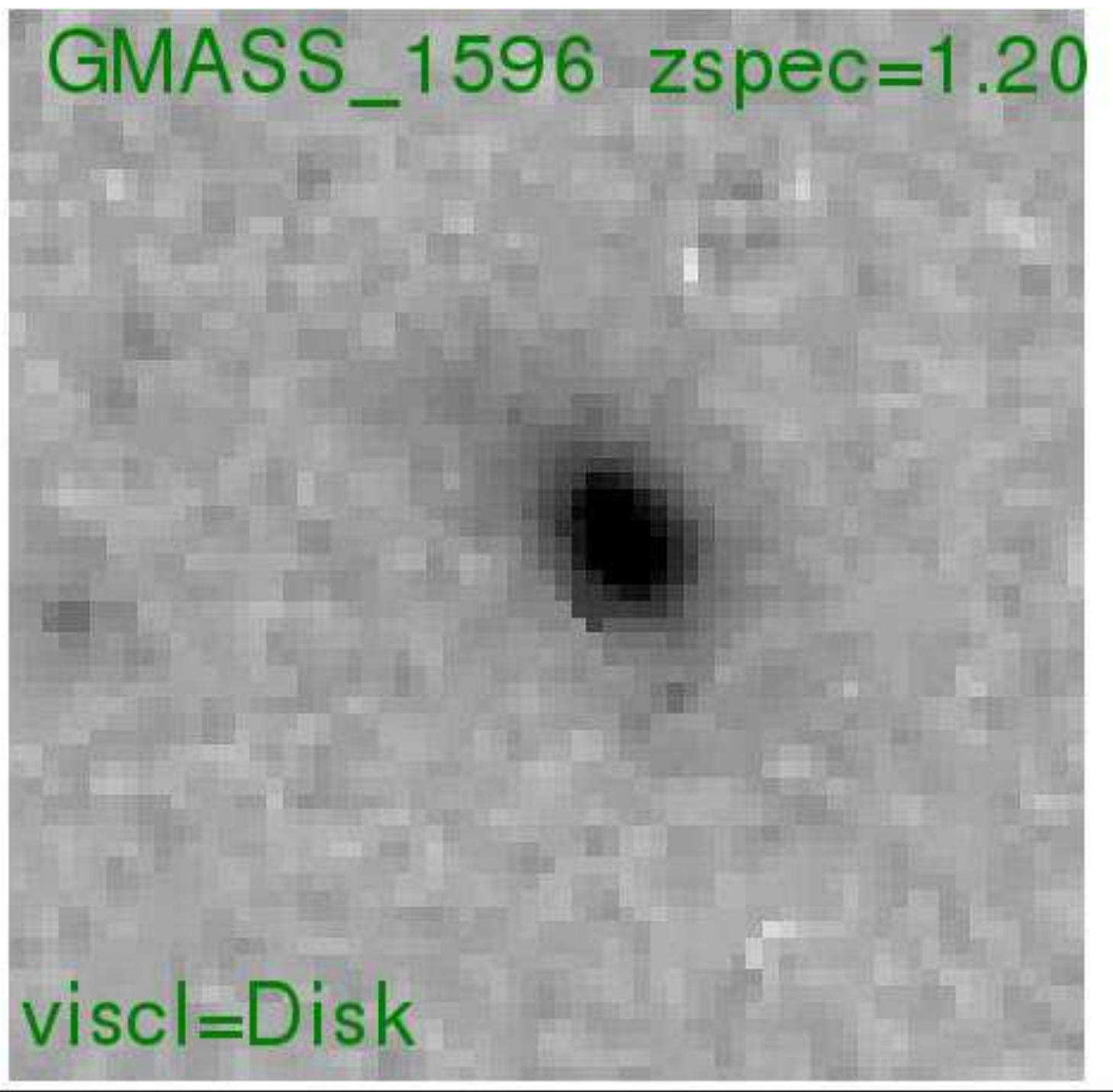}	
\end{figure*}
\begin{figure*}
\centering   
\includegraphics[trim=100 40 75 390, clip=true, width=30mm]{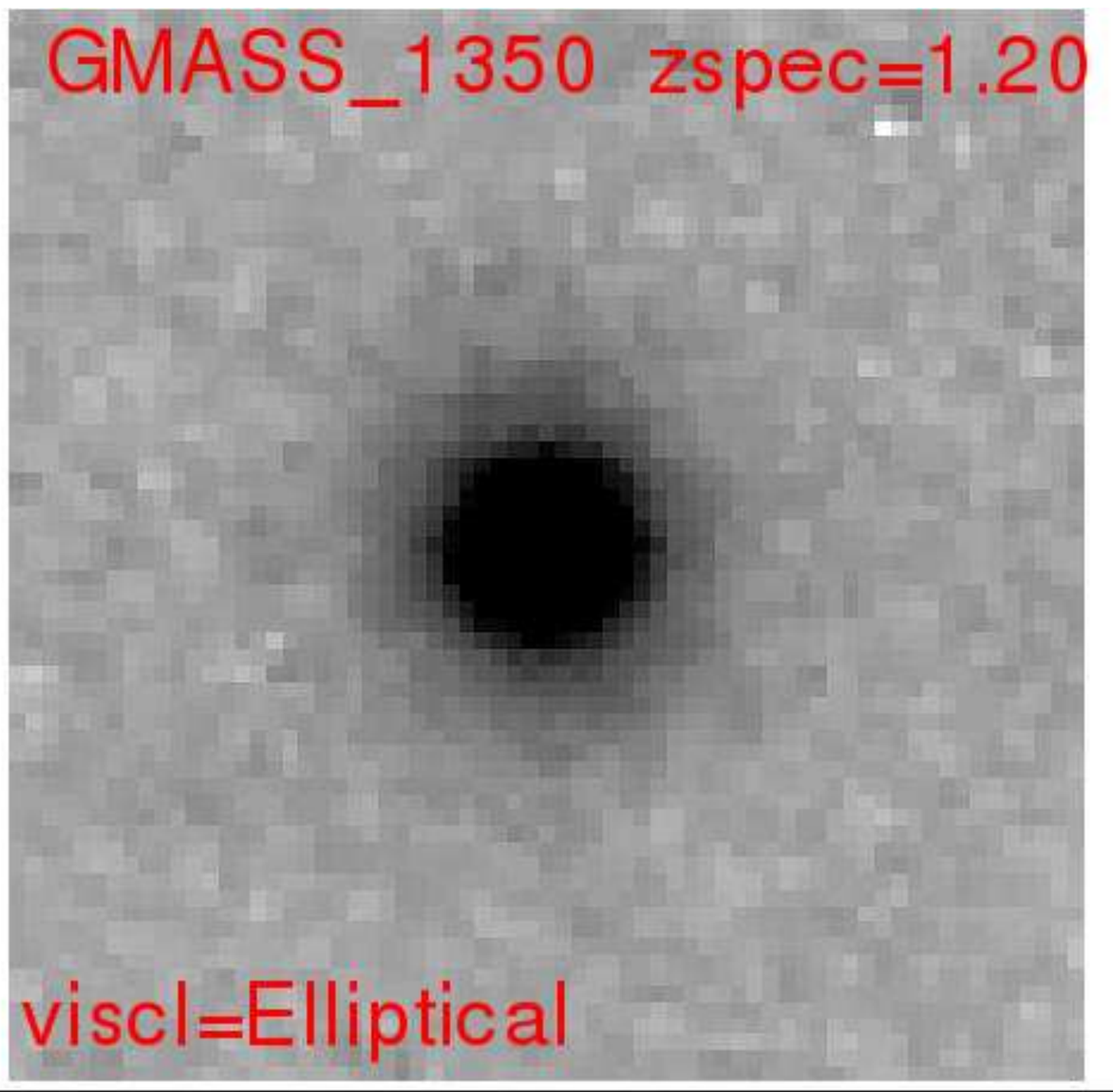}			     
\includegraphics[trim=100 40 75 390, clip=true, width=30mm]{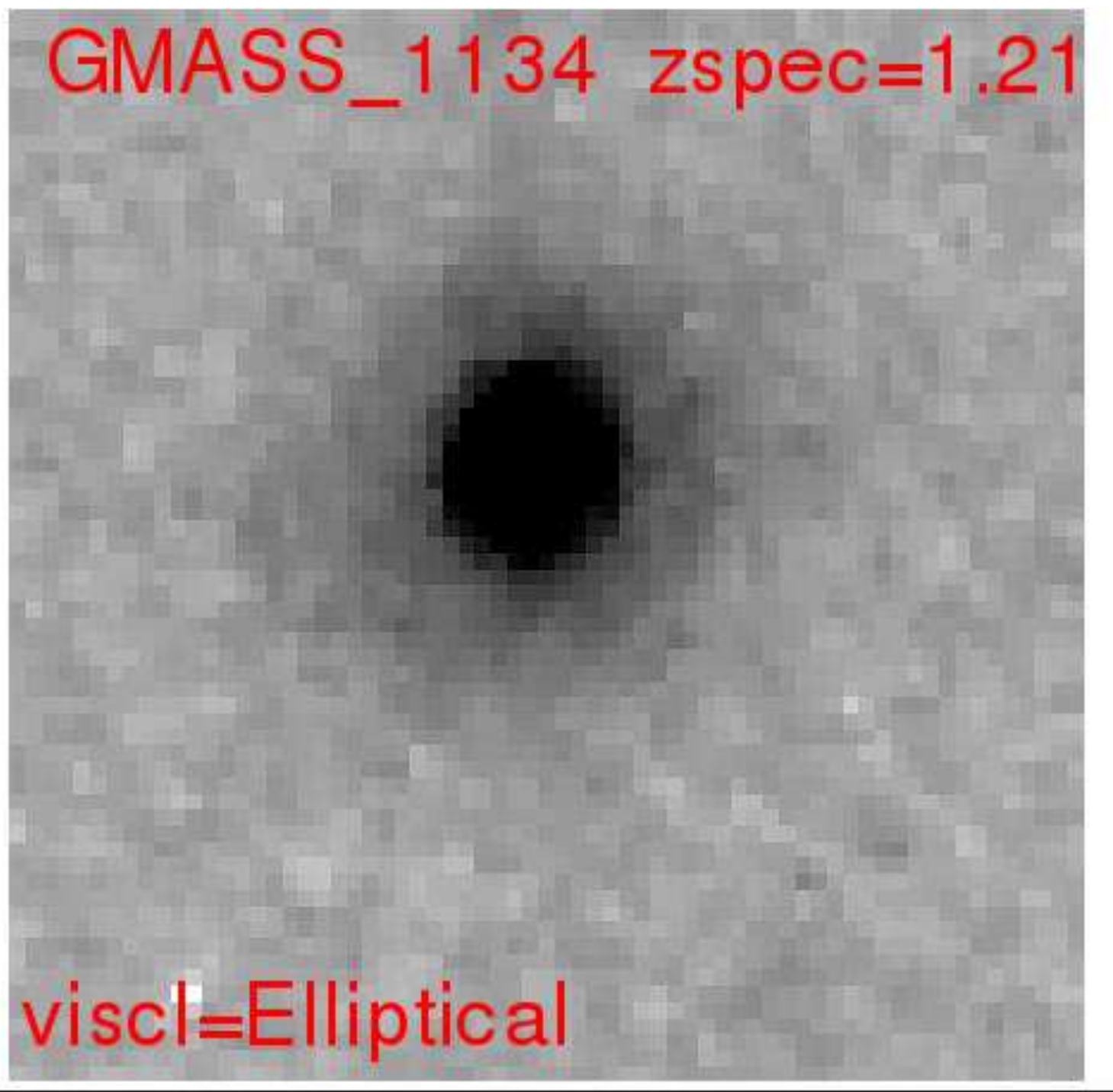}			     
\includegraphics[trim=100 40 75 390, clip=true, width=30mm]{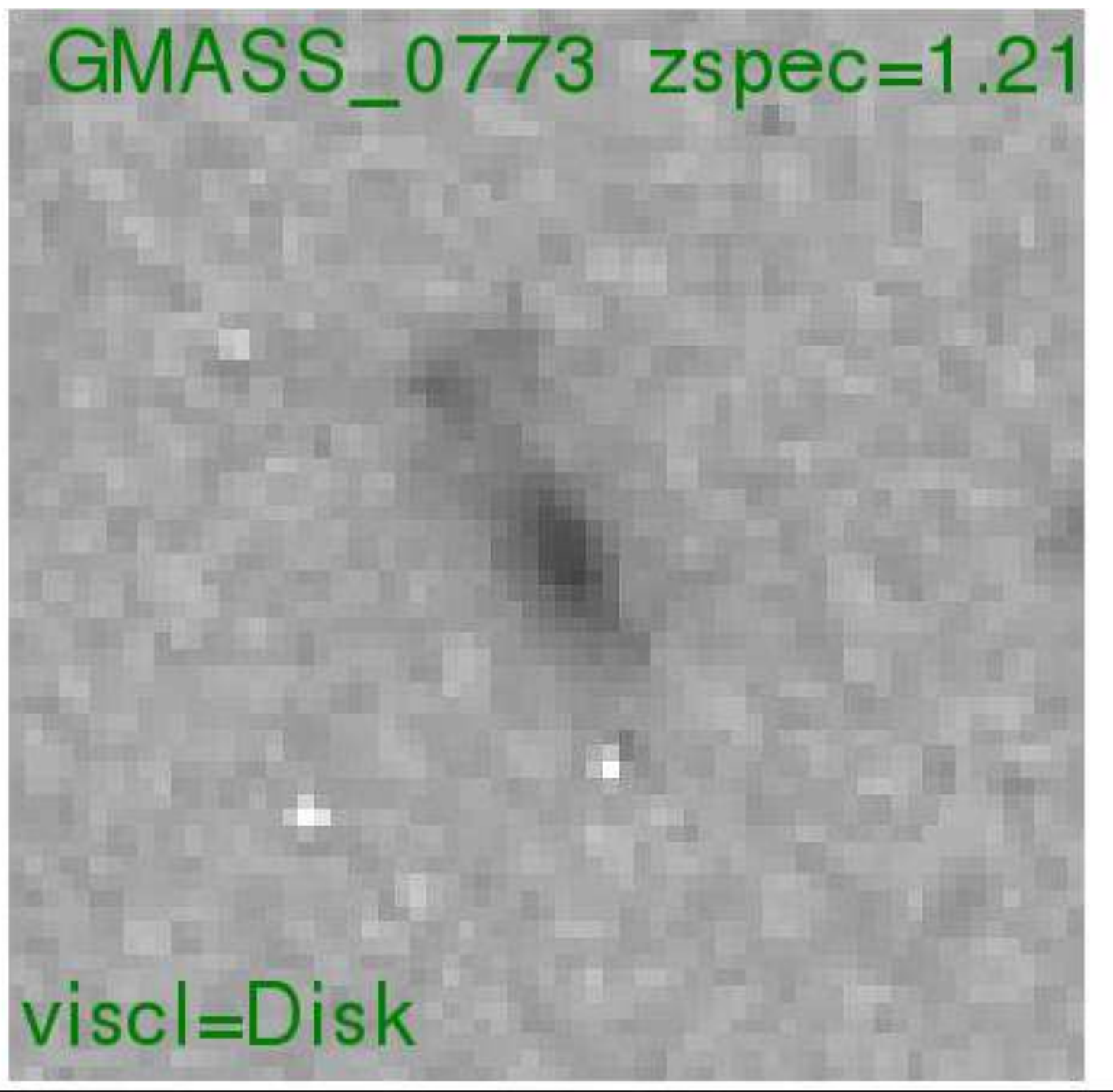}			     
\includegraphics[trim=100 40 75 390, clip=true, width=30mm]{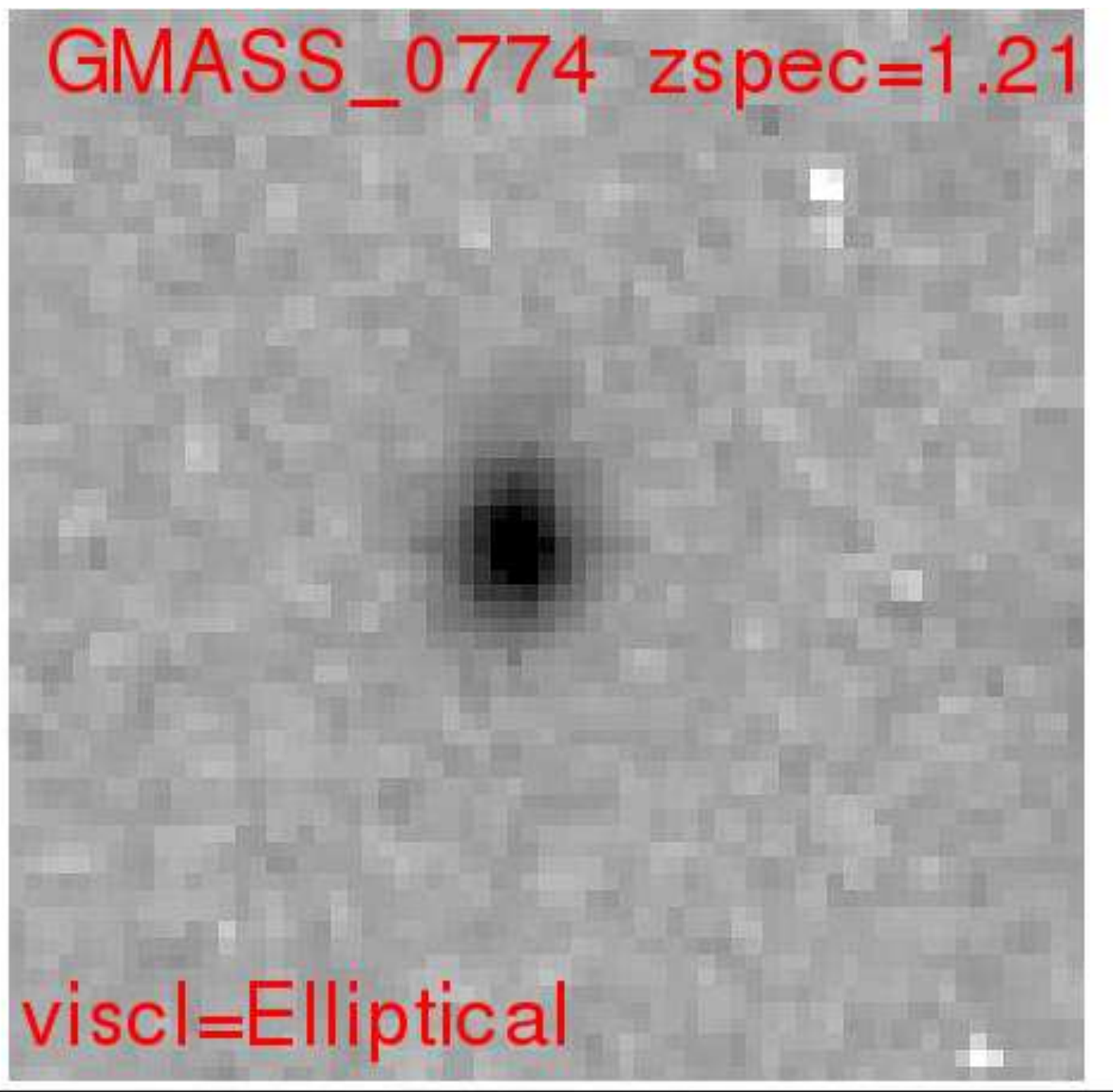}			     
\includegraphics[trim=100 40 75 390, clip=true, width=30mm]{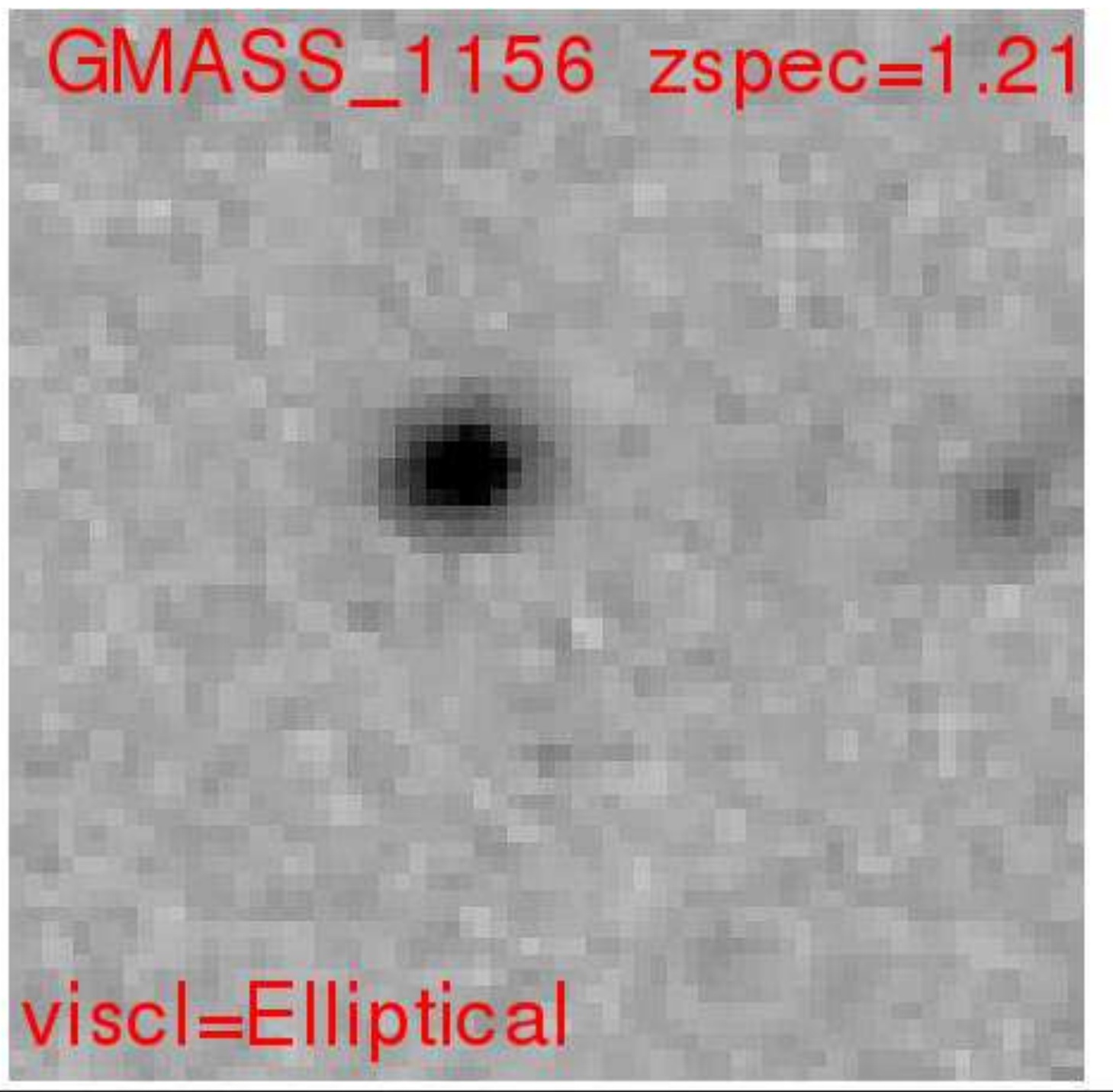}			     
\includegraphics[trim=100 40 75 390, clip=true, width=30mm]{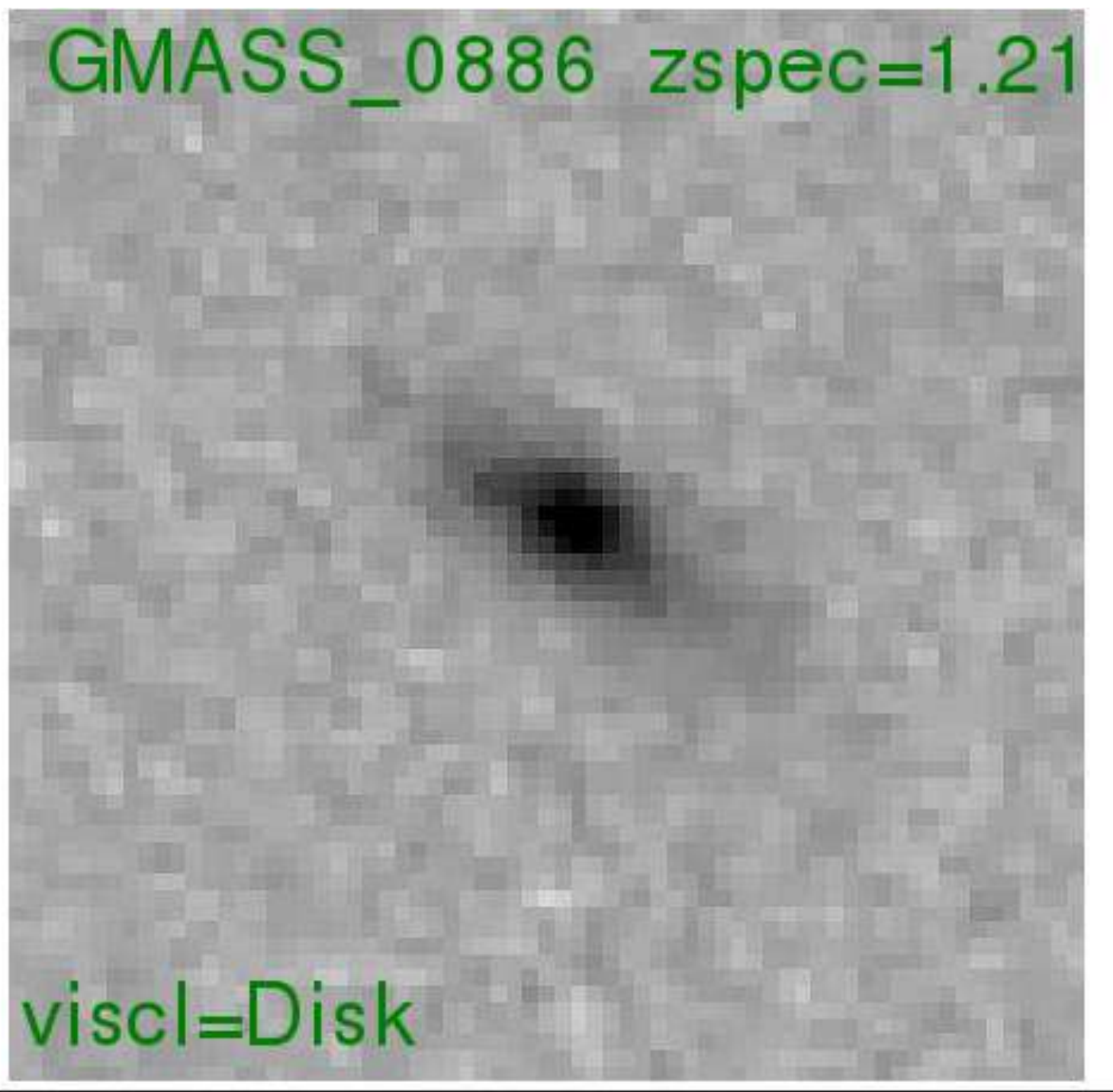}			     

\includegraphics[trim=100 40 75 390, clip=true, width=30mm]{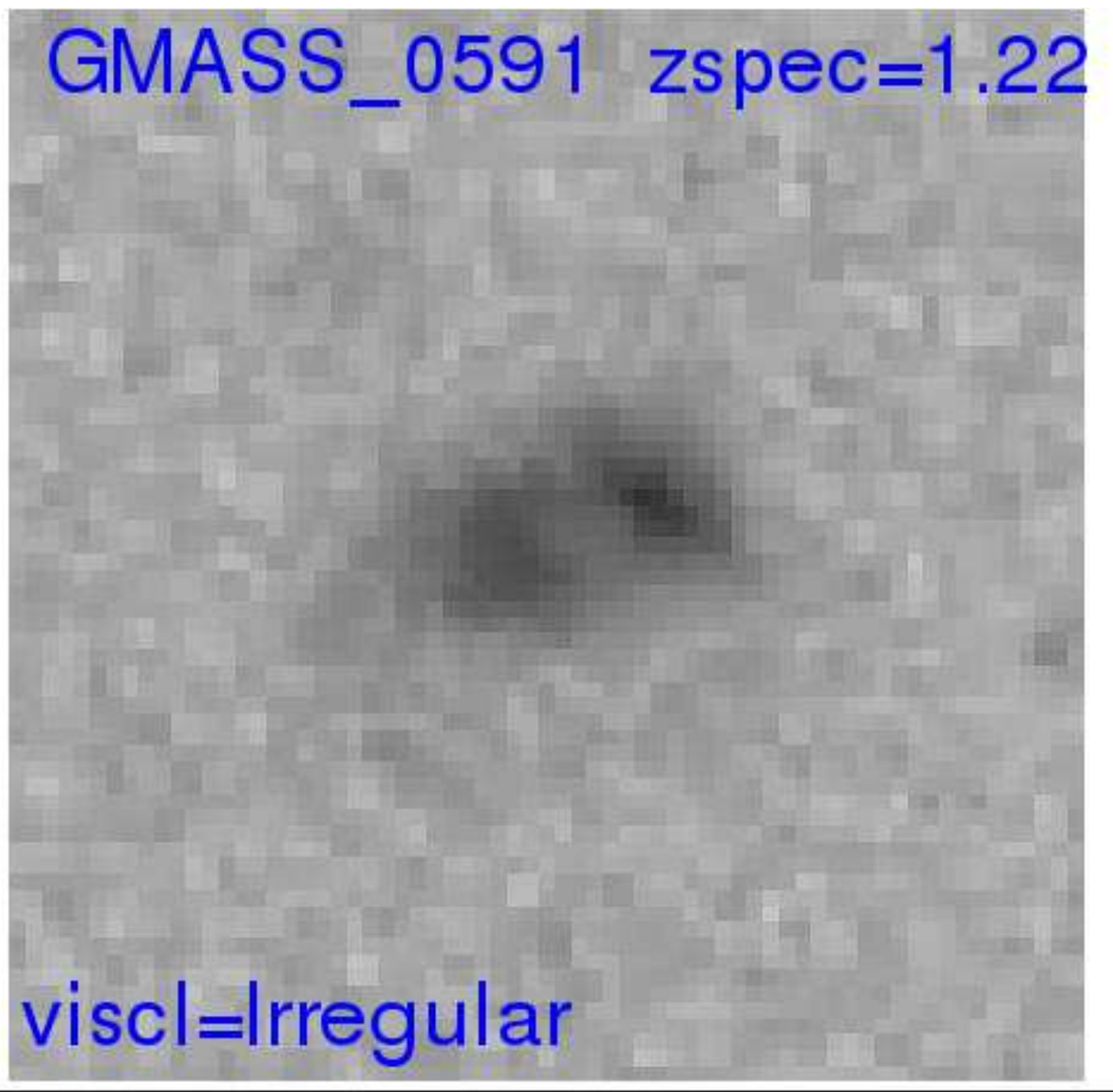}			     
\includegraphics[trim=100 40 75 390, clip=true, width=30mm]{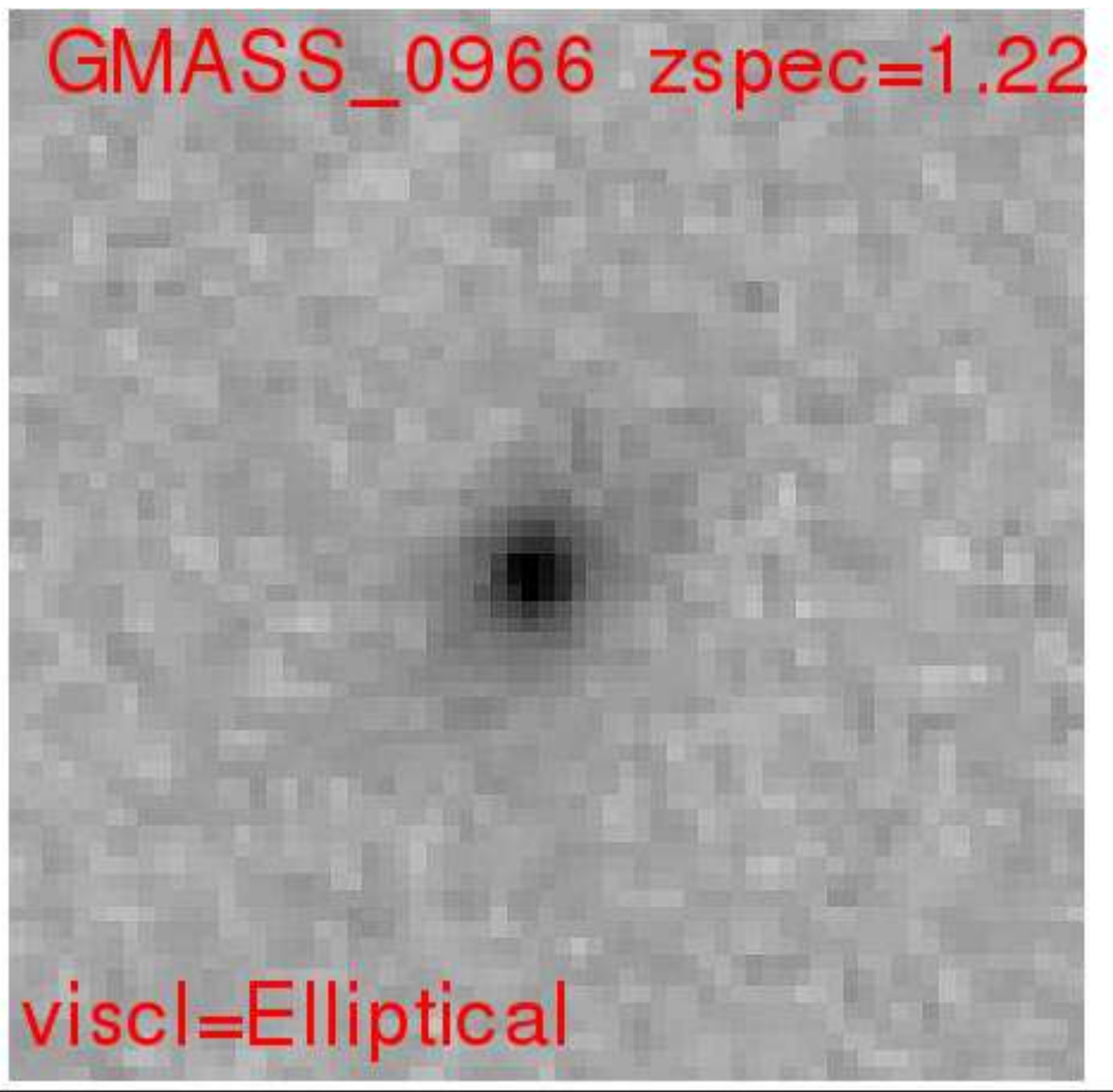}		     
\includegraphics[trim=100 40 75 390, clip=true, width=30mm]{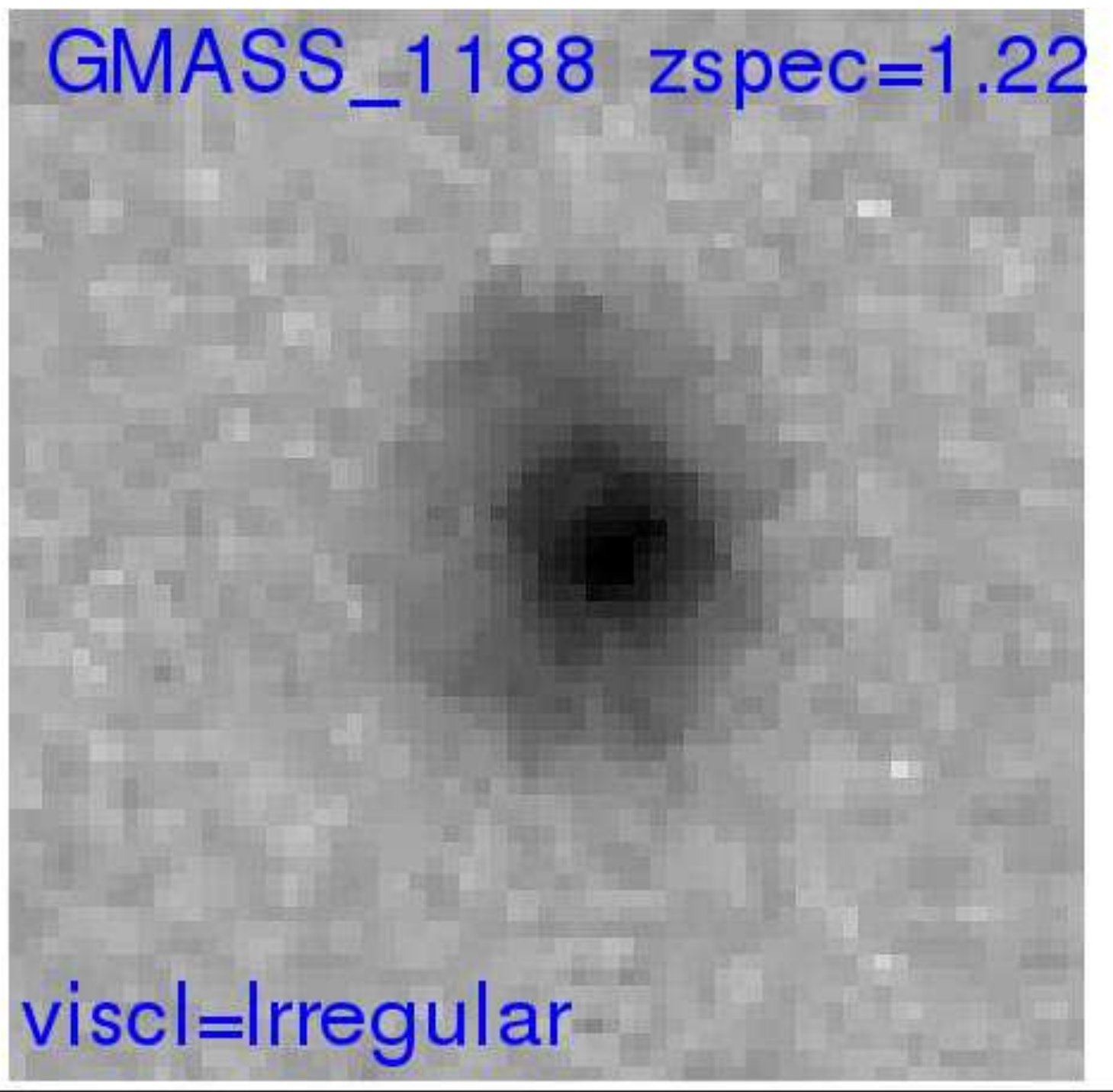}			     
\includegraphics[trim=100 40 75 390, clip=true, width=30mm]{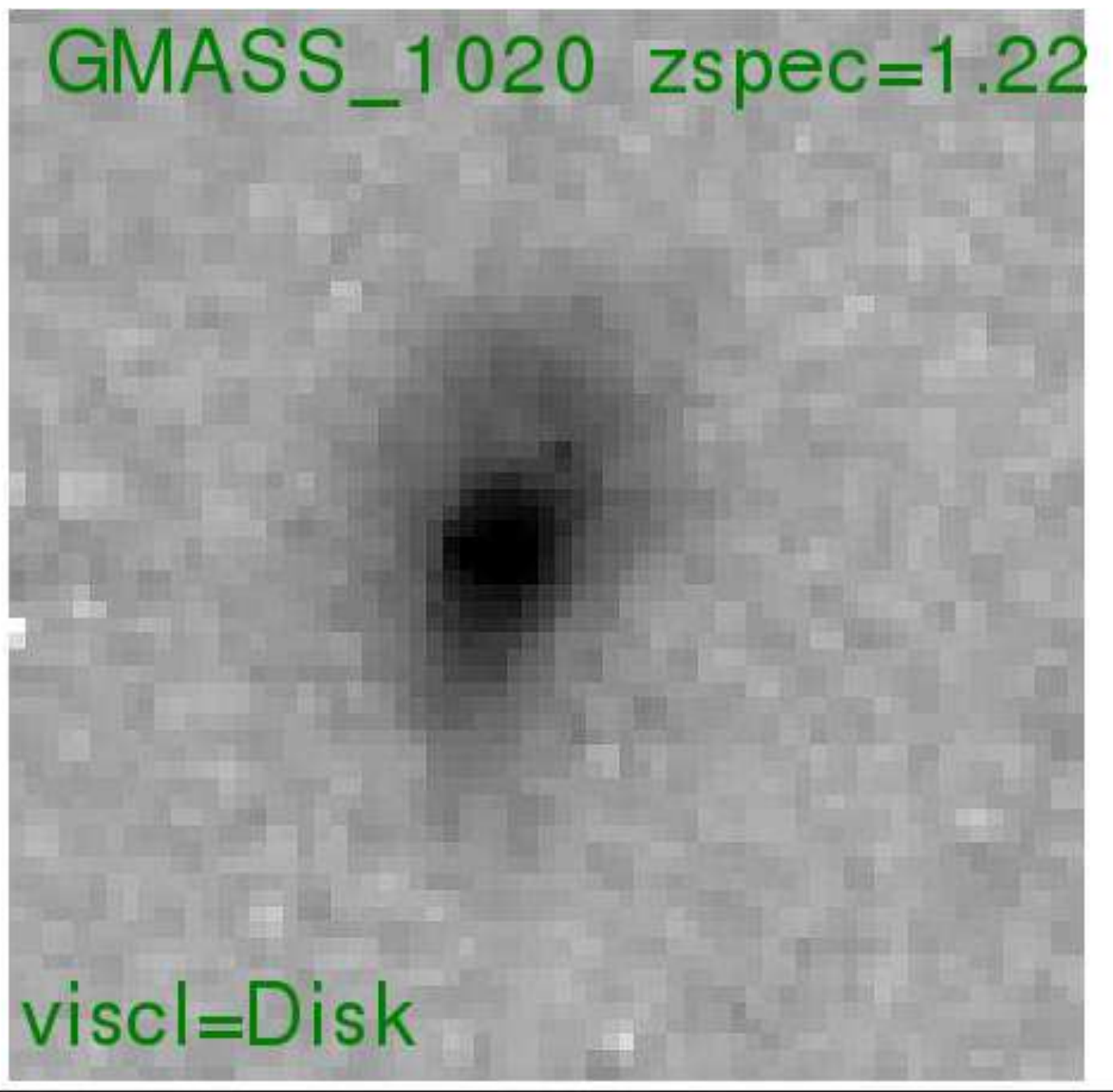}			     
\includegraphics[trim=100 40 75 390, clip=true, width=30mm]{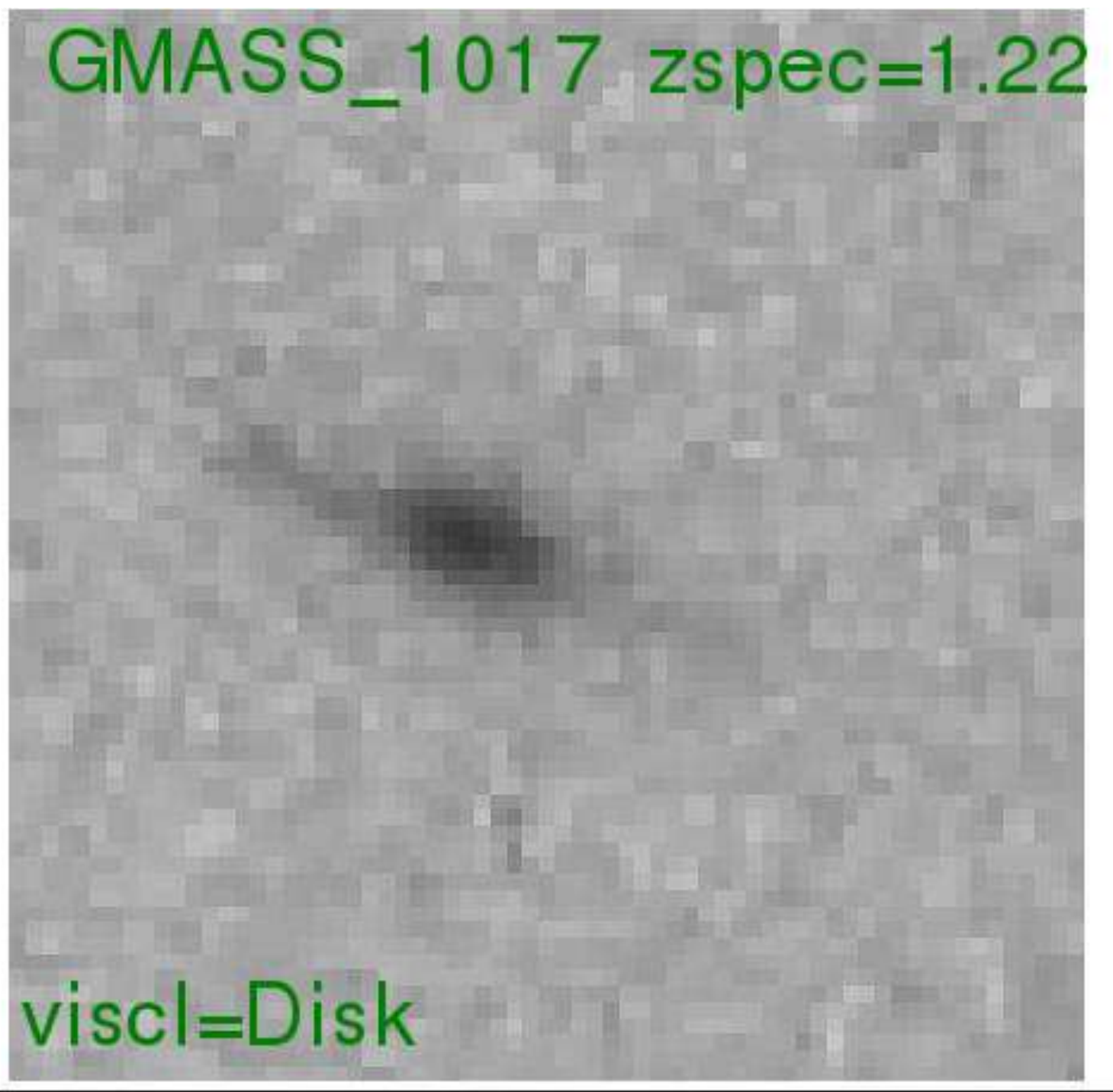}		     
\includegraphics[trim=100 40 75 390, clip=true, width=30mm]{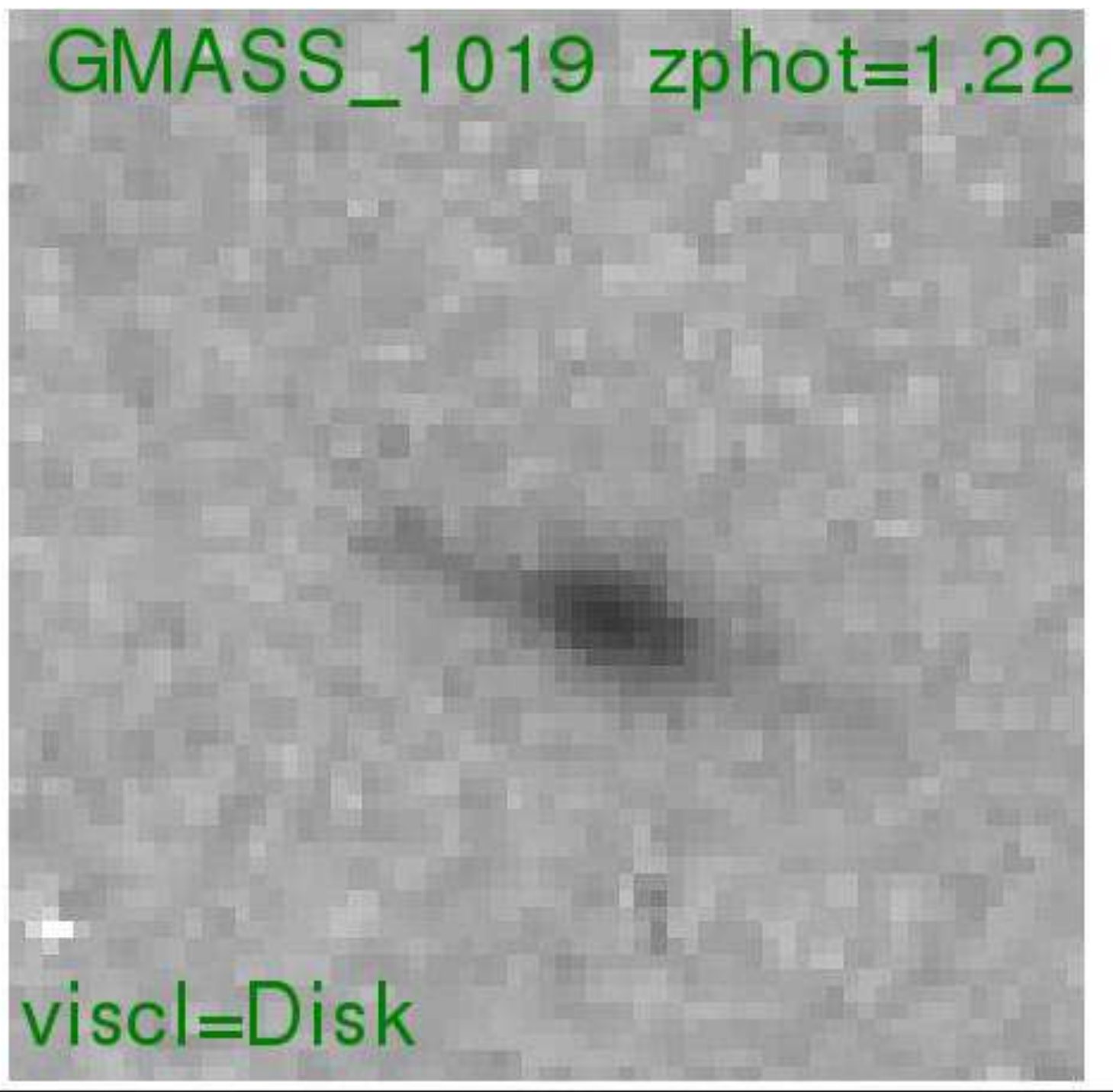}			     

\includegraphics[trim=100 40 75 390, clip=true, width=30mm]{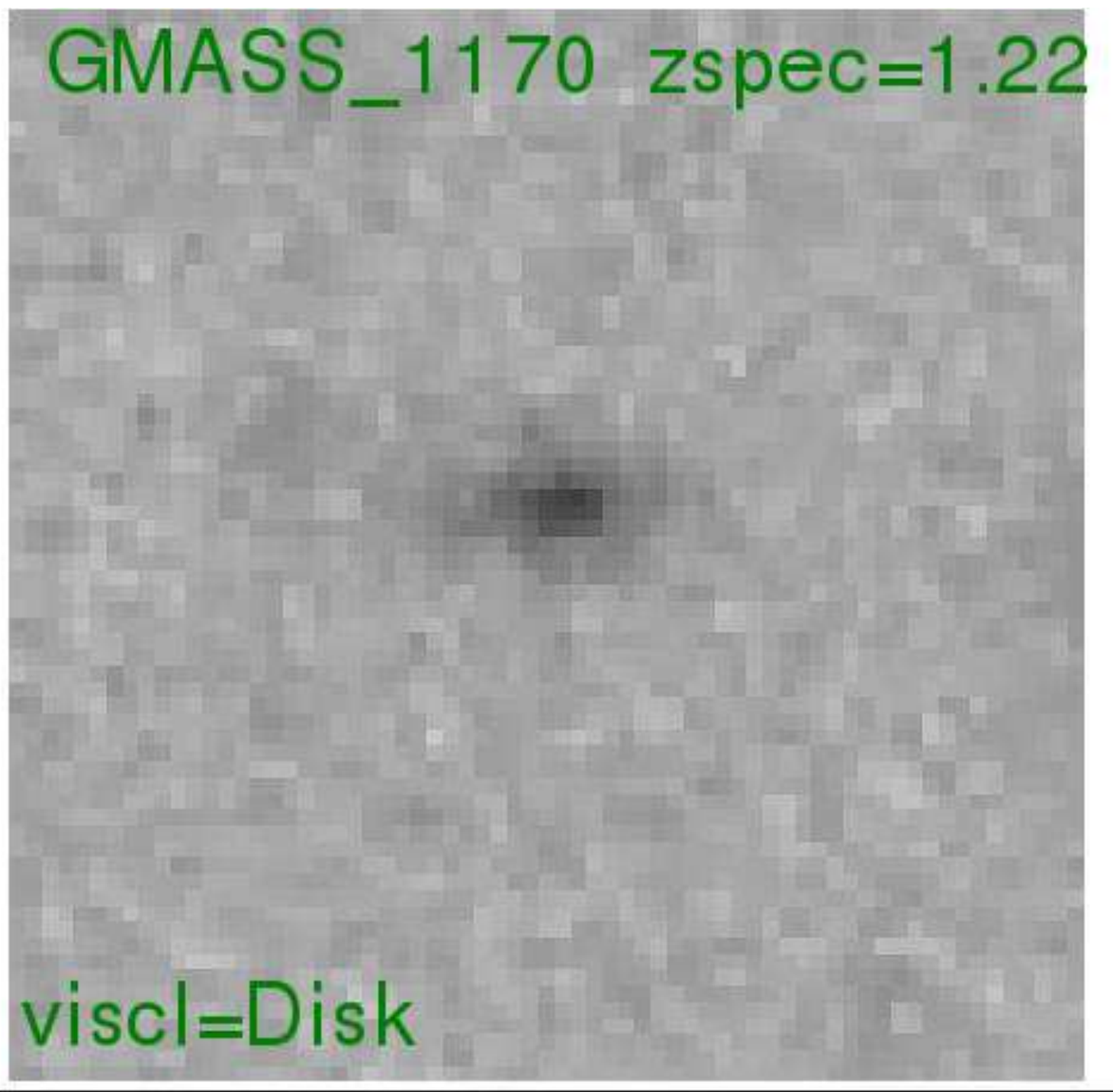}		     
\includegraphics[trim=100 40 75 390, clip=true, width=30mm]{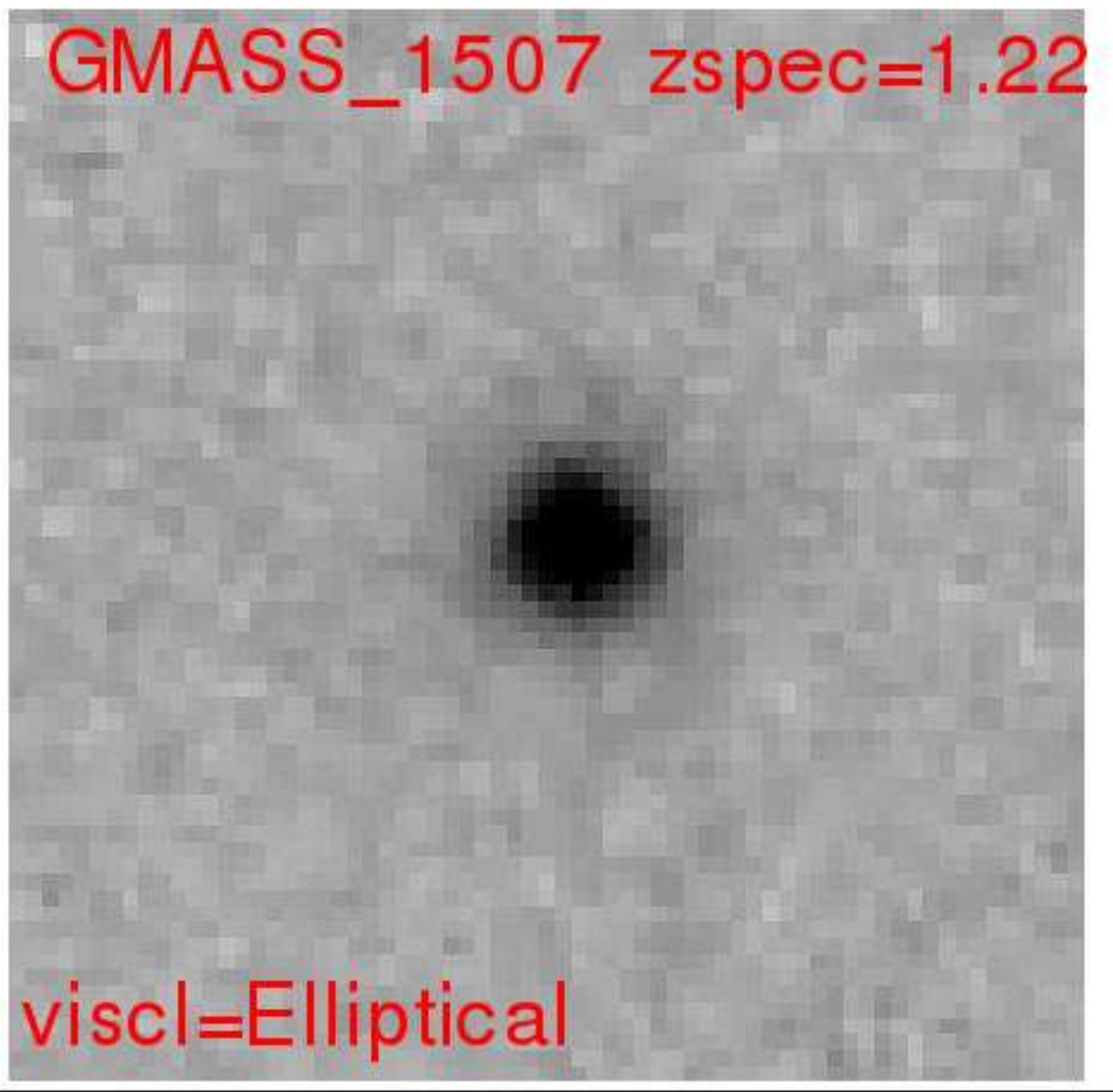}			     
\includegraphics[trim=100 40 75 390, clip=true, width=30mm]{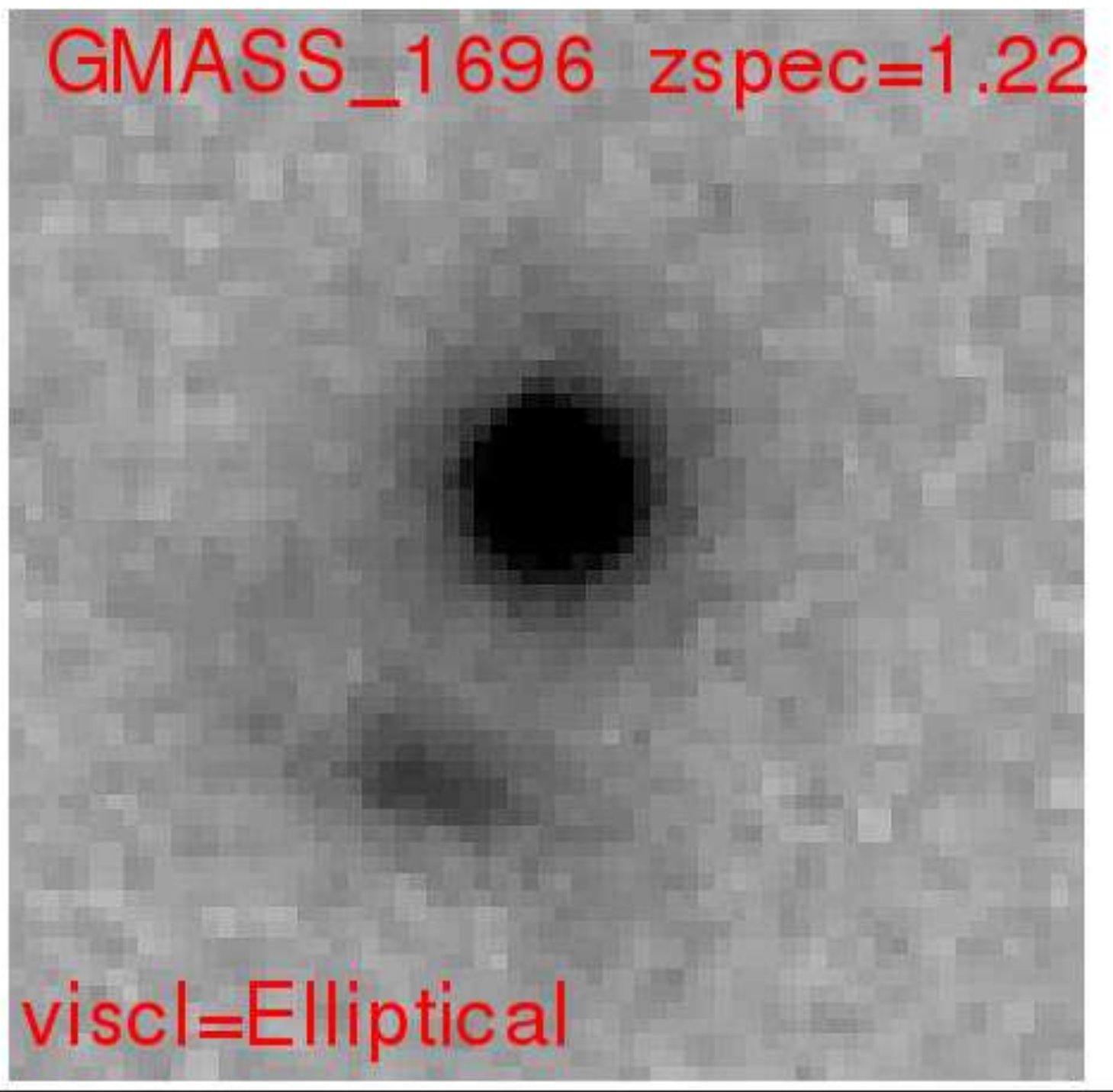}			     
\includegraphics[trim=100 40 75 390, clip=true, width=30mm]{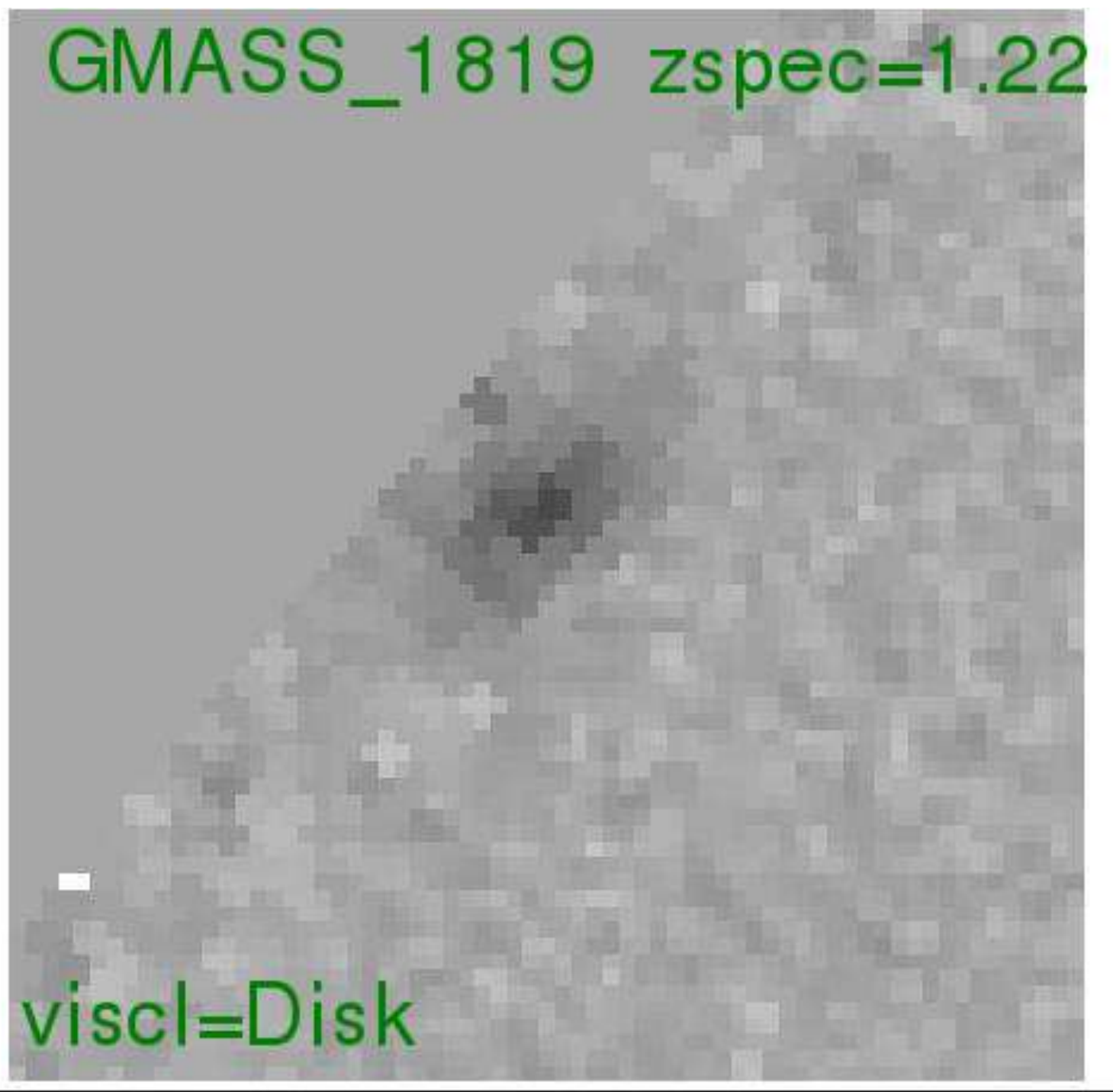}			     
\includegraphics[trim=100 40 75 390, clip=true, width=30mm]{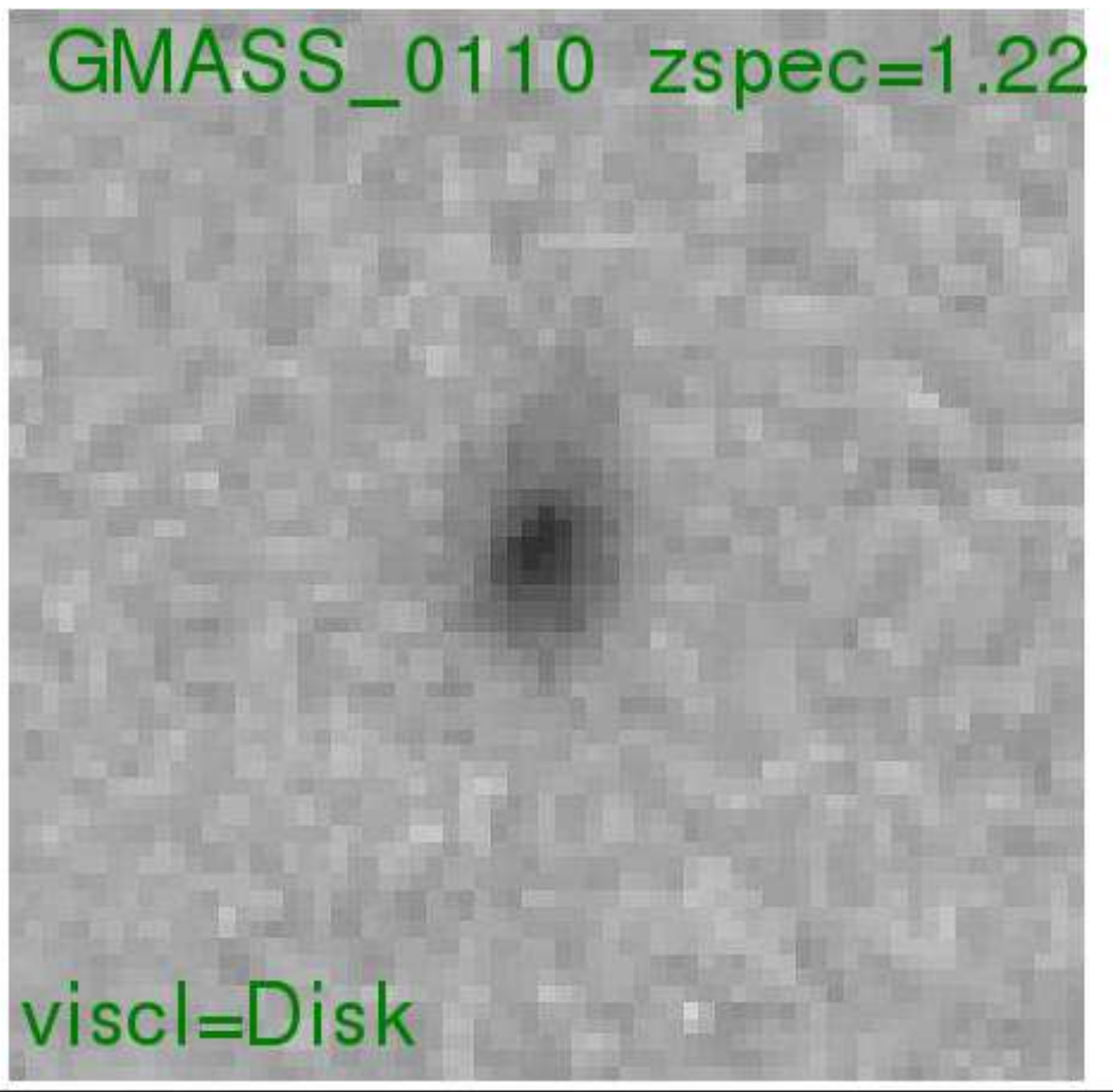}			     
\includegraphics[trim=100 40 75 390, clip=true, width=30mm]{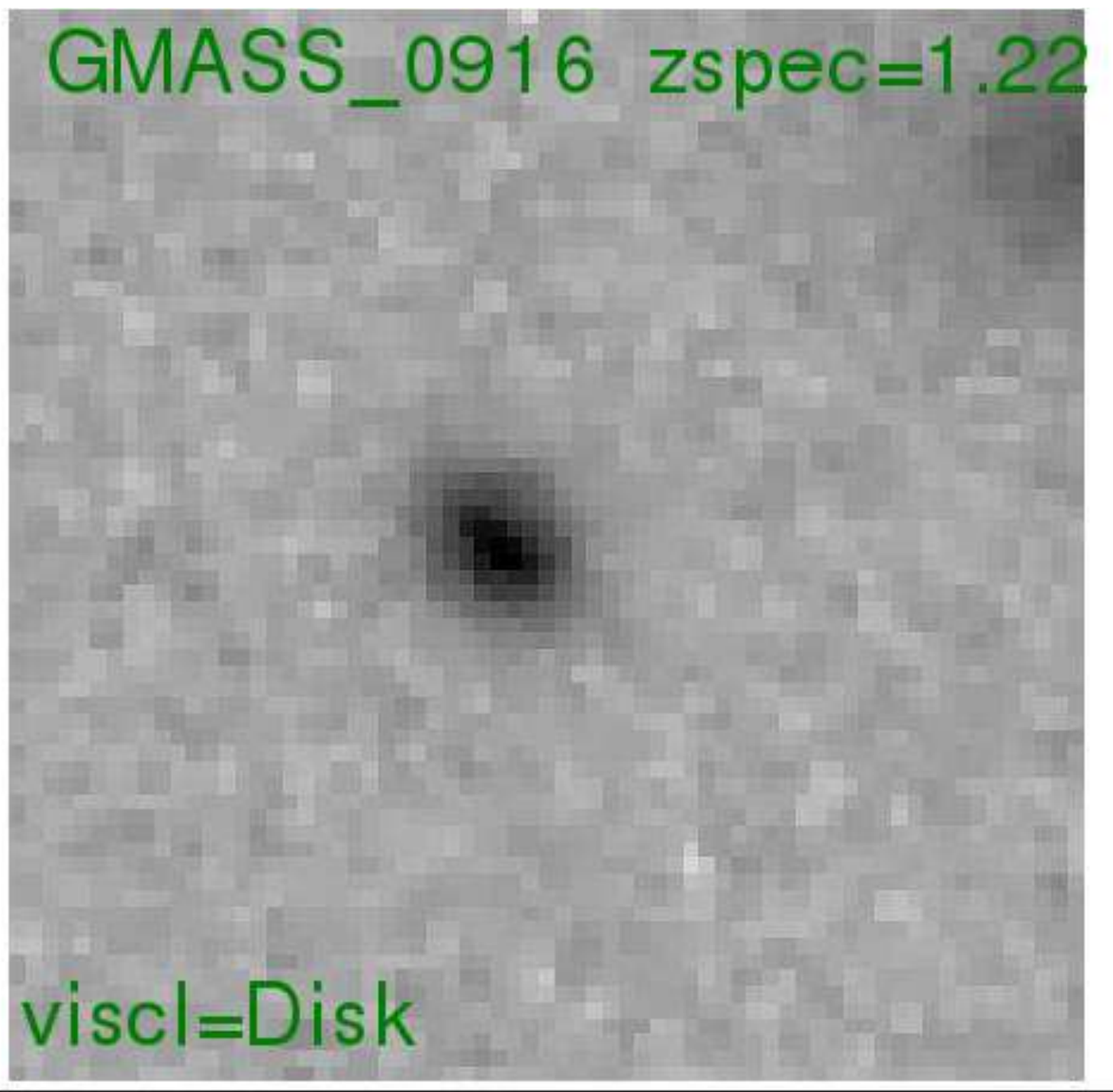}			     

\includegraphics[trim=100 40 75 390, clip=true, width=30mm]{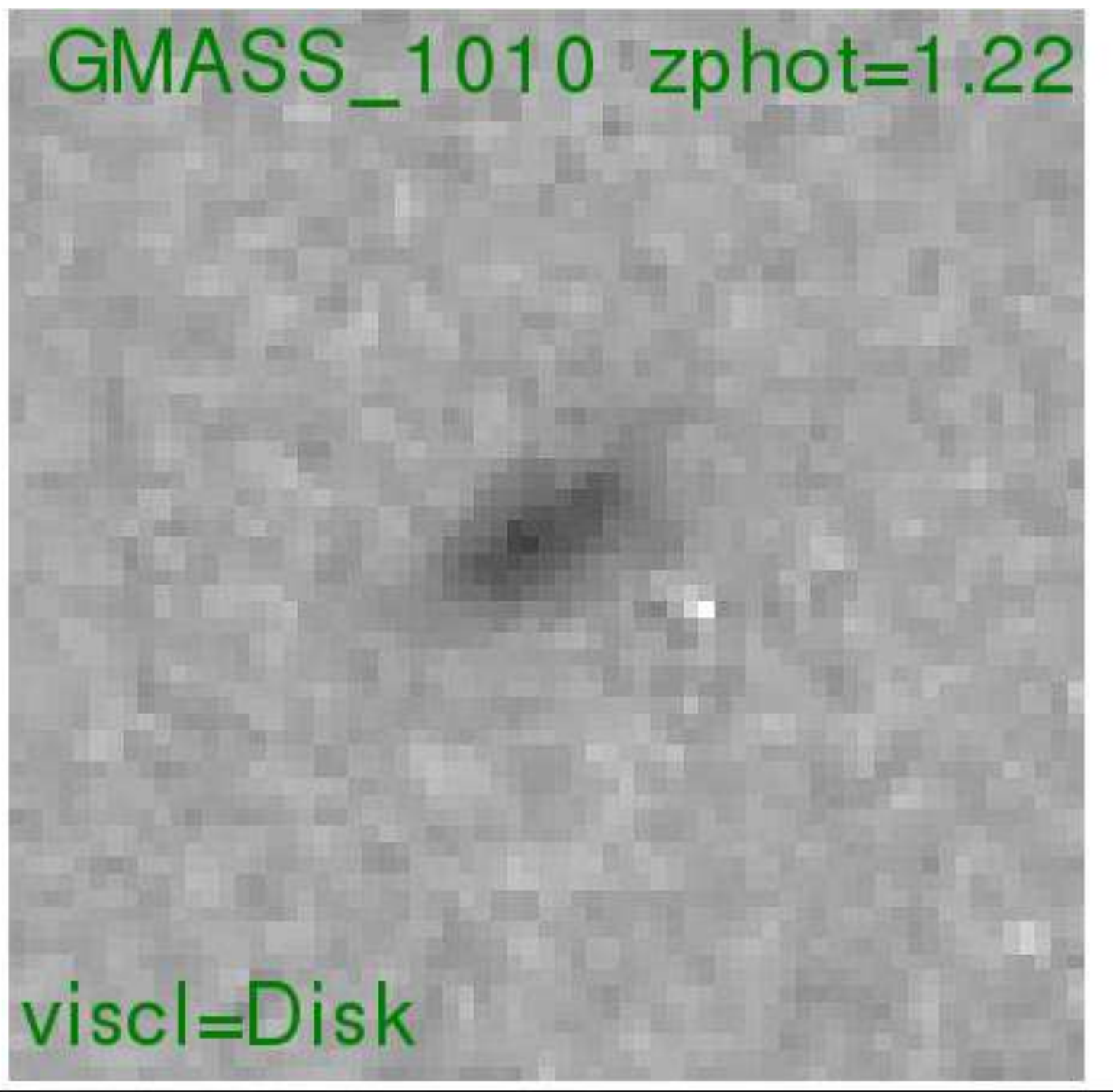}			     
\includegraphics[trim=100 40 75 390, clip=true, width=30mm]{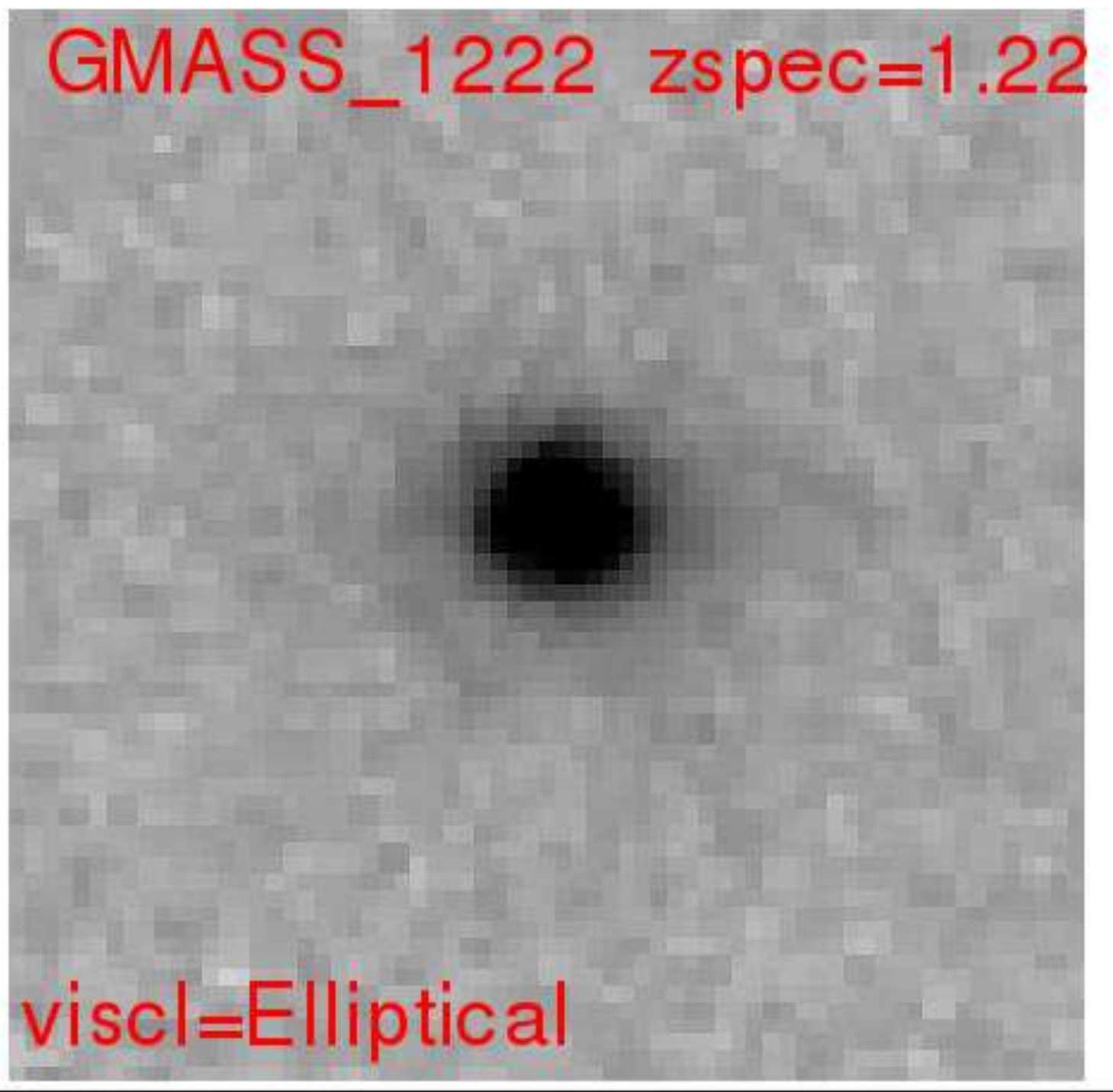}		     
\includegraphics[trim=100 40 75 390, clip=true, width=30mm]{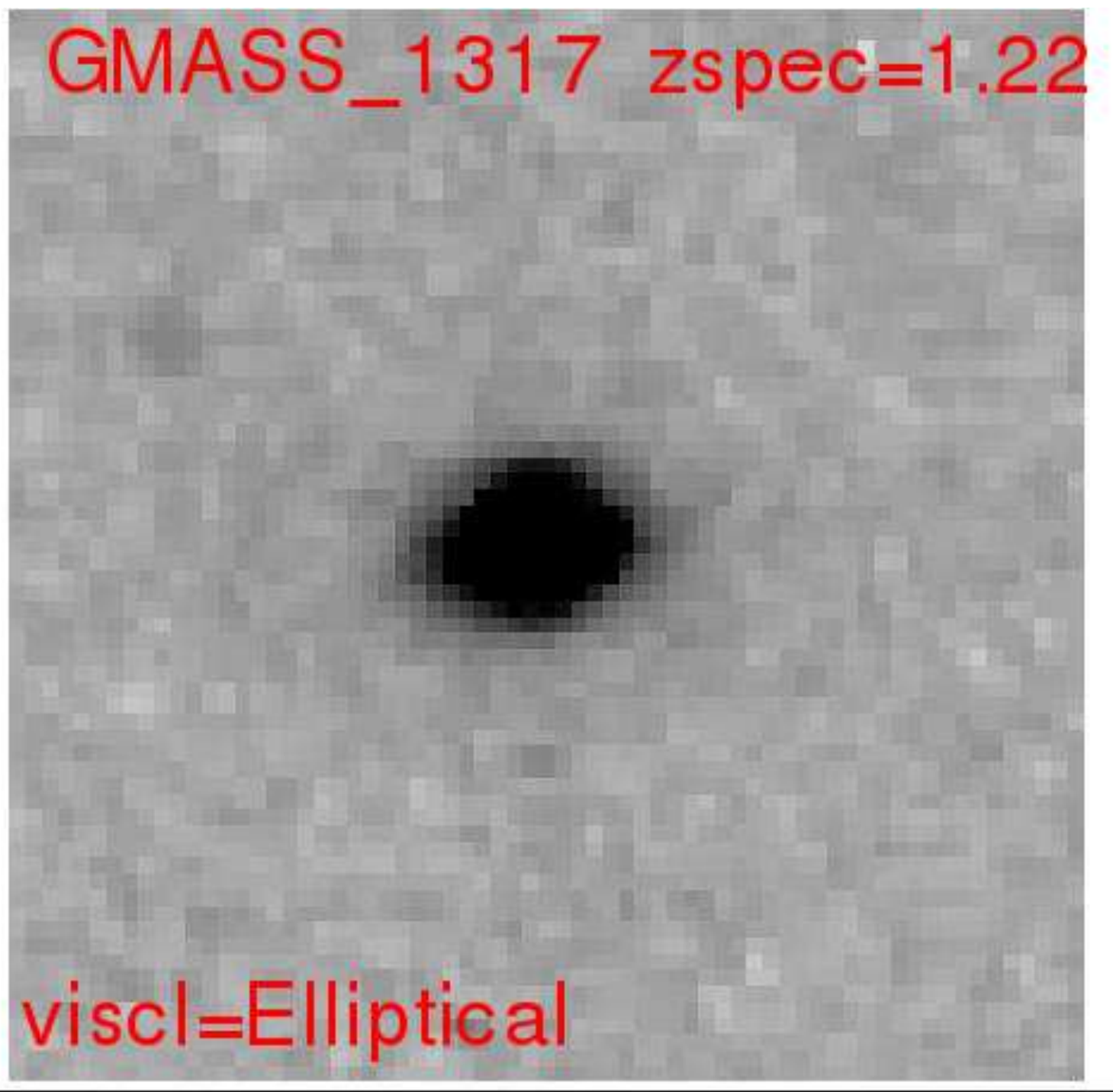}			     
\includegraphics[trim=100 40 75 390, clip=true, width=30mm]{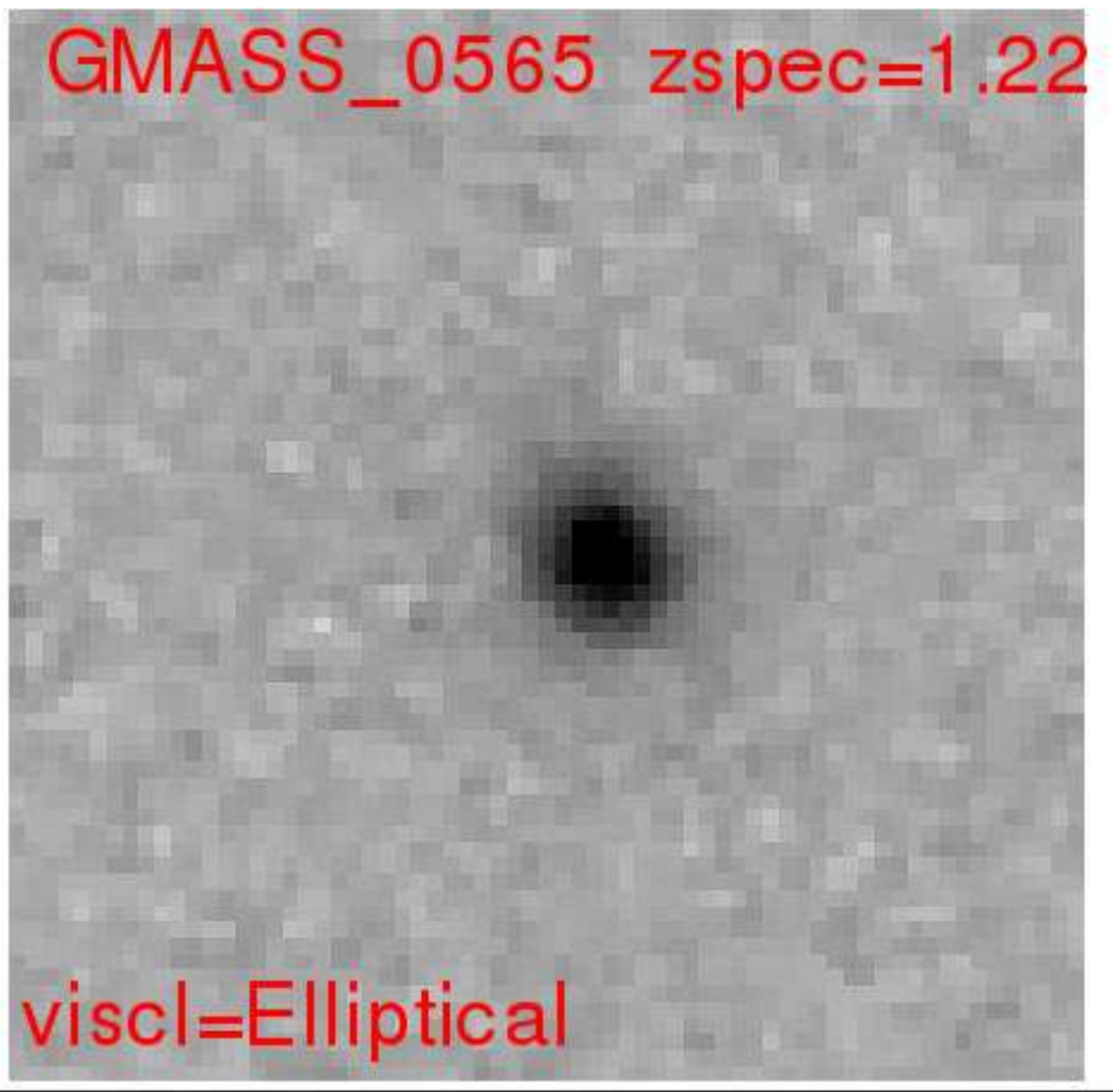}		     
\includegraphics[trim=100 40 75 390, clip=true, width=30mm]{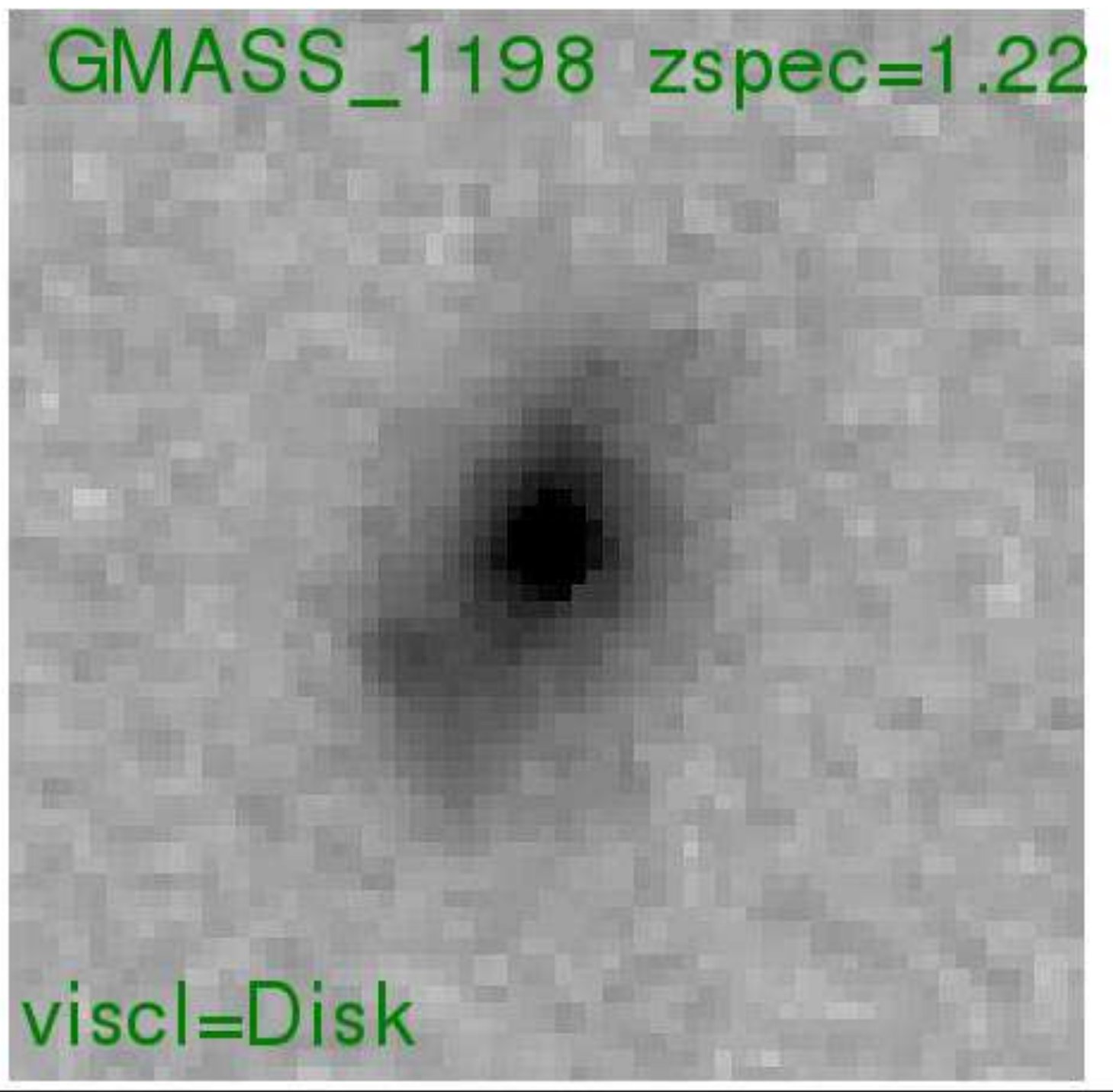}			     
\includegraphics[trim=100 40 75 390, clip=true, width=30mm]{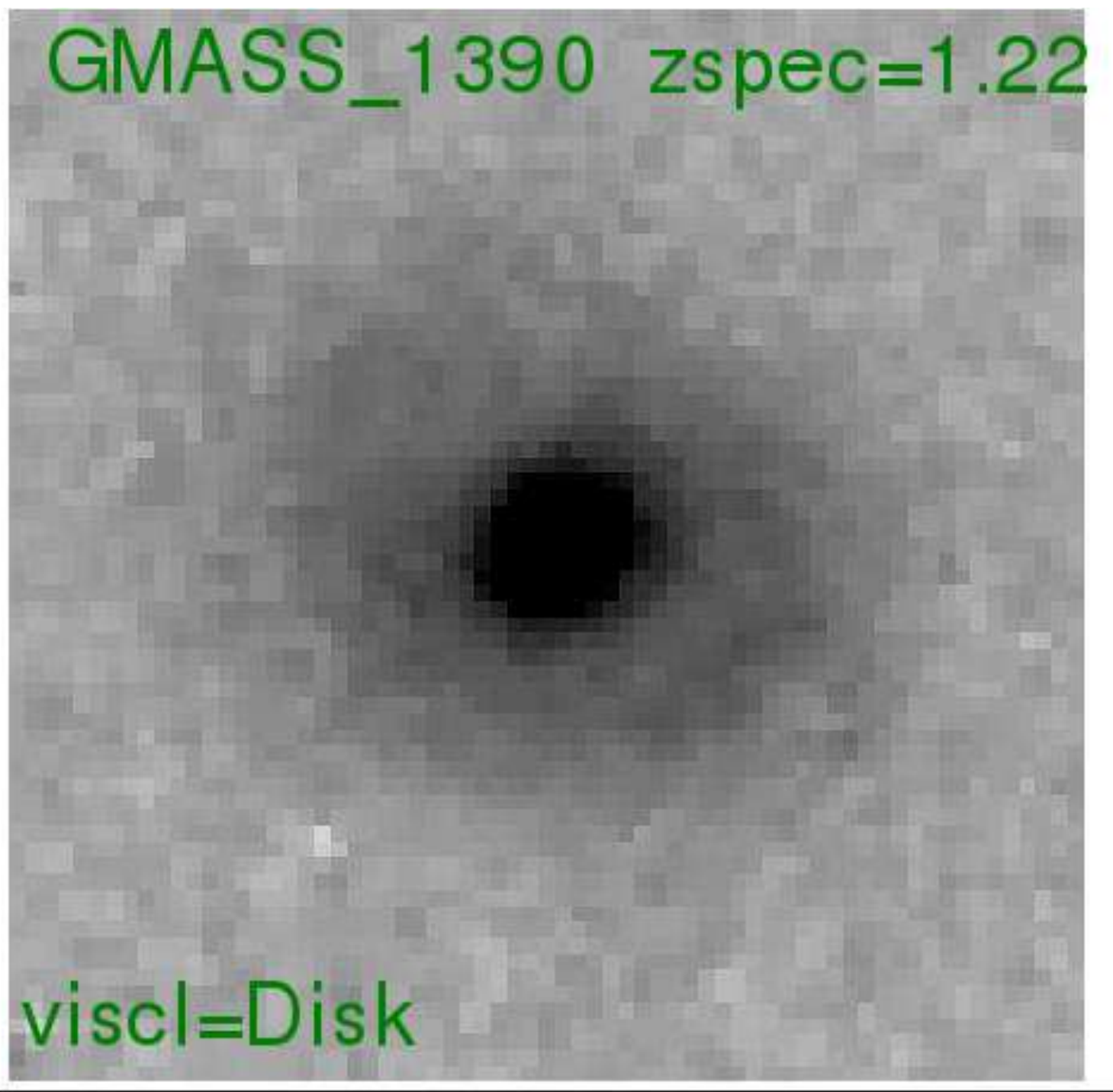}			     

\includegraphics[trim=100 40 75 390, clip=true, width=30mm]{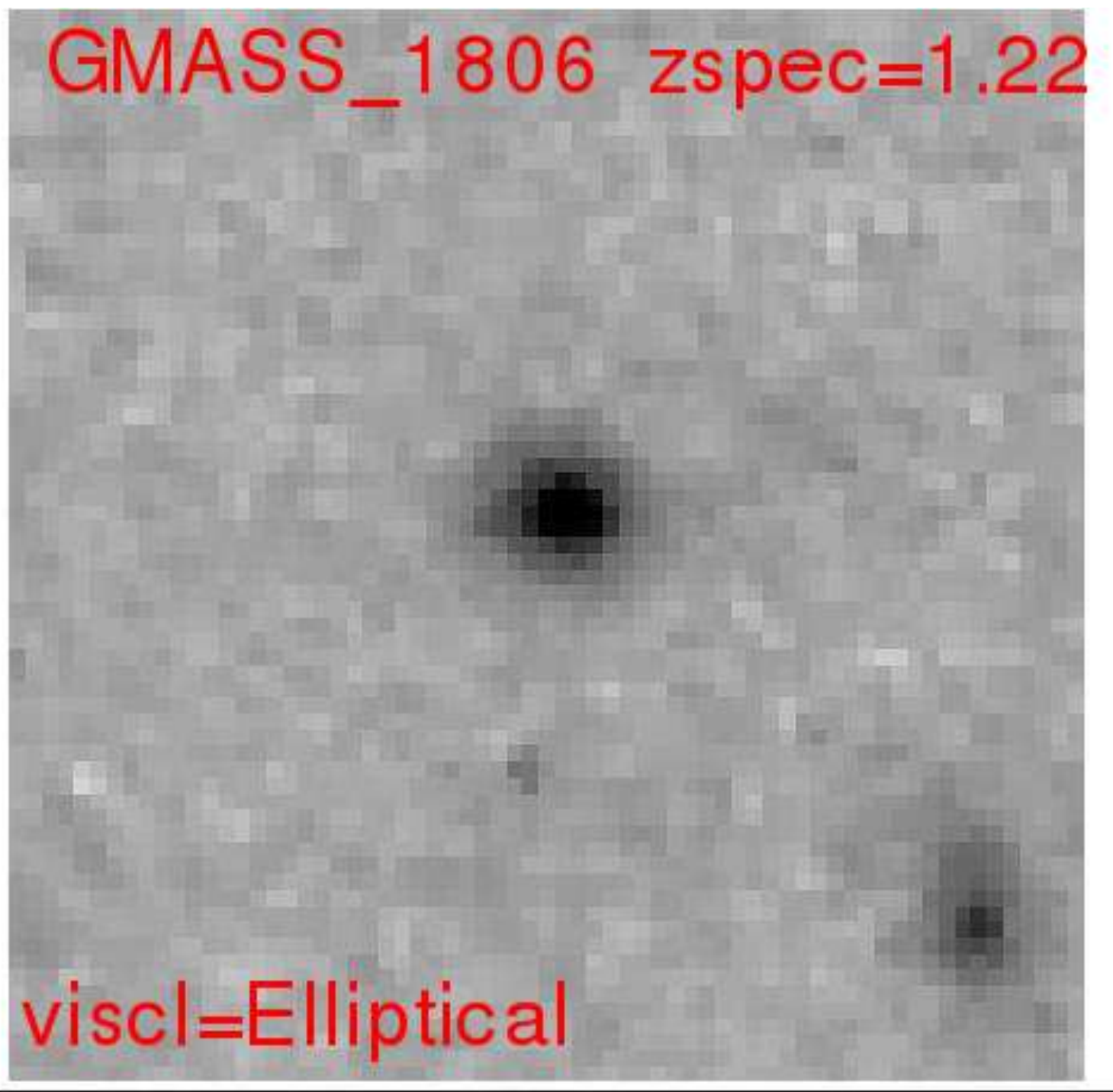}			     
\includegraphics[trim=100 40 75 390, clip=true, width=30mm]{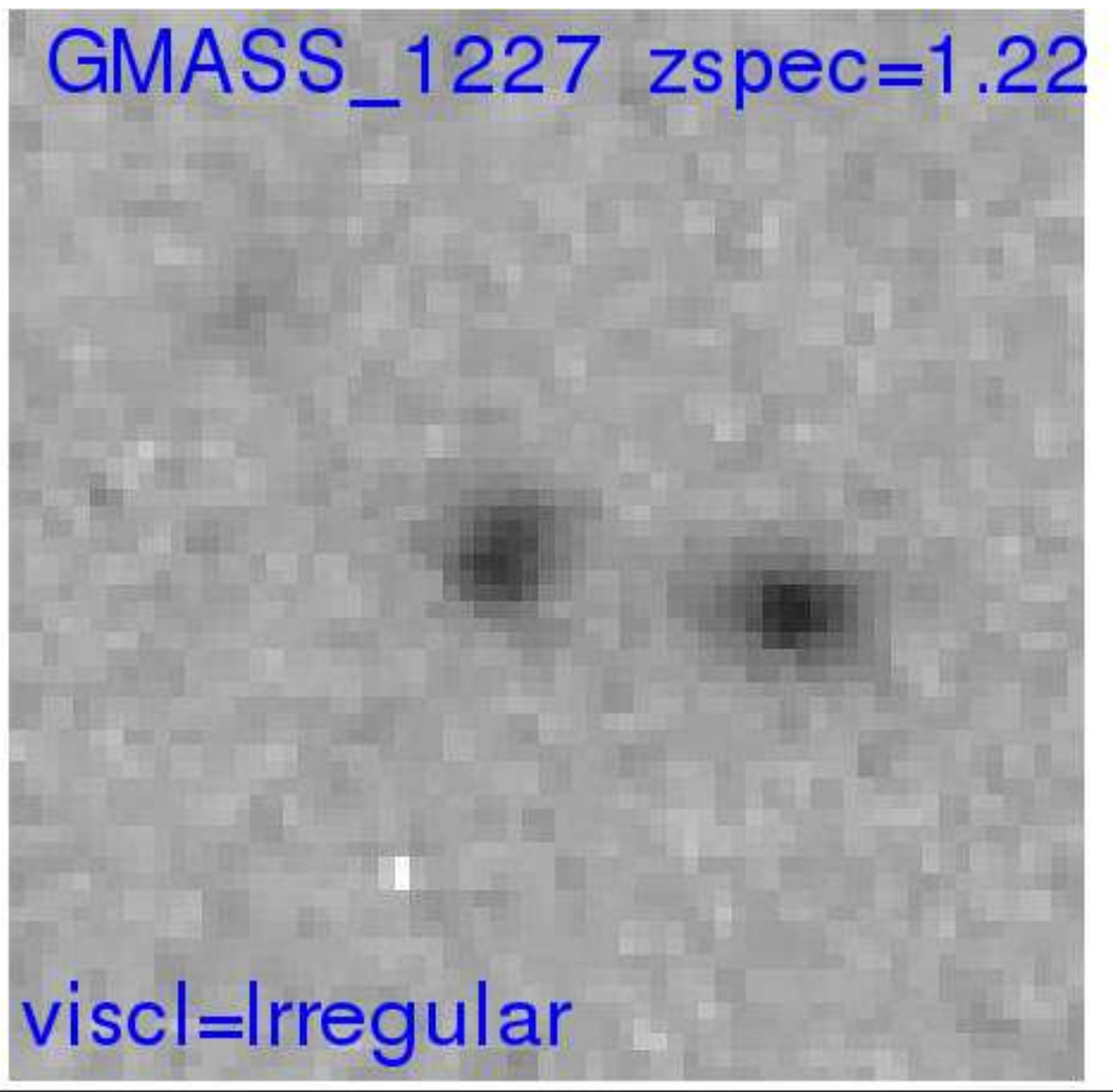}			     
\includegraphics[trim=100 40 75 390, clip=true, width=30mm]{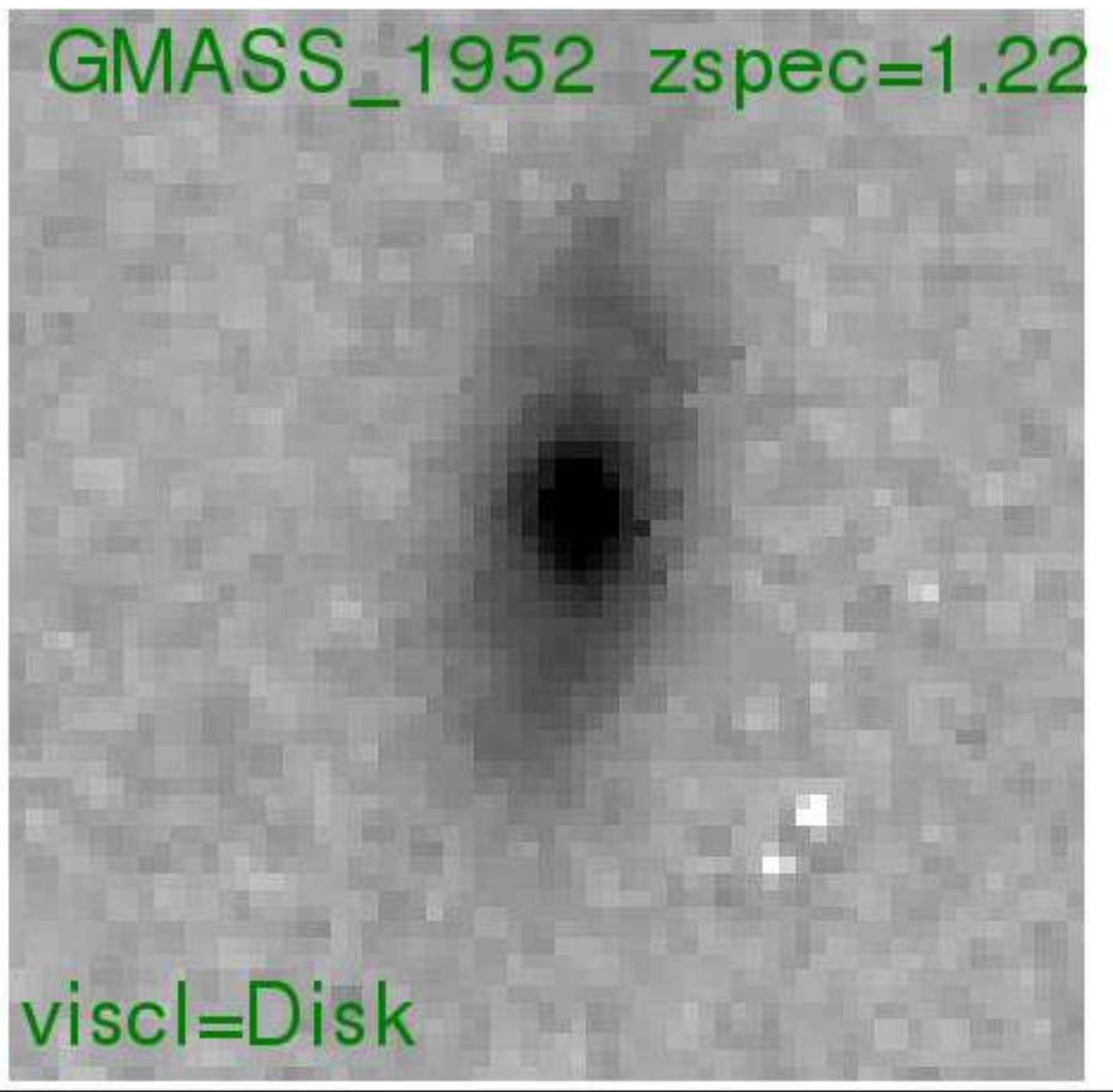}		     
\includegraphics[trim=100 40 75 390, clip=true, width=30mm]{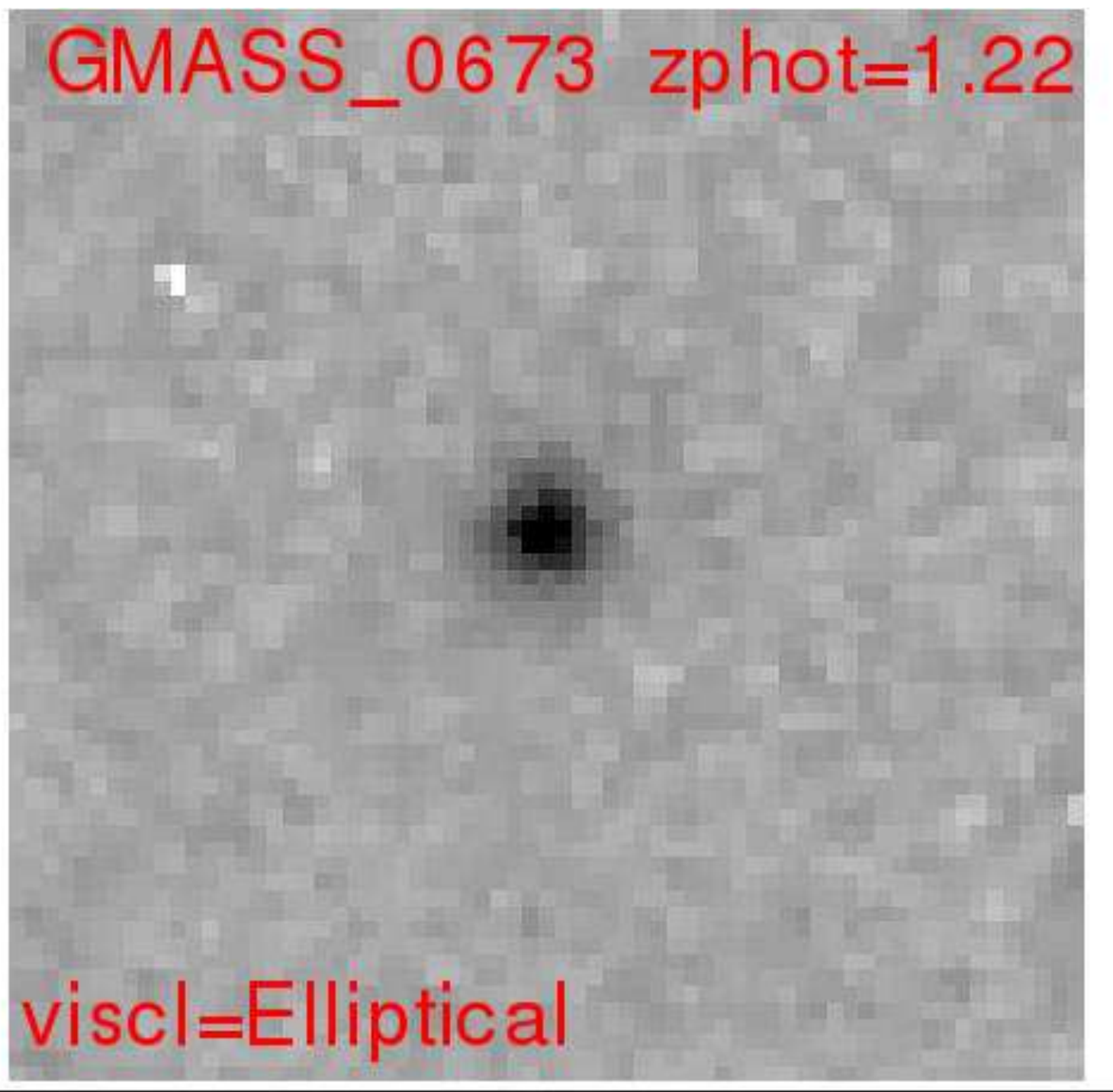}			     
\includegraphics[trim=100 40 75 390, clip=true, width=30mm]{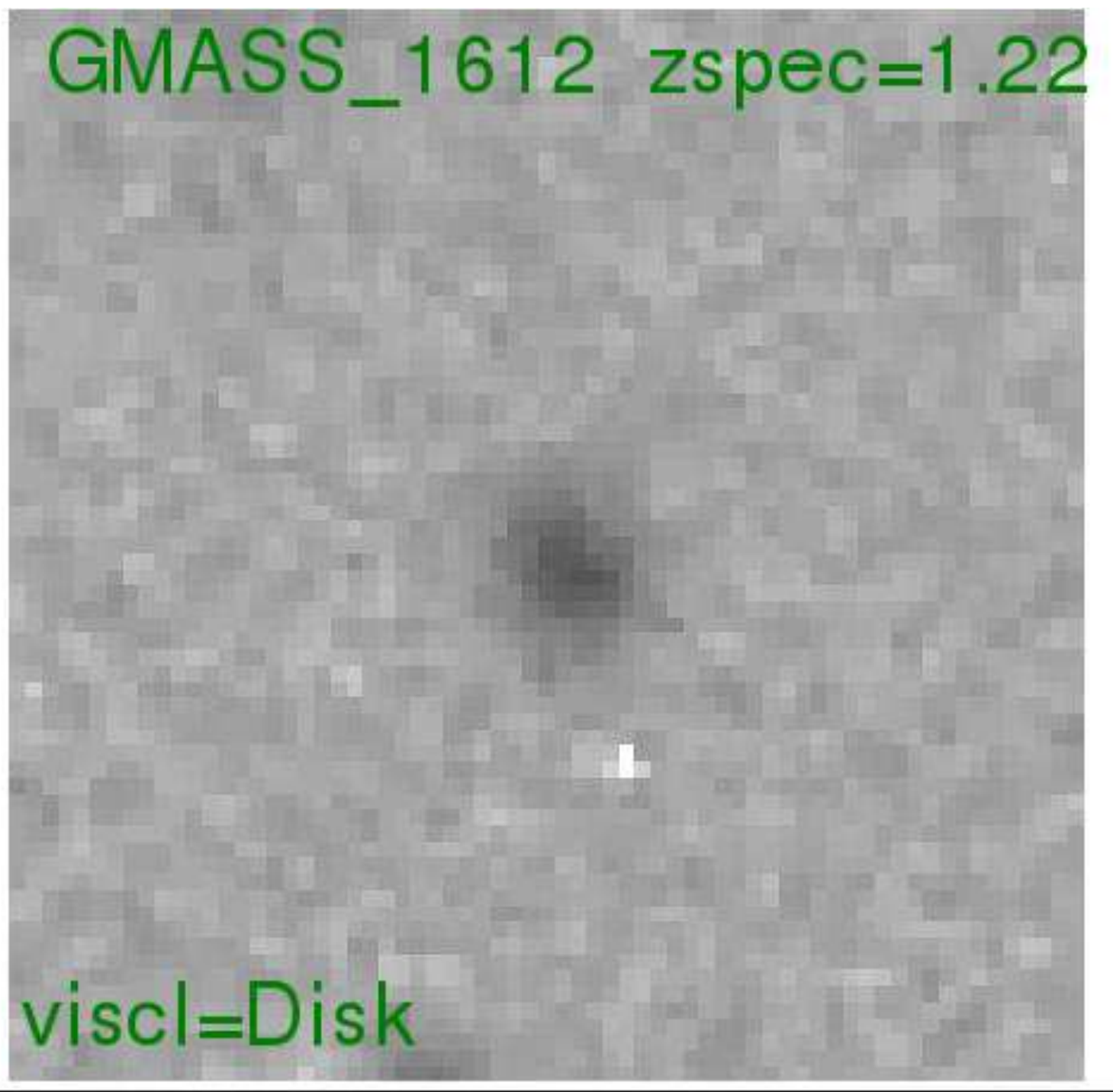}			     
\includegraphics[trim=100 40 75 390, clip=true, width=30mm]{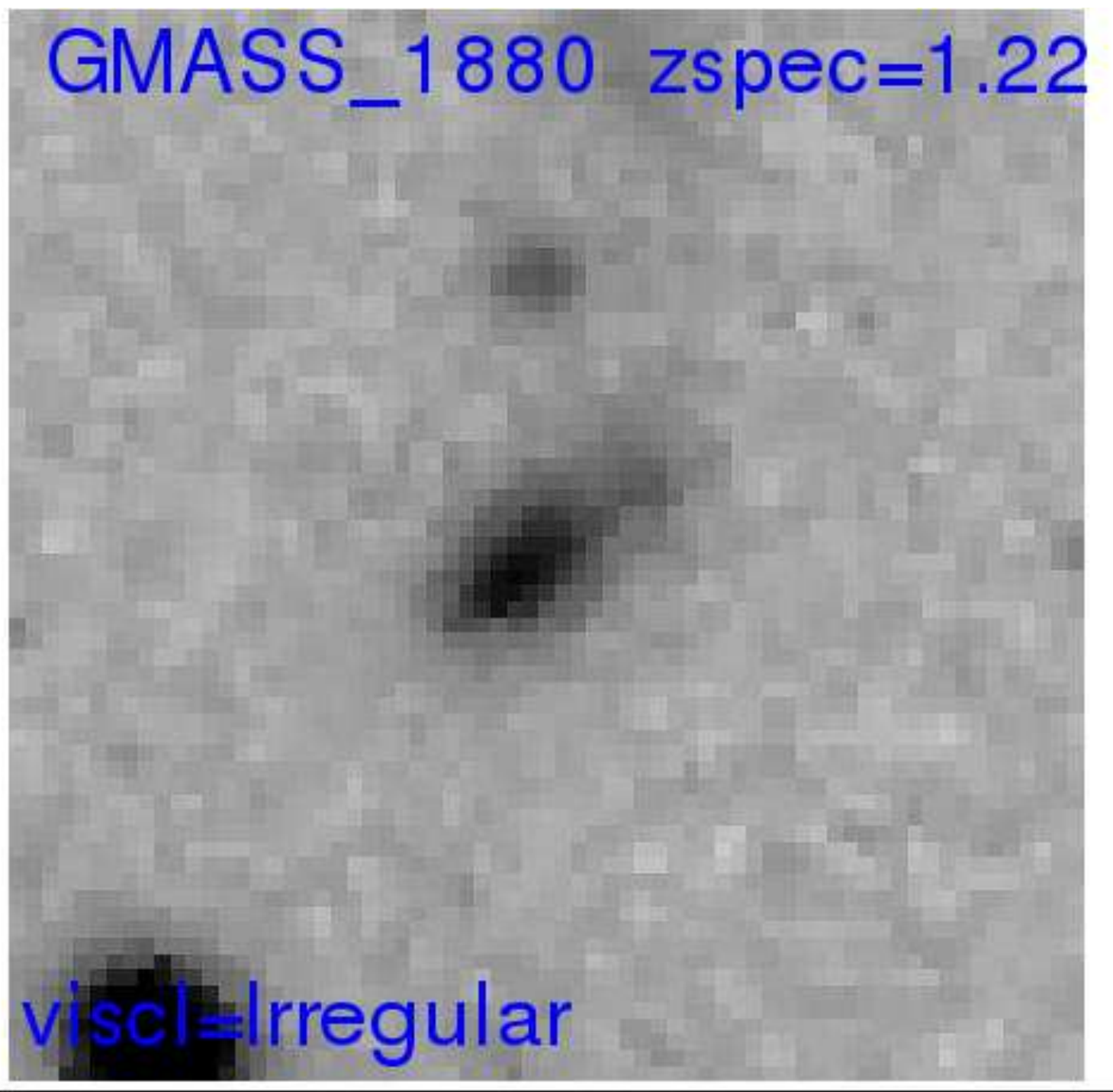}			     

\includegraphics[trim=100 40 75 390, clip=true, width=30mm]{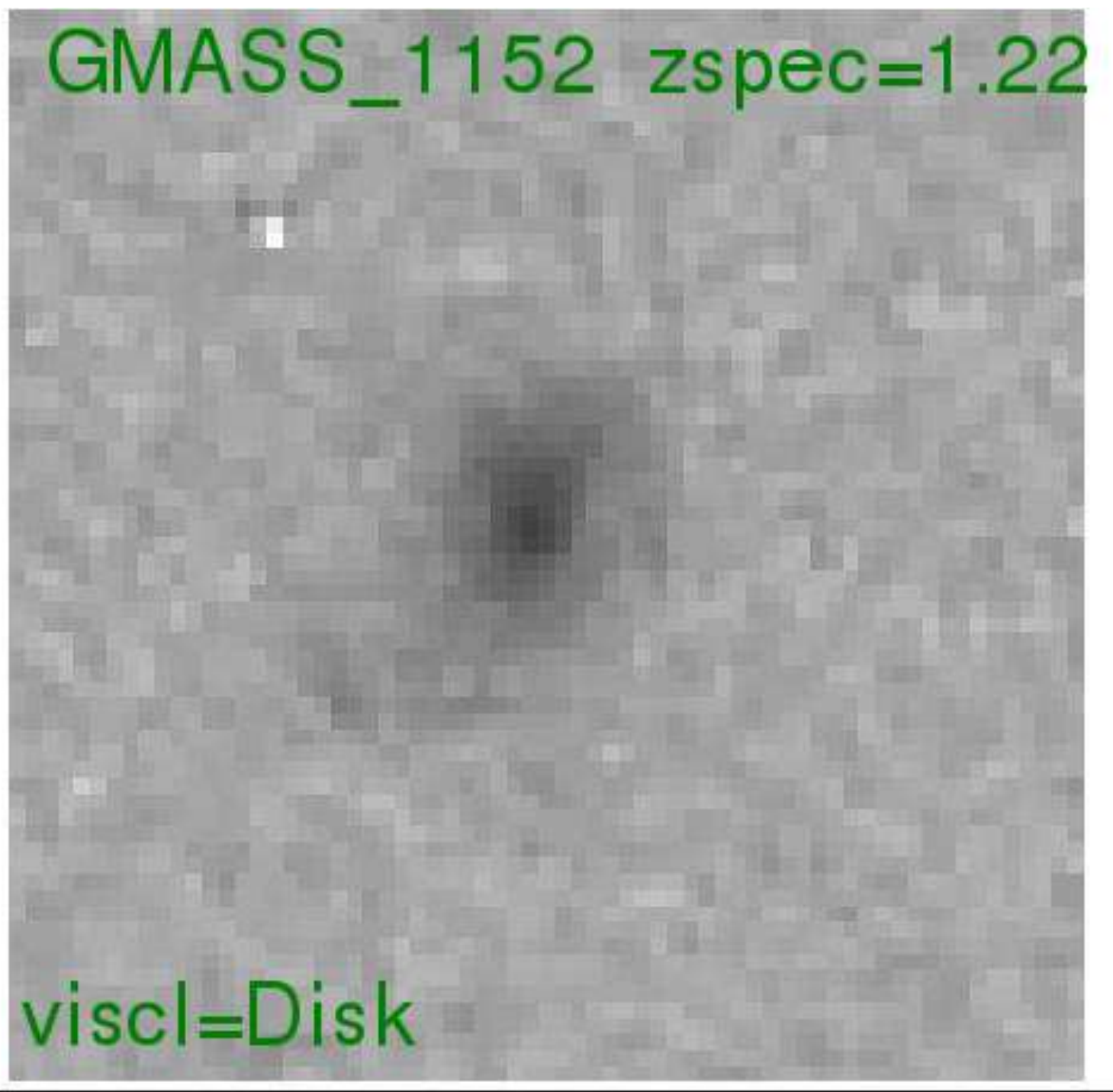}			     
\includegraphics[trim=100 40 75 390, clip=true, width=30mm]{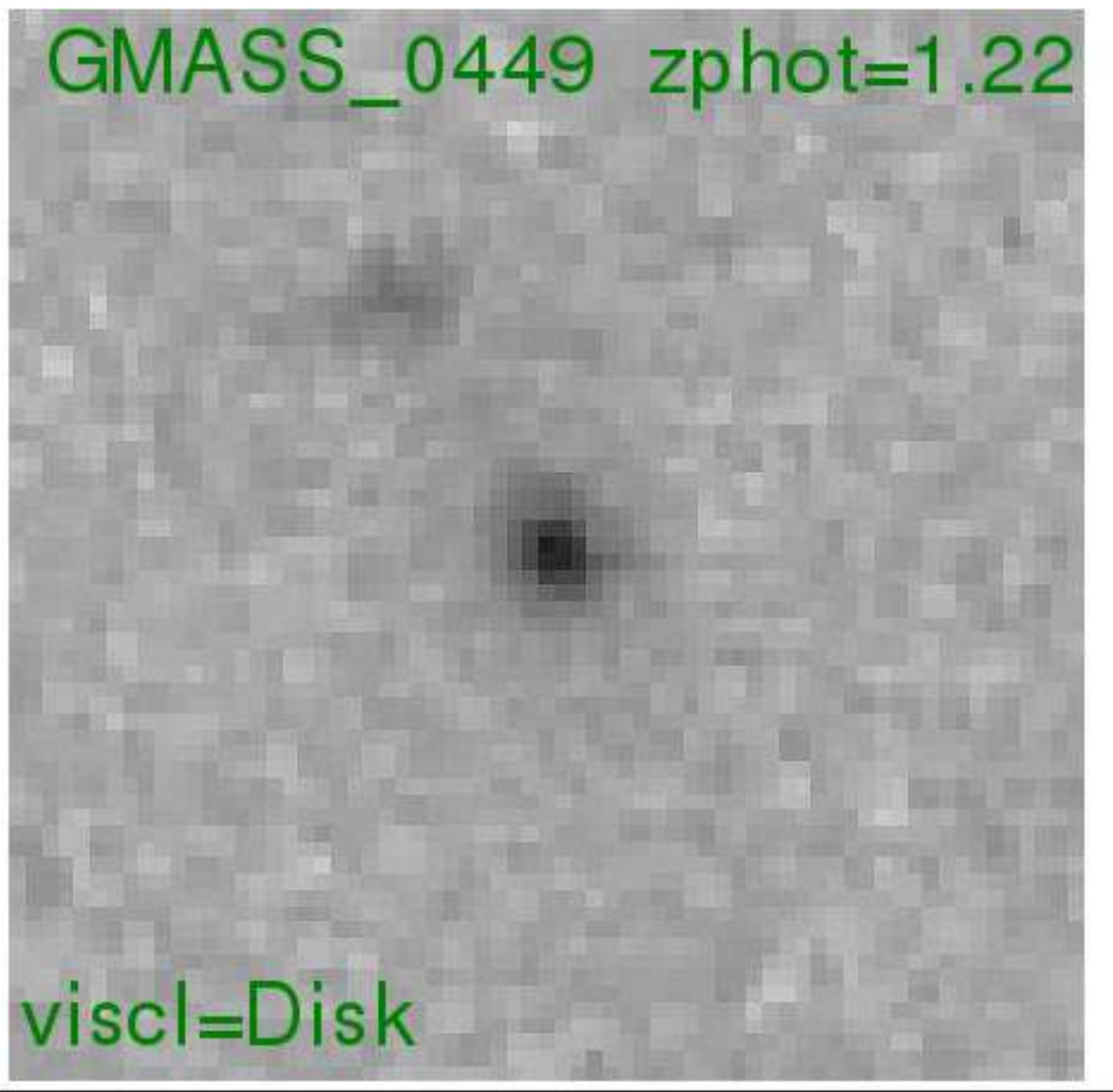}		     
\includegraphics[trim=100 40 75 390, clip=true, width=30mm]{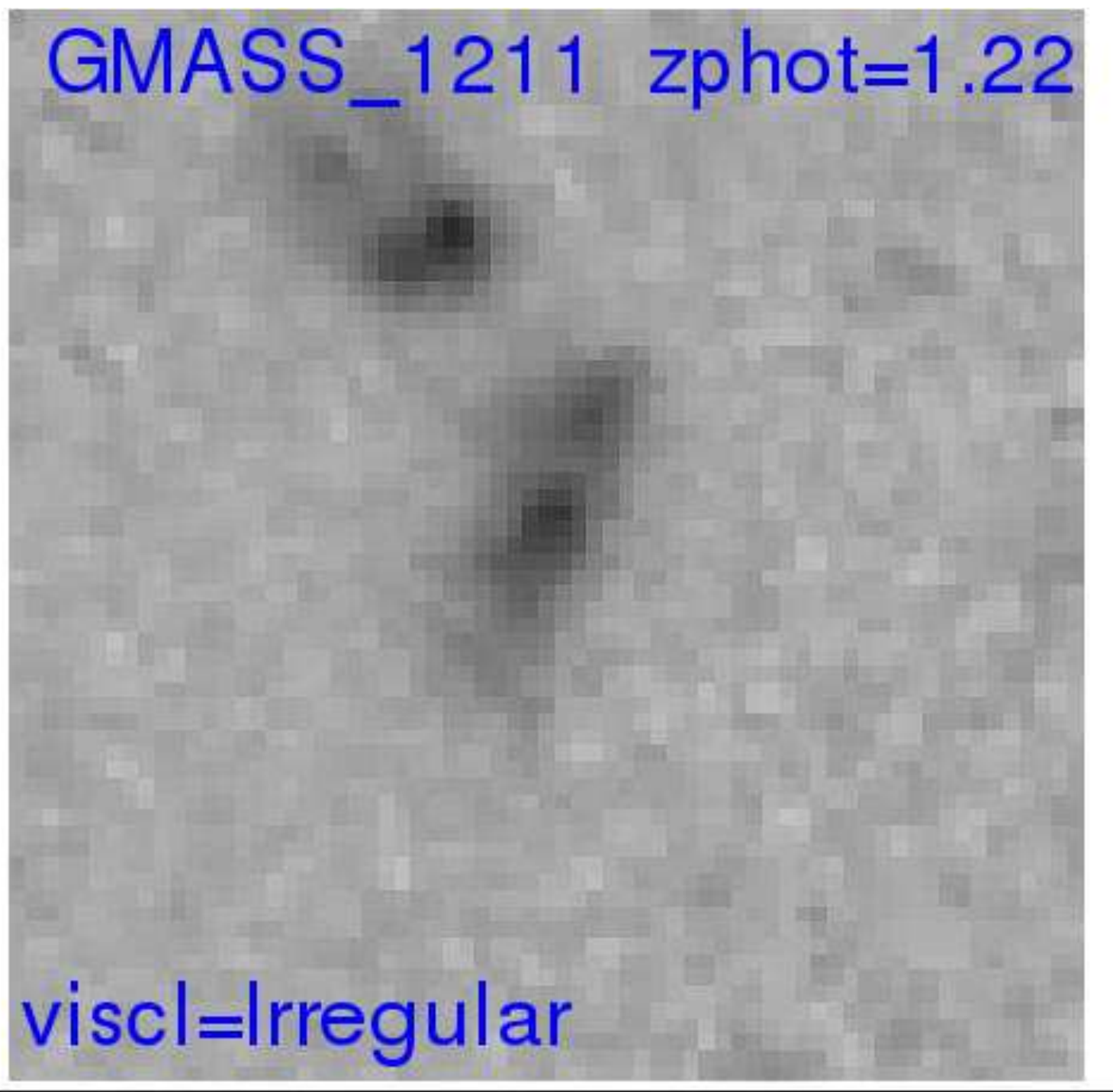}			     
\includegraphics[trim=100 40 75 390, clip=true, width=30mm]{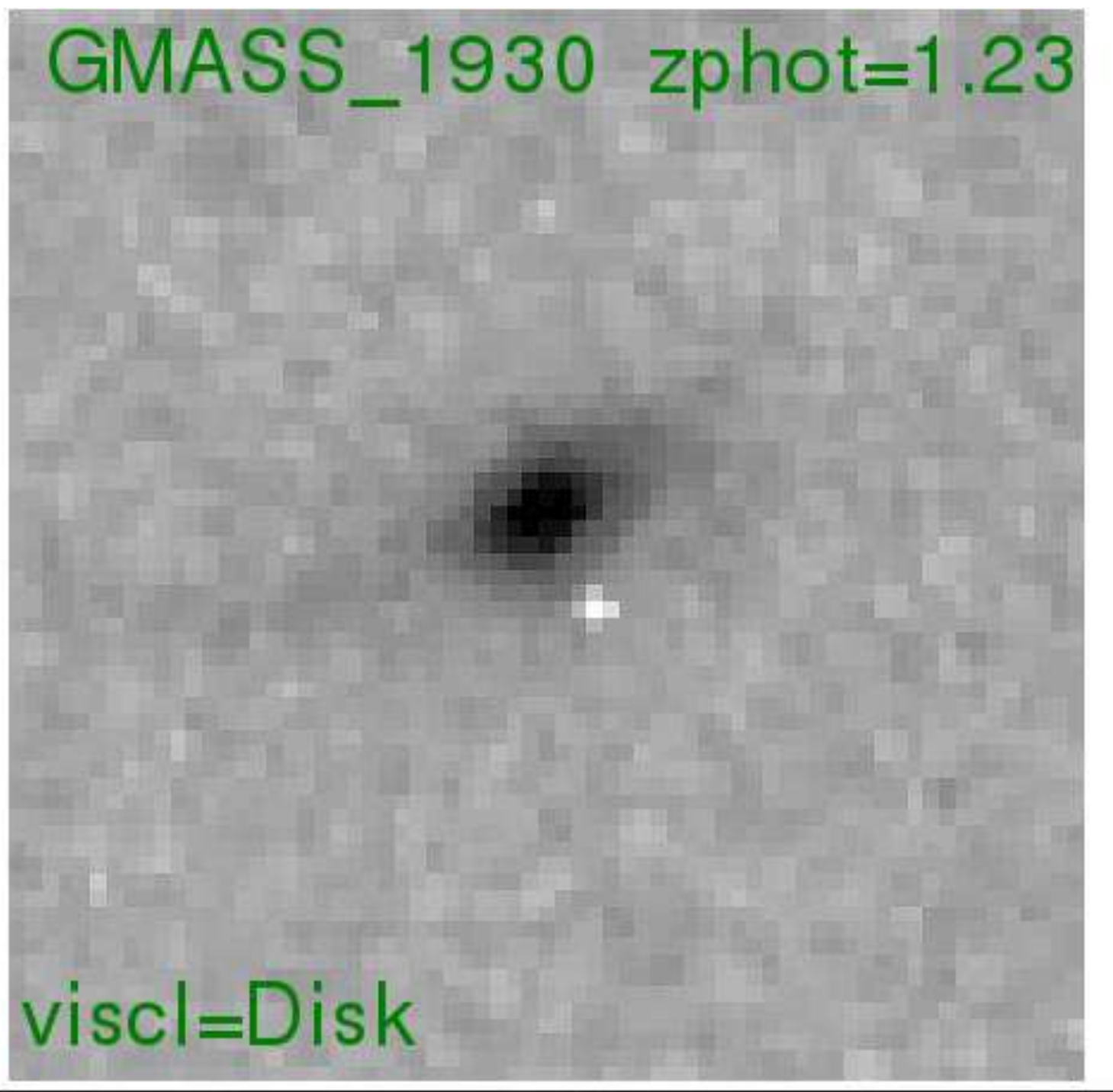}			     
\includegraphics[trim=100 40 75 390, clip=true, width=30mm]{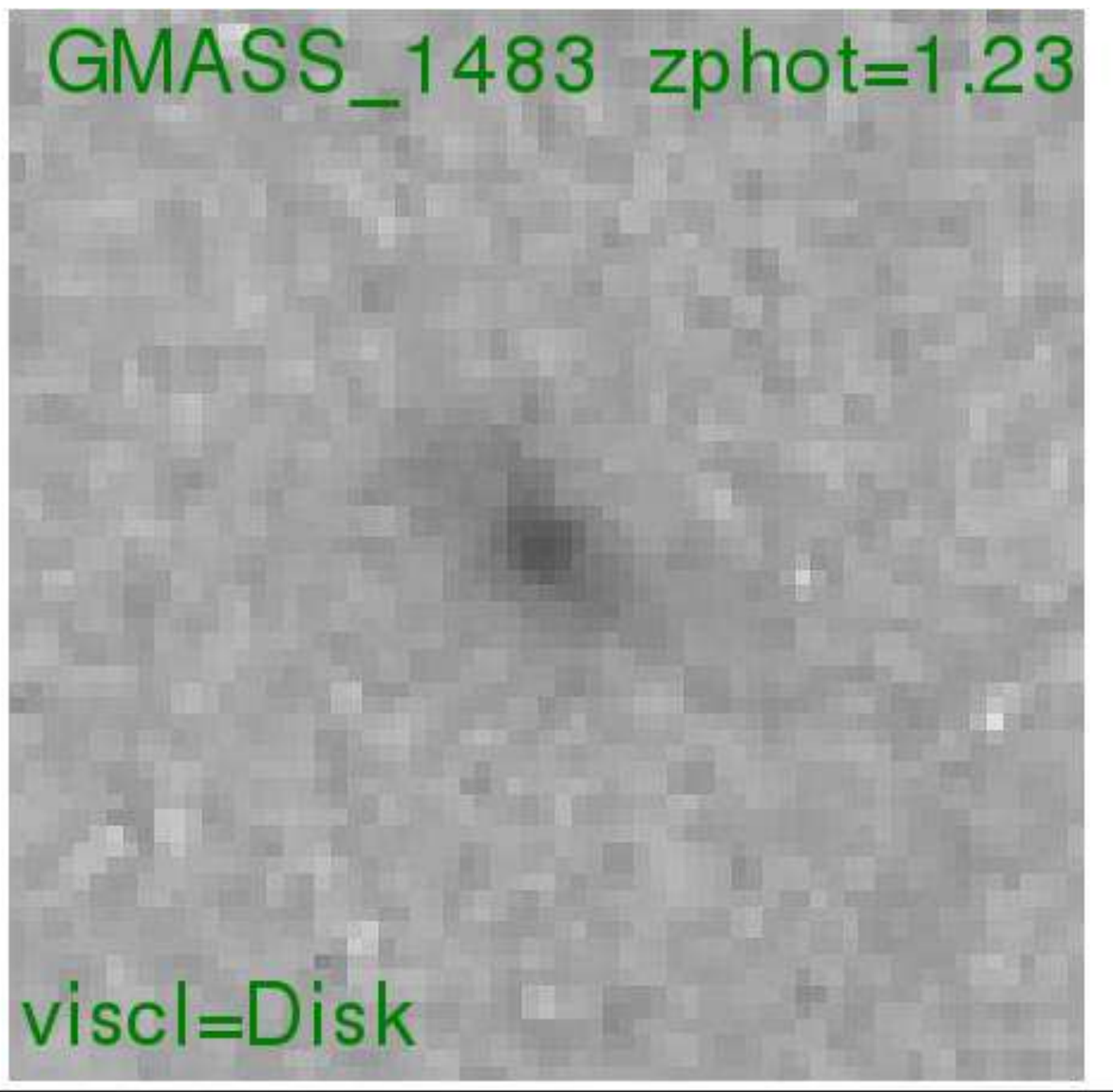}		     
\includegraphics[trim=100 40 75 390, clip=true, width=30mm]{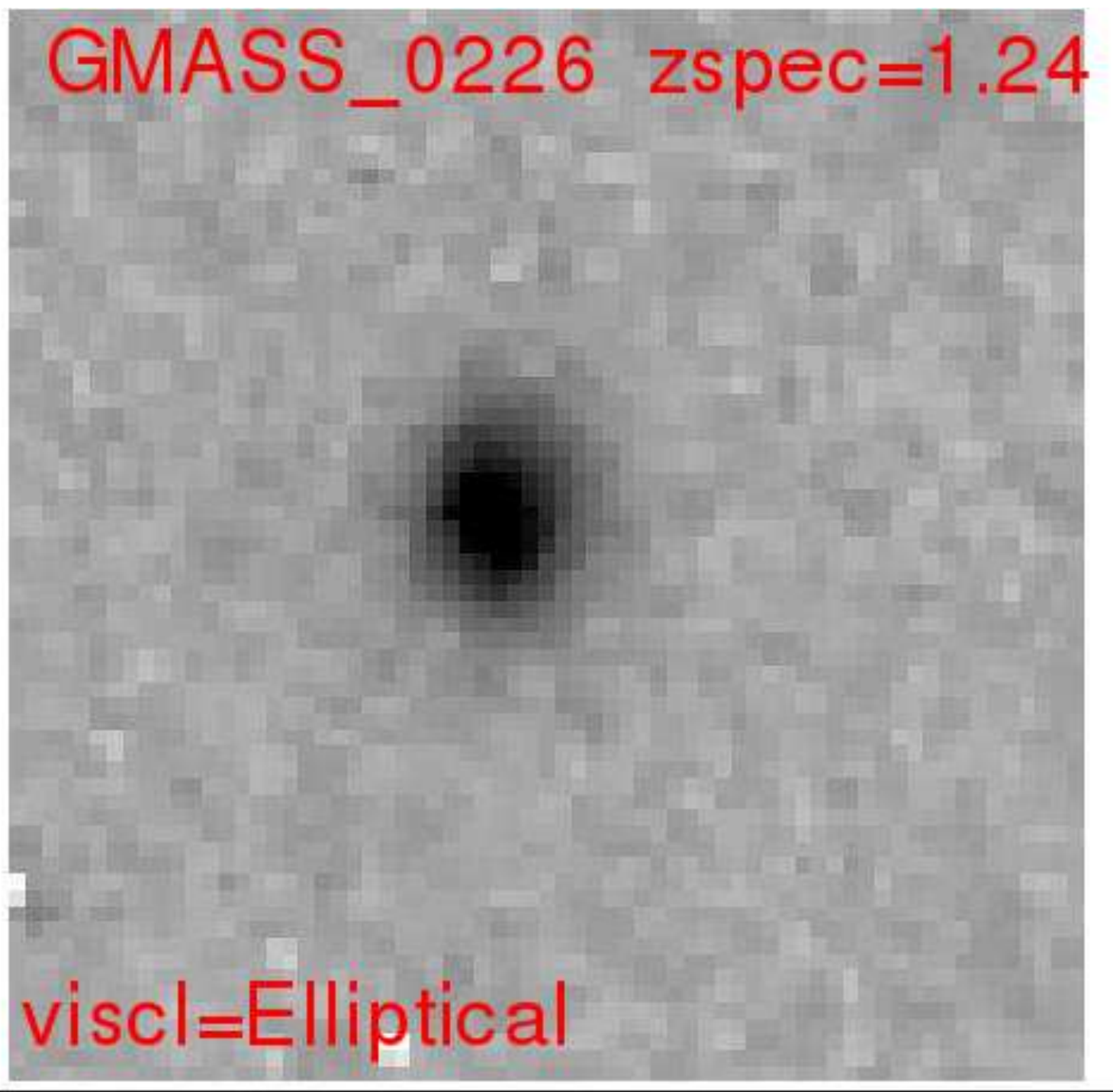}			     

\includegraphics[trim=100 40 75 390, clip=true, width=30mm]{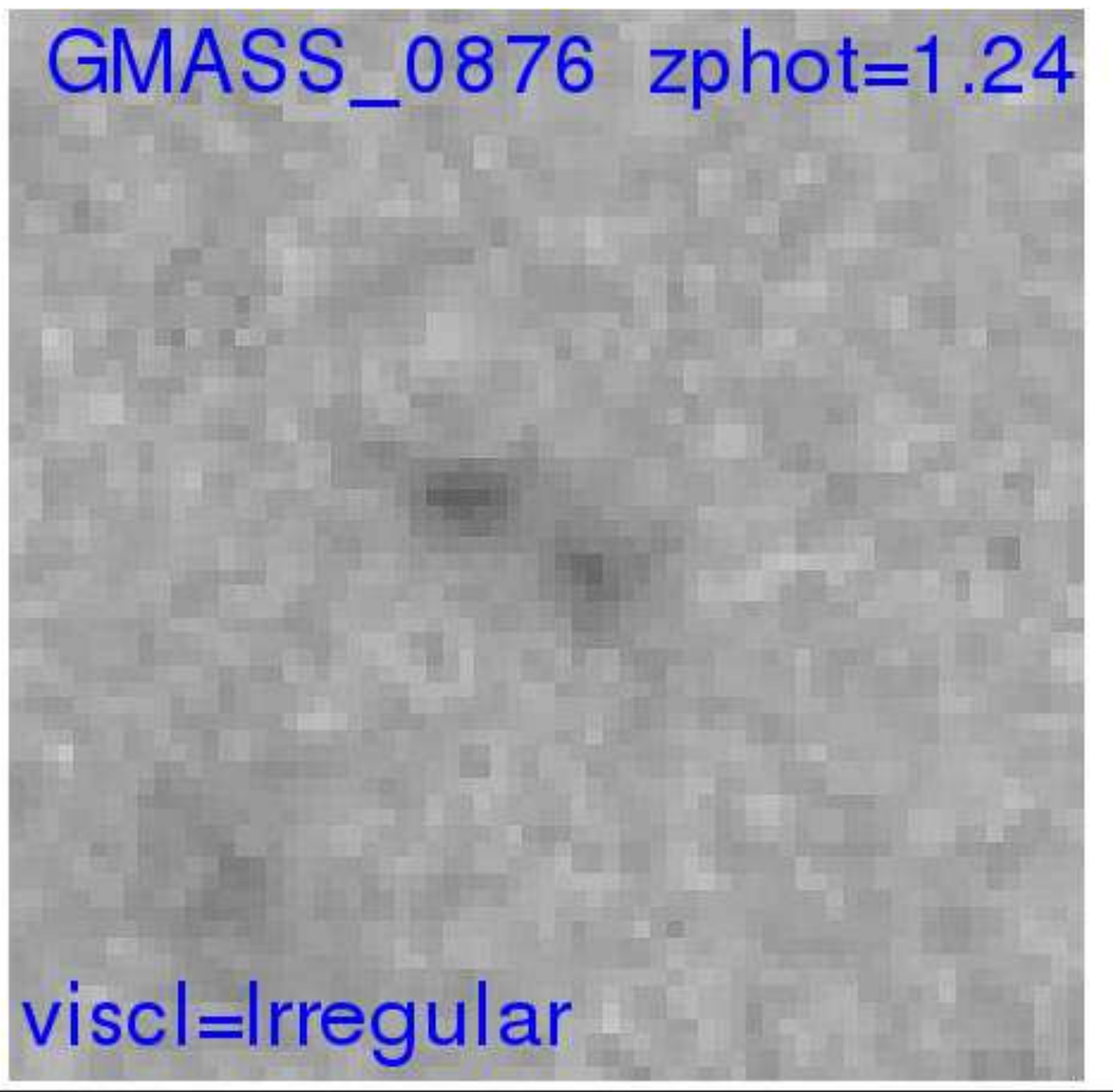}		     
\includegraphics[trim=100 40 75 390, clip=true, width=30mm]{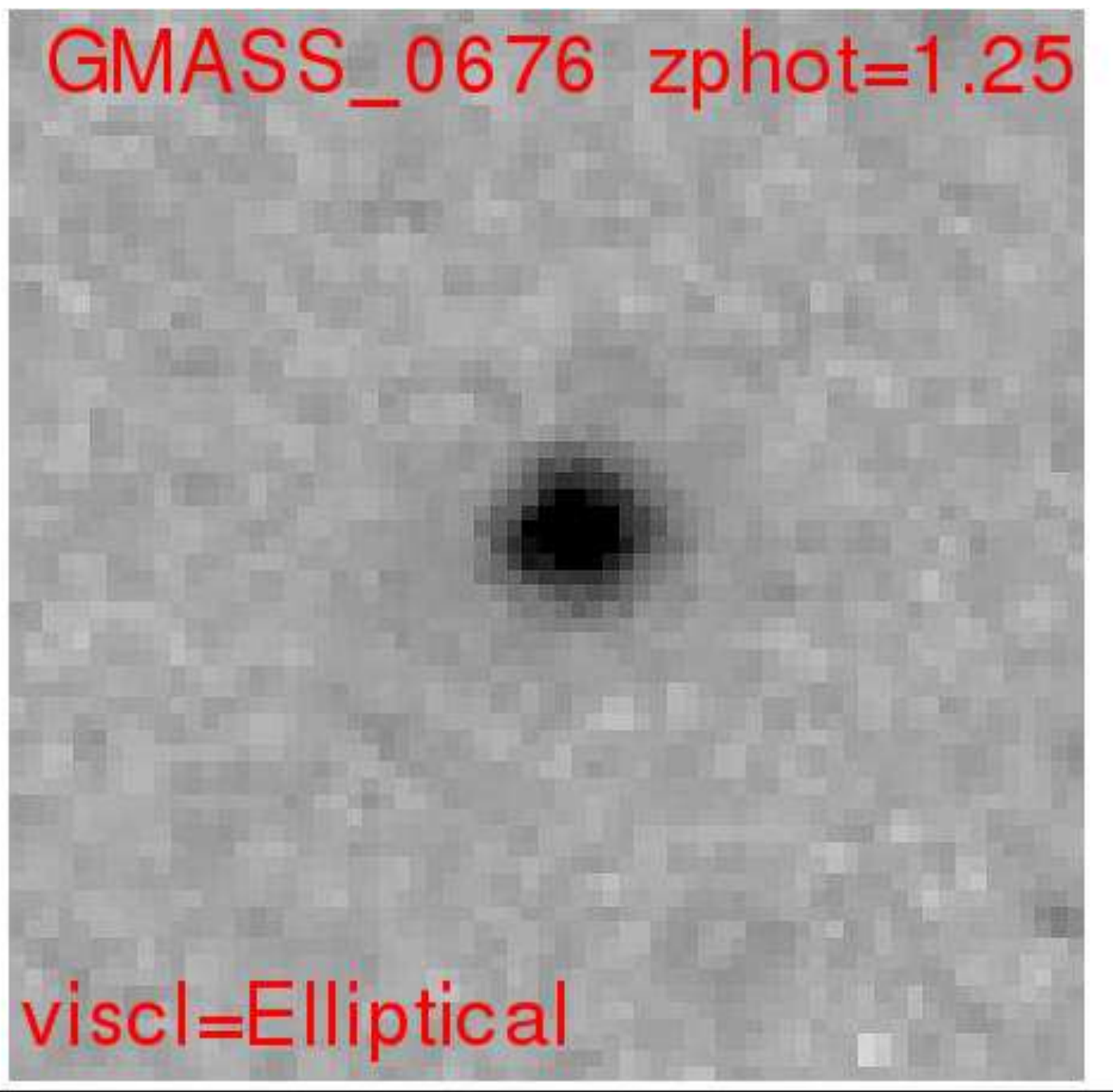}			     
\includegraphics[trim=100 40 75 390, clip=true, width=30mm]{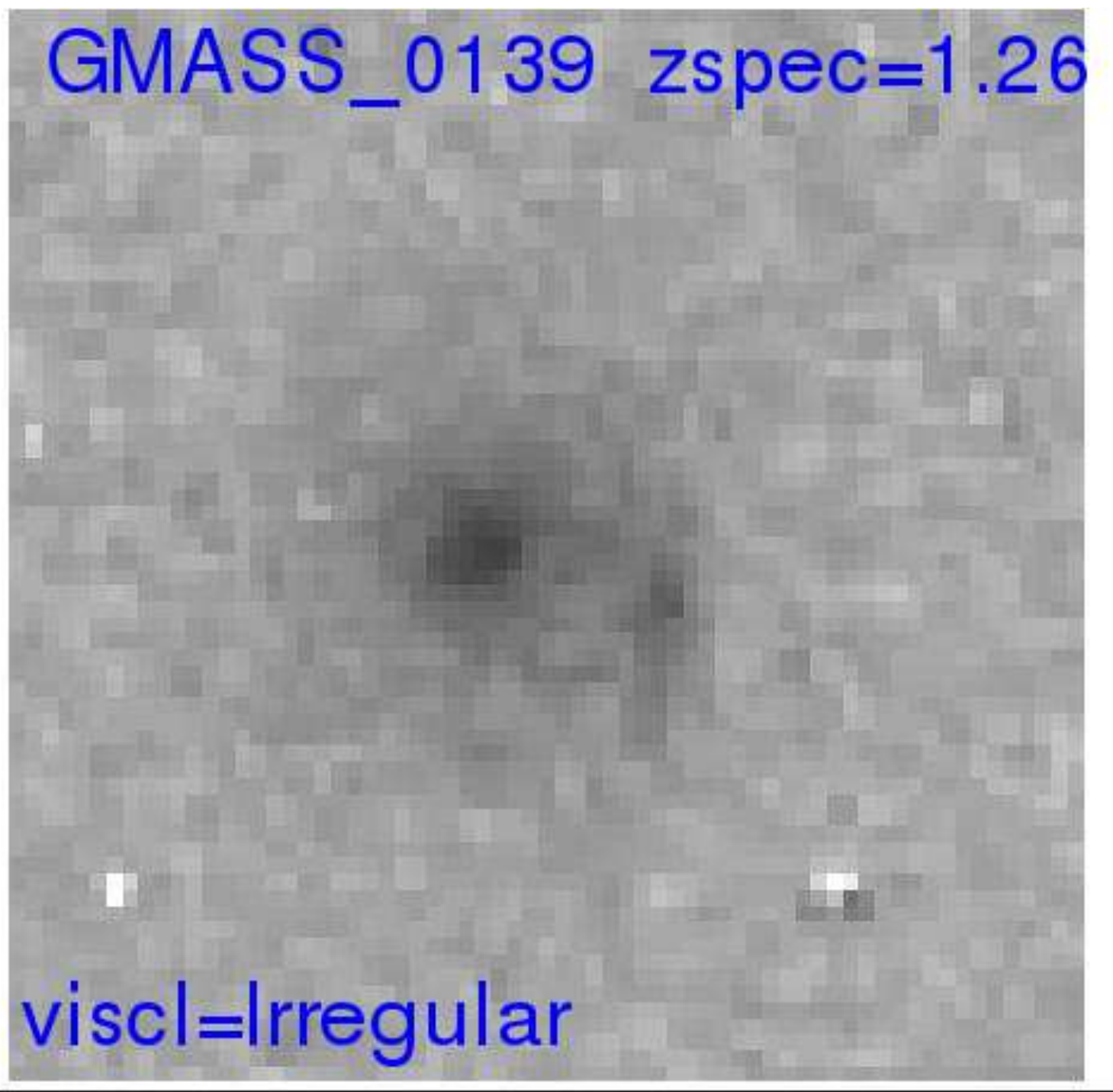}			     
\includegraphics[trim=100 40 75 390, clip=true, width=30mm]{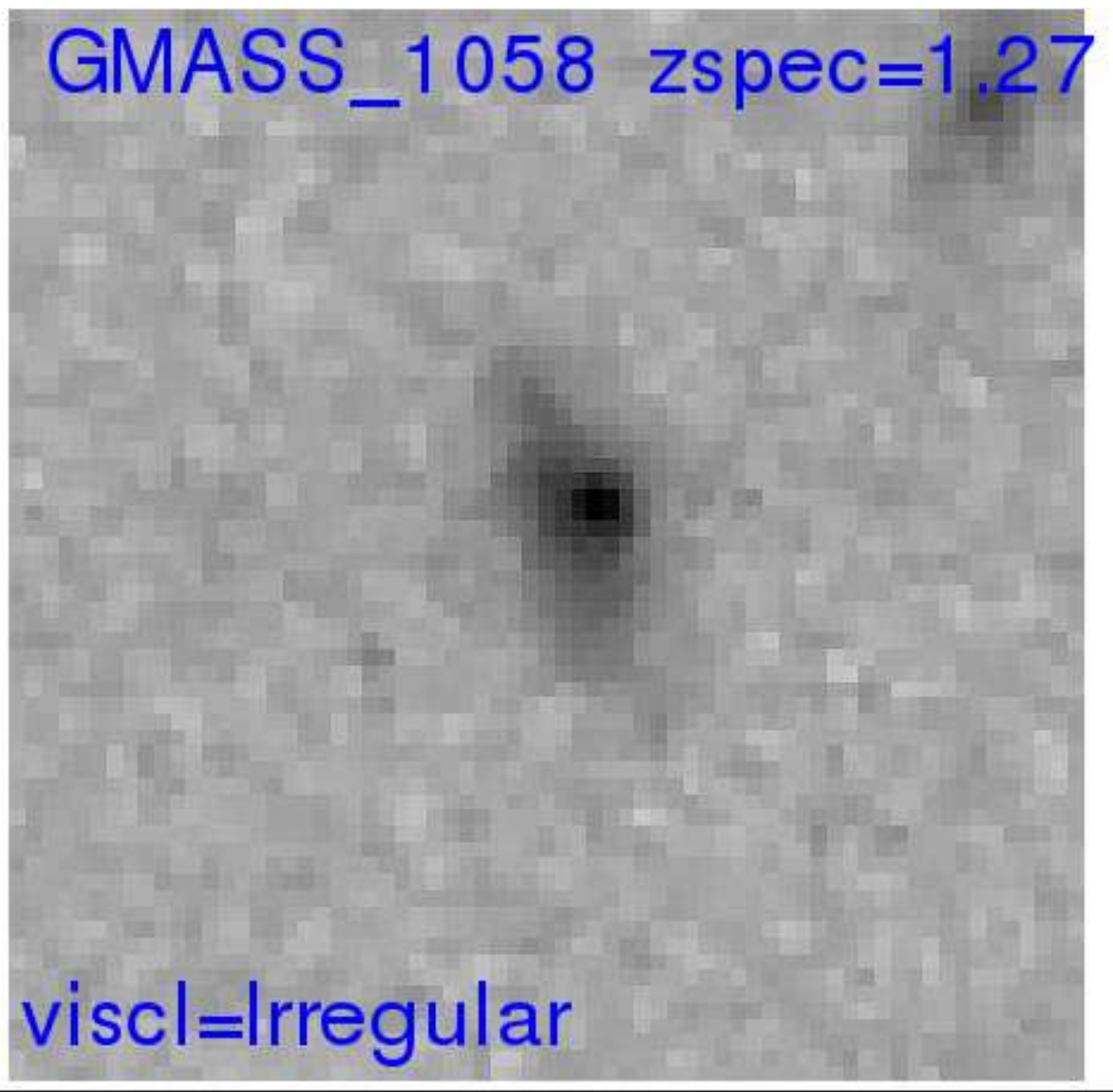}		     
\includegraphics[trim=100 40 75 390, clip=true, width=30mm]{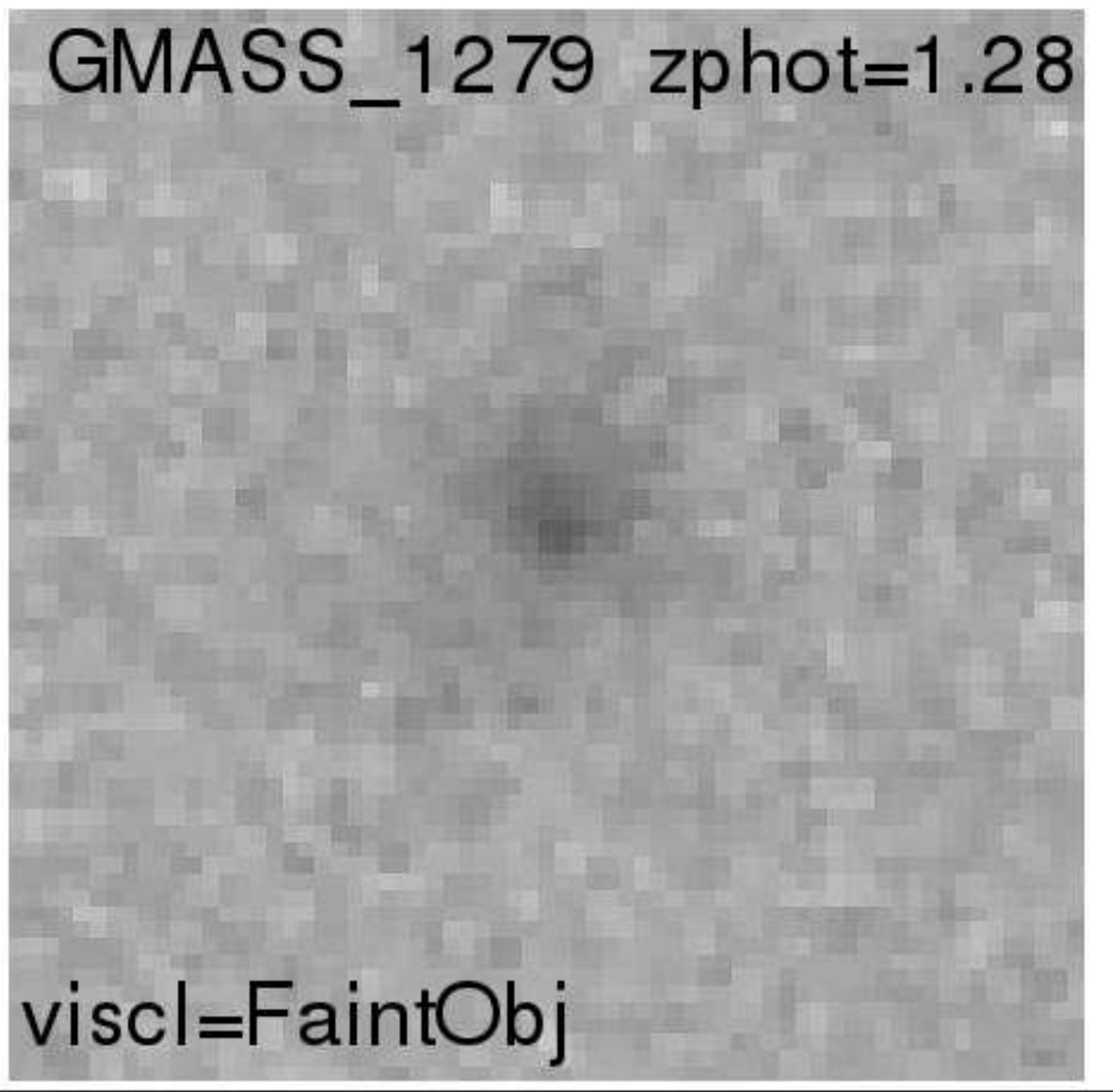}			     
\includegraphics[trim=100 40 75 390, clip=true, width=30mm]{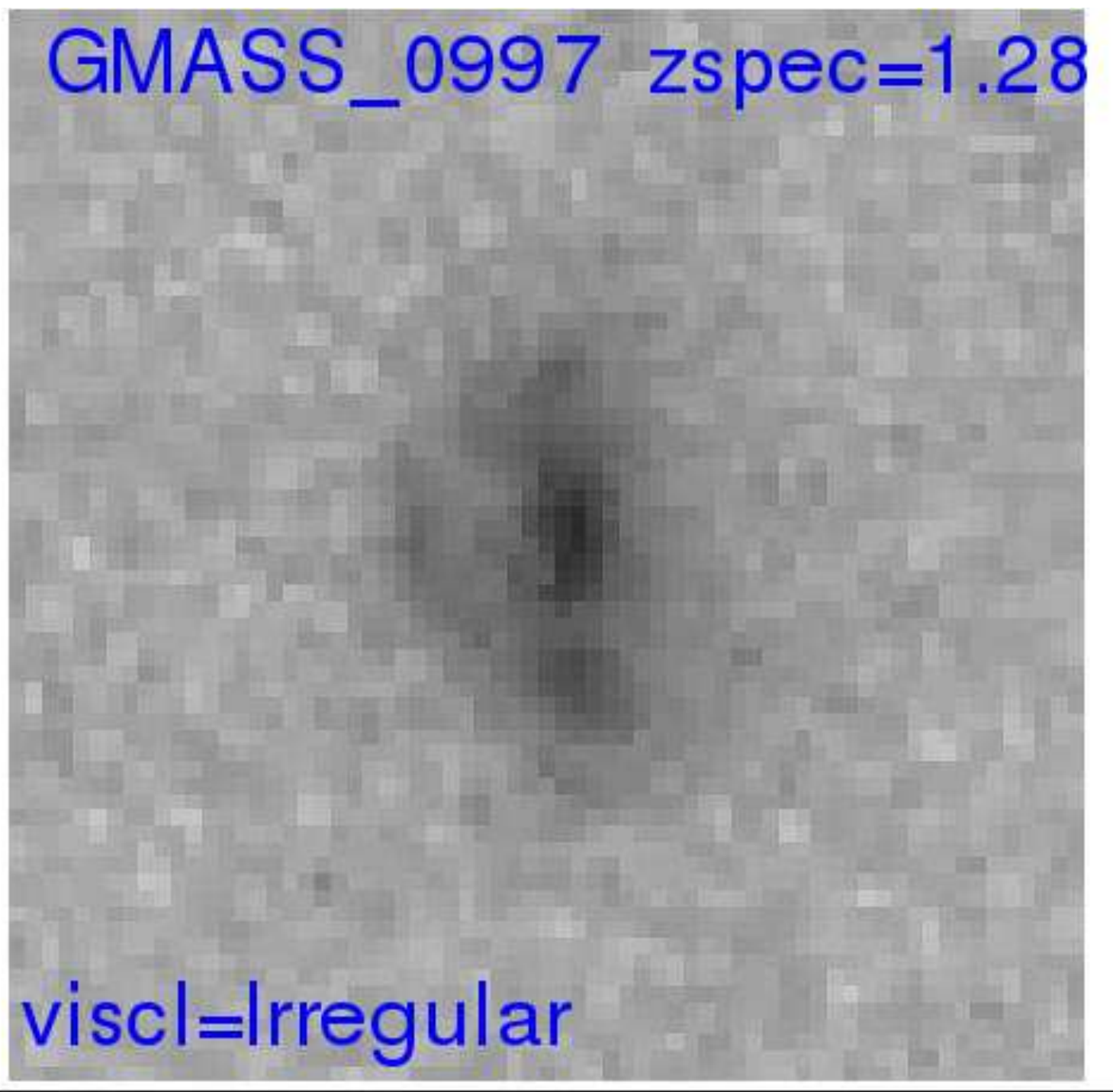}			     

\includegraphics[trim=100 40 75 390, clip=true, width=30mm]{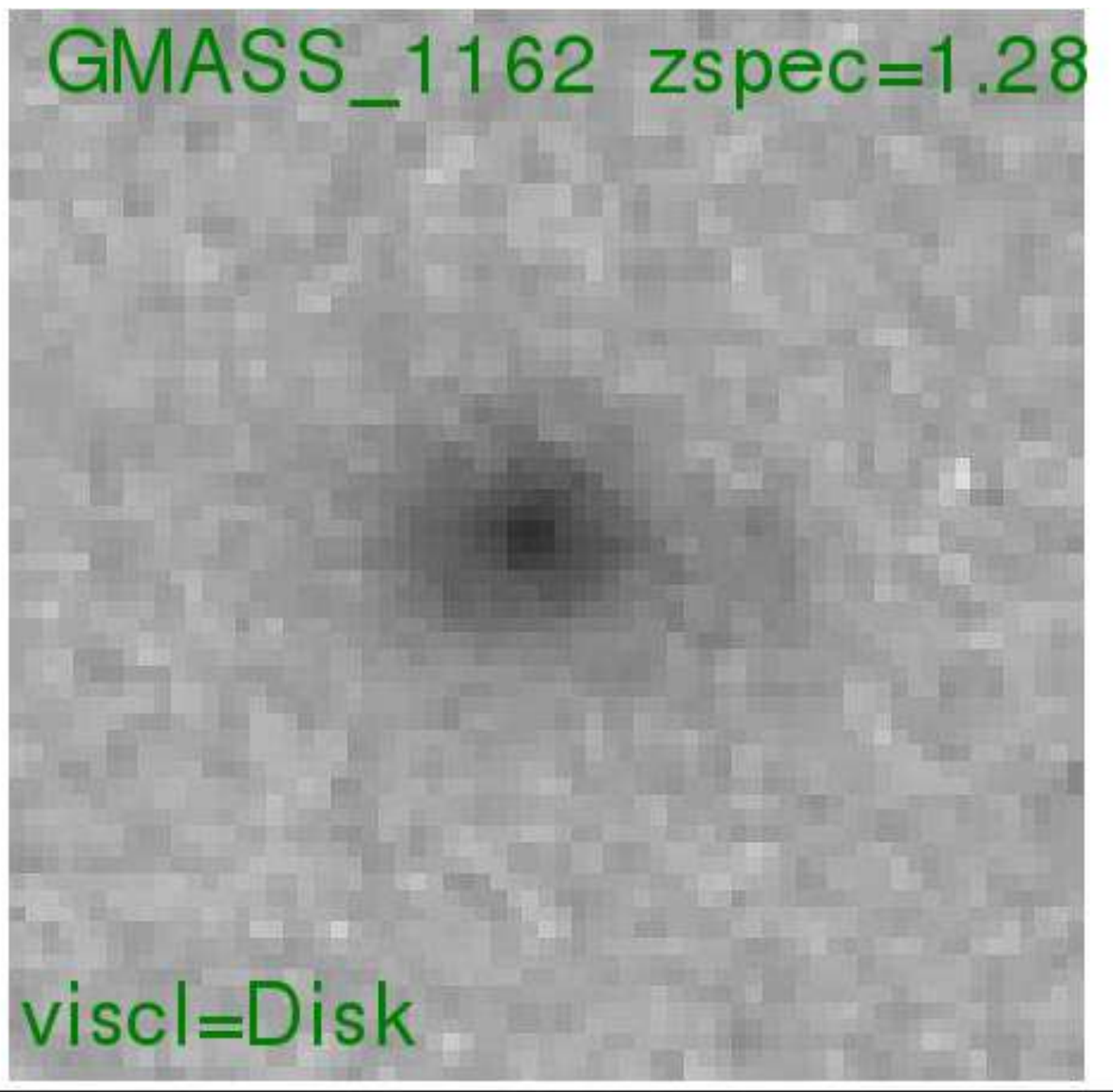}			     
\includegraphics[trim=100 40 75 390, clip=true, width=30mm]{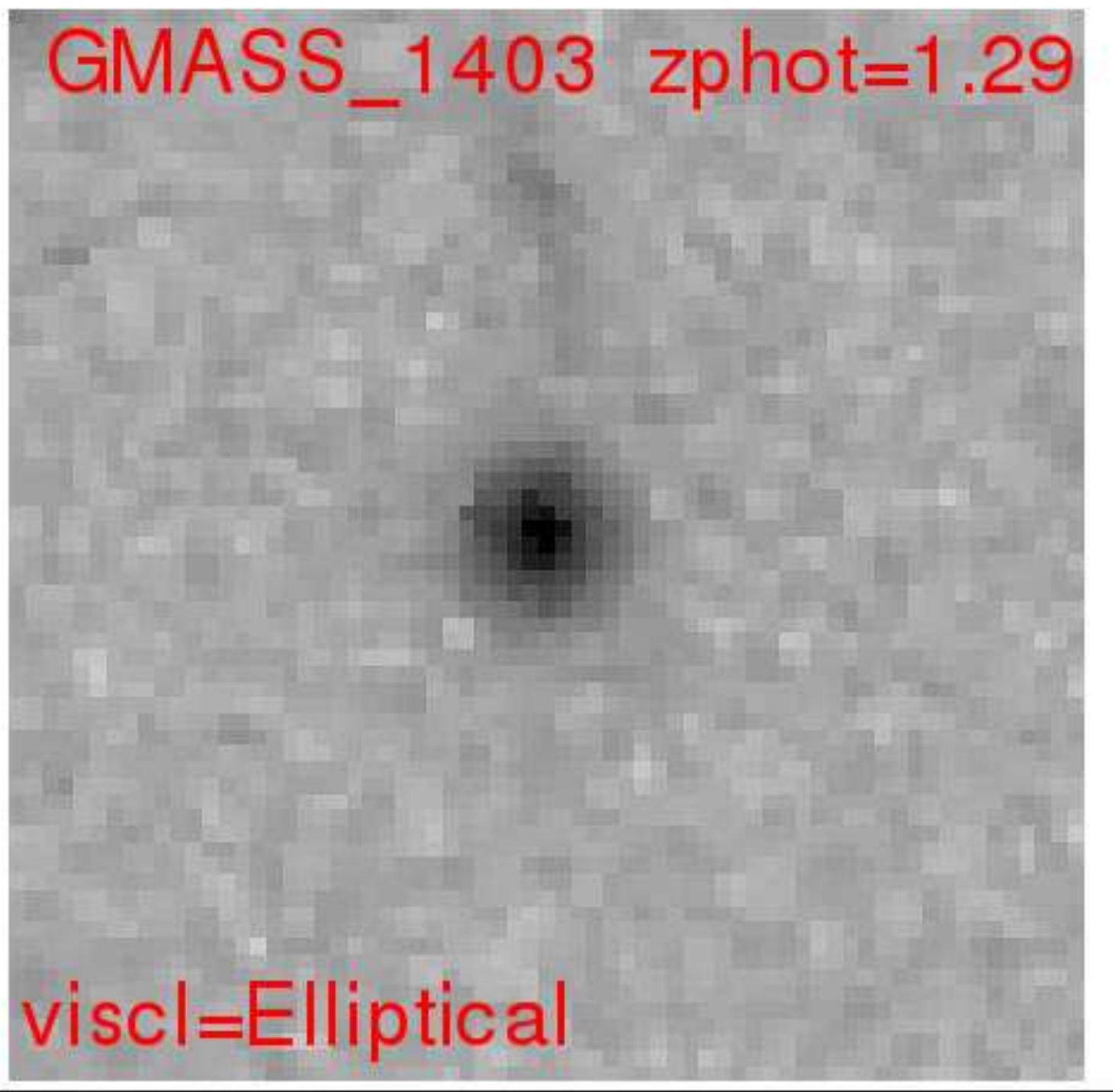}			     
\includegraphics[trim=100 40 75 390, clip=true, width=30mm]{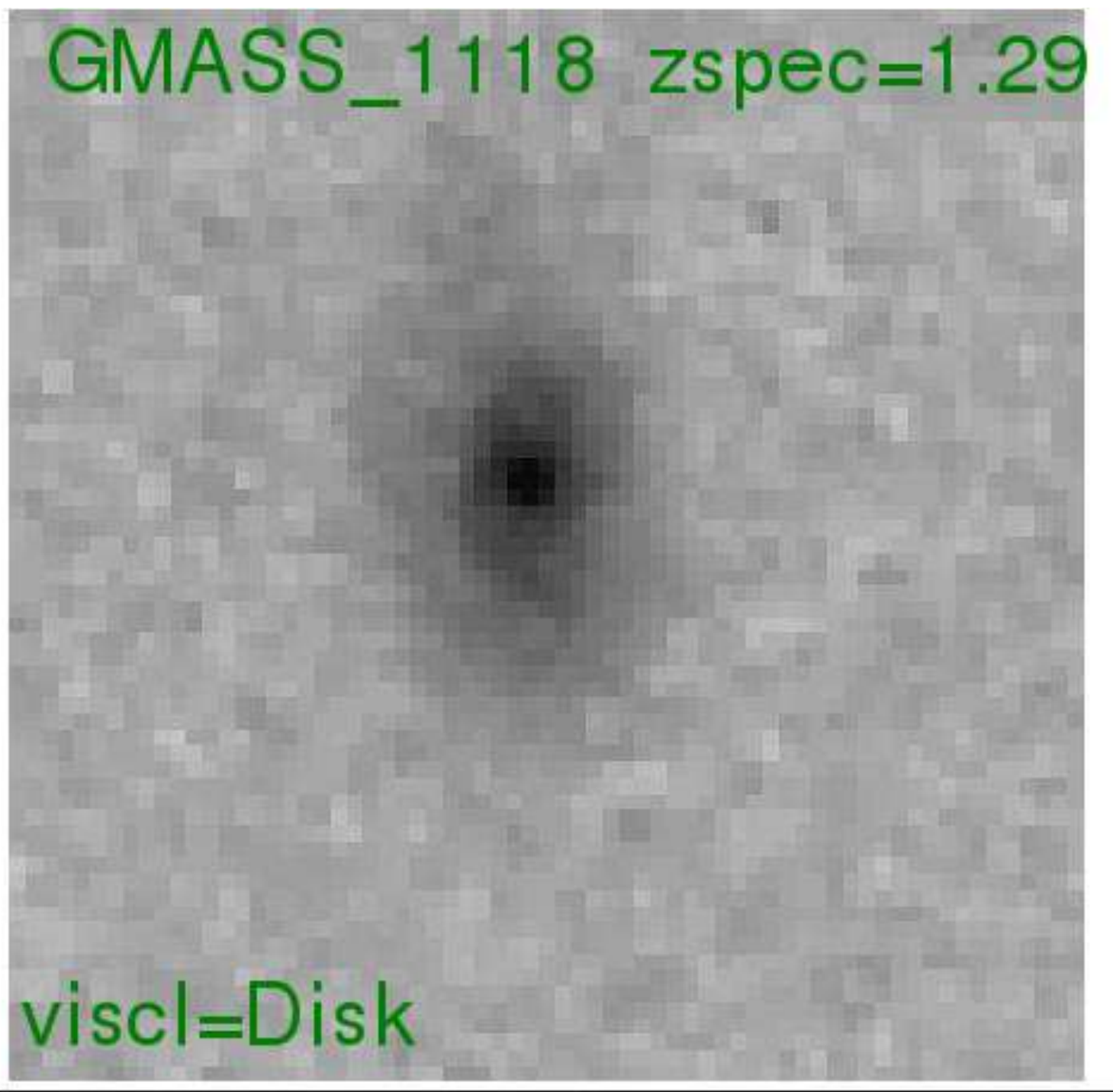}		     
\includegraphics[trim=100 40 75 390, clip=true, width=30mm]{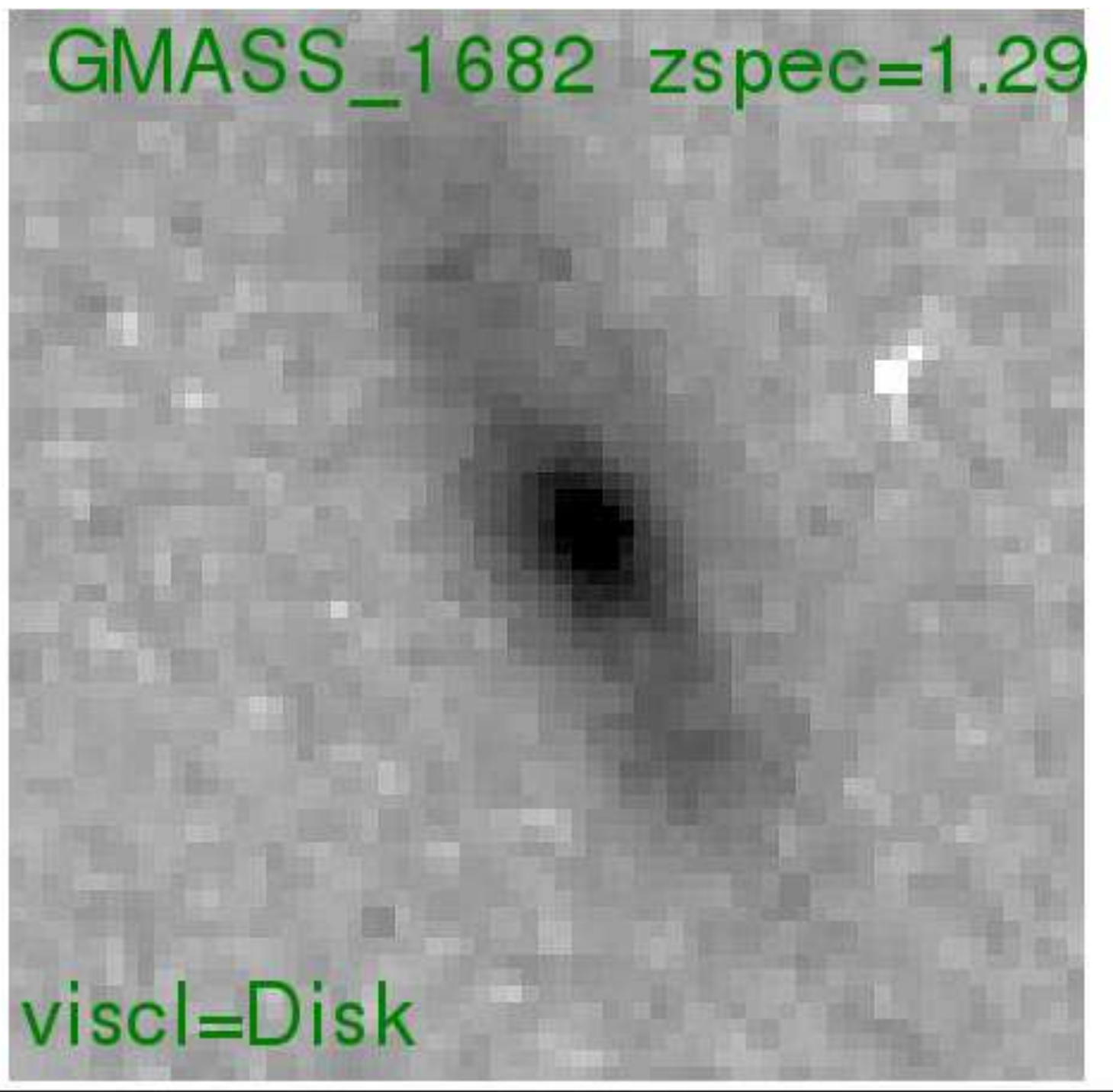}			     
\includegraphics[trim=100 40 75 390, clip=true, width=30mm]{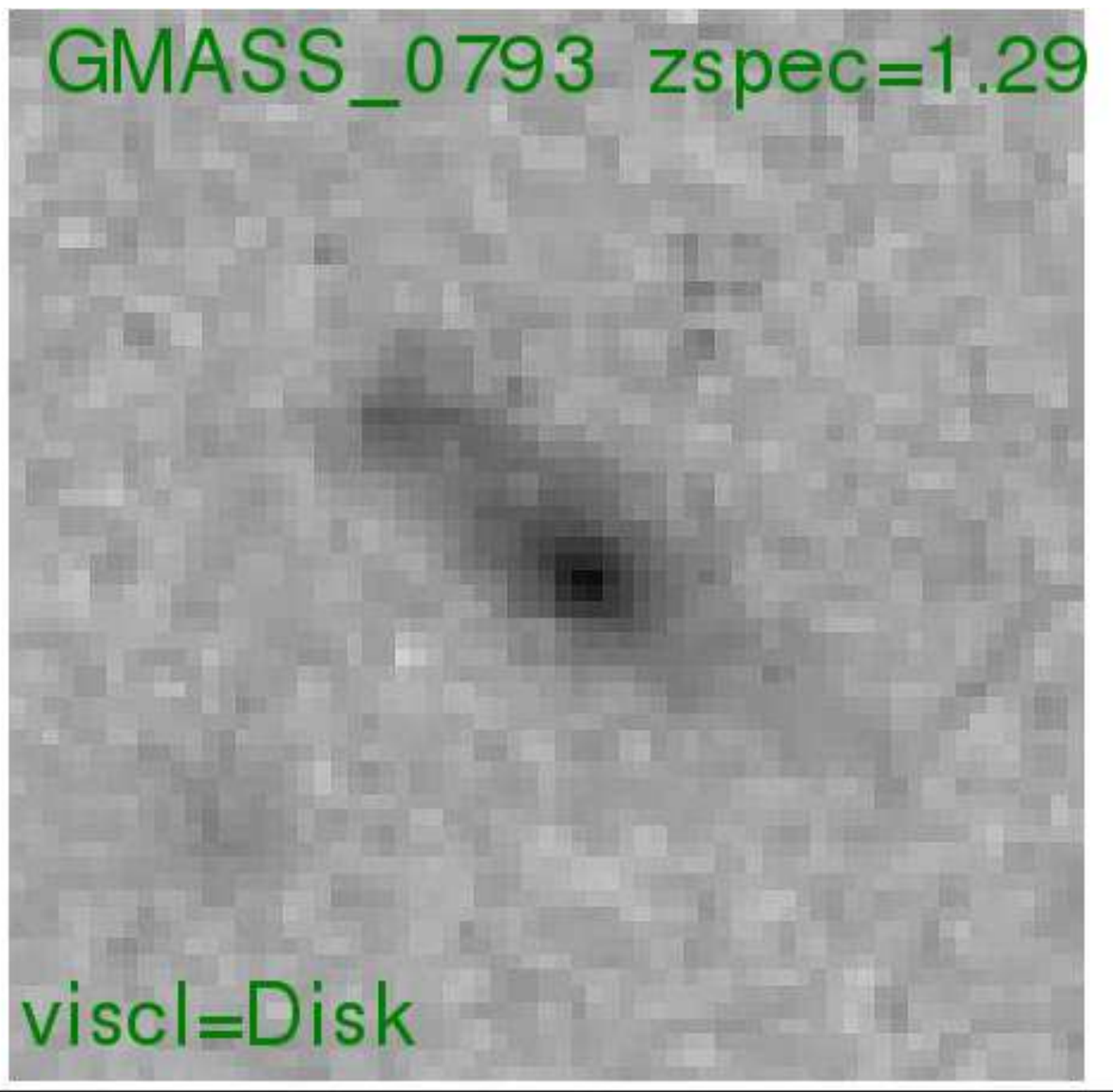}			     
\includegraphics[trim=100 40 75 390, clip=true, width=30mm]{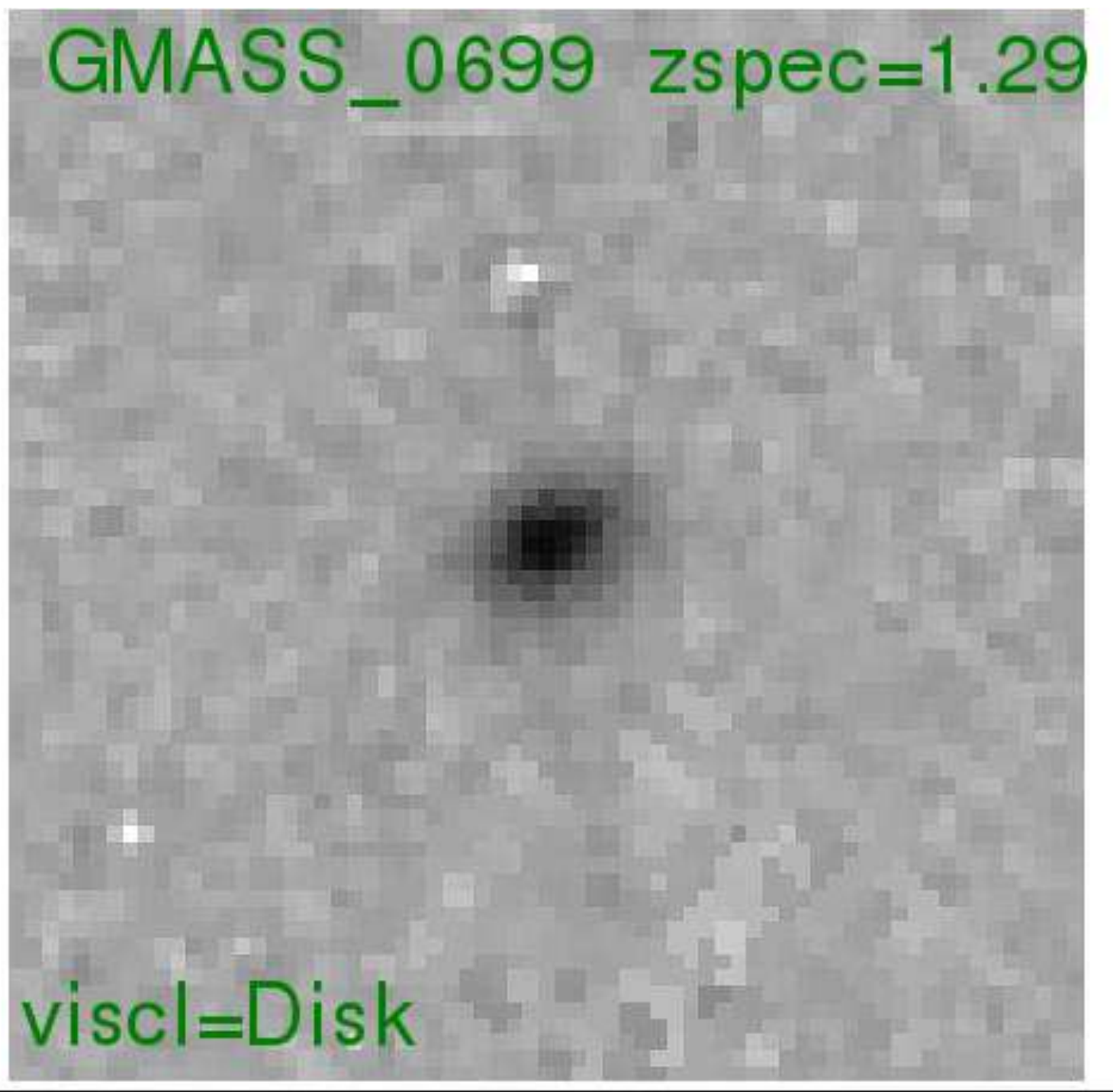}
\end{figure*}
\begin{figure*}
\centering   
\includegraphics[trim=100 40 75 390, clip=true, width=30mm]{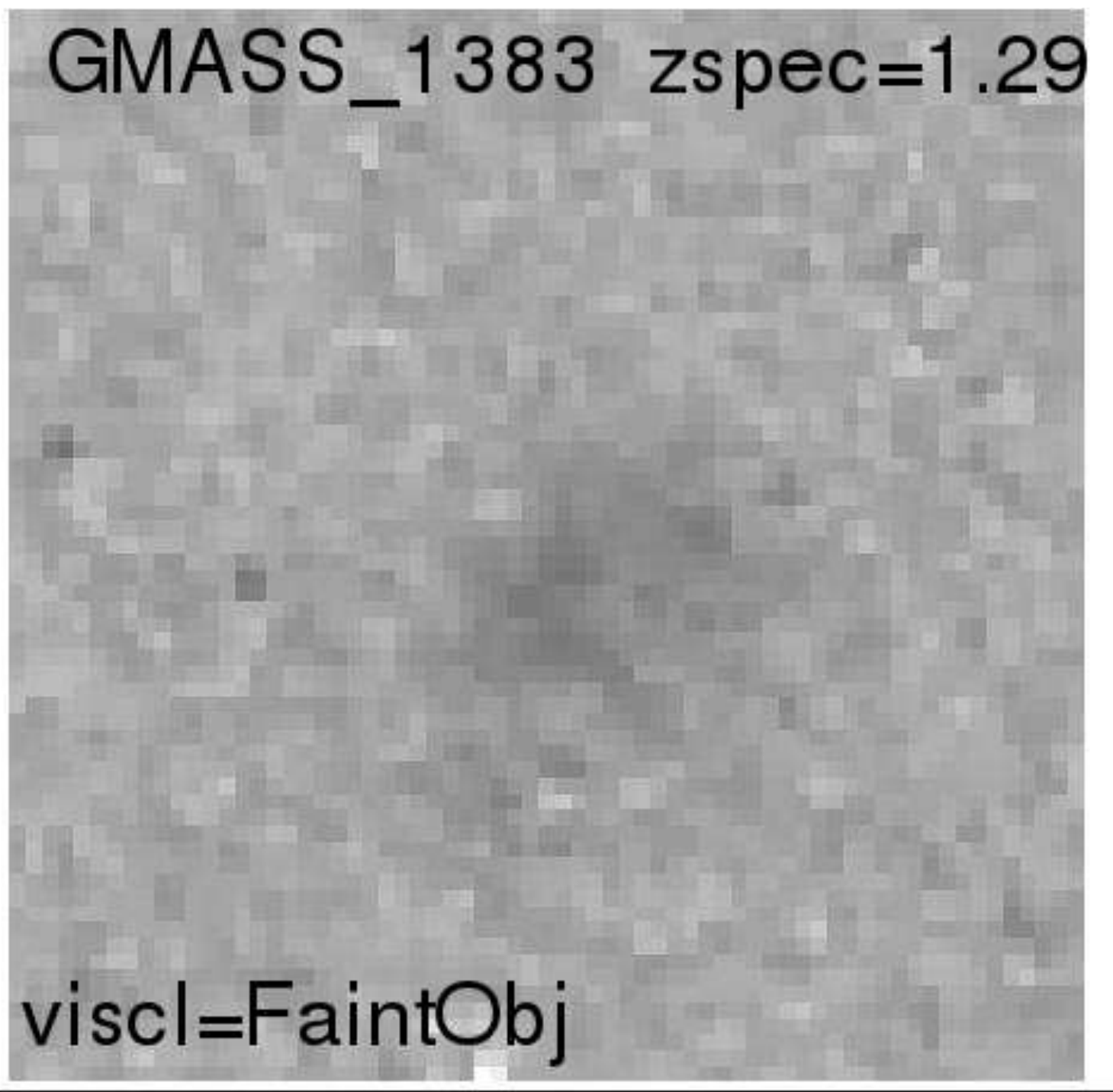}		     
\includegraphics[trim=100 40 75 390, clip=true, width=30mm]{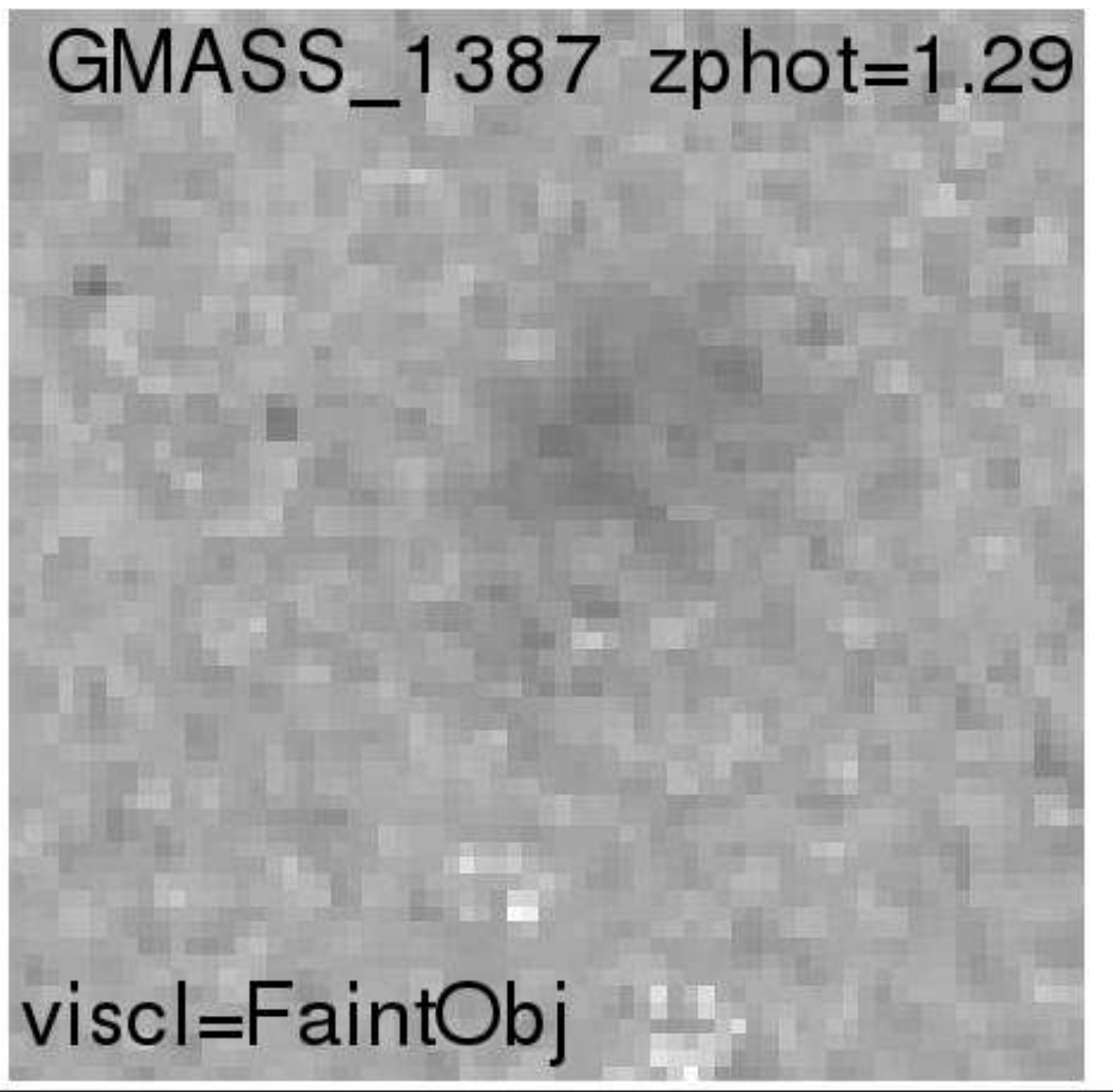}			     
\includegraphics[trim=100 40 75 390, clip=true, width=30mm]{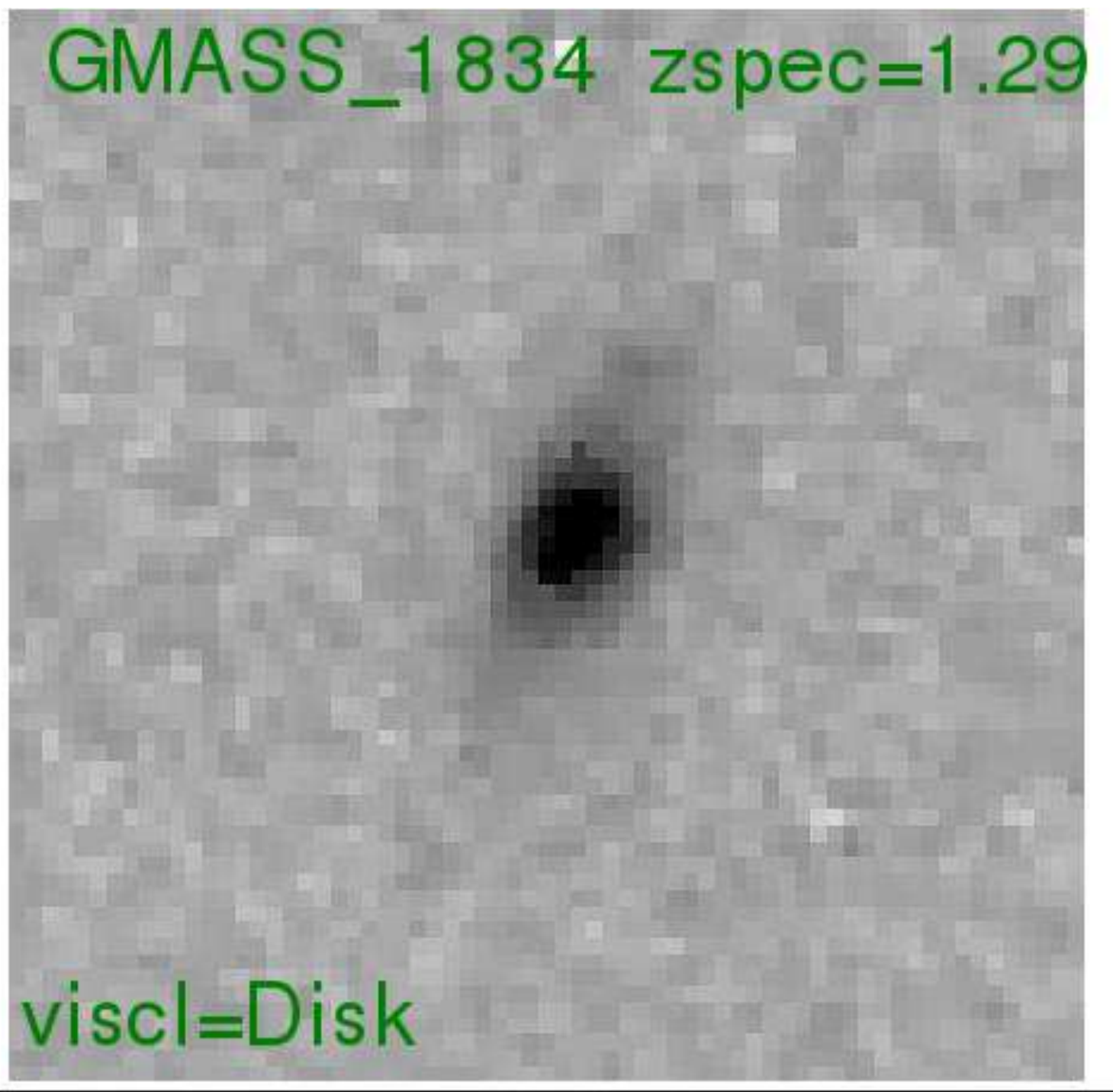}			     
\includegraphics[trim=100 40 75 390, clip=true, width=30mm]{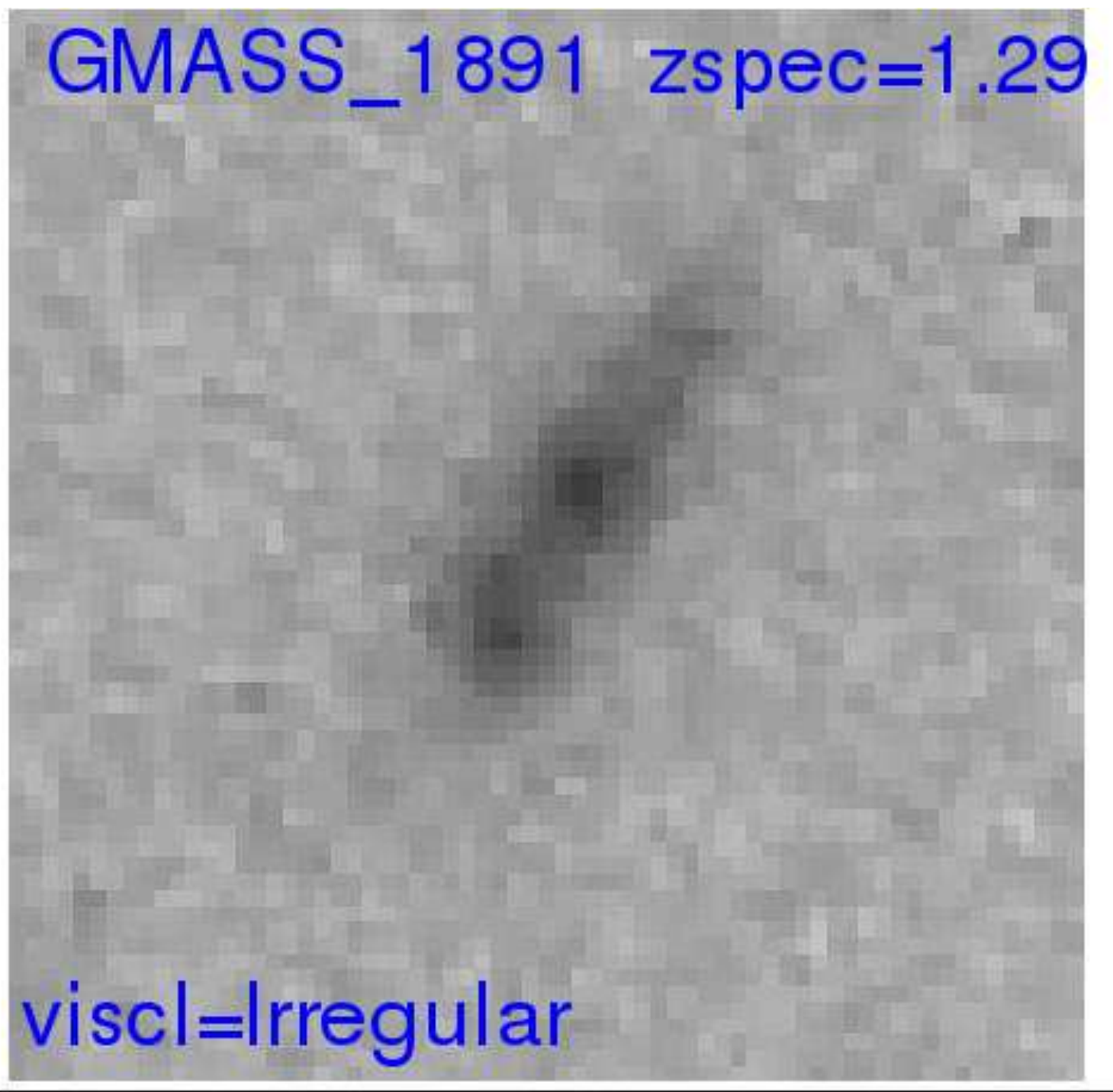}			     
\includegraphics[trim=100 40 75 390, clip=true, width=30mm]{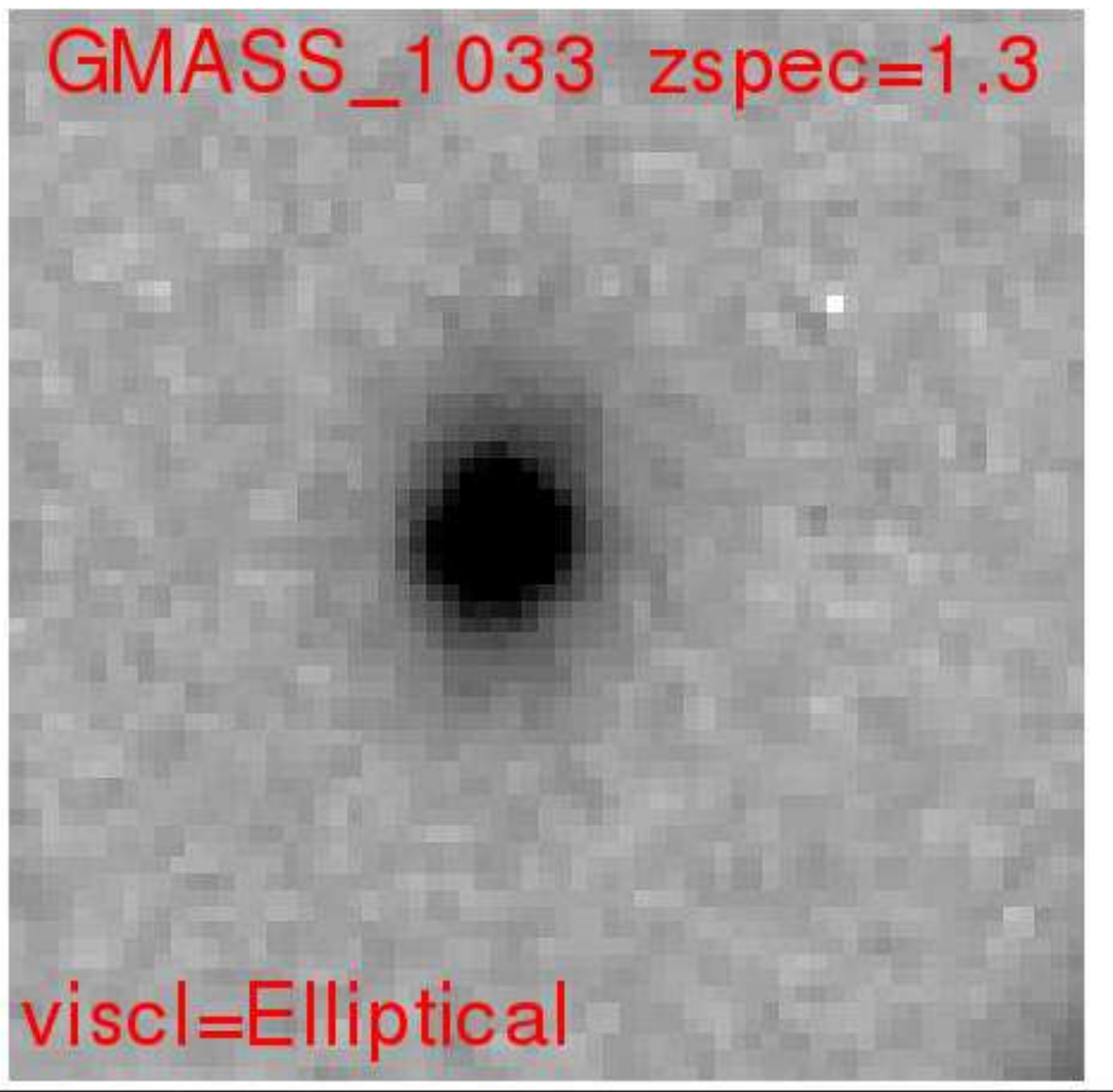}		     
\includegraphics[trim=100 40 75 390, clip=true, width=30mm]{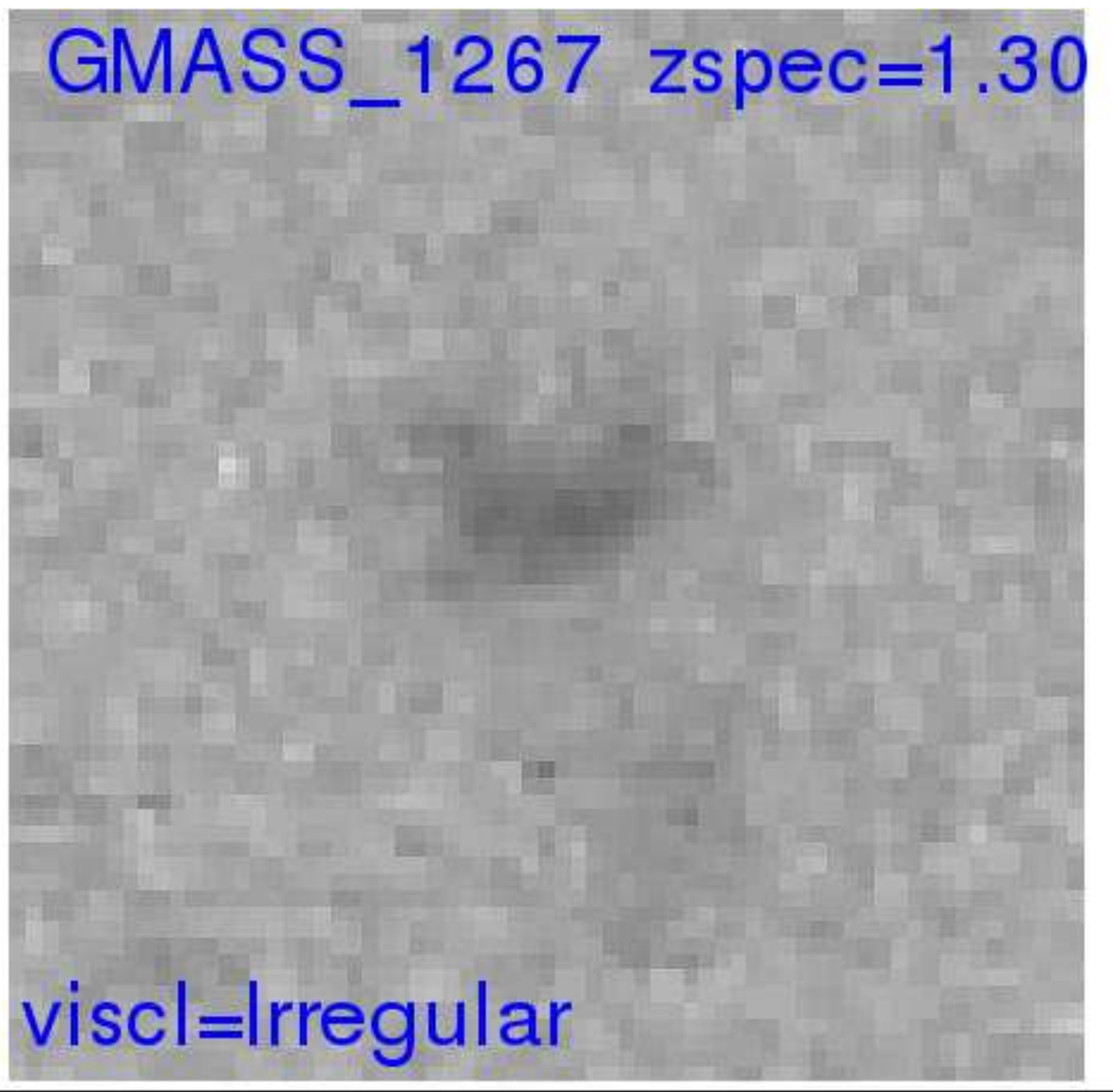}			     

\includegraphics[trim=100 40 75 390, clip=true, width=30mm]{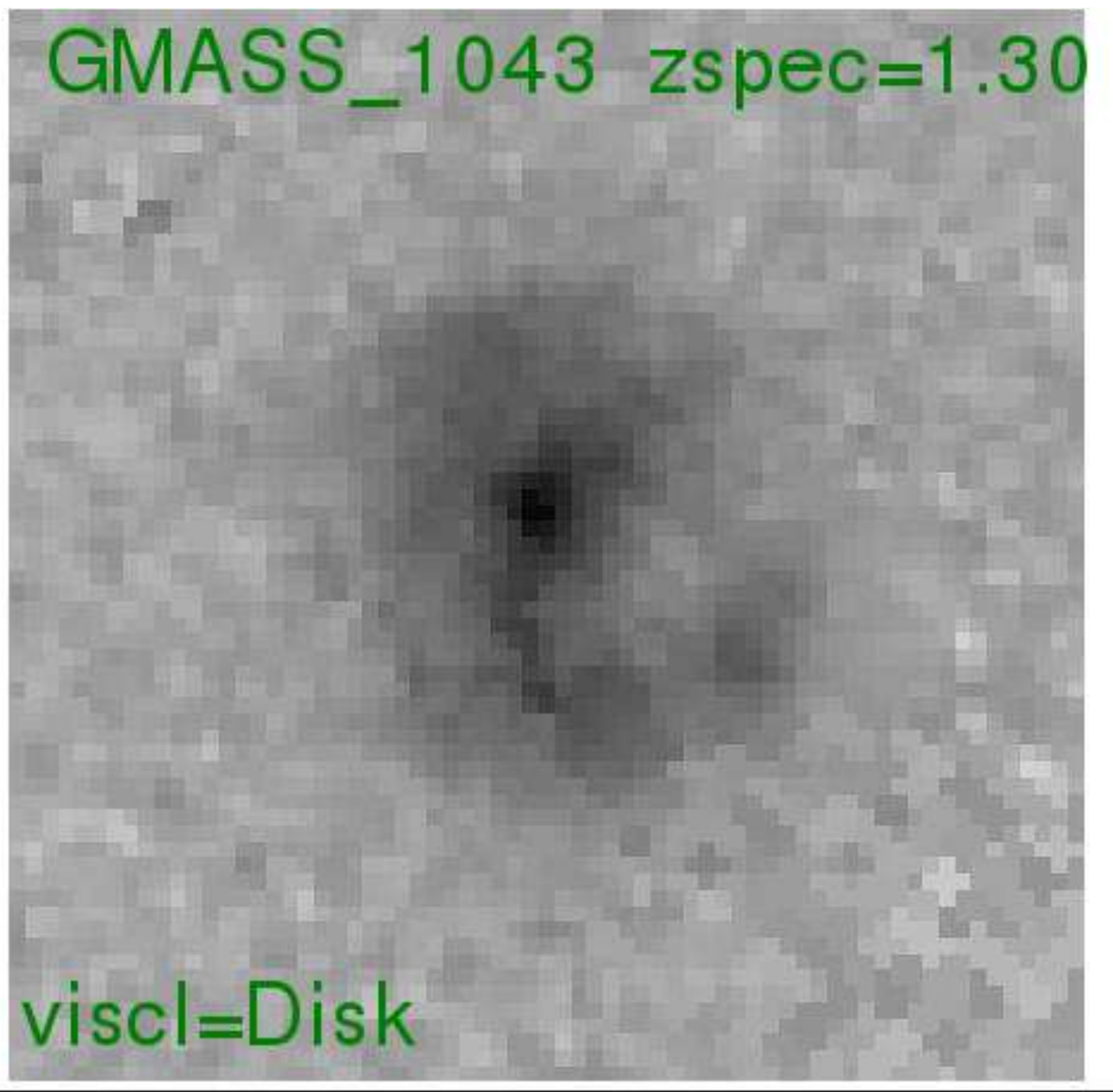}			     
\includegraphics[trim=100 40 75 390, clip=true, width=30mm]{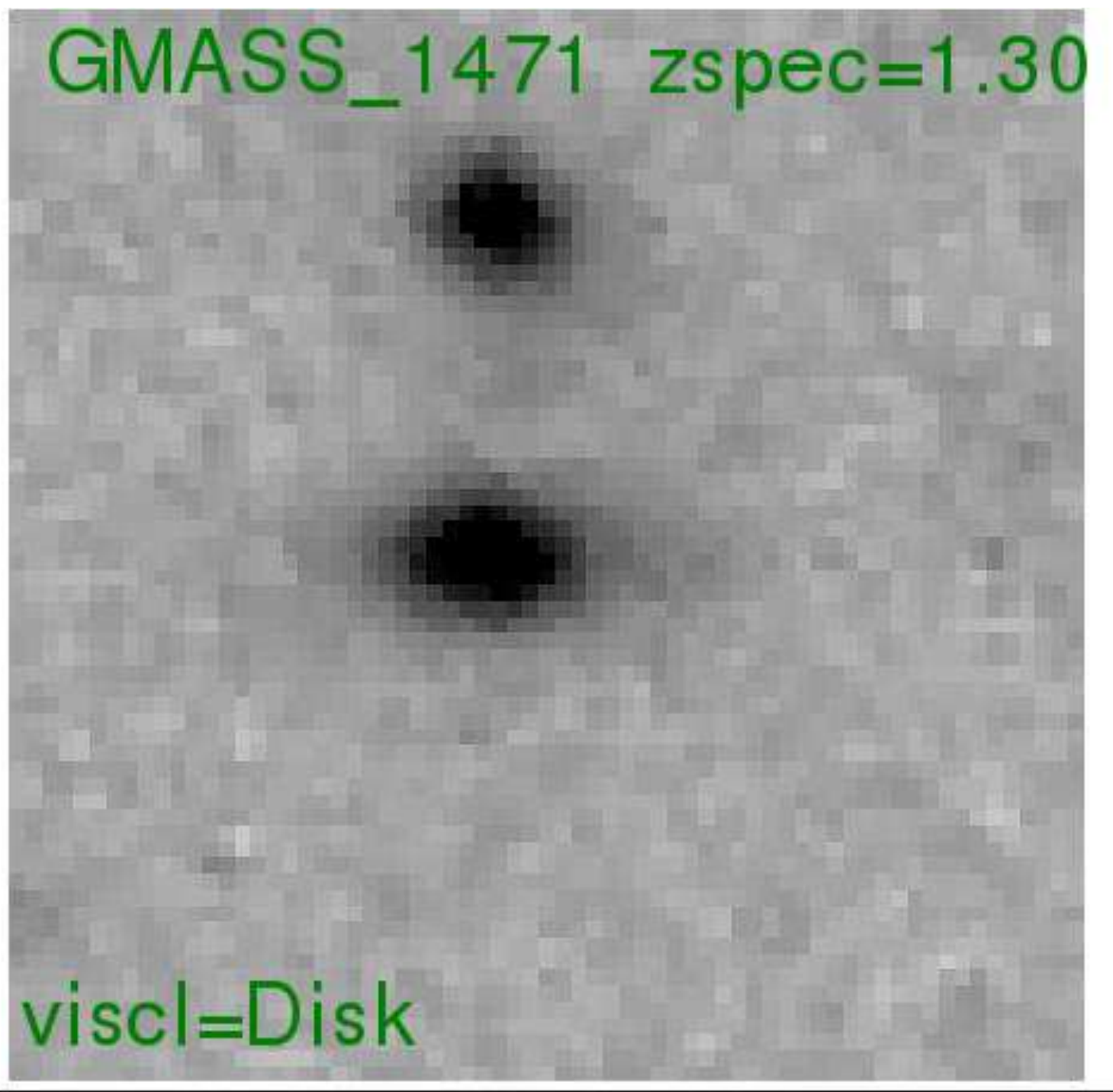}			     
\includegraphics[trim=100 40 75 390, clip=true, width=30mm]{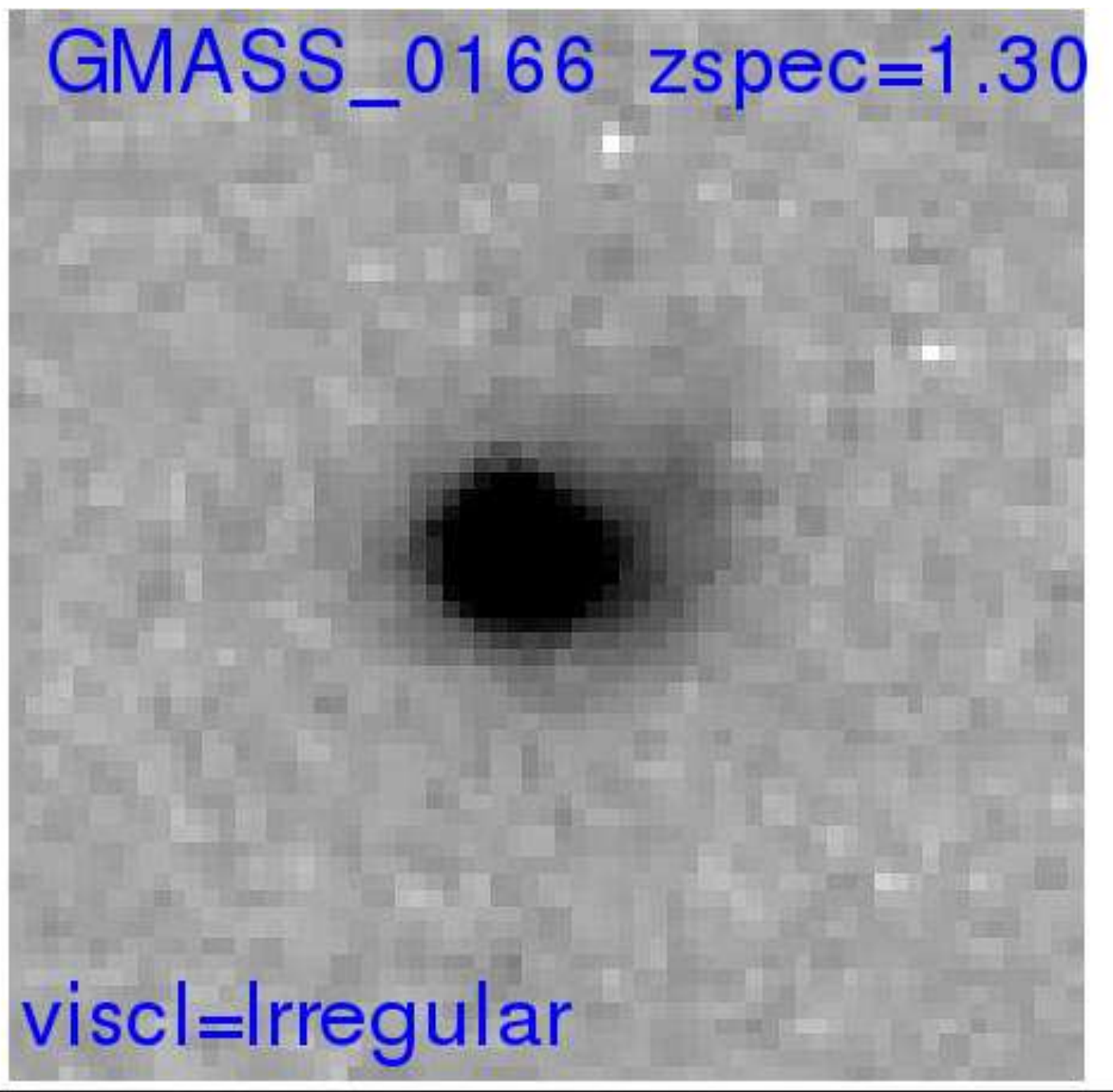}			     
\includegraphics[trim=100 40 75 390, clip=true, width=30mm]{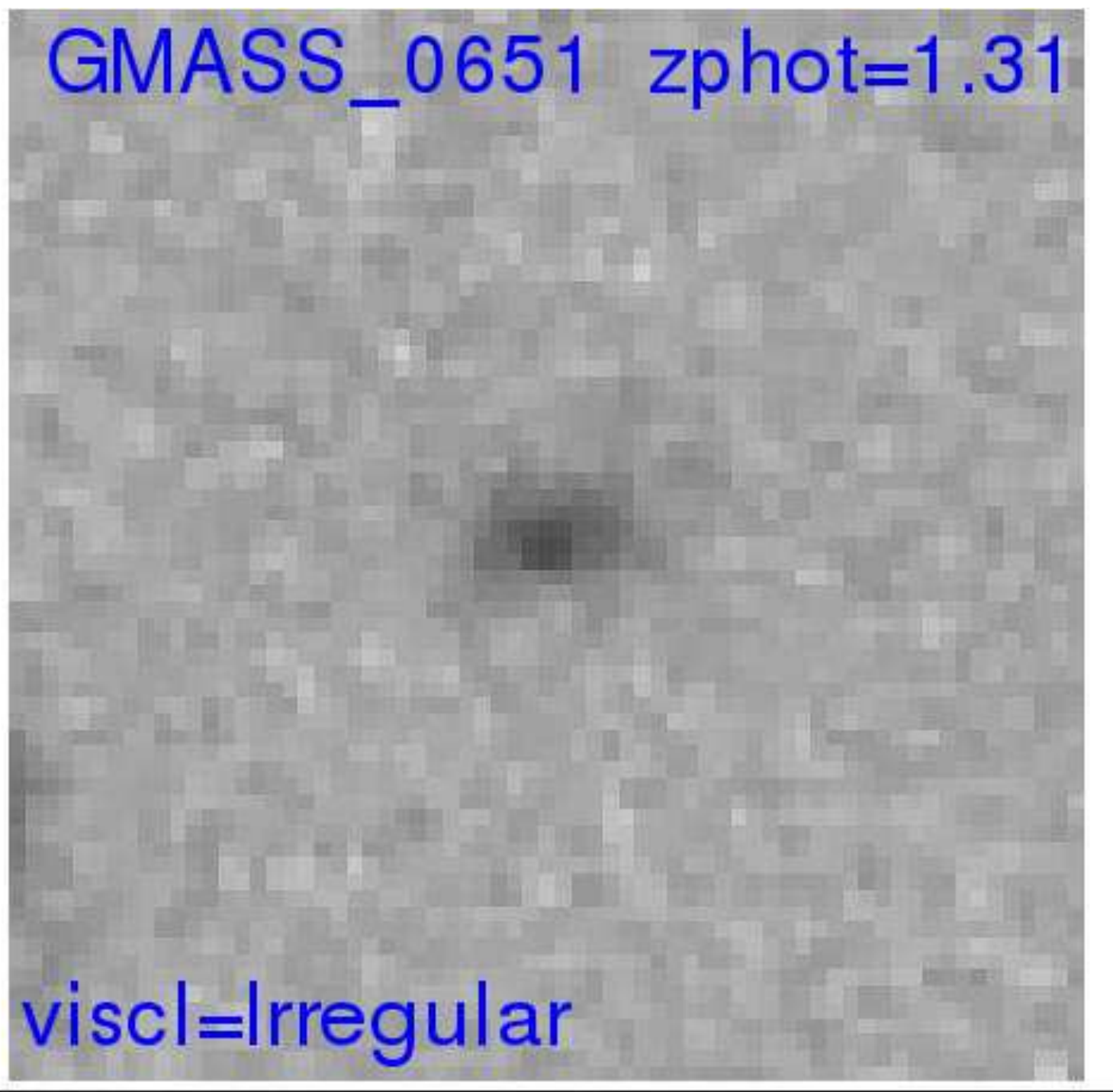}			     
\includegraphics[trim=100 40 75 390, clip=true, width=30mm]{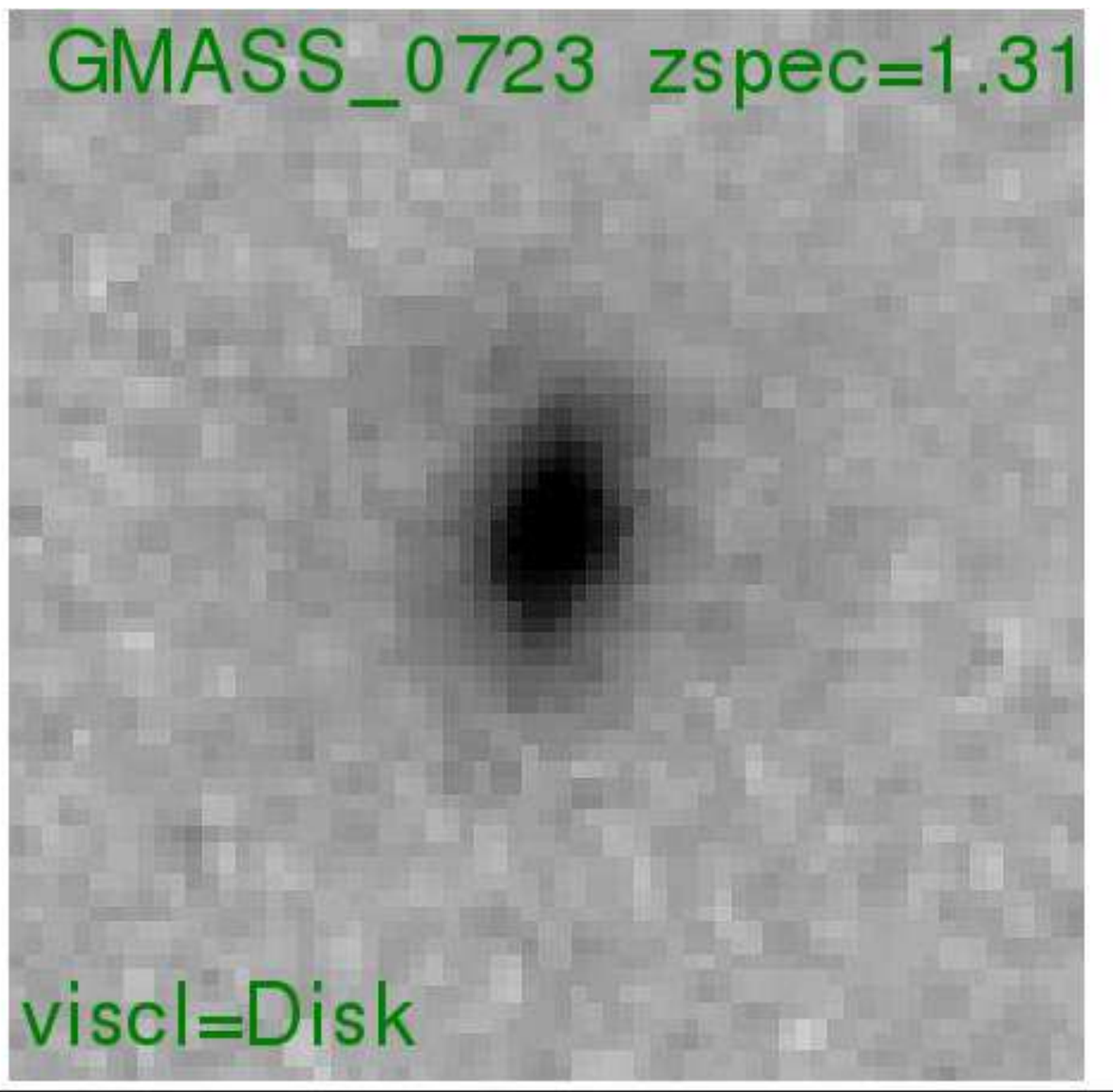}		     
\includegraphics[trim=100 40 75 390, clip=true, width=30mm]{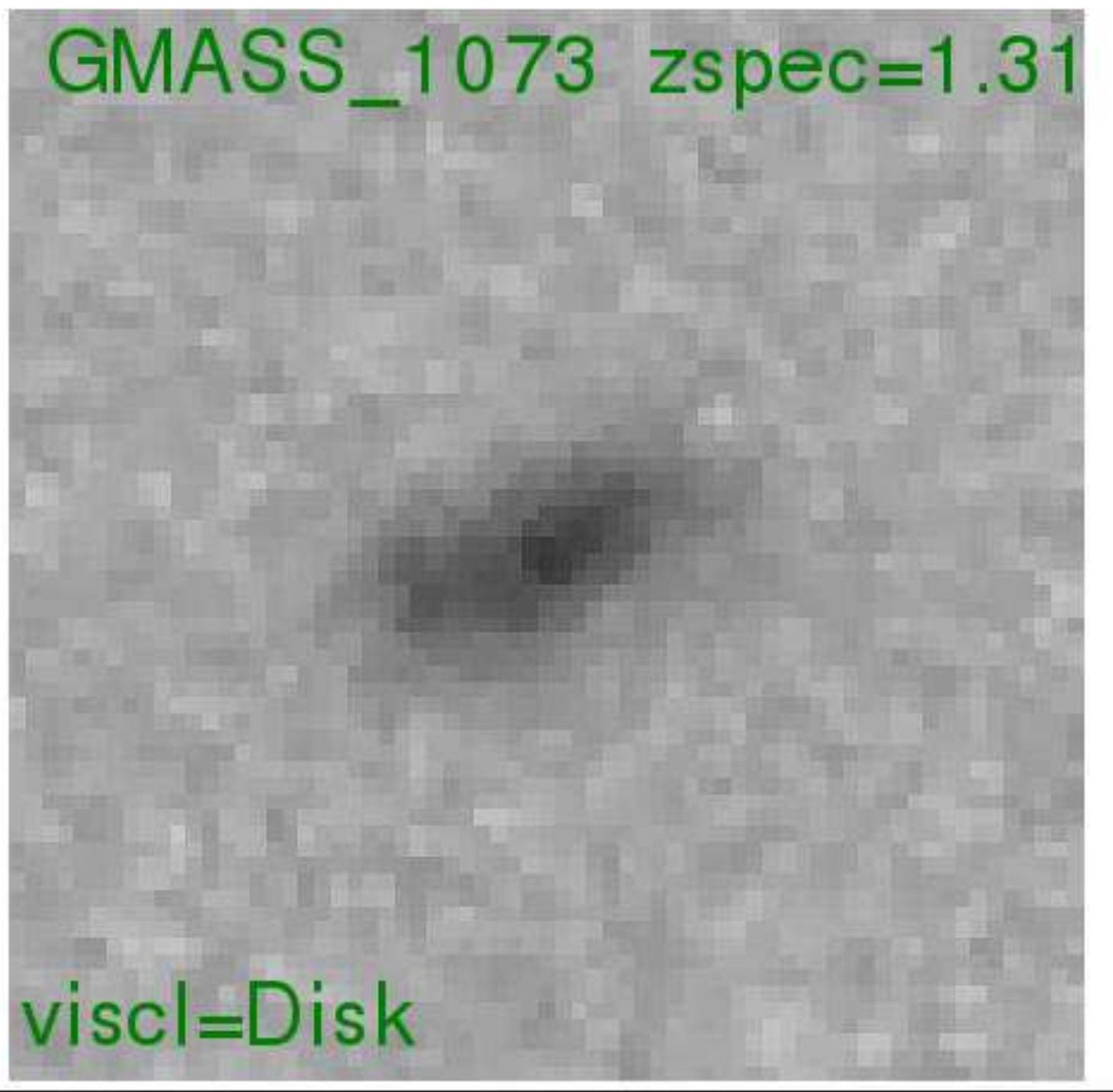}			     

\includegraphics[trim=100 40 75 390, clip=true, width=30mm]{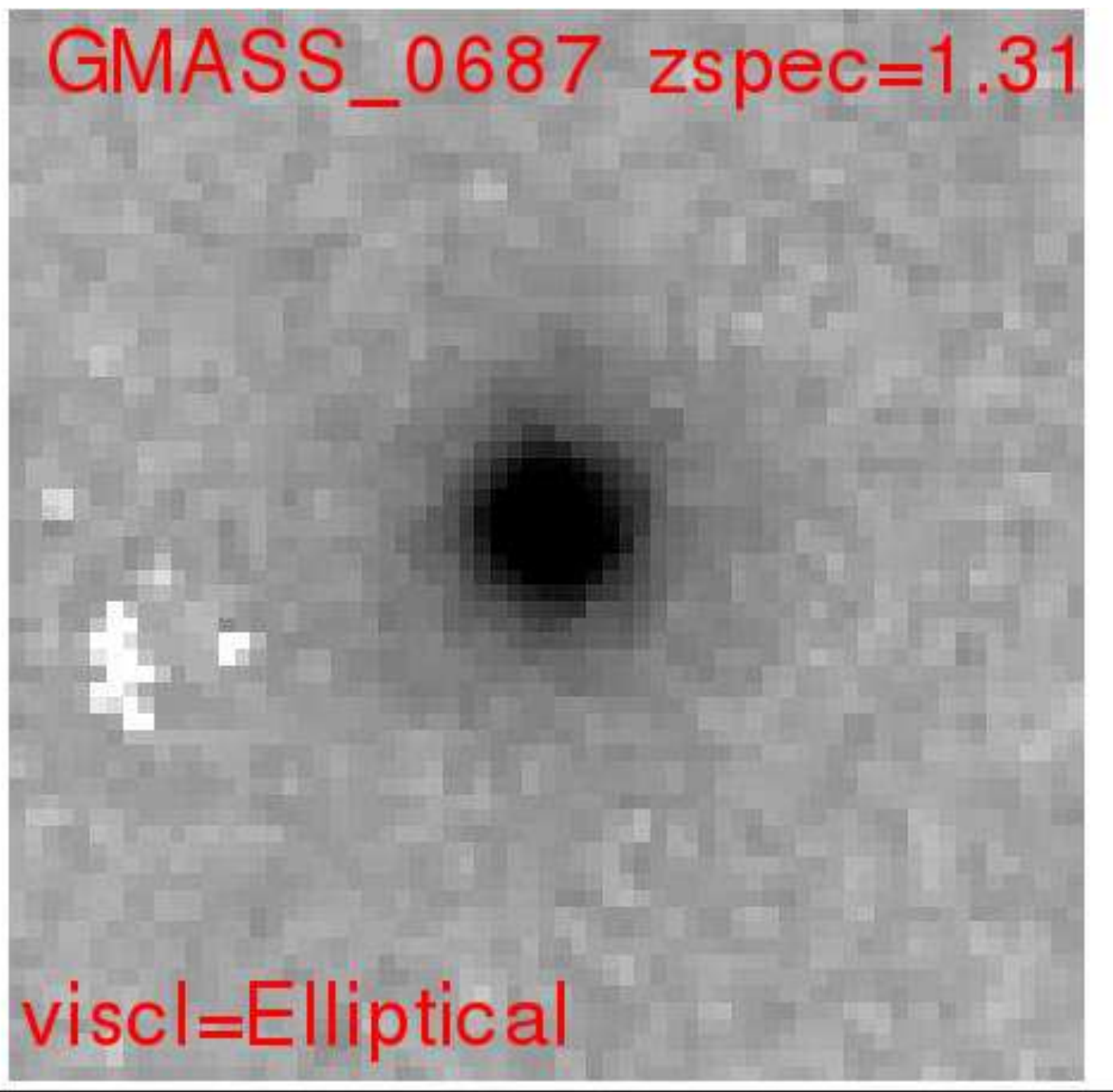}			     
\includegraphics[trim=100 40 75 390, clip=true, width=30mm]{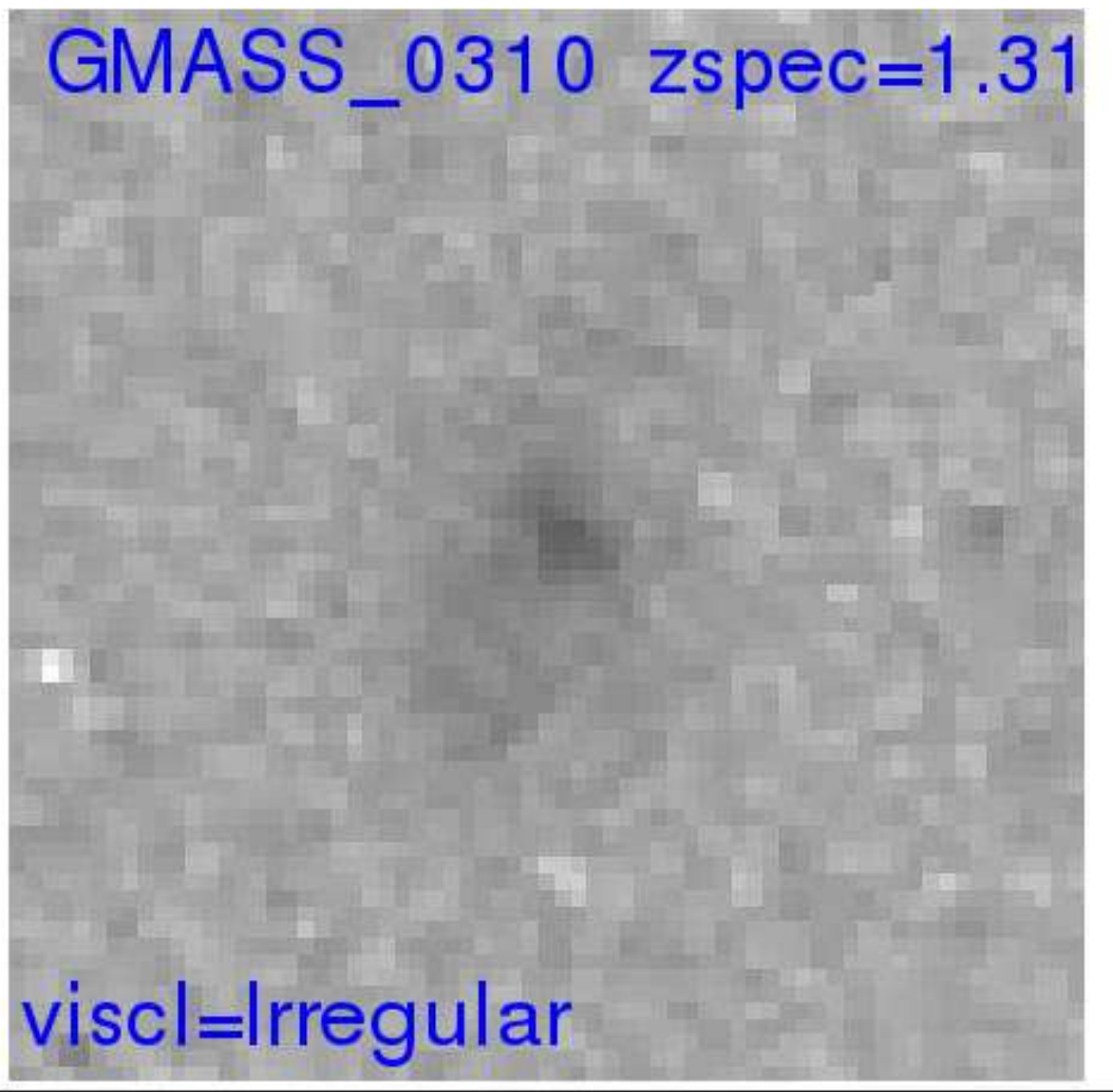}			     
\includegraphics[trim=100 40 75 390, clip=true, width=30mm]{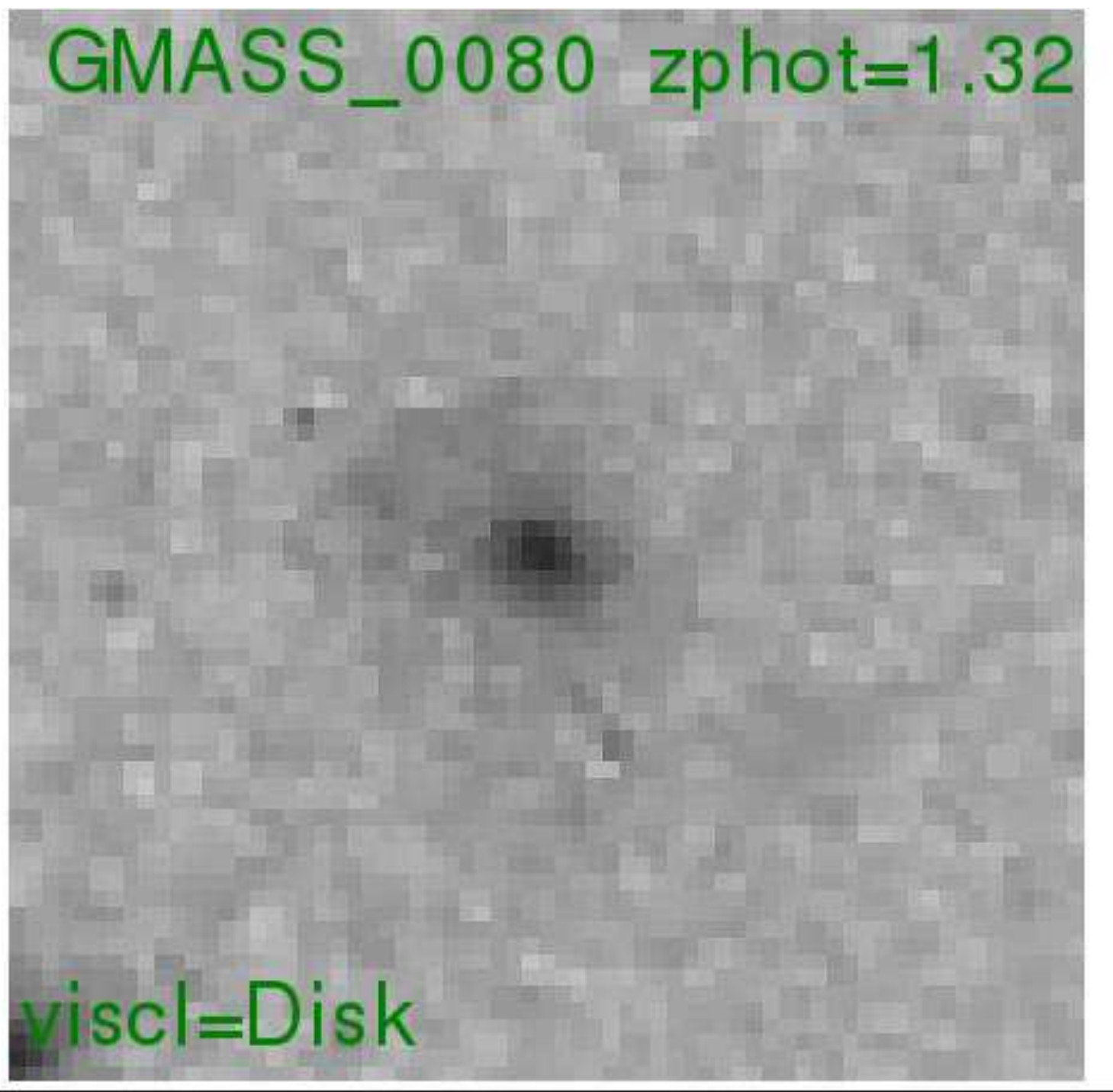}			     
\includegraphics[trim=100 40 75 390, clip=true, width=30mm]{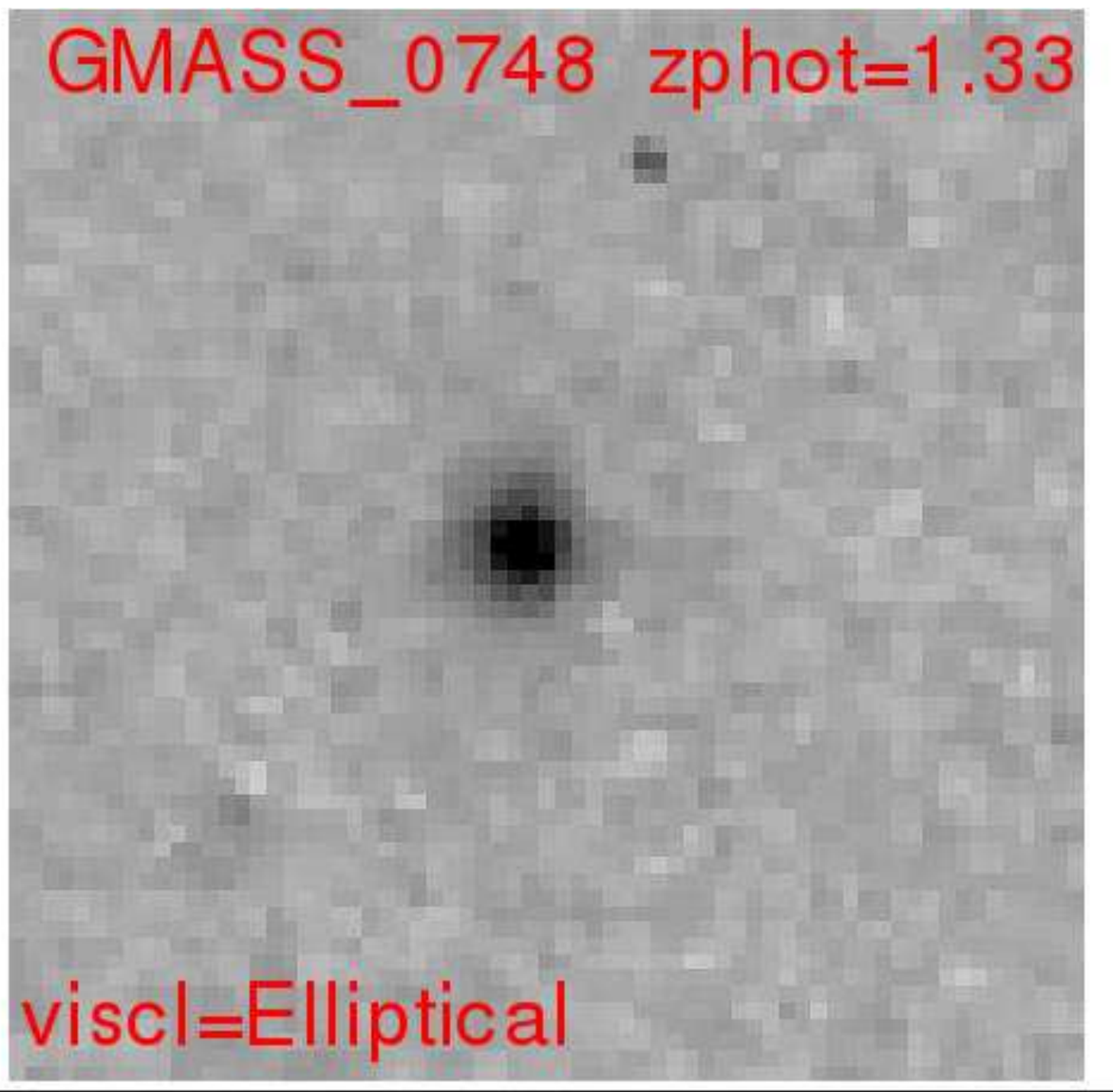}			     
\includegraphics[trim=100 40 75 390, clip=true, width=30mm]{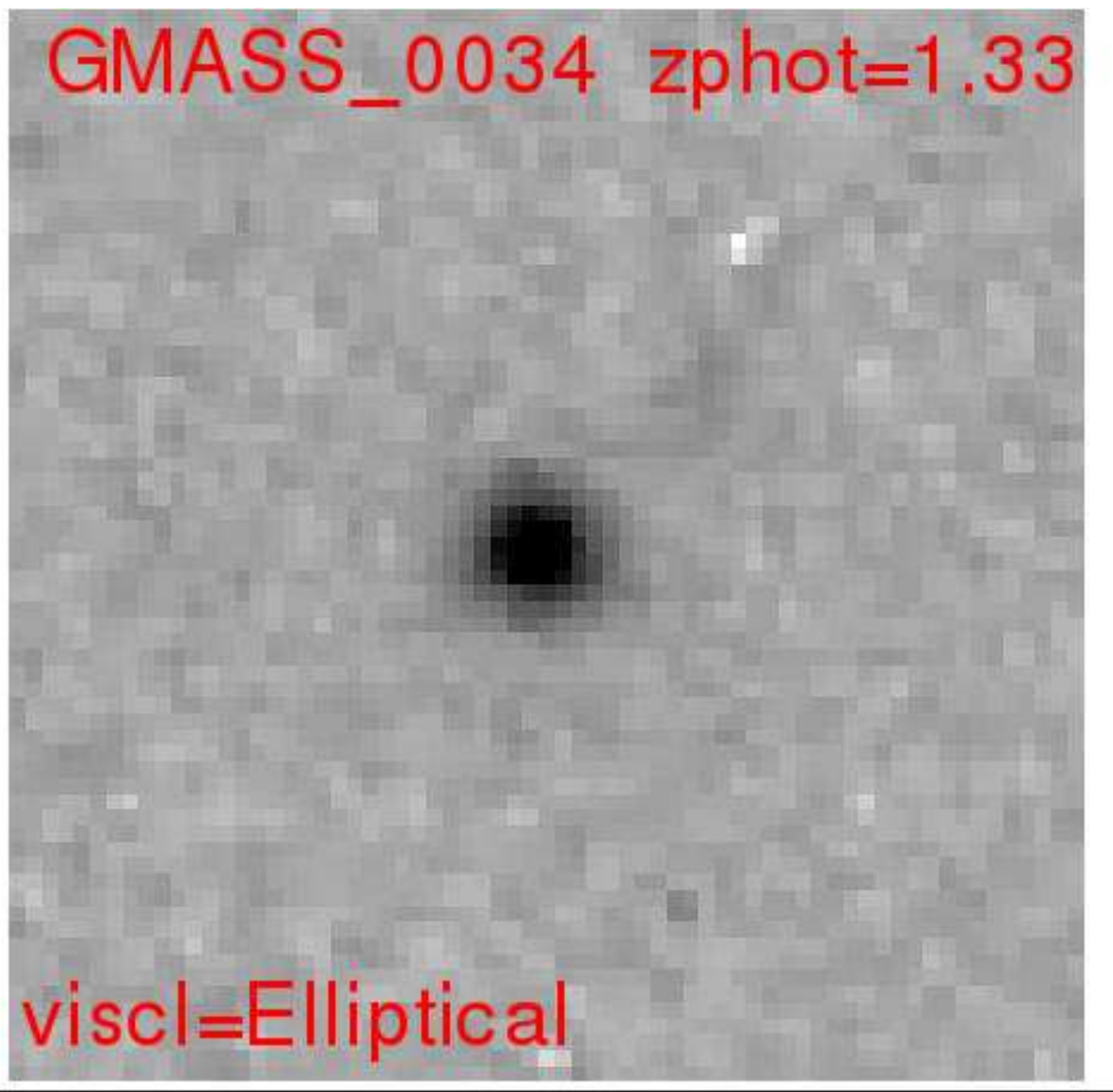}			     
\includegraphics[trim=100 40 75 390, clip=true, width=30mm]{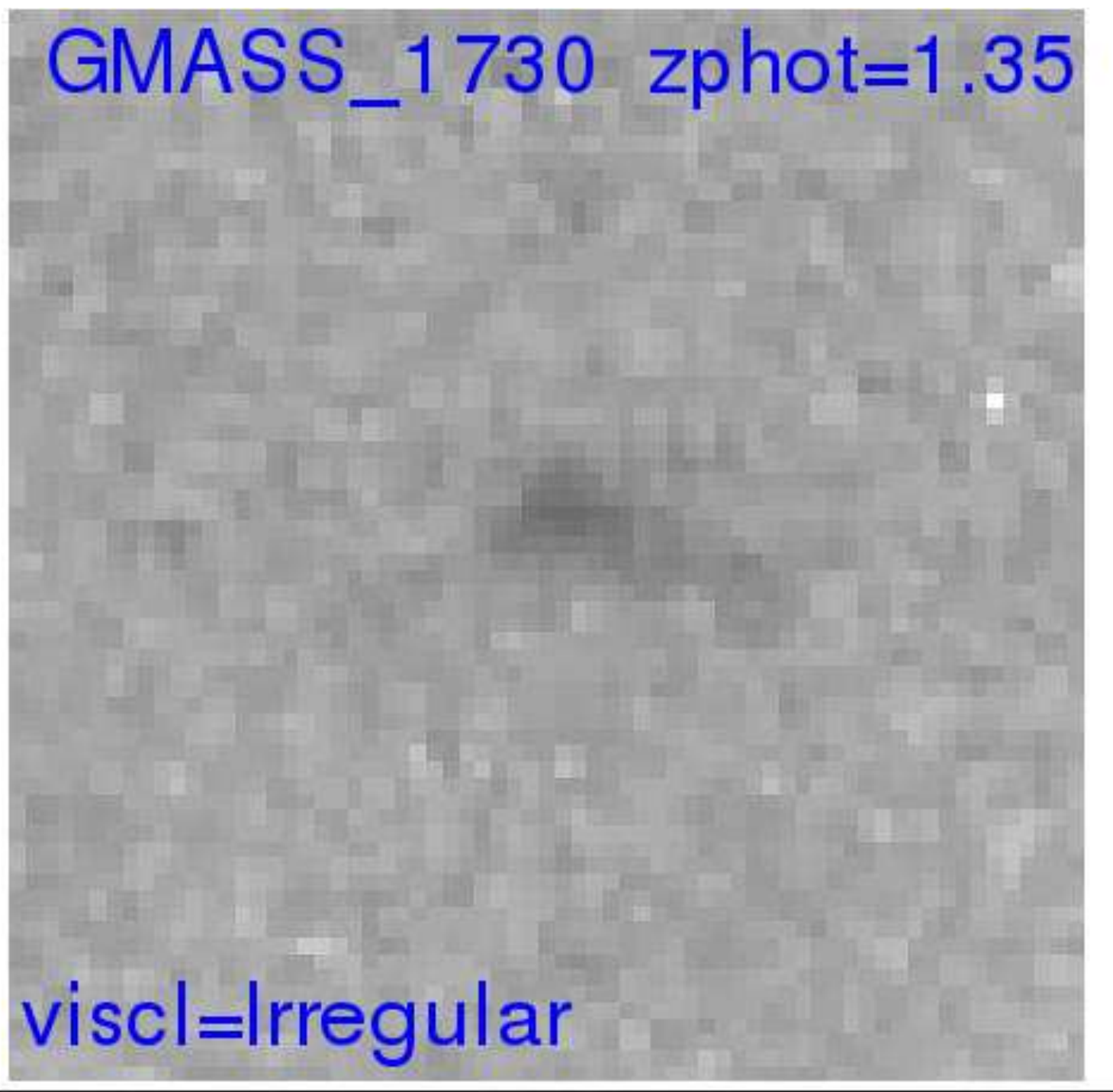}			     

\includegraphics[trim=100 40 75 390, clip=true, width=30mm]{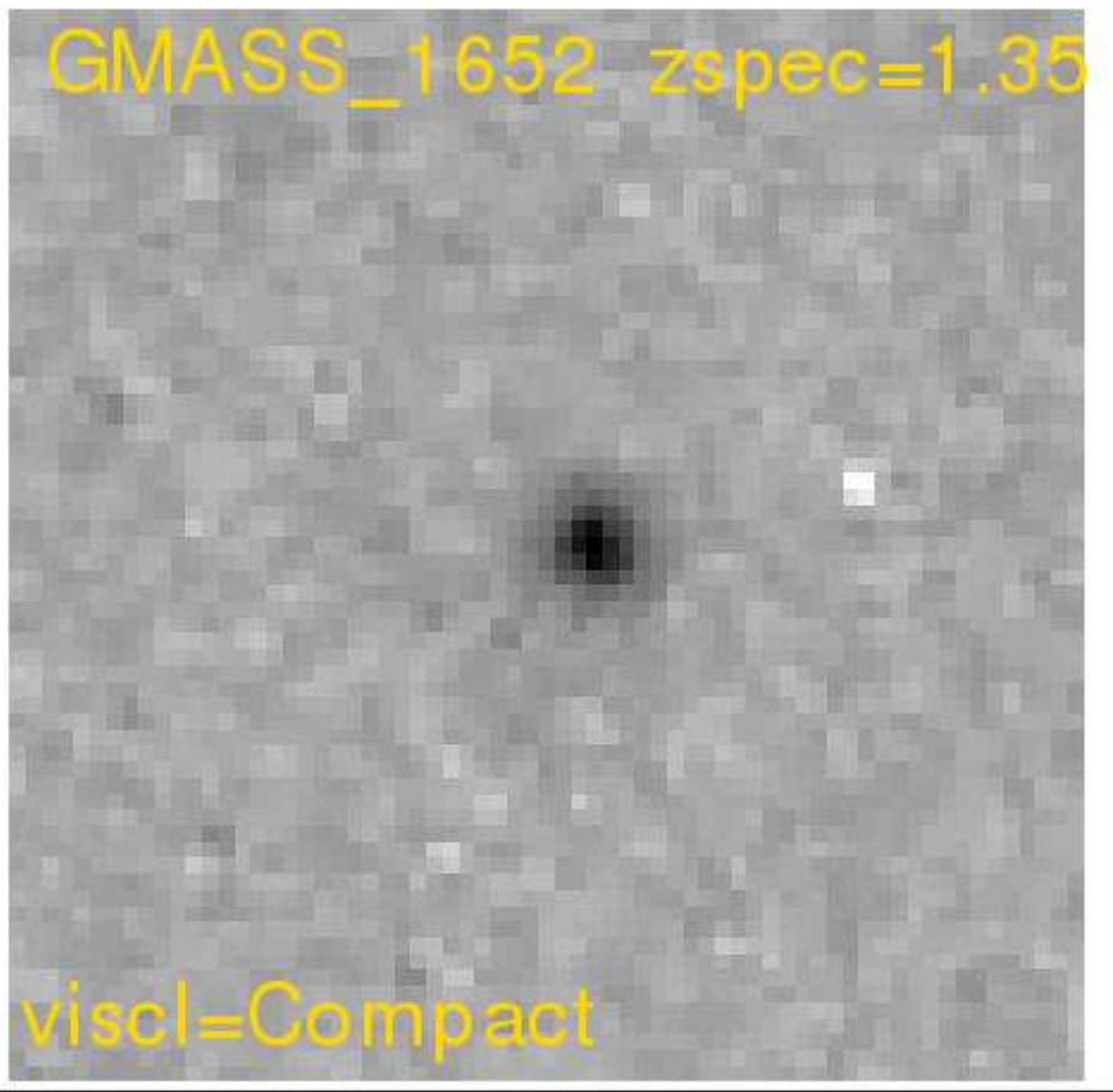}		     
\includegraphics[trim=100 40 75 390, clip=true, width=30mm]{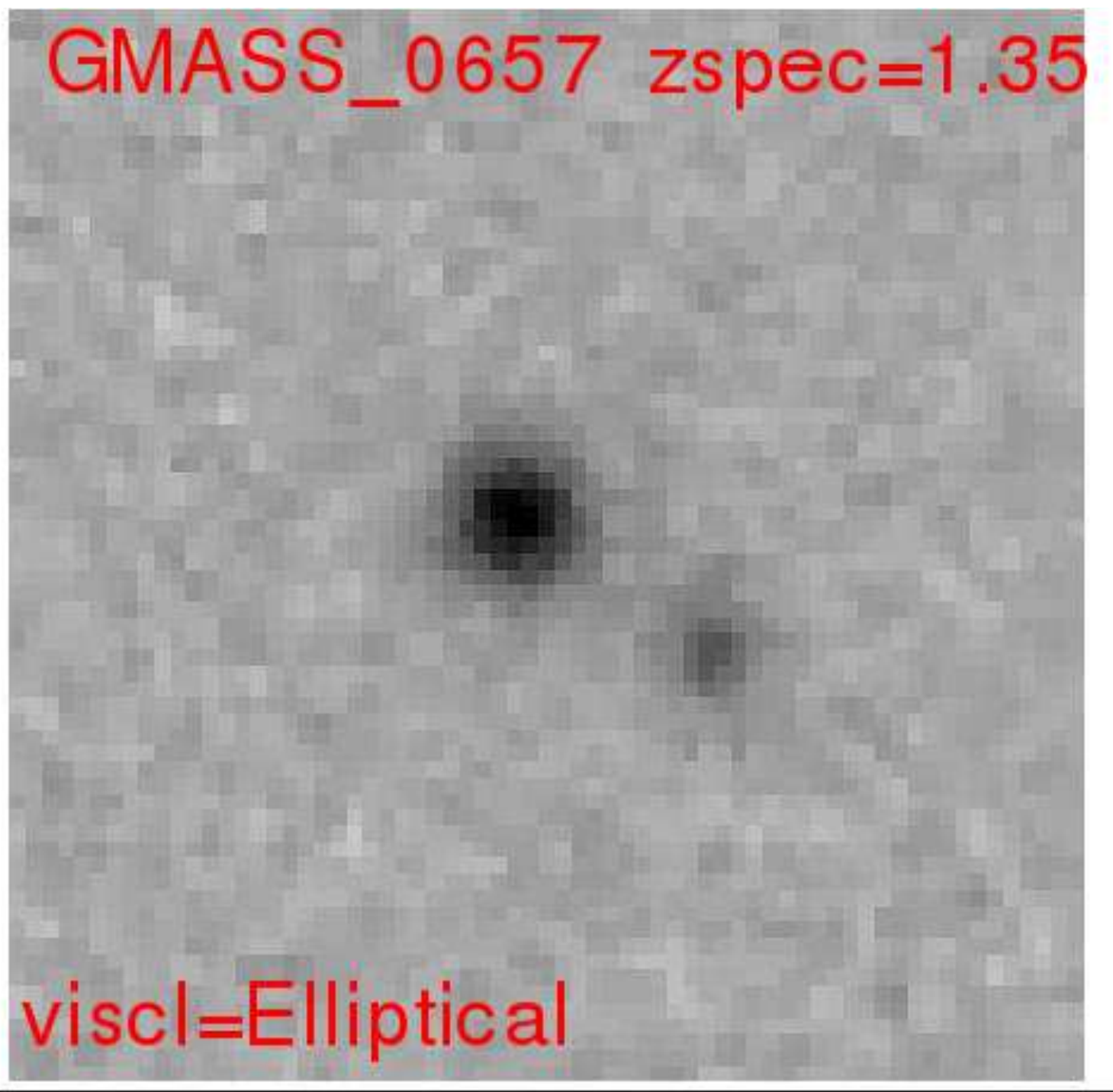}			     
\includegraphics[trim=100 40 75 390, clip=true, width=30mm]{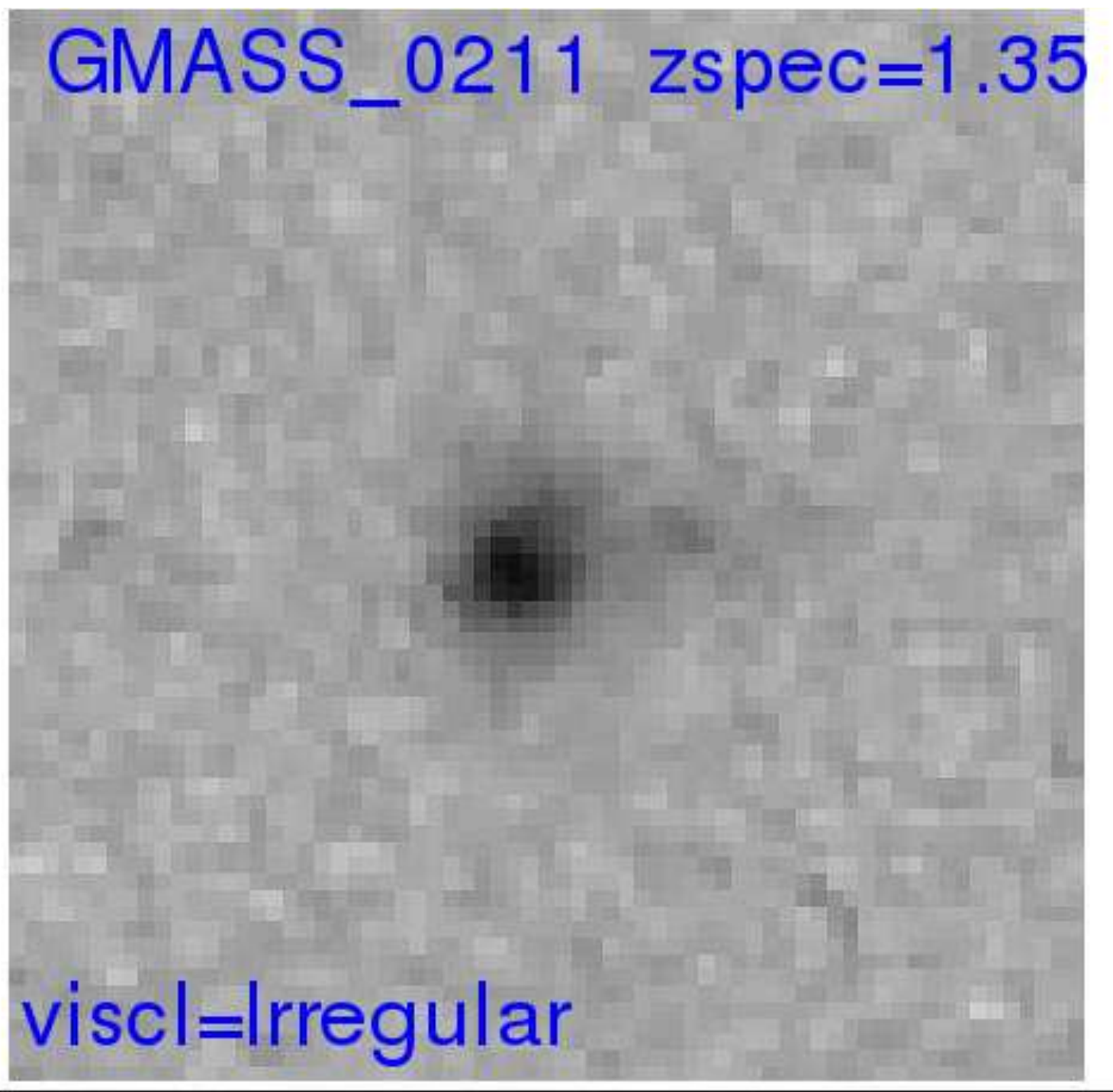}		     
\includegraphics[trim=100 40 75 390, clip=true, width=30mm]{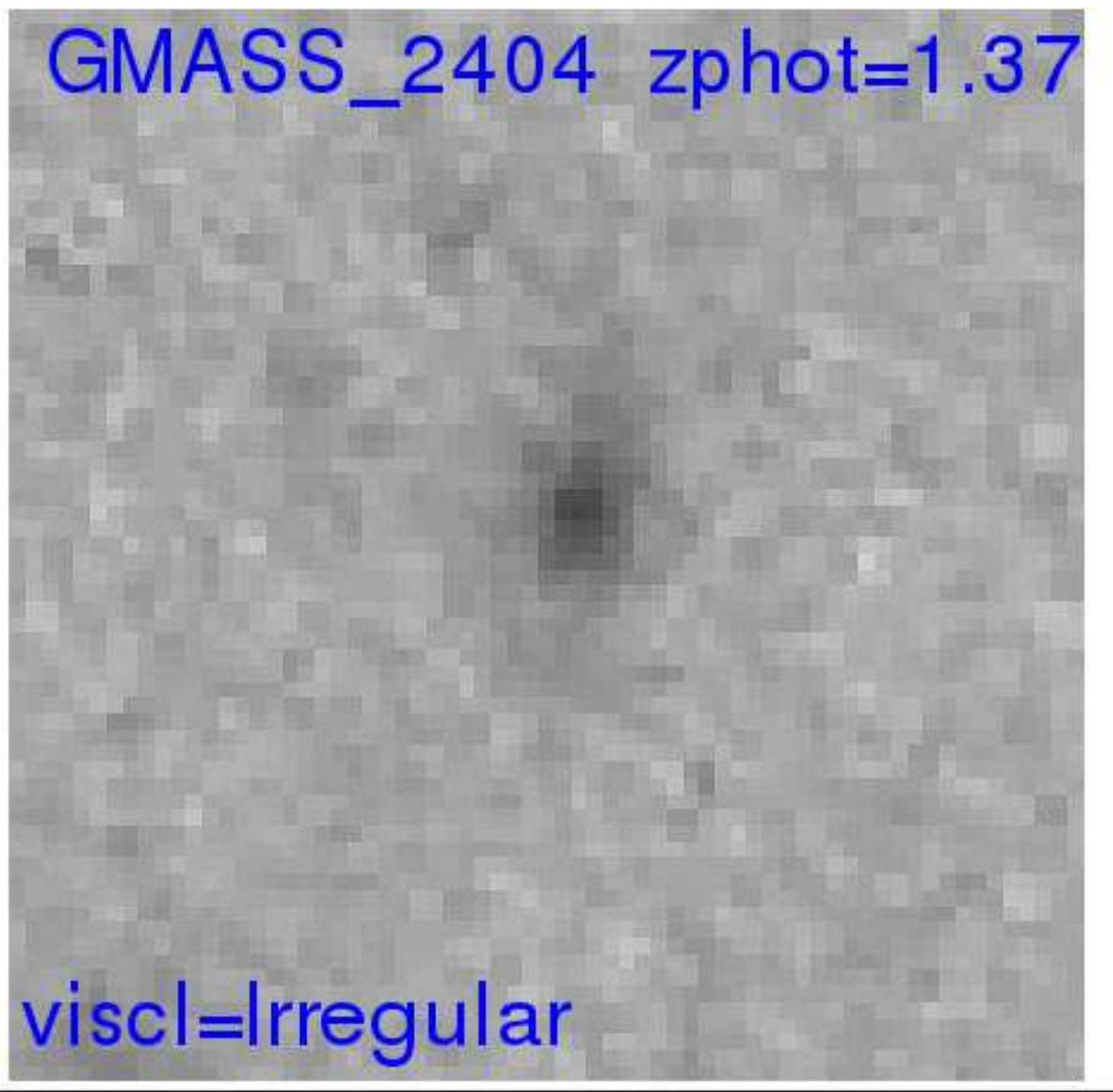}			     
\includegraphics[trim=100 40 75 390, clip=true, width=30mm]{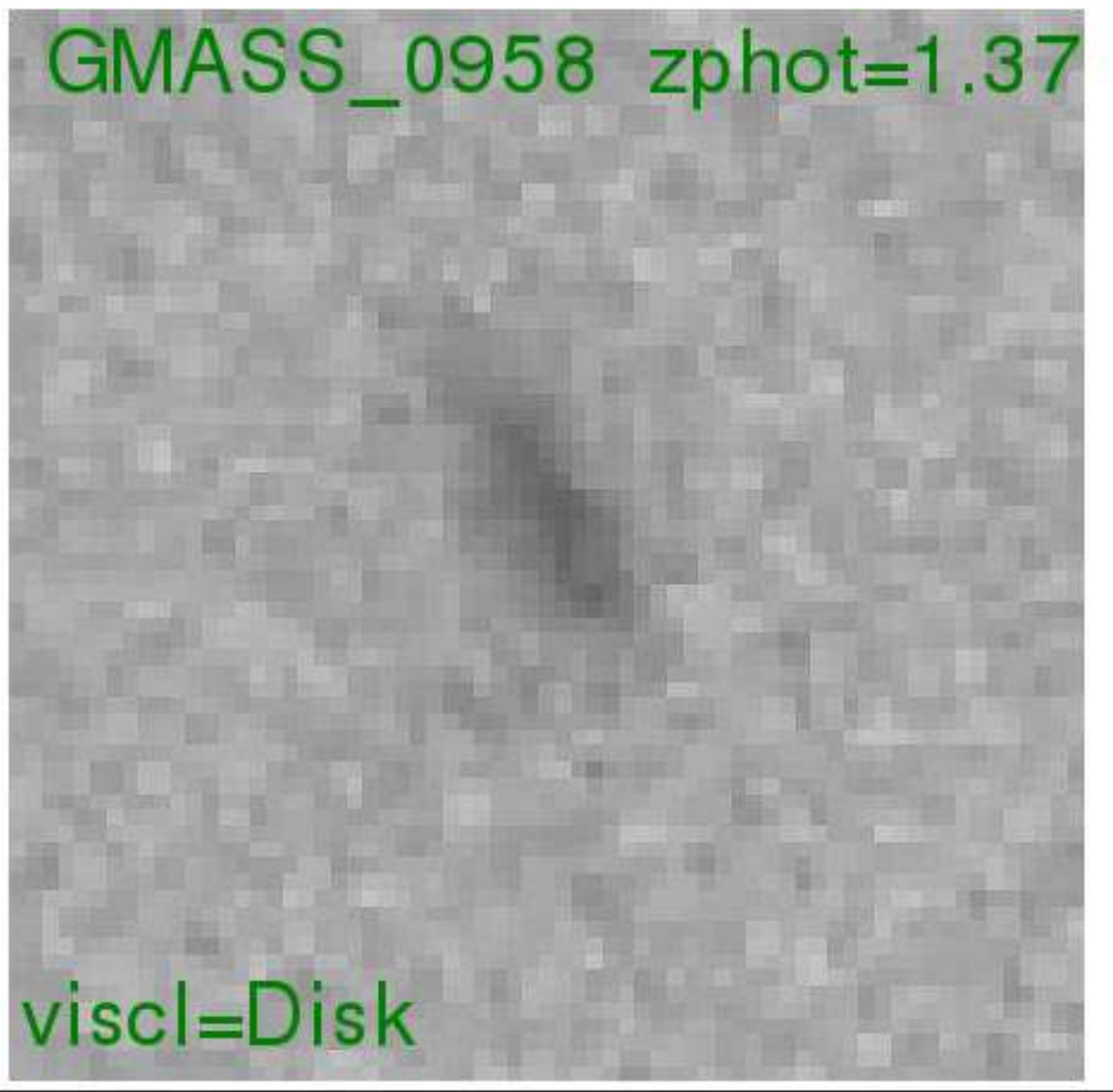}			     
\includegraphics[trim=100 40 75 390, clip=true, width=30mm]{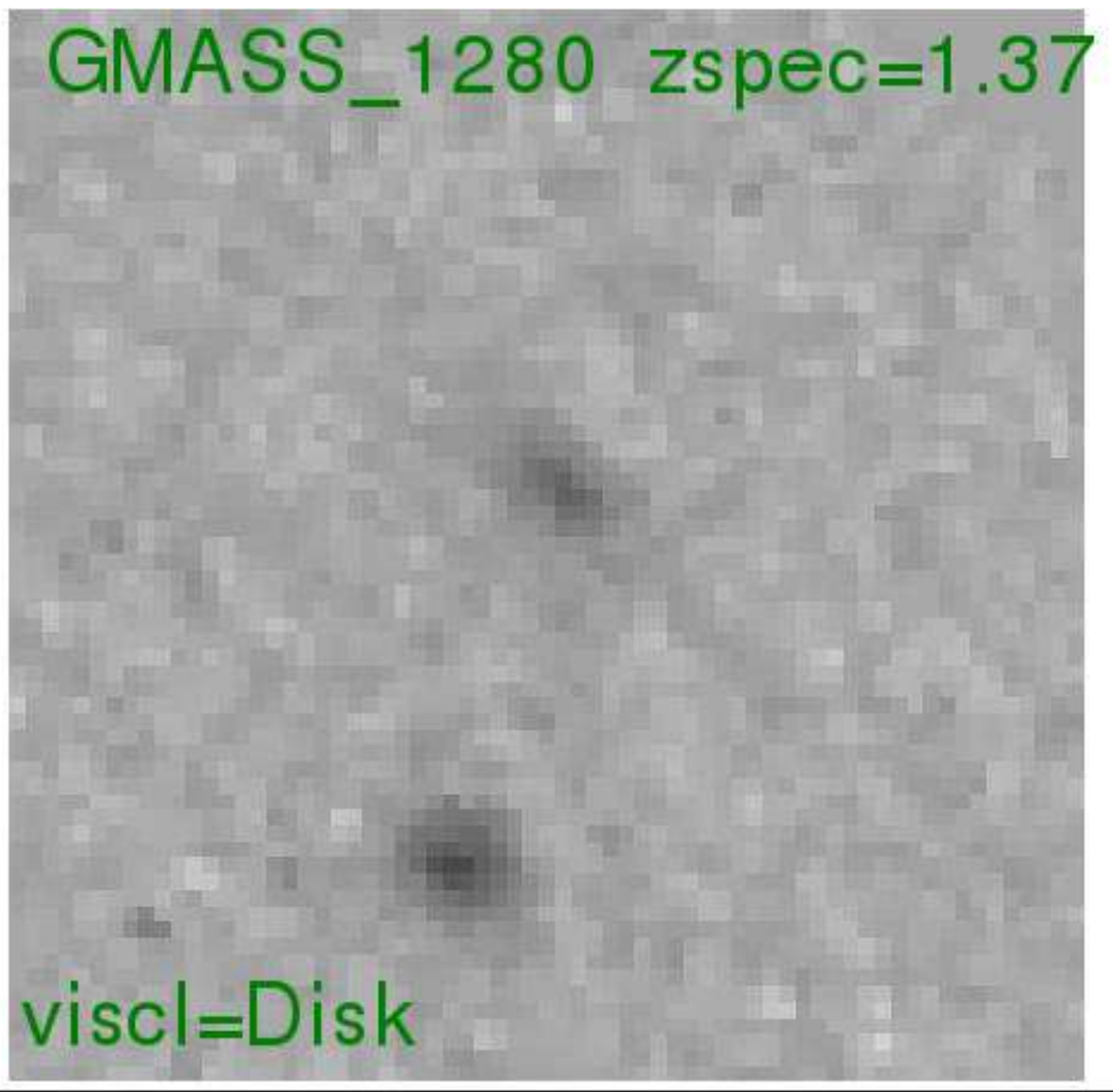}			     

\includegraphics[trim=100 40 75 390, clip=true, width=30mm]{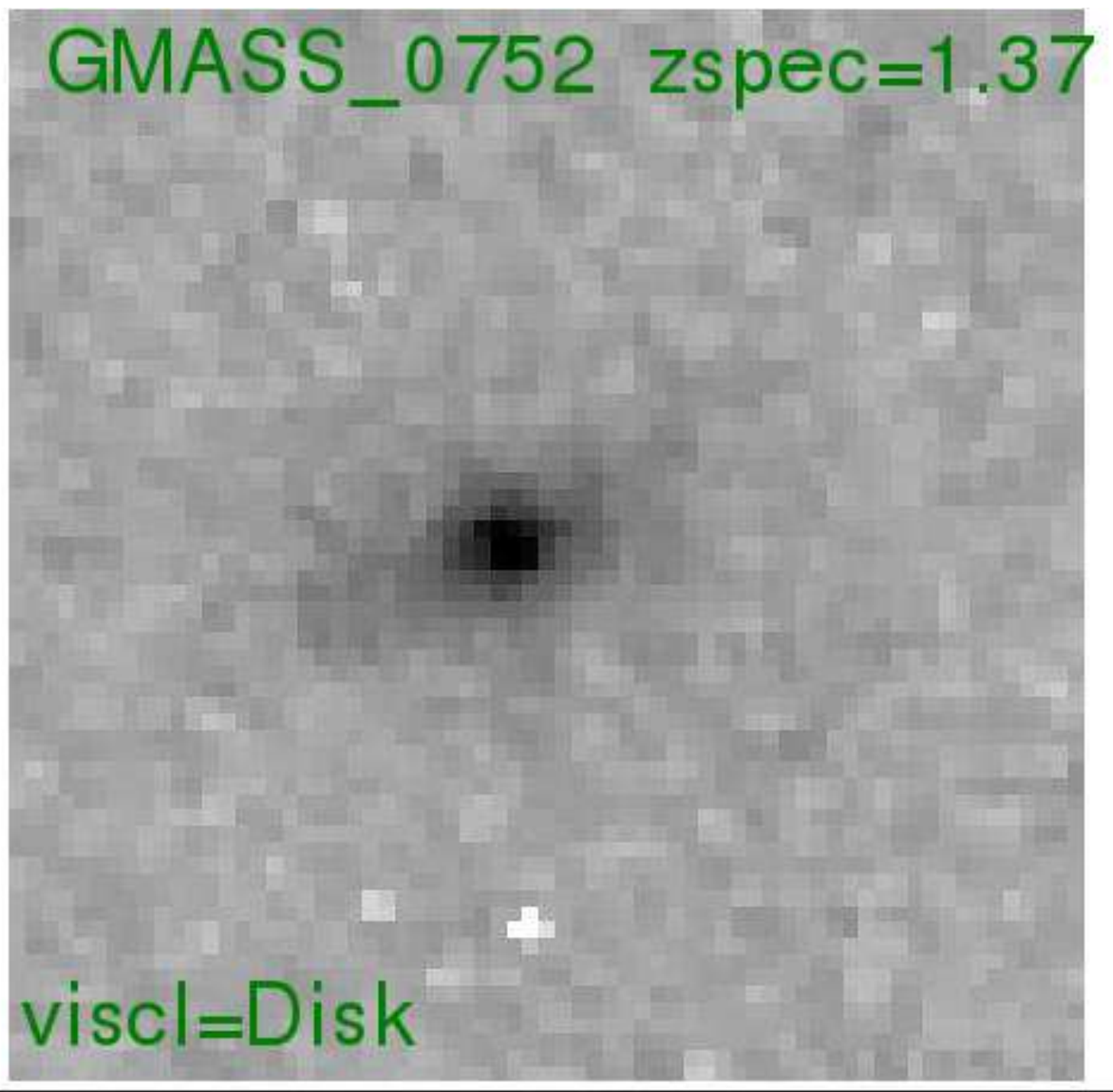}			     
\includegraphics[trim=100 40 75 390, clip=true, width=30mm]{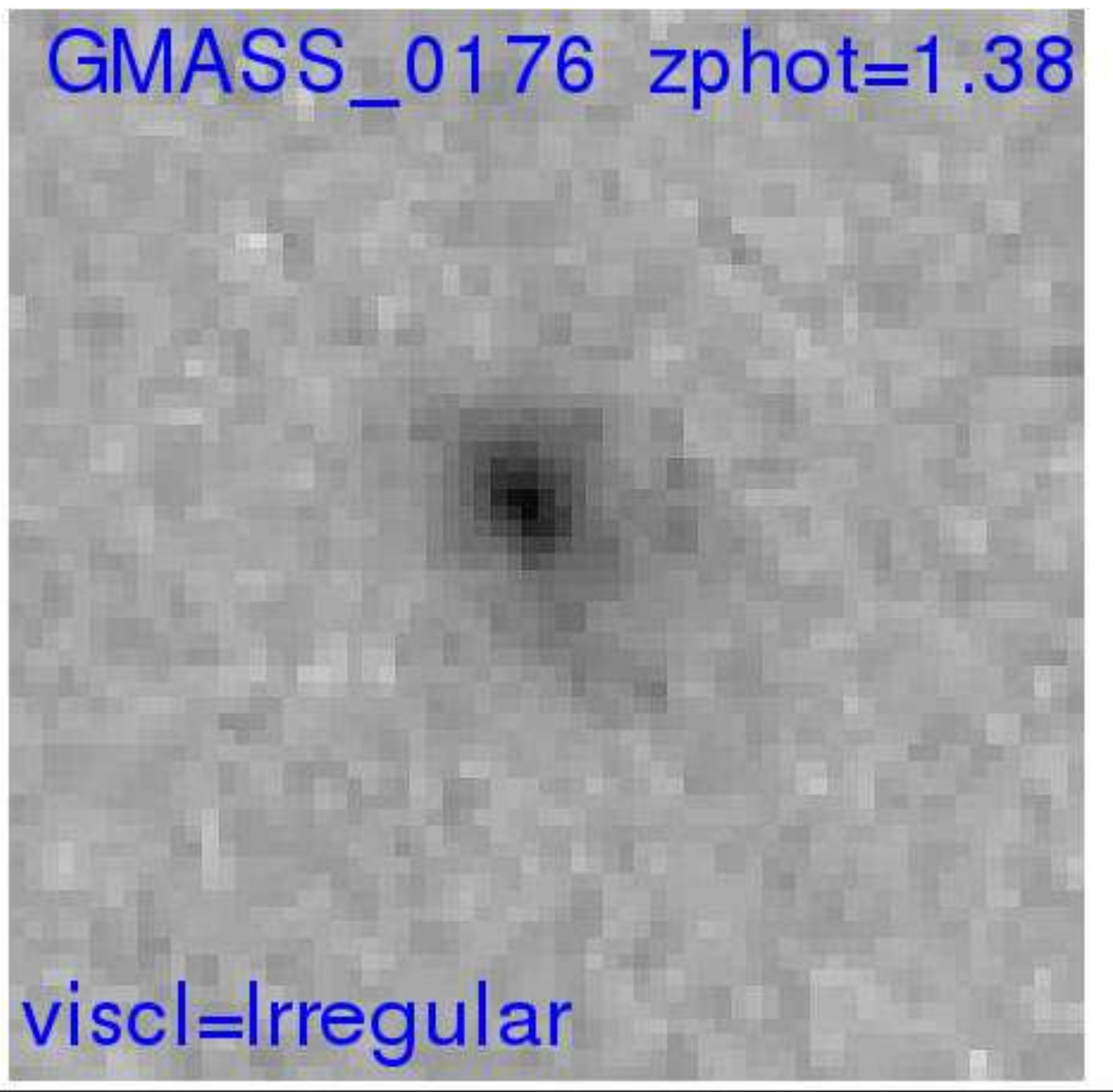}			     
\includegraphics[trim=100 40 75 390, clip=true, width=30mm]{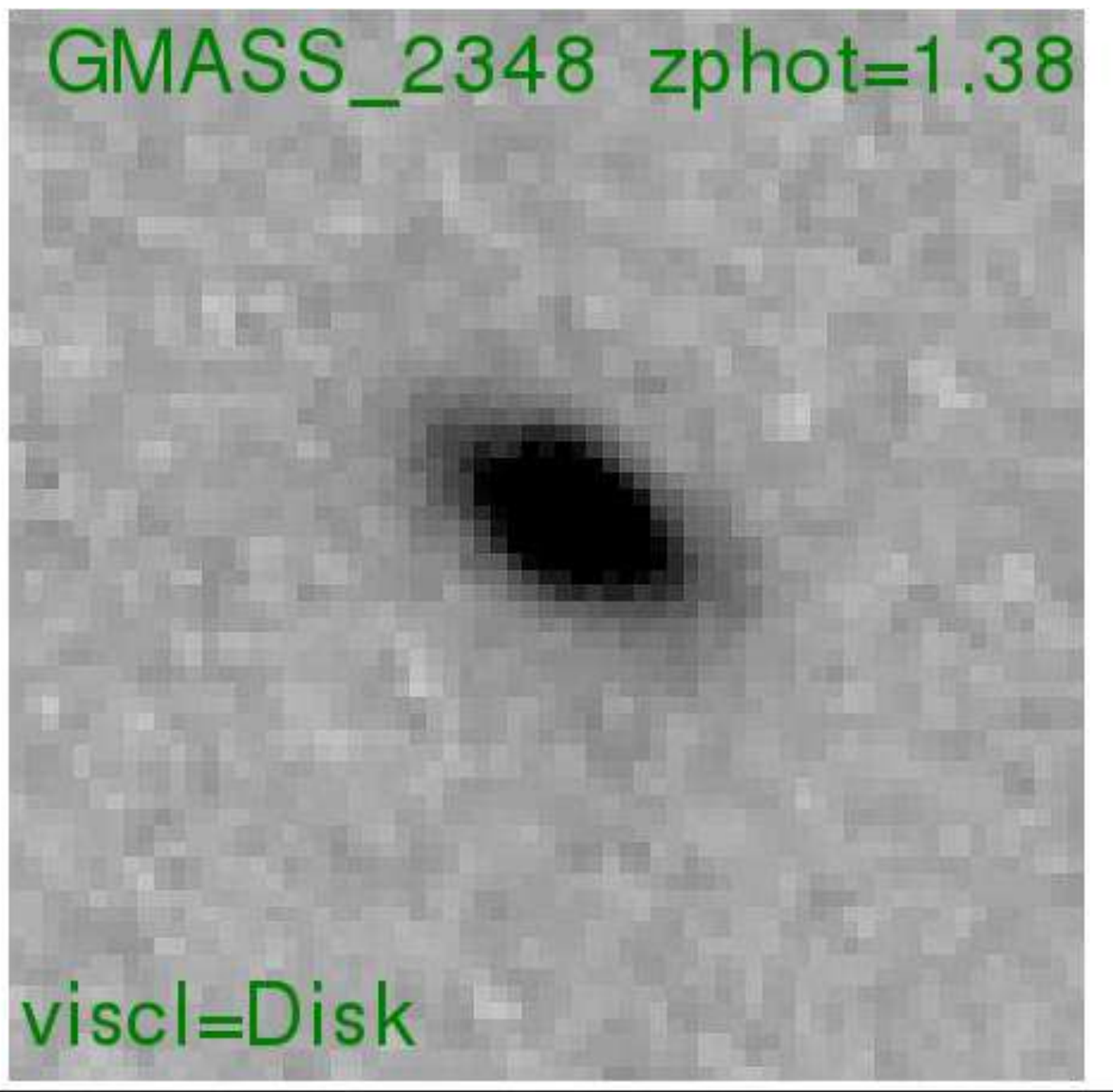}			     
\includegraphics[trim=100 40 75 390, clip=true, width=30mm]{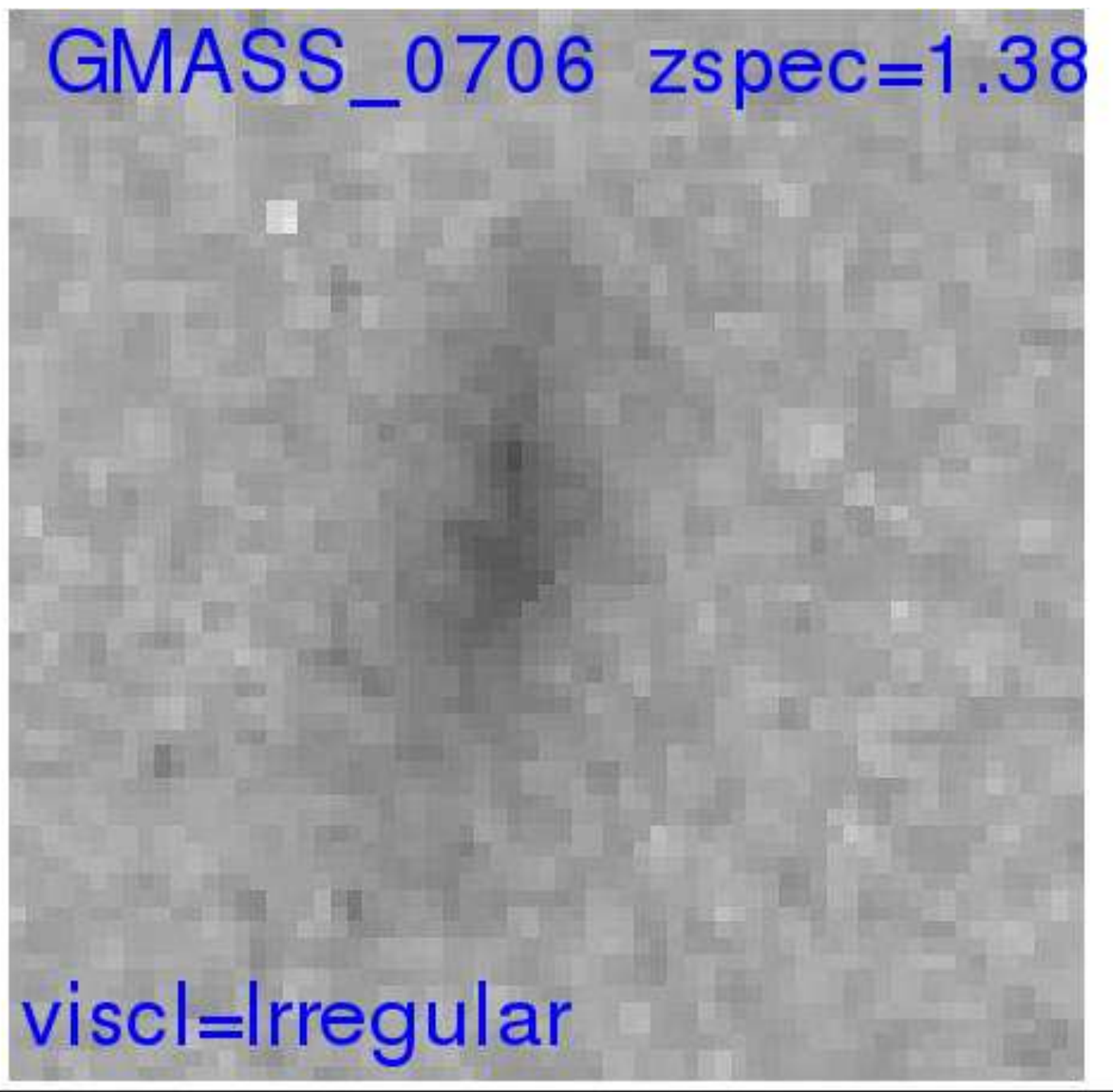}			     
\includegraphics[trim=100 40 75 390, clip=true, width=30mm]{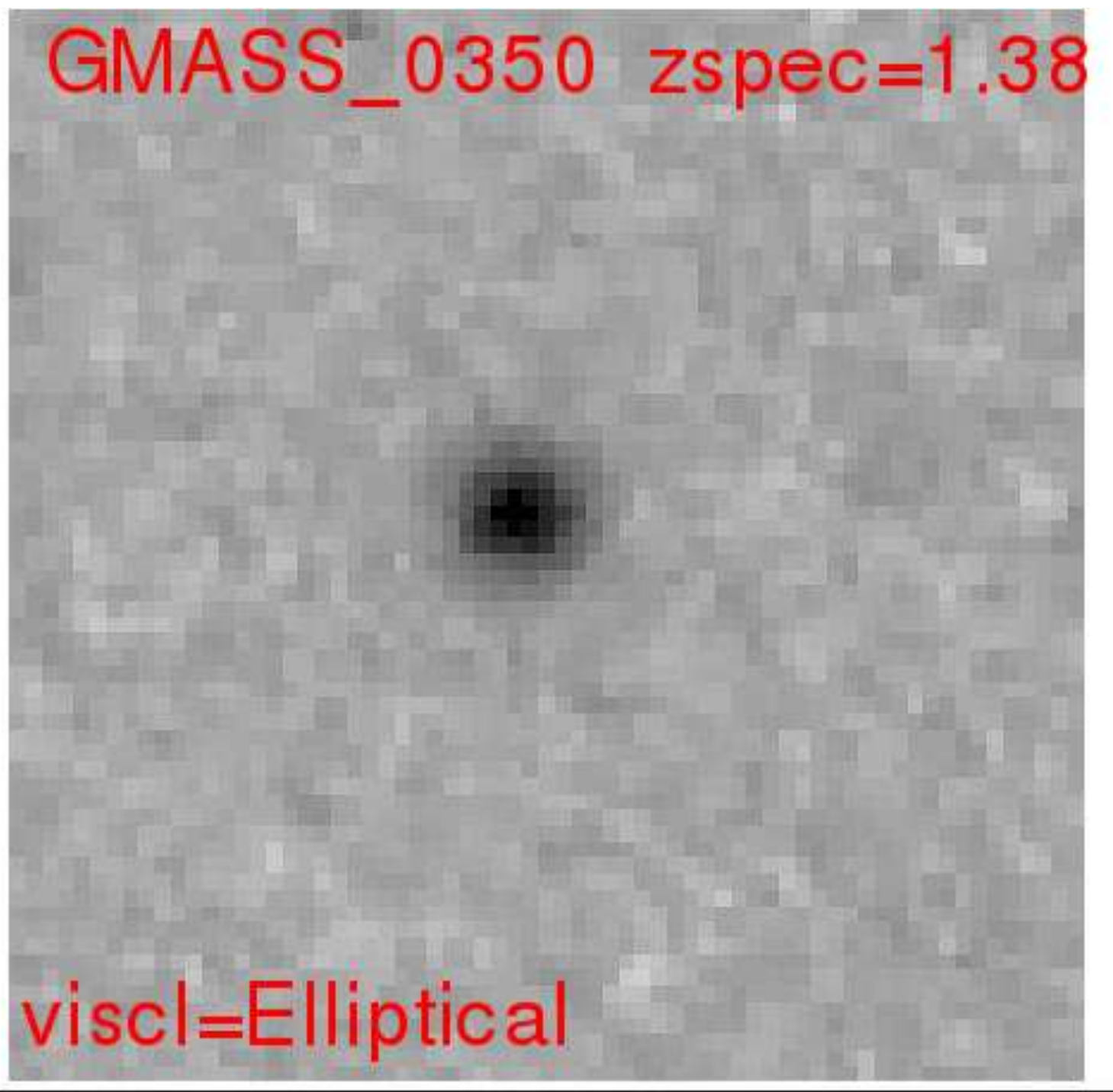}			     
\includegraphics[trim=100 40 75 390, clip=true, width=30mm]{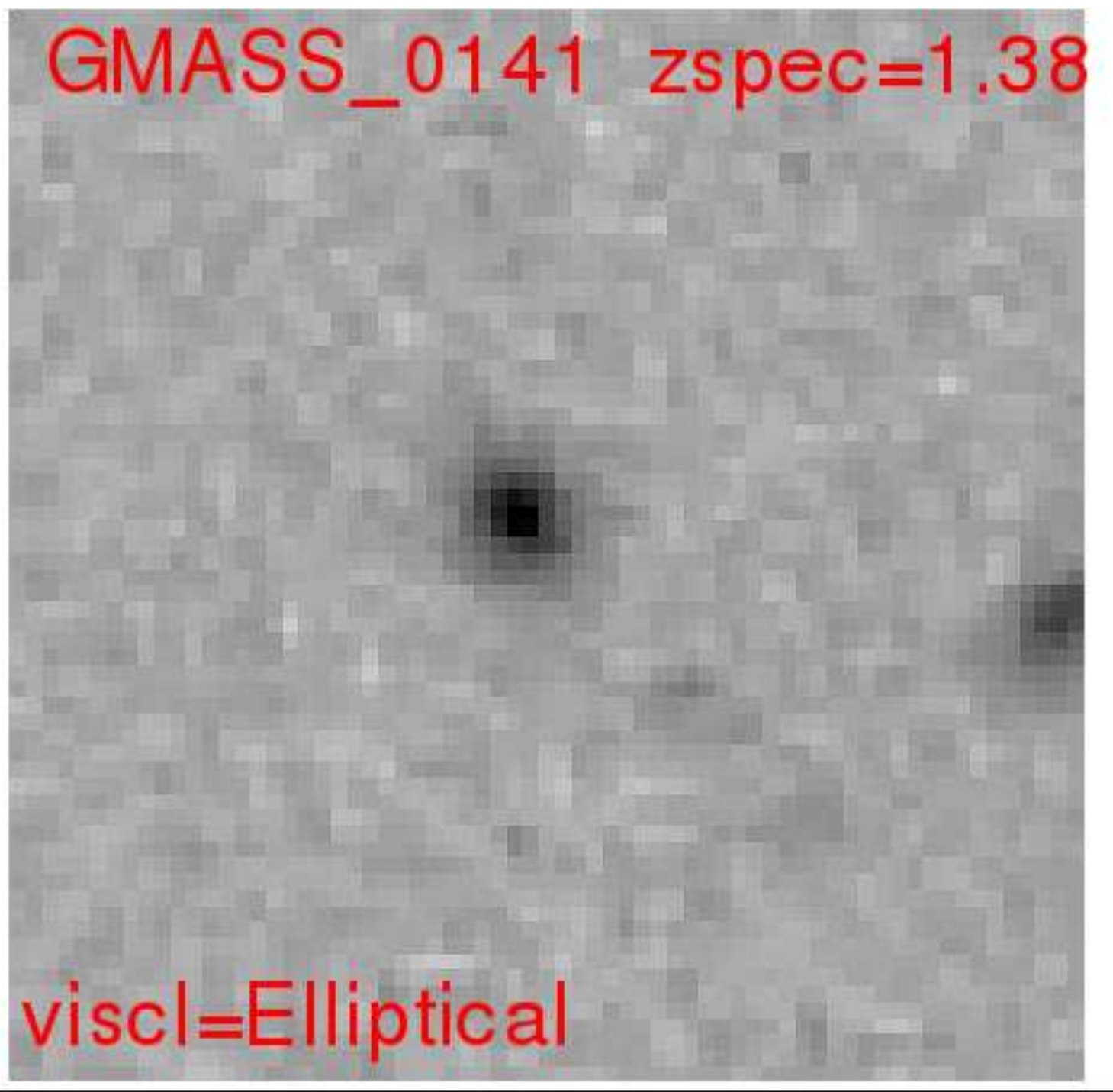}			     

\includegraphics[trim=100 40 75 390, clip=true, width=30mm]{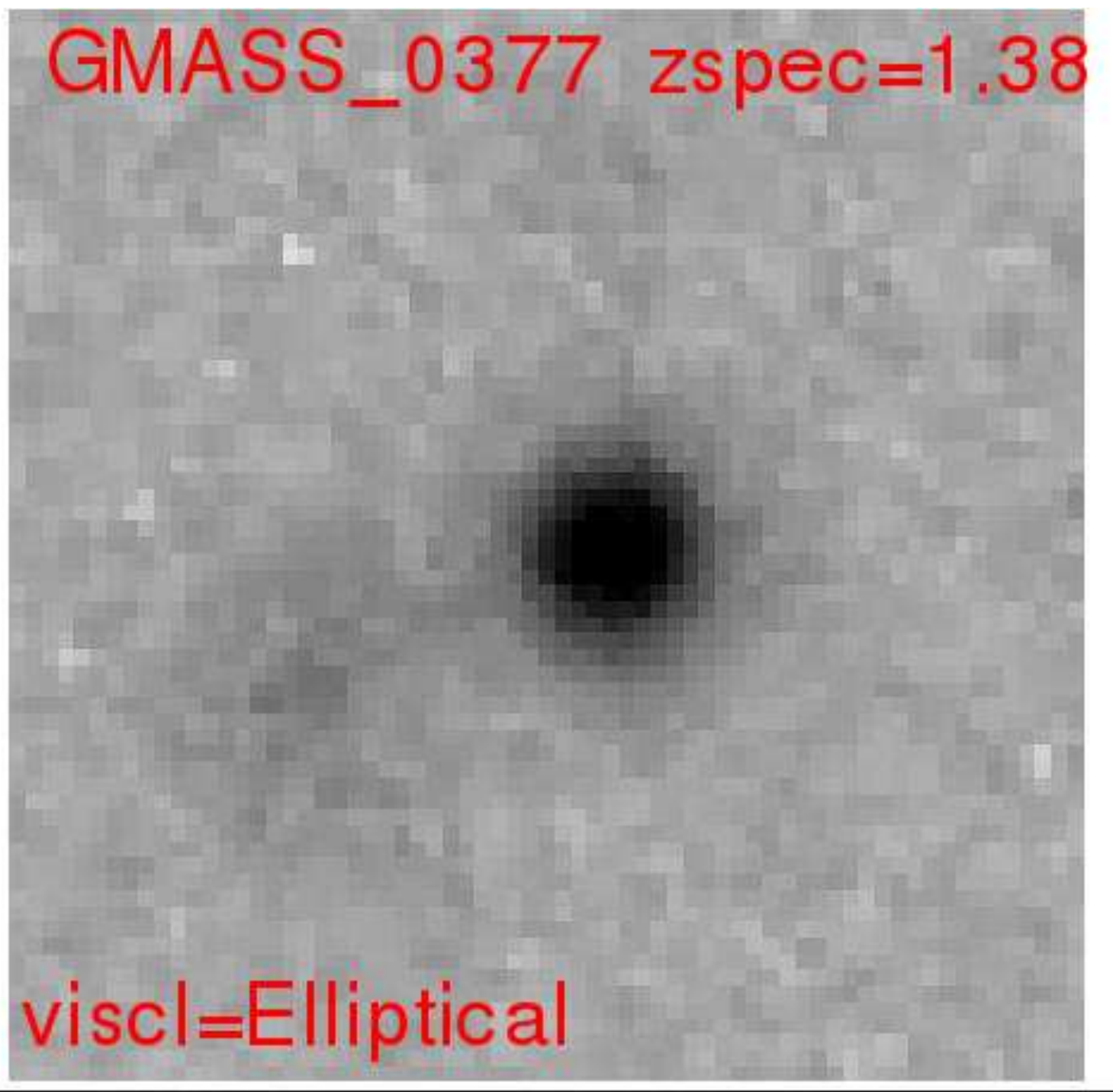}		     
\includegraphics[trim=100 40 75 390, clip=true, width=30mm]{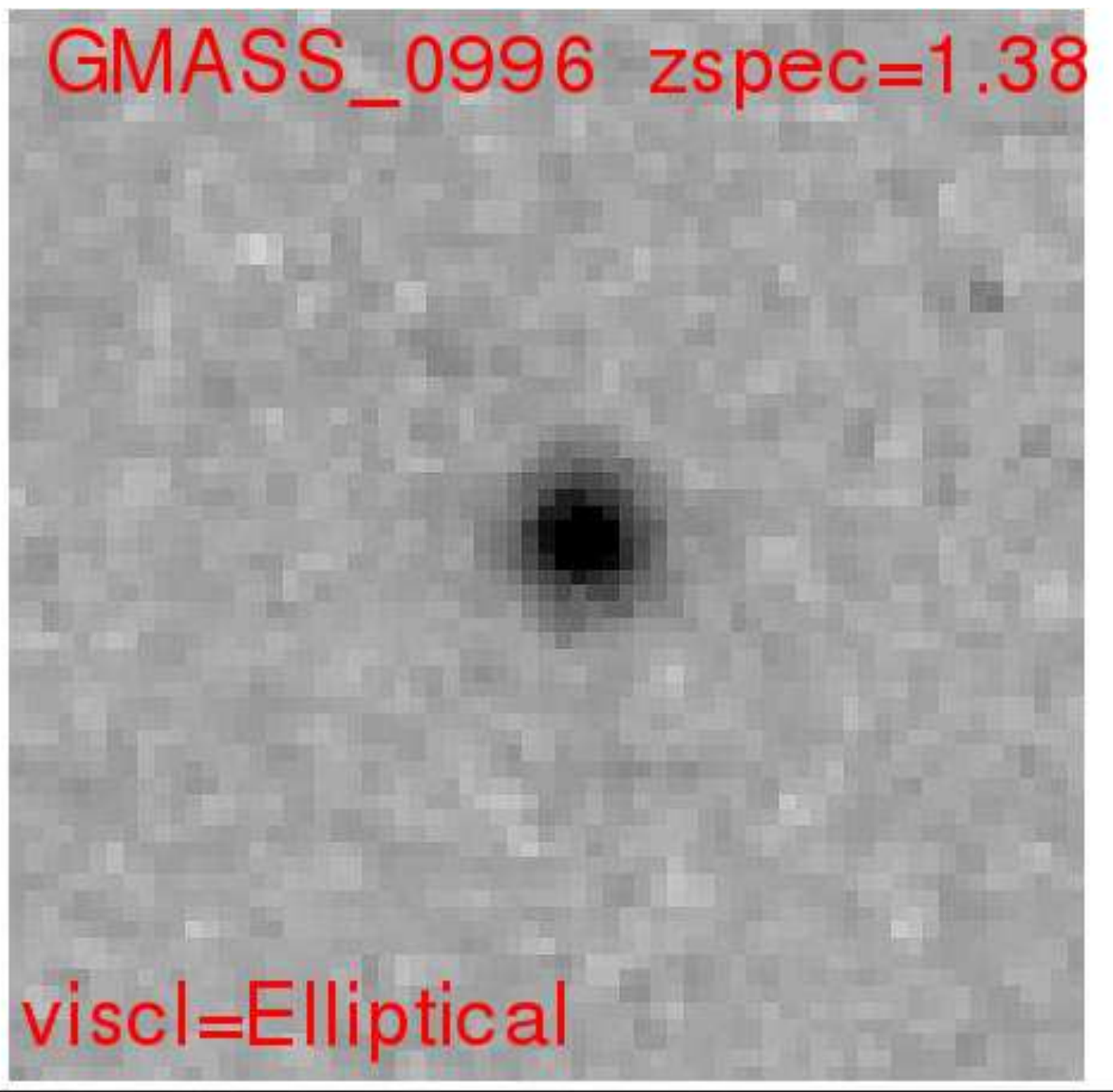}			     
\includegraphics[trim=100 40 75 390, clip=true, width=30mm]{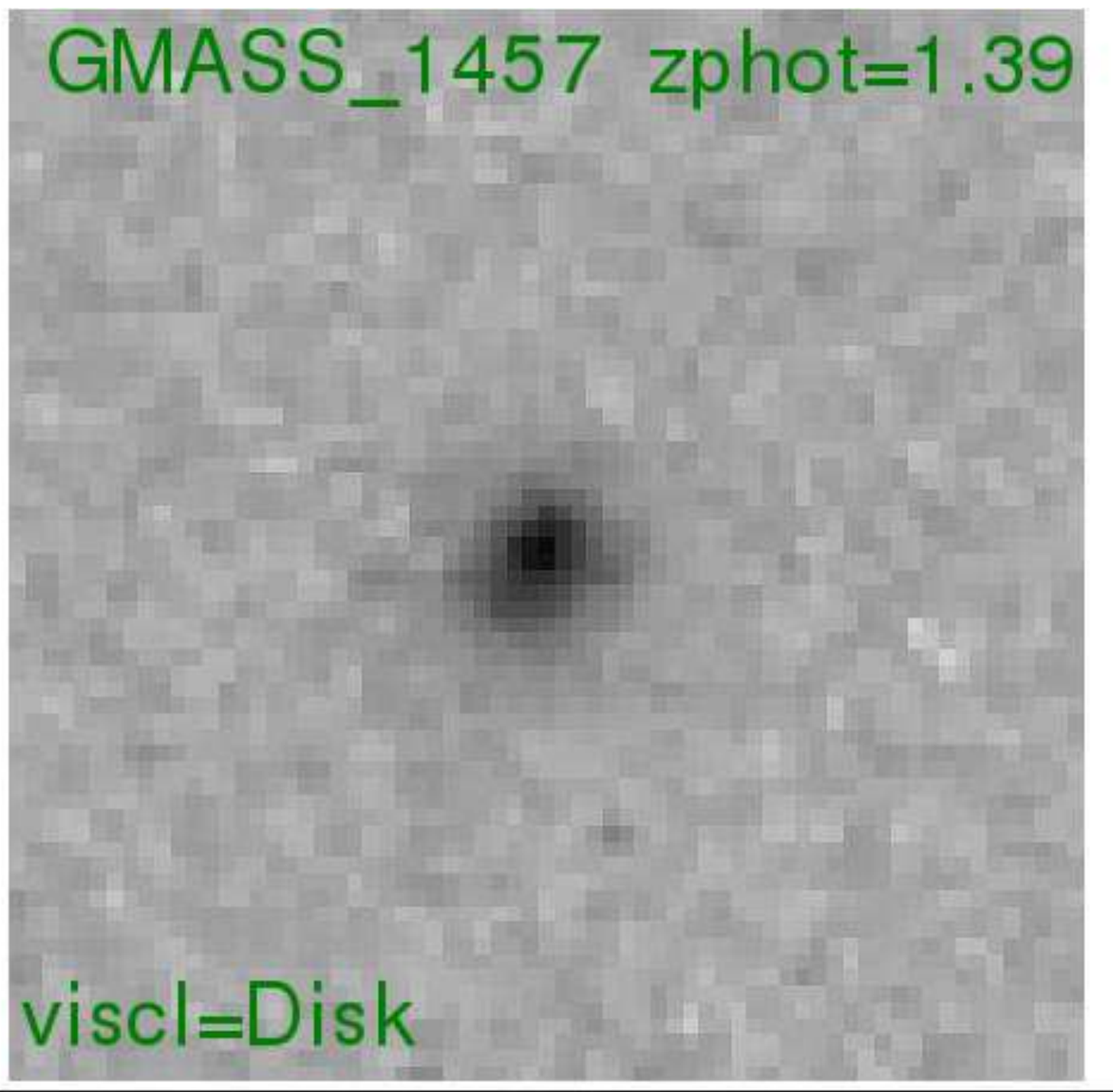}			     
\includegraphics[trim=100 40 75 390, clip=true, width=30mm]{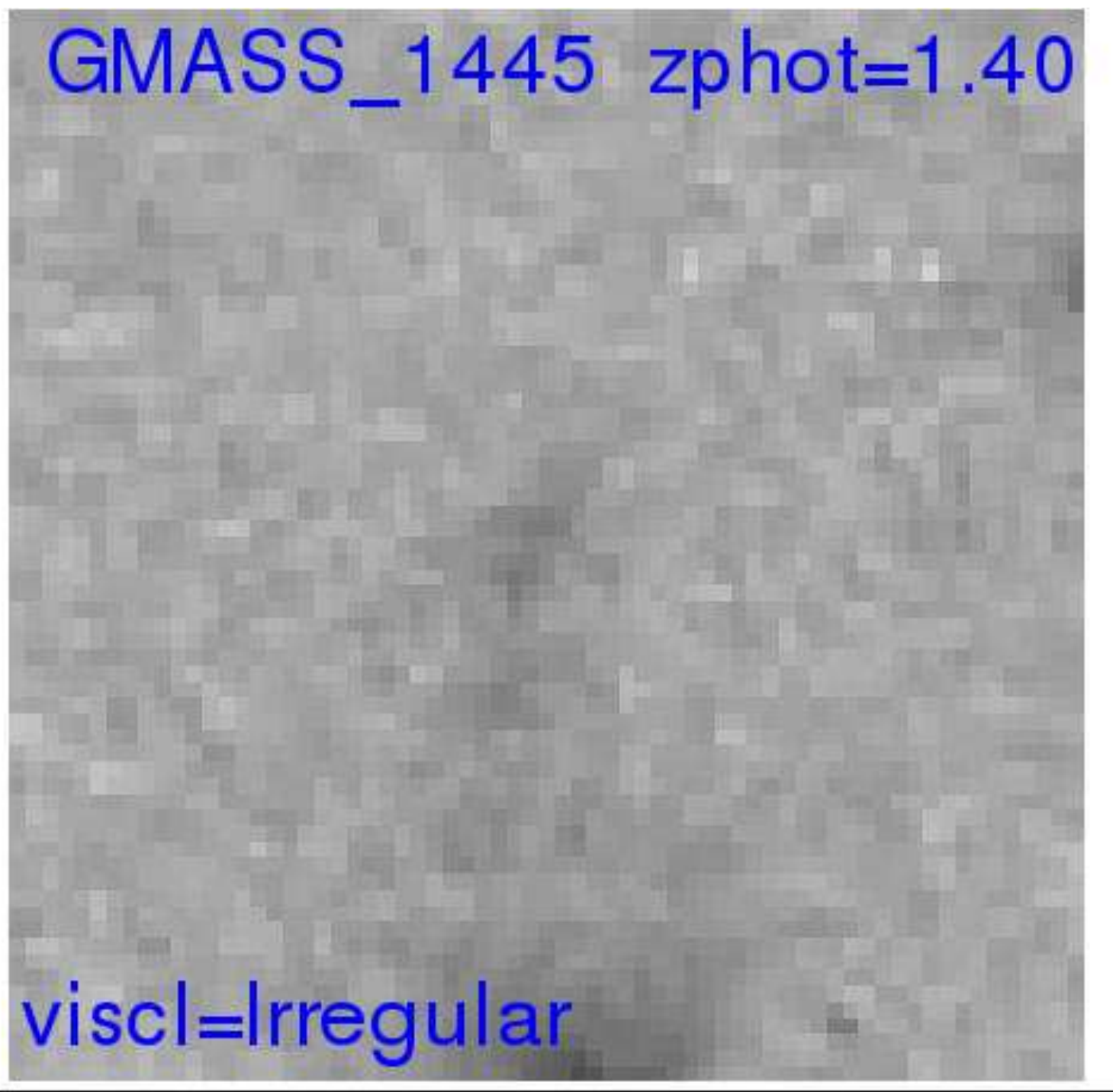}			     
\includegraphics[trim=100 40 75 390, clip=true, width=30mm]{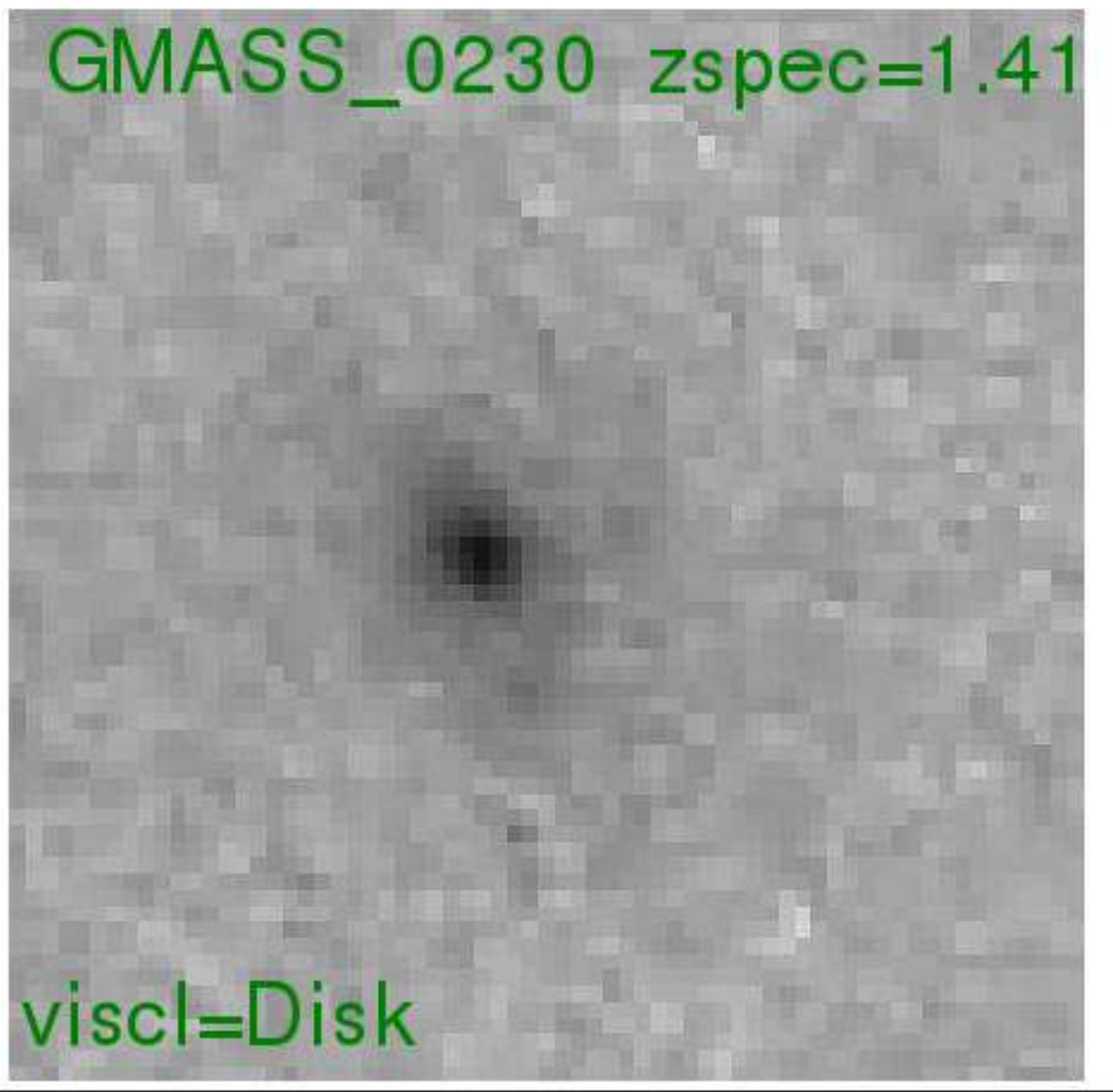}			     
\includegraphics[trim=100 40 75 390, clip=true, width=30mm]{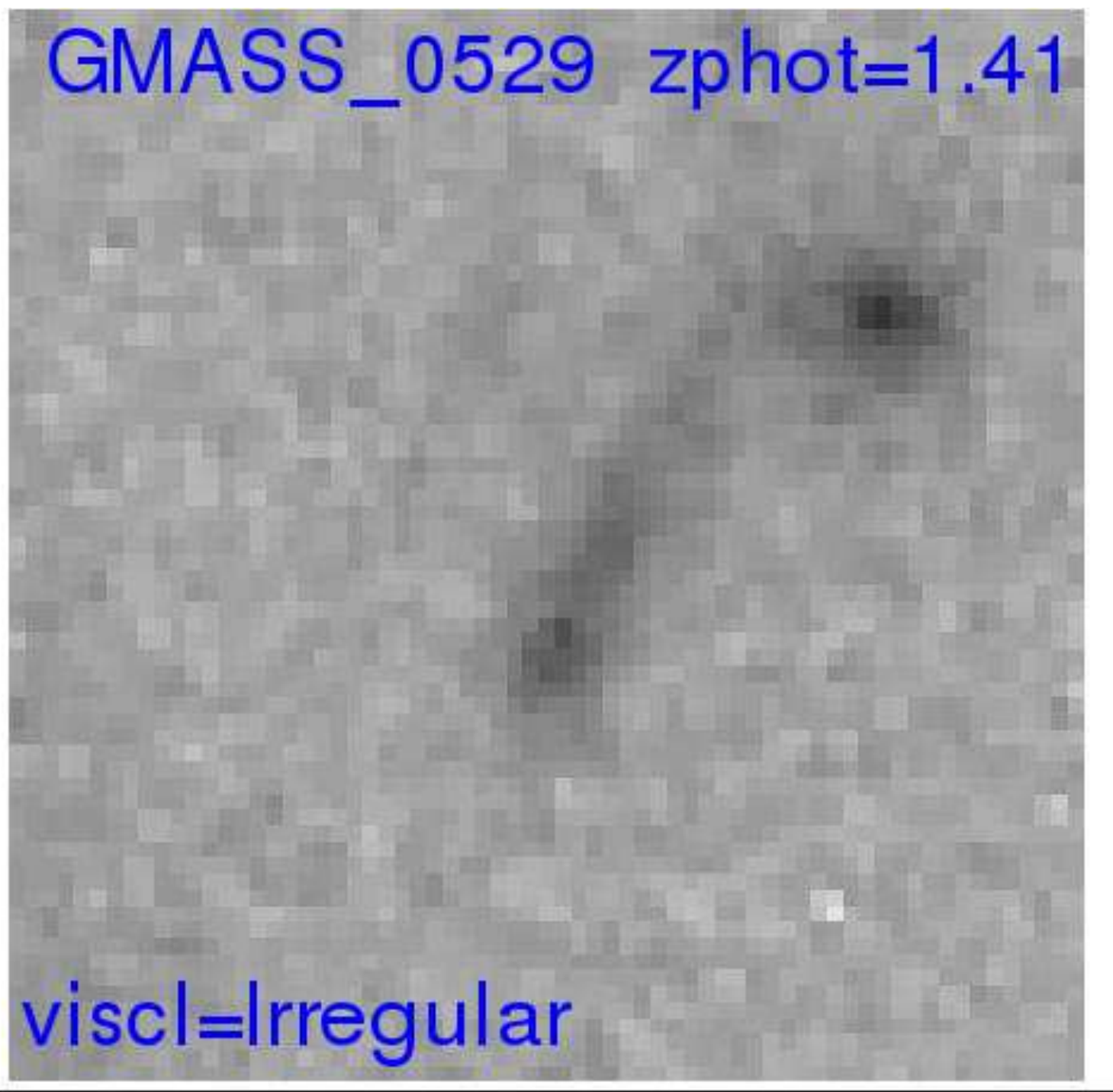}		     

\includegraphics[trim=100 40 75 390, clip=true, width=30mm]{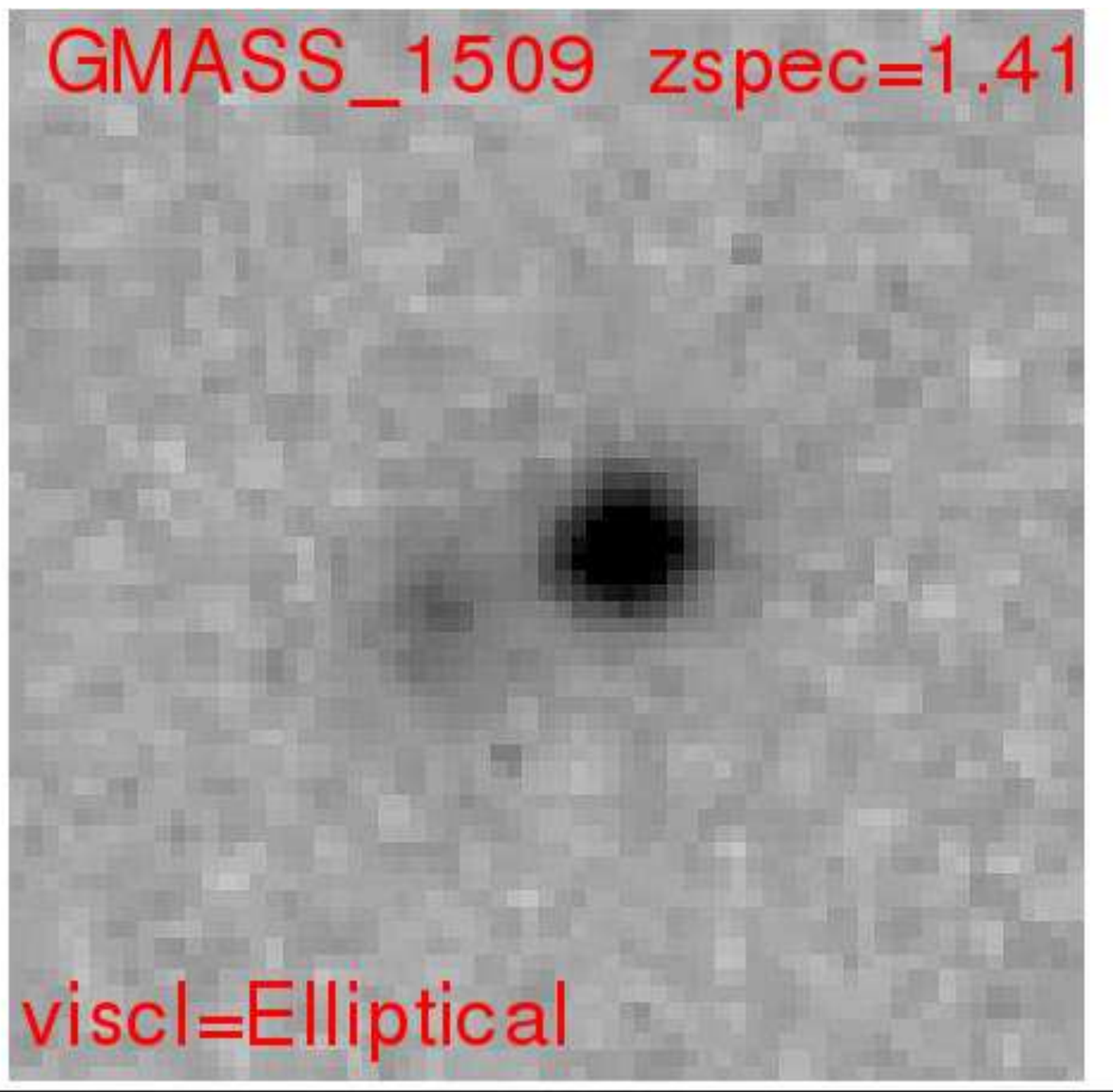}			     
\includegraphics[trim=100 40 75 390, clip=true, width=30mm]{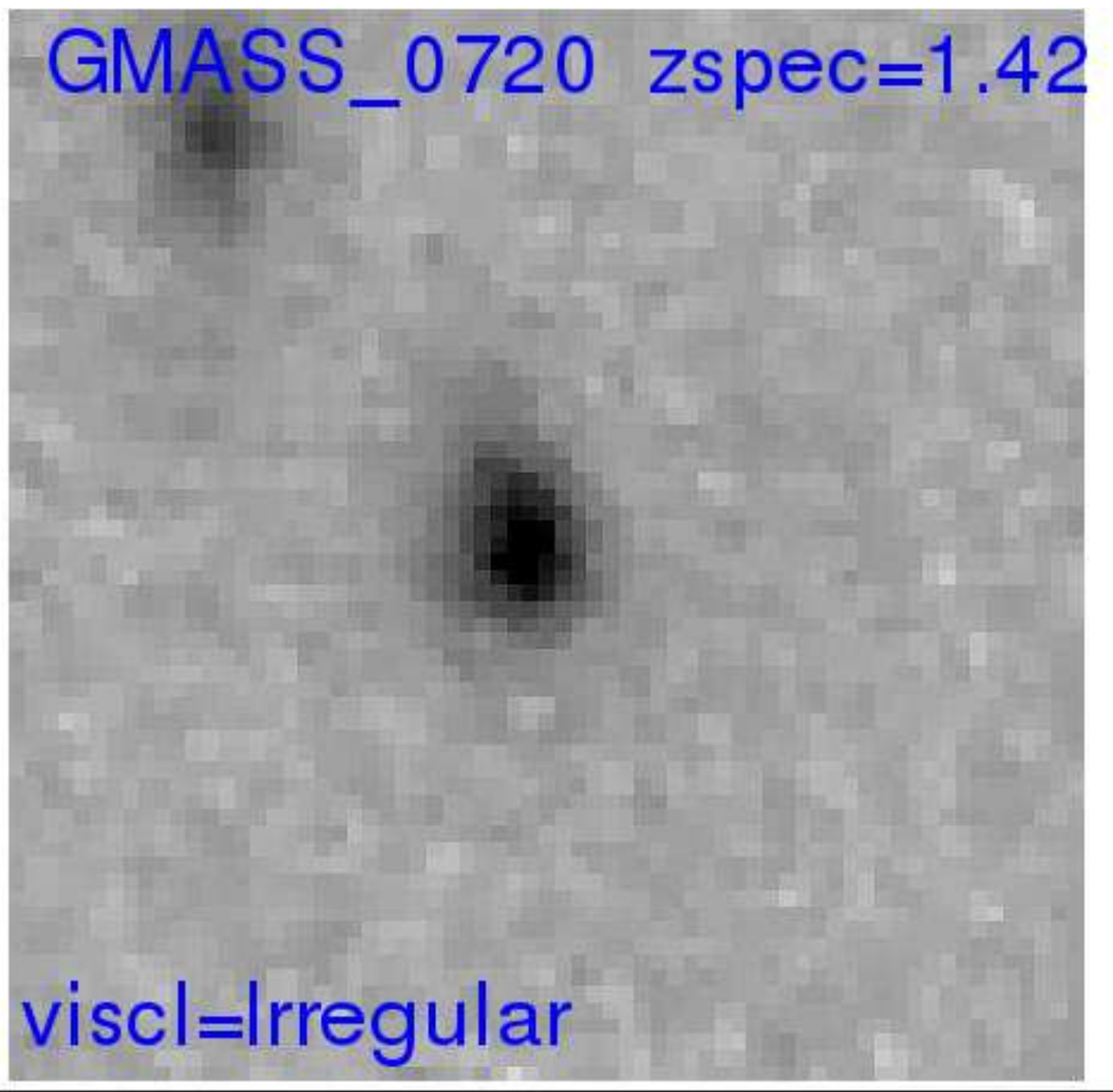}			     
\includegraphics[trim=100 40 75 390, clip=true, width=30mm]{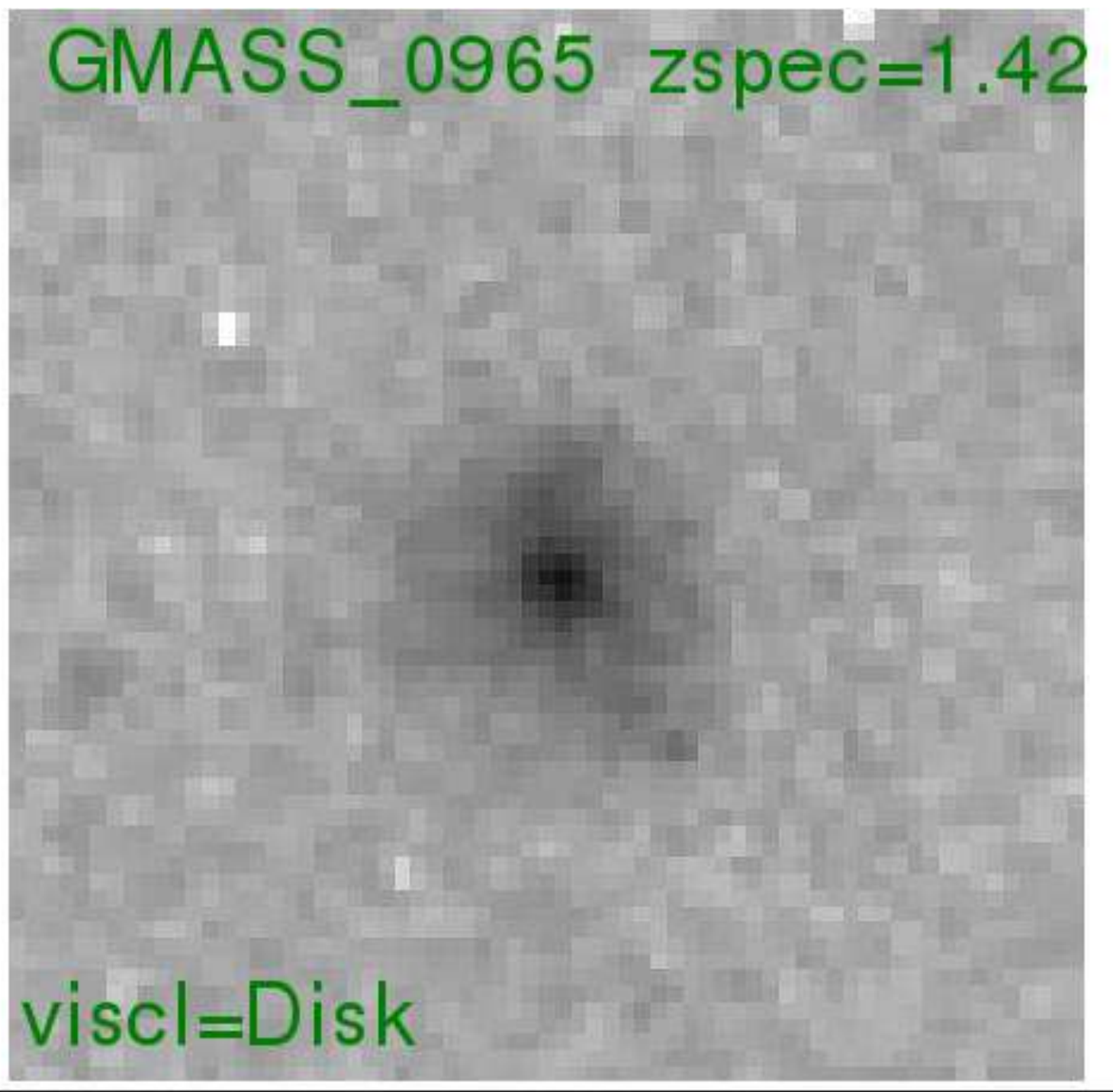}			     
\includegraphics[trim=100 40 75 390, clip=true, width=30mm]{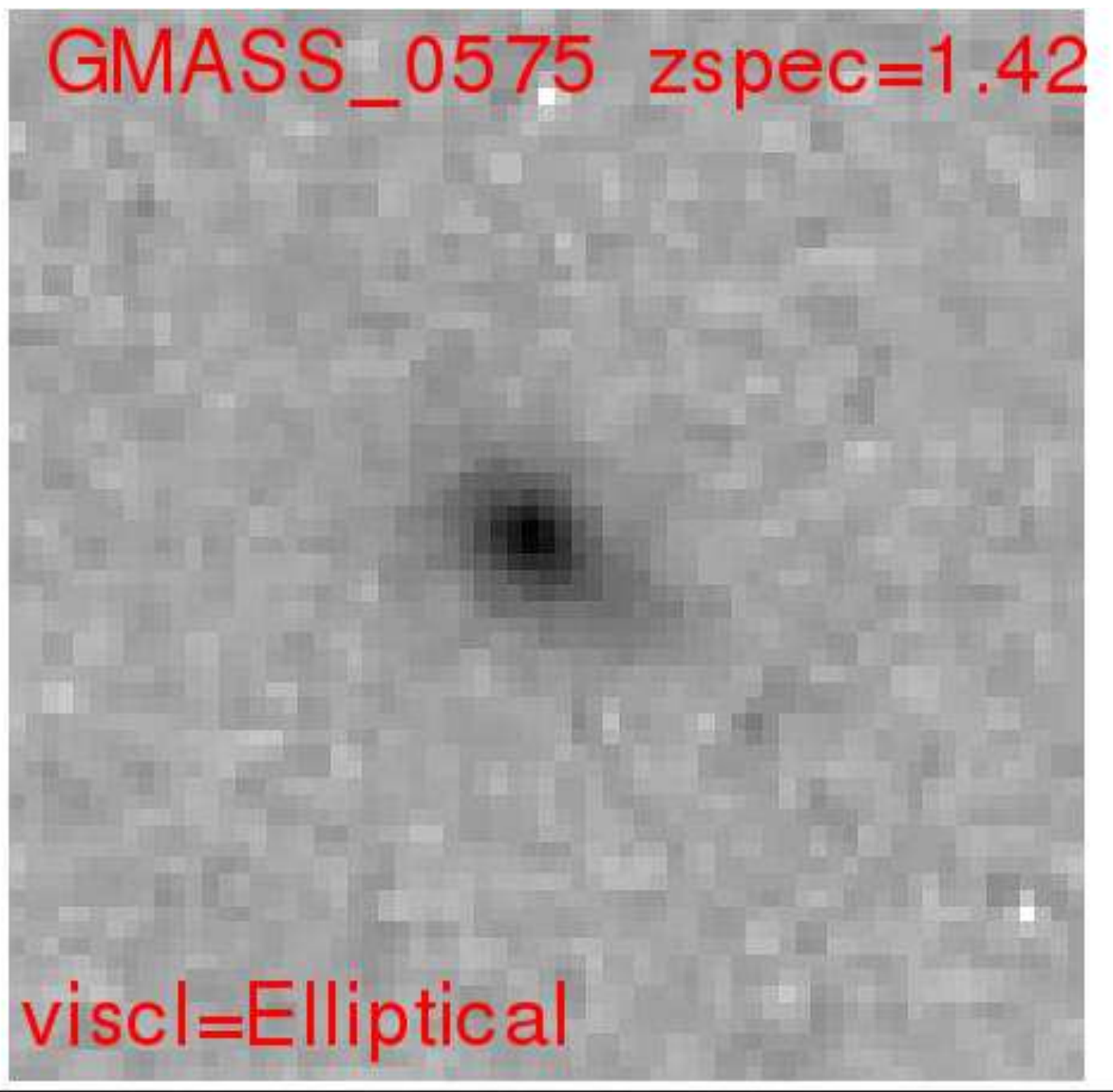}			     
\includegraphics[trim=100 40 75 390, clip=true, width=30mm]{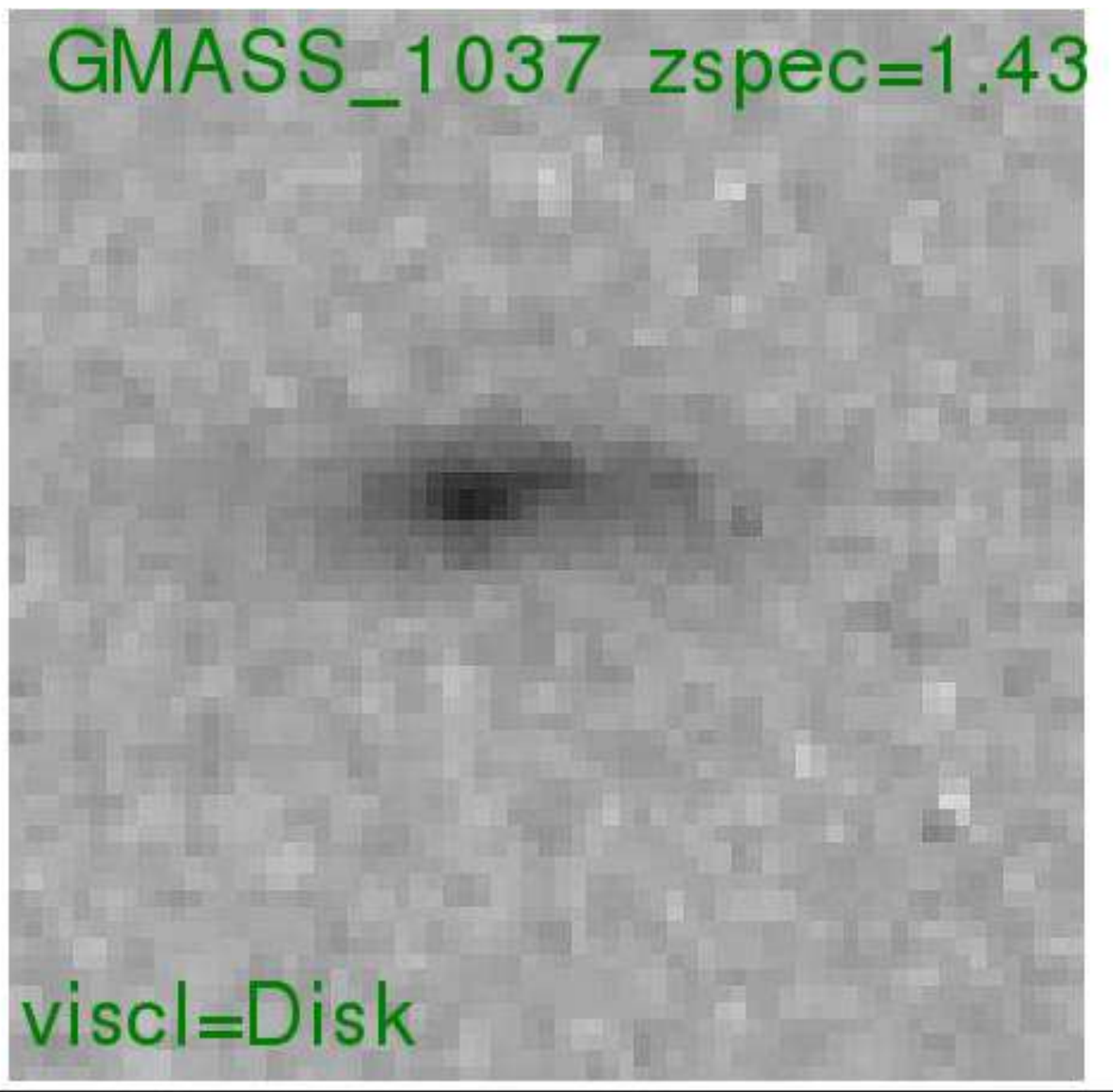}			     
\includegraphics[trim=100 40 75 390, clip=true, width=30mm]{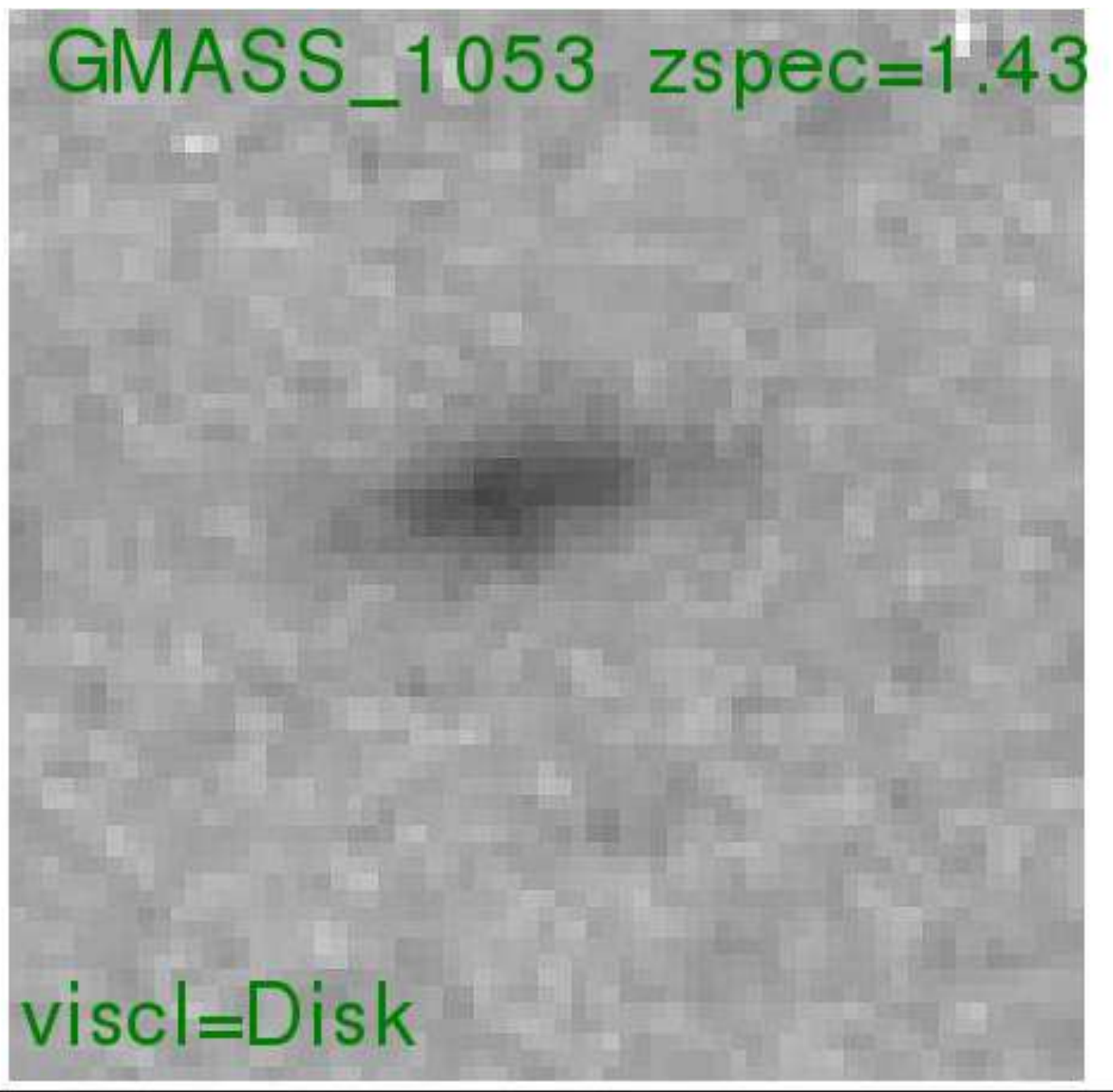}			     

\includegraphics[trim=100 40 75 390, clip=true, width=30mm]{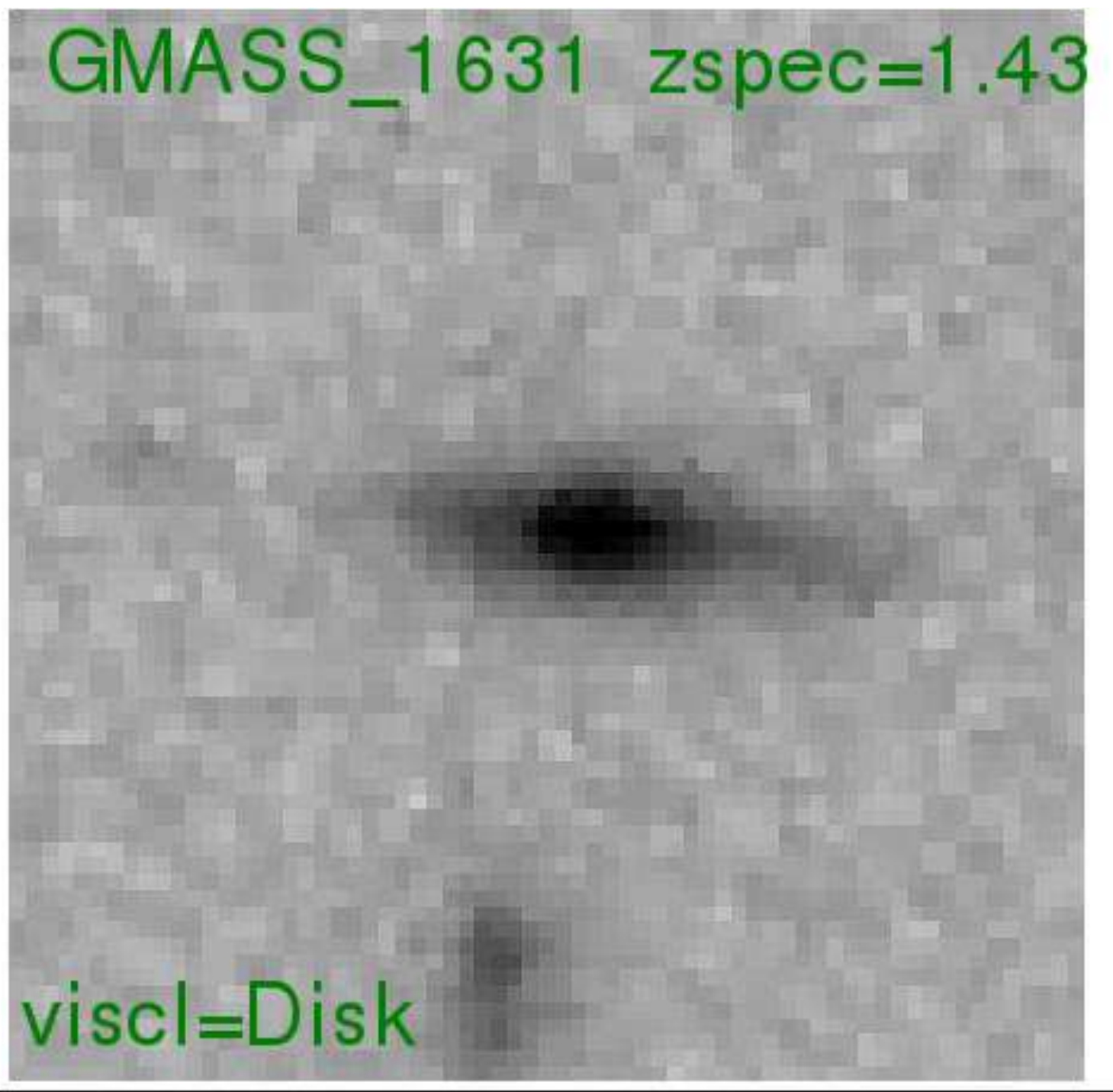}		     
\includegraphics[trim=100 40 75 390, clip=true, width=30mm]{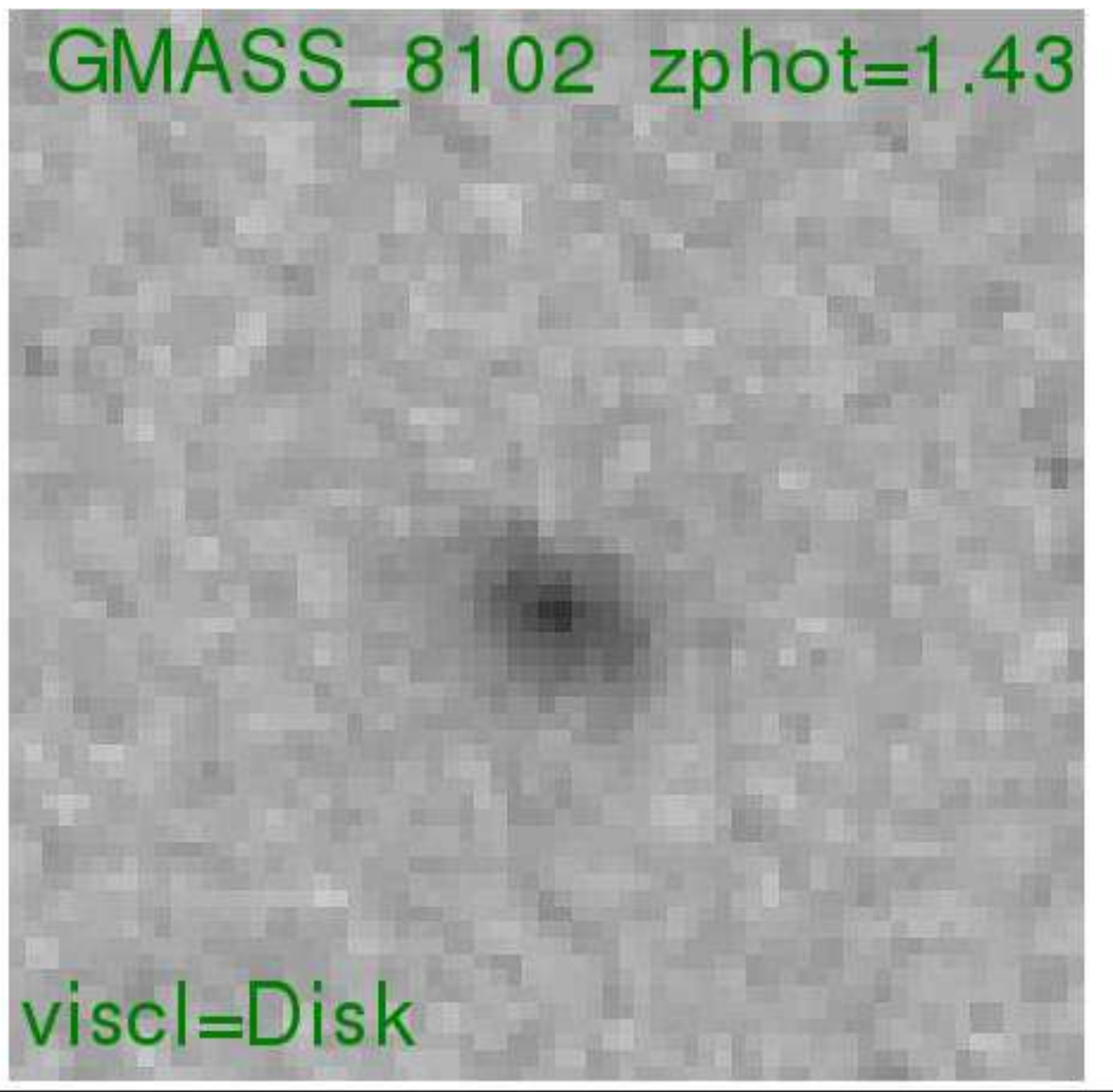}			     
\includegraphics[trim=100 40 75 390, clip=true, width=30mm]{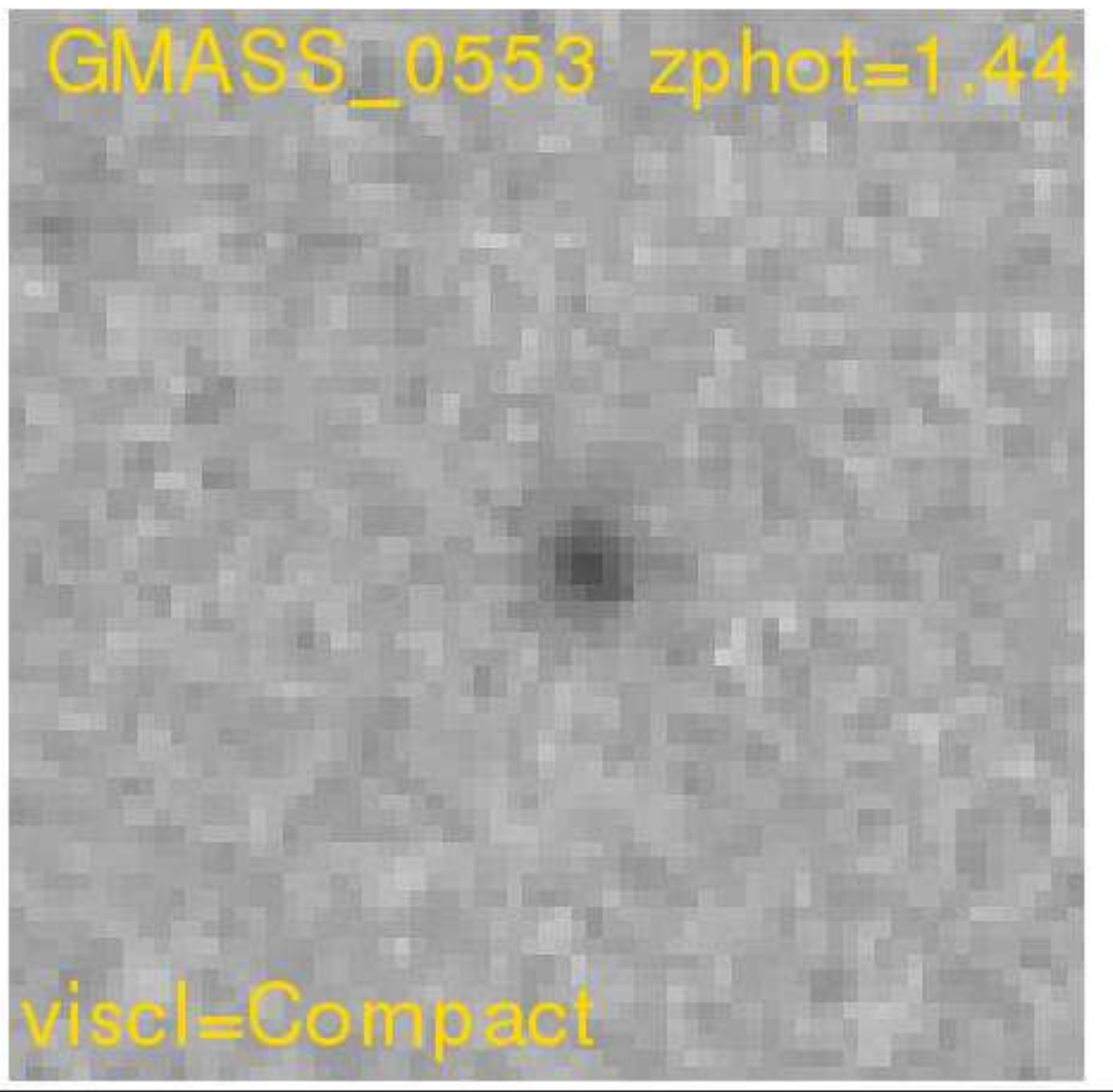}			     
\includegraphics[trim=100 40 75 390, clip=true, width=30mm]{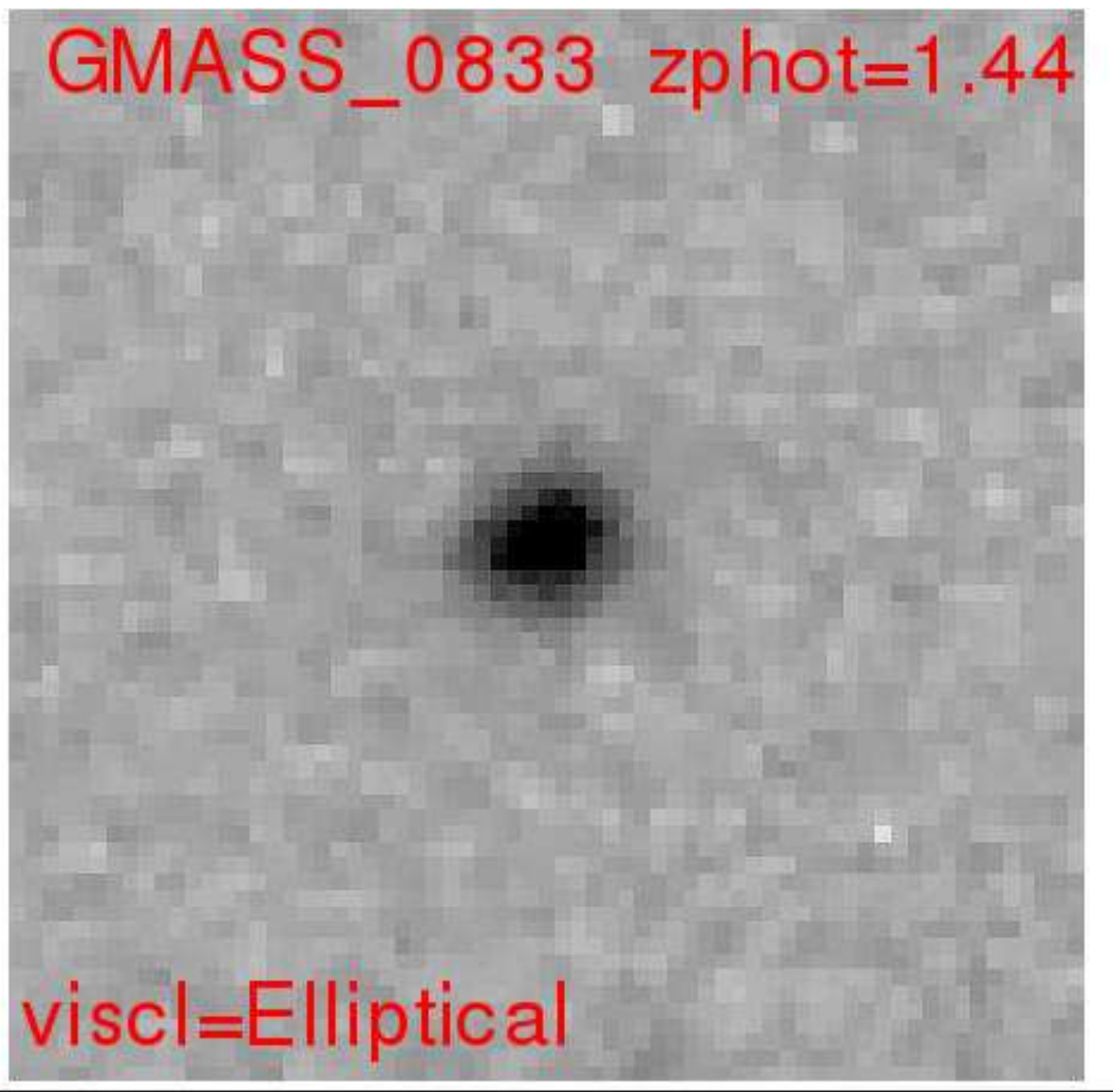}			     
\includegraphics[trim=100 40 75 390, clip=true, width=30mm]{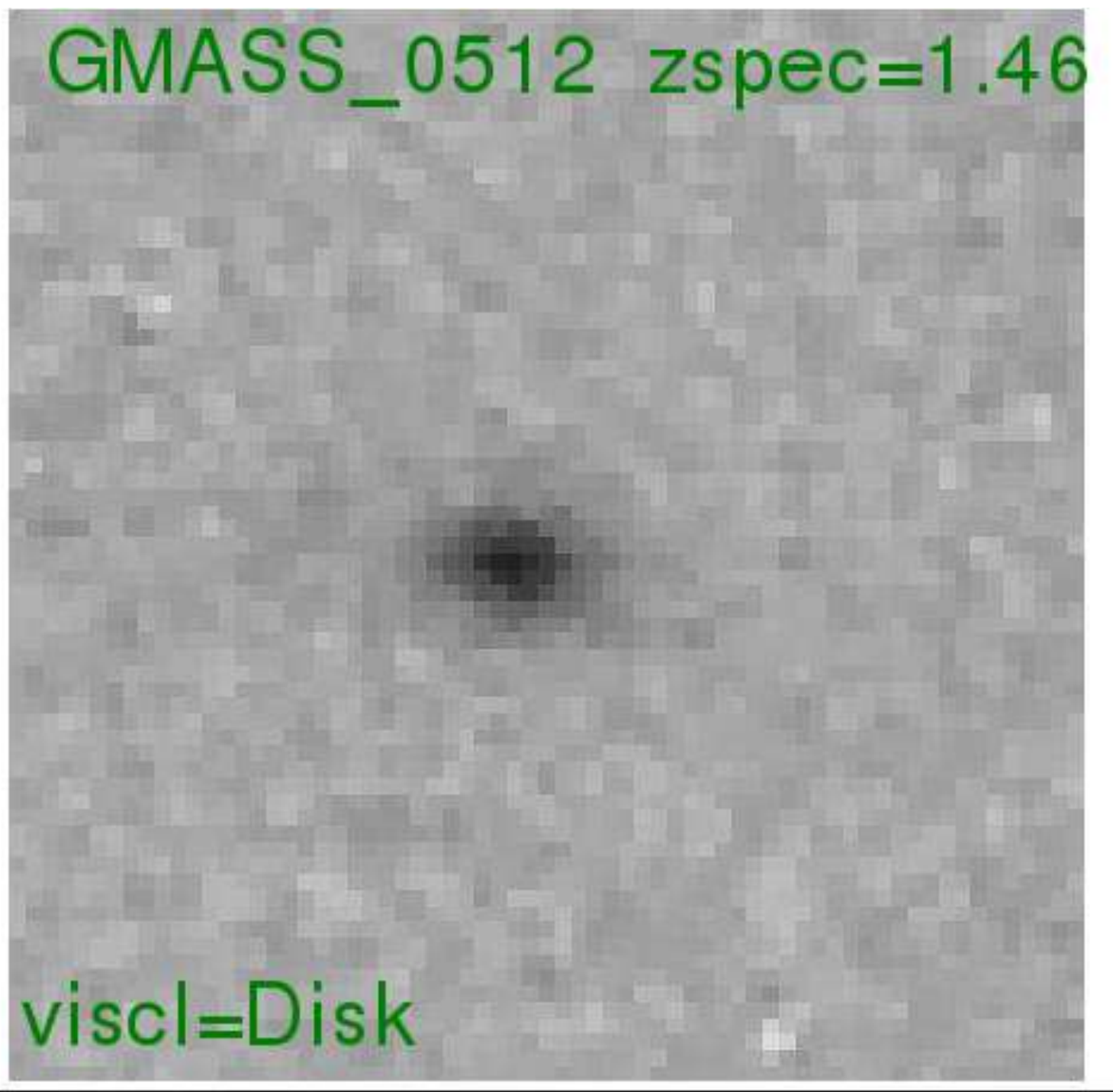}			     
\includegraphics[trim=100 40 75 390, clip=true, width=30mm]{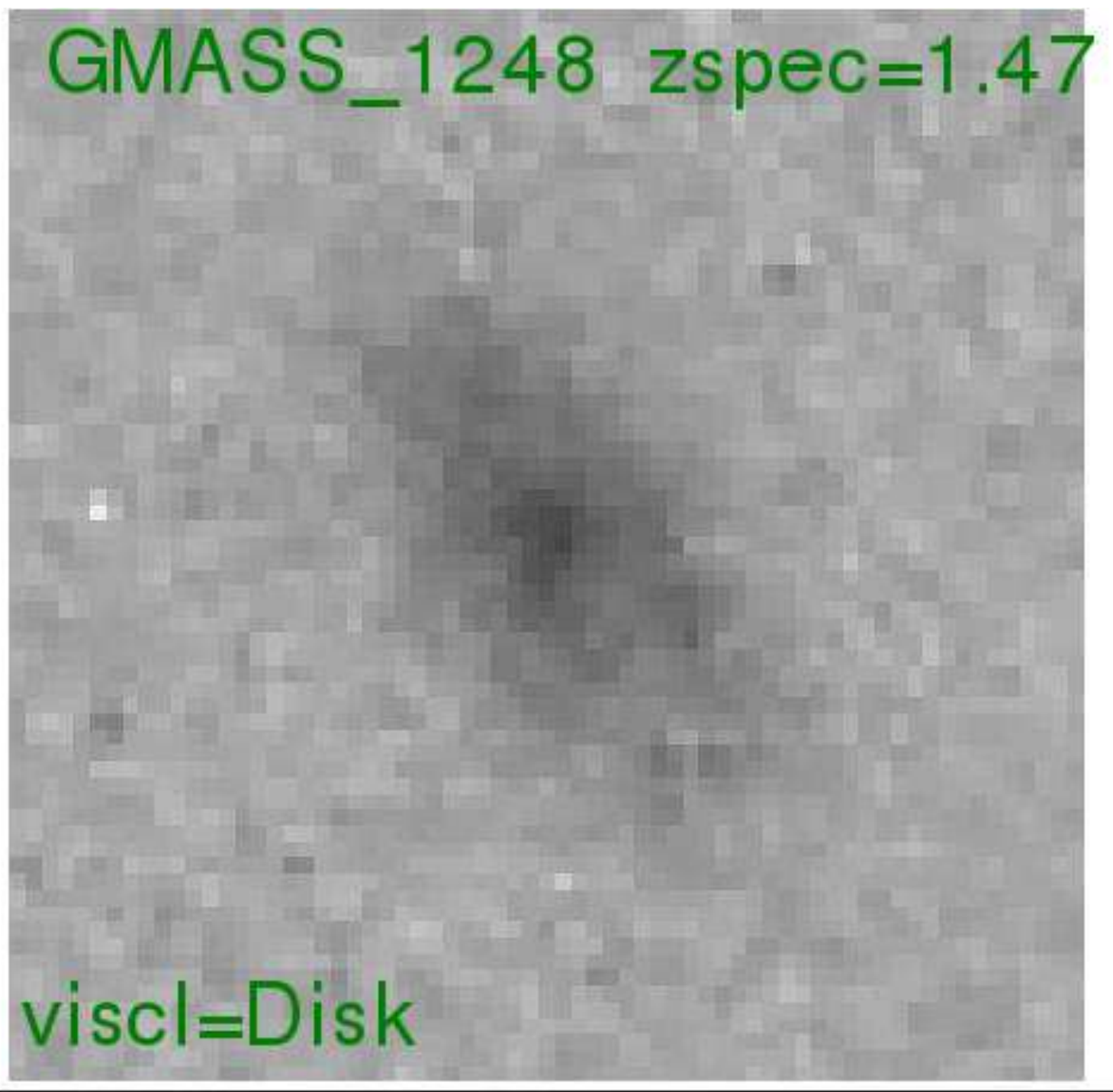}
\end{figure*}
\begin{figure*}
\centering      
\includegraphics[trim=100 40 75 390, clip=true, width=30mm]{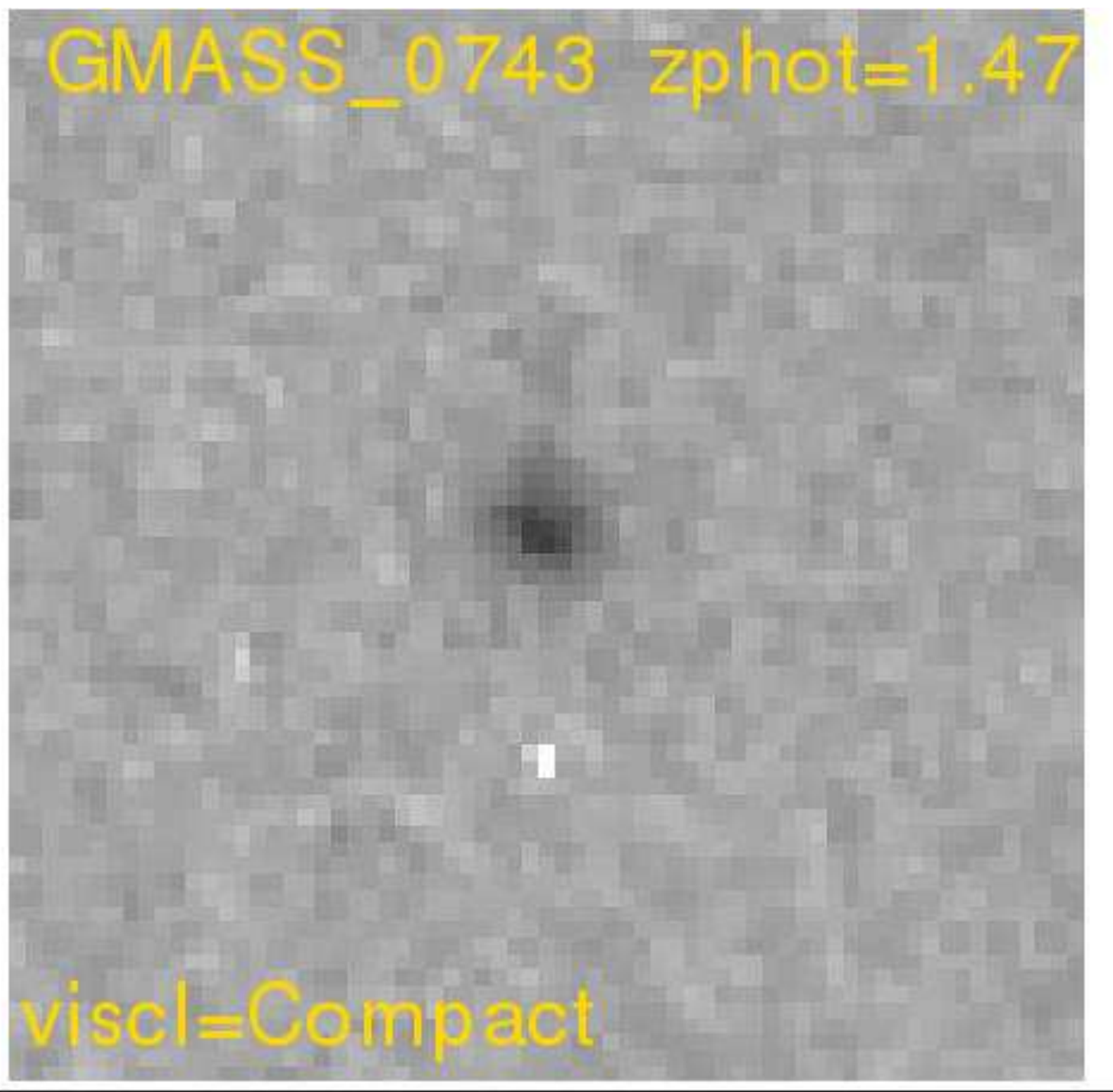}			     
\includegraphics[trim=100 40 75 390, clip=true, width=30mm]{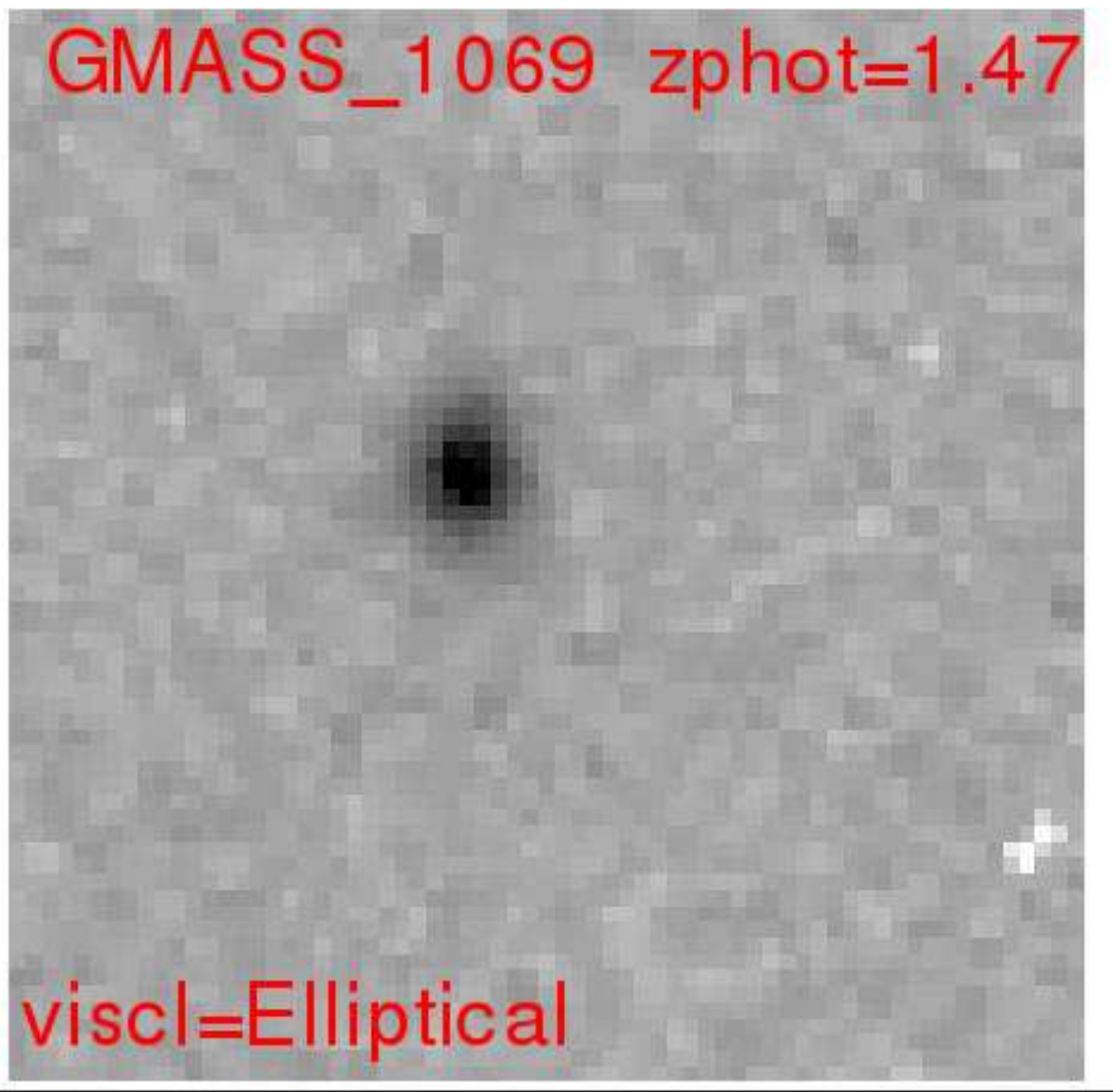}			     
\includegraphics[trim=100 40 75 390, clip=true, width=30mm]{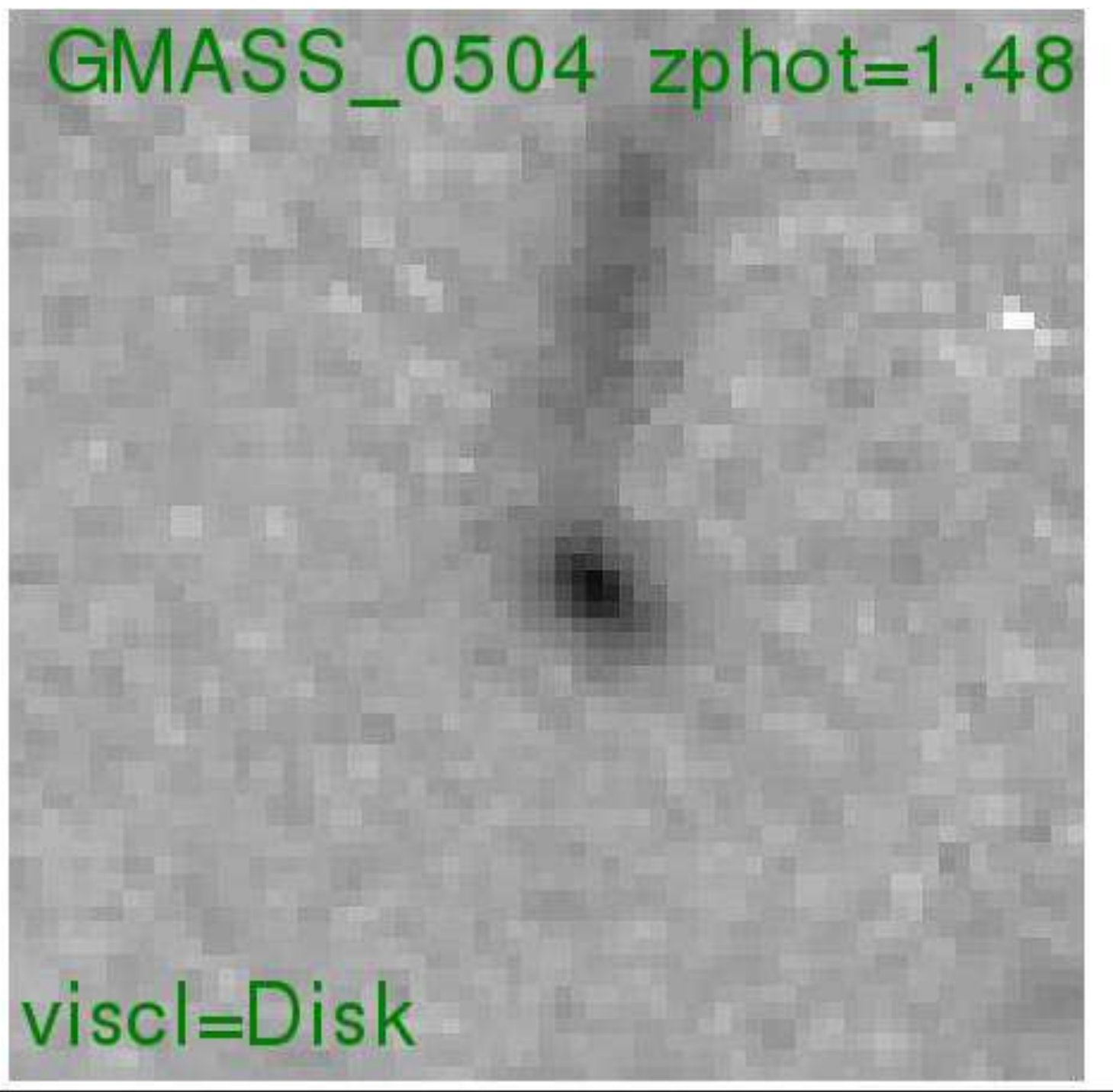}			     
\includegraphics[trim=100 40 75 390, clip=true, width=30mm]{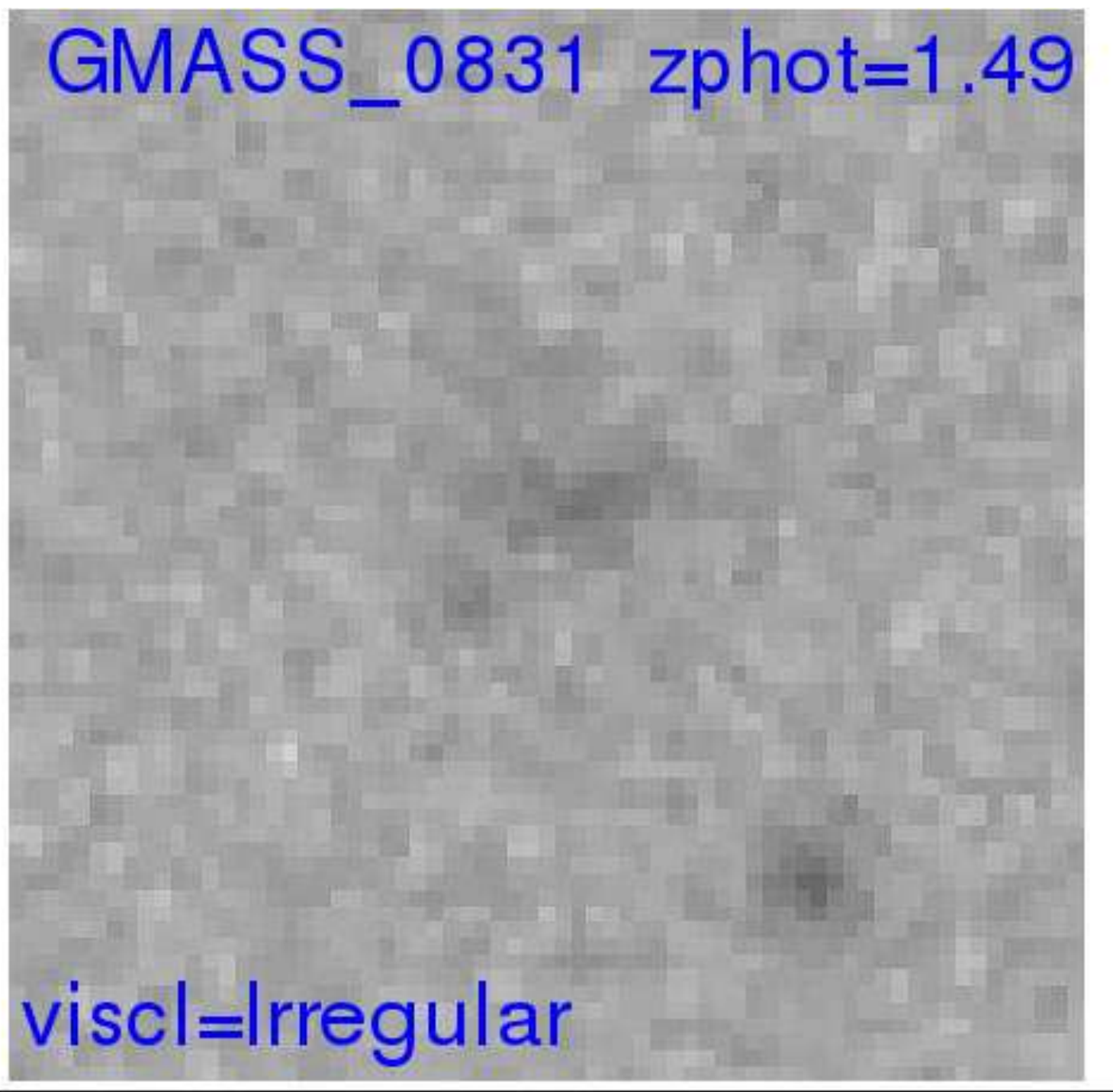}			     
\includegraphics[trim=100 40 75 390, clip=true, width=30mm]{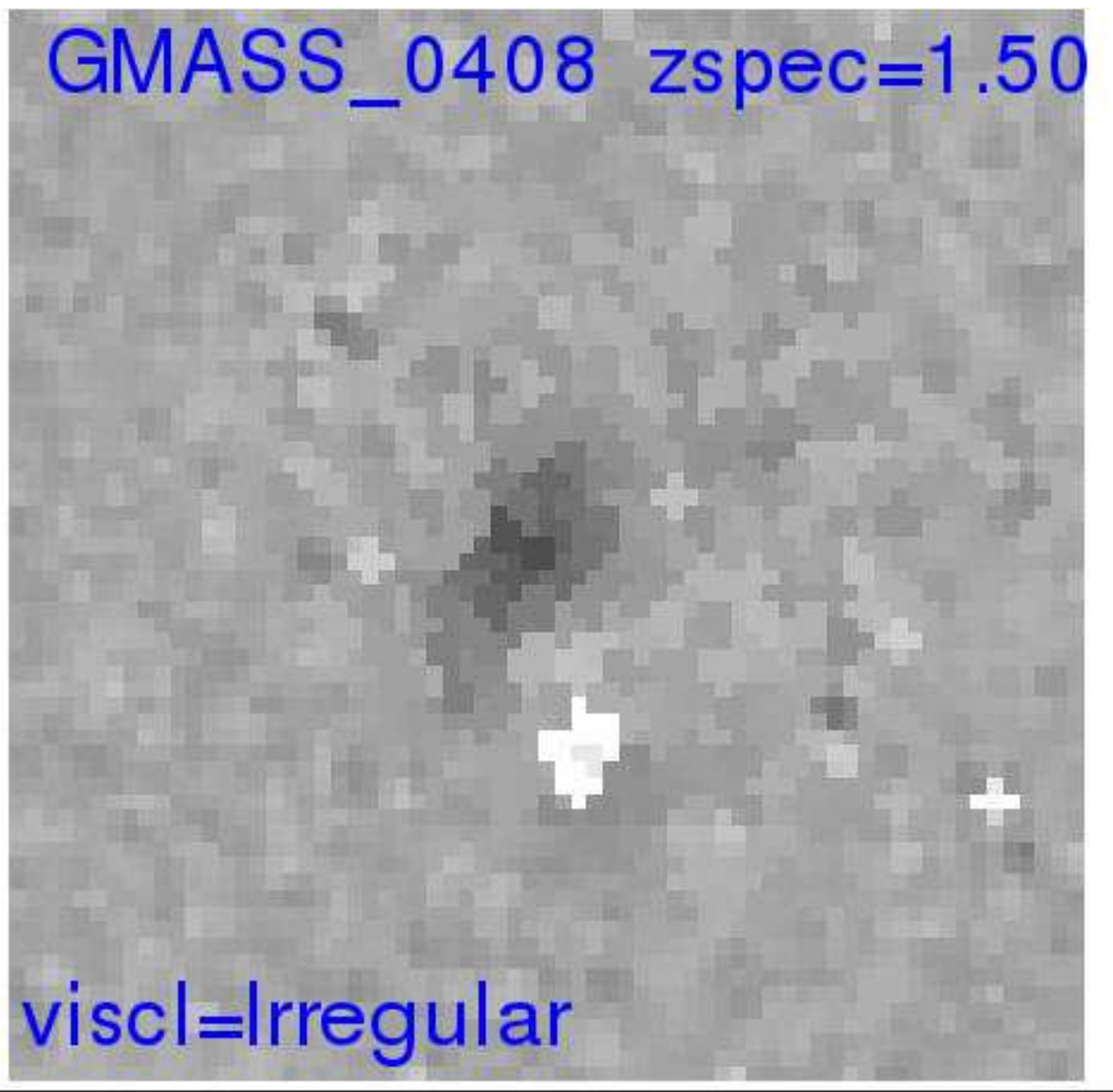}			     
\includegraphics[trim=100 40 75 390, clip=true, width=30mm]{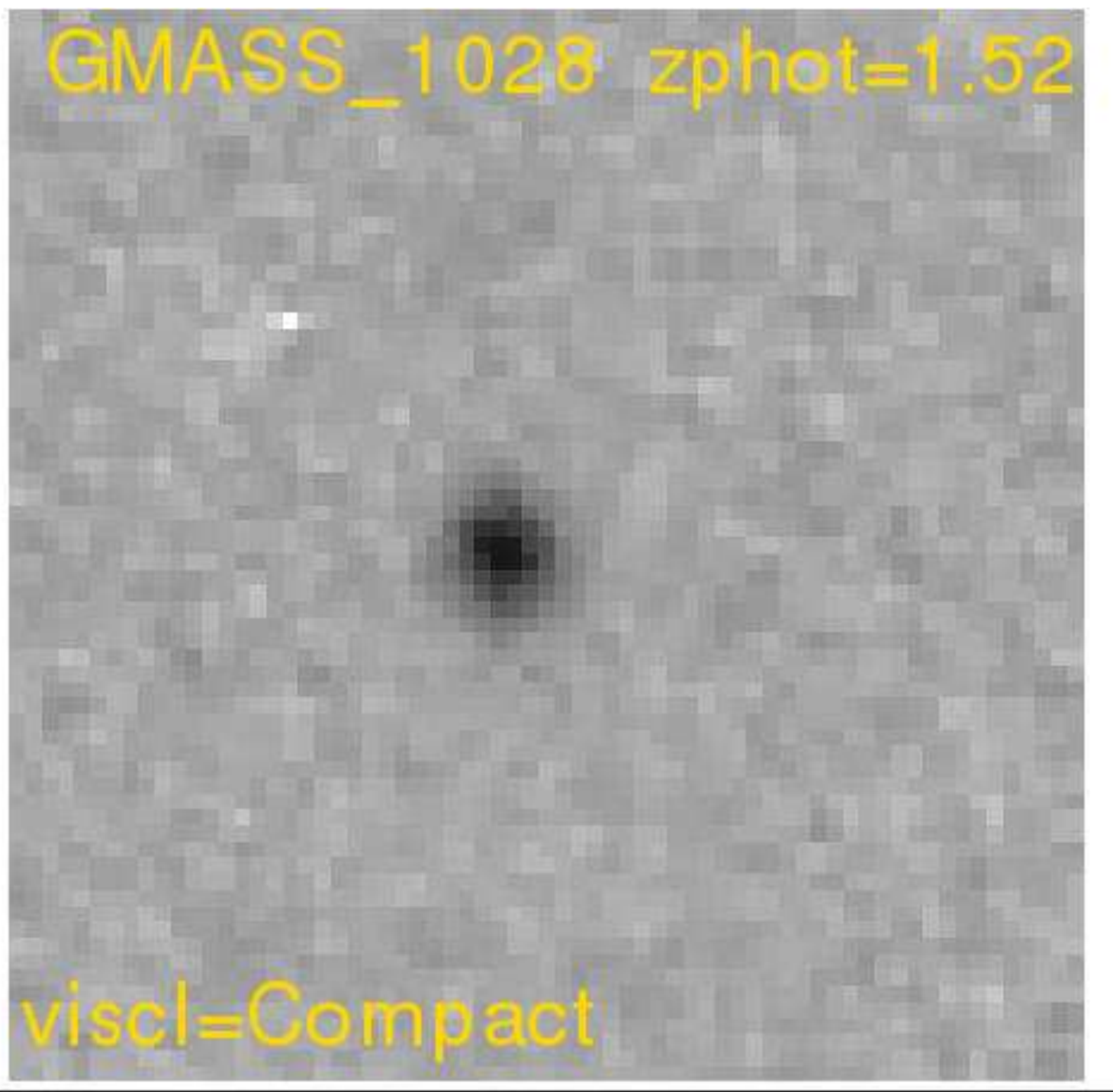}			     

\includegraphics[trim=100 40 75 390, clip=true, width=30mm]{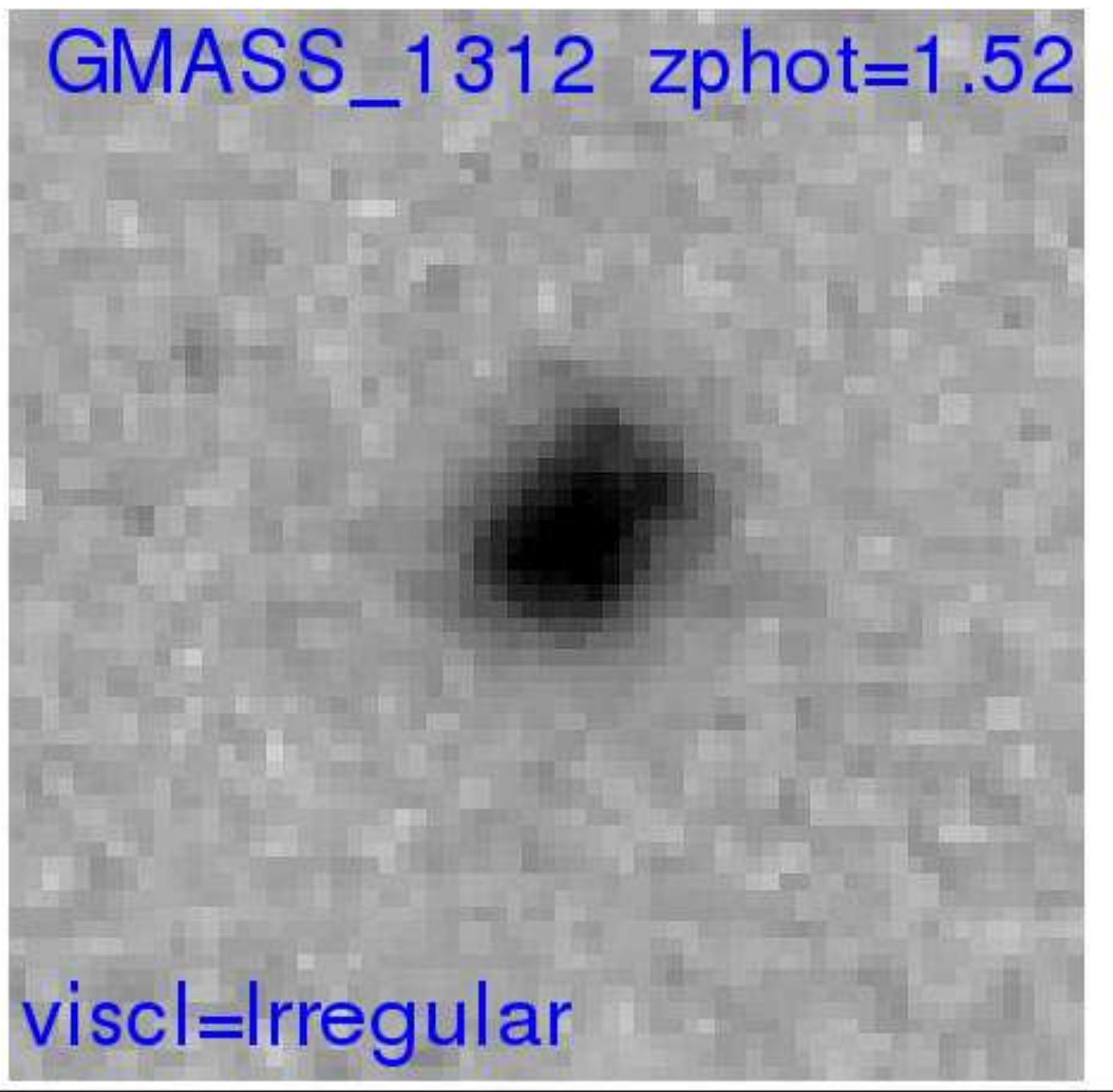}			     
\includegraphics[trim=100 40 75 390, clip=true, width=30mm]{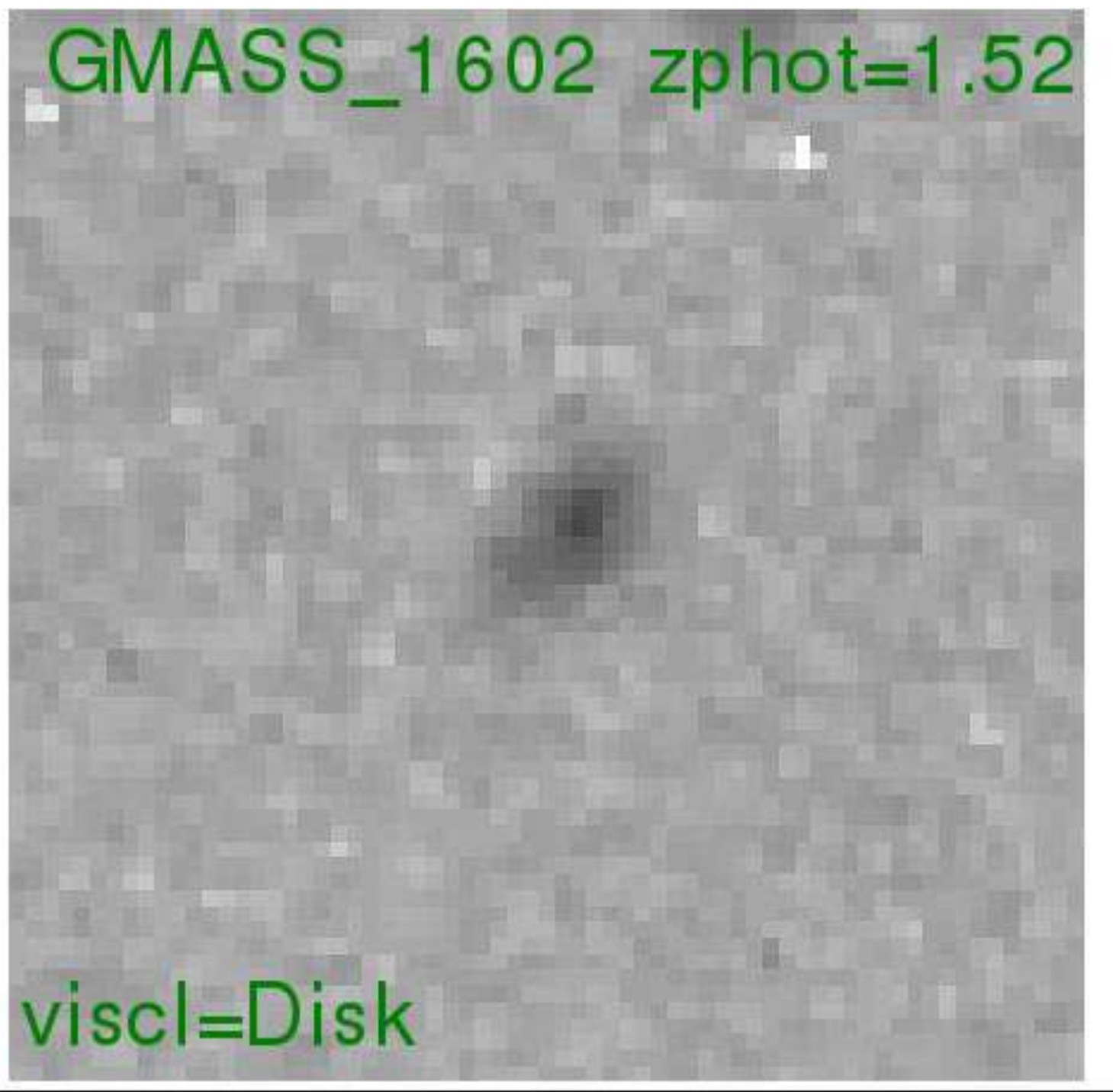}			     
\includegraphics[trim=100 40 75 390, clip=true, width=30mm]{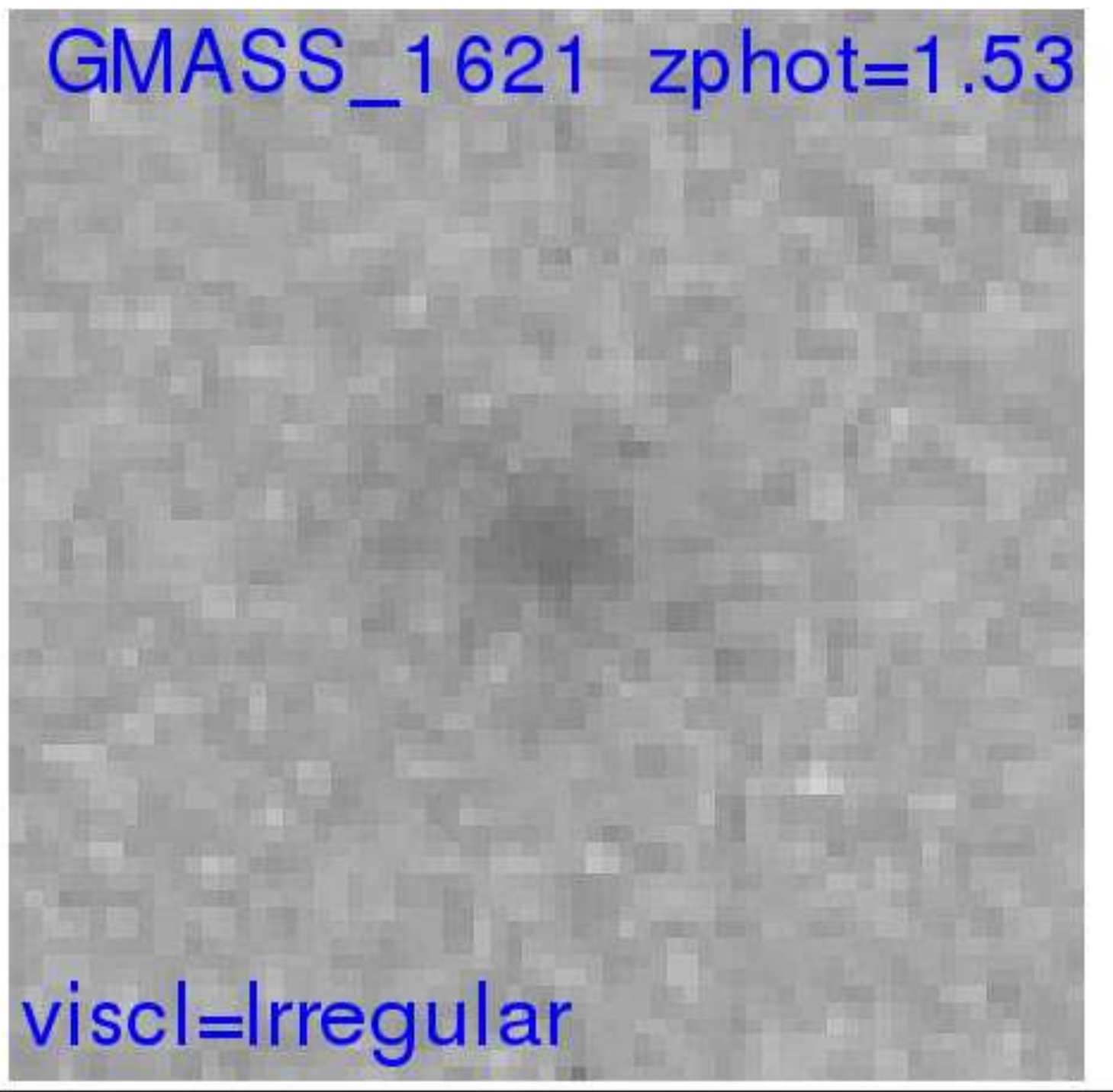}			     
\includegraphics[trim=100 40 75 390, clip=true, width=30mm]{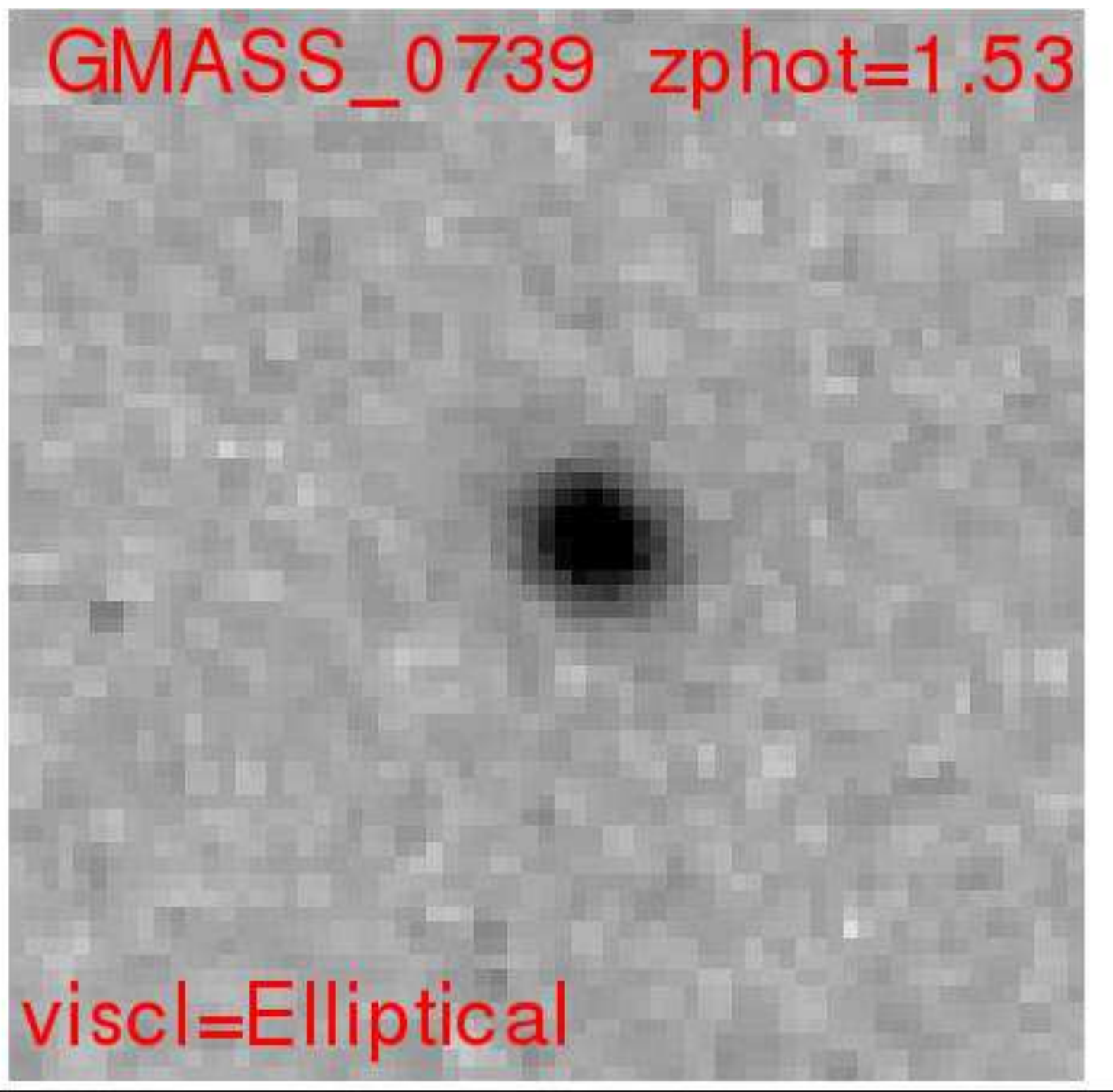}			     
\includegraphics[trim=100 40 75 390, clip=true, width=30mm]{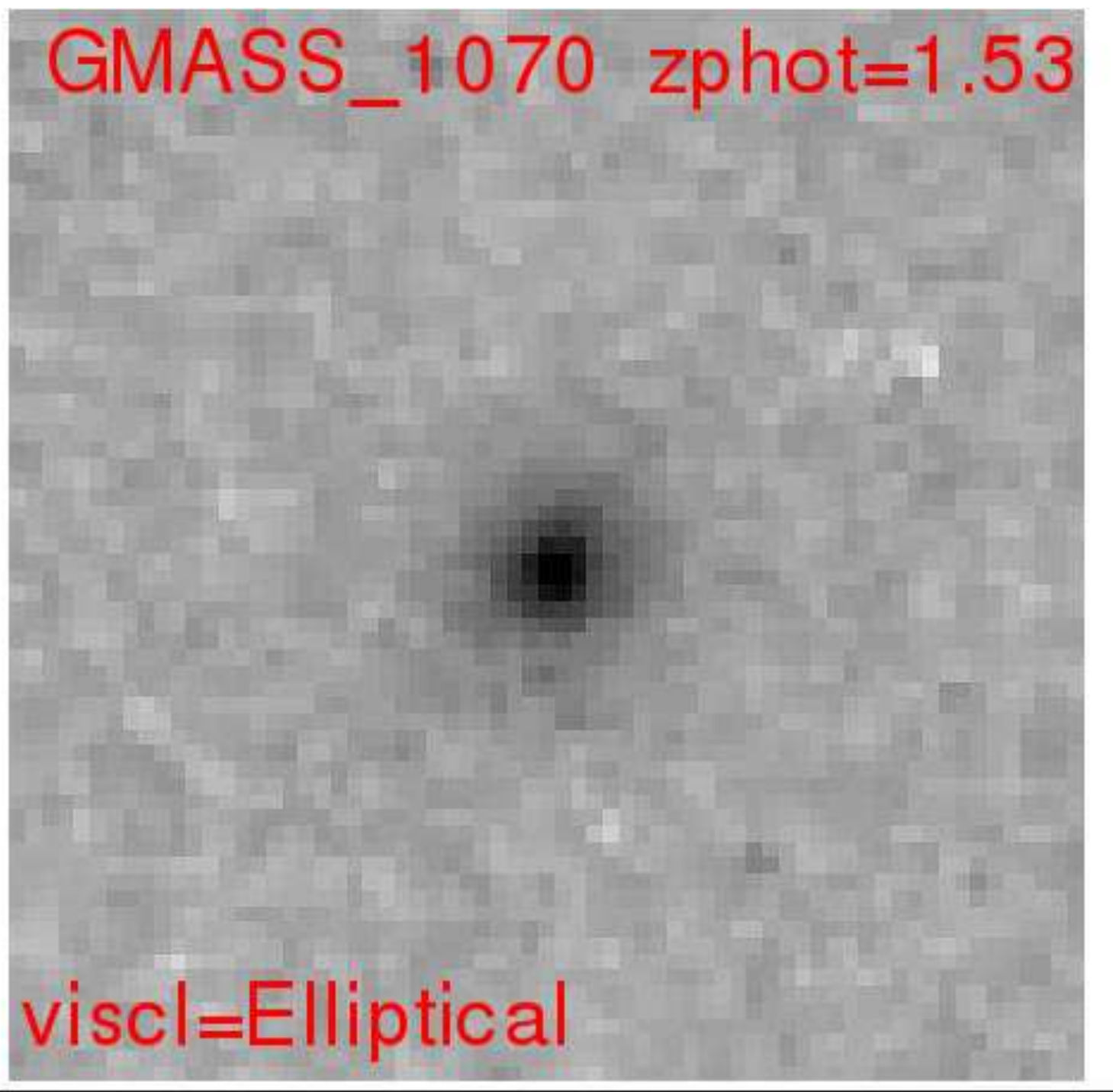}			     
\includegraphics[trim=100 40 75 390, clip=true, width=30mm]{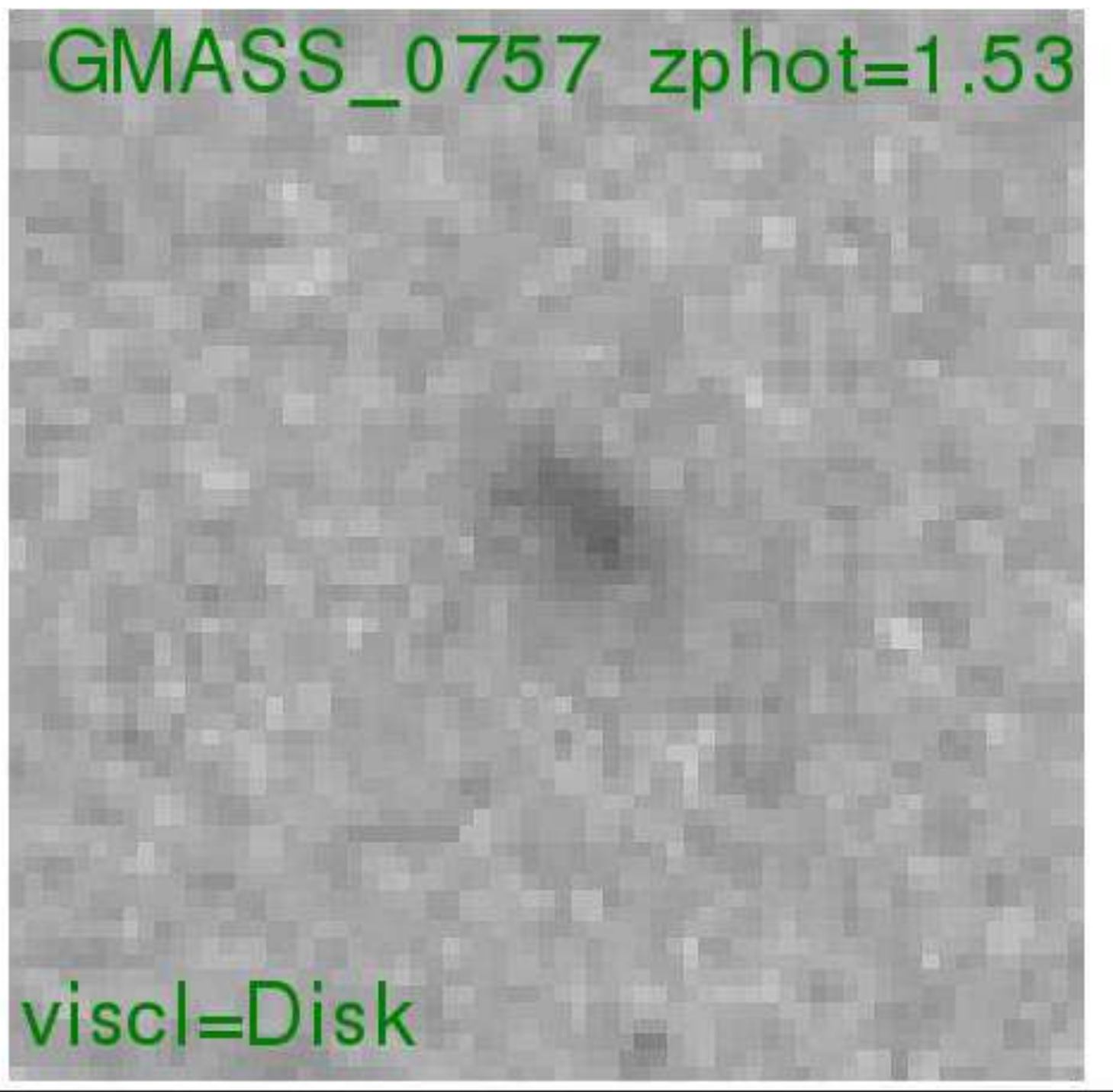}		     

\includegraphics[trim=100 40 75 390, clip=true, width=30mm]{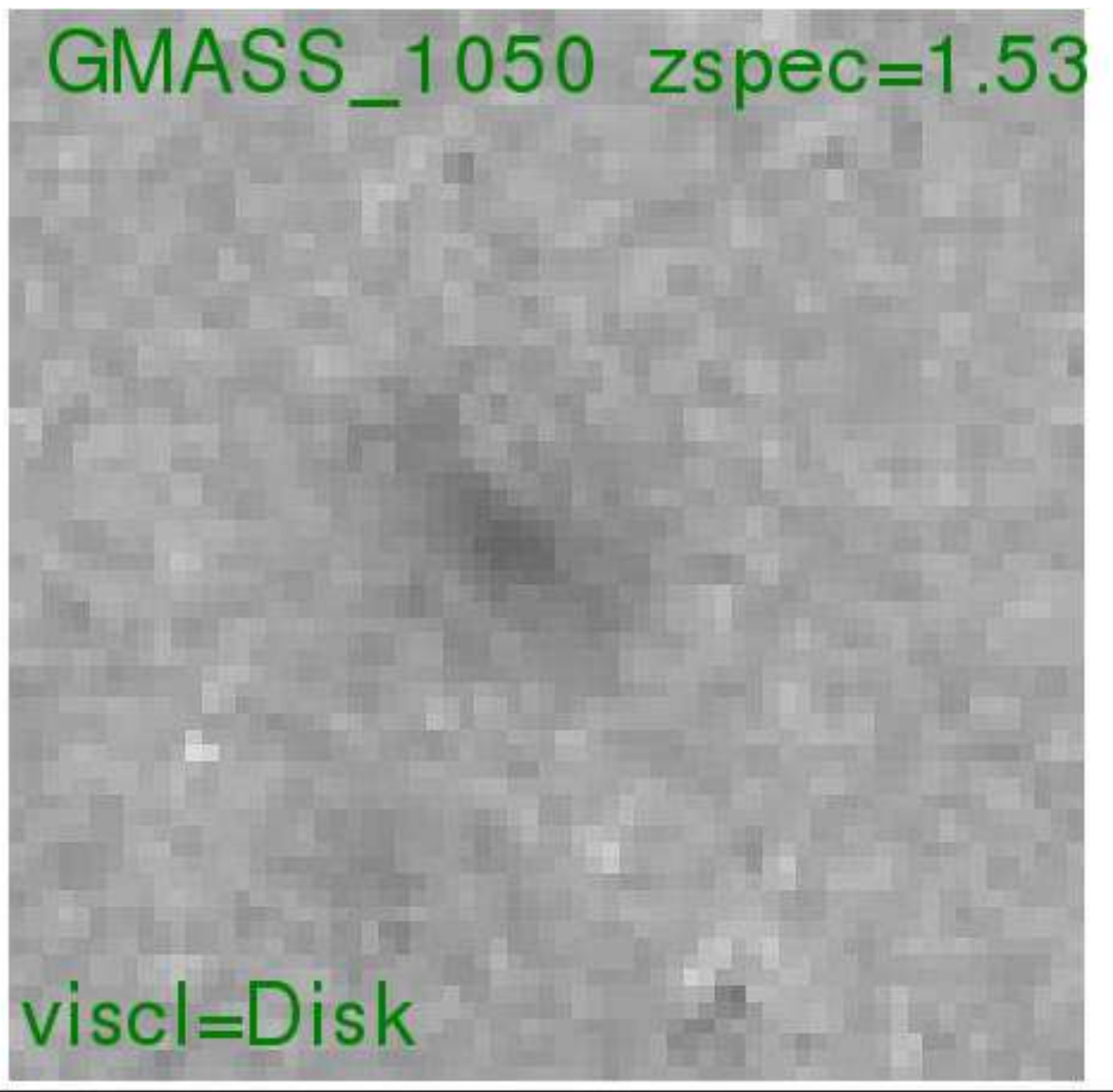}		     
\includegraphics[trim=100 40 75 390, clip=true, width=30mm]{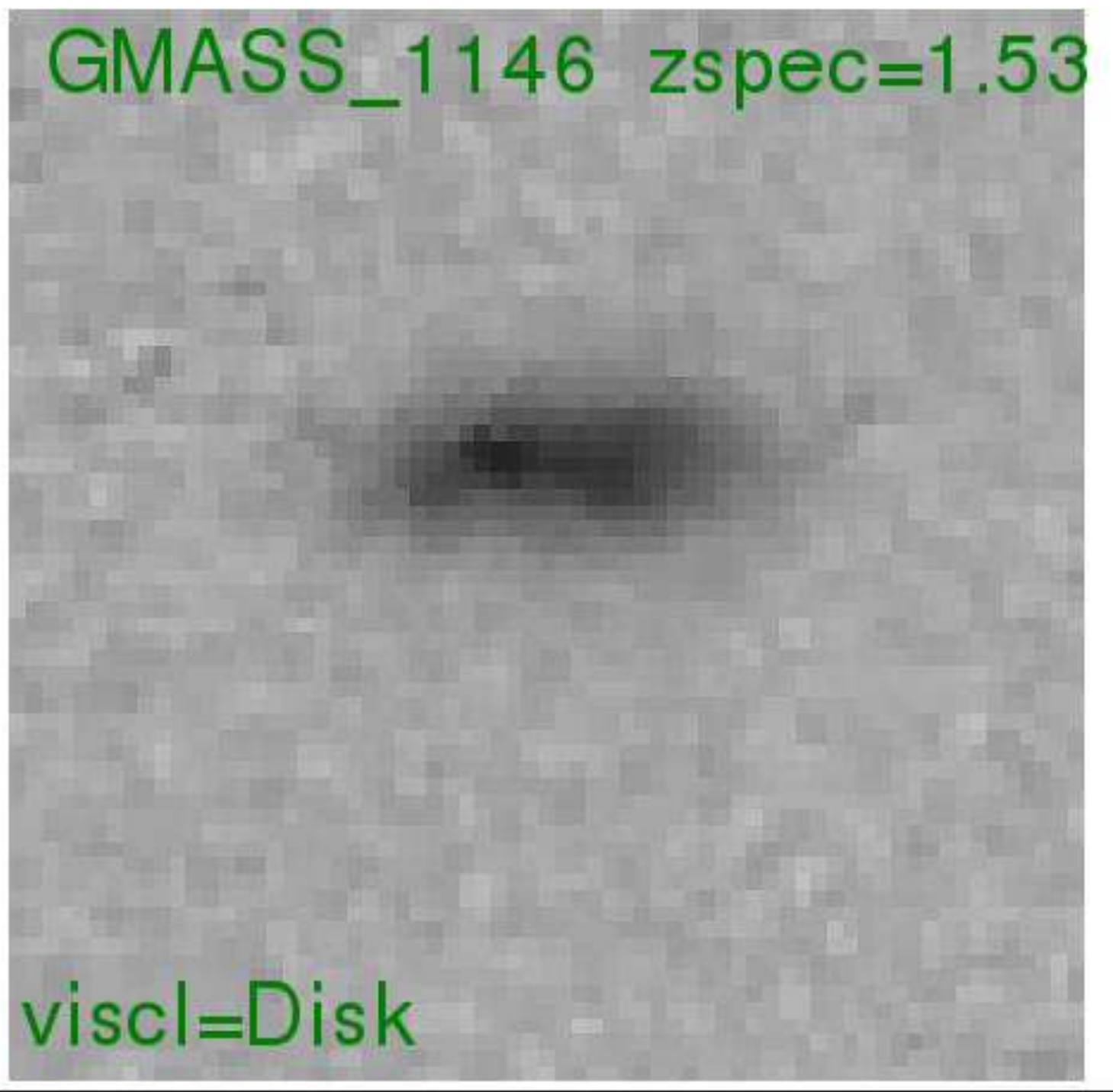}			     
\includegraphics[trim=100 40 75 390, clip=true, width=30mm]{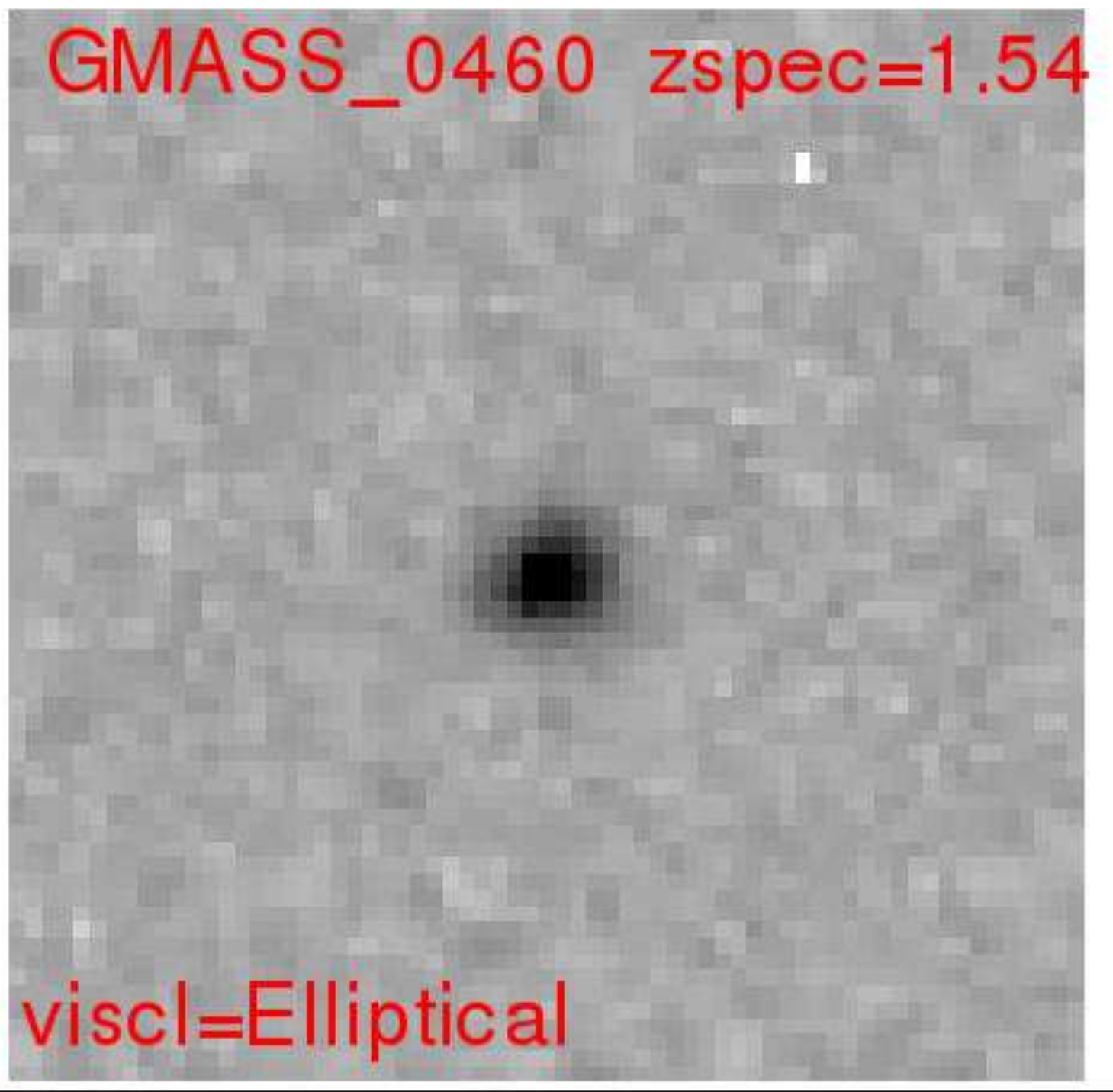}			     
\includegraphics[trim=100 40 75 390, clip=true, width=30mm]{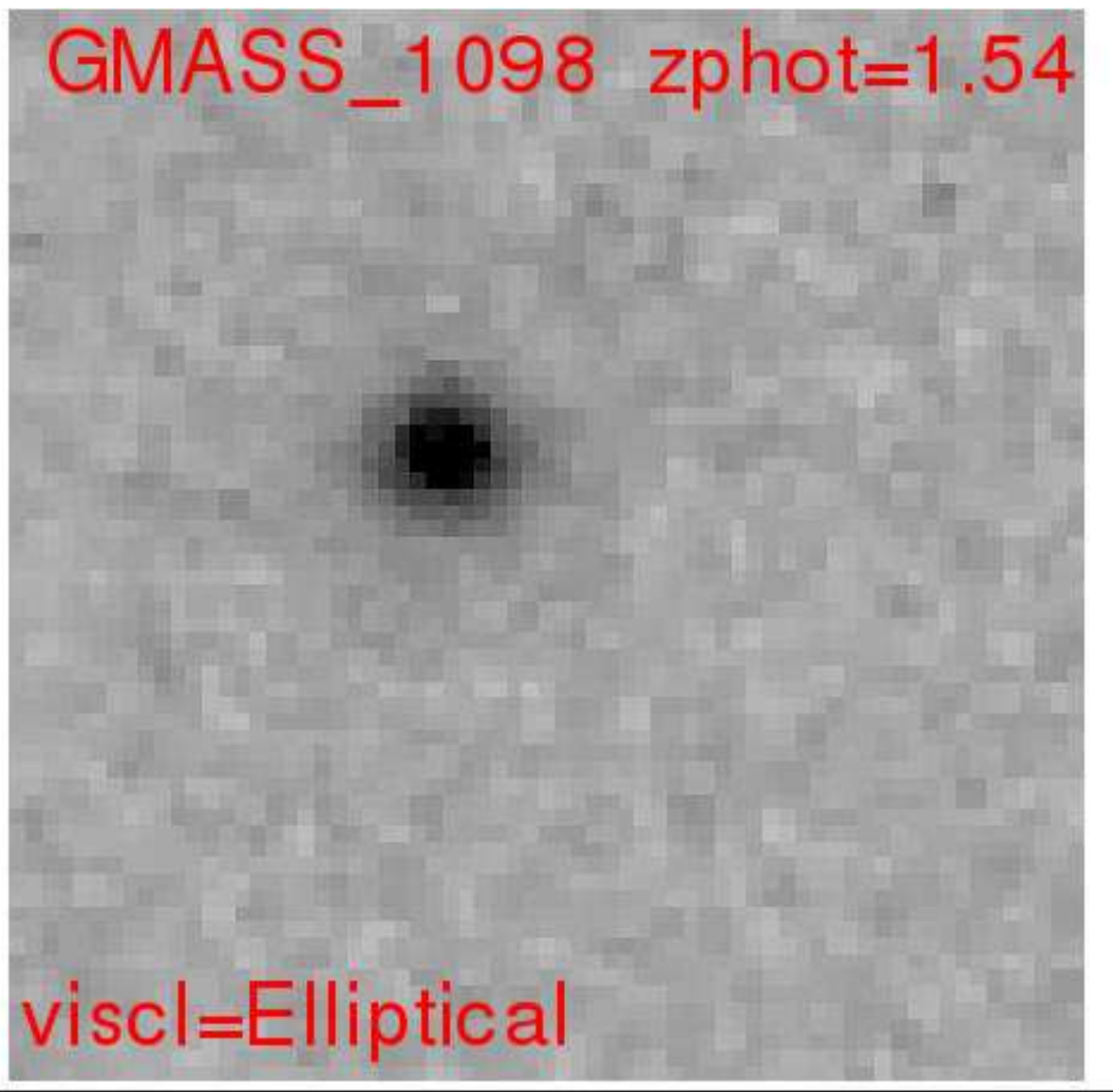}		     
\includegraphics[trim=100 40 75 390, clip=true, width=30mm]{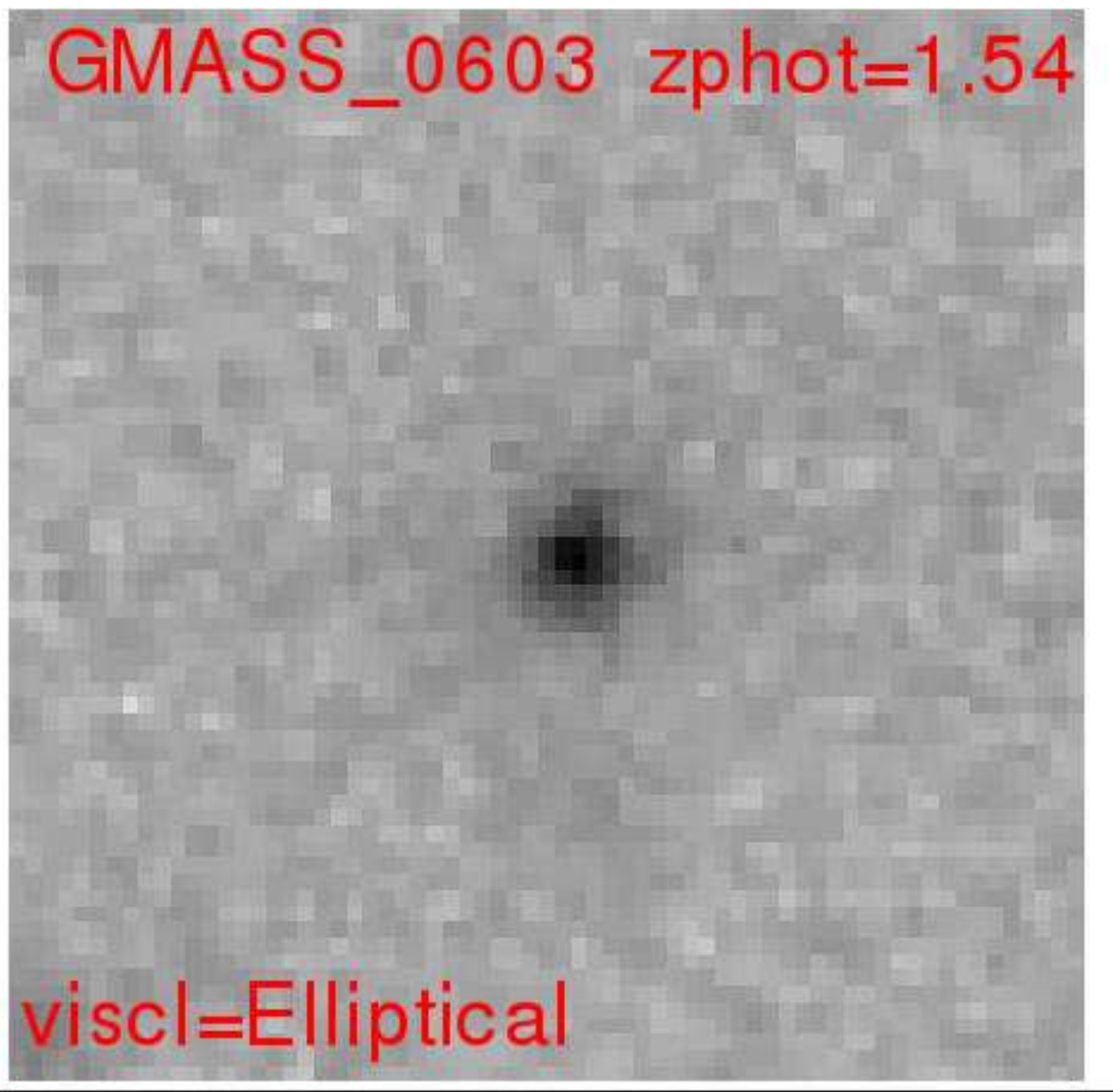}			     
\includegraphics[trim=100 40 75 390, clip=true, width=30mm]{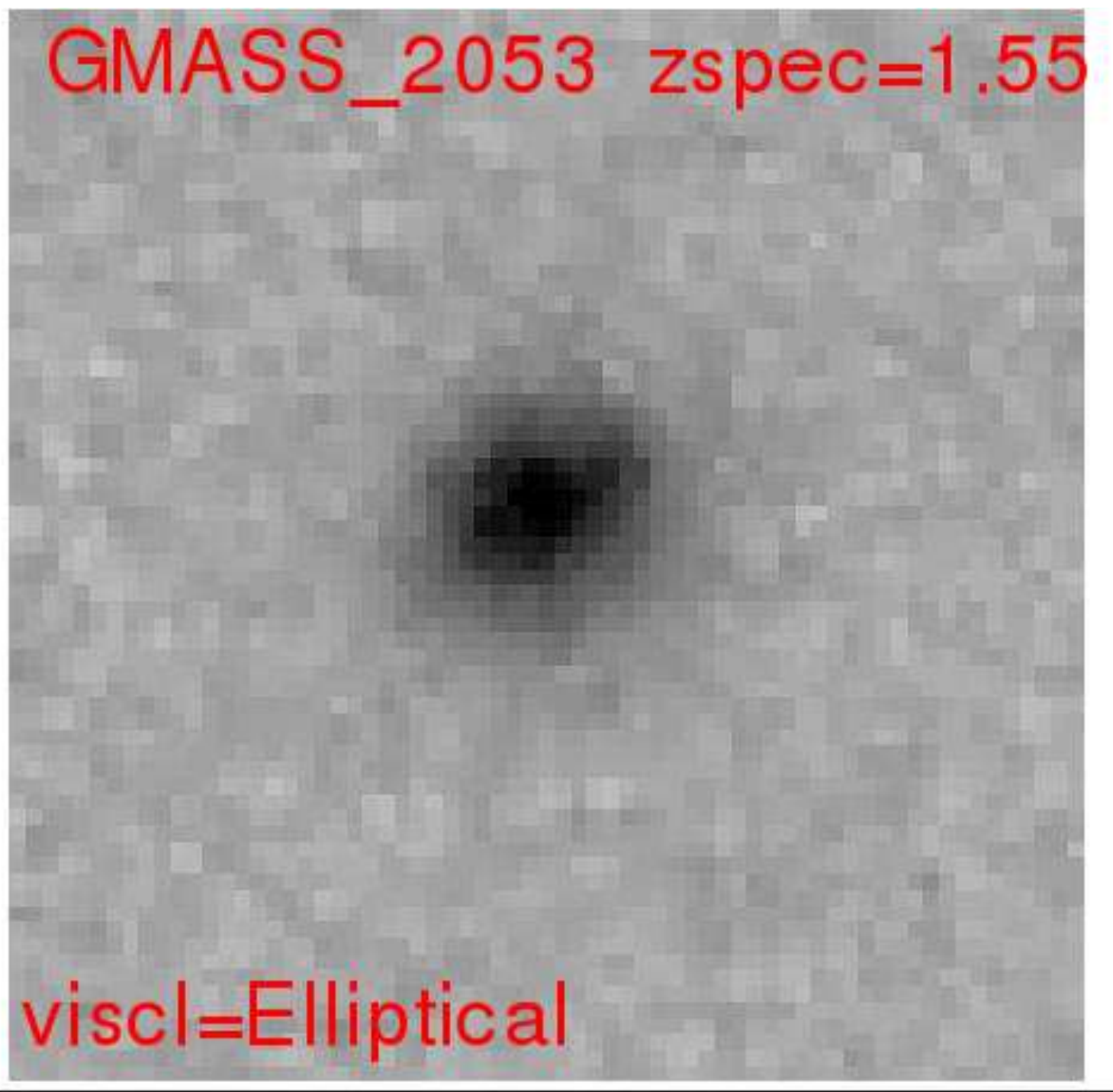}			     

\includegraphics[trim=100 40 75 390, clip=true, width=30mm]{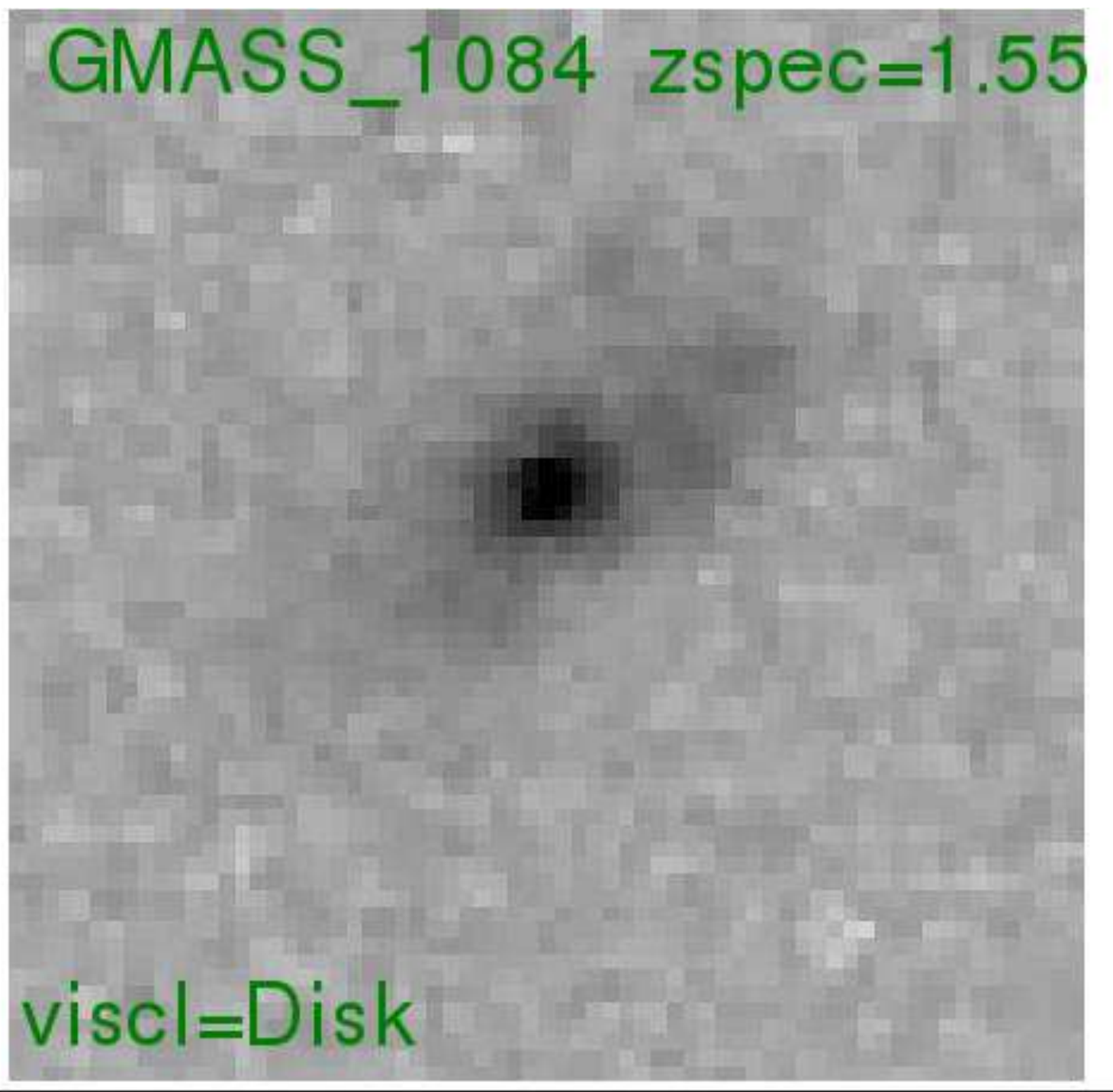}			     
\includegraphics[trim=100 40 75 390, clip=true, width=30mm]{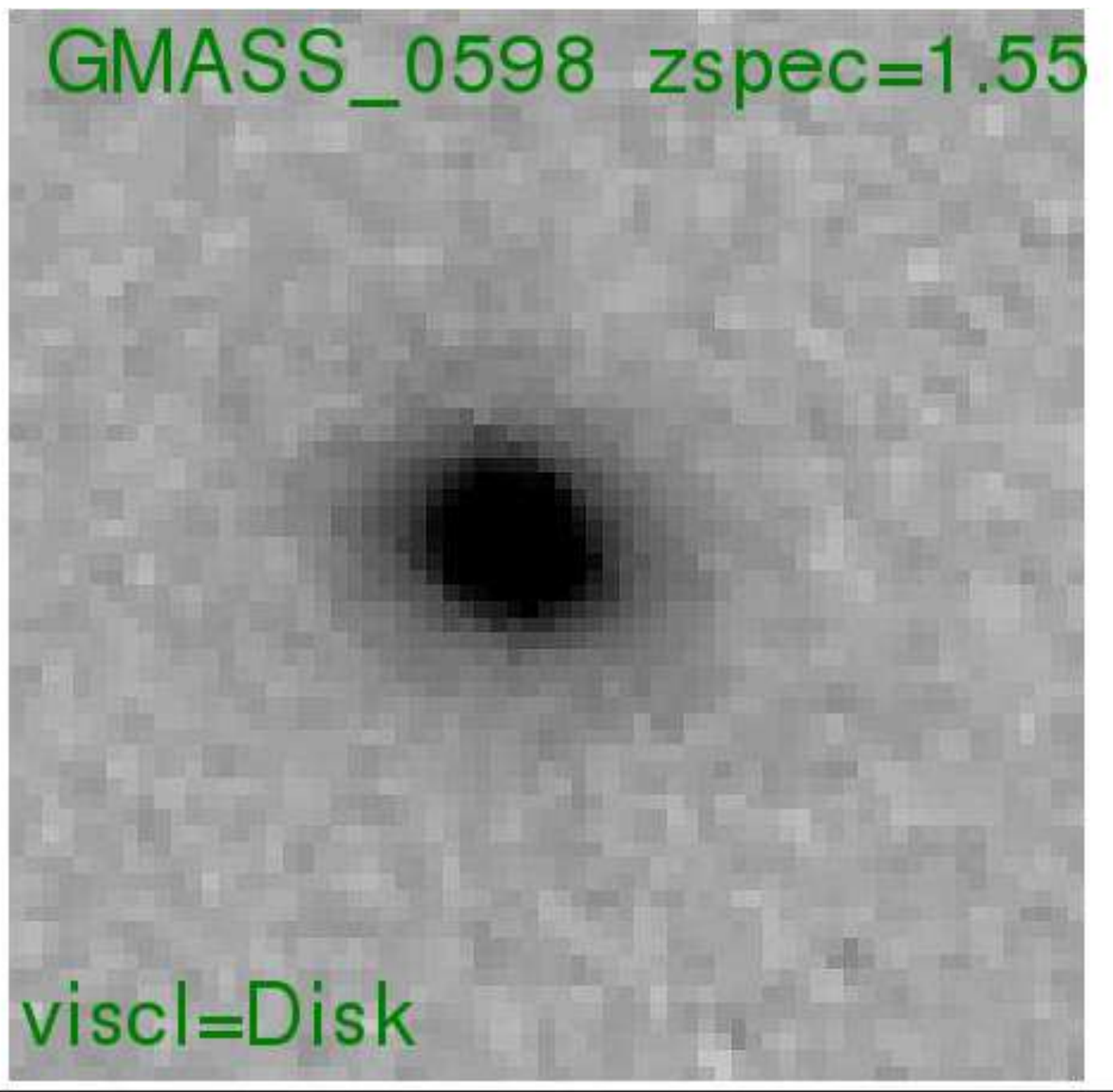}			     
\includegraphics[trim=100 40 75 390, clip=true, width=30mm]{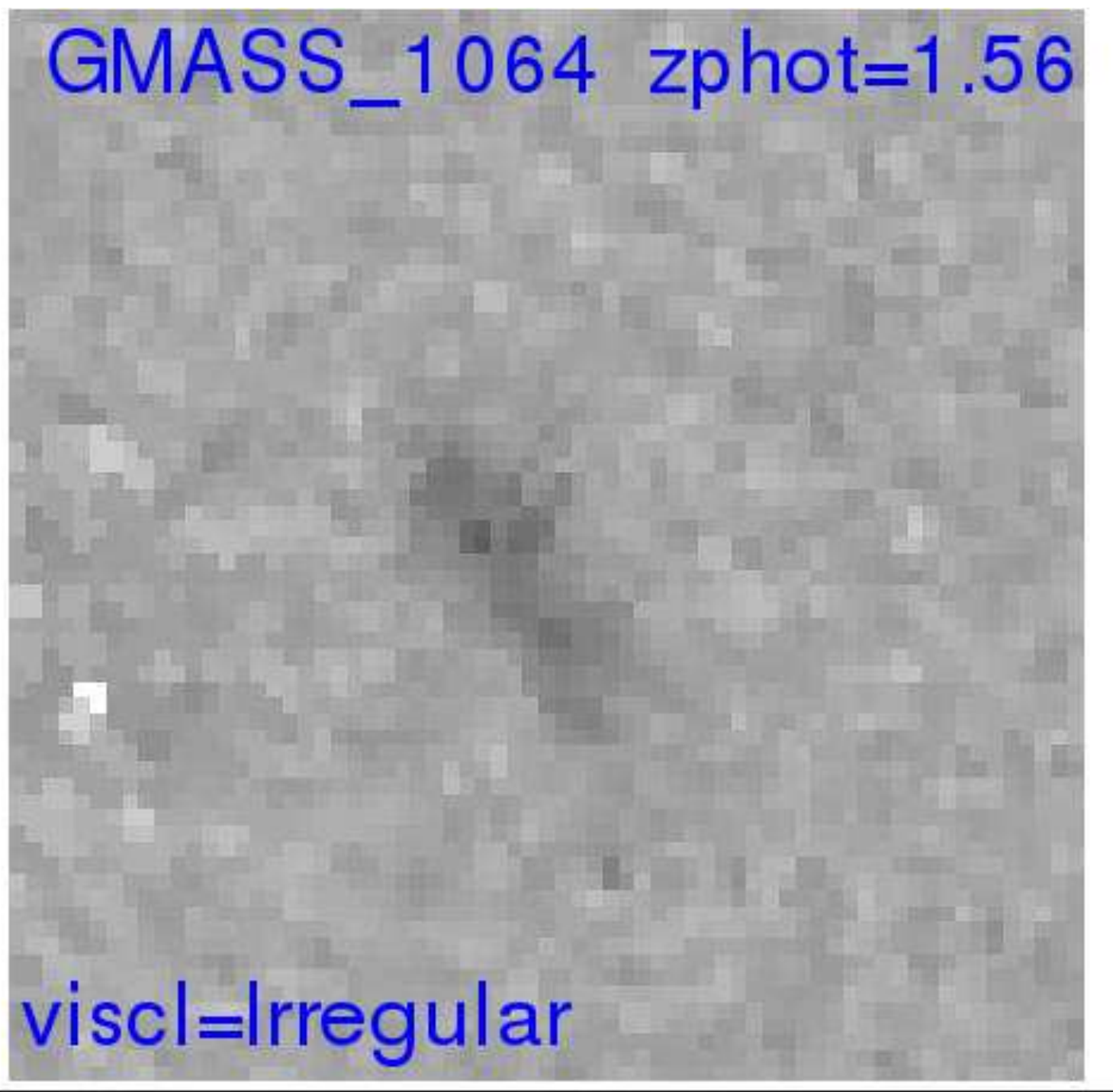}			     
\includegraphics[trim=100 40 75 390, clip=true, width=30mm]{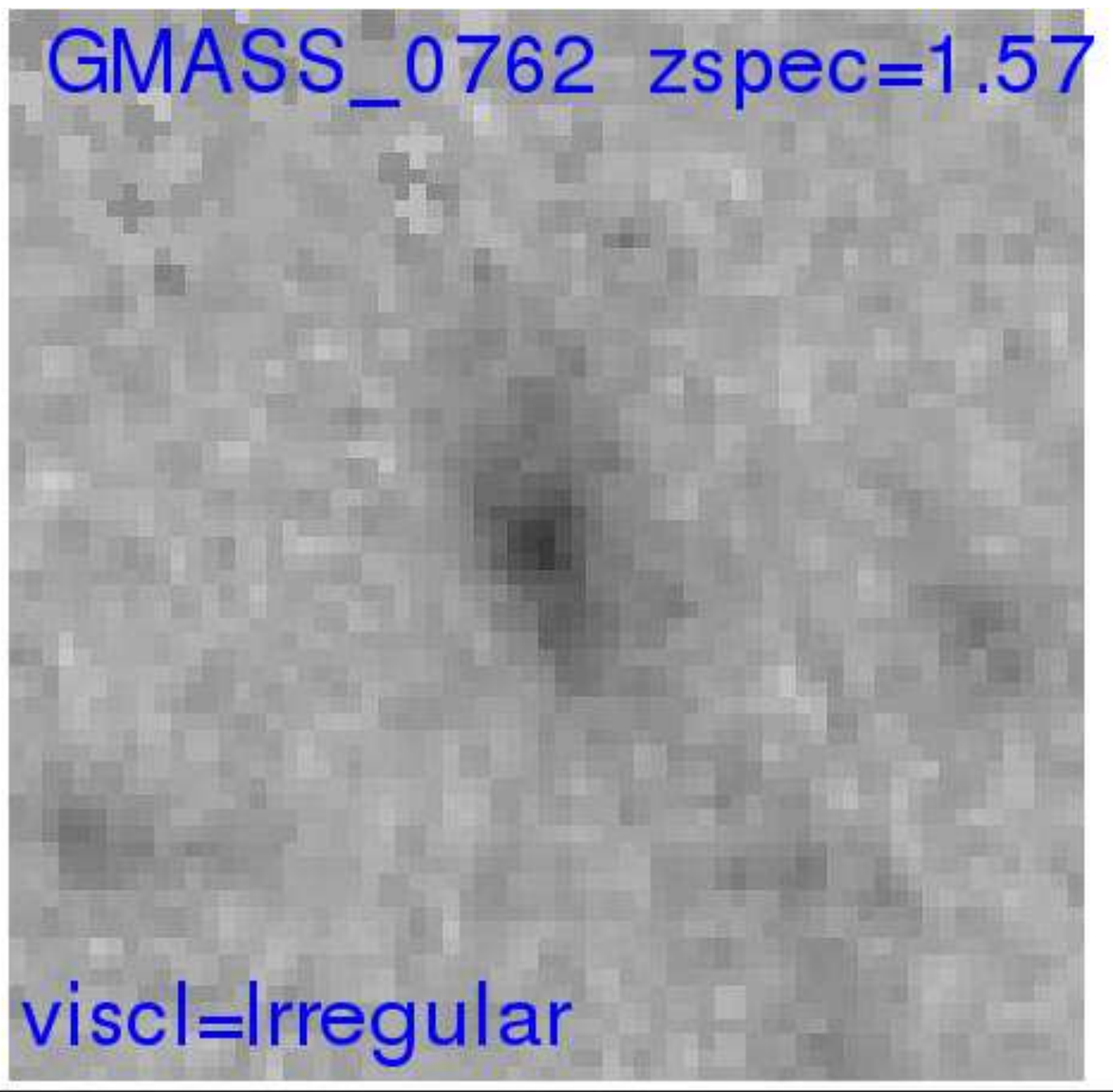}			     
\includegraphics[trim=100 40 75 390, clip=true, width=30mm]{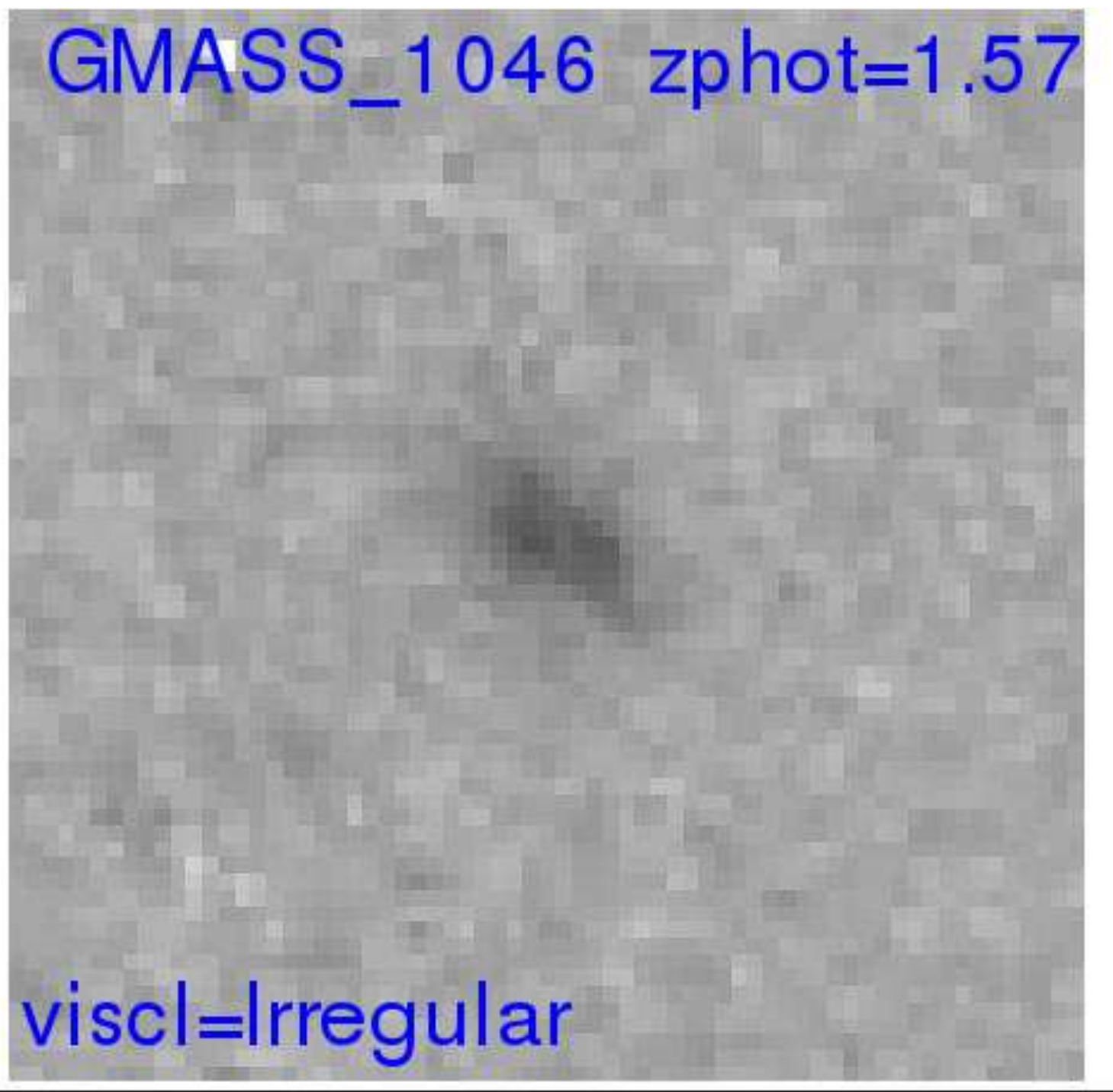}			     
\includegraphics[trim=100 40 75 390, clip=true, width=30mm]{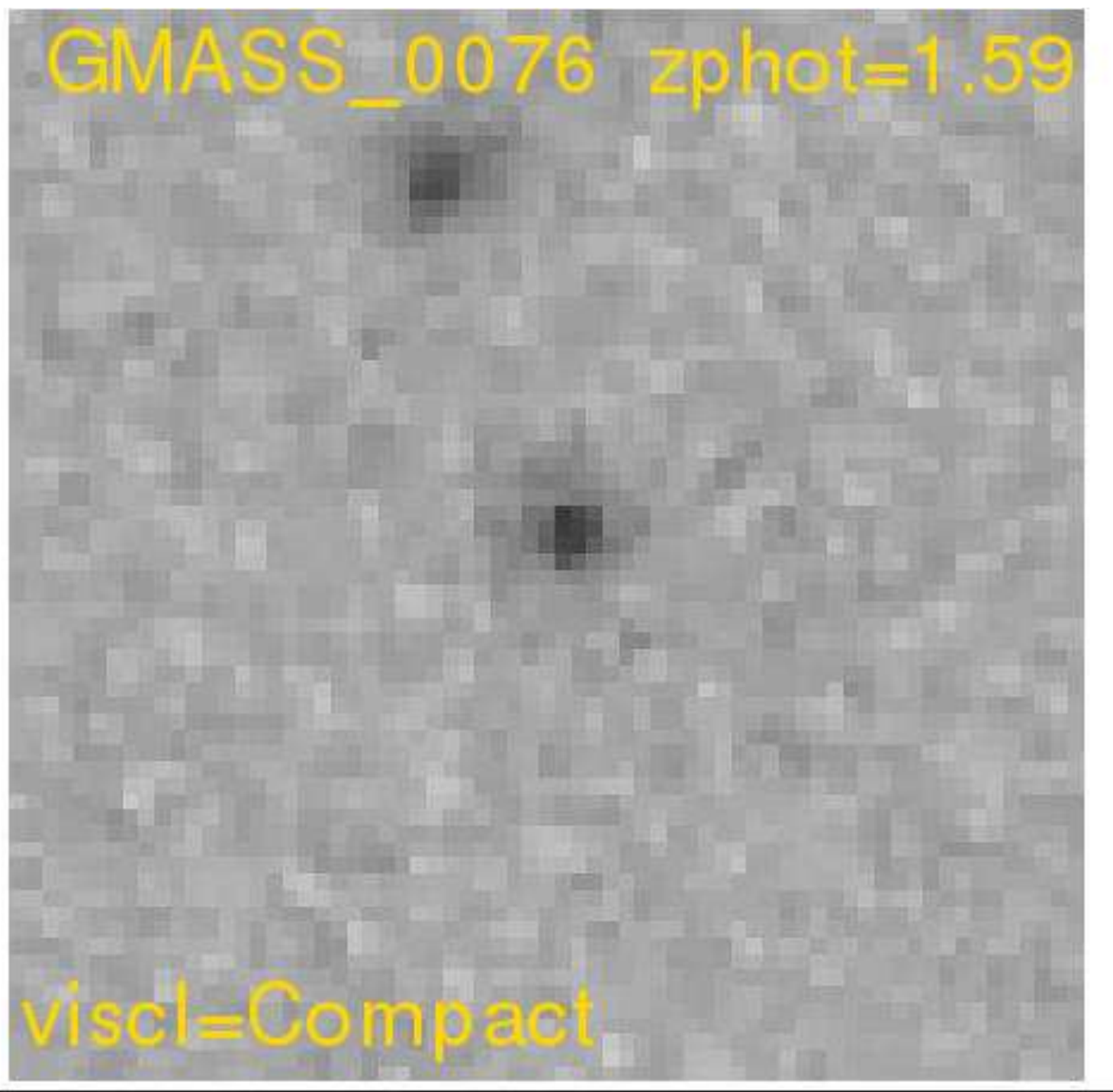}			     

\includegraphics[trim=100 40 75 390, clip=true, width=30mm]{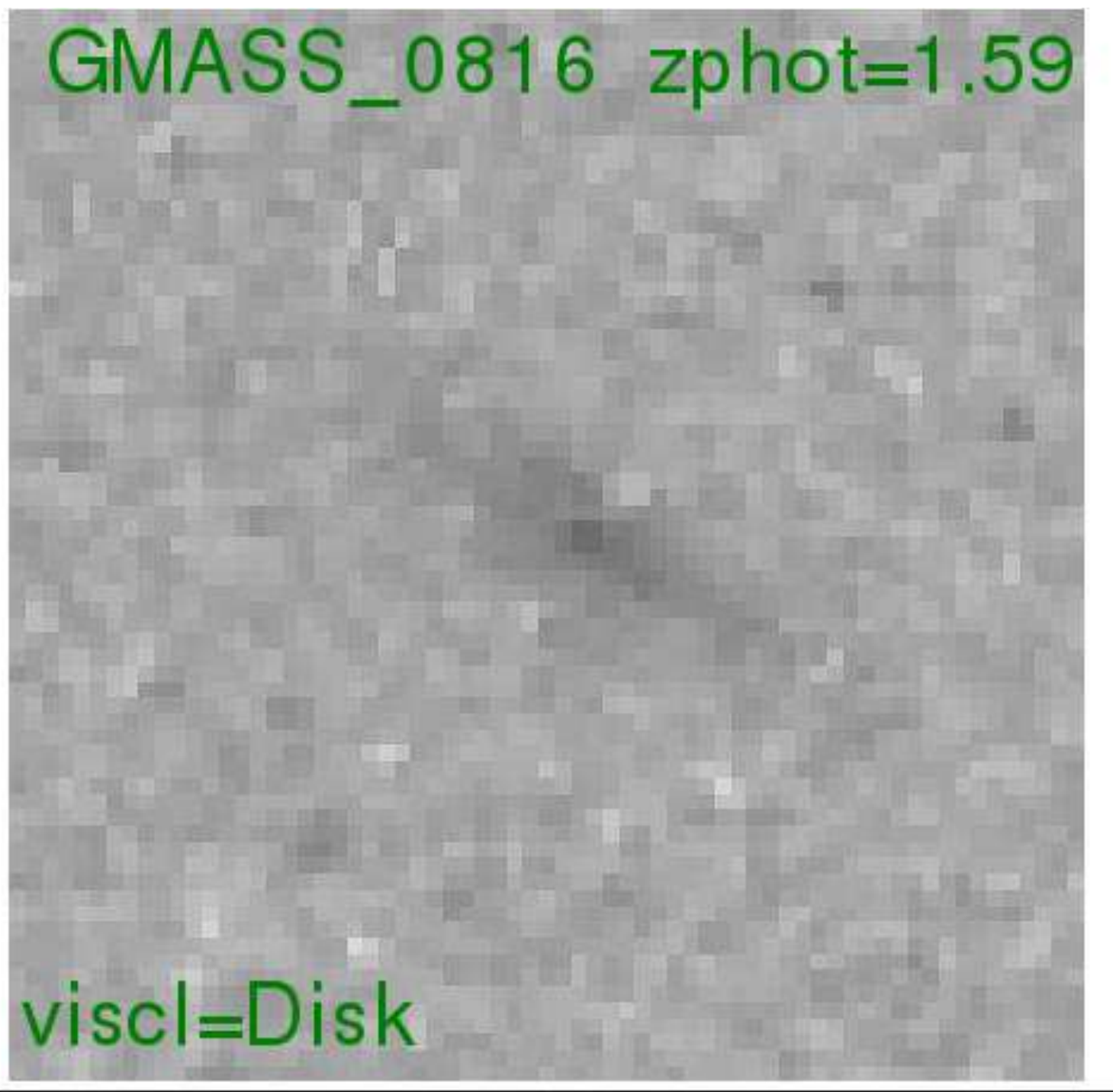}			     
\includegraphics[trim=100 40 75 390, clip=true, width=30mm]{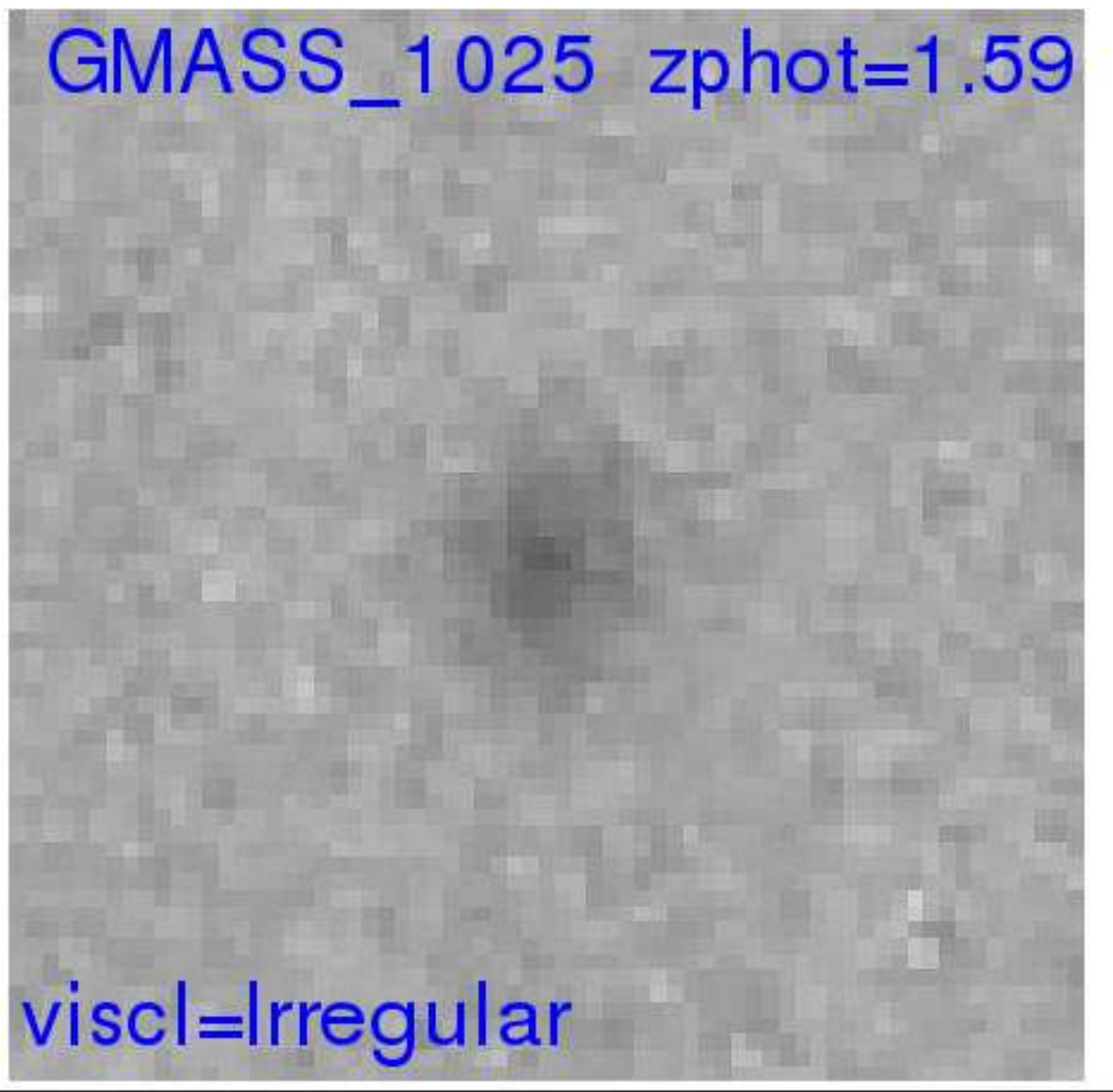}			     
\includegraphics[trim=100 40 75 390, clip=true, width=30mm]{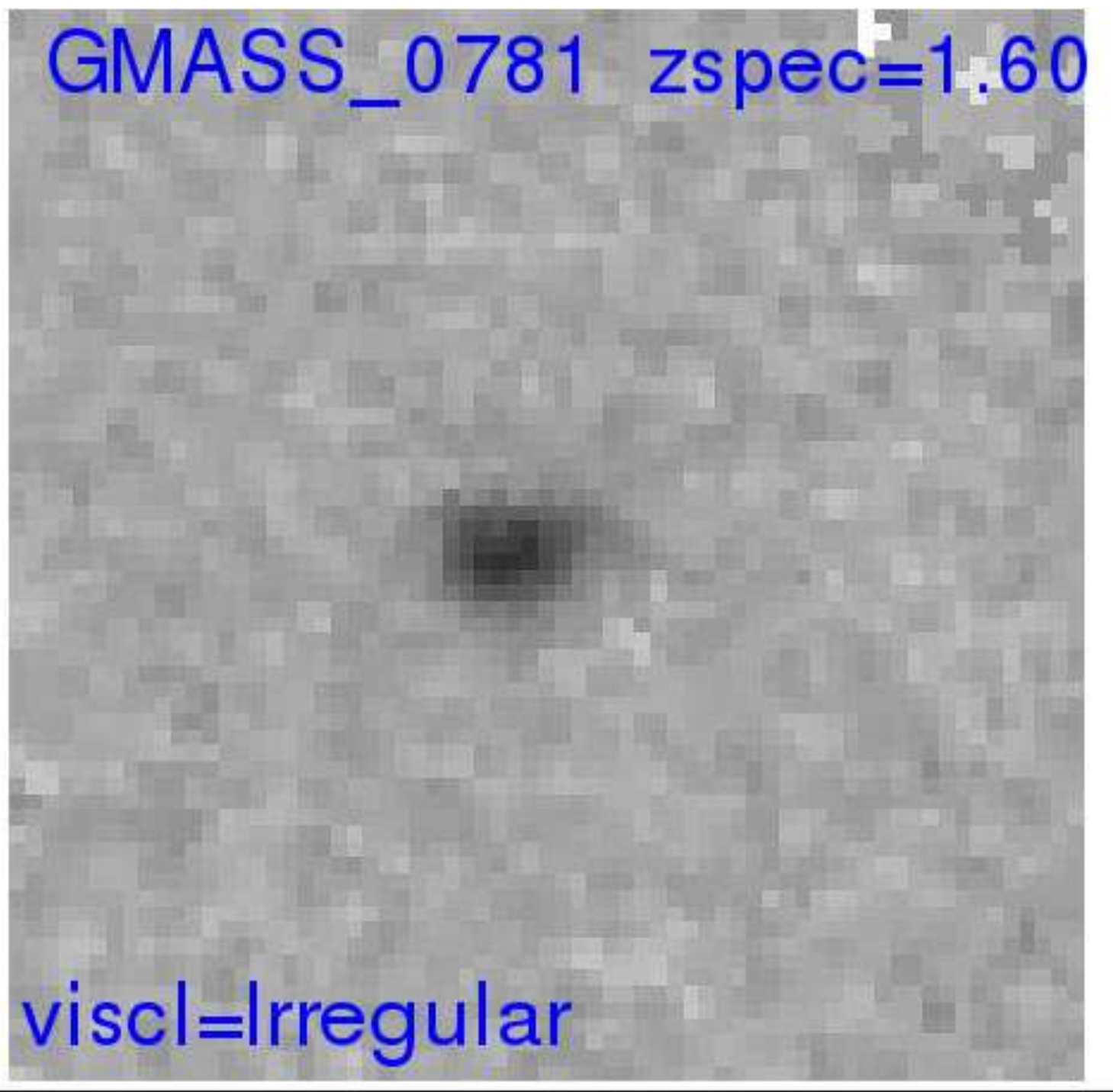}			     
\includegraphics[trim=100 40 75 390, clip=true, width=30mm]{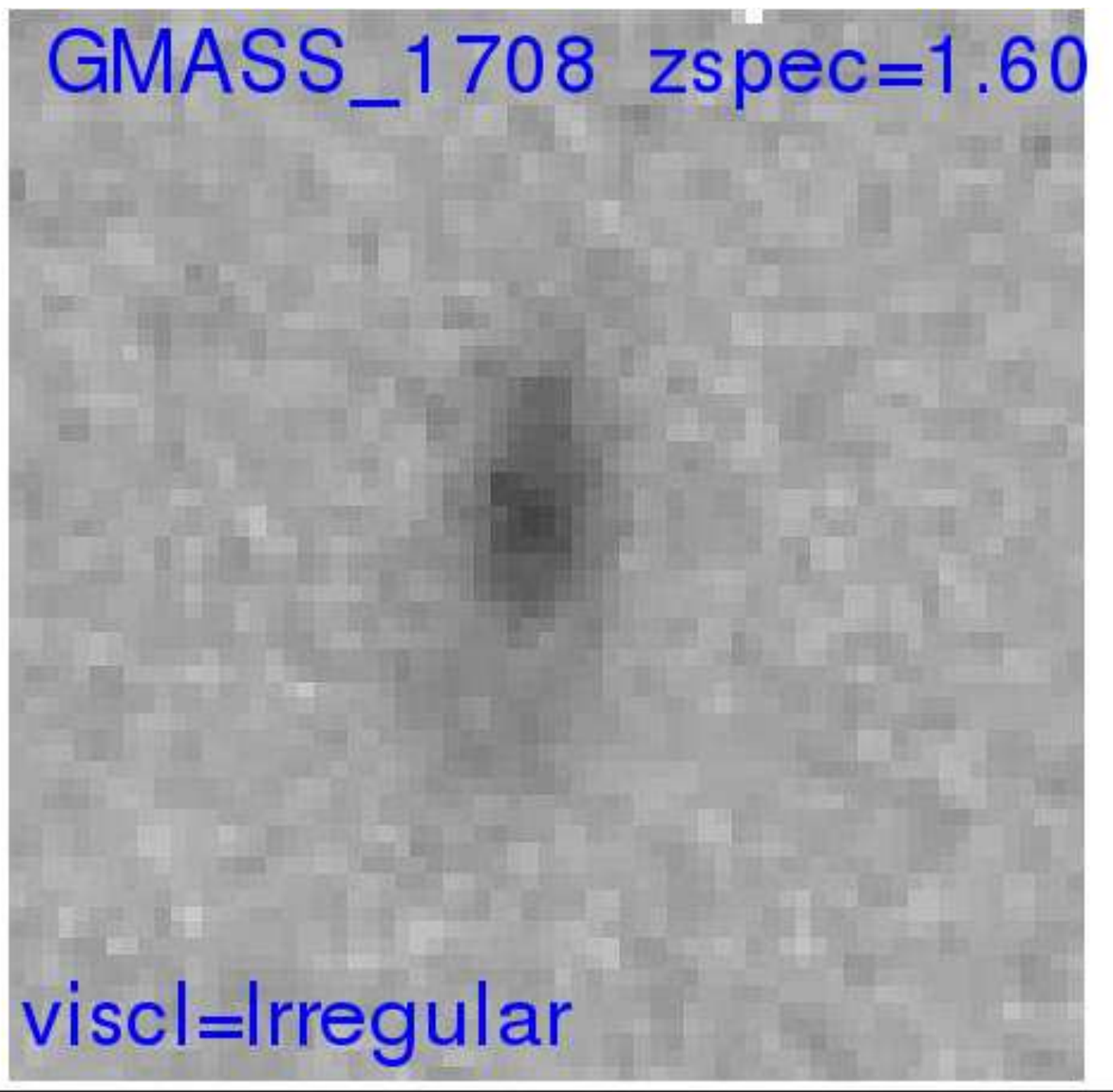}			     
\includegraphics[trim=100 40 75 390, clip=true, width=30mm]{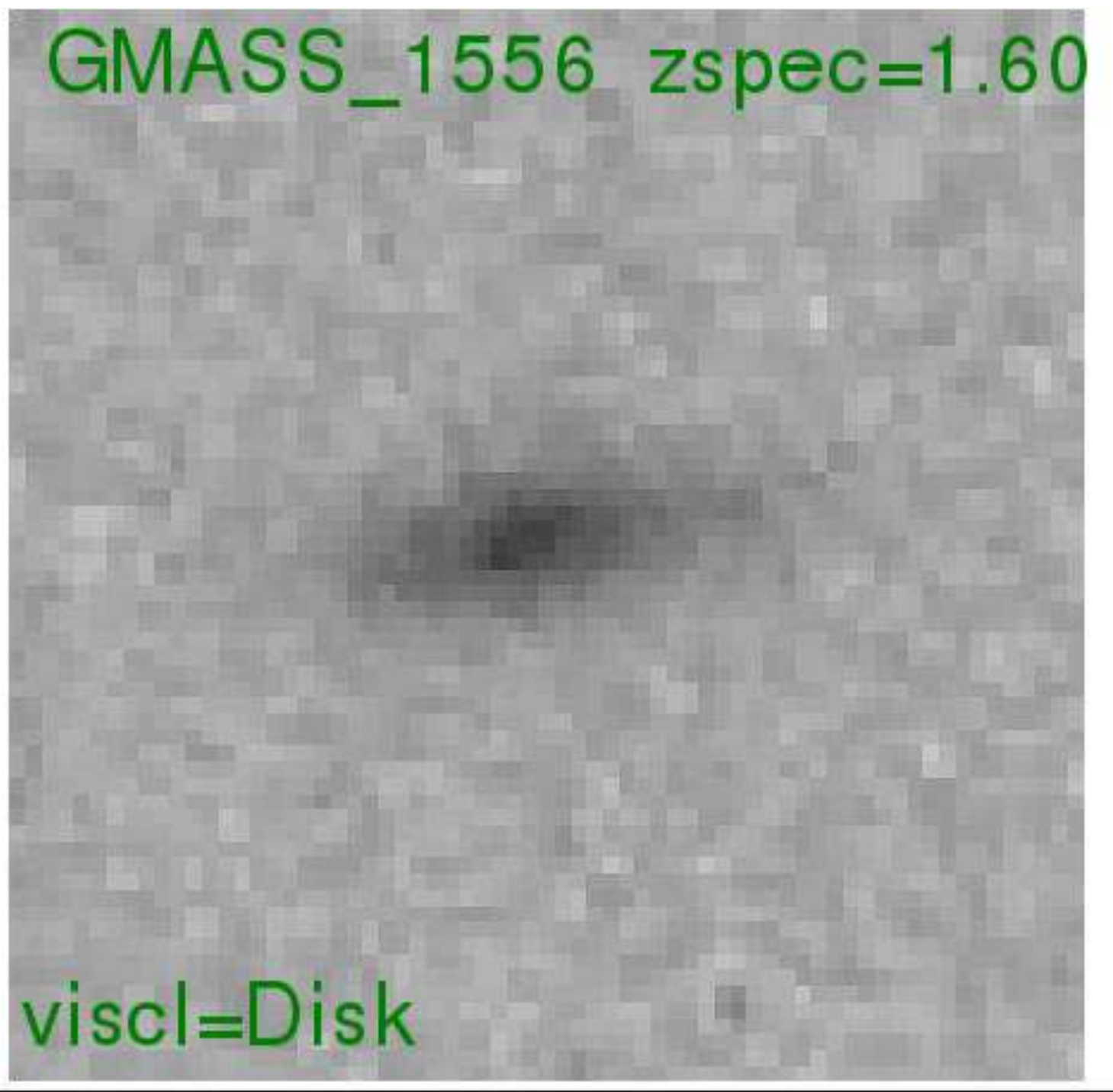}			     
\includegraphics[trim=100 40 75 390, clip=true, width=30mm]{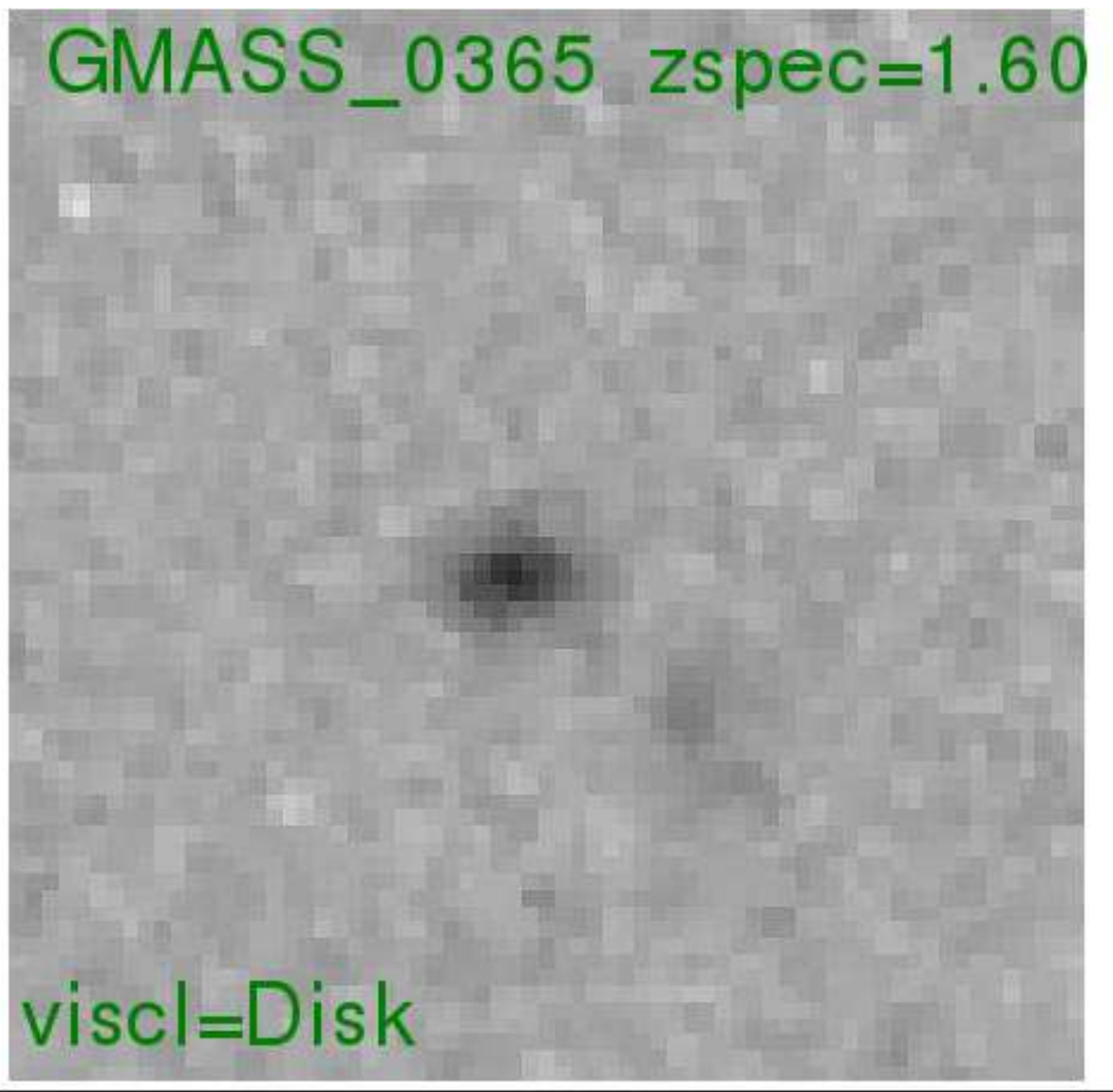}			     

\includegraphics[trim=100 40 75 390, clip=true, width=30mm]{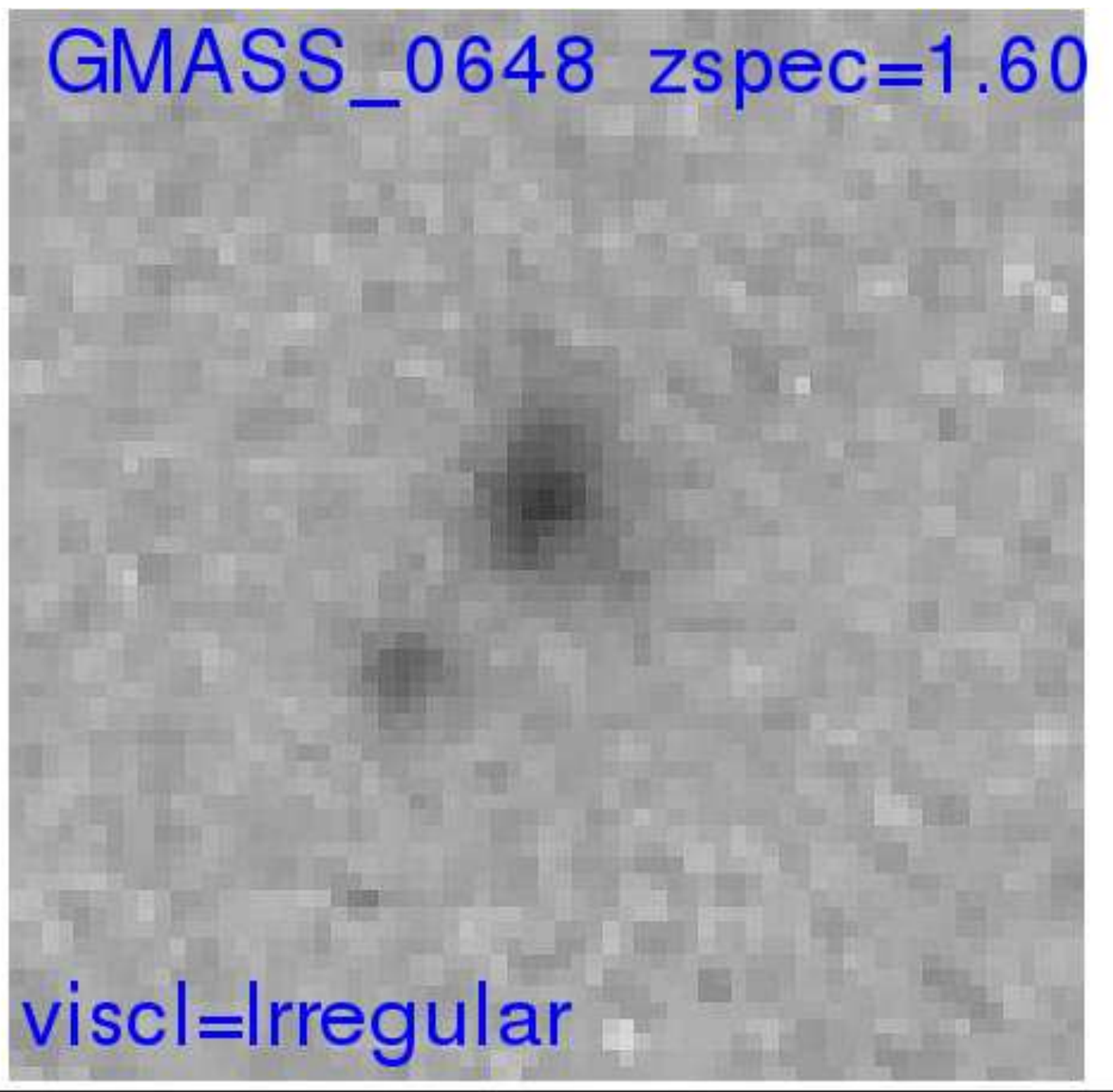}			     
\includegraphics[trim=100 40 75 390, clip=true, width=30mm]{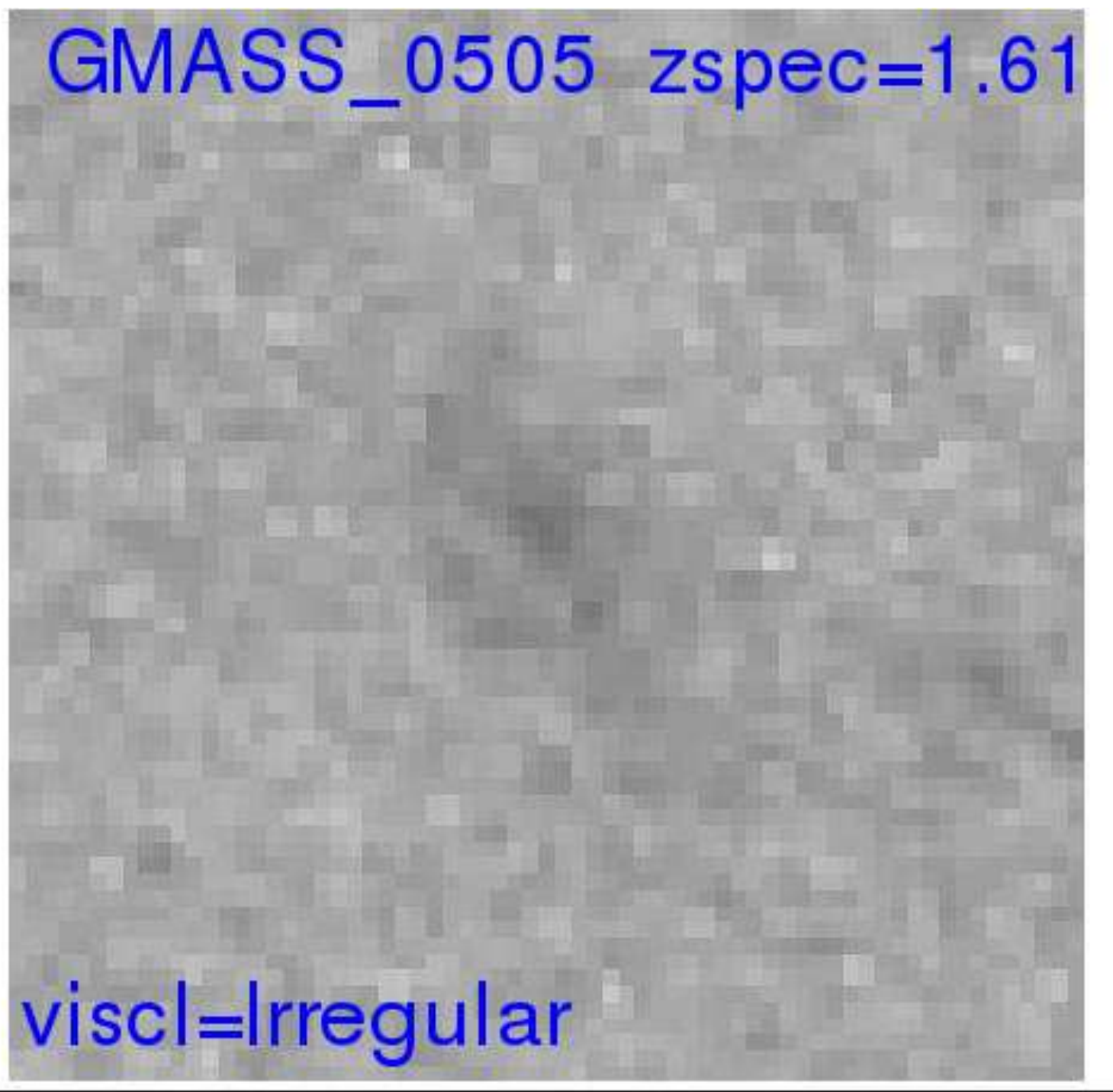}			     
\includegraphics[trim=100 40 75 390, clip=true, width=30mm]{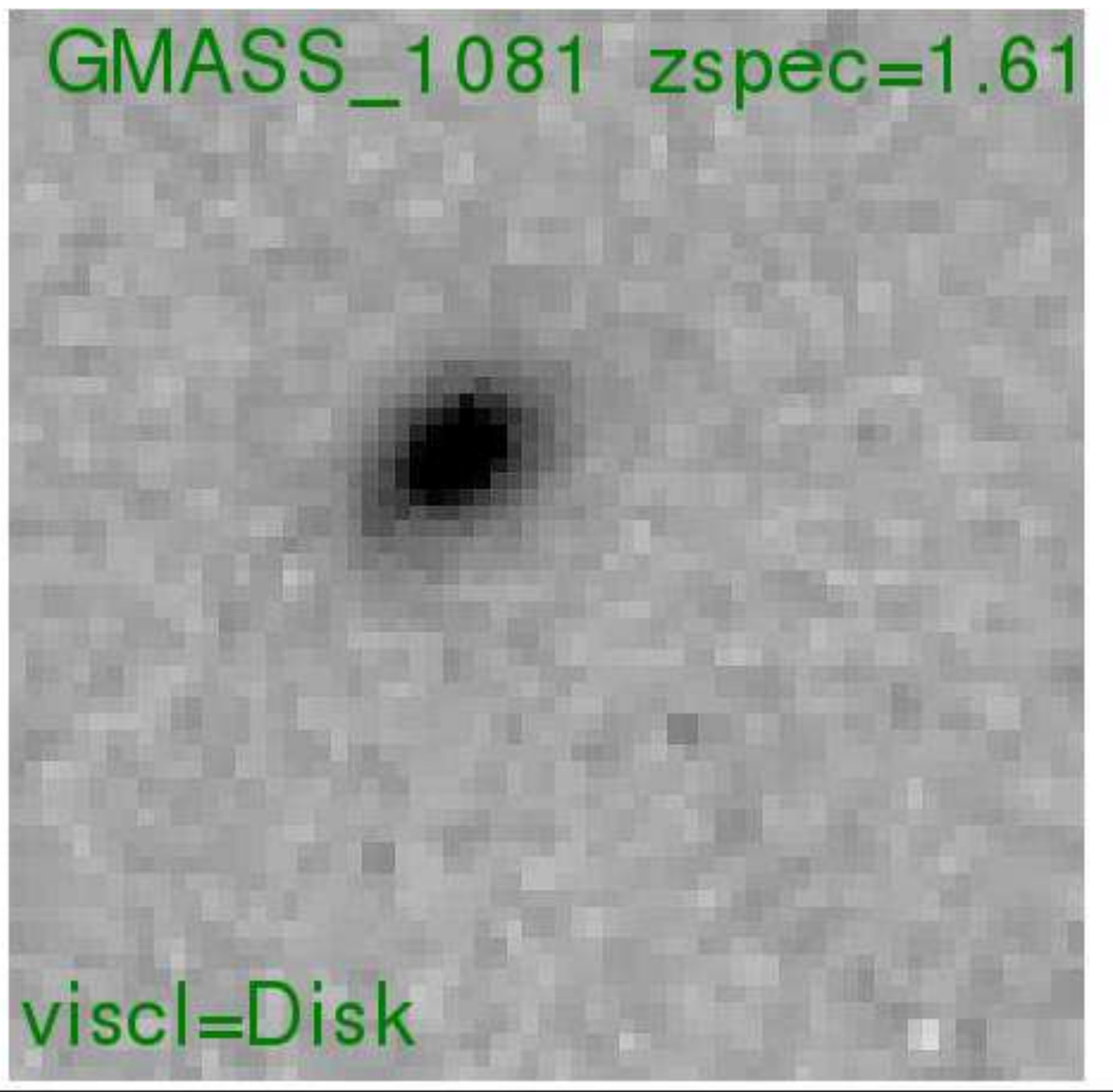}			     
\includegraphics[trim=100 40 75 390, clip=true, width=30mm]{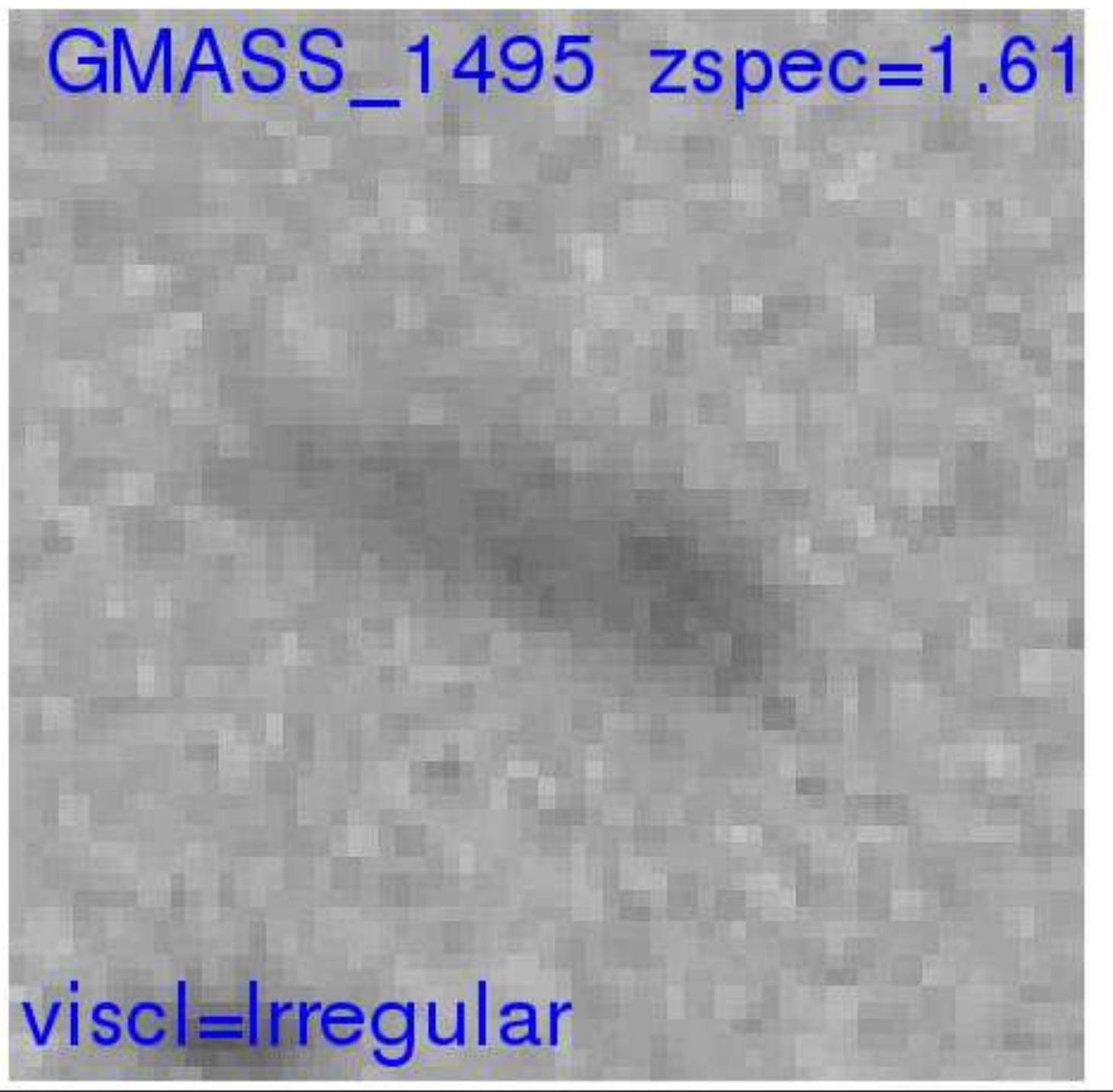}		     
\includegraphics[trim=100 40 75 390, clip=true, width=30mm]{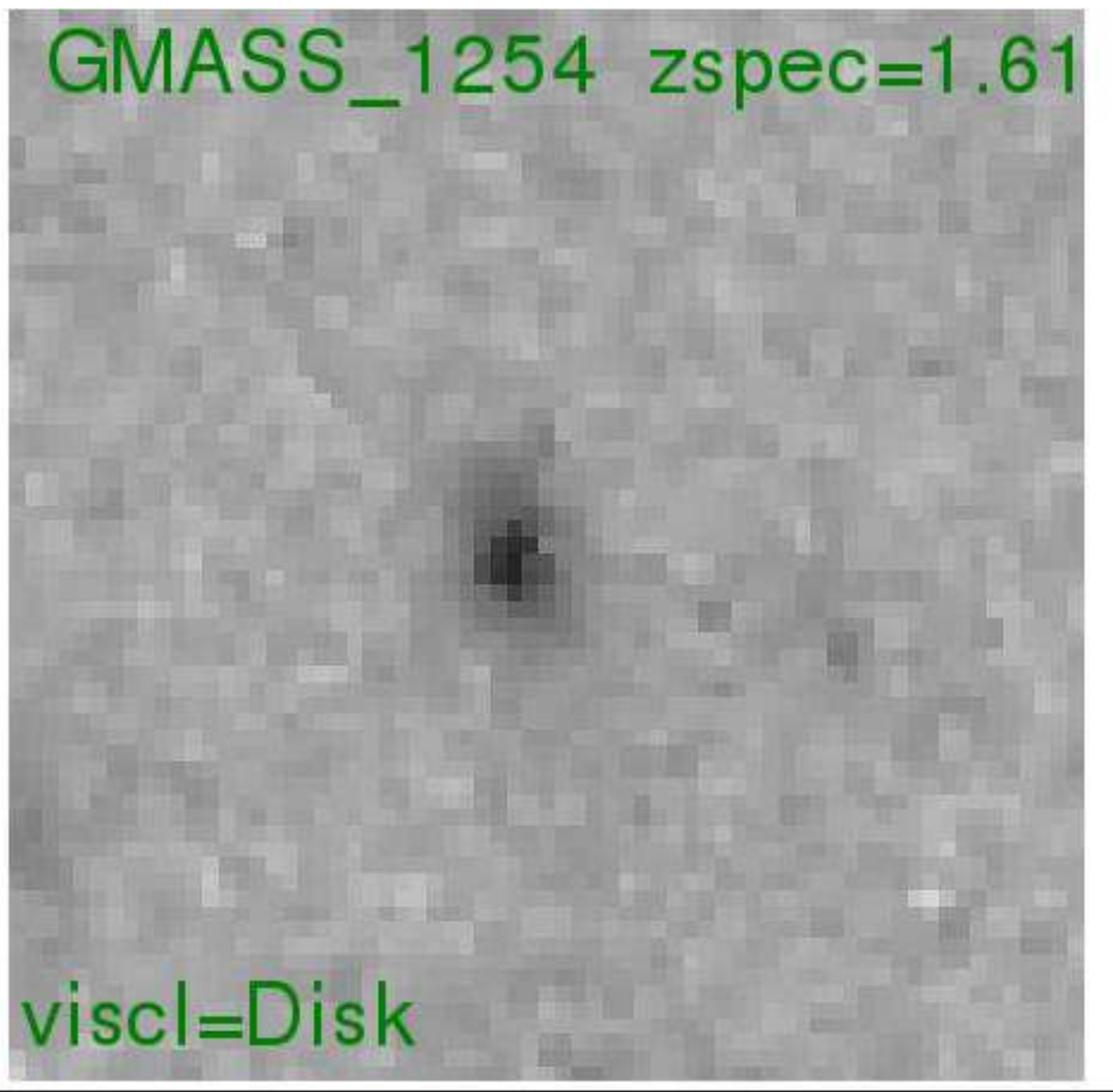}			     
\includegraphics[trim=100 40 75 390, clip=true, width=30mm]{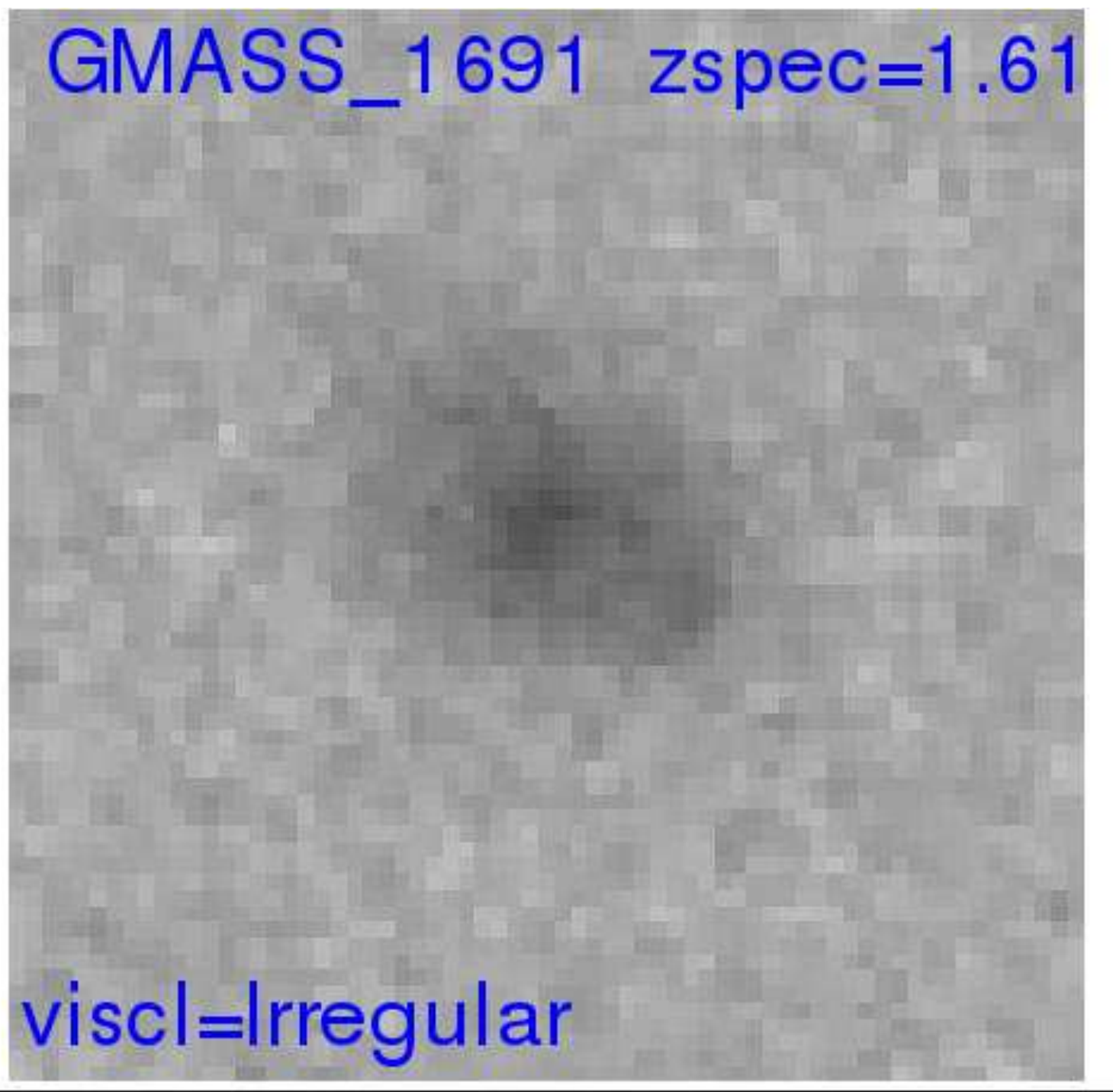}			     

\includegraphics[trim=100 40 75 390, clip=true, width=30mm]{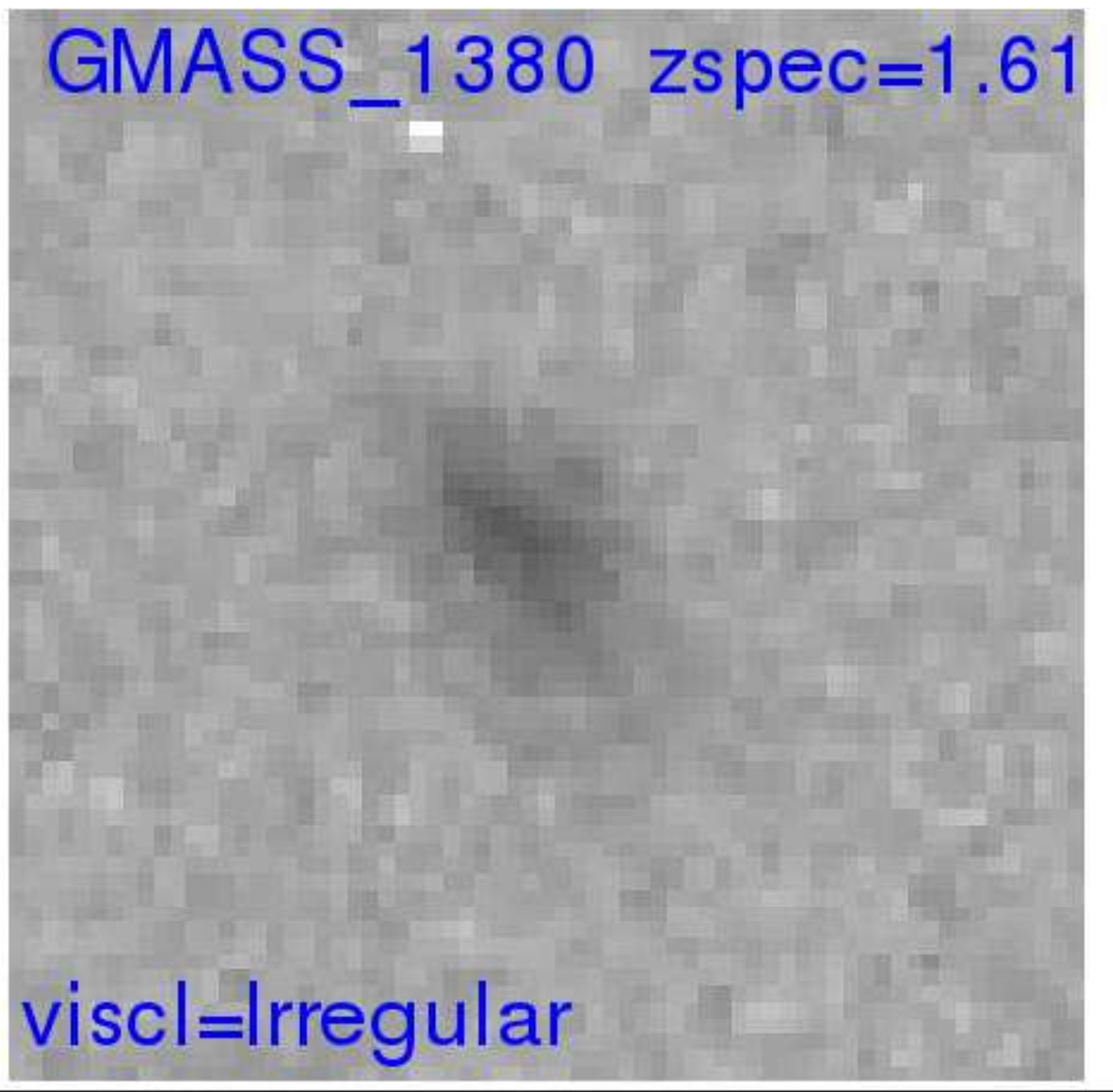}			     
\includegraphics[trim=100 40 75 390, clip=true, width=30mm]{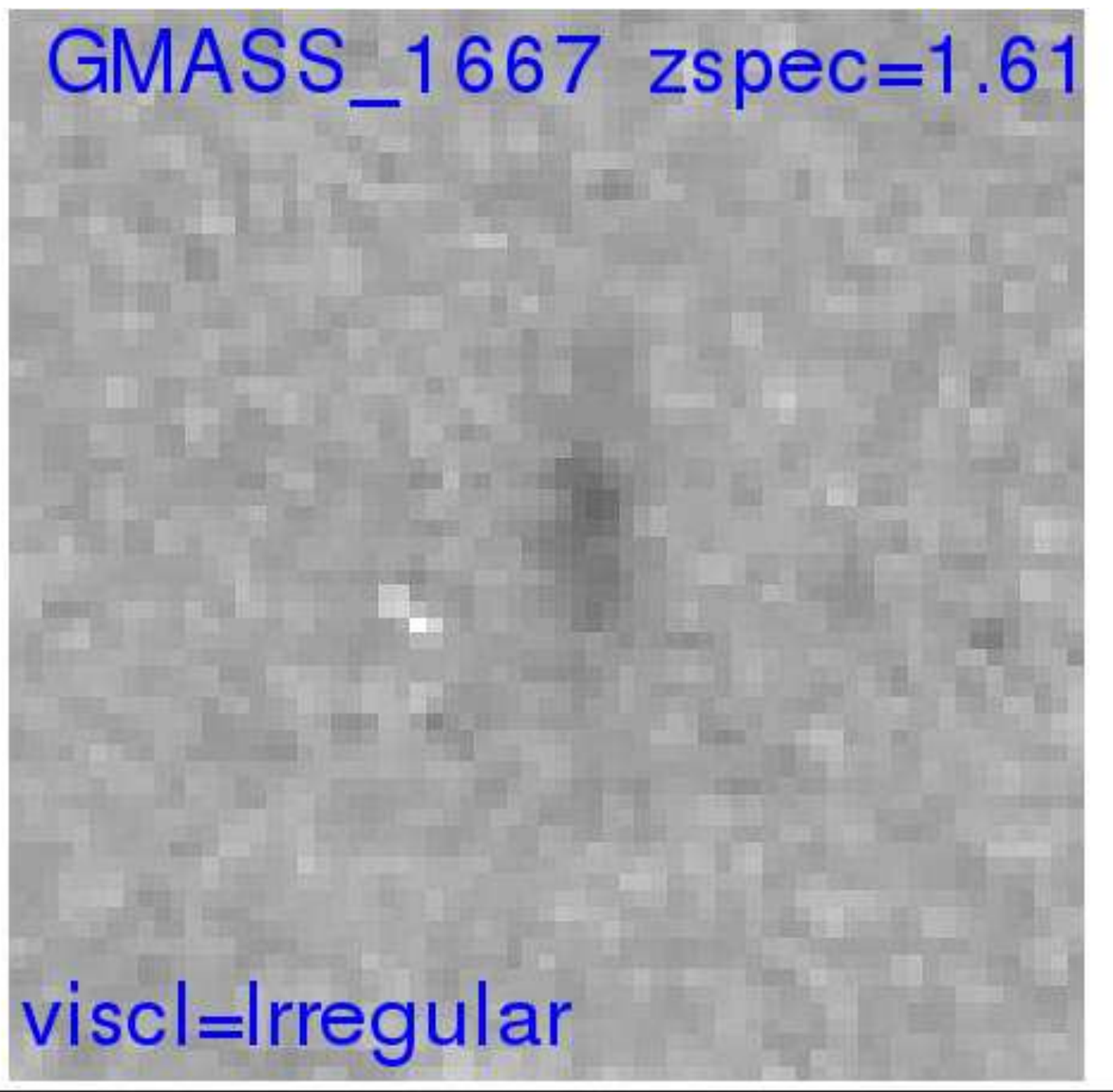}			     
\includegraphics[trim=100 40 75 390, clip=true, width=30mm]{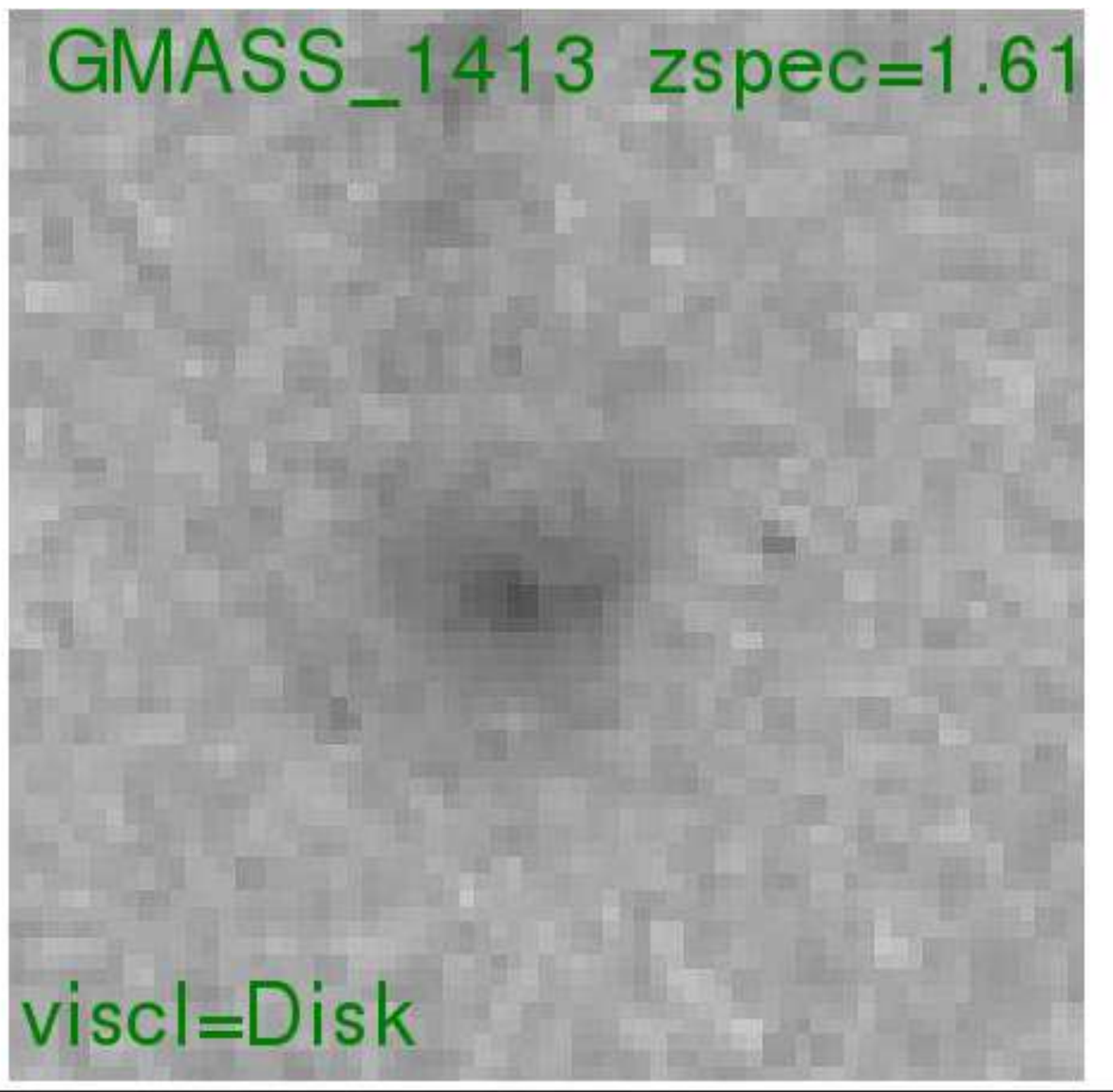}			     
\includegraphics[trim=100 40 75 390, clip=true, width=30mm]{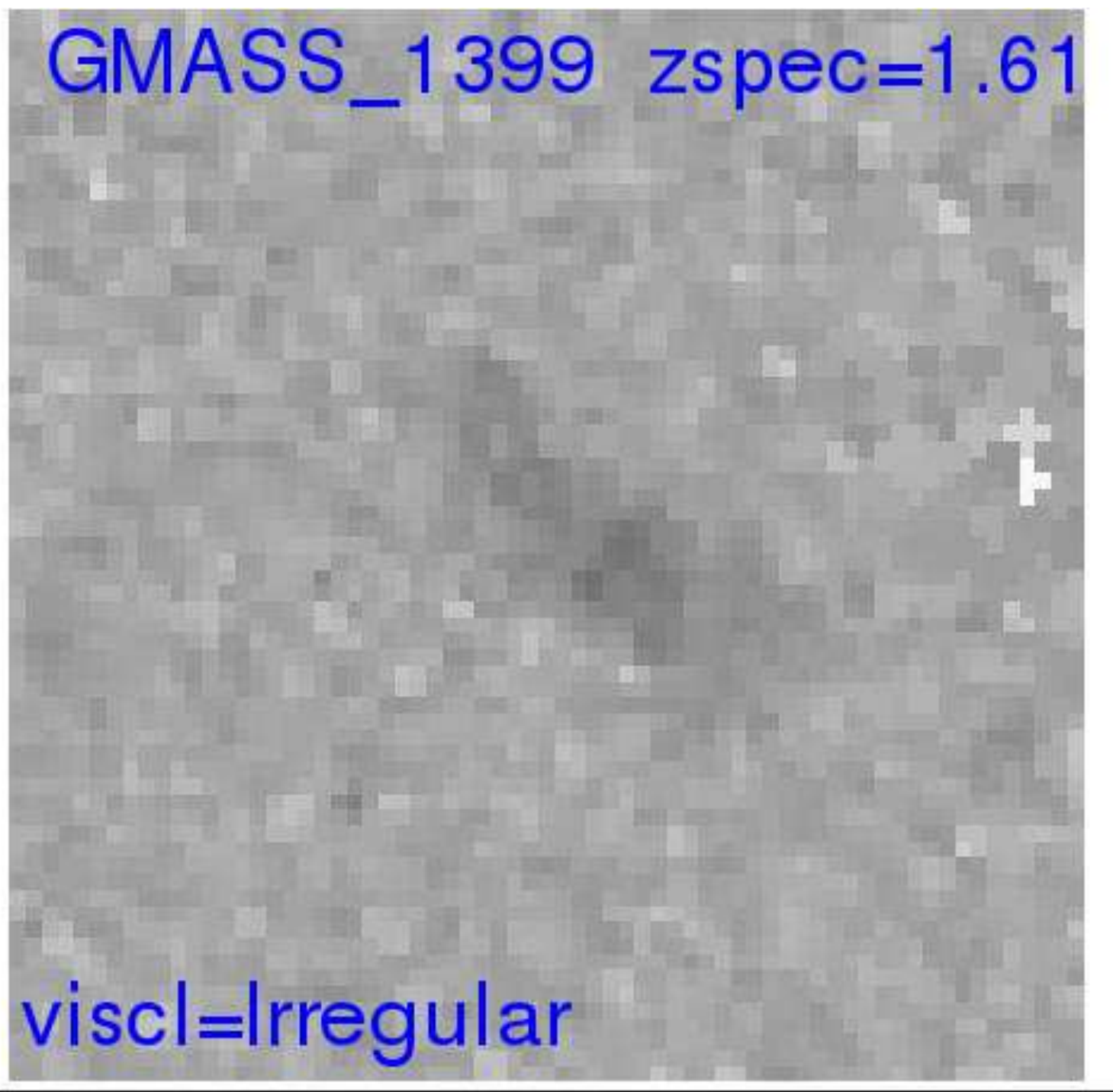}			     
\includegraphics[trim=100 40 75 390, clip=true, width=30mm]{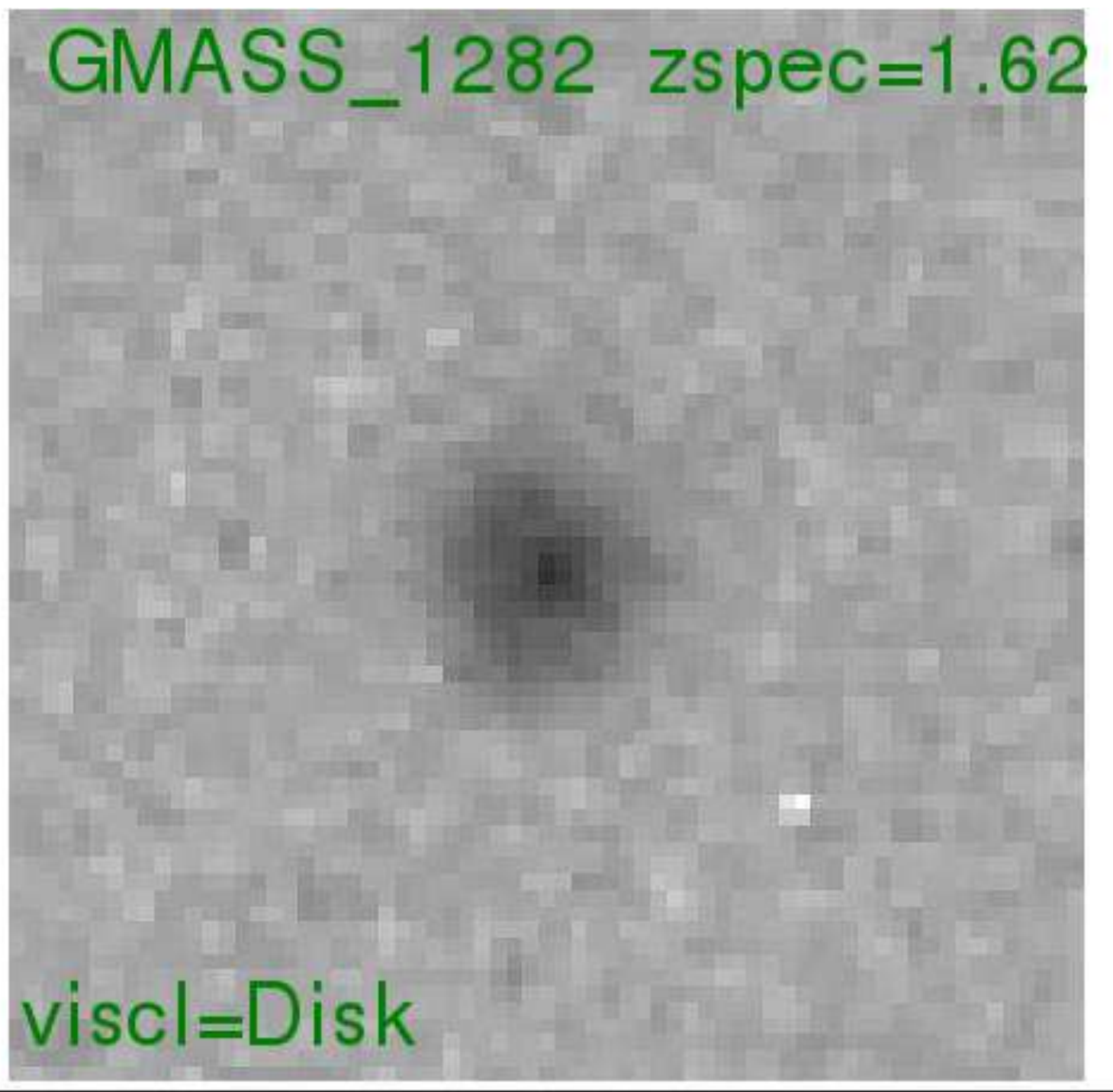}			     
\includegraphics[trim=100 40 75 390, clip=true, width=30mm]{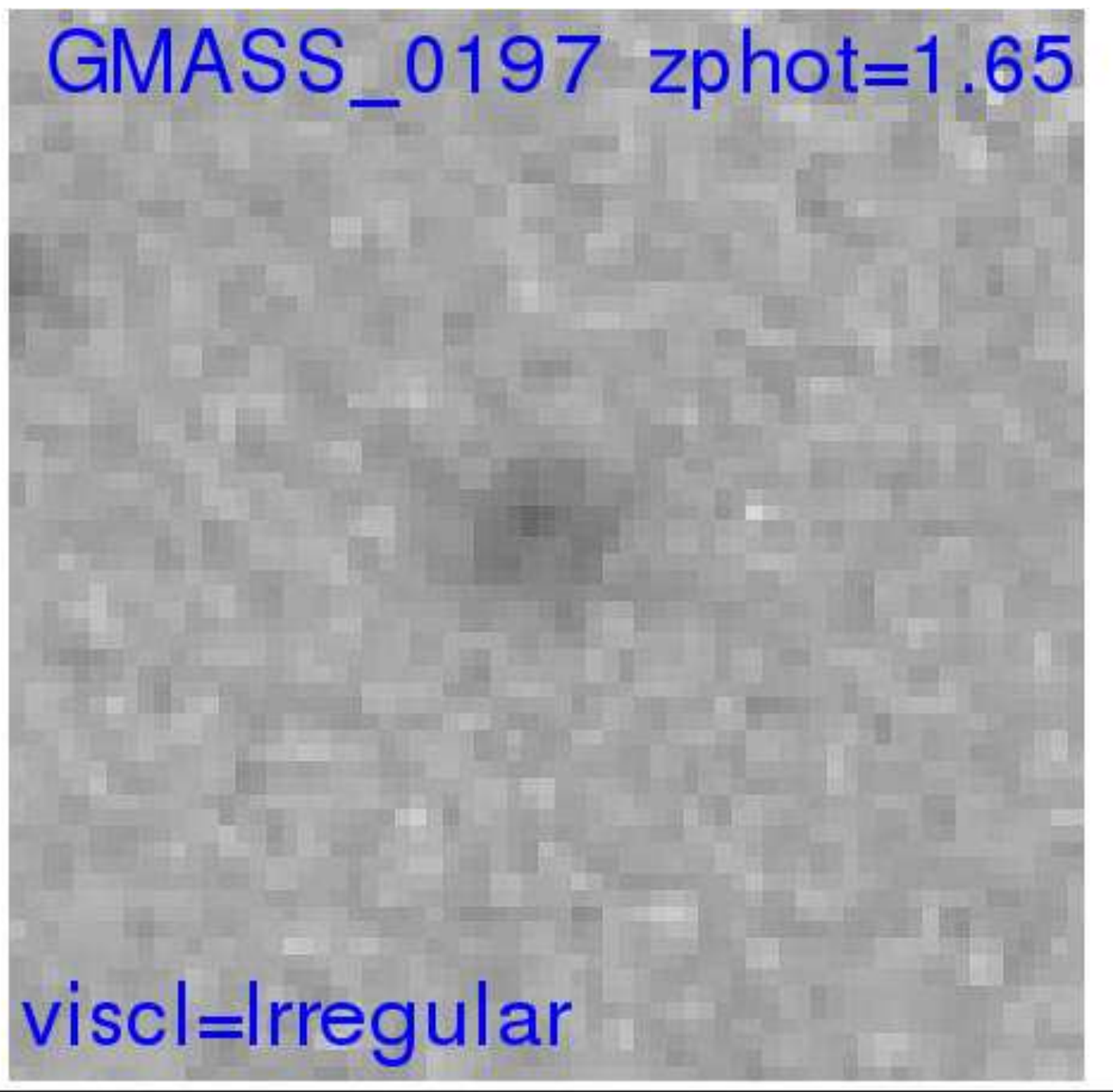}			     

\includegraphics[trim=100 40 75 390, clip=true, width=30mm]{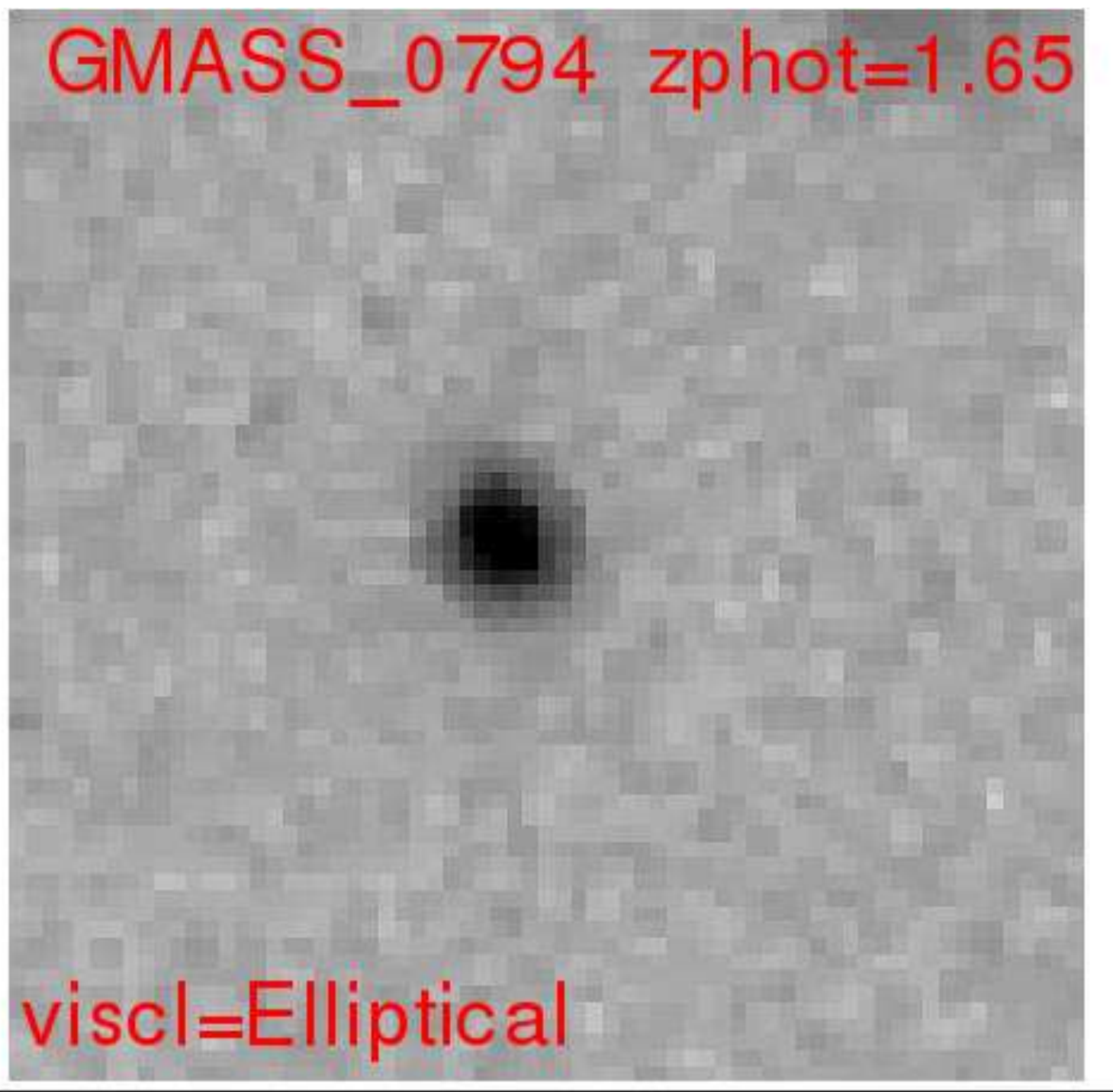}			     
\includegraphics[trim=100 40 75 390, clip=true, width=30mm]{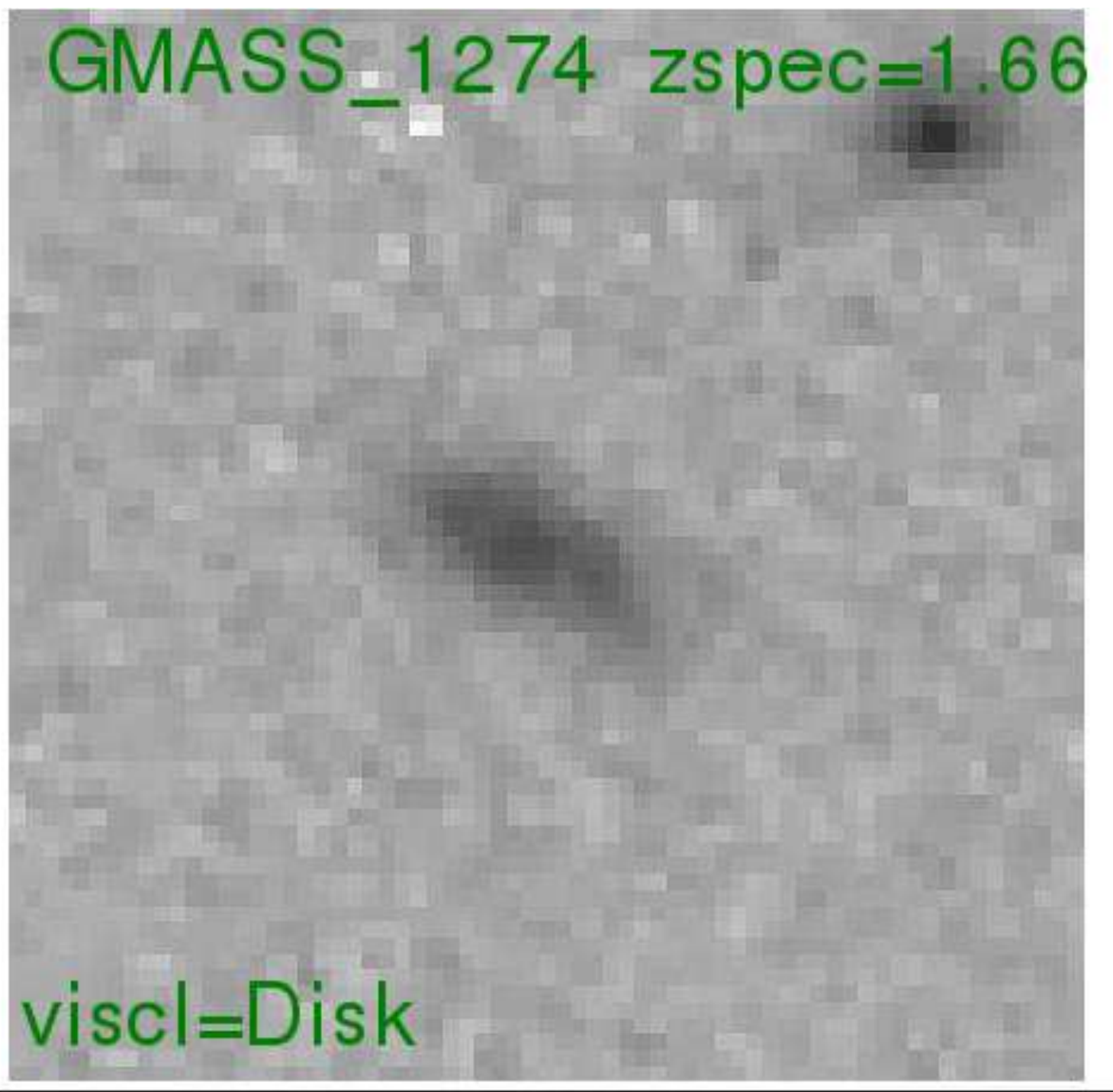}			     
\includegraphics[trim=100 40 75 390, clip=true, width=30mm]{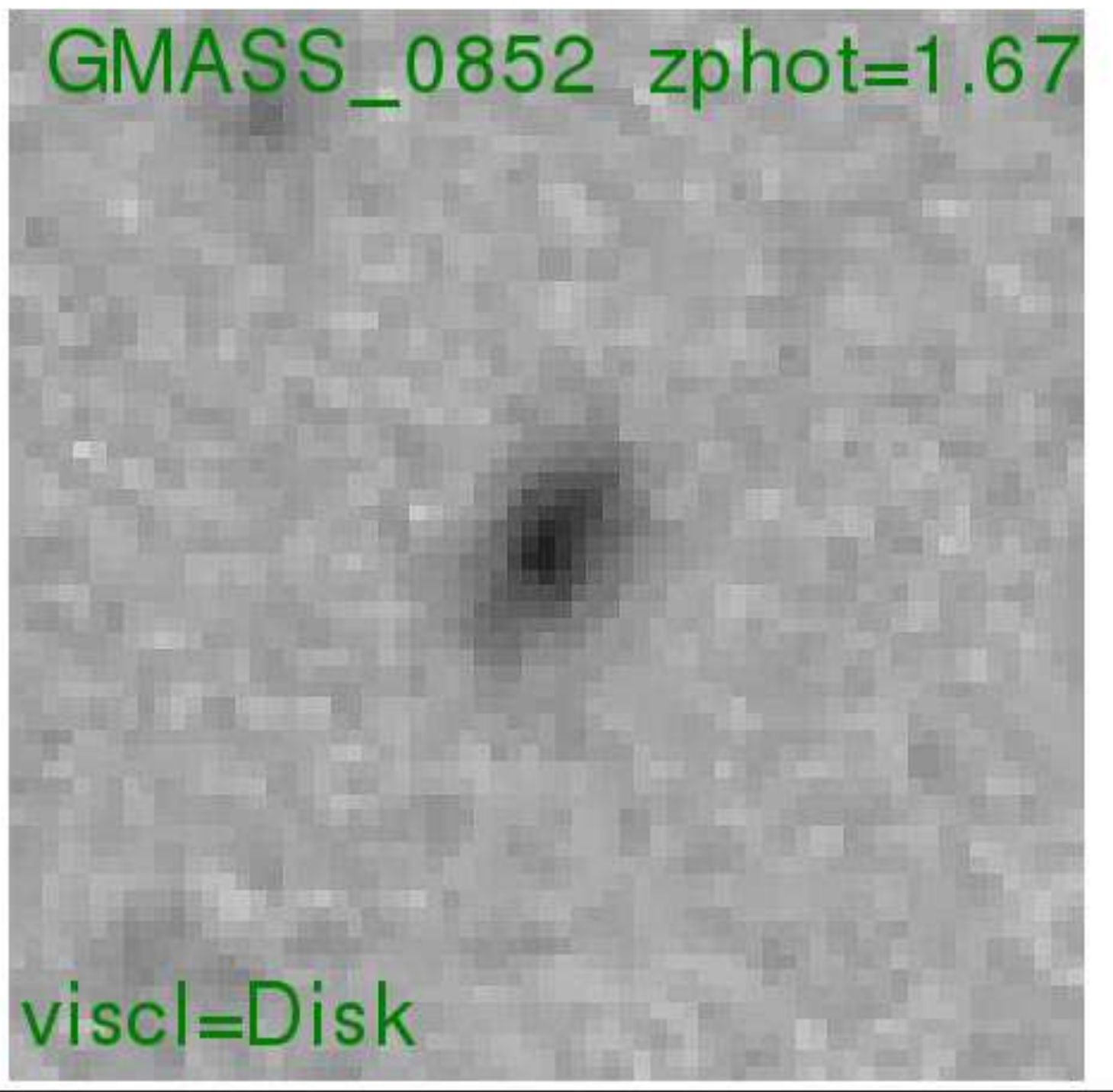}			     
\includegraphics[trim=100 40 75 390, clip=true, width=30mm]{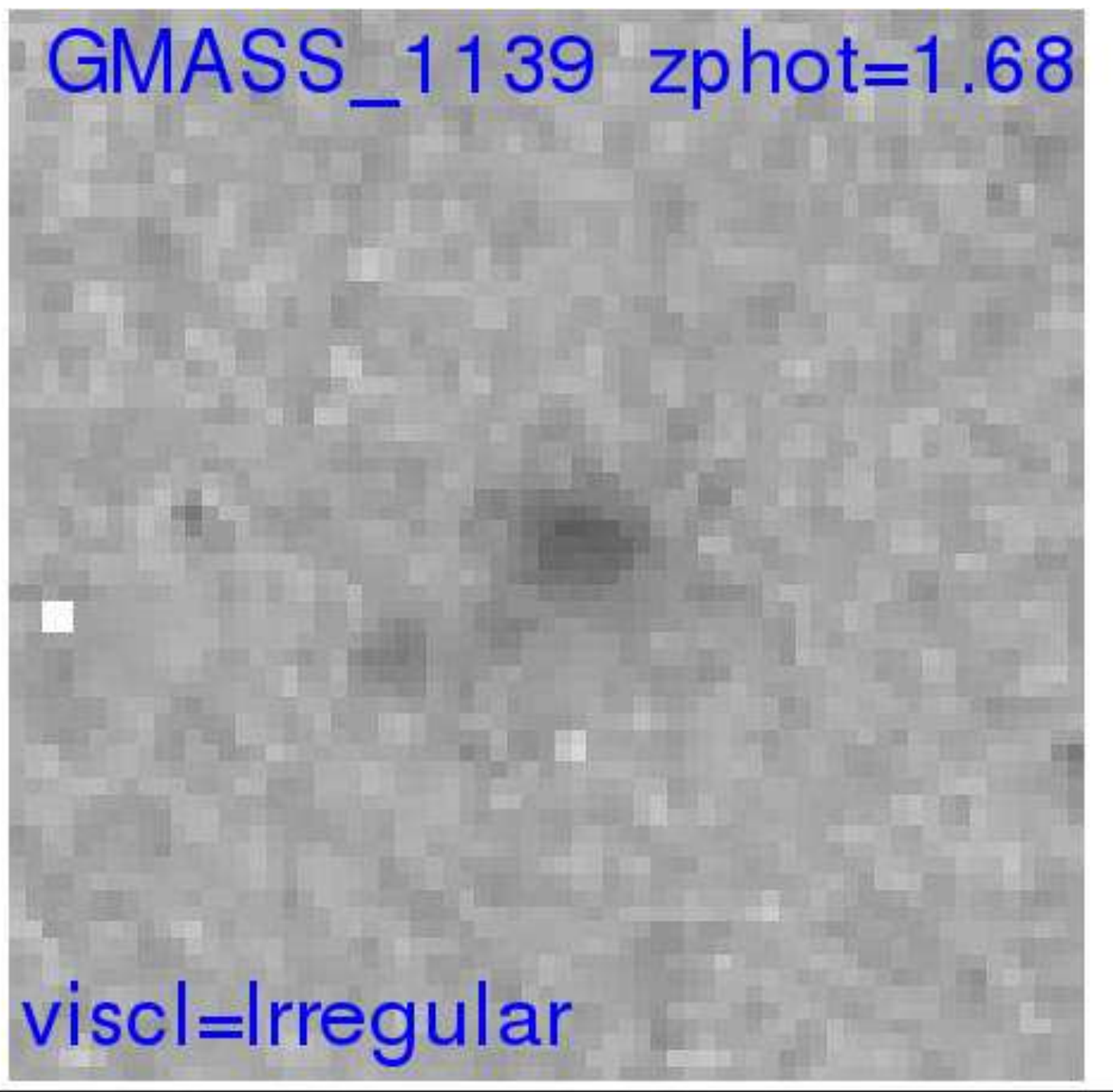}		     
\includegraphics[trim=100 40 75 390, clip=true, width=30mm]{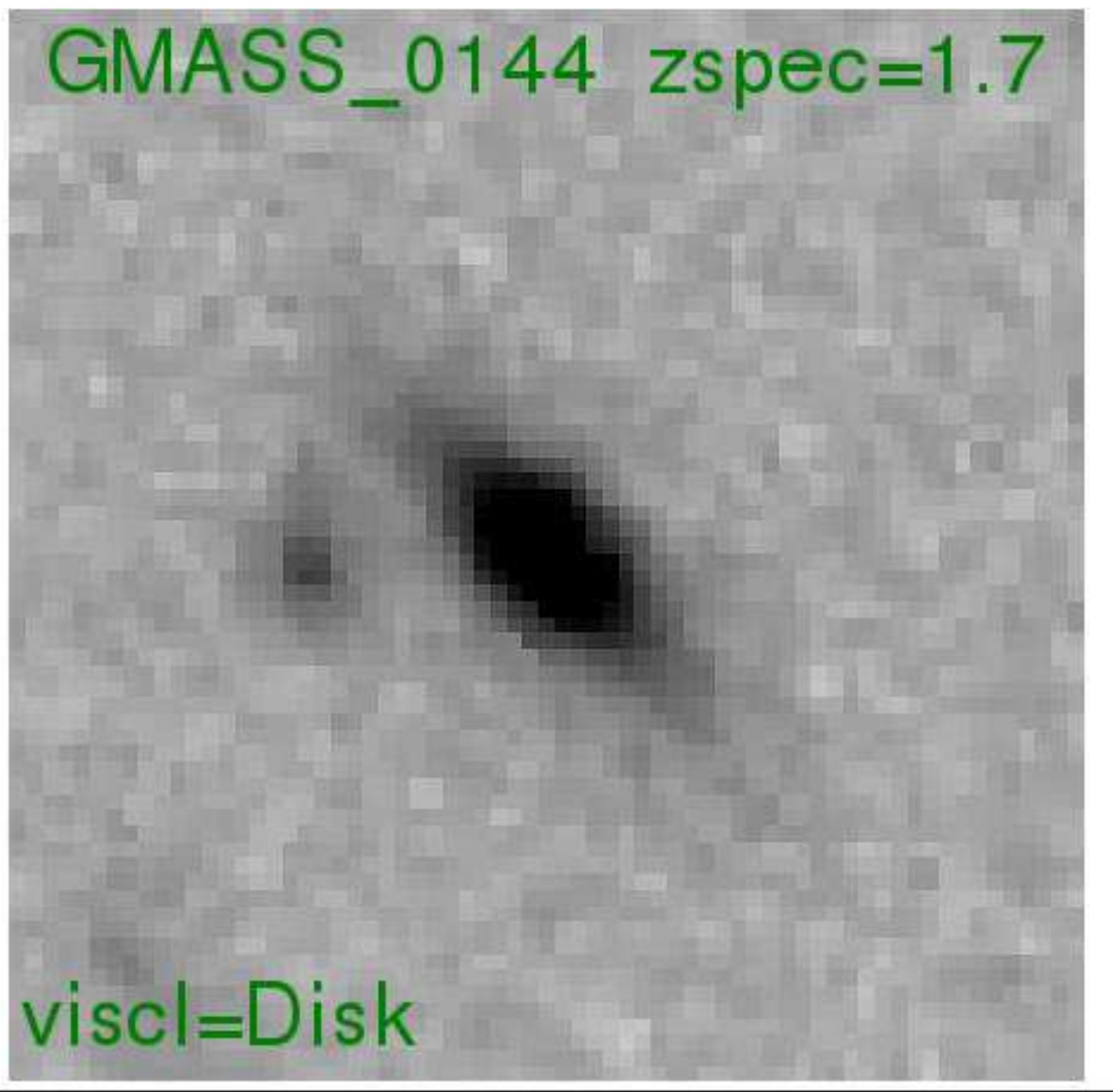}			     
\includegraphics[trim=100 40 75 390, clip=true, width=30mm]{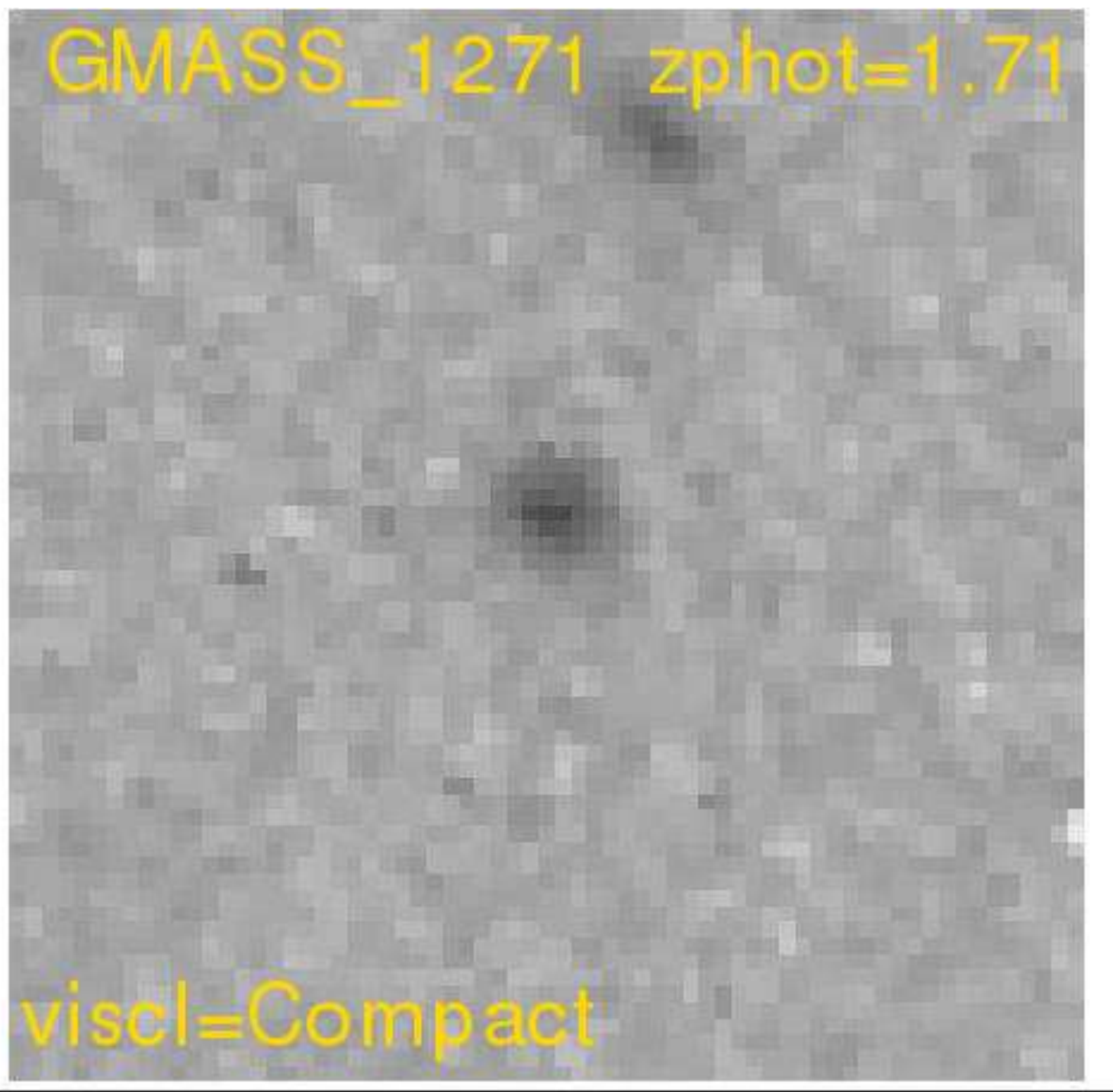}
\end{figure*}
\begin{figure*}
\centering   
\includegraphics[trim=100 40 75 390, clip=true, width=30mm]{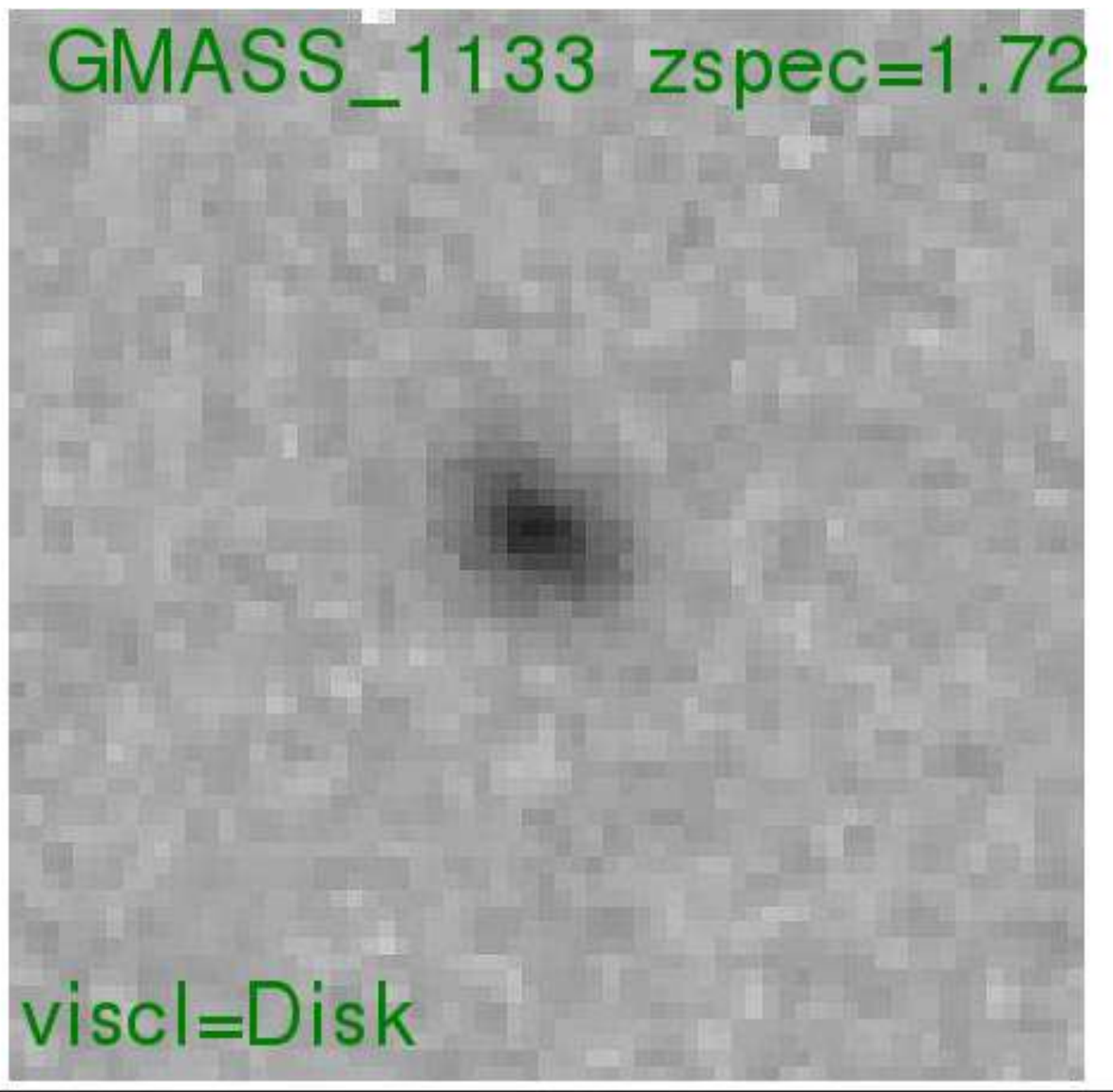}		     
\includegraphics[trim=100 40 75 390, clip=true, width=30mm]{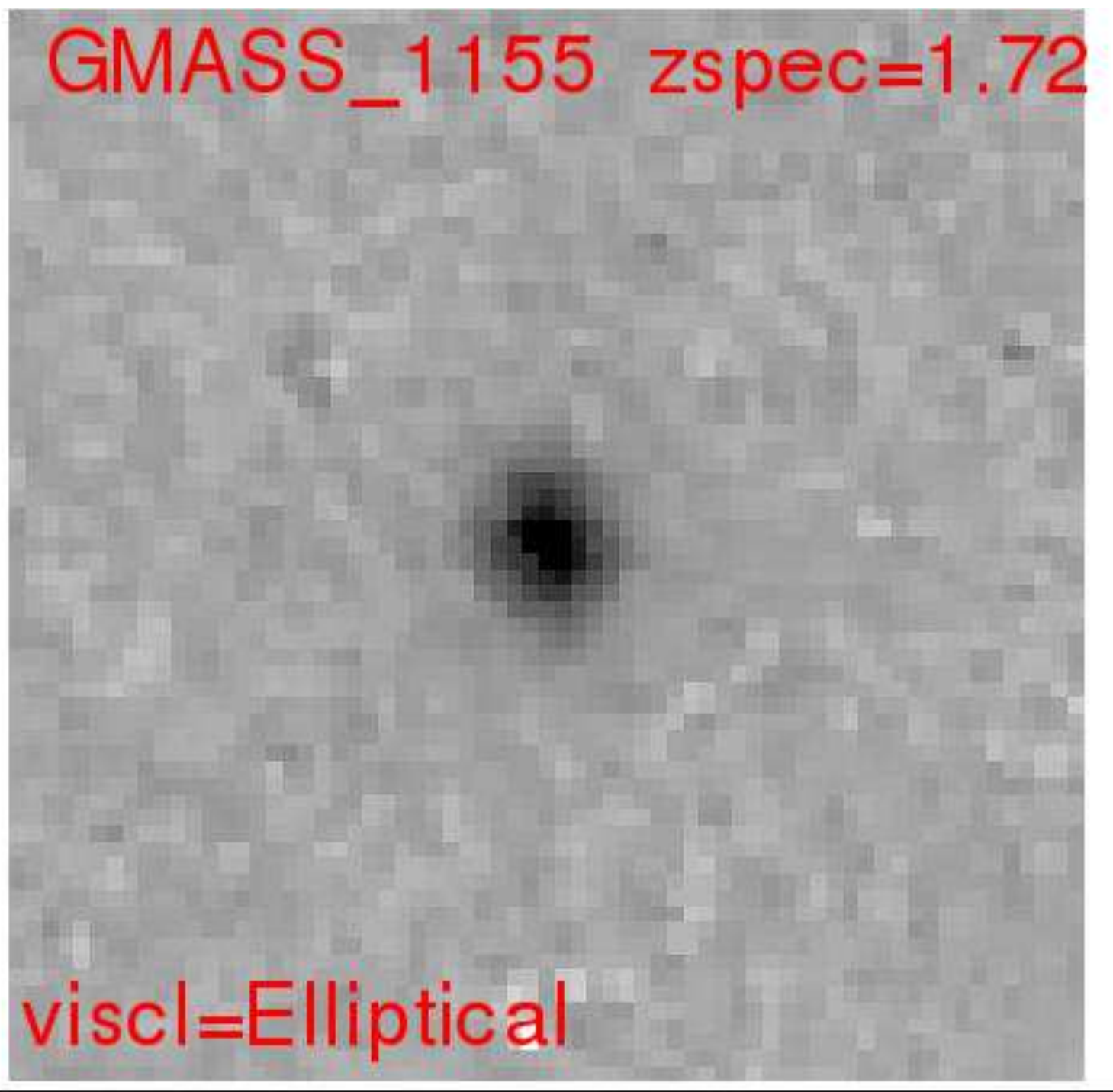}			     
\includegraphics[trim=100 40 75 390, clip=true, width=30mm]{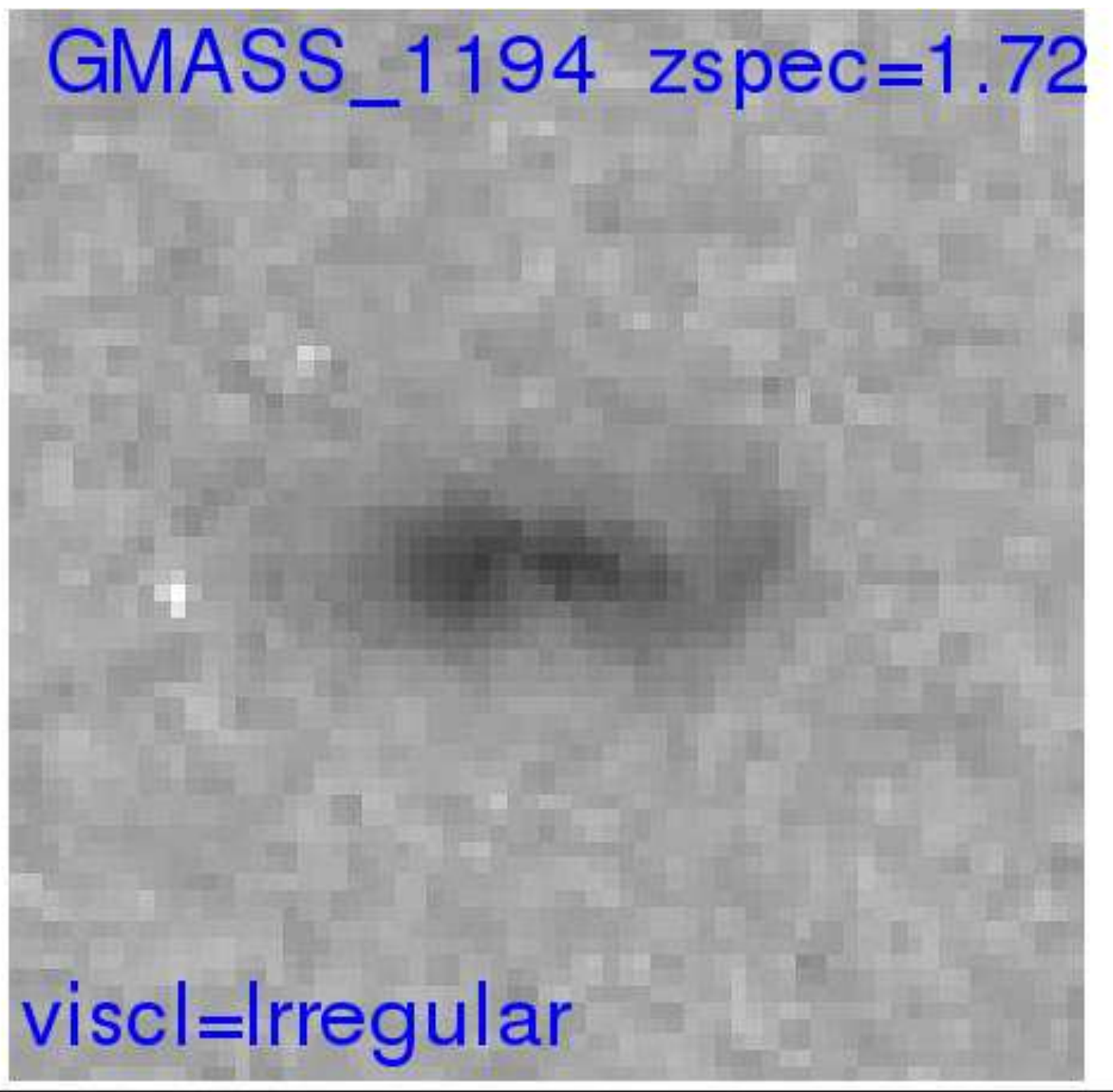}			     
\includegraphics[trim=100 40 75 390, clip=true, width=30mm]{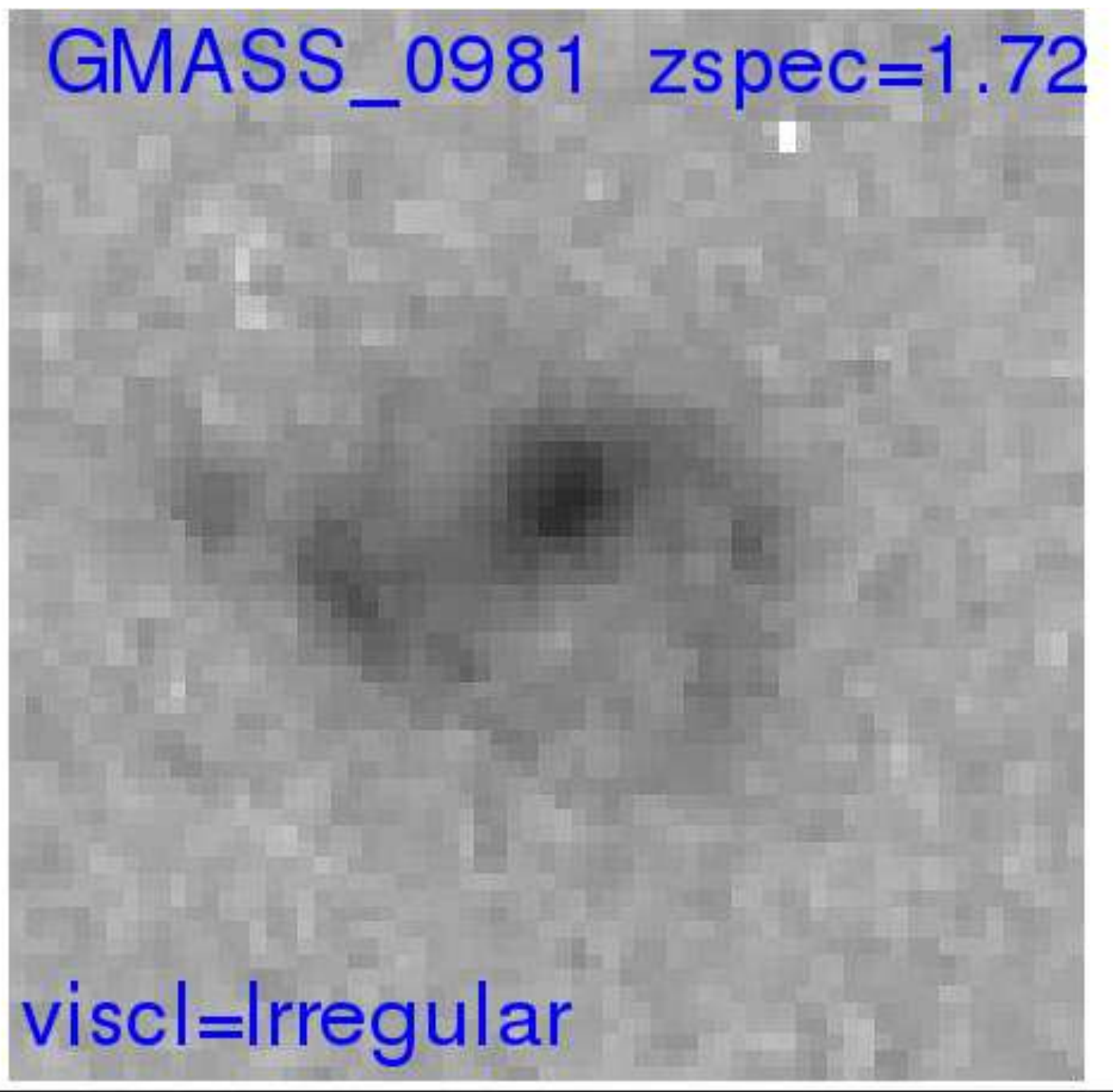}			     
\includegraphics[trim=100 40 75 390, clip=true, width=30mm]{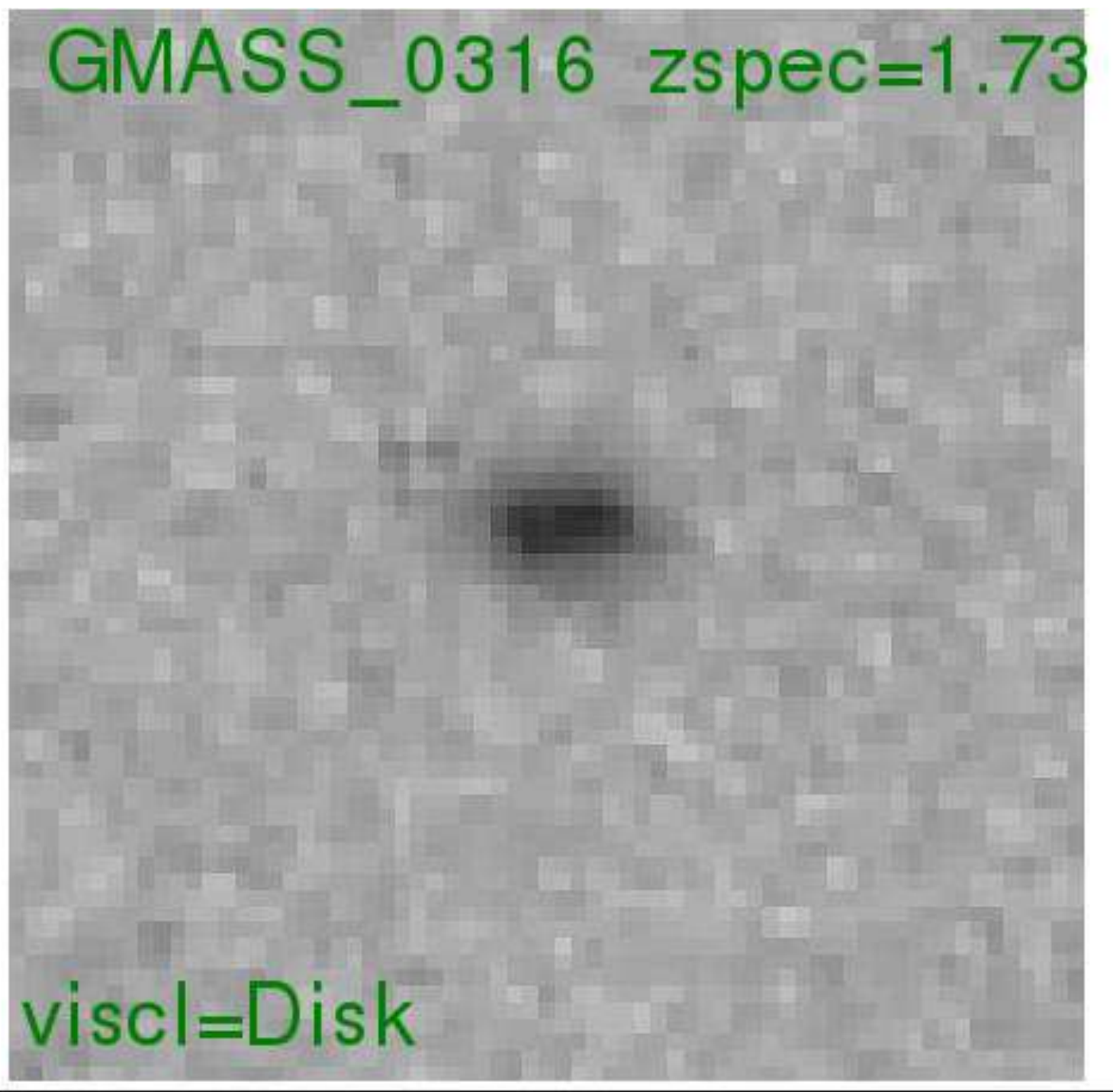}			     
\includegraphics[trim=100 40 75 390, clip=true, width=30mm]{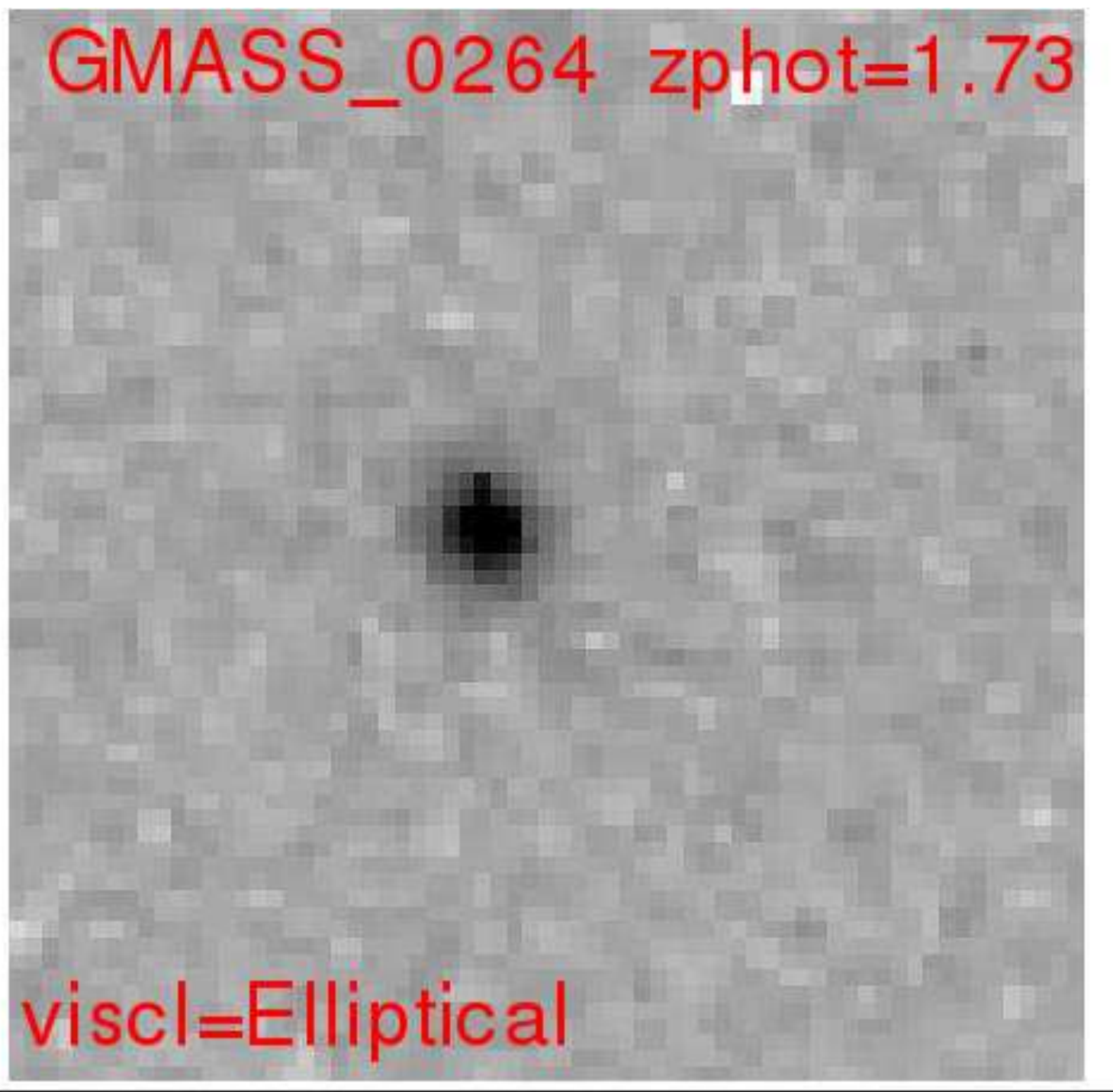}			     

\includegraphics[trim=100 40 75 390, clip=true, width=30mm]{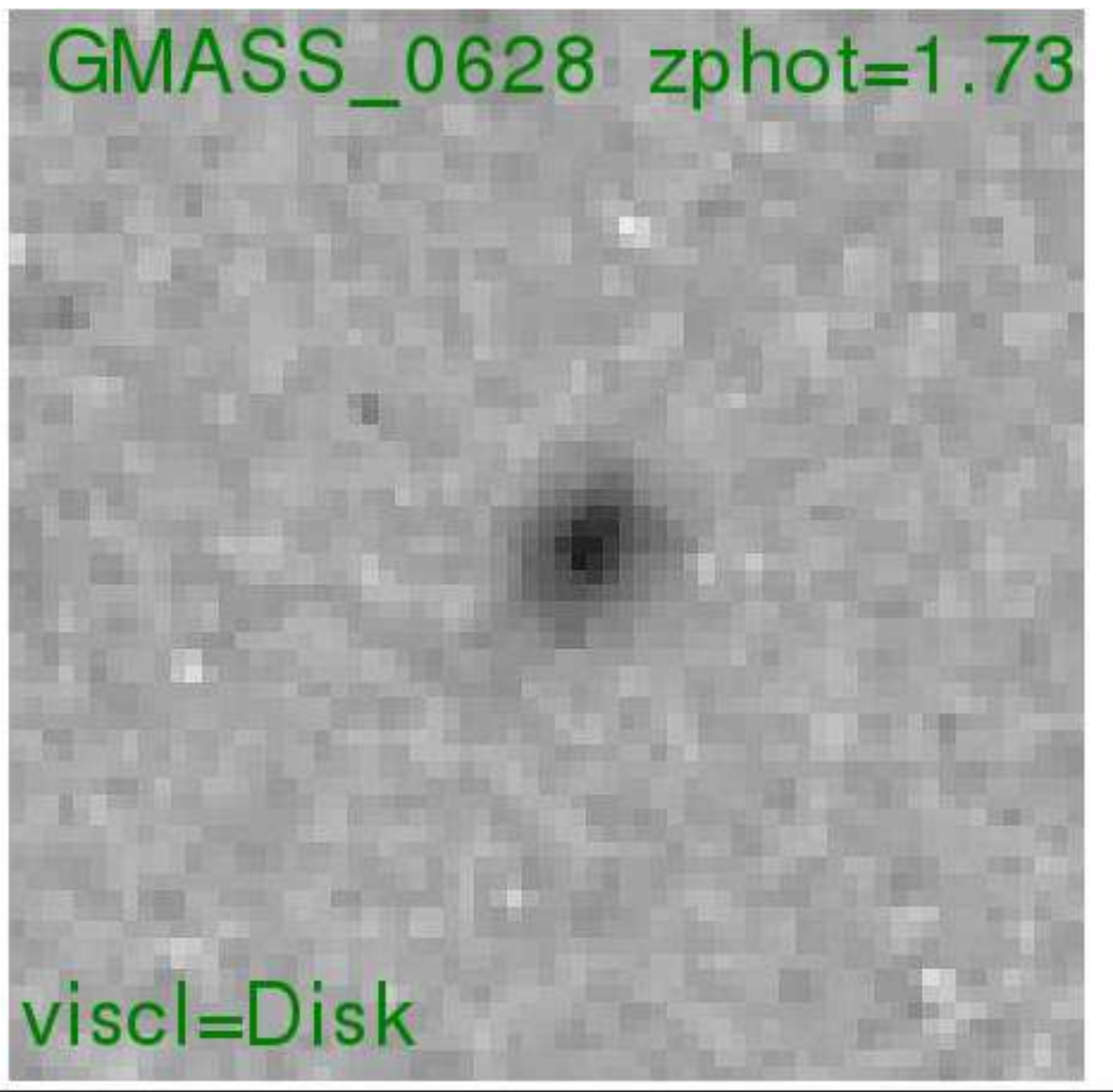}		     
\includegraphics[trim=100 40 75 390, clip=true, width=30mm]{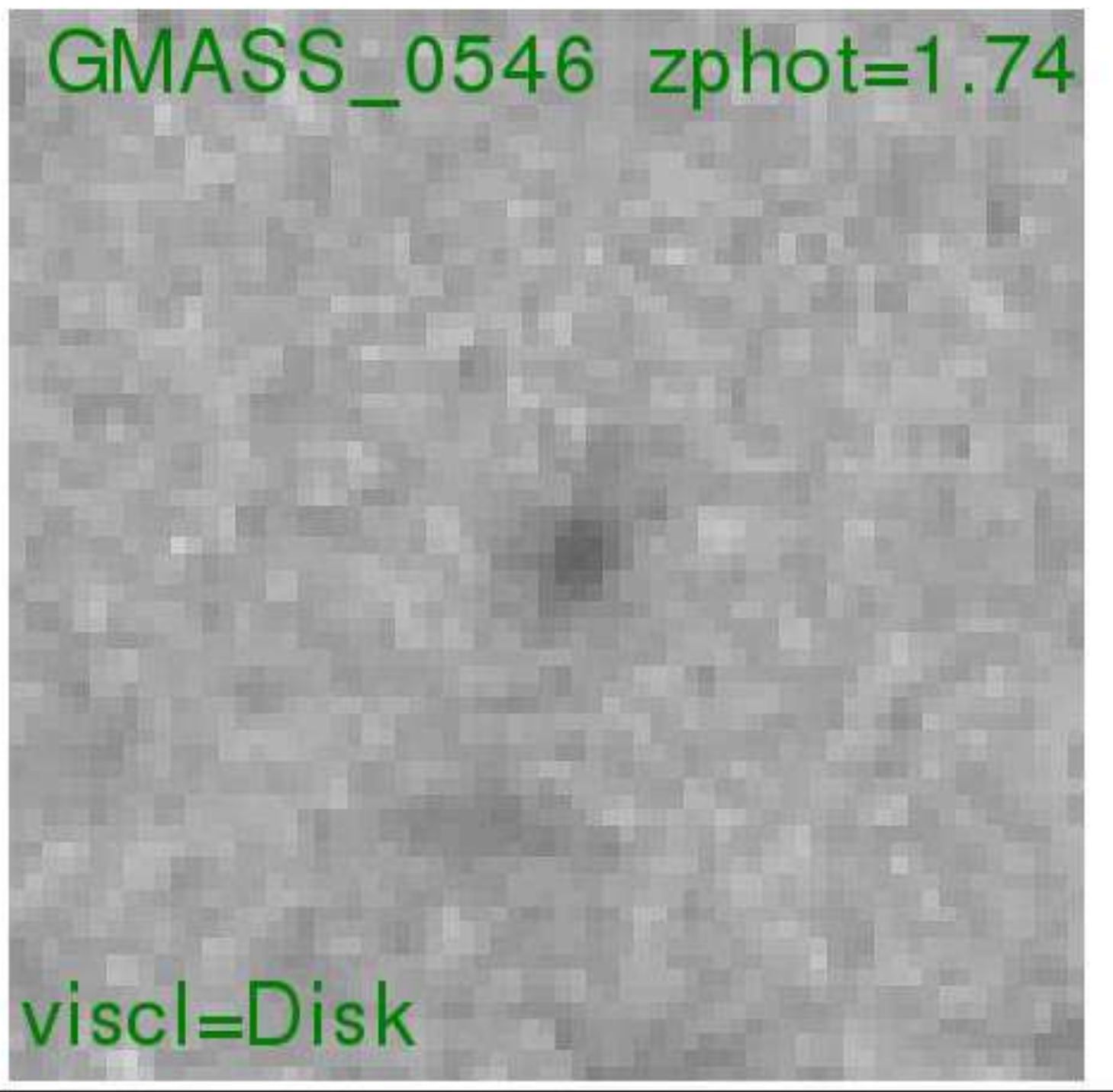}		     
\includegraphics[trim=100 40 75 390, clip=true, width=30mm]{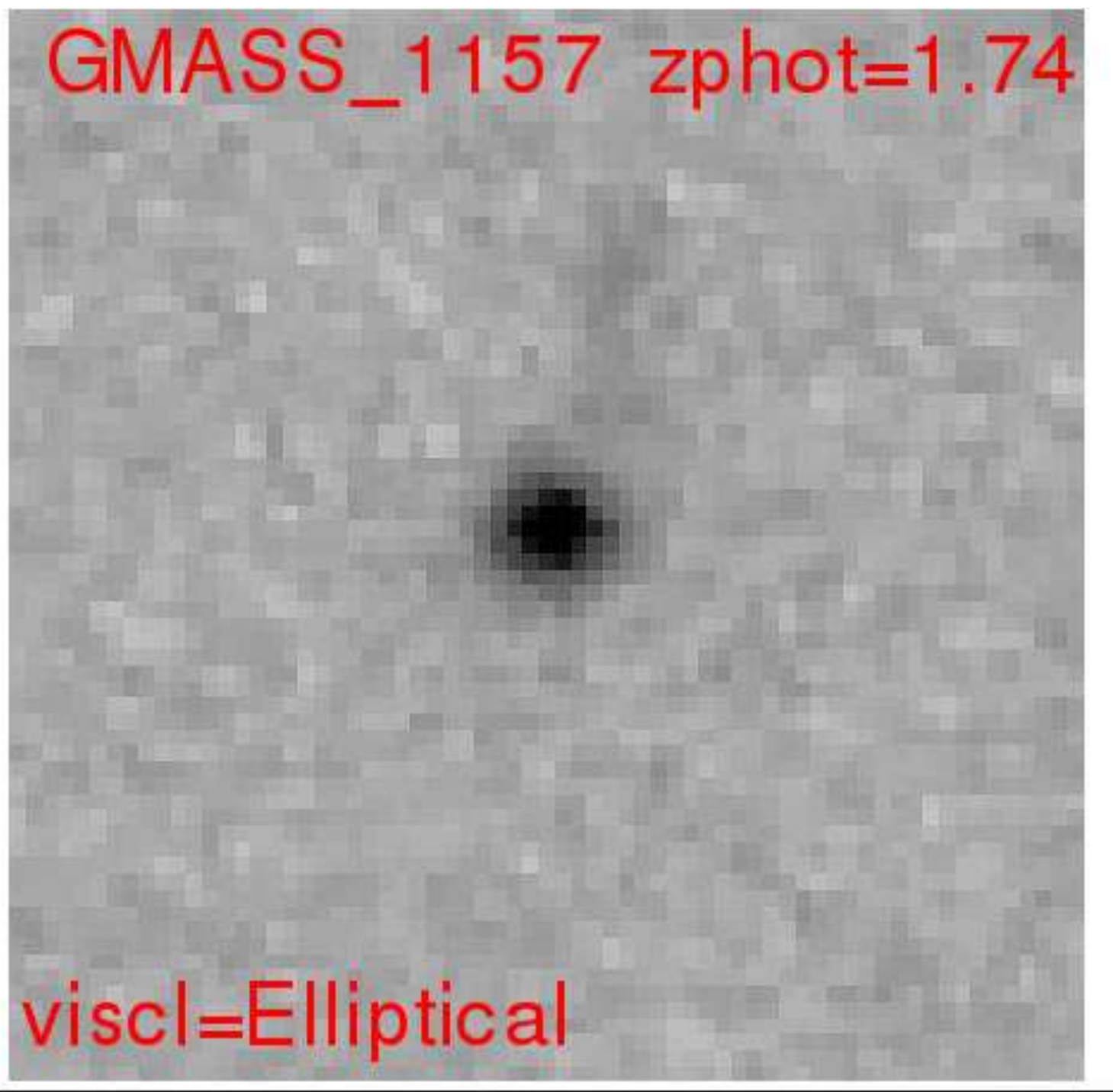}		     
\includegraphics[trim=100 40 75 390, clip=true, width=30mm]{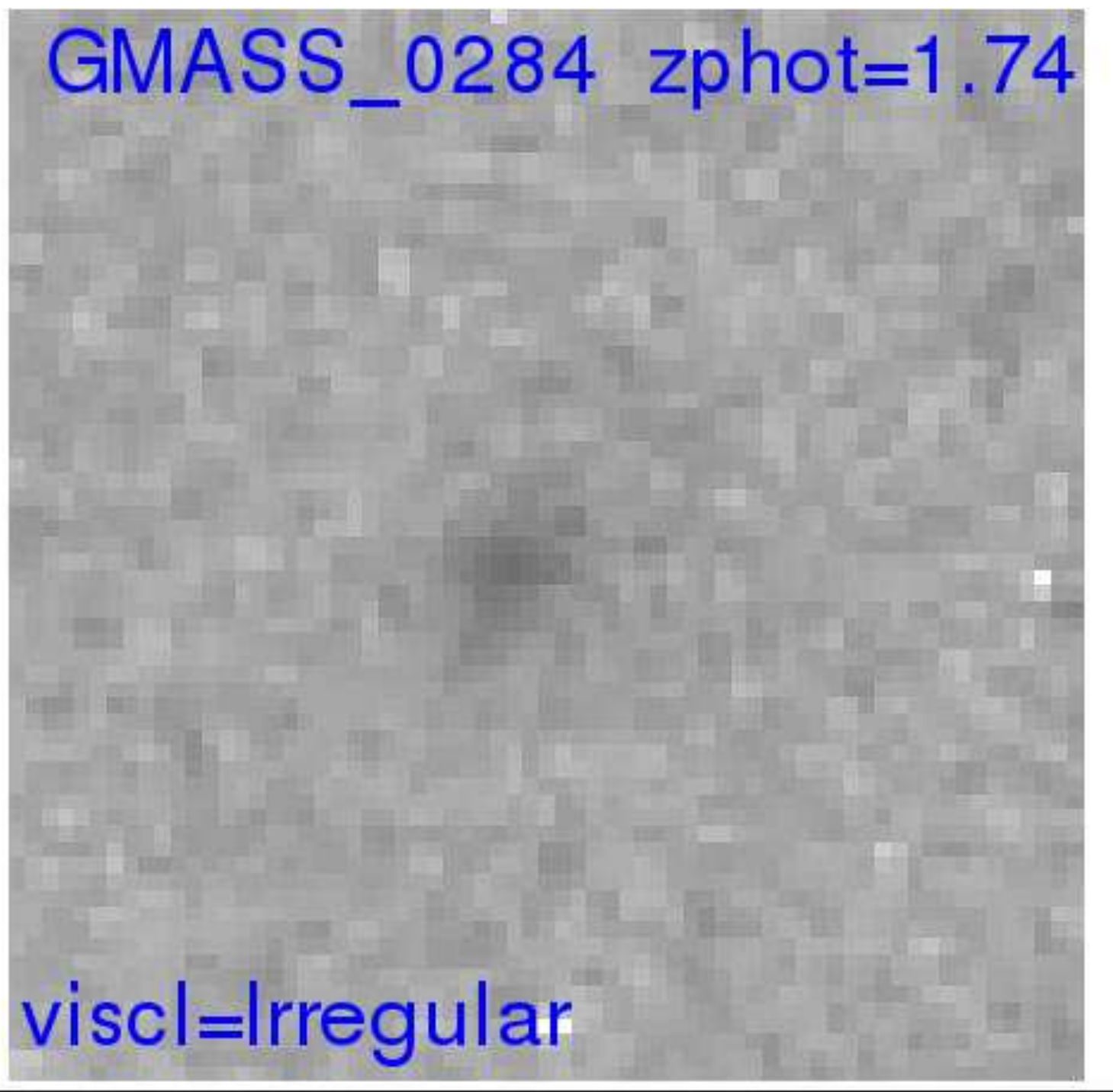}			     
\includegraphics[trim=100 40 75 390, clip=true, width=30mm]{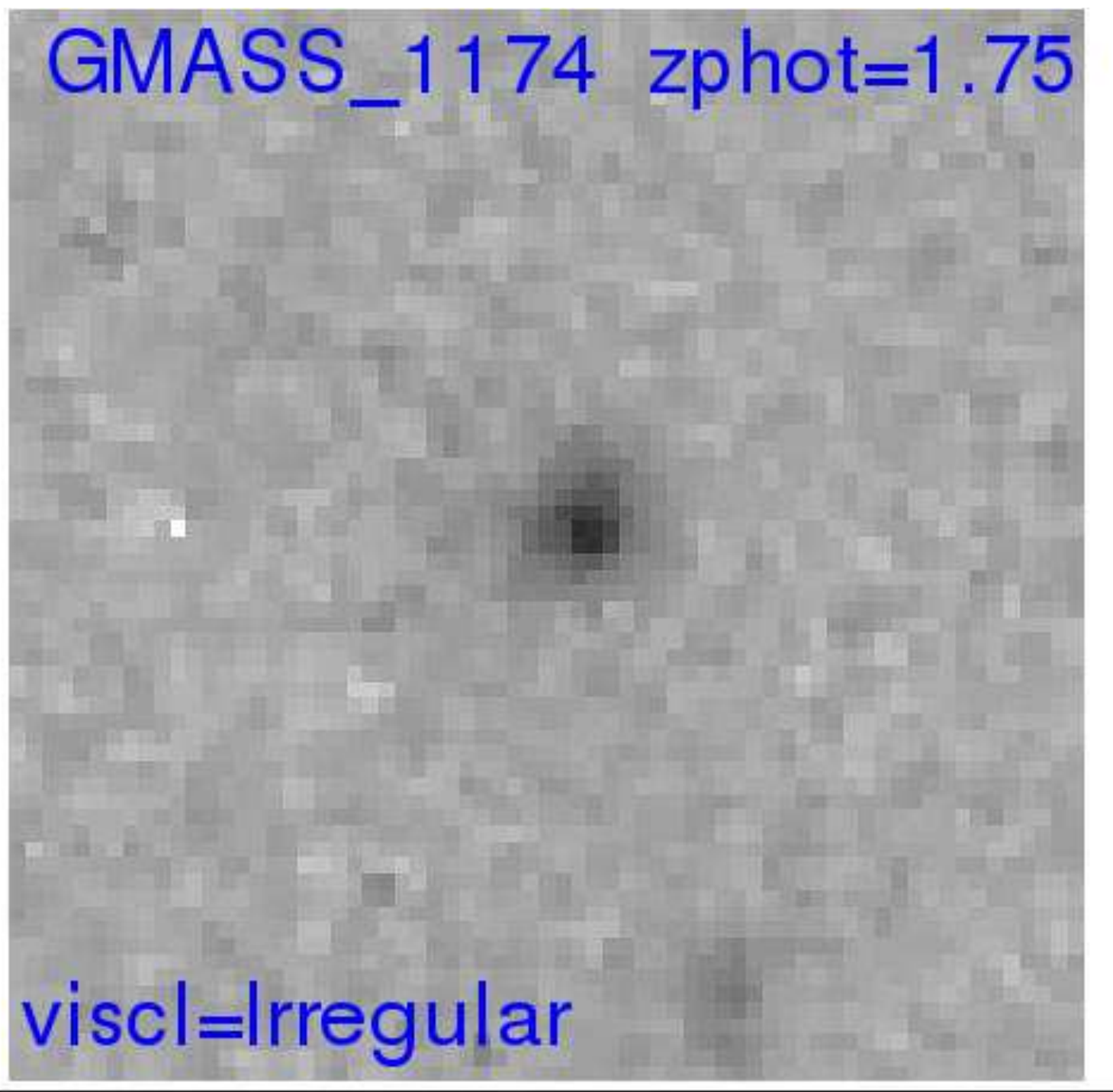}			     
\includegraphics[trim=100 40 75 390, clip=true, width=30mm]{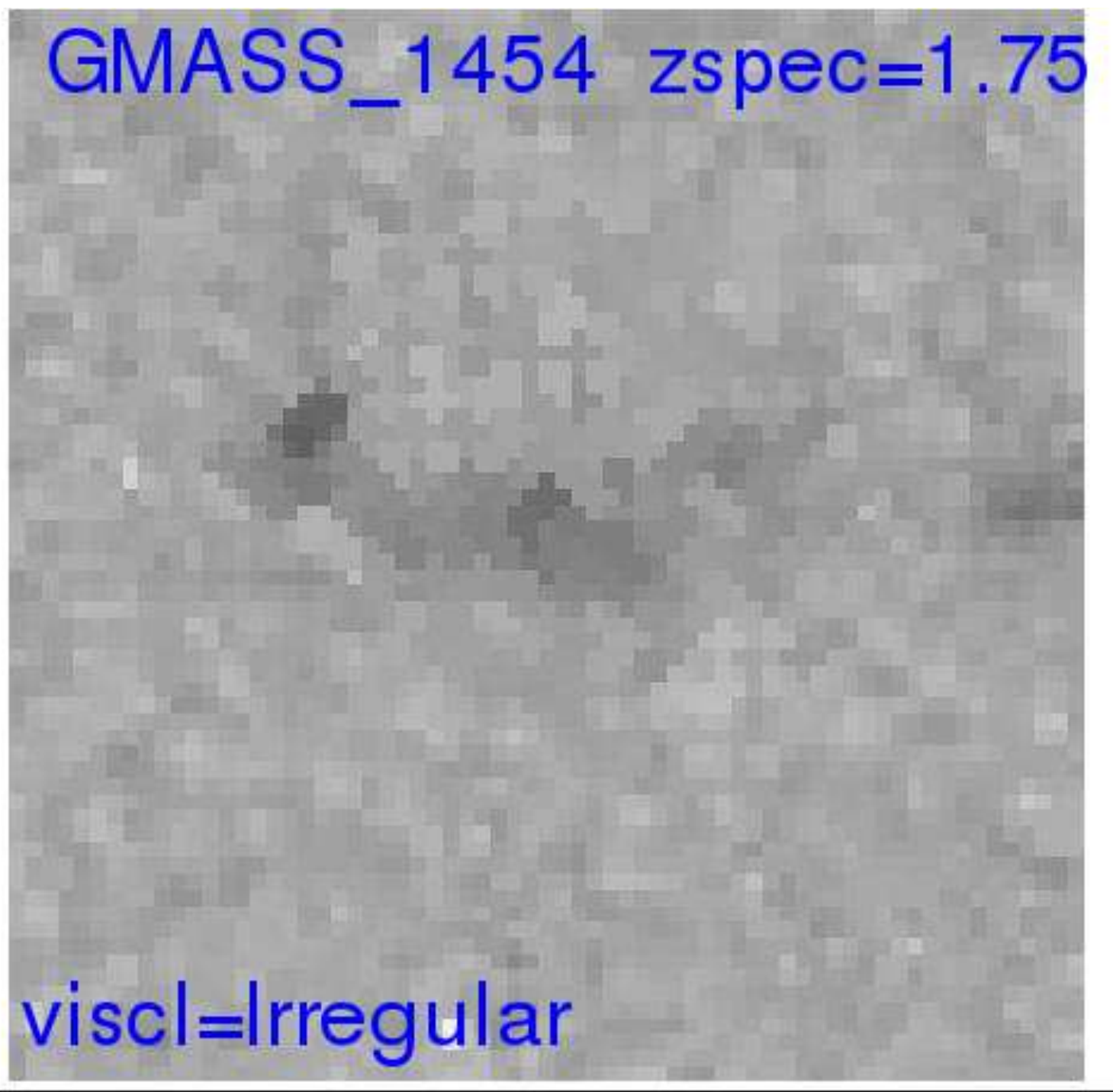}			     

\includegraphics[trim=100 40 75 390, clip=true, width=30mm]{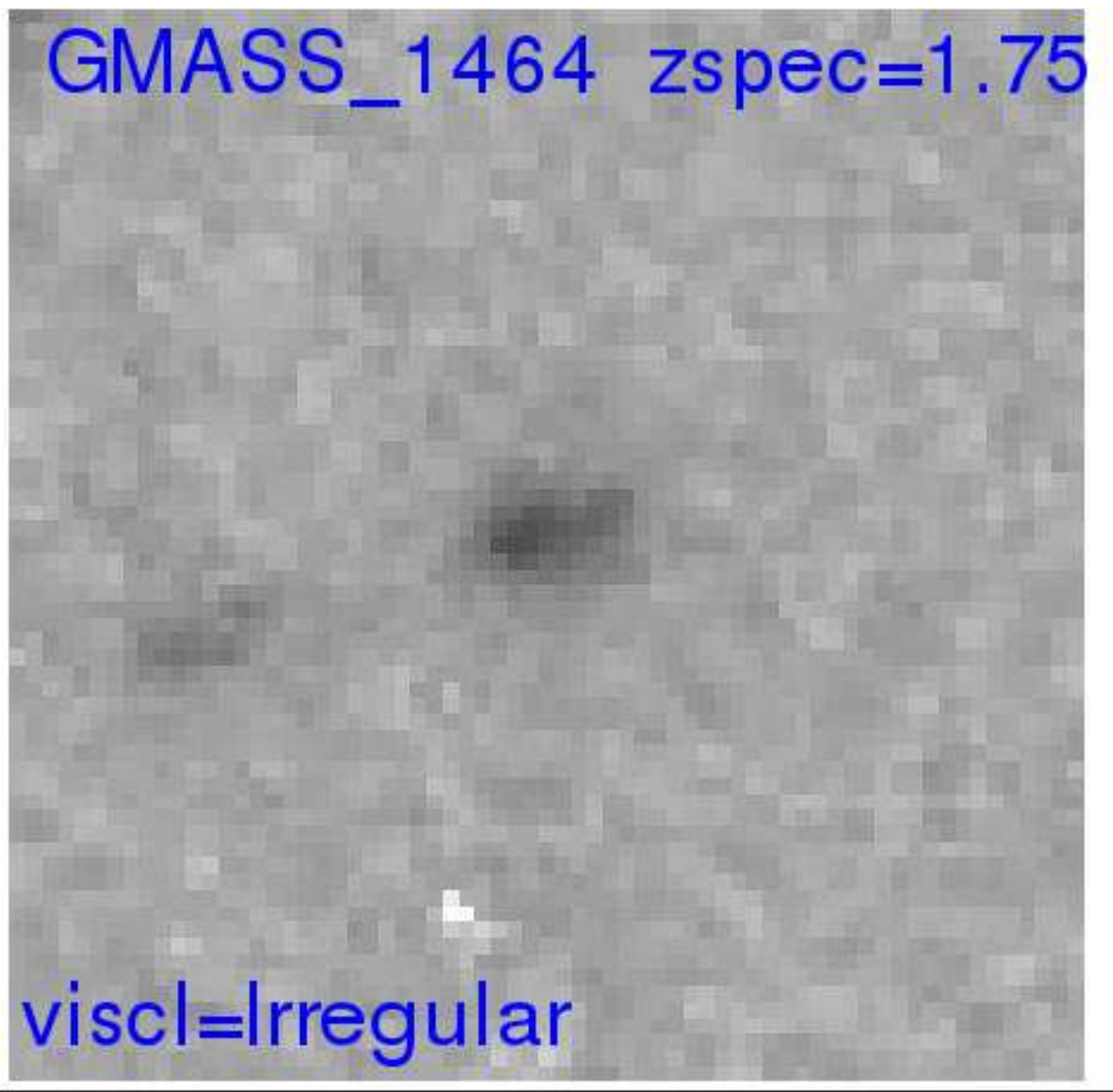}		     
\includegraphics[trim=100 40 75 390, clip=true, width=30mm]{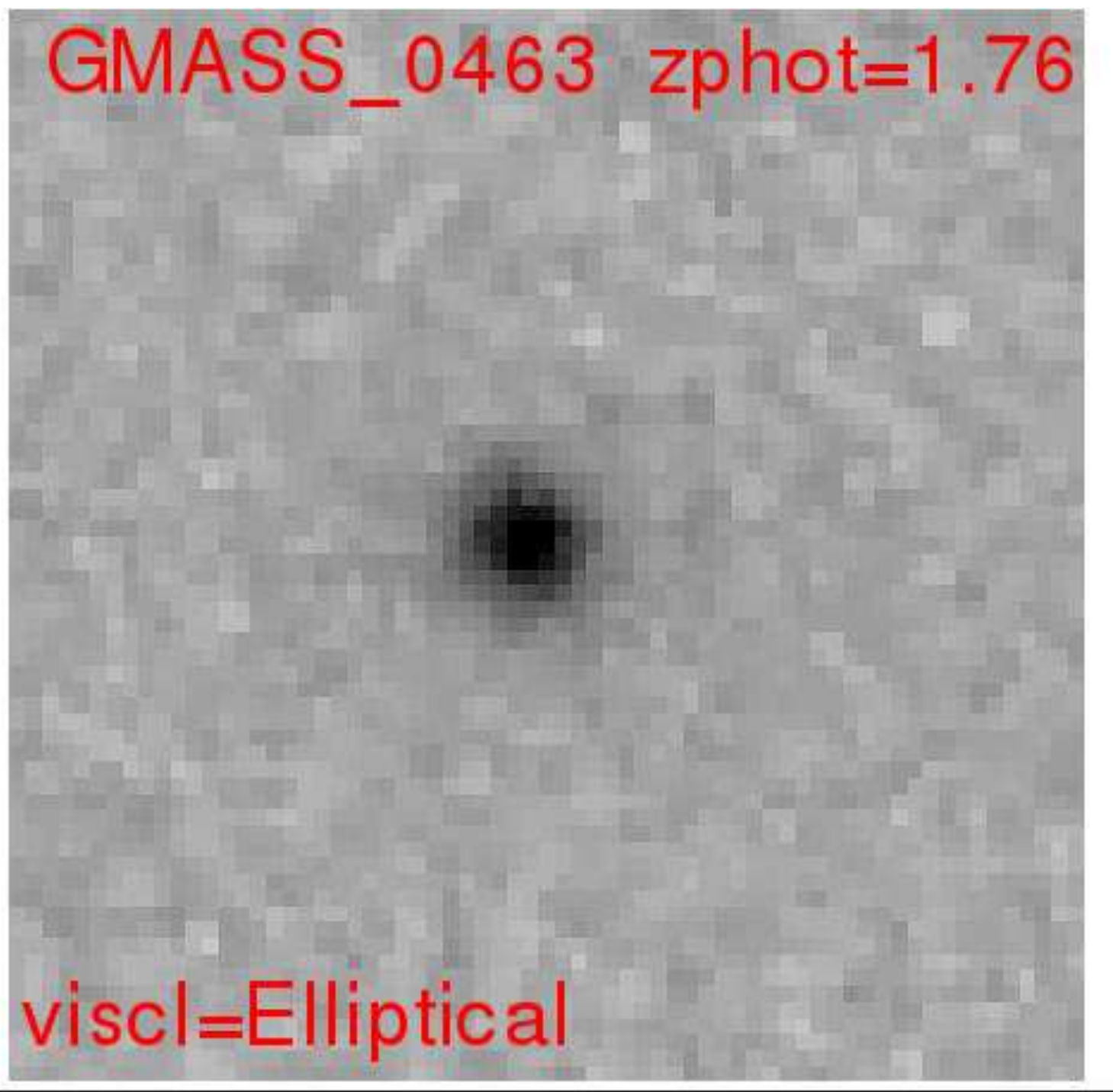}		     
\includegraphics[trim=100 40 75 390, clip=true, width=30mm]{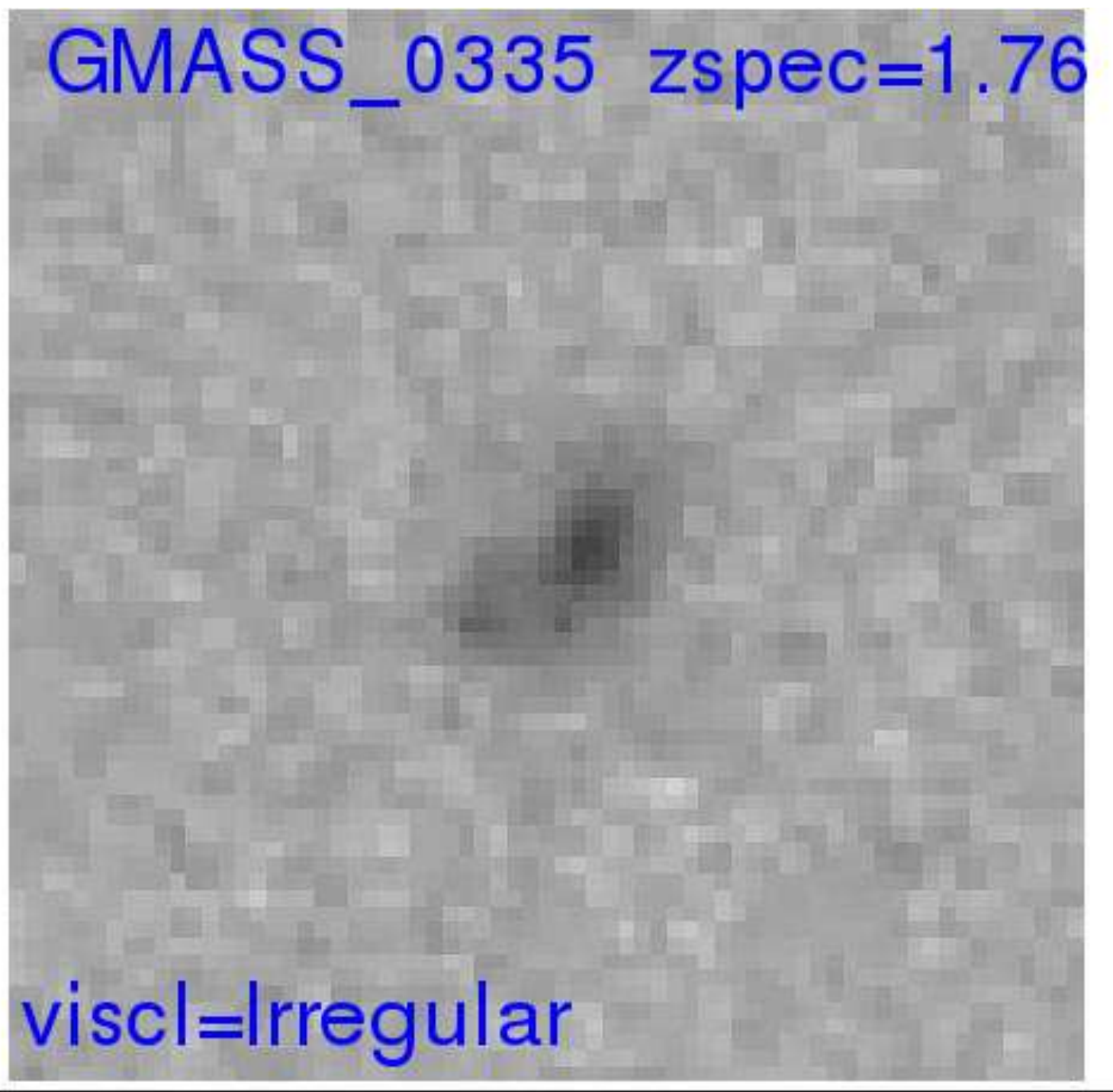}			     
\includegraphics[trim=100 40 75 390, clip=true, width=30mm]{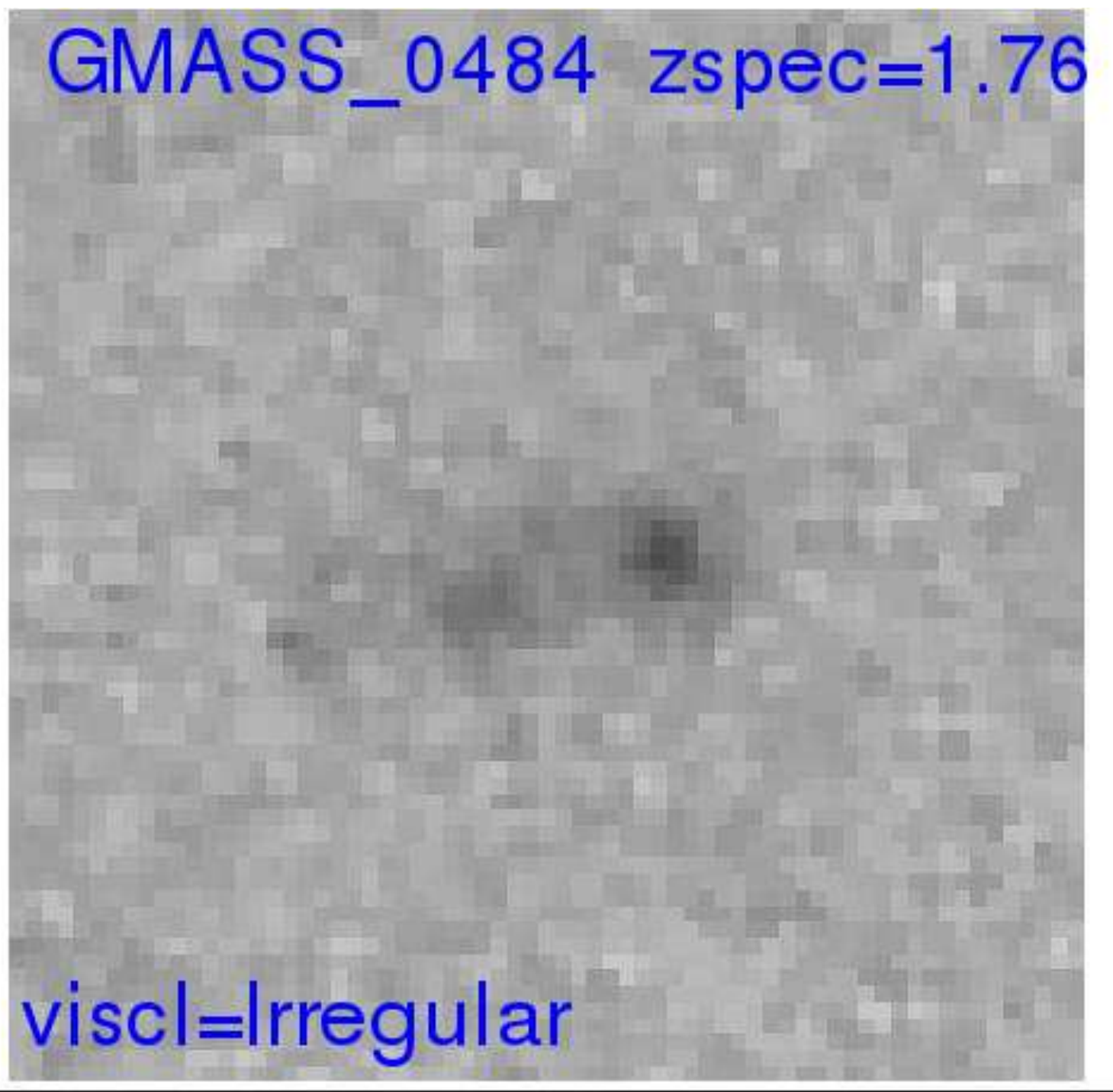}			     
\includegraphics[trim=100 40 75 390, clip=true, width=30mm]{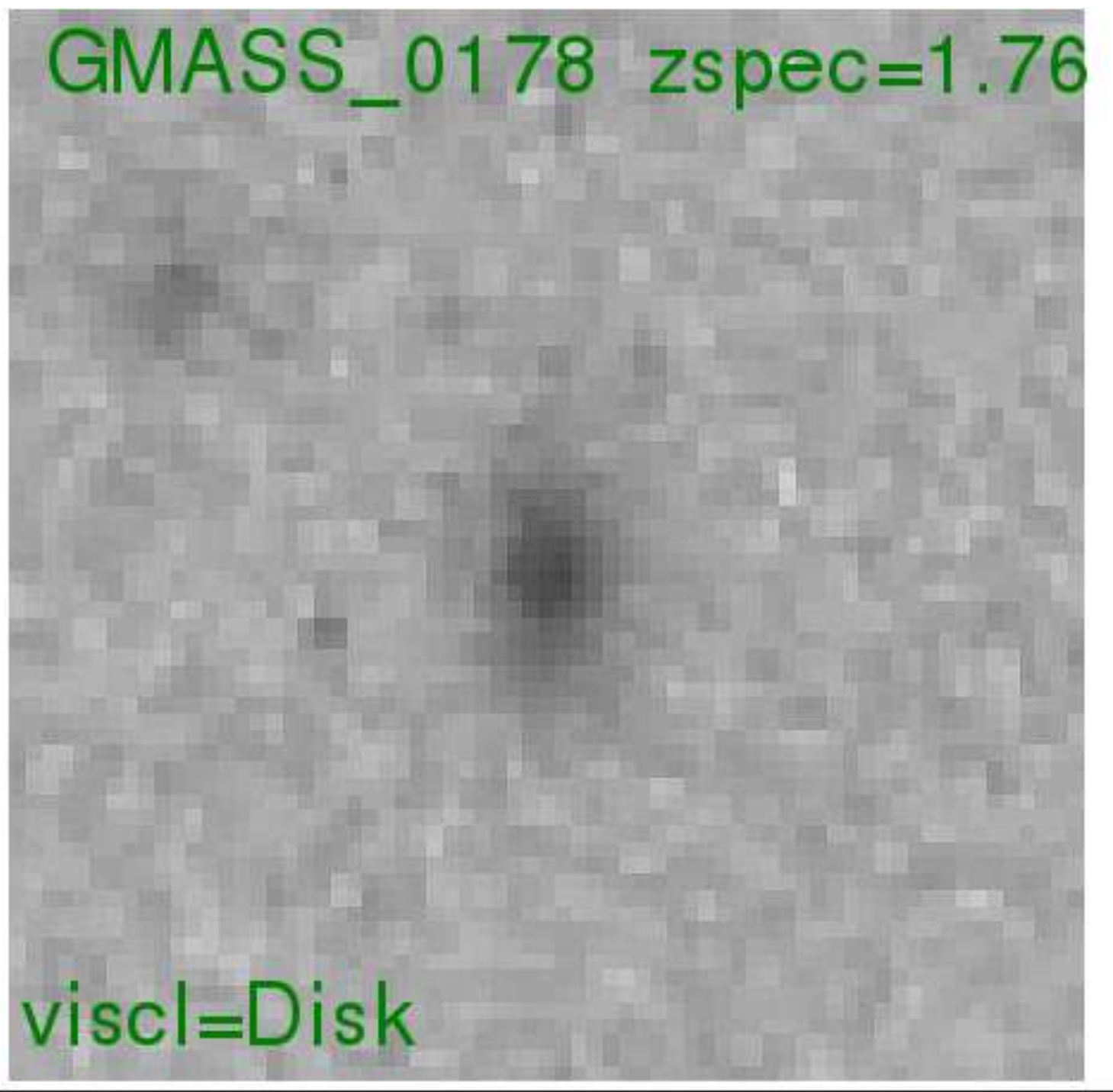}			     
\includegraphics[trim=100 40 75 390, clip=true, width=30mm]{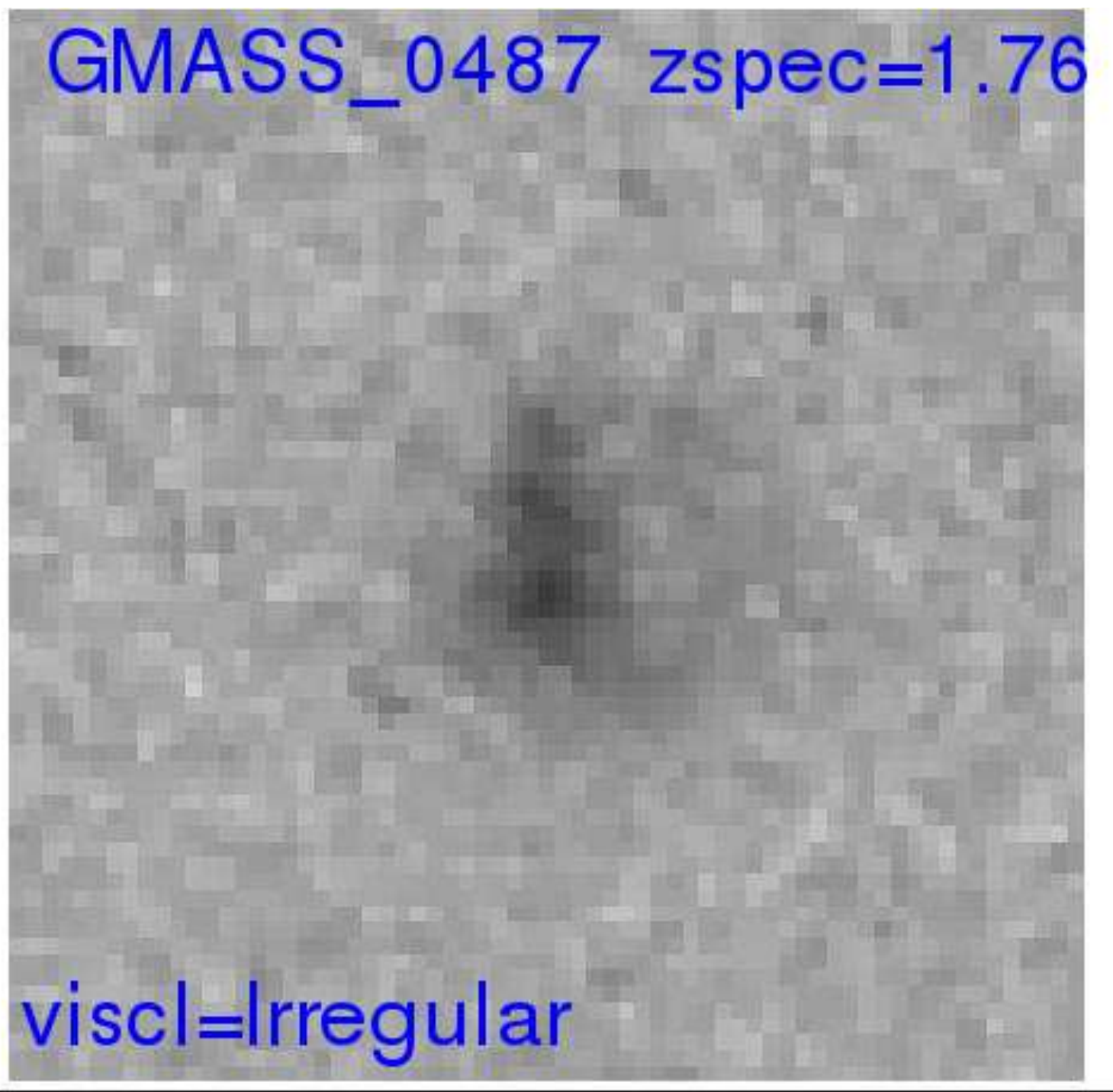}		     

\includegraphics[trim=100 40 75 390, clip=true, width=30mm]{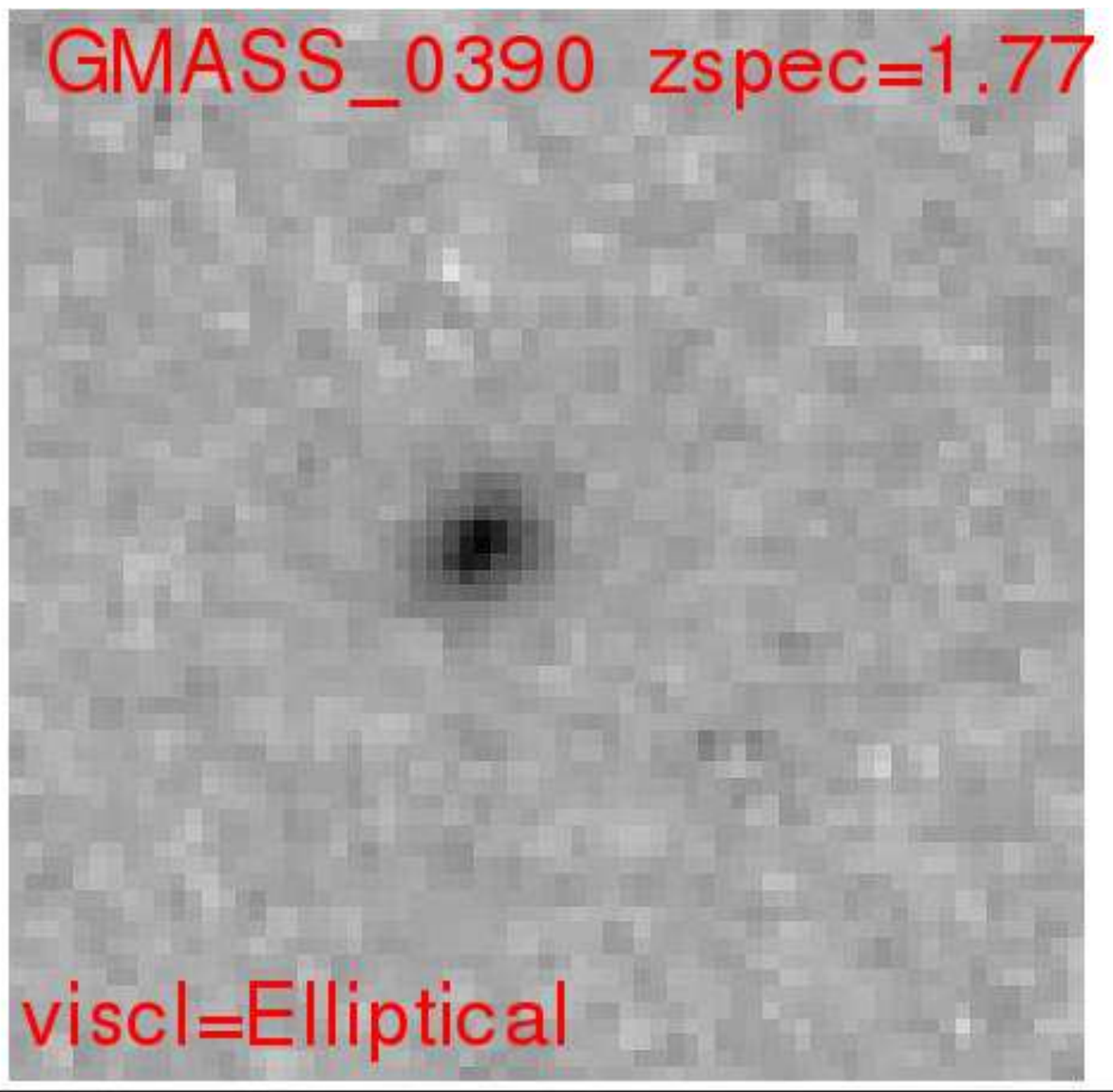}			     
\includegraphics[trim=100 40 75 390, clip=true, width=30mm]{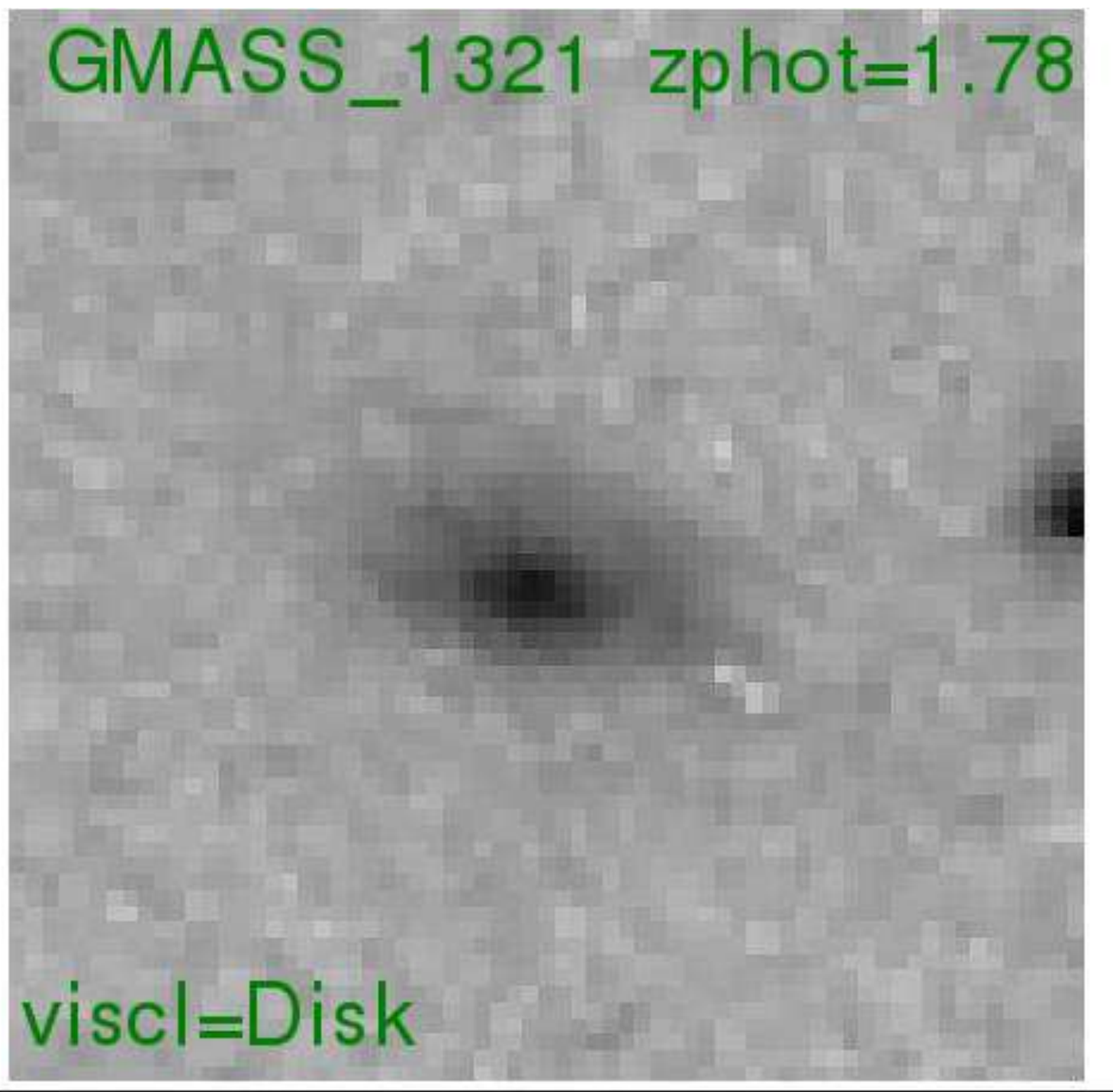}			     
\includegraphics[trim=100 40 75 390, clip=true, width=30mm]{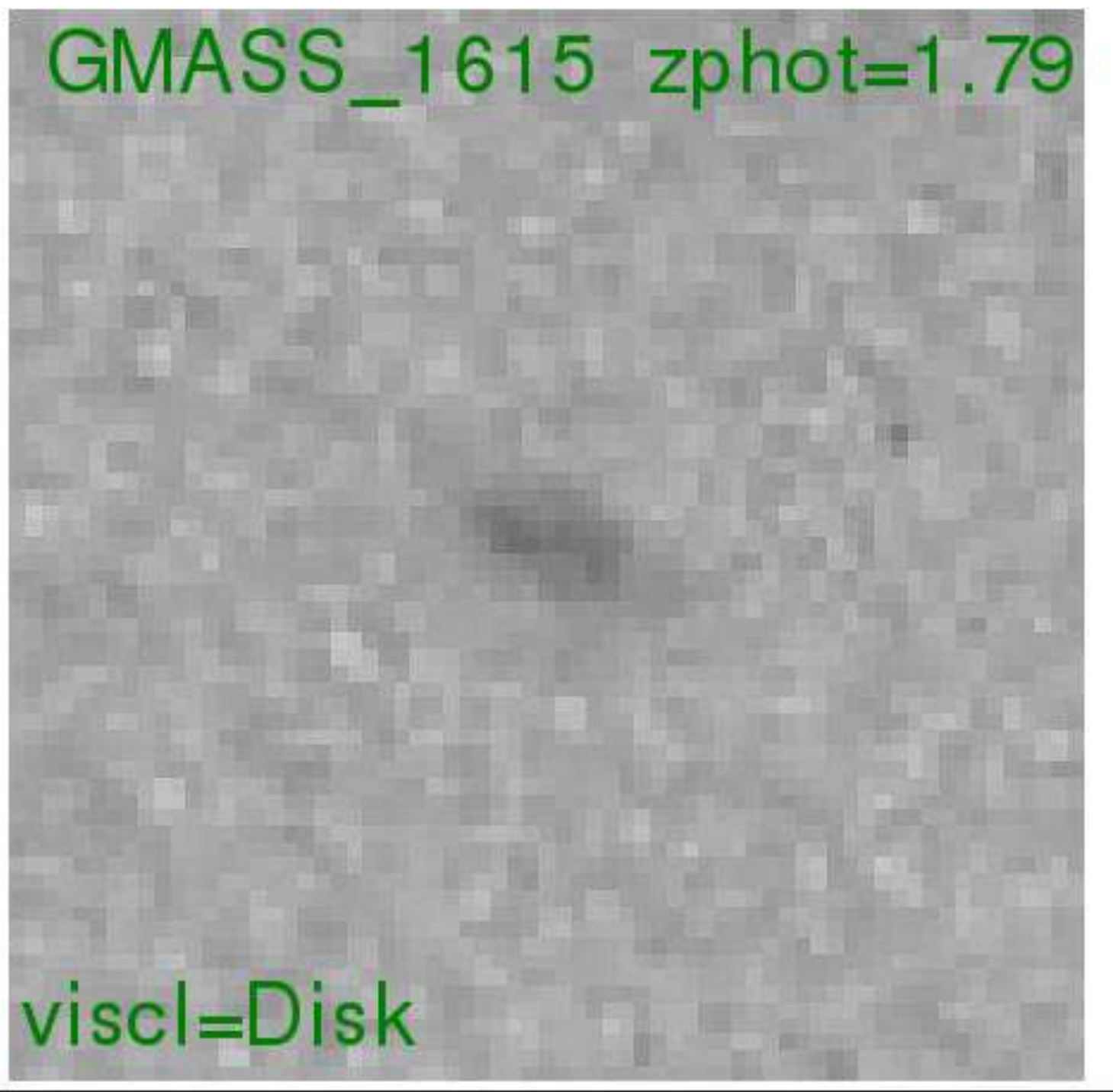}			     
\includegraphics[trim=100 40 75 390, clip=true, width=30mm]{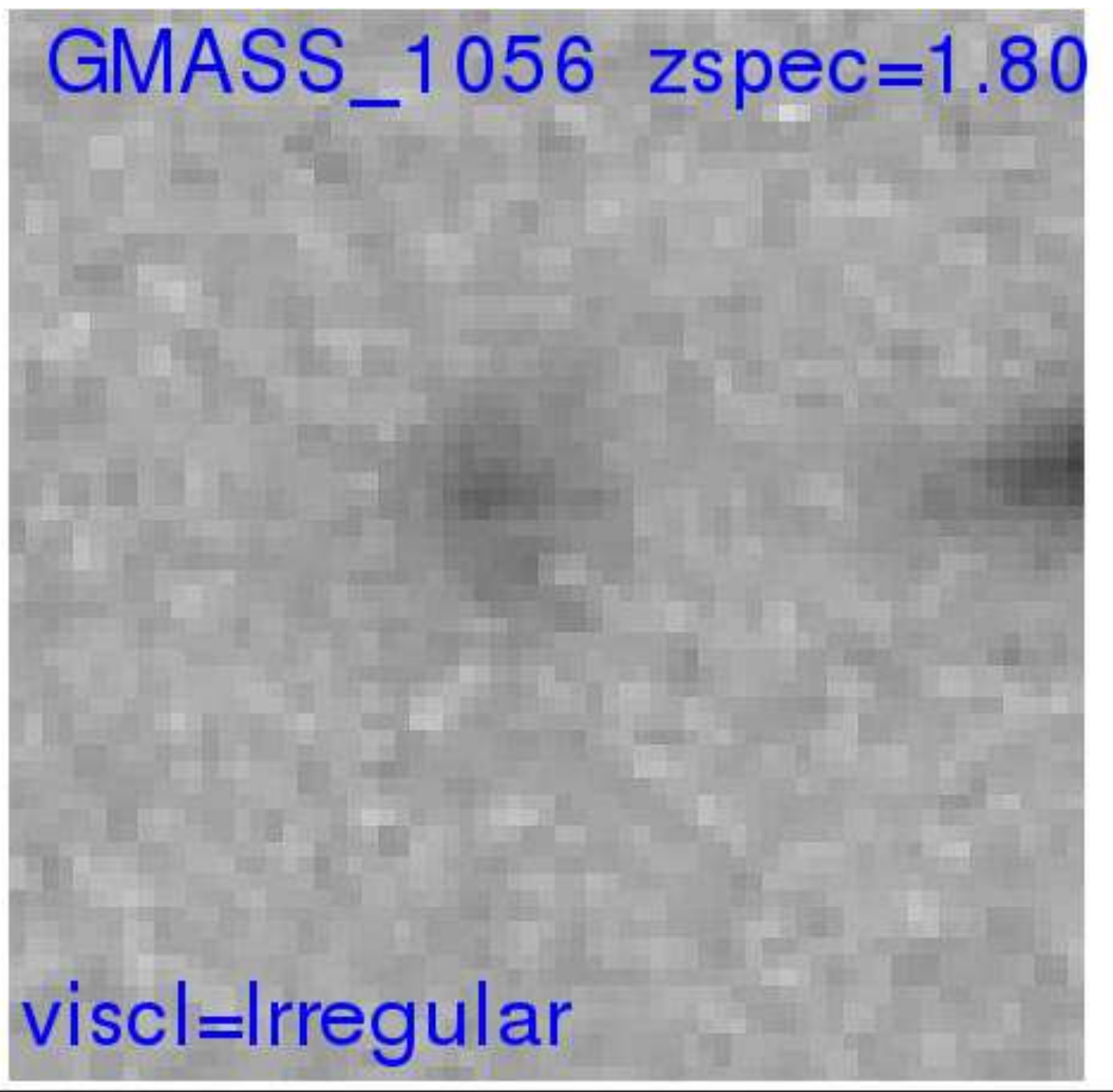}		     
\includegraphics[trim=100 40 75 390, clip=true, width=30mm]{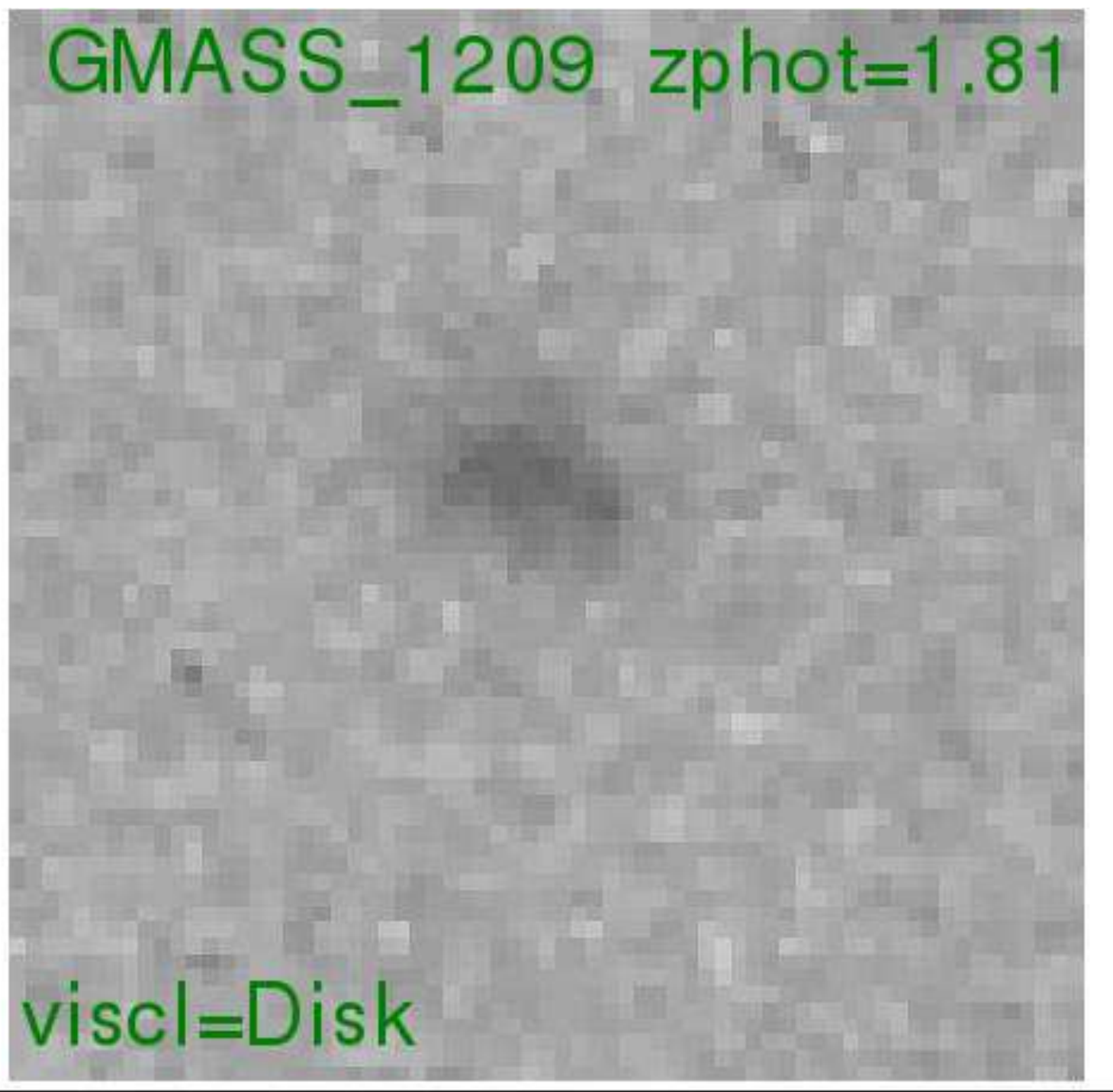}		     
\includegraphics[trim=100 40 75 390, clip=true, width=30mm]{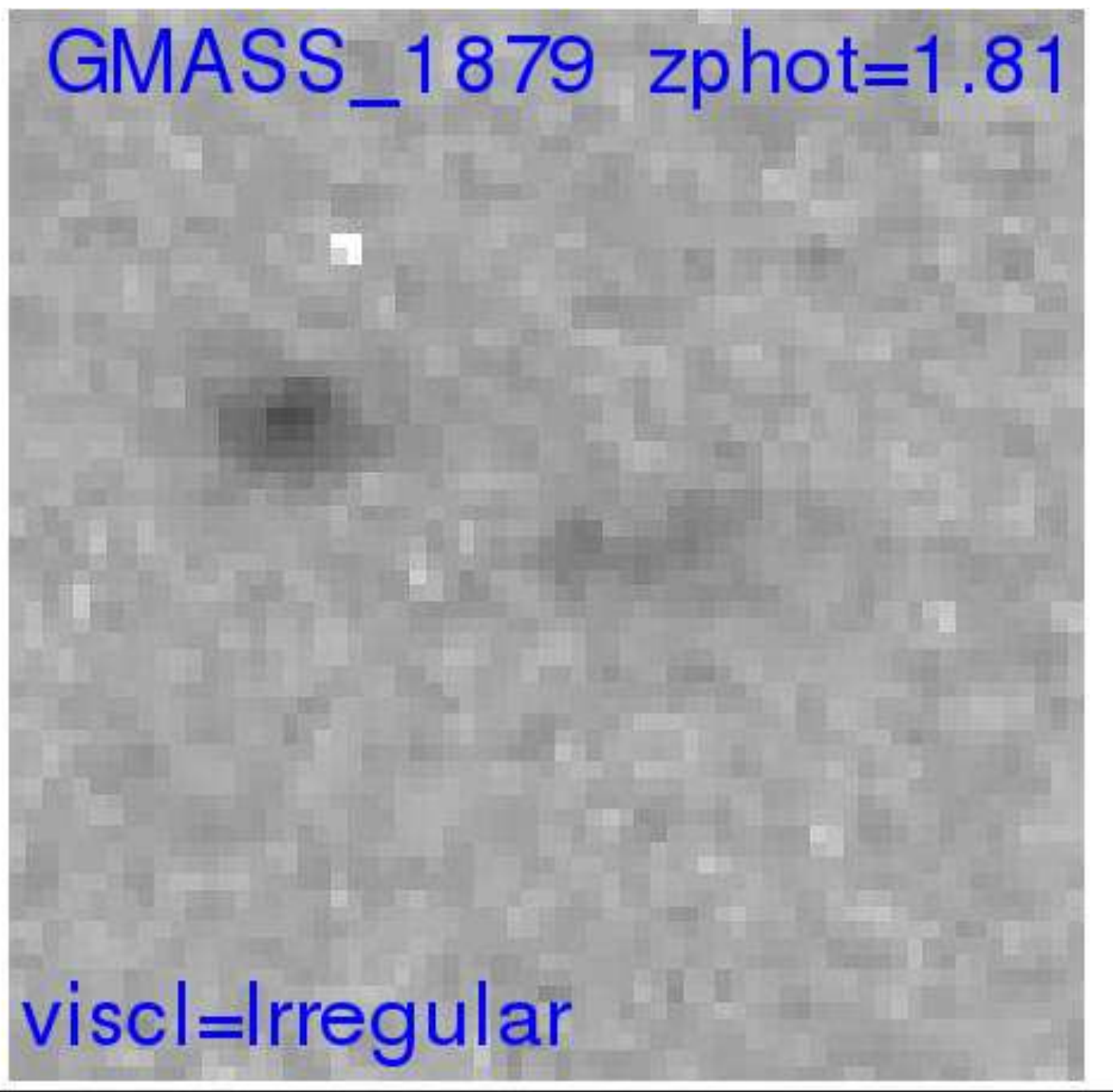}			     

\includegraphics[trim=100 40 75 390, clip=true, width=30mm]{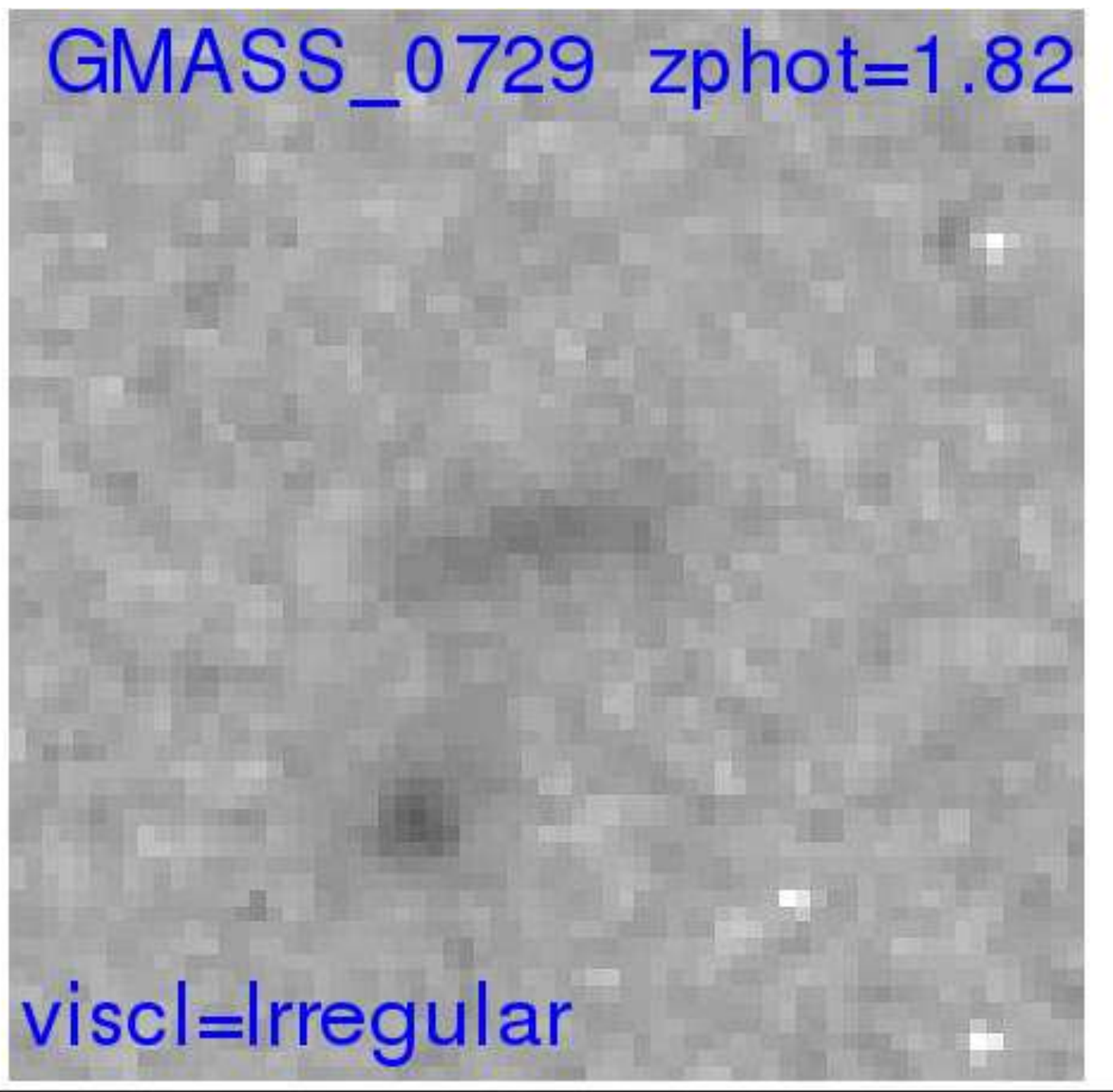}		     
\includegraphics[trim=100 40 75 390, clip=true, width=30mm]{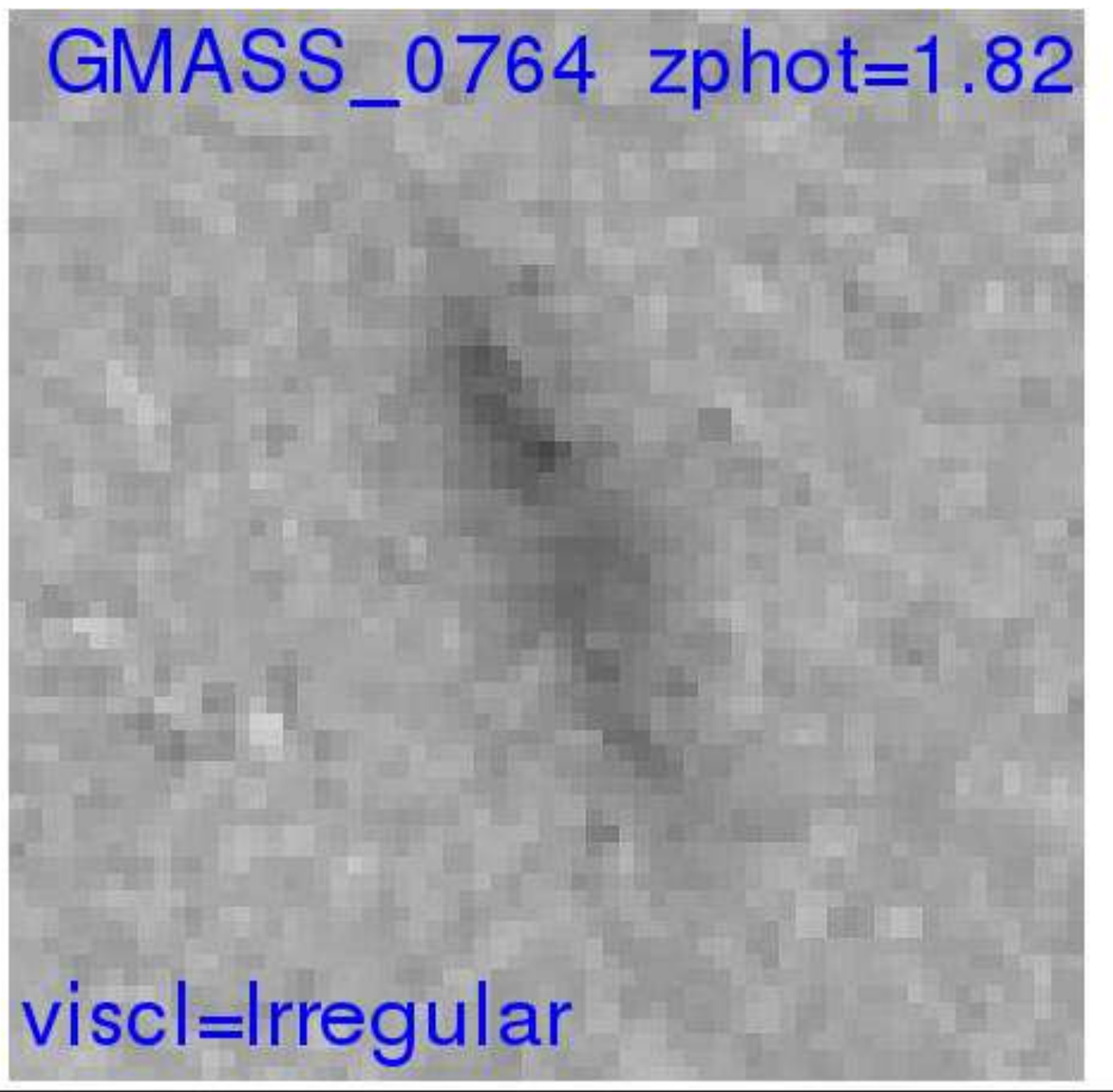}			     
\includegraphics[trim=100 40 75 390, clip=true, width=30mm]{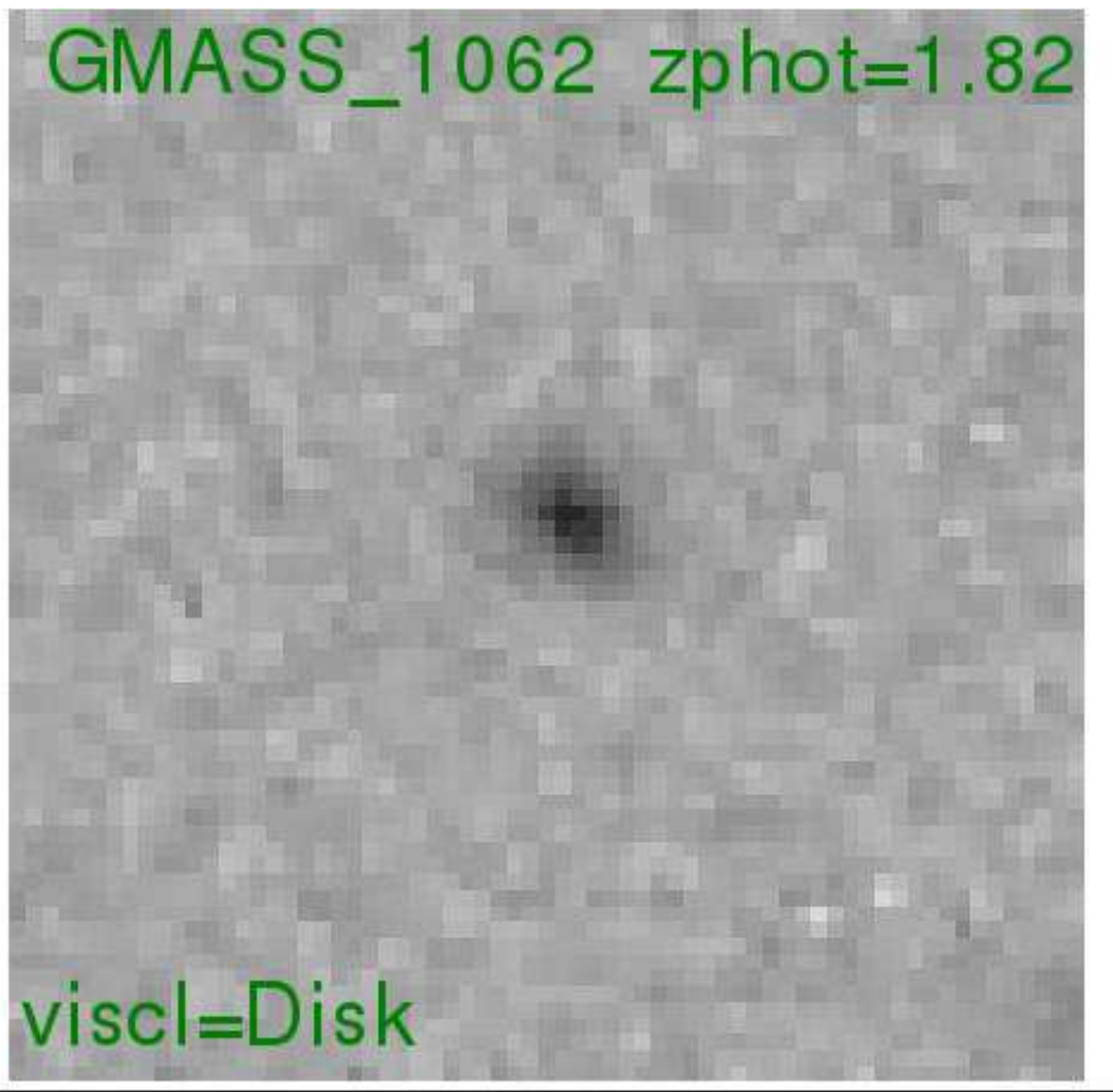}		     
\includegraphics[trim=100 40 75 390, clip=true, width=30mm]{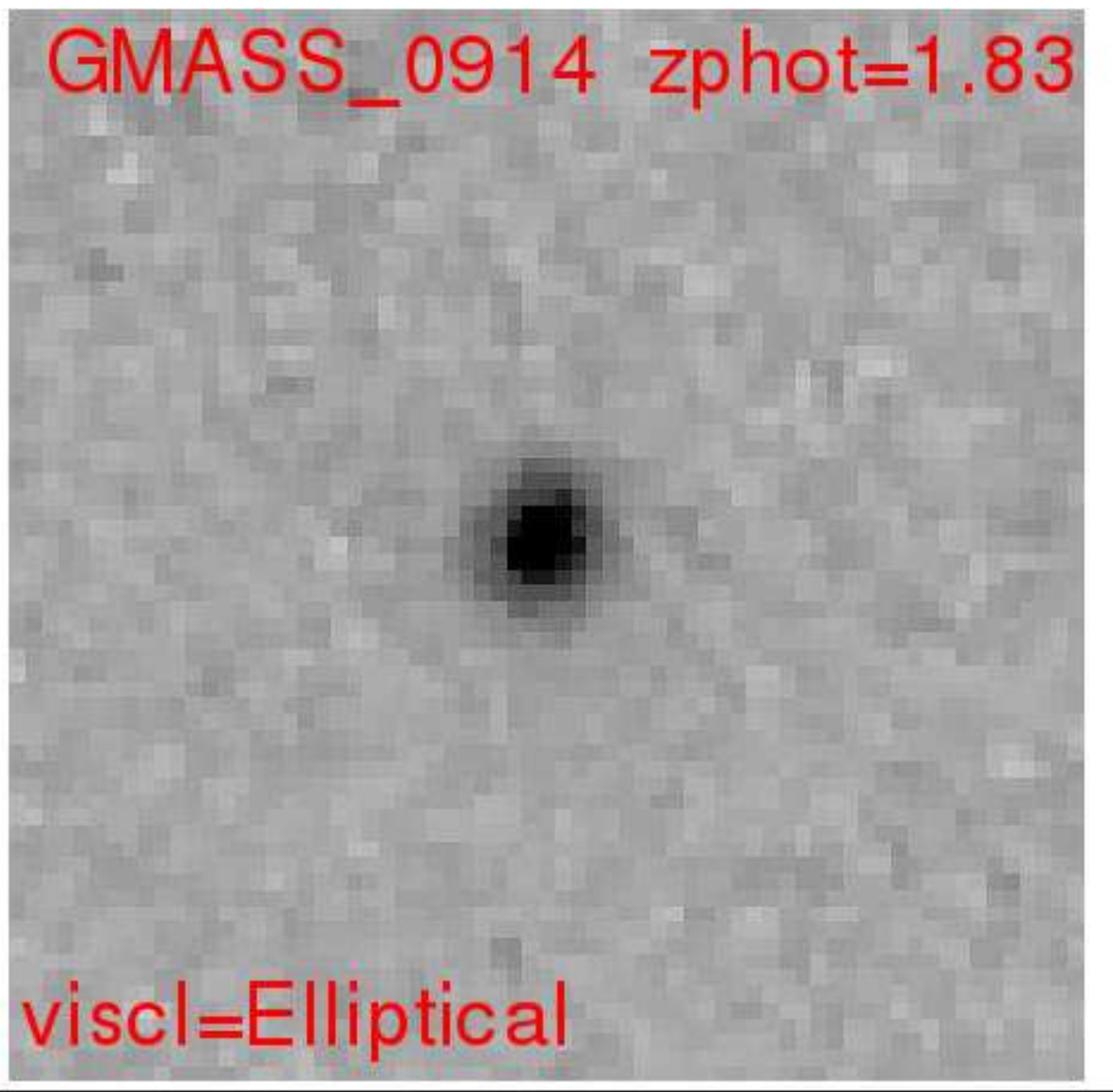}		     
\includegraphics[trim=100 40 75 390, clip=true, width=30mm]{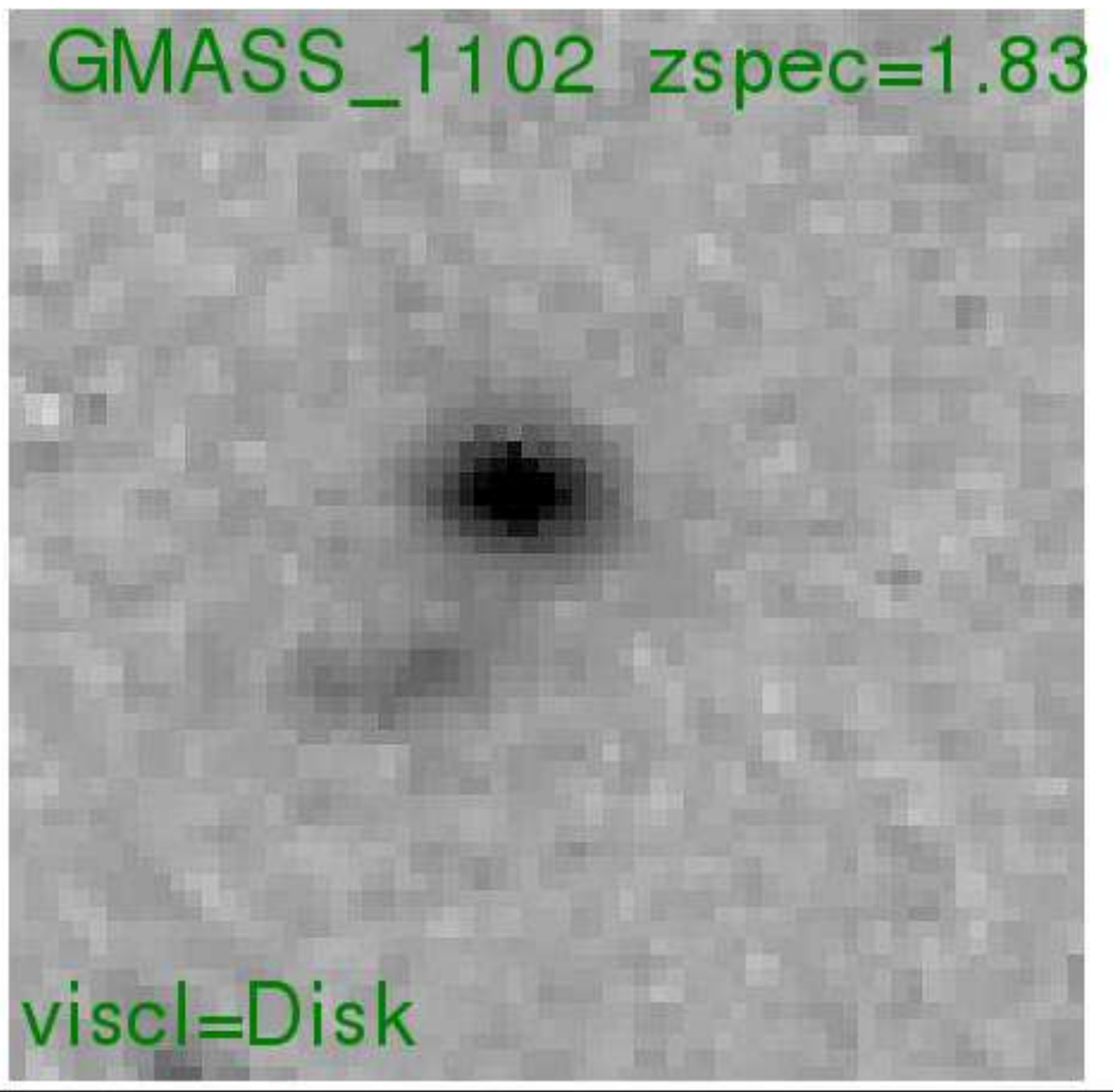}			     
\includegraphics[trim=100 40 75 390, clip=true, width=30mm]{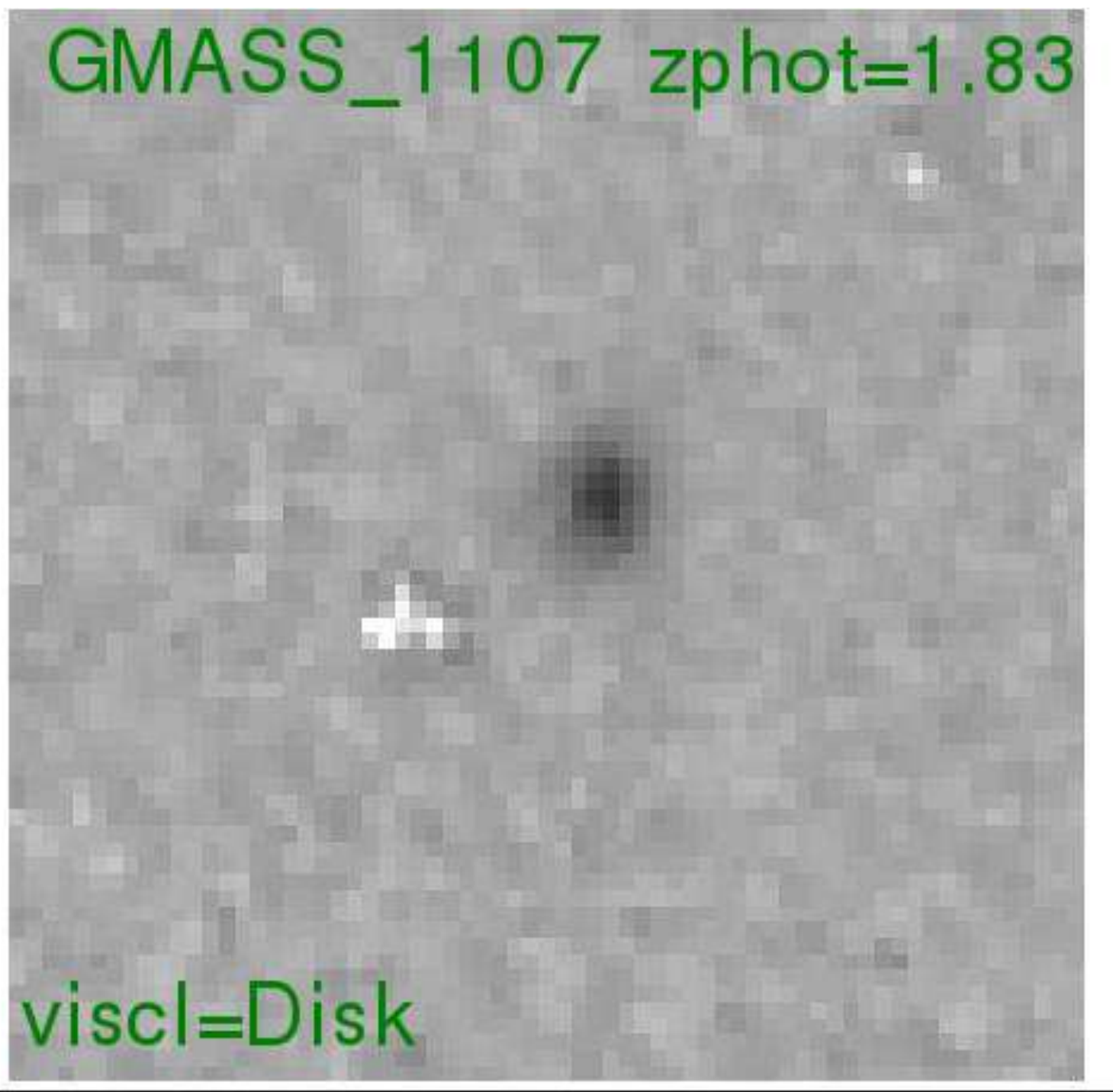}			     

\includegraphics[trim=100 40 75 390, clip=true, width=30mm]{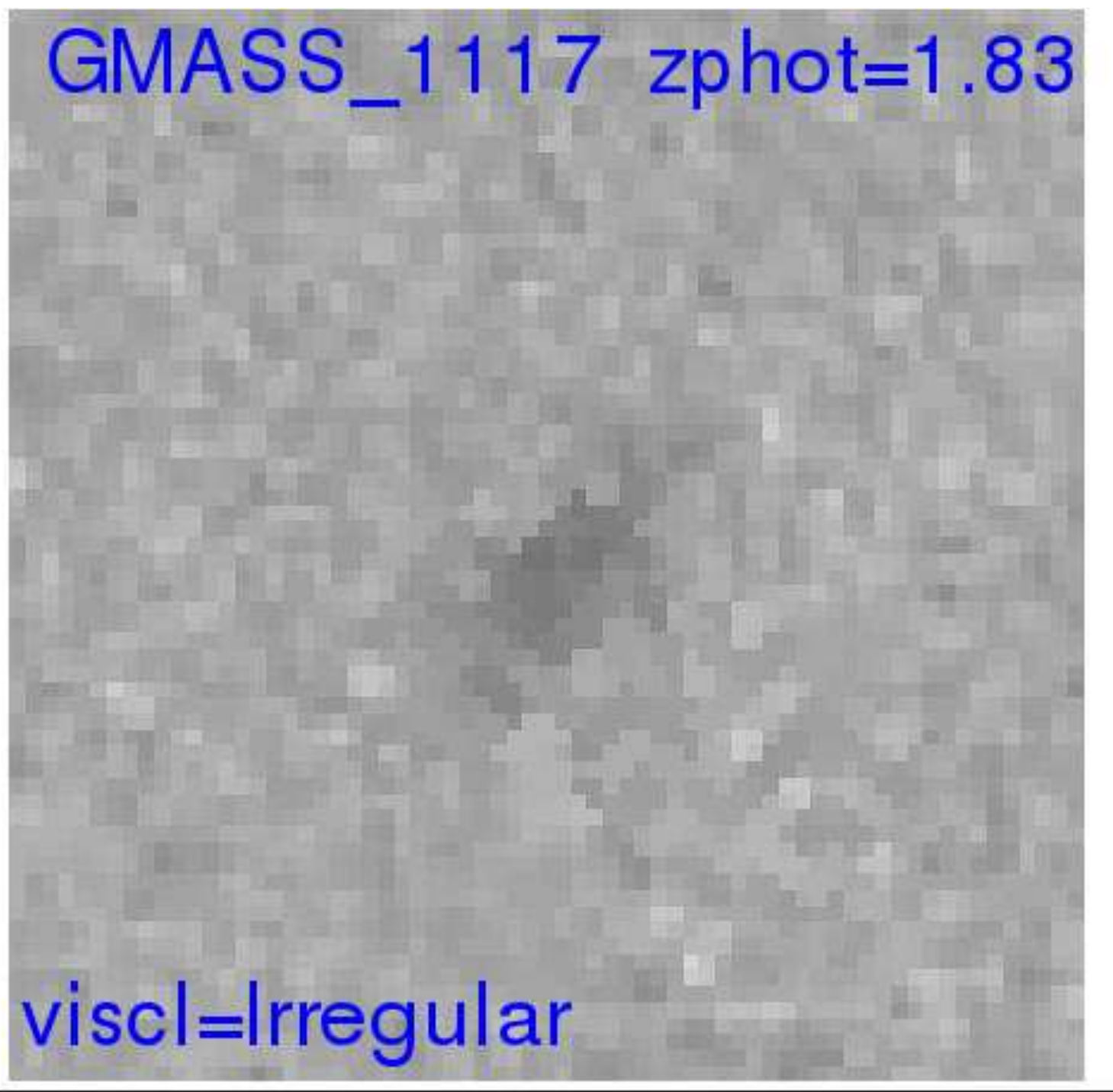}			     
\includegraphics[trim=100 40 75 390, clip=true, width=30mm]{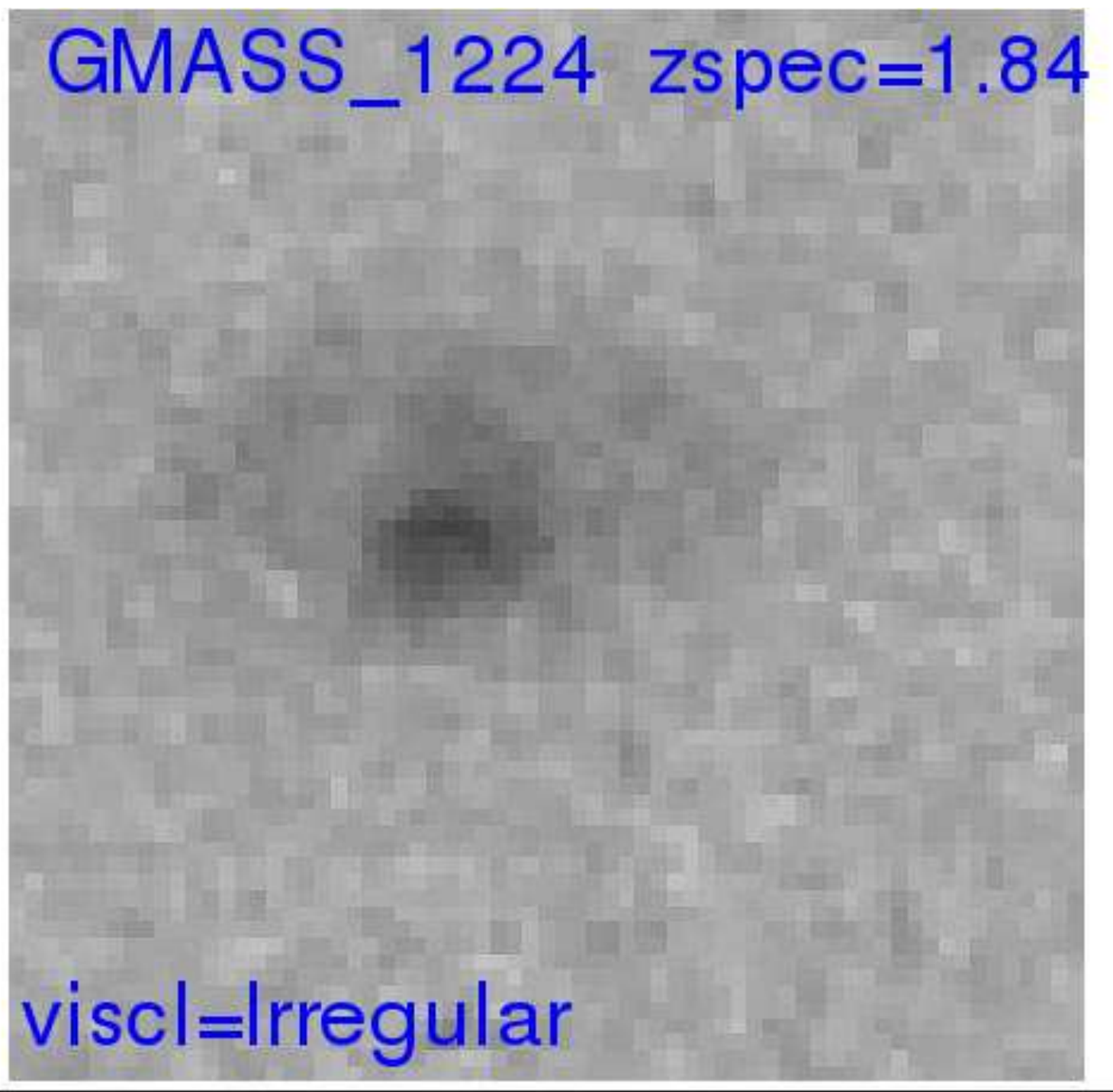}			     
\includegraphics[trim=100 40 75 390, clip=true, width=30mm]{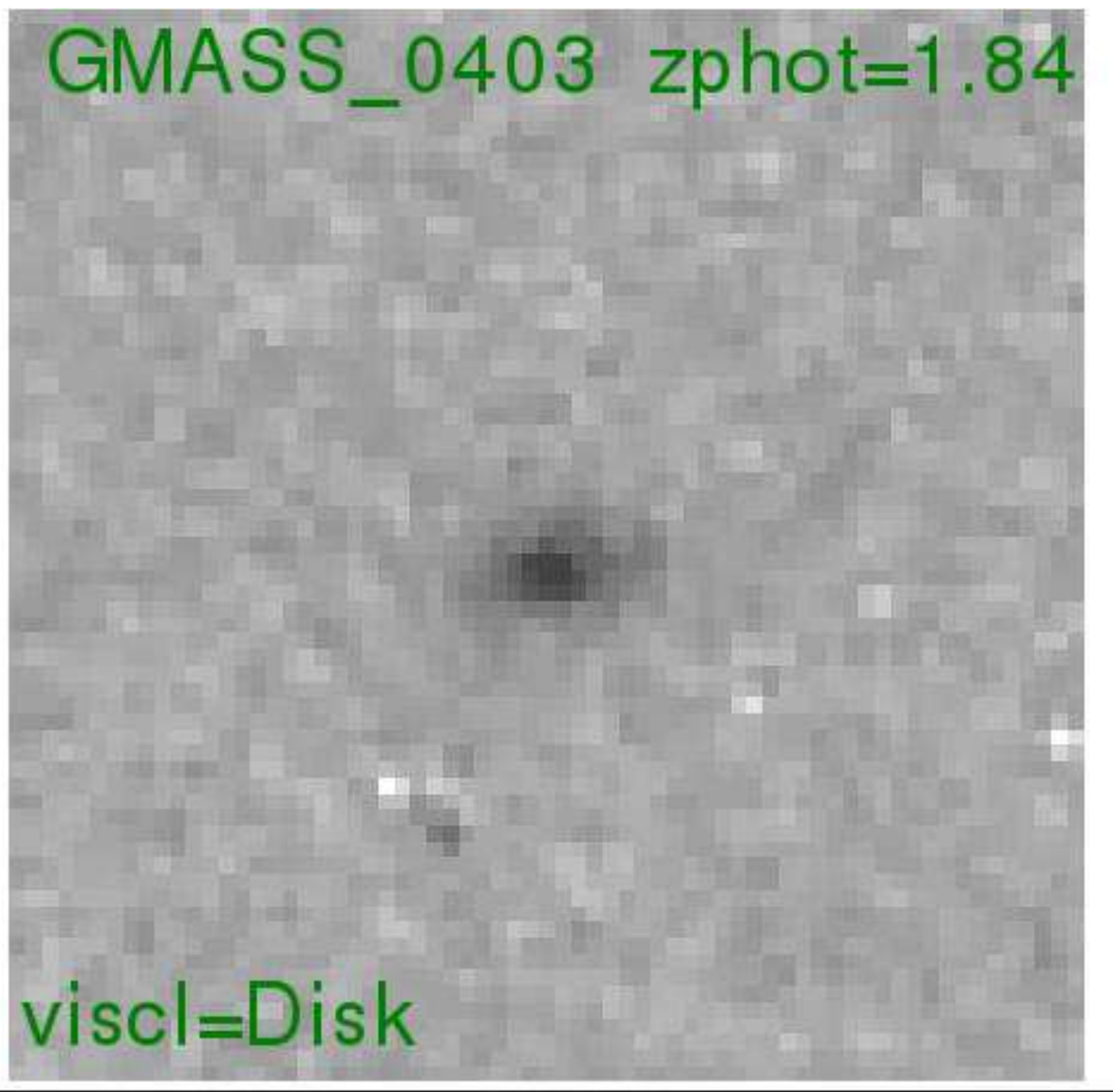}			     
\includegraphics[trim=100 40 75 390, clip=true, width=30mm]{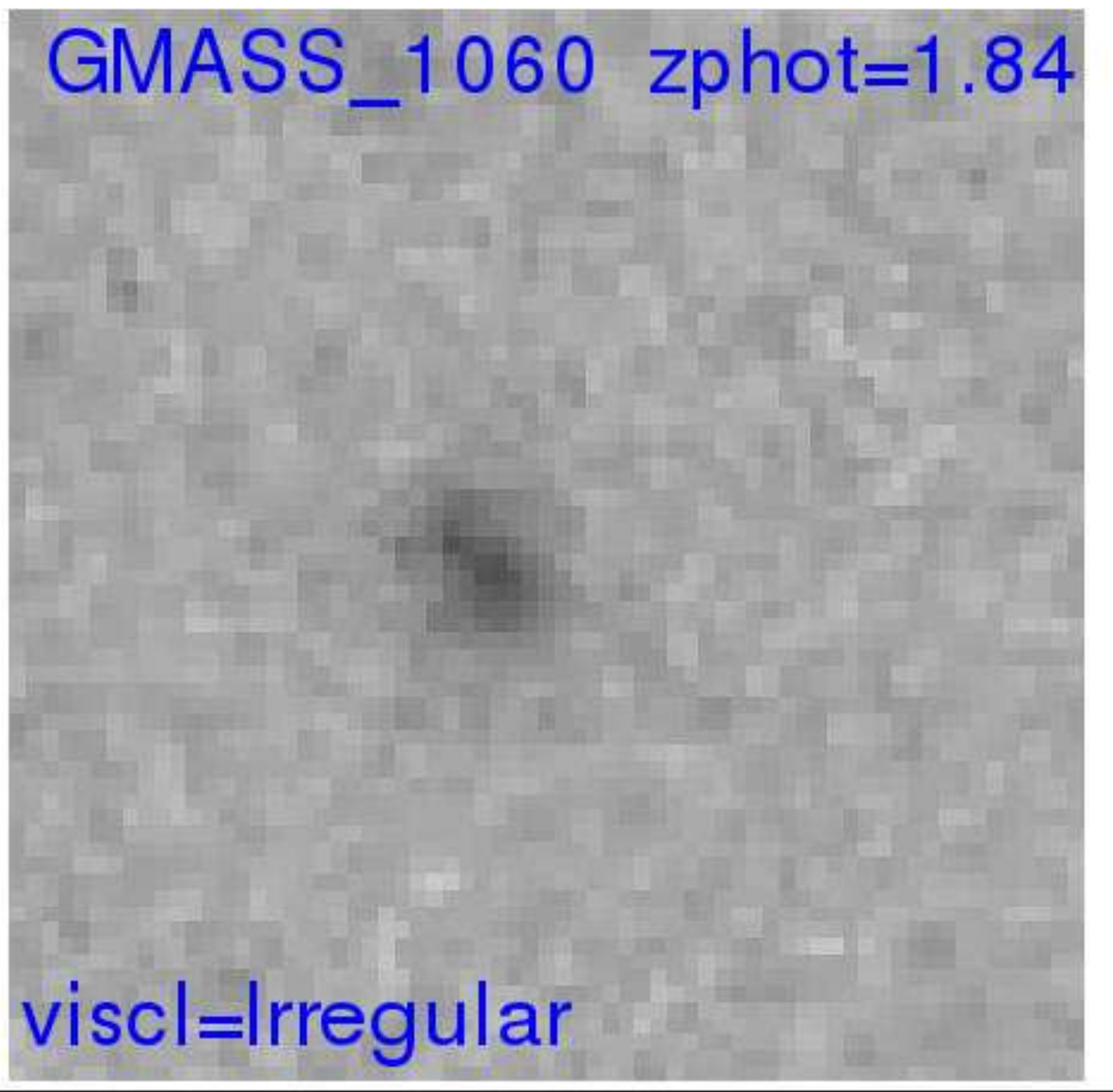}			     
\includegraphics[trim=100 40 75 390, clip=true, width=30mm]{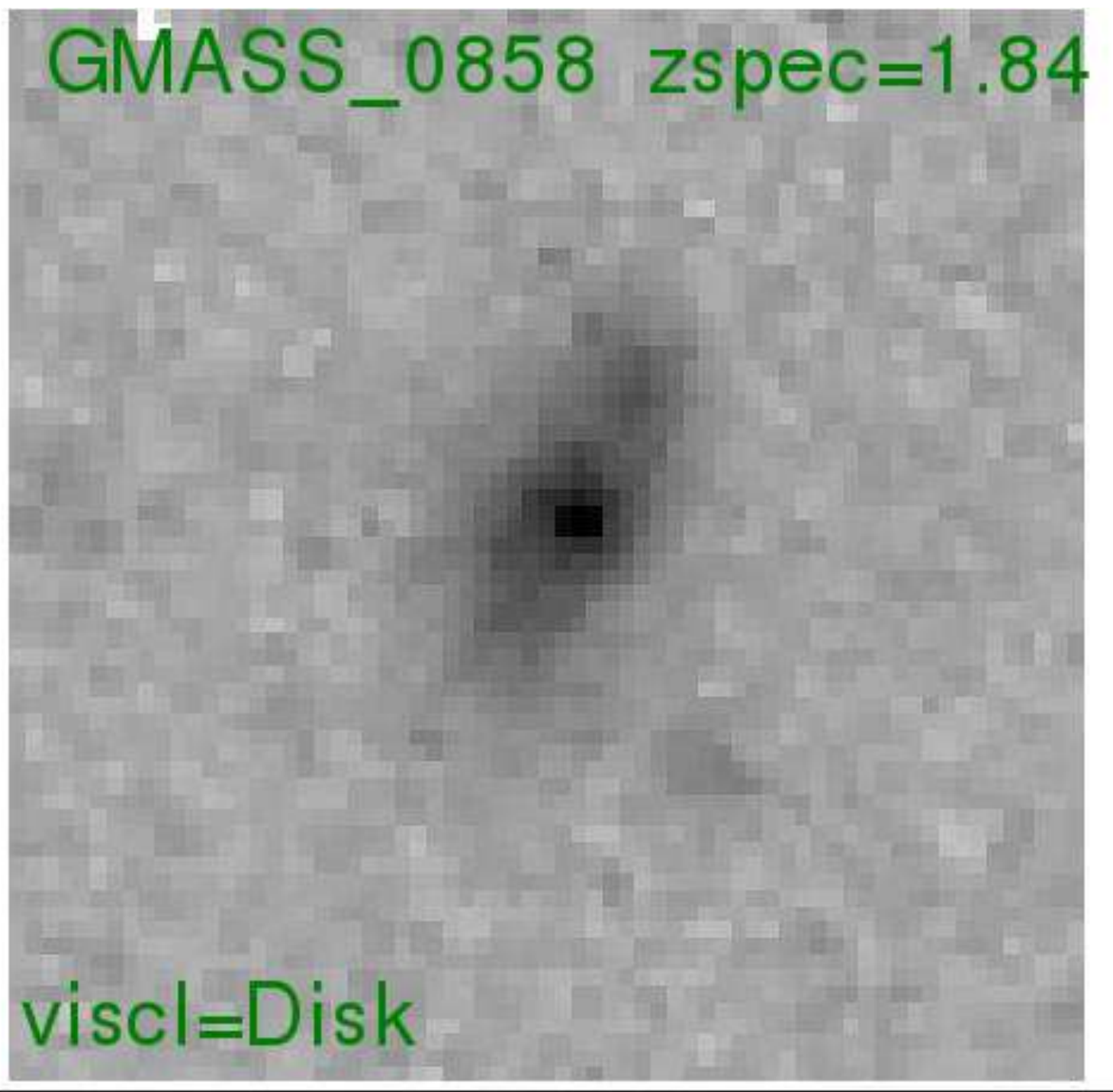}			     
\includegraphics[trim=100 40 75 390, clip=true, width=30mm]{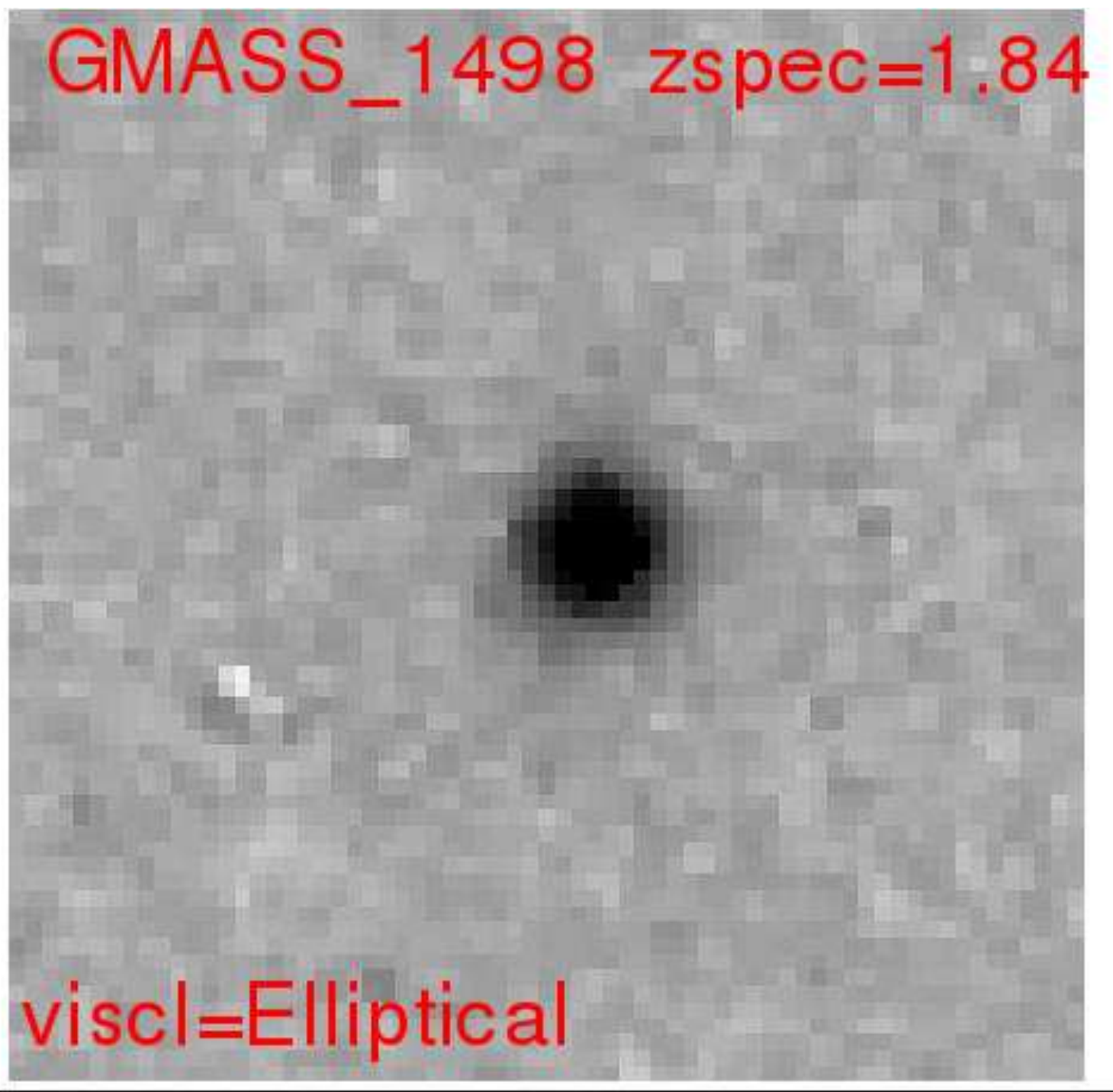}			     

\includegraphics[trim=100 40 75 390, clip=true, width=30mm]{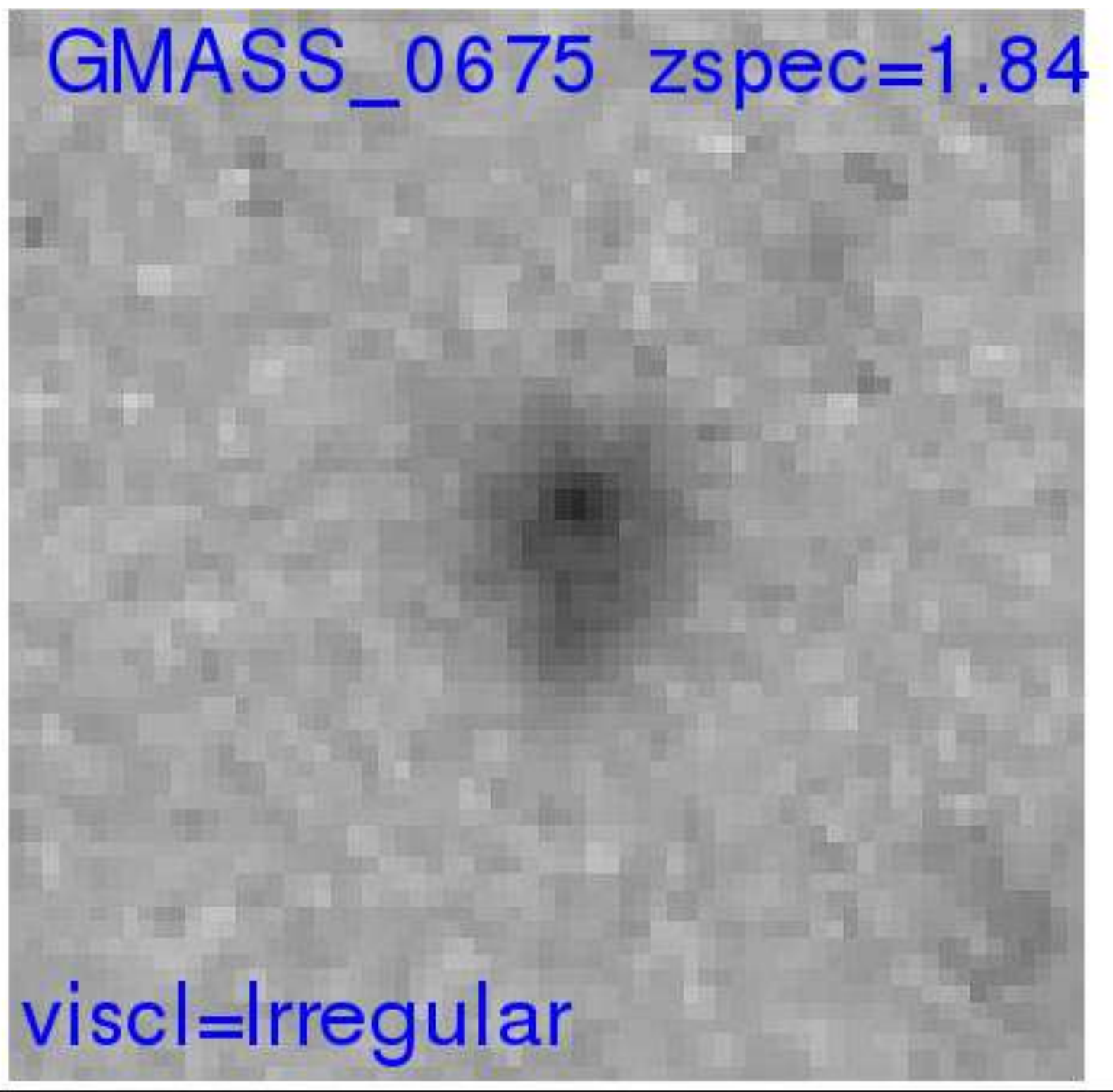}			     
\includegraphics[trim=100 40 75 390, clip=true, width=30mm]{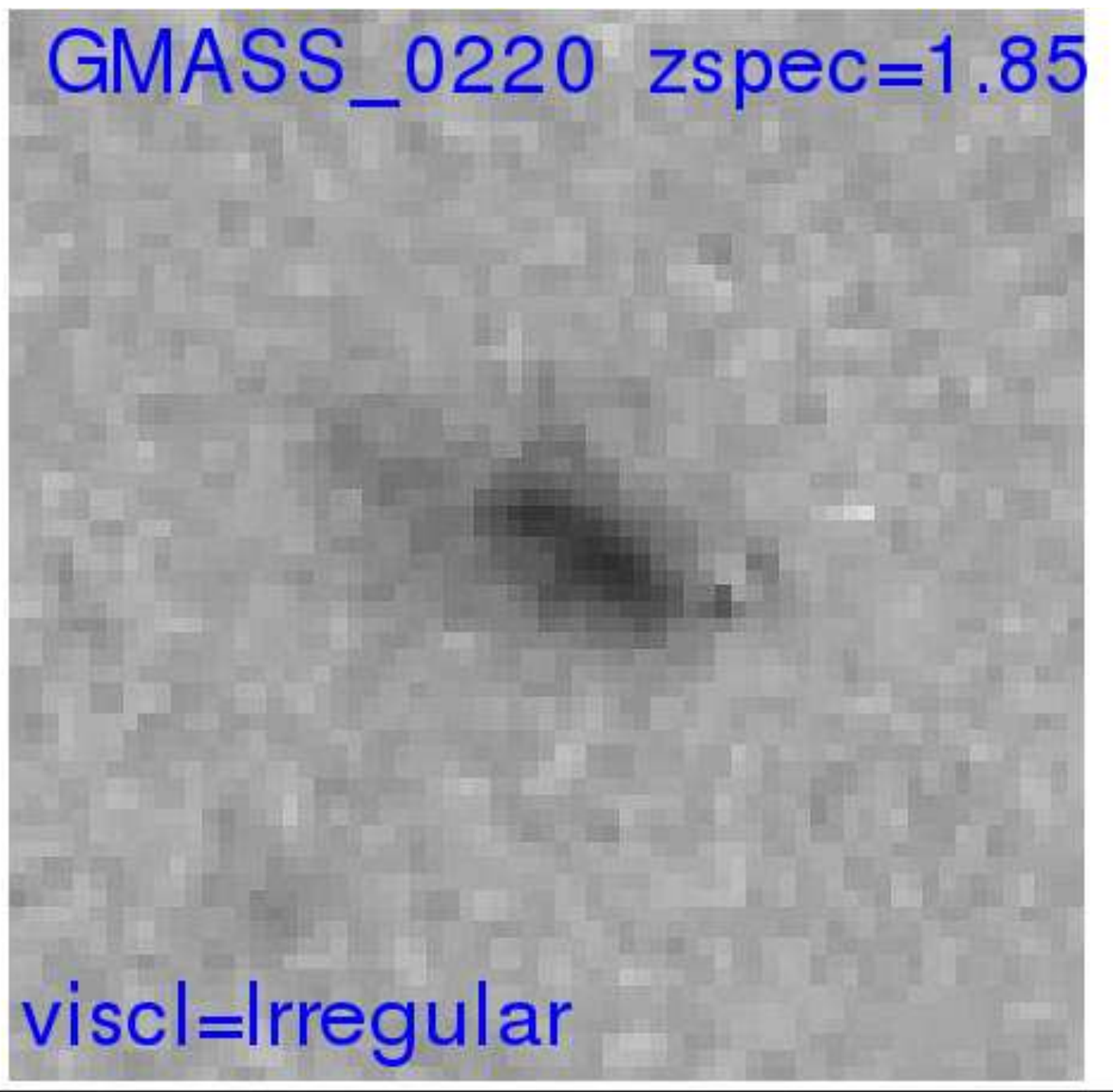}			     
\includegraphics[trim=100 40 75 390, clip=true, width=30mm]{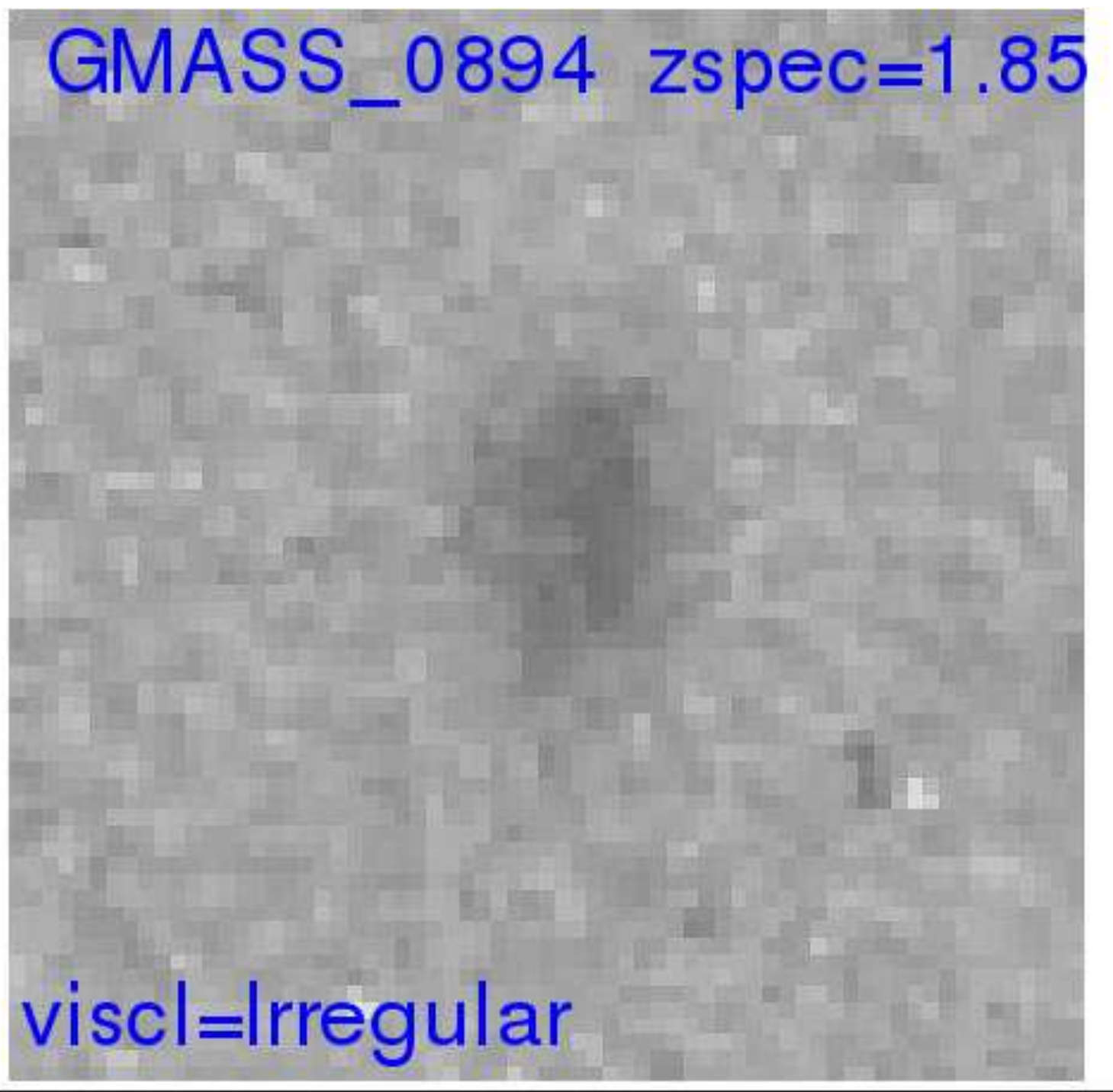}			     
\includegraphics[trim=100 40 75 390, clip=true, width=30mm]{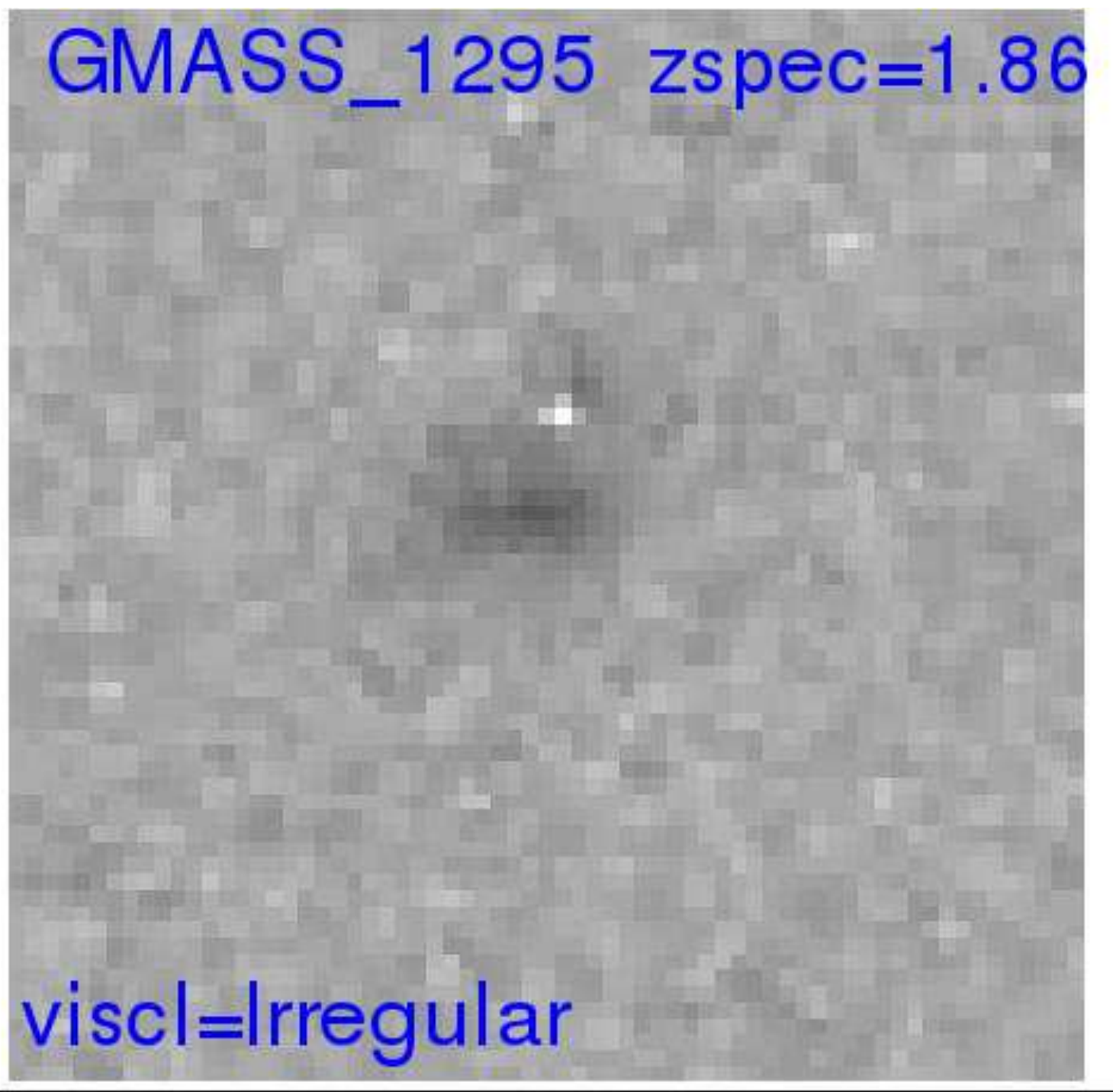}			     
\includegraphics[trim=100 40 75 390, clip=true, width=30mm]{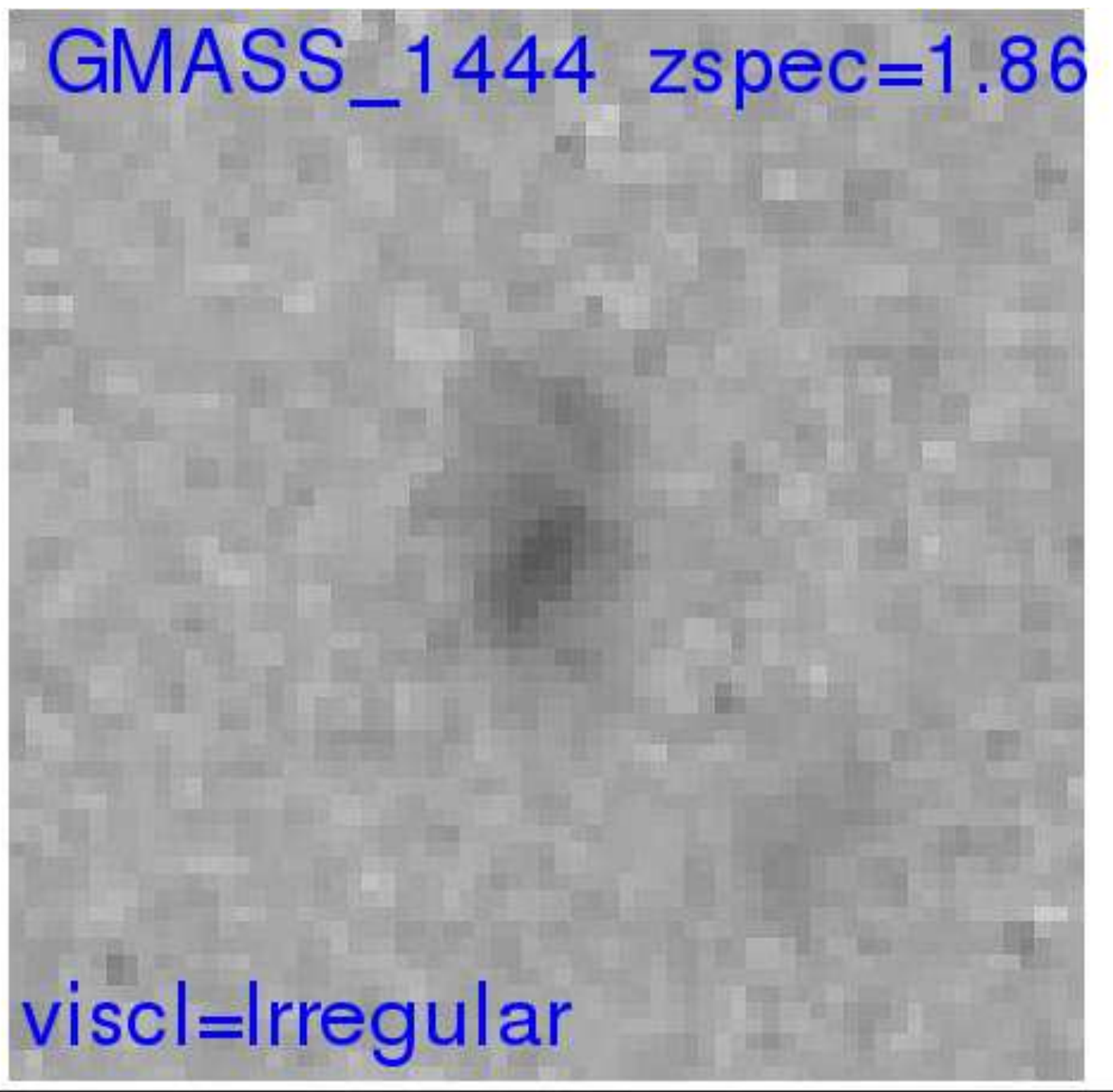}			     
\includegraphics[trim=100 40 75 390, clip=true, width=30mm]{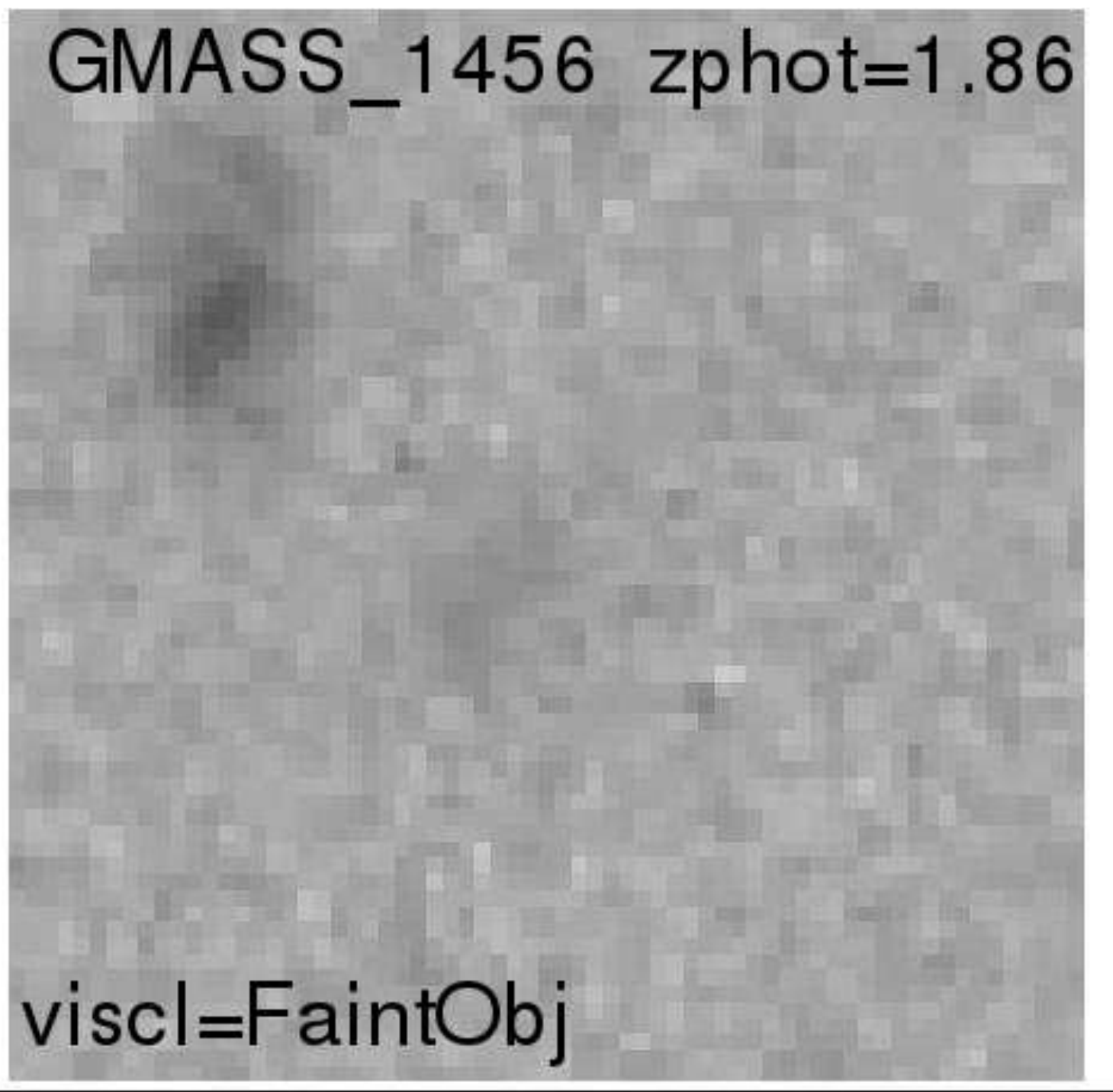}			     

\includegraphics[trim=100 40 75 390, clip=true, width=30mm]{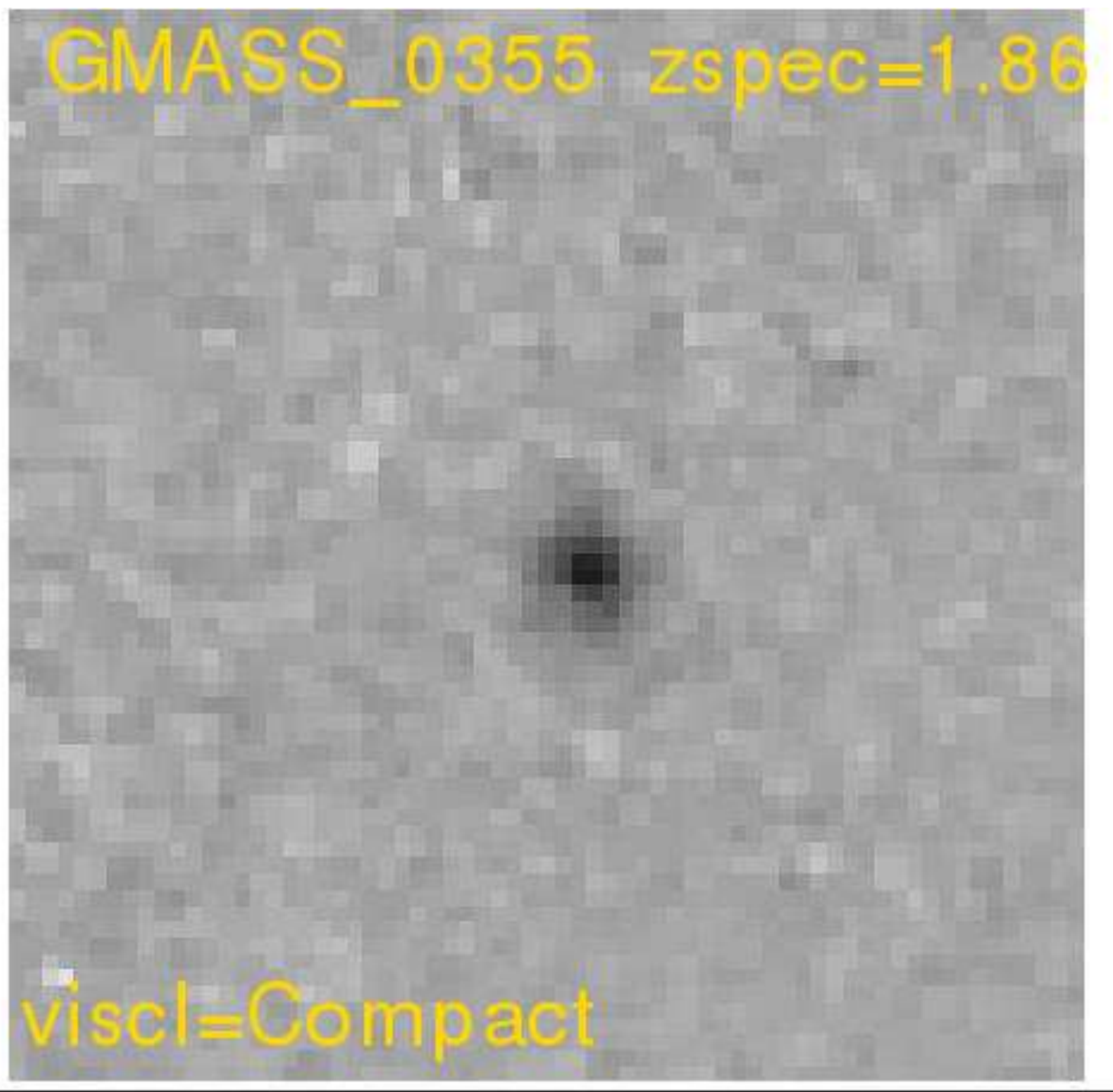}			     
\includegraphics[trim=100 40 75 390, clip=true, width=30mm]{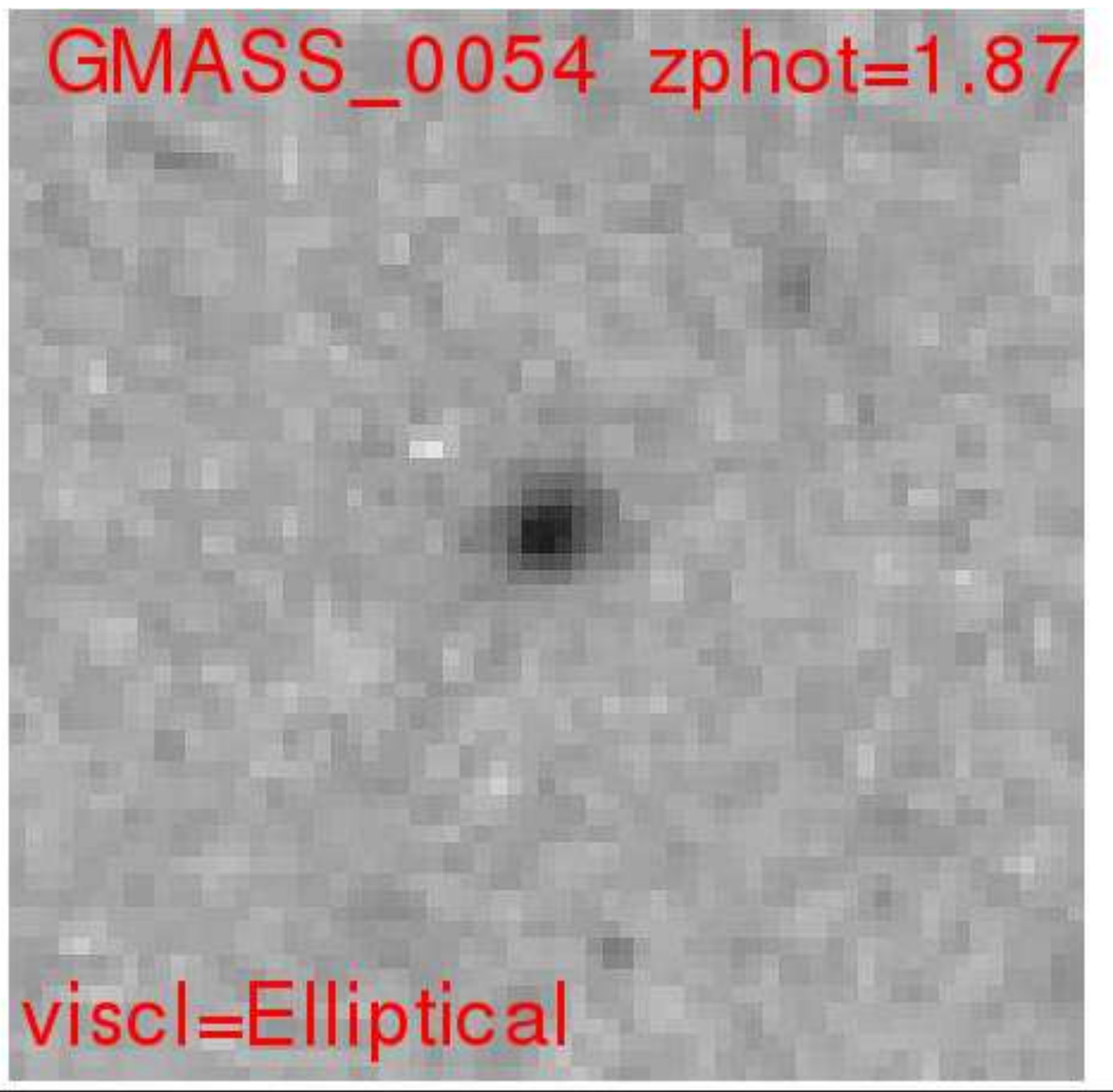}		     
\includegraphics[trim=100 40 75 390, clip=true, width=30mm]{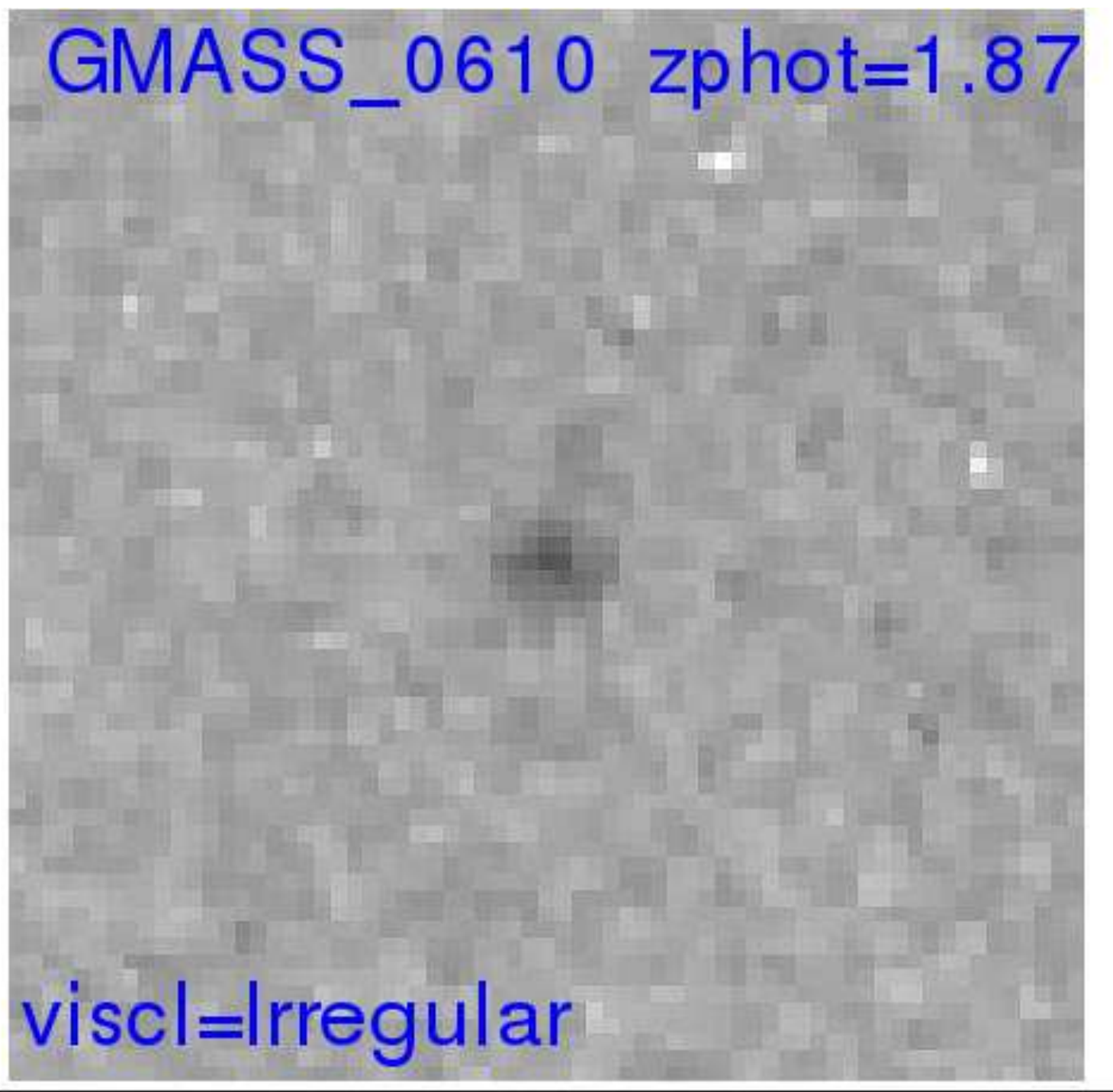}		     
\includegraphics[trim=100 40 75 390, clip=true, width=30mm]{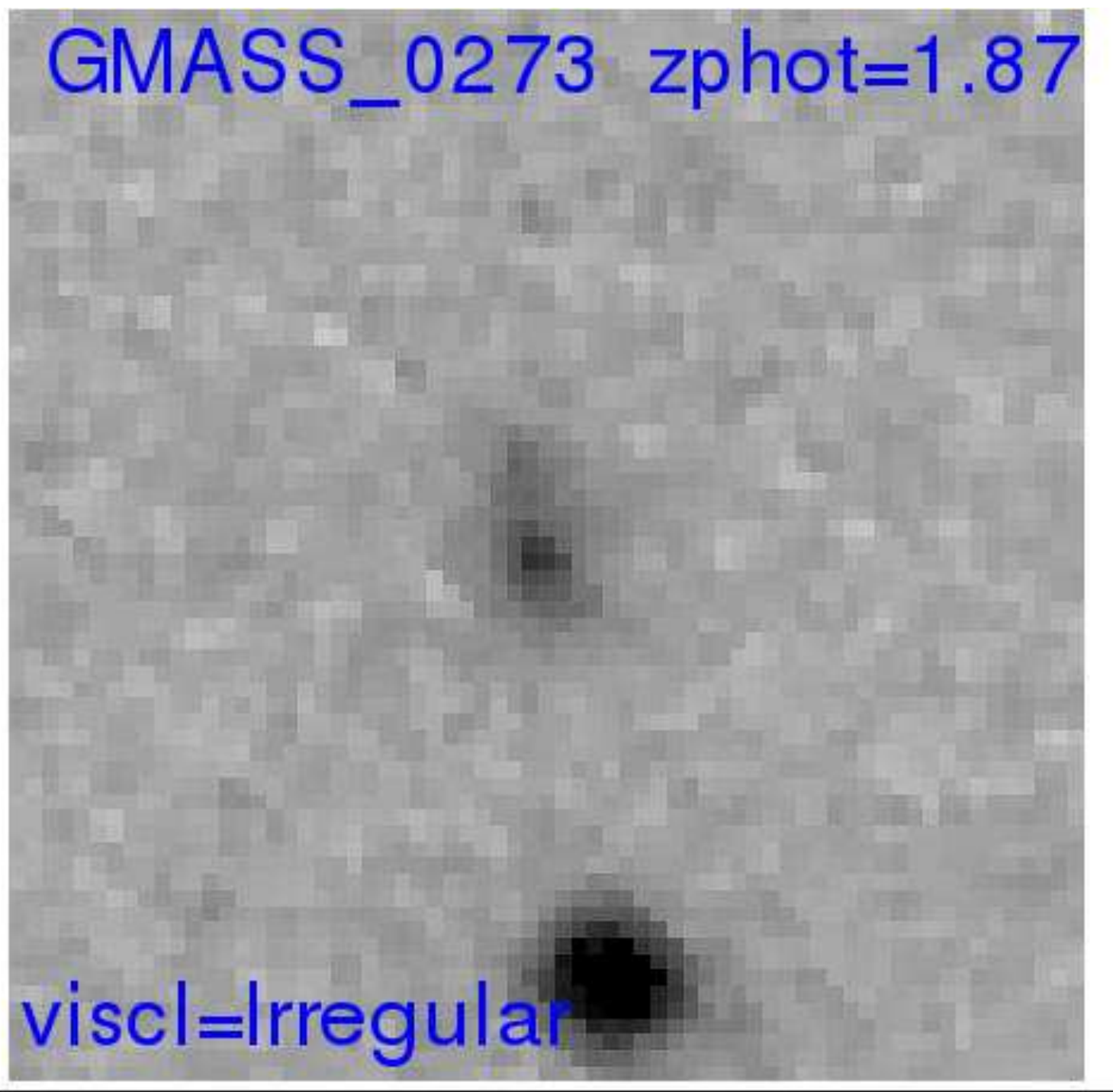}			     
\includegraphics[trim=100 40 75 390, clip=true, width=30mm]{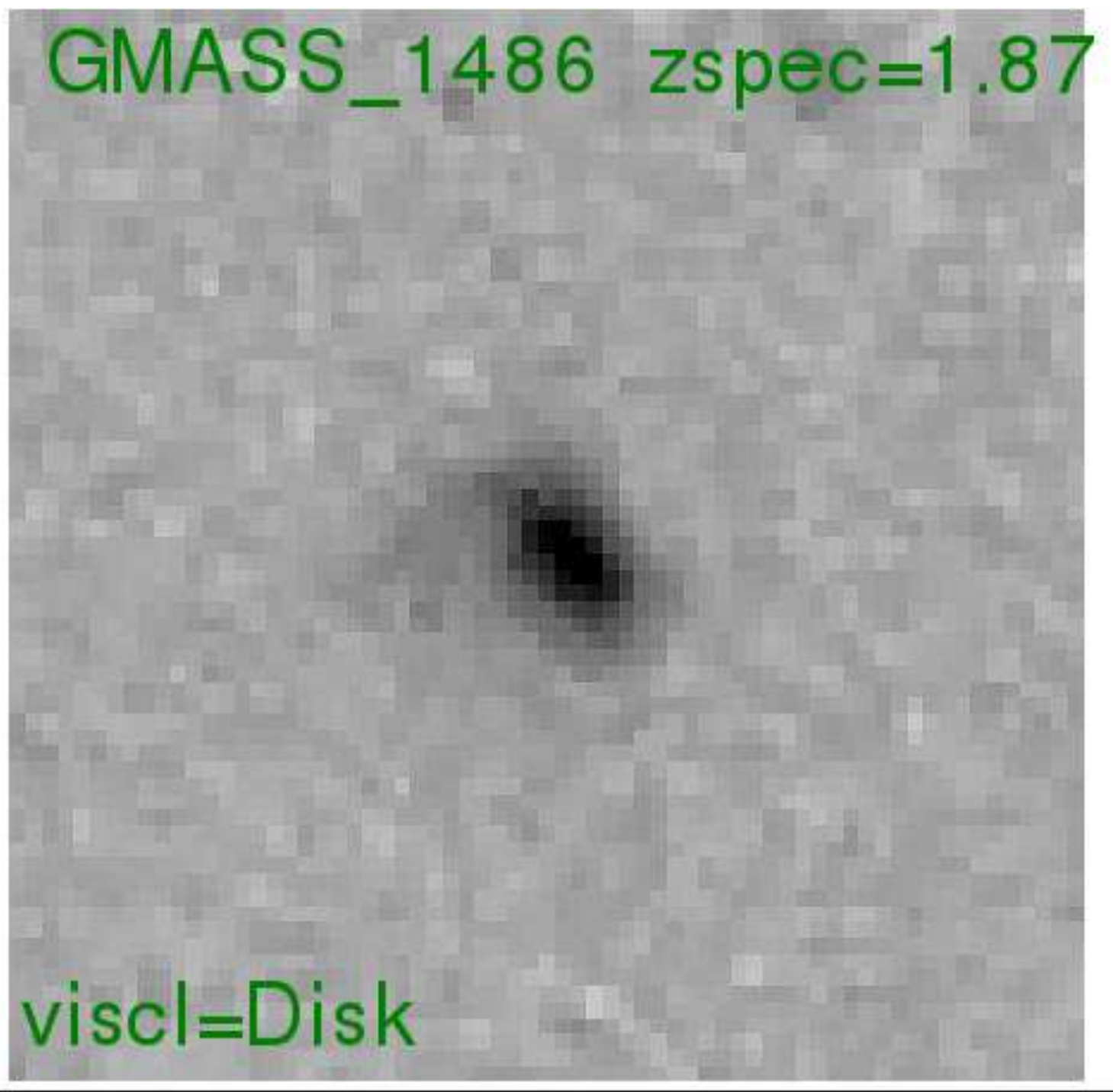}			     
\includegraphics[trim=100 40 75 390, clip=true, width=30mm]{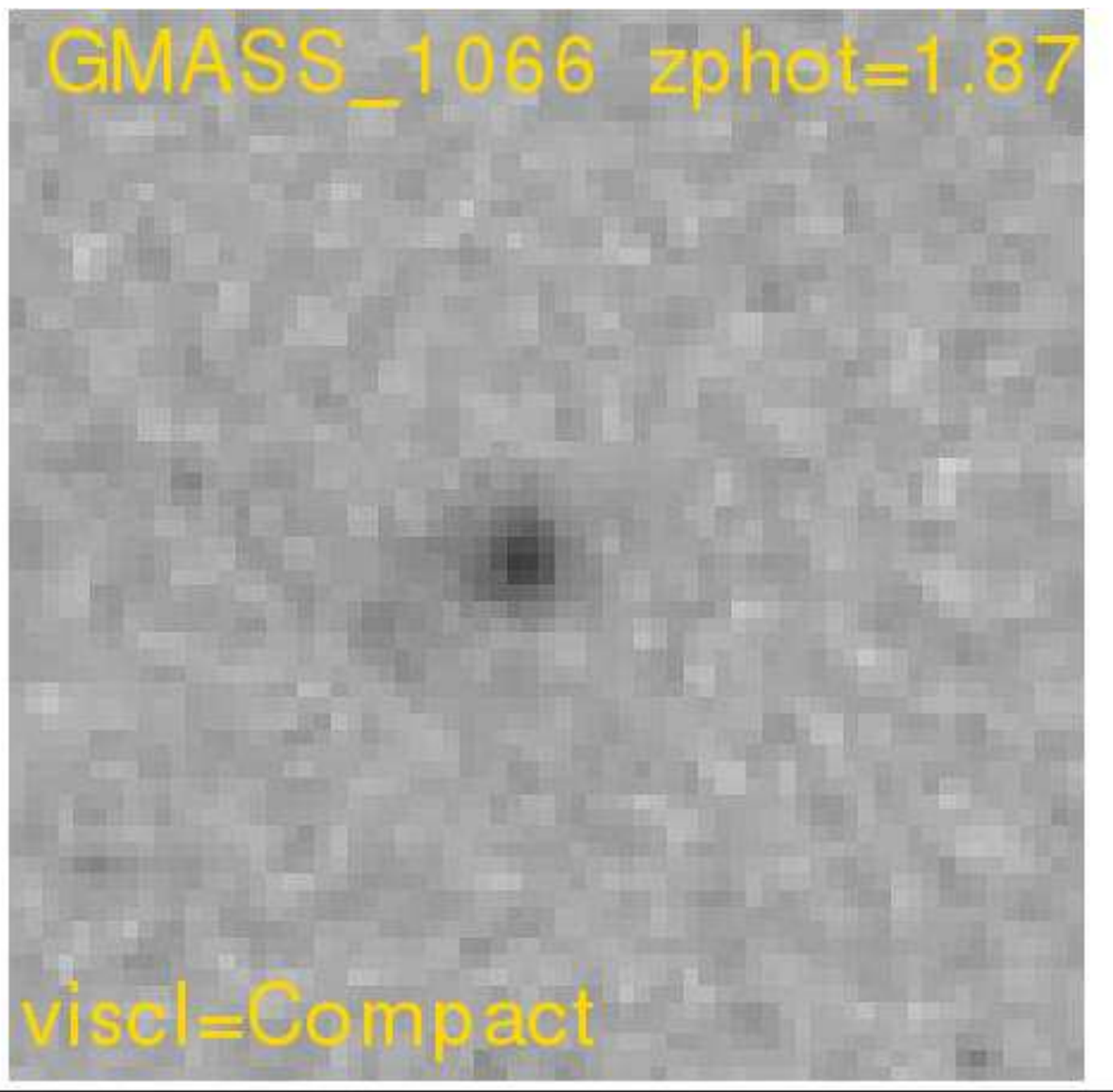}	
\end{figure*}
\begin{figure*}
\centering   
\includegraphics[trim=100 40 75 390, clip=true, width=30mm]{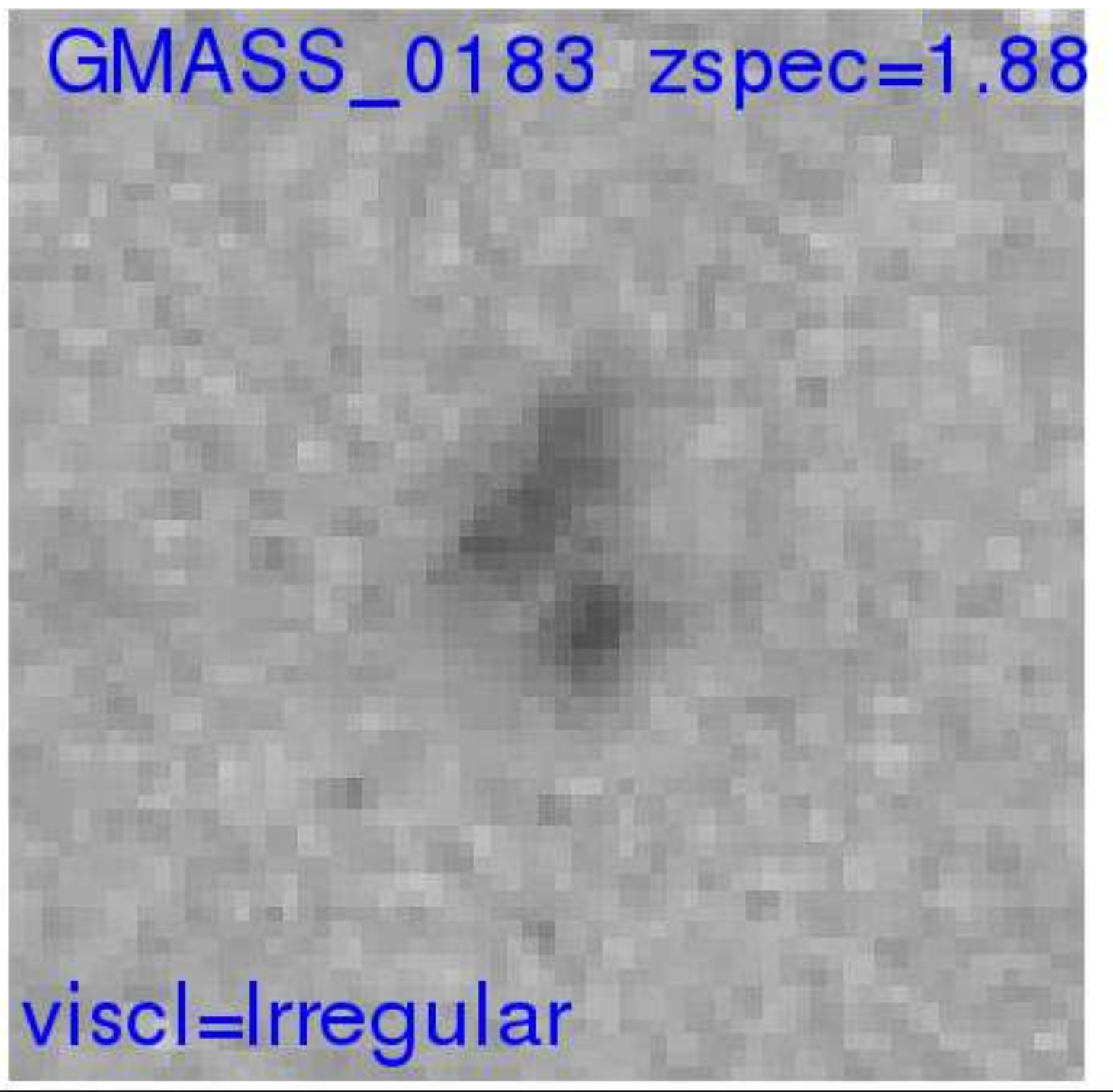}			     
\includegraphics[trim=100 40 75 390, clip=true, width=30mm]{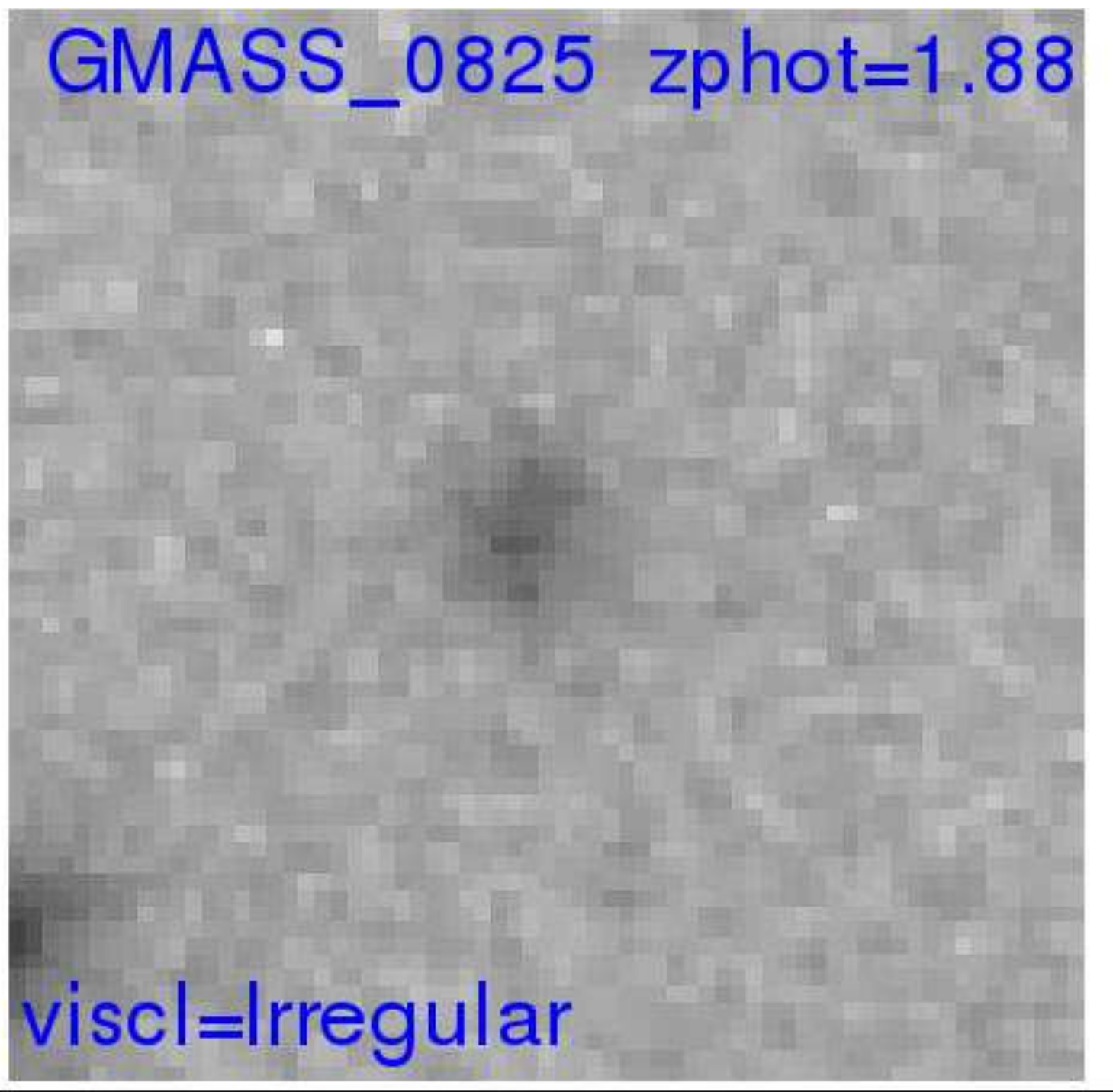}			     
\includegraphics[trim=100 40 75 390, clip=true, width=30mm]{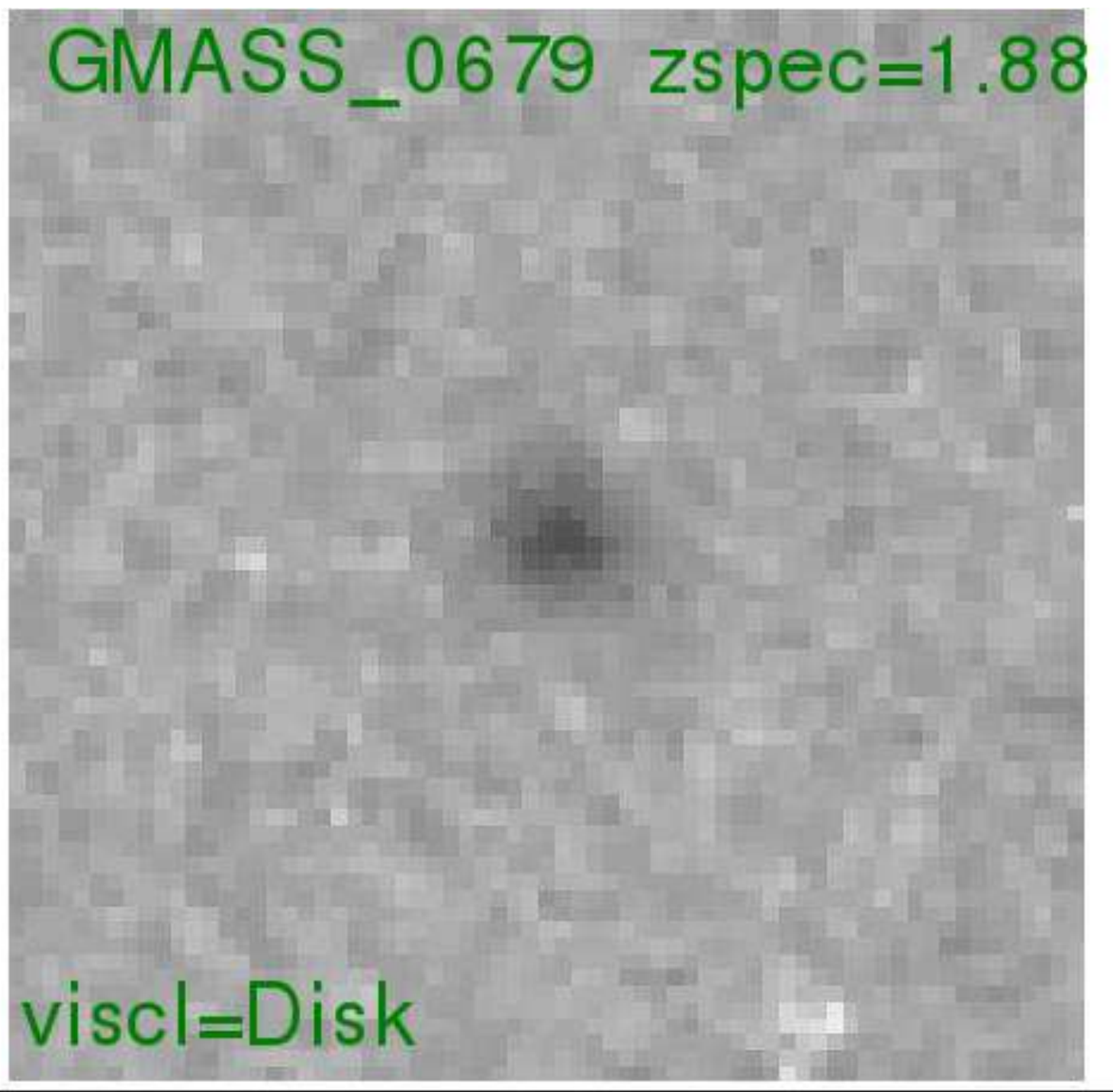}			     
\includegraphics[trim=100 40 75 390, clip=true, width=30mm]{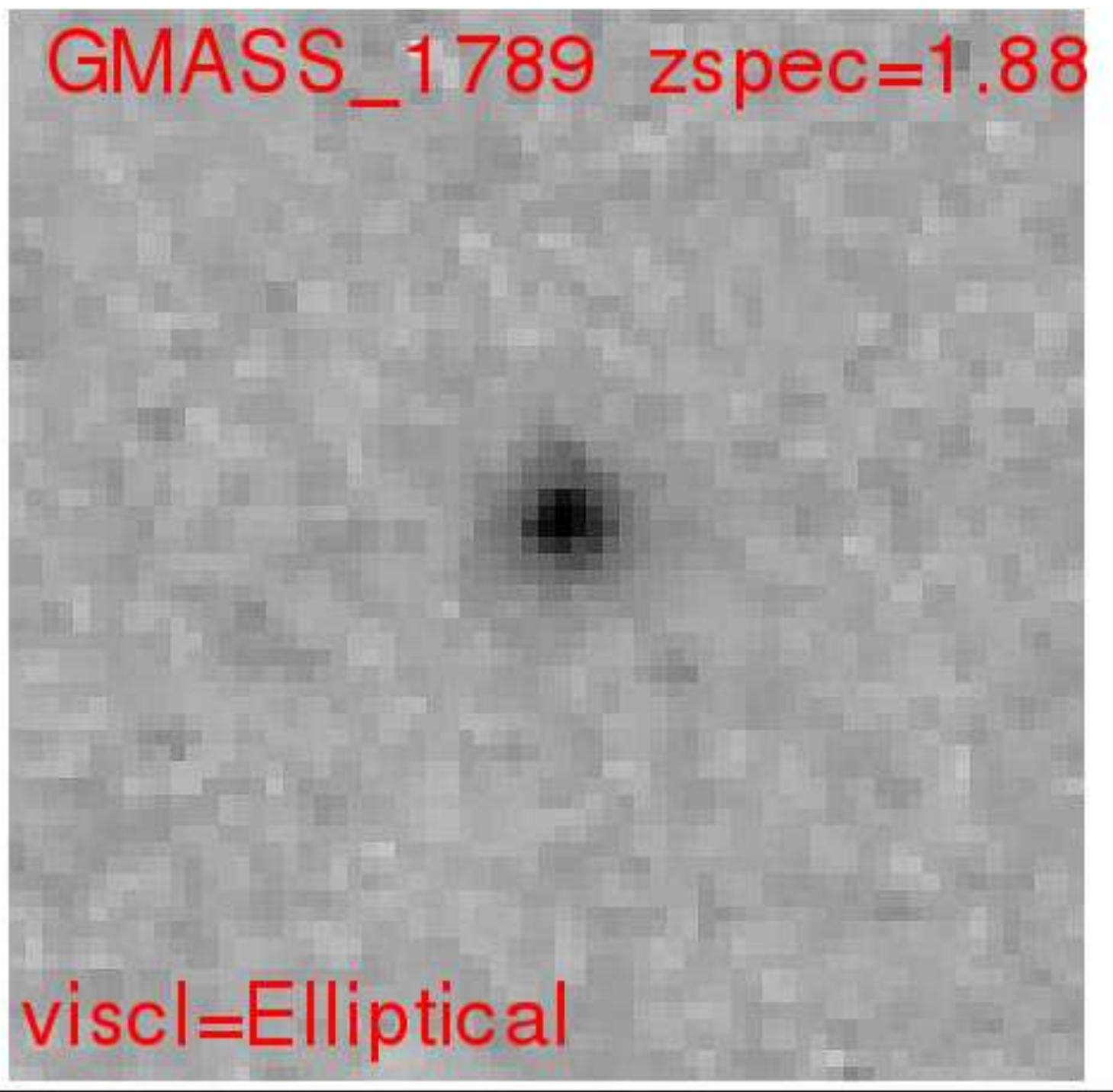}			     
\includegraphics[trim=100 40 75 390, clip=true, width=30mm]{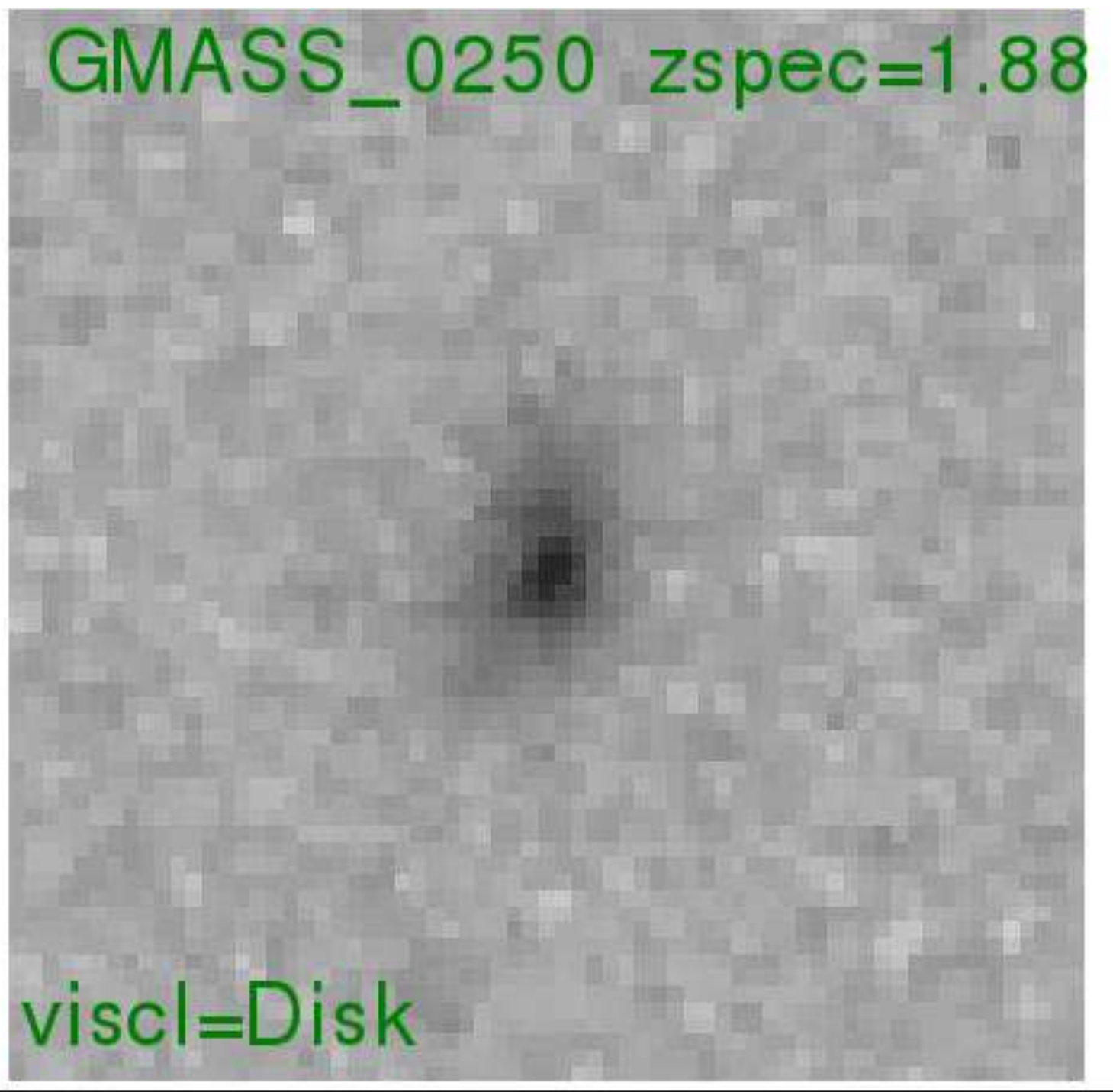}			     
\includegraphics[trim=100 40 75 390, clip=true, width=30mm]{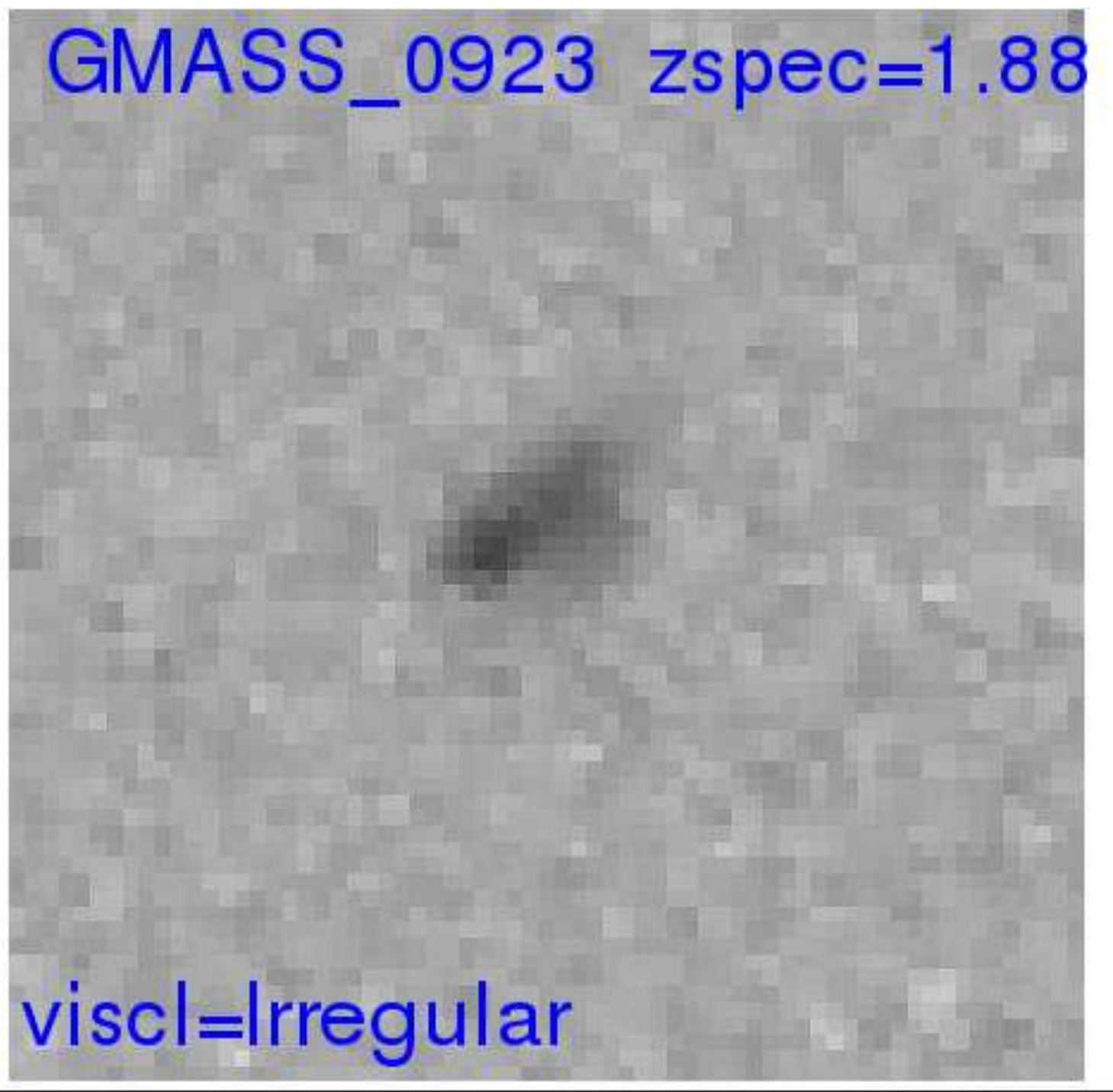}			     

\includegraphics[trim=100 40 75 390, clip=true, width=30mm]{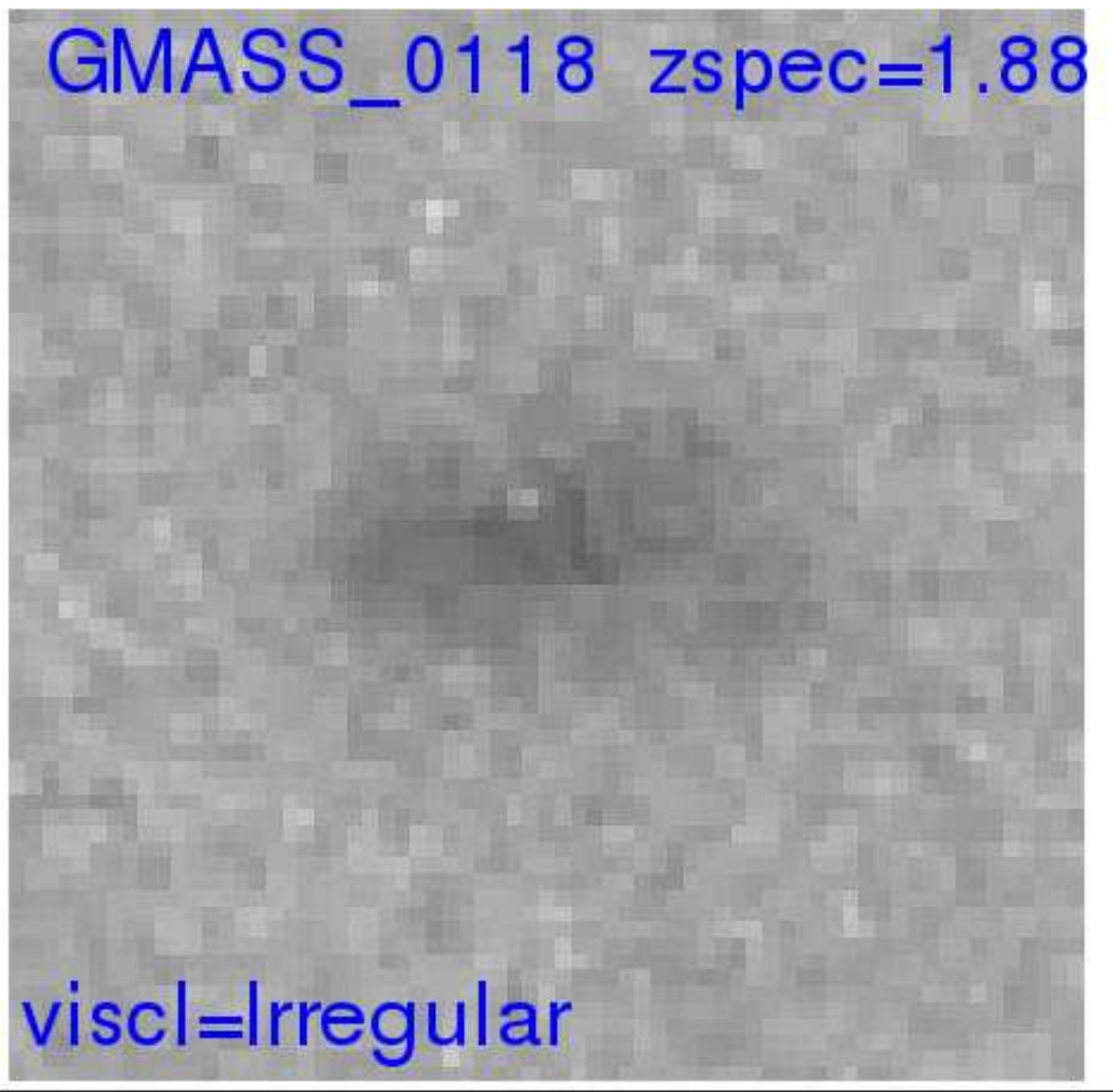}			     
\includegraphics[trim=100 40 75 390, clip=true, width=30mm]{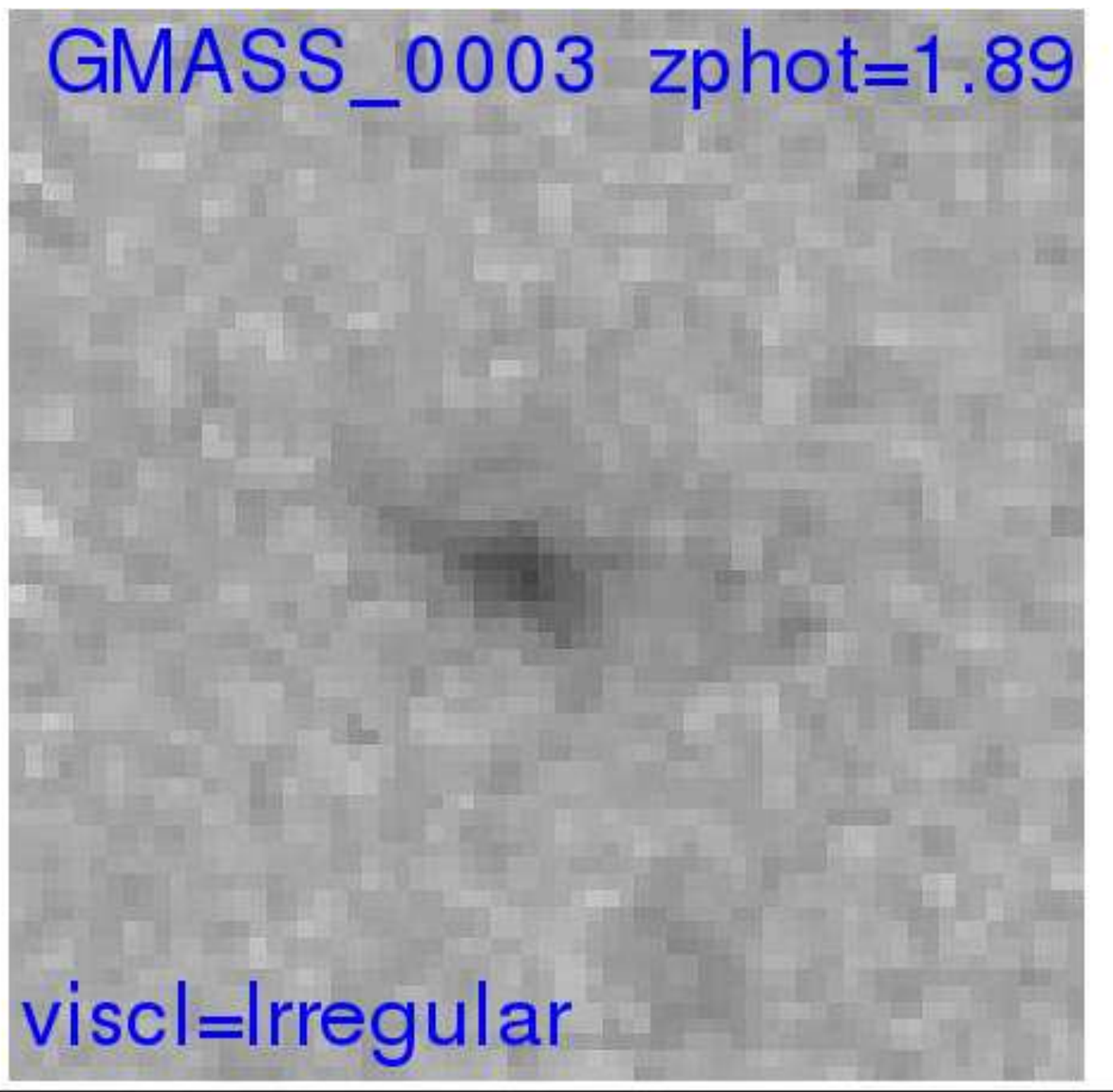}			     
\includegraphics[trim=100 40 75 390, clip=true, width=30mm]{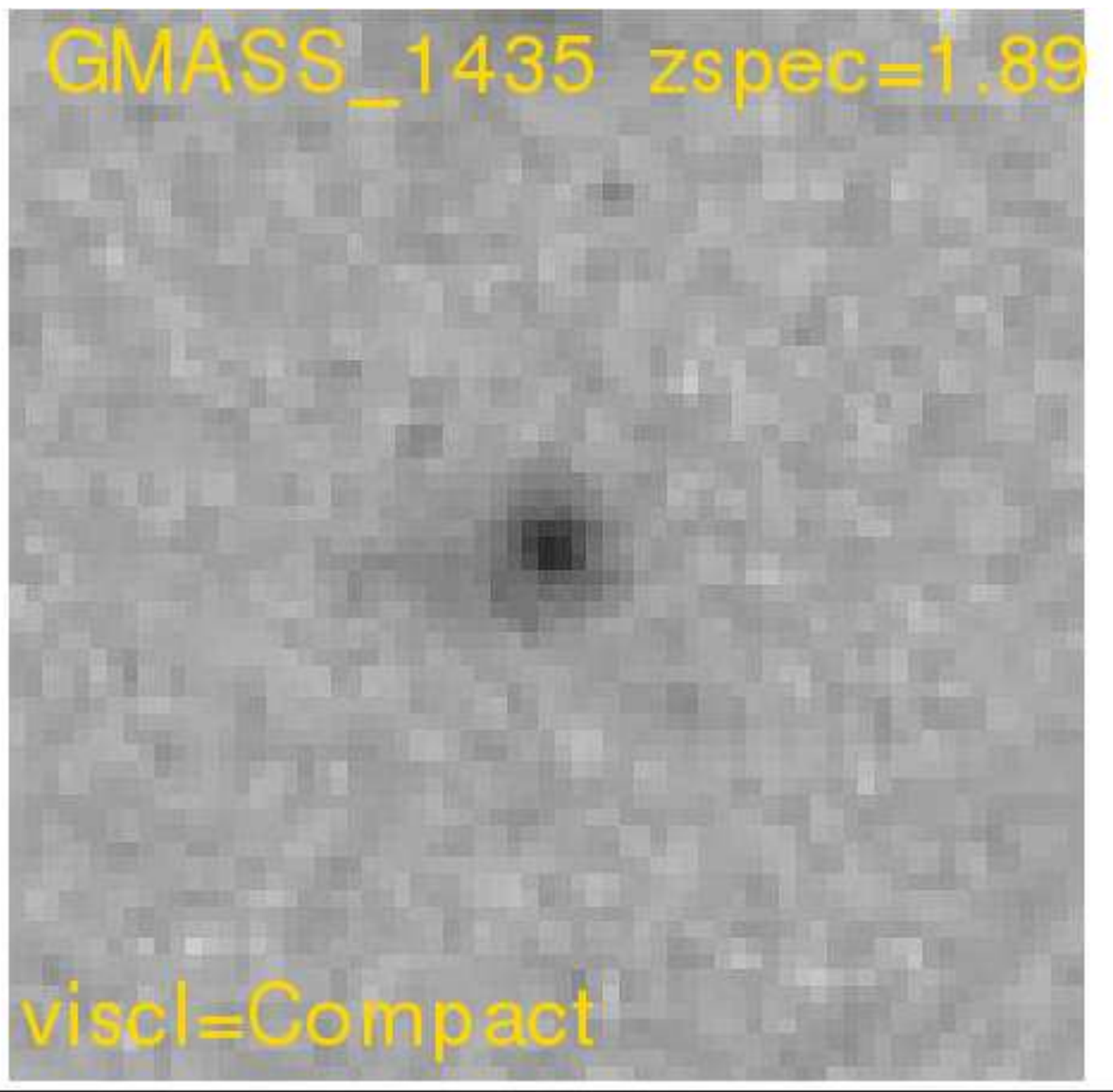}			     
\includegraphics[trim=100 40 75 390, clip=true, width=30mm]{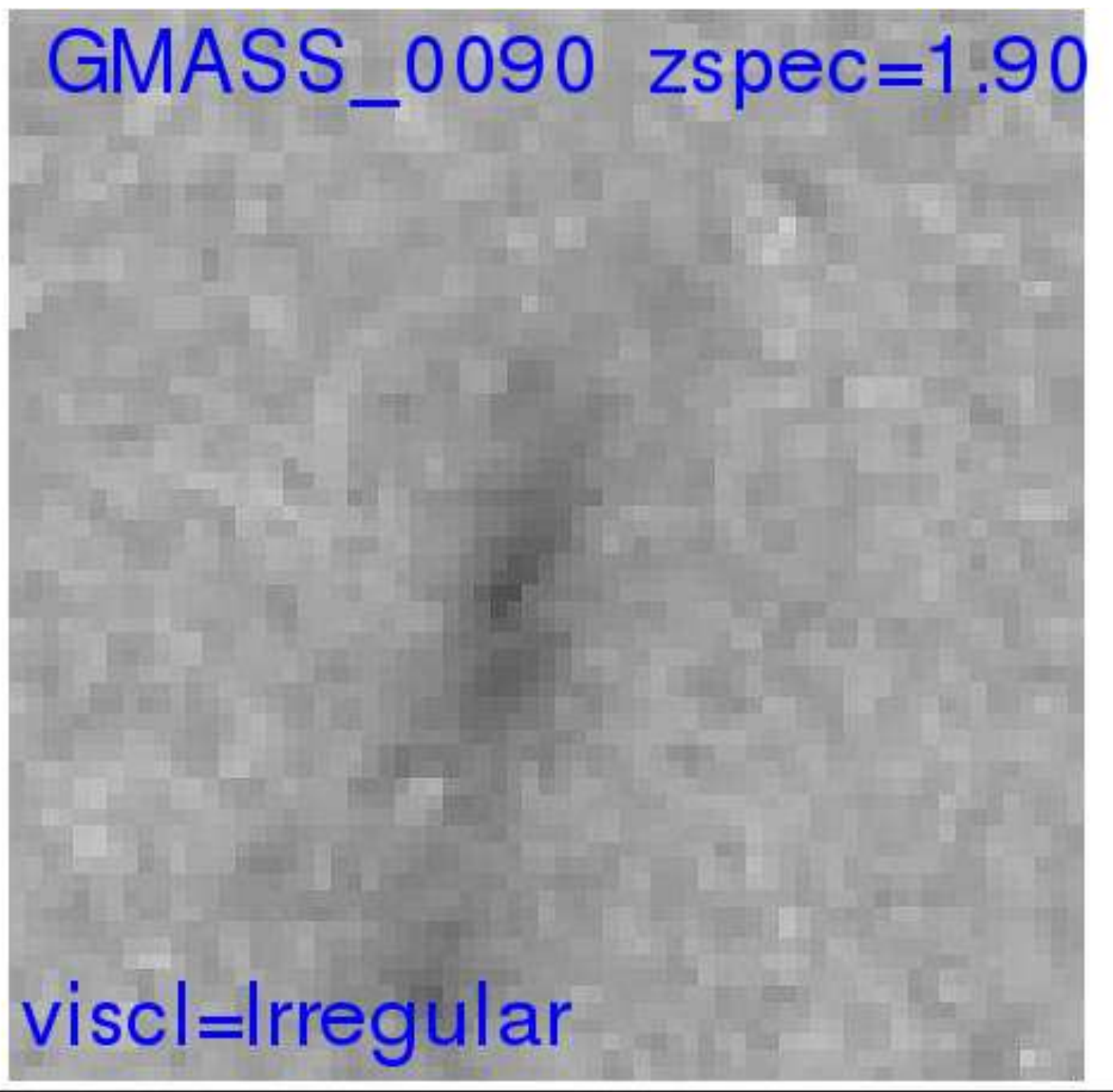}			     
\includegraphics[trim=100 40 75 390, clip=true, width=30mm]{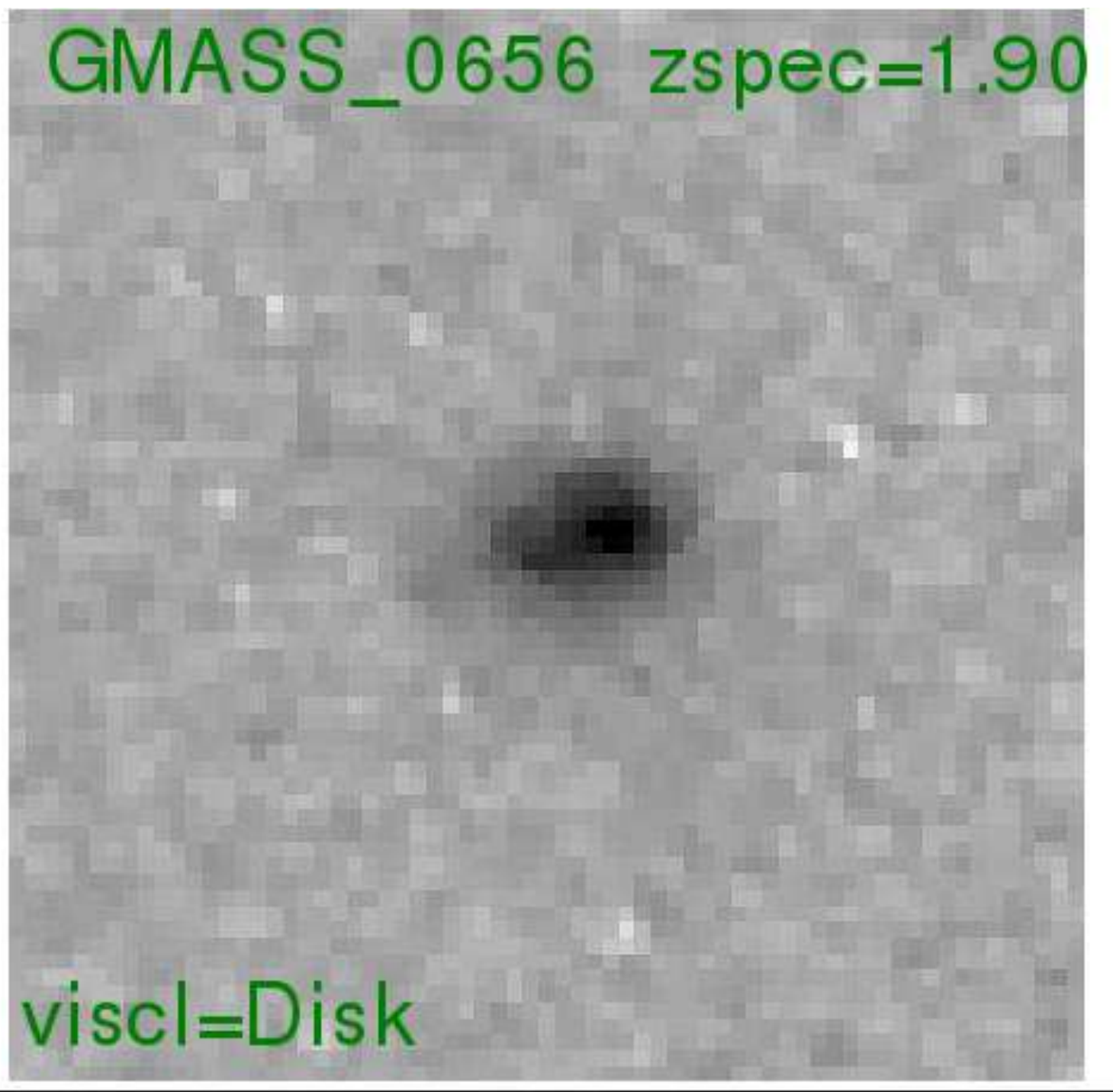}			     
\includegraphics[trim=100 40 75 390, clip=true, width=30mm]{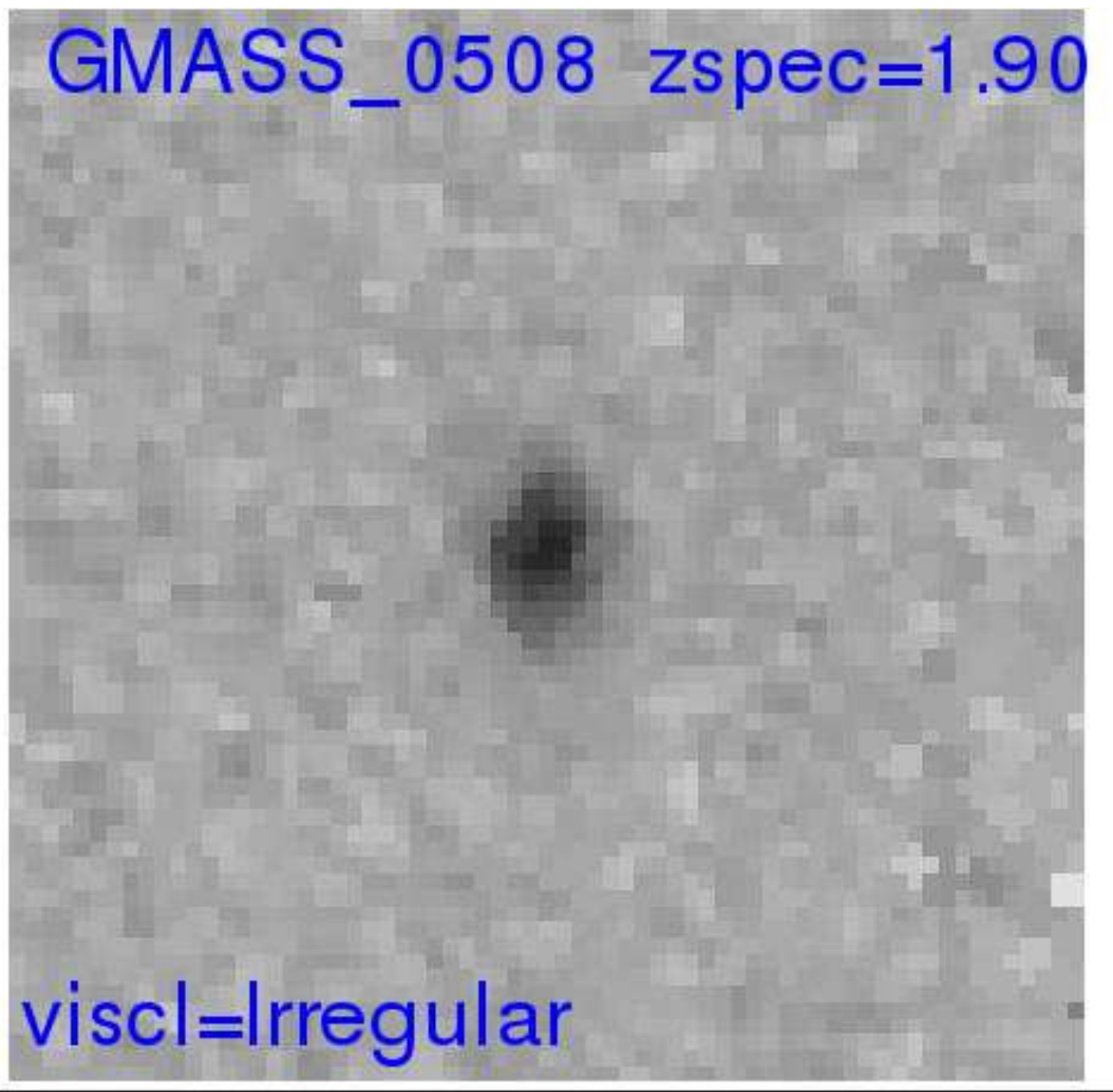}			     

\includegraphics[trim=100 40 75 390, clip=true, width=30mm]{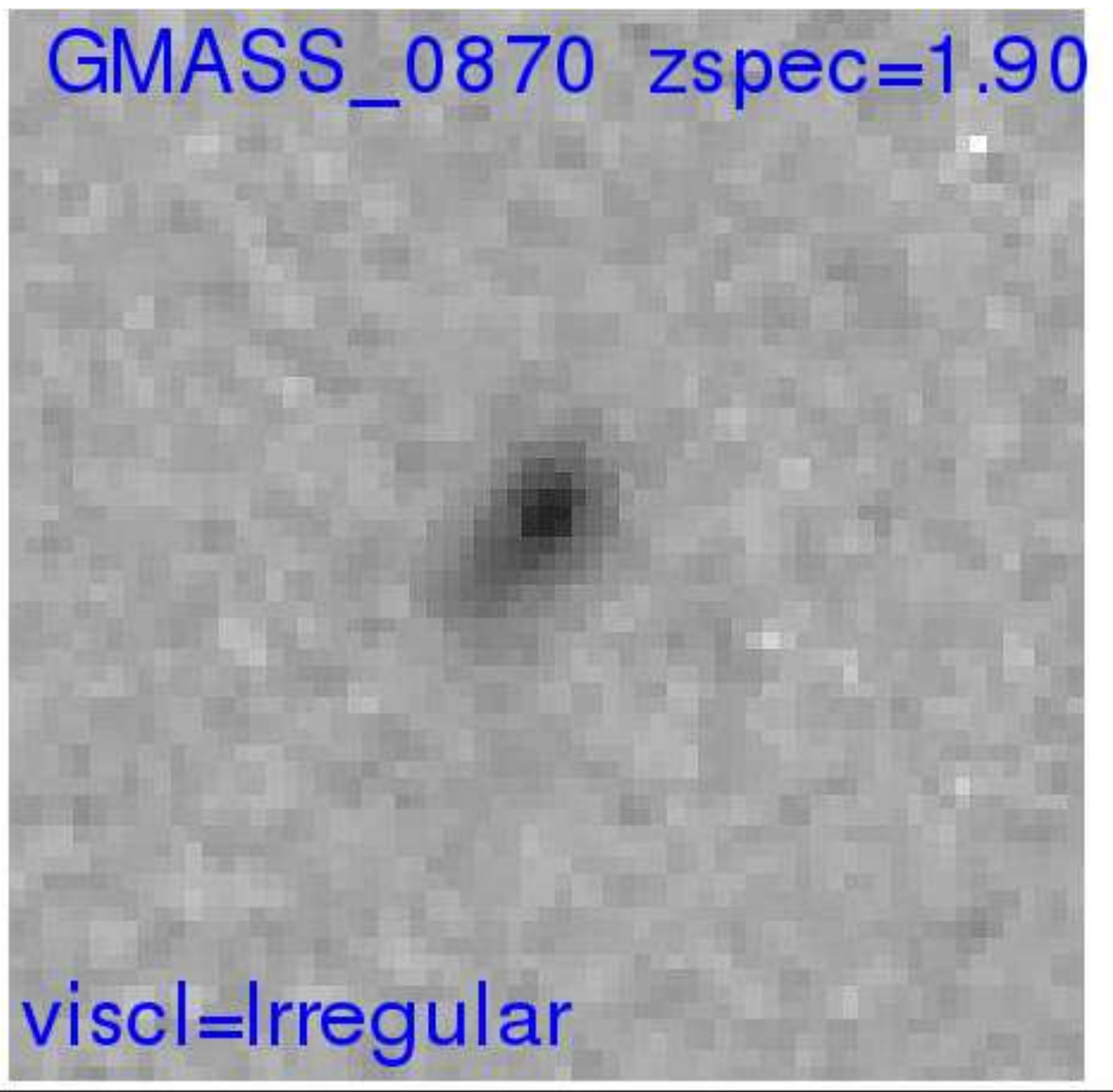}			     
\includegraphics[trim=100 40 75 390, clip=true, width=30mm]{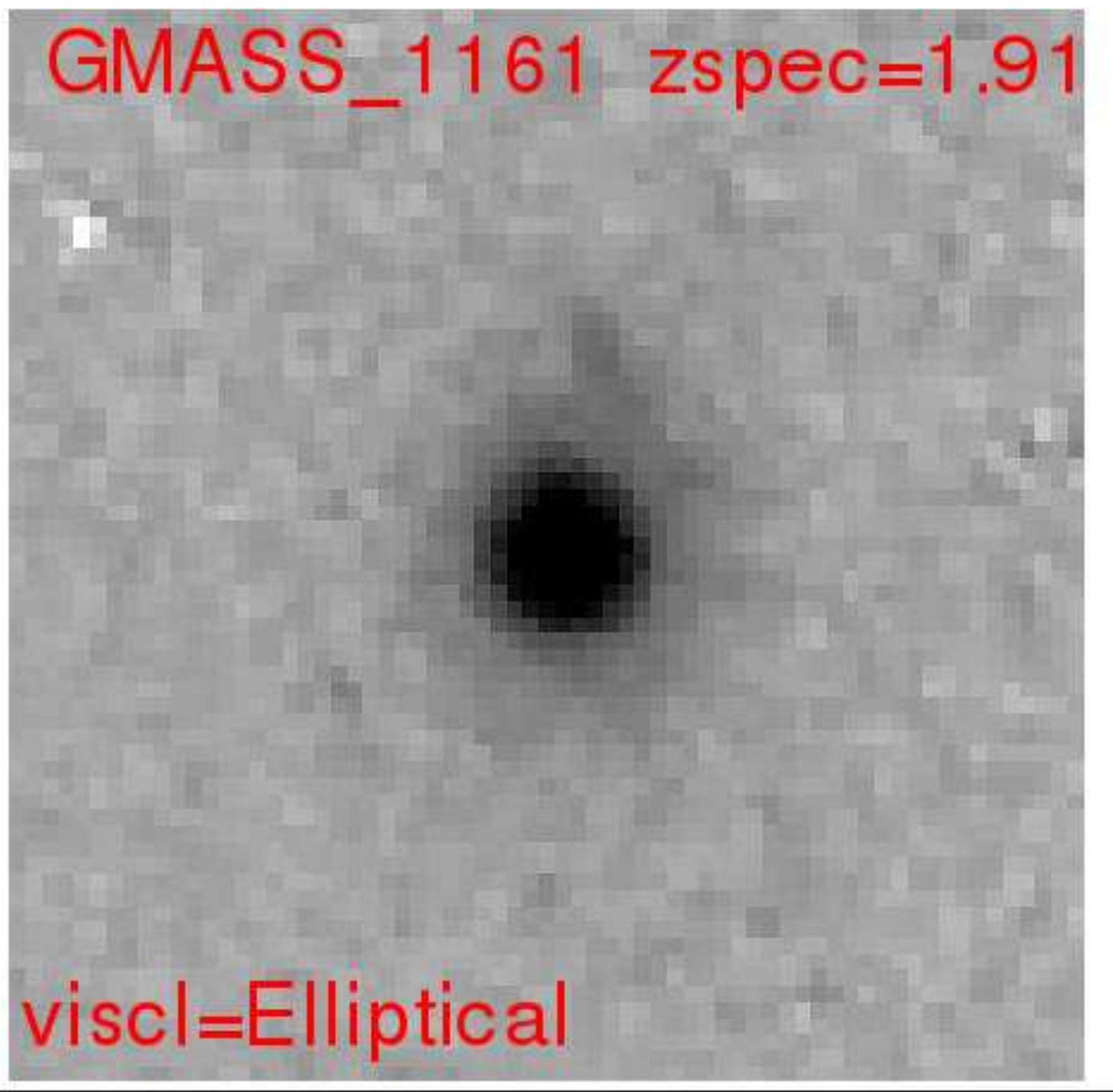}			     
\includegraphics[trim=100 40 75 390, clip=true, width=30mm]{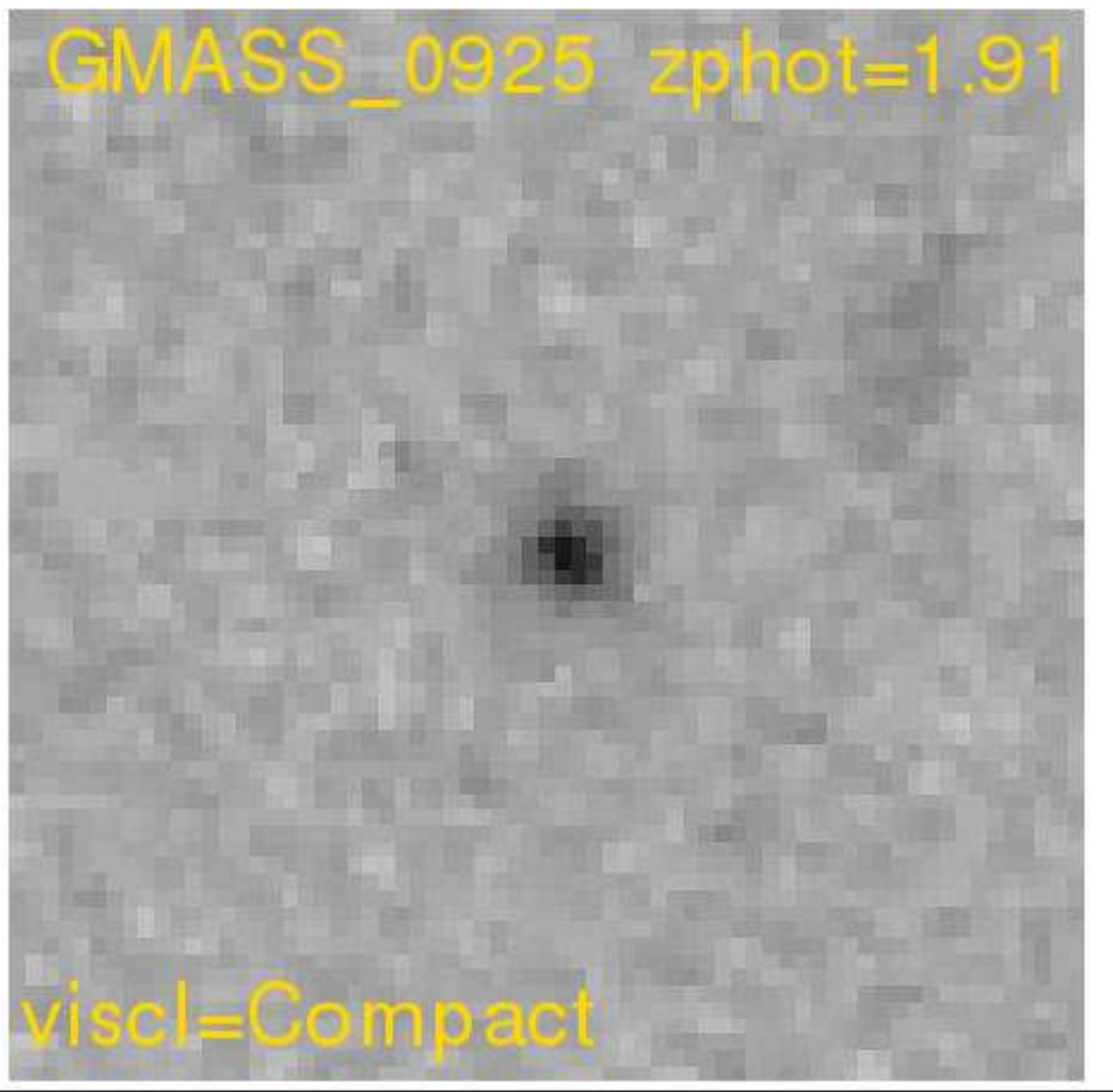}			     
\includegraphics[trim=100 40 75 390, clip=true, width=30mm]{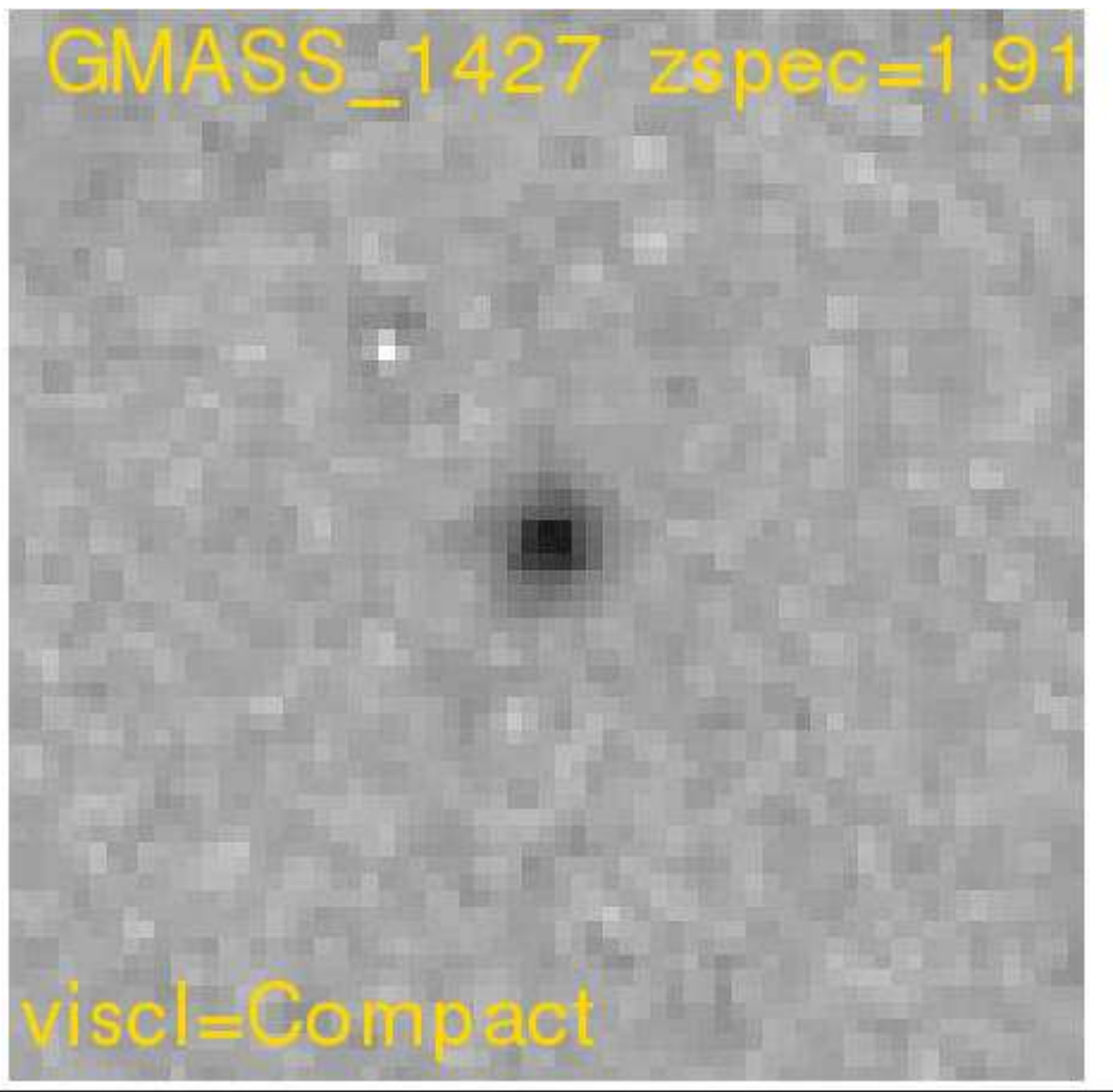}			     
\includegraphics[trim=100 40 75 390, clip=true, width=30mm]{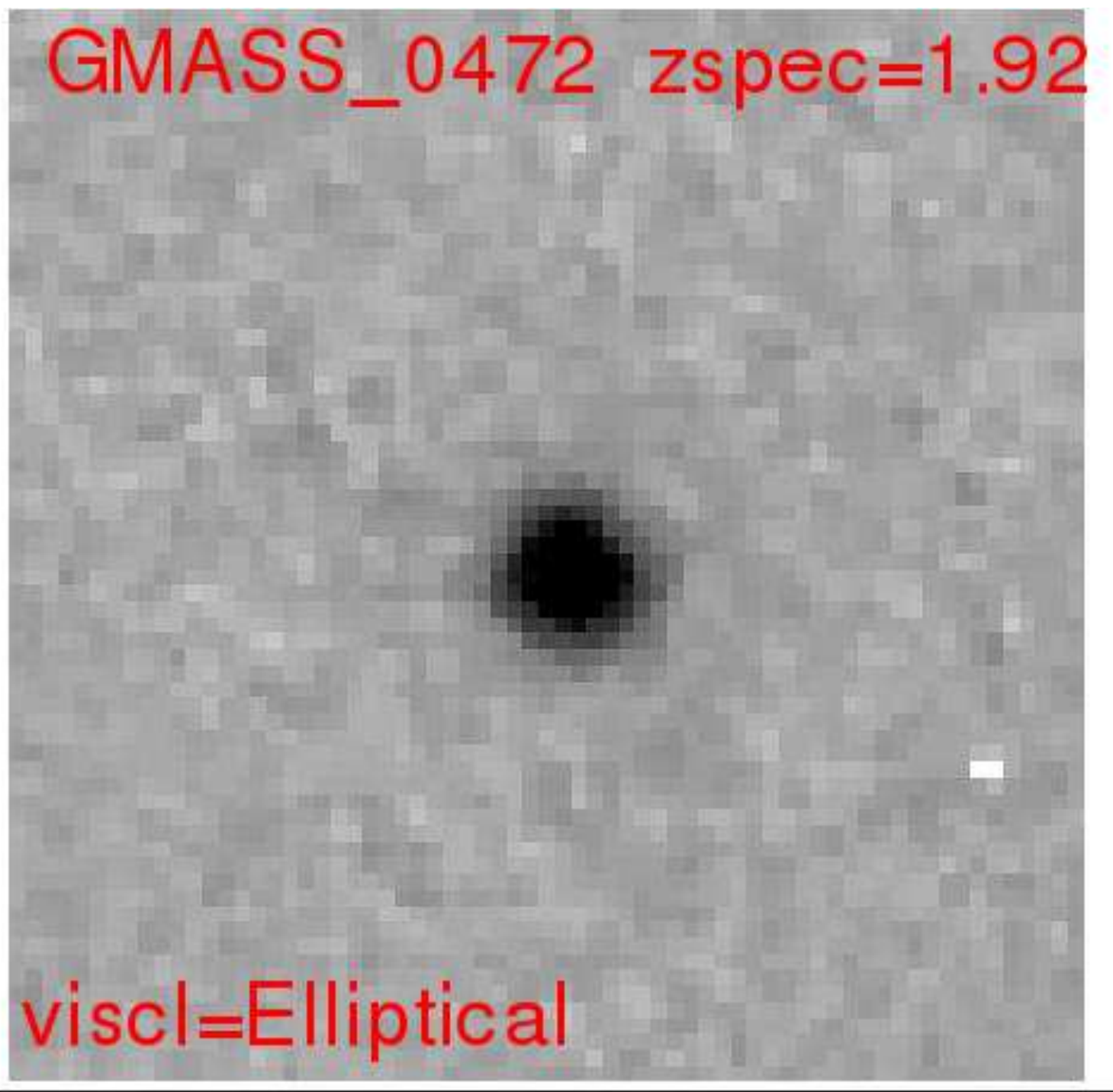}			     
\includegraphics[trim=100 40 75 390, clip=true, width=30mm]{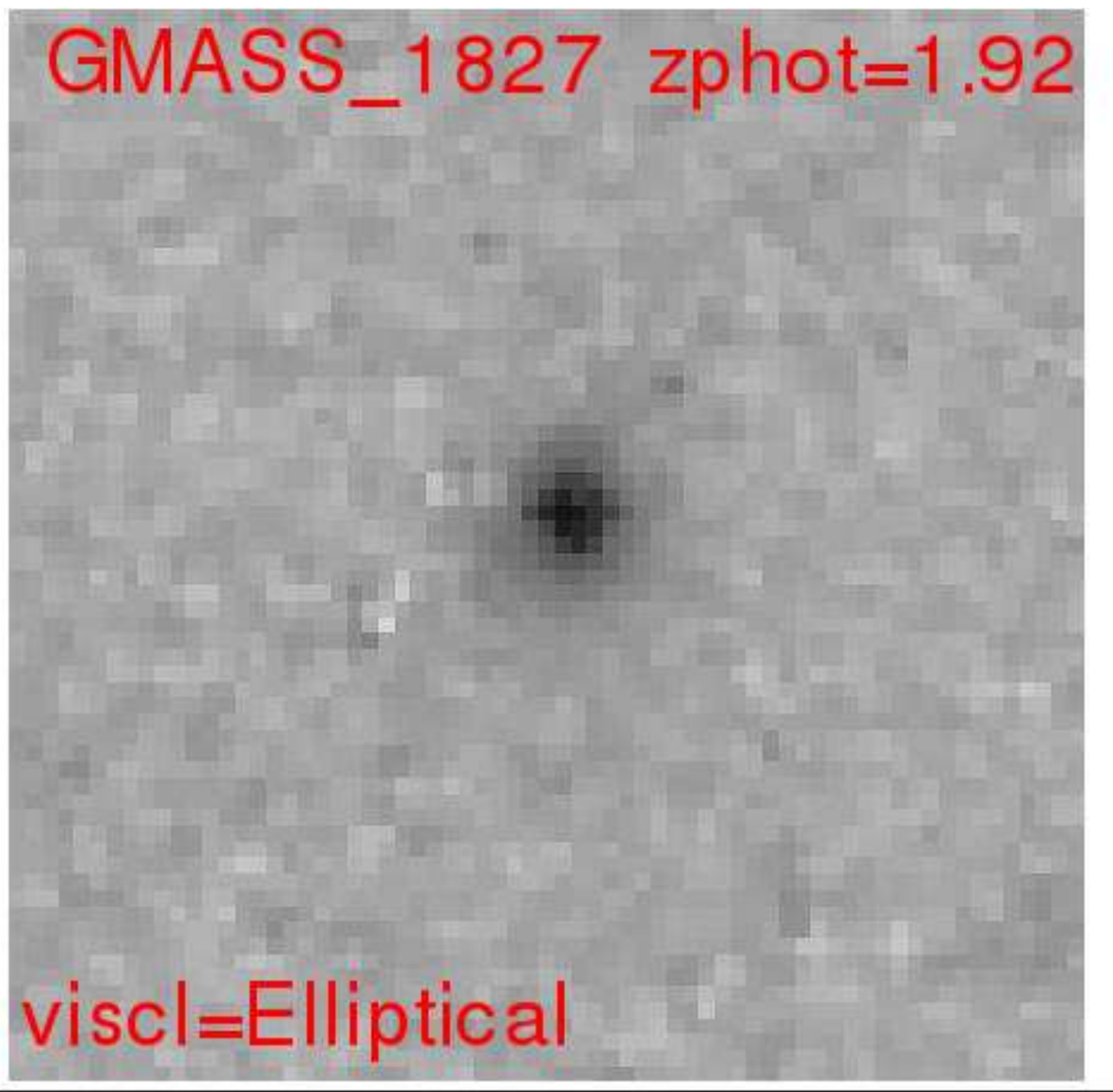}		     

\includegraphics[trim=100 40 75 390, clip=true, width=30mm]{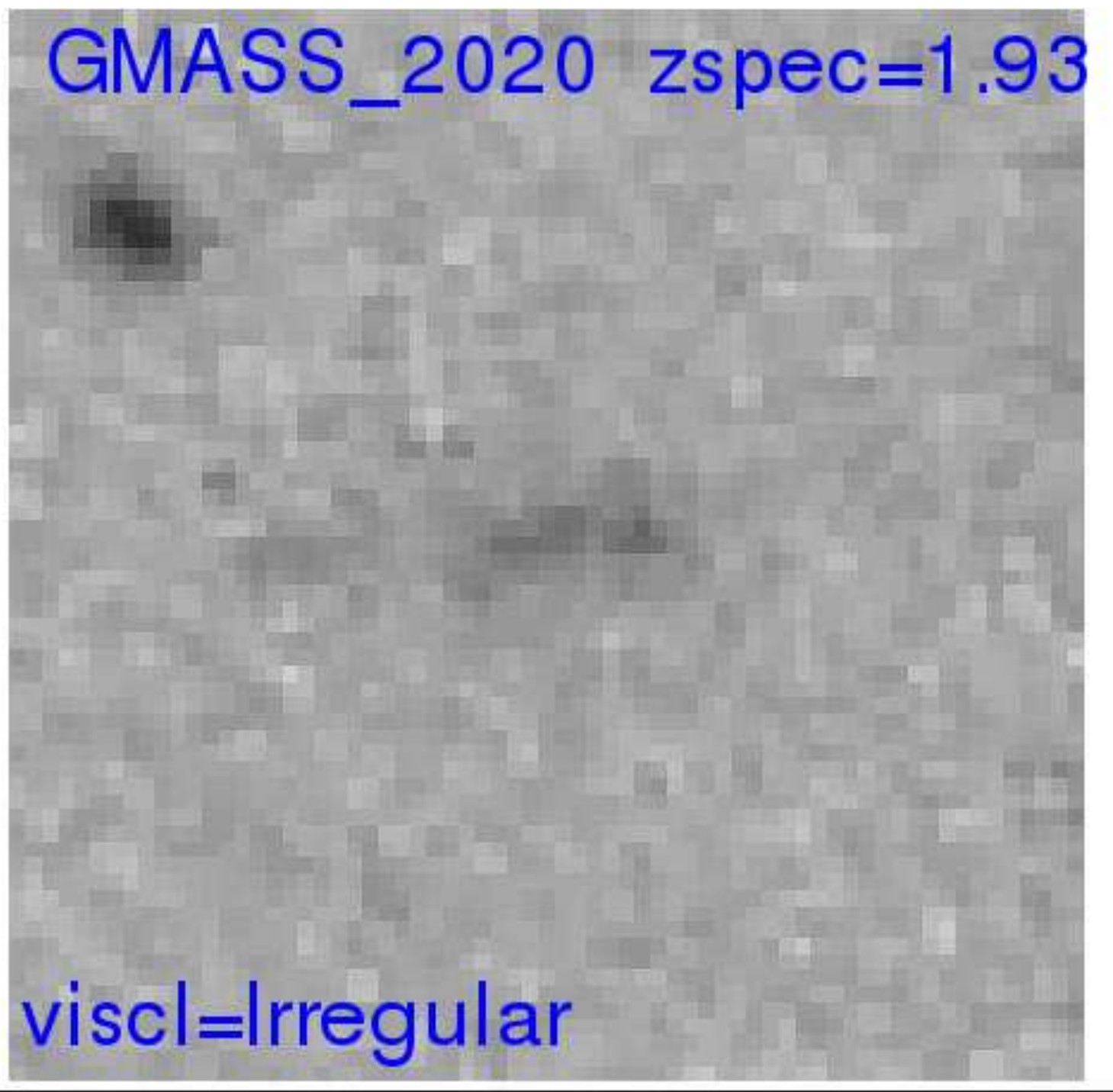}			     
\includegraphics[trim=100 40 75 390, clip=true, width=30mm]{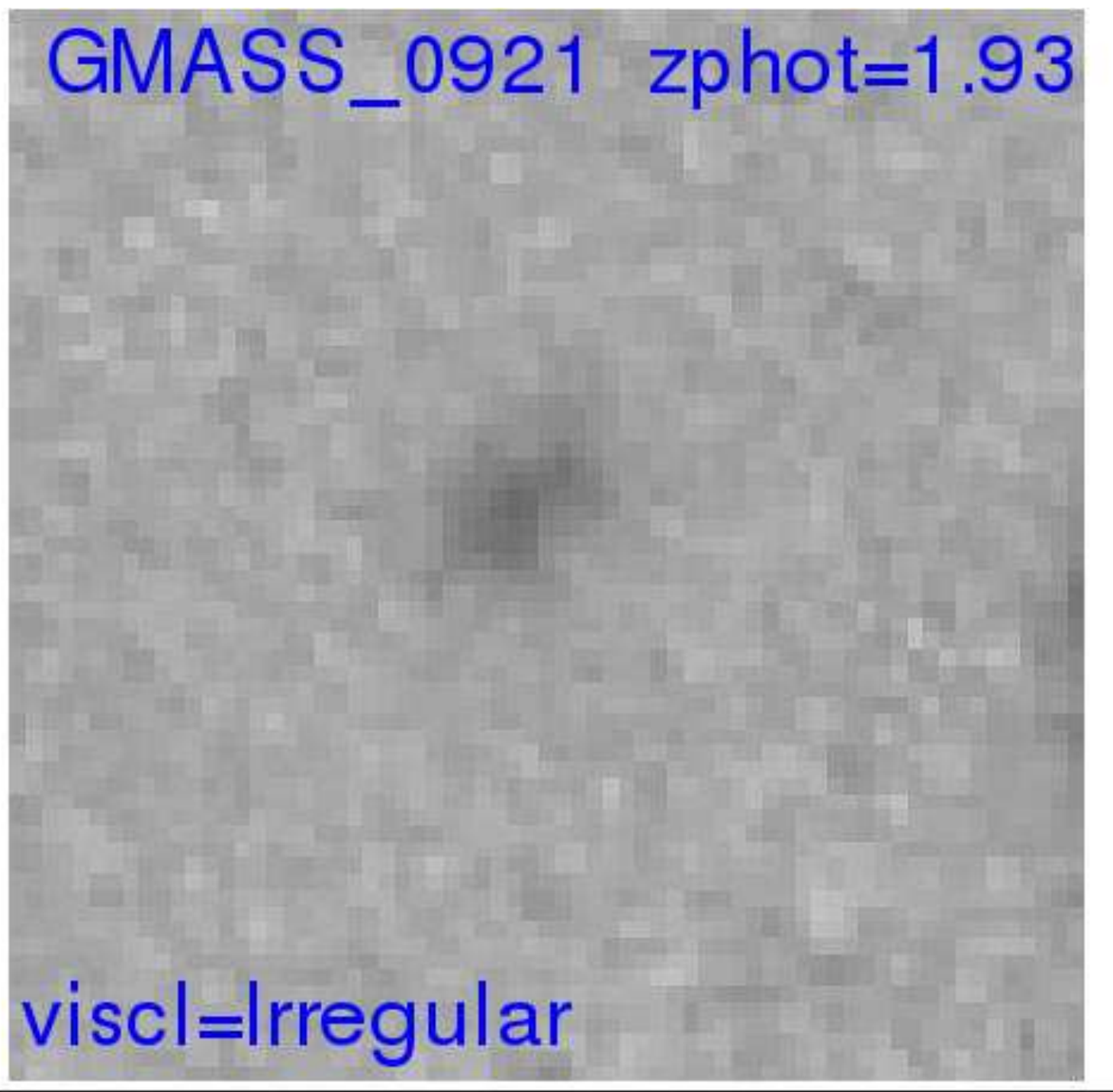}			     
\includegraphics[trim=100 40 75 390, clip=true, width=30mm]{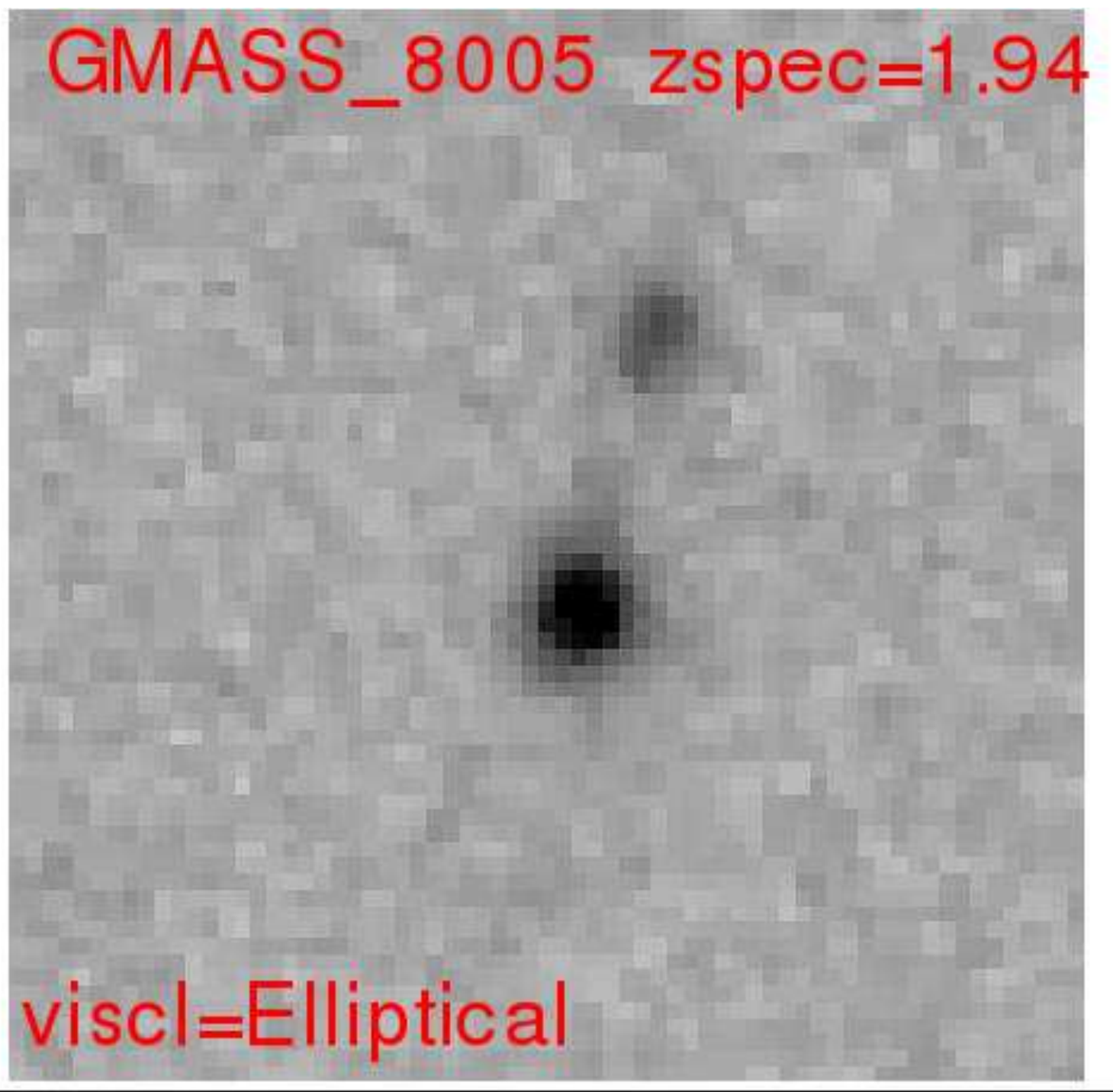}			     
\includegraphics[trim=100 40 75 390, clip=true, width=30mm]{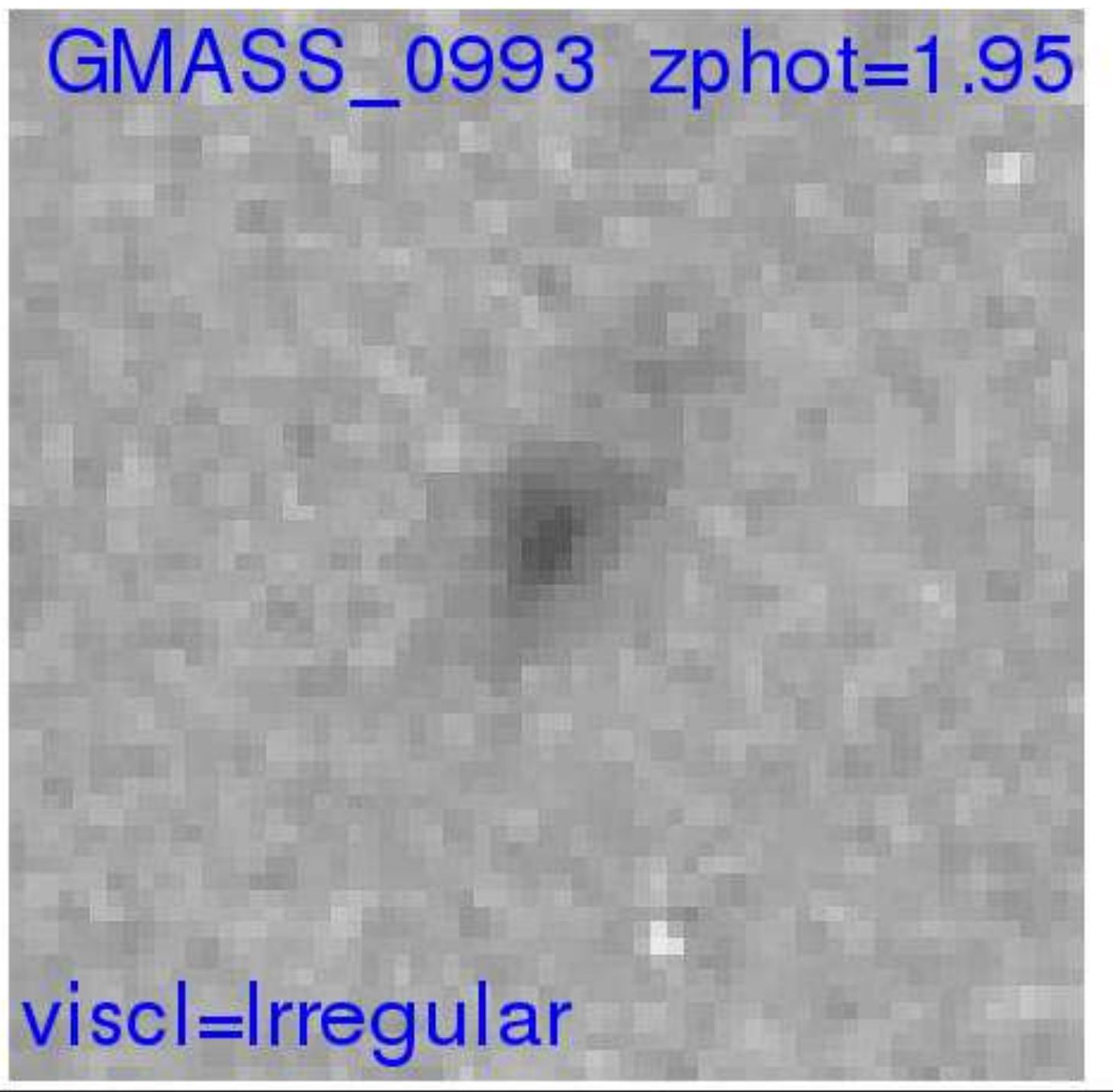}			     
\includegraphics[trim=100 40 75 390, clip=true, width=30mm]{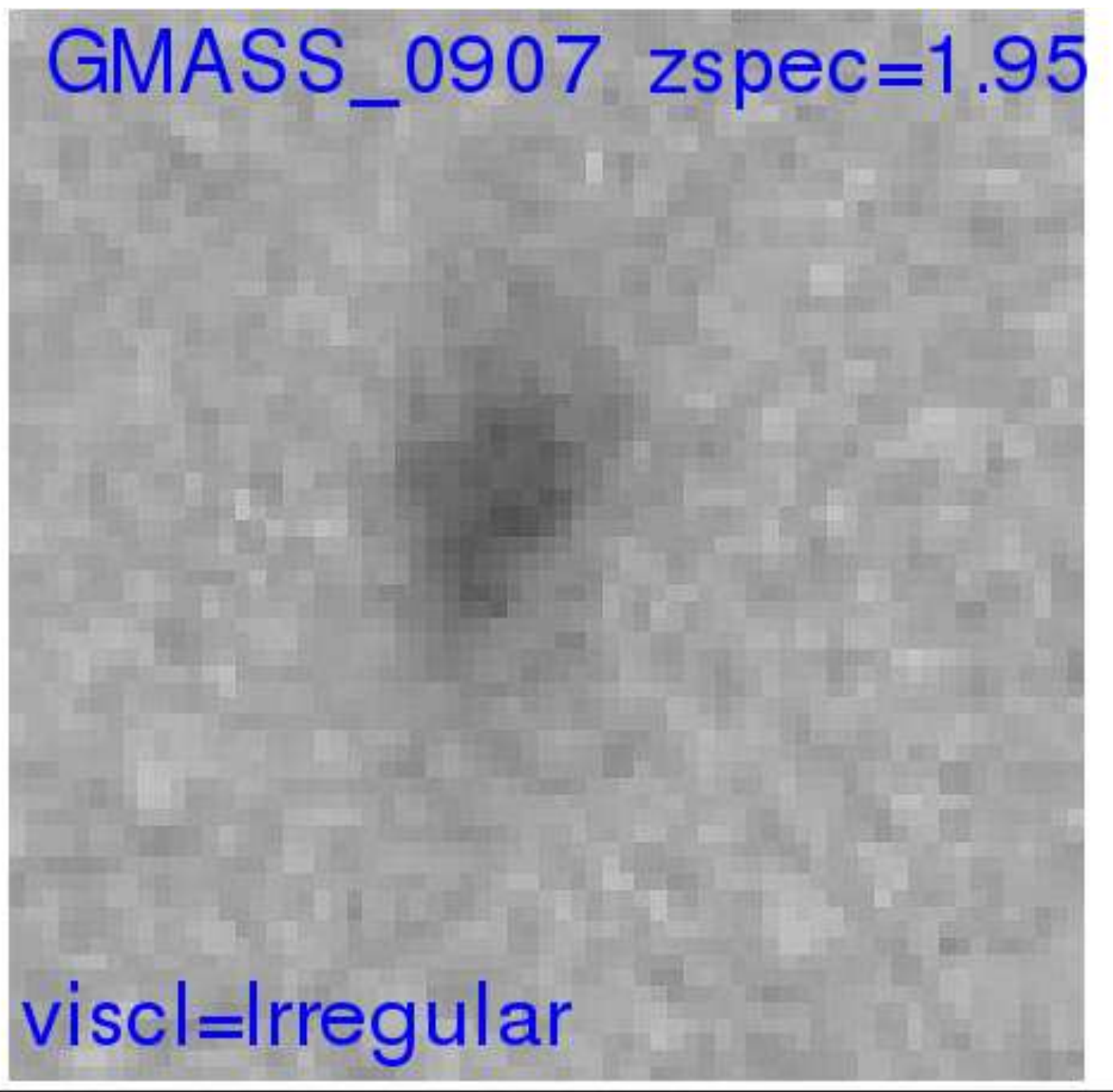}			     
\includegraphics[trim=100 40 75 390, clip=true, width=30mm]{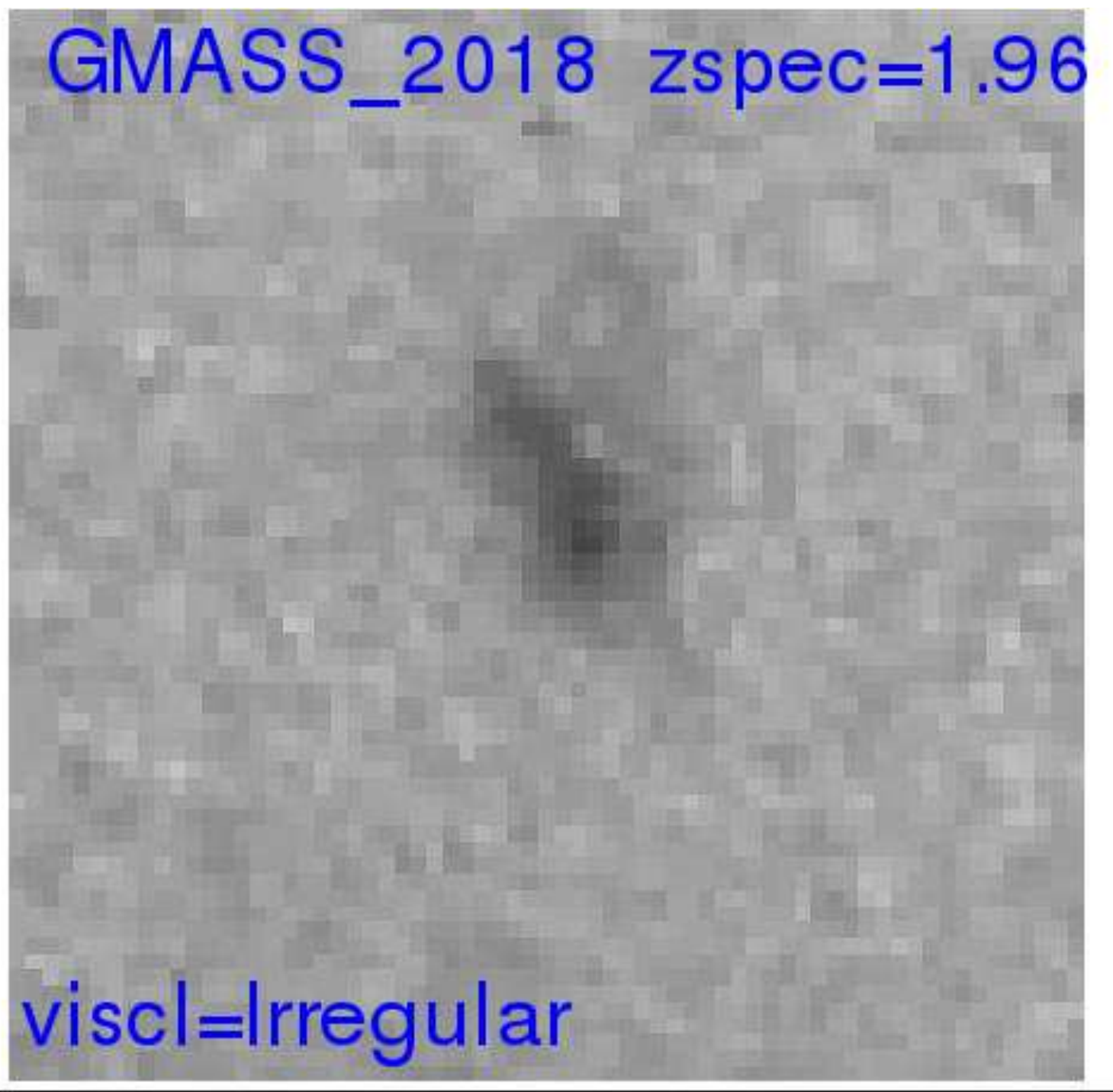}			     

\includegraphics[trim=100 40 75 390, clip=true, width=30mm]{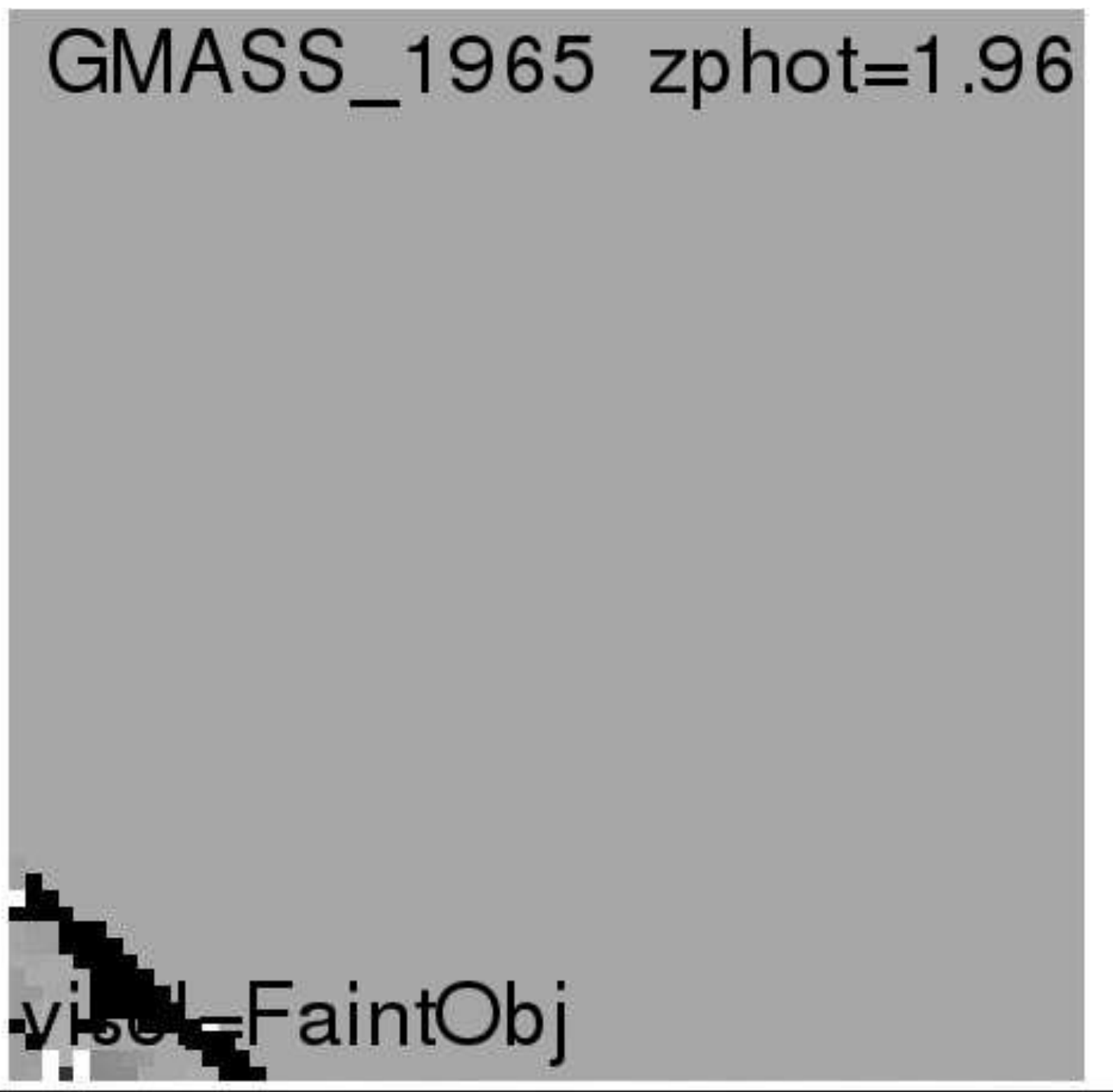}			     
\includegraphics[trim=100 40 75 390, clip=true, width=30mm]{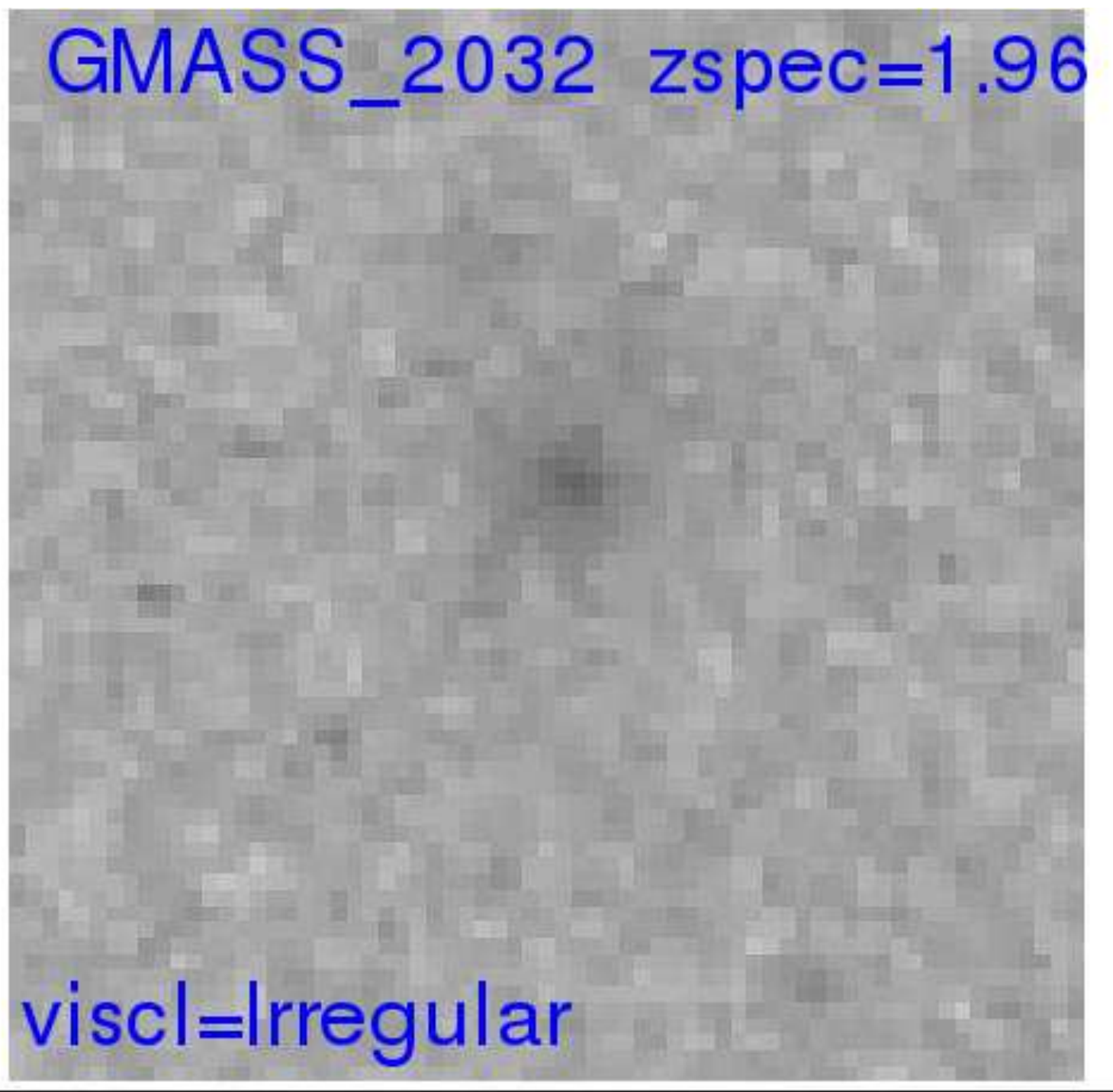}			     
\includegraphics[trim=100 40 75 390, clip=true, width=30mm]{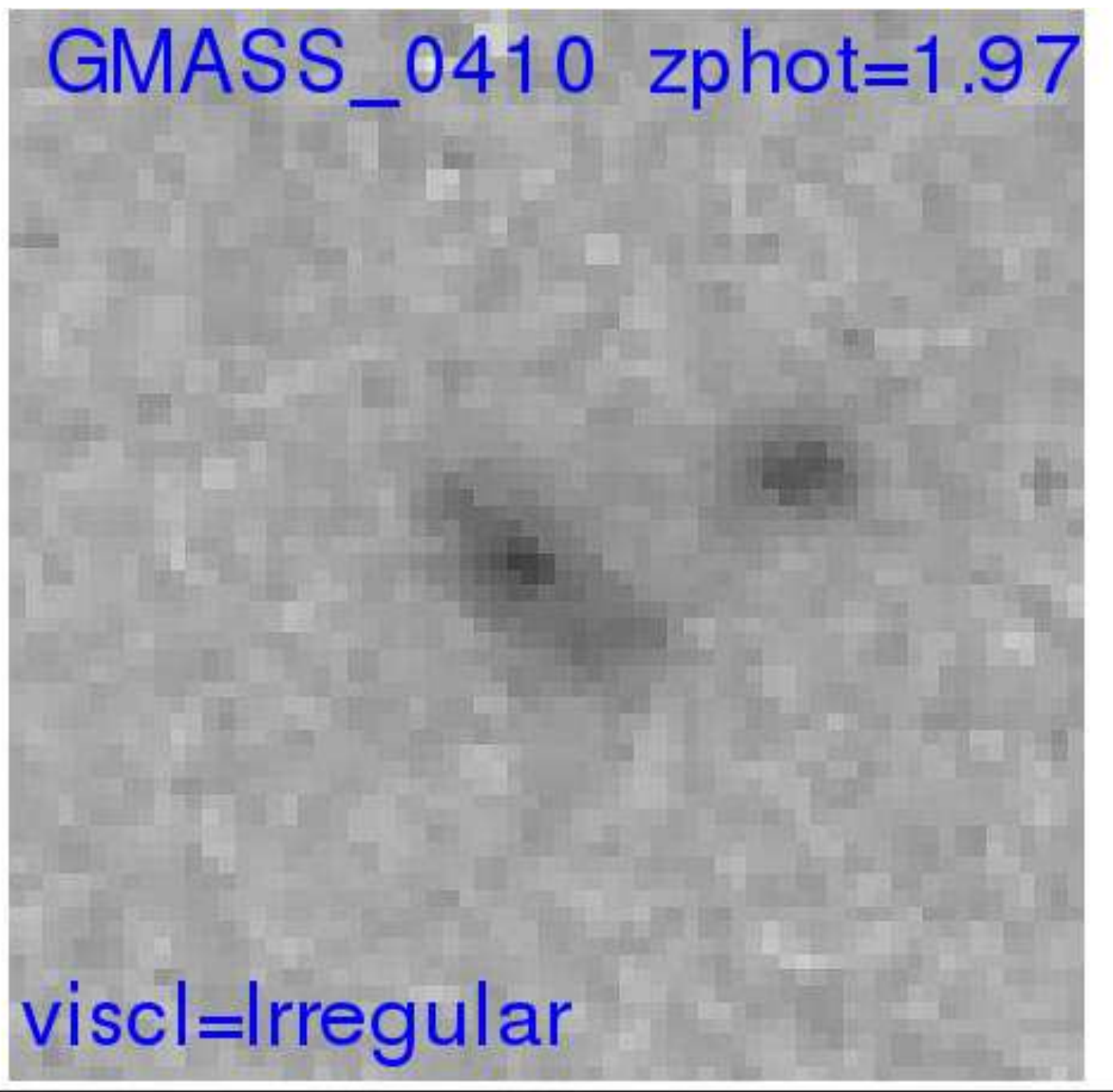}			     
\includegraphics[trim=100 40 75 390, clip=true, width=30mm]{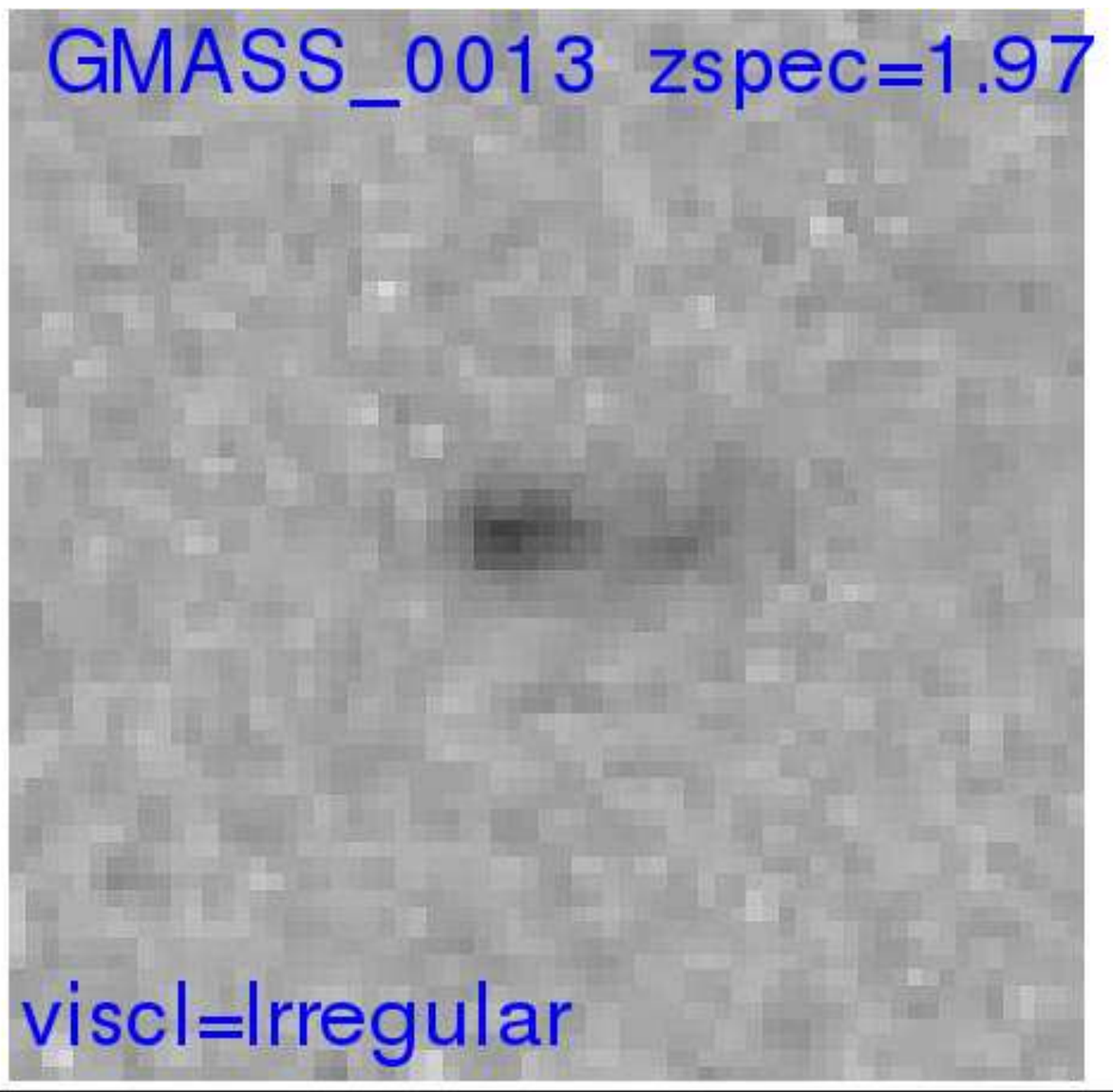}		     
\includegraphics[trim=100 40 75 390, clip=true, width=30mm]{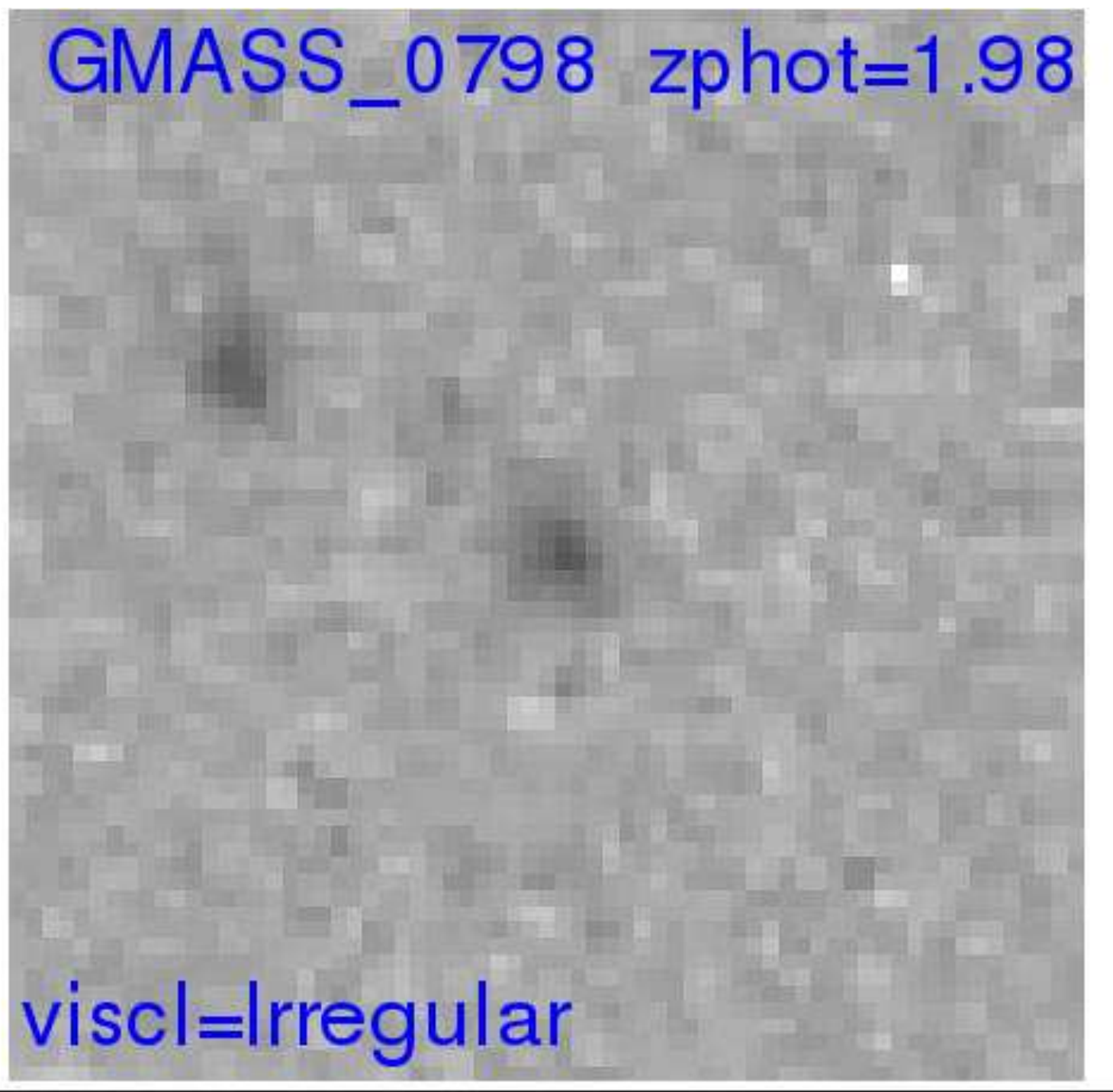}			     
\includegraphics[trim=100 40 75 390, clip=true, width=30mm]{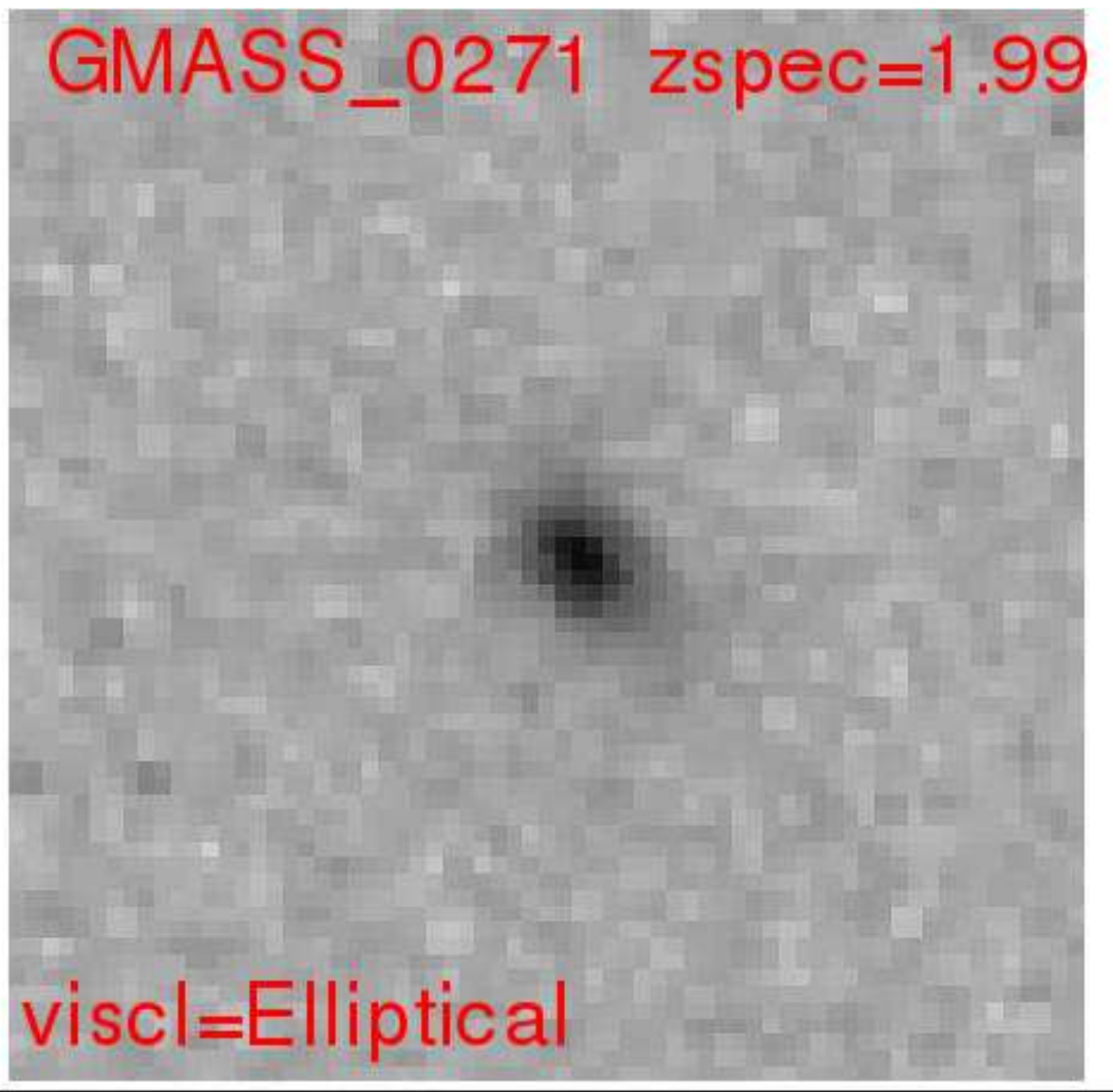}			     

\includegraphics[trim=100 40 75 390, clip=true, width=30mm]{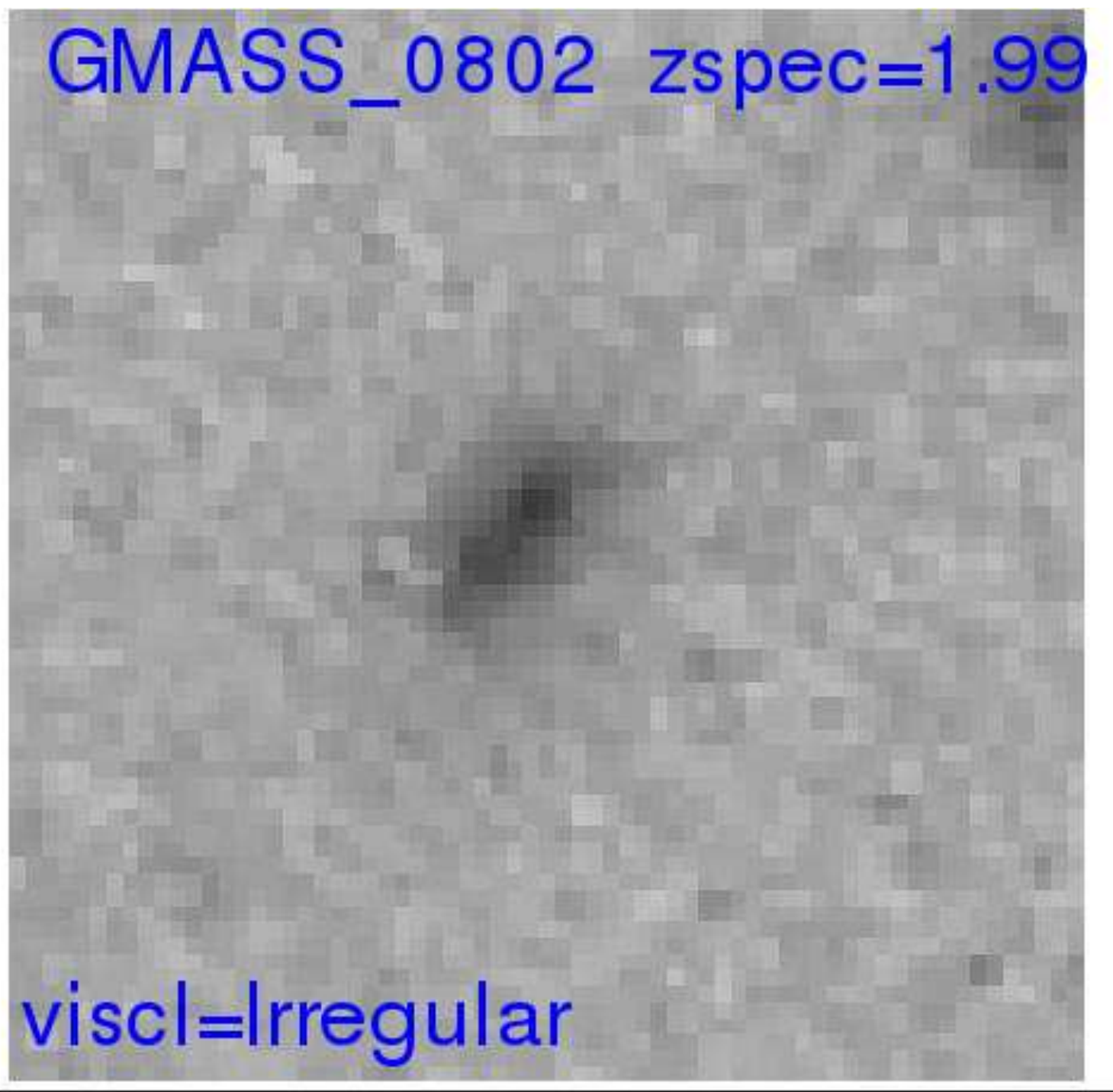}			     
\includegraphics[trim=100 40 75 390, clip=true, width=30mm]{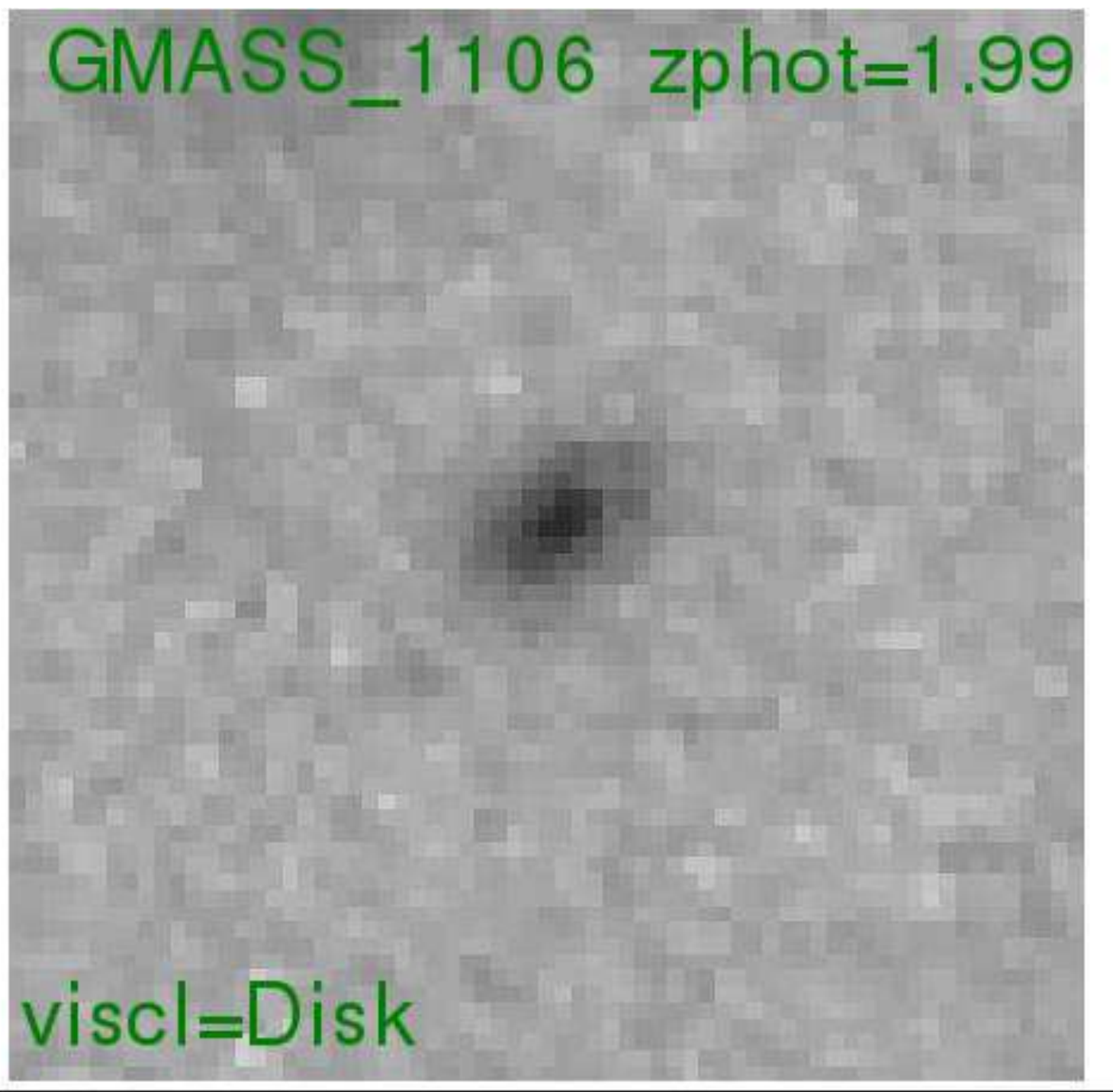}			     
\includegraphics[trim=100 40 75 390, clip=true, width=30mm]{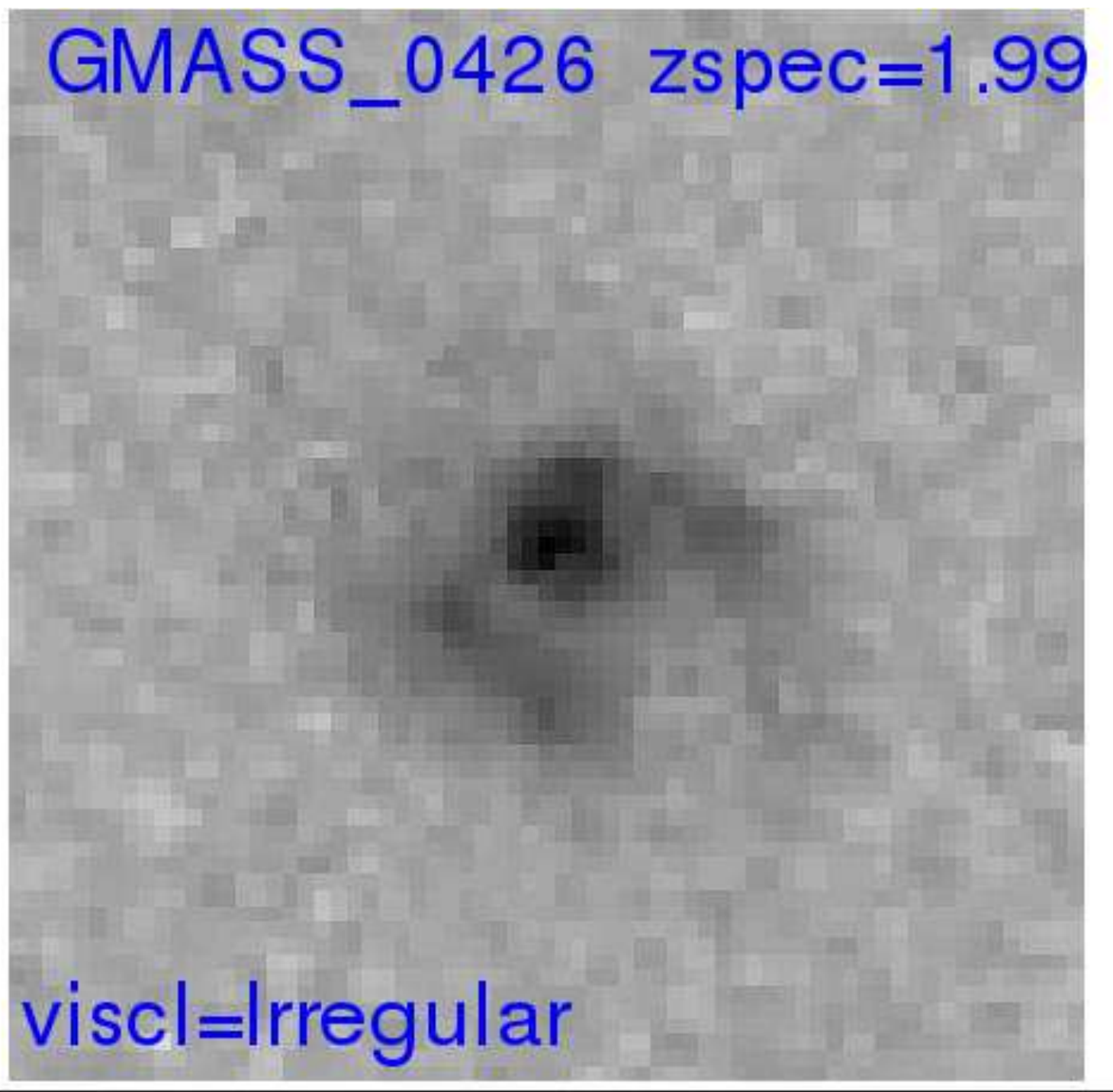}			     
\includegraphics[trim=100 40 75 390, clip=true, width=30mm]{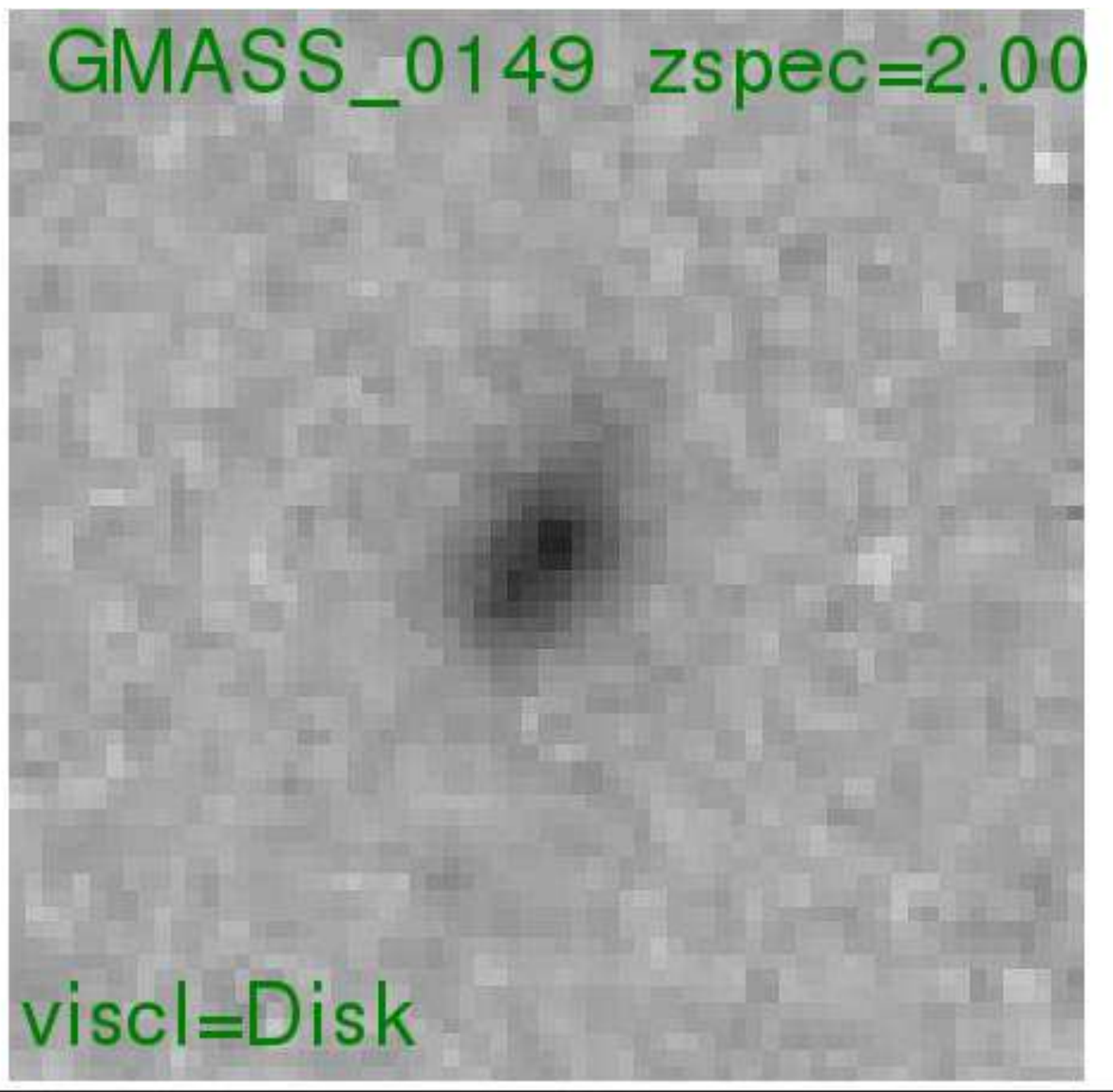}			     
\includegraphics[trim=100 40 75 390, clip=true, width=30mm]{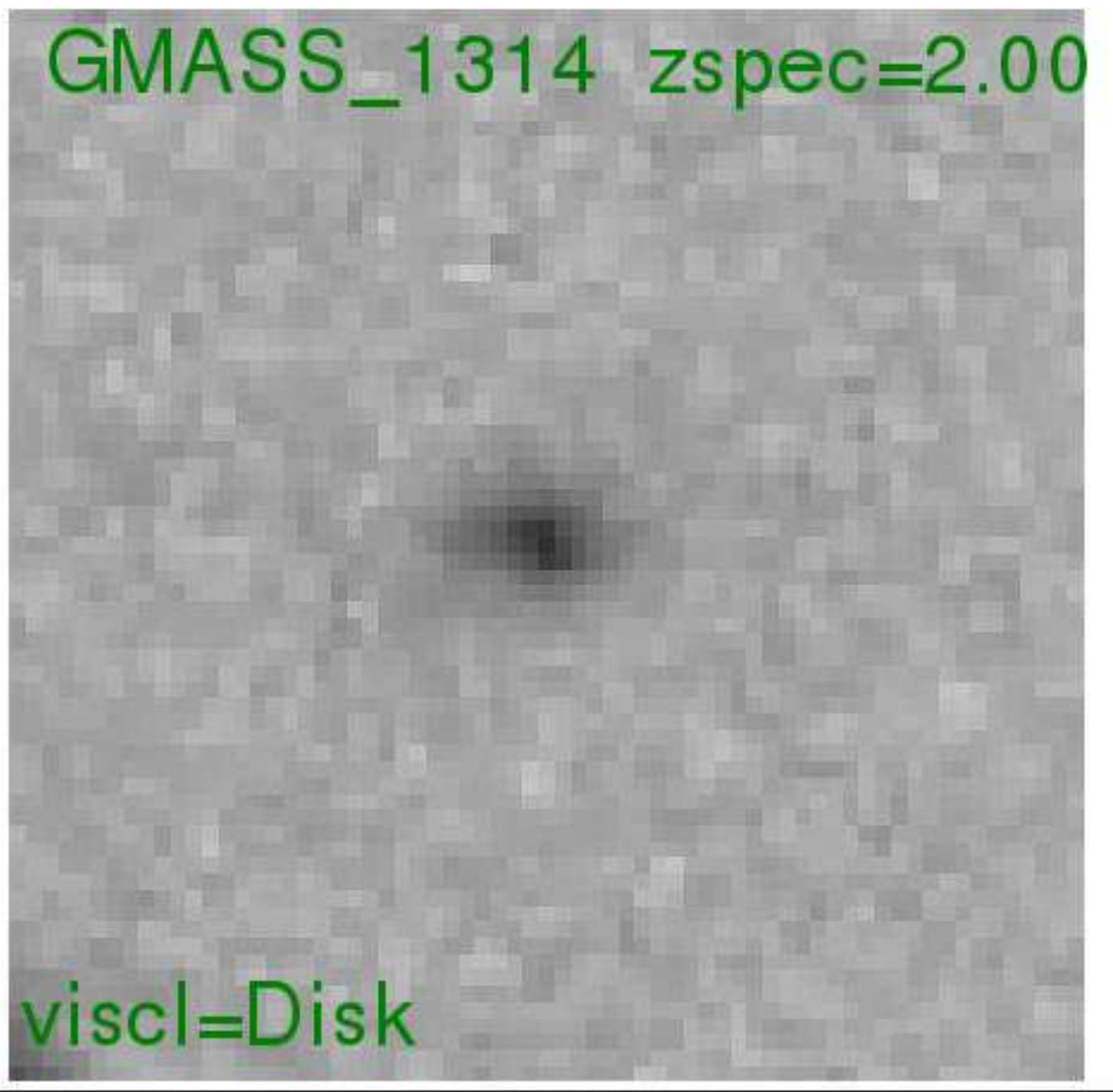}		     
\includegraphics[trim=100 40 75 390, clip=true, width=30mm]{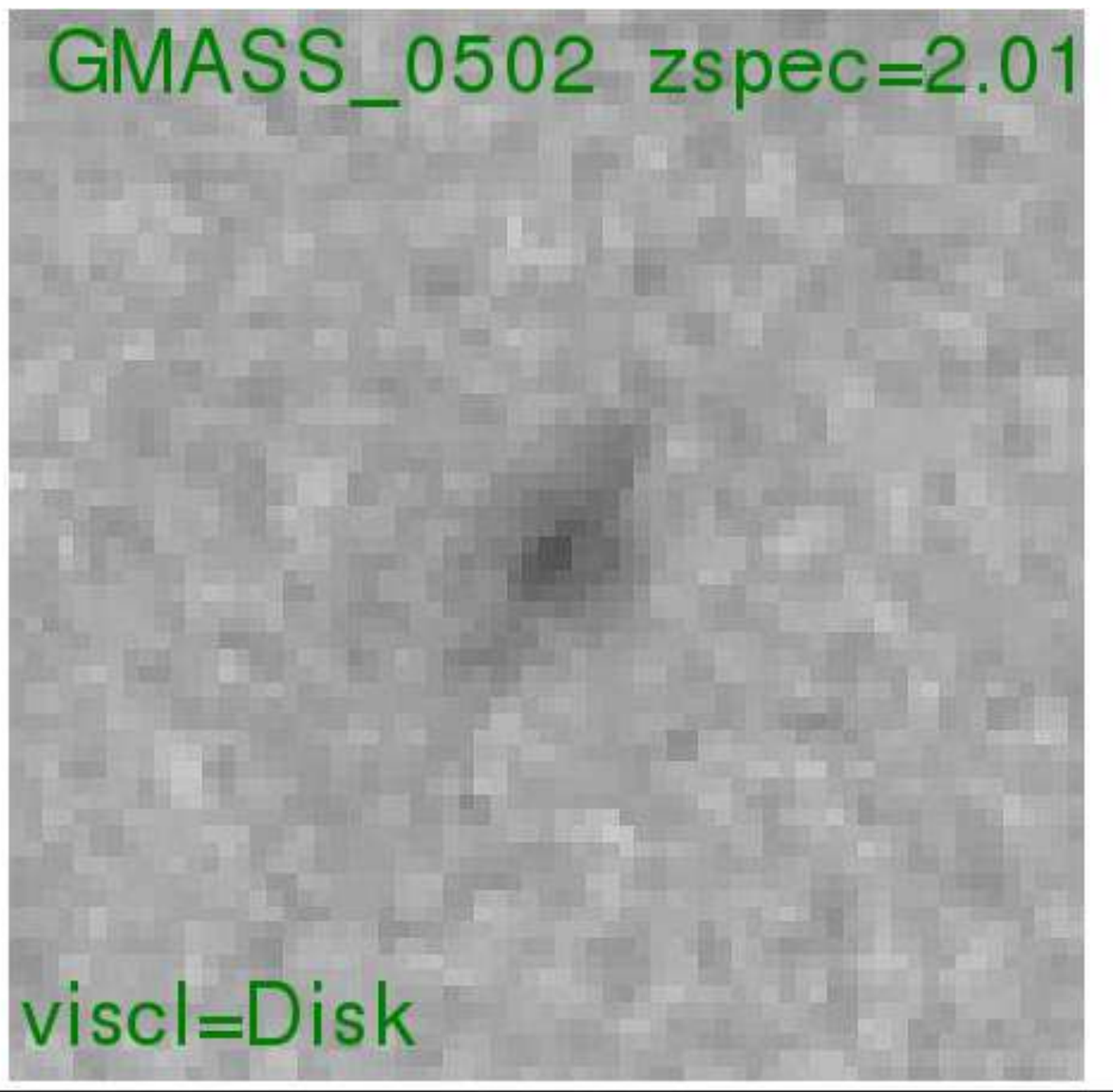}			     

\includegraphics[trim=100 40 75 390, clip=true, width=30mm]{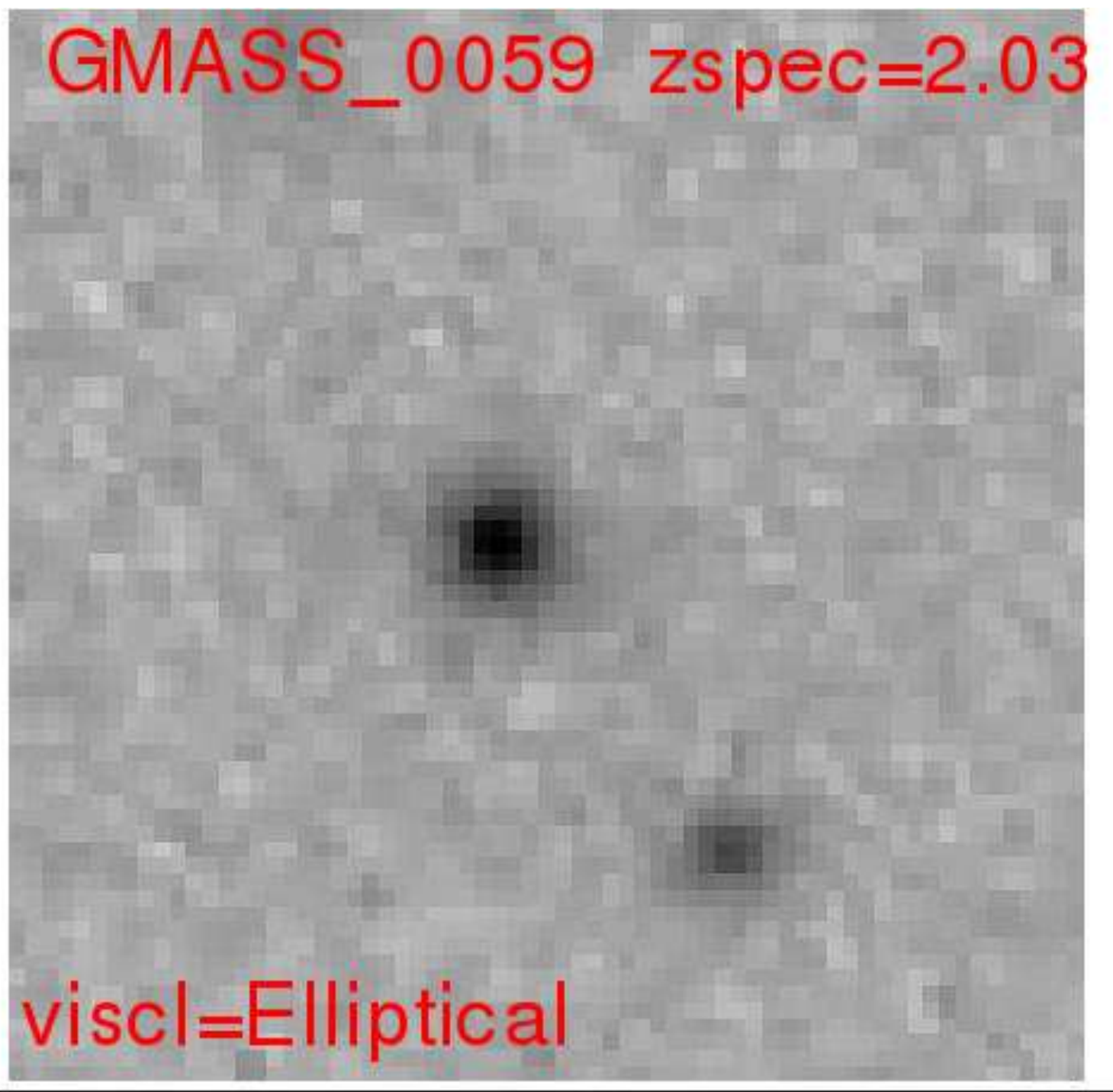}			     
\includegraphics[trim=100 40 75 390, clip=true, width=30mm]{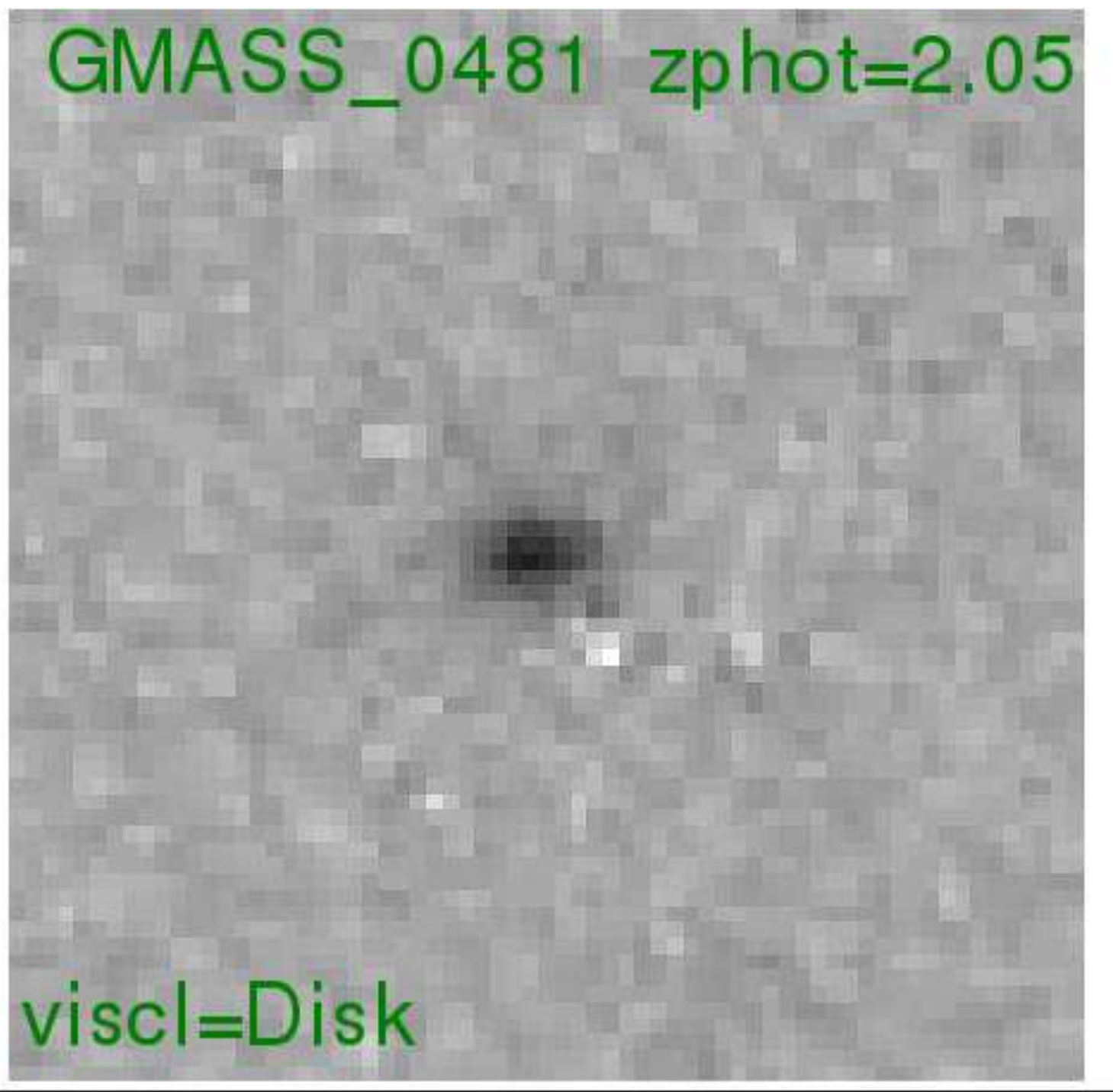}			     
\includegraphics[trim=100 40 75 390, clip=true, width=30mm]{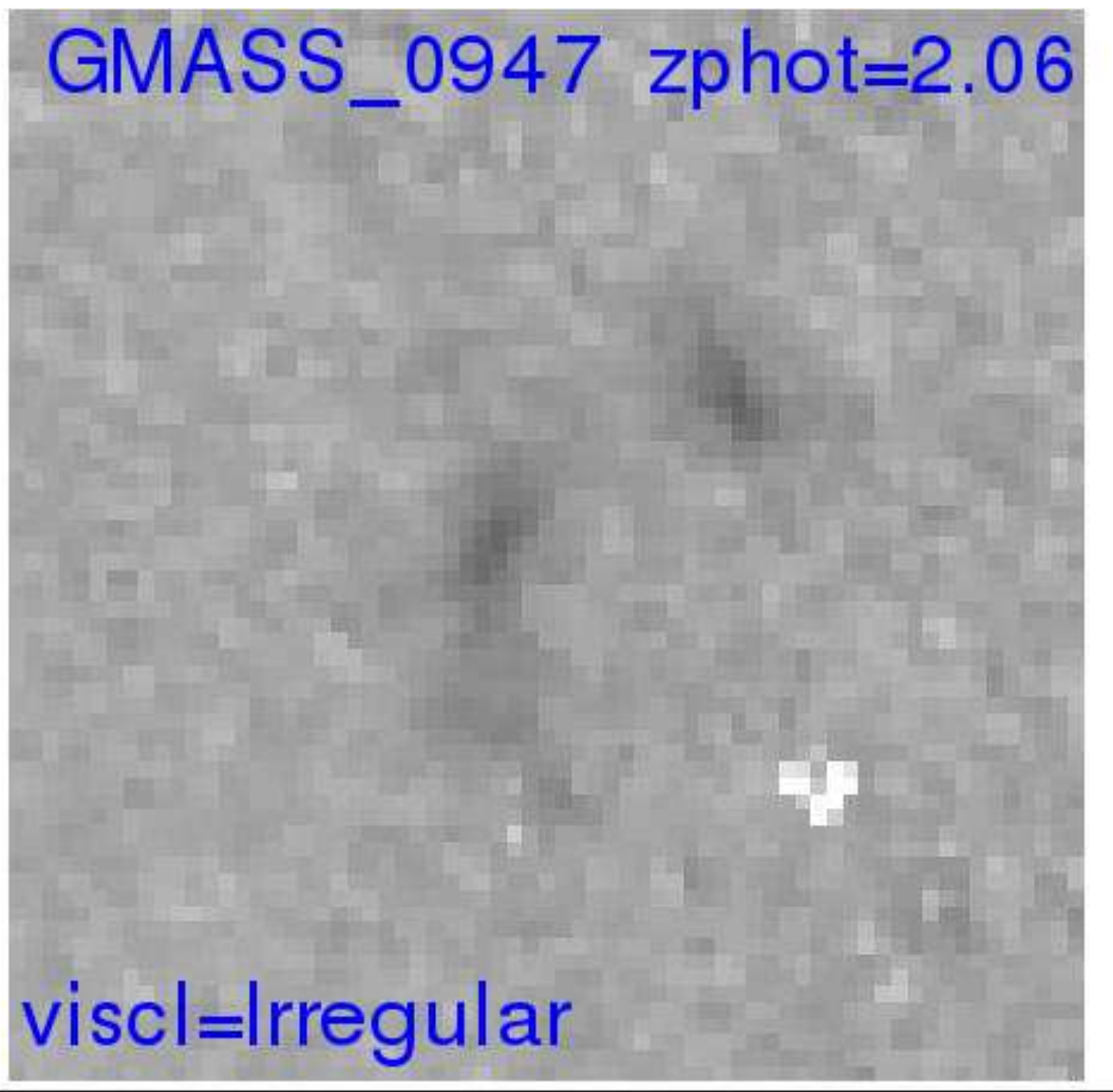}		     
\includegraphics[trim=100 40 75 390, clip=true, width=30mm]{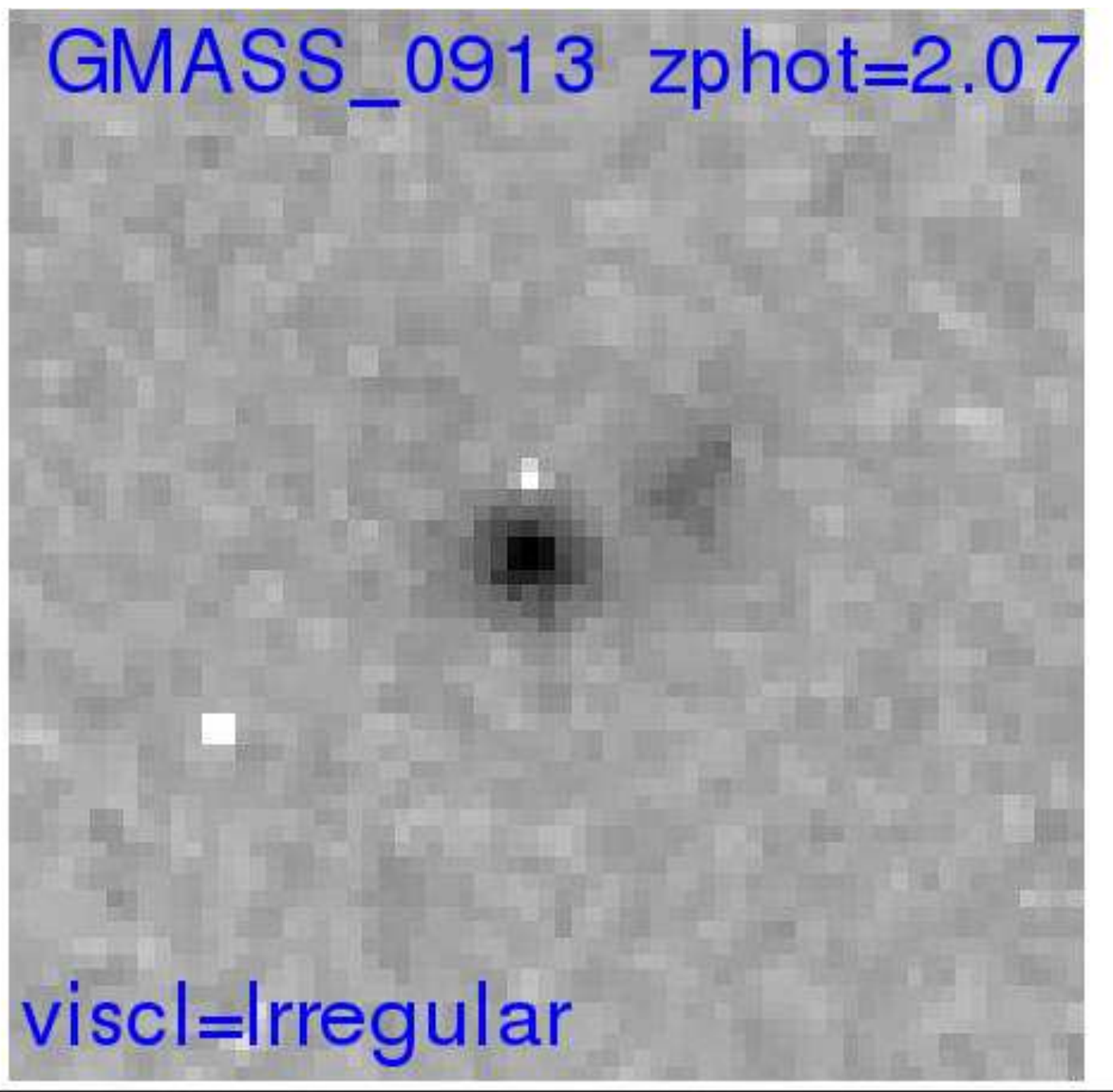}			     
\includegraphics[trim=100 40 75 390, clip=true, width=30mm]{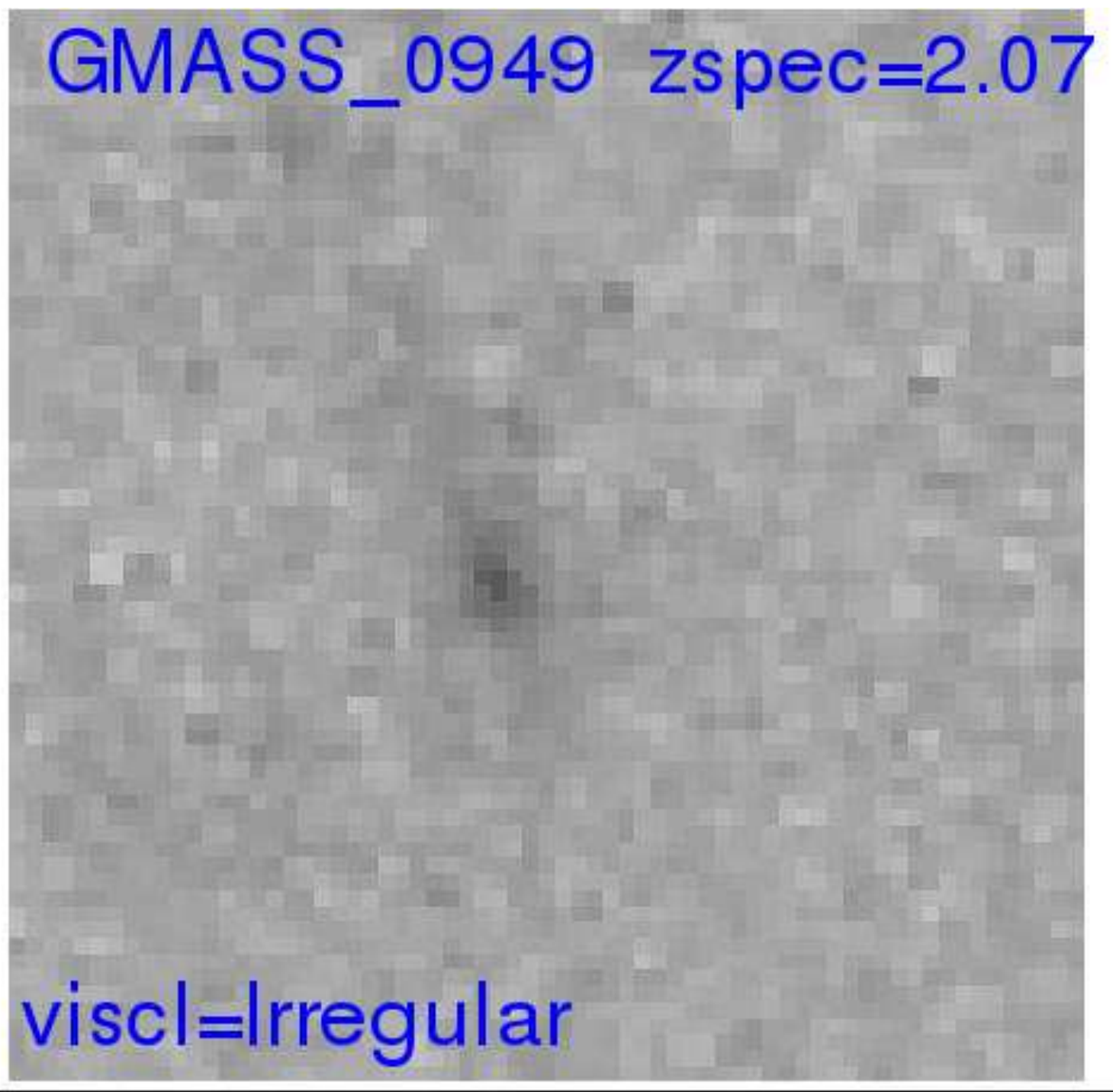}			     
\includegraphics[trim=100 40 75 390, clip=true, width=30mm]{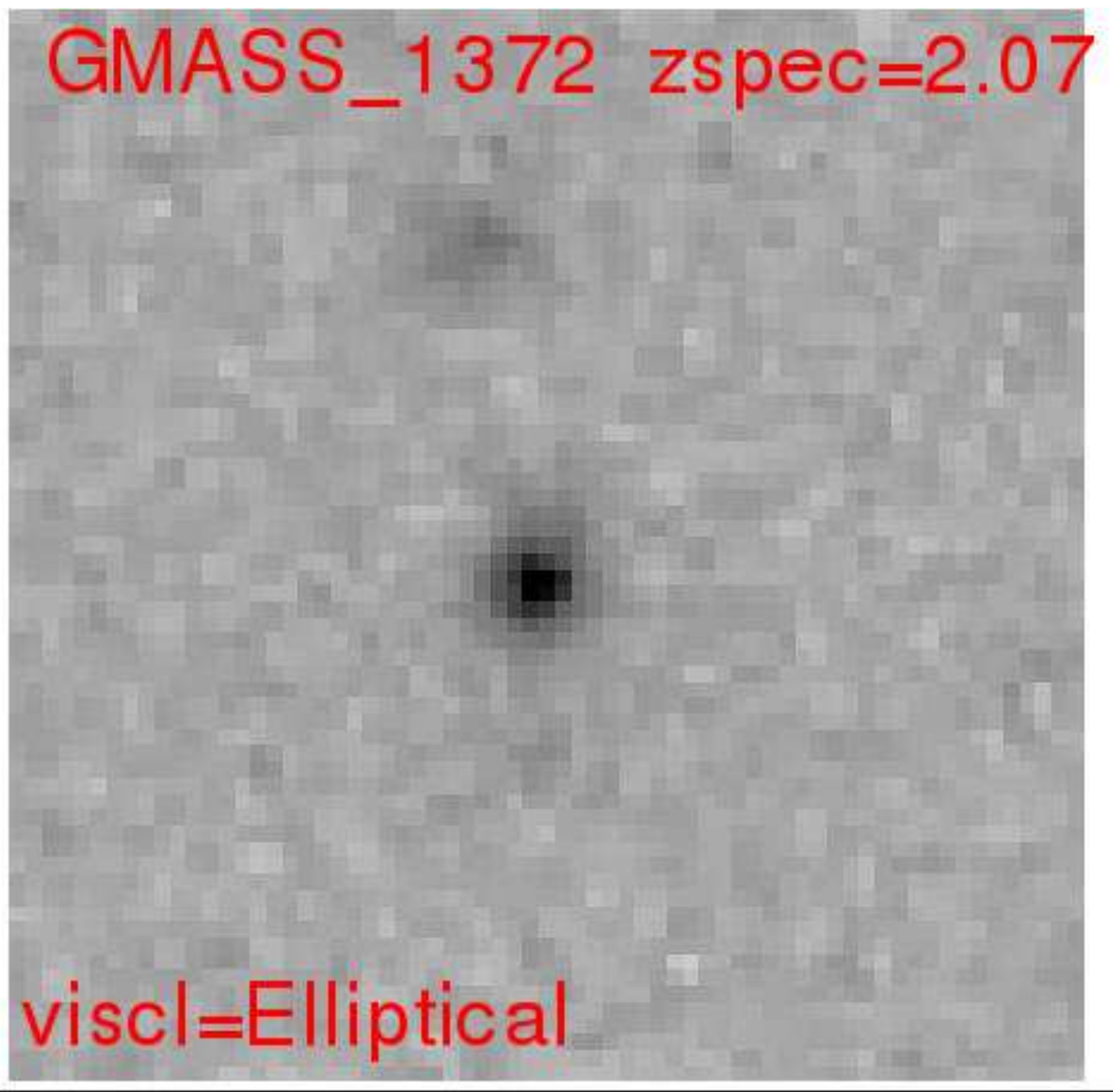}			     

\includegraphics[trim=100 40 75 390, clip=true, width=30mm]{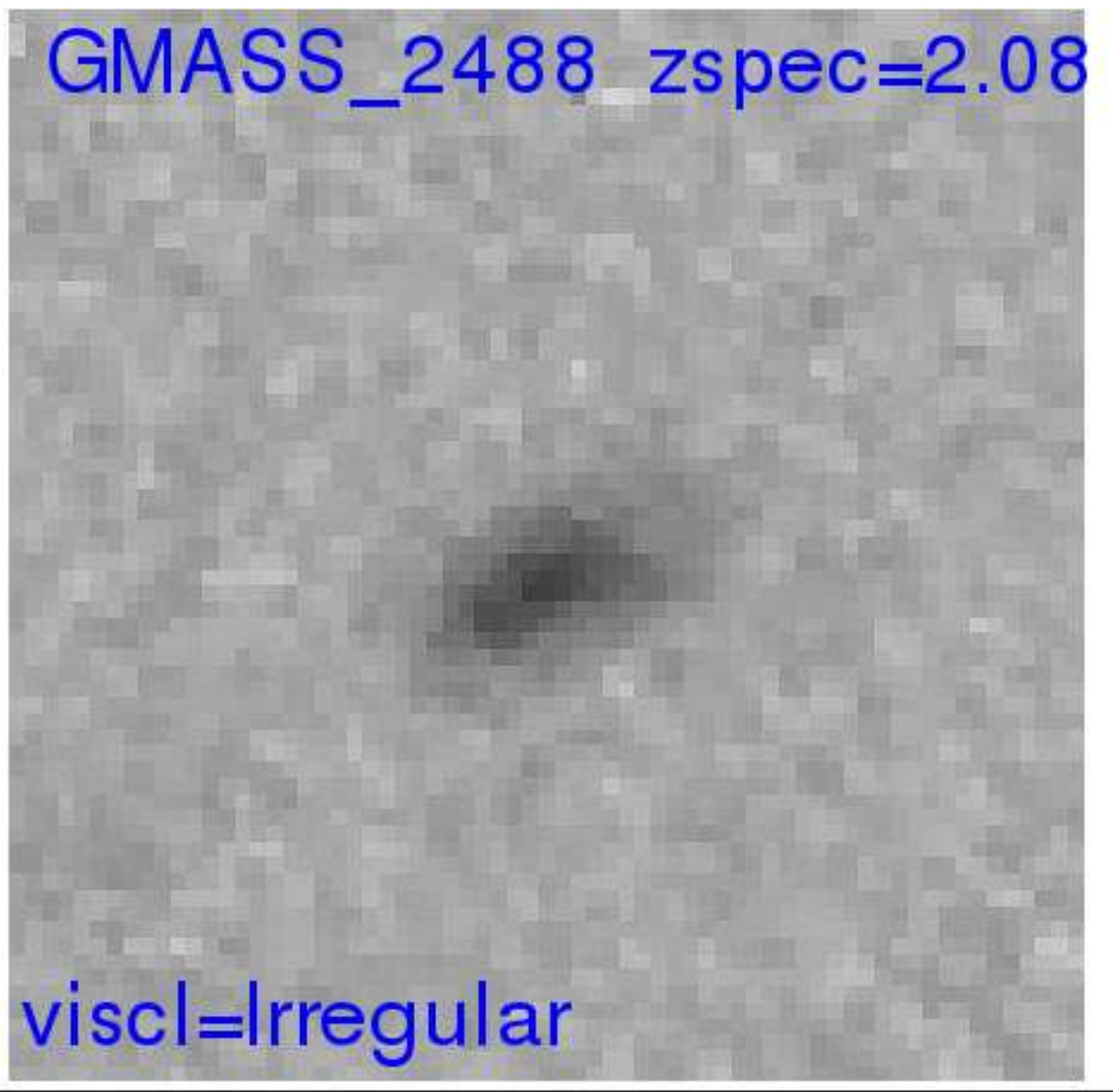}			     
\includegraphics[trim=100 40 75 390, clip=true, width=30mm]{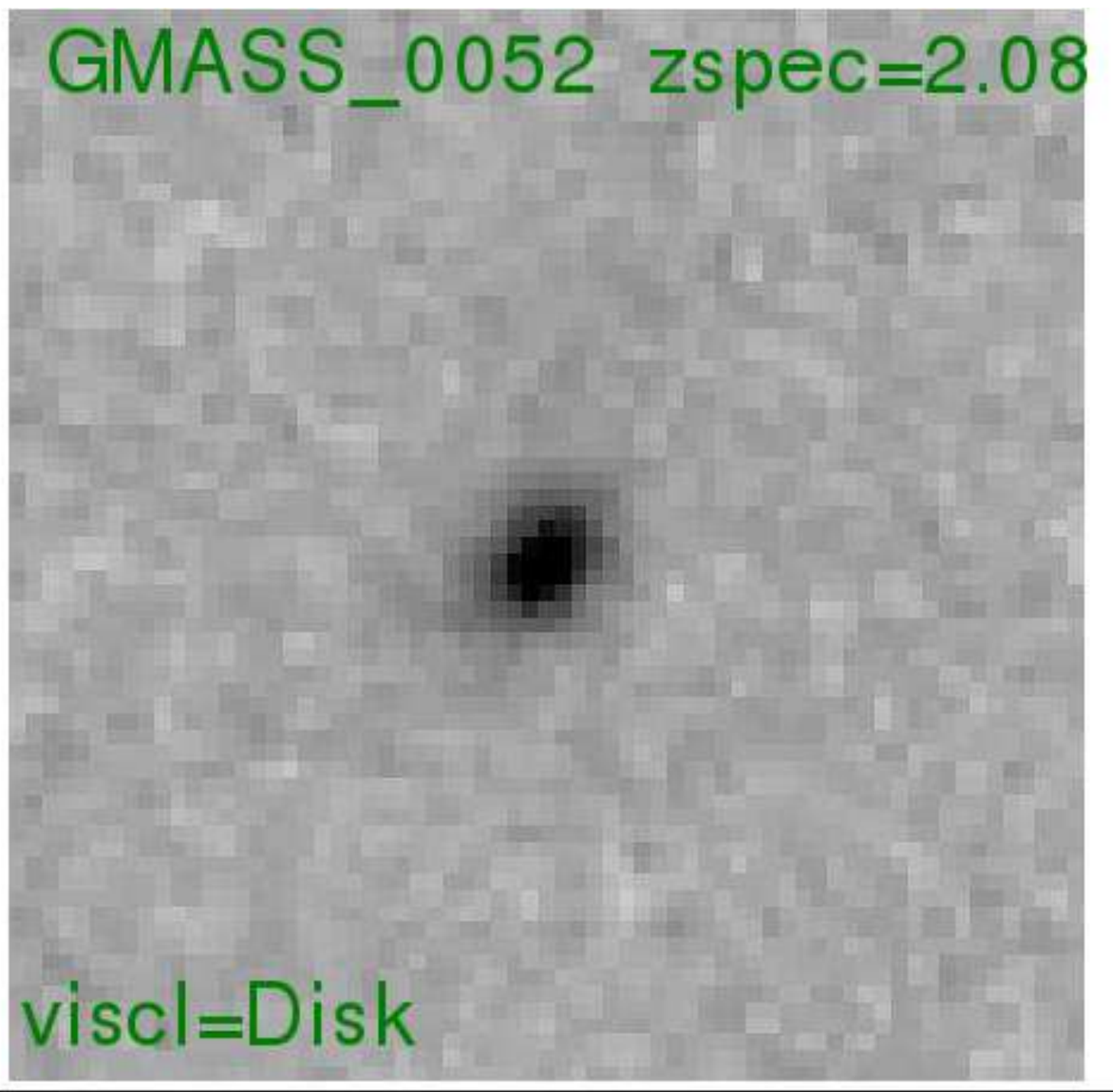}			     
\includegraphics[trim=100 40 75 390, clip=true, width=30mm]{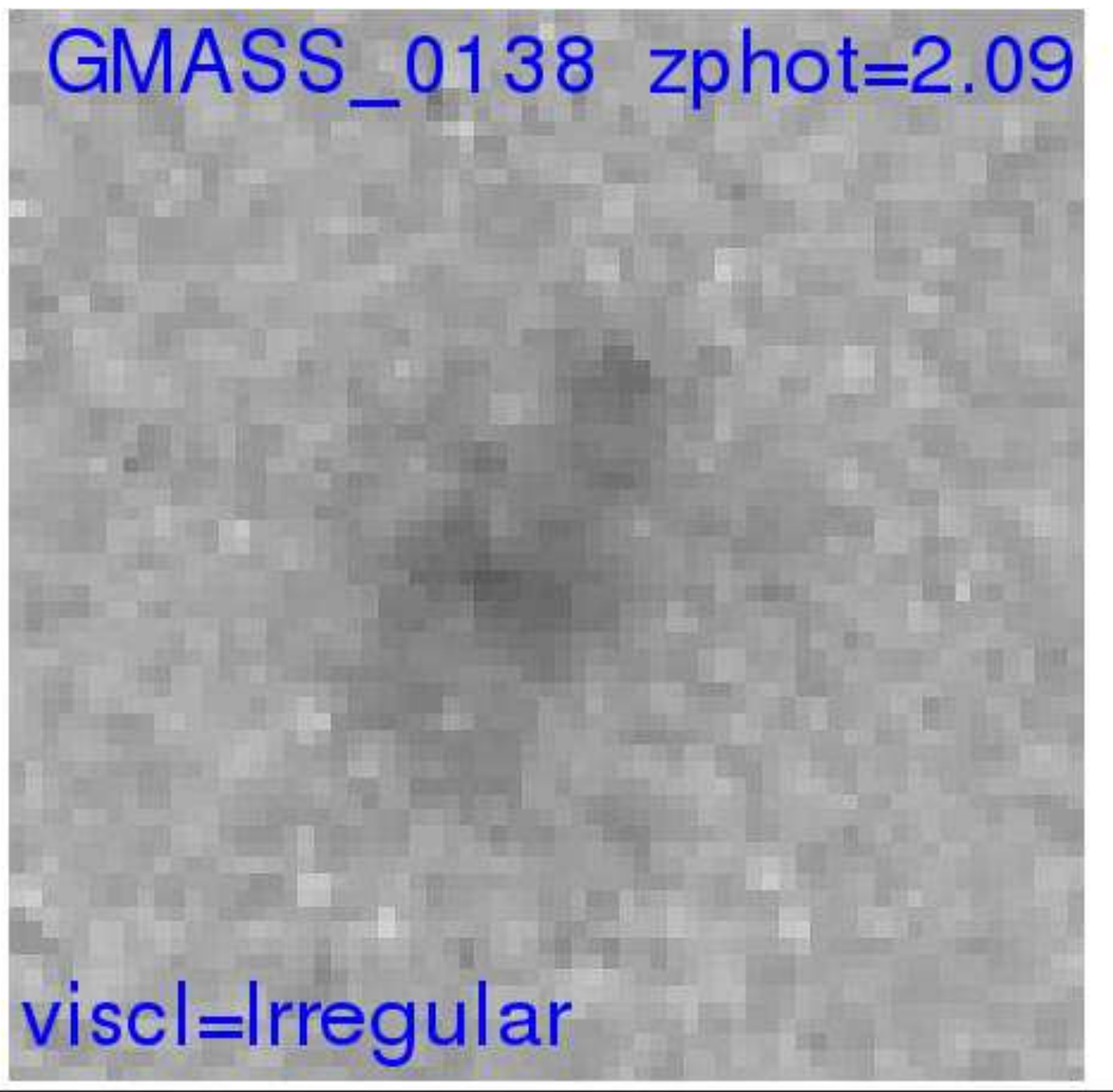}			     
\includegraphics[trim=100 40 75 390, clip=true, width=30mm]{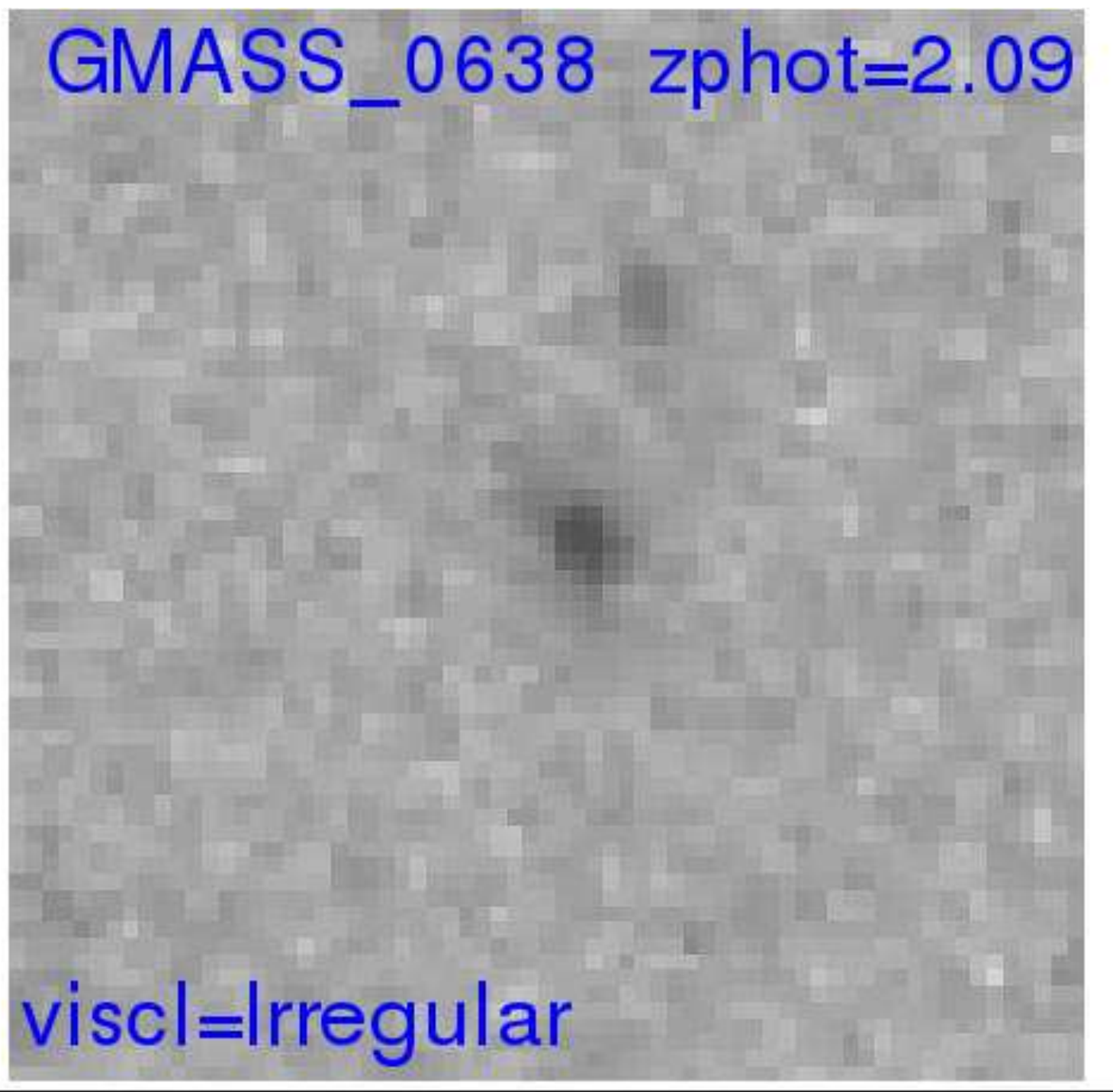}			     
\includegraphics[trim=100 40 75 390, clip=true, width=30mm]{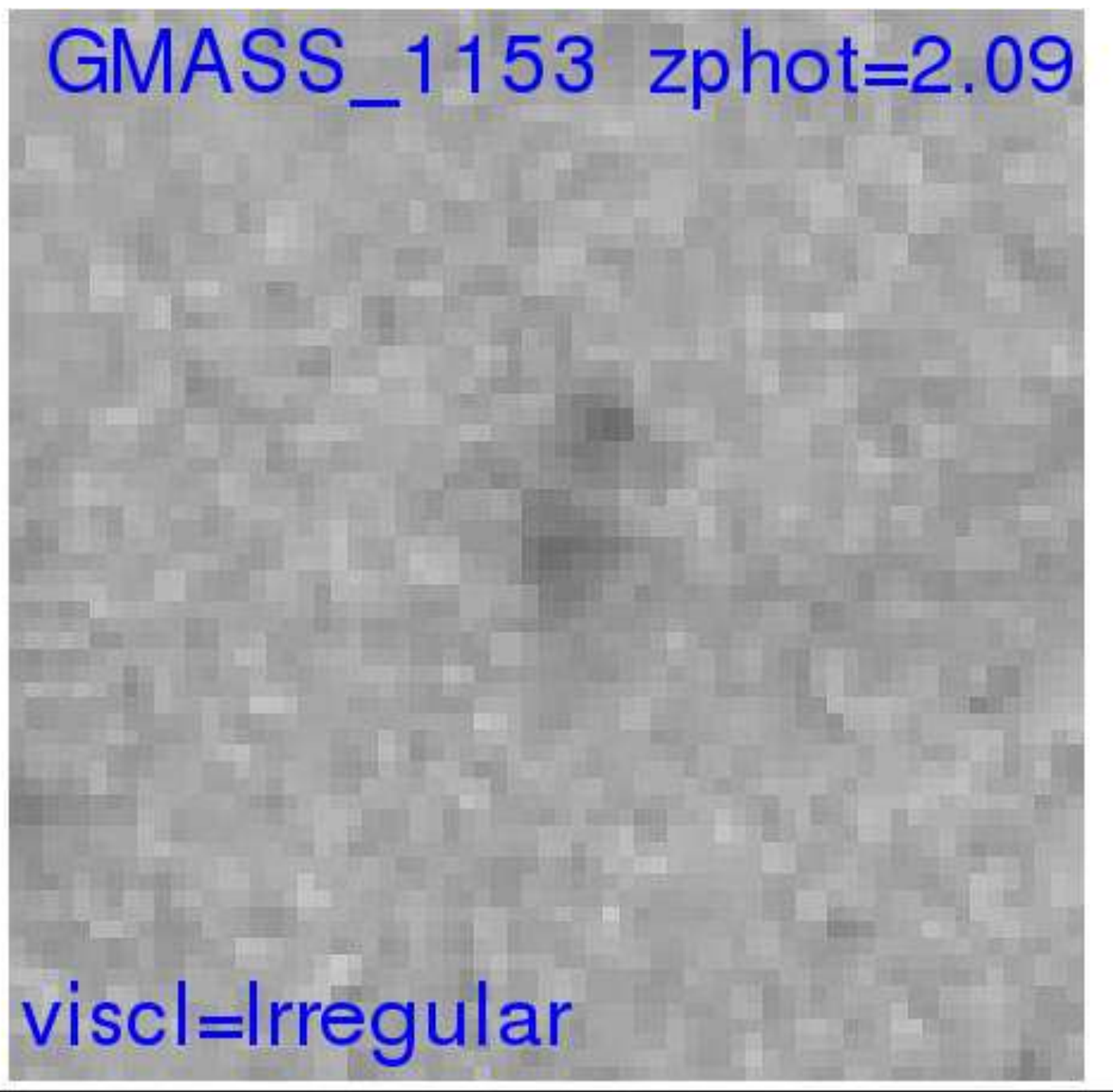}			     
\includegraphics[trim=100 40 75 390, clip=true, width=30mm]{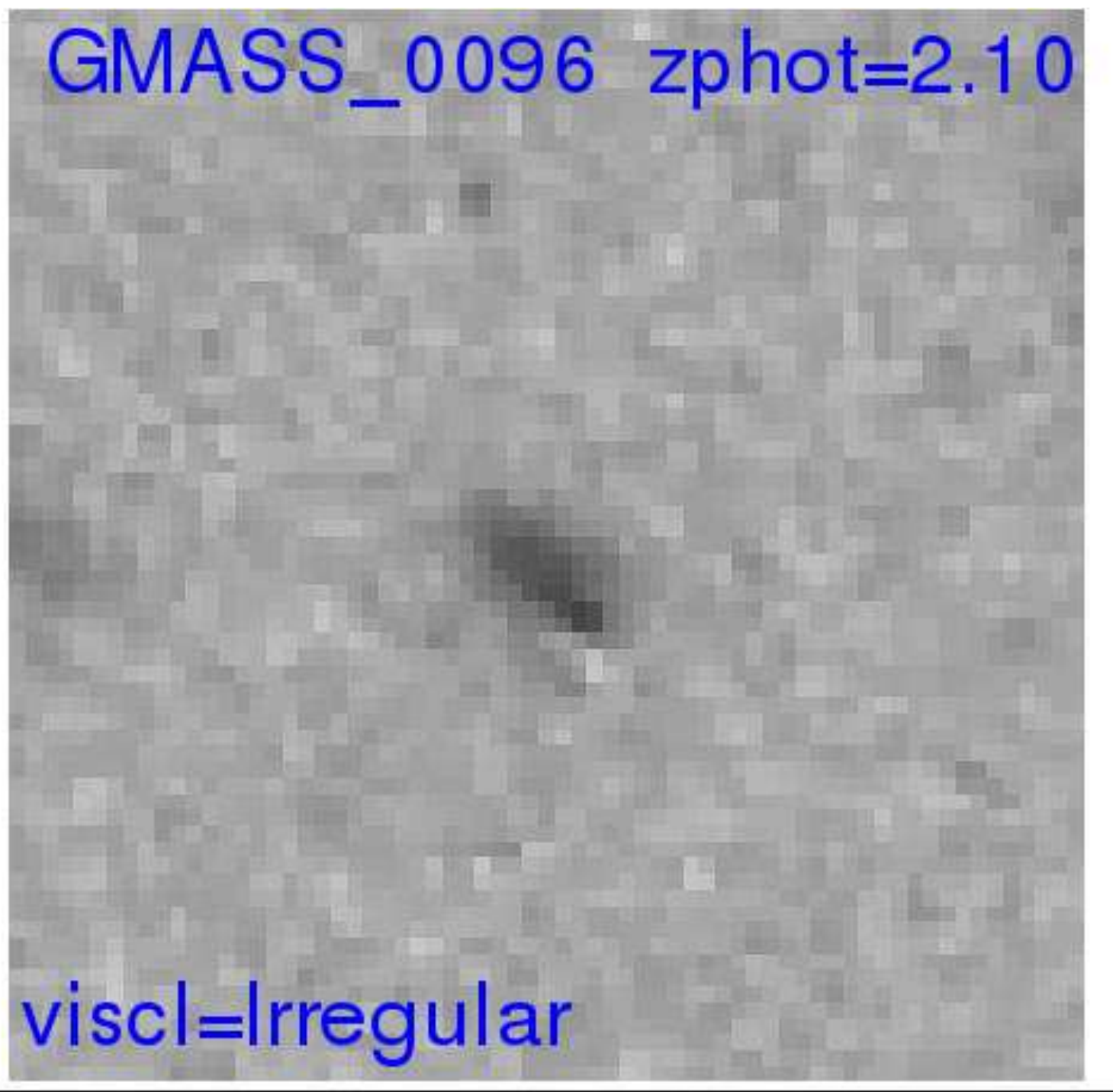}
\end{figure*}
\begin{figure*}
\centering   	 
\includegraphics[trim=100 40 75 390, clip=true, width=30mm]{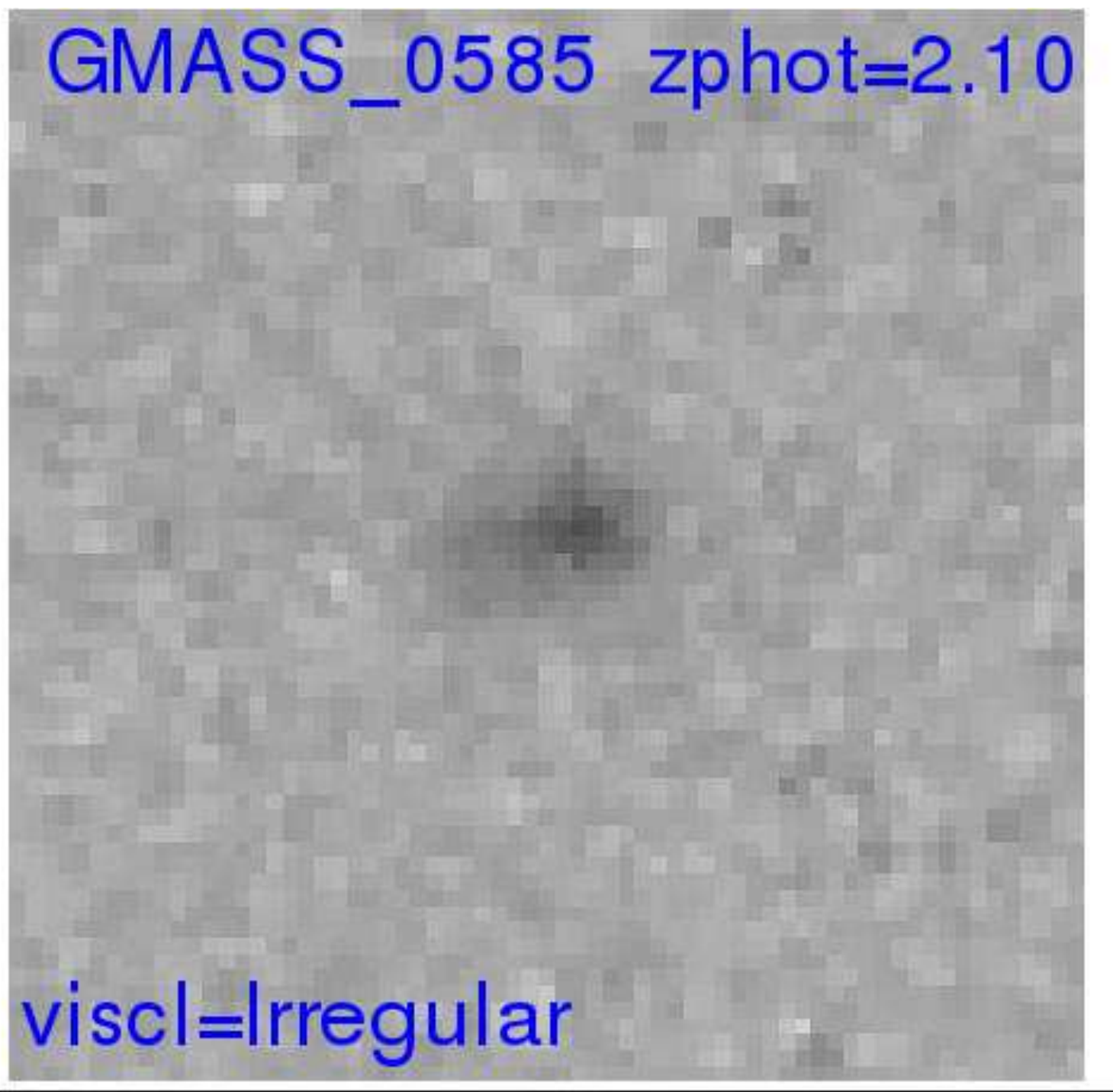}			     
\includegraphics[trim=100 40 75 390, clip=true, width=30mm]{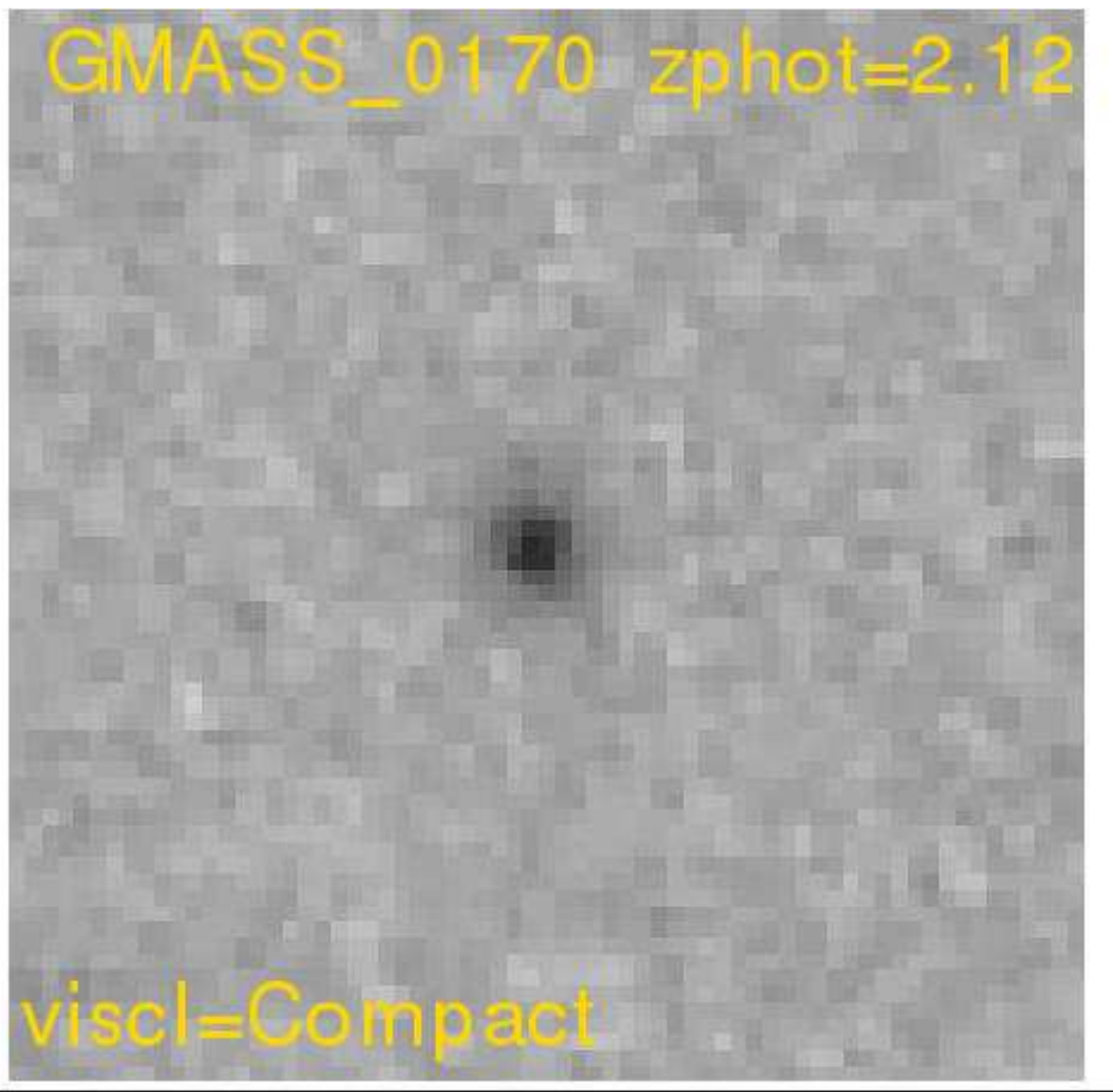}			     
\includegraphics[trim=100 40 75 390, clip=true, width=30mm]{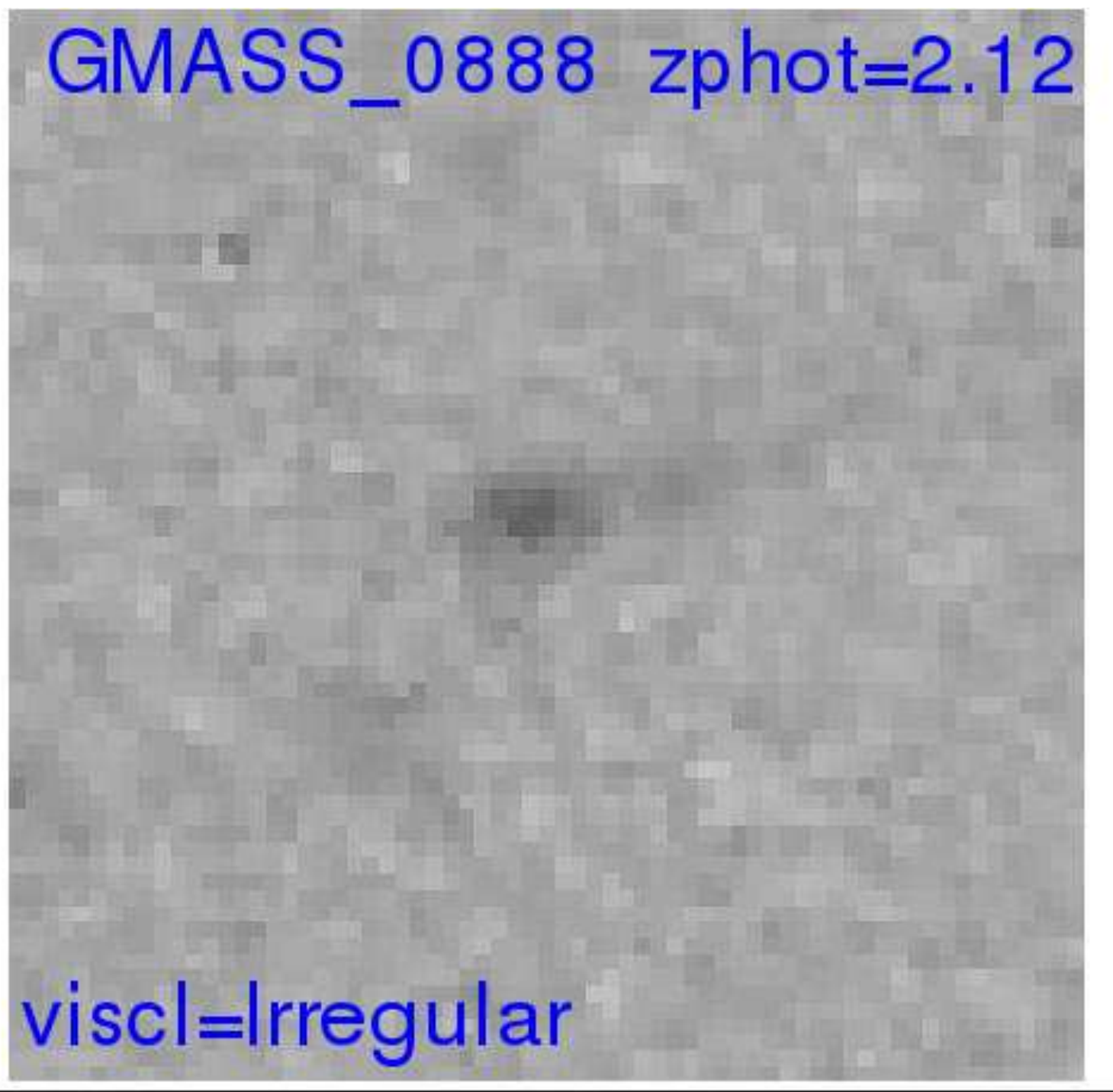}		     
\includegraphics[trim=100 40 75 390, clip=true, width=30mm]{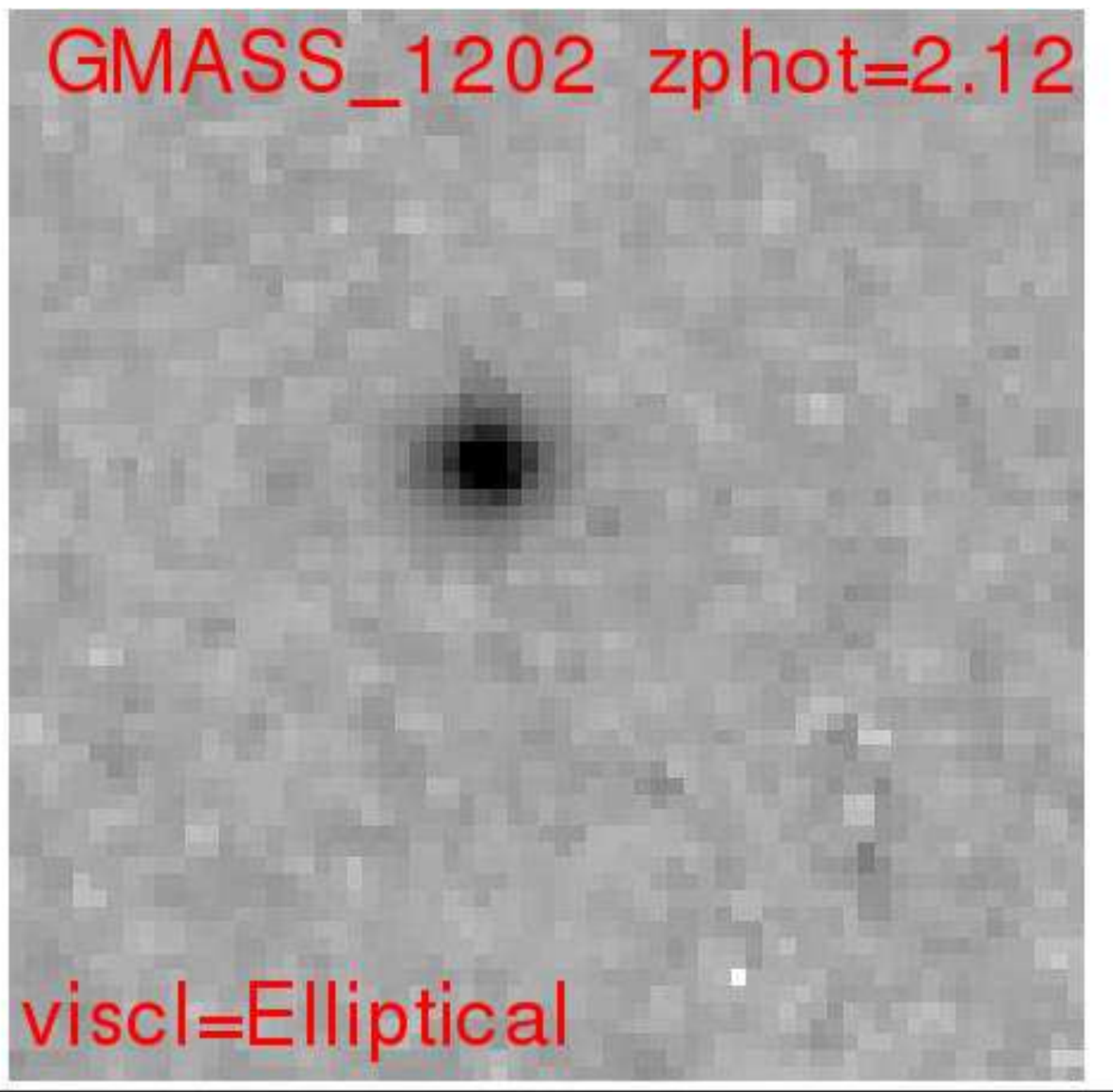}			     
\includegraphics[trim=100 40 75 390, clip=true, width=30mm]{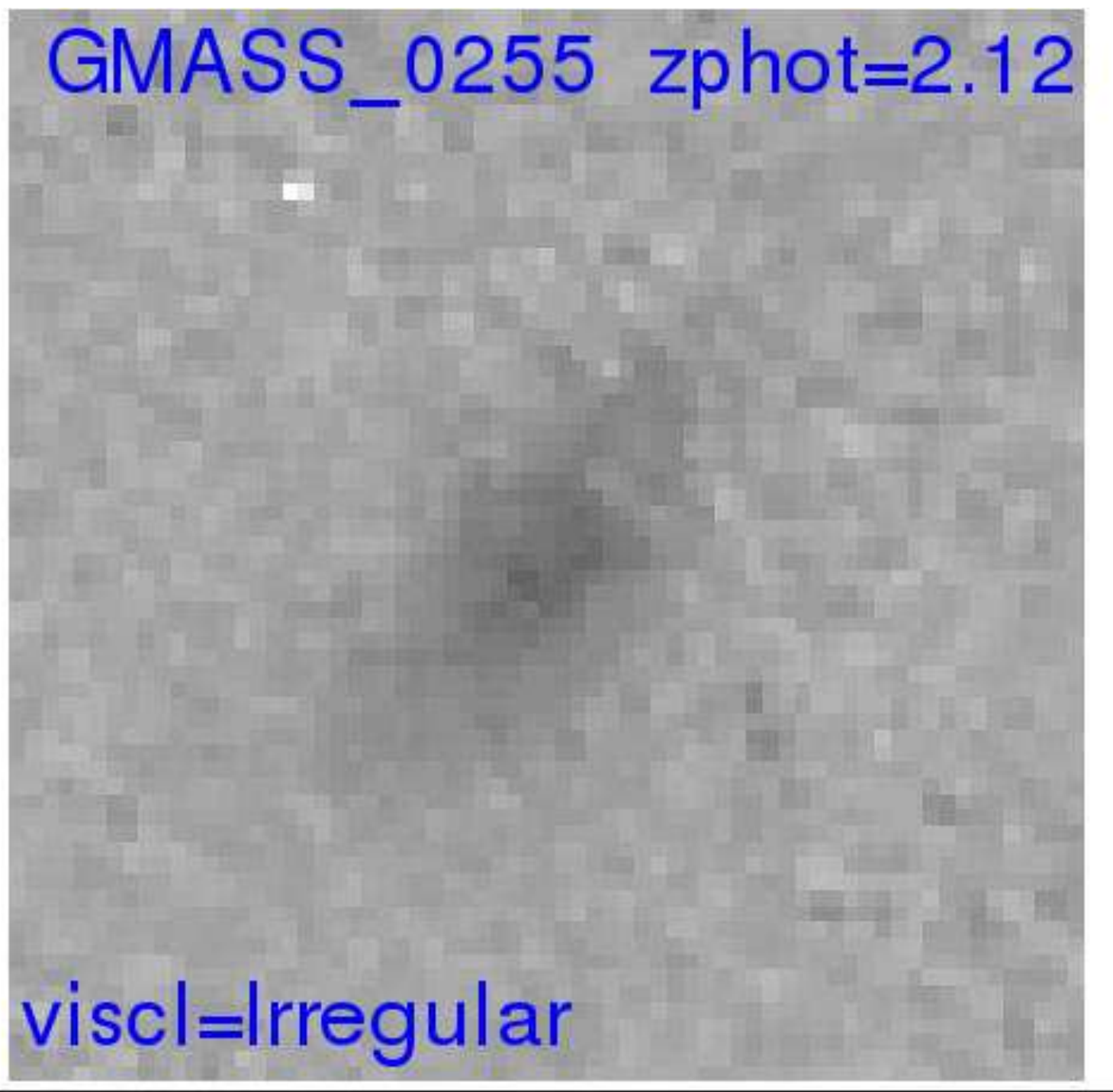}			     
\includegraphics[trim=100 40 75 390, clip=true, width=30mm]{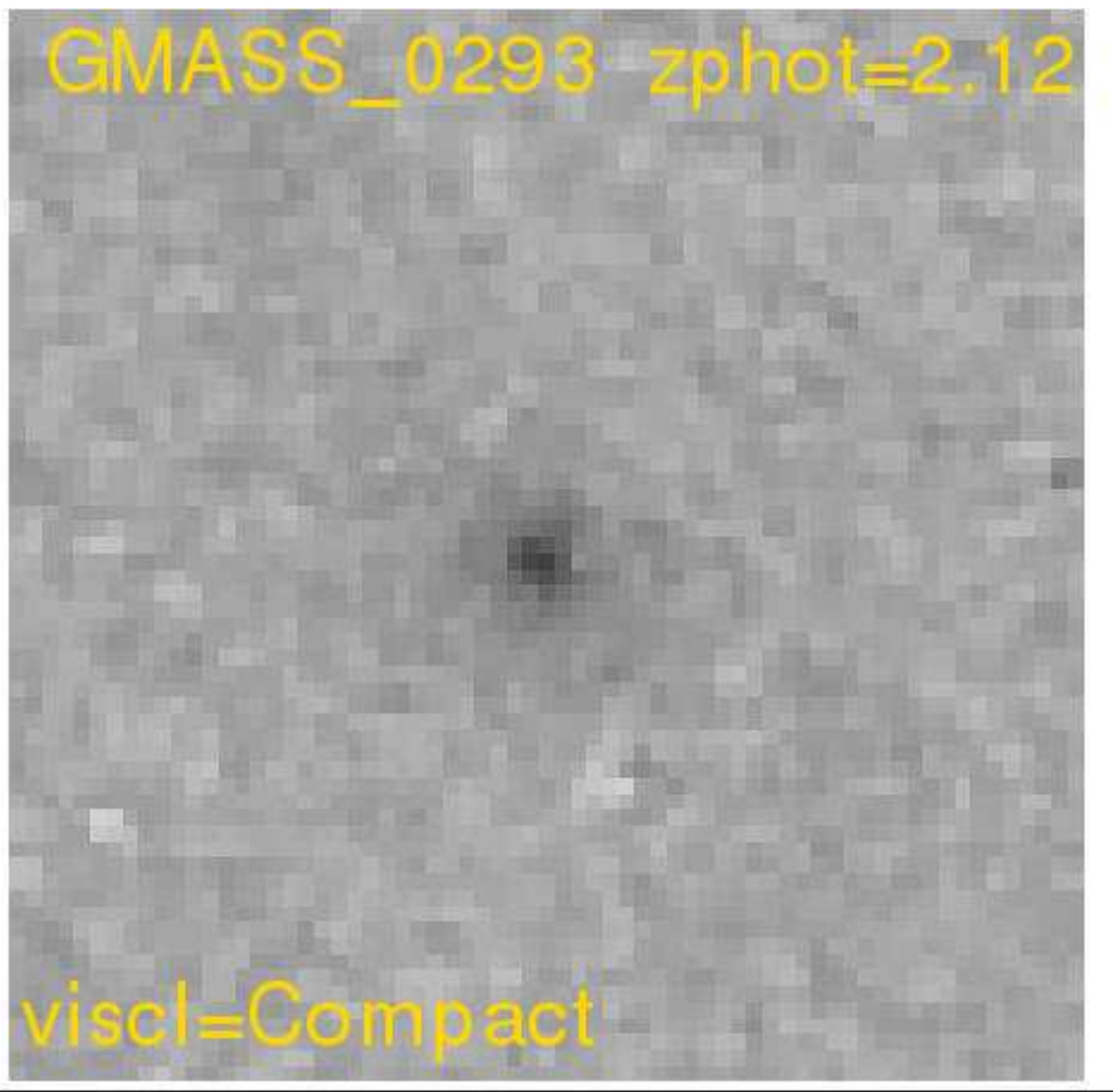}			     

\includegraphics[trim=100 40 75 390, clip=true, width=30mm]{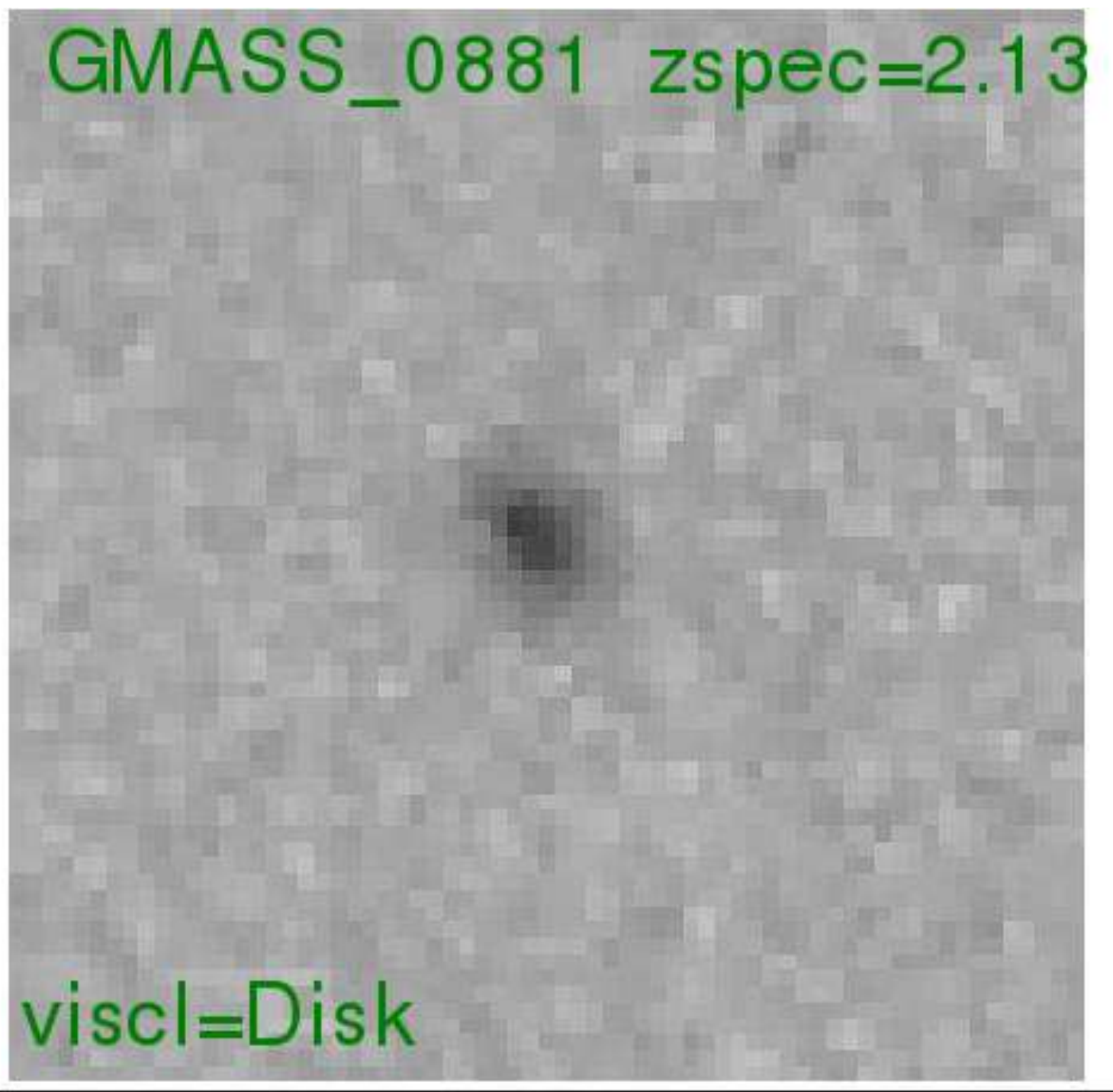}			     
\includegraphics[trim=100 40 75 390, clip=true, width=30mm]{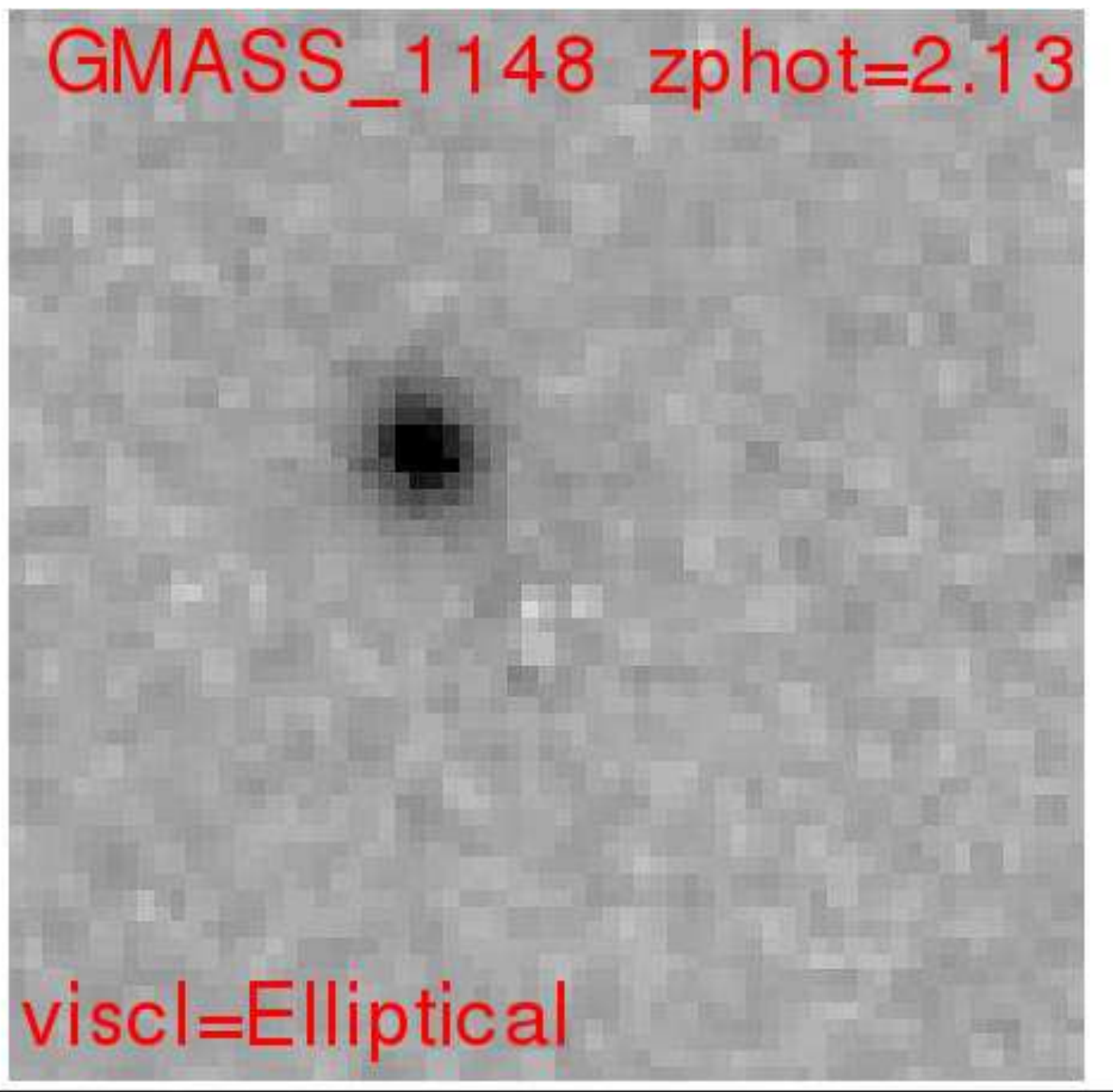}			     
\includegraphics[trim=100 40 75 390, clip=true, width=30mm]{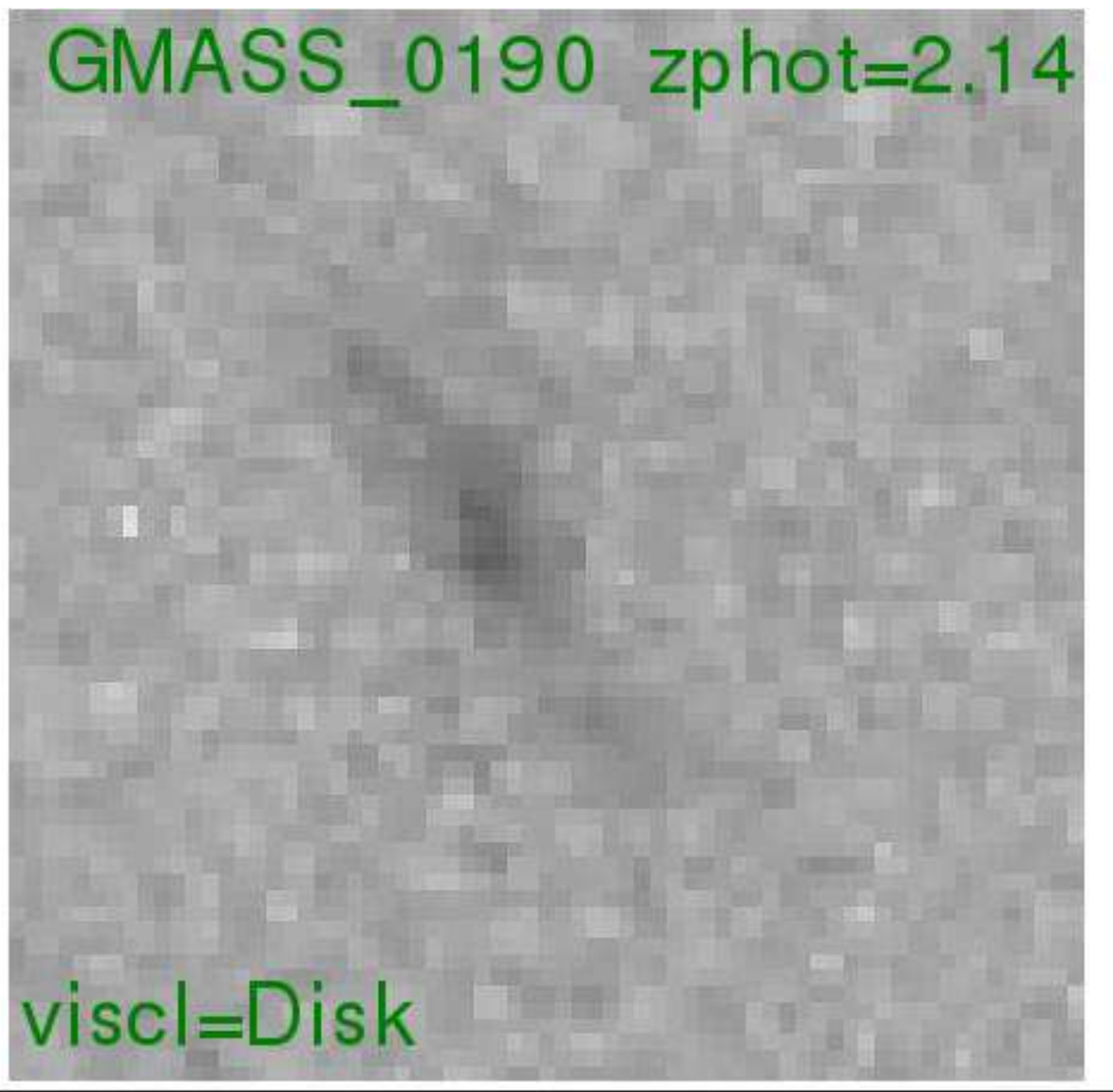}			     
\includegraphics[trim=100 40 75 390, clip=true, width=30mm]{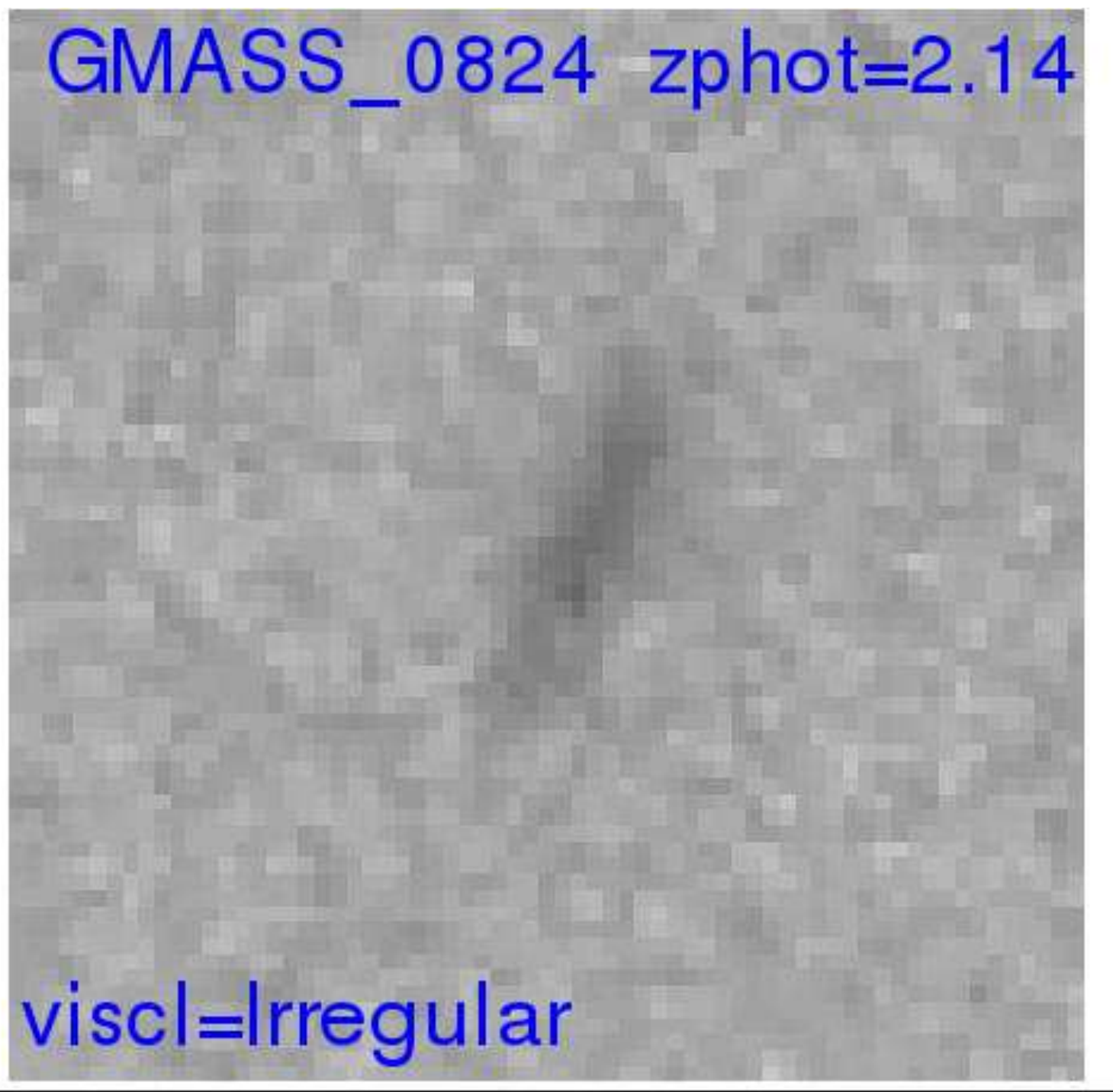}			     
\includegraphics[trim=100 40 75 390, clip=true, width=30mm]{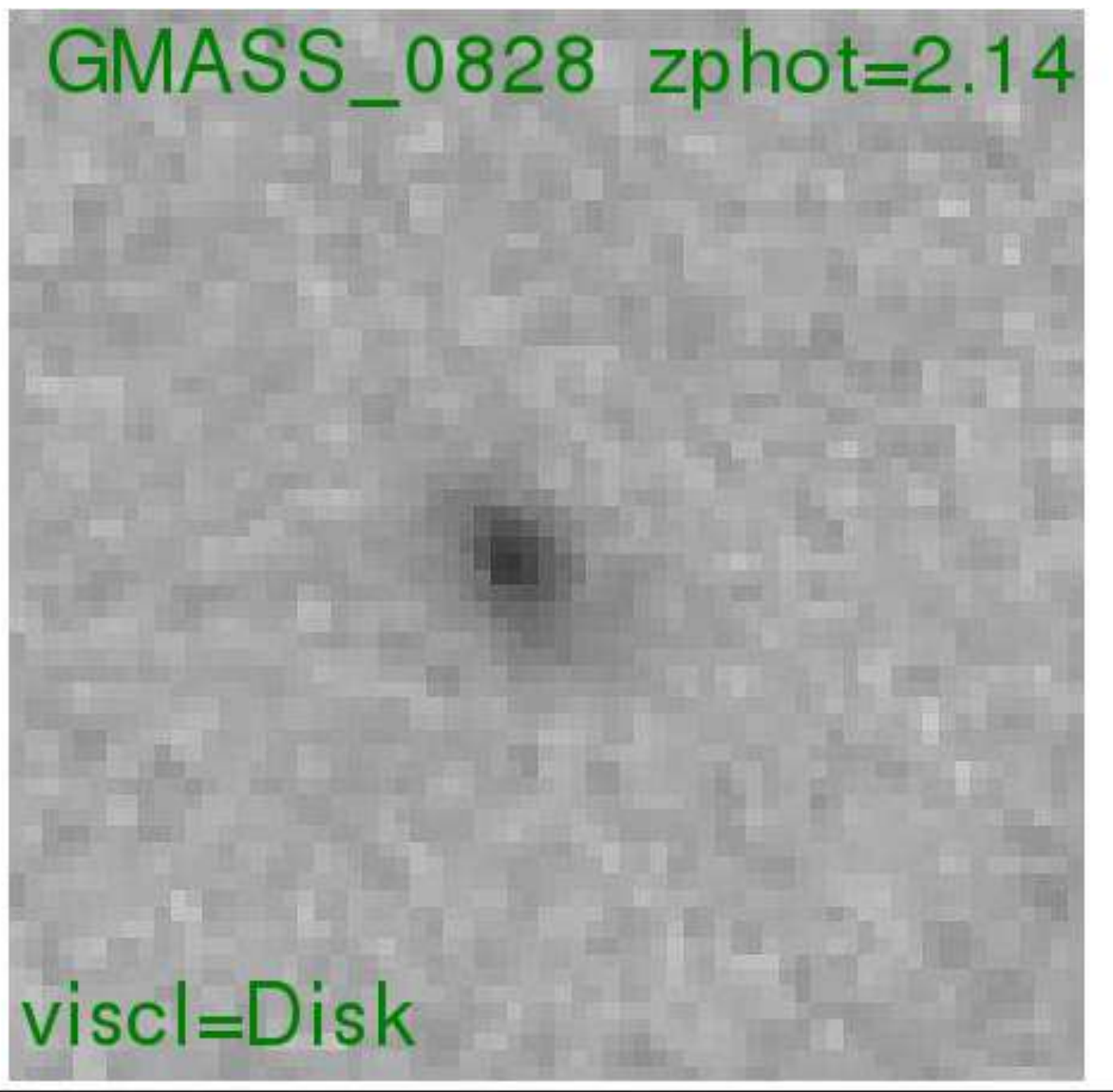}		     
\includegraphics[trim=100 40 75 390, clip=true, width=30mm]{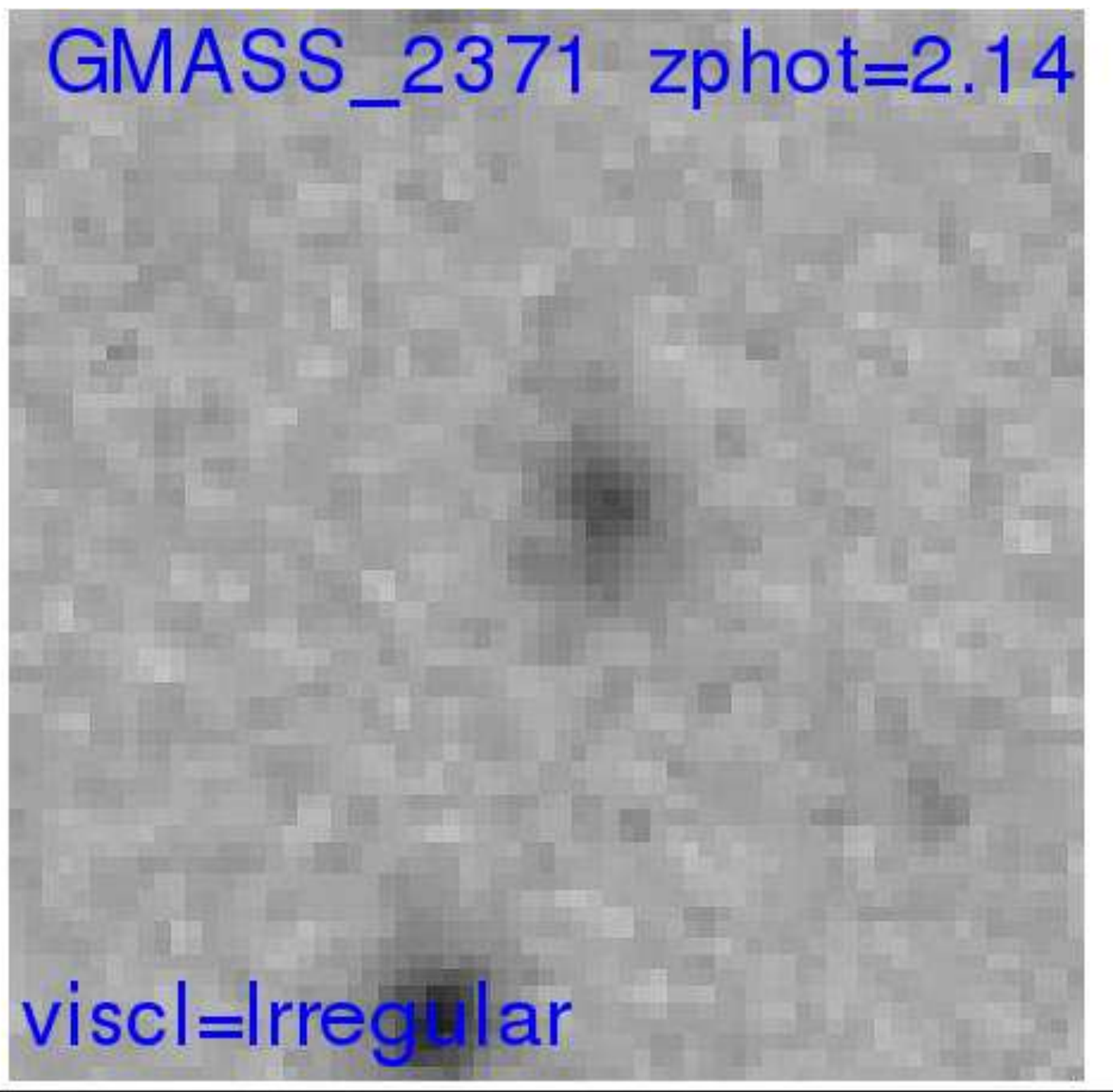}			     

\includegraphics[trim=100 40 75 390, clip=true, width=30mm]{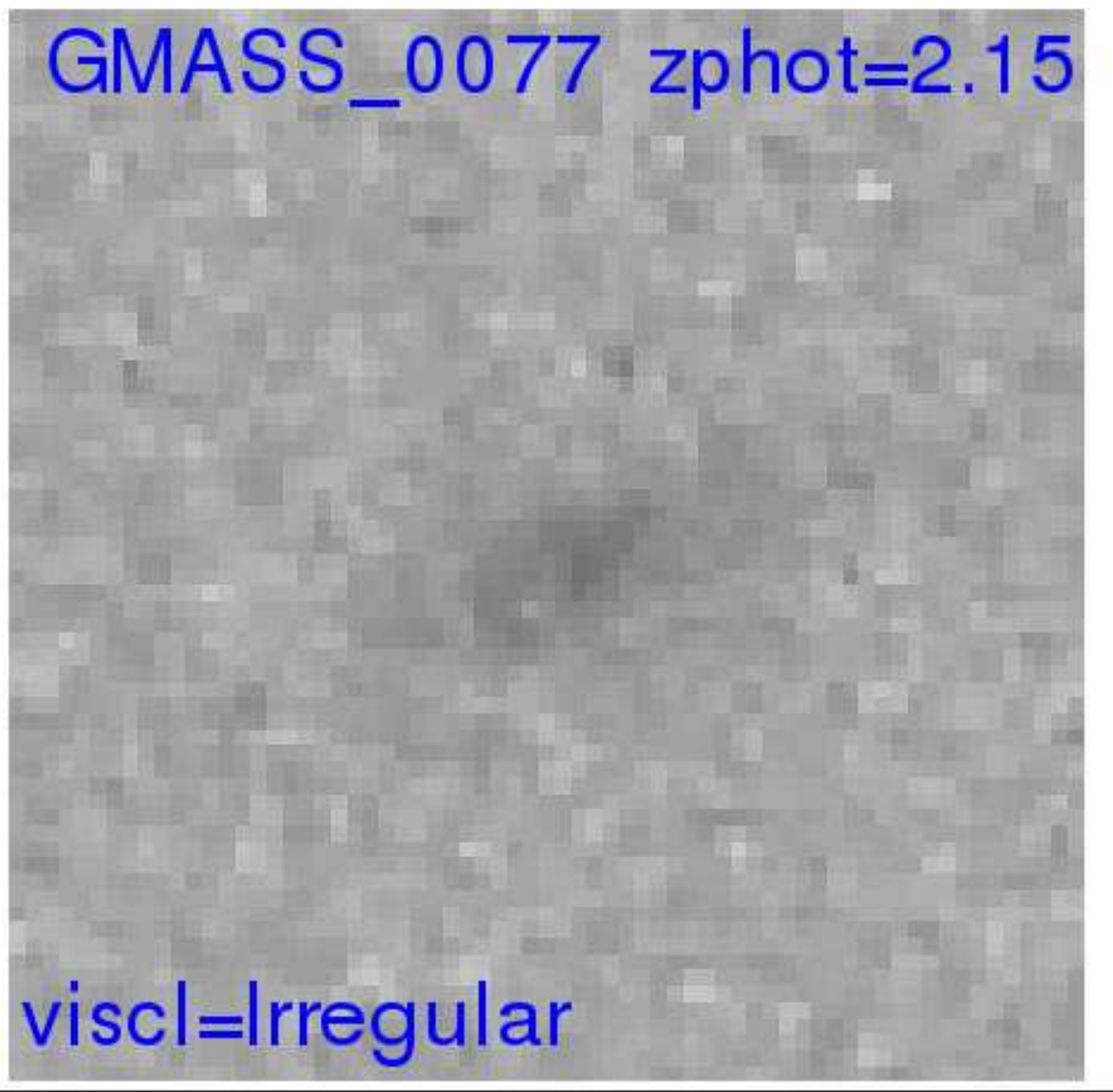}			     
\includegraphics[trim=100 40 75 390, clip=true, width=30mm]{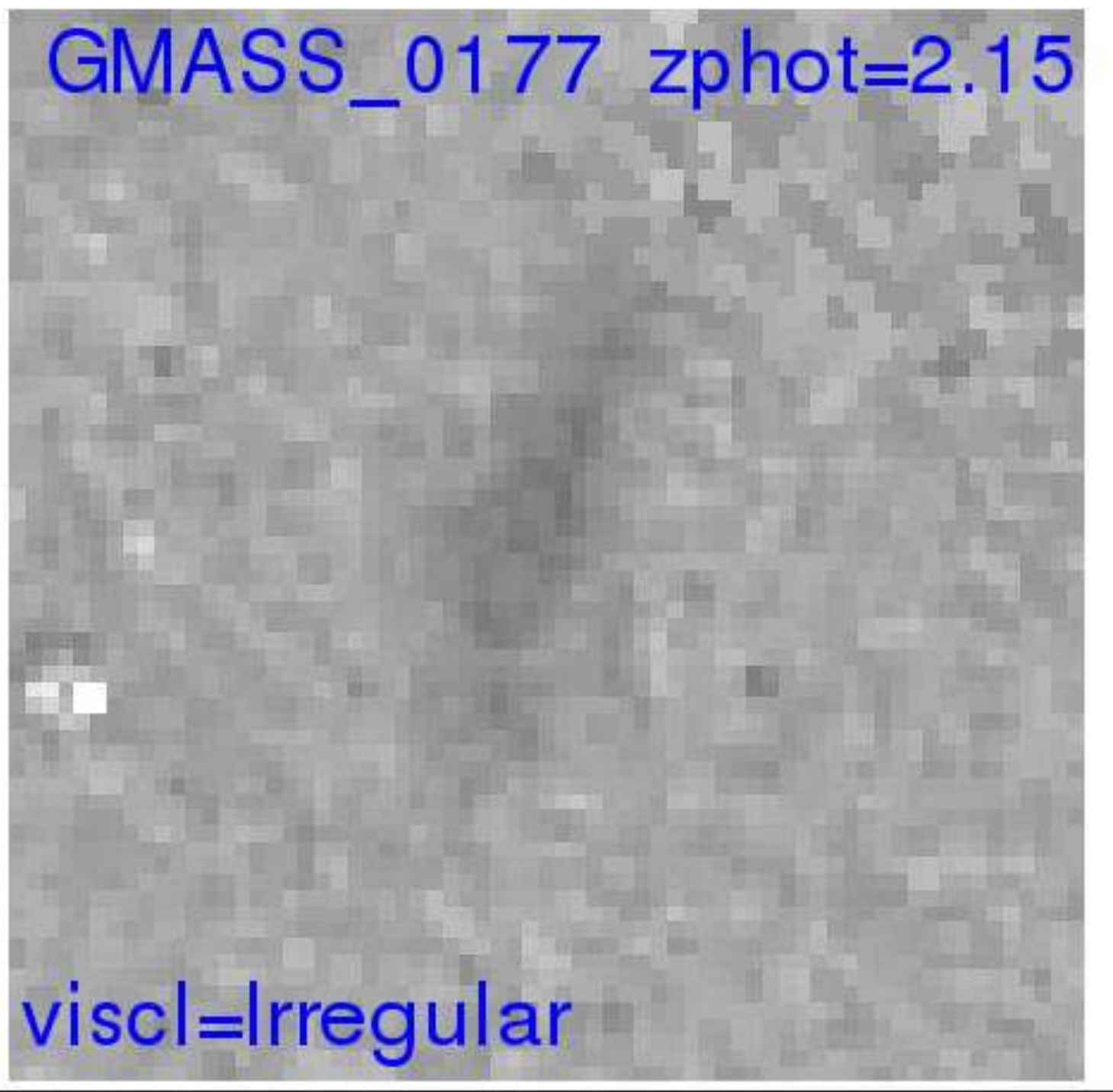}			     
\includegraphics[trim=100 40 75 390, clip=true, width=30mm]{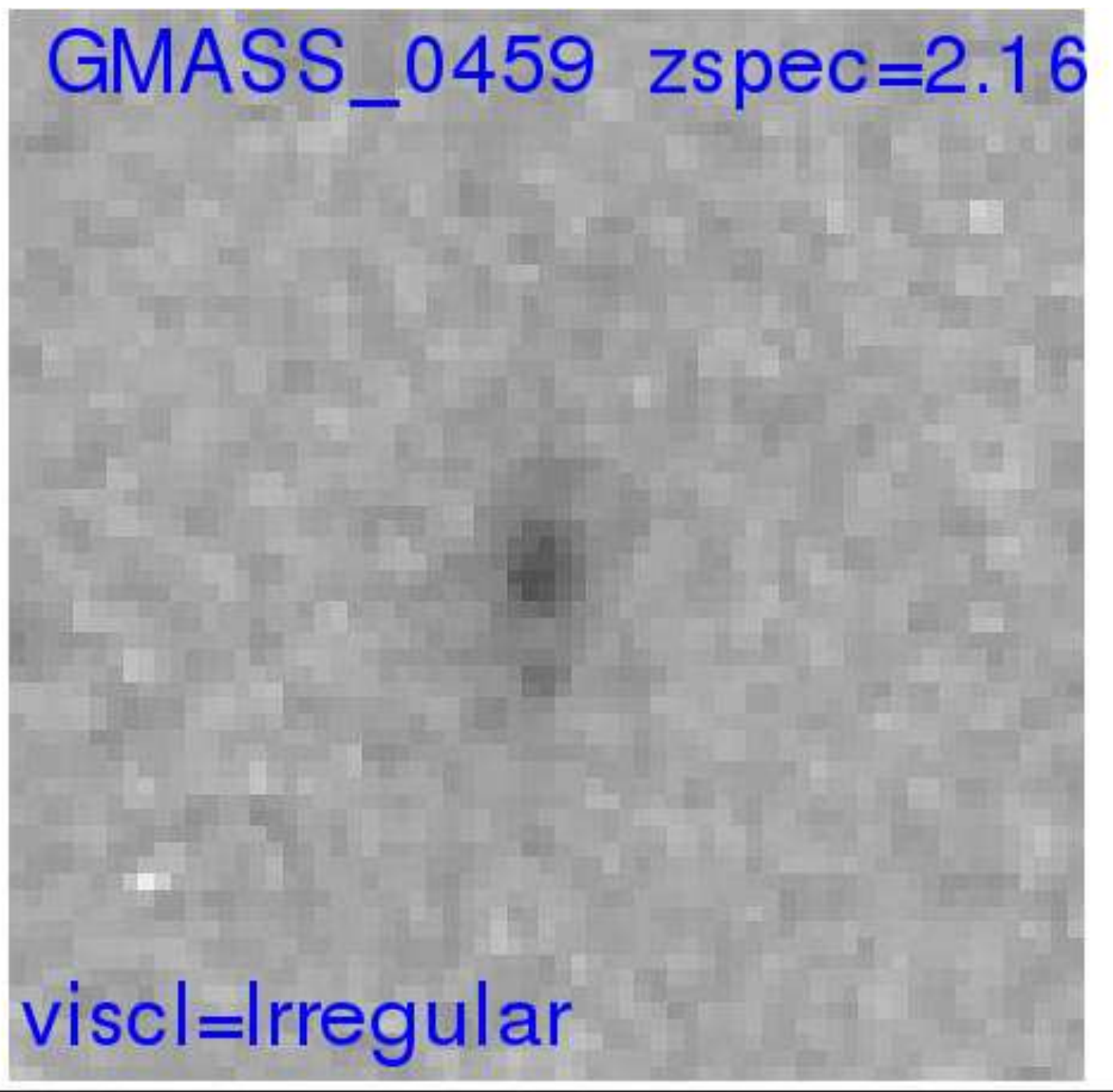}		     
\includegraphics[trim=100 40 75 390, clip=true, width=30mm]{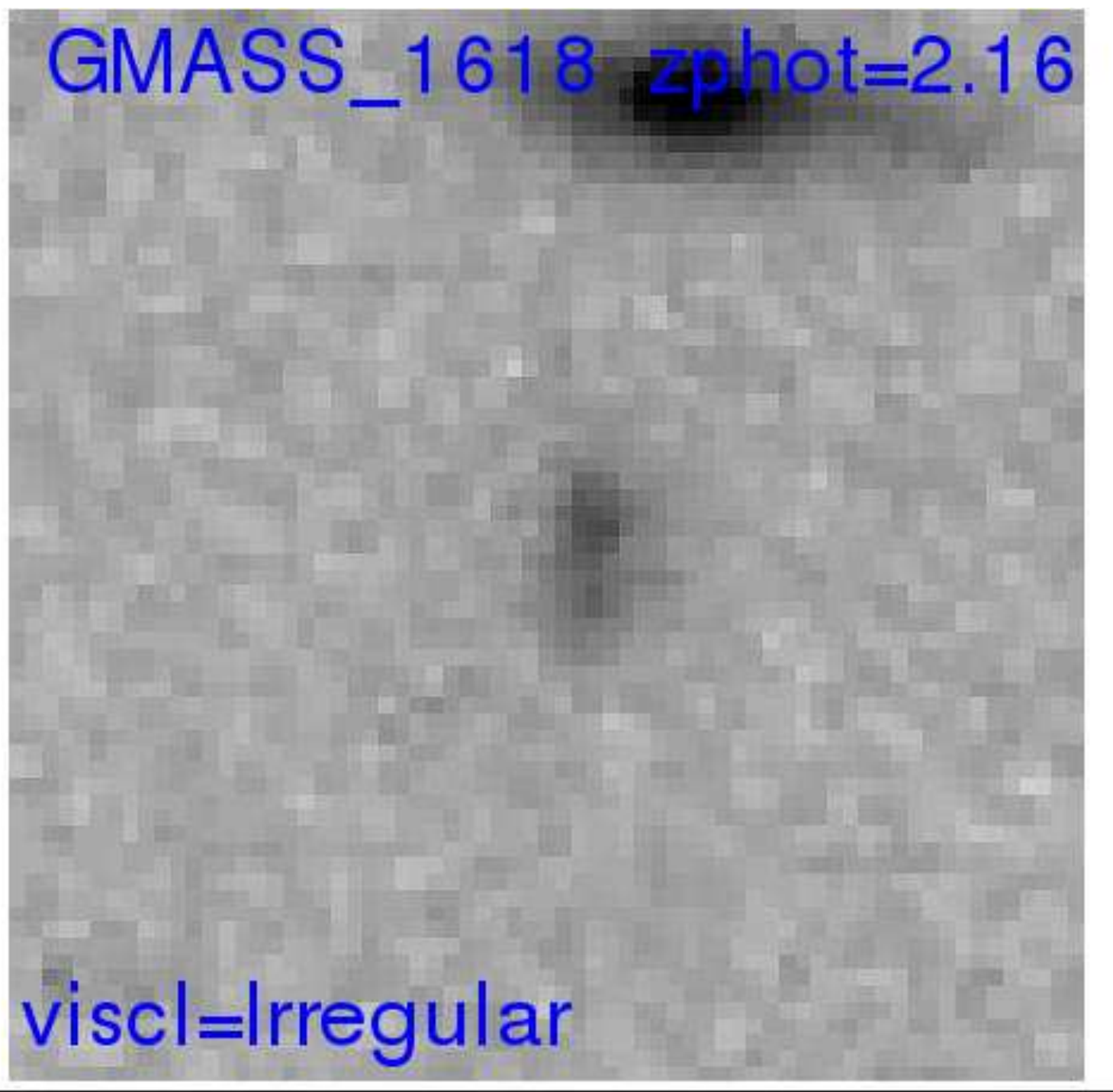}			     
\includegraphics[trim=100 40 75 390, clip=true, width=30mm]{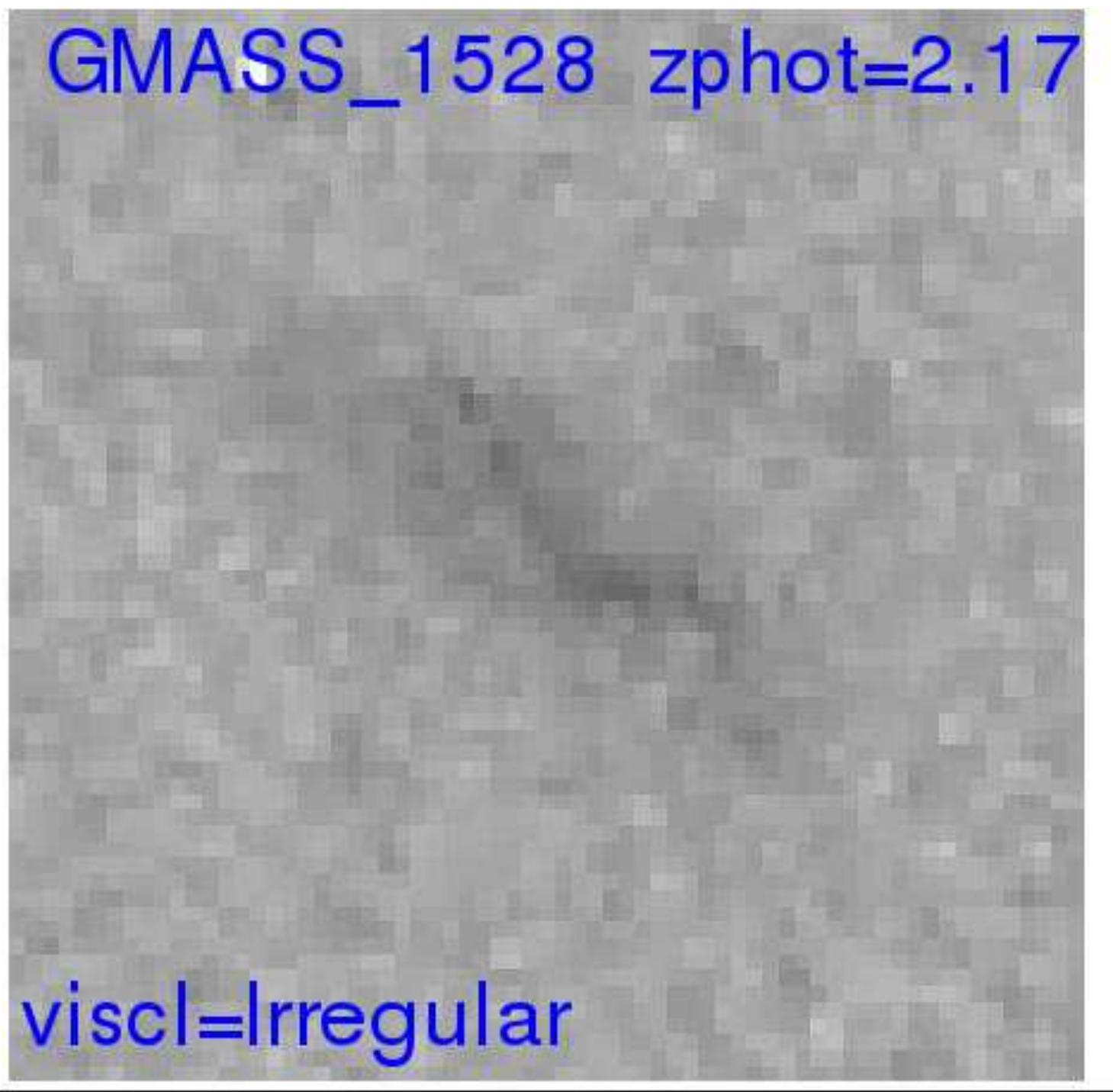}			     
\includegraphics[trim=100 40 75 390, clip=true, width=30mm]{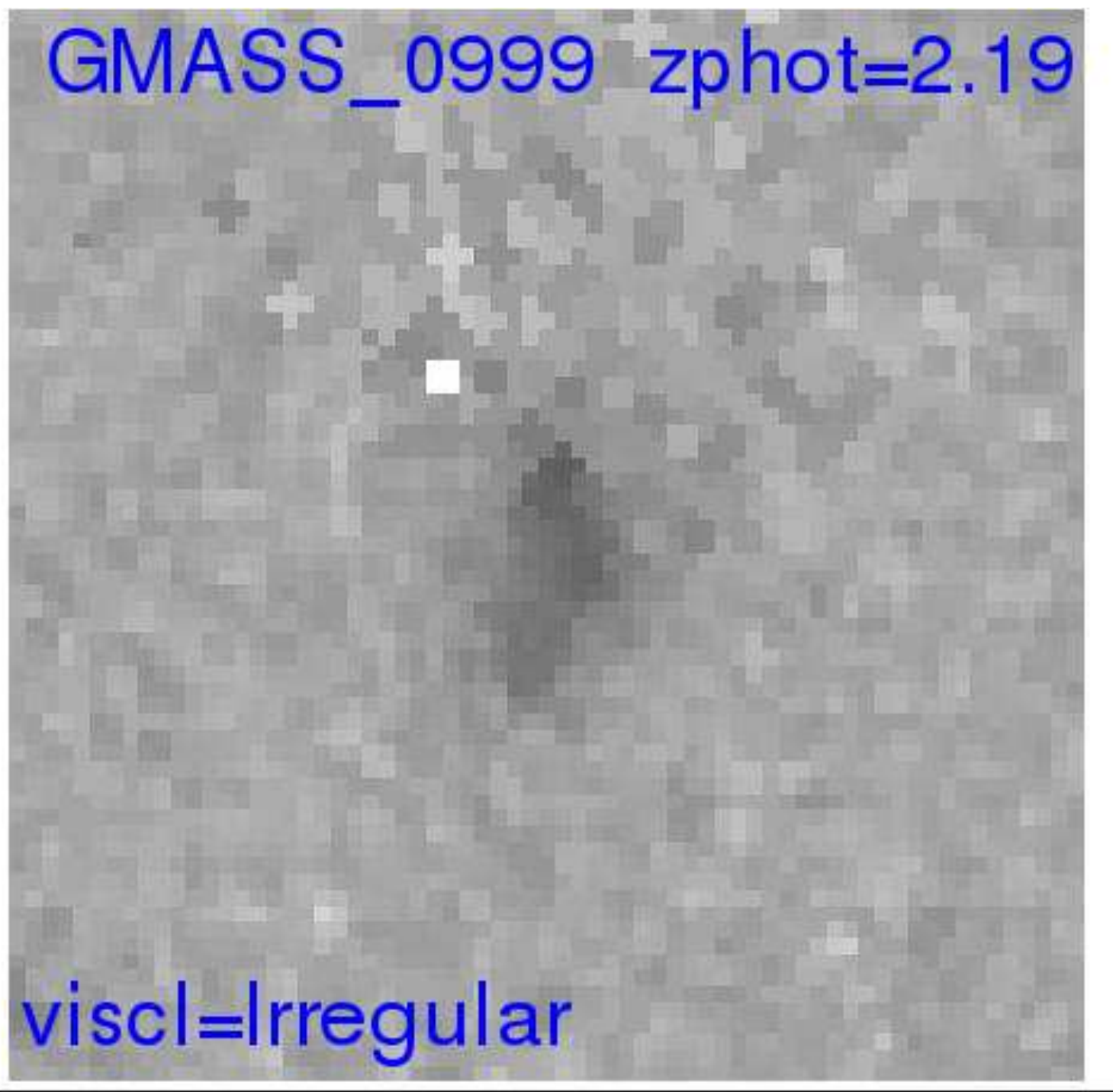}			     

\includegraphics[trim=100 40 75 390, clip=true, width=30mm]{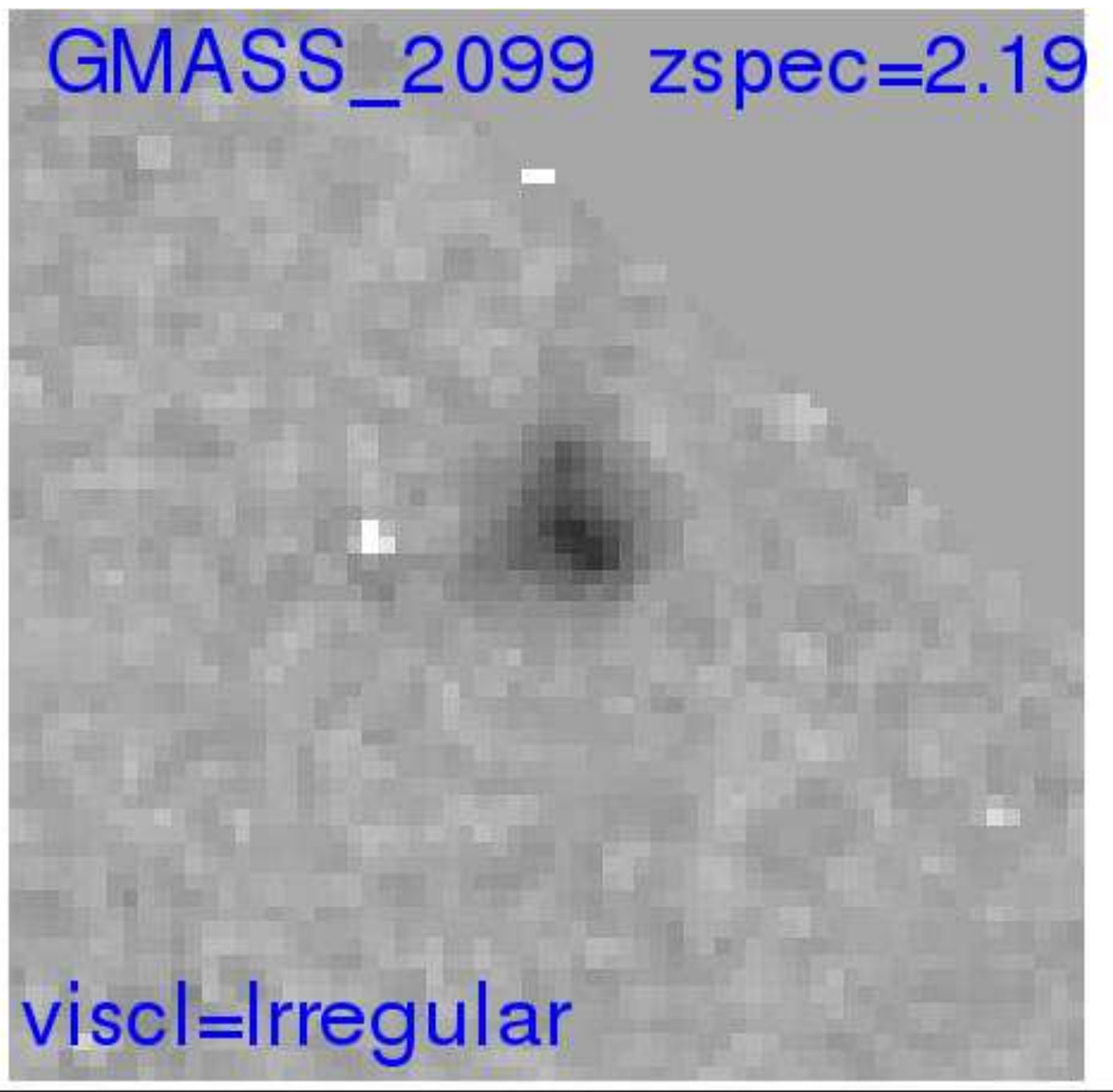}			     
\includegraphics[trim=100 40 75 390, clip=true, width=30mm]{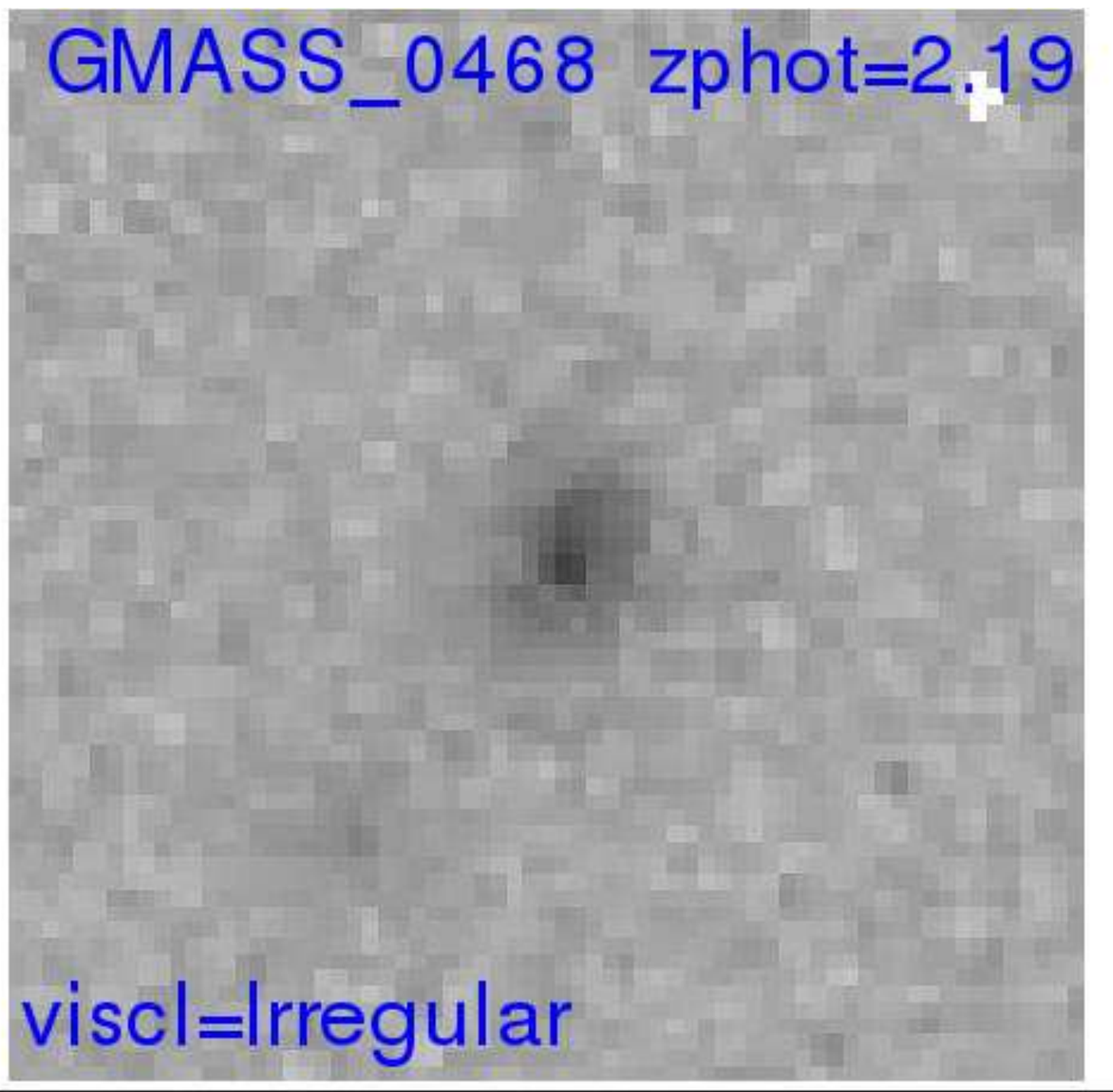}			     
\includegraphics[trim=100 40 75 390, clip=true, width=30mm]{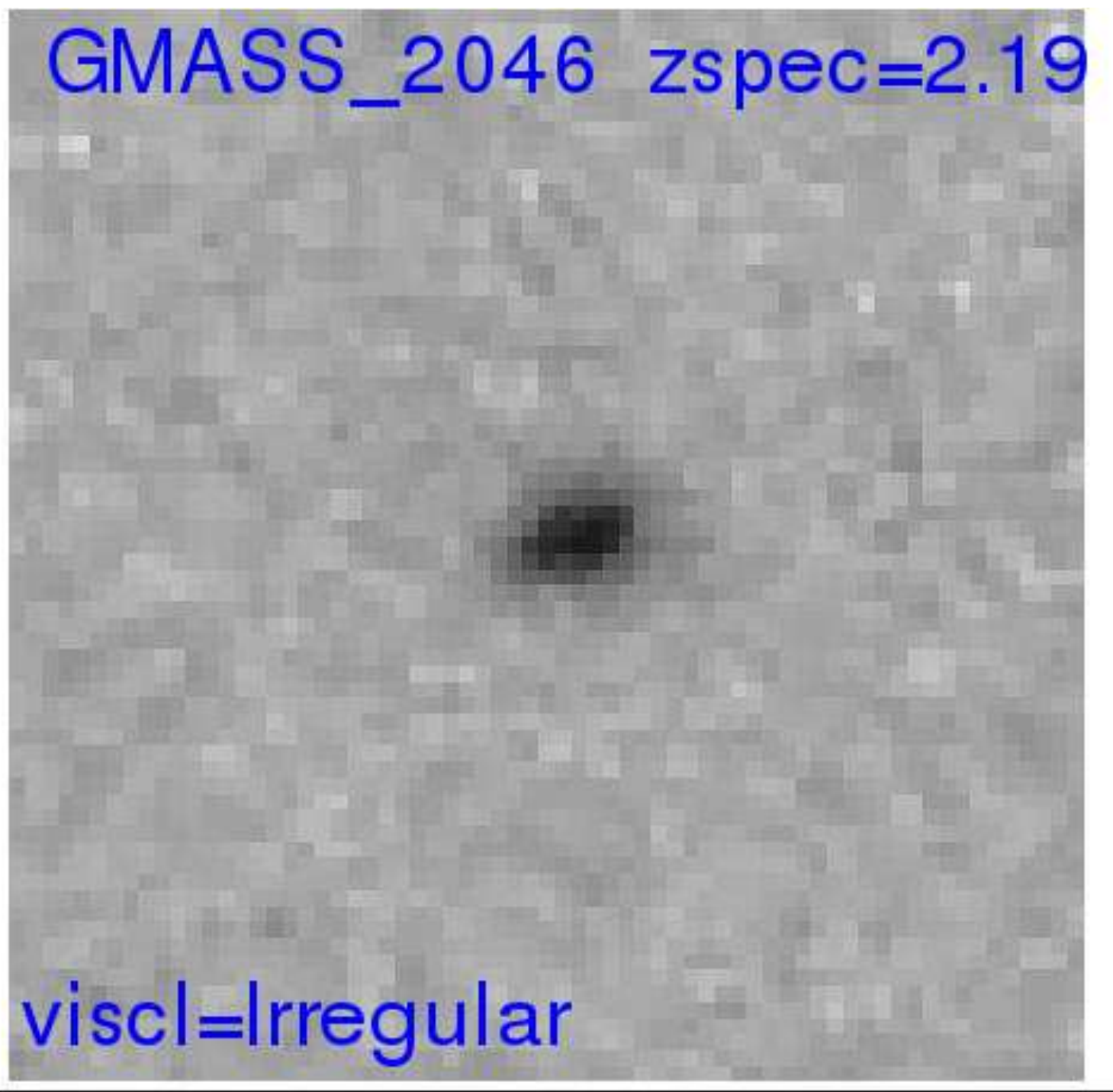}			     
\includegraphics[trim=100 40 75 390, clip=true, width=30mm]{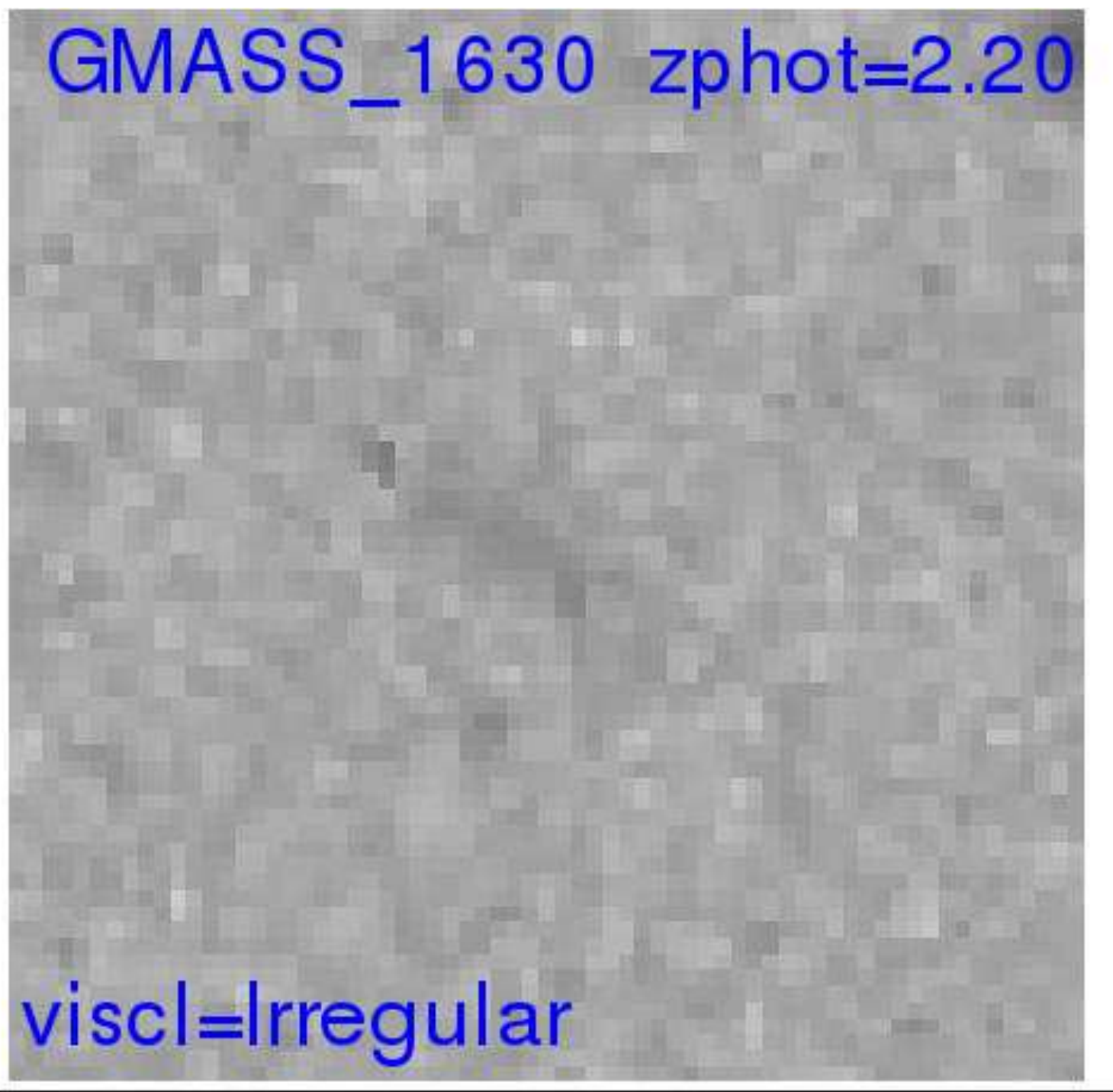}			     
\includegraphics[trim=100 40 75 390, clip=true, width=30mm]{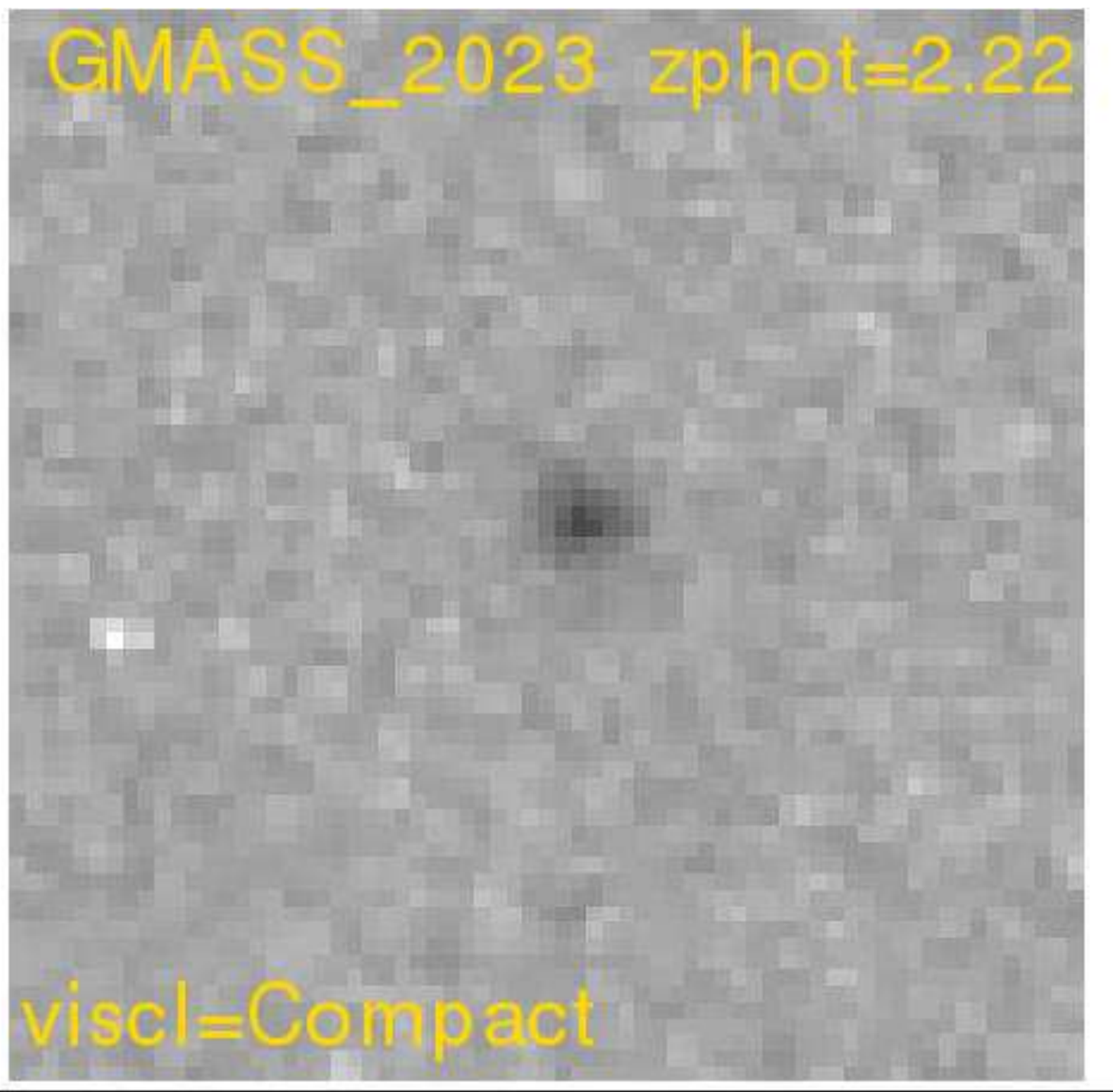}			     
\includegraphics[trim=100 40 75 390, clip=true, width=30mm]{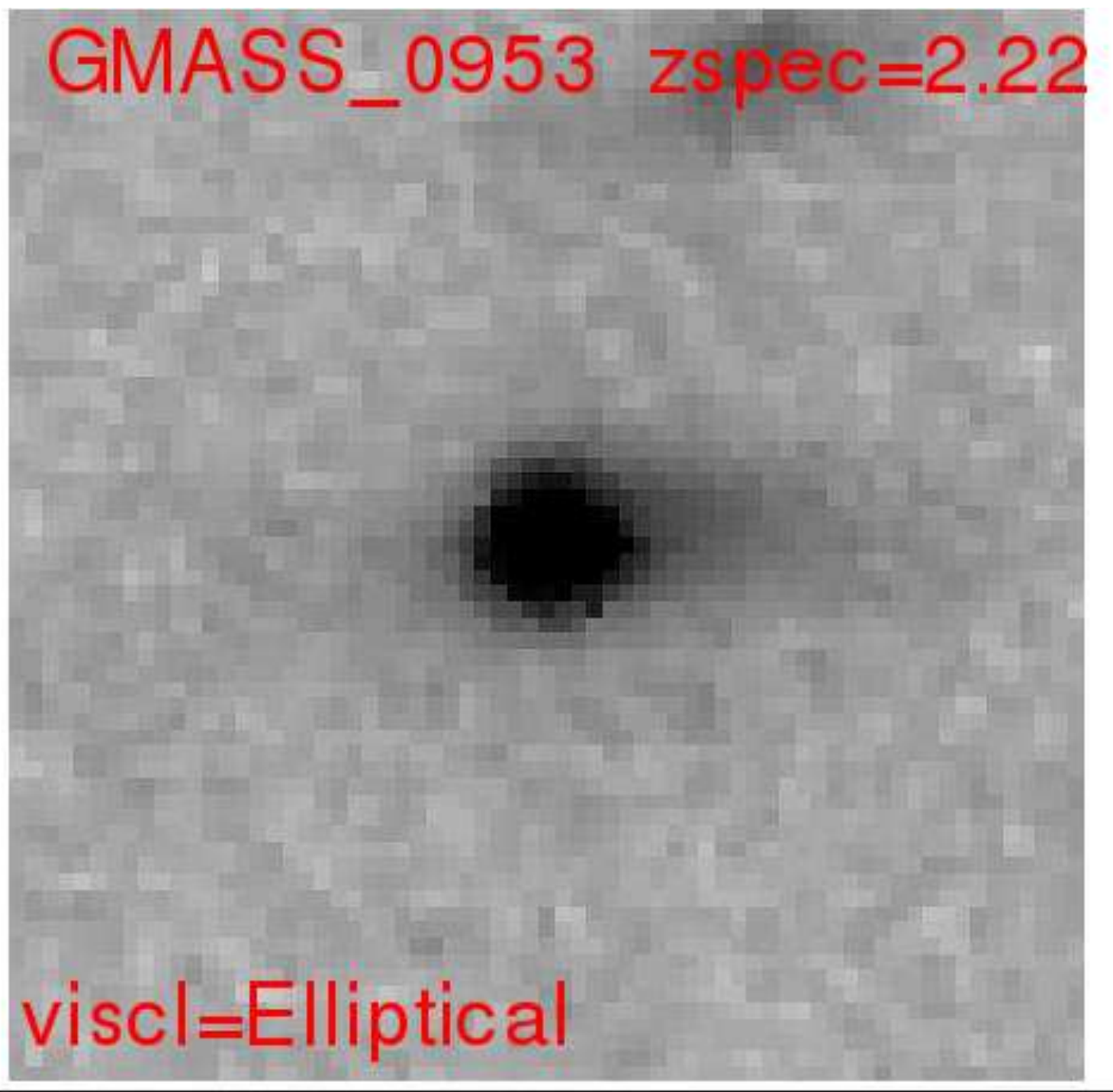}			     

\includegraphics[trim=100 40 75 390, clip=true, width=30mm]{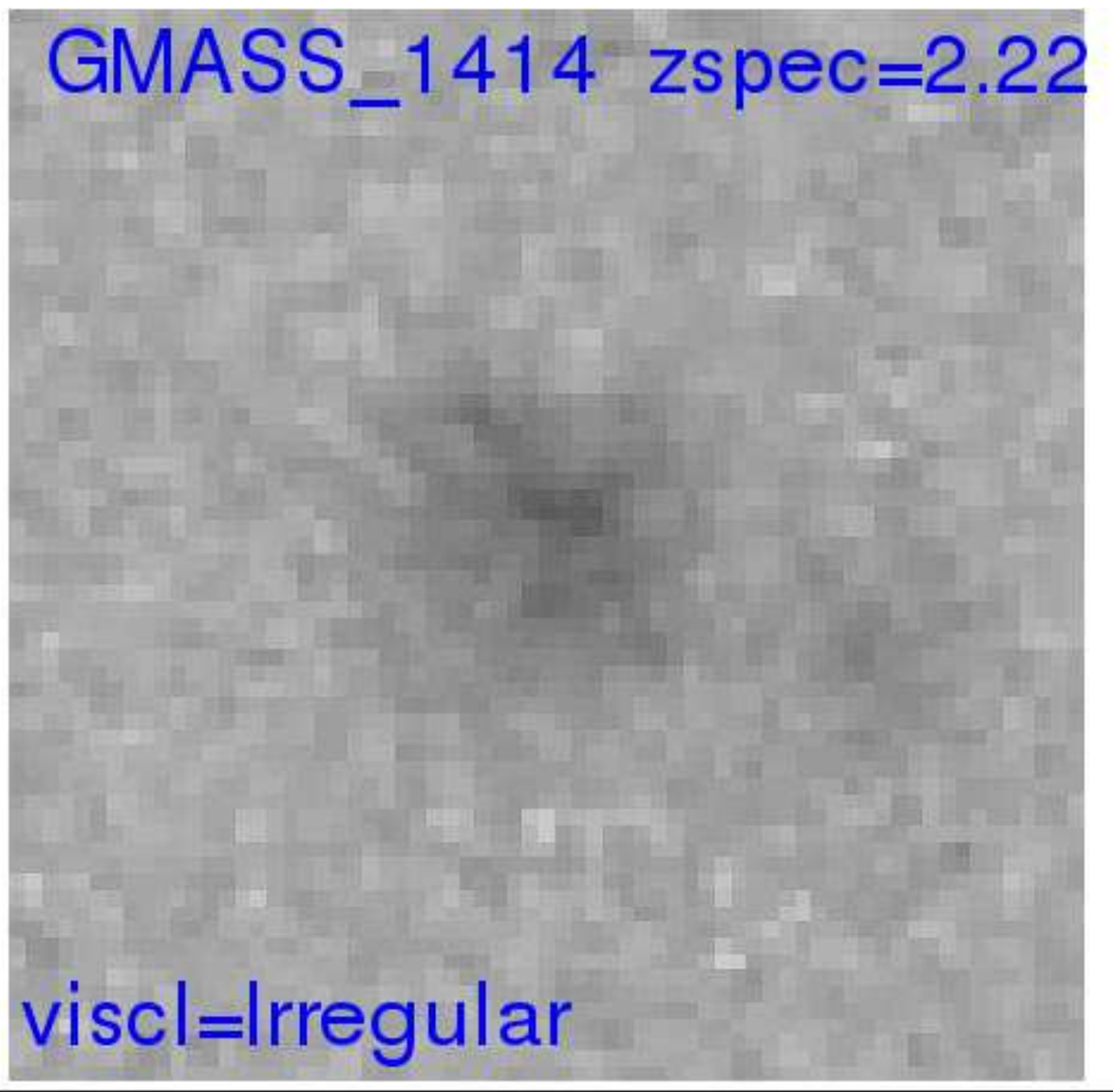}		     
\includegraphics[trim=100 40 75 390, clip=true, width=30mm]{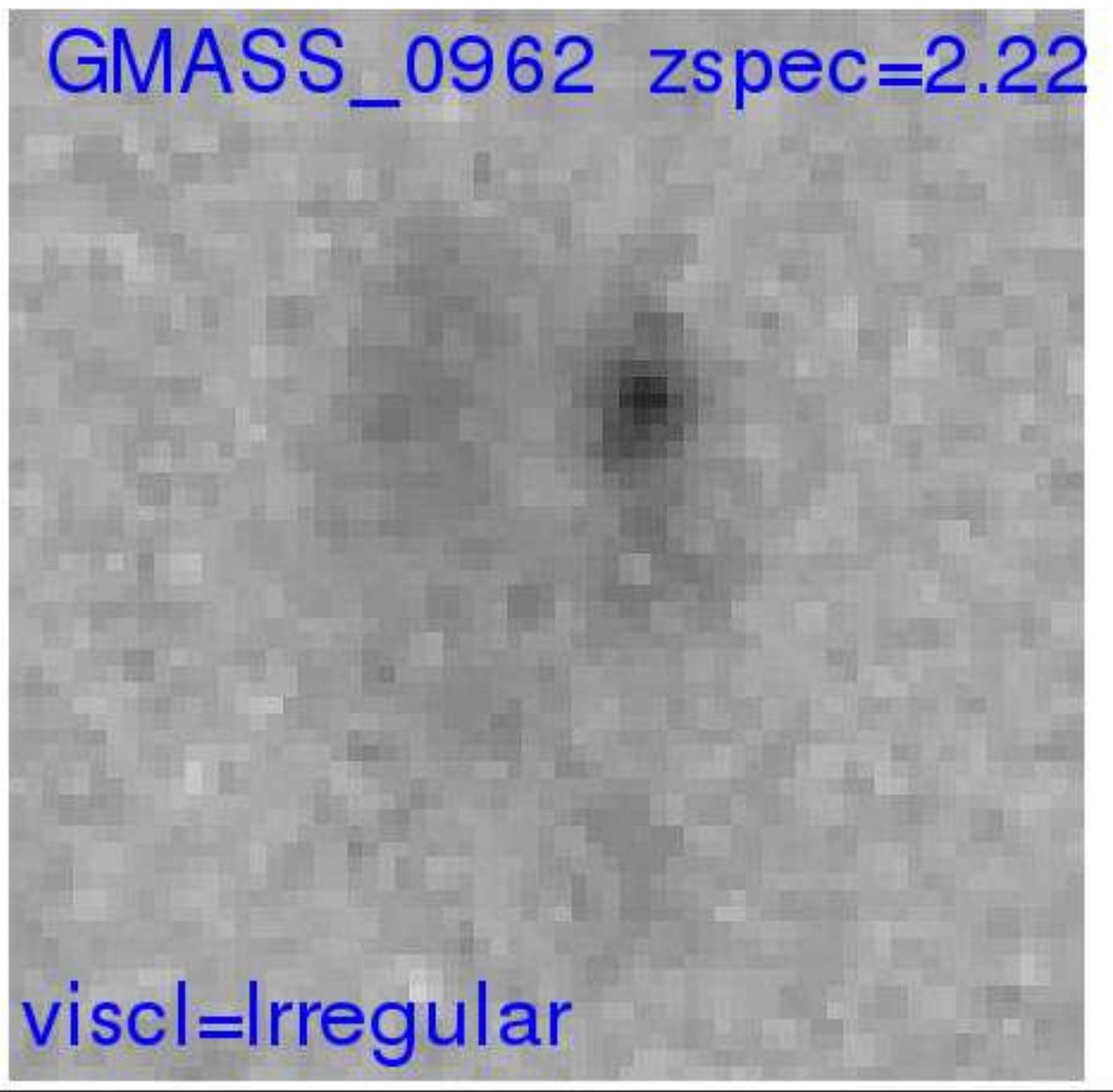}			     
\includegraphics[trim=100 40 75 390, clip=true, width=30mm]{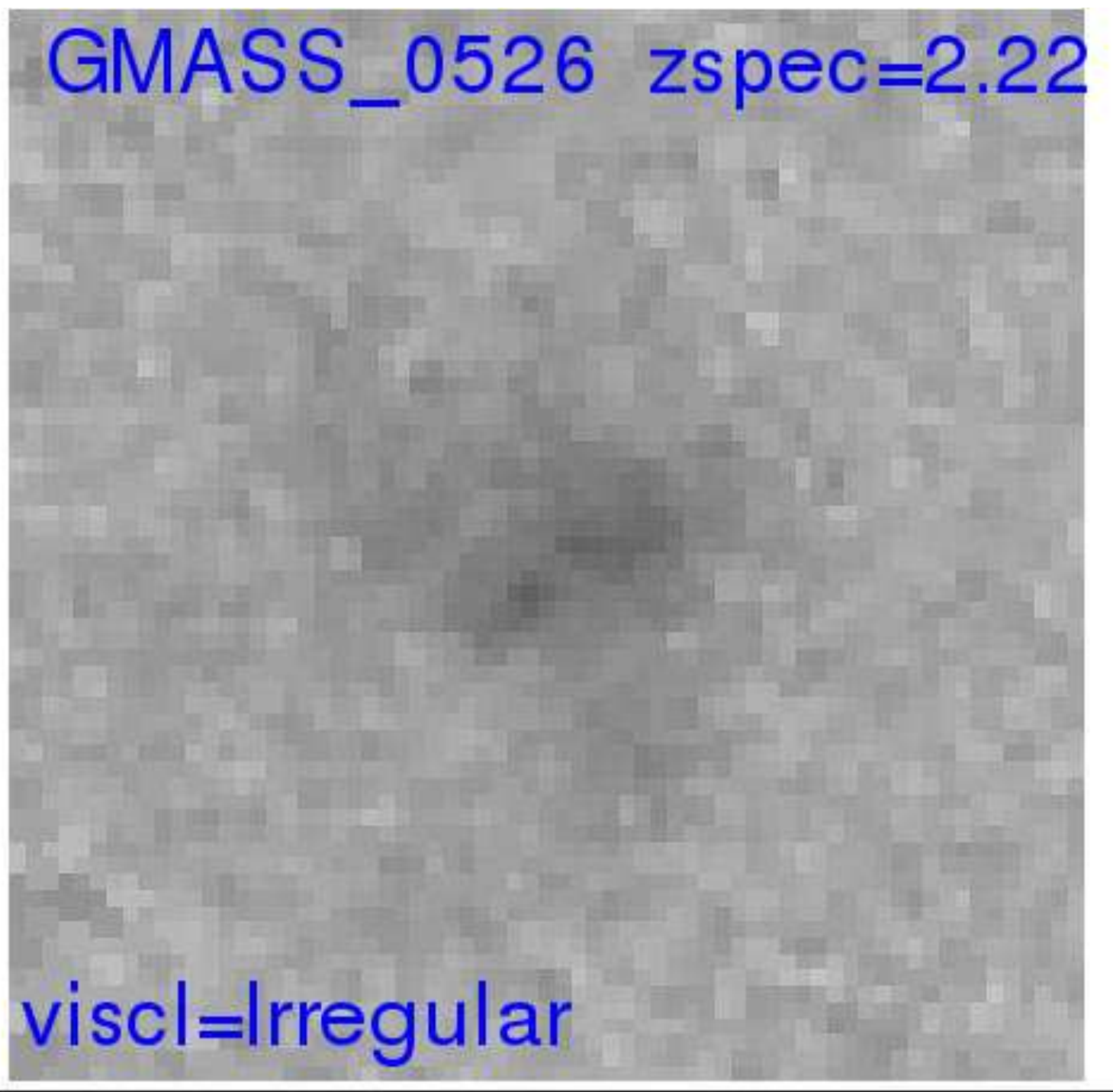}			     
\includegraphics[trim=100 40 75 390, clip=true, width=30mm]{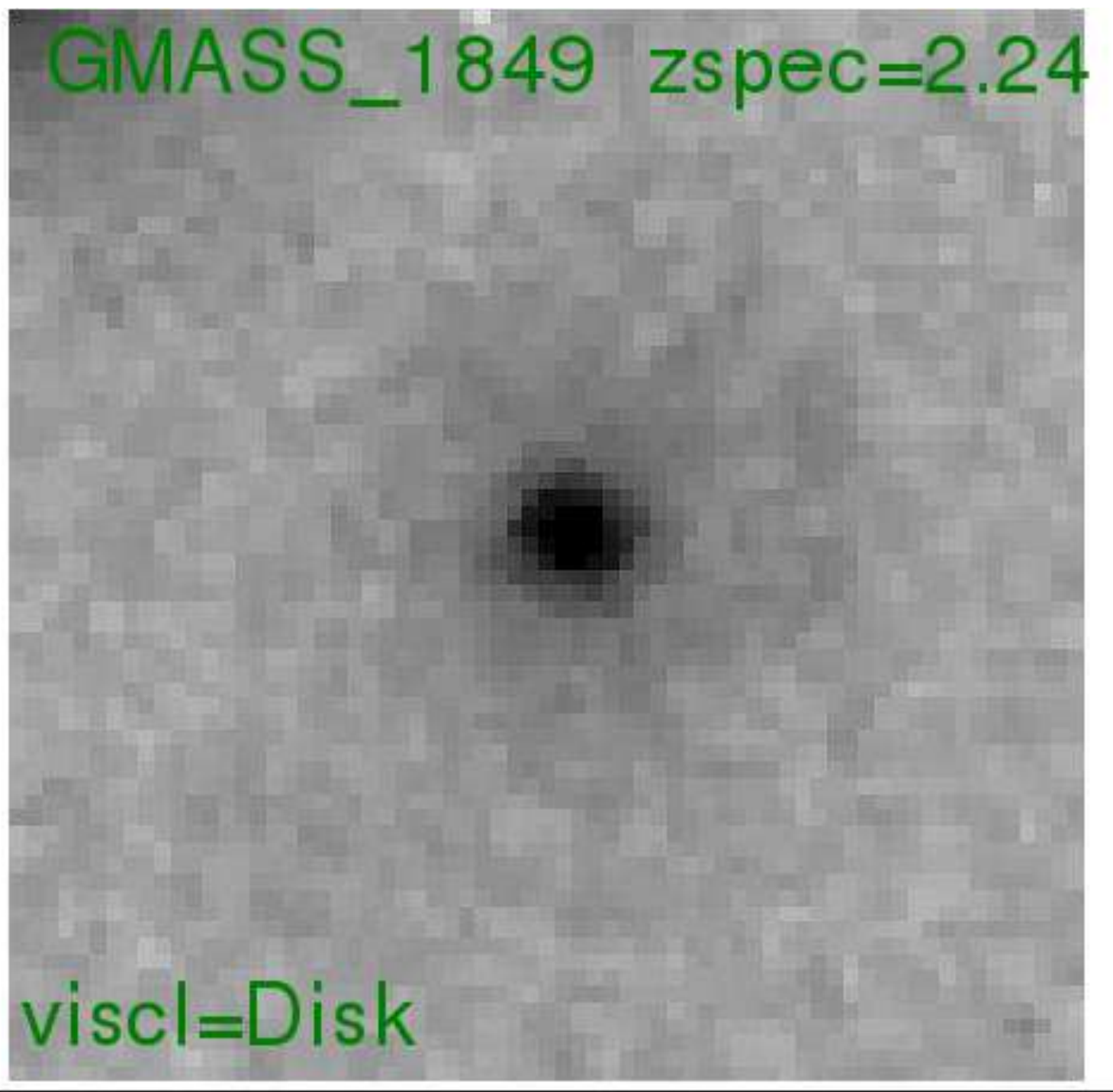}			     
\includegraphics[trim=100 40 75 390, clip=true, width=30mm]{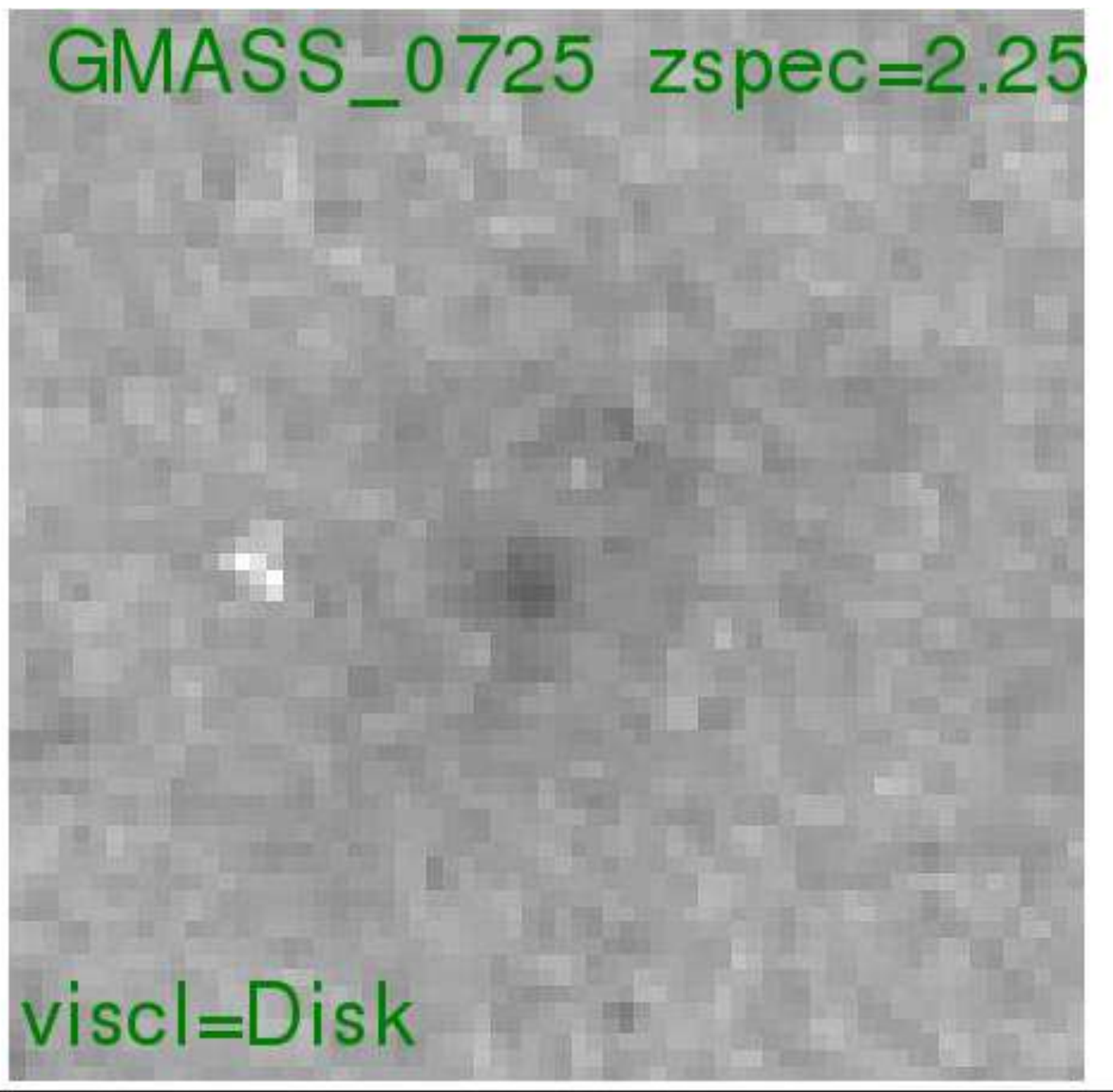}			     
\includegraphics[trim=100 40 75 390, clip=true, width=30mm]{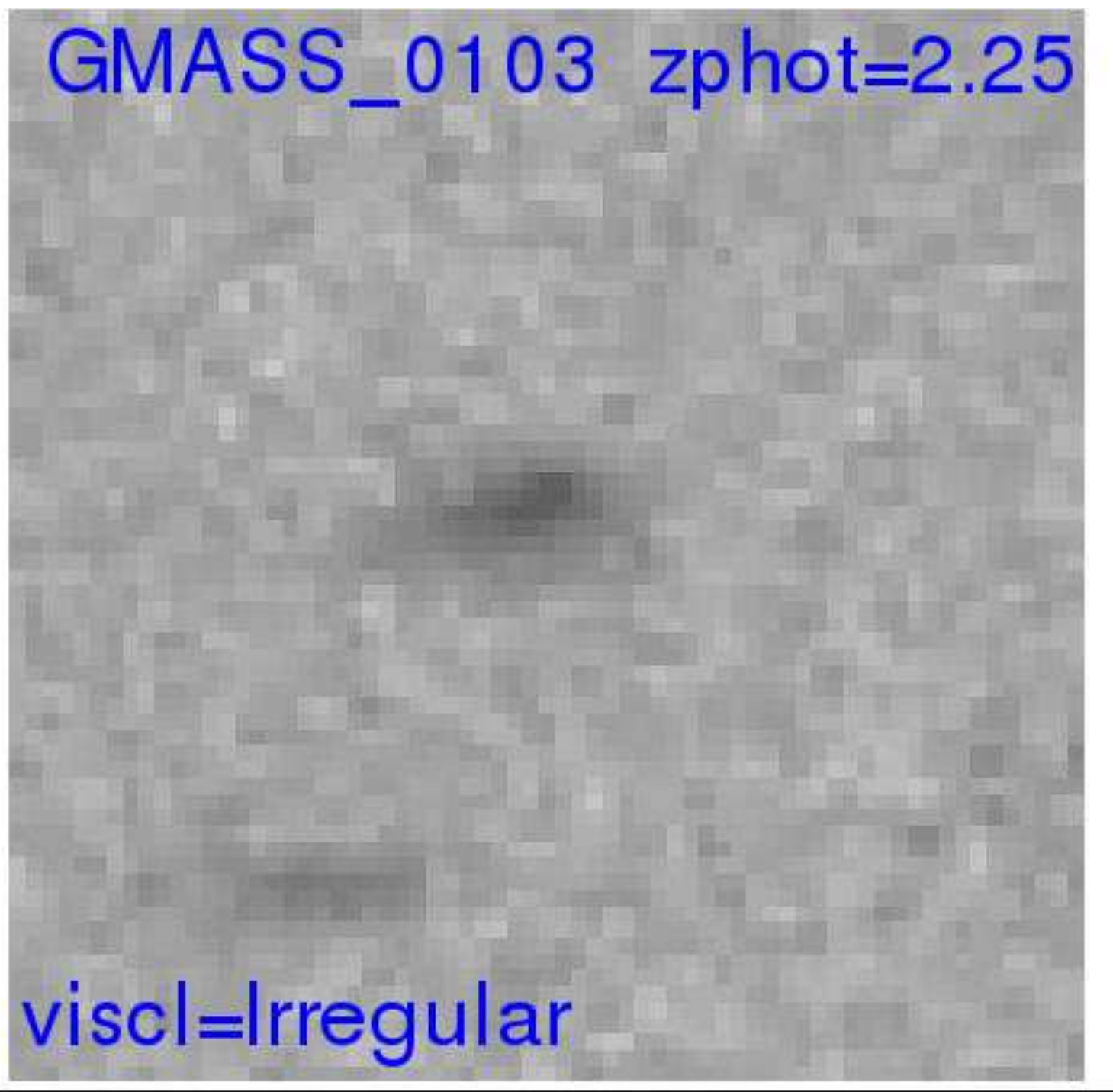}			     

\includegraphics[trim=100 40 75 390, clip=true, width=30mm]{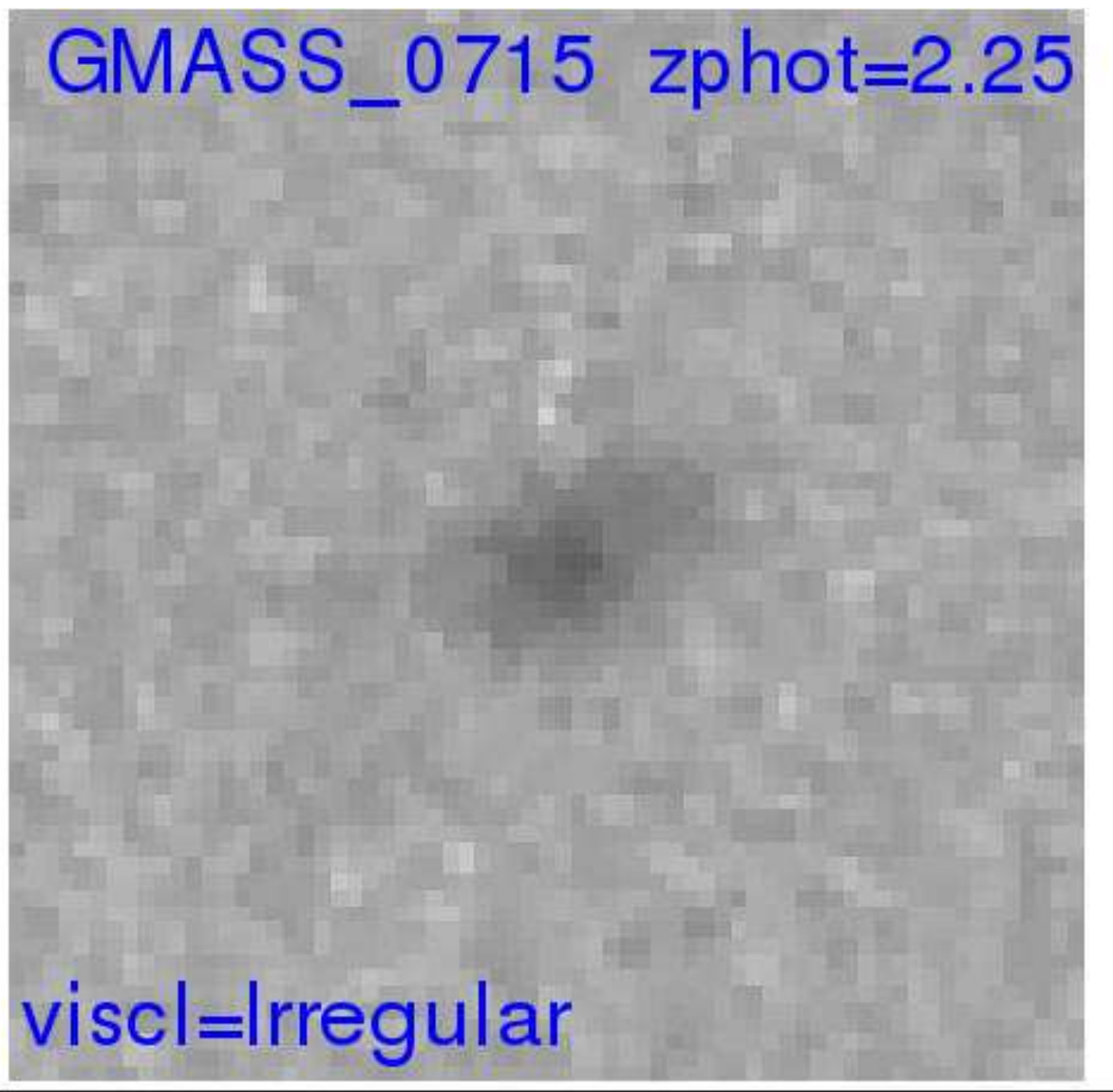}			     
\includegraphics[trim=100 40 75 390, clip=true, width=30mm]{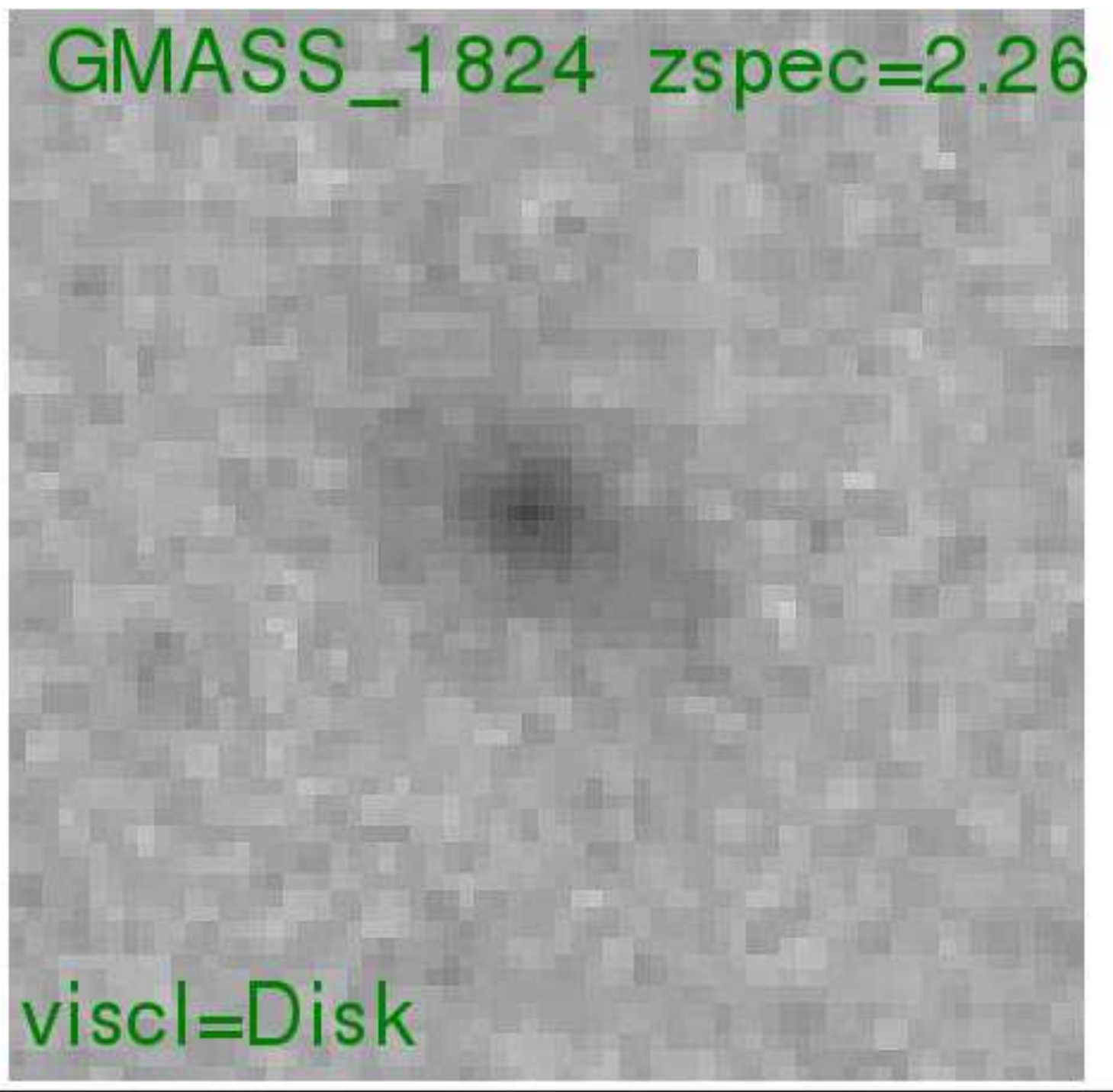}			     
\includegraphics[trim=100 40 75 390, clip=true, width=30mm]{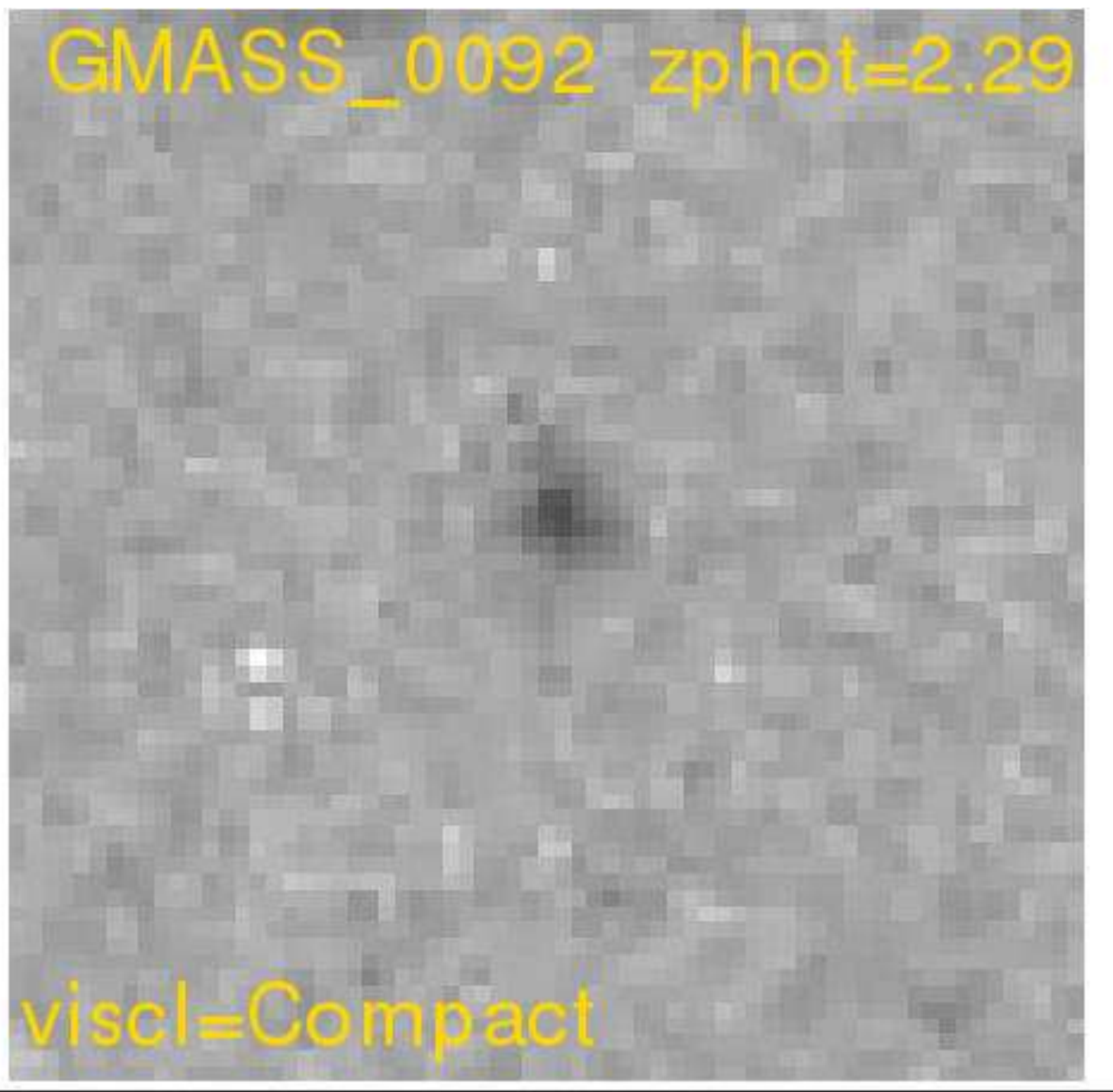}		     
\includegraphics[trim=100 40 75 390, clip=true, width=30mm]{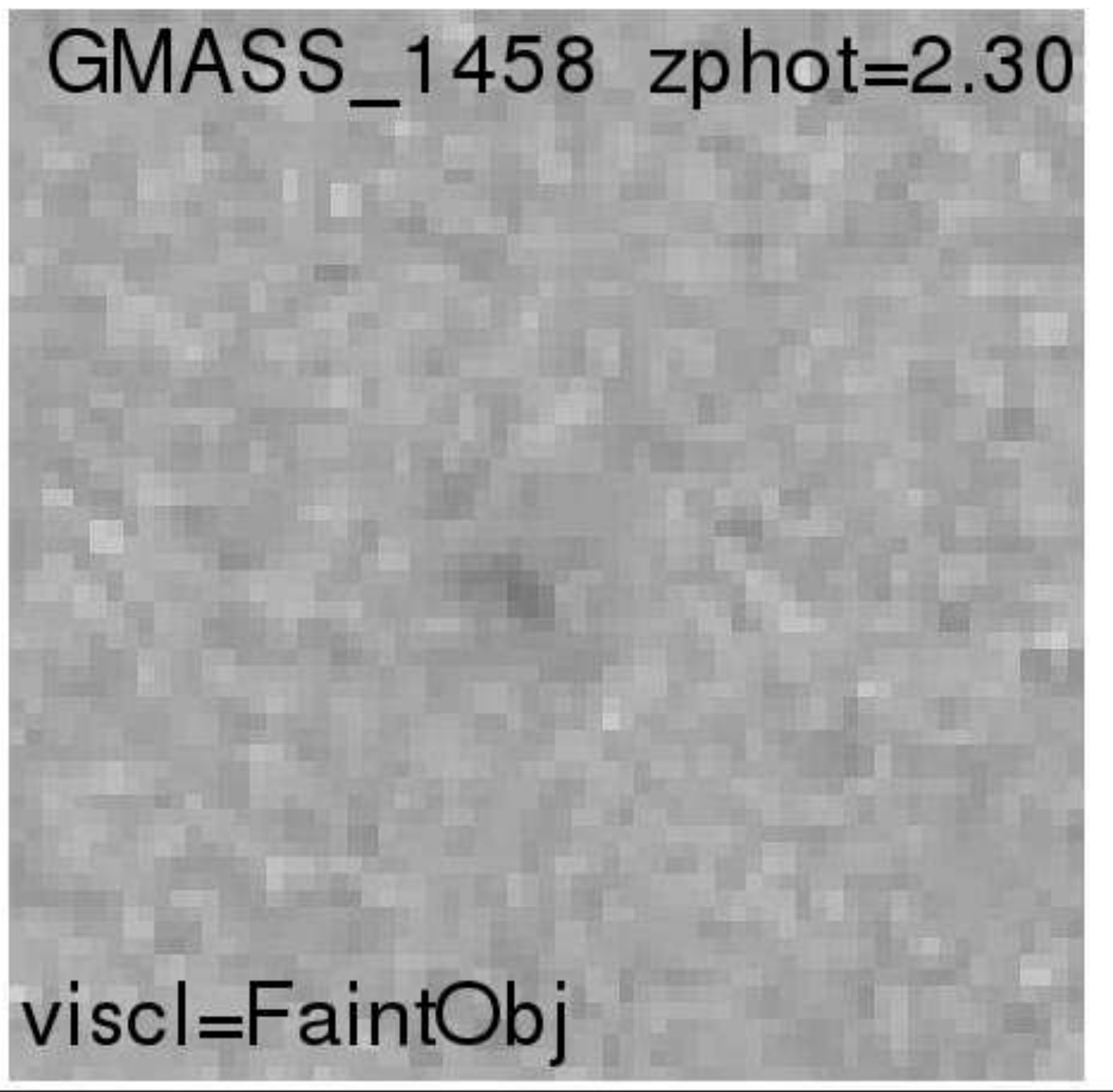}			     
\includegraphics[trim=100 40 75 390, clip=true, width=30mm]{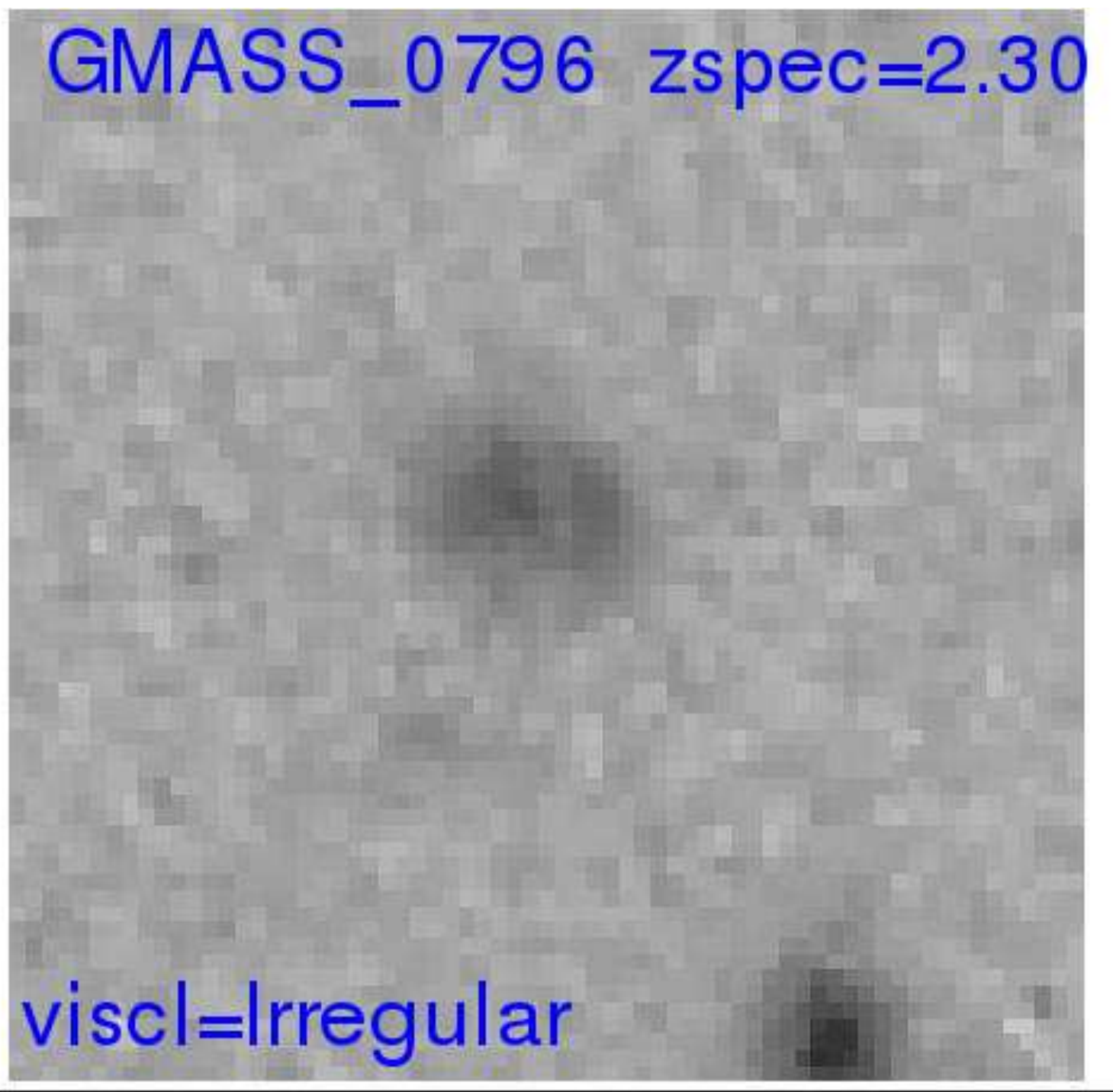}			     
\includegraphics[trim=100 40 75 390, clip=true, width=30mm]{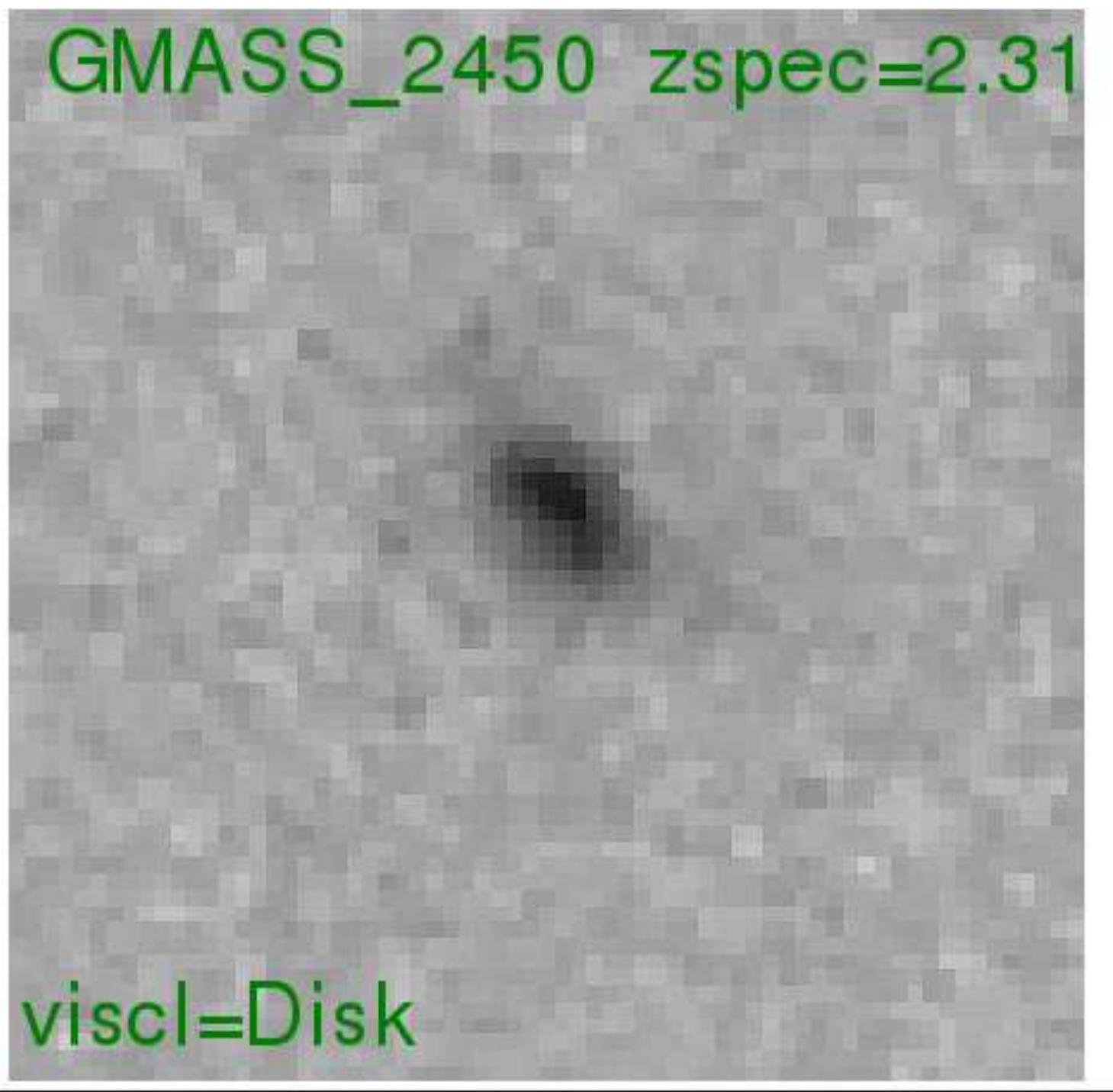}			     

\includegraphics[trim=100 40 75 390, clip=true, width=30mm]{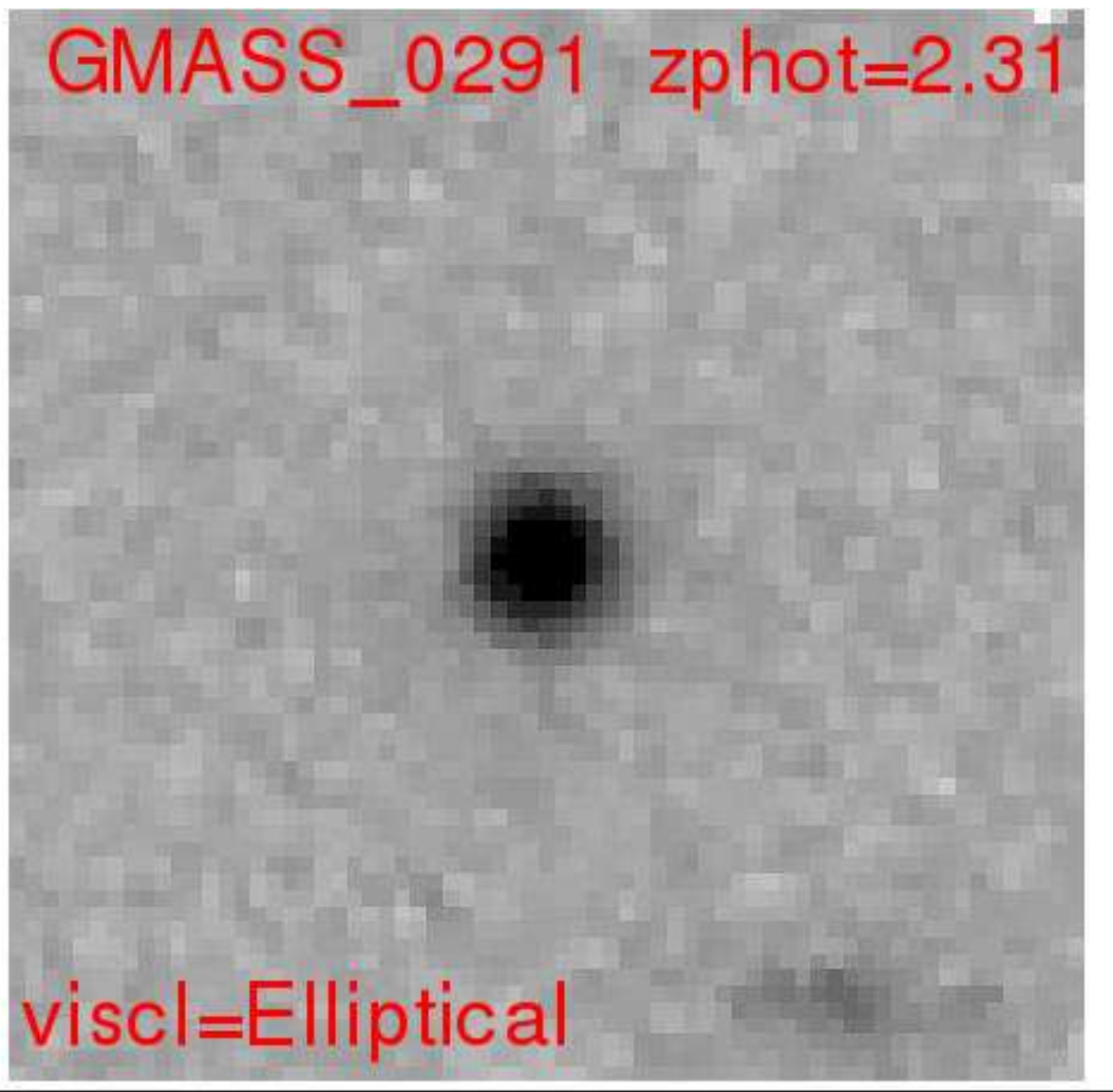}			     
\includegraphics[trim=100 40 75 390, clip=true, width=30mm]{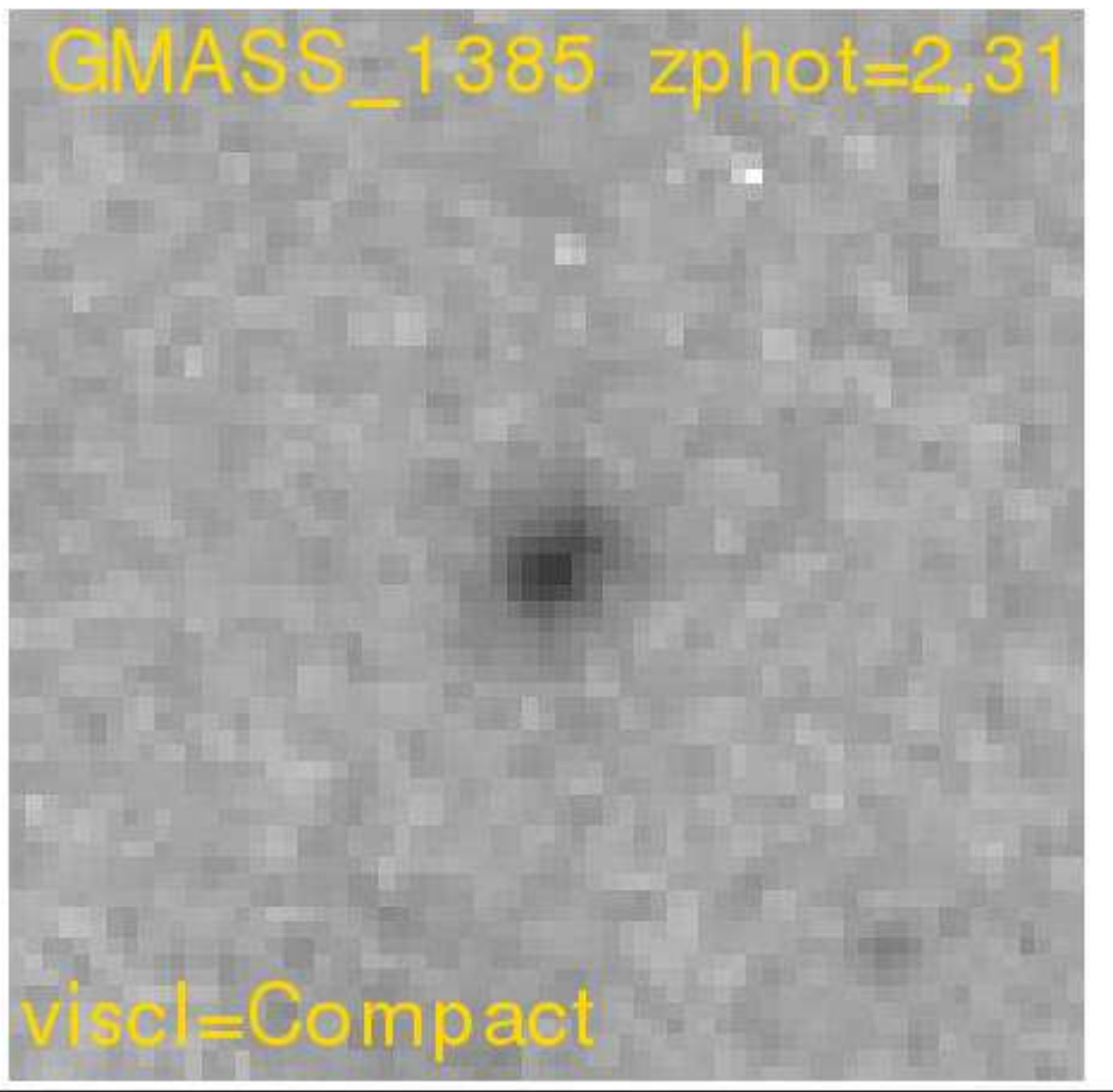}			     
\includegraphics[trim=100 40 75 390, clip=true, width=30mm]{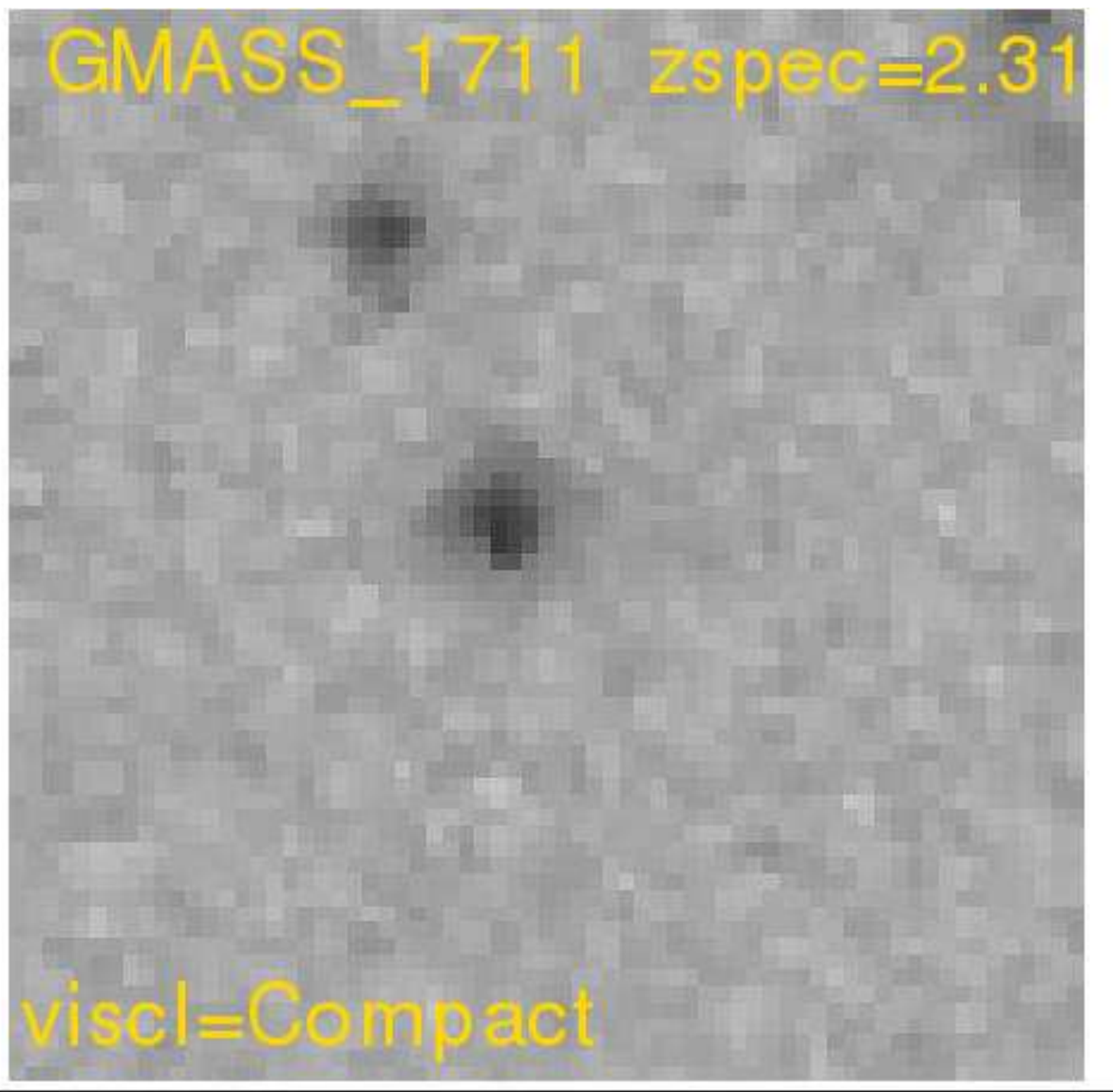}			     
\includegraphics[trim=100 40 75 390, clip=true, width=30mm]{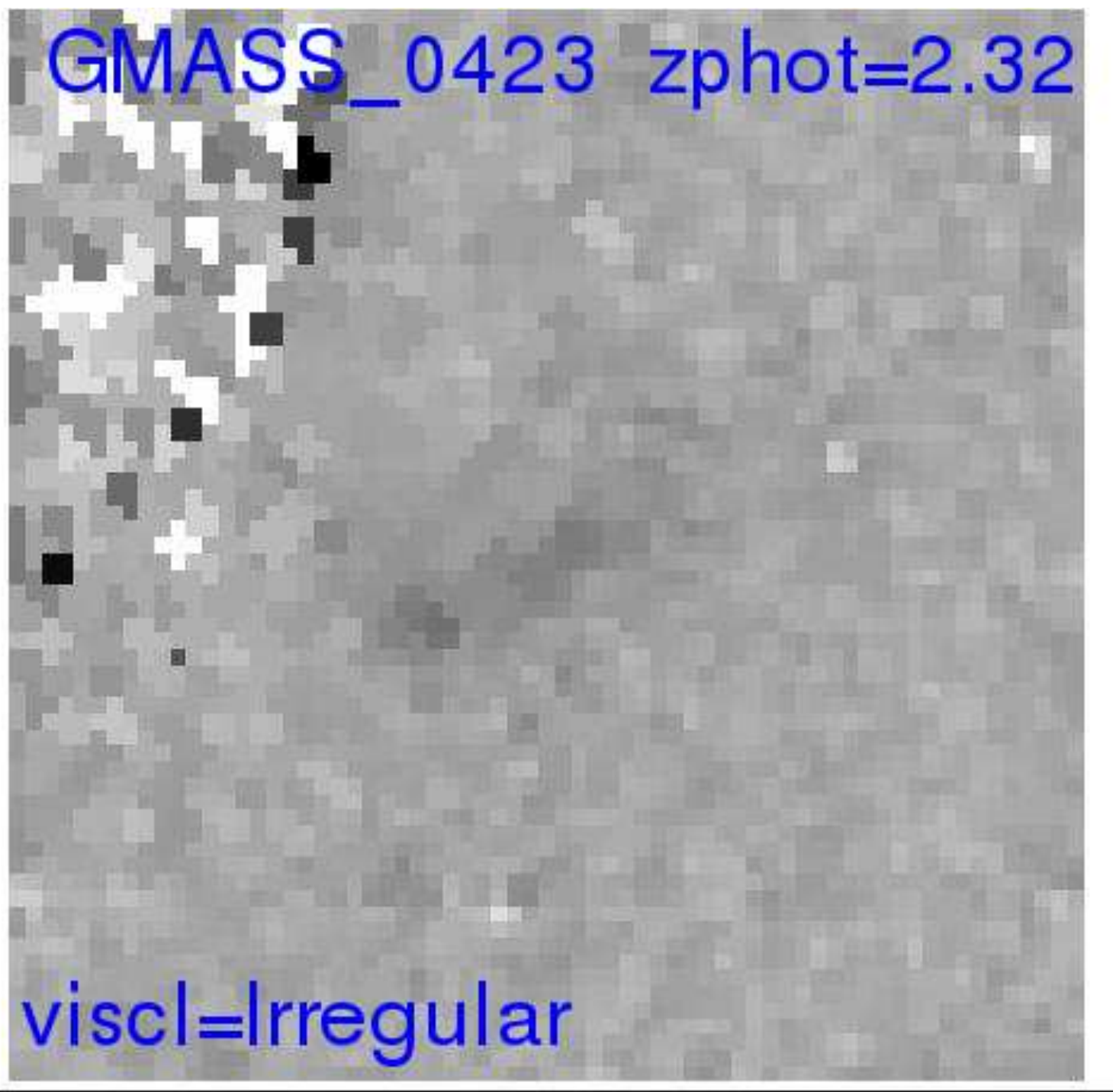}			     
\includegraphics[trim=100 40 75 390, clip=true, width=30mm]{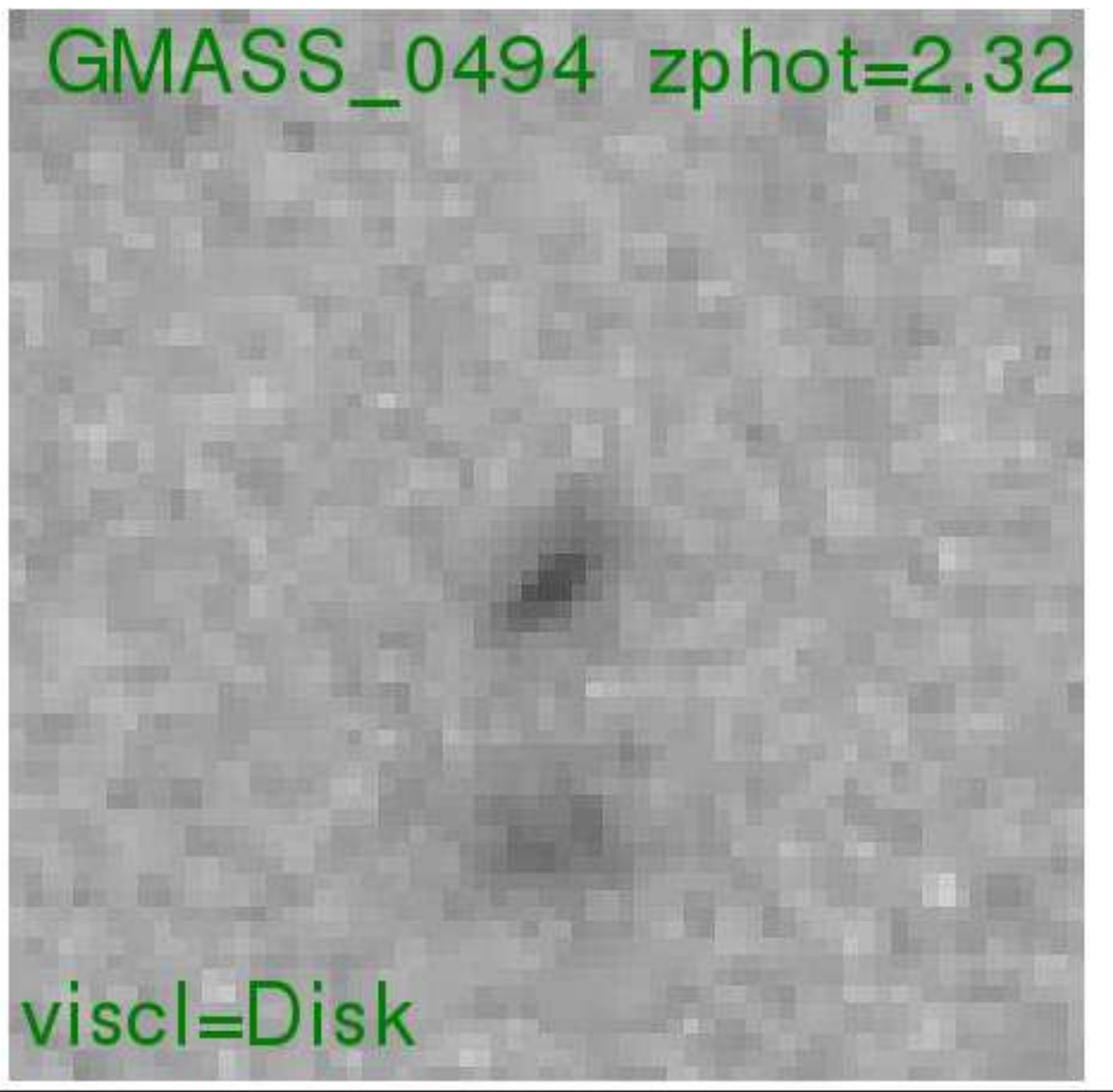}			     
\includegraphics[trim=100 40 75 390, clip=true, width=30mm]{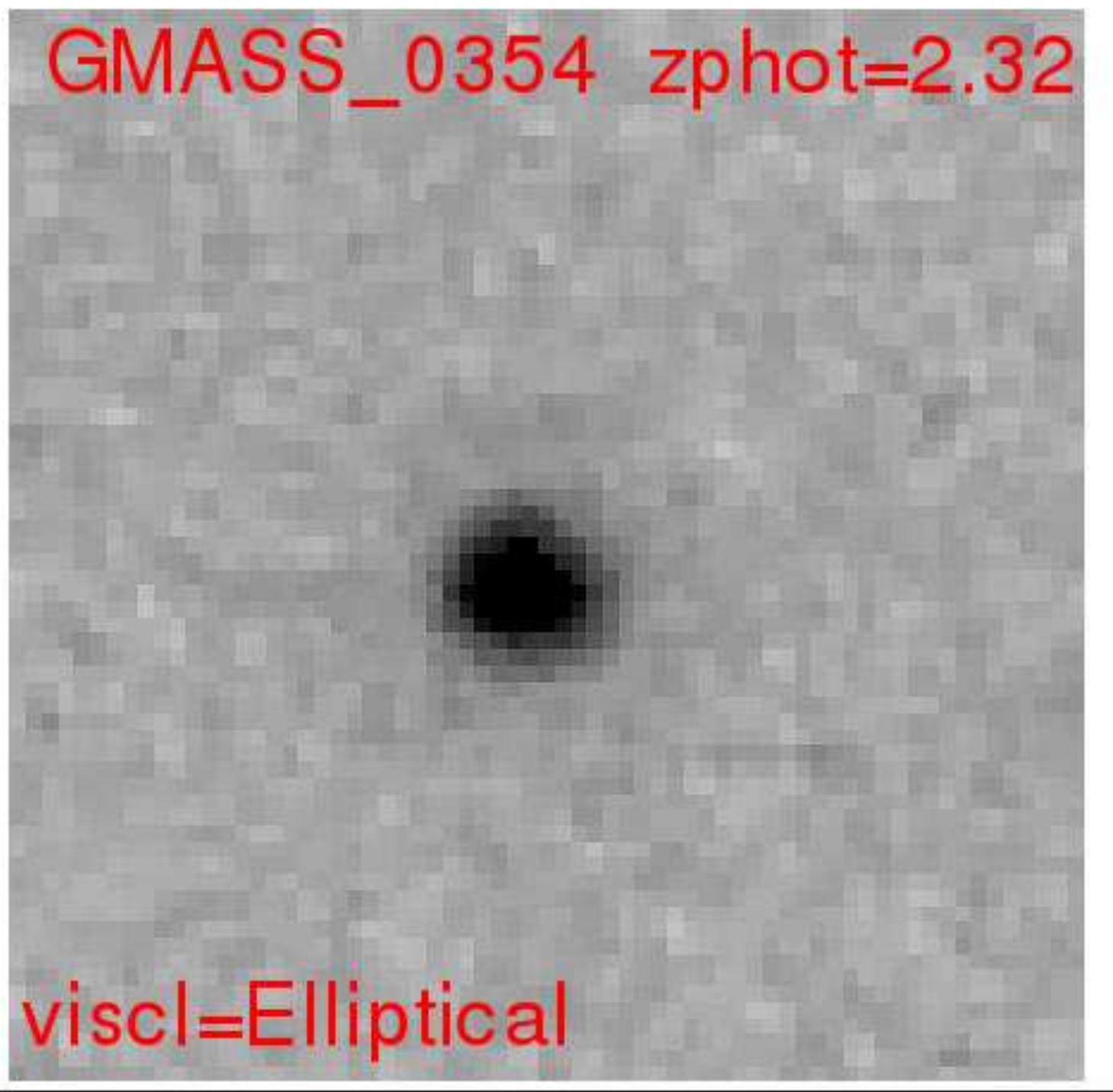}			     

\includegraphics[trim=100 40 75 390, clip=true, width=30mm]{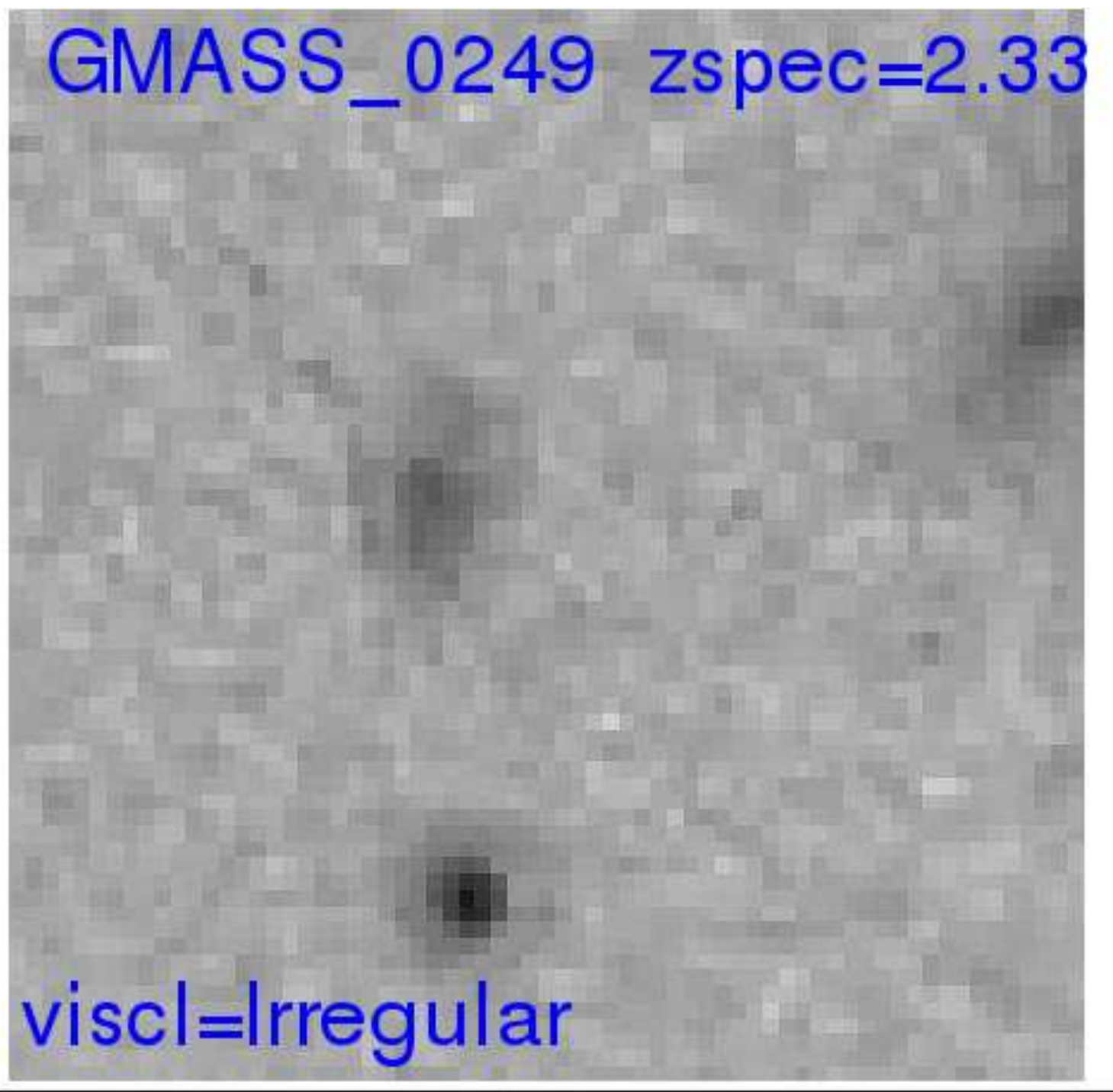}			     
\includegraphics[trim=100 40 75 390, clip=true, width=30mm]{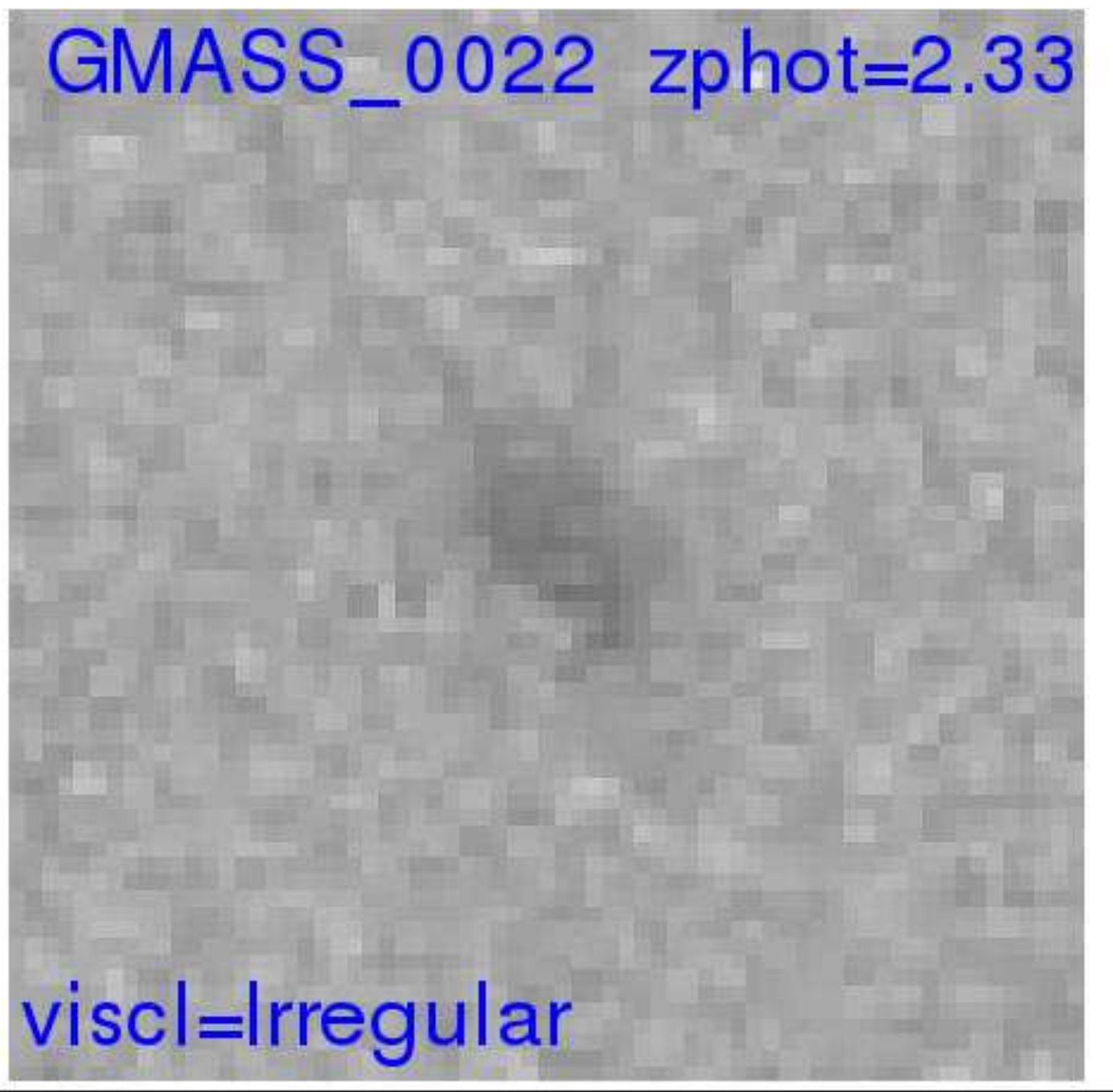}			     
\includegraphics[trim=100 40 75 390, clip=true, width=30mm]{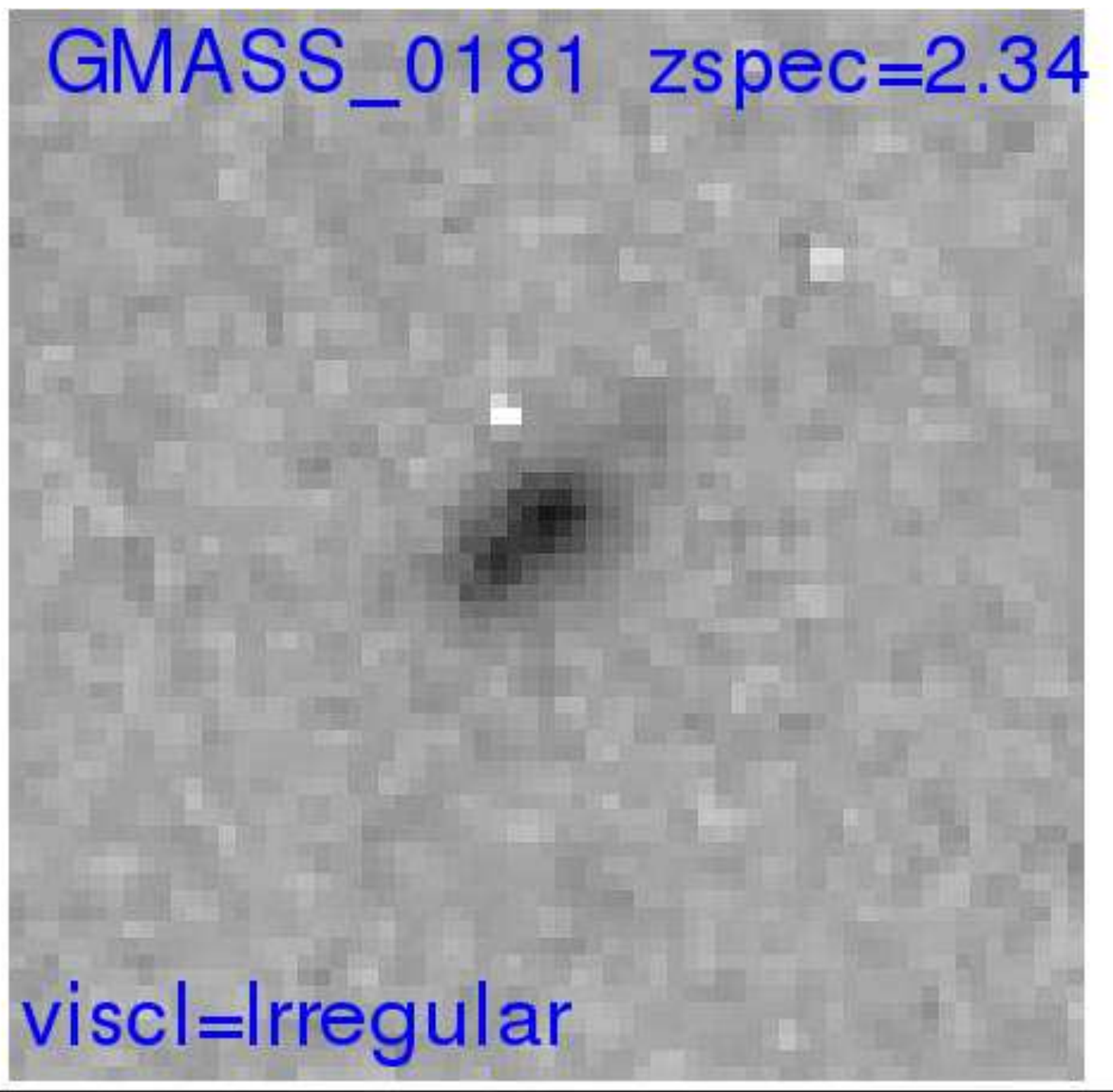}			     
\includegraphics[trim=100 40 75 390, clip=true, width=30mm]{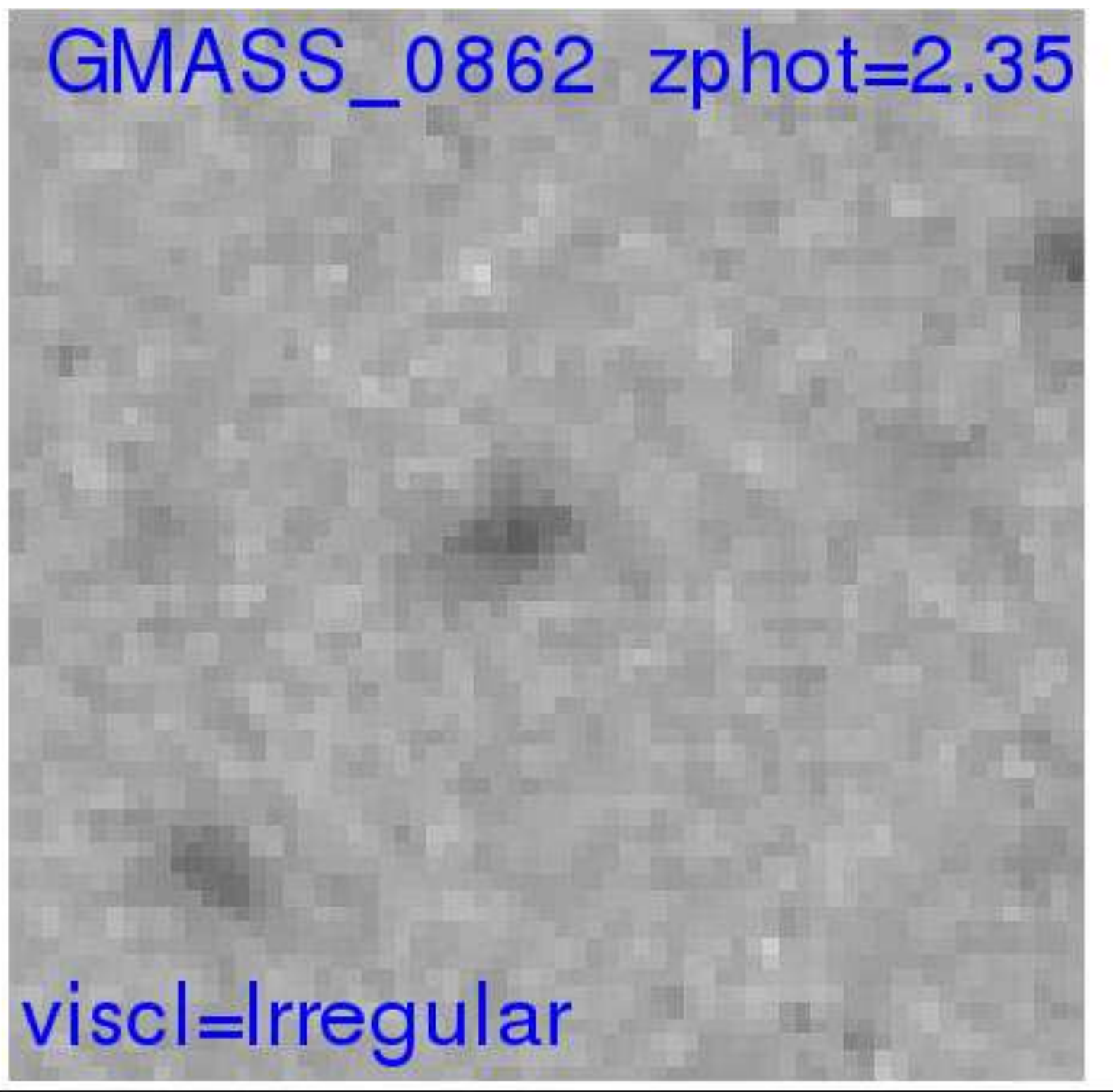}			     
\includegraphics[trim=100 40 75 390, clip=true, width=30mm]{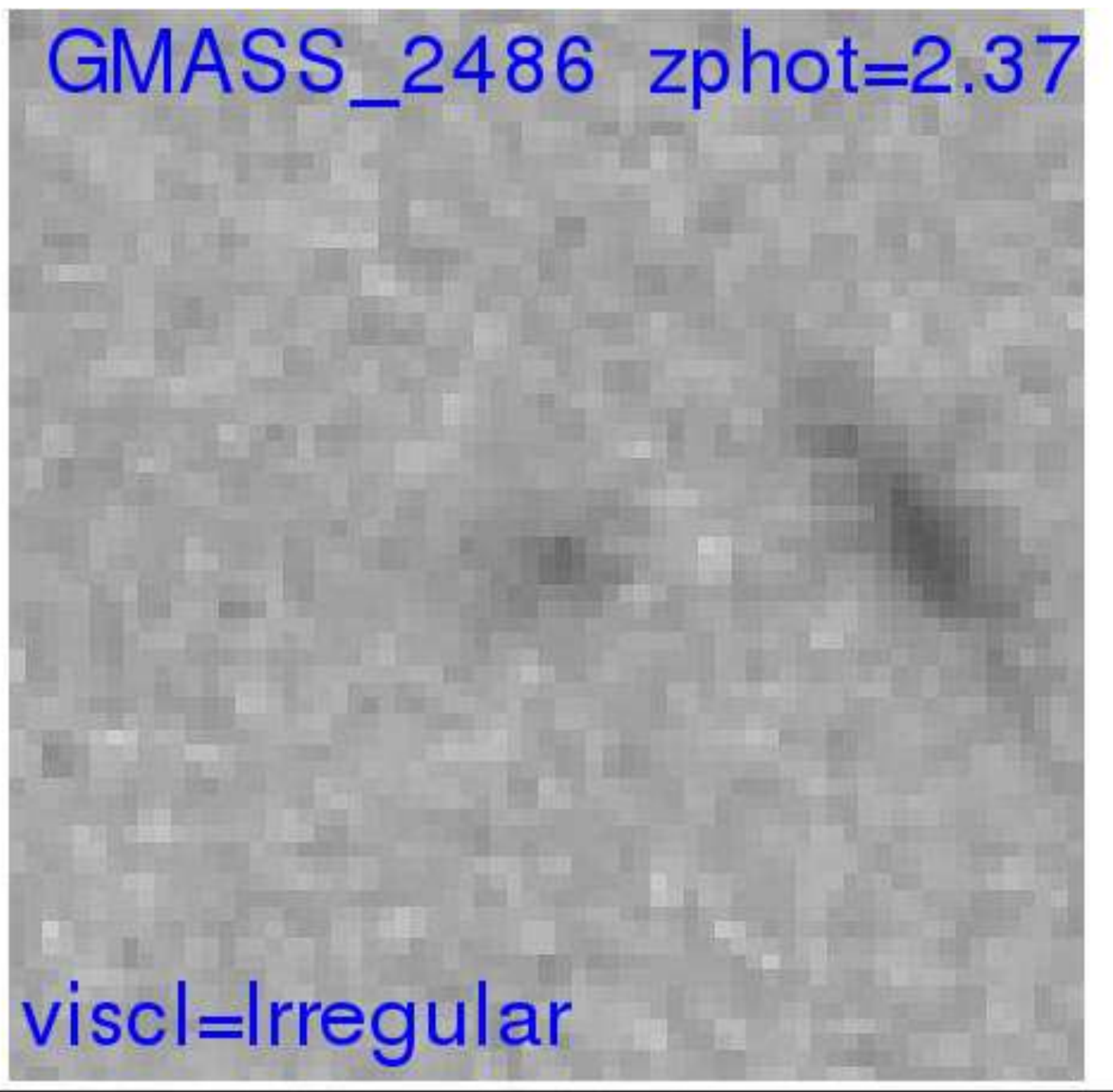}			     
\includegraphics[trim=100 40 75 390, clip=true, width=30mm]{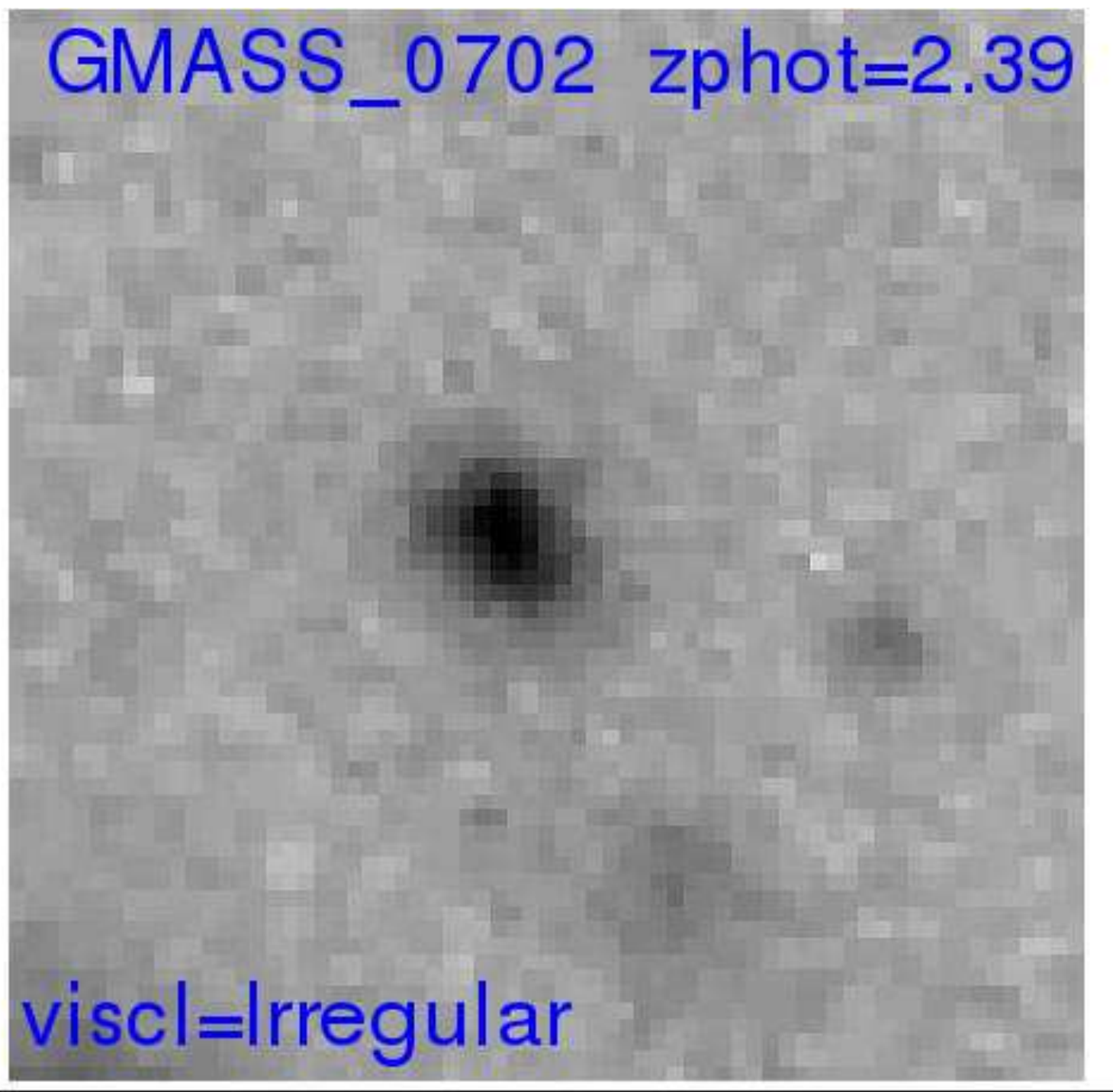}
\end{figure*}
\begin{figure*}
\centering  
\includegraphics[trim=100 40 75 390, clip=true, width=30mm]{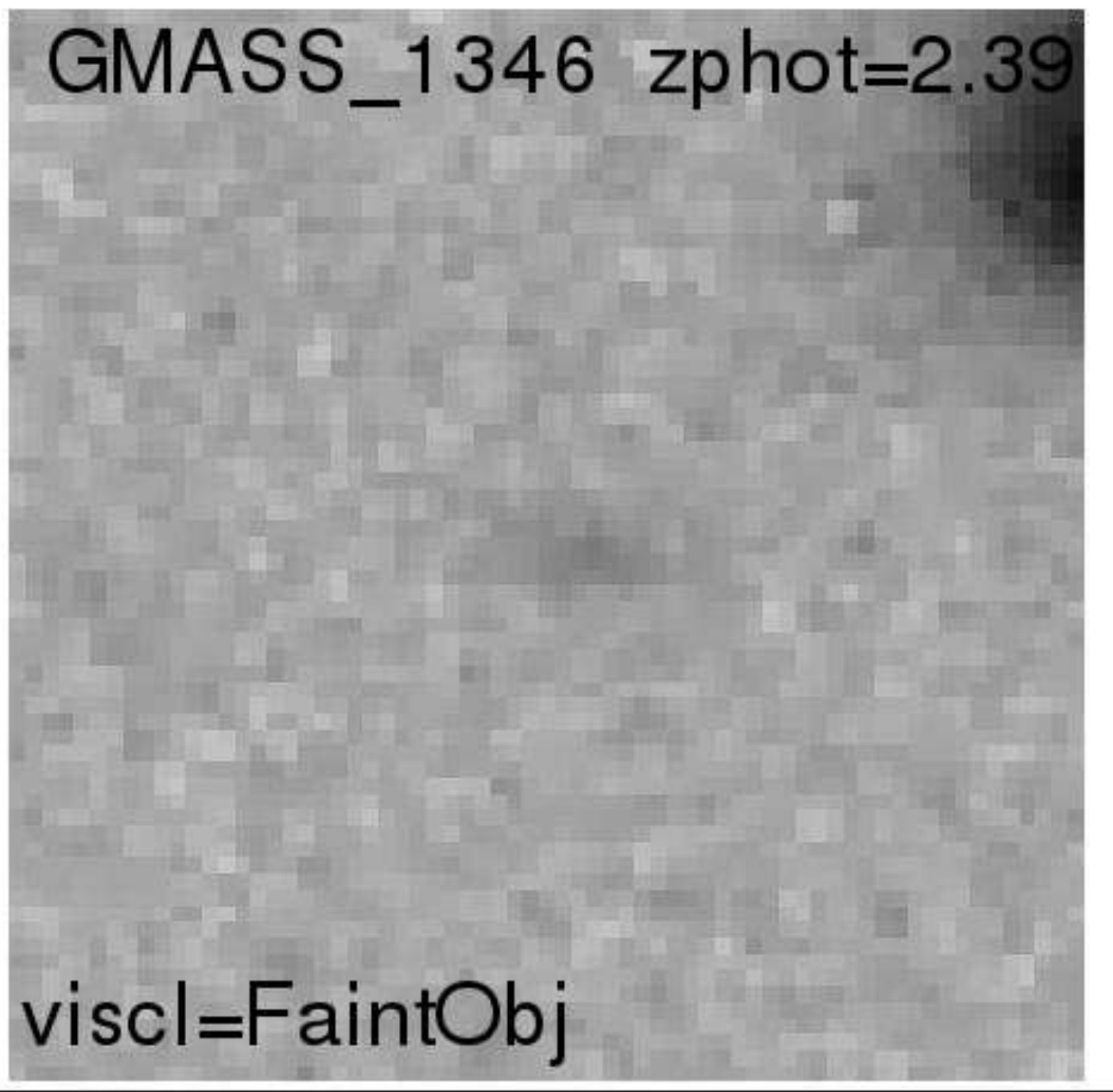}			     
\includegraphics[trim=100 40 75 390, clip=true, width=30mm]{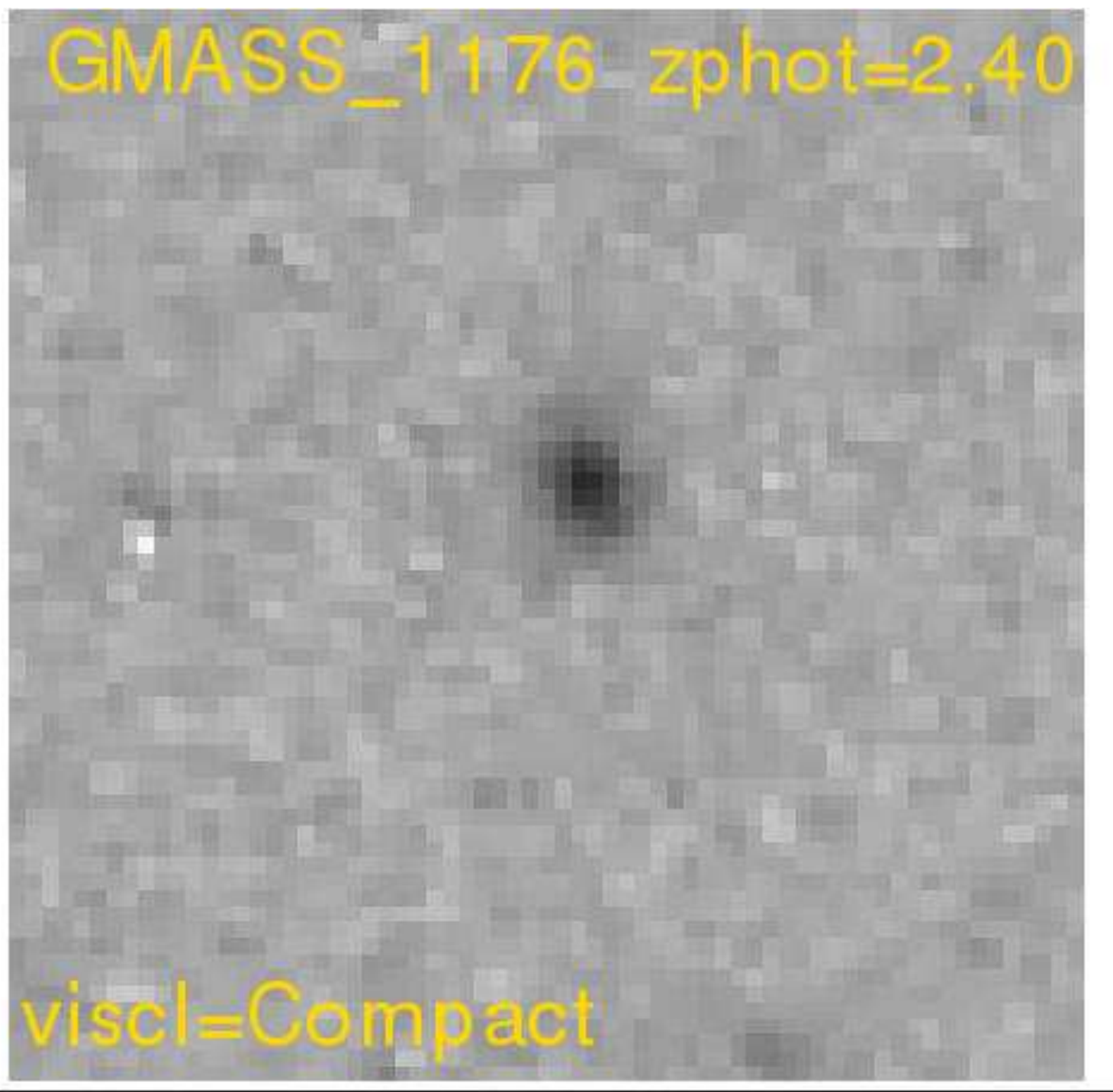}			     
\includegraphics[trim=100 40 75 390, clip=true, width=30mm]{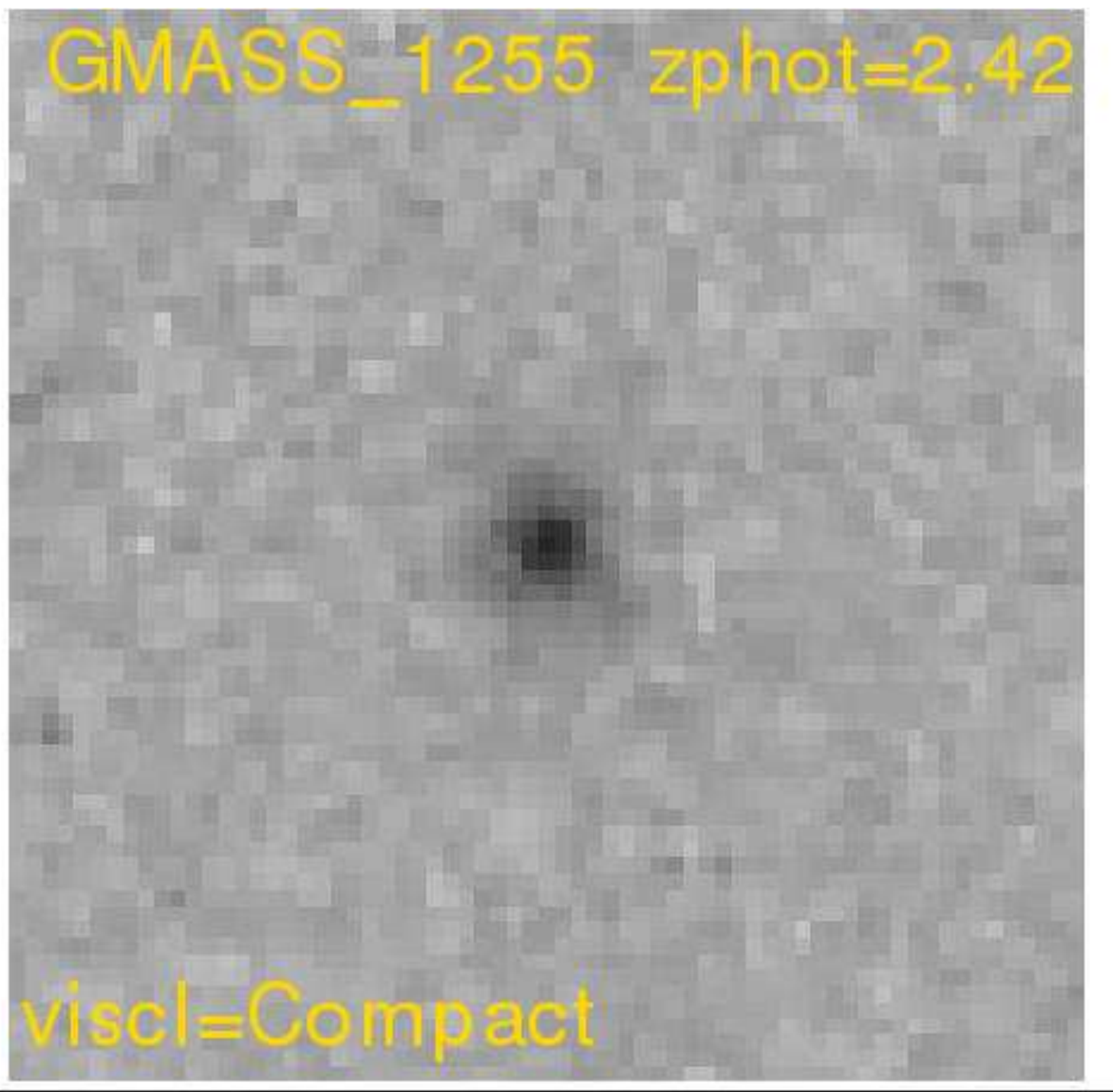}			     
\includegraphics[trim=100 40 75 390, clip=true, width=30mm]{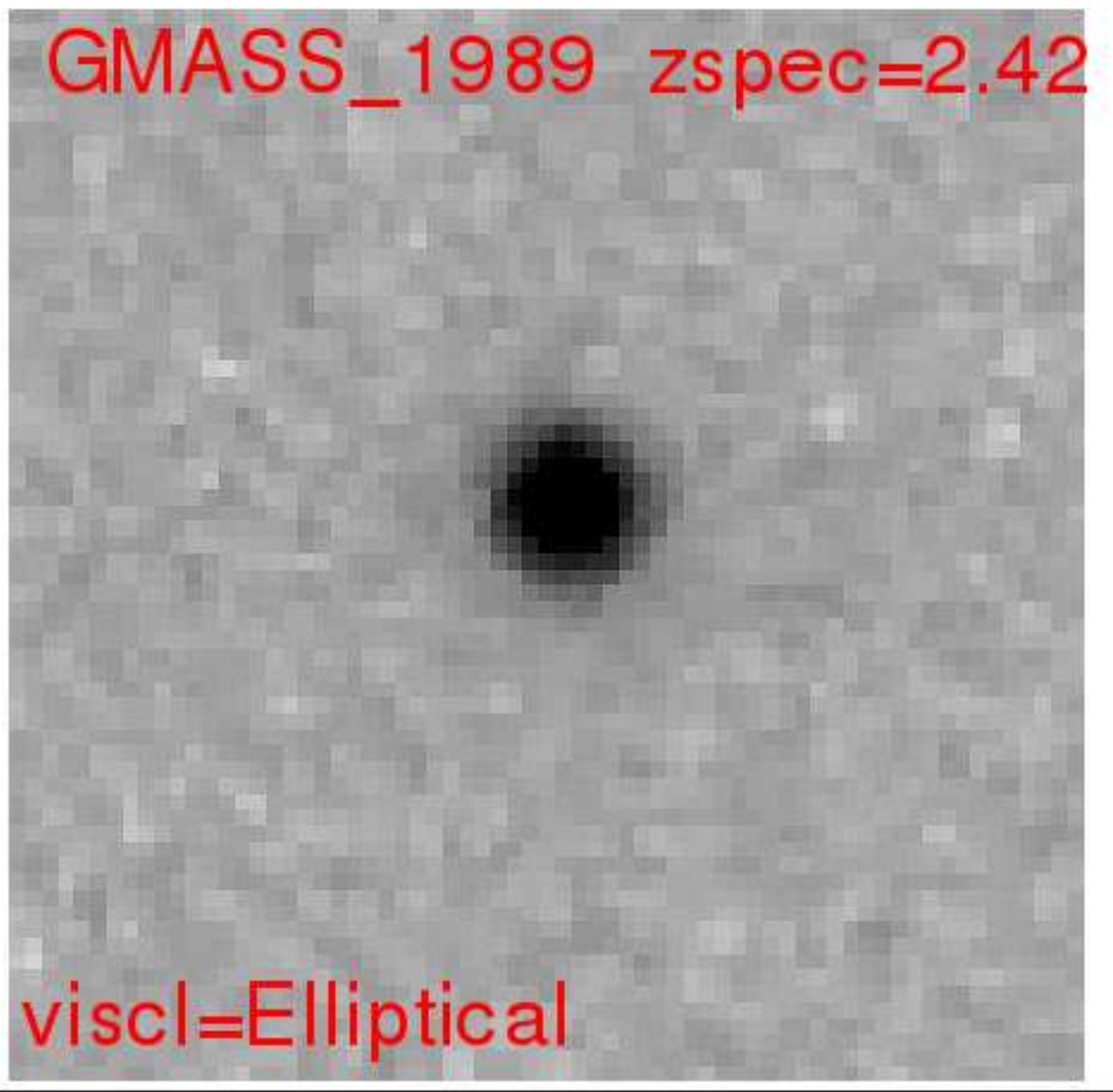}			     
\includegraphics[trim=100 40 75 390, clip=true, width=30mm]{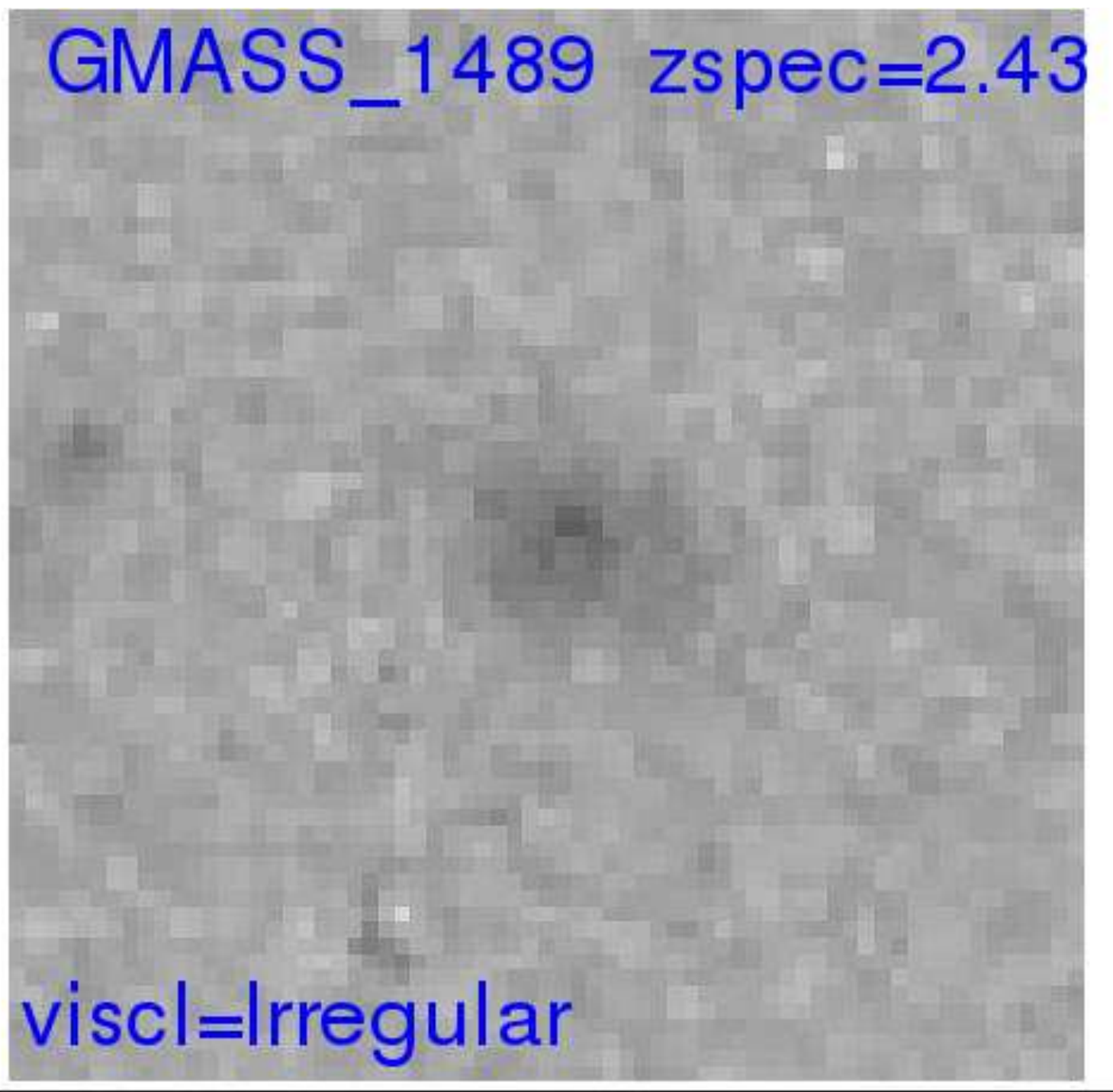}			     
\includegraphics[trim=100 40 75 390, clip=true, width=30mm]{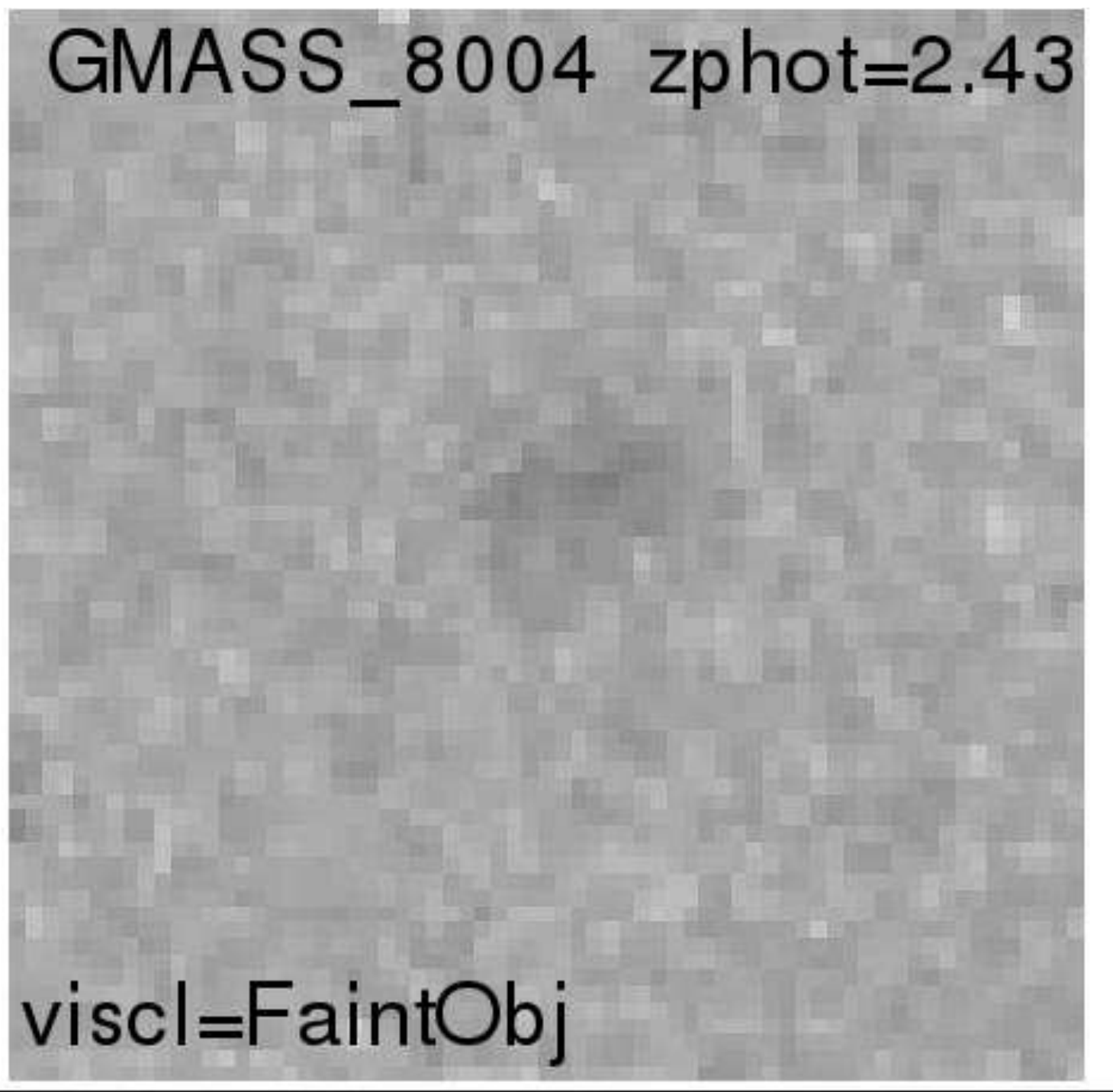}			     

\includegraphics[trim=100 40 75 390, clip=true, width=30mm]{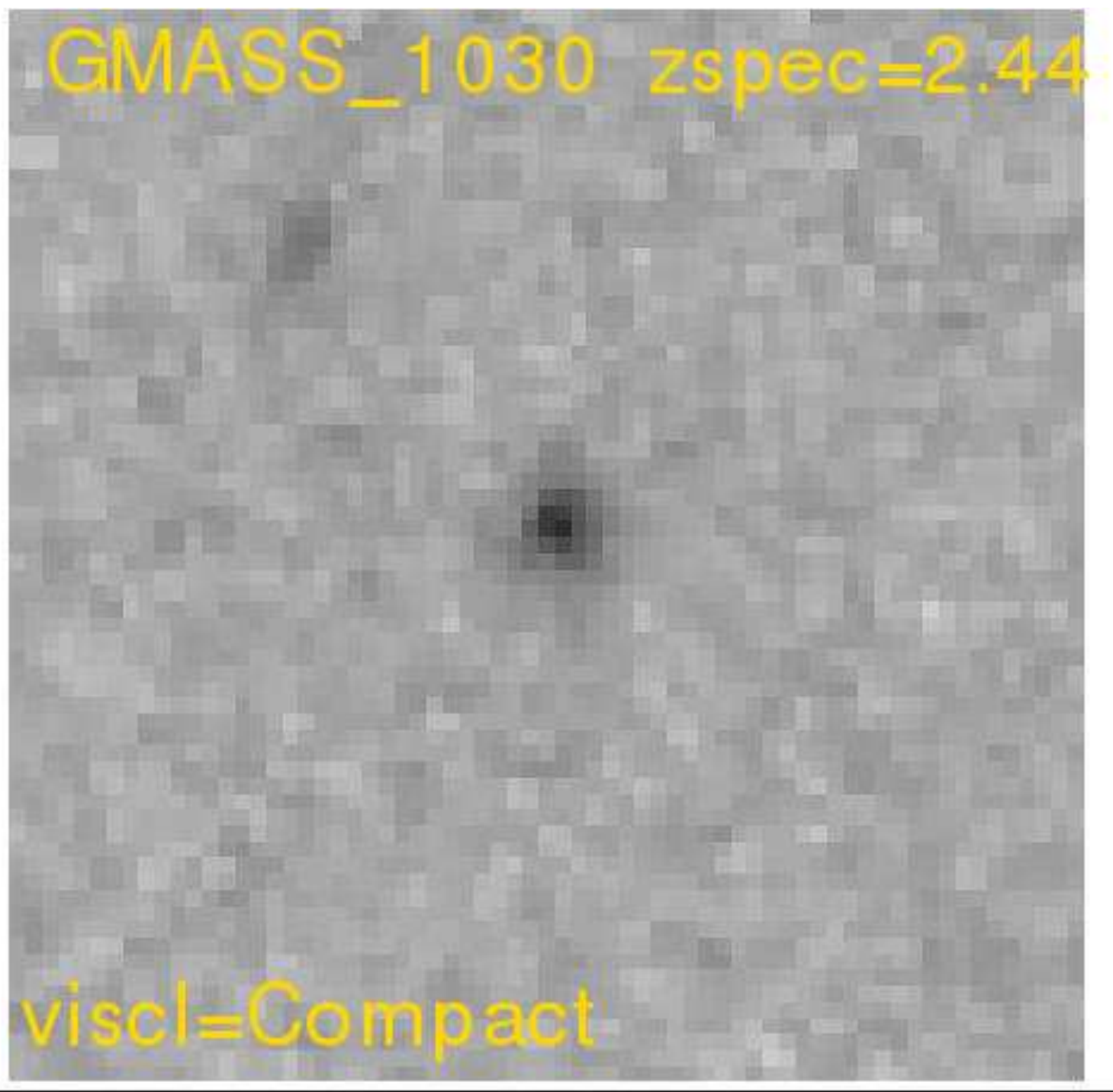}			     
\includegraphics[trim=100 40 75 390, clip=true, width=30mm]{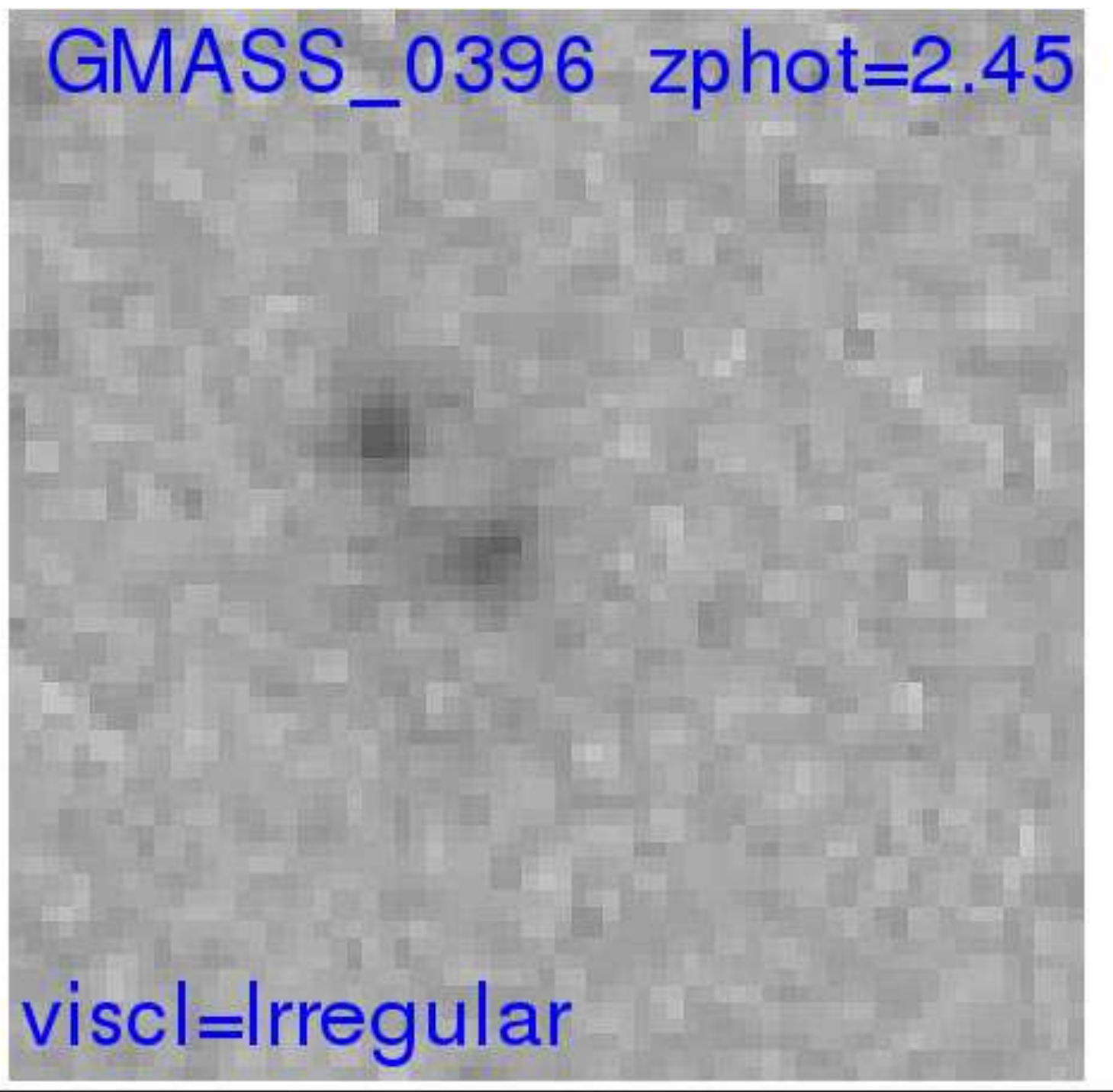}			     
\includegraphics[trim=100 40 75 390, clip=true, width=30mm]{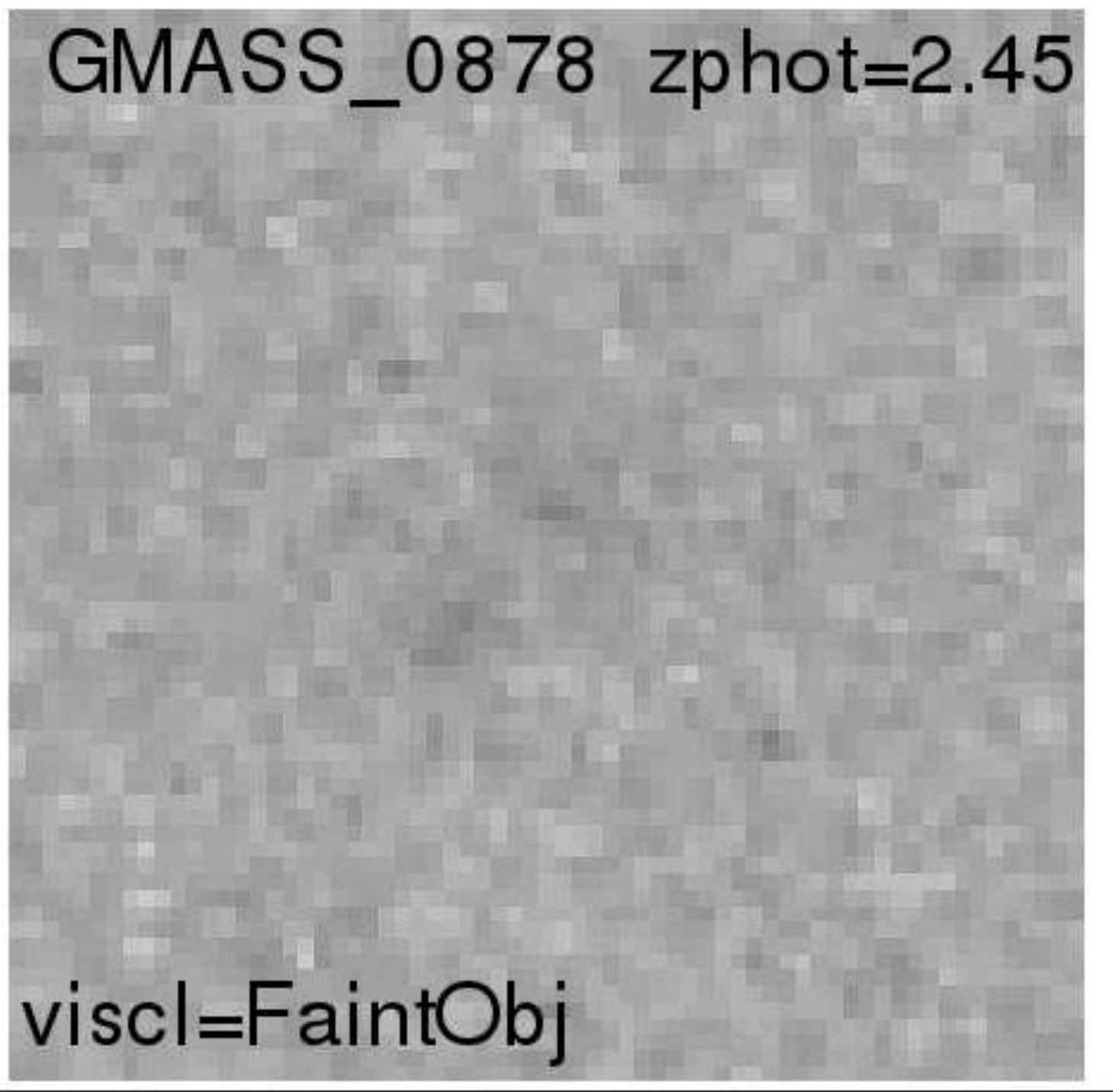}			     
\includegraphics[trim=100 40 75 390, clip=true, width=30mm]{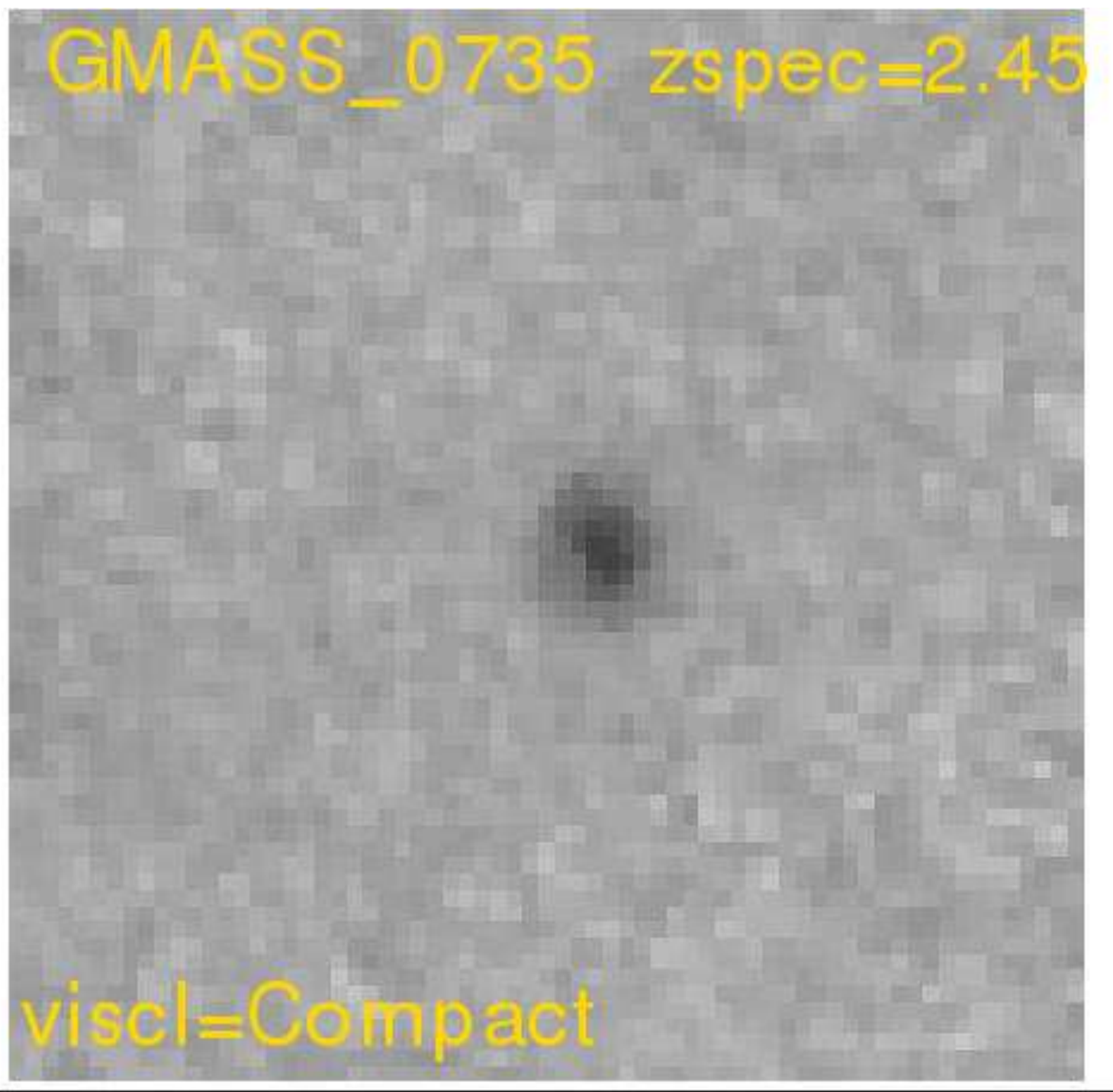}			     
\includegraphics[trim=100 40 75 390, clip=true, width=30mm]{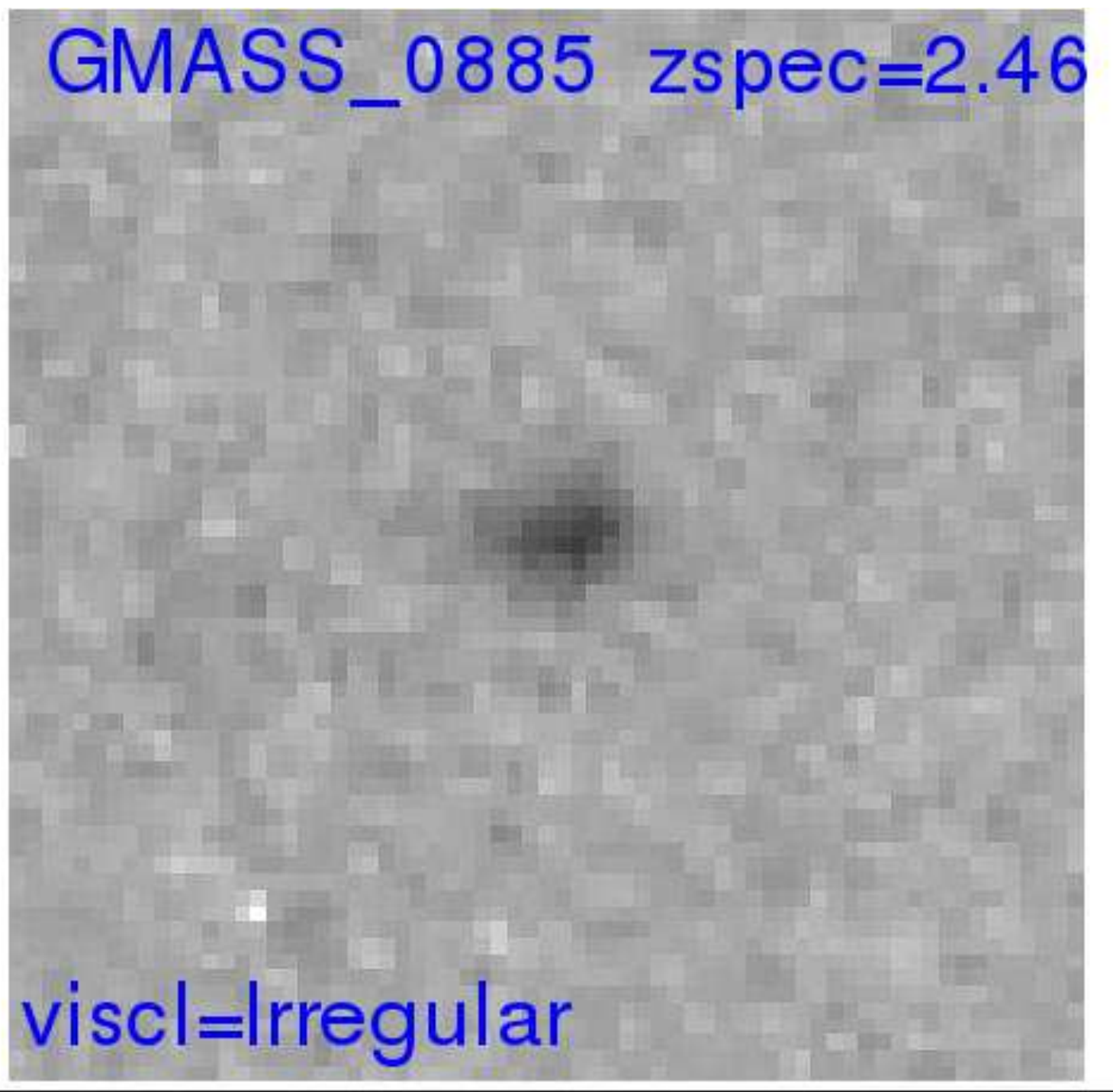}			     
\includegraphics[trim=100 40 75 390, clip=true, width=30mm]{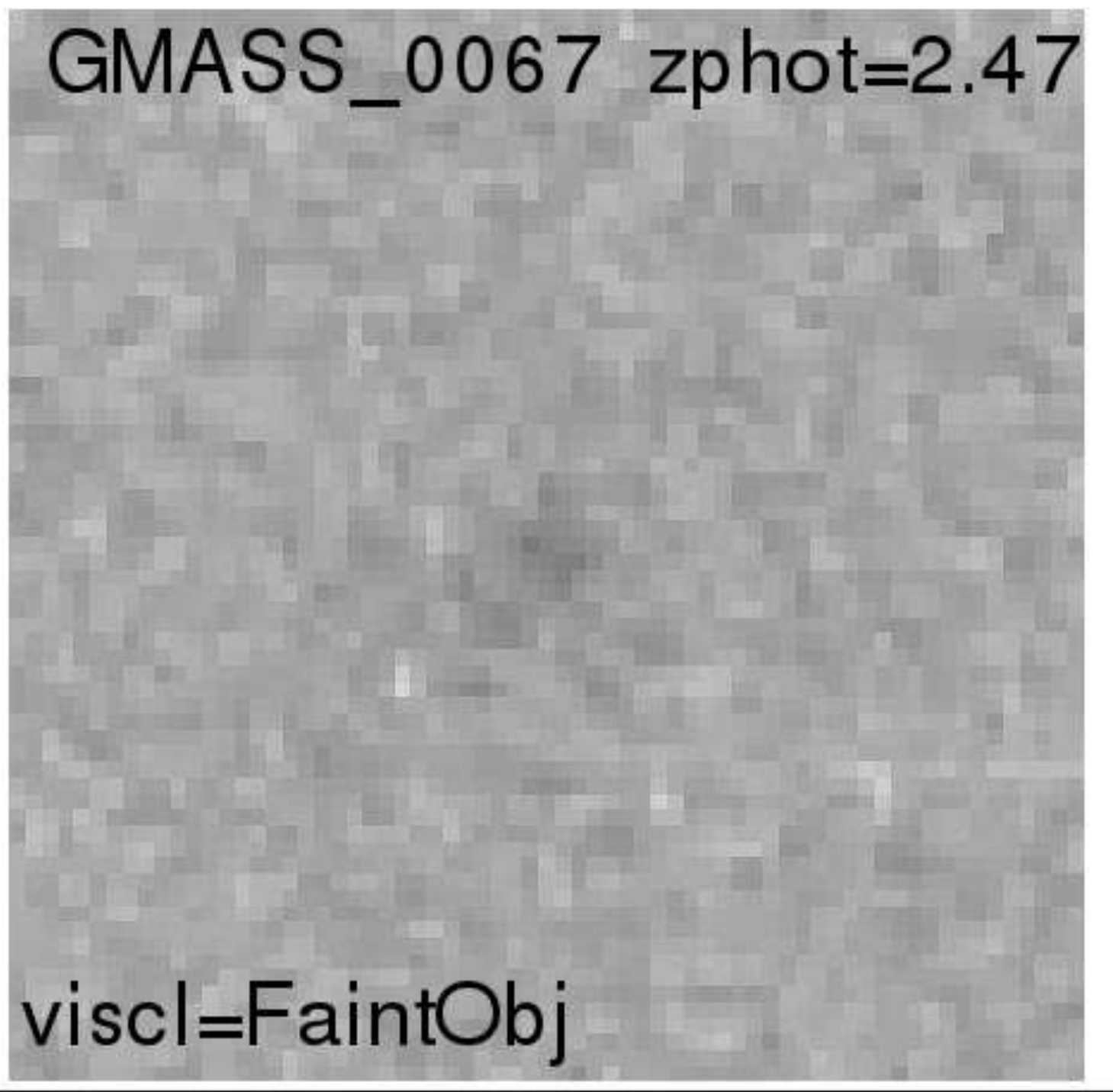}			     

\includegraphics[trim=100 40 75 390, clip=true, width=30mm]{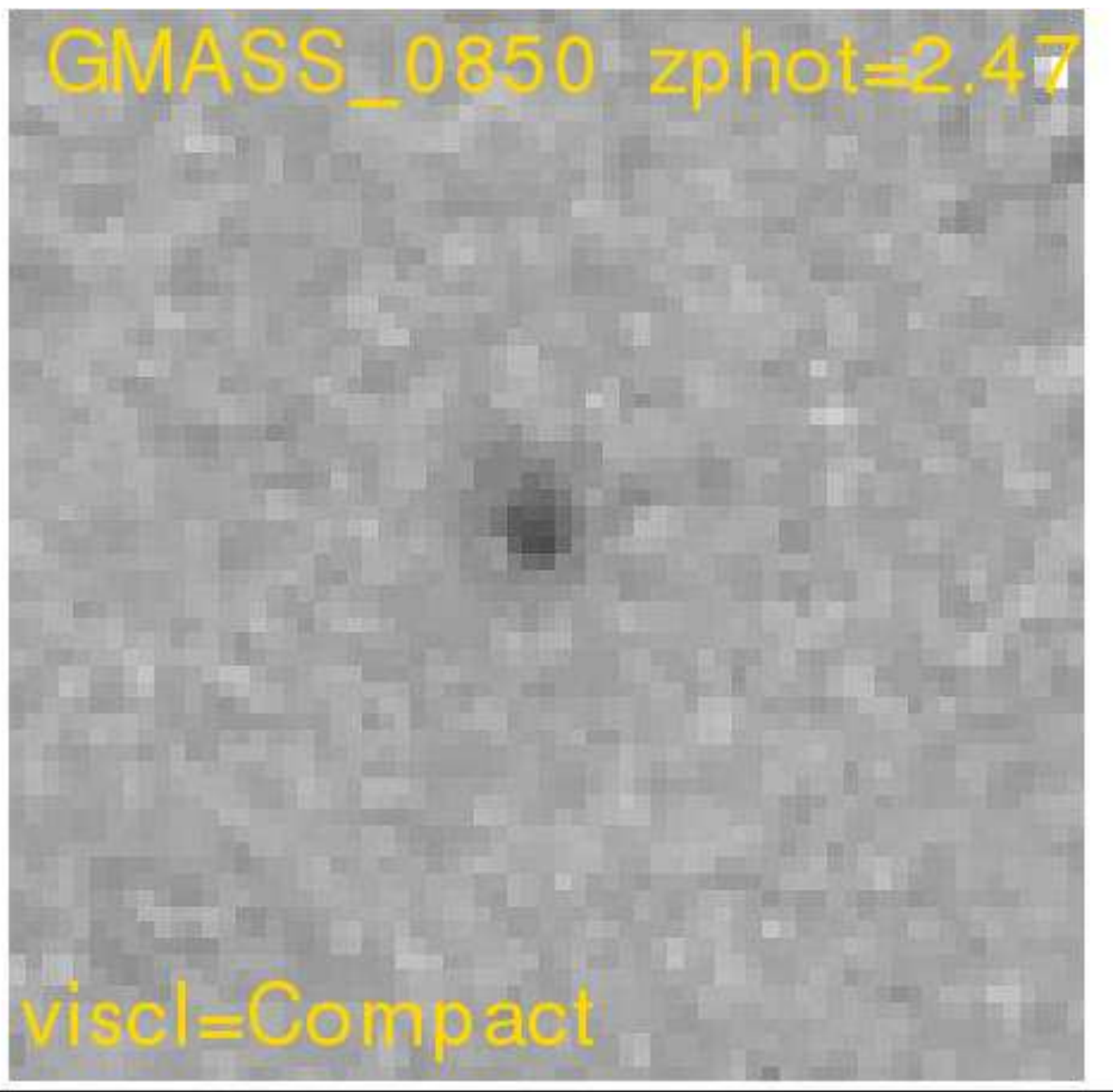}		     
\includegraphics[trim=100 40 75 390, clip=true, width=30mm]{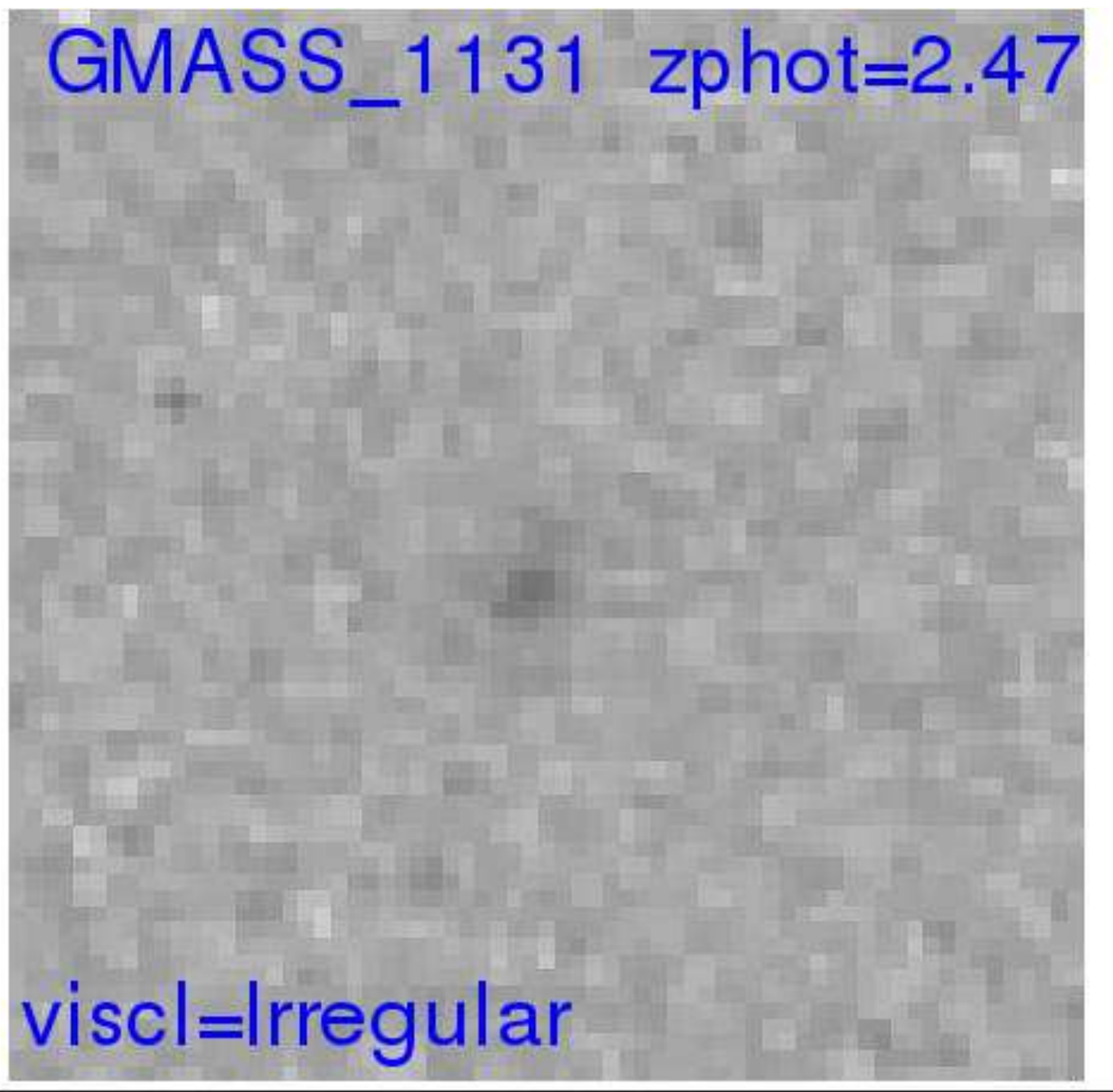}			     
\includegraphics[trim=100 40 75 390, clip=true, width=30mm]{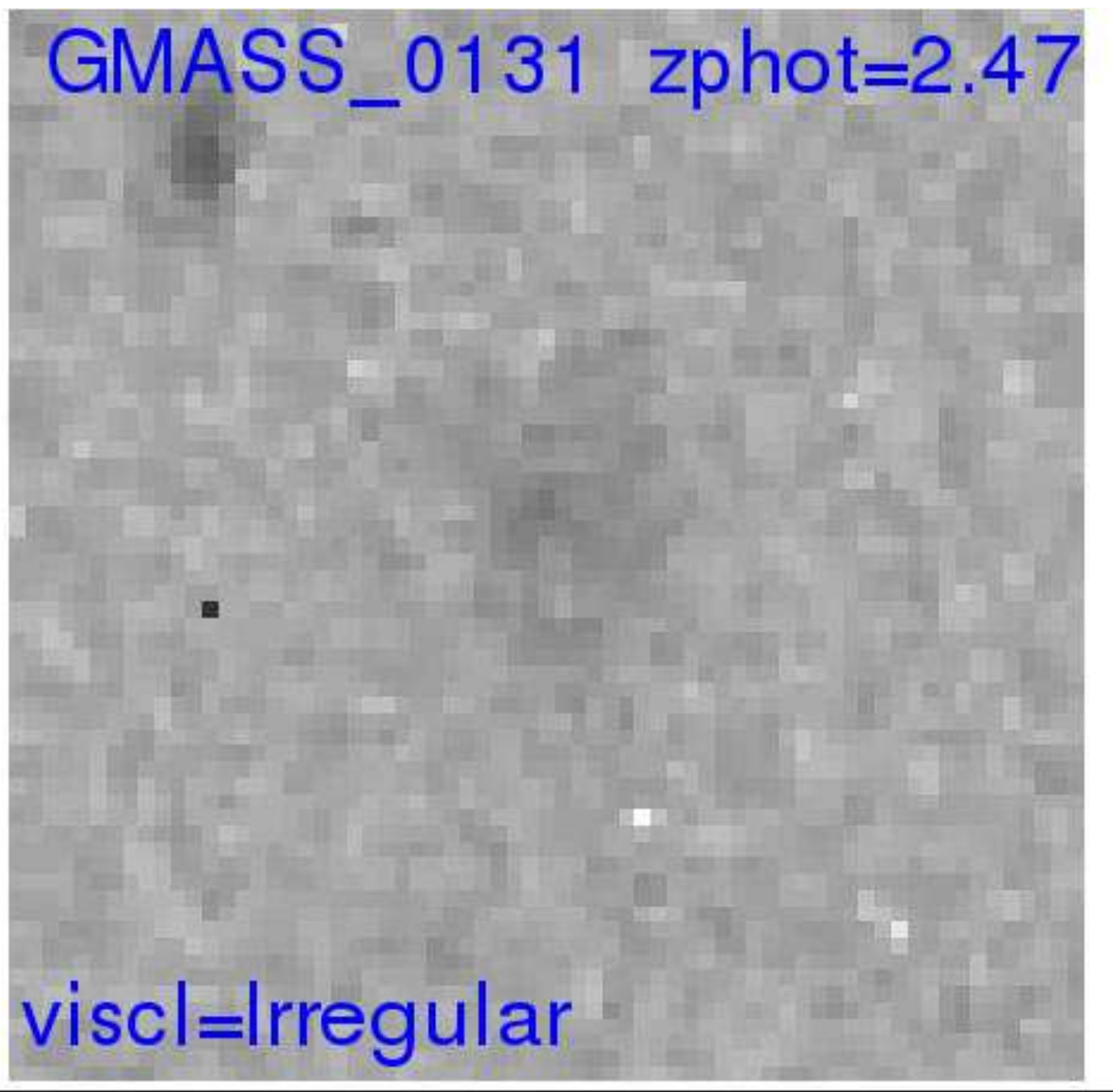}			     
\includegraphics[trim=100 40 75 390, clip=true, width=30mm]{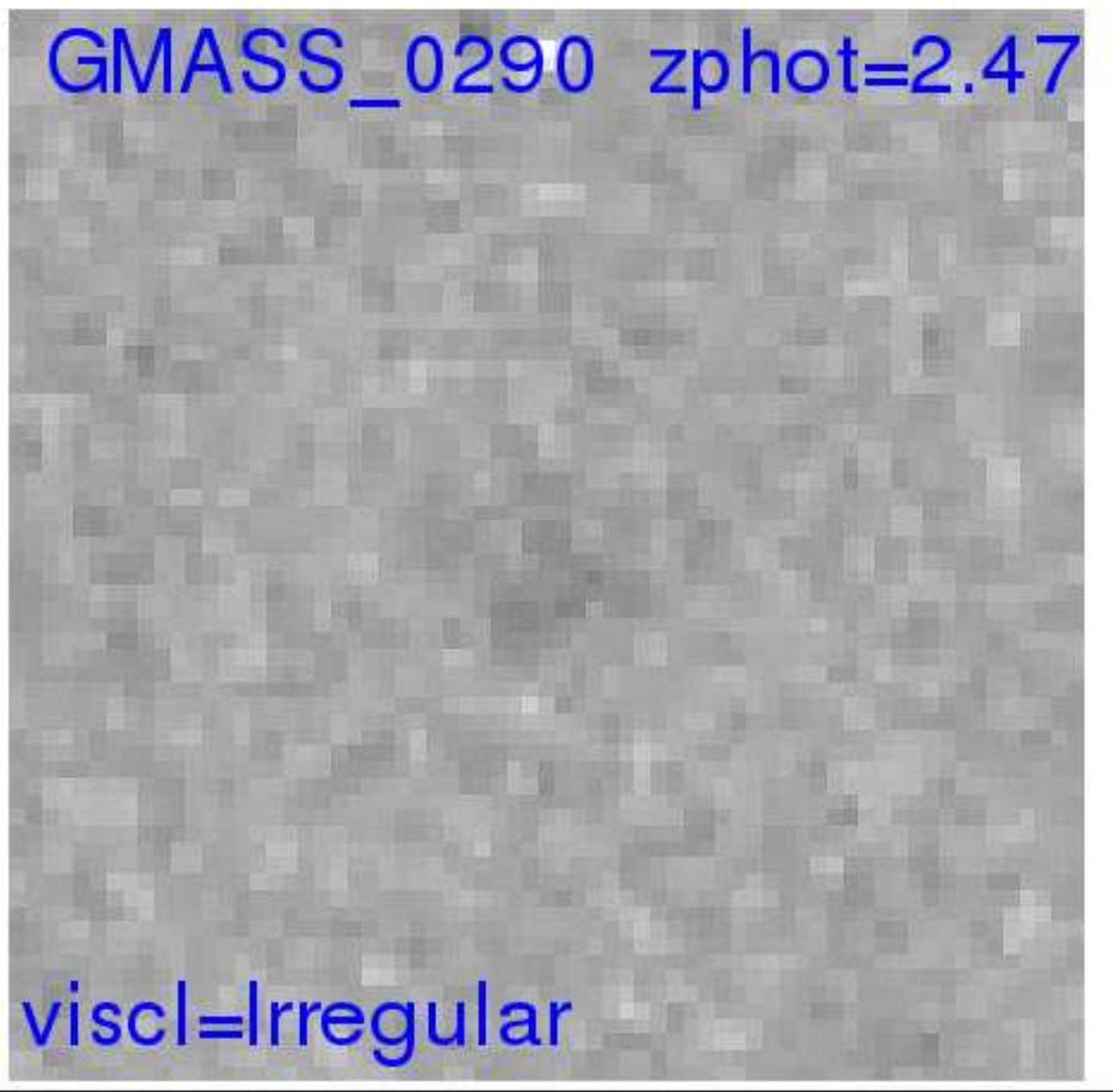}			     
\includegraphics[trim=100 40 75 390, clip=true, width=30mm]{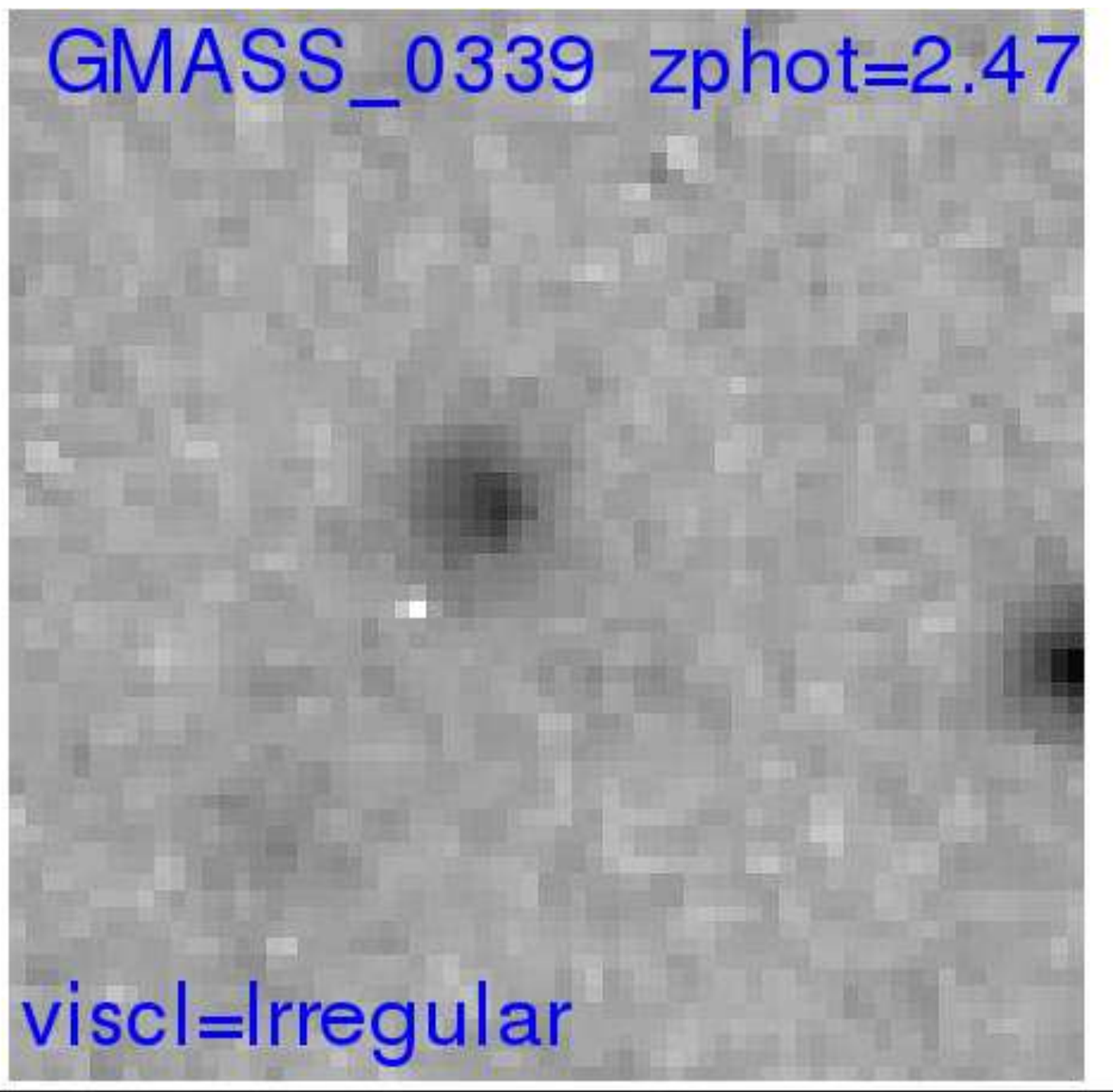}			     
\includegraphics[trim=100 40 75 390, clip=true, width=30mm]{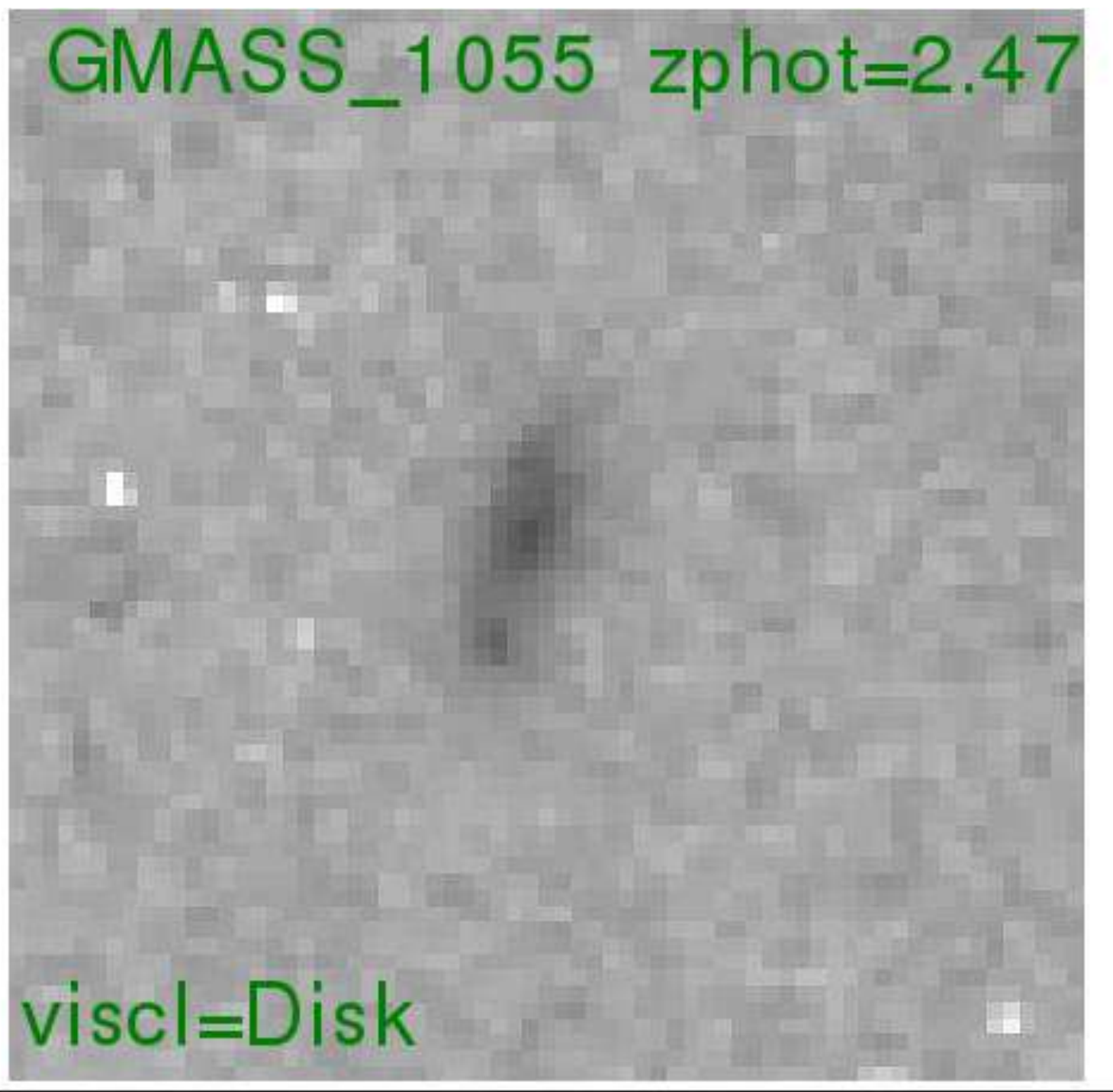}			     

\includegraphics[trim=100 40 75 390, clip=true, width=30mm]{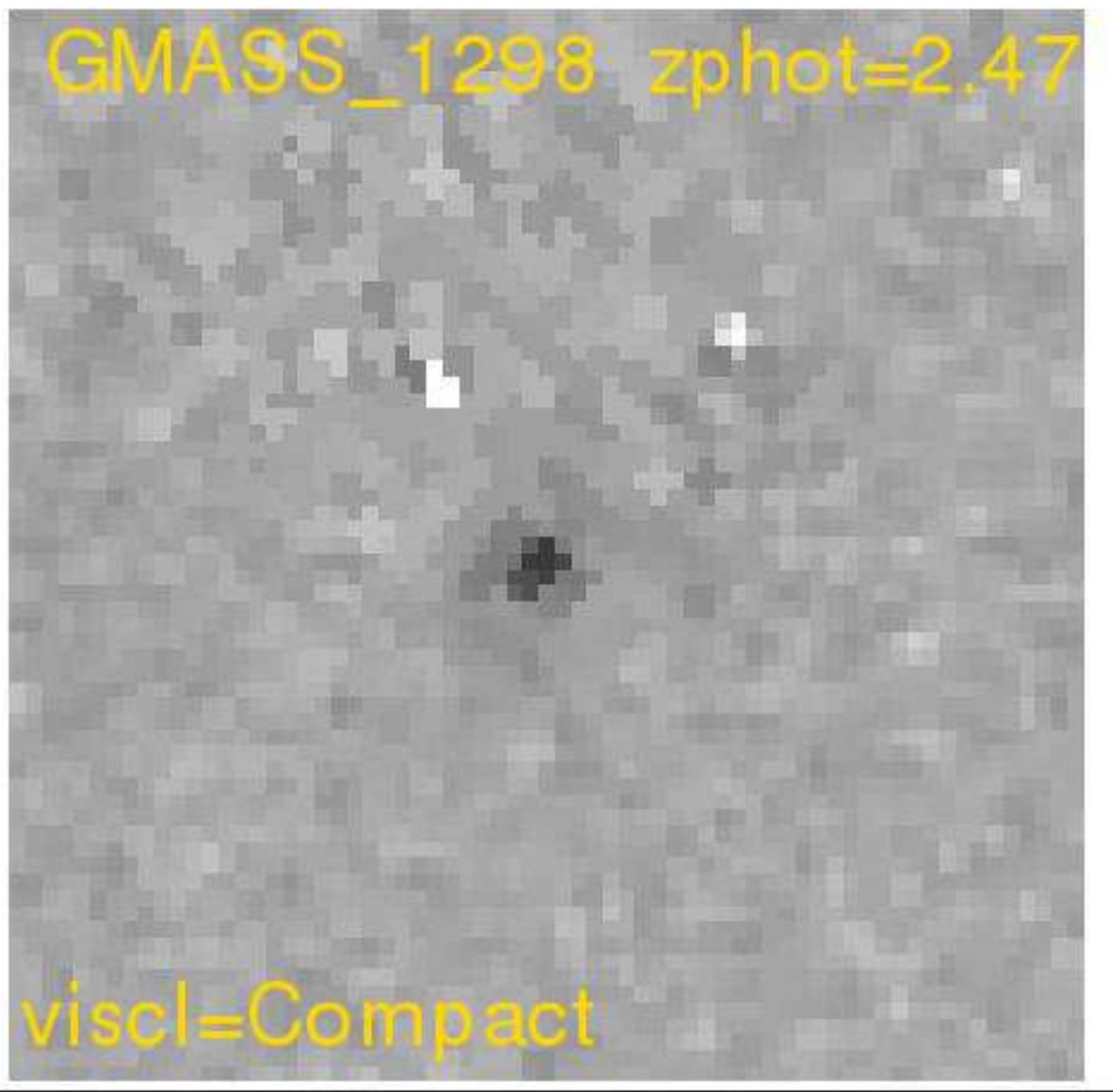}			     
\includegraphics[trim=100 40 75 390, clip=true, width=30mm]{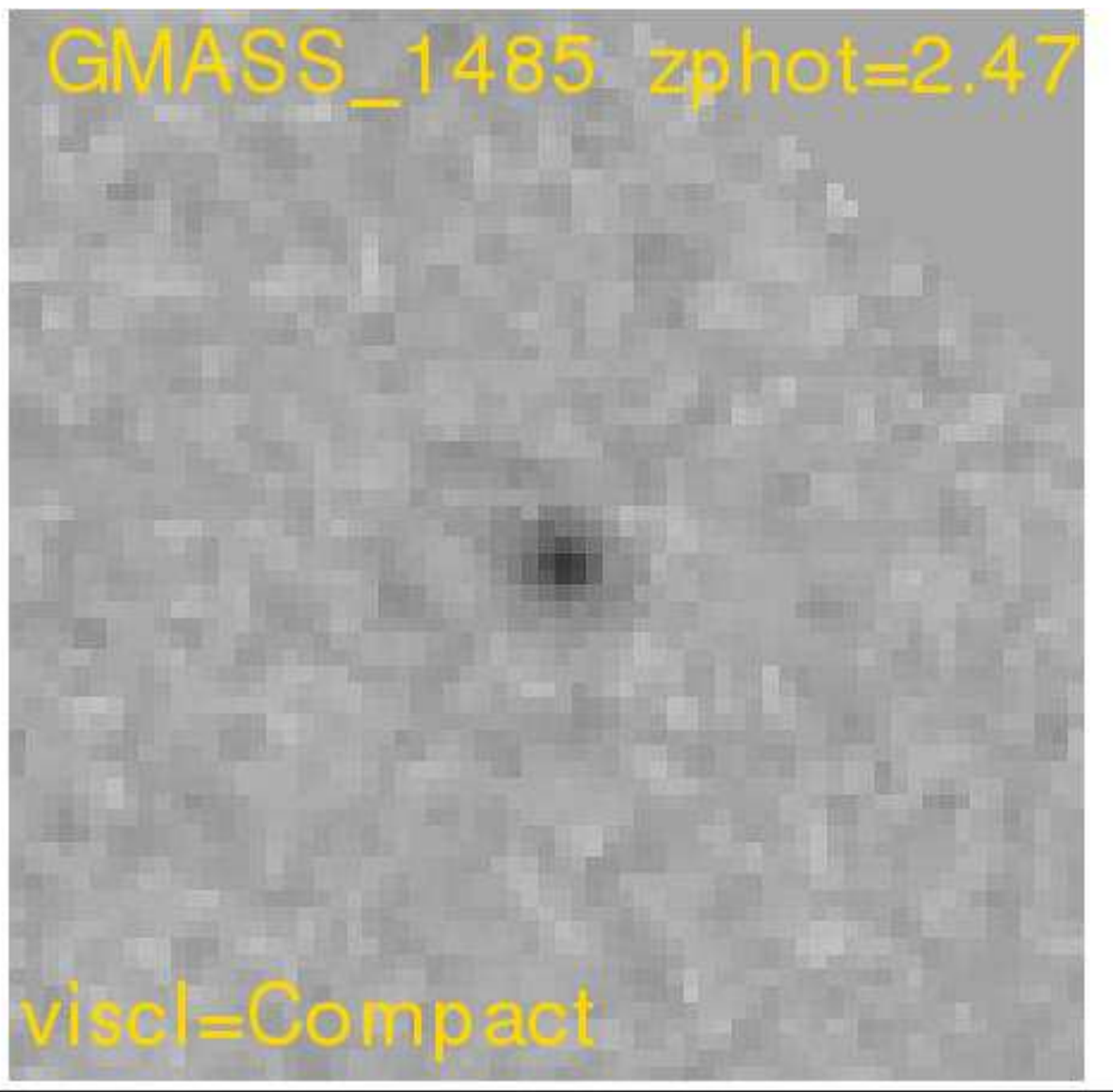}			     
\includegraphics[trim=100 40 75 390, clip=true, width=30mm]{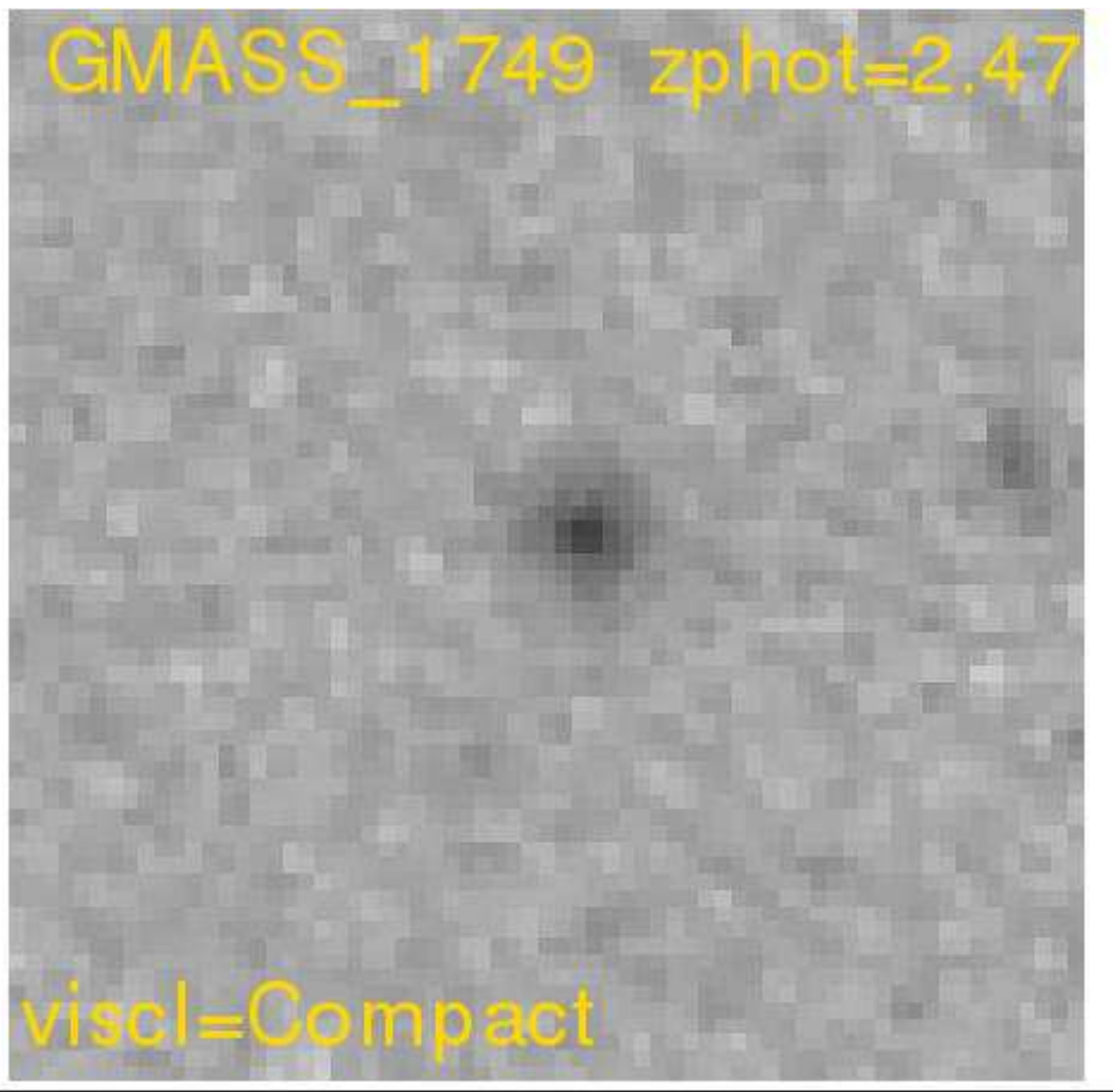}			     
\includegraphics[trim=100 40 75 390, clip=true, width=30mm]{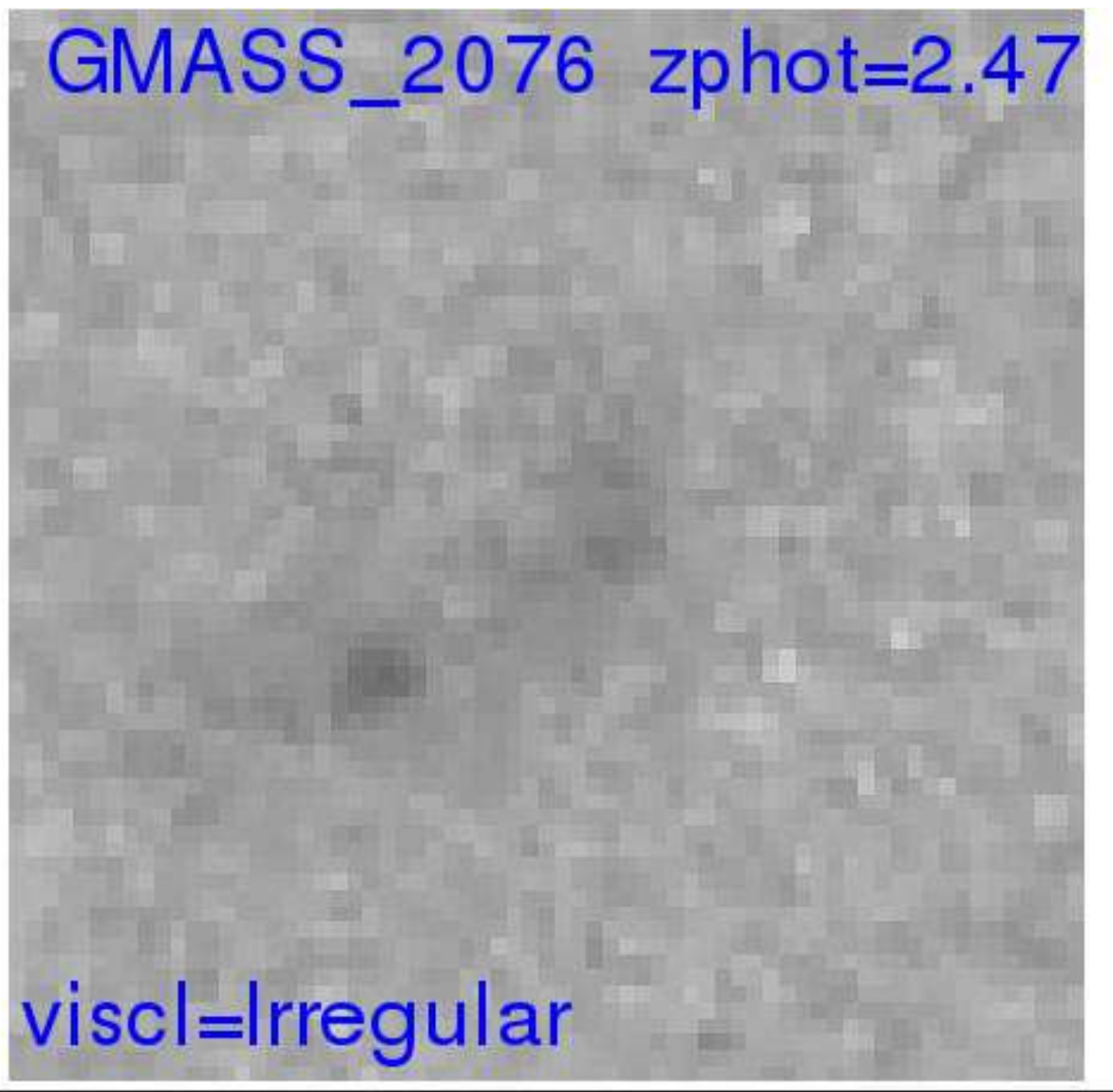}			     
\includegraphics[trim=100 40 75 390, clip=true, width=30mm]{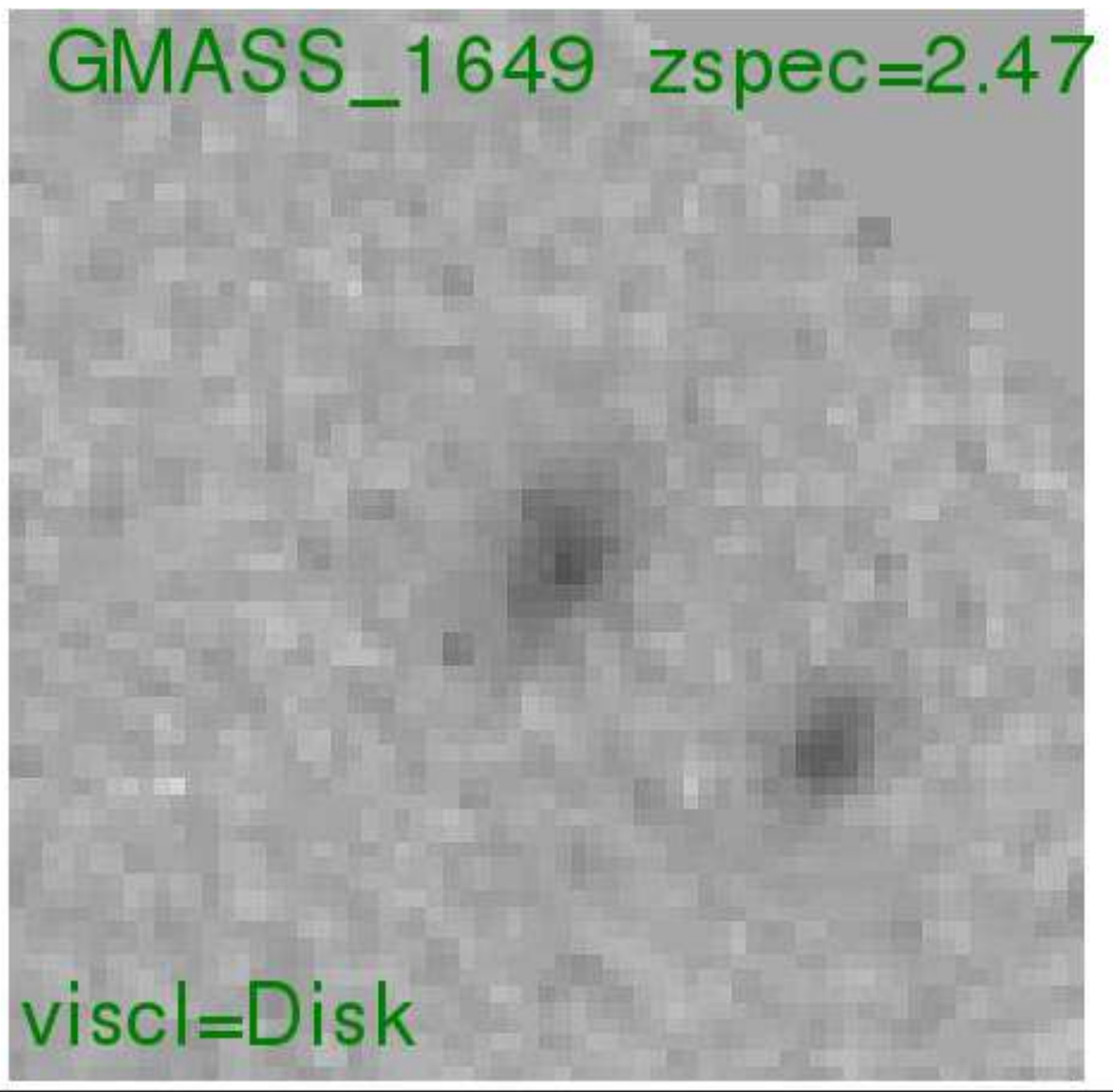}			     
\includegraphics[trim=100 40 75 390, clip=true, width=30mm]{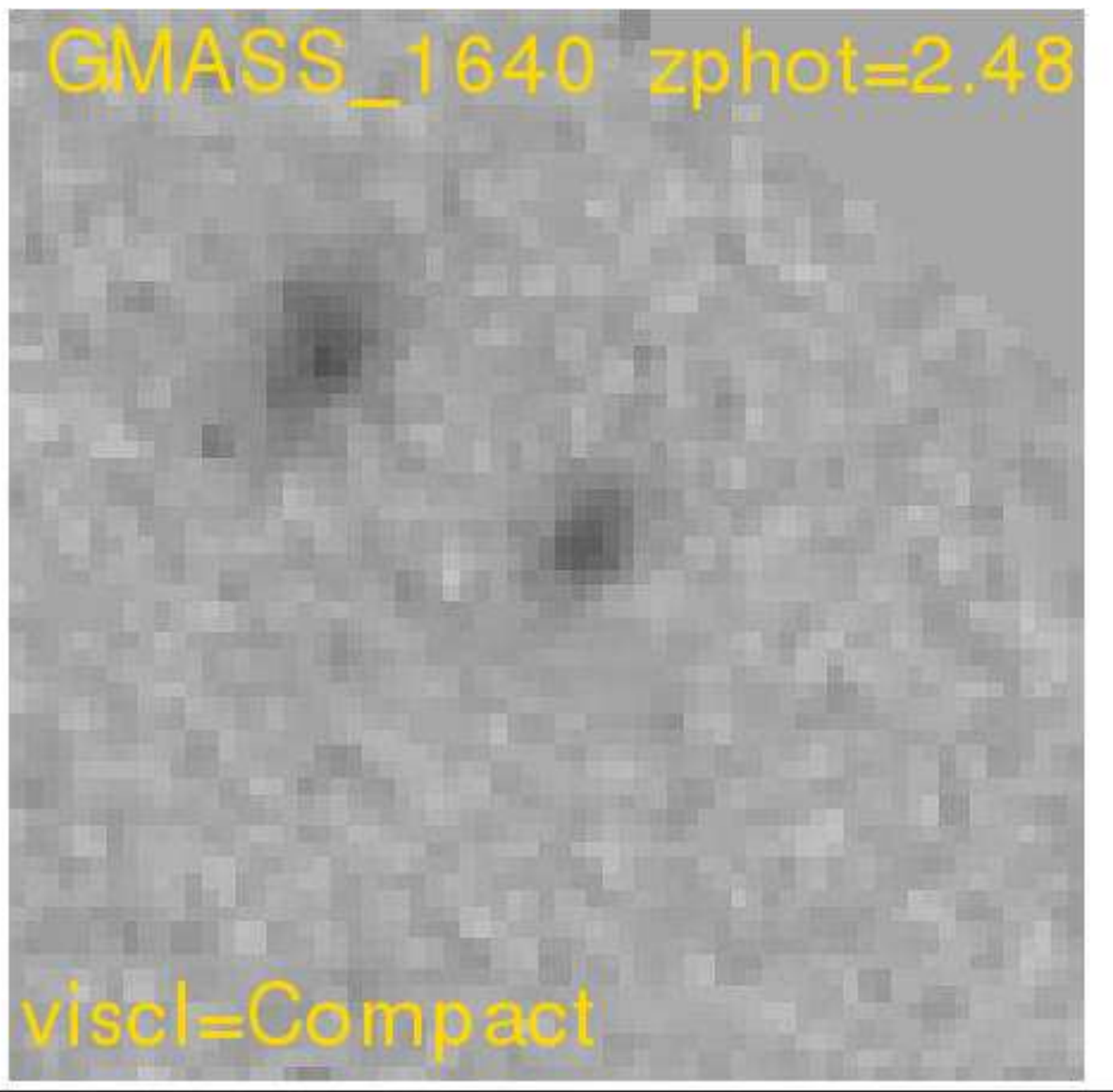}			     

\includegraphics[trim=100 40 75 390, clip=true, width=30mm]{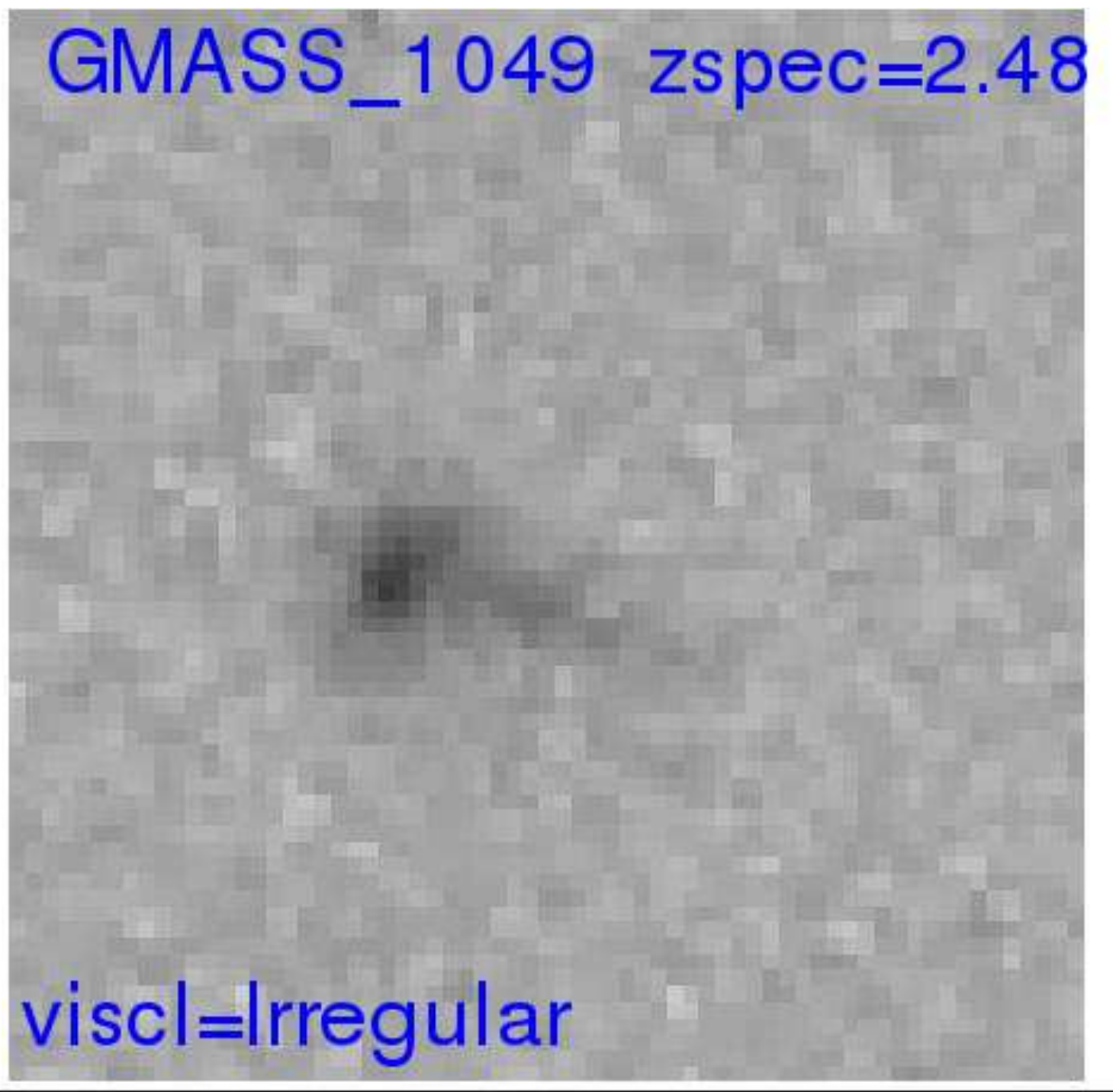}			     
\includegraphics[trim=100 40 75 390, clip=true, width=30mm]{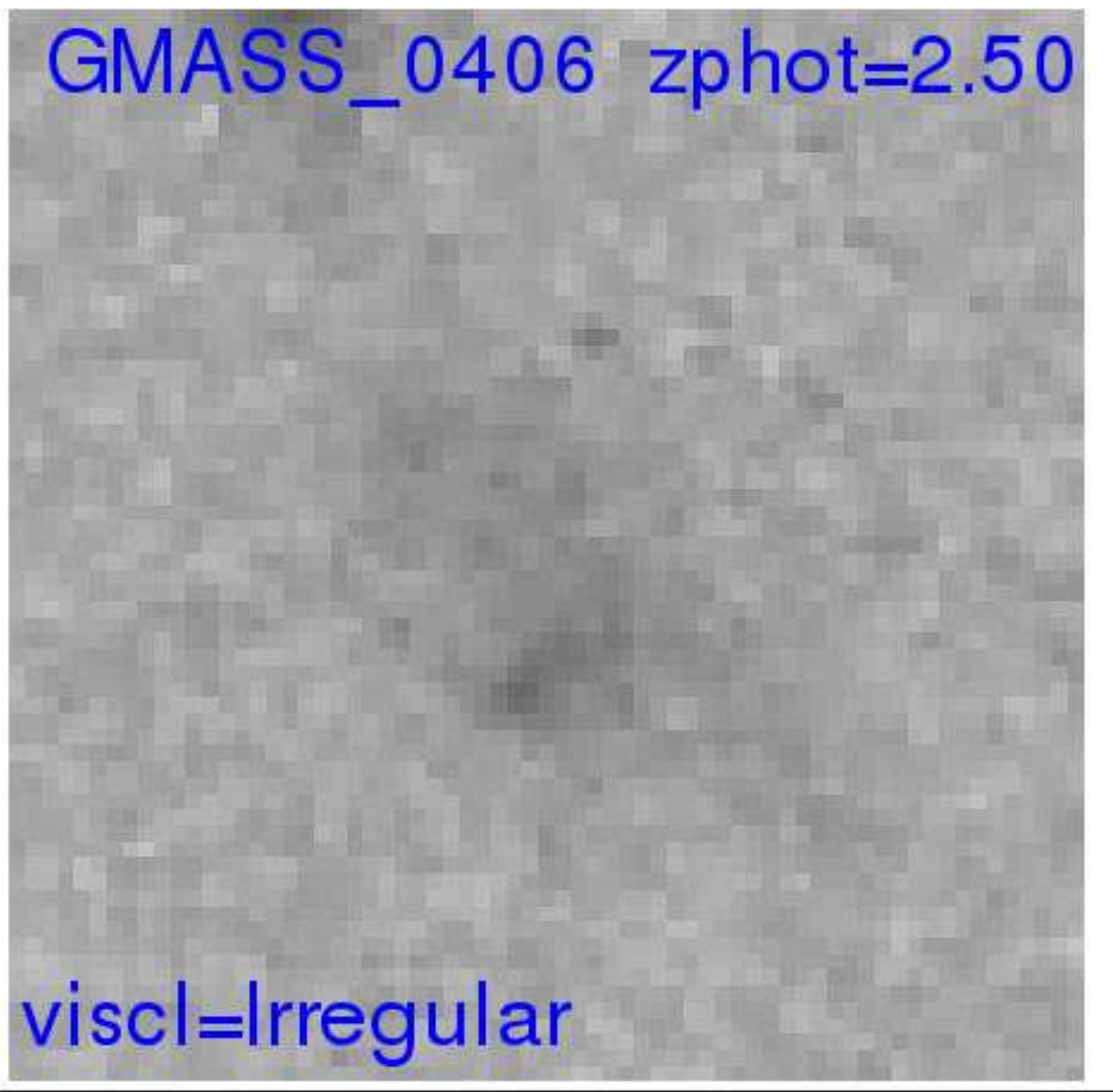}		     
\includegraphics[trim=100 40 75 390, clip=true, width=30mm]{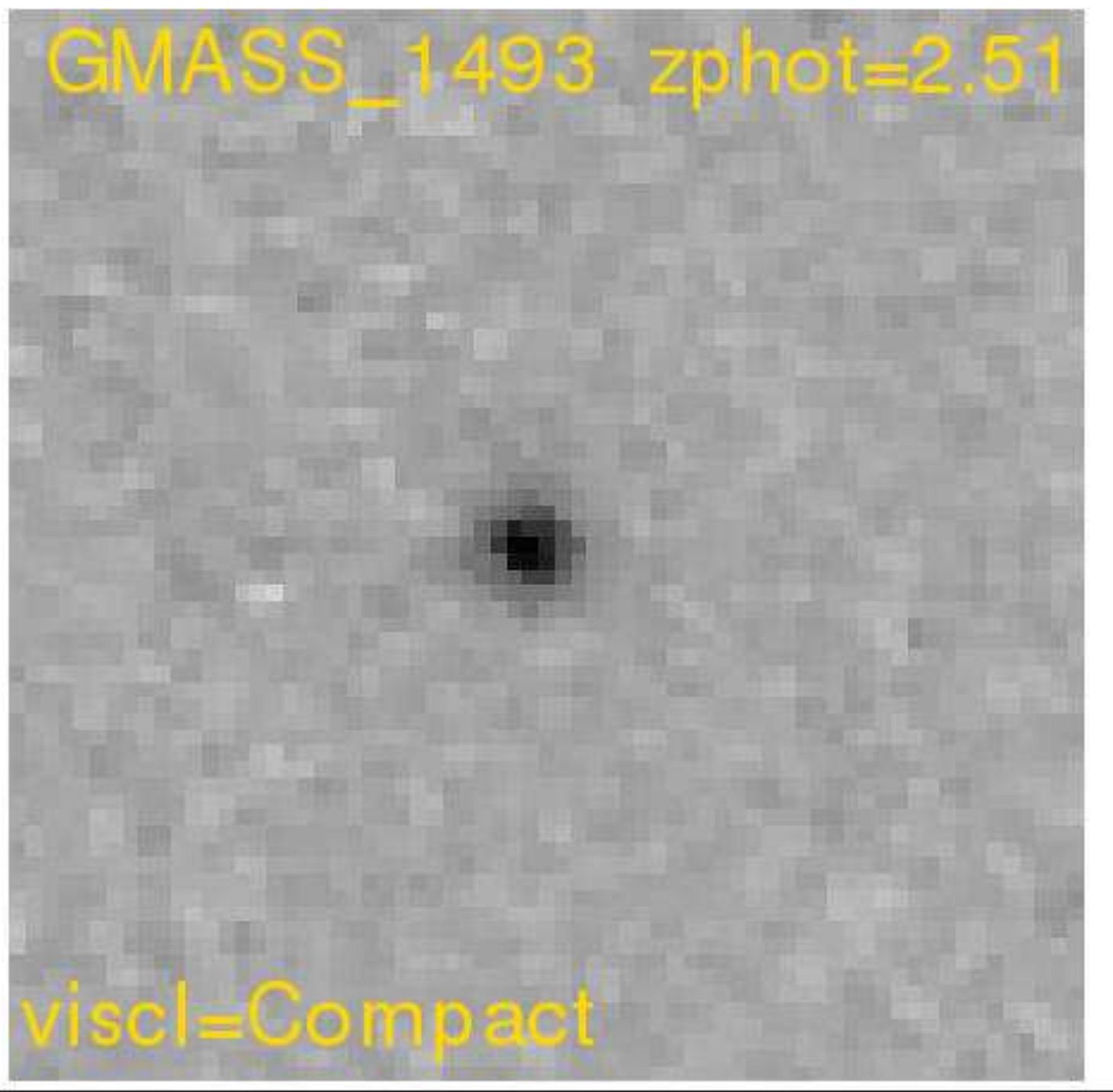}			     
\includegraphics[trim=100 40 75 390, clip=true, width=30mm]{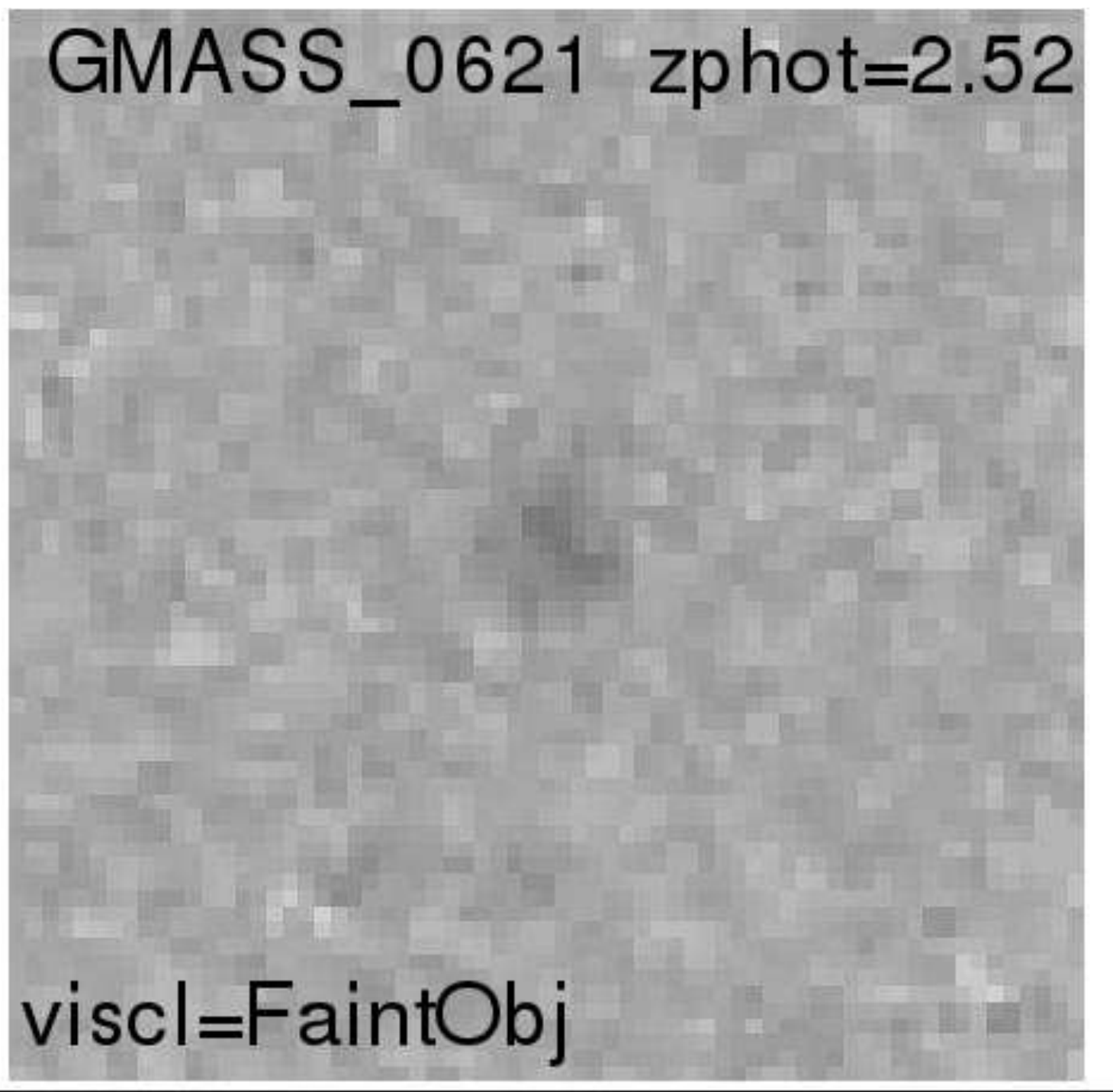}			     
\includegraphics[trim=100 40 75 390, clip=true, width=30mm]{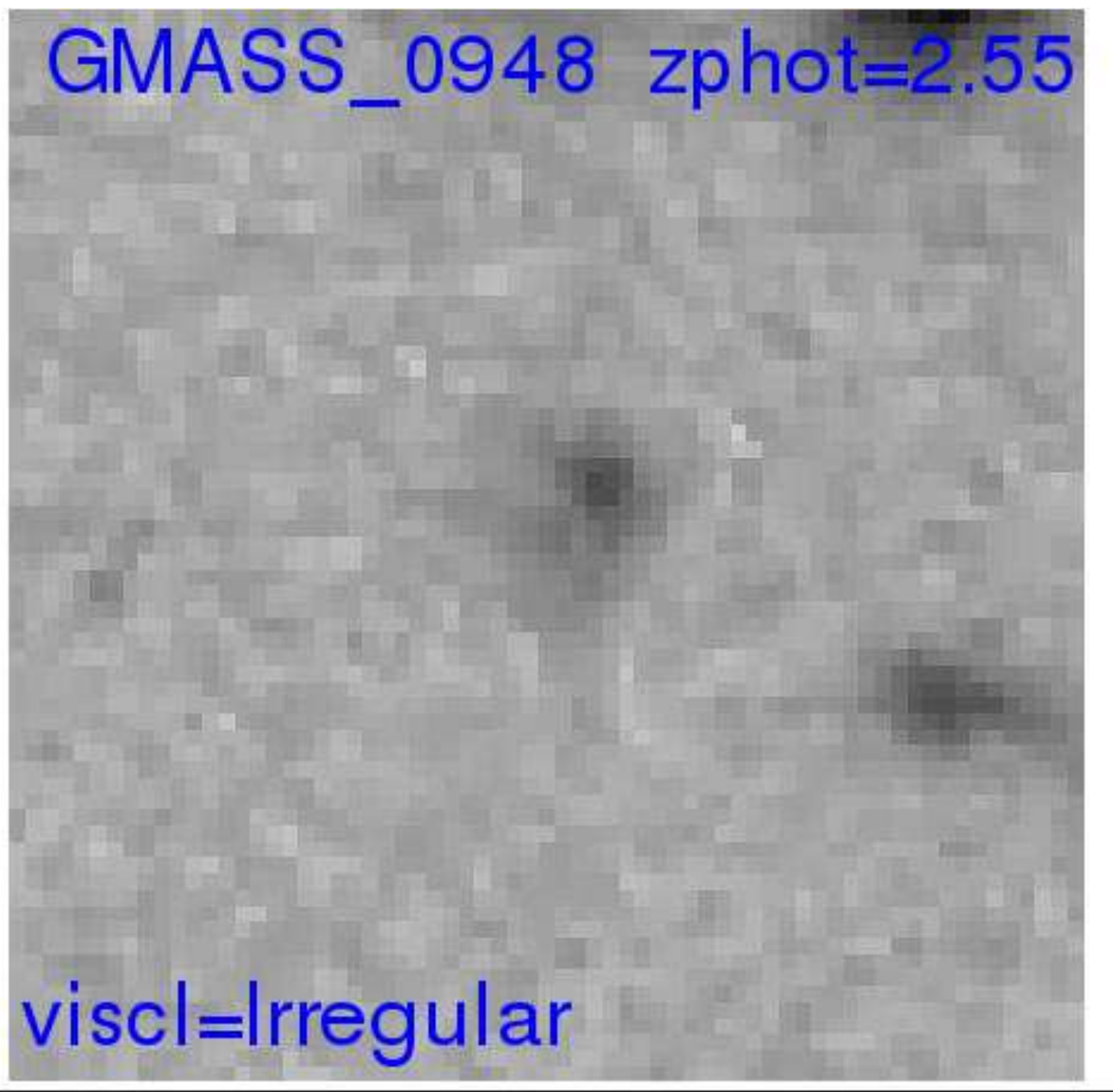}			     
\includegraphics[trim=100 40 75 390, clip=true, width=30mm]{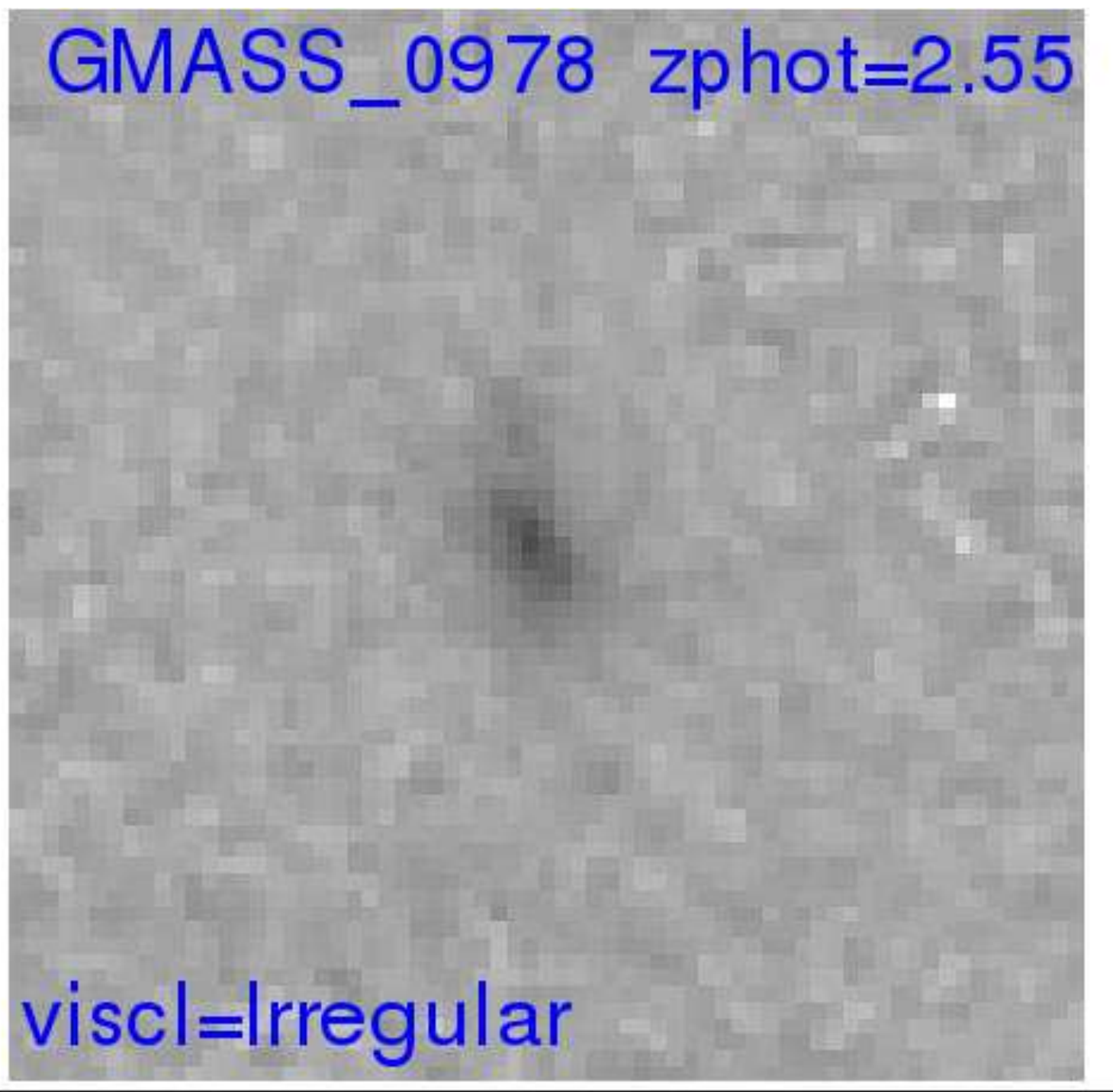}			     

\includegraphics[trim=100 40 75 390, clip=true, width=30mm]{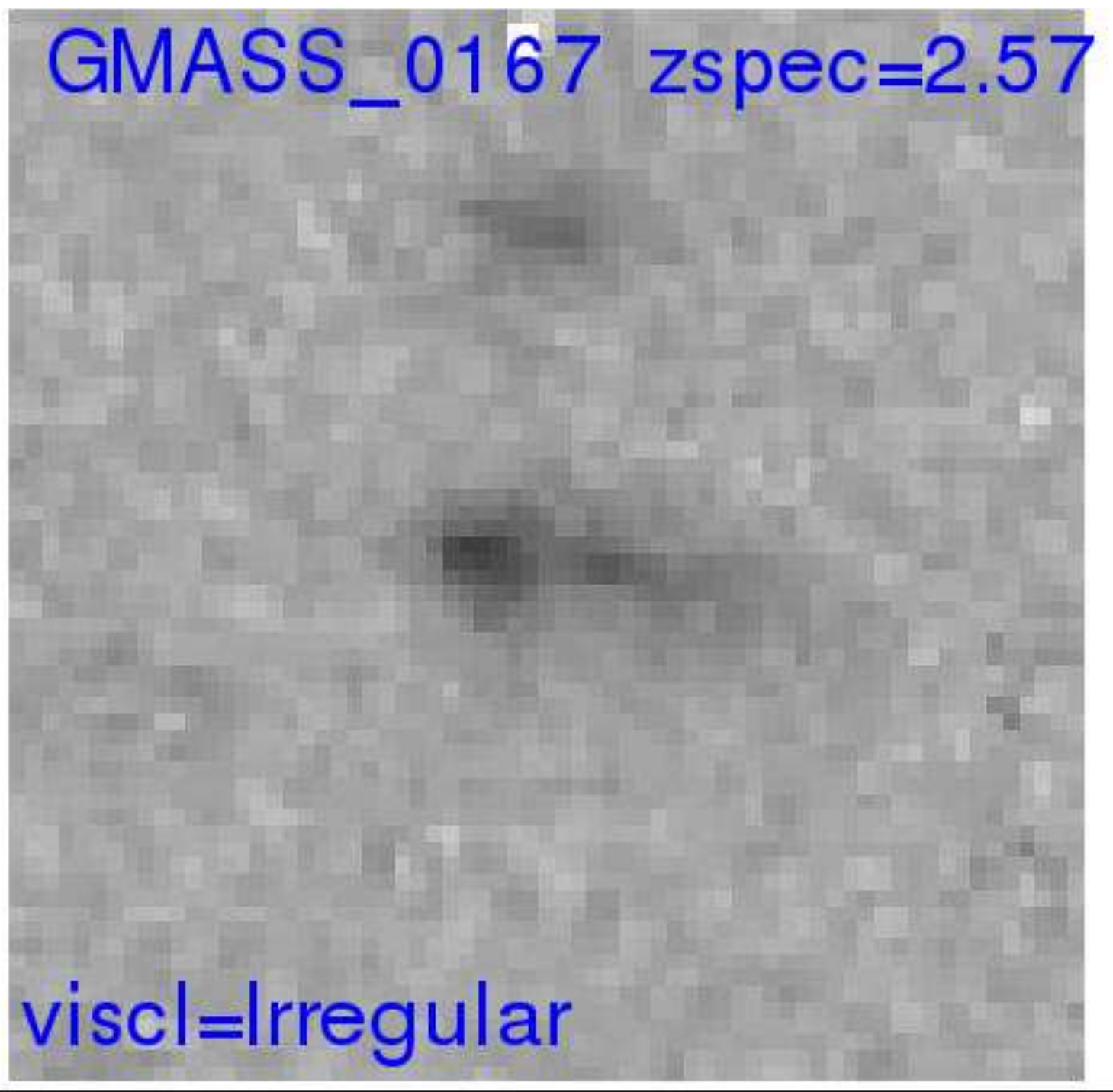}			     
\includegraphics[trim=100 40 75 390, clip=true, width=30mm]{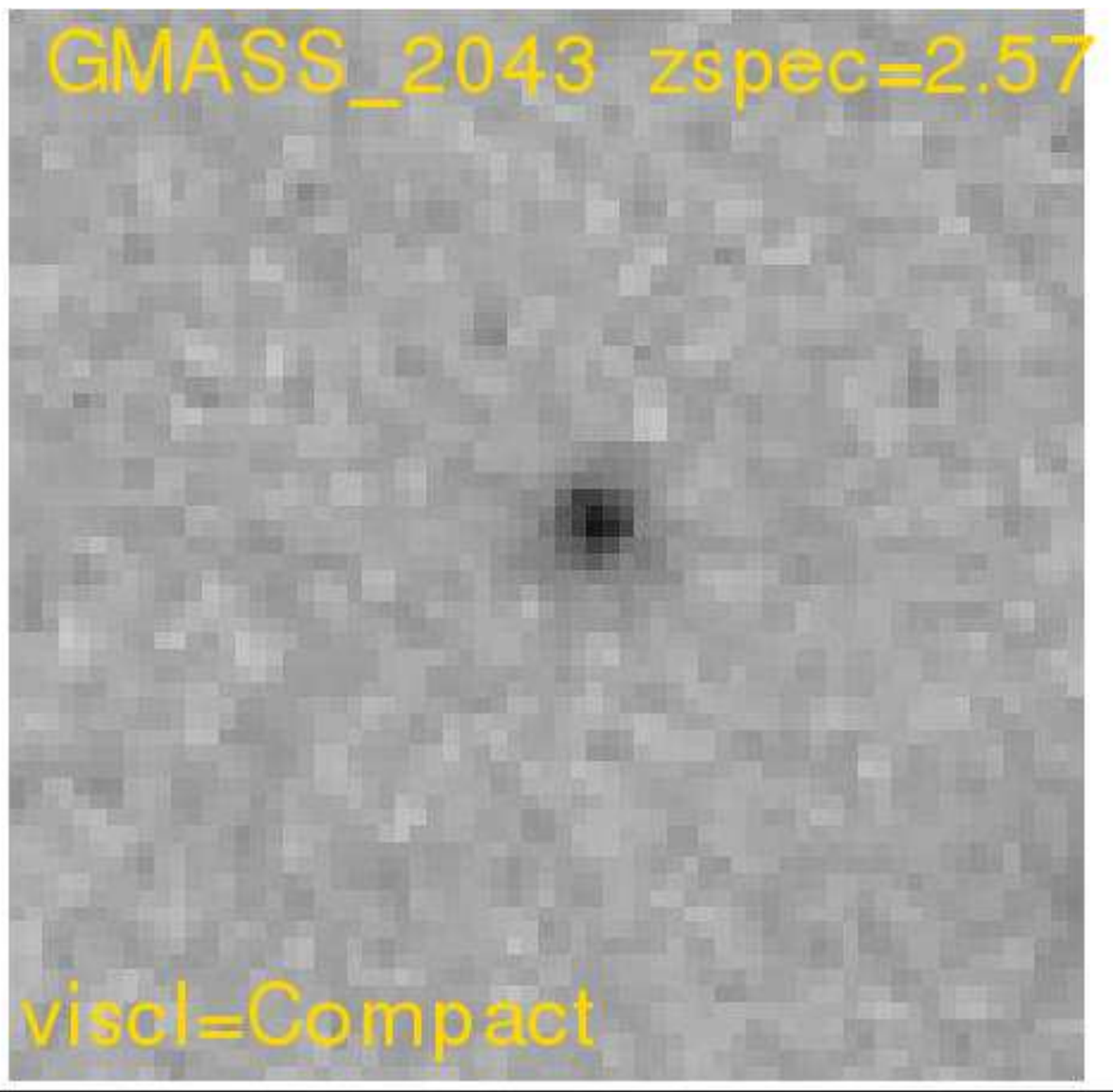}			     
\includegraphics[trim=100 40 75 390, clip=true, width=30mm]{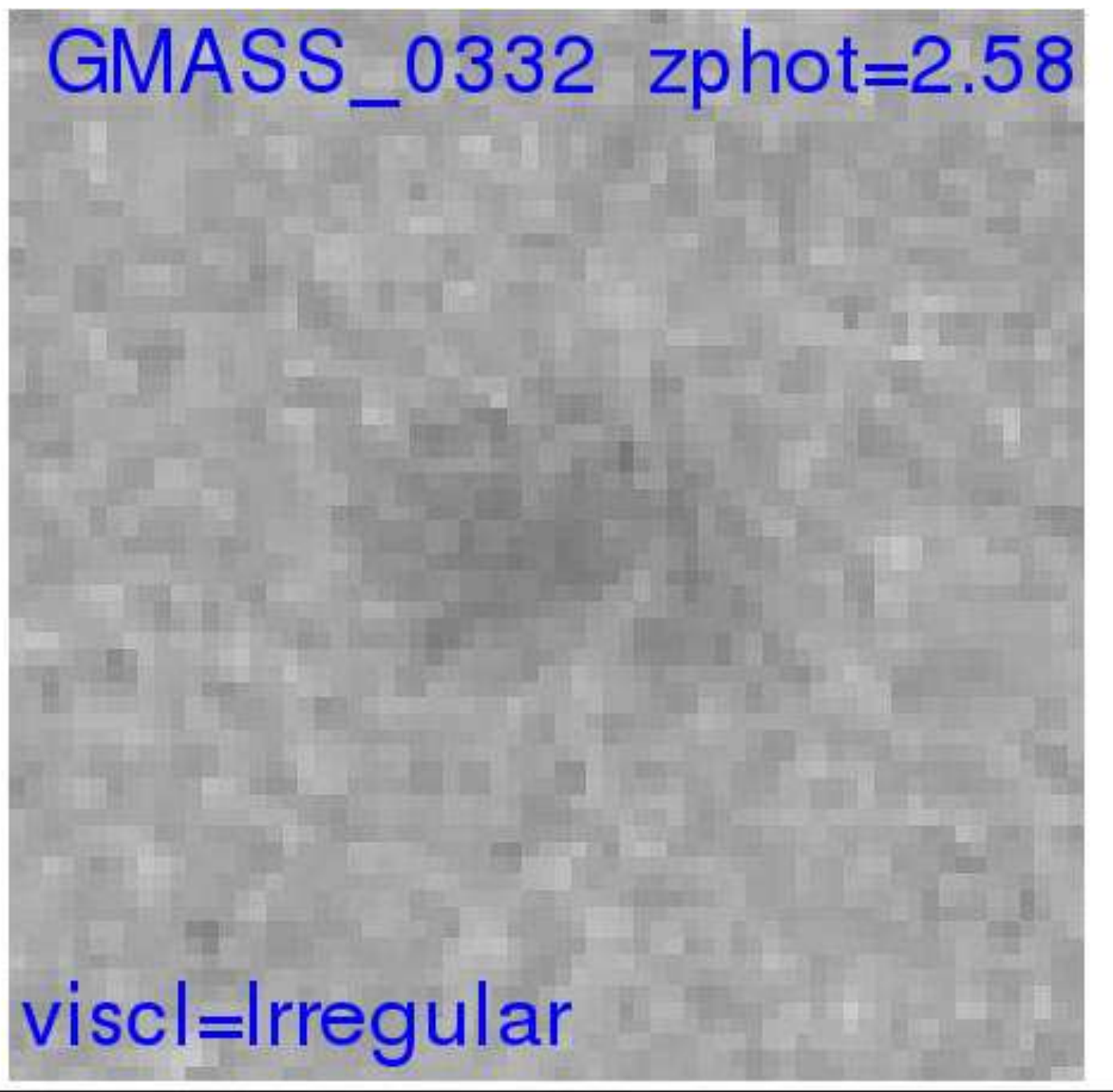}			     
\includegraphics[trim=100 40 75 390, clip=true, width=30mm]{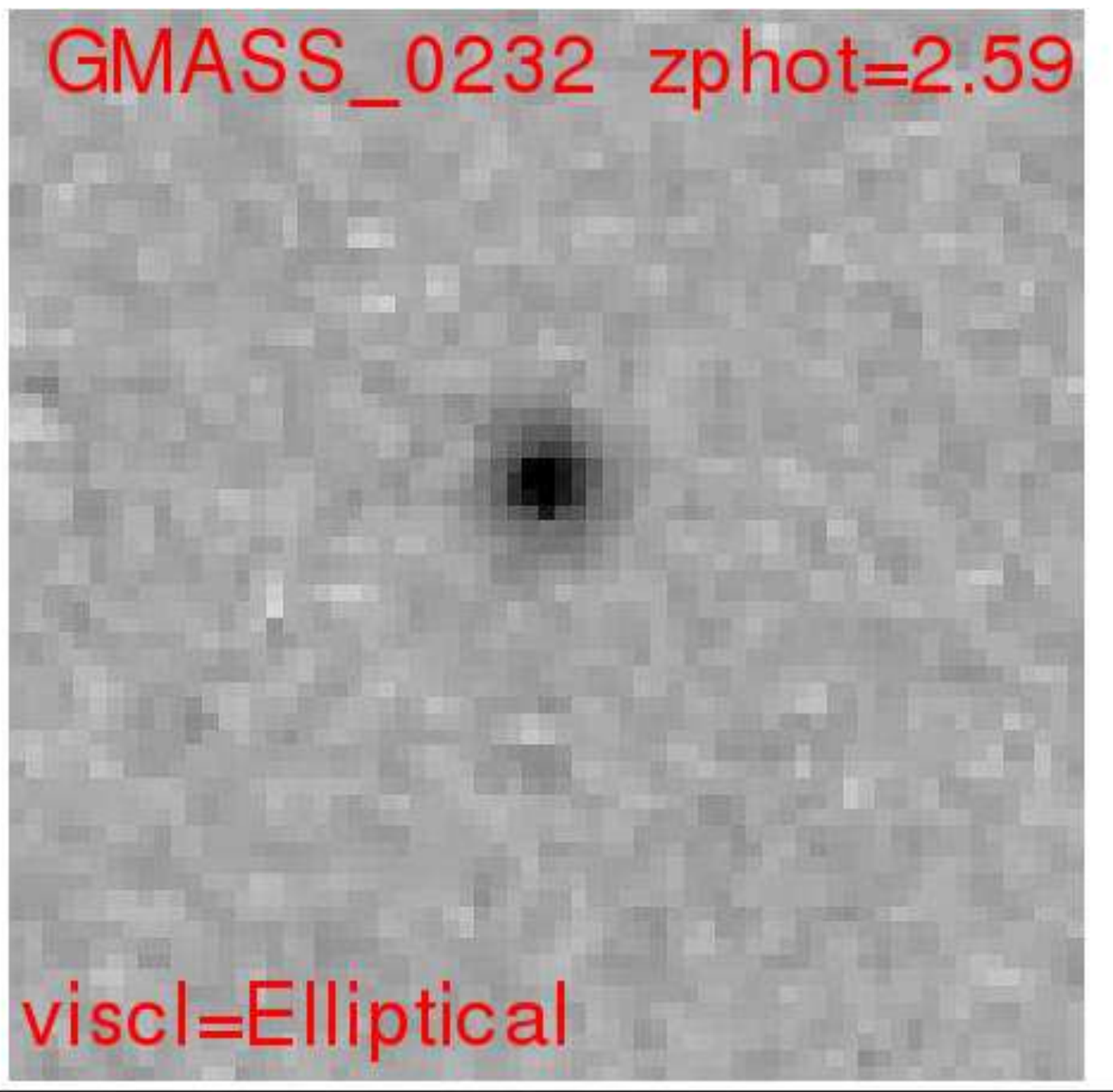}			     
\includegraphics[trim=100 40 75 390, clip=true, width=30mm]{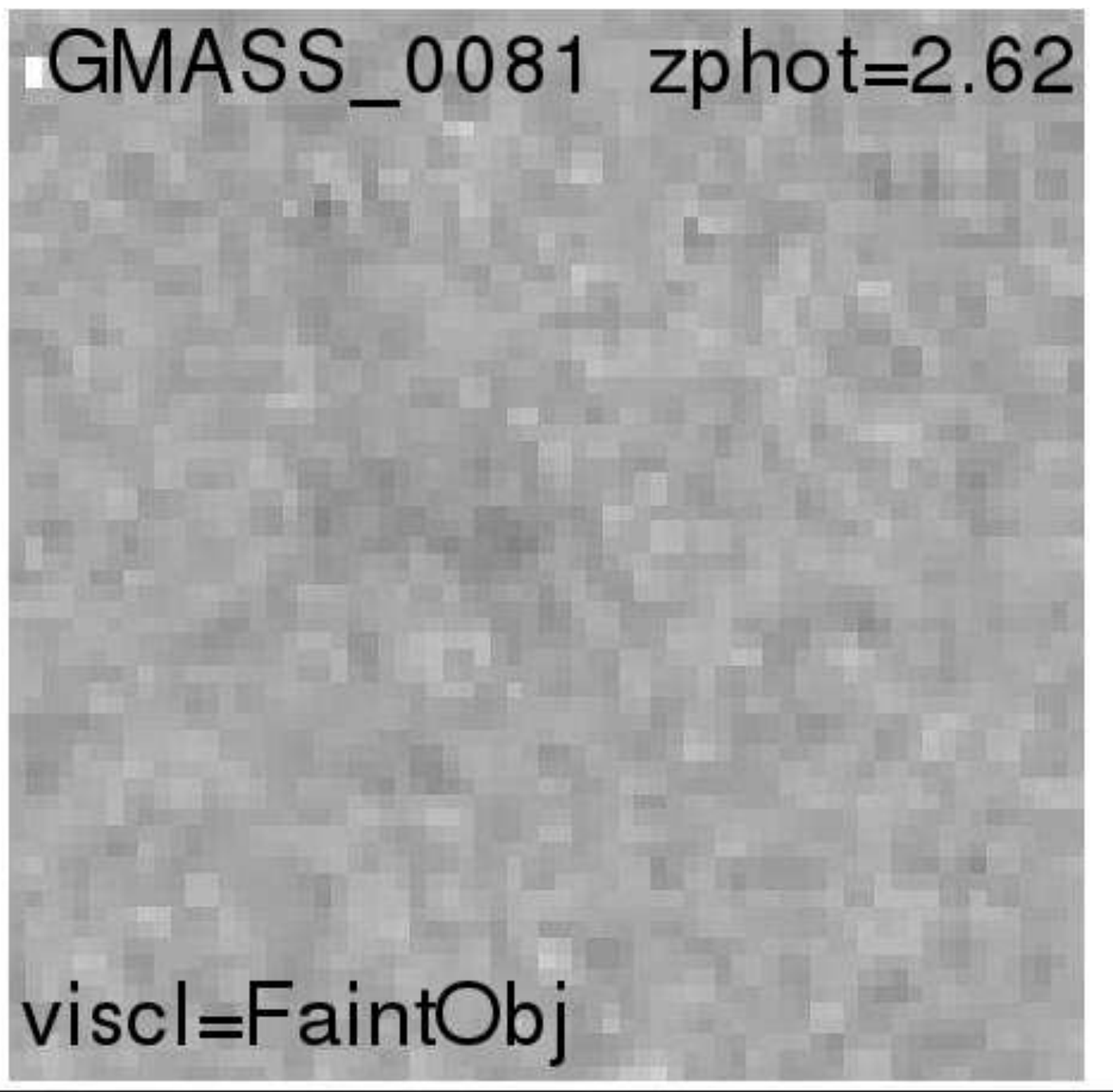}			     
\includegraphics[trim=100 40 75 390, clip=true, width=30mm]{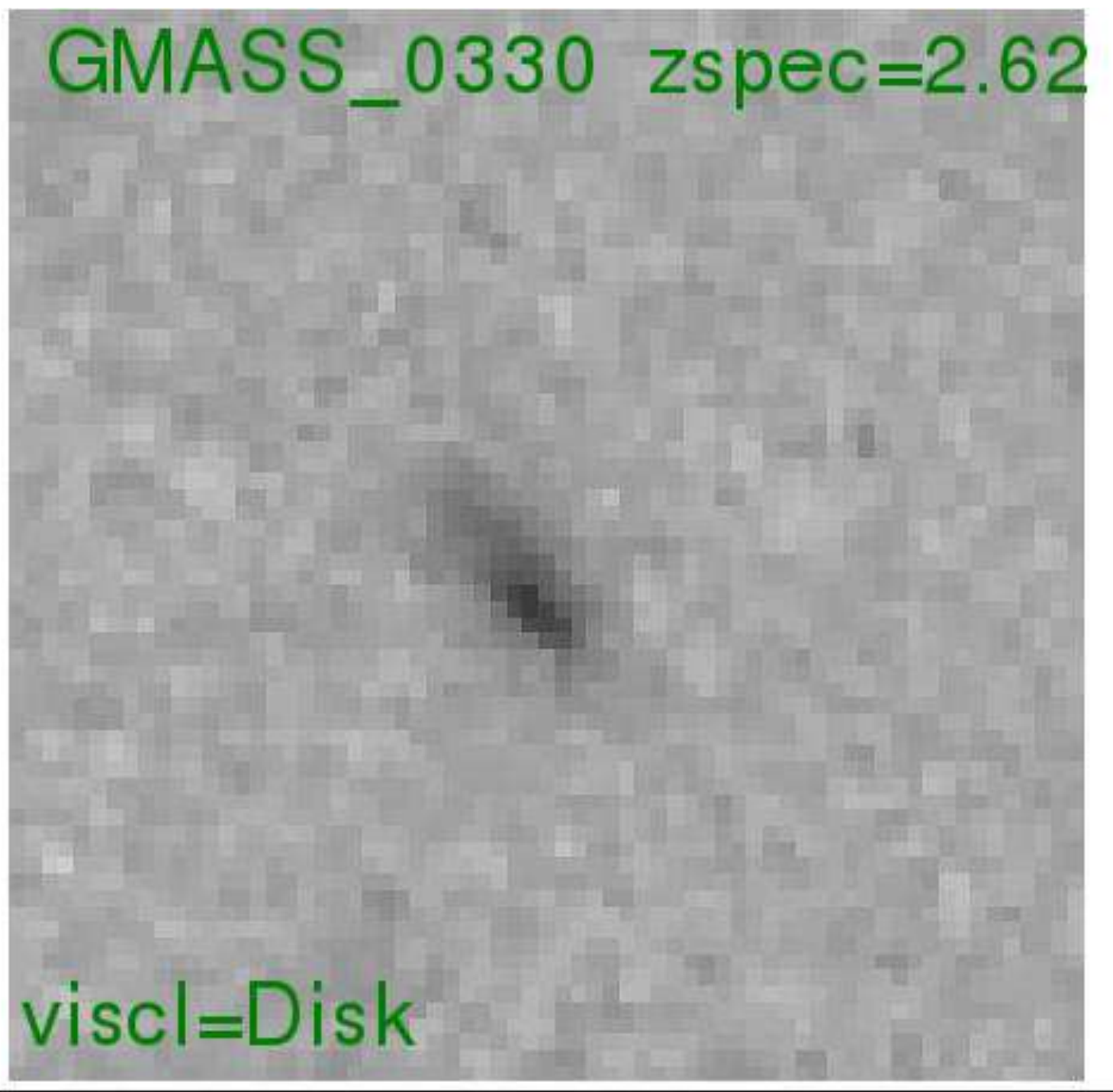}		     

\includegraphics[trim=100 40 75 390, clip=true, width=30mm]{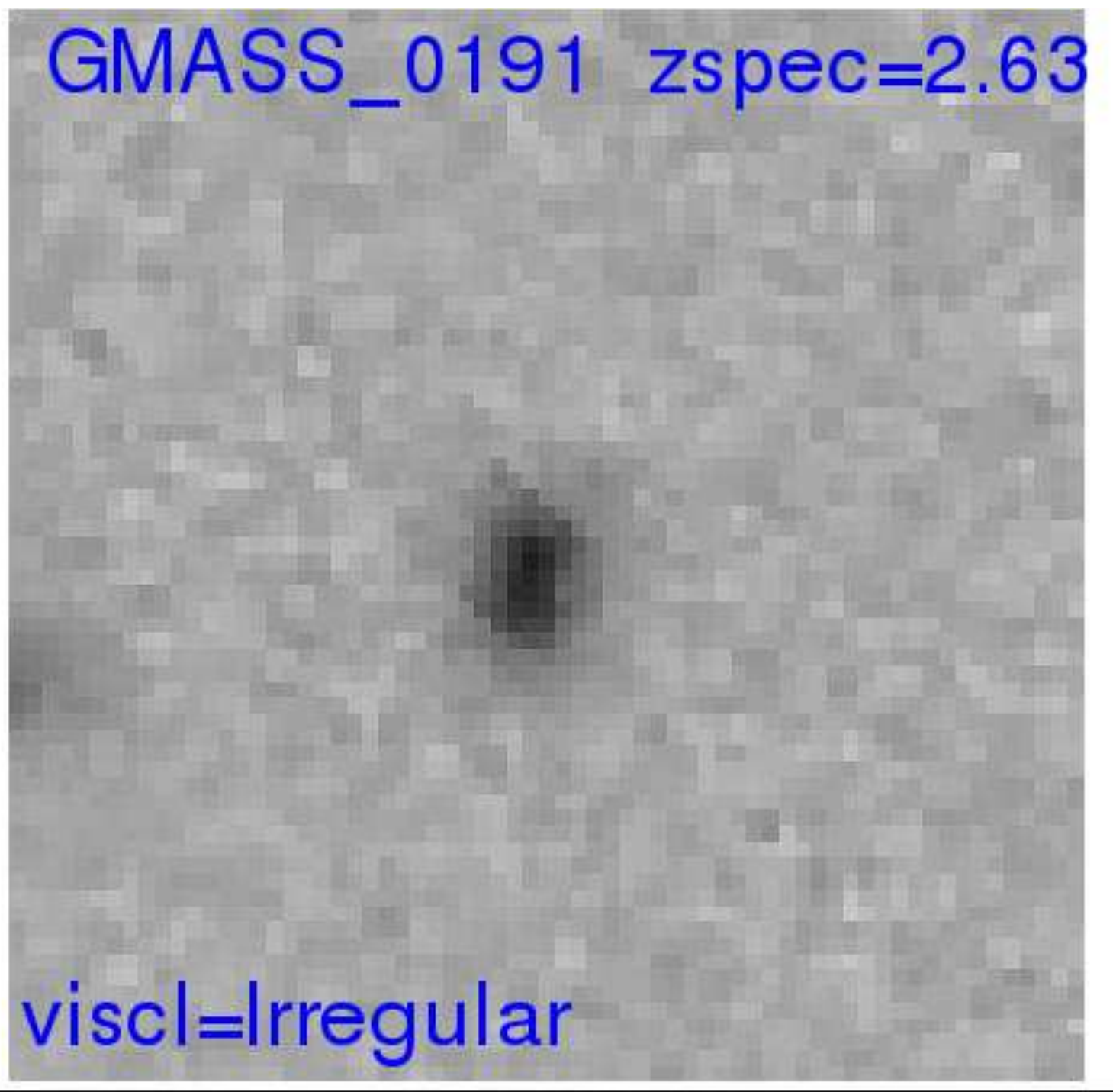}			     
\includegraphics[trim=100 40 75 390, clip=true, width=30mm]{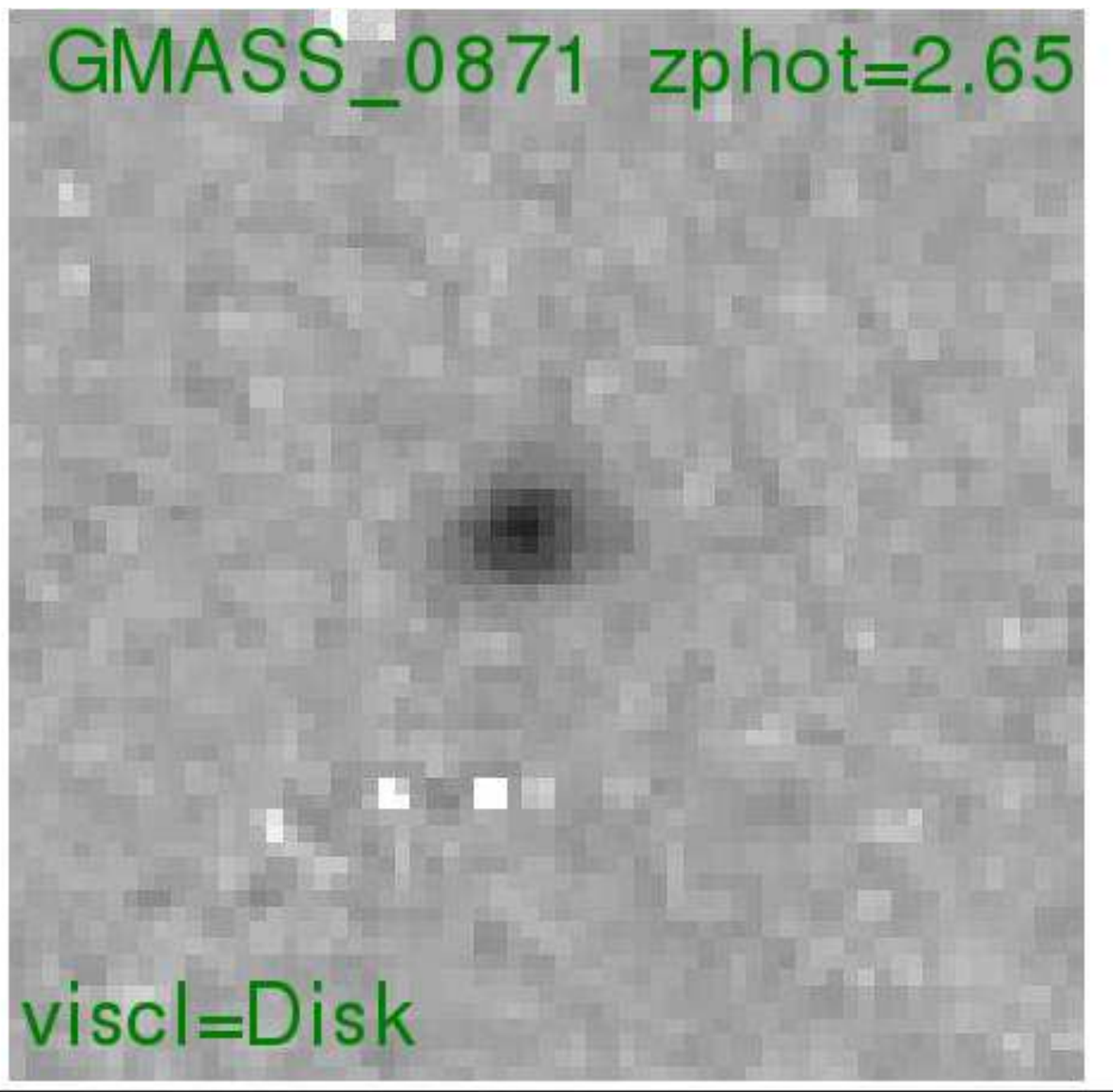}			     
\includegraphics[trim=100 40 75 390, clip=true, width=30mm]{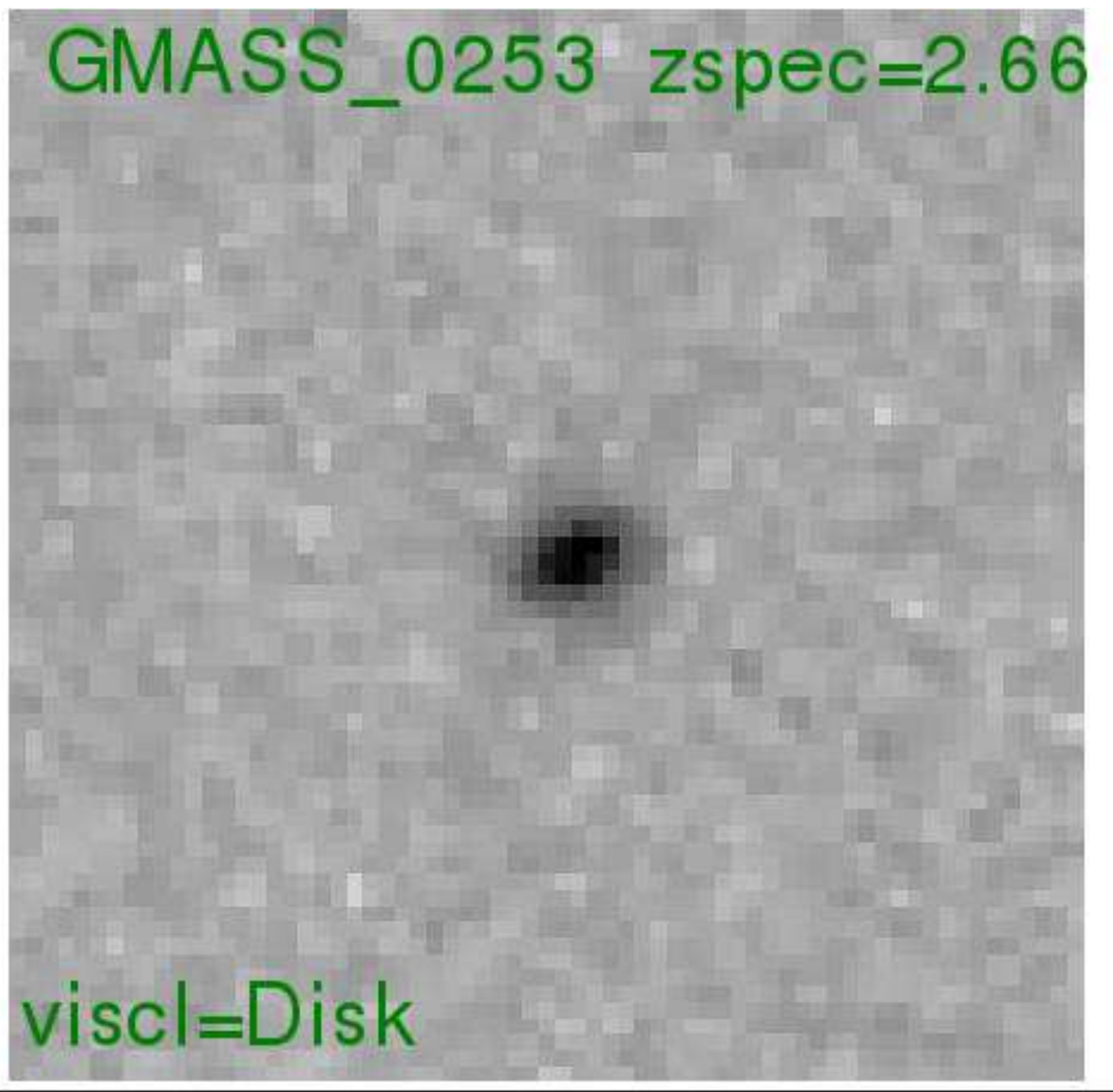}			     
\includegraphics[trim=100 40 75 390, clip=true, width=30mm]{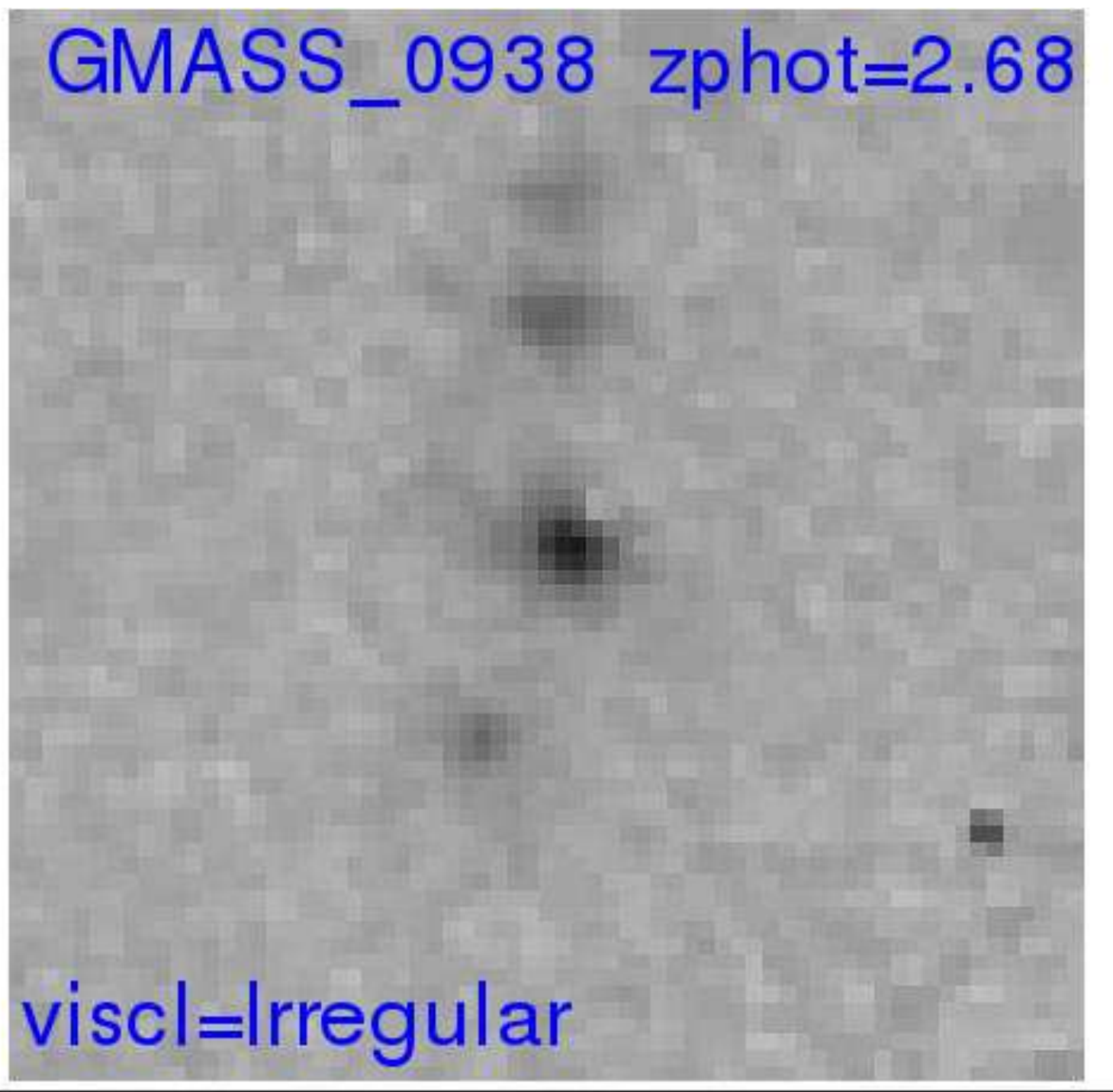}		     
\includegraphics[trim=100 40 75 390, clip=true, width=30mm]{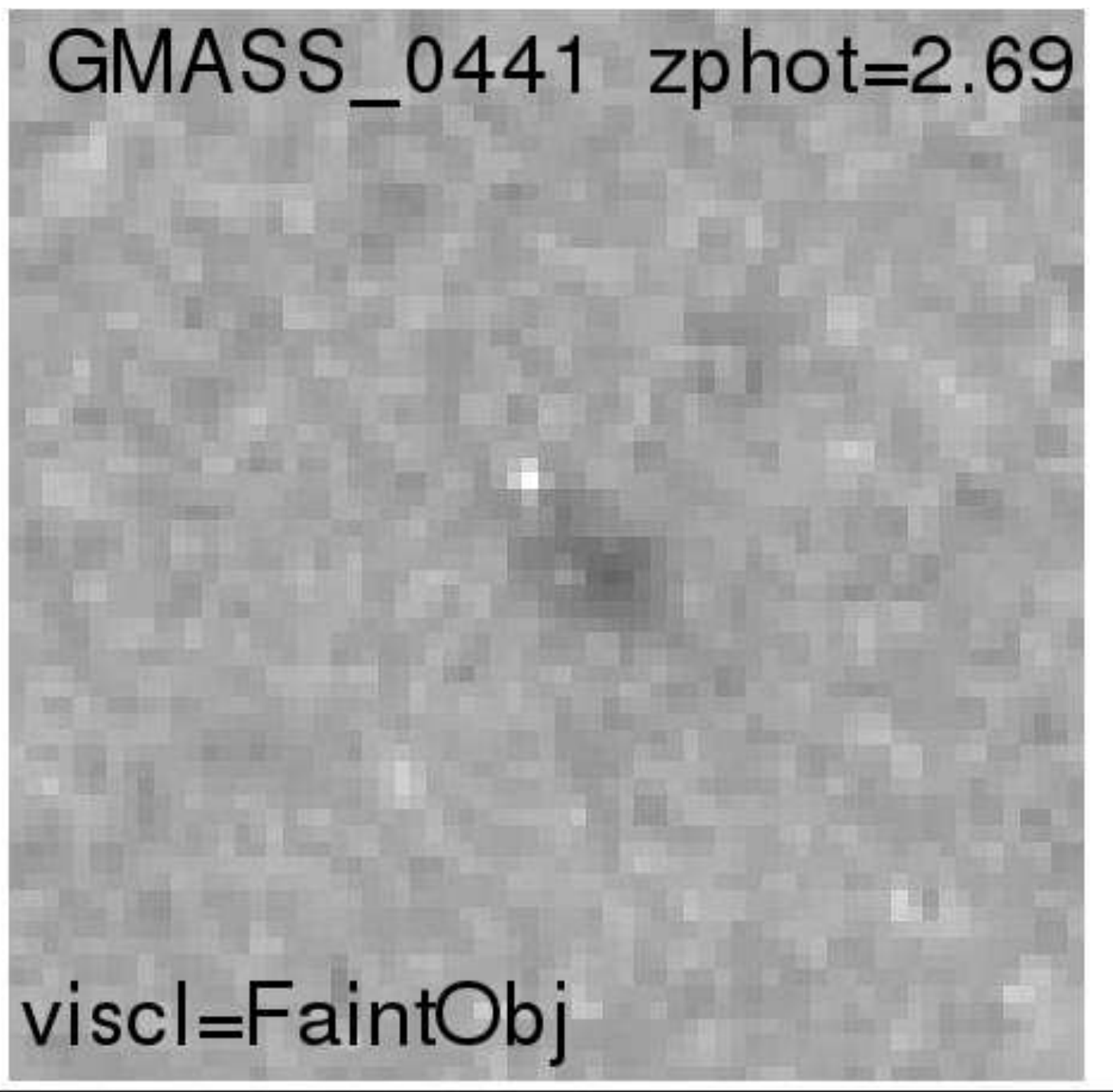}			     
\includegraphics[trim=100 40 75 390, clip=true, width=30mm]{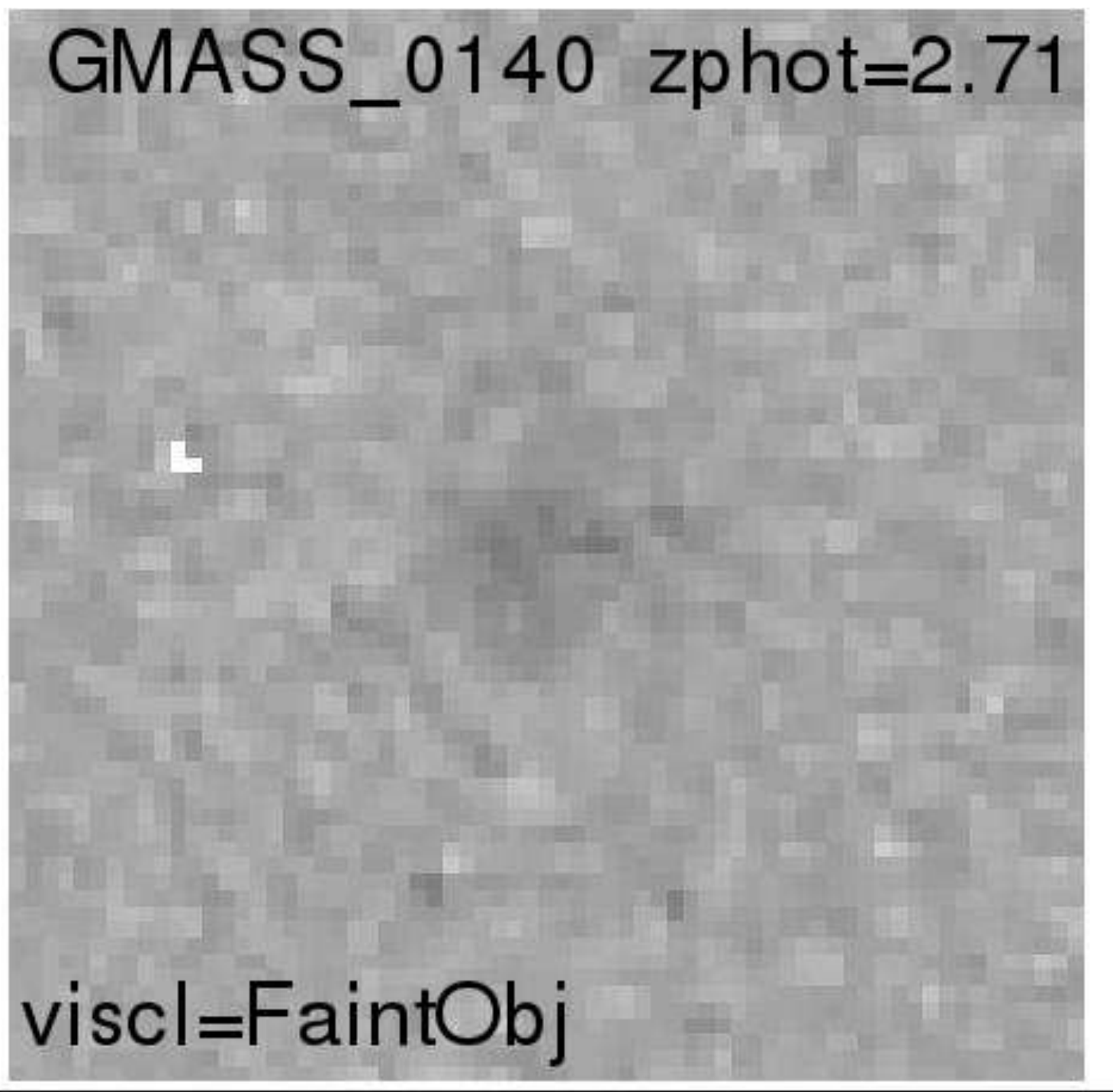}			     

\includegraphics[trim=100 40 75 390, clip=true, width=30mm]{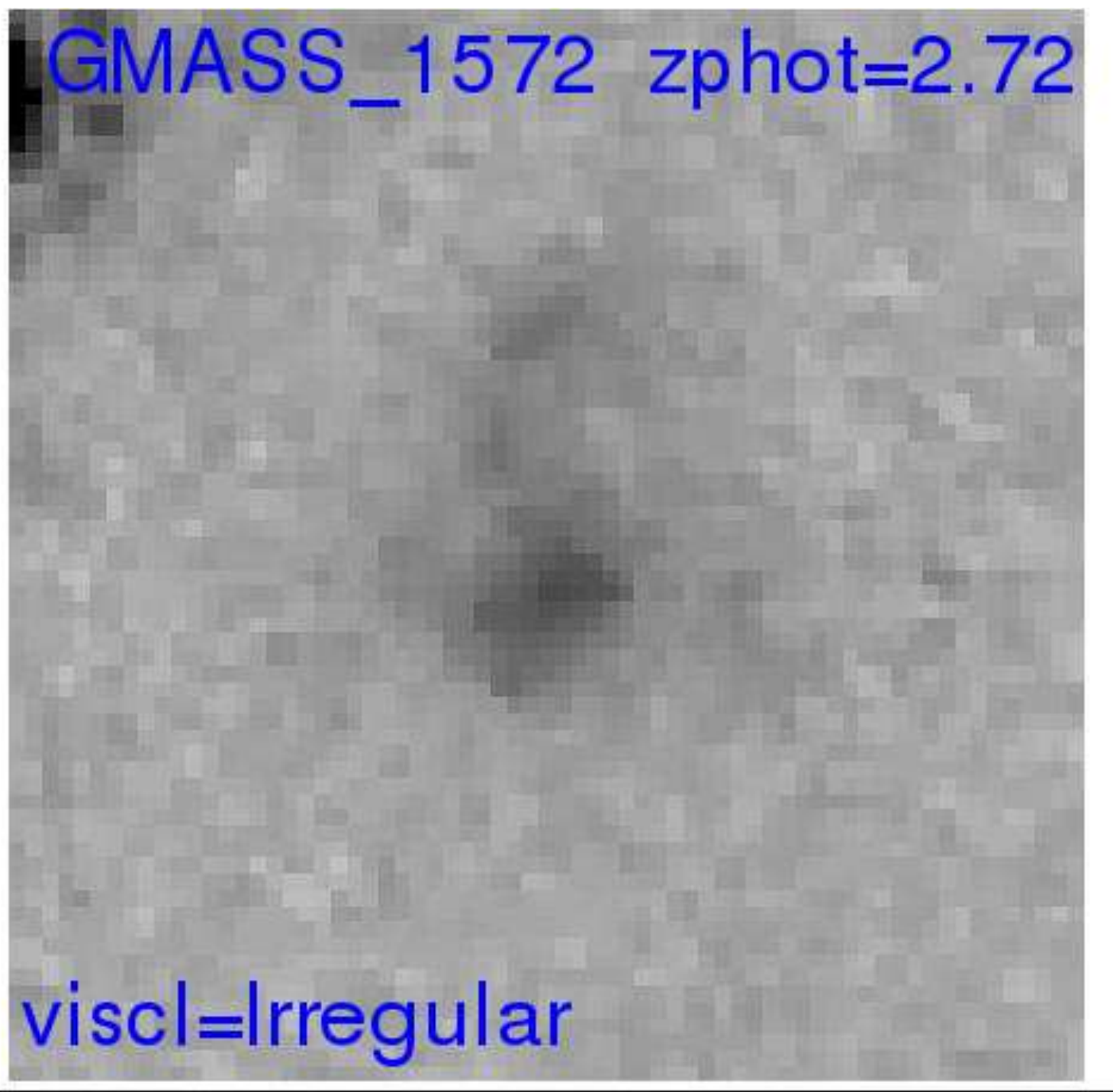}			     
\includegraphics[trim=100 40 75 390, clip=true, width=30mm]{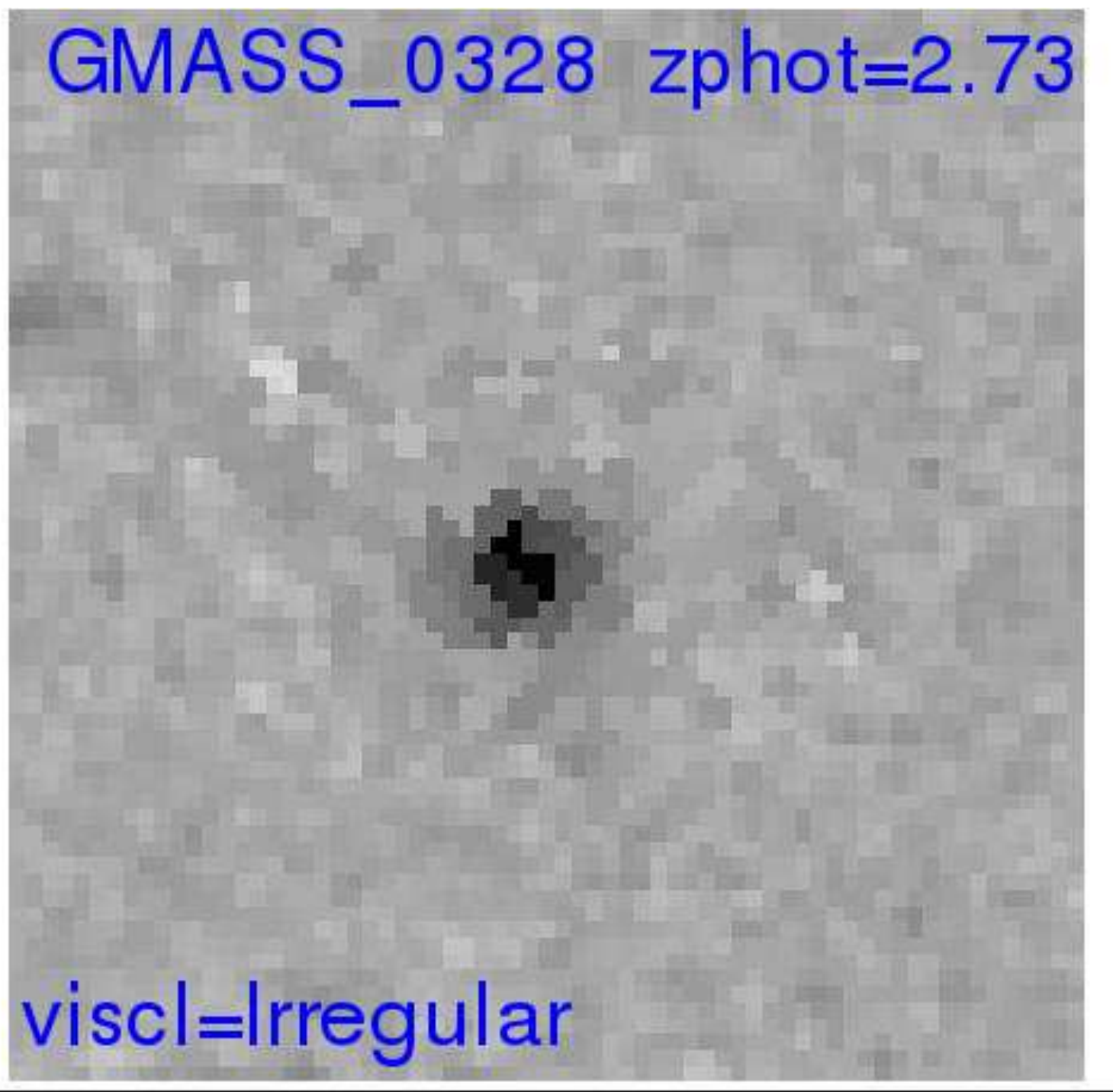}			     
\includegraphics[trim=100 40 75 390, clip=true, width=30mm]{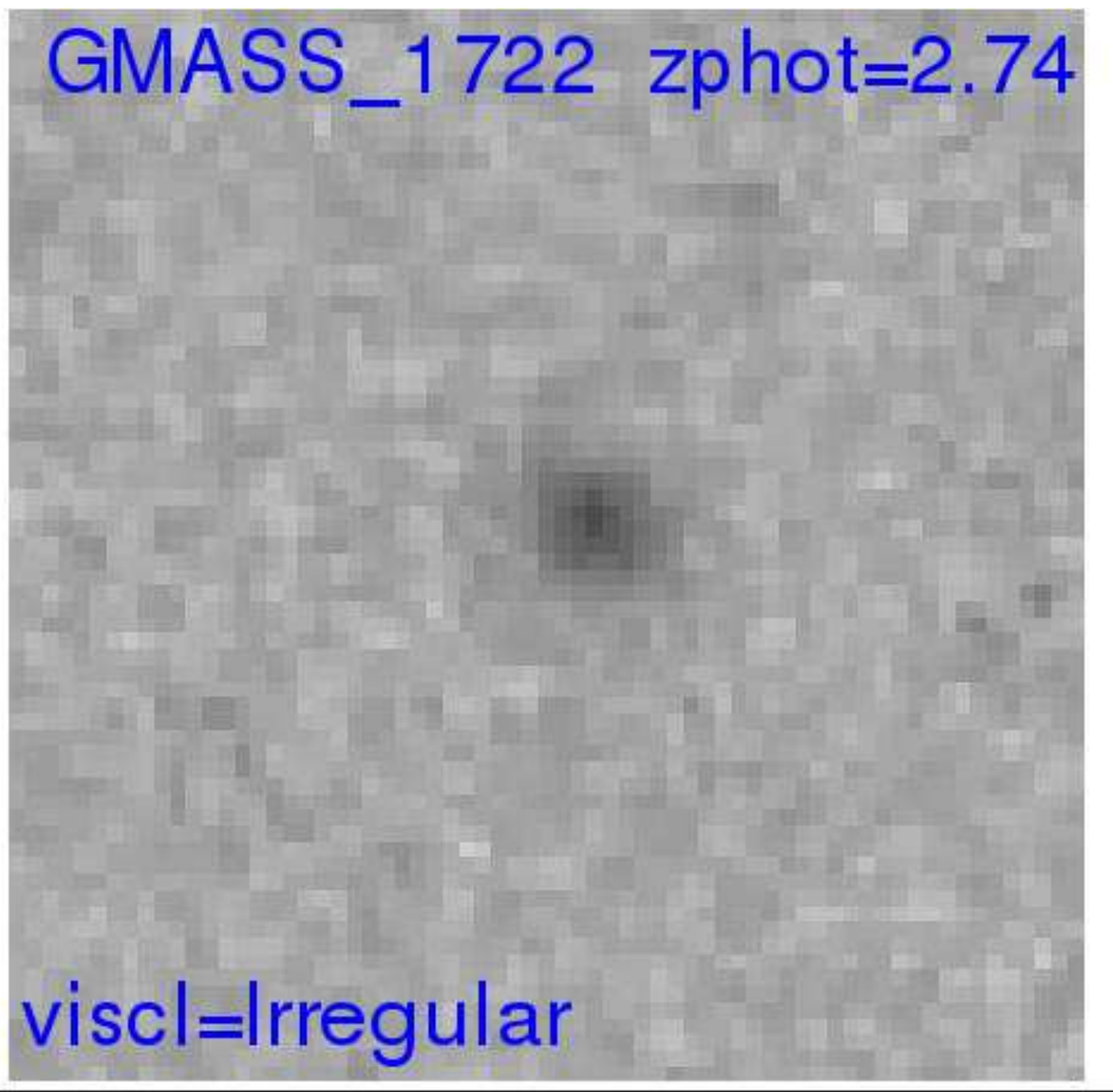}			     
\includegraphics[trim=100 40 75 390, clip=true, width=30mm]{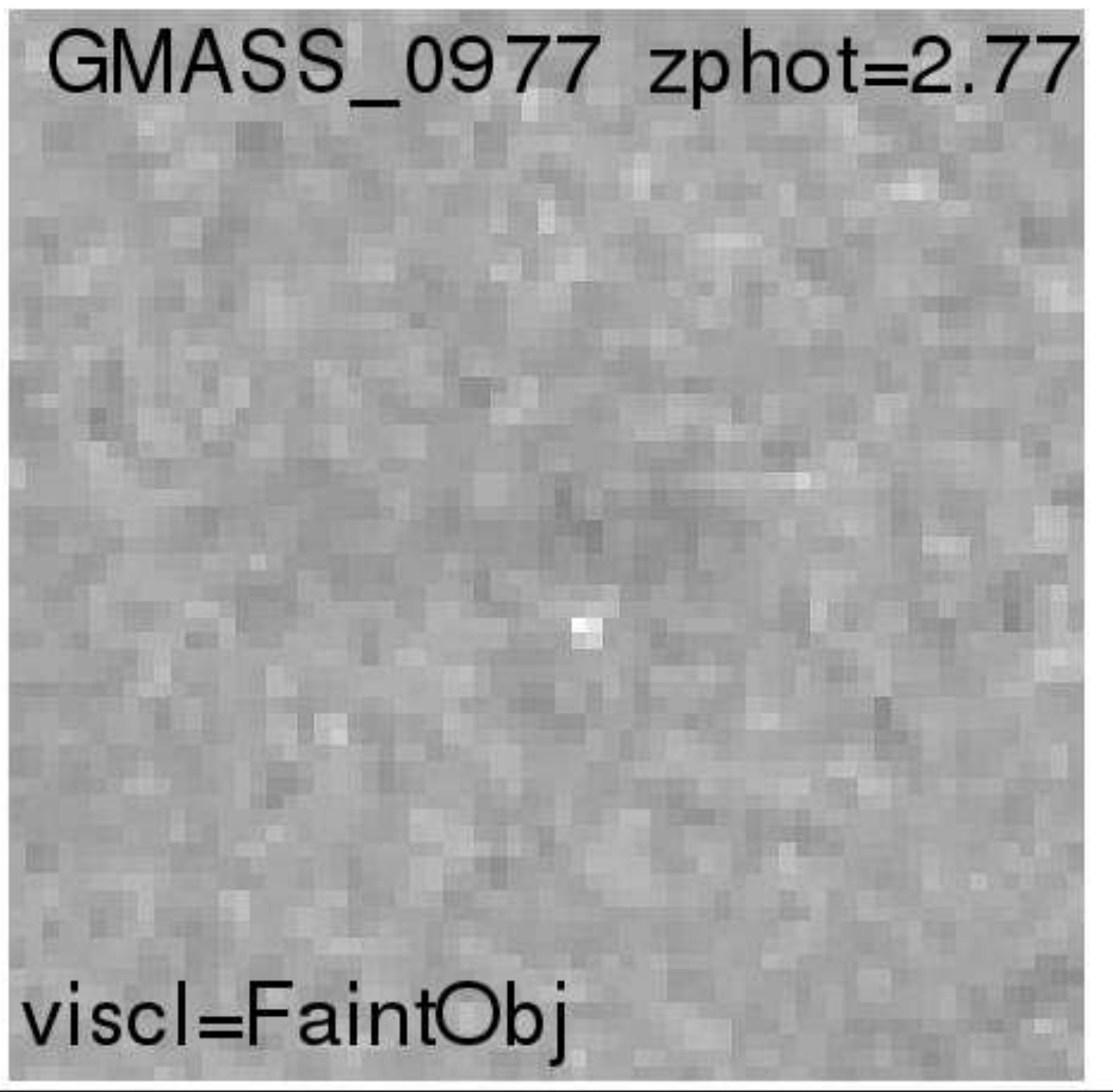}			     
\includegraphics[trim=100 40 75 390, clip=true, width=30mm]{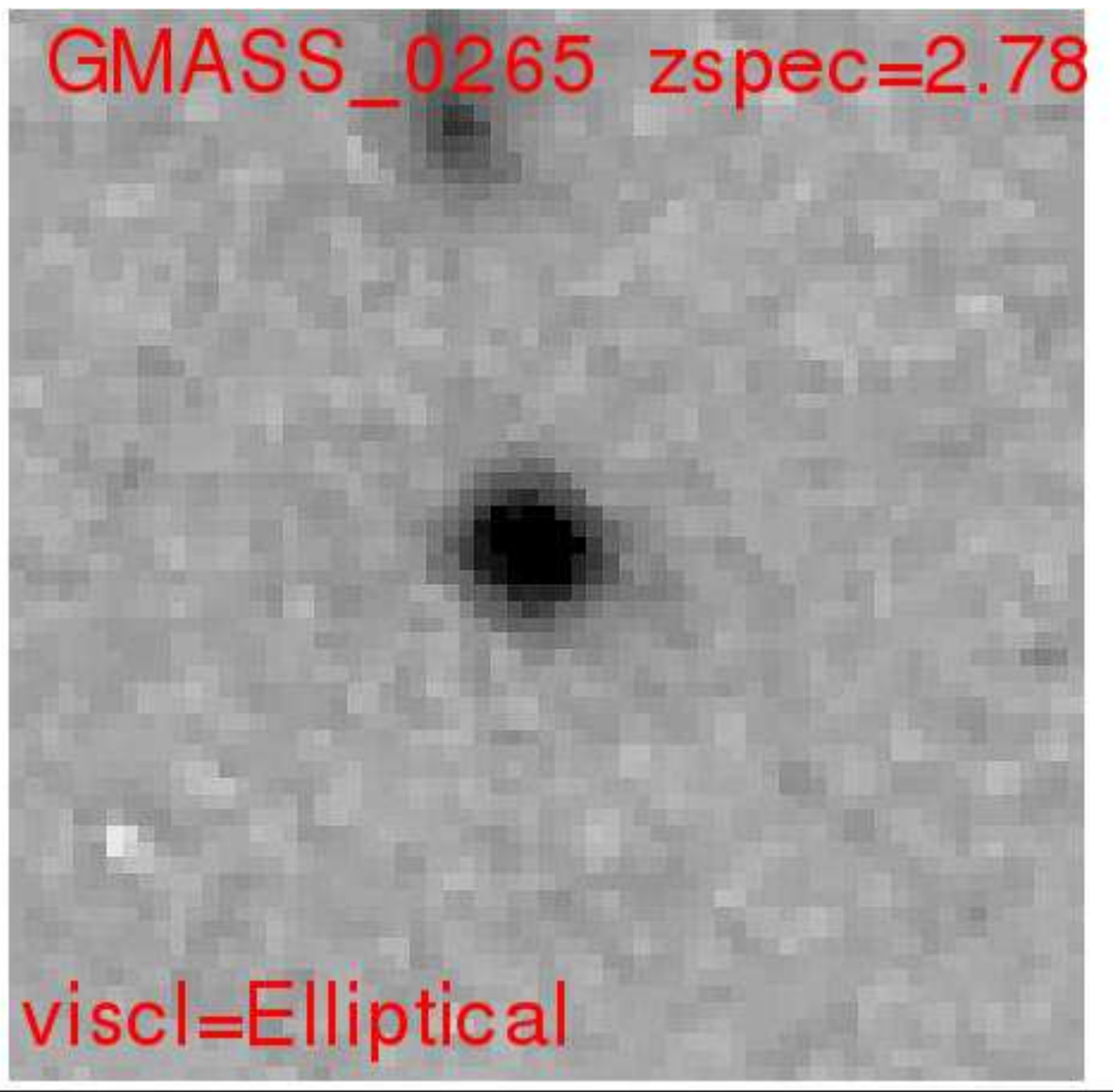}			     
\includegraphics[trim=100 40 75 390, clip=true, width=30mm]{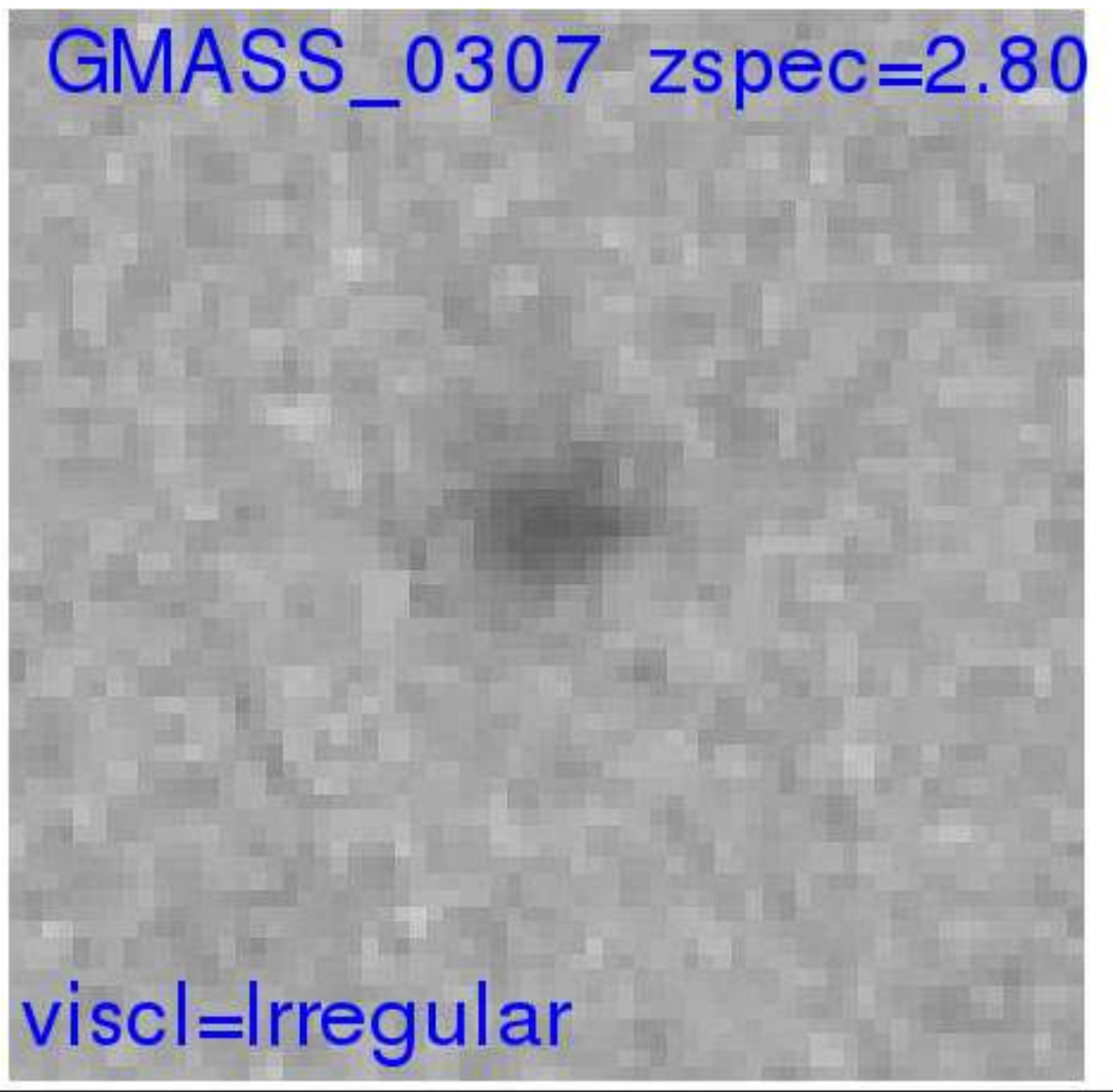}
\end{figure*}
\begin{figure*}
\centering   
\includegraphics[trim=100 40 75 390, clip=true, width=30mm]{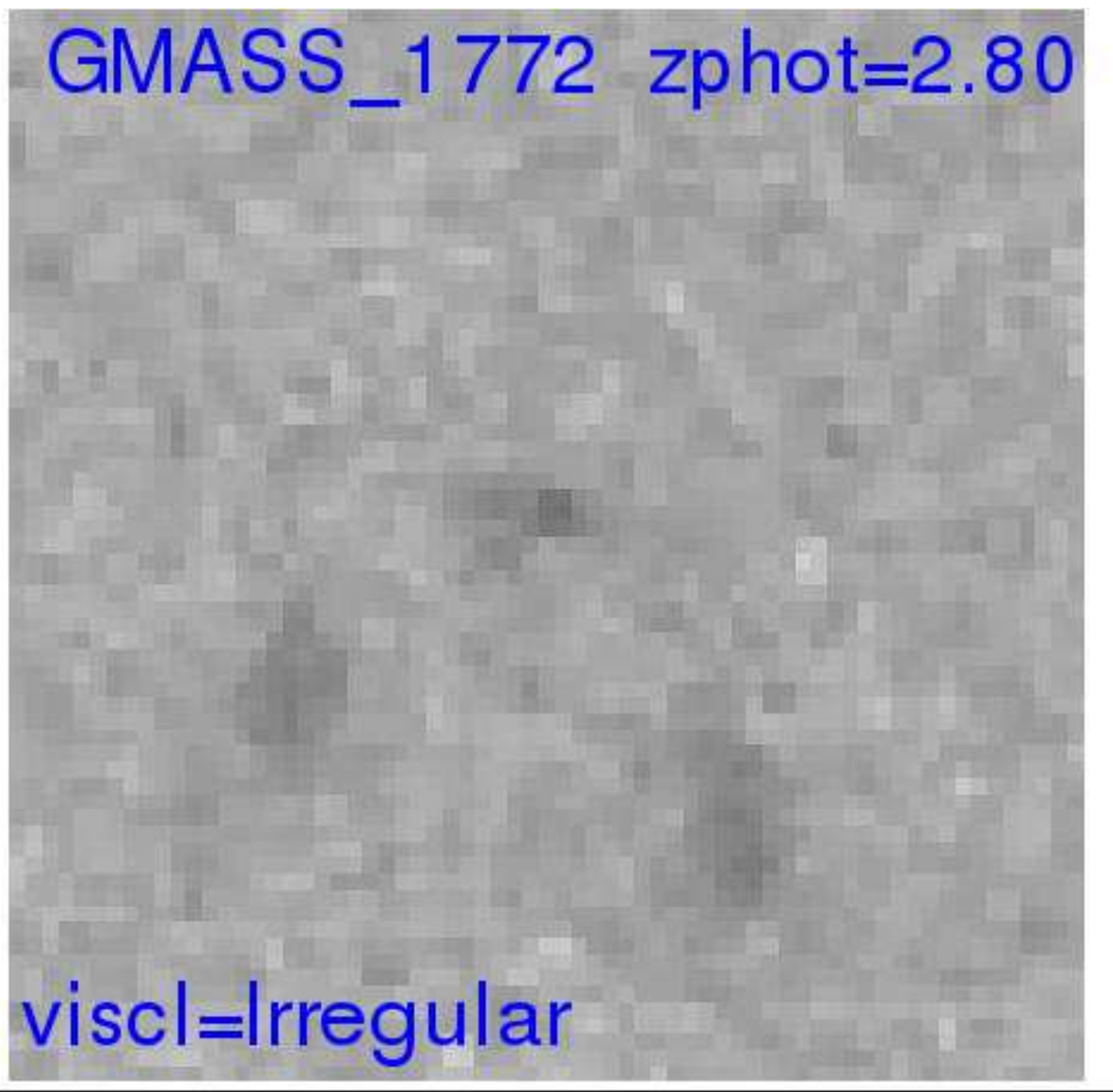}			     
\includegraphics[trim=100 40 75 390, clip=true, width=30mm]{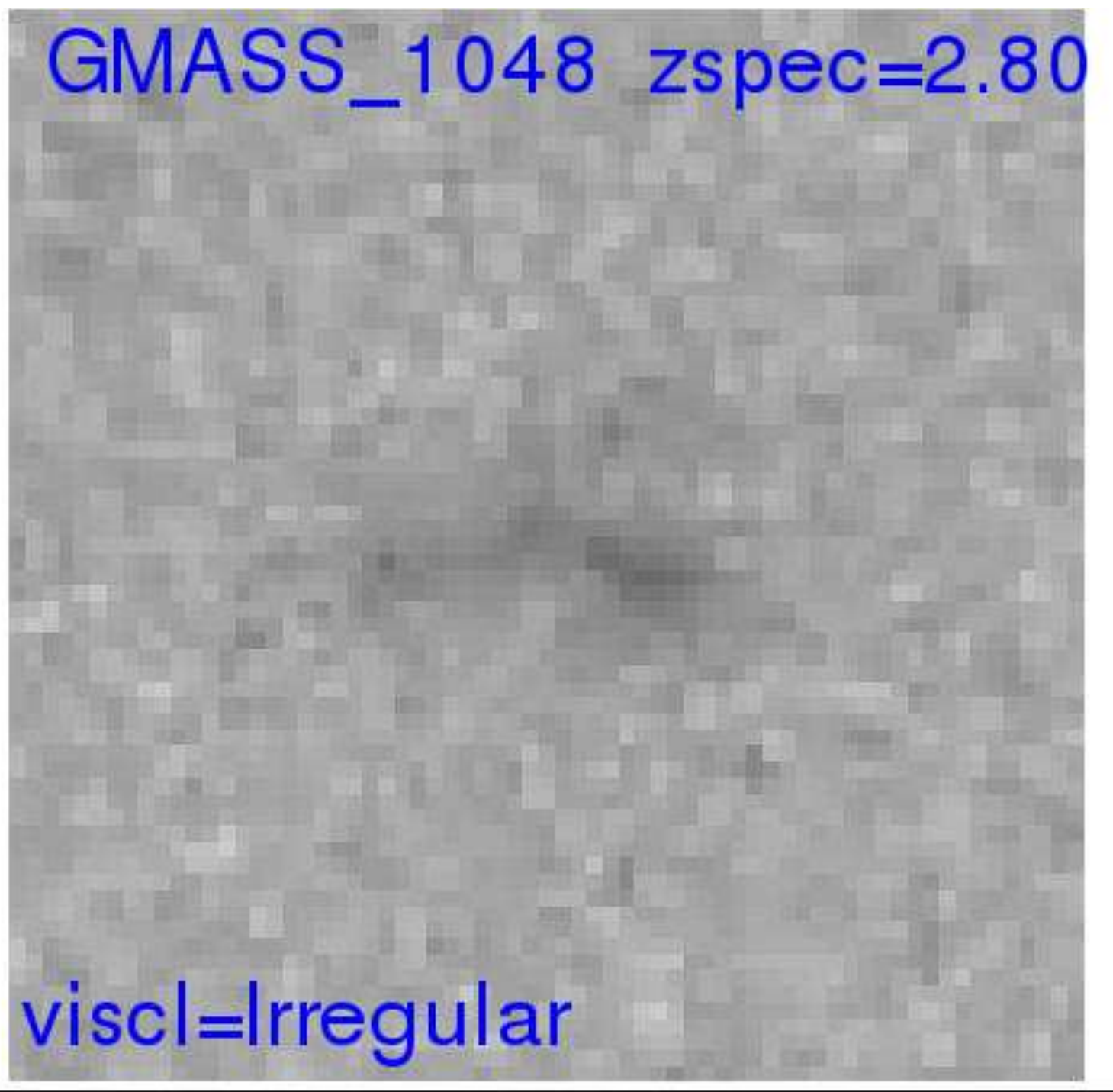}			     
\includegraphics[trim=100 40 75 390, clip=true, width=30mm]{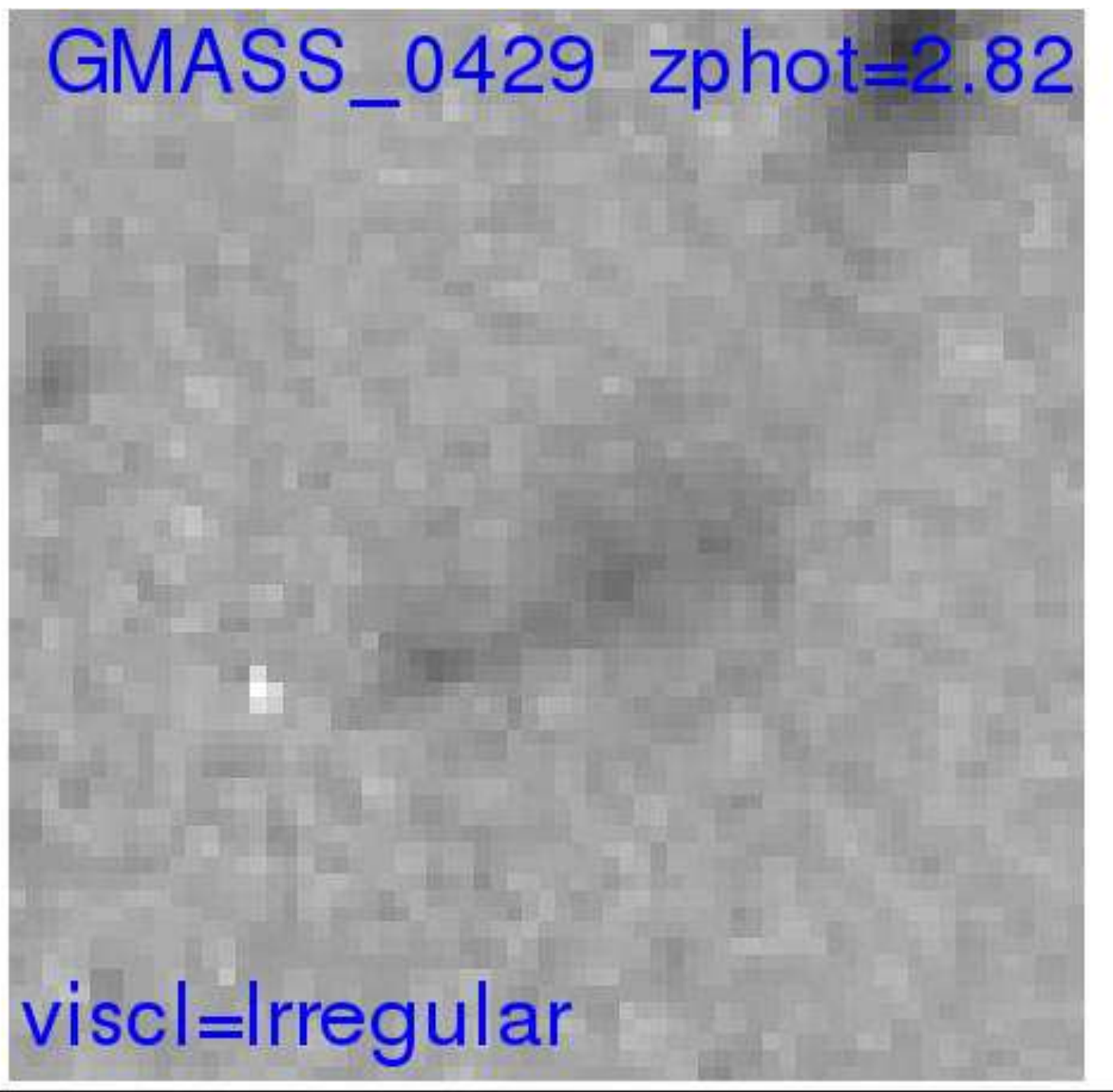}			     
\includegraphics[trim=100 40 75 390, clip=true, width=30mm]{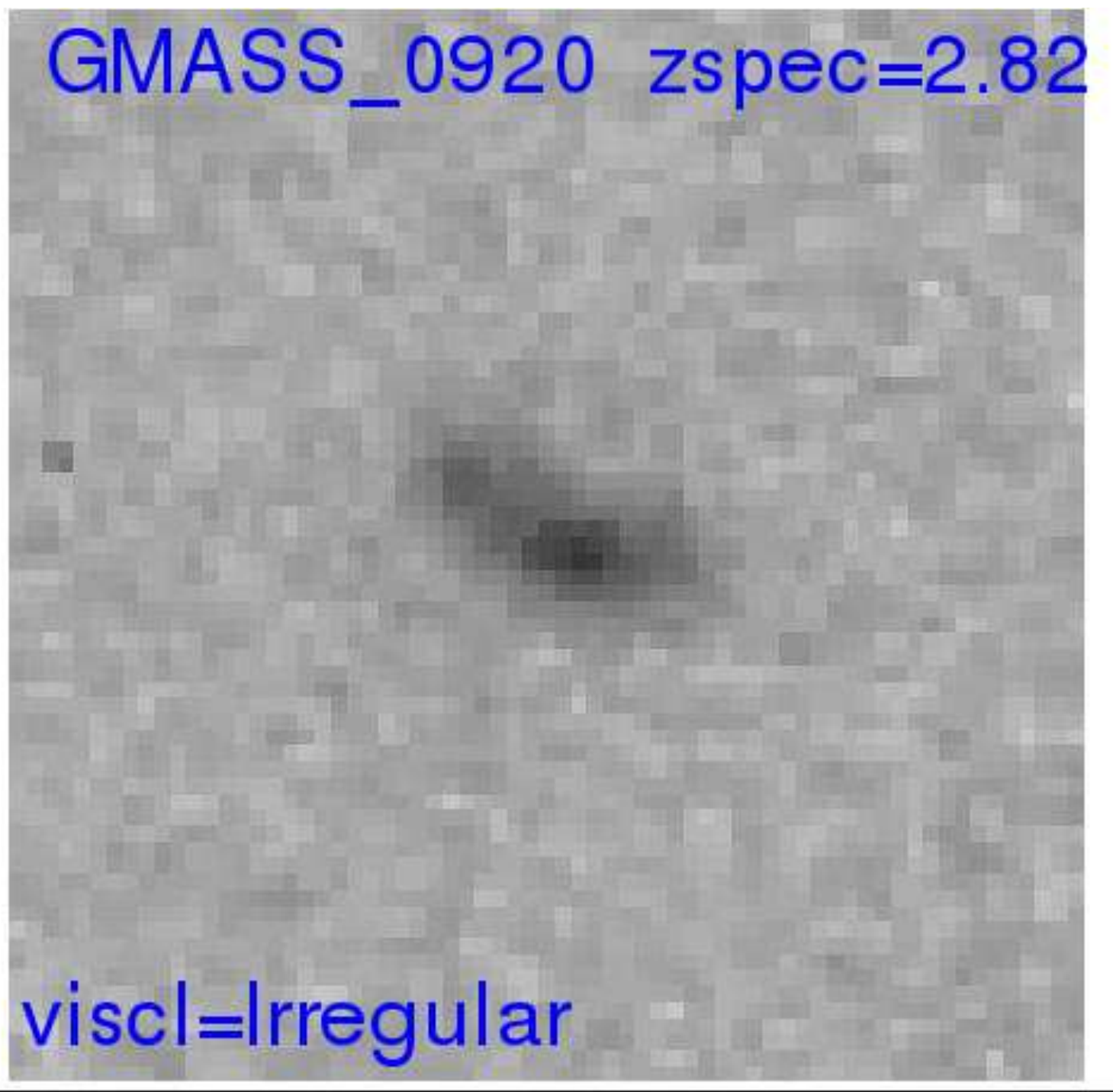}			     
\includegraphics[trim=100 40 75 390, clip=true, width=30mm]{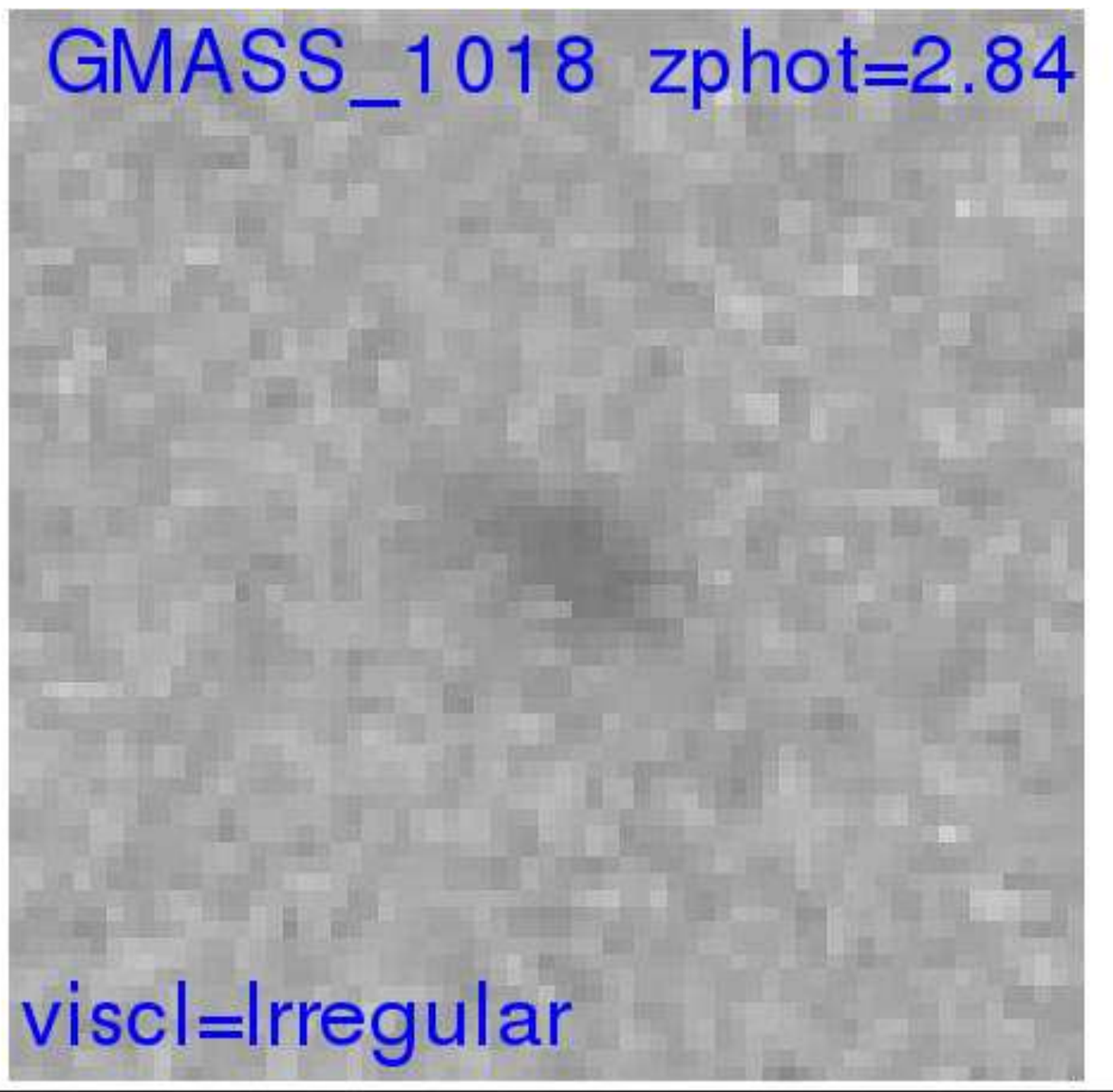}			     
\includegraphics[trim=100 40 75 390, clip=true, width=30mm]{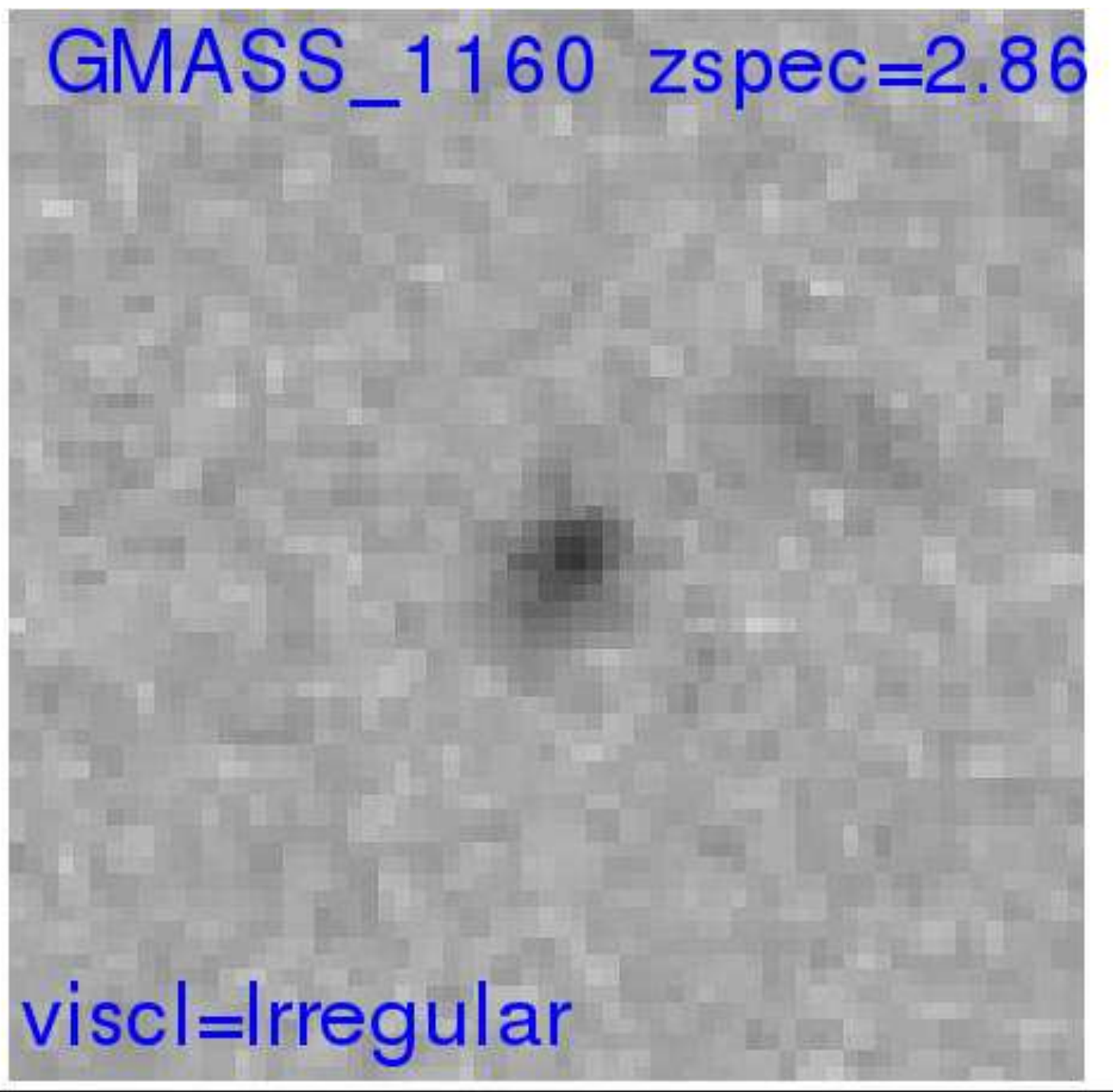}			     

\includegraphics[trim=100 40 75 390, clip=true, width=30mm]{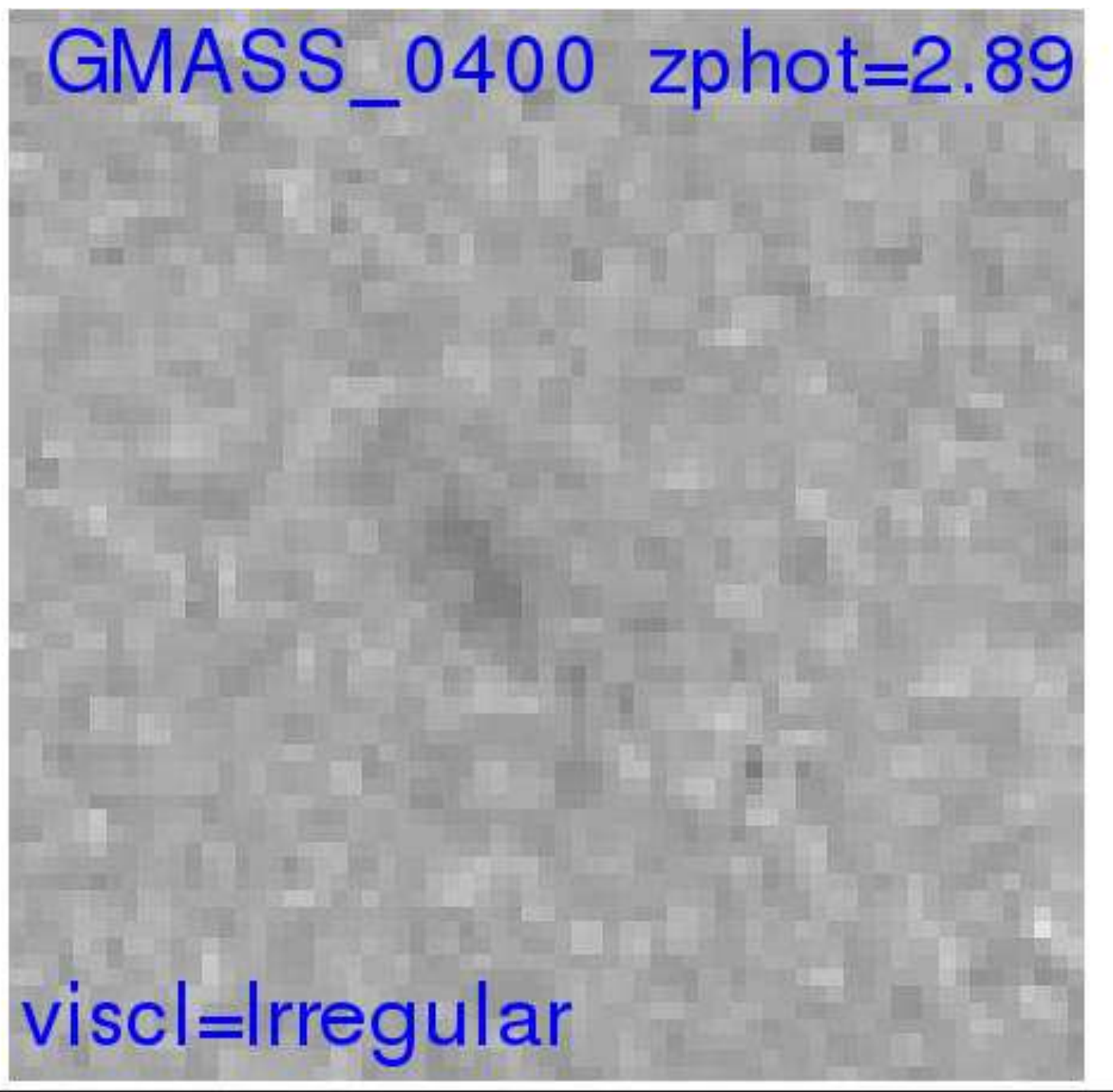}			     
\includegraphics[trim=100 40 75 390, clip=true, width=30mm]{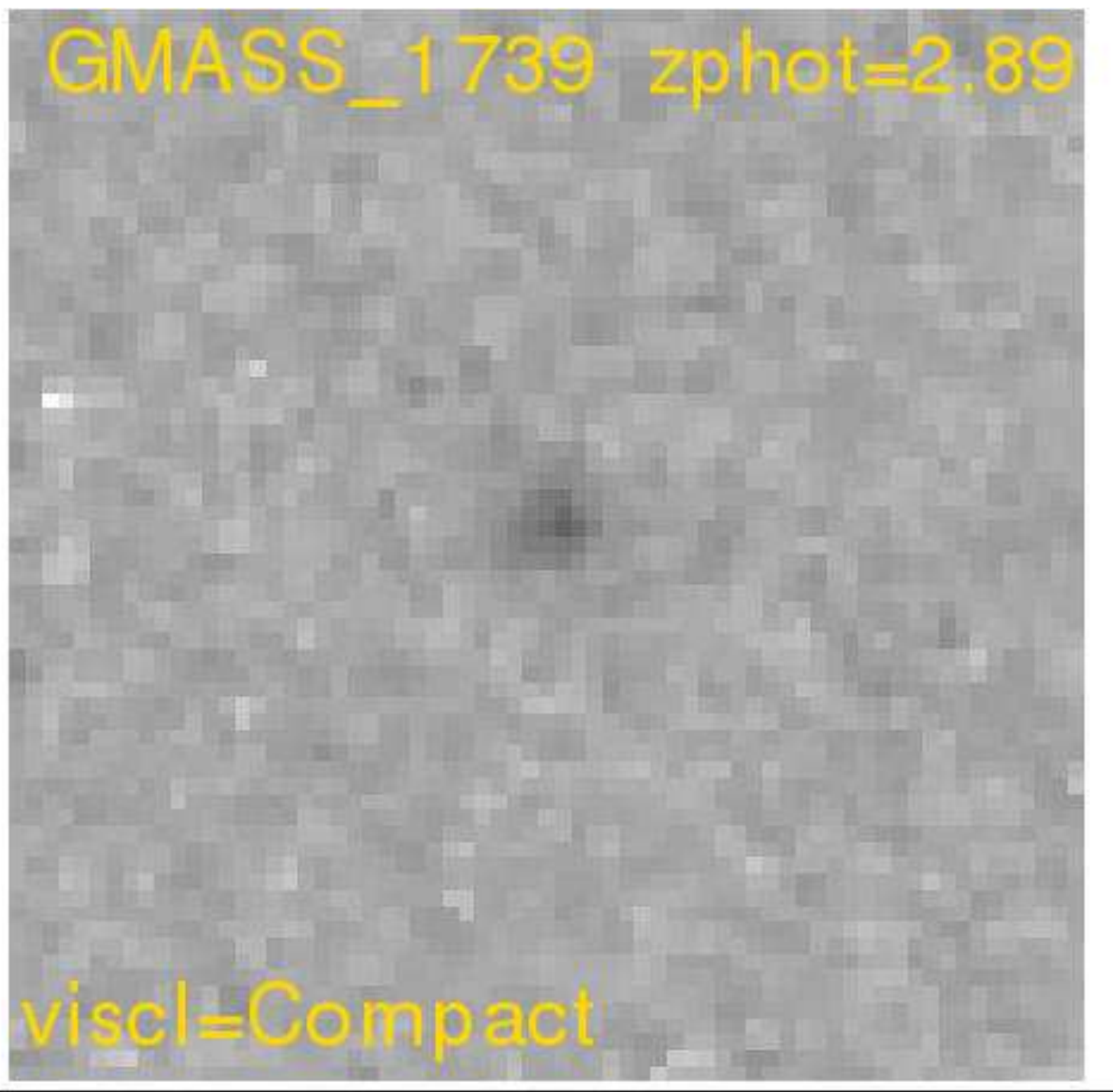}		     
\includegraphics[trim=100 40 75 390, clip=true, width=30mm]{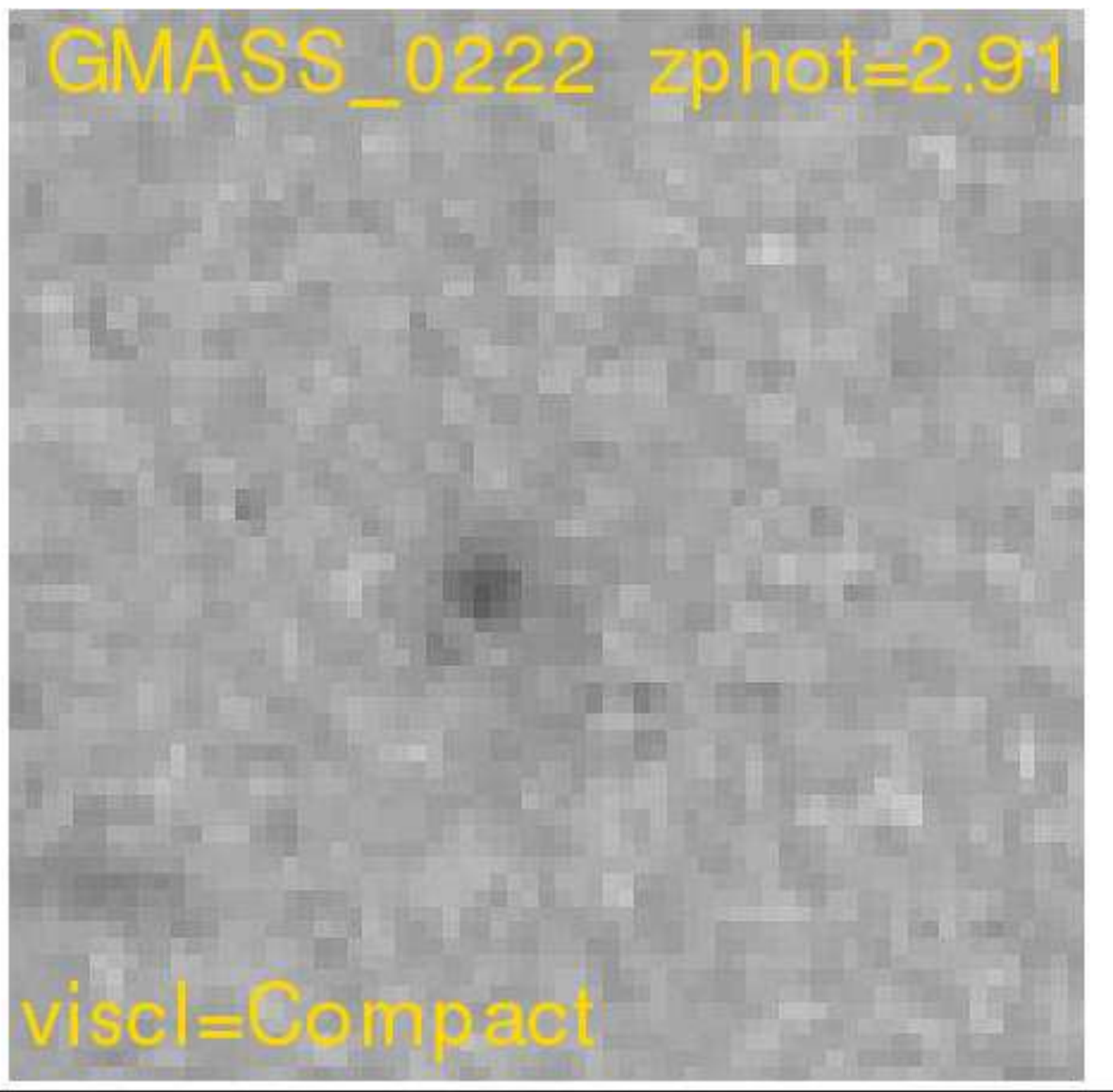}			     
\includegraphics[trim=100 40 75 390, clip=true, width=30mm]{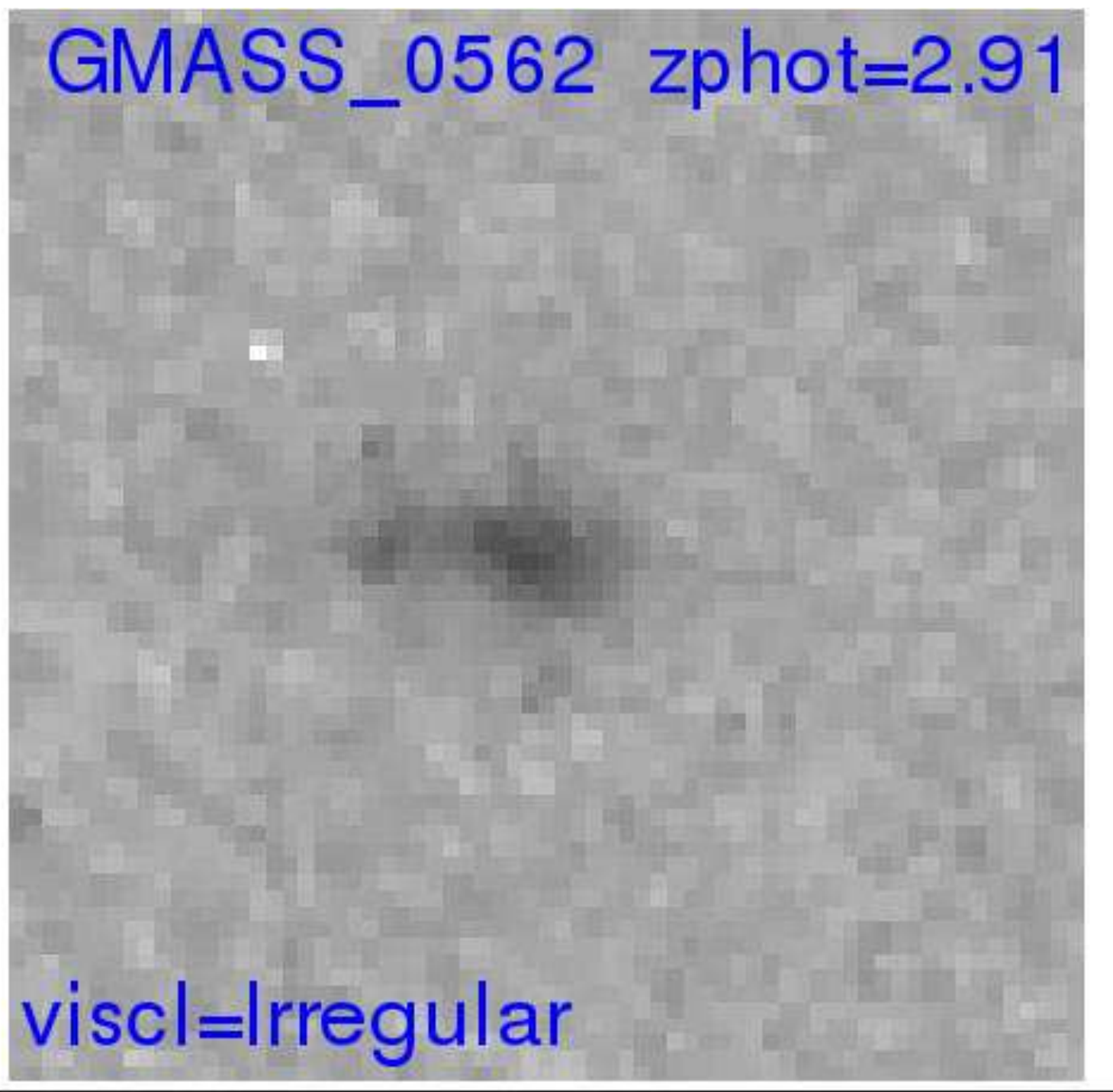}			     
\includegraphics[trim=100 40 75 390, clip=true, width=30mm]{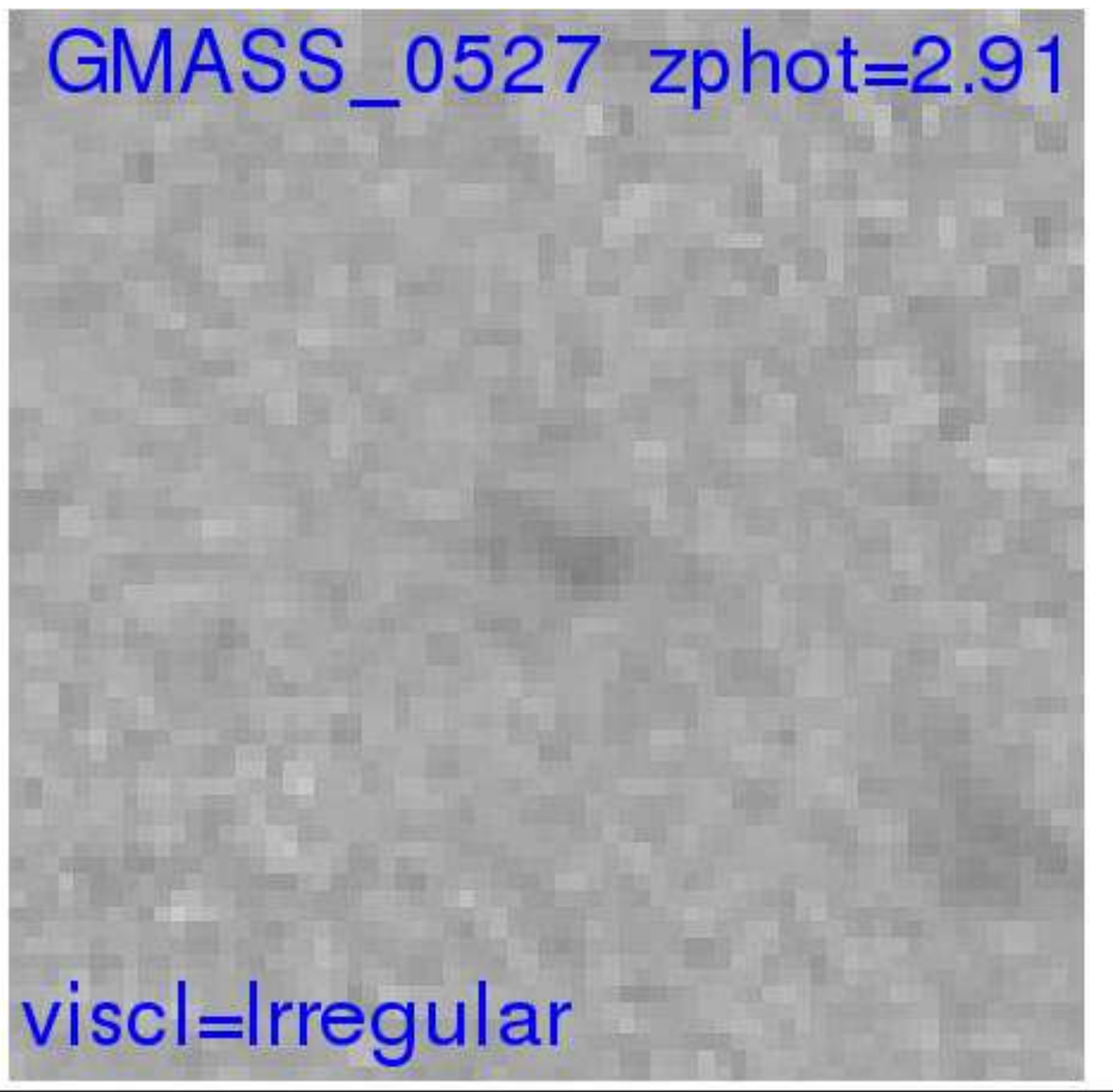}			     
\includegraphics[trim=100 40 75 390, clip=true, width=30mm]{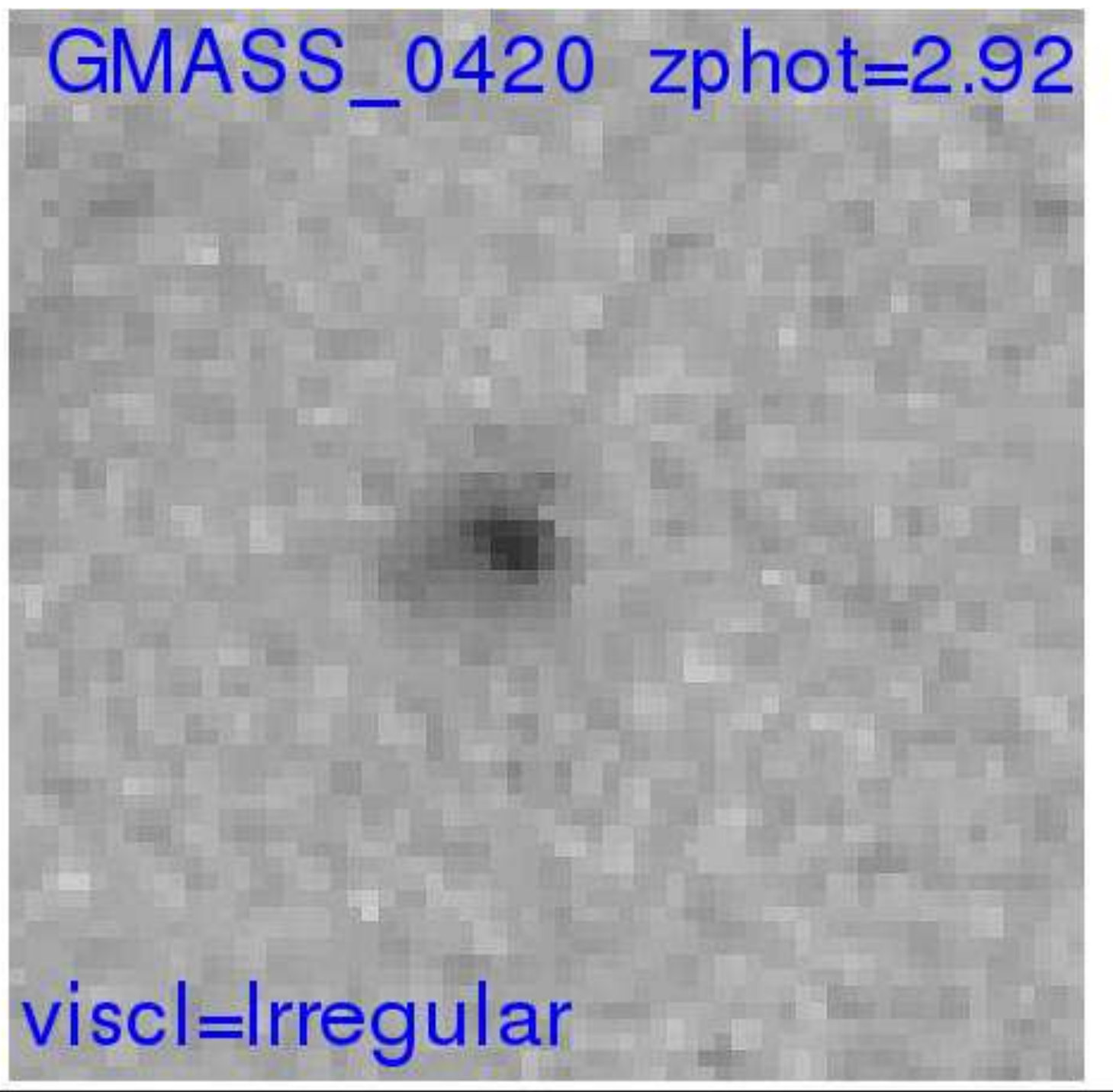}			     

\includegraphics[trim=100 40 75 390, clip=true, width=30mm]{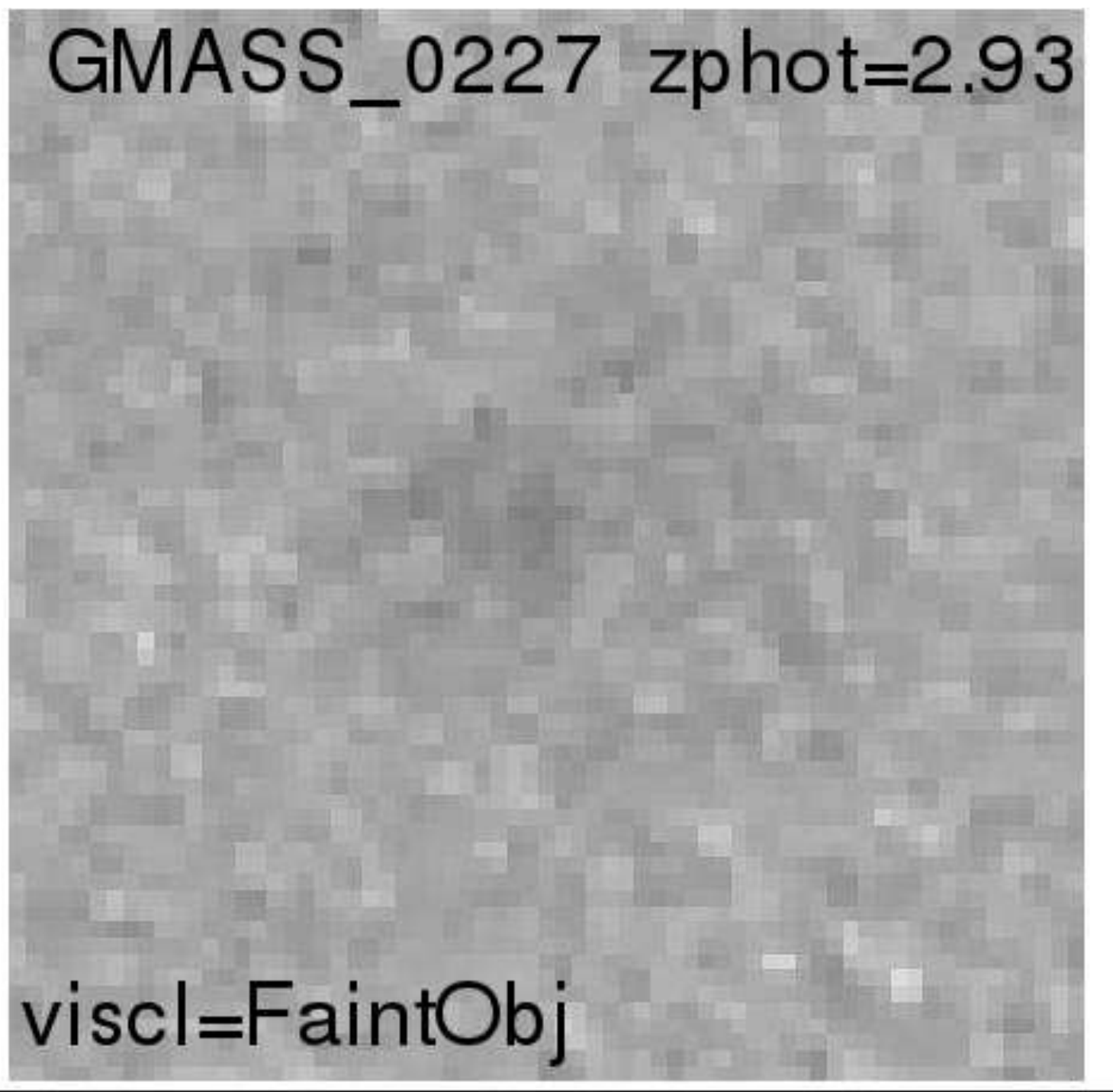}			     
\includegraphics[trim=100 40 75 390, clip=true, width=30mm]{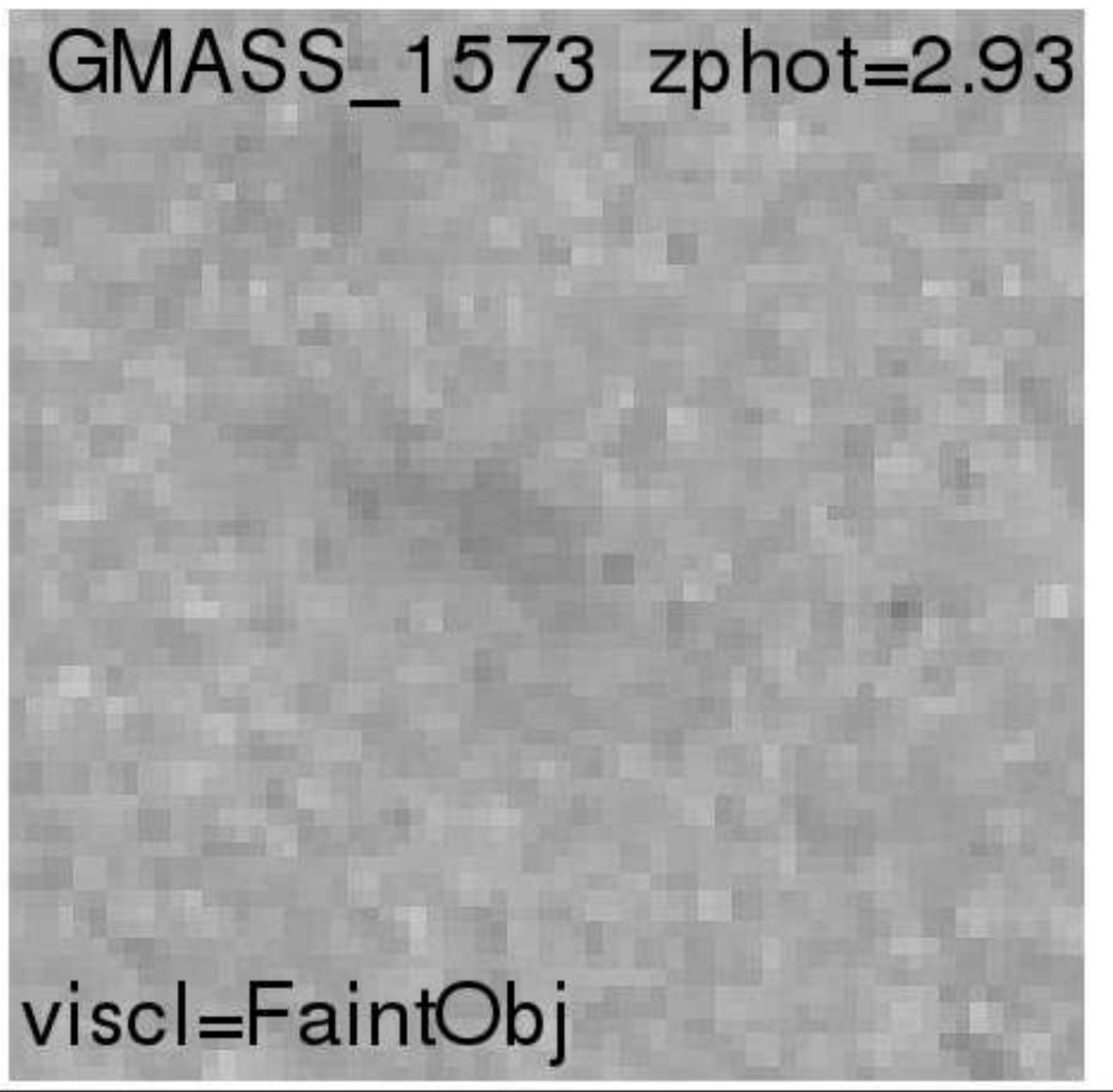}			     
\includegraphics[trim=100 40 75 390, clip=true, width=30mm]{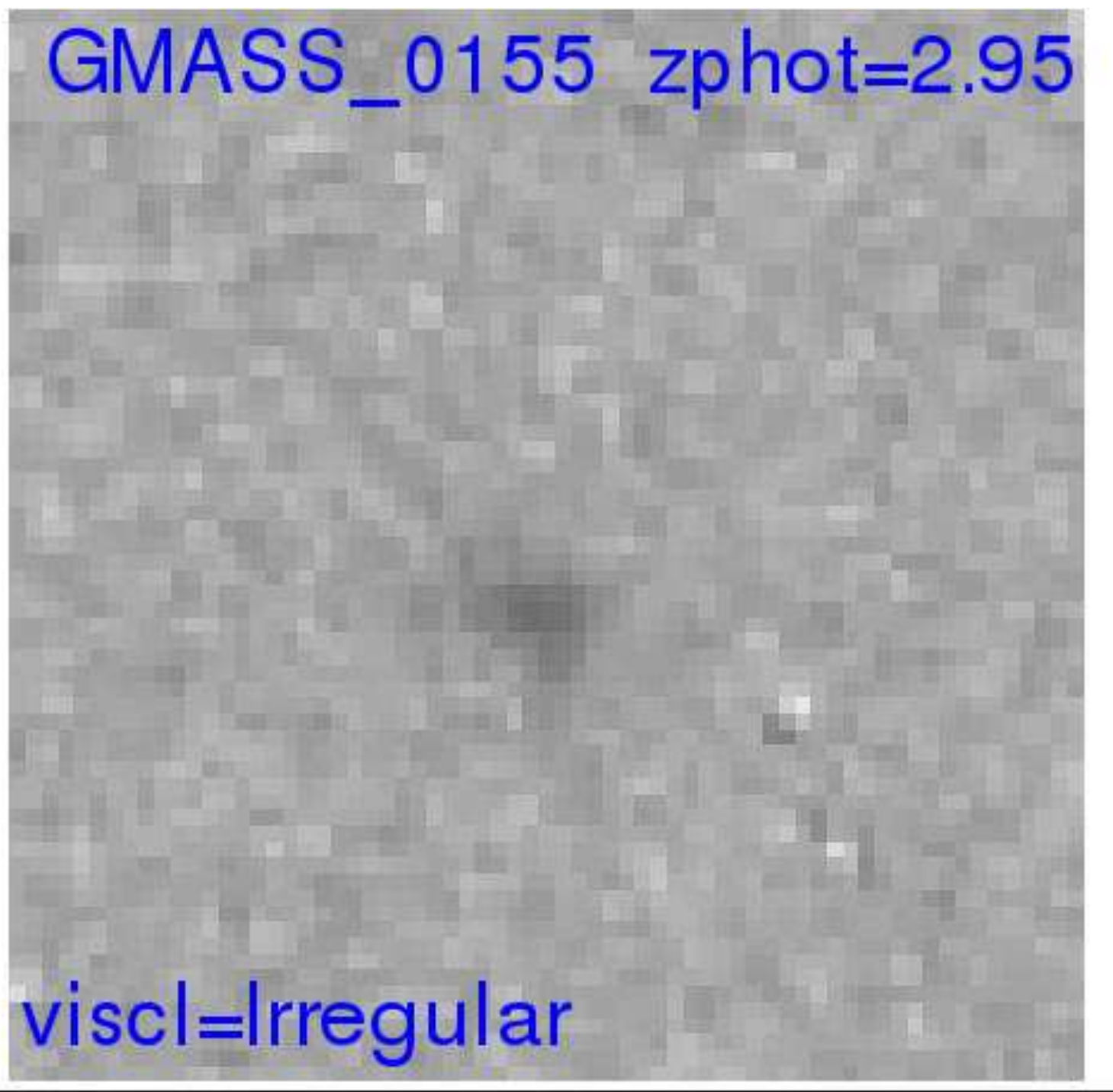}			     
\includegraphics[trim=100 40 75 390, clip=true, width=30mm]{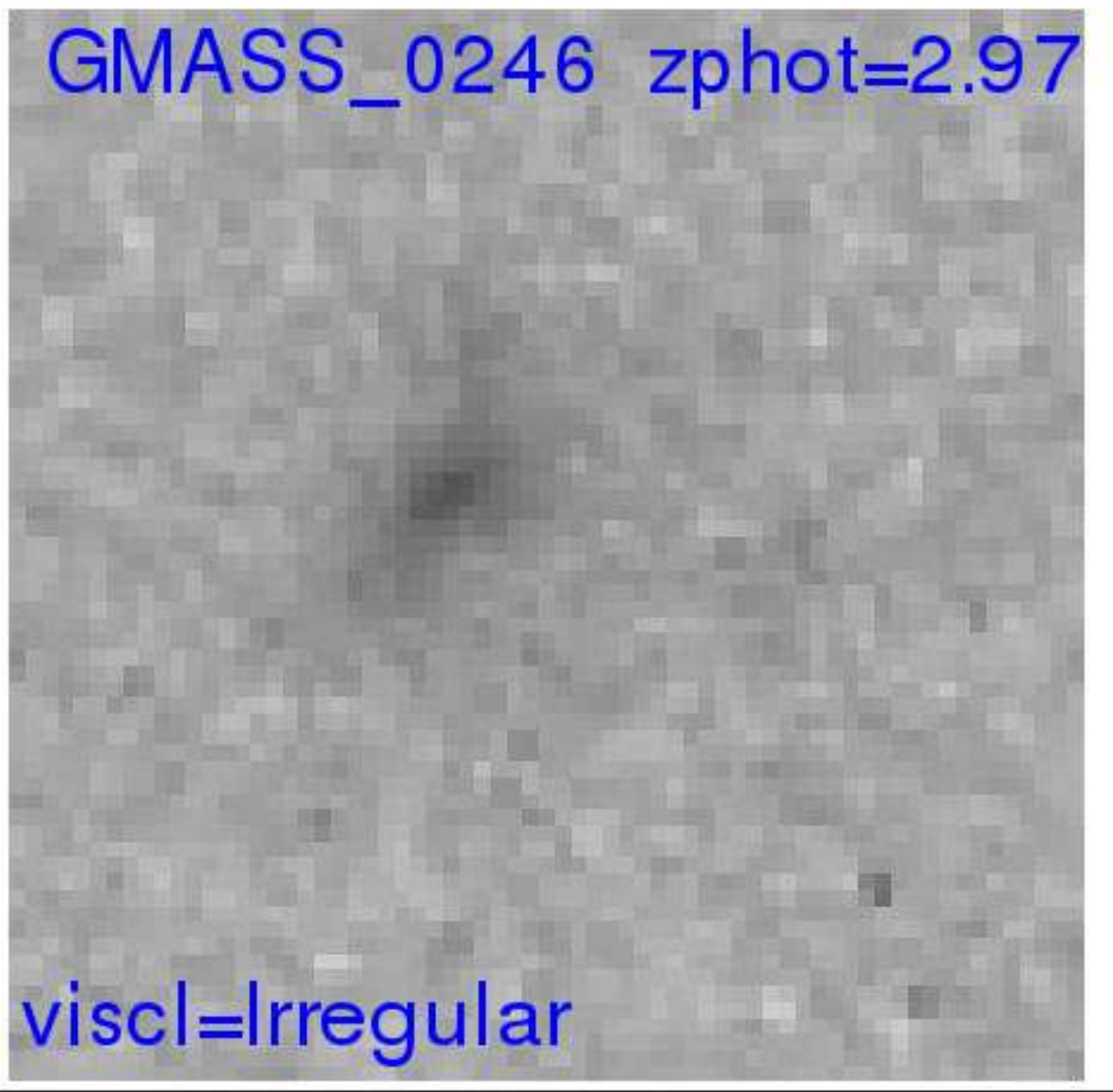}			     
\includegraphics[trim=100 40 75 390, clip=true, width=30mm]{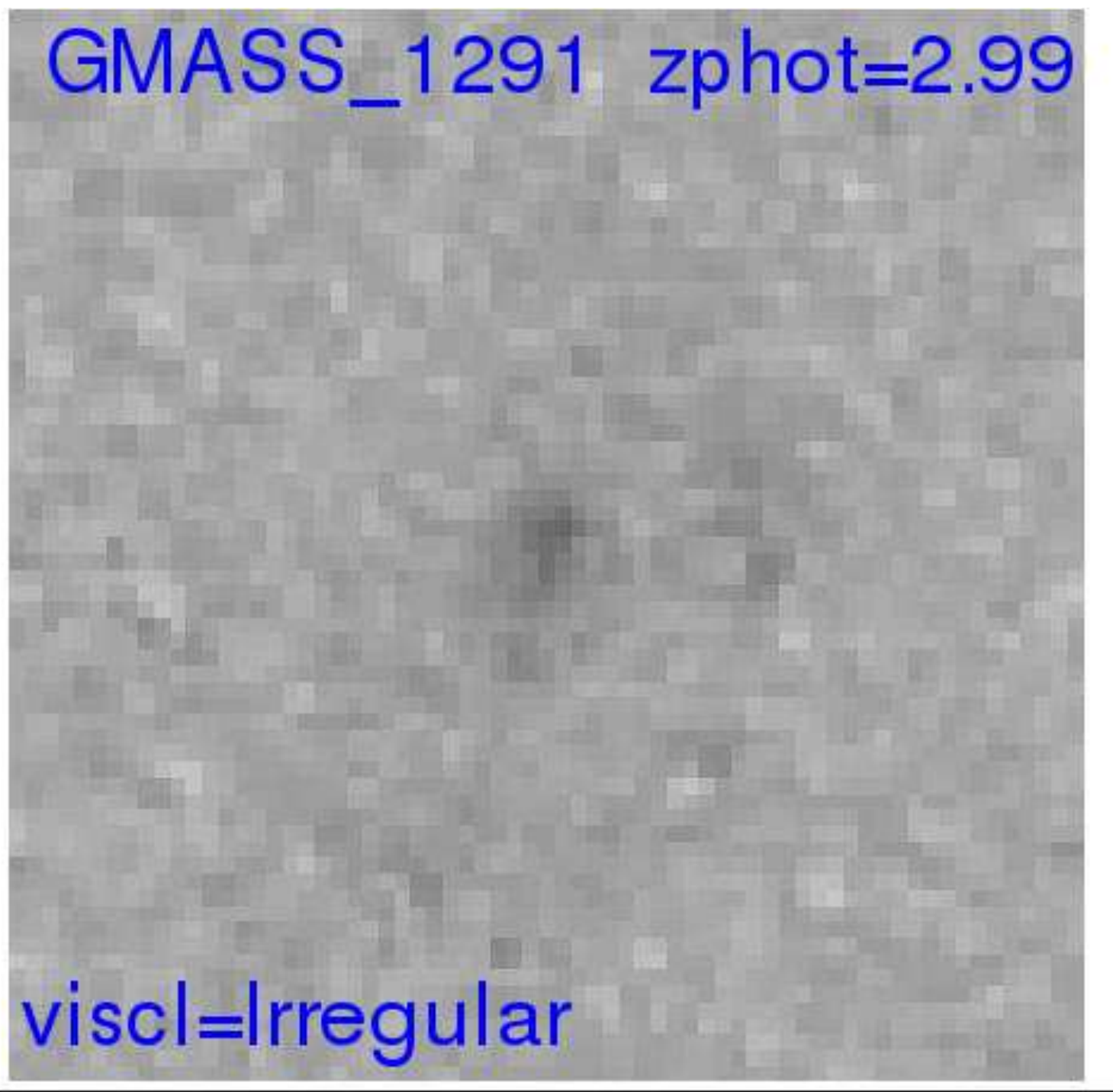}		     
\includegraphics[trim=100 40 75 390, clip=true, width=30mm]{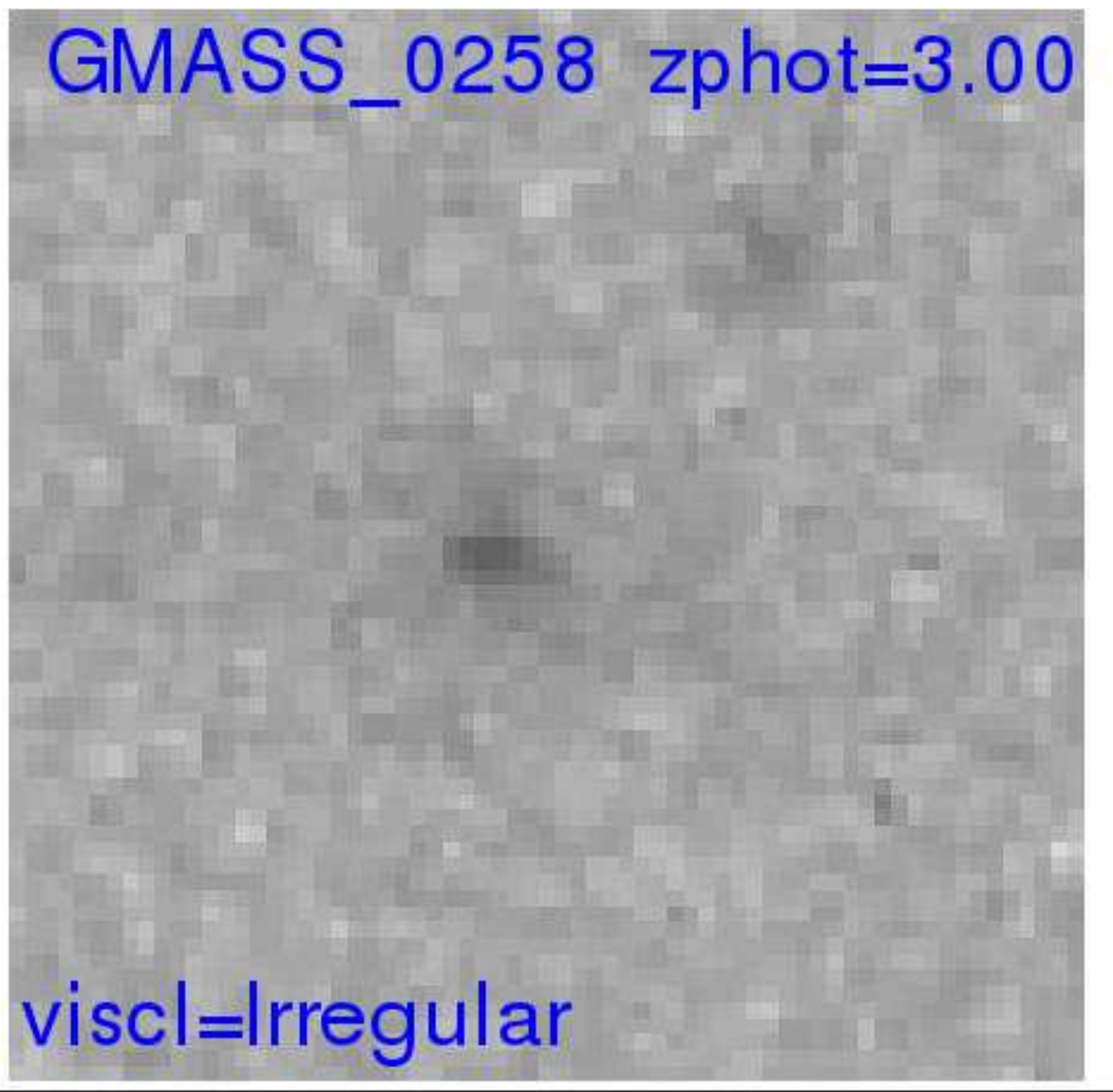}			     

\includegraphics[trim=100 40 75 390, clip=true, width=30mm]{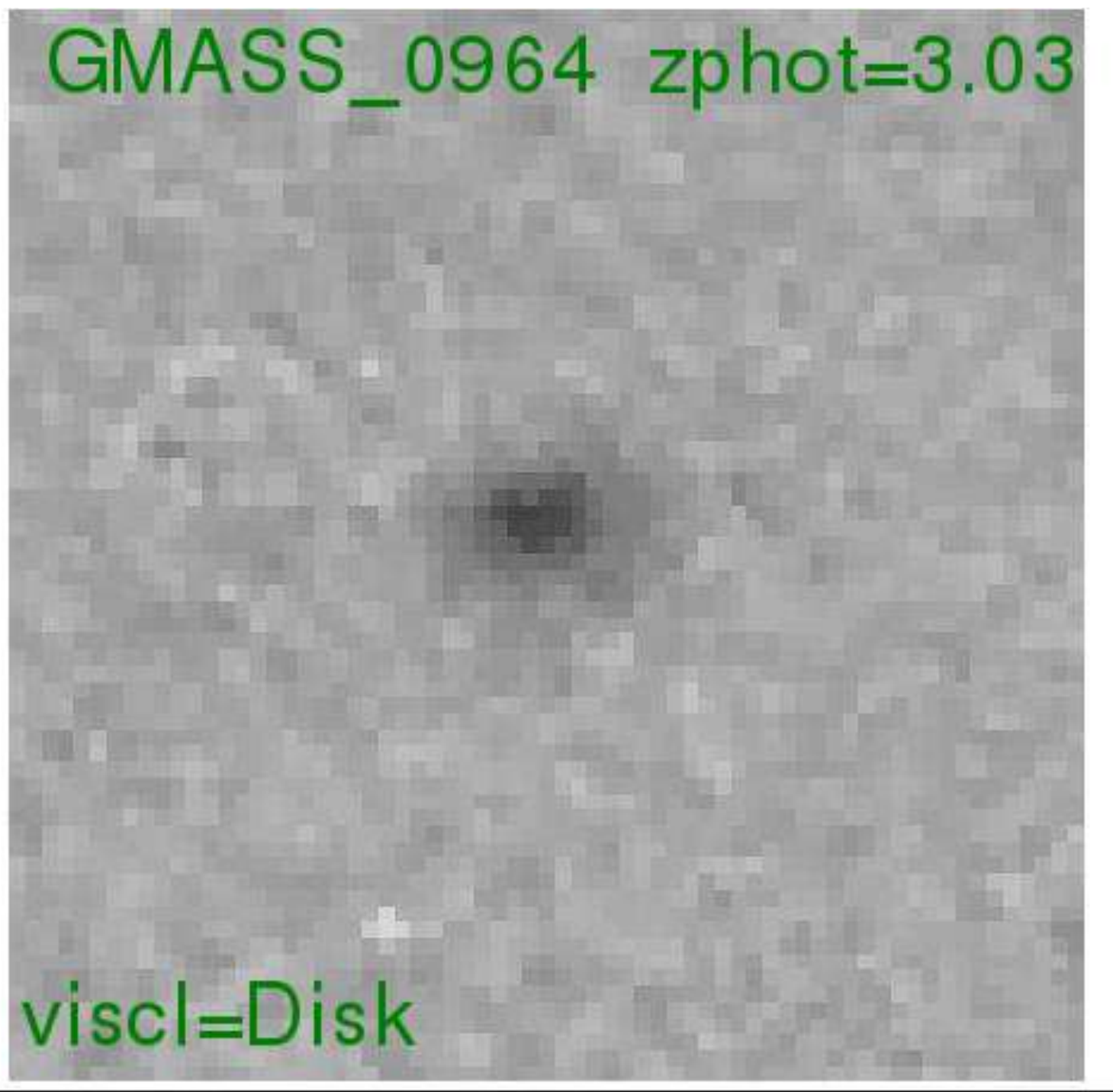}			     
\includegraphics[trim=100 40 75 390, clip=true, width=30mm]{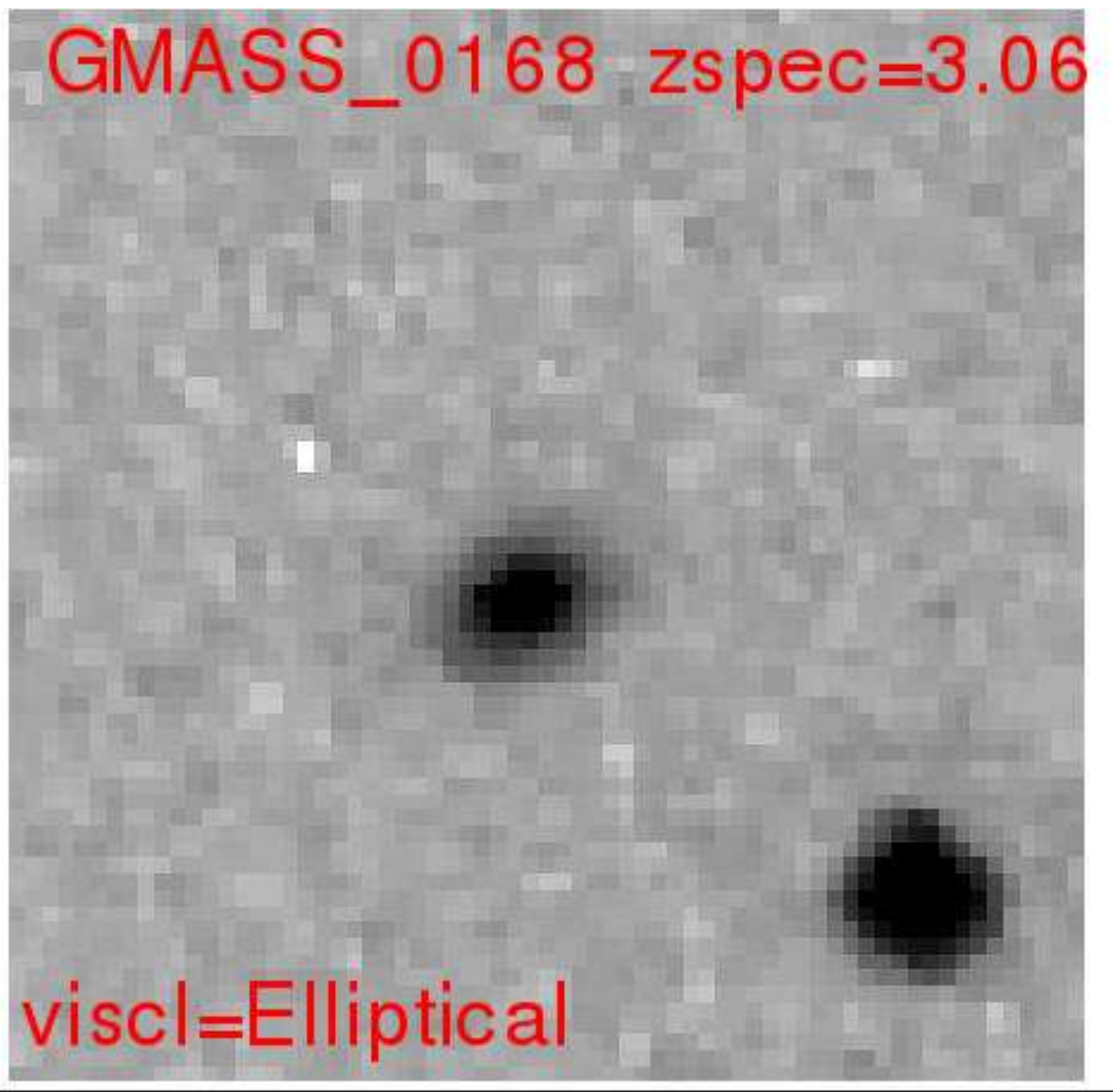}			     
\includegraphics[trim=100 40 75 390, clip=true, width=30mm]{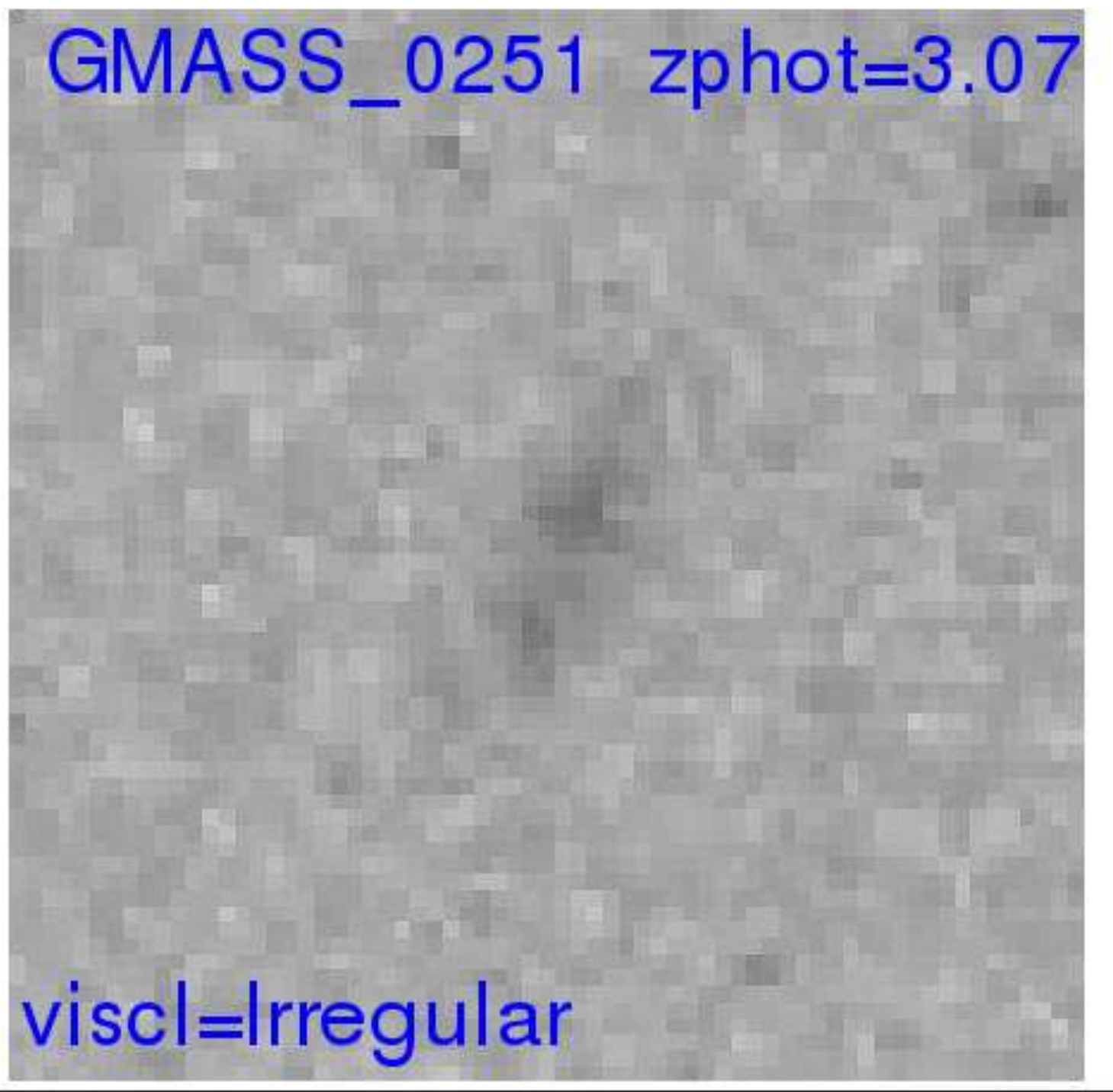}			     
\includegraphics[trim=100 40 75 390, clip=true, width=30mm]{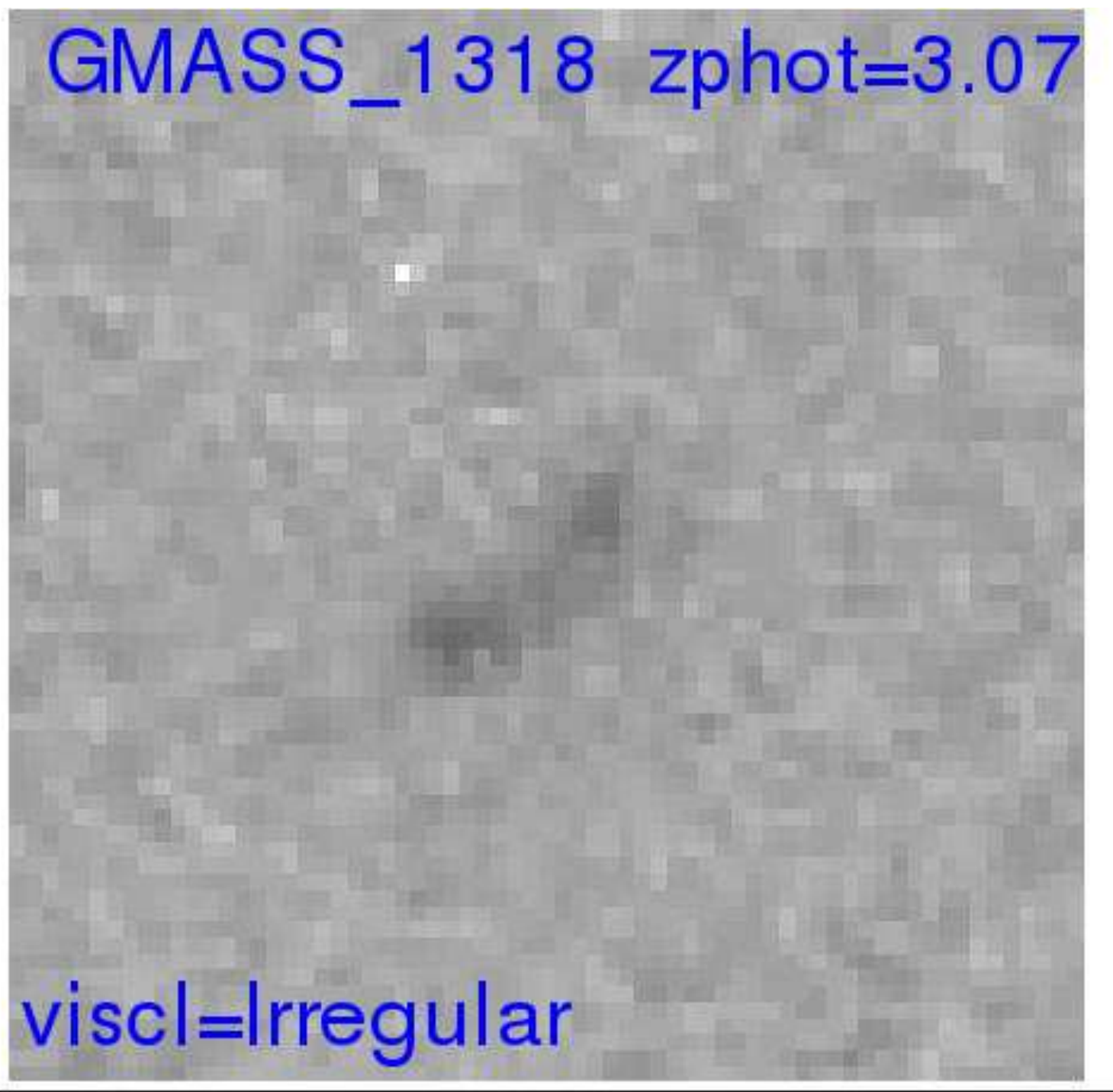}			     
\includegraphics[trim=100 40 75 390, clip=true, width=30mm]{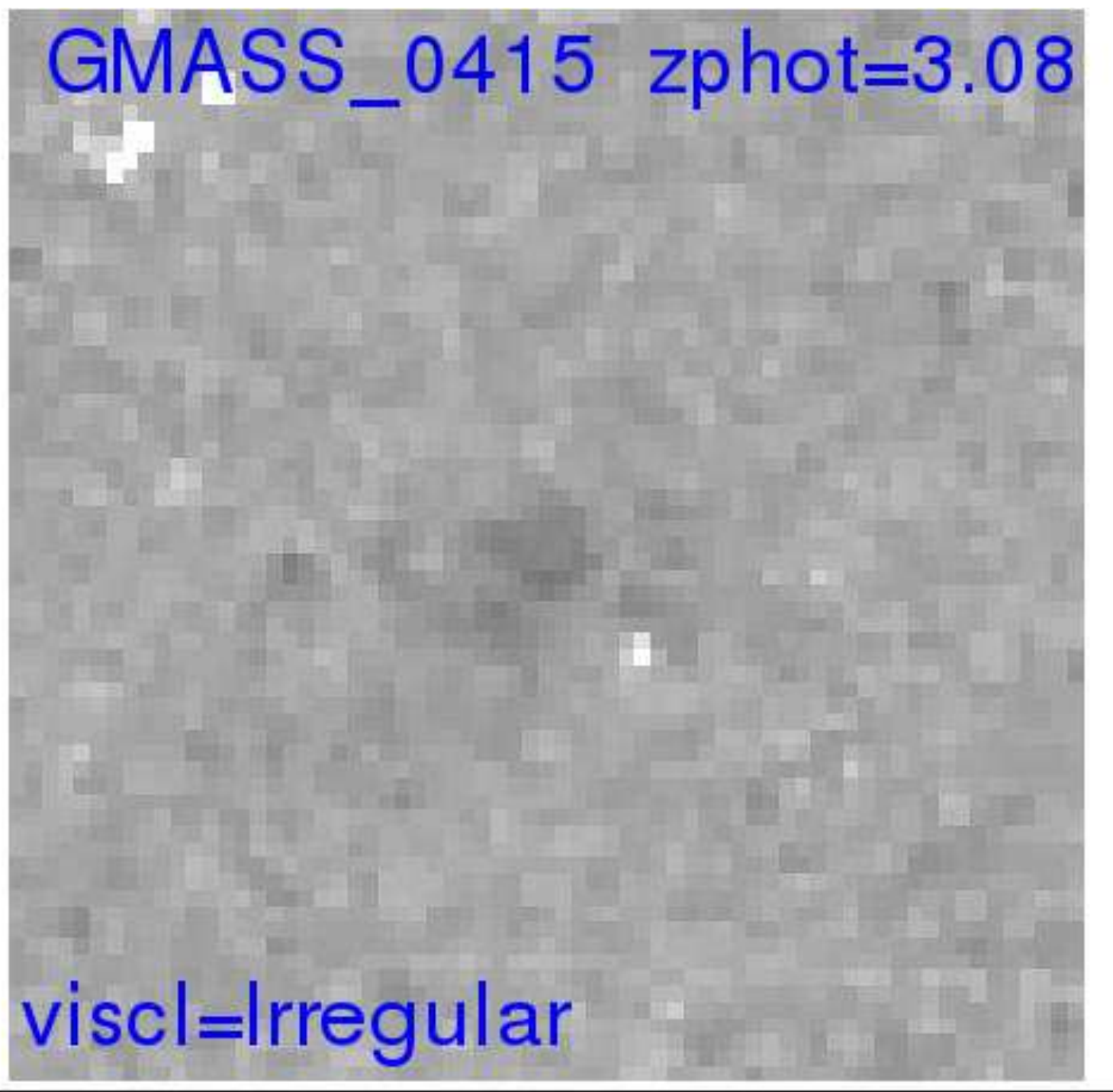}			     
\includegraphics[trim=100 40 75 390, clip=true, width=30mm]{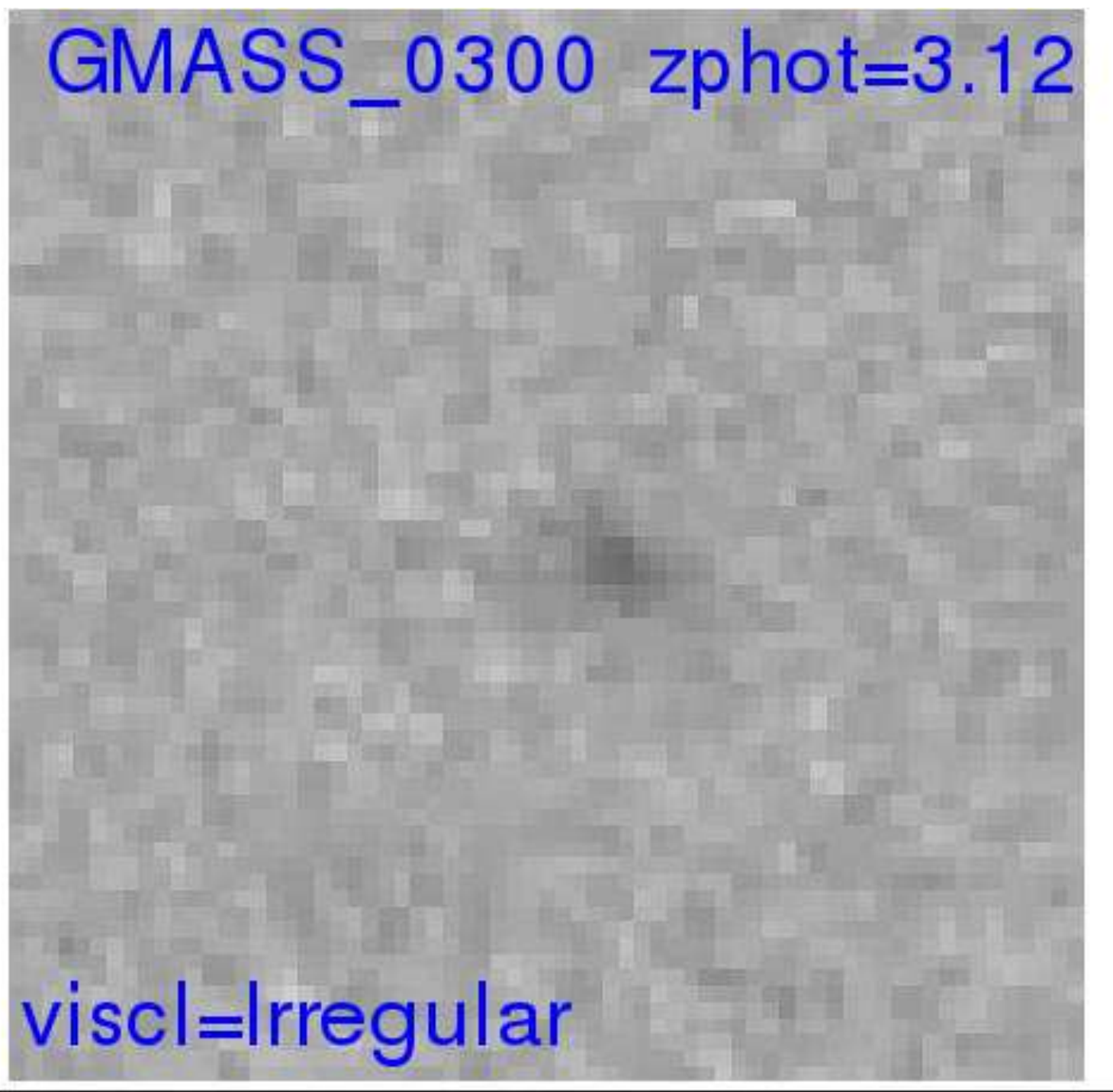}			     

\includegraphics[trim=100 40 75 390, clip=true, width=30mm]{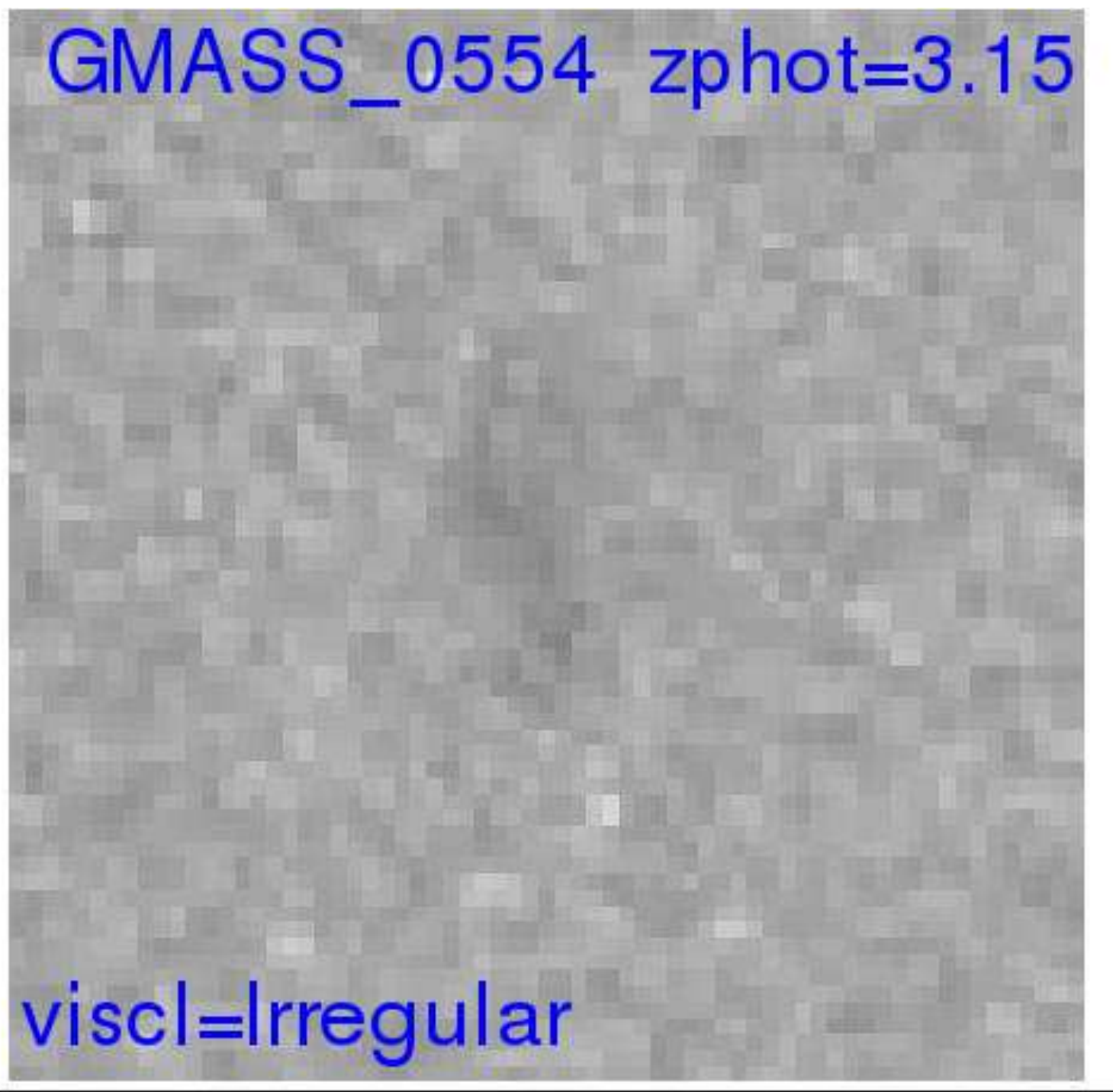}			     
\includegraphics[trim=100 40 75 390, clip=true, width=30mm]{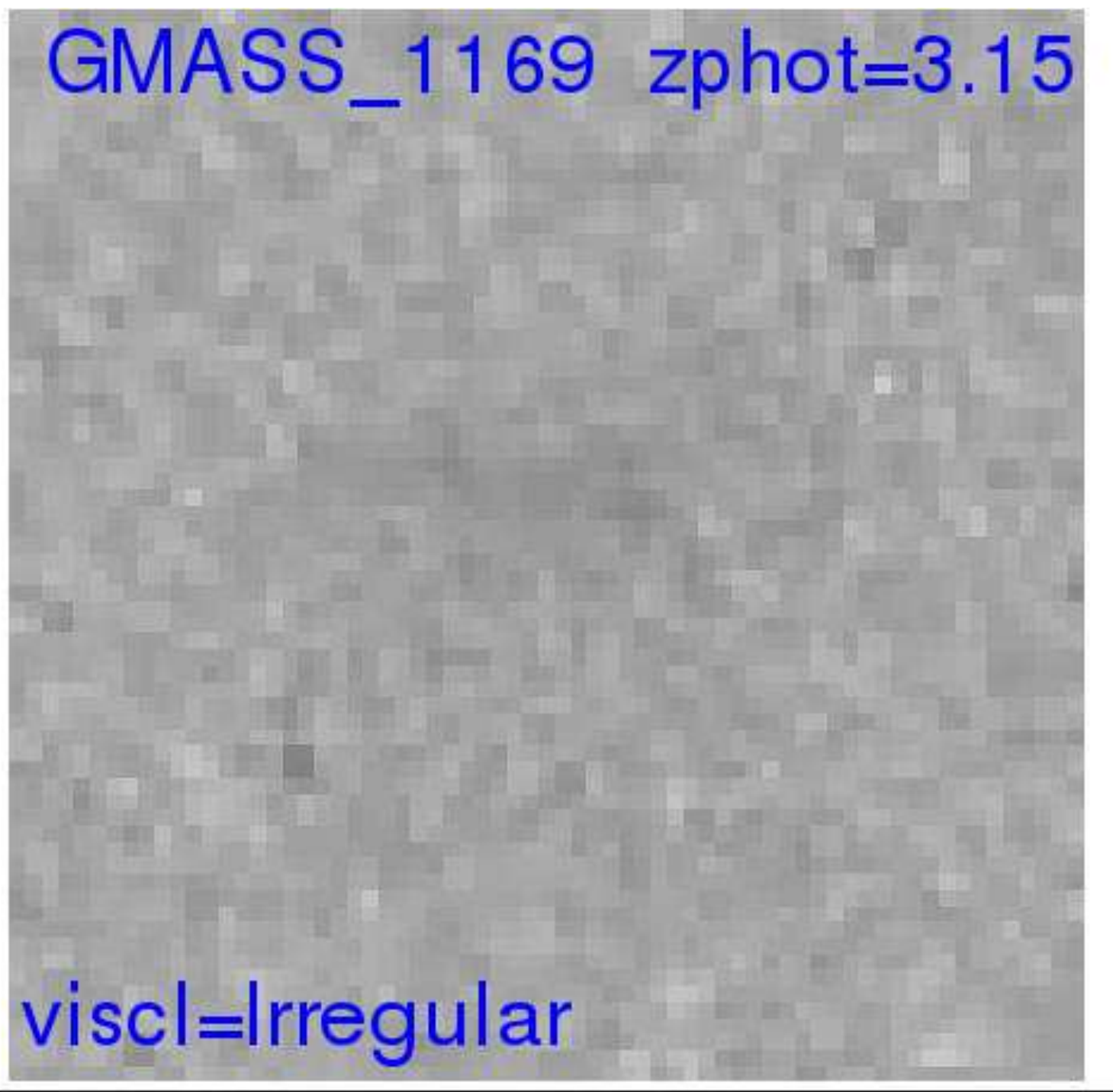}			     
\includegraphics[trim=100 40 75 390, clip=true, width=30mm]{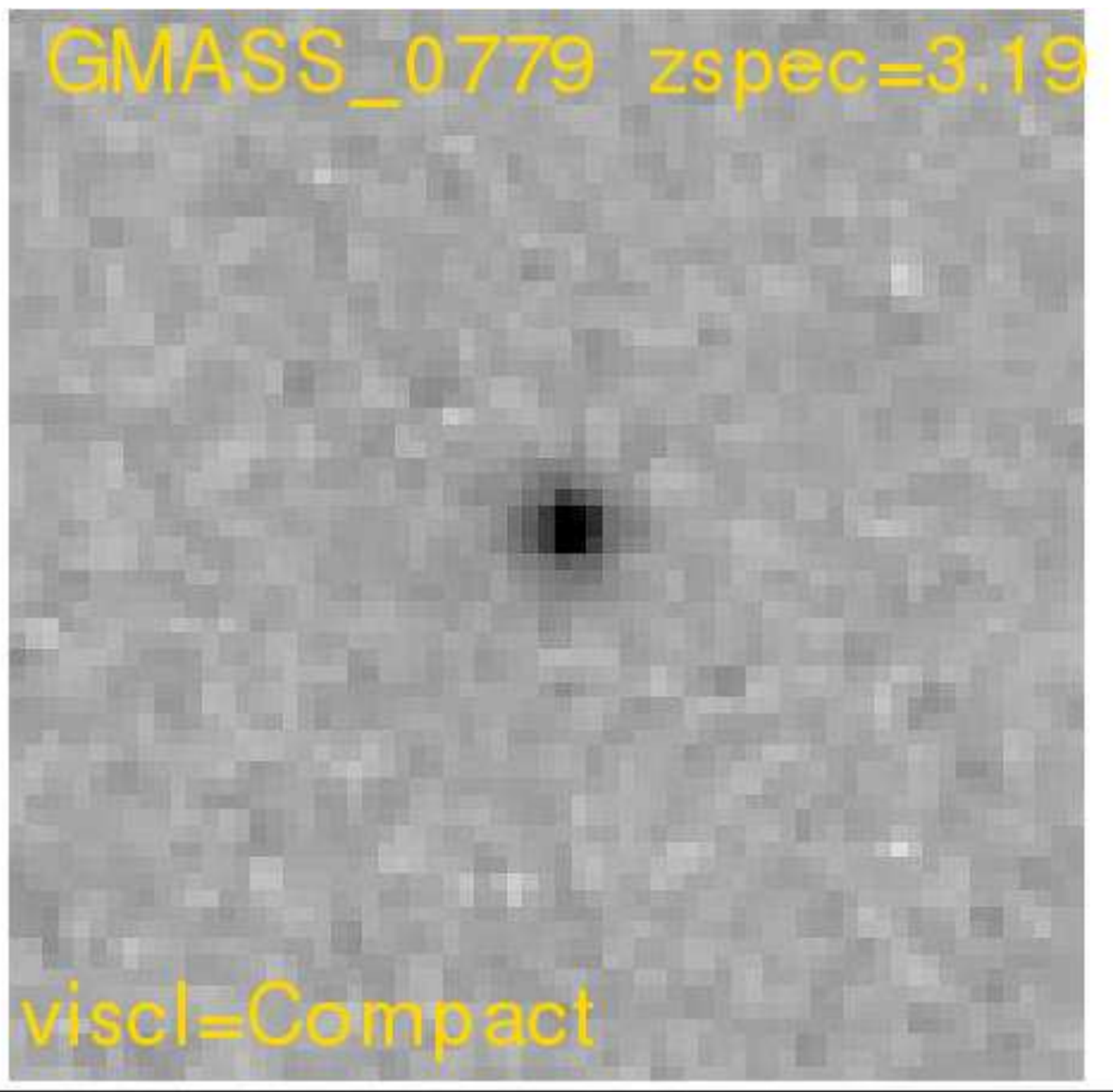}			     
\includegraphics[trim=100 40 75 390, clip=true, width=30mm]{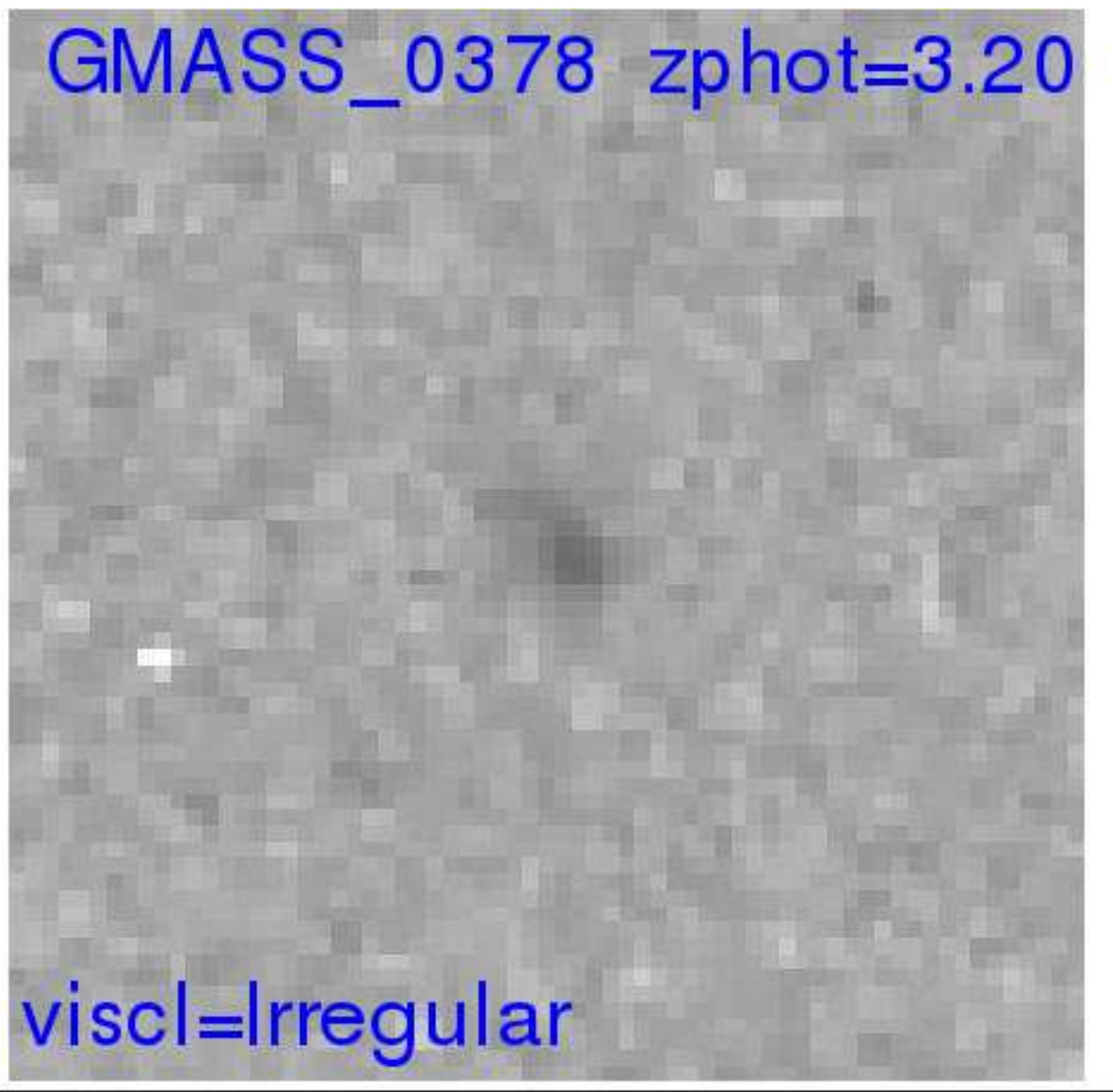}			     
\includegraphics[trim=100 40 75 390, clip=true, width=30mm]{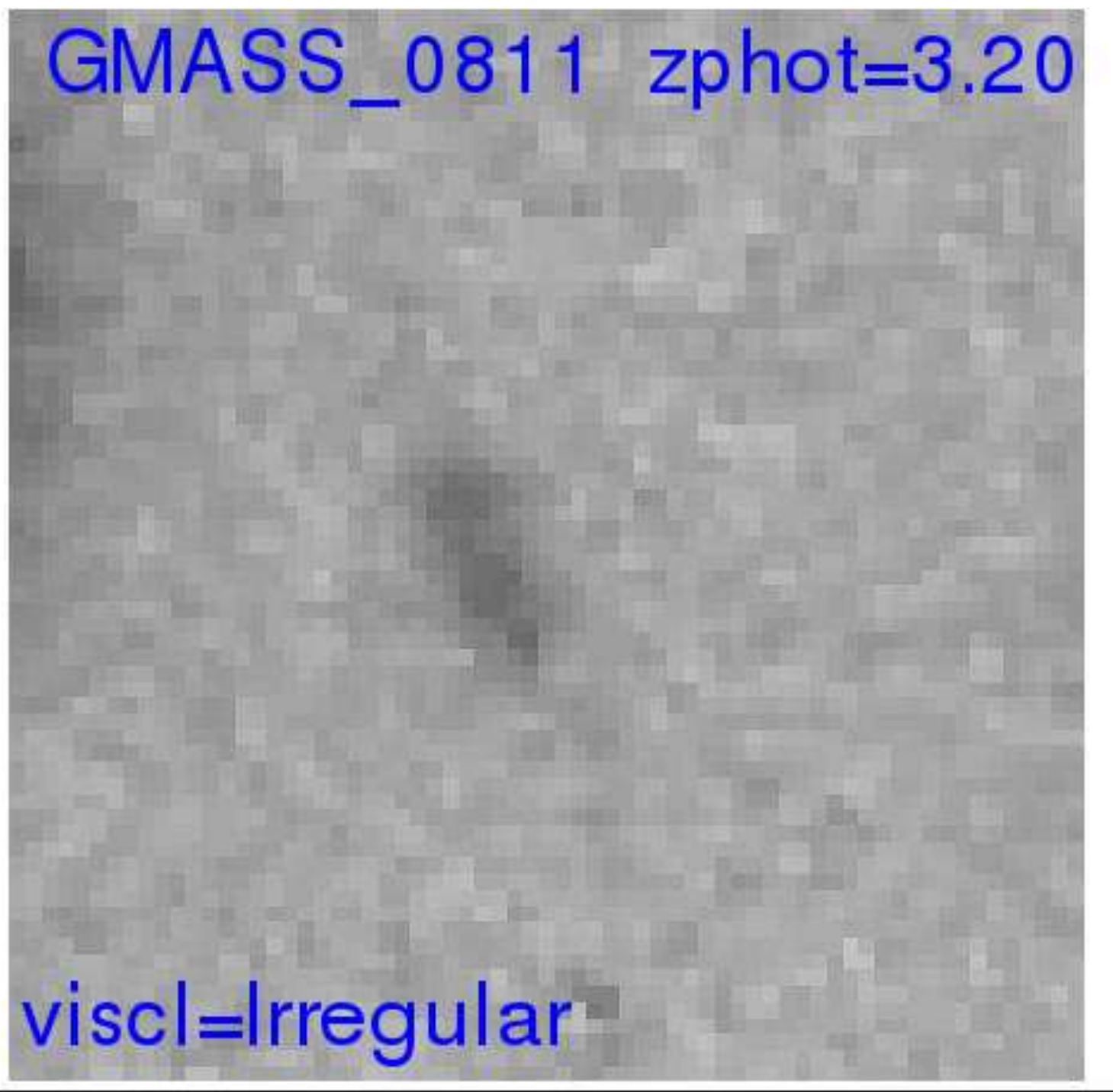}			     
\includegraphics[trim=100 40 75 390, clip=true, width=30mm]{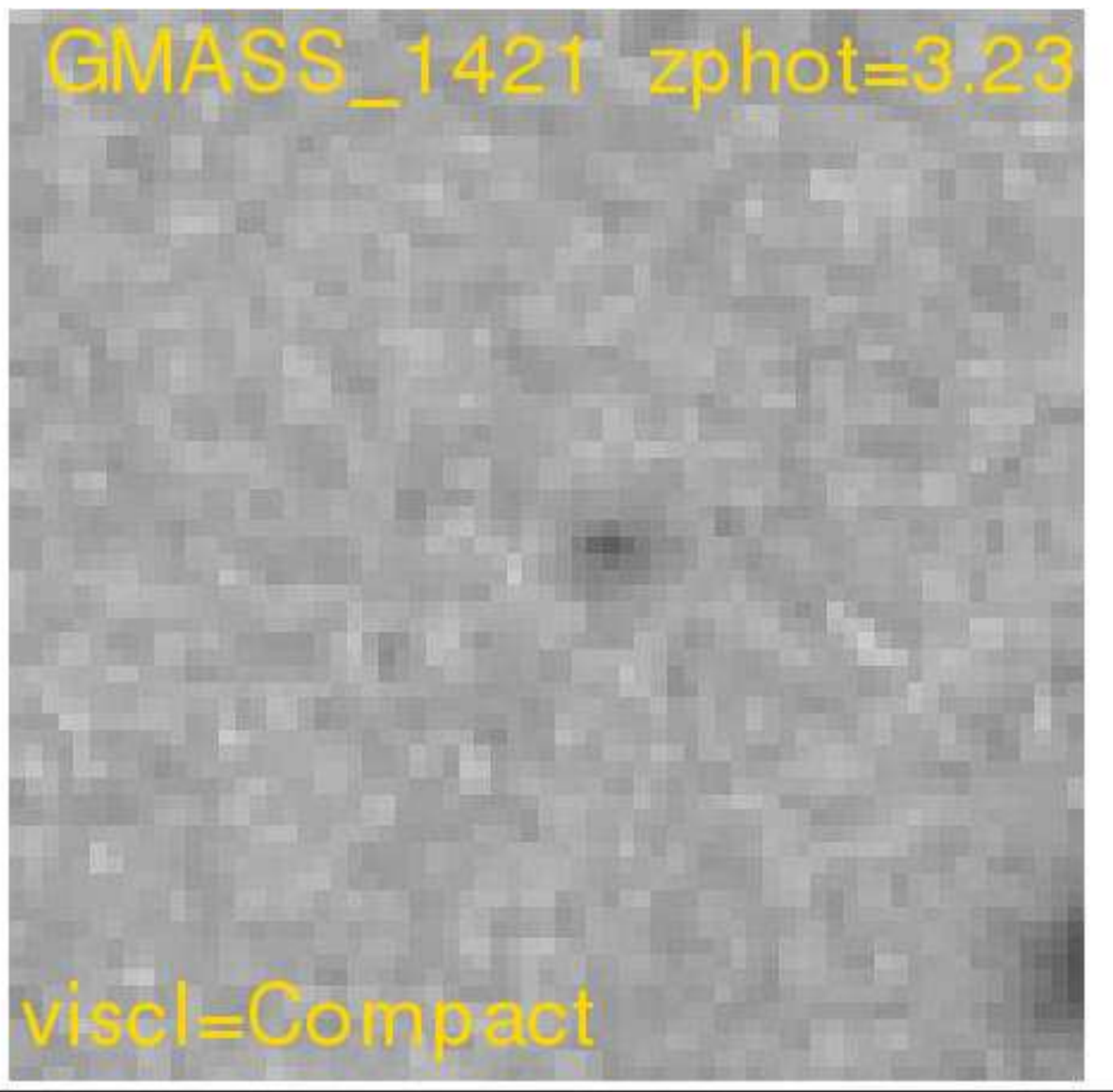}			     

\includegraphics[trim=100 40 75 390, clip=true, width=30mm]{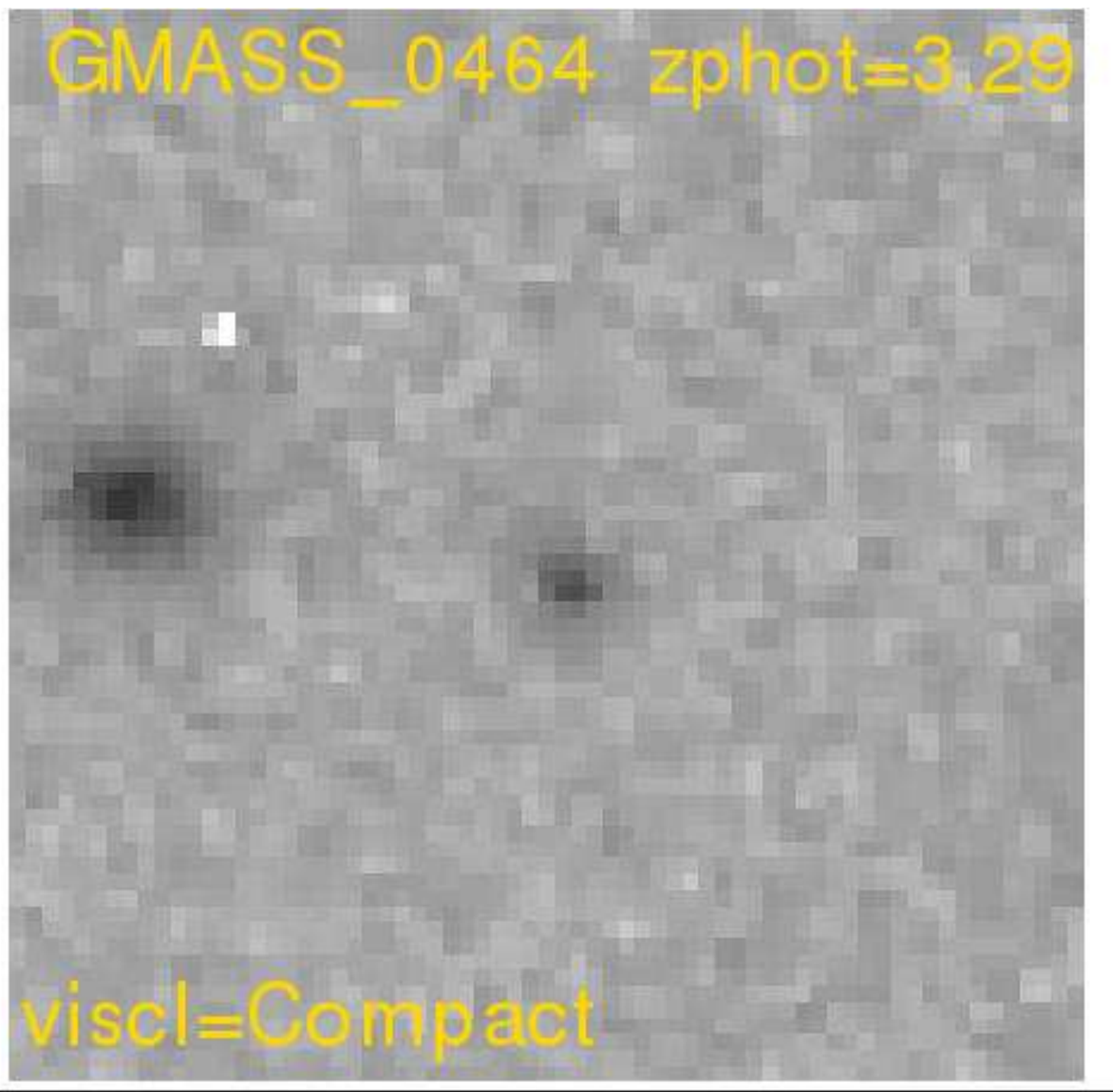}			     
\includegraphics[trim=100 40 75 390, clip=true, width=30mm]{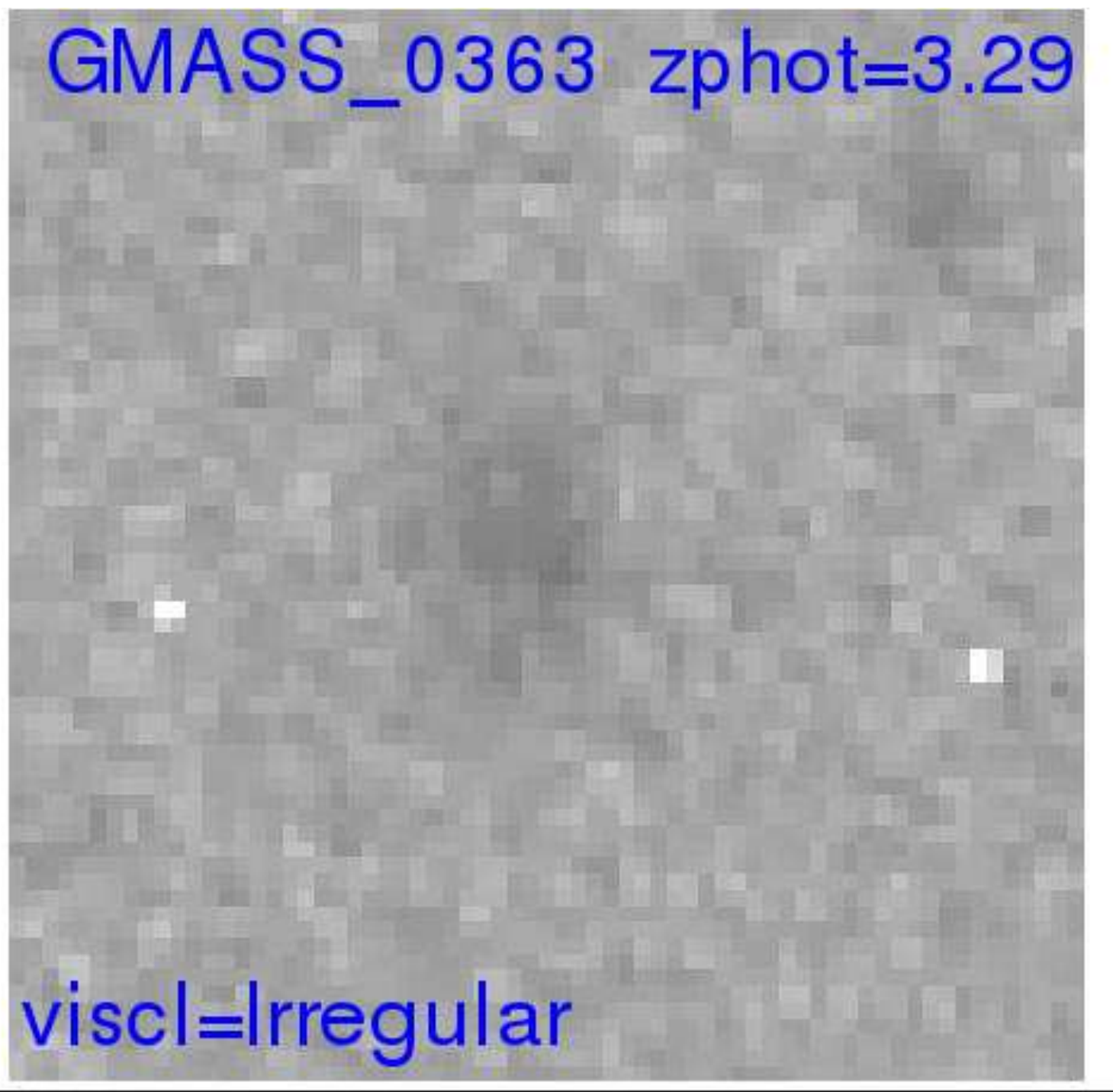}		     
\includegraphics[trim=100 40 75 390, clip=true, width=30mm]{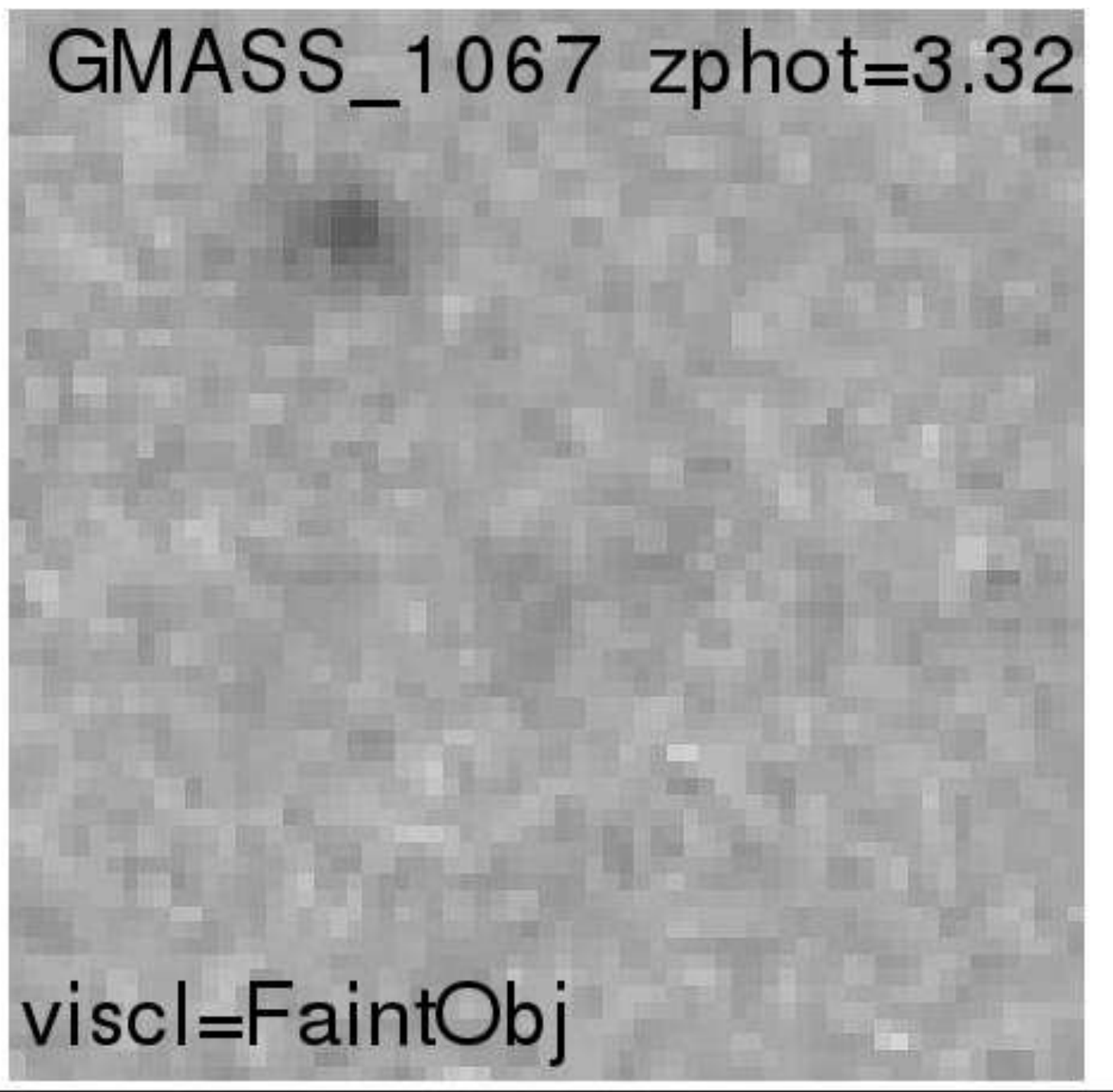}		     
\includegraphics[trim=100 40 75 390, clip=true, width=30mm]{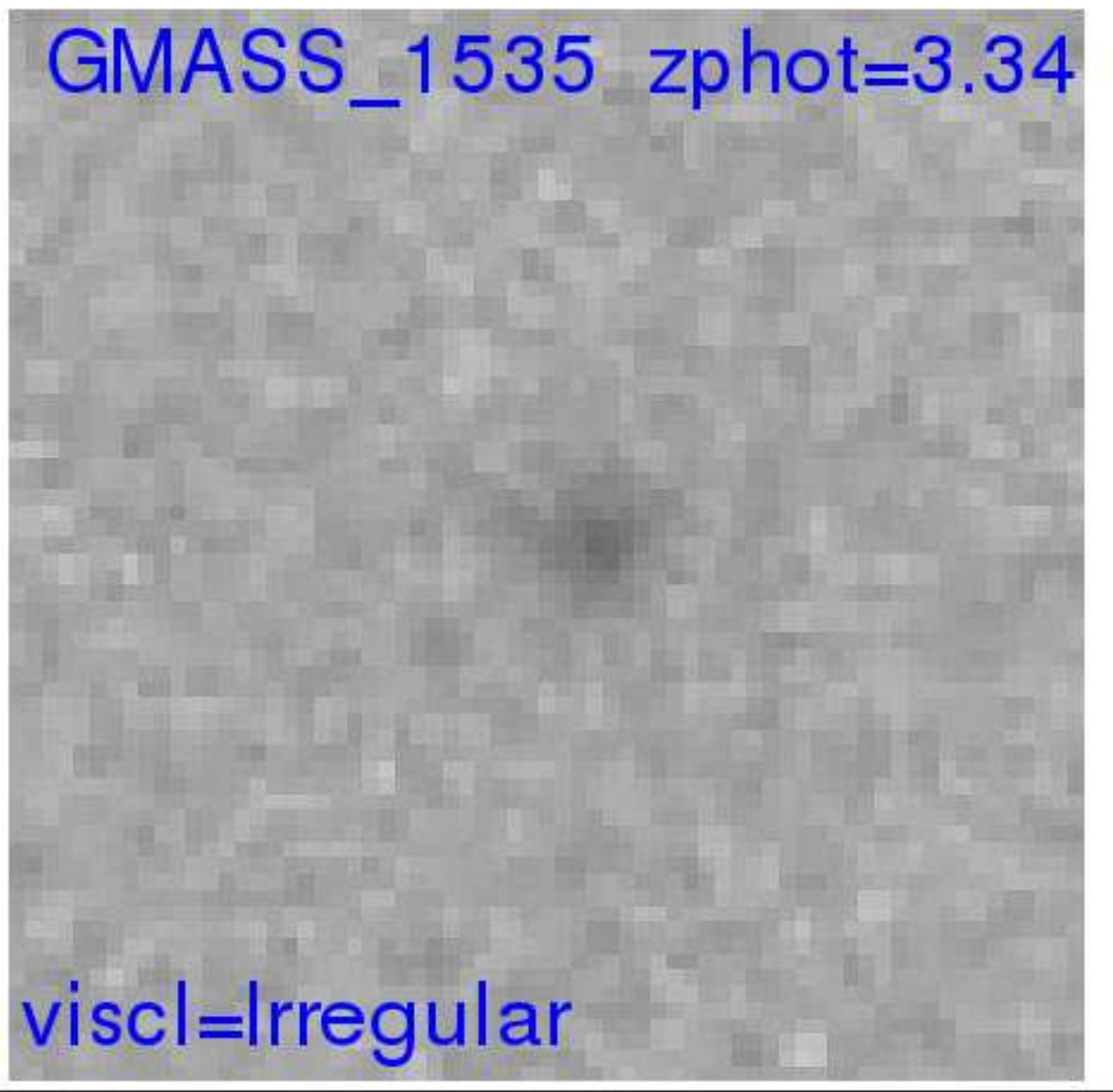}		     
\includegraphics[trim=100 40 75 390, clip=true, width=30mm]{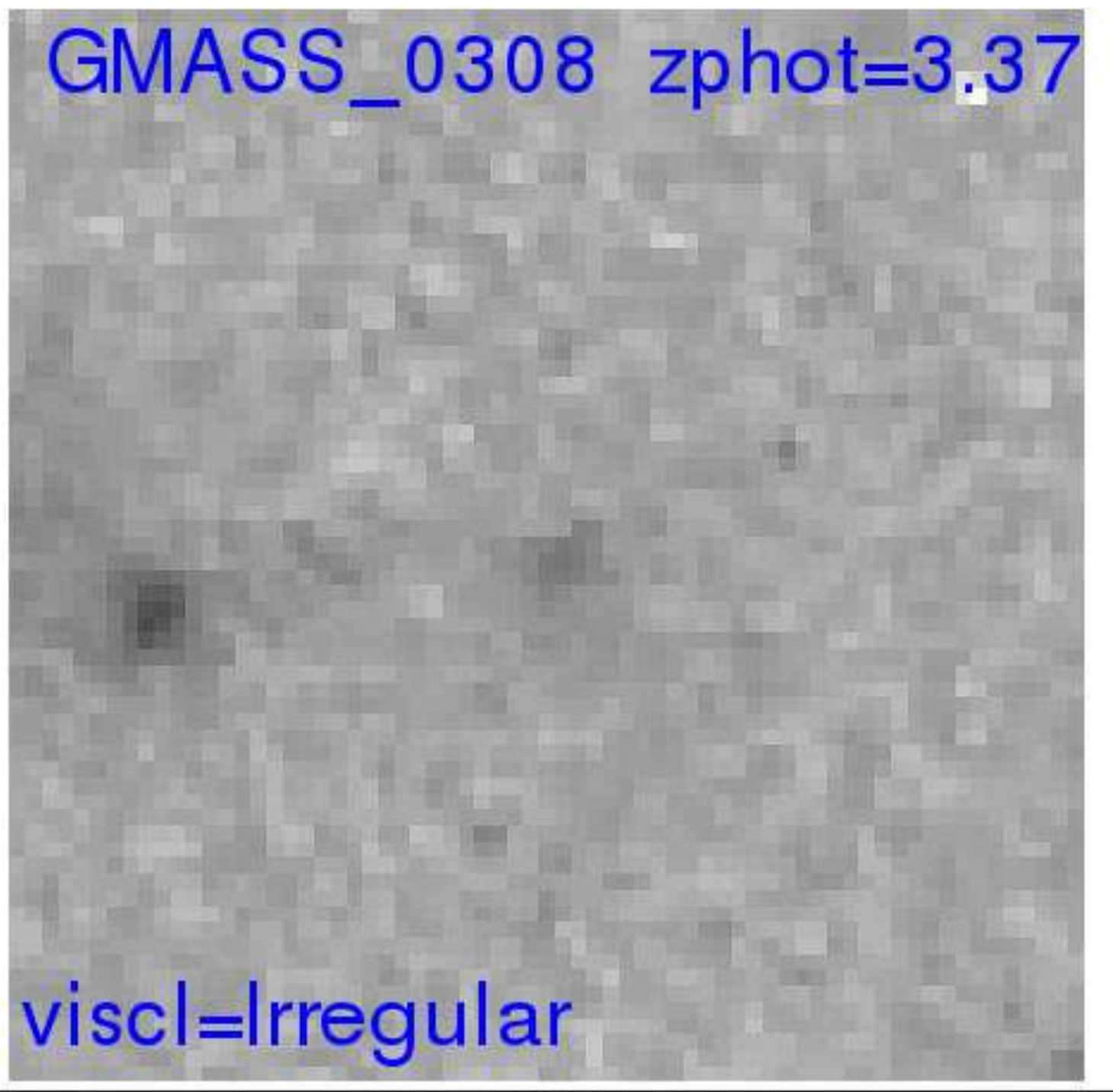}			     
\includegraphics[trim=100 40 75 390, clip=true, width=30mm]{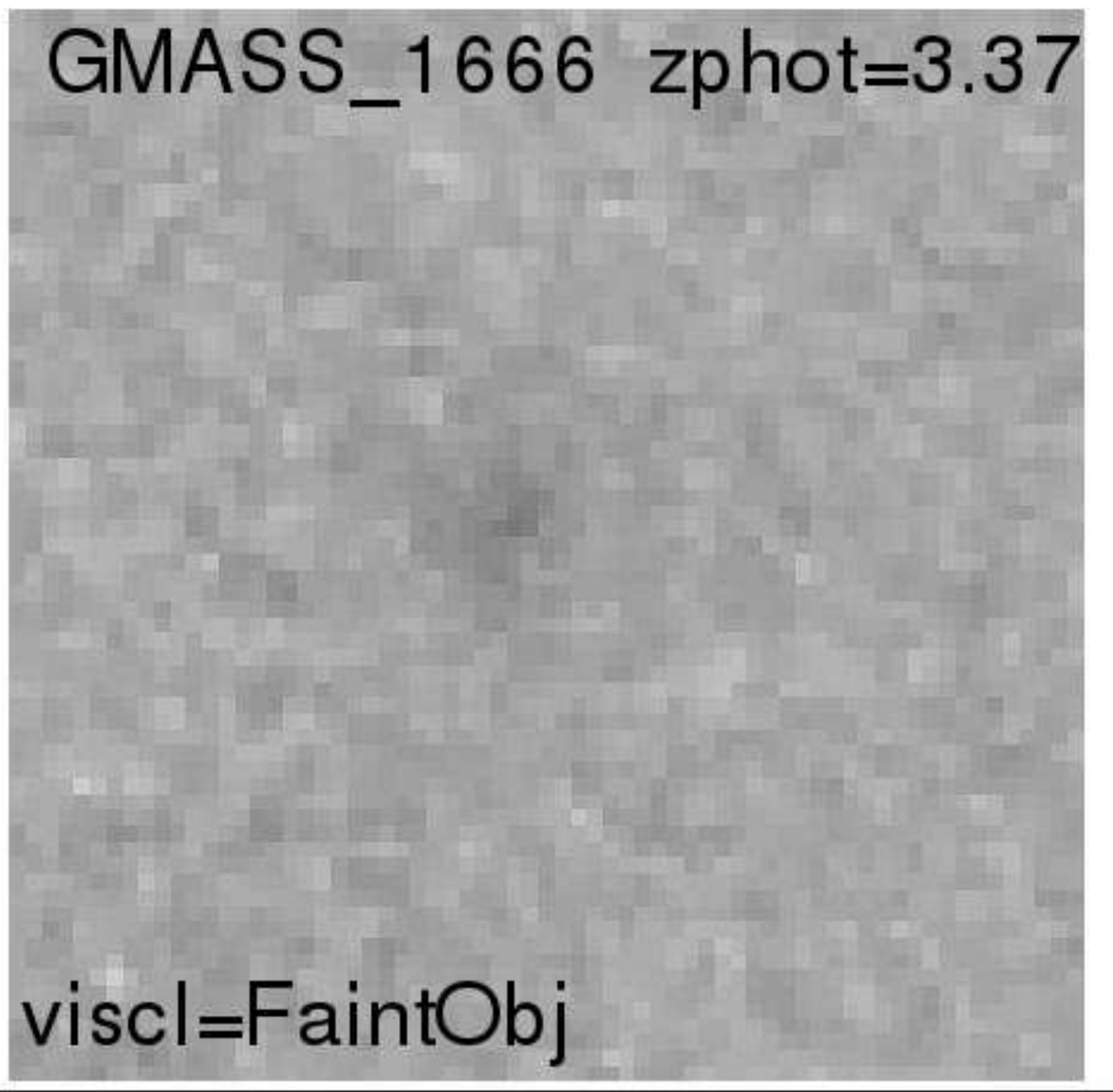}			     

\includegraphics[trim=100 40 75 390, clip=true, width=30mm]{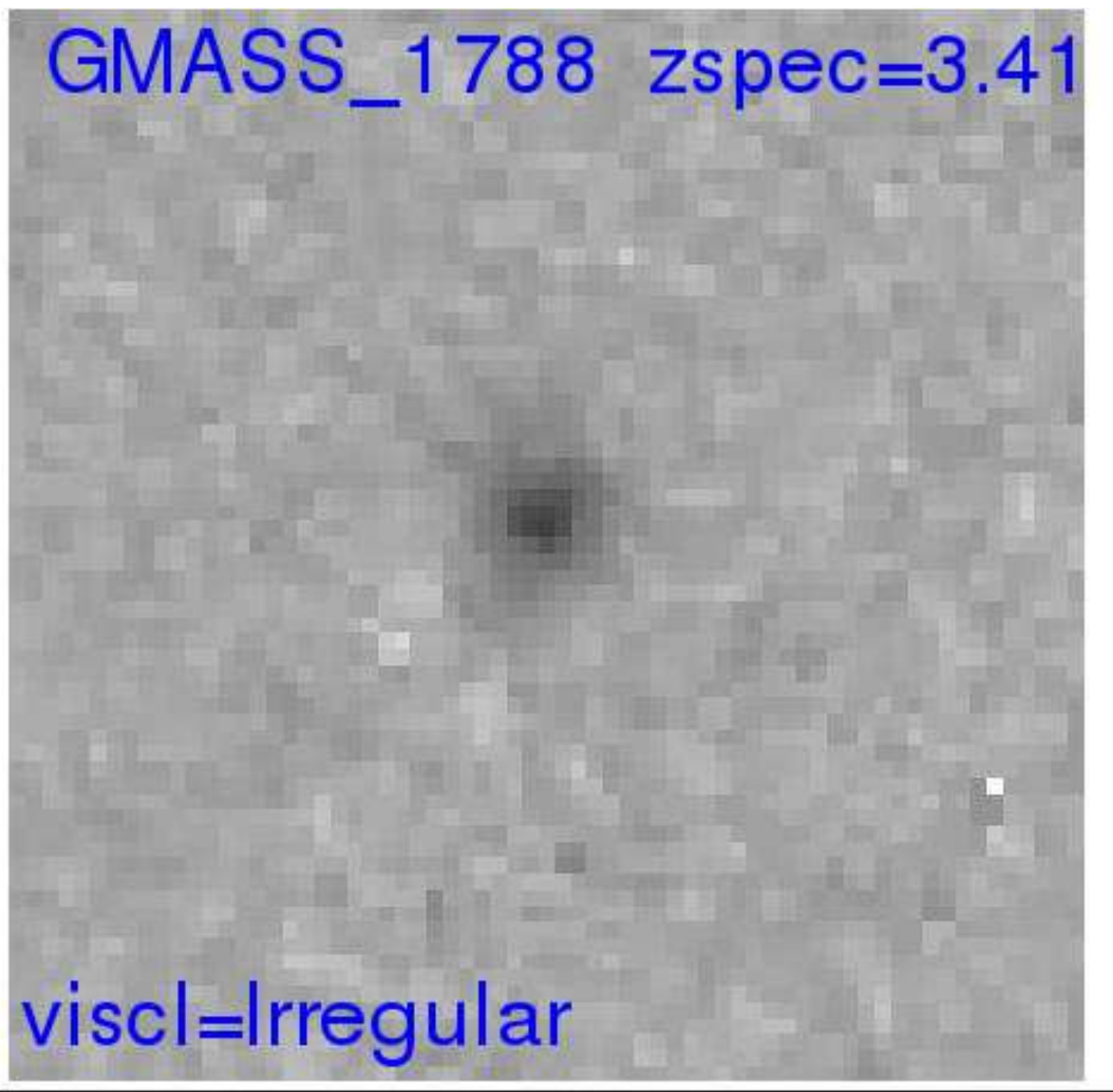}			     
\includegraphics[trim=100 40 75 390, clip=true, width=30mm]{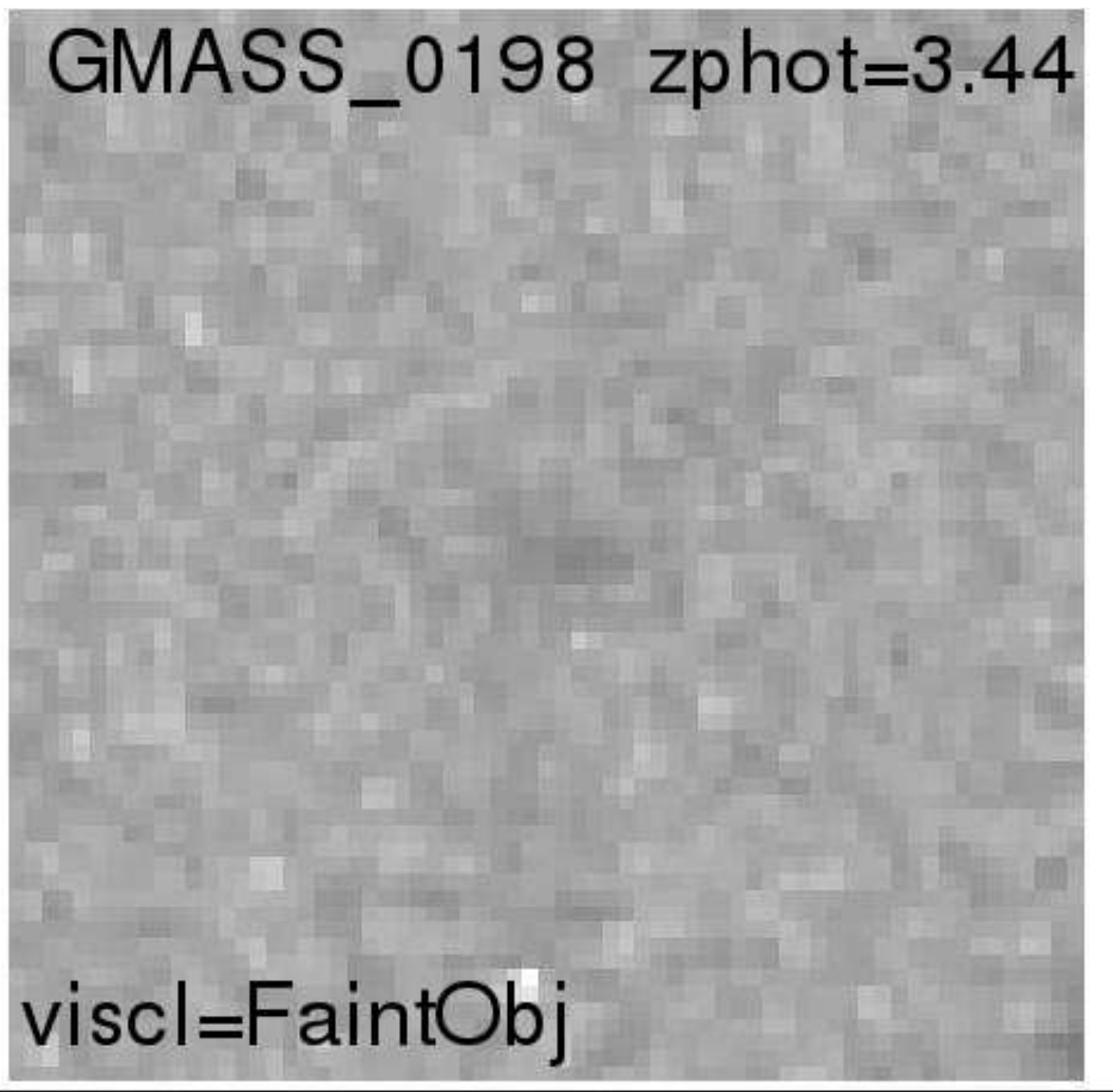}			     
\includegraphics[trim=100 40 75 390, clip=true, width=30mm]{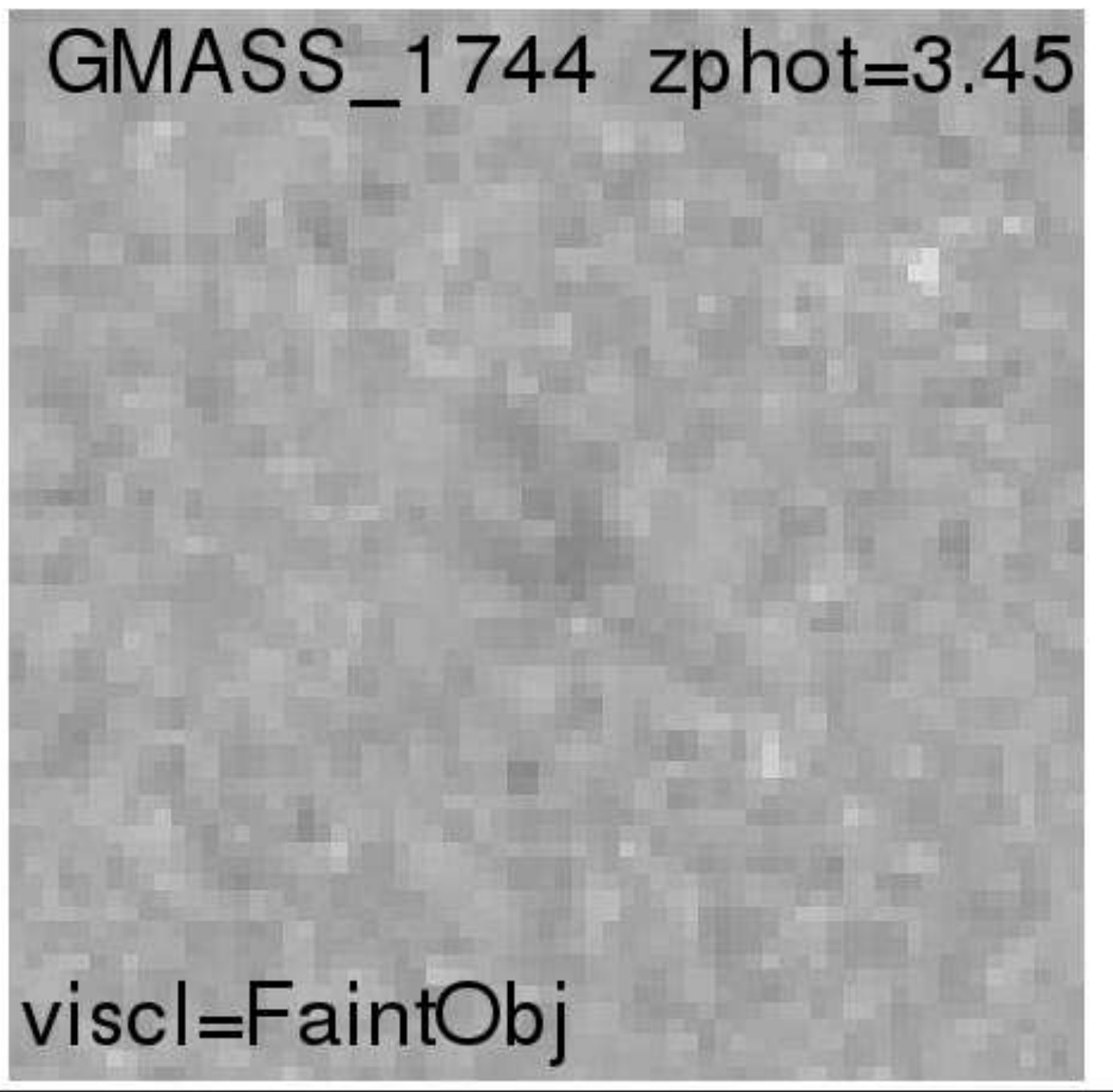}			     
\includegraphics[trim=100 40 75 390, clip=true, width=30mm]{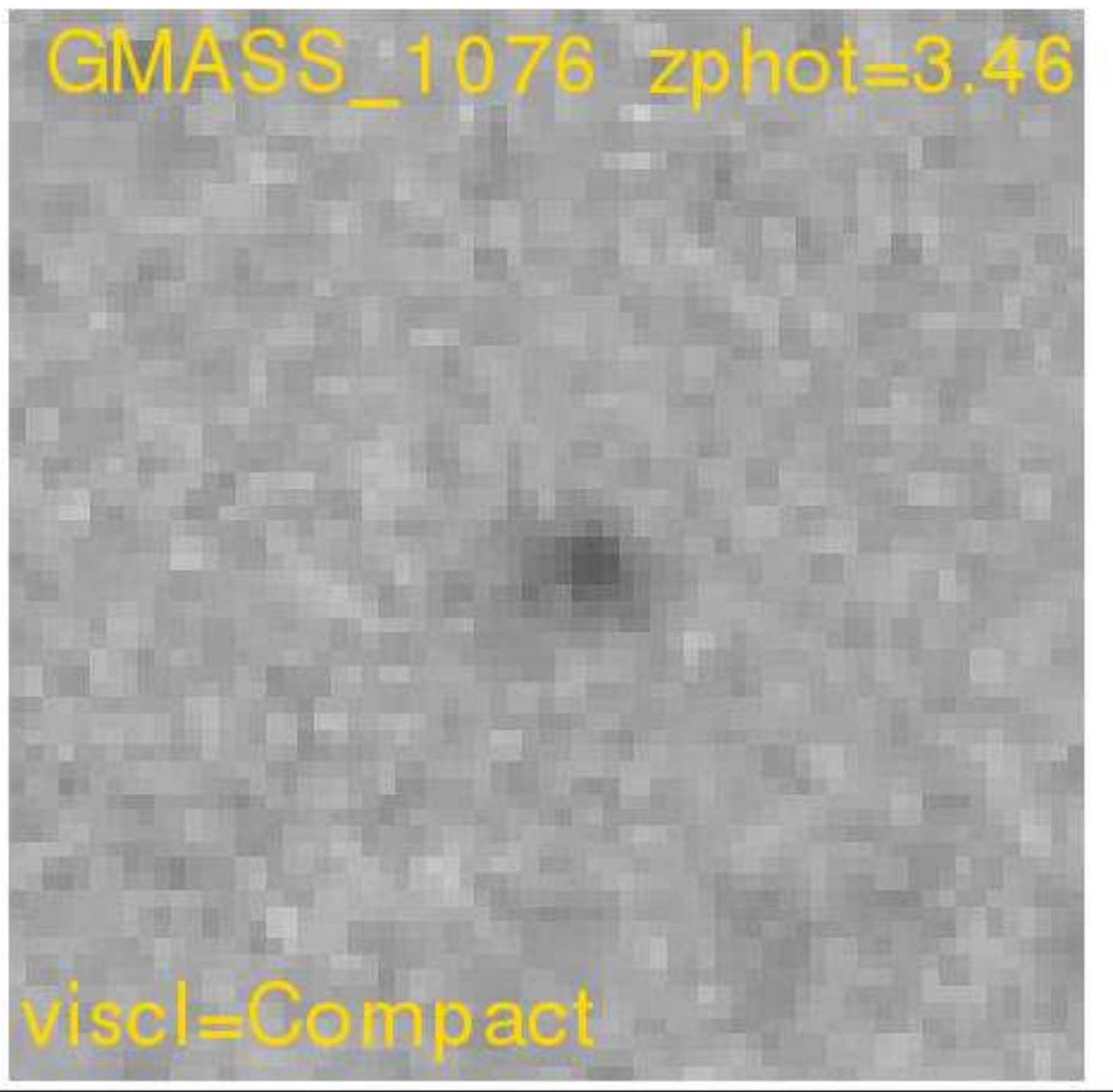}			     
\includegraphics[trim=100 40 75 390, clip=true, width=30mm]{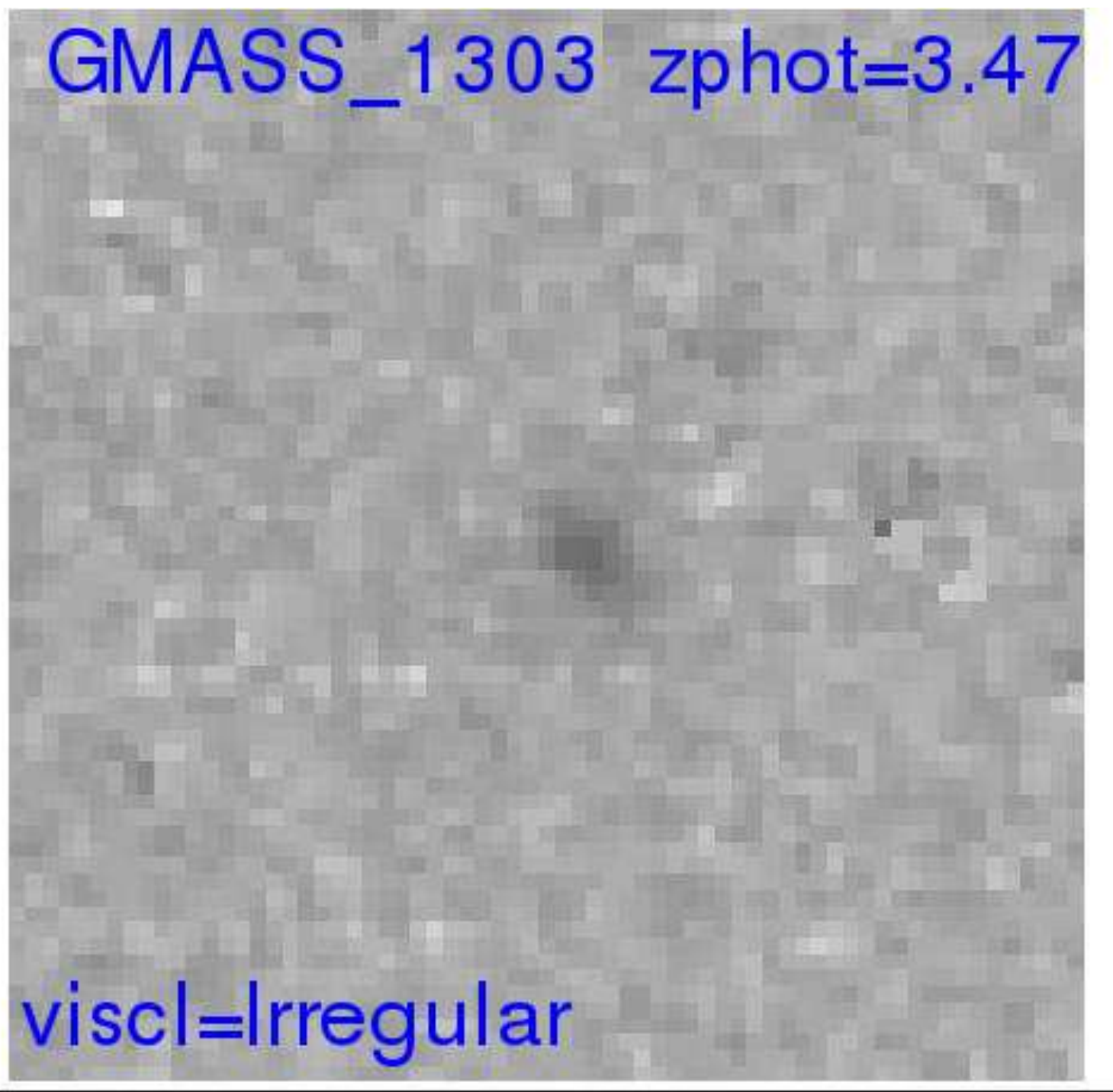}			     
\includegraphics[trim=100 40 75 390, clip=true, width=30mm]{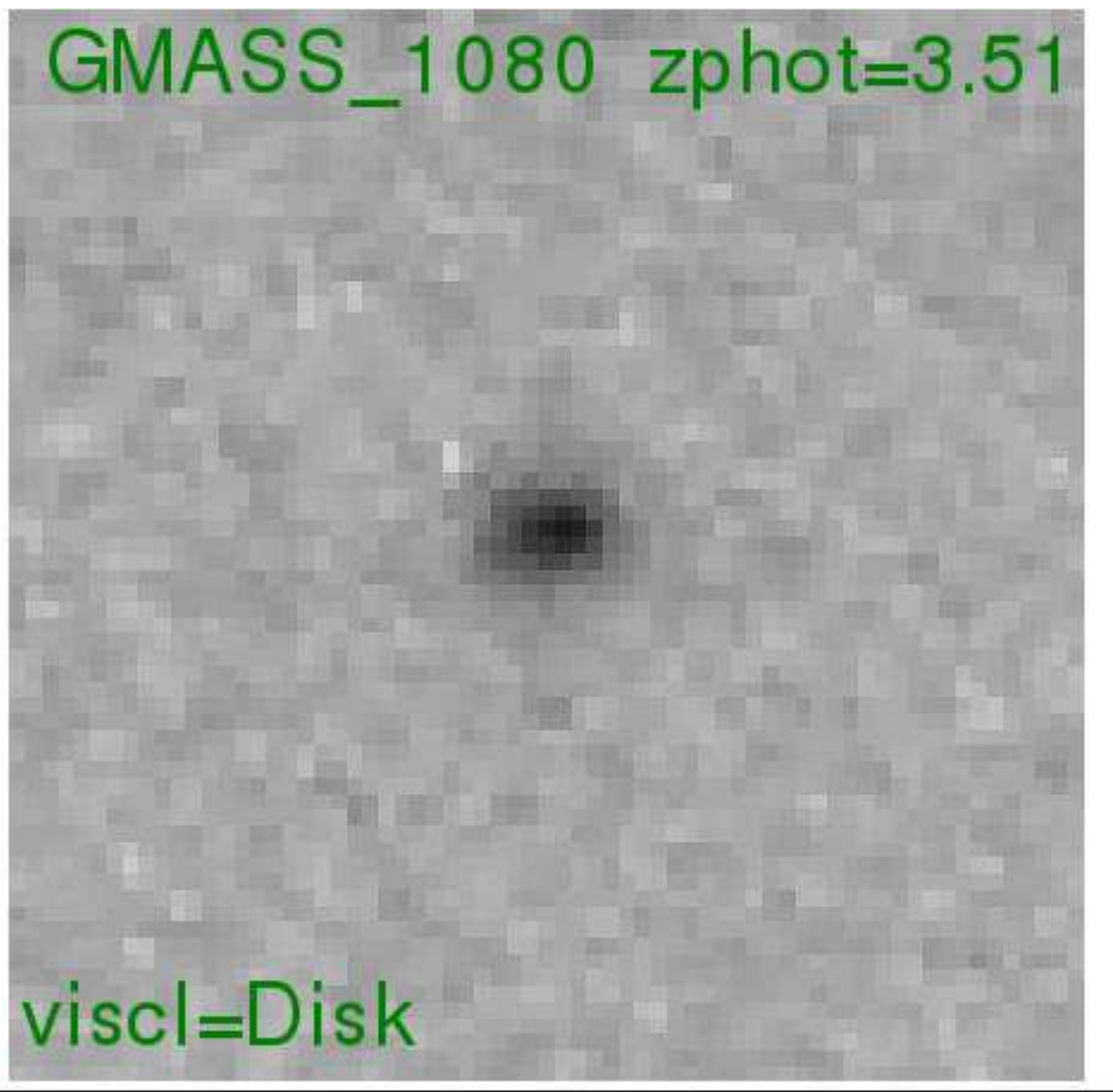}			     

\includegraphics[trim=100 40 75 390, clip=true, width=30mm]{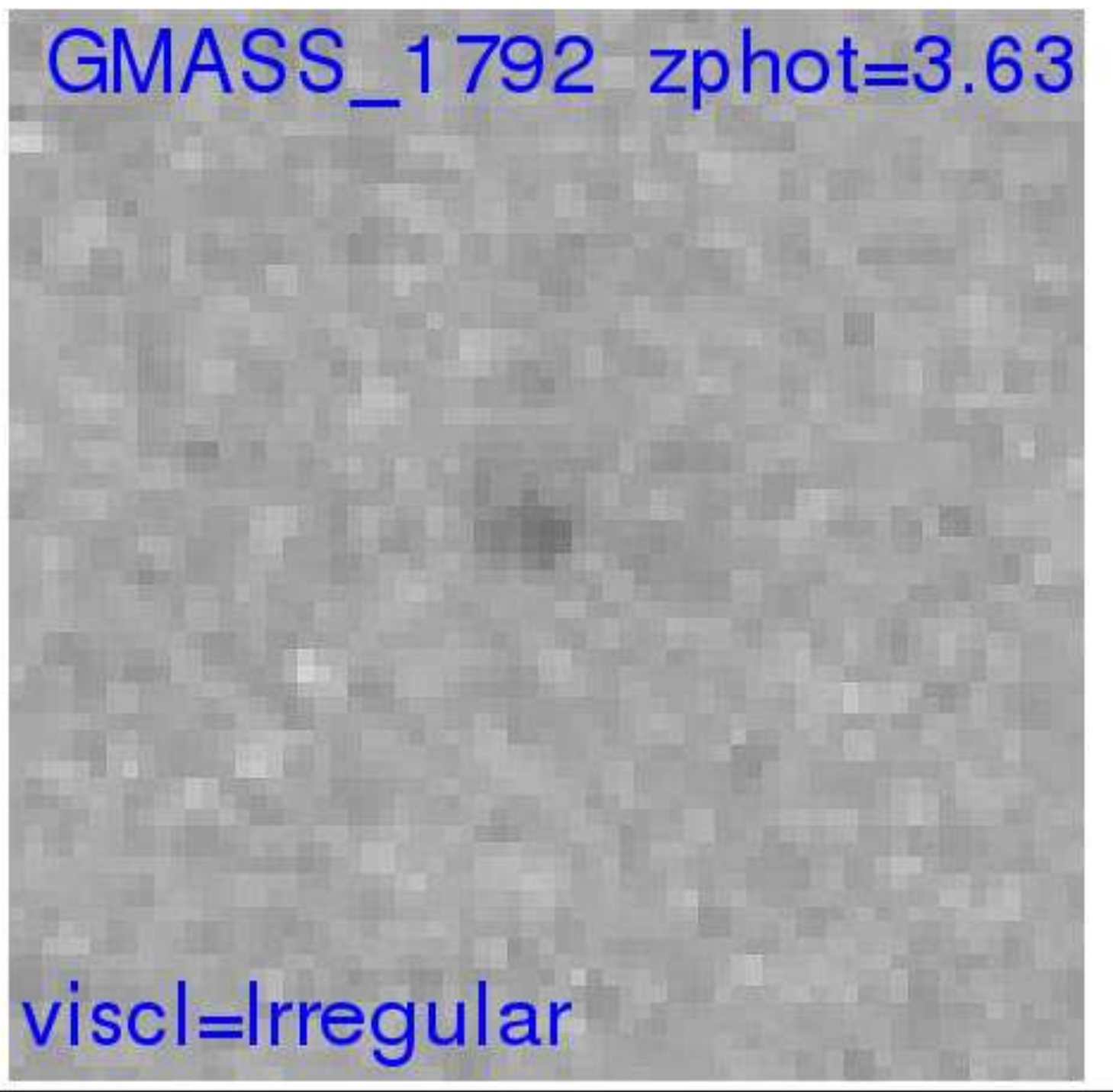}			     
\includegraphics[trim=100 40 75 390, clip=true, width=30mm]{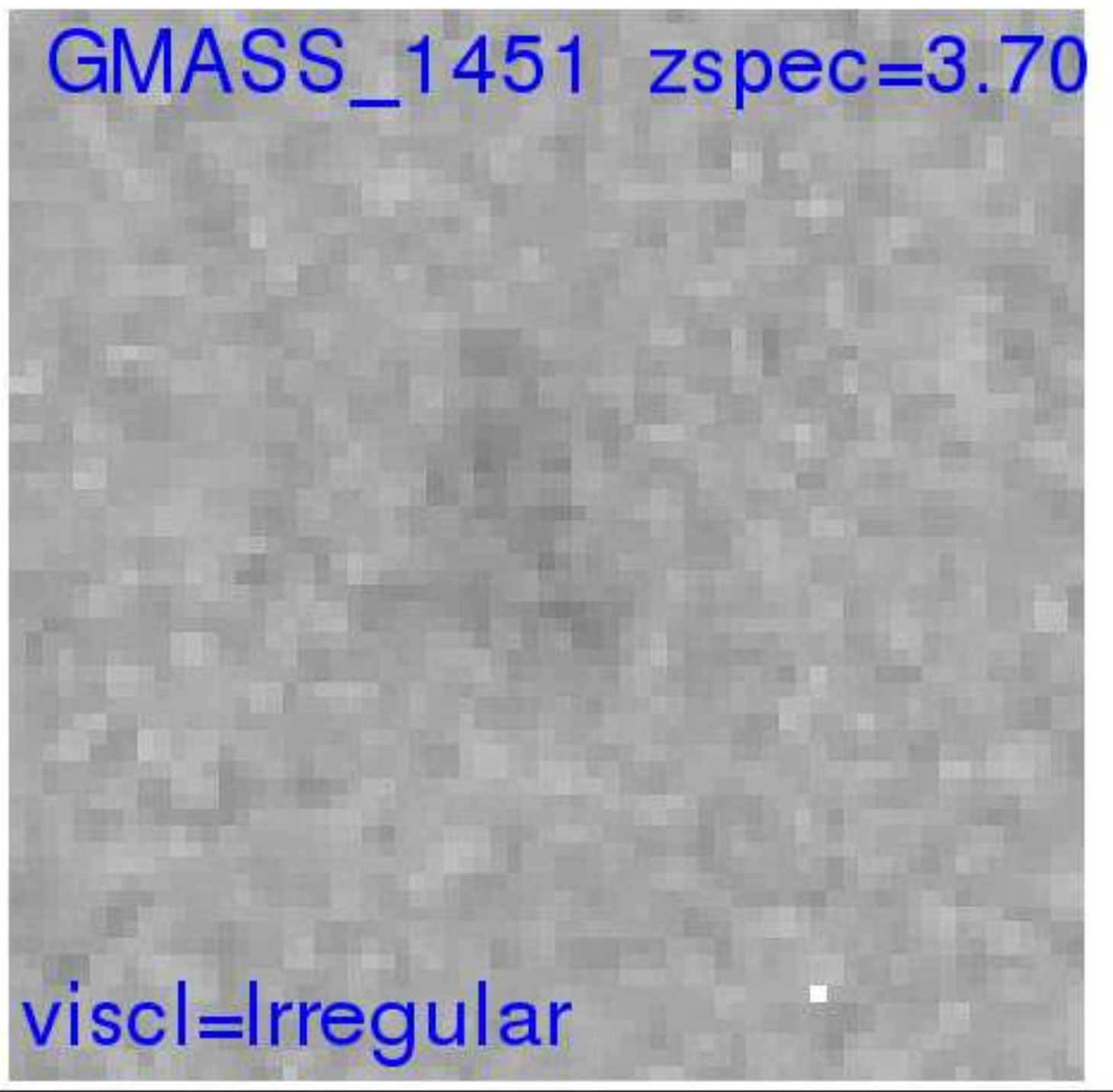}			     
\includegraphics[trim=100 40 75 390, clip=true, width=30mm]{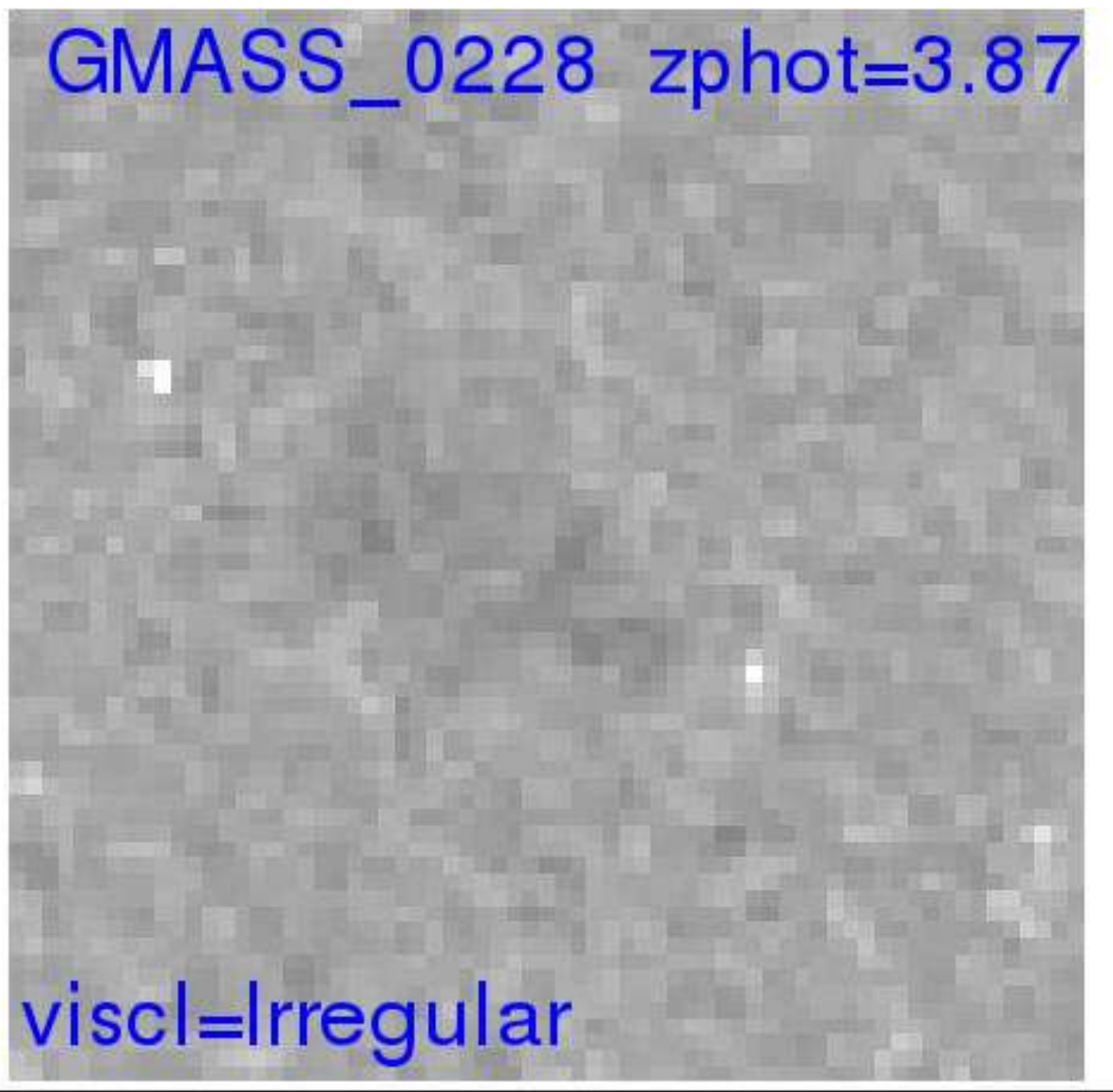}			     
\includegraphics[trim=100 40 75 390, clip=true, width=30mm]{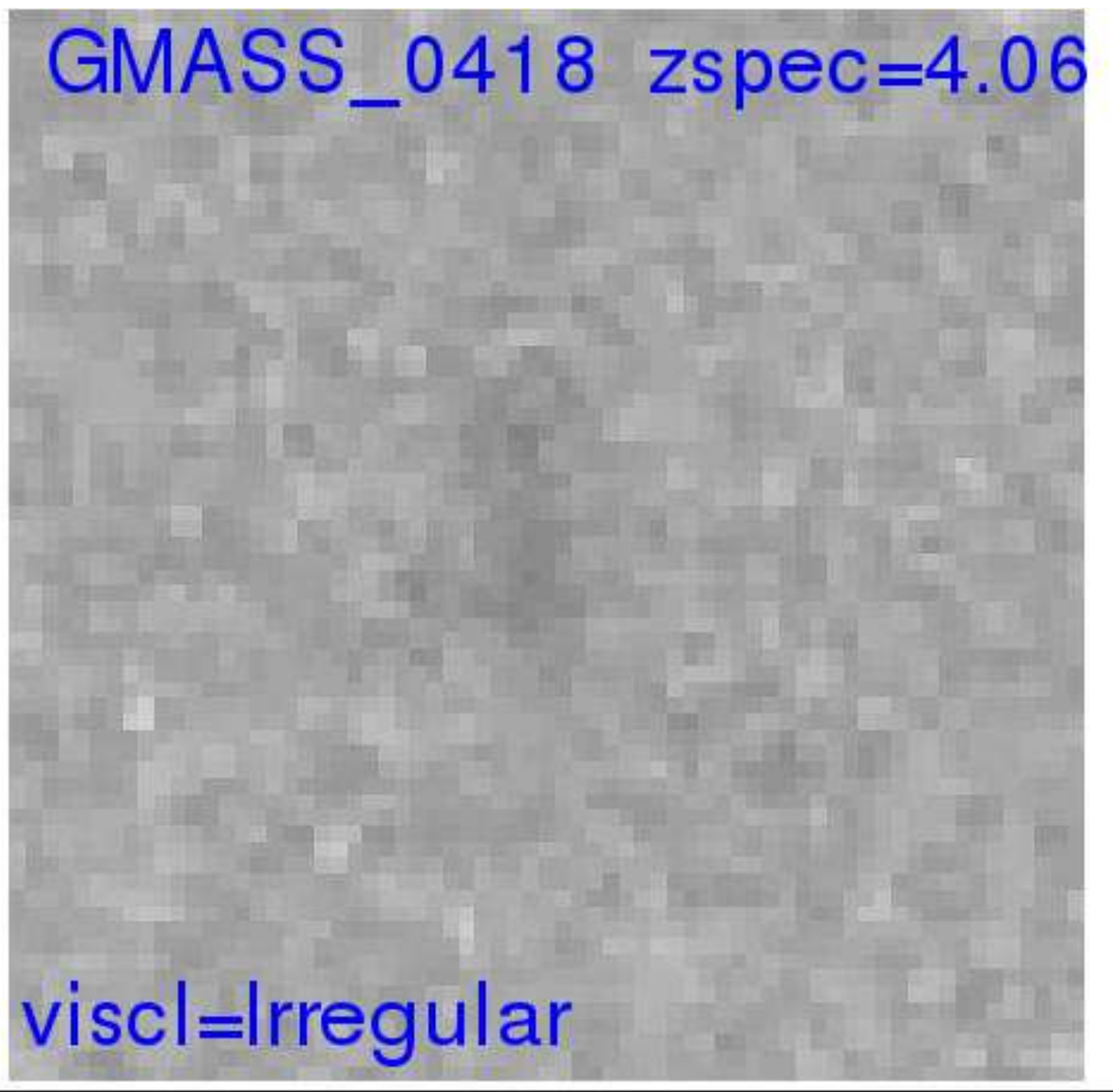}			     
\includegraphics[trim=100 40 75 390, clip=true, width=30mm]{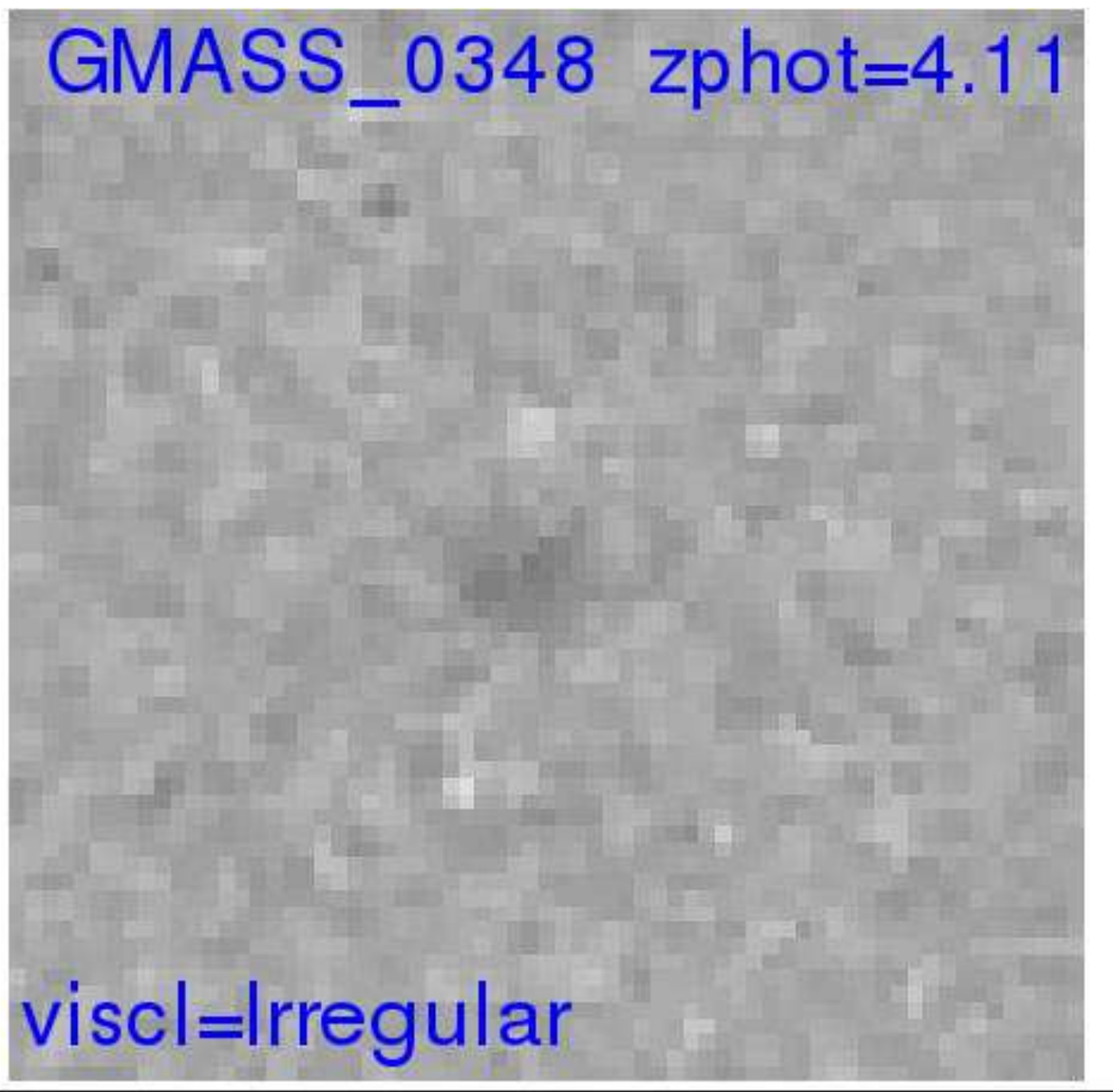}			     
\includegraphics[trim=100 40 75 390, clip=true, width=30mm]{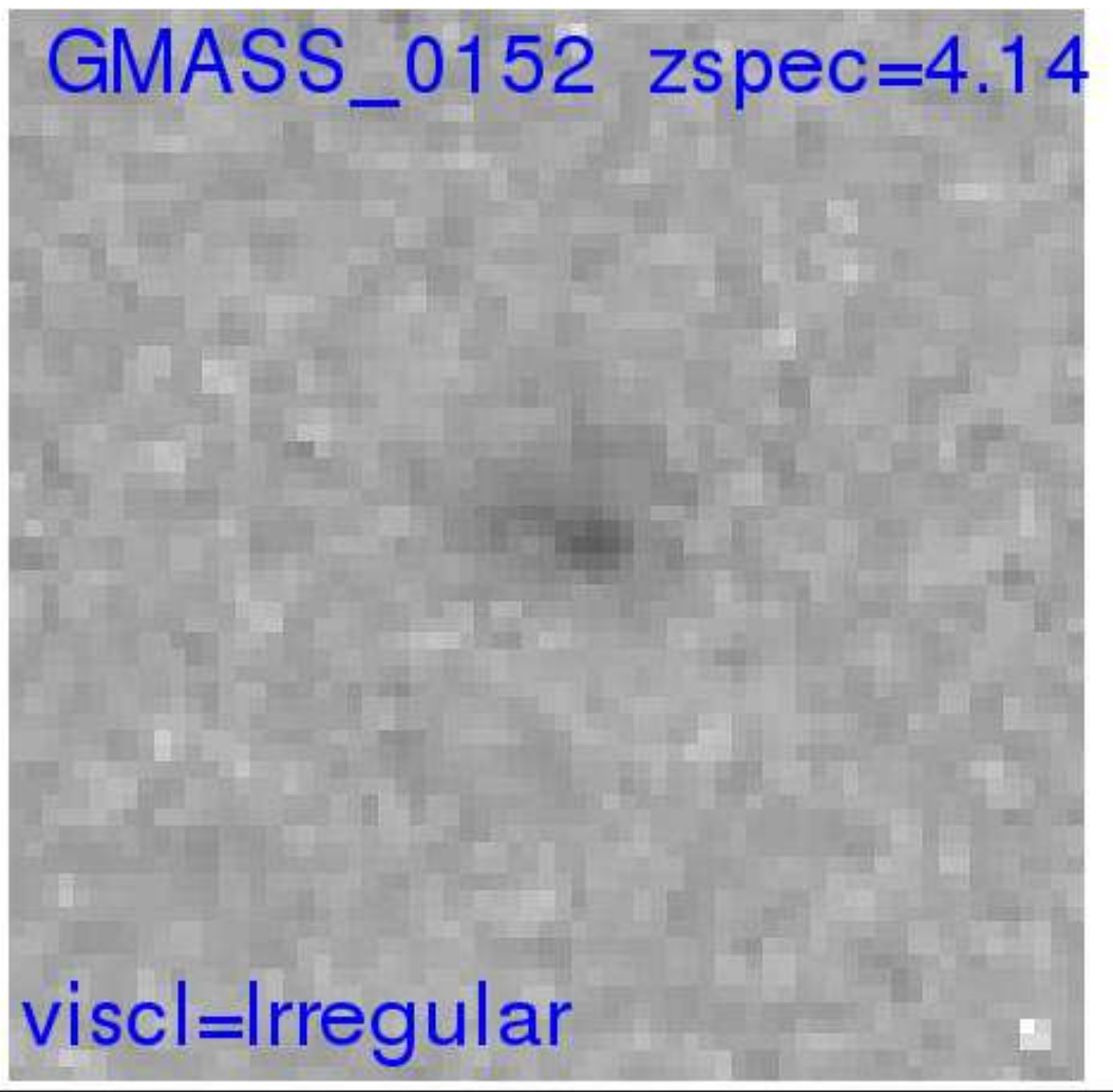}
\end{figure*}
\begin{figure*}
\centering   		
\includegraphics[trim=100 40 75 390, clip=true, width=30mm]{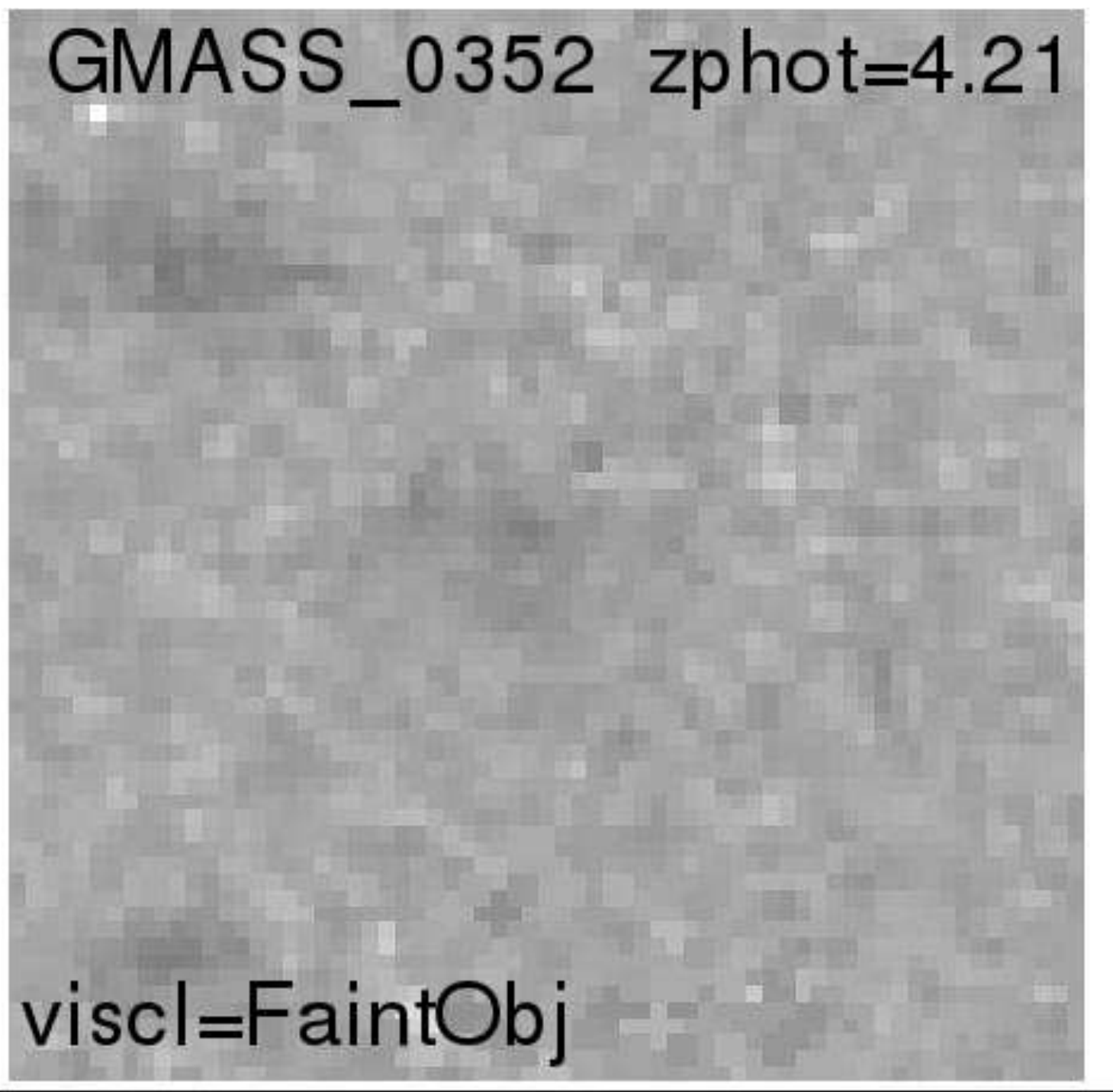}		     
\includegraphics[trim=100 40 75 390, clip=true, width=30mm]{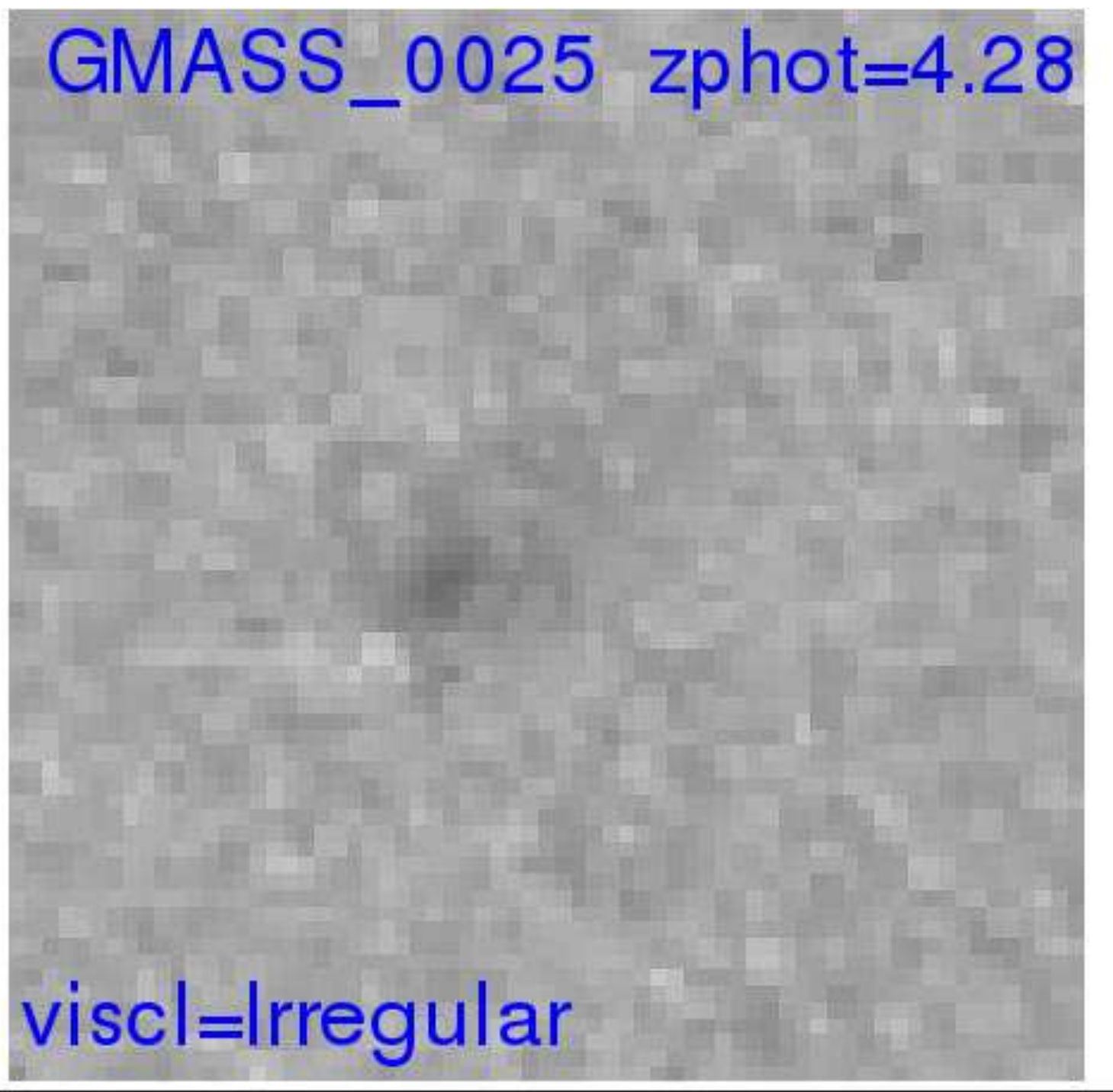}		     
\includegraphics[trim=100 40 75 390, clip=true, width=30mm]{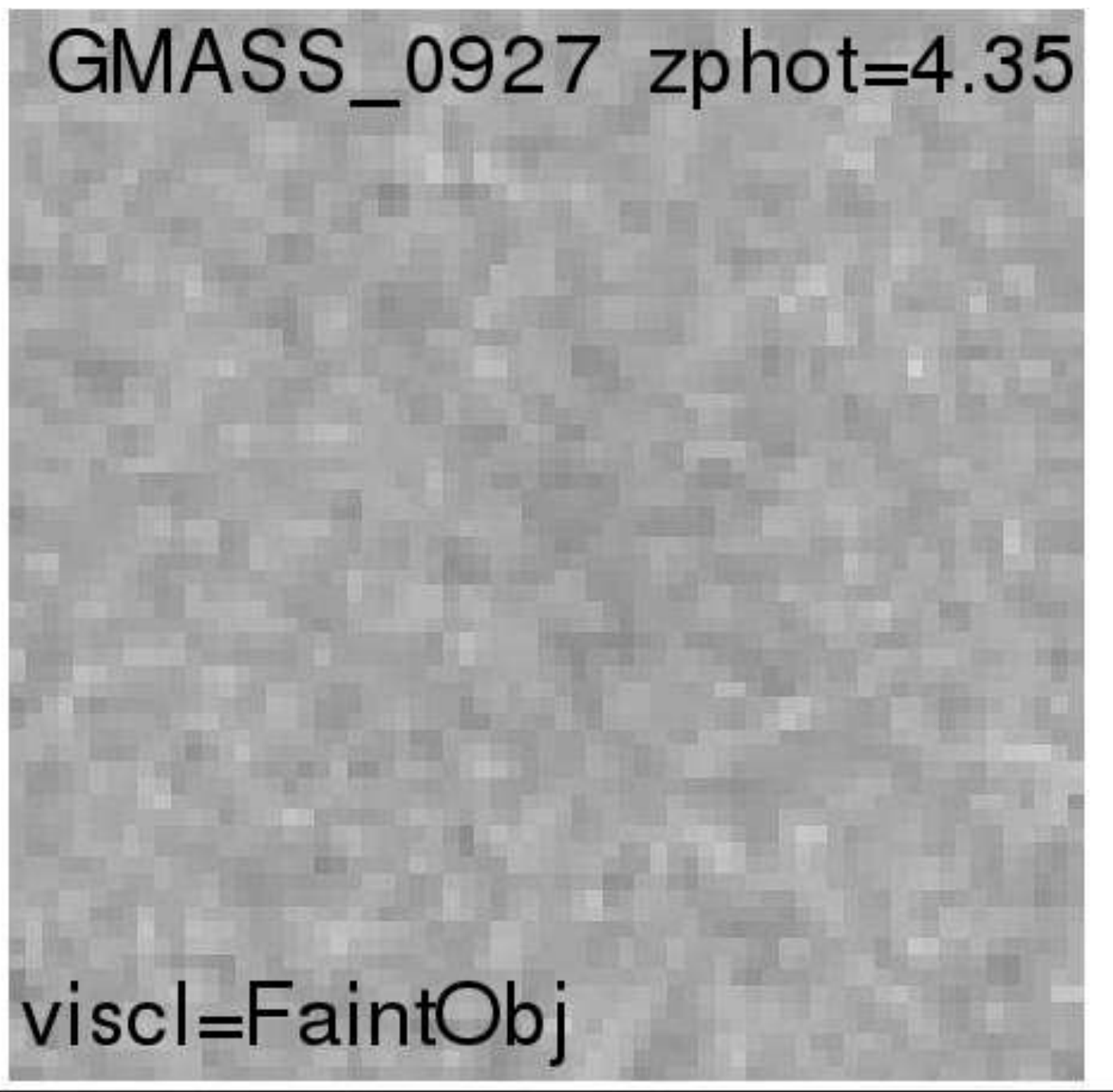}			     
\includegraphics[trim=100 40 75 390, clip=true, width=30mm]{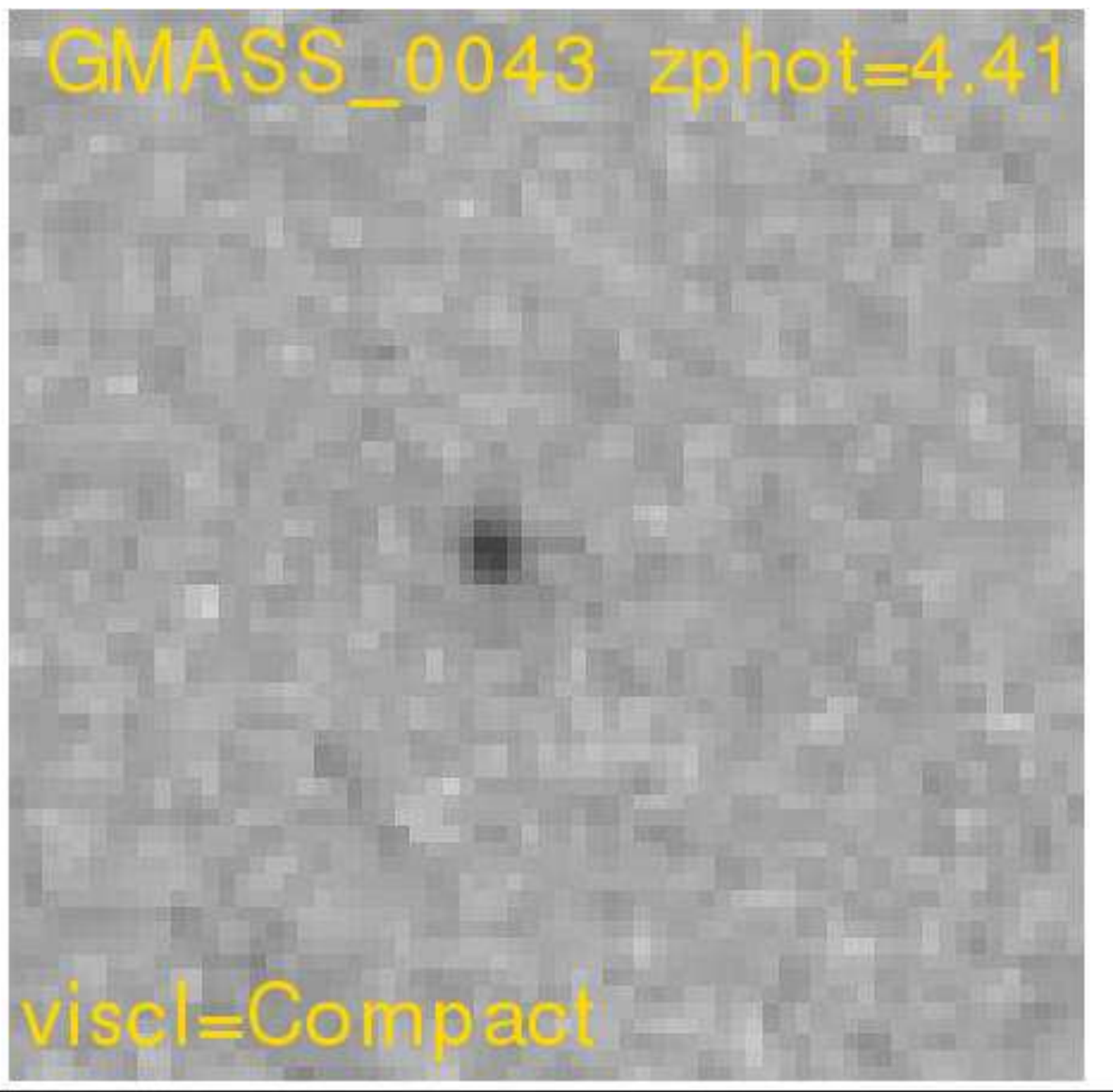}			     
\includegraphics[trim=100 40 75 390, clip=true, width=30mm]{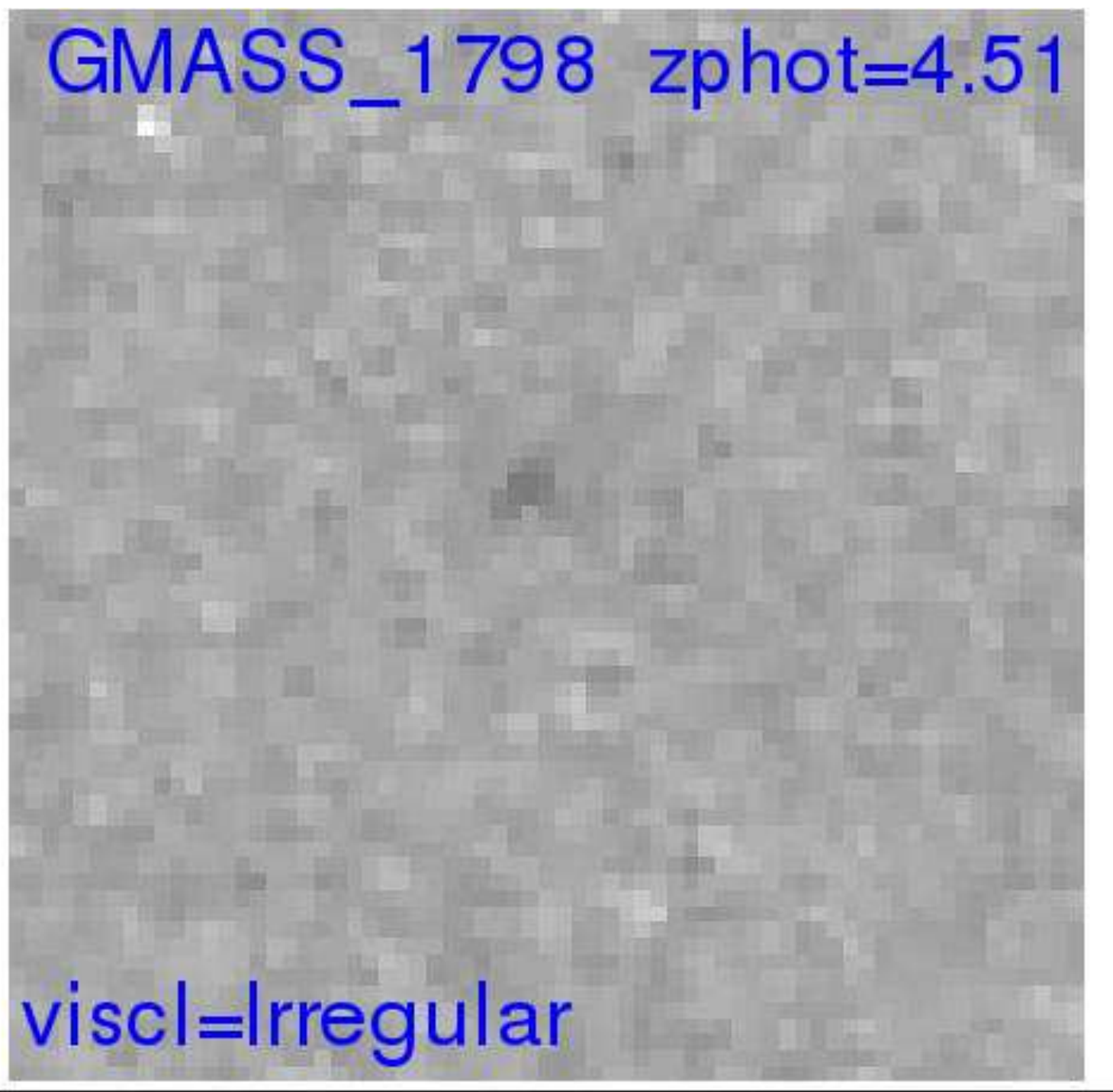}			     
\includegraphics[trim=100 40 75 390, clip=true, width=30mm]{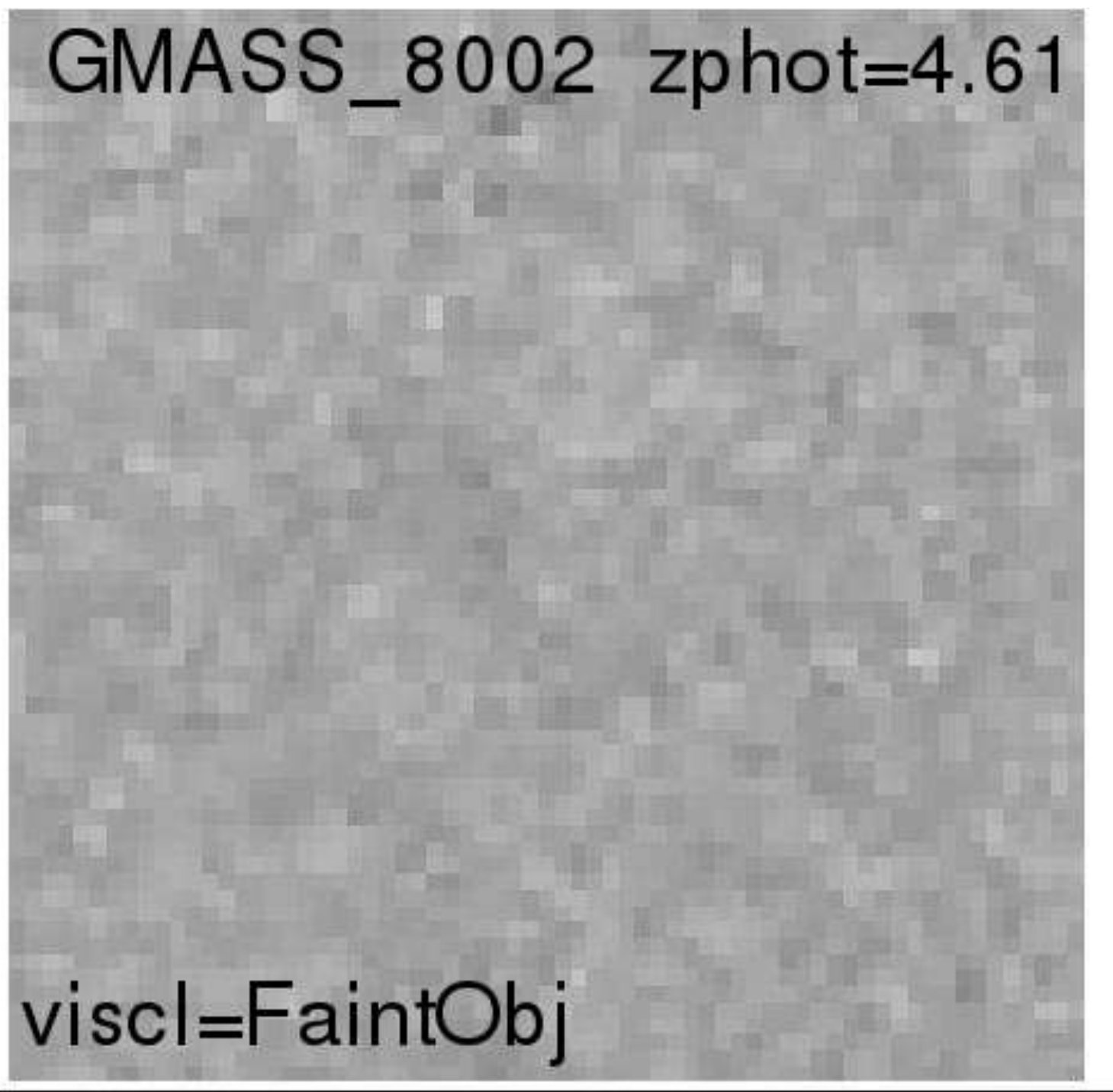}			     

\includegraphics[trim=100 40 75 390, clip=true, width=30mm]{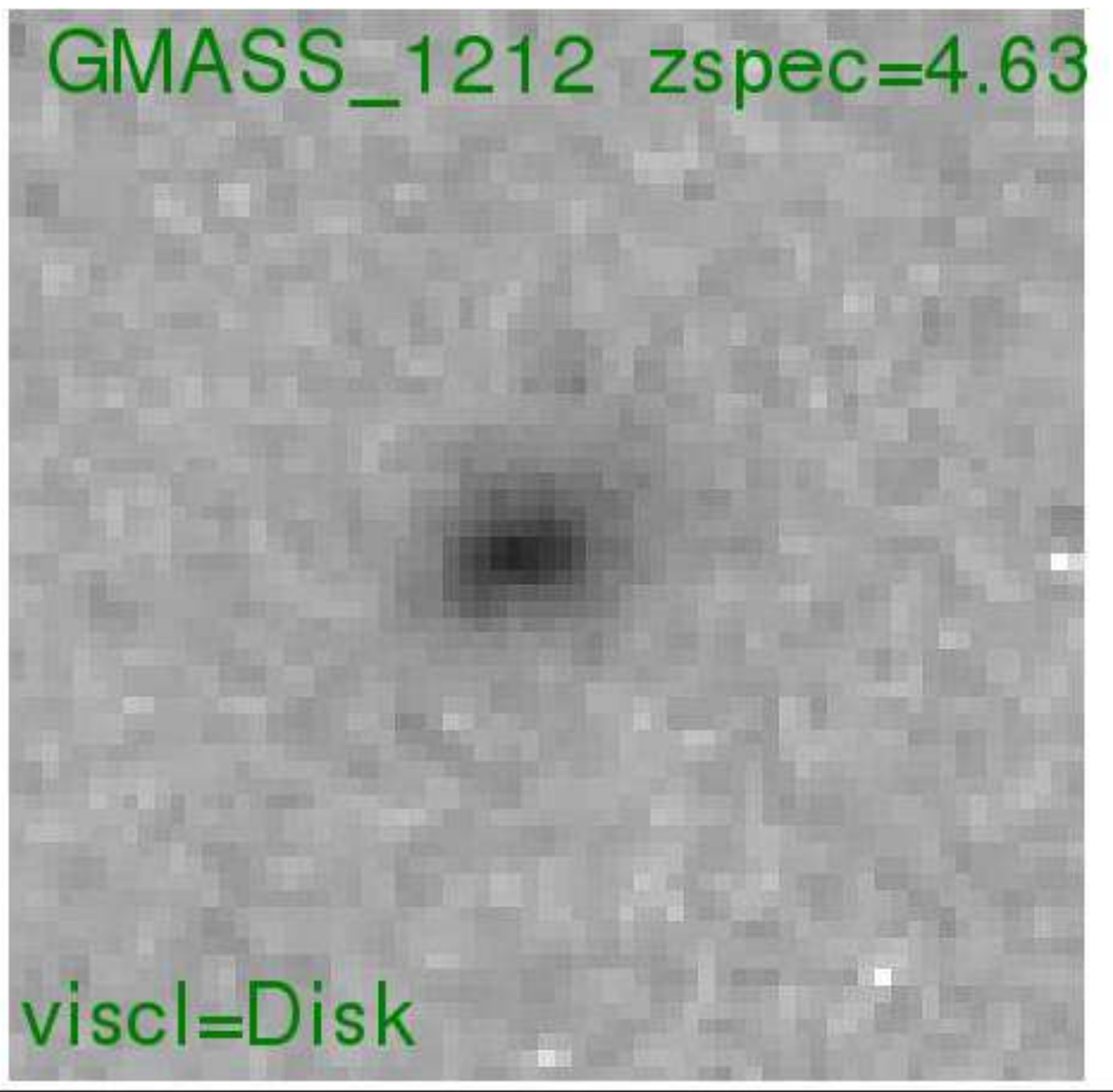}			     
\includegraphics[trim=100 40 75 390, clip=true, width=30mm]{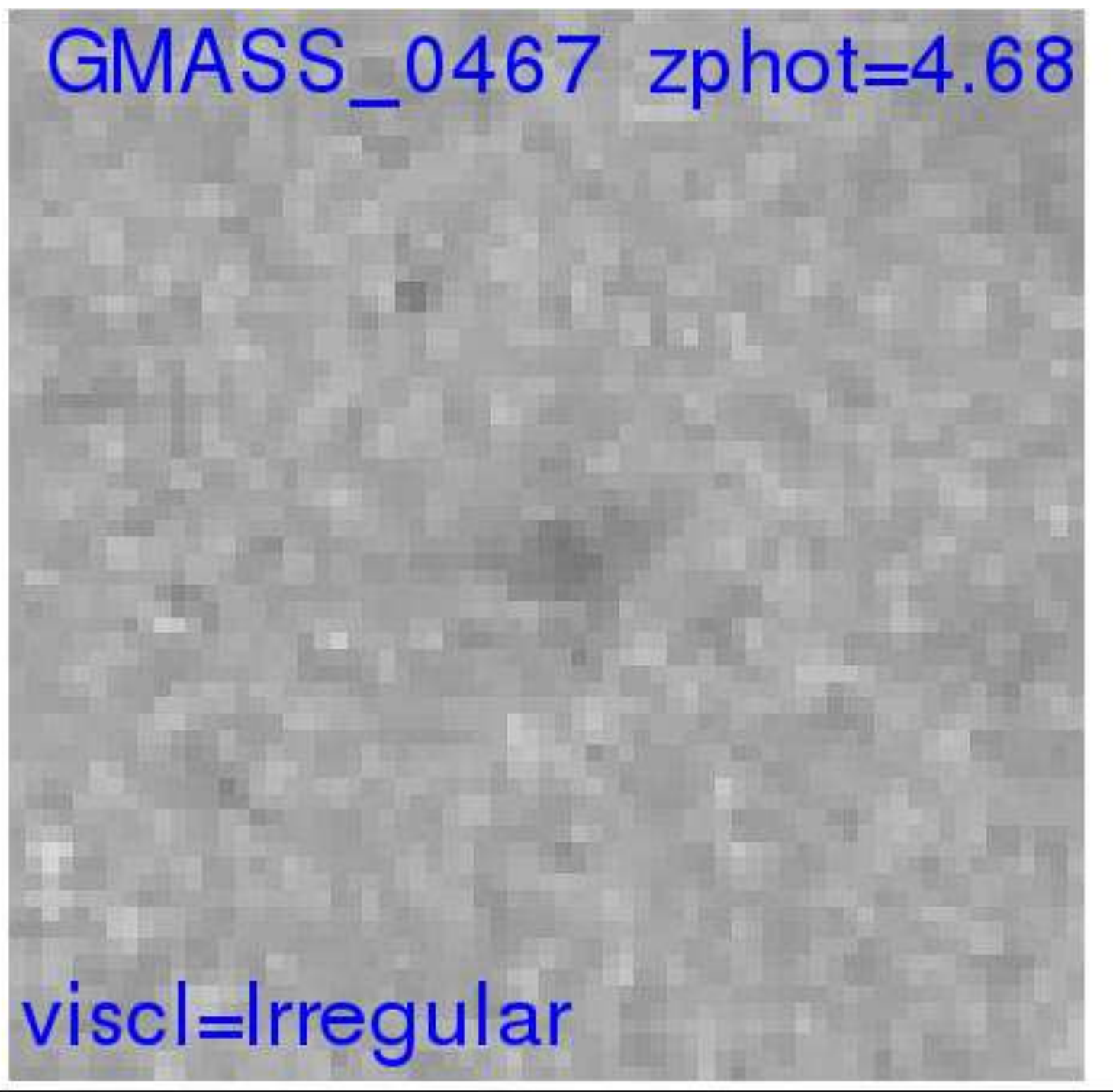}			     
\includegraphics[trim=100 40 75 390, clip=true, width=30mm]{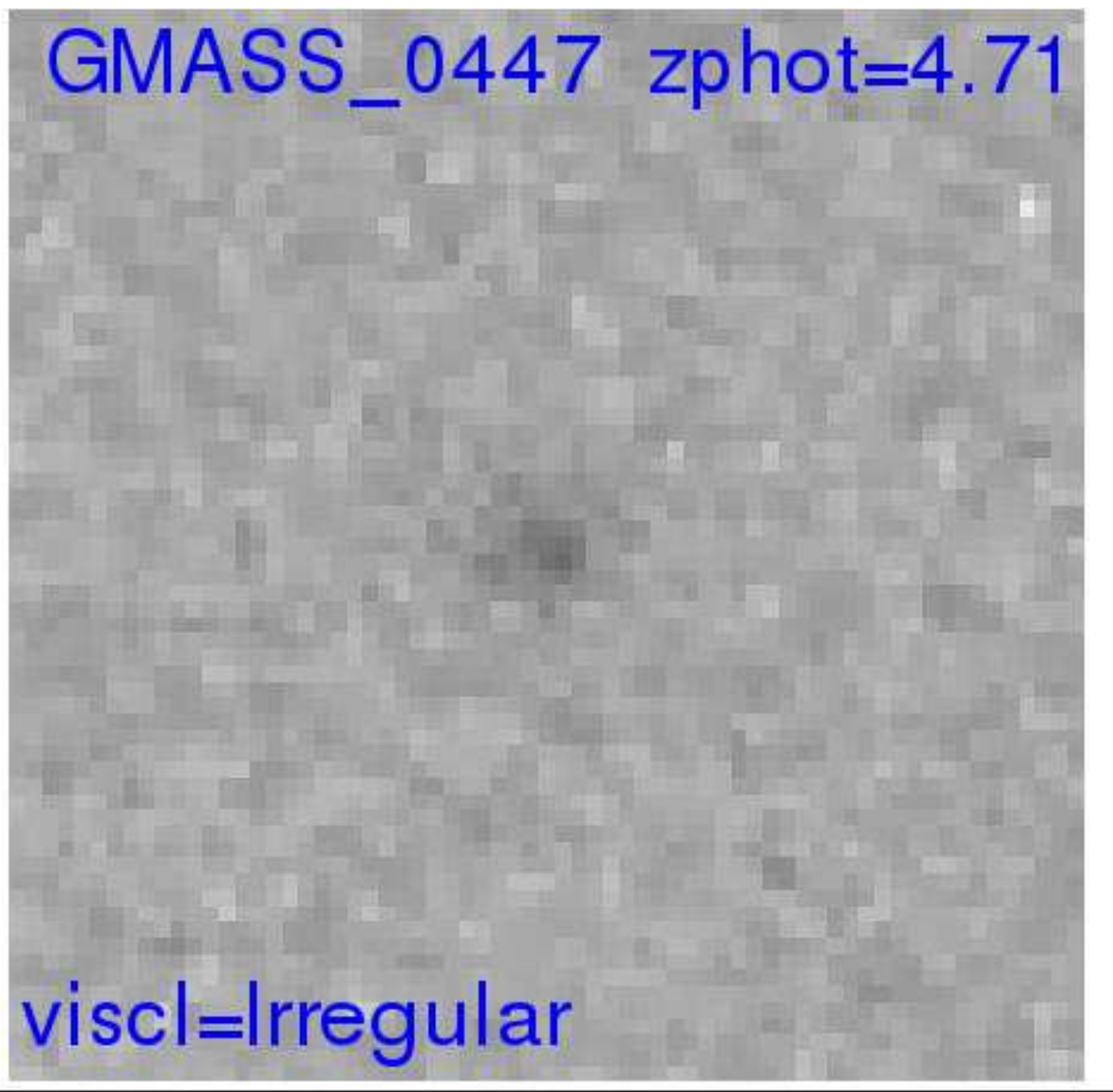}			     
\includegraphics[trim=100 40 75 390, clip=true, width=30mm]{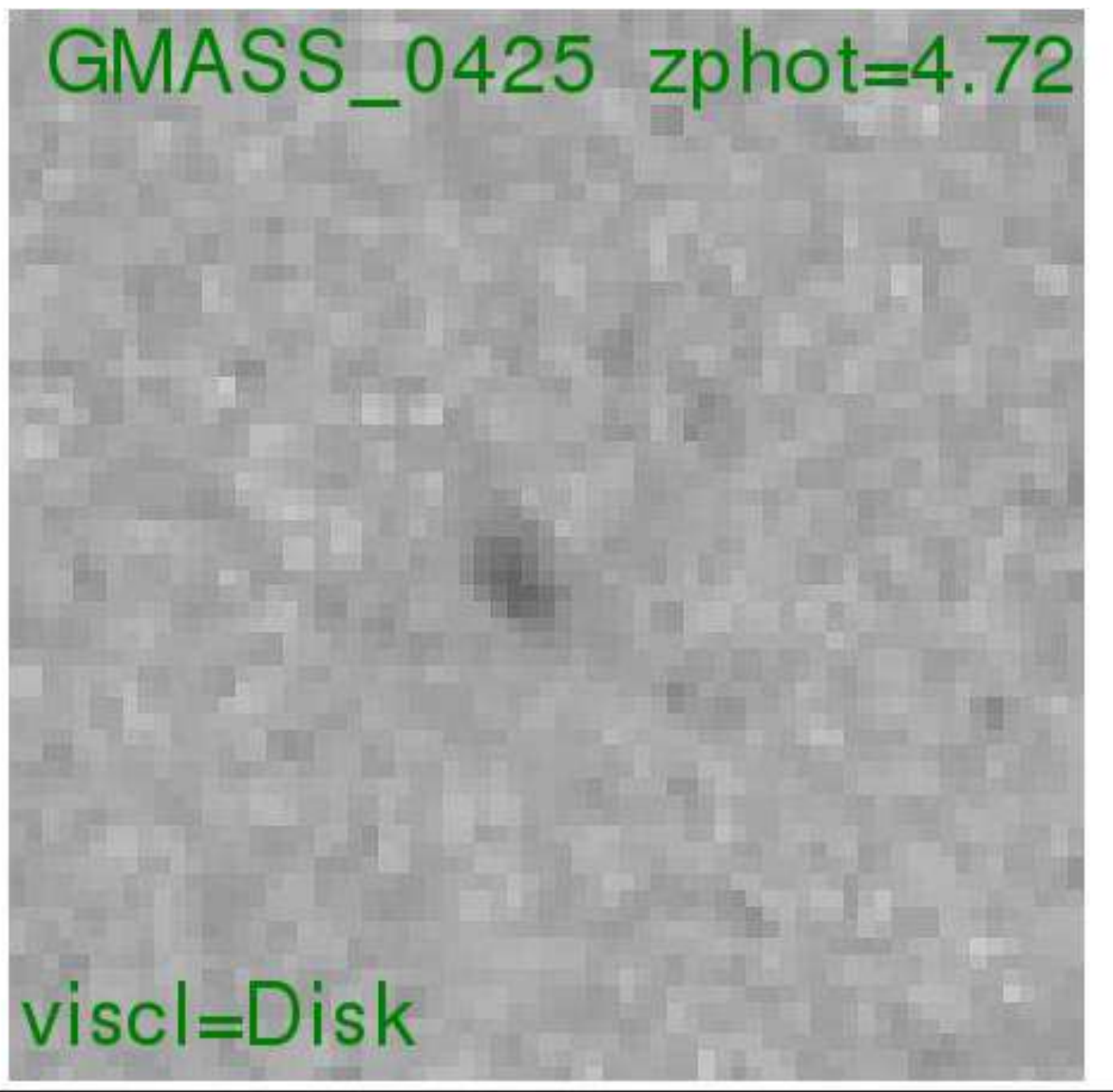}			     
\includegraphics[trim=100 40 75 390, clip=true, width=30mm]{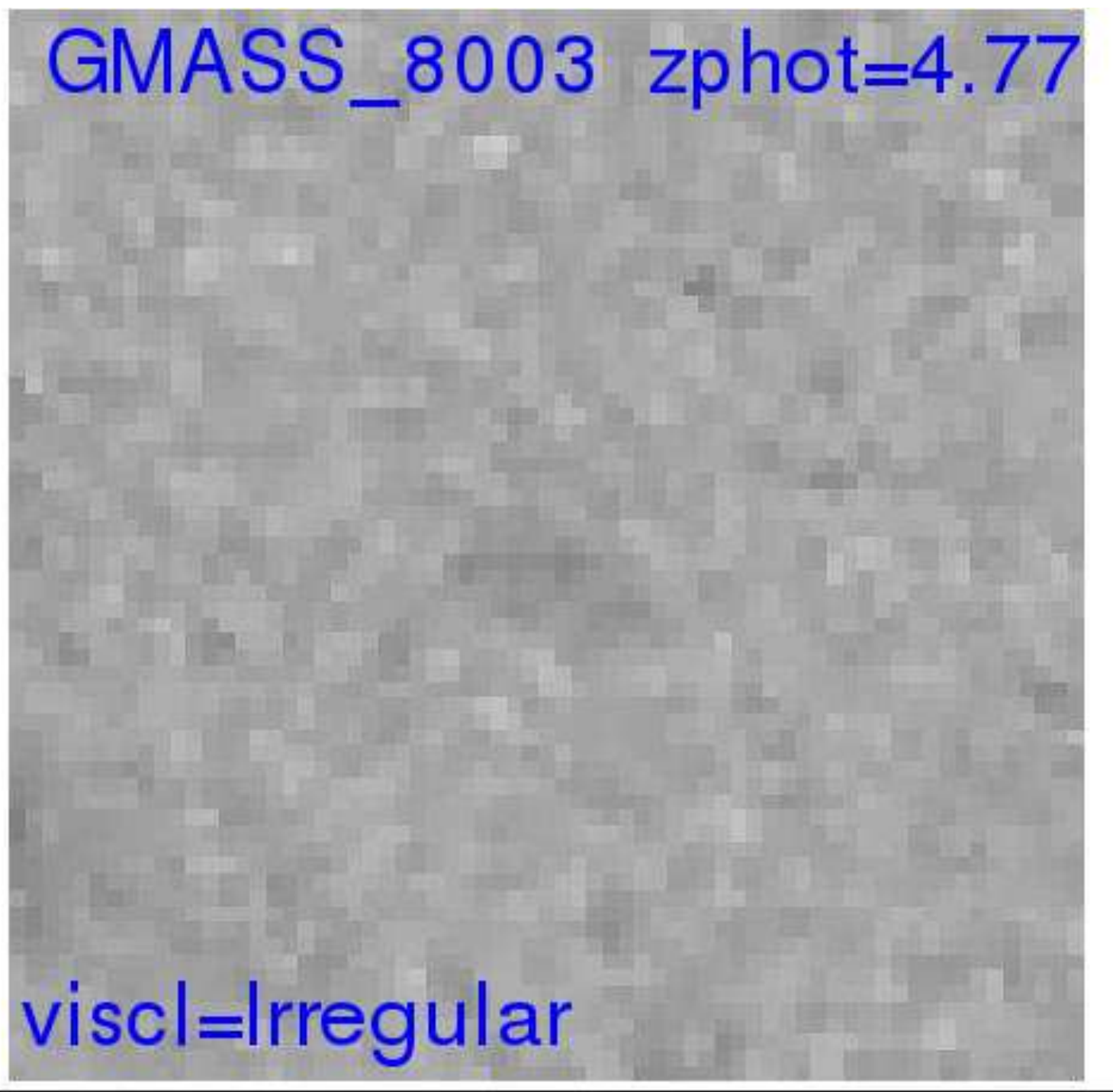}			     
\includegraphics[trim=100 40 75 390, clip=true, width=30mm]{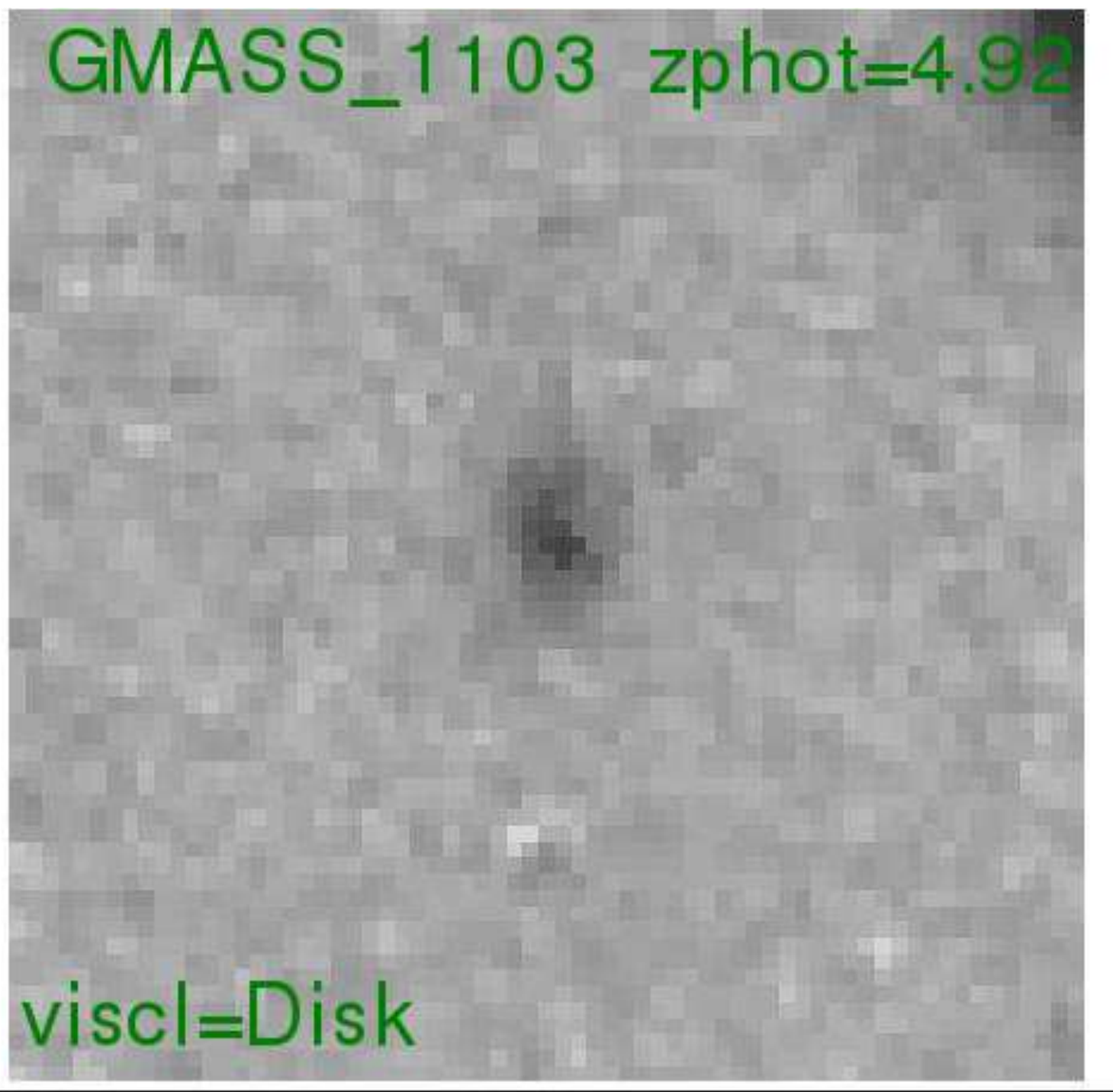}			     

\includegraphics[trim=100 40 75 390, clip=true, width=30mm]{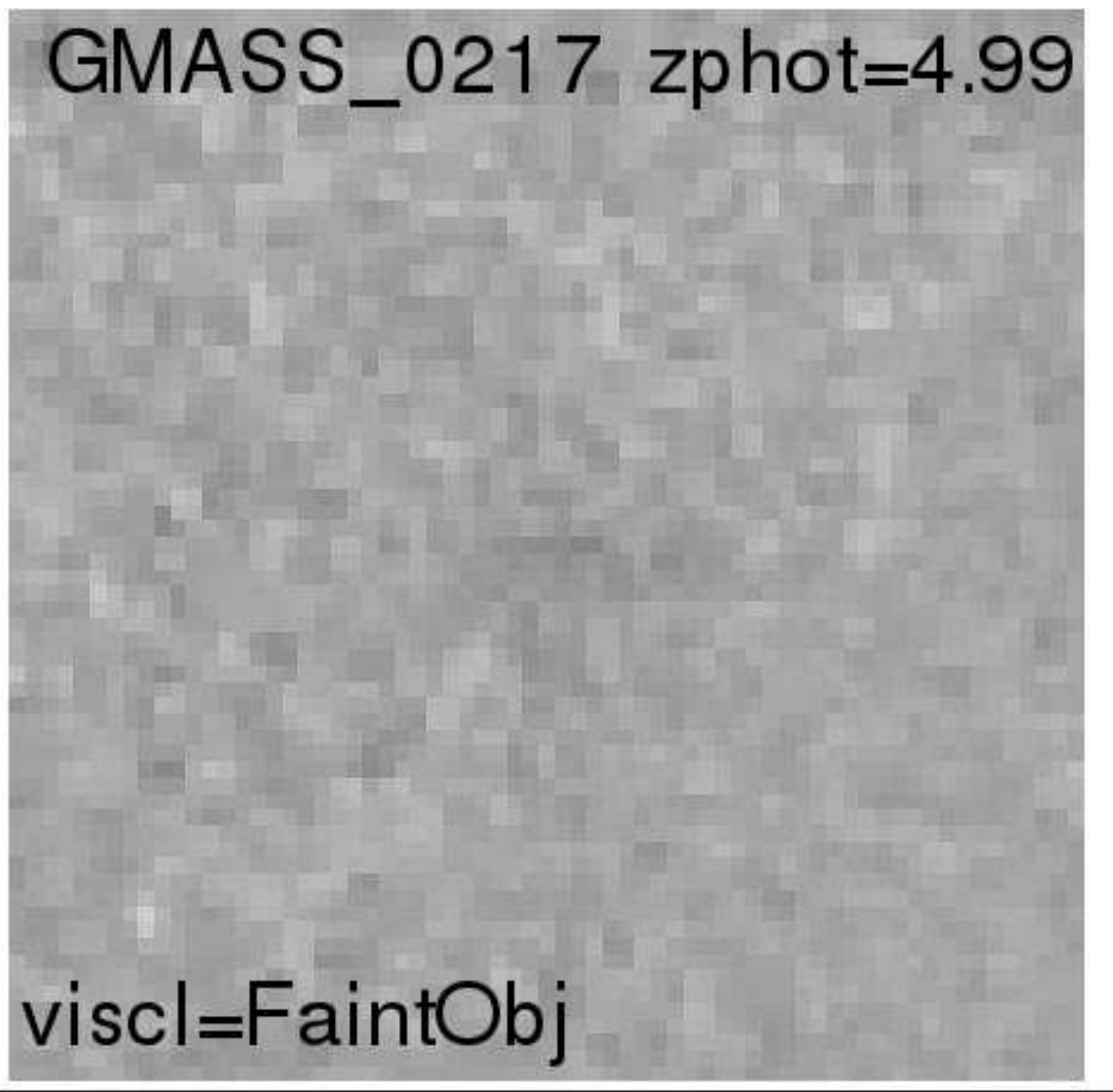}		     
\includegraphics[trim=100 40 75 390, clip=true, width=30mm]{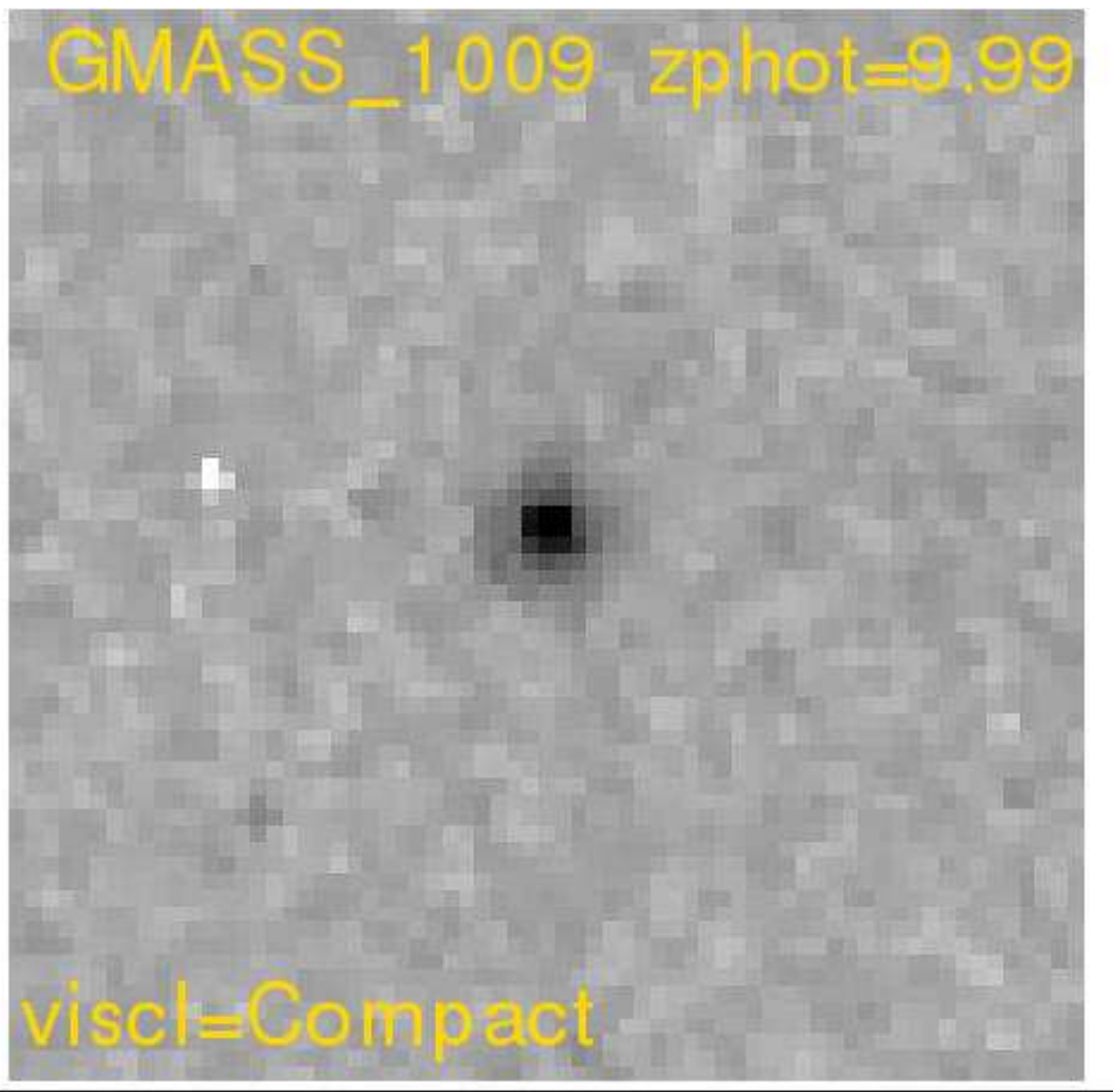}			     
\end{figure*}

\end{document}